\documentclass[final,3p,times]{elsarticle}

\usepackage{graphicx}
\usepackage{makecell}
\usepackage{amssymb}
\usepackage{amsmath}
\usepackage{booktabs}
\usepackage{bm}
\usepackage{xcolor}
\usepackage{multirow}
\usepackage{arydshln}
\usepackage{rotating}
\usepackage{caption}
\usepackage{CJKutf8}
\usepackage{float}
\usepackage[normalem]{ulem}

\usepackage{pifont,amssymb}

\usepackage{subfigure}
\usepackage{threeparttable}

\usepackage{bbm}
\usepackage{multirow}

\usepackage[colorlinks, citecolor=blue,anchorcolor=red,menucolor=red, linkcolor=red,filecolor=red,runcolor=red,urlcolor=blue,frenchlinks=true]{hyperref}

\def\stat{\mathrm{(stat.)}}
\def\syst{\mathrm{(syst.)}}

\journal{Physics Reports}

\newcommand{\changelabel}[1]{#1}
\newcommand{\changelabelb}[1]{#1}

\begin{document}

\begin{frontmatter}



\title{Unquenched Charmonium and Beyond}


\author[taga,tagb,tagc,tagd]{Zi-Yue Bai\footnotemark[1]}

\author[tage,tagb]{Dian-Yong Chen\footnotemark[1]}


\author[tagf,tagb]{Qi Huang\footnotemark[1]}


\author[taga,tagb,tagc,tagd]{Xiang Liu\footnotemark[1]\corref{cor2}}
\cortext[cor2]{Corresponding author}
\ead{xiangliu@lzu.edu.cn}

\author[taga,tagb,tagd]{Si-Qiang Luo\footnotemark[1]}


\author[tagg,tagb]{Jun-Zhang Wang\footnotemark[1]}


\address{(Lanzhou Group)}

\address[taga]{School of Physical Science and Technology, Lanzhou University, Lanzhou 730000, China}
\address[tagb]{Lanzhou Center for Theoretical Physics,
        Key Laboratory of Theoretical Physics of Gansu Province,
        Key Laboratory of Quantum Theory and Applications of MoE,
        Gansu Provincial Research Center for Basic Disciplines of Quantum Physics, Lanzhou University, Lanzhou 730000, China}
\address[tagc]{MoE Frontiers Science Center for Rare Isotopes, Lanzhou University, Lanzhou 730000, China}

\address[tagd]{Research Center for Hadron and CSR Physics, Lanzhou University $\&$ Institute of Modern Physics of CAS, Lanzhou 730000, China}

\address[tage]{School of Physics, Southeast University, Nanjing 210094, China} 


\address[tagf]{School of Physics and Technology, Nanjing Normal University, Nanjing 210023, China}

\address[tagg]{Department of Physics and Chongqing Key Laboratory for Strongly Coupled Physics, Chongqing University, Chongqing 401331, China}



\footnotetext[1]{These authors equally contribute to this work and are listed in alphabetical order.}

\begin{abstract}
The year 2024 marked the 50th anniversary of the discovery of the $J/\psi$ particle, which unveiled the charm quark and the charmonium spectrum, instigating the "November Revolution" in particle physics. This discovery catalyzed the development of quenched potential models, most notably the Cornell model, which provided a foundational quantitative description of the hadronic spectrum. However, the landscape of hadron spectroscopy has been profoundly transformed since the turn of the 21st century with the observation of numerous charmonium-like states, such as $X(3872)$, which exhibit properties starkly at odds with quenched model predictions. These discrepancies, exemplified by the "$X(3872)$ low-mass puzzle" and the "$Y$ problem" associated with vector states like $Y(4260)$, underscore the critical limitations of the quenched approximation and signal the necessity for a new theoretical paradigm. This review synthesizes recent advances in hadronic spectroscopy, arguing that the unquenched picture, which incorporates coupled-channel effects such as hadronic loops, is essential for a unified description of these new states and associated anomalies. We demonstrate how unquenched effects provide compelling solutions to long-standing puzzles in charmonium decays (e.g., the "$\rho\pi$ puzzle" and anomalous dipion transitions), predict and explain the existence of exotic charged states like  $Z_c(3900)$ and $Z_b(10610)$ via mechanisms such as Initial Single Pion Emission, and offer a framework for understanding interactions between charmonia and with nucleons. Furthermore, we emphasize the universality of unquenched effects, extending their application to bottomonium and light-flavor sectors. As experimental precision continues to improve, we advocate for the systematic development of unquenched hadronic spectroscopy, which heralds a new era of high-precision theoretical and experimental studies, moving decisively beyond the quenched approximation to achieve a more complete understanding of non-perturbative behavior of the strong interaction.    
\end{abstract}


\end{frontmatter}


\tableofcontents


\section{Introduction}

In the recently passed year of 2024, the academic community worldwide celebrated the 50th anniversary of the discovery of the $J/\psi$ particle \cite{SLAC:2024-11-20,CDS:2935663}. This major breakthrough, hailed as the "November Revolution" of particle physics \cite{E598:1974sol,SLAC-SP-017:1974ind}, not only confirmed the existence of the charm quark-the fourth flavor quark proposed by the \changelabel{Glashow-Iliopoulos-Maiani (GIM)} mechanism \cite{Glashow:1970gm} to suppress flavor-changing neutral currents (FCNCs)-but also revealed a new family of hadrons: charmonium. 

New discoveries often provide crucial opportunities for theoretical advancement. In the 1950s, with the construction of particle accelerators, a large number of light flavor hadrons were discovered. Against this background, SU(3) symmetry was proposed as a theoretical framework for classifying hadrons \cite{Gell-Mann:1961omu,Neeman:1961jhl,Gell-Mann:1964ewy,Zweig:1964ruk}, becoming a milestone in the development of particle physics and marking the beginning of the first phase of hadronic spectroscopy research. This is a classic example of experimental phenomena driving theoretical breakthroughs.

A similar situation emerged with the observation of charmonium states beginning in 1974. The distinctive mass spectrum of charmonium inspired the Cornell group to propose the famous Cornell potential model \cite{Eichten:1974af,Eichten:1978tg,Eichten:1979ms}, which combines a linear potential and a Coulomb potential described by $V(r) = a r + b/r$, thereby achieving for the first time a quantitative description of the hadronic spectrum. It should be noted that the Cornell potential is a typical quenched approximation model. For low-lying hadronic states, such models can adequately address certain specific problems in spectroscopy. Subsequently, inspired by the Cornell model, various potential models were developed, including the Godfrey-Isgur (GI) model \cite{Godfrey:1985xj}, further advancing the description of the hadronic spectrum. However, after 1986, no new charmonium states were reported experimentally, and hadronic spectroscopy entered a period of relative inactivity.

Since the beginning of the 21st century, with the continuous accumulation of experimental data, many new hadronic states have been discovered, particularly the charmonium-like $XYZ$ states \cite{Klempt:2007cp,Brambilla:2010cs,Liu:2013waa,Hosaka:2016pey,Richard:2016eis,Chen:2016qju,Esposito:2016noz,Chen:2016spr,Lebed:2016hpi,Guo:2017jvc,Olsen:2017bmm,Ali:2017jda,Liu:2019zoy,Brambilla:2019esw,Chen:2022asf,Meng:2022ozq,Wang:2021aql,Wang:2025dur}. These discoveries signify the arrival of a revival in hadronic physics, continuing to this day. Based on past experience, we have reason to expect that these new phenomena will drive further theoretical breakthroughs. These developments are closely related to a deeper understanding of the non-perturbative behavior of the strong interaction—one of the core challenges in modern particle physics.

\begin{figure}[htbp]
\centering
\includegraphics[height=0.275\textheight, angle=0]{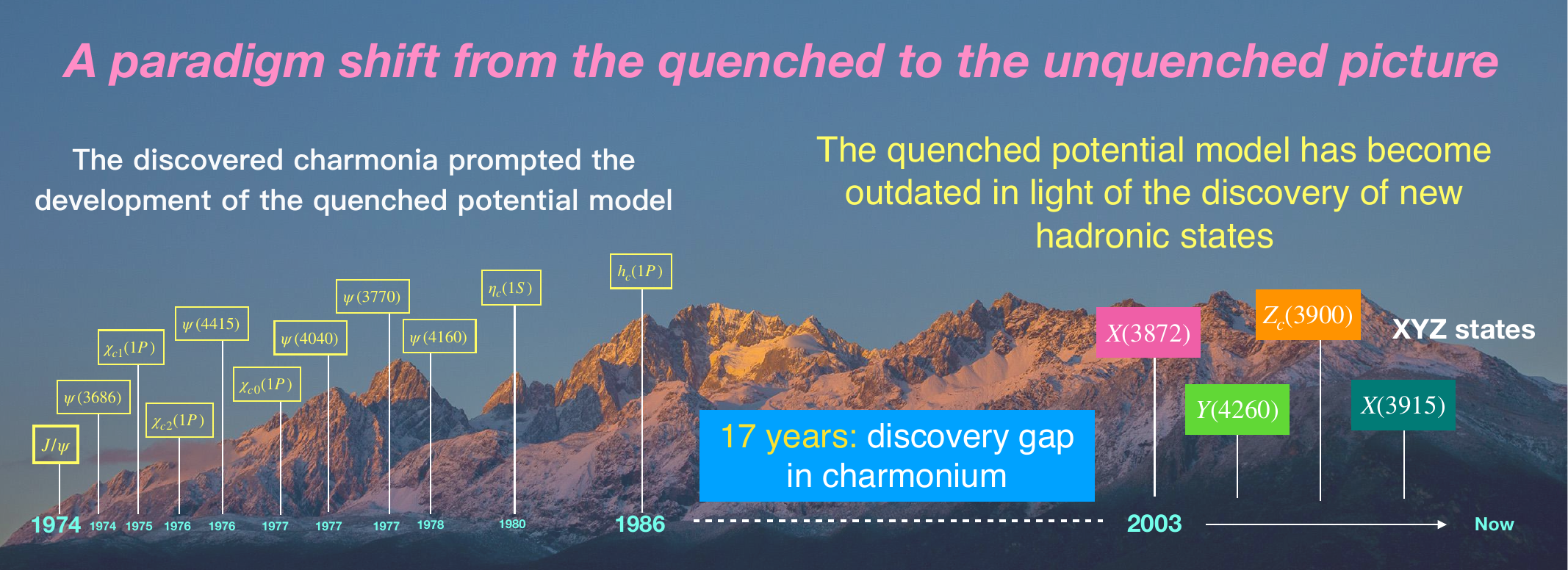}
\caption{A paradigm shift from the quenched to the unquenched picture.}
\label{fig:general}
\end{figure}

In 2003, the Belle Collaboration, through studying the $B \to K J/\psi\pi^+\pi^-$ decay process, first reported the charmonium-like state $X(3872)$ \cite{Belle:2003nnu}. Like a stone thrown into a calm lake, the discovery of $X(3872)$ stirred widespread and far-reaching research waves, promoting extensive exploration of hidden-charm hadronic matter over the past two decades. The charmonium states discussed in this review fall within this category. Comparing the mass of $\chi_{c1}(2P)$ predicted by quenched potential models \cite{Eichten:1974af,Eichten:1978tg,Eichten:1979ms,Godfrey:1985xj} with the actual mass of $X(3872)$ reveals a significant discrepancy, known as the "$X(3872)$ low-mass puzzle." In fact, this puzzle is not limited to $X(3872)$; it also appears in other new hadronic states such as $D_{s0}(2317)$ \cite{BaBar:2003oey}, $D_{s1}(2460)$ \cite{CLEO:2003ggt}, and $\Lambda_c(2940)$ \cite{BaBar:2006itc}. One explanatory approach involves introducing exotic hadronic structures like multiquark states \cite{Klempt:2007cp,Chen:2016qju,Chen:2016spr,Esposito:2016noz,Guo:2017jvc,Ali:2017jda,Liu:2019zoy,Brambilla:2019esw,Chen:2022asf}. Subsequently, the theoretical community gradually recognized that quenched models were insufficient to describe these new states, and the development of unquenched models significantly alleviated these mass discrepancy issues \cite{Silvestre-Brac:1991qqx,vanBeveren:2003kd,Hwang:2004cd,Kalashnikova:2005ui,Li:2009ad,Zhang:2009bv,Ortega:2009hj,Danilkin:2010cc,Song:2015nia,Song:2015nia,Luo:2019qkm,Luo:2021dvj,Zhang:2022pxc,Zhang:2024usz}. Additionally, $X(3915)$ \cite{Belle:2009and} and $Z(3930)$ \cite{Belle:2005rte}, produced via the $\gamma\gamma$ fusion process, have also been included in the candidate list for $2P$ charmonium. This review will use the establishment of $2P$ charmonium as a main thread to systematically introduce recent advances in unquenched charmonium research.

The charmonium-like state $Y(4260)$ is a particularly striking example among these new discoveries, and its study benefited from advances in initial state radiation (ISR) technology. The BaBar Collaboration first observed $Y(4260)$ in the  cross sections for the $e^+e^-\to J/\psi\pi^+\pi^-$ process \cite{BaBar:2005hhc}. Subsequently, \changelabel{the BaBar} Collaboration reported the observation of $Y(4360)$ in the $e^+e^-\to \psi(3686)\pi^+\pi^-$ process \cite{BaBar:2006ait}. This state was later confirmed by the Belle collaboration, as well as another state, $Y(4660)$ \cite{Belle:2007umv}. Near 4.6 GeV, Belle Collaboration also discovered another state $Y(4630)$ through the $e^+e^-\to\Lambda_c\bar{\Lambda}_c$ process \cite{Belle:2008xmh}. With more precise data, the BESIII Collaboration reanalyzed $e^+e^-\to J/\psi\pi^+\pi^-$ in 2017 and proposed that $Y(4260)$ should be resolved into two substructures, $Y(4220)$ and $Y(4320)$ \cite{BESIII:2016bnd}. Furthermore, $Y(4220)$ and $Y(4390)$ were observed in the $e^+e^-\to h_c\pi^+\pi^-$ process \cite{BESIII:2016adj}. The challenge of understanding these $Y$ states constitutes the so-called "$Y$ problem"—a term first appearing in the BESIII white paper \cite{BESIII:2020nme}. Although early attempts interpreted them as exotic hadronic states, comprehensively understanding so many $Y$ states remains highly challenging. This review will introduce a unified theoretical framework based on the unquenched charmonium picture, incorporating interference effects and an $S$-$D$ mixing scheme to provide a solution to the "$Y$ problem." It is worth mentioning that this framework can be naturally extended to the study of bottomonium and light flavor vector meson states in $e^+e^-$ annihilation, demonstrating its universality.

Unquenched effects also play a key role in understanding the anomalous decay behaviors of low-lying charmonia. For example, the "$\rho\pi$ puzzle" in the decays of $J/\psi$ and $\psi(3686)$ to light flavor meson pairs clearly violates expectations based on the "12\% rule" \cite{Harris:1999wn}. In 2006, the hadronic loop mechanism—a direct manifestation of unquenched effects—was first introduced to study $J/\psi \to P + V$ decays (where $P$ and $V$ represent light pseudoscalar and vector mesons, respectively) \cite{Liu:2006dq}. This mechanism was later extended to understand anomalous decays of other charmonia, such as $\psi(3770)$ \cite{Liu:2009dr,Li:2013zcr,Zhang:2009kr} and $\chi_{cJ}(1P)$ ($J=0,1,2$) \cite{Chen:2009ah,Liu:2009vv,Chen:2010re,Huang:2021kfm}. Notable cases include the large non-$D\bar{D}$ decay branching fraction of $\psi(3770)$ observed by the CLEO-c \cite{CLEO:2005mpm} and BES Collaborations \cite{BES:2006fpf,BES:2006dso,BES:2007cev,BES:2008vad}, and the unexpectedly large branching fractions of $\chi_{cJ}(1P)$ in the $\gamma V$ \cite{CLEO:2008sah,BESIII:2011ysp} and $VV$ \cite{BES:2004imp,BESII:2011hcd} decay channels.

Beyond anomalous decay widths, the dipion invariant mass spectrum of $\psi(3770)\to J/\psi\pi^+\pi^-$ \cite{CLEO:2005zky} could not be explained by quenched models such as QCD multipole expansion \cite{Kuang:1989ub}. Introducing an intermediate $D\bar{D}$ hadronic loop connecting the initial state $\psi(3770)$ and the final state $J/\psi\pi^+\pi^-$ successfully reproduced the dipion invariant mass distribution. A similar approach was also applied to describe the dipion spectra in bottomonium decays such as $\Upsilon(4S)\to \Upsilon(nS)\pi^+\pi^-$, where a $B\bar{B}$ hadronic loop must be introduced \cite{Chen:2011jp}. These results provide compelling evidence for the importance of unquenched effects.

The anomalous decays of $\Upsilon(10860)\to \Upsilon(nS)\pi^+\pi^-$ ($n=1,2,3$) and $\Upsilon(10860)\to h_b(mP)\pi^+\pi^-$ ($m=1,2$), reported by the Belle Collaboration \cite{Belle:2007xek,Belle:2011aa}, further promoted the introduction of hadronic loop contributions. In particular, the discovery of two charged bottomonium-like structures, $Z_b(10610)$ and $Z_b(10650)$, led to the proposal of the initial single pion emission (ISPE) mechanism \cite{Chen:2011pv}. This mechanism successfully explained the $Z_b$ states appearing near the $B\bar{B}^*$ and $B^*\bar{B}^*$ thresholds in the $\Upsilon(nS)\pi^\pm, (n=1,2,3)$ and $h_b(mP)\pi^\pm, (m=1,2)$ invariant mass spectra. The ISPE mechanism is essentially a manifestation of the hadronic loop mechanism, a typical unquenched effect.

The ISPE mechanism can also be extended to the hidden-charm dipion decays of highly excited charmonium and charmonium-like $Y$ states, predicting charged charmonium-like structures near the $D\bar{D}^*$ and $D^*\bar{D}^*$ thresholds in the final states such as $J/\psi\pi^\pm$, $\psi(3686)\pi^\pm$, and $h_c\pi^\pm$ \cite{Chen:2011xk}. In 2013, the BESIII Collaboration observed $Z_c(3900)$ \cite{BESIII:2013qmu}, a discovery later selected as one of the highlights of the year by Physics magazine, a publication of the American Physical Society. This review will summarize these advances, highlighting the critical role of unquenched effects.

In recent years, the LHCb, CMS, and ATLAS Collaborations have observed enhanced structures in the invariant mass spectra of double $J/\psi$ and $J/\psi\psi(3686)$ produced in $pp$ collisions \cite{LHCb:2020bwg,ATLAS:2023bft,CMS:2023owd,ATLAS:2025nsd,CMS:2025xwt,CMS:2025vnq,CMS:2025fpt}. Understanding these new phenomena is closely related to interactions between charmonia. Within the unquenched picture, charmonium states are strongly coupled to charmed meson pairs. Therefore, hadronic loops composed of charmed mesons may provide a mechanism for charmonium interactions.

With advances in computing power, lattice QCD studies have effectively derived the interaction potentials between charmonia (such as $J/\psi$ and $\eta_c$) and nucleons ($N$) \cite{Lyu:2024ttm}. These results provide important basis for testing related theoretical schemes, which propose that unquenched charmonia may interact with nucleons through charmed meson loops—a mechanism similar to charmonium-charmonium interactions. This framework can also be applied to study interactions between fully heavy hadrons, such as the recently investigated $\Omega_{QQQ}$-$\Omega_{QQQ}$ interaction via lattice QCD \cite{Lyu:2021qsh}. This review will introduce the latest progress in this field.

As the title indicates, "beyond" is a key theme of this review, emphasizing that unquenched effects are not limited to charmonium. As discussed in Section \ref{section7}, this effect equally applies to bottomonium and light vector meson states, possessing universal significance.

Over the past two decades, numerous new hadronic states and related phenomena have been discovered. Through collaborative efforts between theorists and experimentalists, hadronic spectroscopy research has entered a new developmental stage, marking the advent of the "new era of hadronic spectroscopy." What defines \changelabel{modern} hadronic spectroscopy? It is reflected not only in the improvement of experimental measurement accuracy but also in the advancement of theoretical prediction precision. How can we further enhance the accuracy of theoretical predictions? We believe it is essential to vigorously promote unquenched hadronic spectroscopy research. This review uses charmonium as an example to demonstrate the importance of unquenched effects from multiple perspectives  (see Fig. \ref{fig:general}). Clearly, this effect is universally present in all hadrons.

In the 1970s and 1980s, the discovery of numerous charmonium states drove the development of quenched models, represented by the Cornell potential. Entering the 21st century, with the continuous emergence of new charmonium-like hadronic states, hadronic physics has stepped into a new phase where quenched models can no longer meet the demands of contemporary research. As we celebrate the 50th anniversary of the discovery of the $J/\psi$ particle, we also witness the end of one era and the beginning of another—the era of hadronic spectroscopy characterized by unquenched effects. This new era is undoubtedly full of challenges and significant opportunities.

\section{Anomalous decay behaviors of charmonia and hadronic loop mechanism}\label{section2}\label{sec:hlm}

\subsection{Phenomenology}

\subsubsection{$\rho\pi$ puzzle}

In perturbative QCD, the dominant hadronic decay width of charmonium states, which proceeds primarily via three-gluon annihilation, is proportional to their leptonic width, with both quantities depending on the square of the wave function at the origin \cite{Appelquist:1974zd}. This leads to a simple \changelabel{relation} between the hadronic branching fractions of the ground state $J/\psi$ and its first radial excitation $\psi(3686)$. Specifically, for any hadronic final state $hh^\prime$, one expects \cite{ParticleDataGroup:2024cfk}
\begin{align}
\begin{split}
Q_{hh^\prime}\equiv\frac{BR(\psi(3686)\to hh^\prime)}{BR(J/\psi\to hh^\prime)}&\simeq \frac{BR(\psi(3686)\to e^+e^-)}{BR(J/\psi\to e^+e^-)}=\frac{(7.94\pm0.22)\times10^{-3}}{(5.971\pm0.032)\times10^{-2}}=(13.3\pm0.4)\%.
\end{split}
\end{align}
Experimentally, the ratio of leptonic branching fractions is about 12\%, which gives rise to the so-called "12\% rule". This rule implies that the hadronic decay patterns of $\psi(3686)$ should resemble those of $J/\psi$, with an overall suppression factor of roughly 0.12.

The "12\% rule'' is reasonably well satisfied for most hadronic channels. However, violations are observed in certain exclusive decay modes, most notably in the $\rho\pi$ and $K^+K^*(892)^- + \text{c.c.}$ final states. These are dominant decay modes of $J/\psi$ but are strongly suppressed in $\psi(3686)$, with branching fractions more than two orders of magnitude below the expectation \cite{Harris:1999wn,ParticleDataGroup:2024cfk} 
\begin{align}
\begin{split}
&Q_{\rho\pi}=\frac{BR(\psi(3686)\to \rho\pi)}{BR(J/\psi\to \rho\pi)}= \frac{(3.2\pm1.2)\times10^{-5}}{(1.88\pm0.12)\times10^{-2}}\simeq0.2\%,\\
&Q_{K^+K^*(892)^-+\text{c.c.}}=\frac{BR(\psi(3686)\to K^+K^*(892)^-+\text{c.c.})}{BR(J/\psi\to K^+K^*(892)^-+\text{c.c.})}= \frac{(2.9\pm0.4)\times10^{-5}}{(6.0^{+0.8}_{-1.0})\times10^{-3}}\simeq0.5\%.
\end{split}
\end{align}
Many other vector-pseudoscalar ($VP$) channels show similar suppression, as summarized in Table~\ref{rule12_compare}. This dramatic suppression, observed experimentally and in stark contrast to the "12\% rule", constitutes the longstanding ``$\rho\pi$ puzzle'' in charmonium physics.

 \begin{table}[htbp]
		\centering
		\caption{Measured branching ratios of $\psi(3686)$ and $J/\psi$ decays into
two-light-meson final states $h^\prime$ \cite{ParticleDataGroup:2024cfk}. The corresponding ratios $Q_{h^\prime}$ and their qualitative deviations from the perturbative-QCD 12\% rule are listed, showing cases of strong suppression, enhancement, or consistency with the expectation.}
		\label{rule12_compare}
		\renewcommand\arraystretch{1.8}
		\setlength{\arrayrulewidth}{0.5pt}
		\begin{tabular*}{1.0\textwidth}{l@{\extracolsep{\fill}}cccc}
            \hline
        $hh^\prime$ &$BR(\psi(3686)\to hh^\prime)$ &$BR(J/\psi\to hh^\prime)$ &Ratio $Q_{hh^\prime}$ $(\%)$ &Expected $\sim 12\%$?\\ 
        \hline
        $\rho\pi$ &$(3.2\pm1.2)\times10^{-5}$ &$(1.88\pm0.12)\times10^{-2}$ &$0.2\pm0.1$ &Strong Suppression\\ 
        $K^+K^*(892)^-+\text{c.c.}$ &$(2.9\pm0.4)\times10^{-3}$ &$(6.0^{+0.8}_{-1.0})\times10^{-1}$ &$0.5^{+0.1}_{-0.2}$ &Strong Suppression\\ $\bar{K}^0K^*(892)^0+\text{c.c.}$ &$(1.09\pm0.20)\times10^{-4}$ &$(4.2\pm0.4)\times10^{-3}$ &$2.6\pm0.6$ &Suppression\\ 
        $\phi\eta^\prime$ &$(1.54\pm0.20)\times10^{-5}$ &$(4.6\pm0.5)\times10^{-4}$ &$3.3\pm0.6$ &Suppression\\ $\phi\eta$ &$(3.10\pm0.31)\times10^{-5}$ &$(7.4\pm0.6)\times10^{-4}$ &$4.2\pm1.0$ &Suppression\\ $\rho a_2(1320)$ &$(2.6\pm0.9)\times10^{-4}$ &$(1.09\pm0.22)\times10^{-2}$ &$2.4\pm1.0$ &Suppression\\ $\omega f_2(1270)$ &$(2.2\pm0.4)\times10^{-4}$ &$(4.3\pm0.6)\times10^{-3}$ &$5.1\pm1.2$ &Suppression\\  $\omega\pi^0$ &$(2.1\pm0.6)\times10^{-5}$ &$(4.5\pm0.5)\times10^{-4}$ &$4.7\pm1.5$ &Suppression\\ $\pi^+\pi^-$ &$(7.8\pm2.6)\times10^{-6}$ &$(1.47\pm0.14)\times10^{-4}$ &$5.3\pm1.9$ &Suppression\\ 
        $K^+K^-$ &$(7.5\pm0.5)\times10^{-5}$ &$(3.06\pm0.05)\times10^{-4}$ &$24.5\pm1.7$ &Enhancement\\ $K_S^0K_L^0$ &$(5.34\pm0.33)\times10^{-5}$ &$(1.95\pm0.11)\times10^{-4}$ &$27.4\pm2.3$ &Enhancement\\$b_1(1235)^0\pi^0$ &$(2.4\pm0.6)\times10^{-4}$ &$(2.3\pm0.6)\times10^{-3}$ &$10.4\pm3.8$ &Consistency\\ $\rho\eta$ &$(2.2\pm0.6)\times10^{-5}$ &$(1.93\pm0.23)\times10^{-4}$ &$11.4\pm3.4$ &Consistency\\ $\rho\eta^\prime$ &$(1.9^{+1.7}_{-1.2})\times10^{-5}$ &$(8.1\pm0.8)\times10^{-5}$ &$23.5^{+21.2}_{-15.0}$ &Consistency (Large error)\\ $\phi f_1(1285)$ &$(3.0\pm1.3)\times10^{-5}$ &$(2.6\pm0.5)\times10^{-4}$ &$11.5\pm5.5$ &Consistency\\ $\omega\eta^\prime$ &$(3.2^{+2.5}_{-2.1})\times10^{-5}$ &$(1.89\pm0.18)\times10^{-4}$ &$16.9^{+13.4}_{-11.3}$ &Consistency (Large error)\\ $b_1(1235)^\pm\pi^\mp$ &$(4.0\pm0.6)\times10^{-4}$ &$(3.0\pm0.5)\times10^{-3}$ &$13.3\pm3.0$ &Consistency\\ 
        \hline
		\end{tabular*}
	\end{table}
     
The ``$\rho\pi$ puzzle'' points to the involvement of additional nonperturbative QCD effects and has motivated a wide range of theoretical efforts to account for this anomaly. Hou and Soni proposed a $J/\psi$–glueball mixing mechanism to explain the enhancement of the $J/\psi \to \rho\pi$ and $K\bar{K}^*(892)+\text{c.c.}$ decays \cite{Hou:1982kh}. Brodsky, Lepage, and Tuan suggested that these decays proceed via an intermediate glueball \cite{Brodsky:1987bb}. Brodsky and Karliner argued that the $J/\psi$ decays could be enhanced through quark rearrangement mechanisms involving $|\bar q q \bar c c\rangle$ Fock components present in light vector mesons, while the corresponding $\psi(3686)$ decays are suppressed due to the node in its radial wave function \cite{Brodsky:1997fj}. Feldmann and Kroll further investigated these enhancements through the mixing of $J/\psi$ with light-quark states, notably $\omega$ and $\phi$ \cite{Feldmann:2000hs}. Clavelli and Intemann proposed a vector–meson–mixing model involving the $\rho$, $\omega$, $\phi$, and $J/\psi$, while Guo {\it et al.} suggested the molecular components in $J/\psi$~\cite{Guo:2023igo},  which yields sizable hadronic decay modes of $J/\psi$ \cite{Clavelli:1983rk}. In addition, Li, Bugg, and Zou introduced the final-state interactions (FSI) as an explanation, demonstrating that the $a_2\rho$ loop contribution can enhance the decay rate of $J/\psi \to \rho\pi$ \cite{Li:1996yn}. Collectively, these frameworks focus on resolving the "$\rho\pi$ puzzle'' by enhancing the $J/\psi$ decay rates, while the following approaches instead emphasize the suppression of $\psi(3686)$ decays.

Suzuki suggested that the coupling of $\psi(3686)$ to virtual charmed-meson pairs could generate an amplitude destructively interfering with the perturbative QCD $\psi(3686)\to 3g$ process, thereby suppressing specific channels like $\rho\pi$ and $K\bar{K}^*(892)+\text{c.c.}$ \cite{Suzuki:1998ea,Suzuki:2000yq}. Rosner introduced a $2S$–$1D$ mixing scheme, wherein the $\psi(1D)$ contribution destructively interferes with that of $\psi(3686)$ in these channels~\cite{Rosner:2001nm}. Gérard and Weyers proposed that the three-gluon decay of $\psi(3686)$ is absent or suppressed by treating $J/\psi$ and $\psi(3686)$ as orthogonal states, with $\psi(3686)$ decays proceeding mainly through a two-step process involving an intermediate $h_c(1^{+-})$ state \cite{Gerard:1999uf}. Conversely, Chen and Braaten argued that these decays are dominated by higher color-octet Fock states, attributing the suppression of $\psi(3686)$ to an additional dynamical effect related to the small mass gap between $\psi(3686)$ and the $D\bar D$ threshold \cite{Chen:1998ma}. Other proposed mechanisms include the sequential fragmentation model by Karl and Roberts \cite{Karl:1984en}, the hindered M1 transition scenario by Pinsky \cite{Pinsky:1989ue}, the exponential falloff of hadronic form factors within a nonrelativistic quark model suggested by Chaichian and T\"{o}rnqvist \cite{Chaichian:1988kn}.

In fact, under the unquenched picture, the hadronic loop mechanism also offers a potential explanation for the ``$\rho\pi$'' puzzle, which will be introduced in detail later.

\subsubsection{Large non-$D\bar D$ decays of $\psi(3770)$}

Before discussing the non-$D\bar{D}$ decays of $\psi(3770)$, we first briefly review the Okubo–Zweig–Iizuka (OZI) rule, which plays a crucial role in understanding the suppression of such decay processes. The OZI rule, a consequence of an empirical observation, was independently proposed in the 1960s by Okubo \cite{Okubo:1963fa}, Zweig \cite{Zweig:1964ruk,Zweig:1964jf}, and Iizuka \cite{Iizuka:1966fk}. It states that a decay process involving disconnected quark lines between the initial and final states is strongly suppressed relative to one in which these lines are connected. This rule successfully explained several anomalies encountered in the 1960s in the study of short-lived particles. For instance, certain decay modes expected to occur frequently were found to have significantly smaller widths than predicted, such as $\phi \to \pi^+\pi^-\pi^0$ \cite{Klempt:2007cp,ParticleDataGroup:2024cfk,Lipkin:1991bf,Nomokonov:2002jb}.

\changelabel{Furthermore, the OZI rule accounts for the narrow widths observed in low-lying heavy quarkonium states. In general, for states below the open-flavor meson pair threshold, such as $J/\psi$, $\psi(3686)$, and $\Upsilon(nS)$ $(n=1,2,3)$, their hadronic decays are suppressed by the OZI rule, as well as three gluons emission, leading to relatively small total widths.  As indicated by width of $J/\psi$ and $\psi(3686)$, the widths of the vector charmonia resulted from three gluons annihilation are of order $100$ keV. As for $\psi(3770)$, its width is reported to be $27.2\pm 1.0$ MeV, which is about two orders of magnitude larger than that of $\psi(3686)$ \cite{ParticleDataGroup:2024cfk}. Since the $\psi(3770)$ lies just above the $D\bar{D}$ threshold and the open charm decay process are OZI-allowed, $\psi(3770)$ should dominantly decay into $D\bar{D}$~\cite{LeYaouanc:1977gm,Barnes:2005pb}.} However, experimental measurements have revealed a surprisingly large non-$D\bar{D}$ branching fraction. Before 2005, the branching ratio for $\psi(3770)\to \text{non-}D\bar{D}$ was reported as $(10.9 \pm 6.9 \pm 9.2)\%$ \cite{Rong:2005it}, with subsequent measurements by the BESII Collaboration yielding values of $(16.4 \pm 7.3 \pm 4.2)\%$ \cite{BES:2006fpf}, $(14.5 \pm 1.7 \pm 5.8)\%$ \cite{BES:2006dso}, $(13.4 \pm 5.0 \pm 3.6)\%$ \cite{BES:2007cev}, and $(15.1 \pm 5.6 \pm 1.8)\%$ \cite{BES:2008vad}. Notably, the CLEO Collaboration reported a branching fraction of $BR(\psi(3770)\to D\bar D)=(100.3 \pm 1.4^{+4.8}_{-6.6})\%$ \cite{CLEO:2005mpm}, further deepening the puzzle.

These experimental results, indicating a non-$D\bar{D}$ branching ratio of approximately $10\%$ to $15\%$, are significantly larger than theoretical expectations. Meanwhile, the exclusive non-$D\bar{D}$ decay modes of $\psi(3770)$ listed by the PDG \cite{ParticleDataGroup:2024cfk}—such as $J/\psi\eta$, $J/\psi\pi\pi$, $\phi\eta$, and $\gamma\chi_{cJ}$ ($J=0,1$)—sum to a branching fraction of less than $2\%$, insufficient to account for the observed discrepancy. The unexpectedly large non-$D\bar{D}$ decay rate of $\psi(3770)$ has therefore long stood as a significant puzzle in understanding its decay dynamics, motivating extensive experimental and theoretical investigations over the past two decades.

Theoretical efforts to resolve the $\psi(3770)$ non-$D\bar{D}$ decay puzzle have been extensive \cite{Lipkin:1986av,Kuang:1989ub,Ding:1991vu,Achasov:1990gt,Achasov:1991qp,Achasov:1994vh,Achasov:2005qb,Rosner:2001nm,Rosner:2004wy,Voloshin:2005sd,Eichten:2007qx,He:2008xb,Zhang:2009kr,Liu:2009dr,Li:2013zcr,Qi:2025xkz,Qi:2025xkz}. Kuang and Yan studied the hidden-charm decay $\psi(3770) \to J/\psi\pi\pi$ within the framework of QCD multipole expansion \cite{Kuang:1989ub}, finding results consistent with the corresponding exclusive experimental measurements \cite{ParticleDataGroup:2024cfk}. He, Fan, and Zhao investigated the decay $\psi(3770) \to \text{light hadrons}$ and, by including color-octet contributions, obtained $\Gamma(\psi(3770) \to \text{light hadrons}) = 467^{+338}_{-187}\,\text{keV}$, corresponding to a total non-$D\bar{D}$ branching fraction of about $5\%$ \cite{He:2008xb}.

Given that the dominant decay mode of $\psi(3770)$ is $D\bar{D}$, the hadronic loop mechanism was proposed as a possible pathway for $\psi(3770)$ to access non-$D\bar{D}$ final states \cite{Zhang:2009kr,Liu:2009dr,Li:2013zcr}. Theoretical estimates for the $VP$ final states vary considerably, ranging from $0.2$–$1.1\%$ \cite{Zhang:2009kr,Liu:2009dr} to $(0.04-0.17)\%$ or $(3.38-5.23)\%$ \cite{Li:2013zcr}, depending on the input parameters. These results suggest that accounting for hadronic loop effects can substantially mitigate the discrepancy between theoretical predictions and experimental observations \cite{Liu:2009dr}. Later, we will come back to this point.

\subsubsection{Anomalous transitions in high-lying heavy quarkonia}

For a charmonium state $\psi$ decaying into two light mesons $h_1$ and $h_2$, perturbative QCD predicts that the branching ratio behaves asymptotically as
\begin{align}
BR[\psi(\lambda)\to h_1(\lambda_1)h_2(\lambda_2)]\sim\left(\frac{\Lambda_{\text{QCD}}^2}{m_c^2}\right)^{|\lambda_1+\lambda_2|+2},
\end{align}
where $\lambda$, $\lambda_1$, and $\lambda_2$ denote the helicities of the respective mesons. It is evident that the leading contribution arises when $\lambda_1+\lambda_2=0$, while configurations violating this condition are power-suppressed. This phenomenon is known as the helicity selection rule \cite{Chernyak:1981zz}. The rule can equivalently be expressed in terms of the naturalness quantum number, $\sigma\equiv P(-1)^J$, as $\sigma(\psi)=\sigma(h_1)\sigma(h_2)$, where $P$ and $J$ are the parity and spin of the corresponding meson. The helicity selection rule has been widely applied in analyzing exclusive decay processes of heavy quarkonia. In Refs.~\cite{Feldmann:2000hs,Chernyak:1981zz,Chernyak:1983ej}, specific examples of helicity-allowed and helicity-suppressed decay modes are explicitly discussed. However, with the accumulation of experimental data, an increasing number of observations have revealed substantial deviations from the rule’s predictions. For instance, the well-known $J/\psi \to VP$ decay mentioned earlier, as well as the $\eta_c \to VV$ process, should be strongly suppressed according to the rule, yet they constitute important decay channels of the $J/\psi$ and $\eta_c$, respectively \cite{ParticleDataGroup:2024cfk}.

For the processes $\chi_{cJ}\to VV$, the decay of $\chi_{c1}$ violates the helicity selection rule, whereas those of $\chi_{c0}$ and $\chi_{c2}$ do not. Consequently, the decays of $\chi_{c1}$ into two light vector mesons are expected to be strongly suppressed compared with the corresponding decays of $\chi_{c0}$ and $\chi_{c2}$. In addition, $\chi_{c1}\to\omega\phi$ is a \changelabel{doubly OZI–suppressed} process. Therefore, the branching ratios of the channels $\chi_{c1}\to\phi\phi$, $\omega\omega$, and $\omega\phi$ are expected to be small, making them difficult to observe experimentally, especially in the case of $\chi_{c1}\to\omega\phi$.
However, in 2011, the BESIII Collaboration reported the first observations of $\chi_{c1}\to\phi\phi$, $\omega\omega$, and $\omega\phi$, with the corresponding branching ratios measured to be $(4.4\pm0.3\pm0.5)\times10^{-4}$, $(6.0\pm0.3\pm0.7)\times10^{-4}$, and $(2.2\pm0.6\pm0.2)\times10^{-5}$, respectively \cite{BESII:2011hcd}. These sizable branching fractions, comparable to those of $\chi_{c0}$ and $\chi_{c2}$, indicate the presence of significant nonperturbative contributions.

\begin{figure}
\centering
\includegraphics[width=0.6\textwidth]{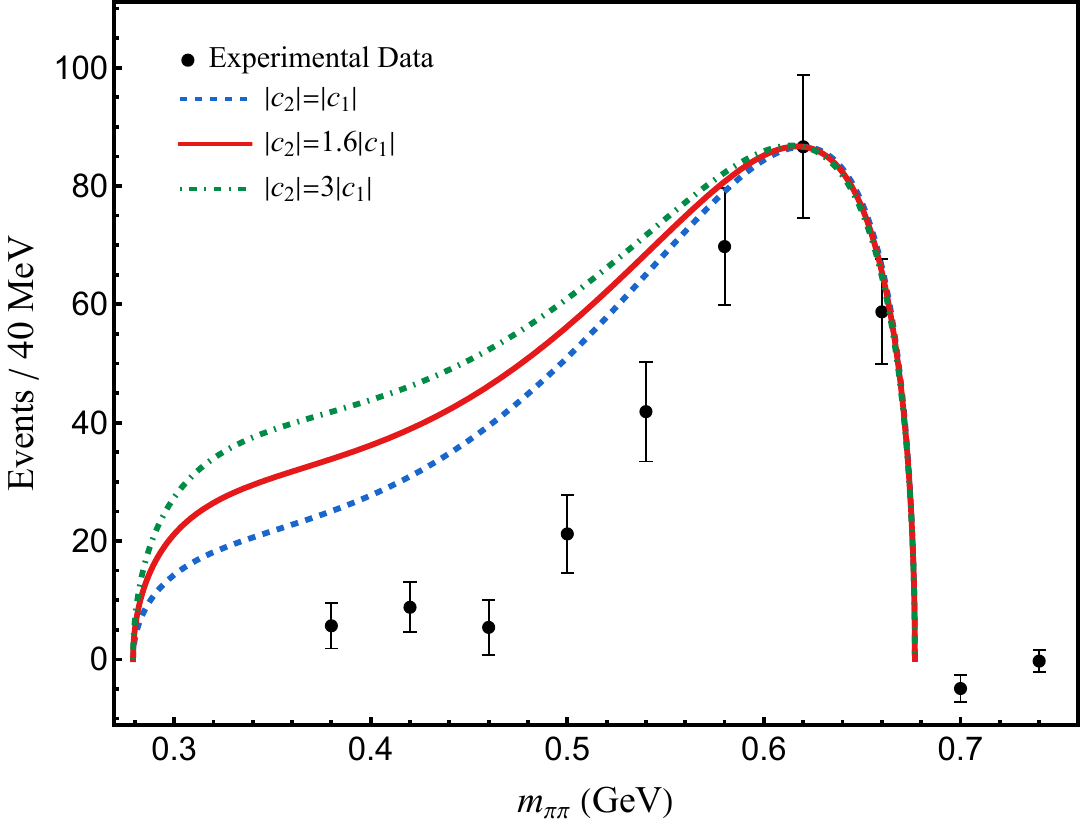}
\caption{Comparison of the dipion invariant-mass distribution between the experimental measurement \cite{CLEO:2005zky} and the result from the QCD multipole expansion \cite{Kuang:1989ub}. The $2S$–$1D$ mixing angle is taken as $\theta_{2S-1D} = -10^\circ$, and several representative values of $|c_1|$ and $|c_2|$ suggested in Ref.~\cite{Kuang:1989ub} are illustrated.}
\label{fig:3770dipion}
\end{figure}

For the decay $\psi(3770)\to J/\psi\pi^+\pi^-$, the dipion invariant mass spectrum distribution has been measured experimentally \cite{CLEO:2005zky}. A clear discrepancy is evident when comparing these data with theoretical predictions based on the QCD multipole expansion (a quenched method) \cite{Kuang:1989ub}, as shown in Fig. \ref{fig:3770dipion}. \changelabel{ A similar discrepancy had also been noticed in the $\Upsilon(3S)\to \Upsilon(1S) \pi^+\pi^- $, and the double-bump structure in the dipion invariant mass distributions lead to the predictions of a new isovector hidden bottom tetraquark state with the mass in the region $10.4-10.8$ GeV~\cite{Anisovich:1995zu, Guo:2004dt}. As for $\Upsilon(4S)\to \Upsilon(nS)\pi^+\pi^-$ $(n=1,2)$, the measured dipion invariant mass spectrum distributions also cannot be described by the QCD multipole expansion approach \cite{Belle:2006wip,BaBar:2006udk,Kuang:1981se}, as shown in Fig. \ref{fig:Mpipi-4S-1S-Pap2-Model}. It should be noted that both the $\psi(3770)$ and $\Upsilon(4S)$ are heavy quarkonia with masses above the open-flavor thresholds ($D\bar{D}$ and $B\bar{B}$, respectively). Therefore, effects beyond the quenched approximation must be considered \cite{Chen:2011jp, Guo:2006ai}.}

Anomalies have also been observed in the radiative decays of heavy quarkonia. In particular, the radiative transitions
$J/\psi\to \gamma \eta_c$ and $\psi^\prime\to\gamma\eta_c^{(\prime)}$ are expected to be dominantly governed by magnetic dipole (M1) transitions, which involve a quark spin flip in the conventional $q\bar q$ picture. However, as studied in Ref.~\cite{Barnes:2005pb}, both the NR model and the GI model exhibit significant discrepancies in their predictions for these M1 transitions when compared with the experimental data \cite{ParticleDataGroup:2024cfk}. Specifically, for $J/\psi \to \gamma \eta_c$, the predicted partial decay widths are approximately a factor of two larger than the measured value, while for $\psi^\prime \to \gamma \eta_c$, the theoretical predictions exceed the experimental results by nearly an order of magnitude (see Table~\ref{tab:etac_gamma} for details).

Besides, Gao, Zhang, and Chao studied the radiative decays of charmonia $J/\psi$ and $\chi_{cJ}$ into light mesons within the framework of perturbative QCD, performing a complete numerical calculation of the quark–gluon loop diagrams~\cite{Gao:2006bc}. Their theoretical results for $J/\psi\to\gamma\eta$ and $\gamma\eta^\prime$ are in good agreement with the experimental data. Moreover, they predicted the branching ratios of $\chi_{c1}\to\gamma\rho^0$, $\gamma\omega$, and $\gamma\phi$ to be $1.4\times10^{-5}$, $1.6\times10^{-6}$, and $3.6\times10^{-6}$, respectively. Nevertheless, the corresponding measurements reported by the CLEO Collaboration are $(2.43\pm0.19\pm0.22)\times10^{-4}$, $(8.5\pm1.5\pm1.2)\times10^{-5}$, and $(1.28\pm0.76\pm0.15)\times10^{-5}$~\cite{CLEO:2008sah}, which are about an order of magnitude larger than the theoretical predictions. Furthermore, the measured branching ratios for the radiative decay processes $h_b(1P)\to\eta_b(1S)\gamma$, $h_b(2P)\to\eta_b(1S)\gamma$, and $h_b(2P)\to\eta_b(2S)\gamma$ \cite{Belle:2012fkf} are found to be larger than the corresponding theoretical expectations by a factor of about 1.2–2.5 \cite{Godfrey:2002rp}.

These deviations indicate that additional unquenched effects may play an important role in these radiative decay processes.

\subsection{Hadronic loop \changelabel{mechanism}}

In prior studies \cite{Eichten:1978tg,Kinnunen:1978qm,Tornqvist:1979hx,Ono:1983rd,Li:1996yn,Ding:1993uy,Ding:1995he,Suzuki:2000yq,Badalian:1981xj,Heikkila:1983wd,Zhou:1990ik,Geiger:1989yc,Geiger:1991qe,Geiger:1991ab,Geiger:1992va,Isgur:1998kr,Maiani:2004qj}, the limitations of the quenched approximation for charmonium physics were recognized. For instance, Kuang attempted to extend the QCD multipole expansion method by incorporating coupled-channel effects \cite{Zhou:1990ik}. However, progress in this direction remained limited, primarily due to constraints imposed by the experimental data available at the time. The situation began to change around 2006, when the observations of anomalous behaviors in certain charmonium states \cite{Franklin:1983ve,Zhu:2006uc,Rong:2005it,BES:2004imp,CLEO:2005mpm,BES:2006fpf,BES:2006dso,BES:2007cev,BES:2008vad,CLEO:2008sah,BESII:2011hcd,BESIII:2011ysp}, as mentioned above, motivated theorists to re-examine this issue. This renewed effort involved introducing the hadronic loop mechanism \cite{Liu:2006dq,Guo:2007up,Wang:2012mf,Zhang:2009kr,Liu:2009dr,Li:2013zcr,Chen:2009ah,Liu:2009vv,Chen:2010re,Wang:2012mf} into the description of heavy quarkonium transitions.

A key aspect of the perturbative QCD framework is that heavy quarkonium annihilation into gluons occurs at short distances on the order of $r \sim 1/m_Q$, where the strong coupling constant $\alpha_s(m_Q)$ is sufficiently small to justify a perturbative expansion. However, for high-lying charmonia and bottomonia, their spatial radii increase. For sufficiently excited states, these radii can even exceed the interaction range of the soft gluon field. In such cases, long-distance contribution may become significant, potentially violating the power-counting rules of perturbative QCD \cite{Meng:2007tk,Meng:2008bq}.
Another contributing factor is that, although the charm quark is heavy, its mass may still be insufficient to fully satisfy the requirements of a reliable perturbative QCD treatment, thereby undermining the validity of the perturbative framework \cite{Liu:2009vv}. When strong interactions occur at long distances, the momentum transfer is small, and the running coupling constant of QCD becomes large. Consequently, perturbative QCD is no longer applicable. Thus, long-distance contributions constitute an important reason for the failure of rules deduced from perturbative QCD.

\begin{figure}[htb]
\begin{center}
\scalebox{0.6}{\includegraphics{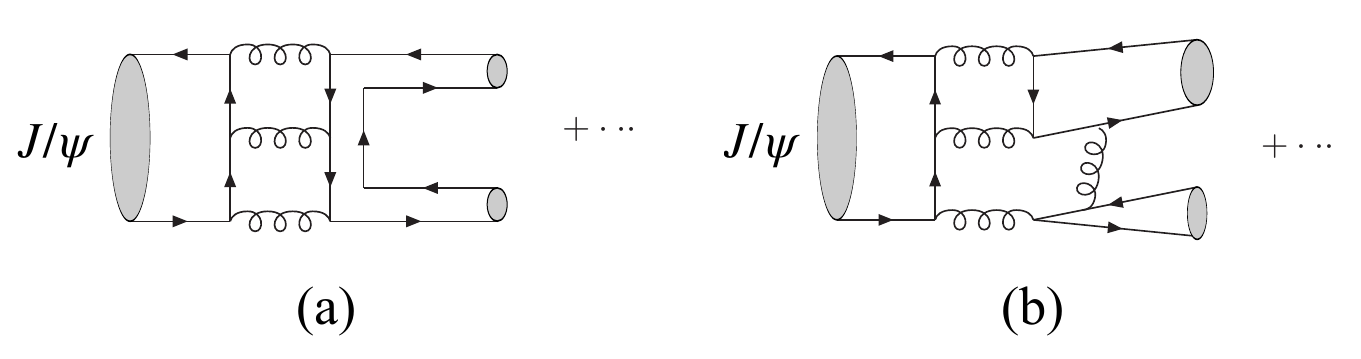}}
\end{center}
\caption{Quark-level short-distance contribution for the $J/\psi\to VP$ decay. Here, (a) and (b) represent the \changelabel{singly OZI-suppressed} (SOZI) and \changelabel{doubly OZI-suppressed} (DOZI) processes for the decays of $J/\psi \to VP$, respectively. Figure adapted from Ref.~\cite{Liu:2006dq}.}
\label{fig:three-gluon}
\end{figure}

By taking the $J/\psi\to VP$ decay as an example, we may explicitly illustrate this issue. The decays of $J/\psi \to \rho\pi$ and other $VP$ final states are OZI-suppressed. The corresponding diagrams for $J/\psi$ decays into $VP$ via three-gluon annihilation are shown in Fig.~\ref{fig:three-gluon}, where Fig.~\ref{fig:three-gluon}(a) represents the \changelabel{singly OZI-suppressed} (SOZI) process and Fig.~\ref{fig:three-gluon}(b) corresponds to the \changelabel{doubly OZI-suppressed} (DOZI) process. In the SOZI process, the quark lines between the initial and final states are disconnected once, whereas in the DOZI process they are disconnected twice.

\begin{figure}[htbp]
    \centering
    \includegraphics[width=0.8\textwidth]{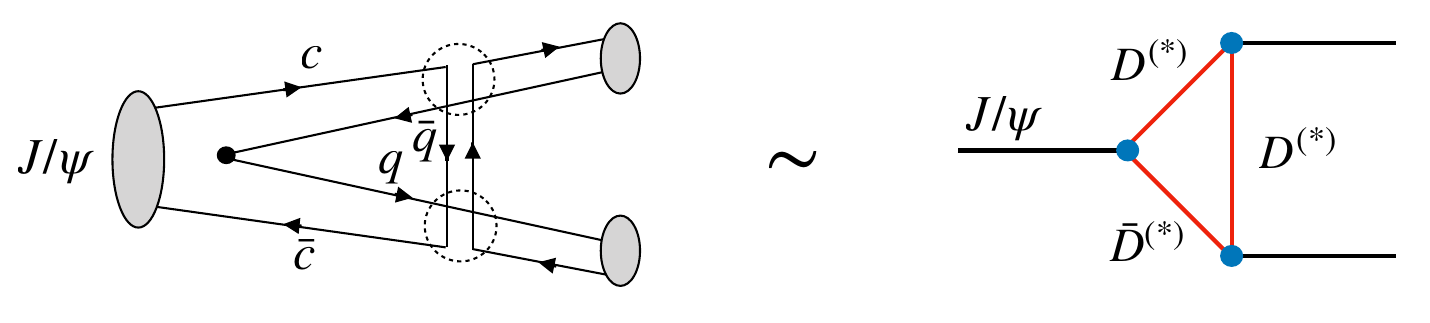}
    \caption{Quark-level (left) and hadron level (right) long-distance contribution for the $J/\psi\to VP$ decay. The left figure adapted from Ref.~\cite{Liu:2006dq}.}
    \label{fig:hadronicloop_quarklevel}
\end{figure}

There exists a long-distance contribution to the $J/\psi \to VP$ decay, as illustrated at the quark level in Fig.~\ref{fig:hadronicloop_quarklevel}. However, it is difficult to directly evaluate such a process at the quark level using quantum field theory due to the nonperturbative nature of strong interactions in the low-energy regime. Therefore, a phenomenological approach must be developed to quantitatively describe this process.

At the hadron level, the long-distance contribution to the $J/\psi \to VP$ decay can be described equivalently. In this framework, the $J/\psi$ meson first dissociates into two virtual charmed mesons, which then interact via the exchange of another charmed meson carrying appropriate charge, flavor, spin, and isospin quantum numbers. Thus, charmed-meson hadronic loops serve as a bridge connecting the initial and final states, as shown in Fig. \ref{fig:hadronicloop_quarklevel}. It is a phenomenological framework known as the hadronic loop mechanism, which provides a tractable method for calculating non-perturbative, long-distance effects in hadronic transitions.

Focusing on a general transition $A\to BC$ mediated by a hadronic loop composed of hadrons $M_i$ ($i=1,2,3$), the process involves three interaction vertices $V_j$ ($j=1,2,3$), which can be described using an effective Lagrangian approach (ELA). Consequently, the decay amplitude can be expressed as
\begin{align}
\raisebox{-0.5\totalheight}{\includegraphics[width=3.5 cm]{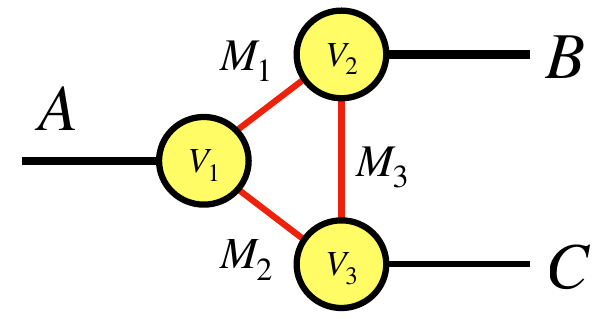}}:\qquad
\mathcal{M}_{A\to BC} = \int \frac{d^4 q}{(2\pi)^4}
\frac{i}{q_{M_1}^2 - m_{M_1}^2}
\frac{i}{q_{M_2}^2 - m_{M_2}^2}
\frac{i}{q_{M_3}^2 - m_{M_3}^2} \,\, \times
\mathcal{V}_{1} \mathcal{V}_{2} \mathcal{V}_{3} \,\, \times\mathcal{FF},
\label{eq:hadronicloop}
\end{align}
where $\mathcal{FF}$ denotes a monopole or dipole form factor. This form factor is introduced primarily to account for the off-shell effects of the exchanged hadron in the subprocess $M_1 + M_2 \to B + C$, and to incorporate the internal structure effects of the exchanged meson. Additionally, $\mathcal{FF}$ serves to suppress high-momentum contributions and regularize the ultraviolet divergence of the loop integral, analogous to the role of a Pauli-Villars regulator.

In the following three subsections, we illustrate how the hadronic loop mechanism plays a crucial role in addressing several long-standing puzzles in quarkonium physics: the $\rho\pi$ puzzle in $J/\psi\to VP$ decays, the unexpectedly large non-$D\bar{D}$ decay width of the $\psi(3770)$, and anomalous transitions among high-lying heavy quarkonia. It should be emphasized that the hadronic loop mechanism effectively embodies the coupled-channel effect, thereby representing a physically motivated and universal unquenched picture of these processes.

\subsubsection{Decoding the $\rho \pi$ puzzle through the hadronic loop mechanism}

The $J/\psi$ decays into a vector ($V$) and a pseudoscalar ($P$) meson provide a critical testing ground for understanding the interplay between perturbative and nonperturbative QCD dynamics. The widely discussed ``$\rho\pi$ puzzle''—the significant suppression of $J/\psi \to \rho\pi$ relative to naive perturbative QCD expectations—highlights the importance of nonperturbative mechanisms. A coherent description requires the inclusion of both the short-distance OZI-suppressed three-gluon annihilation, and the long-distance hadronic loop effects. Since the former cannot be reliably calculated \textit{ab initio}, it is treated via a phenomenological effective coupling $|\mathcal{G}_{S}^{PV}|$, which is assumed universal across all $J/\psi \to PV$ channels due to the flavor-blind nature of gluon interactions.

\begin{figure}[htbp]
    \centering
    \includegraphics[width=0.8\textwidth]{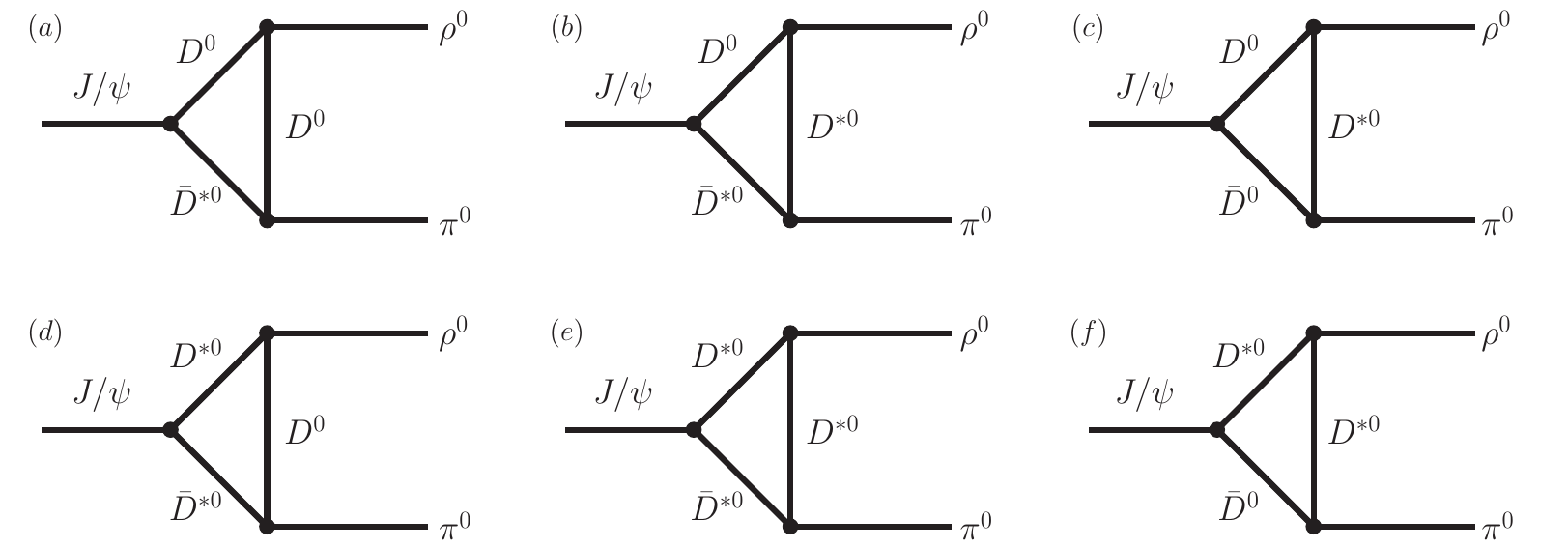}
    \caption{Feynman diagrams for $J/\psi\to \rho^0\pi^0$ via $D^{(*)}$-meson loops. Diagrams (a)--(f) involve neutral charmed mesons; charged-meson loops follow by replacement $D^{(*)0} \to D^{(*)\pm}$. Adapted from Ref.~\cite{Liu:2006dq}.}
    \label{fig:rho-pi}
\end{figure}

The dominant long-distance contribution is modeled by the hadronic loop mechanism, where the $J/\psi$ couples to the final state via virtual intermediate charmed mesons ($D^{(*)}$, $D_s^{(*)}$). In Fig. \ref{fig:rho-pi}, the contribution of the hadronic loop mechanism to $J/\psi \to \rho^0\pi^0$ is illustrated as an example. The total transition amplitude for a channel $J/\psi \to VP$ can be expressed in a compact Lorentz structure \cite{Liu:2006dq}:
\begin{equation}
\mathcal{M}^{PV} = \mathcal{G}^{PV}\,
\varepsilon_{\mu\nu\alpha\beta}\,
\epsilon_{J/\psi}^{\mu}\,\epsilon_{V}^{*\nu}\,p_{3}^{\alpha}\,p_{4}^{\beta},
\label{eq:total_amp_structure}
\end{equation}
where the effective coupling $\mathcal{G}^{PV}$ encapsulates contributions from both the short-distance ($\mathcal{G}^{PV}_S$) and long-distance ($\mathcal{G}^{PV}_{H}(\alpha)$) processes:
\begin{eqnarray}
\mathcal{G}^{PV} = \mathcal{G}^{PV}_{S} \, \beta^{PV} + \mathcal{G}^{PV}_{H}(\alpha).
\label{eq:coupling_decomposition}
\end{eqnarray}
Here, $\beta^{PV}$ are the SU(3) flavor factors obtained from $\mathrm{Tr}(PV)$, and $\alpha$ is a dimensionless parameter governing the hadronic form factor cutoff, $\Lambda(m_i) = m_i + \alpha \Lambda_{\text{QCD}}$ ($\Lambda_{\text{QCD}}=220\;\text{MeV}$)~\cite{Cheng:2004ru}. The loop amplitudes $\mathcal{G}^{PV}_{H}(\alpha)$ are calculated explicitly via Feynman diagrams involving $D^{(*)}$ loops (for light non-strange final states) or $D_s^{(*)}$ loops (for final states involving strangeness). A monopole form factor, $\mathcal{F}(m_i, q^2) = (\Lambda^2 - m_i^2)/(\Lambda^2 - q^2)$, is employed to regulate the loop integrals and account for the off-shell effects of the exchanged mesons, where $\Lambda$ plays a role analogous to the cutoffs used in the Pauli–Villars renormalization scheme.

The model is constrained using the experimental branching ratios for two reference channels: $J/\psi \to \rho^0\pi^0$ and $J/\psi \to K^{*+}K^{-} + \text{c.c.}$. A global fit yields the parameter values:
$\alpha = 0.13$, $|\mathcal{G}^{PV}_{S}| = 4.51 \times 10^{-3}\ \text{GeV}^{-1}$.
With these parameters fixed, predictions are made for all other $J/\psi \to PV$ channels. The results, summarized in Table~\ref{pv-table}, reveal several key insights:
\begin{itemize}
    \item The hadronic loop contribution is \textit{sizable}, comparable in magnitude to the short-distance OZI amplitude.
    \item A \textit{destructive interference} between the two amplitudes is essential to reproduce the observed branching ratios, particularly for the suppressed $\rho\pi$ channel.
    \item The predictions for channels such as $\phi\eta$, $\phi\eta'$, and $\omega\eta'$ show satisfactory agreement with experimental data, validating the model's consistency.
\end{itemize}

\begin{table}[htbp]
\centering
\caption{Comparison between the calculated results and experimental data. The first two decay modes are used to fix the parameters $\alpha$ and $|\mathcal{G}^{PV}_{S}|$. The table is adapted from Ref.~\cite{Liu:2006dq}.}
\renewcommand\arraystretch{1.3}
\setlength{\tabcolsep}{4pt}
\begin{tabular*}{1.0\textwidth}{l@{\extracolsep{\fill}}cccccc}
            \hline
Decay
mode&$\rho^{0}\pi^{0}$&$K^{*+}K^{-}+\mathrm{c.c.}$&$\phi\eta$&$\phi\eta'$&$\omega\eta$&$\omega\eta'$\\
\hline
$BR\times10^{-3}$~(Experiment)~\cite{ParticleDataGroup:2006fqo}&$4.2\pm0.5$&$5.0\pm0.4$&$0.65\pm0.07$&$0.33\pm0.04$
&$1.58\pm0.16$&$0.167\pm0.025$\\\hline $\mathcal{G}^{PV} (10^{-3}$
GeV$^{-1})$&$2.08\pm0.25$&$1.65\pm0.26$&$0.89\pm0.096$&$0.71\pm0.086$&$1.27\pm0.13$&$0.46\pm0.069$\\\hline
$\mathcal{G}^{PV}_{H} (10^{-3}
$GeV$^{-1})$~(Theory)&6.59&6.16&3.55&5.13&5.46&4.07\\\hline
$\mathcal{G}^{PV}
(10^{-3}$GeV$^{-1})$~(Theory)&2.08~(fitting)&1.65~(fitting)&0.92&0.62&0.95&0.40\\\hline
BR~($\times10^{-3}$)~(Theory)&4.2~(fitting)&5.0~(fitting)&0.70&0.25&0.86&0.14\\
\hline
\end{tabular*}
\label{pv-table}
\end{table}

Ref. \cite{Wang:2012mf} also pointed out that the hadronic loop effect is important for understanding the “$\rho\pi$ puzzle”. In that work, the authors investigated the long-distance hadronic loop effects to the decays of $J/\psi(\psi^\prime)\to VP$, together with the short-distance contributions, and the electromagnetic (EM) \changelabel{transition} amplitudes which dominate the isospin-violating channels. By fixing the relevant parameters with the experimental data for $J/\psi$, $\psi^\prime\to \rho\pi$ and $K^*\bar{K}+\text{c.c.}$, they reached a similar conclusion: the destructive interference between the long-distance charmed-meson loops and the short-distance amplitudes in $\psi^\prime$ decays leads to the observed deviations from the ``12\% rule". Ref. \cite{Wang:2012mf} further suggested that the hadronic loop mechanism constitutes a general nonperturbative effect in the charmonium energy region and may be involved in an even wider range of decay modes.

In conclusion, the combined framework incorporating both the short-distance three-gluon annihilation and the long-distance hadronic loop mechanism provides a consistent and quantitative description of the $J/\psi/\psi^\prime \to VP$ decays.
The significant interference between the short-distance three-gluon and long-distance hadronic-loop amplitudes offers a compelling explanation for the long-standing ``$\rho\pi$ puzzle'', underscoring the critical role of nonperturbative charmed-meson loops in charmonium decays.

\subsubsection{\changelabel{Explaining the }large non-$D\bar D$ decays of $\psi(3770)$}
The unexpectedly large non-$D\bar{D}$ decay rates observed for $\psi(3770)$ indicate the presence of additional mechanisms. Among the possible explanations, the hadronic loop mechanism has been regarded as a plausible non-perturbative contribution capable of accounting for these anomalously large non-$D\bar{D}$ decay modes.

\begin{figure}[htbp]
    \centering
    \includegraphics[width=0.4\textwidth]{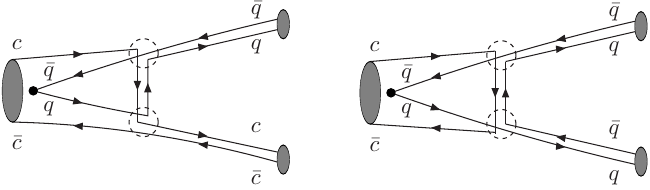}\qquad
\raisebox{0.1\totalheight}{\includegraphics[width=0.4\textwidth]{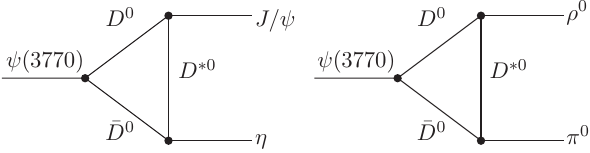}}
    \caption{Quark- and hadron-level Feynman diagrams illustrating the hadronic loop transitions $\psi(3770)\to J/\psi\eta$ and $\psi(3770)\to\rho^{0}\pi^{0}$. Figure adapted from Ref.~\cite{Liu:2009dr}.}
    \label{fig:quark-hadron-level}
\end{figure}

The large non-$D\bar{D}$ decays of $\psi(3770)$ include both hidden-charm and light-hadron (L–H) decay channels. The corresponding Feynman diagrams in the hadronic loop framework are shown in Fig.~\ref{fig:quark-hadron-level}, with the left panel depicting the quark level and the right panel the hadron level. In this mechanism, $\psi(3770)$ first couples to an intermediate pair of charmed mesons, $D\bar{D}$, which subsequently convert into the final states via a charmed meson exchange \cite{Liu:2009dr}. It should be noted that since the mass of $\psi(3770)$ lies above the $D\bar{D}$ threshold, the intermediate $D\bar{D}$ pair can go on-shell and become real. Consequently, the hadronic loop effect in this case is also referred to as final-state interaction (FSI). The absorptive part of the decay amplitudes for $\psi(3770)\to D(k_{1})+\bar{D}(k_{2})\to J/\psi(k_{3})+\eta(k_{4})$ and $\psi(3770)\to D(k_{1})+\bar{D}(k_{2})\to \mathbb{V}(k_{3})+\mathbb{P}(k_{4})$ corresponding to Fig. \ref{fig:quark-hadron-level} are given by \cite{Liu:2009dr}

\begin{eqnarray}
{A}_{J/\psi
\eta}&=&\mathcal{G}_{J/\psi\eta}\,\mathcal{Q}_{J/\psi\eta}\,\frac{|\mathbf{p}|}{32\pi^2
M_{\psi}}\int \mathrm{d}\Omega\left[ig_{{\psi \mathcal{DD}}}
(k_{1}-k_{2})\cdot
\epsilon_{\psi}\right]\left[i\,g_{{J/\psi}\mathcal{DD}^{*}}
\varepsilon_{\kappa\xi\lambda\tau}\epsilon_{J/\psi}^{*\kappa}(-i\,k_{1}^{\xi})(iq^{\tau})\right]
\left[g_{\mathcal{D}^{*}\mathcal{D}\mathbb{P}}(i\,k_{4\mu})\right]\nonumber\\
&&\times\frac{i}{q^2 -m_{D^*}^2
}\left(-g^{\mu\lambda}+\frac{q^{\mu}q^{\lambda}}{m_{D^*}^2}\right)\,\mathcal{F}^2\left[m_{D^*}^2
,q^2\right],
\label{aaaa1}
\end{eqnarray}
and
\begin{eqnarray}
{A}_{\mathbb{PV}}&=&\mathcal{G}_{\mathbb{PV}}\,\mathcal{Q}_{\mathbb{PV}}
\,\frac{|\mathbf{p}|}{32\pi^2 M_{\psi}}\int
\mathrm{d}\Omega\left[i\,g_{\psi \mathcal{DD}} (k_{1}-k_{2})\cdot
\epsilon_{\psi}\right]
\left[-2i\,f_{\mathcal{D}^{*}\mathcal{D}\mathbb{V}}\varepsilon_{\kappa\xi\tau\lambda}(i\,k_{3}^{\kappa})
\epsilon_{\mathbb{V}}^{*\xi}(-ik_{1}^{\tau}-iq^{\tau})\right]\left[g_{\mathcal{D}^*\mathcal{D}\mathbb{P}}
(i\,k_{4\mu})\right]\nonumber\\
&&\times
\left(-g^{\mu\lambda}+\frac{q^{\mu}q^{\lambda}}{m_{D^*}^2}\right)\,\,
\frac{i}{q^2 -m_{D^*}^2
}\,\,\mathcal{F}^2\left[m_{D^*}^2,q^2\right]\,,
\label{aaaa2}
\end{eqnarray}
respectively, where $\mathcal{F}^2\left[m_{D^*}^2,q^2\right]$ denotes the monopole form factor. The full decay amplitude is ultimately obtained via the dispersion relation technique~\cite{Liu:2009dr,Meng:2007cx,Liu:2007qi,Zhang:2007su,Pennington:2007xr,vanBeveren:2007cb}.

\begin{figure}[htbp]
\centering
\includegraphics[width=0.31\textwidth]{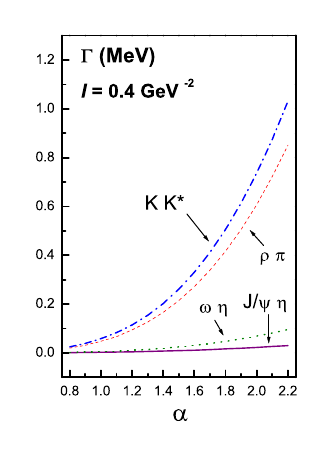}\hspace{0.02\textwidth}
\raisebox{0.07\totalheight}{\includegraphics[width=0.52\textwidth]{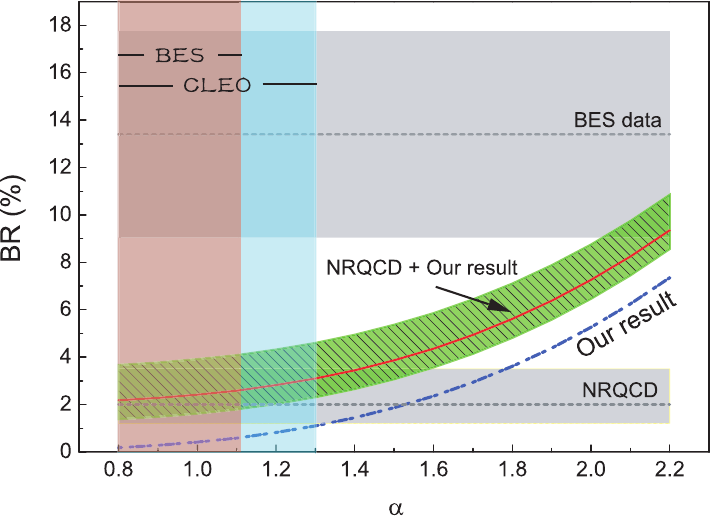}}
\caption{Left panel: The decay widths for the non-$D\bar D$ channels of $\psi(3770)$ on the dependence of parameter $\alpha$.
Right panel: a comparison of the branching ratios predicted by the hadronic loop mechanism (blue dash-dotted line) with the BES measurement of the excess non-$D\bar D$ component in inclusive $\psi(3770)$ decays (dashed line with shaded band), as well as with the NRQCD prediction for $\psi(3770)\to\mathrm{Light~Hadrons}$ obtained via the color-octet mechanism and calculated up to next-to-leading order (dash-dotted line with shaded band)~\cite{He:2008xb}. In the right panel, the red line with the green shaded band denotes the total prediction that includes both the NLO NRQCD contribution and the hadronic loop effect. The green band represents the uncertainty range estimated in Ref.~\cite{He:2008xb}. The orange and light-blue shaded regions indicate the allowed ranges of $\alpha$ inferred from the BES~\cite{BES:2005sch} and CLEO~\cite{CLEO:2005zrs} measurements of $\psi(3770)\to\rho\pi$, respectively. Figure adapted from Ref.~\cite{Liu:2009dr}.}
\label{fig:compare}
\end{figure}

The $\alpha$ parameter, introduced in the form factors, has been constrained to the ranges of $0.8<\alpha<1.1$ and $0.8<\alpha<1.3$, based on the BES \cite{BES:2005sch} and CLEO measurements \cite{CLEO:2005zrs} of $BR(\psi(3770)\to\rho\pi)$, respectively \cite{Liu:2009dr}. Within these ranges, the predicted branching fraction for $\psi(3770)\to\mathrm{non}\text{-}D\bar{D}$ arising from the hadronic loop mechanism is estimated to be $BR_{\mathrm{non}\text{-}D\bar{D}}^{\text{Hadronic~Loop}}=(0.2\text{–}1.1)\%$, as shown in Fig. \ref{fig:compare}. When combined with the NRQCD contribution, the total non-$D\bar{D}$ branching ratio can reach a maximum value of approximately
4.6\%.

\begin{figure}[htbp]
\centering
\includegraphics[width=0.9\textwidth]{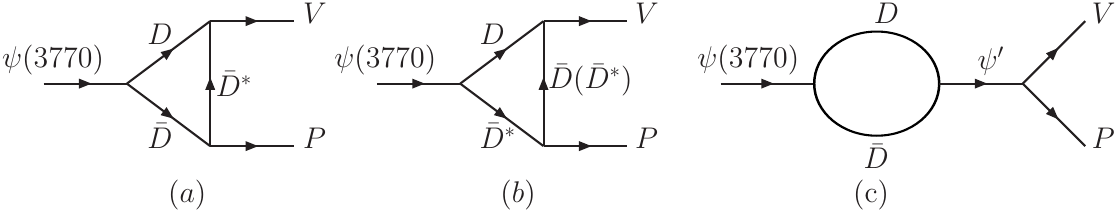}
\caption{$t$-channel (a, b) and $s$-channel (c) hadronic loop of $\psi(3770)\to VP$. Figure adapted from Ref.~\cite{Zhang:2009kr}.}
\label{fig:st}
\end{figure}

More comprehensive analyses has been presented, incorporating both the long-distance $t$-channel and $s$-channel hadronic loop contributions (see Fig.~\ref{fig:st}), together with the short-distance SOZI  contribution \cite{Zhang:2009kr} and EM contribution \cite{Li:2013zcr}, to the decay processes of $\psi(3770)\to VP$. The total decay amplitude can be expressed as \cite{Zhang:2009kr}

\begin{align}
\mathcal{M}_{fi} = \mathcal{M}^{L} + e^{i\delta}\mathcal{M}^{SOZI}
\equiv i \left( g_{L} + e^{i\delta} g_{S} \mathcal{F}_{S}(\vec{p}_{V}) \right)
\, \varepsilon_{\alpha\beta\mu\nu} P_{\psi}^{\alpha} \epsilon_{\psi}^{\beta}
P_{V}^{\mu} \epsilon_{V}^{*\nu} / M_{\psi(3770)}.
\end{align}
Here, an exponential form factor, ${\mathcal F}_S^2({\vec P}_V) \equiv \exp(-{\vec P}V^2/{8\beta^2})$ with $\beta = 0.5~\text{GeV}$, is applied to depicting the flavor-blind SOZI transition \cite{Close:2000yk,Li:2007ky,Li:2013zcr,Zhang:2009kr}. The $s$-channel contributions are evaluated within the on-shell approximation, while the $t$-channel contributions follow similar treatment as in Ref.~\cite{Liu:2009dr} discussed above, where the parameter $\alpha$ is determined either by the measured branching ratio $BR(\psi(3770)\to J/\psi\eta)=(9.0\pm4.0)\times10^{-4}$~\cite{ParticleDataGroup:2008zun,Li:2013zcr,Zhang:2009kr} or from the $\psi(3686)$ decays \cite{Zhang:2009kr}.

Both Refs.~\cite{Zhang:2009kr} and \cite{Li:2013zcr} find that the $t$-channel transitions appear to play a dominant role in the $\psi(3770)\to VP$ decays, whereas the $s$-channel and EM contributions are generally small and negligible. Ref.~\cite{Zhang:2009kr} obtained the total branching ratio of about 0.64\% for all non-$D\bar D$ $VP$ channels of $\psi(3770)$, consistent with the result of $(0.2\text{–}1.1)\%$ calculated in Ref.~\cite{Liu:2009dr}. In Ref.~\cite{Li:2013zcr}, the first scenario yields a total branching ratio in the range of $(0.04\text{–}0.17)\%$, which is close to the results of Refs.~\cite{Liu:2009dr,Zhang:2009kr}, while the second scenario leads to a substantially larger value of $(3.38\text{–}5.23)\%$.

In summary, even with the inclusion of color-octet contributions, the NRQCD prediction for the branching ratio of $\psi(3770)\to \mathrm{non}\text{-}D\bar D$, calculated up to next-to-leading order, still fails to reproduce the BES data~\cite{He:2008xb}, as shown in Fig.~\ref{fig:compare}. The hadronic loop mechanism is expected to play an essential role. When the hadronic loop effects are incorporated, the discrepancy between theoretical predictions and experimental measurements is substantially reduced. 

\subsubsection{\changelabel{Understanding the} anomalous transitions in high-lying heavy quarkonia}

\paragraph{$\chi_{c1}\to VV$} The observations of $\chi_{c1}\to VV$ processes \cite{BES:2004imp,BESII:2011hcd}, which exhibit pronounced violations of the helicity selection rule \cite{Chernyak:1981zz,Chernyak:1983ej}, together with the sizable discrepancies between the measured branching ratios of $\chi_{c1}\to \gamma V$ \cite{CLEO:2008sah,BESIII:2011ysp} and the corresponding pQCD predictions \cite{Gao:2006bc}, provide compelling evidence for the presence of nonperturbative effects. As shown below, these issues can be naturally understood by incorporating the long-distance hadronic loop contributions.

\begin{figure}[htb]
\centering
\begin{tabular}{@{}c c c c c c@{}} 
\scalebox{0.40}{\includegraphics{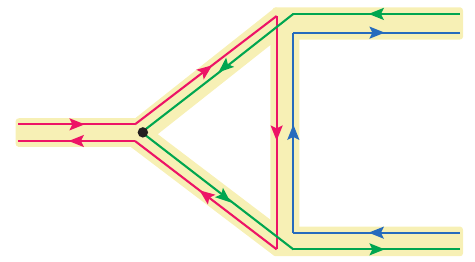}}
& \raisebox{5.5ex}{$\Longrightarrow$}
& \scalebox{0.60}{\includegraphics{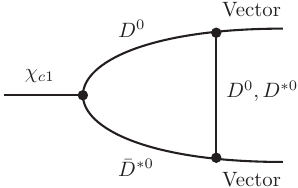}}
& \scalebox{0.60}{\includegraphics{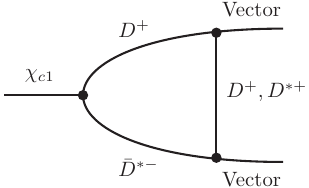}}
& \scalebox{0.60}{\includegraphics{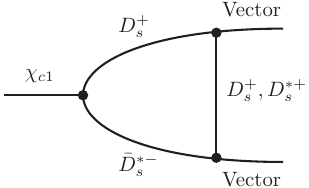}}
& \raisebox{5.5ex}{$+\,\cdots$}
\end{tabular}
\vspace{4pt} 
\caption{Hadronic loop contributions to $\chi_{c1}\to VV$ at the quark and hadron levels. The ellipsis indicates additional hadron level diagrams obtained via the charge-conjugation operation $D_{(s)}^{(*)}\rightleftharpoons \bar{D}_{(s)}^{(*)}$. Figure adapted from Ref. \cite{Chen:2009ah}}
\label{fig:BW}
\end{figure}

The processes $\chi_{c1}\to VV$ proceed via the hadronic loop mechanism as illustrated in Fig.~\ref{fig:BW}, $\chi_{c1}$ first dissociates into a pair of virtual charmed mesons, $D_{(s)}\bar{D}^{*}_{(s)}+\text{h.c.}$, through an $S$-wave coupling. The intermediate charmed mesons then transition into two on-shell light vector mesons by exchanging an appropriate charmed meson. Through this mechanism at long distance, the decays of $\chi_{c1}\to VV$ can evade the helicity selection rule. In addition, an $\omega$-$\phi$ mixing scheme
\begin{align}
\left(
  \begin{array}{c}
| \phi^{p} \rangle  \\| \omega^{p}  \rangle\\
  \end{array}
\right)
 =
\left(
  \begin{array}{cc}
    \cos \theta & \sin \theta \\
    -\sin \theta &  \cos \theta
  \end{array}
\right) \left(
  \begin{array}{c}
    | \phi^{I} \rangle \\ | \omega^{I} \rangle \\
  \end{array}
\right)\label{eq0}
\end{align}
is introduced to prevent the DOZI amplitude of $\chi_{c1}\to \omega\phi$ from vanishing. In this scheme, $|\phi^{p}\rangle$ ($|\omega^{p}\rangle$) and $|\phi^{I}\rangle$ ($|\omega^{I}\rangle$) denote the physical states and the ideally mixed states, respectively, and the mixing angle $\theta=(3.4\pm0.2)^\circ$ is suggested \cite{Dolinsky:1991vq,Benayoun:1999fv,Kucukarslan:2006wk,Chen:2009ah}.

\begin{figure}
\centering
\includegraphics[width=0.32\textwidth]{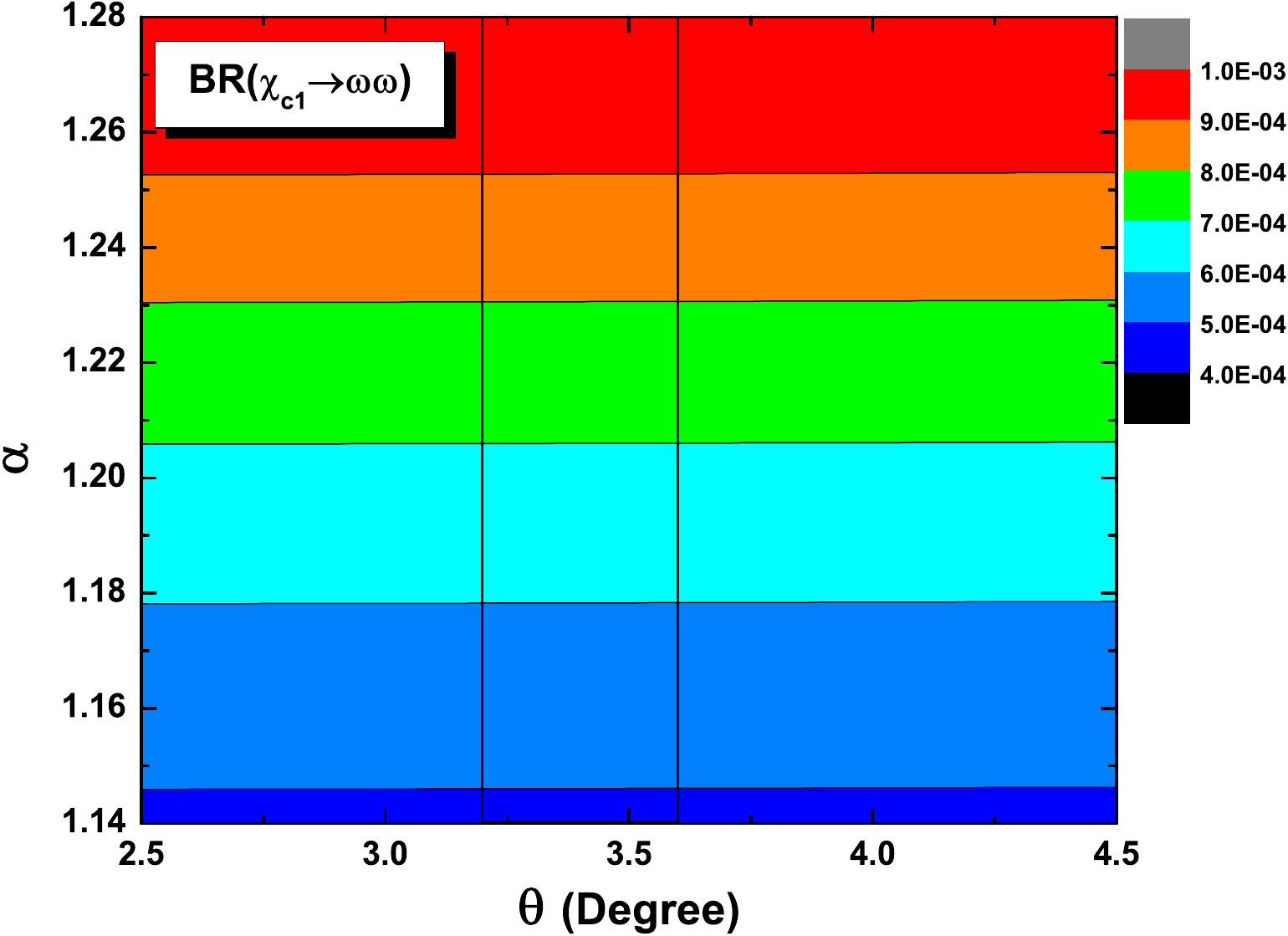}
\includegraphics[width=0.32\textwidth]{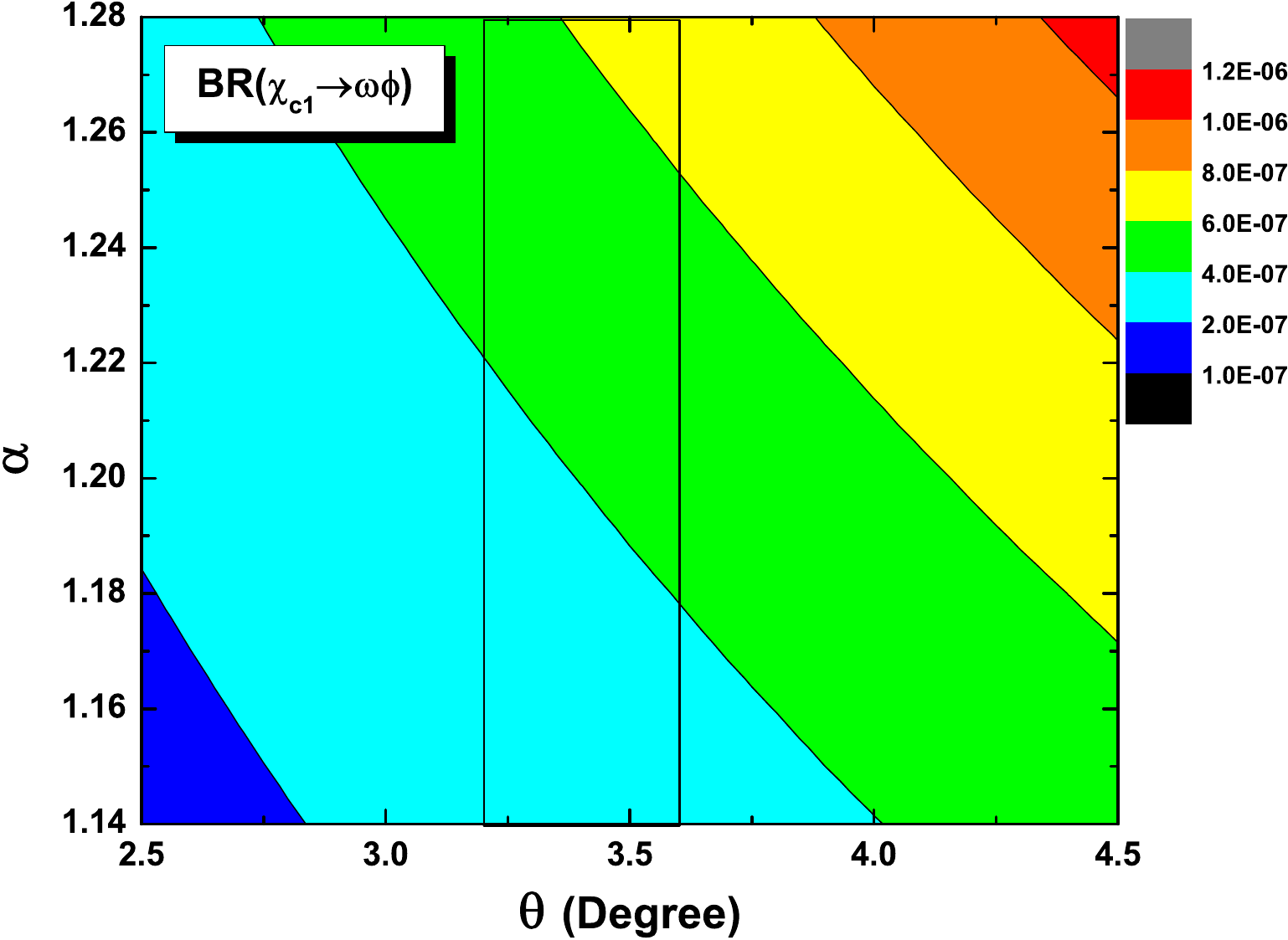}
\includegraphics[width=0.32\textwidth]{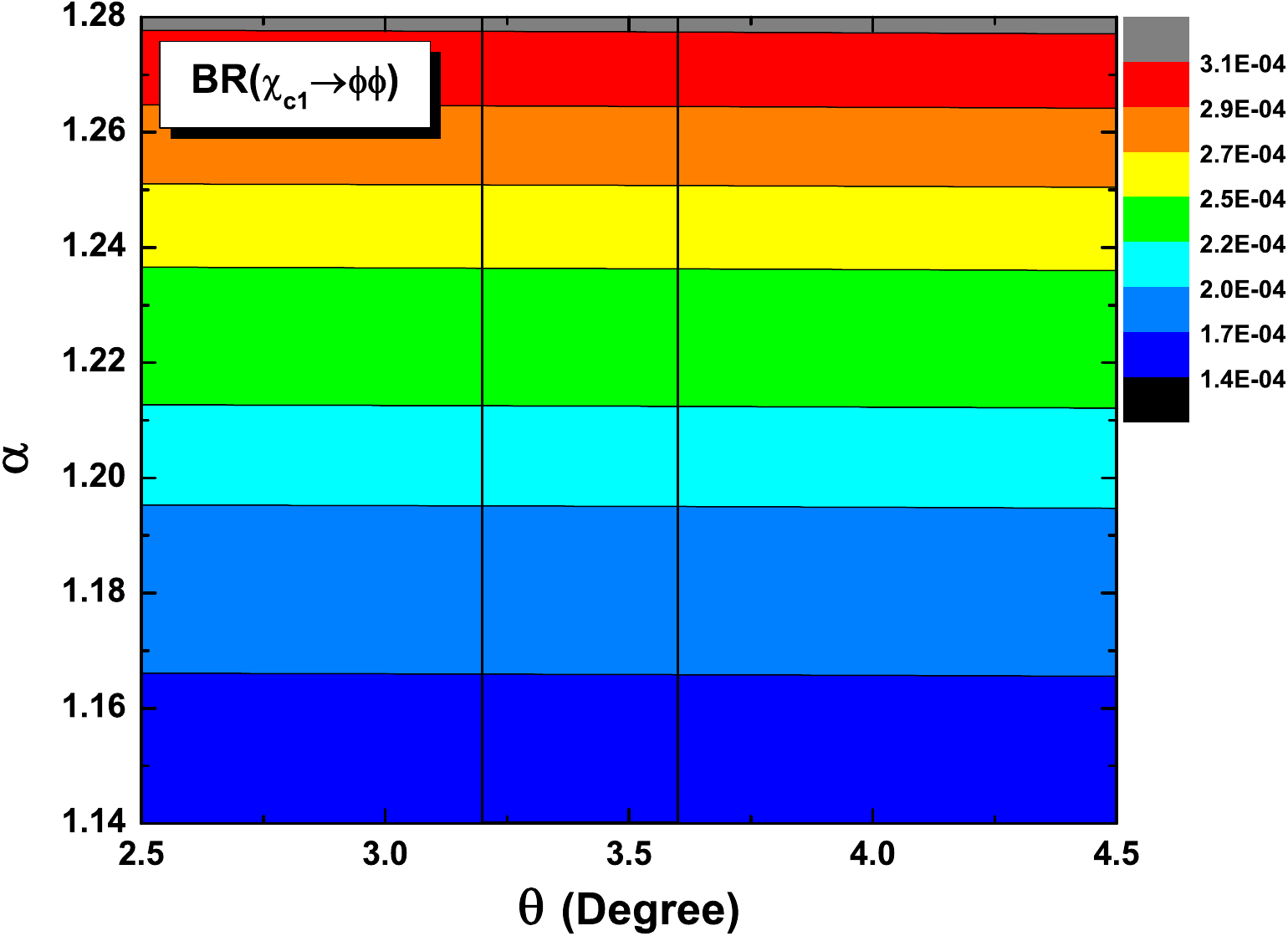}
\caption{Contour plots showing the dependence of $BR(\chi_{c1}\to \omega\omega)$, $BR(\chi_{c1}\to \phi\phi)$, and $BR(\chi_{c1}\to \omega\phi)$ on the parameters $\theta$ and $\alpha$. The region between the two vertical solid lines denotes the range allowed by the $1\sigma$ uncertainty of the mixing angle $\theta$. The figure is adapted from Ref.~\cite{Chen:2009ah}.}
\label{fig:Contour}
\end{figure}

The amplitudes corresponding to pseudoscalar and vector charmed meson exchange, as depicted in Fig.~\ref{fig:BW}, can be written as \cite{Chen:2009ah}
\begin{align}
\mathcal{M}^{(P)}=&\int\frac{d^4 q}{(2\pi)^4}[ig_{\chi_{c1}\mathcal{DD}^*} \epsilon_\sigma][-ig_{\mathcal{DDV}}(p_1+q)\cdot \epsilon_3^*][2if_{\mathcal{D^*DV}}\varepsilon_{\mu\nu\alpha\beta}p_4^\mu \epsilon_4^{*\nu} (q^\alpha-p_2^\alpha)]\nonumber\\&\times
\frac{i}{p_1^2-m^2_\mathcal{D}}\frac{i}{p_2^2-m^2_{\mathcal{D}^*}}\Big(-g^{\sigma\beta}+\frac{p_2^\sigma p_2^\beta}{m^2_{\mathcal{D}^*}}\Big)\frac{i}{q^2-m^2_\mathcal{D}}\,\mathcal{F}^2(q^2,m_{\mathcal{D}}^2),
\end{align}
and
\begin{align}
\mathcal{M}^{(V)}=&\int\frac{d^4 q}{(2\pi)^4}[ig_{\chi_{c1}\mathcal{DD}^*} \epsilon_\sigma][2if_{\mathcal{D^*DV}}\varepsilon_{\mu\nu\alpha\beta}p_3^\mu \epsilon_3^{*\nu} (p_1^\alpha+q^\alpha)]\nonumber\\&\times
\Big[ig_{\mathcal{D^*D^*V}}(q-p_2)\cdot \epsilon_4^* g_{\lambda\kappa}-4i f_{\mathcal{D^*D^*V}}(p_{4\kappa}\epsilon_{4\lambda}^*-p_{4\lambda}\epsilon_{4\kappa}^*)\Big]\nonumber\\&\times
\frac{i}{p_1^2-m^2_\mathcal{D}}\frac{i}{p_2^2-m^2_{\mathcal{D}^*}}\Big(-g^{\sigma\kappa}+\frac{p_2^\sigma p_2^\kappa}{m^2_{\mathcal{D}^*}}\Big)\frac{i}{q^2-m^2_{\mathcal{D}^*}}
\Big(-g^{\beta\lambda}+\frac{q^\beta q^\lambda}{m^2_{D^*}}\Big)\mathcal{F}^2(q^2,m_{\mathcal{D}^*}^2)
,
\end{align}
respectively, in which the dipole form factor $\mathcal{F}^2(q^2,m_{E}^2)$ is adopted. The $\alpha$ parameter introduced in the form factor is constrained into the range of $\alpha=1.14\sim 1.28$, by reproducing the measured branching ratio of $BR[\chi_{c1}\to K^{*0}\bar{K}^{*0}]=(1.67 \pm 0.32 \pm 0.31)\times10^{-3}$ \cite{BES:2004imp}. With this constraint, the branching ratios for $\chi_{c1}\to\omega\omega$, $\phi\phi$, and $\omega\phi$ can be predicted. The results are summarized in Table \ref{tab:typical}, along with the subsequently measured decay branching ratios. The dependence of the calculated branching ratios on the parameter $\alpha$ and the $\omega$–$\phi$ mixing angle $\theta$ is illustrated in Fig.~\ref{fig:Contour}.

\begin{table}[htbp]
\centering
\caption{Predicted ranges of the branching ratios for $\chi_{c1}\to VV$ within the hadronic loop framework in Refs.~\cite{Chen:2009ah} and \cite{Liu:2009vv}, shown together with the subsequently measured decay branching ratios.}
\renewcommand\arraystretch{1.3}
\begin{tabular*}{1.0\textwidth}{l@{\extracolsep{\fill}}cccc}
\hline
\multicolumn{2}{c}{Channel} &Ref. \cite{Chen:2009ah} &Ref. \cite{Liu:2009vv} &Experimental value \\
\hline
\multirow{5}{*}{$\chi_{c1}\to VV$} &$K^{* 0} \bar{K}^{* 0}$& $(1.1 \sim 2.3)\times10^{-3}$ &$(1.2\sim2.0)\times10^{-3}$ &$(1.67 \pm 0.32 \pm 0.31)\times10^{-3}$ \cite{BES:2004imp}\\
{}&$\omega\omega$& $(4.8 \sim 10.4)\times 10^{-4}$ &$(8.7\sim18.0)\times10^{-4}$ &$(6.0\pm0.3\pm0.7)\times10^{-4}$ \cite{BESII:2011hcd}\\
{}&$\phi\phi$& $(1.5 \sim 3.2)\times10^{-4} $ &$(2.7\sim4.6)\times10^{-4}$ &$(4.4\pm0.3\pm0.5)\times10^{-4}$ \cite{BESII:2011hcd}\\
{}&$\omega\phi$& $(2.5 \sim 6.9) \times 10^{-7}$&$\cdots$&($2.2\pm0.6\pm0.2)\times10^{-5}$ \cite{BESII:2011hcd}\\
{}&$\rho\rho$ &$\cdots$ &$(2.6\sim5.4)\times10^{-3}$ &$\cdots$\\
\hline
\multirow{3}{*}{$\chi_{c2}\to VP$} &$K^{*0}\bar{K}^0 +\text{c.c.}$ &$\cdots$&$(4.0\sim 6.7)\times10^{-5}$ &$\cdots$\\
{}& $K^{*+}{K}^- +\text{c.c.}$ &$\cdots$&$(4.0\sim 6.7)\times10^{-5}$\\
{}&$\rho^+\pi^- +\text{c.c.}$&$\cdots$&$(1.2\sim 2.0)\times 10^{-7}$ &$\cdots$\\
\hline
\end{tabular*}
\label{tab:typical}
\end{table}

\begin{figure}
\centering
\begin{minipage}{0.47\textwidth}
  \centering
  \includegraphics[width=\textwidth]{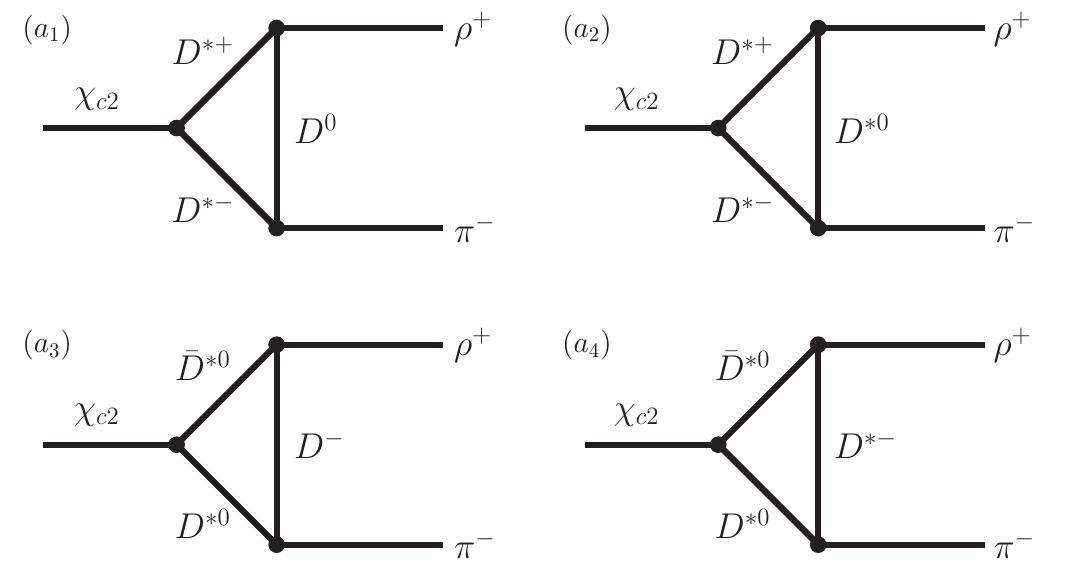}
\end{minipage}
\hspace{0.01\textwidth}
\begin{minipage}{0.01\textwidth}
  \centering
  \rule{0pt}{0.9\linewidth}\\[-0.9\linewidth]
  \rule{0.4pt}{22\linewidth}
\end{minipage}
\hspace{0.01\textwidth}
\begin{minipage}{0.47\textwidth}
  \centering
  \includegraphics[width=\textwidth]{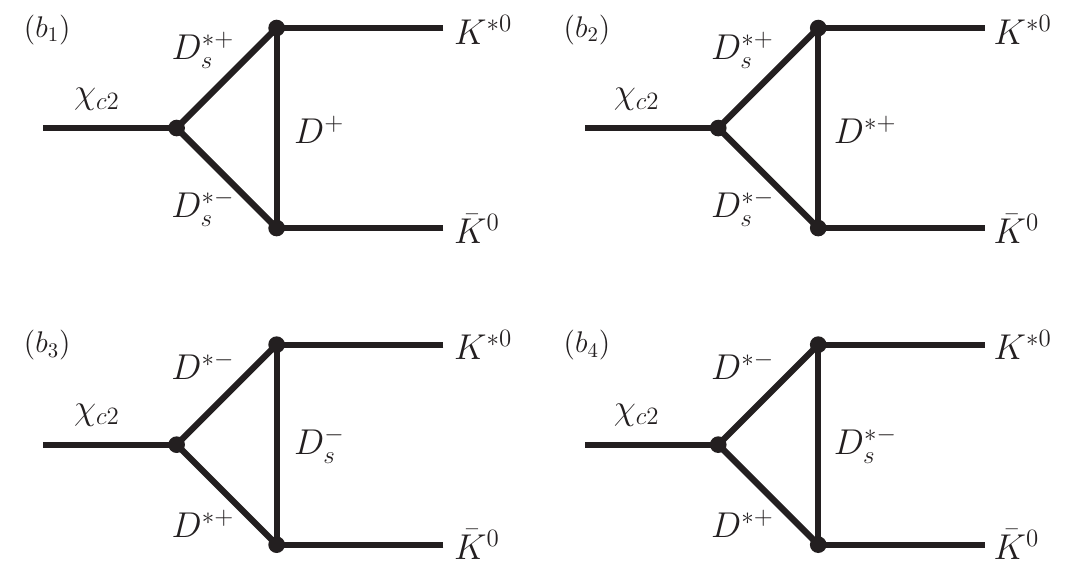}
\end{minipage}

\caption{Hadronic loop diagrams illustrating the long-distance contributions of
$\chi_{c2}\to \rho^+\pi^-$ (left panel) and $K^{*0}\bar K^0$ (right panel) as studied
in Ref.~\cite{Liu:2009vv}. The diagrams for $\chi_{c2}\to \rho^-\pi^+$ and
$\bar K^{*0}K^0$ are implicated.}
\label{fig:chic2rho}
\end{figure}

Ref.~\cite{Liu:2009vv} also studies the $\chi_{c1}\to VV$ channels, and in addition, presents an analysis of the $\chi_{c2}\to VP$ decays (see these diagrams in Fig.~\ref{fig:chic2rho}), both of them violate the helicity selection rule. The calculations employ procedures similar to those of Ref.~\cite{Chen:2009ah} discussed above, and the corresponding numerical results are summarized in Table~\ref{tab:typical}.

From Table~\ref{tab:typical}, we see that the predicted branching ratios for $\chi_{c1} \to \omega\omega$ and $\chi_{c1} \to \phi\phi$ in Refs.~\cite{Chen:2009ah,Liu:2009vv} are close, and both are comparable to the subsequently measured values. This indicates that the long-distance hadronic loop mechanism provides a reasonable explanation for the evasion of the helicity selection rule. Furthermore, the $\omega$--$\phi$ mixing scheme may offer a possible source for the DOZI forbidden process $\chi_{c1} \to \omega\phi$, as the corresponding branching ratio increases rapidly with the mixing angle $\theta$, as shown in Fig. \ref{fig:Contour}. However, with the currently adopted value $\theta=(3.4 \pm 0.2)^\circ$~\cite{Chen:2009ah}, the calculated branching ratio remains about two orders of magnitude smaller than the experimental measurement, suggesting that further investigation is required. The predicted branching ratio for $\chi_{c2} \to K^{*}\bar{K} + \text{c.c.}$ is at the level of $10^{-5}$, indicating that this decay mode may be observable in future experiments.

\paragraph{Radiative decays}
In Refs.~\cite{Li:2007xr,Li:2011ssa}, Li and Zhao showed that hadronic loop effects provide explicit corrections to the M1 transition amplitudes. Their interference with the conventional M1 amplitudes can naturally explain the discrepancies between the GI model predictions and the experimental measurements for the $J/\psi$ and $\psi^\prime \to \gamma \eta_c$ transitions. The total amplitude can be parameterized as~\cite{Li:2011ssa}
\begin{eqnarray}
{\cal M}_{fi}=(g_{V\gamma P}+
\tilde{g}_{tri}e^{i\delta})\varepsilon_{\alpha\beta\mu\nu}p_\gamma^\alpha
\varepsilon_\gamma^\beta p_i^\mu \varepsilon_i^\nu \ ,
\end{eqnarray}
where the first term represents the M1 transition amplitude, while the second denotes the charm-meson loop contributions. In Ref.~\cite{Li:2011ssa}, a destructive phase $\delta=\pi$ is adopted, since the M1 transition amplitude in the quenched quark model has overestimated the experimental data.

\begin{figure}
\centering
\includegraphics[width=0.55\textwidth]{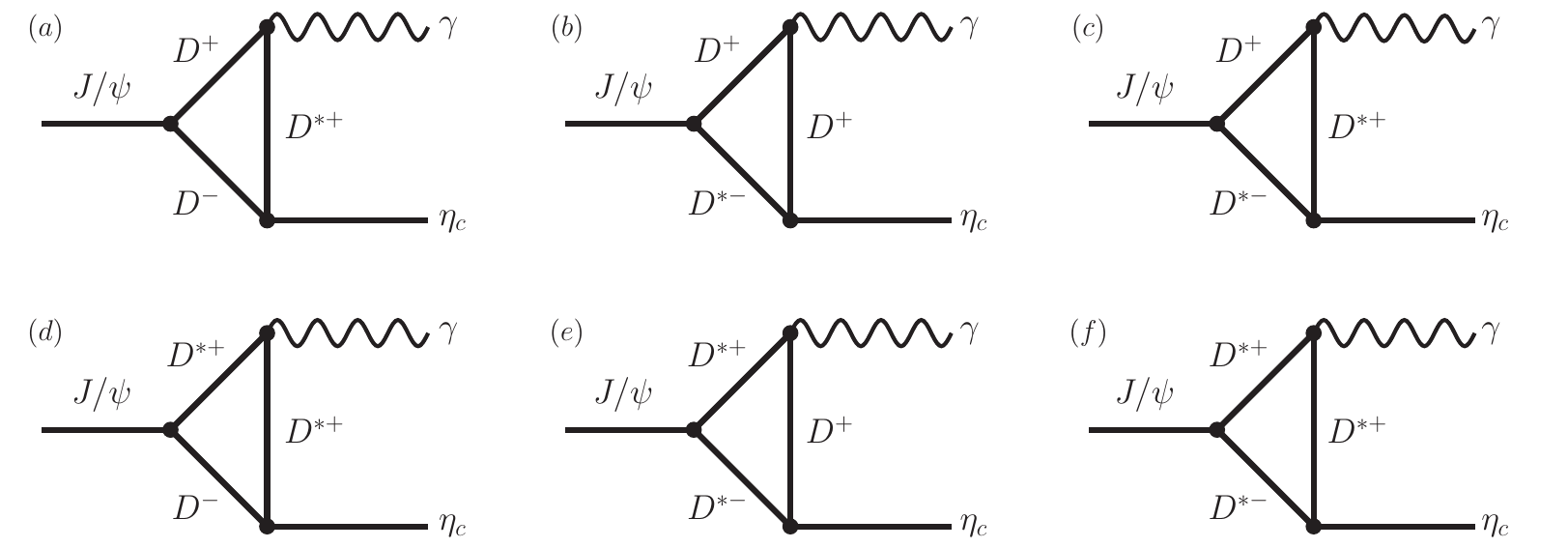}
\hspace{0.01\textwidth} \vrule width 0.5pt height 28mm \hspace{0.01\textwidth}
\includegraphics[width=0.4\textwidth]{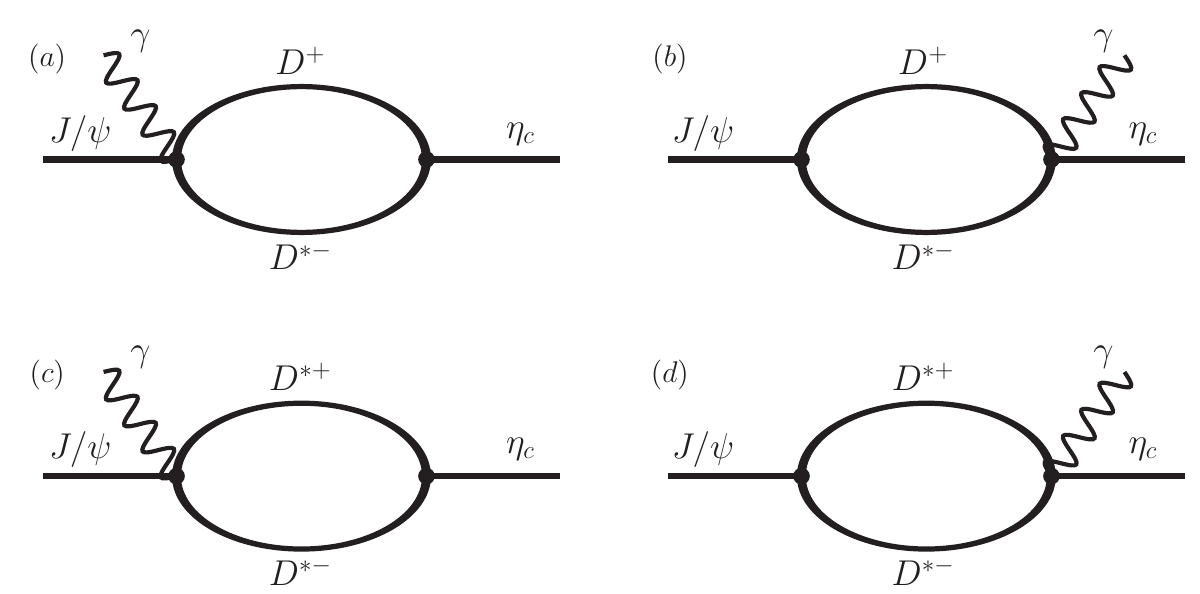}
\caption{Feynman diagrams for the $J/\psi\to\gamma\eta_c$ process in the hadronic loop mechanism. The left panel shows the triangle diagram contributions, while the right panel shows the contact terms. Similar diagrams also occur in $\psi^\prime\to\gamma\eta_c$ and $\gamma\eta_c^\prime$~\cite{Li:2011ssa}.}
\label{fig:Jpsi-etac}
\end{figure}

The concrete diagrams for $J/\psi\to \gamma \eta_c$ and $\psi^\prime\to \gamma \eta_c^{(\prime)}$ via hadronic loops are shown in Fig.~\ref{fig:Jpsi-etac}. As discussed in Ref.~\cite{Li:2011ssa}, the contact diagrams (right panel of Fig.~\ref{fig:Jpsi-etac}) do not contribute to the transition matrix elements in the effective Lagrangian approach. Therefore, it is sufficient to concentrate on the triangle diagrams (left panel of Fig.~\ref{fig:Jpsi-etac}).

The explicit transition amplitude for the first triangle diagram shown in the left panel of Fig.~\ref{fig:Jpsi-etac} can be written as
\begin{align}
{\cal M}_T^{(a)}=&(i^3)\int \frac{d^4p_2} {(2\pi)^4}[g_{\psi
DD}\varepsilon_\psi^\rho (p_{1\rho}-p_{3\rho})] [-g_{\eta_c
D^*D}(p_{f\sigma}+p_{3\sigma})][-eg_{D^*D\gamma}\varepsilon_{\alpha\beta\mu\nu}
p_\gamma^\alpha  \varepsilon_\gamma^{*\beta} p_2^\mu]\\
&\times\frac {i} {p_1^2-m_1^2} \frac {i(-g^{\nu\sigma}+p_2^\nu
p_2^\sigma/m_2^2)} {p_2^2-m_2^2} \frac {i} {p_3^2-m_3^2}{\cal
F}(m_2,p_2^2),
\end{align}
where ${\cal F}(m_2,p_2^2)$ is the dipole form factor. The amplitudes for the remaining triangle diagrams can be obtained analogously.

With the destructive phase $\delta=\pi$, the parameter $\alpha=0.98\pm 0.27$ \cite{Li:2011ssa} is fixed by combining the GI model result \cite{Barnes:2005pb} with the hadronic loop mechanism so as to reproduce the experimental partial width $\Gamma_{\rm exp}(J/\psi \to \gamma \eta_c)=(1.58\pm 0.37)\,\text{keV}$~\cite{ParticleDataGroup:2010dbb}. The same values of $\alpha$ and $\delta$ are subsequently used to predict the partial widths of $\psi^\prime$ and $\psi(3770)\to \gamma \eta_c$ and $\gamma \eta_c^\prime$, as shown in Table \ref{tab:etac_gamma}.

\begin{table}[htbp]
\centering
\caption{Radiative partial decay widths from different mechanisms. $\Gamma^{\mathrm{NR}}_{\mathrm{M1}}$ and $\Gamma^{\mathrm{GI}}_{\mathrm{M1}}$ denote the M1 transition widths obtained in the NR and GI quark models, respectively~\cite{Barnes:2005pb}. $\Gamma_{\mathrm{IML}}$ represents the contributions from the intermediate meson loop (IML) transitions shown in Fig.~\ref{fig:Jpsi-etac}, while $\Gamma_{\mathrm{all}}$ denotes the coherent results including both the M1 transition amplitude from GI model and the IML contributions. The experimental values are taken from the PDG~\cite{ParticleDataGroup:2010dbb} and the measurements reported by the BESIII~\cite{BES:2012uhz} and CLEO~\cite{CLEO:2008pln} Collaborations. The LQCD results are also listed for comparison~\cite{Dudek:2009kk}. The Table is adapted from Ref. \cite{Li:2011ssa}.}
\renewcommand\arraystretch{1.3}
\begin{tabular*}{1.0\textwidth}{l@{\extracolsep{\fill}}ccccc}
 \hline
 Initial meson                 & $J/\psi(1^3S_1)$   &\multicolumn{2}{c}      {$\psi^\prime(2^3S_1)$}       & \multicolumn{2}{c} {$\psi{''}(1^3 D_1$)}  \\
 Final meson                   & $\eta_c(1^1S_0)$   & $\eta_c'(2^1S_0)$   &  $\eta_c(1^1S_0)$         & $\eta_c'(2^1S_0)$   &  $\eta_c(1^1S_0)$\\
 \hline
 $\Gamma^{\text{NR}}_{\text{M1}}$ (keV)      & 2.9                & 0.21                      &  9.7                &$\cdots$ & $\cdots$      \\
 $\Gamma^{\text{GI}}_{\text{M1}}$ (keV)      & 2.4                & 0.17                      &  9.6                &$\cdots$&$\cdots$    \\
 $\Gamma_{\text{IML}}$     (keV)      & $0.08_{-0.06}^{+0.13}$   & $0.02_{-0.01}^{+0.02}$        & $2.78_{-1.96}^{+5.73}$        & $1.82_{-1.19}^{+1.95}$  & $17.14_{-12.03}^{+22.93}$\\
 $\Gamma_{\text{all}}$     (keV)      & $1.58^{+0.37}_{-0.37}$    & $0.08^{+0.03}_{-0.03}$             & $2.05^{+2.65}_{-1.75}$    & $1.82_{-1.19}^{+1.95}$  & $17.14_{-12.03}^{+22.93}$     \\
 $\Gamma_{\text{exp}}$     (keV)      & $1.58\pm 0.37$~\cite{ParticleDataGroup:2010dbb}     & $0.143\pm 0.027\pm 0.092$~\cite{BES:2012uhz}                      & $0.97\pm 0.14$~\cite{CLEO:2008pln}      &$\cdots$&$\cdots$     \\
 $\Gamma_{\text{LQCD}}$ (keV)      & $2.51\pm 0.08$     & $\cdots$                     & $0.4\pm 0.8$      &$\cdots$& $10\pm 11$     \\
 \hline
\end{tabular*}
\label{tab:etac_gamma}
\end{table}

As shown in Table \ref{tab:etac_gamma}, the inclusion of unquenched hadronic loop contributions, which destructively interfere with the M1 amplitude, effectively suppresses the M1 transition strength. Consequently, the predicted partial widths for $\psi^\prime \to \gamma \eta_c\,(\gamma \eta_c^\prime)$ are consistent with the experimental measurements. Moreover, the hadronic loop effects in $\psi^\prime \to \gamma \eta_c\ (\gamma \eta_c^\prime)$ are much more significant than those in $J/\psi \to \gamma \eta_c$ and become comparable to the M1 contributions. This behavior highlights the necessity of the unquenched quark model scenario, especially for transitions occurring near open-charm thresholds.

Furthermore, \changelabel{ the LQCD calculations of charmonium radiative transitions in the quenched approximation, listed in Table~\ref{tab:etac_gamma}, provide a valuable benchmark. For the $M1$ transitions $J/\psi\to\gamma\eta_c$, the quenched LQCD prediction is close to the tranisition widths obtained from the quark model \cite{Barnes:2005pb}, and sightly larger than the experimental value \cite{ParticleDataGroup:2010dbb}, indicating that unquenched effects in this channel are likely small. For both $\psi(3686)\to\gamma\eta_c$ and $\psi(3770)\to\gamma\eta_c$, the quenched LQCD results suffer from relatively large uncertainties, which hinders a definitive assessment of the role played by unquenched contributions. Notably, the quenched LQCD result for $\psi(3686)\to\gamma\eta_c$ is significantly smaller than the quark model prediction~\cite{Barnes:2005pb}, yet remains consistent with the experimental measurement within errors~\cite{BES:2012uhz}. In addition, the quenched LQCD estimate for $\psi(3770)\to\gamma\eta_c$ is found to be of the same order of magnitude as the hadronic loop contribution, suggesting that the interference between quenched and unquenched amplitudes may be important in this channel. This interplay is closely connected to the non-$D\bar{D}$ decay modes of the $\psi(3770)$.}

\begin{center}
\begin{figure}[htb]
\begin{tabular}{cccc}
Quark level&Hadron level\\
\raisebox{5ex}{\scalebox{0.5}{\includegraphics{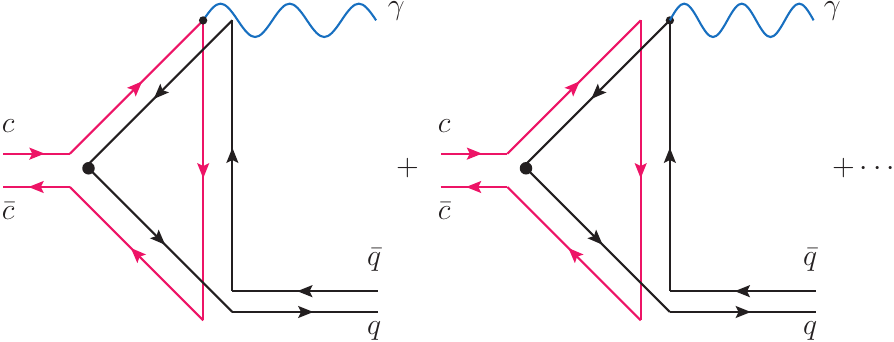}}}&
\scalebox{0.5}{\includegraphics{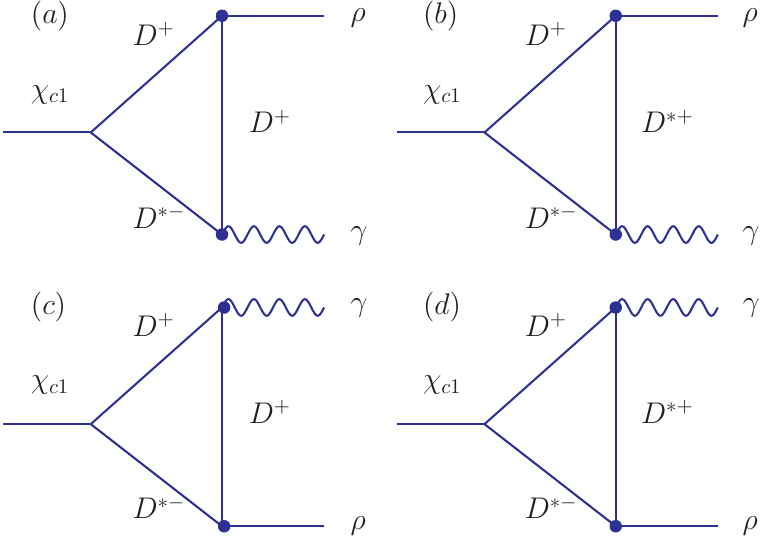}}\\
\end{tabular}
\caption{The left panel shows the typical quark level diagrams that describe the hadronic loop contributions to the processes $\chi_{c1}\to \gamma V$, while the right panel illustrates the corresponding hadronic level schematic diagram for $\chi_{c1}\to \gamma\rho^0$. The red and black lines represent the charm and light quarks, respectively. The photon can be emitted either from the charm-quark line or from the light-quark line. By applying the charge-conjugation transformation $D^{(*)+}\rightleftharpoons D^{(*)0}$ and $D^{(*)-}\rightleftharpoons \bar D^{(*)0}$, the rest two diagrams for $\chi_{c1}\to D\bar{D}^*+\text{h.c.}\to\gamma\rho^0$ can be obtained from diagrams (a) and (d). Figure adapted from Ref. \cite{Chen:2010re}.} 
\label{Fig:gpept-1}
\end{figure}
\end{center}

\begin{figure}
\centering
\includegraphics[width=0.6\textwidth]{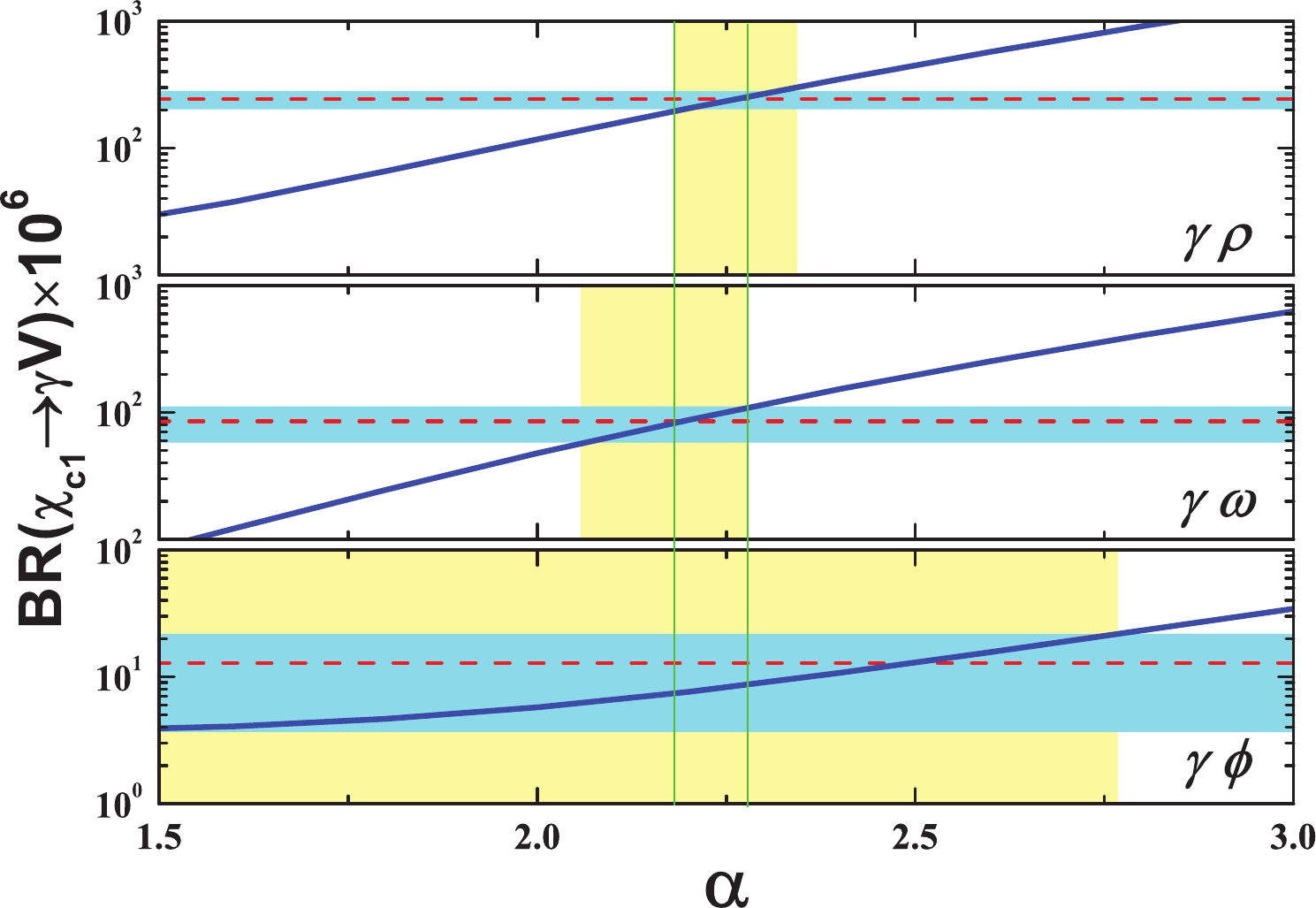}
\caption{The branching ratios of $\chi_{c1} \to \gamma \rho^0,\, \gamma \omega,\, \gamma \phi$ as functions of the parameter $\alpha$.
The red dashed lines with blue bands represent the experimental measurements~\cite{CLEO:2008sah}, while the blue solid curves denote the theoretical predictions incorporating both the hadronic loop contribution and the pQCD calculation~\cite{Gao:2006bc}. The vertical yellow bands indicate the regions where the theoretical results overlap with the corresponding experimental data. Within the same $\alpha$ interval marked by the two green vertical lines, the obtained branching ratios are consistent with the measurements.}
\label{fig:radiative}
\end{figure}

For the radiative decays $\chi_{c1}\to \gamma V$, the typical quark-level and hadron-level Feynman diagrams within the hadronic loop mechanism are illustrated in Fig.~\ref{Fig:gpept-1}. The transition element can be written as \cite{Chen:2010re}
\begin{align}
\mathcal{M}[\chi_{c1} \to \gamma V] = \sum_i \langle \gamma V | \mathcal{H}^{(2)} | i \rangle \langle i | \mathcal{H}^{(1)} | \chi_{c1} \rangle,
\end{align}
in which $\mathcal{H}^{(1)}$ denotes the interaction between $\chi_{c1}$ and $D\bar D^*+\text{h.c.}$, while $\mathcal{H}^{(2)}$ describes the transition $D\bar D^*+\text{h.c.}\to\gamma V$ mediated by the exchange of an appropriate charmed meson. In Fig.~\ref{fig:radiative}, the $\alpha$–dependence of the branching ratios for $\chi_{c1} \to \gamma \rho^{0},\, \gamma \omega,\, \gamma \phi$ are presented. For comparison with the experimental measurements~\cite{CLEO:2008sah}, the theoretical predictions shown here include both the hadronic loop contributions obtained in Ref. \cite{Chen:2010re} and the pQCD estimation calculated in Ref.~\cite{Gao:2006bc}.  As shown in Fig.~\ref{fig:radiative}, an overlap exists between the numerical results obtained in Ref. \cite{Liu:2009vv} and the experimental data reported by CLEO. The corresponding $\alpha$ ranges for $\chi_{c1} \to \gamma \rho^0,\, \gamma \omega,\, \gamma \phi$ are $2.18 < \alpha < 2.35$, $2.06 < \alpha < 2.28$, and $1.16 < \alpha < 2.77$, respectively, all of which lie within a reasonable parameter space. Notably, a common $\alpha$ range of $2.18 < \alpha < 2.28$ is found for all three radiative decay channels. Thus, the hadronic loop mechanism can serve as the underlying source that reduces the discrepancy between the pQCD calculations and the experimental measurements for $\chi_{c1} \to \gamma V$.

\begin{figure}[htbp]
\centering
\includegraphics[width=0.40\textwidth]{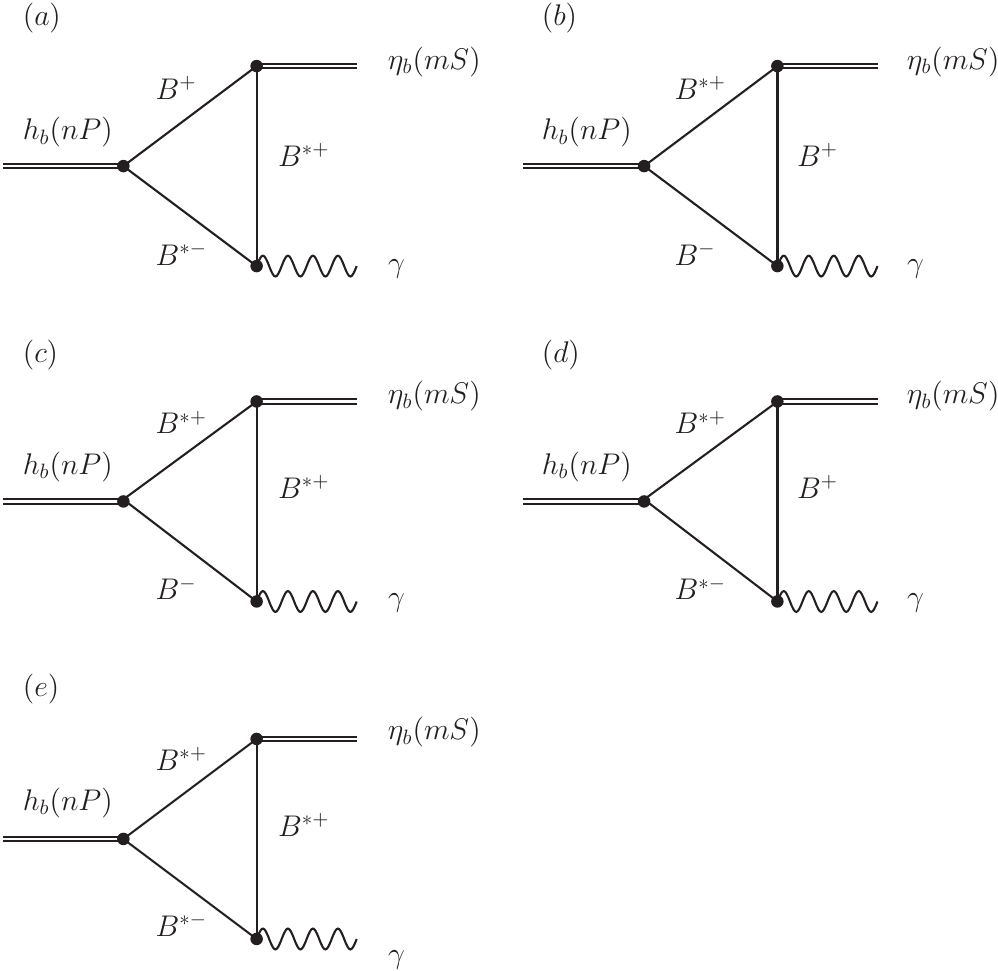}
\hspace{0.05\textwidth}
\vrule width 0.5pt height 60mm
\hspace{0.05\textwidth}
\raisebox{25mm}{
    \includegraphics[width=0.40\textwidth]{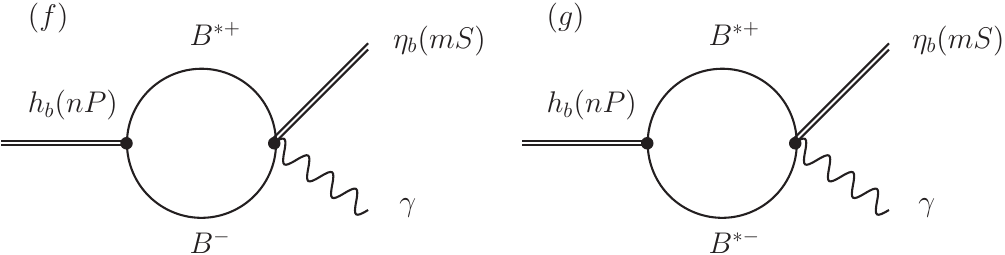}
}
\caption{Feynman diagrams for the hadronic loop mechanism in the radiative transitions $h_b(nP)\to\gamma\eta_b(mS)$ ($n,m=1,2$). The triangle-loop and contact-term contributions are shown in the left and right panels, respectively. Adapted from Ref.~\cite{Chen:2013cpa}.}
\label{fig:hb-etab}
\end{figure}

\begin{figure}
\centering
\includegraphics[width=0.32\textwidth]{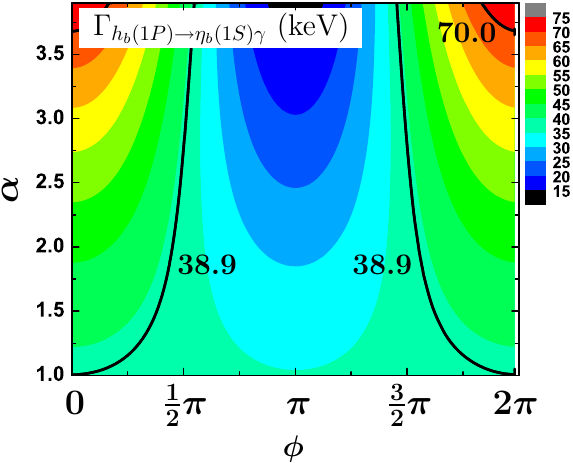}
\includegraphics[width=0.32\textwidth]{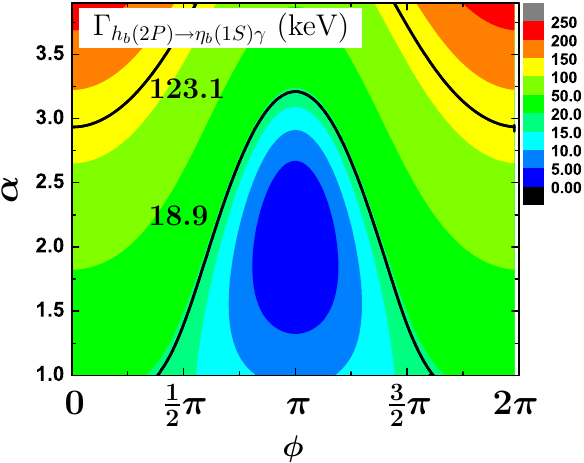}
\includegraphics[width=0.32\textwidth]{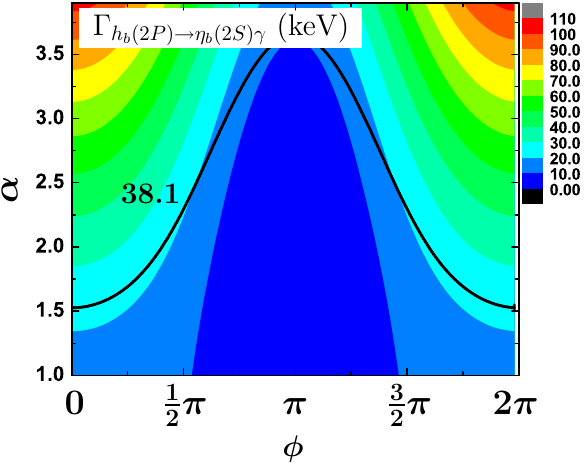}
\caption{Partial decay widths for $h_b(nP)\to\gamma\eta_b(mS)$: $h_b(1P)\to\gamma\eta_b(1S)$ (left panel), $h_b(2P)\to\gamma\eta_b(1S)$ (middle panel), and $h_b(2P)\to\gamma\eta_b(2S)$ (right panel), shown as functions of the parameters $\alpha$ and $\phi$. The black curves denote the experimental upper and lower bounds measured by the Belle Collaboration \cite{Belle:2012fkf}. Adapted from Ref.~\cite{Chen:2013cpa}.}
\label{fig:hbetabt}
\end{figure}

\changelabel{
Motivated by heavy-quark symmetry, the hadronic loop mechanism has been employed to address the discrepancies between experimental measurements \cite{Belle:2012fkf} and quenched quark-model predictions \cite{Godfrey:2002rp} for the radiative transitions \(h_b(nP)\to\gamma\eta_b(mS)\) (\(n,m=1,2\)) in the bottomonium sector. The corresponding hadronic-loop diagrams are depicted in Fig.~\ref{fig:hb-etab}. To compare the coherent decay widths, arising from the interference between the quenched quark-model and hadronic-loop amplitudes, with those inferred from the experimental branching fractions, a broad parameter space was explored, with $\alpha=1$--$4$ and $\phi=0$--$2\pi$, where \(\phi\) denotes the relative phase between the two amplitudes. The resulting decay widths are summarized in Table~\ref{tab:hb_etab_widths}. It was found that the hadronic-loop contributions can significantly enhance the radiative transition rates of \(h_b(nP)\to\gamma\eta_b(mS)\), thereby substantially reducing the discrepancies with experimental data. Moreover, common regions in the \((\alpha,\phi)\) parameter space were identified, in which all measured transitions \(h_b(nP)\to\eta_b(mS)\gamma\) can be simultaneously described (see Fig.~\ref{fig:hbetabt} for details), suggesting a common underlying dynamical mechanism. Future high-precision measurements will be valuable for further constraining the model parameters and testing the role of hadronic-loop effects.}

\begin{table}[htbp]
\centering
\caption{Partial decay widths for $h_b(nP)\to\gamma\eta_b(mS)$ within different mechanisms. $\Gamma_{\mathrm{QM}}$ denotes the results obtained from the quenched quark model \cite{Godfrey:2002rp}, while $\Gamma_{\mathrm{ML}}$ represents the hadronic-loop contributions. $\Gamma_{\mathrm{Tot}}$ is the coherent result including both contributions. The experimental ranges $\Gamma_{\mathrm{Exp}}$ are taken from Belle Collaboration \cite{Belle:2012fkf}. Table adapted from Ref. \cite{Chen:2013cpa}.}
\renewcommand\arraystretch{1.3}
\begin{tabular*}{1.0\textwidth}{l@{\extracolsep{\fill}}ccccc}
\hline
Initial states & Final states & $\Gamma_{\mathrm{QM}}$ [keV] & $\Gamma_{\mathrm{ML}}$ [keV] & $\Gamma_{\mathrm{Tot}}$ [keV] & $\Gamma_{\mathrm{Exp}}$ [keV] \\
\hline
$h_b(1P)$ & $\eta_b(1S)$ & 37.0 & $0.1$--$6.4$ & $12.6$--$74.2$ & $38.9$--$70.0$ \\
\multirow{2}{*}{$h_b(2P)$} & $\eta_b(1S)$ & 15.4 & $0.9$--$123.3$ & $0$--$203.5$ & $18.9$--$123.1$ \\
{} & $\eta_b(2S)$ & 10.0 & $0.5$--$48.8$ & $0$--$119.0$ & $38.1$--$267.9$ \\
\hline
\end{tabular*}
\label{tab:hb_etab_widths}
\end{table}

In summary, the long-distance hadronic loop mechanism provides a natural explanation for processes that violate the helicity selection rule, such as $\chi_{c1}\to VV$, and reconciles the significant discrepancies between quenched theoretical predictions and experimental measurements in several radiative transitions, including $J/\psi\,(\psi^\prime)\to\gamma\eta_c\,(\gamma\eta_c^\prime)$, $\chi_{c1}\to \gamma V$, and $h_b(nP)\to\gamma\eta_b(mS)$. Experimental searches for promising channels, such as $\chi_{c2}\to K^{*}\bar{K}+\text{c.c.}$ and $\psi(3770)\to \gamma\eta_c\,(\gamma\eta_c^\prime)$, would be particularly valuable for confirming the presence of this mechanism.

\changelabel{\subsection{The nonrelativistic effective field theory and power counting for loop integrals}

The above discussion of the hadronic loop mechanism is based on the ELA, and a form factor with a model parameter is introduced in the amplitudes in order to make the loop integral finite in the ultraviolet region, which makes ELA actually model dependent on the effective Lagrangian adopted, model parameter selected, coupling constant determination method, and the kind of form factor used. For example, it had been indicated in Ref.~\cite{Cheng:2004ru} that, if a monopole-type form factor $\mathcal{F}(q^2) = \frac{\Lambda^2-m^2}{\Lambda^2-q^2}$ is adopted with cut-off parameterized as $\Lambda=m+\alpha\Lambda_{\mathrm{QCD}}$, the parameter $\alpha$ should be of order unity and its concrete value can be determined by comparing the experimental data with the corresponding estimations. In fact, for transitions between heavy quarkonia described by Eq.~\eqref{eq:hadronicloop}, when the mass of the initial state lies close to the corresponding threshold of the intermediate heavy-meson pair, the intermediate mesons will propagate with small velocities and can therefore be treated nonrelativistically. This allows one to formulate a model-independent nonrelativistic effective field theory (NREFT), in which loop contributions can be systematically organized through a power counting scheme~\cite{Guo:2017jvc,Guo:2009wr,Guo:2010ak}.

In any effective field theory, the key ingredient is a small dimensionless quantity that serves as the basis for an ordered expansion. This quantity not only dictates the power counting of various contributions, but also enables an estimate of the theoretical uncertainty incurred by truncating the expansion at a given order. In the case of heavy quarkonium transitions near open-flavor thresholds, the intermediate heavy mesons are typically nonrelativistic, as the mass difference satisfies 
\begin{equation}
m_{Q\bar Q}-2m_{Q\bar q}\ll m_{Q\bar q},
\end{equation}
where $m_{Q\bar{Q}}$ and $m_{Q\bar{q}}$ denote the masses of the heavy quarkonium and the open-flavor heavy meson, respectively. Consequently, the velocity of the intermediate heavy meson satisfies
\begin{equation}
v \sim \sqrt{\frac{\left| m_{Q\bar{Q}} - 2 m_{Q\bar{q}} \right|}{m_{Q\bar{q}}}} \ll 1,
\end{equation}
and thus can serve as the desired small expansion quantity. Thus, for loop diagram depicted in Eq.~\eqref{eq:hadronicloop}, the power is counted in the nonrelativistic velocity $v$ as following: The three-momentum of the intermediate nonrelativistic particles counts as $|\bm{q}|\sim\mathcal O(v)$, while the nonrelativistic energy counts as $E\sim\mathcal O(v^2)$, thus the loop integration counts as
\begin{equation}
\int d^4 q \sim \mathcal O(v^5),
\end{equation}
whereas each nonrelativistic heavy-meson propagator behaves as
\begin{equation}
\frac{i}{E-\vec{q}^2/(2m_{Q\bar{q}})+i\epsilon}
\sim \mathcal O(v^{-2}),
\end{equation}
therefore, a scalar triangle diagram containing three nonrelativistic propagators scales as
\begin{equation}
\mathcal{M}_{\rm loop}
\sim
\mathcal O\left(v^5(v^{-2})^3\right)
\sim \mathcal O(v^{-1}),
\label{eq:scalar_loop_counting}
\end{equation}
this power-counting estimate is consistent with explicit calculations of the loop integrals, as demonstrated in Refs.~\cite{Guo:2017jvc,Guo:2009wr,Guo:2010ak}.

In addition to the scaling of the loop integral as discussed above, the momentum dependence of the interaction vertices should also be taken into account in order to determine the overall scaling behavior of a specific transition amplitude. Such momentum-dependent couplings may introduce additional factors that related with the external momentum $p$. In general, the external momentum $p$ can not be assigned with a specific scaling law in powers of the velocity $v$. Thus, it is kept explicitly in the power-counting analysis. The power counting of the NREFT provides a systematic way to control 
theoretical uncertainties.

An illustrative example for the above power counting rule is provided by the decay processes $\psi^\prime\to J/\psi \pi^0$ and $\psi^\prime\to J/\psi \eta$,
which probe isospin and SU(3) flavor symmetry breaking,
respectively. These processes were once regarded as ideal channels for extracting the light-quark mass ratio $m_u/m_d$, since the dominant source of symmetry breaking is provided by the light-quark mass differences \cite{Ioffe:1979rv,Ioffe:1980mx,Donoghue:1989sj,Meissner:1993ah,Leutwyler:1996eq}. However, a significant discrepancy was found between the value of $m_u/m_d$ extracted from these decays and that obtained from the light pseudoscalar meson masses \cite{Weinberg:1977hb,Leutwyler:1996sa,Leutwyler:1996qg}. This puzzle was addressed in Ref.~\cite{Guo:2009wr}, where it was shown that, according to the power counting rule, the intermediate charmed-meson loop amplitude is enhanced relative to the tree-level one by a factor of $1/v\approx 2$. As a consequence, these loop contributions can even dominate over the direct transition amplitudes, which ultimately undermines the reliability of the $\psi^\prime\to J/\psi \pi^0 (\eta)$ decays for a precise extraction of the light-quark mass ratio. Other applications of the NREFT can be found in Refs. \cite{Guo:2010zk,Guo:2010ca,Guo:2011dv,Guo:2012tg,Mehen:2011tp,Guo:2014qra,Cleven:2011gp,Cleven:2013sq,Guo:2013zbw,Esposito:2014hsa,Mehen:2015efa,Abreu:2016xlr,Huo:2015uka,Wu:2016dws,Chen:2016mjn,Guo:2016yxl}.

A comparison between the ELA and NREFT has been carried out in several
charmonium transitions~\cite{Guo:2010ak}. It was found that the
two approaches yield qualitatively consistent predictions when the transitions are governed by long-range meson-loop dynamics. Noticeable deviations typically emerge only when the ELA results become sensitive to the particular choice of form factors or when short-distance mechanisms contribute significantly. It should be emphasized that the validity of this NREFT framework is restricted to transitions where the mass difference between the initial and final charmonium states is much smaller than the typical hadronic scale $\Lambda_\chi \approx 1$ GeV, i.e., $|\Delta E| \lesssim 600$ MeV. With this qualification, the observed level of agreement between the two approaches provides further support for the power counting of the NREFT, and at the same time, offers additional confidence for applying the ELA to reactions where the NREFT is no longer applicable.}

\subsection{Other applications of hadronic loop mechanism}

Based on the preceding analysis, the introduction of a hadronic loop composed of charm mesons can effectively link the initial charmonium state with the final-state particles, providing an explanation for the anomalous charmonium decay phenomena observed experimentally. This mechanism clearly reflects the unquenched effect, whereby the initial charmonium couples to a charm meson pair following the OZI rule—a typical manifestation of coupled-channel dynamics.

In heavy-flavor physics, the decays of $B$ mesons have drawn considerable interest. While the pQCD factorization framework offers a valuable approach, it remains applicable only to certain typical processes and exhibits notable limitations~\cite{Colangelo:1989gi,Beneke:1999br,Beneke:2000ry,Cheng:2000kt,Keum:2000wi,Keum:2000ph,Lu:2000em,Beneke:2001ev,Song:2002mh,Beneke:2003zv}. Consequently, achieving quantitative descriptions of a broader range of $B$-meson decay modes continues to pose a significant challenge.

To address such challenges, alternative non-perturbative mechanisms such as final-state interactions (FSI) have been introduced. As discussed in Ref.~\cite{Colangelo:2002mj,Colangelo:2003sa,Cheng:2004ru}, the FSI mechanism can be employed to study $B$-meson decays. In this approach, one first identifies experimentally well-measured main decay channels of the $B$ meson to constrain the weak interaction vertex. The resulting on-shell intermediate mesonic states then rescatter via strong interactions into the observed final state. It is worth noting that, while both the FSI and the hadronic loop mechanisms utilize hadronic loops to connect initial and final configurations, their physical origins differ: the hadronic loop mechanism emerges from intrinsic coupled-channel effects, whereas FSI describes post-weak-decay rescattering.

In recent years, the FSI framework has found broad application—for example, in exploring additional sources of $CP$ violation in $B_c$ meson decays~\cite{Liu:2007qs}, investigating $D$ meson decays and related $CP$ puzzles~\cite{Fajfer:2003ag,Yu:2021euw,Hsiao:2019ait,Ling:2021qzl,Bediaga:2022sxw,Pich:2023kim,Geng:2024uxp,Wang:2025mdn}, analyzing $\Lambda_c$ decays~\cite{Yu:2020vlt,Jia:2024pyb,He:2024unv}, studying dominant decay processes and $CP$ violation in $\Lambda_b$ decays~\cite{Wang:2024oyi,Duan:2024zjv,Hsiao:2024szt}, and assisting in the identification of the double-charm baryon $\Xi_{cc}(3620)$~\cite{Yu:2017zst,Li:2020qrh,Han:2021azw,Hu:2024uia,Geng:2025gtd}.

Notably, FSI has also been applied to charmonium(-like) production in $B$-meson decays (see Refs.~\cite{Colangelo:2002mj,Colangelo:2003sa,Xu:2016kbn,Duan:2021bna,Yuan:2025pnt}), including processes such as $B \to K + \chi_{cJ}$. In these cases, the initial $B$ meson decays via the weak interaction into a charmed meson pair, which then transforms into a final-state charmonium and a kaon through charm meson exchange. This transition is facilitated by the coupled-channel mechanism linking the charmonium to charmed mesons. It is important to emphasize that, within the conventional pQCD factorization approach, decays of the type $B \to \text{charmonium} + K$ are forbidden; however, the inclusion of hadronic loop or FSI mechanisms renders such processes viable.

Additionally, the hadronic loop mechanism has also been applied in the past two decades to study the strong decays of the observed charmonium-like $XYZ$ states \cite{Liu:2010um,Chen:2010nv,Guo:2010zk,Guo:2010ak,Chen:2011kc,Wang:2011yh,Wang:2012wj,Chen:2012nva,Guo:2012tj,Li:2013jma,Li:2013xia,Li:2013yla,Chen:2013yxa,Chen:2014sra,Dong:2013kta,Li:2014gxa,Chen:2015bma,Wang:2015xsa,Guo:2016iej,Qian:2021neg,Qian:2021gby,Wang:2022jxj,Liu:2023sdw,Qian:2023taw,Peng:2024xui,Gao:2024qth,Zheng:2024eia,Bai:2024lps,Qi:2025xkz,Qi:2025hwd}, shedding light on their properties.

\section{Low-mass puzzle of $X(3872)$ and the construction of $2P$ charmonia}
\changelabel{
\subsection{The unquenched quark model in early stage}

\begin{figure}
\centering
\begin{tabular*}{\textwidth}{@{\extracolsep{\fill}}cc}
\includegraphics[height=0.275\textheight]{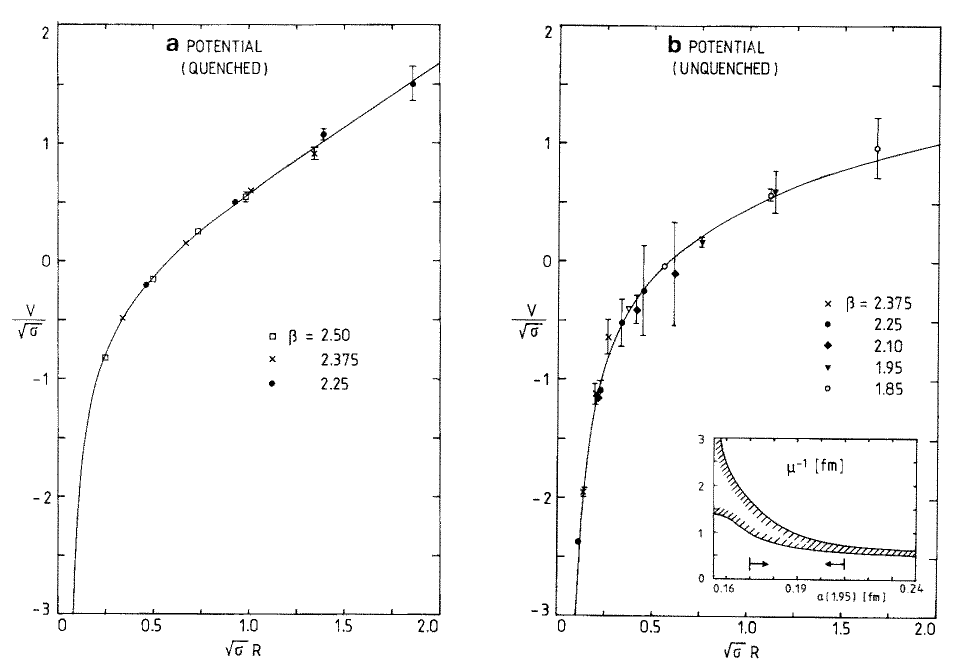}&
\includegraphics[height=0.275\textheight]{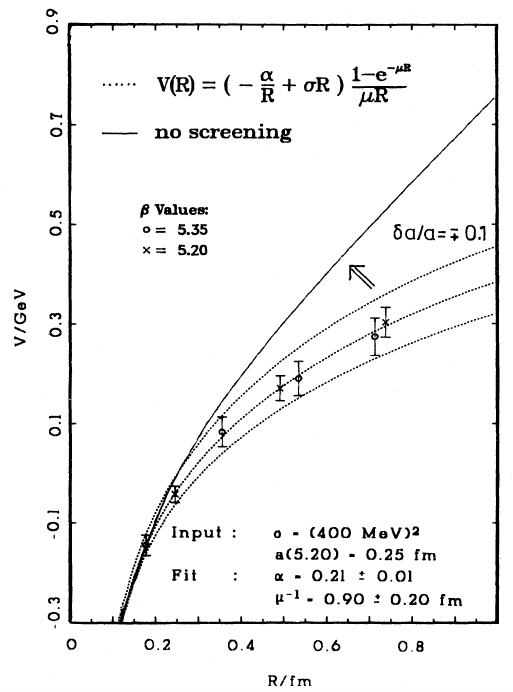}\\
(a)&(b)
\end{tabular*}
\caption{The comparison of the quenched and unquenched potentials (a)~\cite{Laermann:1986pu}, and the screened potential (b)~\cite{Born:1989iv}.}
\label{fig:screened_potential}
\end{figure}

In a long peroid, the quenched quark model was treated as an effective approach to depict the hadron spectroscopy. In 1970s, based on the observation of a series of charmonia, Eichten {\it et al.} established the early potential model, i.e., Cornell potential \cite{Eichten:1974af, Eichten:1978tg, Eichten:1979ms} that contains the Coulomb potential and a linear potential used to phenomenologically describe the color confinement as
\begin{equation}
V(r) = -\frac{\kappa}{r} + \frac{r}{a^2}.
\end{equation}
It turns out that the Cornell potential could be well applied to the spectra of the charmonia in this era, which matched the early lattice QCD calculations~\cite{Otto:1984qr,Barkai:1984ca,Barkai:1984pz}. However, this potential only contains central part, which could not describe the spin splits. Thus, the Cornell potential model was gradually refined in subsequent work, where Isgur and Karl introduced hyperfine interactions to the spectroscopy of the baryons~\cite{Isgur:1977ef,Isgur:1978wd,Isgur:1978xj}, while Godfrey and Isgur~\cite{Godfrey:1985xj} employed relativistic correction to potential model to cover meson spectra from lightest $\pi$ to heaviest bottominum, i.e., the famous GI model, and Capstick and Isgur also successfully employed GI model to baryon systems~\cite{Capstick:1986ter}. Both historically and in the present day, the potential model represents an unequivocal success, accounting for most observed mesons and baryons.

Despite the great successes of the quenched potential models in explaining the experimentally observed mesons and baryons at that time, stimulation from the following lattice QCD~\cite{Laermann:1986pu,Born:1989iv} pointed out that once considering into the dynamic behavior as the creation and annihilation of virtual quark-antiquark pairs, at long distance the potential would be depressed as presented in Fig.~\ref{fig:screened_potential}. Such behavior can be stimulated by a potential formed as~\cite{Laermann:1986pu,Born:1989iv}
\begin{eqnarray}
    V(r)=\left(-\frac{\alpha}{r}+\sigma r\right) \frac{1-e^{-\mu r}}{\mu r},
\end{eqnarray}
with $\mu^{-1}$ defined as the screen length~\cite{Laermann:1986pu,Born:1989iv}. Thus, the linear confinement in Cornell potential model might be correlated to
\begin{equation}
r\to \frac{1-e^{-\mu r}}{\mu},
\end{equation}
which is vividly called as screened confinement. Ding, Chao, and Qin employed this form to the calculations of $c\bar{c}$ and $b\bar{b}$, the results implied that the masses of highly excited states are obviously depressed, while low-lying states could still locate around the original positions, indicating that predictions from quenched potential models on the highly excited states might possibly be questionable if screened correction was not introduced~\cite{Ding:1993uy,Ding:1995he}. 

\begin{figure}
\centering
\includegraphics[width=\textwidth]{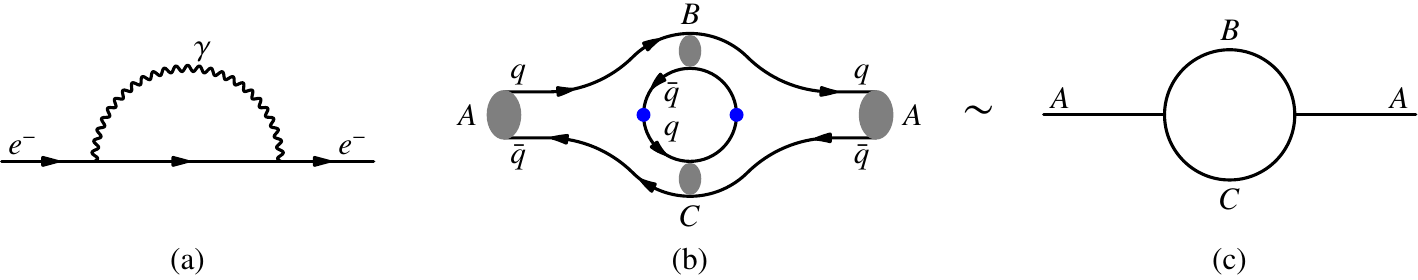}
\caption{Loop diagrams: (a) electron-photon, (b) quark-level hadron, (c) hadron-level hadron.}
\label{fig:loop_electron_photon}
\end{figure}

Combining Refs.~\cite{Ding:1993uy,Ding:1995he,Laermann:1986pu,Born:1989iv}, we may conclude that long distance effect will play more important roles in shifting the spectra when excitation becomes higher. Actually, this one-time creation and annihilation of the virtual quark-antiquark pair, combined with the valence quarks in hadrons, can produce a quark loop diagram like Fig.~\ref{fig:loop_electron_photon}~(b), which can be viewed in hadron scale as hadronic loop mechanism as presented in Fig.~\ref{fig:loop_electron_photon}~(c)\footnote{In Fig.~\ref{fig:loop_electron_photon}~(b), intermediate $q\bar{q}-q\bar{q}$ loop with color octet-octet configuration is also permitted, but according to Ref.~\cite{Tan:2024pqs}, color octet structure can be effectively replaced by a set of relatively complete color singlet bases. While for $qq-\bar{q}\bar{q}$ configuration, Fierz transformation can transit it into $q\bar{q}-q\bar{q}$ structure~\cite{Hu:2022zdh}. Thus, it is already complete to consider only a set of hadronic loop diagrams as Fig.~\ref{fig:loop_electron_photon}~(c), although in practical calculation only a few intermediate channels will dominate due to thresholds. Moreover, when the constituents of $A$ in Fig.~\ref{fig:loop_electron_photon}~(b) are heavy quarks like $c\bar{c}$, additional consideration of the OZI rule will suppress the contributions from $c\bar{c}-q\bar{q}$ loops.}. Thus, it indicates that an introduction of hadronic loop diagram will modify the masses of hadrons. Such phenomena are very analogous to the concept in quantum electrodynamics, where further inclusion of intermediate elctron-photon loop as Fig.~\ref{fig:loop_electron_photon}~(a) makes the electron physical. Thus, similarly, states obtained from the quenched quark model can be seen as ``bare" ones, further interactions with the continuum component through Fig.~\ref{fig:loop_electron_photon}~(b-c) then make them ``physical". In conclusion, the screened potential can be treated as a phenomenological description on this loop mechanism, whose equivalence is later proved by Chao {\it et al.}~\cite{Li:2009ad}, and similar results are also obtained in studying high radial excited $\chi_{cJ}$ states~\cite{Duan:2021alw}. Anyway, the unquenched quark model can provide a systematic way to incorporate these continuum effects, it can not only shift the spectra, but also give rise to another strong decay mechanism as introduced in the last section with states above the thresholds, which thus being valuable to require further investigations.

Actually, early in 1978, Eichten {\it et al.} had already built a coupled-channel framework in charmonia sector to describe the unquenched effect, where a instantaneous four fermion interaction was assumed~\cite{Eichten:1978tg}
\begin{eqnarray}
    H_I=\frac{1}{2} \sum_{a=1}^8 \int \rho_a(\mathbf{r}) V\left(\mathbf{r}-\mathbf{r}^{\prime}\right) \rho_a\left(\mathbf{r}^{\prime}\right) d^3 \mathbf{r} d^3 \mathbf{r}^{\prime}.
\end{eqnarray}
Here, $V$ is the static potential, $\rho_a(\mathbf{r})=\bar{\psi}(\mathbf{r}) \frac{1}{2} \lambda_a \psi(\mathbf{r})$ is the octet of color densities of the quark field $\psi$, which is claimed to contain both the static valence quarks and the dynamical creation of a light quark pair~\cite{Eichten:1978tg}. Then, the oscillatory unquenched effect could be achieved by the projection of a Lippmann-Schwinger-like equation, with a total Hamiltonian that contains the $c\bar{q}$ and $\bar{c}q$ components, onto the $c\bar{c}$ subspace~\cite{Eichten:1978tg}. While in the same year, Kinnunen and T\"ornqvist started to employ the loop diagram as shown in Fig.~\ref{fig:loop_electron_photon}~(b) to study the mass difference and mixing in meson systems~\cite{Kinnunen:1978qm}. Later in 1980, Van Beveren, Dullemond, and Rupp proposed a transition operator from $Q\bar{Q}$ to open-flavor meson pair, where light $q\bar{q}$ was created from vacuum. After substituting it into the coupled-channel equation, the spectra in addition with the strong decays of $c\bar{c}$ and $b\bar{b}$~\cite{vanBeveren:1979bd}, and such method was also adopted later in low-lying scalar meson nonet~\cite{vanBeveren:1986ea}. Also, T\"ornqvist {\it et al.} developed the unitary quark model that naturally contains the coupled-channel effects, with which corrections to the spectra of mesons~\cite{Tornqvist:1979hx,Ono:1983rd,Heikkila:1983wd,Ono:1985eu} and baryons~\cite{Tornqvist:1984fy,Tornqvist:1985fi,Zenczykowski:1985uh} were calculated. Importantly, such this coupled-channel equation could interpret the masses of the light flavor scalar nonet~\cite{vanBeveren:1986ea,Tornqvist:1995kr,Tornqvist:1982yv}. Thus, in the review ``Quarkonium and Quark Loops"~\cite{Tornqvist:1984xz}, T\"ornqvist systematically discussed how non-perturbative quark loops can be included phenomenologically in quarkonium calculations via unitarization. In light hadron sector, the unquenched effect was employed by Silvestre-Brac and Gignoux~\cite{Silvestre-Brac:1991qqx} to the light flavor $N$, $\Delta$, $\Lambda$, and $\Sigma$ baryons, the calculation implied that coupled channel effects could play significant roles in understanding $\Lambda(1405)$. Geiger and Isgur also~\cite{Geiger:1989yc,Isgur:1998kr} pointed out that in most situations, the mass shift from coupled-channel effects could be absorbed into the parameters in potential model, but if a state is nearby an $S$-wave threshold, the mass shift exists a cusp like behavior nearby the threshold, which could not be absorbed into the parameters in potential model. That is why there exists significant coupled-channel effects in light flavor scalar states like $f_0(980)$ and $a_0(980)$, and light flavor baryon $\Lambda(1405)$.

In addition to mass spectra corrections, the coupled-channel effect was also introduced into the decay processes. As we known, although the di-pion emissions between heavy-quarkonium systems are OZI-suppressed processes, as we have metioned in the introduction, one approach to achieve this is the QCD multipole expansion~\cite{Kuang:1989ub}. Thus, such decay operator then can be additionally contained into the total Hamiltonian that already contained the unquenched effect, after expanding the Lippmann-Schwinger-like equation~\cite{Eichten:1978tg} with this Hamiltonian, the di-pion transition will also be affected by the coupled channel effects. Such procedure was discussed by Zhou and Kuang~\cite{Zhou:1990ik}, and it turned out that low-$m_{\pi\pi}$ distribution could really be enhanced by this approach when comparing with the pure QCD multipole expansion.

}

\changelabel{Therefore, it can be seen that in the early stage, the unquenched quark model had already been developed a lot. However, in our view, owing to the limitations of that time, its further development unavoidably hit a plateau. In the light hadron sector, although the unquenched effect can be used to interpret many phenomena, due to the too complex non-perturbative effect in this region, it is too hard to properly use a relatively simple and in development unquenched model to give definite conclusions. While in the heavy quark region, lack of lattice QCD calculations in accompany with the most importantly absence of experimental measurements make the discussions also difficult. The main reason is that the unquenched effect is suppressed within low-lying states, making the quenched approximation sufficiently enough at that time. Thus, a further development on unquenched effect is not so urgent. Even if assuming it had been done, missing of scaling points would also make the developed models unpredictable when extrapolating to higher excited energies. Therefore, more experimental data is thirsty for further studies, especially the measurements performed in the heavy quark sector, since as the transition zone between perturbative and non-perturbative regions, it provides the most ideal environment for refinements.}

\subsection{Low-mass puzzle in new hadronic states}

Since 2003, with the accumulation of experimental data, an increasing number of new hadronic states have been reported. Thus, 2003 marks a renaissance in hadron spectroscopy. Interested readers may consult the review articles~\cite{Klempt:2009pi,Liu:2013waa,Cheng:2015iom,Chen:2016qju,Hosaka:2016pey,Richard:2016eis,Chen:2016spr,Lebed:2016hpi,Guo:2017jvc,Olsen:2017bmm,Yuan:2018inv,Liu:2019zoy,Brambilla:2019esw,Dong:2021juy,Cheng:2021qpd,Chen:2022asf,Meng:2022ozq,Gross:2022hyw,Liu:2024uxn,Wang:2025sic} for further details. These abundant observations have not only stimulated extensive discussions on exotic hadronic states, including multiquark configurations, but also provided opportunities to refine existing theoretical models and frameworks. Over the past two decades, we have witnessed these new advances.

Among these newly observed hadronic states, several typical states—such as $D_{s0}^*(2317)$, $D_{s1}^\prime(2460)$, $X(3872)$, and $\Lambda_c(2940)$—have attracted widespread attentions within the research community due to the emergence of a low-mass puzzle associated with their properties. In the following, we first introduce their experimental observations individually, and then summarize the nature of this low-mass puzzle:
\begin{itemize}

\item {$D_{s0}^*(2317)$ and $D_{s1}^\prime(2460)$:} In 2003, the BaBar Collaboration observed a new charmed-strange state, $D_{s0}^*(2317)$, in the $D_s^+\pi^0$ channel~\cite{BaBar:2003oey}. Its mass was measured to be $m=2316.8\pm 0.4~\text{MeV}$ for $D_s^+\to K^+K^-\pi^+$ and $m=2317.6\pm 1.3~\text{MeV}$ for $D_s^+\to K^+K^-\pi^+\pi^0$, and the width was measured as $\Gamma\lesssim 10~\text{MeV}$. Subsequent experiments, including those by the CLEO, Belle, BESIII, and BaBar collaborations, confirmed its existence~\cite{Belle:2003guh,BaBar:2006eep,CLEO:2003ggt,Belle:2003kup,BESIII:2017vdm,BaBar:2004yux}. If $D_{s0}^*(2317)$ is classified within the charmed-strange meson family, a significant discrepancy arises: its mass is approximately 180 MeV lower than the prediction of quenched models such as the GI model~\cite{Godfrey:1985xj}. This discrepancy provides the main motivation for introducing exotic multiquark interpretations \cite{Barnes:2003dj,Cheng:2003kg,Kolomeitsev:2003ac,Hayashigaki:2004st,Chen:2004dy,Dmitrasinovic:2005gc,Kim:2005gt,Terasaki:2006wm,Guo:2006fu,Guo:2006rp,Faessler:2007gv,Xie:2010zza}, \changelabel{in which, especially, the hadronic molecular picture has nowadays got support from lattice calculations~\cite{Guo:2023wkv}.}
    
Later, the CLEO Collaboration also reported a new charmed-strange state, $D_{s1}^\prime(2460)$, in the $D_s^{*+}\pi^0$ channel, with mass and width $m=2463.4\pm1.7({\rm stat})\pm1.0({\rm syst})~\text{MeV}$ and $\Gamma<7~\text{MeV}$ at 90\% C.L.~\cite{CLEO:2003ggt}. A similar low-mass phenomenon also exists for this state, as its mass is about 80 MeV below the prediction of the GI model~\cite{Godfrey:1985xj}, a representative quenched approach.

\item {$X(3872)$:} Here, we introduce the charmonium-like state $X(3872)$, a prominent figure among the newly observed hadronic states. The Belle Collaboration first observed $X(3872)$ in the $J/\psi \pi^+\pi^-$ channel of the decay $B\to K J/\psi \pi^+\pi^-$~\cite{Belle:2003nnu}. The measured mass and width are $m = 3872.0 \pm 0.6(\text{stat}) \pm 0.5(\text{syst})~\text{MeV}$ and $\Gamma < 2.3~\text{MeV}$ at 90\% C.L., respectively. A clear discrepancy exists between the mass of $X(3872)$ and the theoretical prediction for the charmonium state $\chi_{c1}(2P)$ from Ref.~\cite{Godfrey:1985xj}, i.e., the experimental mass is about 80 MeV below the calculation in Ref.~\cite{Godfrey:1985xj}, presenting another instance of the low-mass puzzle. \changelabel{Given that $X(3872)$ lies very close to the $D\bar{D}^*$ threshold, a hadronic molecular interpretation has been proposed to resolve this discrepancy~\cite{Close:2003sg,Voloshin:2003nt,Wong:2003xk,Swanson:2003tb,Tornqvist:2004qy,AlFiky:2005jd,Liu:2008fh,Thomas:2008ja,Liu:2009qhy,Lee:2009hy,Gamermann:2009uq,Braaten:2010mg,Wang:2013kva,Baru:2013rta}.}

\item $\Lambda_c(2940)$: Besides the mesonic states mentioned above, the low-mass puzzle also occurs in the baryonic sector, as exemplified by $\Lambda_c(2940)$. This state was first reported by the BaBar Collaboration in the $D^0p$ channel~\cite{BaBar:2006itc} and later confirmed by the Belle Collaboration via the $\Sigma_c(2455)\pi$ invariant mass spectrum~\cite{Belle:2006xni}. In 2017, the LHCb Collaboration further analyzed $\Lambda_c(2940)$ in the decay $\Lambda_b^0\to D^0p\pi^-$ and suggested its most likely spin-parity to be $3/2^-$~\cite{LHCb:2017jym}. If it is interpreted as a conventional charm baryon, the mass of $\Lambda_c(2940)$ is about 60–100 MeV lower than the one predicted by \changelabel{Capstick-Isgur model~\cite{Capstick:1986ter,Yu:2022ymb}, diquark model~\cite{Ebert:2007nw,Ebert:2011kk,Chen:2014nyo,Chen:2016iyi}, and Regge trajectory~\cite{Cheng:2017ove}.}

\end{itemize}

\begin{figure}
\centering
\includegraphics[width=\textwidth]{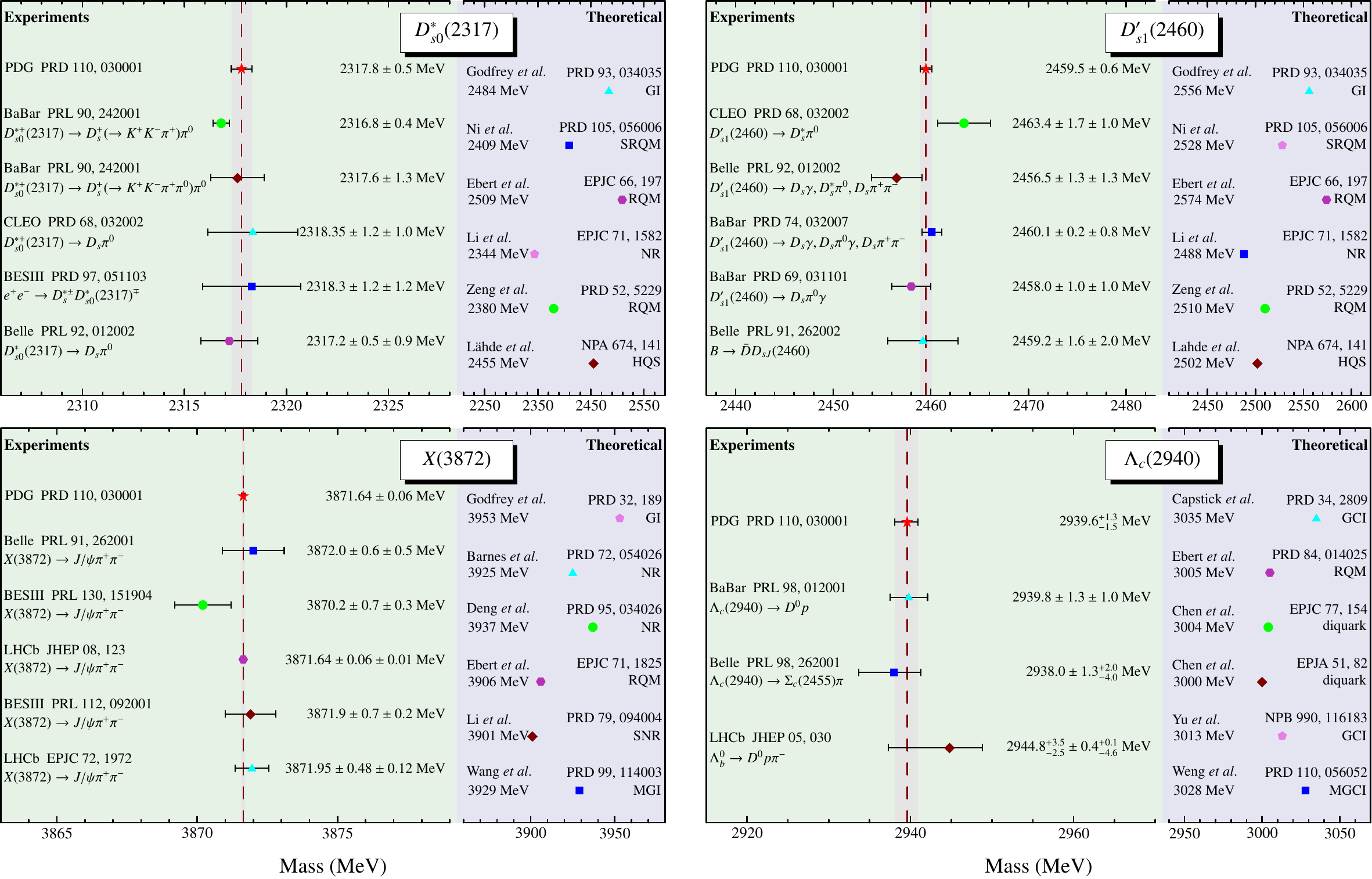}
\caption{The comparison between experimental and theoretical masses is shown in each subfigure. The experimental values are positioned on the left, and the theoretical values on the right. The experimental data are taken from Refs.~\cite{ParticleDataGroup:2024cfk,BaBar:2003oey,CLEO:2003ggt,BESIII:2017vdm,Belle:2003kup,CLEO:2003ggt,Belle:2003kup,BaBar:2006eep,BaBar:2003cdx,Belle:2003guh,Belle:2003nnu,BESIII:2022bse,LHCb:2020fvo,BESIII:2013fnz,LHCb:2011zzp,BaBar:2006itc,Belle:2006xni,LHCb:2017jym}, while the theoretical results are from Refs.~\cite{Godfrey:2015dva,Ni:2021pce,Ebert:2009ua,Li:2010vx,Zeng:1994vj,Lahde:1999ih,Godfrey:1985xj,Barnes:2005pb,Deng:2016stx,Ebert:2011jc,Li:2009zu,Wang:2019mhs,Capstick:1986ter,Ebert:2011kk,Chen:2016iyi,Chen:2014nyo,Yu:2022ymb,Weng:2024roa}. The GI, SRQM, RQM, NR, GCI, HQS, SNR, MGI, MGCI, and 'diquark' imply the theoretical methods of Godfrey-Isgur model, semi-relativistic quark model, relativistic quark model, non-relativistic quark model, Godfrey-Capstick-Isgur model, heavy quark symmetry, screened non-relativistic quark model, modified (screened) Godfrey-Isgur model, modeified (screened) Godfrey-Capstick-Isgur model, and diquark model, respectively.}
\label{fig:comparemass}
\end{figure}

\begin{figure}[htpb]
\centering
\includegraphics[width=0.75\textwidth]{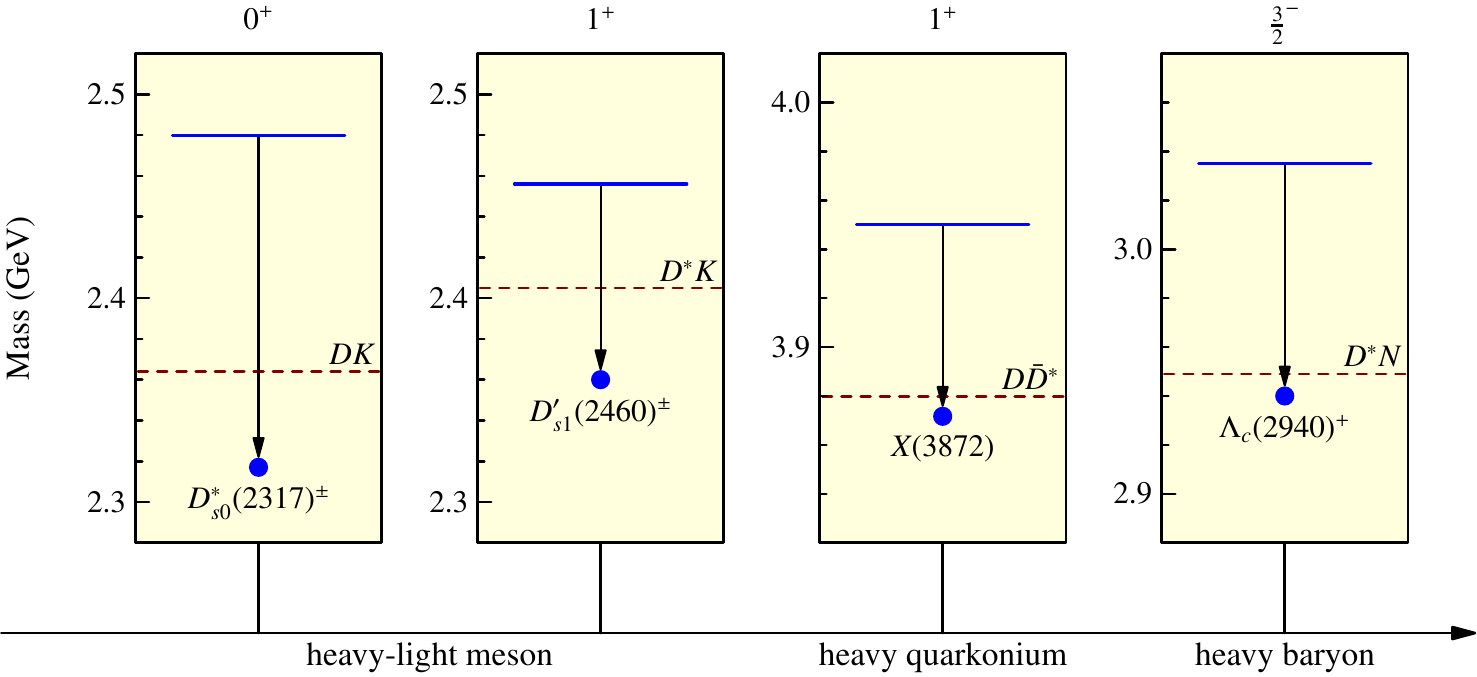}
\caption{The low-mass puzzle for selected hadrons. This figure is updated from Ref.~\cite{Luo:2019qkm}. The theoretical masses (shown as solid blue lines) are taken from Refs.~\cite{Godfrey:2015dva,Godfrey:1985xj,Capstick:1986ter}. The blue points represent experimental results from Ref.~\cite{ParticleDataGroup:2024cfk}. The dashed lines indicate the thresholds of the relevant decay channels.}
\label{fig:low_mass_hadrons}
\end{figure}

\changelabel{
In Figs.~\ref{fig:comparemass}-\ref{fig:low_mass_hadrons}, we compare the experimental masses with theoretical results of $D_{s0}^*(2317)$, $D_{s1}^\prime(2460)$, $X(3872)$, and $\Lambda_c(2940)$. On the one hand, as the discussions above, exotic states such as compact multiquark configurations or hadronic molecular structures are a kind of interpretation, which matches the properties of these states. On the other hand, in fact, this low-mass puzzle phenomenon extends across multiple systems, including heavy-light mesons, heavy quarkonia, and singly heavy-flavor baryons. Turning to the spectroscopy of light-flavor hadrons, we note that $\Lambda(1405)$, $a_0(980)$, and $f_0(980)$ may also exhibits a low-mass phenomenon~\cite{Silvestre-Brac:1991qqx,vanBeveren:1986ea,Tornqvist:1995kr,Tornqvist:1982yv}. While earlier work suggested that such loop effects are generally negligible in most systems~\cite{Geiger:1989yc,Isgur:1998kr}, but they become crucial for states close to relevant thresholds. In the past, limited experimental data restricted such analyses to only a few candidates. However, the discovery of $X(3872)$ and numerous other new hadronic states in recent years has provided renewed impetus to systematically study these unquenched dynamics, offering a clearer window into how open-channel effects reshape hadron spectroscopy.
}


In the study of $D_{s0}^*(2317)$, van Beveren and Rupp found that the $c\bar{s}$ configuration can strongly couple to the OZI-allowed $S$-wave $DK$ channel~\cite{vanBeveren:2003kd}. In Ref.~\cite{vanBeveren:2005ha}, light and heavy scalar mesons were connected by pole trajectories. Simonov and Tjon proposed that the mass of a $P$-wave $c\bar{s}$ state could be lowered via coupling to the $DK$ channel~\cite{Simonov:2004ar}. Ortega {\it et al.} used a coupled-channel framework to analyze the mixing between the $c\bar{s}(0^+)$ core and a $DK$ channel for $D_{s0}^*(2317)$, as well as $c\bar{s}(1^+)$-$D^*K$ for $D_{s1}^\prime(2460)$, showing that such core–meson-meson mixing can significantly alter the properties of these states~\cite{Ortega:2016mms}. Cheng and Yu calculated the self-energy modifications for charm-strange and bottom-strange systems within heavy meson chiral perturbation theory (HMChPT)~\cite{Cheng:2014bca,Cheng:2017oqh}. In Ref.~\cite{Dai:2003yg}, Dai {\it et al.} interpreted $D_{s0}^*(2317)$ and $D_{s1}^\prime(2460)$ with with sum rules in heavy quark effective theory. The results implied that $D_{s0}^*(2317)$ and $D_{s1}^\prime(2460)$ could be included in $j_\ell=\frac{1}{2}$ doublets with $0^+$ and $1^+$, respectively. In addition, the analysis of the decay modes of $D_{s0}^*(2317)$ in Ref.~\cite{Colangelo:2003vg} also implied that $D_{s0}^*(2317)$ may be a scalar $c\bar{s}$ $j_\ell=\frac{1}{2}$ state. Ref.~\cite{Coito:2011qn} systematically studied the axial-vector charmed mesons ($D_1(2420)$, $D_1(2430)$, $D_{s1}(2536)$, and $D_{s1}^\prime(2460)$) with resonance-spectrum-expansion model, where the mixing angle between $^3P_1$ and $^1P_1$ could be naturally obtained.

Alongside the $D_{s0}^*(2317)$ and $D_{s1}^\prime(2460)$ states, $X(3872)$ provides another prominent example of the critical role played by coupled-channel effects. In 2005, Kalashnikova applied the coupled-channel approach to charmonium and identified a mechanism that generates a structure at the $D\bar{D}^*$ threshold, crucial for interpreting $X(3872)$~\cite{Kalashnikova:2005ui}. Barnes and Swanson established general theorems for the unquenched quark model in charmonium~\cite{Barnes:2007xu}. They found that for low-lying $c\bar{c}$ states, the mass shifts induced by hadron loops are identical within a given $NL$ multiplet. Although these shifts are sizable, they can be absorbed into the parameters of potential models. A similar conclusion was reached in $b\bar{b}$ systems~\cite{Liu:2011yp}. The method has since been extended to bottomonium with more refined potential models, wave functions, and inclusion of higher excited states~\cite{Ni:2025gvx,Sultan:2025dfe,Lu:2016mbb,Lu:2017hma,Ferretti:2012zz,Ferretti:2013vua}. Coito {\it et al.} interpreted the $X(3872)$ with coupled-channel effect~\cite{Coito:2010if}, and the $X(3872)$ is not a pure hadronic molecular state~\cite{Coito:2012vf}, where the coupling with $c\bar{c}$ is necessary. Under this framework, Cardoso {\it et al.} calculated the electromagnetic decays of $X(3872)$~\cite{Cardoso:2014xda}. Ortega {\it et al.} performed a coupled-channel calculation for the $1^{++}$ $c\bar{c}$–$D\bar{D}^*$ system, successfully explaining the mass of $X(3872)$ and indicating a large $D\bar{D}^*$ component in its wave function~\cite{Ortega:2009hj,Ortega:2012rs}. This interpretation was supported by Ferretti, who also emphasized the important role of the $D\bar{D}^*$ continuum~\cite{Ferretti:2013faa}.

Pennington and Wilson developed subtracted dispersion relations to handle meson loops~\cite{Pennington:2007xr}, a technique later adopted in several works~\cite{Zhou:2011sp,Zhou:2013ada,Duan:2020tsx,Duan:2021alw,Man:2025zfu,Man:2025vmm}. Danilkin and Simonov demonstrated within a coupled-channel framework that $X(3872)$ emerges from the charmonium $2^3P_1$ state, whose pole is shifted to the $D\bar{D}^*$ threshold by strong coupling to open-charm channels, thereby accounting for its narrow width and precise threshold location~\cite{Danilkin:2009hr,Danilkin:2010cc}. Li {\it et al.} compared coupled-channel and screening effects, finding that while both give consistent results for some excited states, the coupled-channel effect is significantly stronger for $X(3872)$~\cite{Li:2009ad}. This conclusion was later confirmed for higher excited $P$-wave charmonium states~\cite{Duan:2021alw}. Wang {\it et al.} analyzed $X(3872)$ with coupled-channel dynamics, the results concluded that $X(3872)$ originates from the $D\bar{D}^*$ pole with over 99.7\% confidence~\cite{Wang:2024ytk}. However, some works also researched the $X(3872)$ within the conventional charmonium. For example, Ref.~\cite{Colangelo:2025uhs} calculated the branching ratio of $X(3872)\to \psi(2S)\gamma$ to $X(3872)\to \psi\gamma$ and the obtained result is $1.7\pm 0.3$, which is close to the measurement of the LHCb collaboration~\cite{LHCb:2024tpv}.

\subsection{The role of coupled-channel effects in understanding low-mass puzzle}


\begin{figure}
\centering
\begin{tabular*}{\textwidth}{@{\extracolsep{\fill}}cc}
\includegraphics[width=0.45\textwidth]{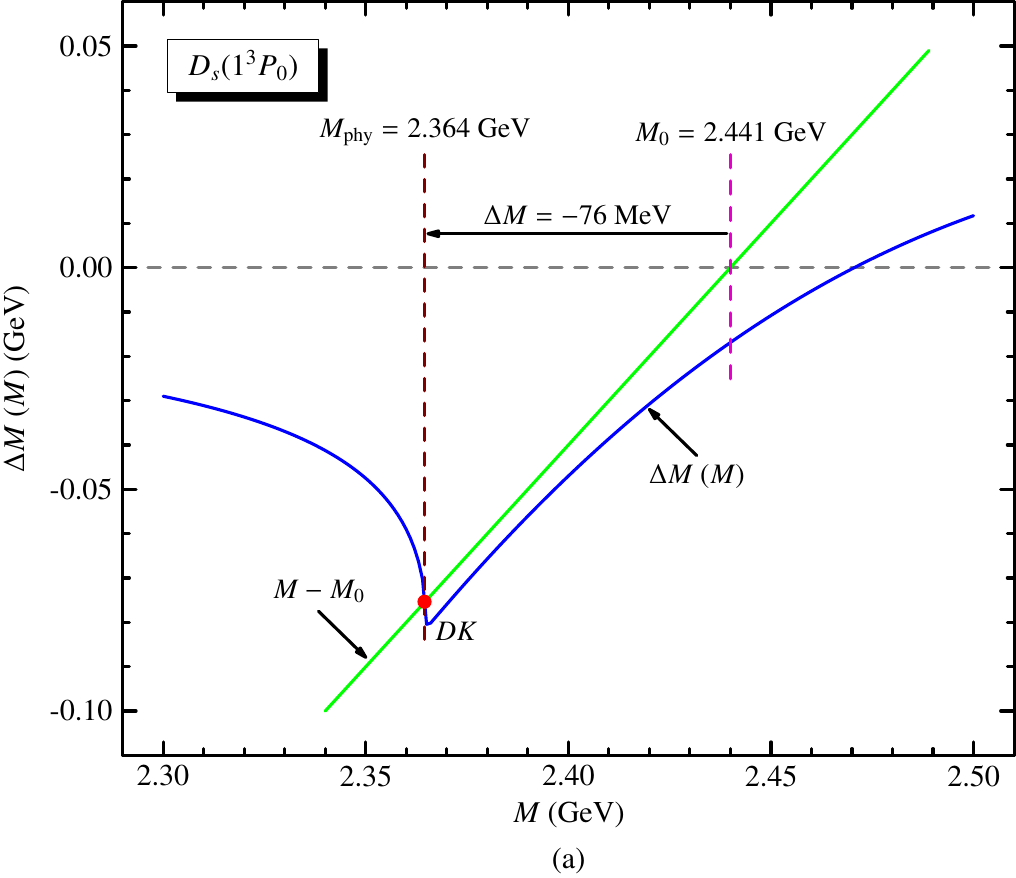}&
\includegraphics[width=0.45\textwidth]{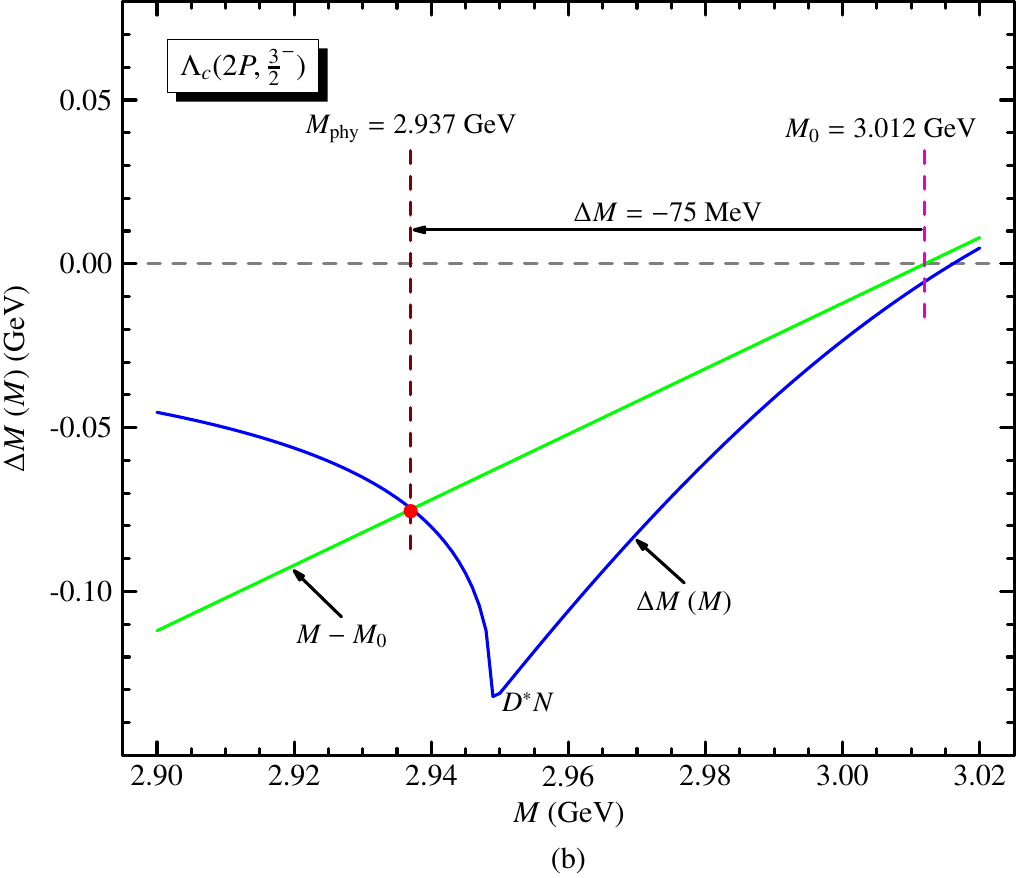}
\end{tabular*}
\caption{Solutions of the coupled-channel equation for $D_{s0}^*(2317)$ (a) and $\Lambda_c(2940)$ (b). Subfigures (a) and (b) are adapted from Ref.~\cite{Luo:2021dvj} and Ref.~\cite{Luo:2019qkm}, respectively.}
\label{fig:coupledchannelsolutions_a_b}
\end{figure}


\changelabelb{As discussed above, under the framework of coupled-channel effects, theorists developed many approached. However, in essence, they could be grouped under a similar method with different expressions. In this review, we mainly introduce two basic and popular expressions to describe the coupled-channel effects deduced from the unquenched quark model.} The first one is based on a quantum mechanical framework, \changelabel{in which the key equation is the coupled channel Schrodinger equation in matrix form as
\begin{equation}\label{eq:cpm}
\left(\begin{array}{cc}
\hat{H}_0&\hat{H}_I\\
\hat{H}_I&\hat{H}_{BC}
\end{array}\right)\left(\begin{array}{c}c_0\left|\psi_A\right\rangle\\\sum\limits_{BC}c_{BC}({\bf p})\left|\psi_{BC}({\bf p})\right\rangle\end{array}\right)=M\left(\begin{array}{c}c_0\left|\psi_A\right\rangle\\\sum\limits_{BC}c_{BC}({\bf p})\left|\psi_{BC}({\bf p})\right\rangle\end{array}\right).
\end{equation}
Here, $\left|\psi_A\right\rangle$ and $\left|\psi_{BC}({\bf p})\right\rangle$ are the wave functions of bare and intermediate states, respectively. $c_0$ and $c_{BC}({\bf p})$ are probability amplitudes of the corresponding components. $H_0$ is the Hamiltonian of the bare state, i.e., the conventional quenched quark model. $H_{BC}$ is the Hamiltonian of the intermediate loop. $H_I$ is the transition Hamiltonian between the bare and intermediate loop. 

By expanding Eq.~(\ref{eq:cpm}), one will obtain
\begin{equation}\label{eq:cpe}
\left\{\begin{array}{l}
\hat{H}_0c_0\left|\psi_0\right\rangle+\sum\limits_{BC}\hat{H}_Ic_{BC}({\bf p})\left|\psi_{BC}({\bf p})\right\rangle=Mc_0\left|\psi_0\right\rangle,\\
\hat{H}_Ic_0\left|\psi_0\right\rangle+\sum\limits_{BC}\hat{H}_{BC}c_{BC}({\bf p})\left|\psi_{BC}({\bf p})\right\rangle=\sum\limits_{BC}Mc_{BC}({\bf p})\left|\psi_{BC}({\bf p})\right\rangle.
\end{array}
\right.
\end{equation}
After defining the bare mass $M_0$ and transition amplitude $\mathcal{M}_{A\to BC}({\bf p})$ as
\begin{equation}
\begin{split}
                 M_0=&\left\langle\psi_A\right|\hat{H}_0\left|\psi_A\right\rangle,\\
\mathcal{M}_{A\to BC}({\bf p})=&\left\langle\psi_{BC}({\bf p})\right|\hat{H}_I\left|\psi_A\right\rangle,
\end{split}
\end{equation}
respectively, we can act $\left\langle\psi_0\right|$ and $\left\langle\psi_{BC}({\bf p})\right|$ to the Eq.~(\ref{eq:cpe}) to obtain
\begin{equation}\label{eq:cper}
\left\{\begin{array}{l}
M_0c_0+\sum\limits_{BC}\int c_{BC}({\bf p})\mathcal{M}_{A\to BC}^*({\bf p}){\rm d}^3{\bf p}=Mc_0,\\
c_0\mathcal{M}_{A\to BC}({\bf p})+c_{BC}({\bf p})E_{BC}(p)=Mc_{BC}({\bf p}),
\end{array}
\right.
\end{equation}
where $E_{BC}(p)=\sqrt{p^2+M_B^2}+\sqrt{p^2+M_C^2}$. Thus, from Eq.~(\ref{eq:cper}), $c_{BC}({\bf p})$ and $c_0$ have the following relationship
\begin{equation}\label{eq:cBCp}
c_{BC}({\bf p})=\frac{\mathcal{M}_{A\to BC}({\bf p})}{M-E_{BC}(p)}c_0,
\end{equation}
with an additional normalization condition as
\begin{equation}
|c_0|^2+\sum\limits_{BC}\int |c_{BC}({\bf p})|^2 {\rm d}^3{\bf p}=1.
\end{equation}
With the combination of Eqs. (\ref{eq:cper})-(\ref{eq:cBCp}), the coupled channel equation finally becomes
\begin{equation}
M_0+\sum\limits_{BC}\int \frac{|\mathcal{M}_{A\to{BC}}({\bf p})|^2}{M-E_{BC}(p)}{\rm d}^3{\bf p}=M,
\end{equation}
which could be simplified as
\begin{equation}
M_0+\Delta M(M)=M,\label{eq:coupled_channel_equation}
\end{equation}
where $\Delta M(M)$ is just the mass shift
\begin{equation}
\Delta M(M)=\sum\limits_{BC}\int \frac{|\mathcal{M}_{A\to{BC}}({\bf p})|^2}{M-E_{BC}(p)}{\rm d}^3{\bf p}.
\end{equation}
If $M>M_B+M_C$, these exist a singularity in the integrand, which lead to imaginary part, i.e.,
\begin{equation}
\begin{split}
{\rm Im}\;\Delta M(M)=&\sum\limits_{BC}-\pi\frac{\sqrt{M_B^2+p^2}\sqrt{M_C^2+p^2}p}{M}\int|{\cal M}_{A\to BC}({\bf p})|^2 {\rm d}\Omega\\
=&-\frac{1}{2}\Gamma_{A}.
\end{split}
\end{equation}
It is related to the decay width. The \changelabelb{corresponding} real part is
\begin{equation}\label{eq:ReDeltaM}
{\rm Re}\;\Delta M(M)=\sum\limits_{BC}{\cal P}\int \frac{|\mathcal{M}_{A\to{BC}}({\bf p})|^2}{M-E_{BC}(p)}{\rm d}^3{\bf p}.
\end{equation}

\changelabelb{In the calculations of the mass shift, the coupling between $A\to BC$ is crucial. Whether the mass shift $\Delta M(M)$ above or self-energy $\Pi (s)$ following, the The UV divergence may occur. In general, if the amplitude is obtained from Lagrangian, the integral may be divergent. However, if we calculate the amplitude ${\cal M}_{A\to BC}$ with quark model, the integral divergence problem could be solved. In quark model, the spatial wave functions of $A$, $B$, and $C$ decays rapidly as distance or momentum increase. In general, the obtained amplitudes $\mathcal{M}_{A\to BC}({\bf p})=\left\langle\psi_{BC}({\bf p})\right|\hat{H}_I\left|\psi_A\right\rangle$ may also decay rapidly as the momentum $p$ increases. For example, in quark model, the QPC model is a popular approach to calculate amplitude ${\cal M}_{A\to BC}({\bf p})$. With the harmonic oscillator wave functions, the obtained lowest $p$-dependence term is~\cite{Roberts:1992esl,Blundell:1996as,Barnes:2007xu,Chen:2017gnu}
\begin{equation}\label{eq:MAtoBC}
{\cal M}_{A\to BC}({\bf p})\sim p^Le^{-\alpha p^2},
\end{equation}
where the $\alpha$ is a factor, highly related to consistent quark masses and spatial wave functions of the hadrons. $L$ in Eq.~(\ref{eq:MAtoBC}) is the orbital momentum between $BC$, for $S$-wave, $L=0$, for $P$-wave, $L=1$, for $D$-wave, $L=2$, and so on. Eq~(\ref{eq:MAtoBC}) gives only the lowest-order term in $p$, while terms of order $p^{L+2}$, $p^{L+4}$, $\cdots$ may also be contained. In this scheme, the integral in Eq.~(\ref{eq:ReDeltaM}) is convergent.}

}

To illustrate the application of this \changelabelb{approach}, we consider the states $D_s(1^3P_0)$ and $\Lambda_c(2P,\,3/2^-)$. As shown in Fig.~\ref{fig:coupledchannelsolutions_a_b}, the bare masses of $D_s(1^3P_0)$ and $\Lambda_c(2P,\,3/2^-)$ from conventional quenched models~\cite{Luo:2021dvj,Luo:2019qkm} are 2.441 GeV and 3.012 GeV, respectively. Using the quark-pair-creation (QPC) model, one can compute the transition amplitudes $\mathcal{M}_{D_s(1^3P_0) \to DK}(p)$ and $\mathcal{M}_{\Lambda_c(2P,3/2^-) \to D^* N}(p)$, from which the mass shifts $\Delta M(M)$ are obtained (blue curves in Fig.~\ref{fig:coupledchannelsolutions_a_b}). The intersection points of the lines $M - M_0$ and the curves $\Delta M(M)$ give the solutions of the coupled-channel equations. The resulting physical masses are 2.364 GeV for $D_s(1^3P_0)$ and 2.937 GeV for $\Lambda_c(2P,\,3/2^-)$, corresponding to mass shifts of $-76$ MeV and $-75$ MeV, respectively. These values agree well with the observed masses of $D_{s0}^*(2317)$ and $\Lambda_c(2940)$. The coupled-channel effects for $D_{s0}^*(2317)$ and $D_{s1}'(2460)$ have been extensively discussed in the literature using various approaches~\cite{vanBeveren:2003kd,Dai:2003yg,Hwang:2004cd,Simonov:2004ar,Ortega:2016mms,Cheng:2017oqh,Cheng:2014bca,Luo:2021dvj,vanBeveren:2020eis,Zhang:2024usz}.

\begin{figure}
\centering
\includegraphics[width=0.55\textwidth]{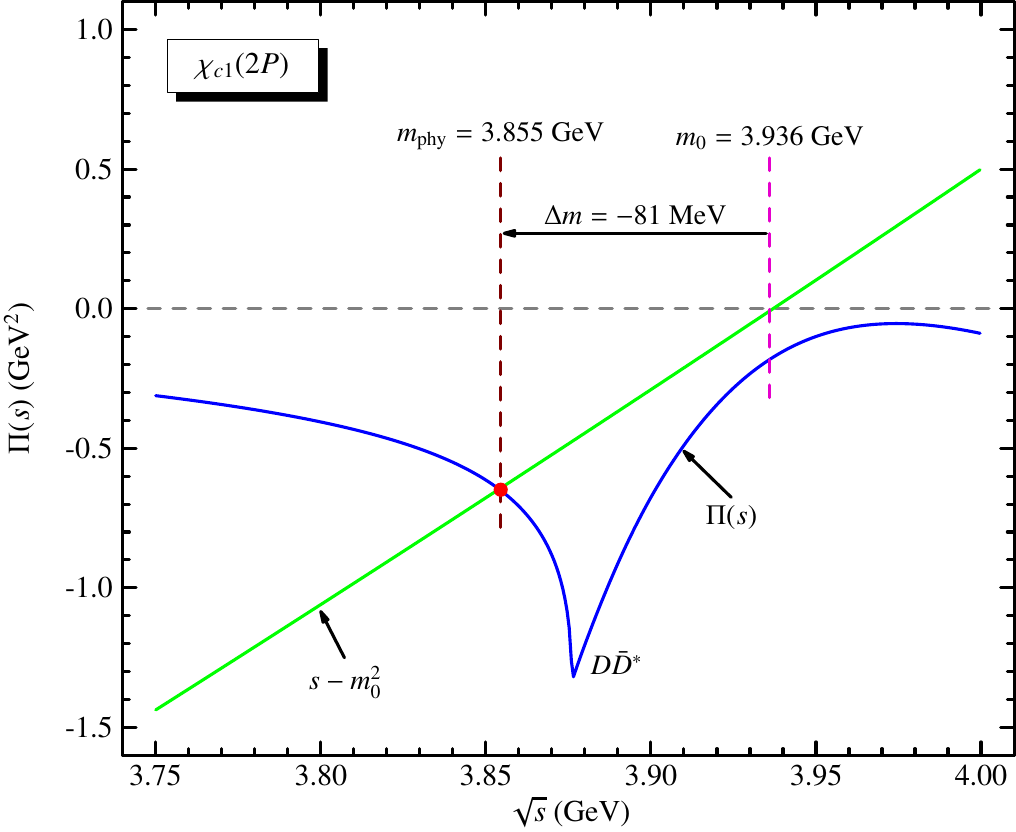}
\caption{Solution of the coupled-channel equation for $\chi_{c1}(2P)$. The data are taken from Ref.~\cite{Duan:2020tsx}.}
\label{fig:coupledchannelsolutions_chi_c1_2P}
\end{figure}

The second \changelabelb{expression} is formulated within quantum field theory, where the normalized propagator of a particle could be written as
\begin{center}
\includegraphics[width=0.8\textwidth]{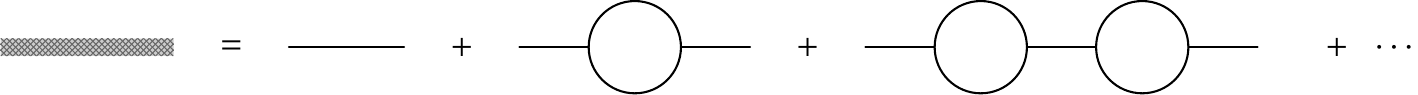},
\end{center}
and thus could be expressed as the following formula
\begin{equation}
\frac{A\bar{A}}{X-\bar{A}YA}=\frac{A\bar{A}}{X}+\frac{A\bar{A}}{X}Y\frac{A\bar{A}}{X}+\frac{A\bar{A}}{X}Y\frac{A\bar{A}}{X}Y\frac{A\bar{A}}{X}+\cdots.
\end{equation}
Here, $A$ is the field of the particle, $X=s-M_0^2$ with $M_0$ being the bare mass, \changelabel{$Y$ is the loop function}. Then, the self-energy function can be defined with $\Pi(s)=\bar{A}YA$, which describes the contribution from one loop. The pole position $s$ of a particle satisfies
\begin{equation}\label{eq:coupled_channel_equation_inf}
s-M_0^2-\Pi(s)=0.
\end{equation}
\changelabel{Usually,} a direct evaluation of $\Pi(s)$ in quantum field theory often leads to divergent integrals~\cite{Cheng:2017oqh,Cheng:2014bca}. \changelabel{However, when doing practical calculations, the loop function is just the same one $\Delta M$ given in the first \changelabelb{approach}. Thus, the introduced damping form factor there will automatically solve this divergence problem.}

For a single unquenched channel $A \to BC \to A \to BC \to A \cdot\cdot\cdot$, according to the quantum field theory, when $\sqrt{s}$ is above the threshold of the intermediate particles in the loop, an imaginary part will emerge in $\Pi_{A \to BC}(s)$, which can be related to the decay width by
\begin{equation}
{\rm Im}\;\Pi_{A \to BC}(s)=-\sqrt{s}\;\Gamma_{A\to BC}.
\end{equation}
Due to the \changelabel{two-body decay width
\begin{equation}
\Gamma_{A\to BC}=2\pi \frac{\sqrt{M_B^2+p^2}\sqrt{M_C^2+p^2}p}{M}\int|{\cal M}_{A\to BC}({\bf p})|^2 {\rm d}\Omega~~(s=M^2),
\end{equation}
}the imaginary part of loop can be expressed as
\begin{equation}
{\rm Im}\;\Pi_{A \to BC}(s)=-2\pi E_B(p)E_C(p)\int\left|{\cal M}_{A\to BC}({\bf p})\right|^2 {\rm d}\Omega,
\end{equation}
with the momentum $p$ given by
\begin{equation}
p=\frac{\sqrt{M_B^4+(M_C^2-s)^2-2M_B^2(M_C^2+s)}}{2\sqrt{s}}.
\end{equation}
Thus, the real part of $\Pi_{A \to BC}(s)$ can be obtained via a dispersion relation,
\begin{equation}\label{eq:RePi}
{\rm Re}\;\Pi_{A \to BC}(s)=\frac{1}{\pi}\mathcal{P}\int_{z_R}^\infty{\rm d}z\frac{{\rm Im}\;\Pi_{A \to BC}(z)}{z-s},
\end{equation}
where $\mathcal{P}$ denotes the principal value. 

\changelabel{ In fact, in the unquenched quark model, since the quark loop diagram given in Fig.~\ref{fig:loop_electron_photon}~(b) only constrains the quark components of the intermediate states, theoretically all the possible channels should be considered. Thus, the $\Pi(s)$ in Eq.~(\ref{eq:coupled_channel_equation_inf}) is actually an infinite summation on the intermediate states as
\begin{eqnarray}
\Pi(s)=\sum_{BC}\Pi_{BC}(s).
\end{eqnarray}
Among these channels, due to the damping form factor we introduced, actually only a few channels with proper thresholds will give considerable contributions. Especially, those channels with thresholds much larger than $\sqrt{s}$ must be suppressed, i.e., their contributions are nearly zero and weakly dependence on the $\sqrt{s}$. Thus, to give a base to the practical calculation, a subtraction is usually performed as~\cite{Pennington:2007xr}}
\begin{equation}
\begin{split}
{\rm Re}\;\Pi(s) \changelabel{- {\rm Re}\;\Pi(s_0)}=&\frac{1}{\pi}\mathcal{P}\int_{z_R}^\infty{\rm d}z\frac{{\rm Im}\;\Pi(z)}{z-s}-\frac{1}{\pi}\mathcal{P}\int_{z_R}^\infty{\rm d}z\frac{{\rm Im}\;\Pi(z)}{z-s_0}\\
=&\frac{s-s_0}{\pi}\mathcal{P}\int_{z_R}^\infty{\rm d}z\frac{{\rm Im}\;\Pi(z)}{(z-s)(z-s_0)}.
\end{split}
\end{equation}
Here, $s_0$ is a subtraction point, which is often chosen as the squared mass of the corresponding ground state. \changelabel{It should be emphasized again here that this subtraction is not to handle ultraviolet divergence, but to transform an infinite consideration on intermediate channels to limited ones~\cite{Pennington:2007xr}. In Fig.~\ref{fig:loop_electron_photon}, we made an analogy between electron-photon and hadron loops. We also emphasize that only the physical mass, i.e., the solution of the coupled-channel equation is observable. But for a hadron, in quark model, the $M_0$, $\Delta M(M)$, and $\Pi(s)$ are computable. In general, for the states which there are no significant coupled-channel effects, the bare masses $M_0$ are close to the experimental measurements in quark model calculations.}


\changelabel{Despite the different expressions between Eqs.~(\ref{eq:coupled_channel_equation}) and (\ref{eq:coupled_channel_equation_inf}), the contributions of these two approaches are not essentially different. Barnes and Swanson also pointed that the results from these two approaches are similar~\cite{Barnes:2007xu}.} We apply the second \changelabelb{approach} to examine the coupled-channel effect on the $\chi_{c1}(2P)$ state. As shown in Fig.~\ref{fig:coupledchannelsolutions_chi_c1_2P}, the bare mass of $\chi_{c1}(2P)$ is calculated to be 3.936 GeV within the GI model with updated parameters~\cite{Duan:2020tsx}. Choosing the subtraction point $s_0 = m_{J/\psi}^2$, the contribution from the $D\bar{D}^*$ channel induces a mass shift of $-81$ MeV, yielding a final mass of 3.855 GeV, which matches the measured mass of $X(3872)$. The coupled-channel effects for $X(3872)$ have also been widely studied~\cite{Ortega:2009hj,Kalashnikova:2005ui,Duan:2020tsx,Li:2009ad}.

As illustrated in Figs.~\ref{fig:coupledchannelsolutions_a_b}--\ref{fig:coupledchannelsolutions_chi_c1_2P}, both the mass shift $\Delta M(M)$ and the self-energy $\Pi(s)$ exhibit cusp-like structures at the relevant thresholds, where their magnitudes increase rapidly. These curves vividly demonstrate how coupled-channel effects lower the hadron masses. A closer examination reveals that such cusp structures arise \changelabel{most explicitly} from $S$-wave couplings between the state and the intermediate channel, \changelabel{although the appearance of threshold cusps is the consequences of unitarity at the level of the relevant discontinuity~\cite{Guo:2019twa}}. This can be understood from the \changelabelb{expression of the amplitude in Eq.~(\ref{eq:MAtoBC}).} For $S$-wave ($L=0$), \changelabel{such an amplitude will lead to the most severe damping behavior away from threshold, which will lead to a very clear presence of a cusp at threshold}. However, if the coupling is not $S$-wave, or even if it is $S$-wave but the bare mass lies far from the threshold, the pronounced effects shown in the figures do not occur, \changelabel{i.e., the contribution to the mass shift will be significantly suppressed}. In Section~\ref{subsubsec:mass_X3915_Z3930}, we will use $\chi_{c0}(2P)$ and $\chi_{c2}(2P)$ as examples to further elaborate on this point.

\subsection{$X(3915)$ and $Z(3930)$  as $2P$ charmonium states}

\subsubsection{Initial research}

\begin{figure}
\begin{tabular*}{\textwidth}{@{\extracolsep{\fill}}ccc}
\includegraphics[height=0.20\textwidth]{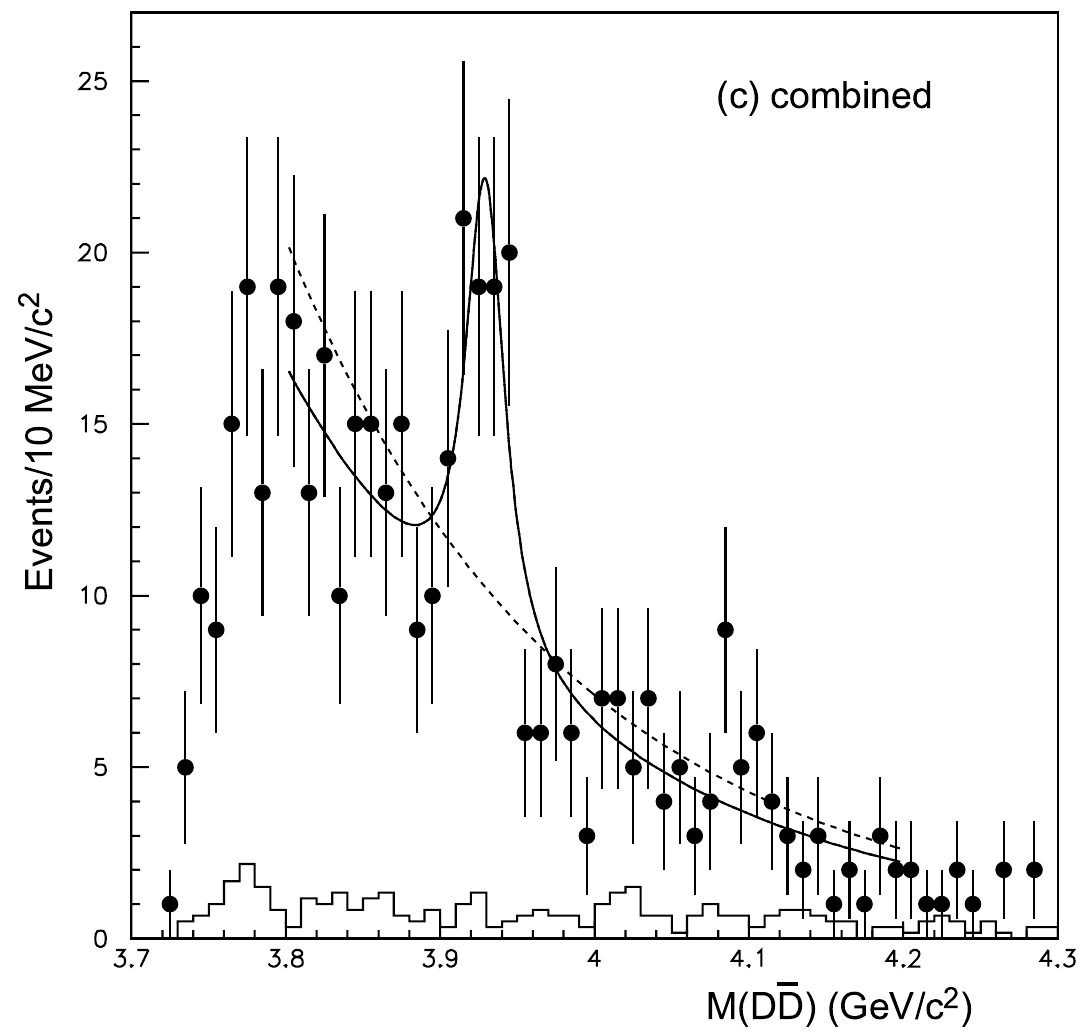}&\includegraphics[height=0.20\textwidth]{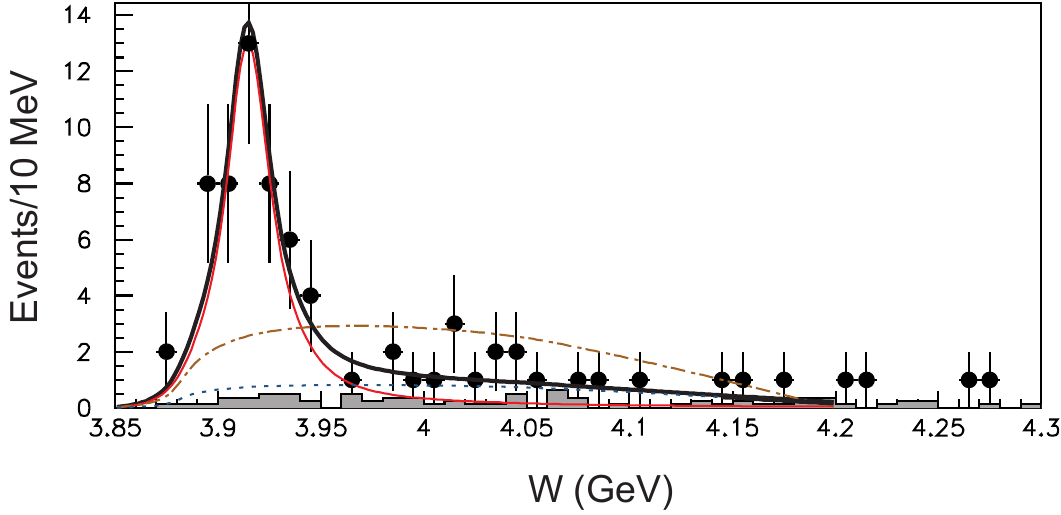}&\includegraphics[height=0.20\textwidth]{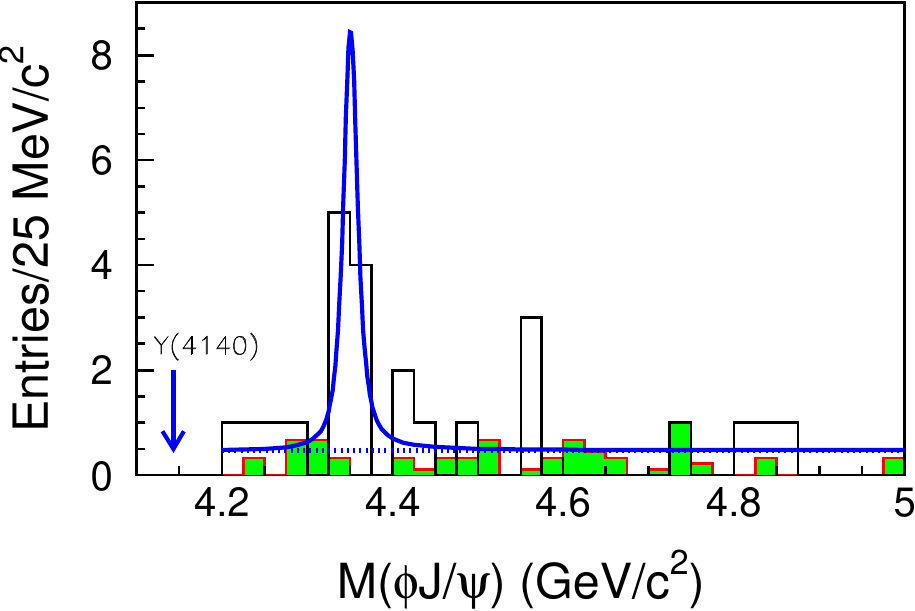}\\
\hspace{1em}(a)&\hspace{2em}(b)&\hspace{2em}(c)
\end{tabular*}
\caption{
(a) The $D\bar{D}$ invariant mass spectrum from $\gamma\gamma\to D\bar{D}$~\cite{Belle:2005rte}.
(b) The $\omega J/\psi$ invariant mass ($W$) distribution from $\gamma\gamma \to \omega J/\psi$ candidate events~\cite{Belle:2009and}.
(c) The $\phi J/\psi$ invariant mass spectrum from $\gamma\gamma\to \phi J/\psi$~\cite{Belle:2009rkh}.
All results are from the Belle Collaboration.}
\label{fig:dd_Jpsiomega_phiJpsi}
\end{figure}

Following the discovery of $X(3872)$, the Belle Collaboration observed a new charmonium-like state around 3.93 GeV in the process $\gamma\gamma\to D\bar{D}$ in 2005~\cite{Belle:2005rte}. This state, denoted $Z(3930)$, is shown in Fig.~\ref{fig:dd_Jpsiomega_phiJpsi} (a) and was reported with a statistical significance of $5.3 \,\sigma$. Its measured resonance parameters are
\begin{equation}
\begin{split}
M&=3929\pm 5({\rm stat})\pm 2({\rm syst})~{\rm MeV},\\
\Gamma&=29\pm 10 ({\rm stat})\pm 2({\rm syst})~{\rm MeV}.
\end{split}
\end{equation}
Furthermore, the product of its two-photon partial decay width and the branching fraction to $D\bar{D}$ was measured as
\begin{equation}
\Gamma_{\gamma\gamma}\times{\cal B}(Z(3930)\to D\bar{D})=0.18 \pm 0.05 ({\rm stat}) \pm 0.03 ({\rm syst})~{\rm keV}~({\rm assuming}\,\, J = 2).
\end{equation}
The angular distribution was found to be consistent with spin $J = 2$. Since $Z(3930)$ is produced in $\gamma\gamma$ fusion, its $C$-parity is positive ($C = +1$). Based on these quantum numbers, $Z(3930)$ was identified as a good candidate for the $\chi_{c2}(2P)$ charmonium state with $J^{PC} = 2^{++}$, an assignment that gained wide acceptance.

In 2009, the Belle Collaboration reported another charmonium-like state, $X(3915)$, observed in $\gamma\gamma\to \omega J/\psi$ as shown in Fig.~\ref{fig:dd_Jpsiomega_phiJpsi} (b)~\cite{Belle:2009and}. The signal had a statistical significance of $7.7\,\sigma$, with mass and width
\begin{equation}
\begin{split}
M=3915\pm 3\pm 2~{\rm MeV},\quad
\Gamma=17\pm 10\pm 3~{\rm MeV}.
\end{split}
\end{equation}
The product of the two-photon decay width and the branching fraction to $\omega J/\psi$ was measured for two possible spin-parity assignments:
\begin{equation}
\renewcommand\arraystretch{1.5}
\Gamma_{\gamma\gamma}\times {\cal B}(X(3915)\to \omega J/\psi)=\left\{\begin{array}{ll}
(61 \pm 17 \pm 8)~{\rm eV} & {\rm for}~J^P=0^+\\
(18 \pm  5 \pm 2)~{\rm eV} & {\rm for}~J^P=2^+
\end{array}\right..
\end{equation}

Motivated by the earlier CDF observation of the $Y(4140)$ in $B^+ \to J/\psi \phi K^+$ decays~\cite{CDF:2009jgo}, the Belle Collaboration also searched for structures in $\gamma\gamma \to \phi J/\psi$ in 2009~\cite{Belle:2009rkh}. While no significant $Y(4140)$ signal was found, a new narrow structure, $X(4350)$, was identified (Fig.~\ref{fig:dd_Jpsiomega_phiJpsi} (c)) with $8.8^{+4.2}_{-3.2}$ signal events and a significance of $3.2\,\sigma$. Its parameters are
\begin{equation}
\begin{split}
M     =4350.6^{+4.6}_{-5.1} ({\rm stat}) \pm 0.7 ({\rm syst})~{\rm MeV},\quad
\Gamma=13^{+18}_{-9} ({\rm stat}) \pm 4 ({\rm syst})~{\rm MeV}.
\end{split}
\end{equation}
The corresponding product of the two-photon decay width and the branching fraction to $\phi J/\psi$ is
\begin{equation}
\renewcommand\arraystretch{1.5}
\Gamma_{\gamma\gamma}\times {\cal B}(X(4350)\to \phi J/\psi)=\left\{\begin{array}{ll}
6.7^{+3.2}_{-2.4}({\rm stat}) \pm 1.1 ({\rm syst})~{\rm eV} & {\rm for}~J^P=0^+\\
1.5^{+0.7}_{-0.6}({\rm stat}) \pm 0.3 ({\rm syst})~{\rm eV} & {\rm for}~J^P=2^+
\end{array}\right. .
\end{equation}

\begin{figure}
\centering
\begin{tabular*}{\textwidth}{@{\extracolsep{\fill}}ccc}
\includegraphics[height=0.15\textheight]{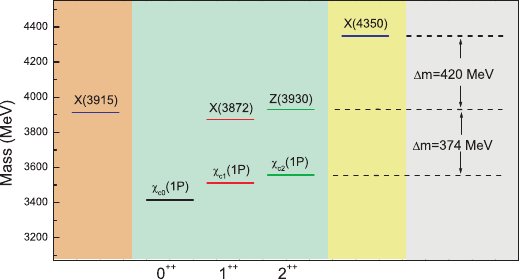}&
\includegraphics[height=0.16\textheight]{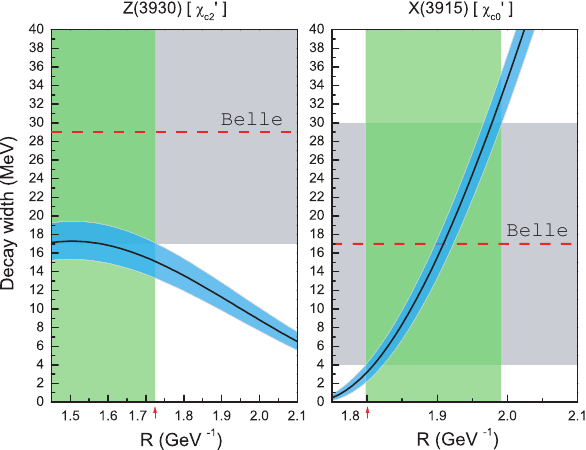}&
\includegraphics[height=0.16\textheight]{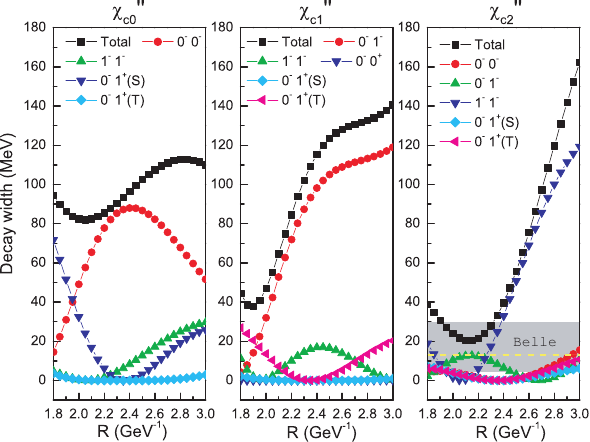}\\
(a)&(b)&\hspace{1em}(c)
\end{tabular*}
\caption{The observed $P$-wave charmonium states and candidates (a), the comparisons of theoretical and experimental decay widths of $Z(3930)$ with assignment of $\chi_{c2}(2P)$ and $X(3915)$ with interpretation of $\chi_{c0}(2P)$, and the theoretical widths of $\chi_{cJ}(3P)$ and the comparisons of theoretical and experimental decay widths of $X(4250)$ with interpretation of $\chi_{c2}(3P)$. The figures are taken from Ref.~\cite{Liu:2009fe}. Here, $\chi_{cJ}^\prime$ and $\chi_{cJ}^{\prime\prime}$ are $\chi_{cJ}(2P)$ and $\chi_{cJ}(3P)$, respectively.}
\label{fig:mesg_3915_3930_4350}
\end{figure}

The $\gamma\gamma$ fusion process is particularly suited for producing mesonic states with positive $C$-parity. According to the Landau--Yang theorem~\cite{Landau:1948kw,Yang:1950rg}, two photons in an $S$-wave can only couple to final states with quantum numbers $J^{PC} = 0^{++}$ or $2^{++}$. Consequently, $\gamma\gamma$ collisions have long been an important way for studying $P$-wave quarkonia, as exemplified by the early observations of the ground-state charmonia $\chi_{c0}(1P)$ and $\chi_{c2}(1P)$~\cite{Belle:2007qae,Belle:2013eck}. It is therefore natural to interpret the charmonium-like states observed in $\gamma\gamma$ fusion at $B$-factories as potential conventional $P$-wave charmonia.

A prime candidate is $X(3915)$, observed in $\gamma\gamma \to \omega J/\psi$ with a mass of $3915 \pm 3 \pm 2~\text{MeV}$ and a narrow width. This state appeared to fill a specific gap in the known spectrum. The first radial excitation multiplet ($n=2$) already contained the $1^{++}$ candidate $X(3872)$ and the well-established $2^{++}$ state $Z(3930)$, but the corresponding scalar partner $\chi_{c0}(2P)$ ($0^{++}$) was missing. The mass of $X(3915)$ aligns well with model predictions for $\chi_{c0}(2P)$. Furthermore, the pattern of mass differences among these $n=2$ states can be understood through coupled-channel effects, which are typically weaker for $0^{++}$ and $2^{++}$ states than for the $1^{++}$ case (see Fig.~\ref{fig:mesg_3915_3930_4350} (a)).

Additional support comes from calculations of open-charm decays using the quark pair creation (QPC) model~\cite{Liu:2009fe}. These calculations reproduce the measured width of $X(3915)$ using parameter values consistent with those needed to describe $Z(3930)$, as shown in Fig.~\ref{fig:mesg_3915_3930_4350} (b). Collectively, this evidence strongly suggested that $X(3915)$ is the conventional $\chi_{c0}(2P)$ charmonium, completing the first radially excited $P$-wave triplet. Its assignment as a $0^{++}$ state, testable via angular distribution analyses, underscores the continuing value of $\gamma\gamma$ fusion in mapping out the conventional quarkonium spectrum.

While states like $X(4350)$ were also observed in similar $\gamma\gamma$ processes and are discussed as higher excitations, the case of $X(3915)$ illustrates how combining systematic spectroscopy with decay calculations can successfully determine the nature of new resonances within the conventional quark model.

In 2012, the BaBar Collaboration performed an angular analysis of $X(3915)$ and confirmed that its quantum numbers are consistent with $J^{PC}=0^{++}$~\cite{BaBar:2012nxg}. This experimental finding agreed with earlier theoretical predictions in Ref.~\cite{Liu:2009fe}, leading to the subsequent listing of $X(3915)$ as $\chi_{c0}(2P)$ in the 2013 edition of the PDG review~\cite{ParticleDataGroup:2012pjm}.

Despite the overall consistency among its measured mass, width, and established quantum numbers—which collectively support the $\chi_{c0}(2P)$ interpretation—several intriguing puzzles remained unresolved:
\begin{enumerate}
\item Quenched quark model calculations typically predict a significantly larger mass splitting between $\chi_{c2}(2P)$ and $\chi_{c0}(2P)$ than the observed $\sim$12 MeV between $Z(3930)$ and $X(3915)$~\cite{Guo:2012tv,Guo:2010ak,Olsen:2014maa}. In addition, the $\chi_{c0}(2P)$ could decay into $D\bar{D}$ via $S$-wave, but the width of $X(3915)$ is \changelabel{too} narrow. What mechanisms are responsible for this anomalously small splitting of $\chi_{c2}(2P)-\chi_{c0}(2P)$ and small width of $X(3915)$?
\item The decay $X(3915) \to J/\psi,\omega$ is an OZI-suppressed process, yet its measured branching fraction is relatively large~\cite{Guo:2012tv,Brambilla:2010cs,Olsen:2014maa}. Why is this hidden-charm mode strongly enhanced?
\item If $X(3915)$ is indeed the conventional $\chi_{c0}(2P)$, its dominant decay channel should be $D\bar{D}$. However, a clear signal of $X(3915)$ was not observed in the $D\bar{D}$ invariant mass spectrum~\cite{Guo:2012tv,Olsen:2014maa}. What suppresses or masks its expected open-charm decay?
\end{enumerate}
These persistent issues rendered the final identification of $X(3915)$ both subtle and challenging. In the following sections, we address each of these problems systematically.

\subsubsection{Anomalous small mass gap bewtween $X(3915)$ and $Z(3930)$}\label{subsubsec:mass_X3915_Z3930}

\begin{figure}[htbp]
\centering
\begin{tabular*}{\textwidth}{@{\extracolsep{\fill}}cc}
\includegraphics[height=0.3\textheight]{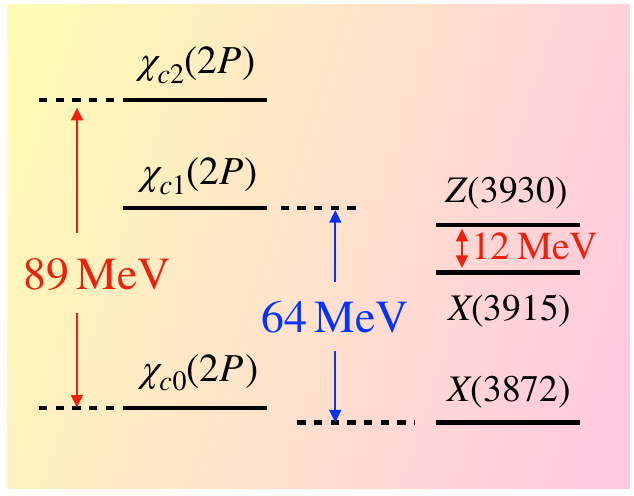}&\includegraphics[height=0.30\textheight]{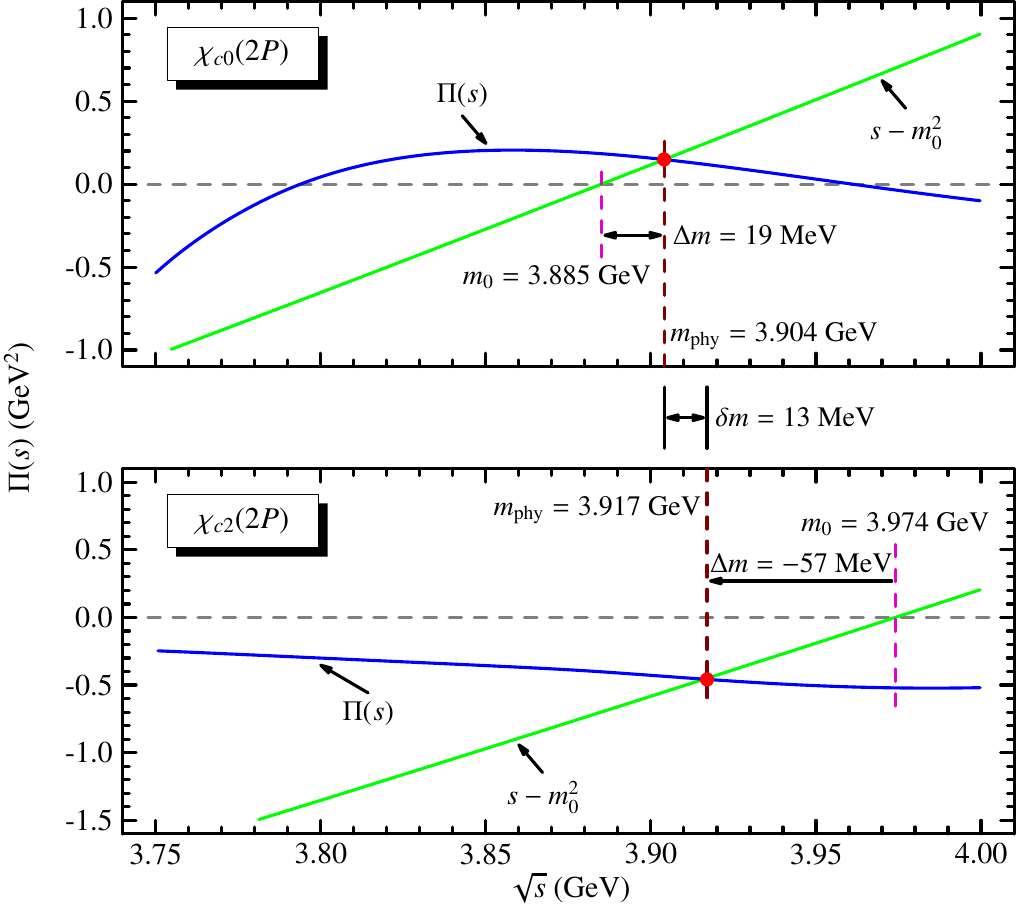}\\
(a)&\hspace{2.5em}(b)
\end{tabular*}
\caption{Comparison of masses between $\chi_{cJ}(2P)$ states from the GI model and the observed $X(3872)$, $X(3915)$, and $Z(3930)$. The figures are taken from Ref.~\cite{Duan:2020tsx}.}
\label{fig:split_02_2P}
\end{figure}

The assignment of $X(3915)$ as the conventional $\chi_{c0}(2P)$ charmonium state faced significant theoretical scrutiny, primarily centered on two seemingly contradictory properties: its anomalously small mass splitting with the established $\chi_{c2}(2P)$ candidate $Z(3930)$ (see Fig.~\ref{fig:split_02_2P} (a)), and its remarkably narrow width despite lying above the $S$-wave $D\bar{D}$ decay threshold. A consistent resolution to both puzzles has emerged from advanced \textit{unquenched} theoretical frameworks that move beyond the static quark model by dynamically incorporating the dressing of the bare $c\bar{c}$ core with virtual hadron loops~\cite{Duan:2020tsx}. These frameworks treat the physical state as a superposition of the bare quark-model configuration and coupled hadronic channels—primarily $D\bar{D}$ and $D\bar{D}^*$—with contributions computed using effective models such as the QPC model.

The mass puzzle is addressed by considering significant and differential mass shifts induced by the coupling to open-charm channels. The numerical results of Ref.~\cite{Duan:2020tsx} are presented in Fig.~\ref{fig:split_02_2P} (b). In a quenched GI model with updated parameters, the bare $\chi_{c0}(2P)$ and $\chi_{c2}(2P)$ states are separated by nearly 89 MeV. However, unquenched calculations reveal a more complex physical picture. The $\chi_{c1}(2P)$ state (associated with $X(3872)$) couples strongly to $D\bar{D}^*$ in an $S$-wave, resulting in a large attractive mass shift of approximately $-80$ MeV, which explains its unusually low mass (discussed previously; see Fig.~\ref{fig:coupledchannelsolutions_chi_c1_2P}). In contrast, the $\chi_{c0}(2P)$ couples to $D\bar{D}$ in an $S$-wave but experiences a modest \textit{positive} shift of about $+20$ MeV\footnote{This positive shift arises when the physical mass lies above the $D\bar{D}$ threshold ($m_{\rm phy} > M_B + M_C$). In this case, principal-value integration is required in the mass shift formulas [Eqs.~(\ref{eq:ReDeltaM}) and (\ref{eq:RePi})], which can yield positive contributions.}. The $\chi_{c2}(2P)$ state, coupling to $D\bar{D}$ and $D\bar{D}^*$ in a $D$-wave, undergoes a moderate negative shift of about $-60$ MeV due to centrifugal suppression. Crucially, the large bare splitting is dramatically compensated: the opposing shifts for the scalar ($+$) and tensor ($-$) states pull their physical masses closer together (see Fig.~\ref{fig:split_02_2P} (b)). The net physical mass difference reduces to approximately $13$ MeV, in striking agreement with the observed $\sim$12 MeV gap between $X(3915)$ and $Z(3930)$. This result demonstrates that the small splitting is a direct signature of coupled-channel effects rather than an anomaly. \changelabel{In addition, Ref.~\cite{Shi:2024llv} also reanalyzed the lattice QCD data of $D\bar{D}$-$D_s\bar{D}_s$ coupled-channel from Ref.~\cite{Prelovsek:2020eiw} and $D\bar{D}^*$ systems from Ref.~\cite{Prelovsek:2013cra}, a series of poles were found. For the $0^{++}$, two poles could be interpreted as mostly molecular states, and one pole is originated from the $\chi_{c0}(2P)$. For the $1^{++}$, one pole could be related to $X(3872)$, and second one may be originated from $\chi_{c1}(2P)$, which is likely corresponds to $\chi_{c1}(4010)$. Two $2^{++}$ poles were found, which correspond to shallow $D^*\bar{D}^*$ bound state and dressed $\chi_{c2}(2P)$.}

\begin{figure}[htbp]
\centering
\begin{tabular*}{\textwidth}{@{\extracolsep{\fill}}cc}
\includegraphics[height=0.230\textheight]{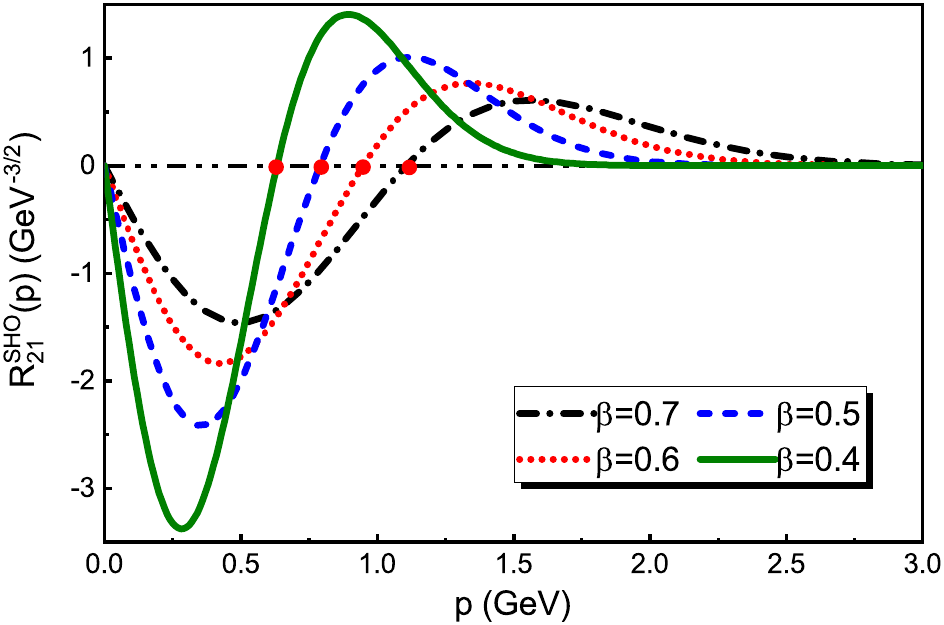}&\includegraphics[height=0.24\textheight]{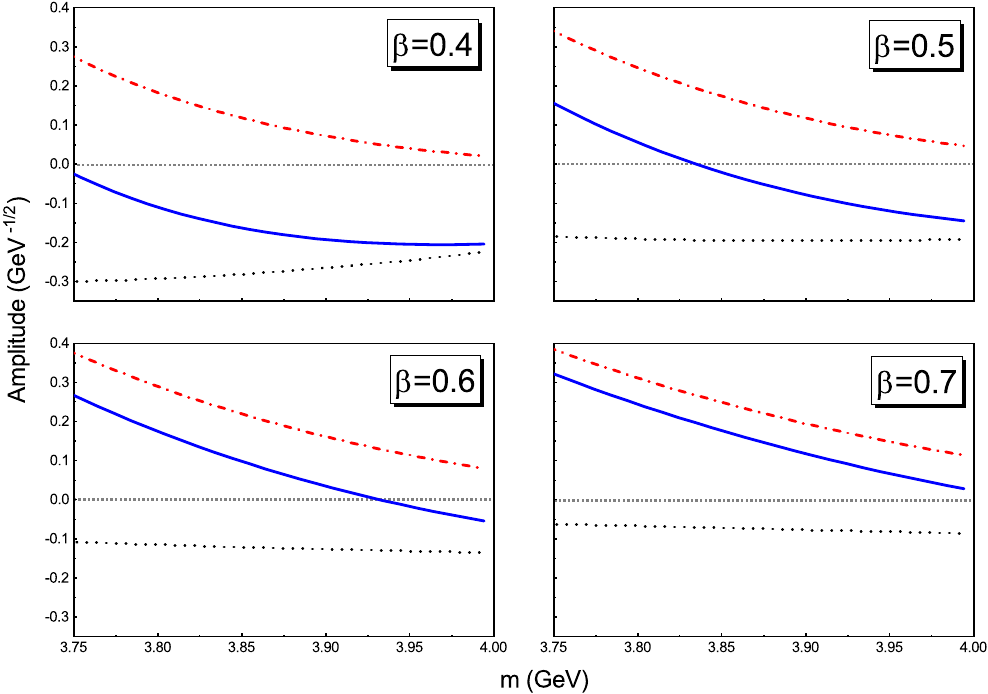}\\
\hspace{1.2em}(a)&\hspace{2em}(b)
\end{tabular*}
\caption{(a) Radial wave functions of $\chi_{cJ}(2P)$ states for different values of the oscillator parameter $\beta$. (b) Decay amplitude for $\chi_{c0}(2P)\to D\bar{D}$ as a function of $\beta$, showing contributions from positive ($R_{nL}(p)>0$, dashed-dotted) and negative ($R_{nL}(p)<0$, dotted) wave function regions, and the total amplitude (solid). Figures from Ref.~\cite{Duan:2020tsx}.}
\label{fig:WF-Amp}
\end{figure}

The width puzzle is resolved through a subtle quantum mechanical phenomenon known as the node effect. As shown in Fig.~\ref{fig:WF-Amp} (a), the radial wave function of a $2P$ state, being the first radial excitation, possesses a single node. The partial width for the dominant open-charm decay $\chi_{c0}(2P) \to D\bar{D}$ is proportional to the square of the decay amplitude $\mathcal{M}$, which is governed by a spatial overlap integral, $I$, between the initial charmonium wave function and the final-state $D$-meson wave functions. This integral can be decomposed as $I = I_{+} + I_{-}$, where $I_{+}$ ($I_{-}$) arises from integrating over the region where the radial wave function is positive (negative). As illustrated in Fig.~\ref{fig:WF-Amp} (b), these two contributions are opposite in sign, leading to significant destructive interference. This near-cancellation drastically suppresses the decay amplitude, resulting in a much narrower total width than naive phase-space estimates would suggest. Detailed calculations yield a total width of $20$--$30$ MeV for $\chi_{c0}(2P)$, aligning perfectly with the experimental width of $X(3915)$. This mechanism not only explains the narrowness of $X(3915)$ but also provides a critical discriminant against alternative interpretations. For instance, a broad state like $X(3860)$\footnote{In 2017, the Belle Collaboration observed a broad state $X(3860)$~\cite{Belle:2017egg} in the $D\bar{D}$ invariant mass spectrum from $e^+e^-\to J/\psi D \bar{D}$, with mass $3862^{+26+40}_{-32-13}$ MeV and width $201^{+154+88}_{-67-82}$ MeV. The $J^{PC}=0^{++}$ hypothesis was favored over $2^{++}$ at the $2.5\sigma$ level. Some theoretical works interpreted $X(3860)$ as $\chi_{c0}(2P)$~\cite{Zhou:2017dwj,Yu:2017bsj,Ortega:2017qmg}, but it is omitted from the PDG summary table~\cite{ParticleDataGroup:2024cfk}.} ($\Gamma \approx 200$ MeV) cannot be the $\chi_{c0}(2P)$ because its large width is incompatible with the node-effect suppression inherent to a conventional radially excited $P$-wave charmonium.

In summary, the initially puzzling spectroscopic features of $X(3915)$—its small mass gap with $Z(3930)$ and its narrow width—find a natural and unified explanation within the unquenched quark model. The mass splitting is determined by state-specific coupled-channel shifts, while the width is suppressed by the nodal structure of the radial wavefunction. Together, these insights provide a robust theoretical foundation for identifying $X(3915)$ as the long-sought $\chi_{c0}(2P)$ state, thereby completing the first radially excited $P$-wave charmonium triplet.

\subsubsection{Anomalous large the branching ratio for the $X(3915)\to J/\psi\omega$ decay}

$X(3915)$ and $Z(3930)$ were both observed in $\gamma\gamma$ collisions. $X(3915)$ was first discovered in the decay $X(3915)\to J/\psi\omega$ by Belle~\cite{Belle:2009and}, with quantum numbers later measured as $J^{PC}=0^{++}$ by BaBar~\cite{BaBar:2012nxg}. $Z(3930)$ was observed in $\gamma\gamma\to D\bar D$ by Belle~\cite{Belle:2005rte} and BaBar~\cite{BaBar:2010jfn} and is a strong candidate for $\chi_{c2}(2P)$. If both states are indeed the first radial excitations of the $P$-wave charmonia $\chi_{c0}$ and $\chi_{c2}$, they should both decay into $J/\psi\omega$. However, experimentally only $X(3915)$ is seen in this channel. This raises the question: why is $Z(3930)$ not observed in $J/\psi\omega$? The answer likely lies in a mechanism that strongly suppresses $\chi_{c2}(2P)\to J/\psi\omega$ relative to $\chi_{c0}(2P)\to J/\psi\omega$, despite the latter being an OZI-suppressed process.

\begin{figure}
    \centering
    \includegraphics[width=\textwidth]{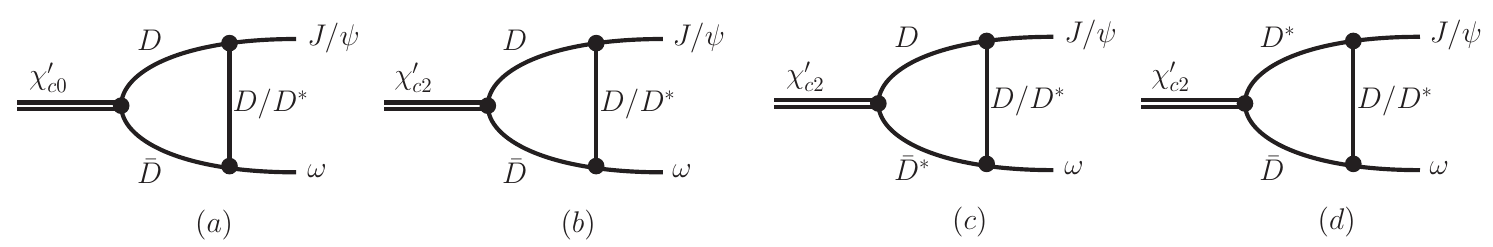}
    \caption{Diagrams for the decays $\chi_{c0}^\prime \to J/\psi\omega$ (a) and $\chi_{c2}^\prime \to J/\psi\omega$ (b-d), taken from Ref.~\cite{Chen:2013yxa}. Here $\chi_{c0}^\prime$ and $\chi_{c2}^\prime$ are $\chi_{c0}(2P)$ and $\chi_{c2}(2P)$, respectively.}
    \label{fig:Chen2013yxa_chicj}
\end{figure}

To address this puzzle, the decay widths for $\chi_{c0}(2P)\to J/\psi\omega$ and $\chi_{c2}(2P)\to J/\psi\omega$ can be computed using a hadronic loop approach~\cite{Chen:2013yxa}. Since both states lie above the $D\bar D$ and $D\bar D^{\ast}$ thresholds but below $D^{}\bar D^{*}$, their dominant decays are to open-charm meson pairs: $X(3915)\to D\bar D$ and $Z(3930)\to D\bar D,\, D\bar D^{*}+\text{c.c.}$ The hidden-charm decays $J/\psi\omega$ are assumed to proceed via rescattering of these dominant modes through intermediate charmed meson loops, as illustrated in Fig.~\ref{fig:Chen2013yxa_chicj}. The loop integrals are evaluated using the Cutkosky cutting rules, with form factors to account for off-shell effects and vertex structure: $\mathcal{F}(q^2)=\left(\frac{m_{\text{E}}^2-\Lambda^2}{q^2-\Lambda^2}\right)^N$, where $\Lambda=m_{\text{E}}+\alpha\Lambda_{\text{QCD}}$, $\Lambda_{\text{QCD}}=220$ MeV, $\alpha\sim\mathcal{O}(1)$, and $N=1$ (monopole) or $N=2$ (dipole).

\begin{figure}
\centering
\begin{tabular*}{0.9\textwidth}{@{\extracolsep{\fill}}cc}
\includegraphics[height=0.230\textheight]{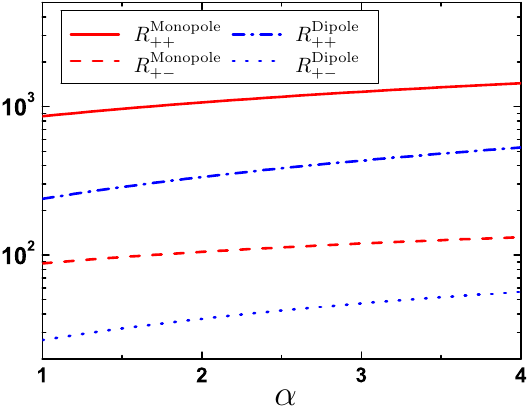}&
\includegraphics[height=0.233\textheight]{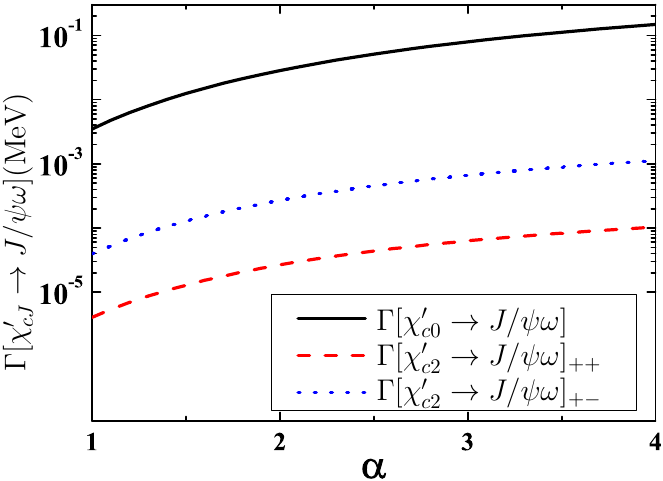}\\
\hspace{1.25em}(a)&\hspace{2.5em}(b)
\end{tabular*}
\caption{Dependence on the parameter $\alpha$ of (a) the ratio of decay widths $\Gamma(\chi_{c0}^\prime\to J/\psi\omega)/\Gamma(\chi_{c2}^\prime\to J/\psi\omega)$, and (b) the partial widths for $\chi_{c0,c2}^\prime\to J/\psi\omega$ (shown for the monopole form factor as an example). Figure adapted from Ref.~\cite{Chen:2013yxa}. Here $\chi_{c0}^\prime$ and $\chi_{c2}^\prime$ are $\chi_{c0}(2P)$ and $\chi_{c2}(2P)$, respectively.}
    \label{fig:width-ratio}
\end{figure}

As presented in Fig.~\ref{fig:width-ratio}, the partial decay widths are calculated as functions of $\alpha$. For $\chi_{c0}(2P)\to J/\psi\omega$, the width ranges from $3.5\times10^{-3}$ MeV to $0.15$ MeV for $\alpha\in[1,4]$ with a monopole form factor. For $\chi_{c2}(2P)\to J/\psi\omega$, the width is much smaller: $4.1\times10^{-6}$ MeV to $1.1\times10^{-4}$ MeV (case $++$) or $4.0\times10^{-5}$ MeV to $1.1\times10^{-3}$ MeV (case $+-$). \changelabel{ Here, $++$ represents that the signs of the coupling constants $g_{\chi_{c2(2P)}DD}$ and $g_{\chi_{c2(2P)}DD}$ are both positive, while $+-$ indicates that the signs of the coupling constants $g_{\chi_{c2(2P)}DD}$ and $g_{\chi_{c2(2P)}DD}$ are positive and negative, respectively ~\cite{Chen:2013yxa}.} The dipole form factor yields widths about an order of magnitude smaller.

More importantly, the ratio $R=\Gamma[\chi_{c0}(2P)\to J/\psi\omega]/\Gamma[\chi_{c2}(2P)\to J/\psi\omega]$ is large and robust:
\[
R_{++}^{\text{Monopole}} \simeq 850\text{--}1400,\quad
R_{++}^{\text{Dipole}} \simeq 239\text{--}558,\quad
R_{+-}^{\text{Monopole}} \simeq 87\text{--}130,\quad
R_{+-}^{\text{Dipole}} \simeq 27\text{--}59.
\]
Thus, $\chi_{c0}(2P)\to J/\psi\omega$ is at least one order of magnitude larger than $\chi_{c2}(2P)\to J/\psi\omega$, regardless of the form factor or coupling sign. \changelabel{However, this ratio is drastically different from the naive heavy quark spin symmetry (HQSS) expectation for the $P$-wave charmonia, which should be of nearly same order~\cite{Guo:2014qra}, indicating a large dynamical spin symmetry breaking or other effects\footnote{It should be mentioned that only the absorptive parts of the meson loops were taken into consideration in the estimations of  Ref.~\cite{Chen:2013yxa}. The dispersive part, which contains contributions arising from $\chi_{c0/2}(2P) D^\ast \bar{D}^\ast$ couplings, is also expected to yield significant contributions to the decays $\chi_{c0/2}(2P)\to J/\psi \omega $ and deserves further investigation.
}.
}

This suppression explains why $Z(3930)$ is not observed in $J/\psi\omega$. Using theoretical estimates for the two-photon widths of $\chi_{c0}(2P)$ (1–5 keV) \cite{Godfrey:1985xj,Munz:1996hb,Ebert:2003mu,Hwang:2010iq,Wang:2007nb} and the measured $\Gamma_{\gamma\gamma}\times BR(\chi_{c0}(2P)\to J/\psi\omega)$ \cite{Belle:2009and,BaBar:2012nxg}, the partial width for $\chi_{c0}(2P)\to J/\psi\omega$ is estimated to be between 200 eV and 1 MeV, consistent with the calculated result (up to 150 keV for $\alpha=4$) \cite{Belle:2009and}.

The hadronic loop mechanism provides a natural explanation for the observed pattern: the decay $Z(3930)\to J/\psi\omega$ is strongly suppressed compared to $X(3915)\to J/\psi\omega$. This supports the identification of $X(3915)$ as $\chi_{c0}(2P)$ and $Z(3930)$ as $\chi_{c2}(2P)$. The study also underscores the importance of coupled-channel effects in charmonium decays, particularly for states above open-charm thresholds.

\subsubsection{Unexpected \changelabel{absence} of $X(3915)$ in $\gamma\gamma\to D\bar{D}$}

As discussed above, the decay $X(3915)\to J/\psi \omega$ is an OZI-suppressed process, where hadron loop mechanisms are crucial. In contrast, the assignment of $Z(3930)$ as $\chi_{c2}(2P)$ appears more straightforward since both Belle~\cite{Belle:2005rte} and BaBar~\cite{BaBar:2010jfn} observed it directly in the OZI-allowed channel $\gamma\gamma\to D\bar{D}$. According to the OZI rule and spin-parity conservation, both $\chi_{c0}(2P)$ ($0^{++}$) and $\chi_{c2}(2P)$ ($2^{++}$) should decay into $D\bar{D}$. In particular, $D\bar{D}$ is expected to be the dominant decay mode for $\chi_{c0}(2P)$~\cite{Chen:2012wy}. However, the experimental data from Belle and BaBar in $\gamma\gamma\to D\bar{D}$ reported only the $Z(3930)$ enhancement without a clear signal for $X(3915)$, raising a puzzle regarding the $\chi_{c0}(2P)$ identification.

To resolve this puzzle, Chen {\it et al.} proposed that the observed $Z(3930)$ enhancement might contain contributions from both $\chi_{c0}(2P)$ and $\chi_{c2}(2P)$~\cite{Chen:2012wy}. Since both states can be produced via $\gamma\gamma$ fusion and decay into $D\bar{D}$, they may appear as a single overlapping structure in the invariant mass spectrum due to their close masses.

\begin{figure}
\centering
\begin{tabular*}{0.66\textwidth}{@{\extracolsep{\fill}}cc}
\includegraphics[height=0.1\textheight]{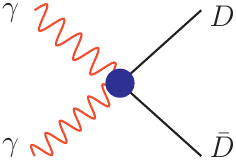}&
\includegraphics[height=0.1\textheight]{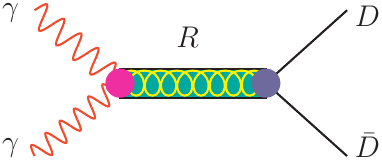}\\
(a)&(b)\\
\end{tabular*}
\caption{Feynman diagrams for (a) the direct non-resonant process and (b) the resonant contributions via \(\chi_{cJ}(2P)\) in \(\gamma\gamma\to D\bar{D}\), adapted from Ref.~\cite{Chen:2012wy}.}
\label{fig:gg-DD}
\end{figure}

Their analysis considered two mechanisms for $\gamma\gamma\to D\bar{D}$, illustrated in Fig.~\ref{fig:gg-DD}: (a) a non-resonant (NOR) direct process providing a smooth background, and (b) resonant contributions via intermediate states $R = \chi_{c0}(2P), \chi_{c2}(2P)$. The total amplitude was constructed as
\begin{equation}
\mathcal{M}_{\mathrm{Total}} = \mathcal{A}_{\mathrm{NOR}} + e^{i\phi_0} \mathcal{A}_{\chi_{c0}(2P)} + e^{i\phi_2} \mathcal{A}_{\chi_{c2}(2P)},
\end{equation}
where $\phi_0$ and $\phi_2$ account for possible interference phases.

By performing a combined fit to the \(D\bar{D}\) invariant mass spectrum and the \(\cos\theta^*\) angular distribution from Belle data, Chen {\it et al.} extracted the resonance parameters for both states. The best-fit results yield:
\begin{equation}
\begin{split}
m_{\chi_{c0}(2P)} &= 3.920 \pm 0.007~\mathrm{GeV}, \quad \Gamma_{\chi_{c0}(2P)} = 8.065 \pm 9.663~\mathrm{MeV}, \\
m_{\chi_{c2}(2P)} &= 3.942 \pm 0.003~\mathrm{GeV}, \quad \Gamma_{\chi_{c2}(2P)} = 11.980 \pm 6.953~\mathrm{MeV}.
\end{split}
\end{equation}
These masses are remarkably close, separated by only about $20~\mathrm{MeV}$, and both widths are narrow, consistent with expectations for $2P$ charmonia.

\begin{figure}
\centering
\begin{tabular*}{0.75\textwidth}{@{\extracolsep{\fill}}cc}
\includegraphics[width=0.4\textwidth]{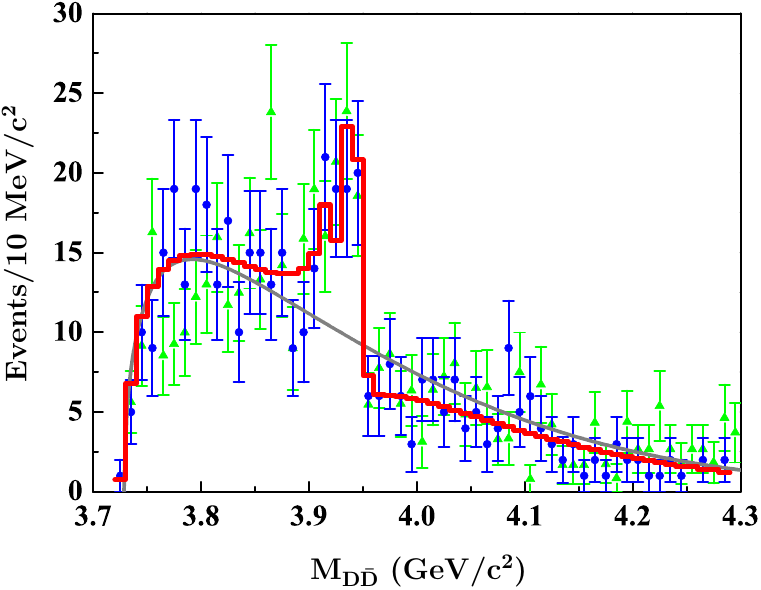}&
\includegraphics[width=0.4\textwidth]{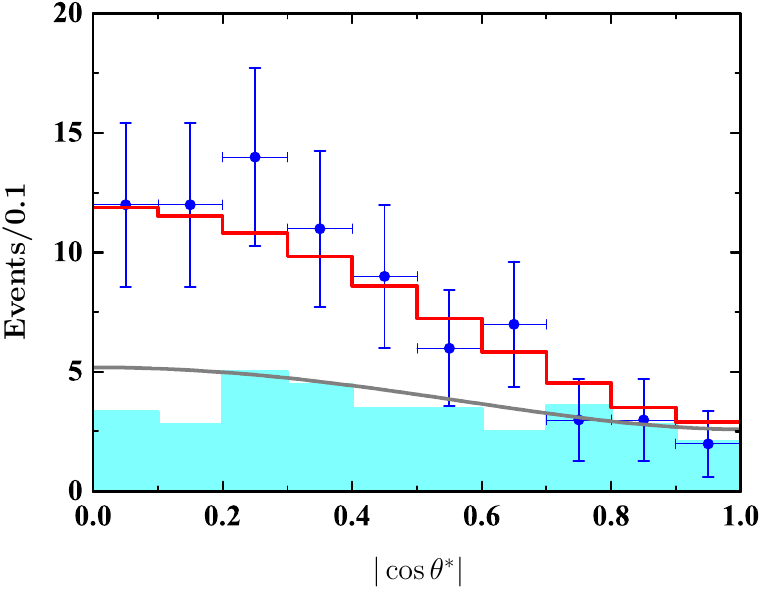}\\
\hspace{1em}(a)&\hspace{1.5em}(b)
\end{tabular*}
\caption{(a) Fitted \(D\bar{D}\) invariant mass spectrum (red histogram) compared with Belle (blue dots) and BaBar (green triangles) data. (b) Fitted \(\cos\theta^*\) distribution (red histogram) compared with Belle data (blue dots) and background (cyan histogram). Figures from Ref.~\cite{Chen:2012wy}.}
\label{fig:mDD-theta}
\end{figure}

The fitting results, shown in Fig.~\ref{fig:mDD-theta}, demonstrate that including both $\chi_{c0}(2P)$ and $\chi_{c2}(2P)$ provides a good description of the experimental data, including the steep drop near $3.95~\mathrm{GeV}$ in the invariant mass spectrum and the $\cos\theta^*$ angular distribution. Although the fitted coupling strength for $\chi_{c0}(2P)$ is numerically smaller than that for $\chi_{c2}(2P)$, its contribution remains significant due to differing dimensional factors. The calculated cross-section ratio $\sigma(\chi_{c2})/\sigma(\chi_{c0}) =1.447$ further supports the necessity of including both resonances.

Thus, the observed $Z(3930)$ enhancement is likely not a single resonance but an overlapping structure of $\chi_{c0}(2P)$ and $\chi_{c2}(2P)$. Their near-degenerate masses make them difficult to resolve in existing data, explaining why only one broad structure was reported. This interpretation resolves the puzzle of the missing $\chi_{c0}(2P)$ signal and provides a coherent picture of the $P$-wave charmonium spectrum. Future high-statistics experiments at Belle II and other facilities will be crucial to test this hypothesis through more precise angular analyses and possibly direct observation of the two separated states as indicated in Ref. \cite{Chen:2012wy} explicitly. \changelabel{ Due to the complex spectral structure around 3.93 GeV, unambiguous assignment of the observed states remains challenging. Besides the above assignment, the authors in Ref.~\cite{Zhou:2015uva} proposed that both $X(3915)$ and $Z(3930)$ may correspond to the same $\chi_{c2}(2P)$ state. In Ref.~\cite{Baru:2017fgv}, the authors also proposed that if the $X(3915)$ is a $D^{(*)}\bar{D}^{(*)}$ hadronic molecule, it may be a scalar state, and if the $X(3915)$ is dominated by the helicity-0 contribution of the nearby tensor state, it may be other exotic interpretations beyond both the regular quarkonia and $D^*\bar{D}^*$ spin partner of the $X(3872)$.}

\subsubsection{The experimental measurement of $B\to KD\bar{D}$ from the LHCb Collaboration}

\begin{figure}
    \centering
    \includegraphics[width=0.8\textwidth]{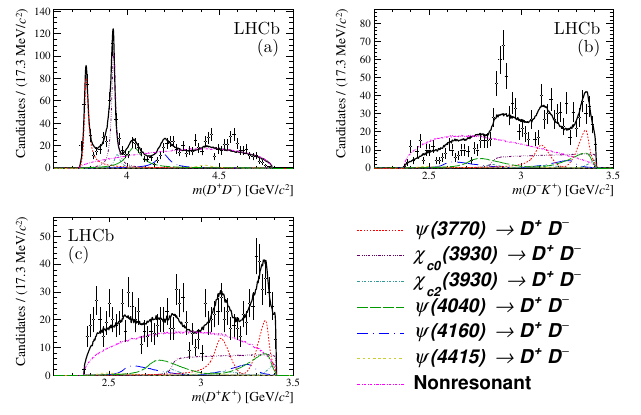}
    \caption{The invariant mass spectra of $D^+D^-$ (a), $D^-K^+$ (b), and $D^+K^+$ (c) in $B^+\to D^+D^-K^+$ process from LHCb~\cite{LHCb:2020pxc}.}
    \label{fig:LHCb_BDDK}
\end{figure}

The decay $B^+ \to D^+ D^- K^+$ proceeds via a $\bar{b} \to c\bar{c}\bar{s}$ transition and offers a clean environment to study charmonium spectroscopy due to low background levels. Resonances in the $D^-K^+$ system would be manifestly exotic, possessing minimal quark content $\bar{c}d u \bar{s}$ (or $c\bar{d} u \bar{s}$ for $D^+K^+$), i.e., open charm and strangeness. While many exotic hadrons containing hidden charm ($c\bar{c}$) have been observed, no significant evidence for exotic states with open flavour existed prior to LHCb's studies.

The LHCb Collaboration reported an analysis of the $B^+ \to K^+ D^+ D^-$ process, with invariant mass spectra shown in Fig.~\ref{fig:LHCb_BDDK}~\cite{LHCb:2020pxc,LHCb:2020bls}. The key results are:
\begin{enumerate}
\item The amplitude analysis confirms several established charmonium resonances in the $D^+D^-$ channel, including $\psi(3770)$, $\psi(4040)$, $\psi(4160)$, and $\psi(4415)$. Notably, both spin-0 and spin-2 components are required in the $\chi_{cJ}(3930)$ region:
 \begin{eqnarray*}
        \chi_{c0}(3930): & M = 3.9238 \pm 0.0015 \pm 0.0004\,\text{GeV}, & \Gamma = 17.4 \pm 5.1 \pm 0.8\,\text{MeV}, \\
        \chi_{c2}(3930): & M = 3.9268 \pm 0.0024 \pm 0.0008\,\text{GeV}, & \Gamma = 34.2 \pm 6.6 \pm 1.1\,\text{MeV}.
    \end{eqnarray*}
\item The analysis requires two new resonances in the $D^-K^+$ system to describe the structure near $2.9,\text{GeV}$:
\begin{eqnarray*}
        X_0(2900): & M = 2.8663 \pm 0.0065 \pm 0.0020\,\text{GeV}, & \Gamma = 57.2 \pm 12.2 \pm 4.1\,\text{MeV}, \\
        X_1(2900): & M = 2.9041 \pm 0.0048 \pm 0.0013\,\text{GeV}, & \Gamma = 110.3 \pm 10.7 \pm 4.3\,\text{MeV}.
    \end{eqnarray*}
Based on their decay patterns, $X_0(2900)$ and $X_1(2900)$ were identified as singly charmed tetraquark candidates~\cite{Yu:2017pmn,Liu:2020orv,Zhang:2020oze,He:2020jna,Chen:2020aos,Lu:2020qmp,Huang:2020ptc,Wang:2020xyc,Hu:2020mxp,Karliner:2020vsi,Agaev:2020nrc,Yang:2020atz,Chen:2020eyu,Mutuk:2020igv,Burns:2020xne,Dong:2020rgs,Xiao:2020ltm,Tan:2020cpu,An:2020vku,Wang:2020prk,Abreu:2020ony,Yang:2021izl,Qi:2021iyv,Albuquerque:2021svg,Wang:2021lwy,Chen:2021tad}, offering new insights into exotic hadrons.
\item No obvious resonant structures are seen in the $D^+K^+$ spectrum.
\end{enumerate}

As reported by LHCb~\cite{LHCb:2020pxc,LHCb:2020bls}, the measured properties of $\chi_{c0}(3930)$ and $\chi_{c2}(3930)$ are very close to those of the previously observed $X(3915)$ and $Z(3930)$, respectively. This is consistent with theoretical predictions that the $\chi_{c0}(2P)$ and $\chi_{c2}(2P)$ states should have close masses and narrow widths~\cite{Duan:2020tsx,Liu:2009fe}. Prior to the LHCb experiment, $X(3915)$ had not been observed in the $D\bar{D}$ decay channel. LHCb's work filled this gap, providing key evidence that ultimately supported identifying $X(3915)$ as the $\chi_{c0}(2P)$ state. Consequently, in the 2022 edition of the PDG~\cite{ParticleDataGroup:2022pth}, $X(3915)$ was listed once again as the $\chi_{c0}(2P)$ charmonium state.

\section{$Y$ Problem and its solution}\label{sec:Y}
\label{section4}

\begin{figure}[htbp]
\centering
\includegraphics[width=\textwidth]{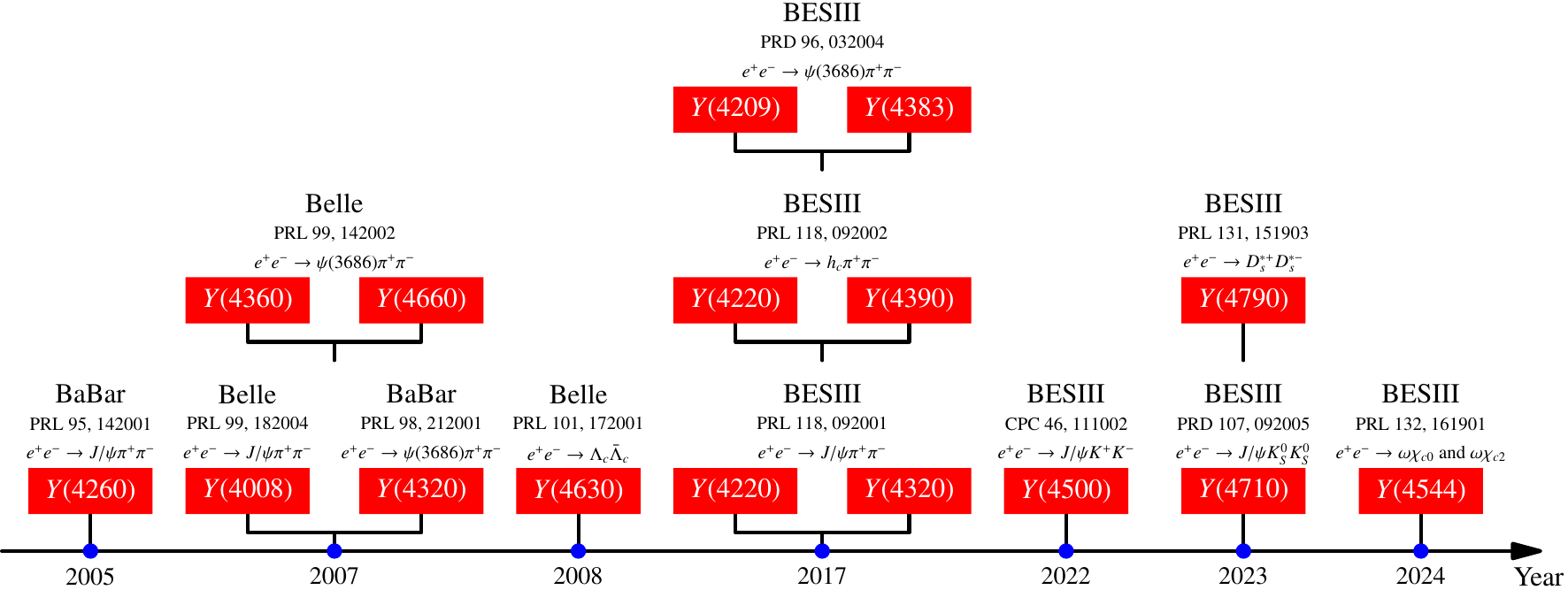}
\caption{The observations of charmonium-like $Y$ states. The data are obtained from Refs.~\cite{BaBar:2005hhc,Belle:2007dxy,BaBar:2006ait,Belle:2007umv,Belle:2008xmh,BESIII:2016adj,BESIII:2016bnd,BESIII:2017tqk,BESIII:2022joj,BESIII:2022kcv,BESIII:2023wsc,BESIII:2024jzg}.}
\label{fig:ObservedY}
\end{figure}

Since the discovery of $X(3872)$ by the Belle collaboration in 2003~\cite{Belle:2003nnu}, an increasing number of $Y$ states have been observed in experiments. Among them, the $Y$ states produced directly via $e^+e^-$ annihilation form a particularly important subgroup. Their current status is summarized in Fig.~\ref{fig:ObservedY}. These observations not only enrich the family of charmonium-like $XYZ$ states~\cite{Klempt:2007cp,Brambilla:2010cs,Liu:2013waa,Hosaka:2016pey,Richard:2016eis,Chen:2016qju,Esposito:2016noz,Chen:2016spr,Lebed:2016hpi,Guo:2017jvc,Olsen:2017bmm,Ali:2017jda,Liu:2019zoy,Brambilla:2019esw,Chen:2022asf,Meng:2022ozq,Wang:2021aql,Wang:2025dur}, but also offer an opportunity to re-examine the $J/\psi$ family, especially the construction of its high-lying states. This, in turn, helps deepen our understanding of the non-perturbative behavior of the strong interaction.

The $Y(4260)$ stands out prominently in the $XYZ$ family, as it was the first $Y$ state discovered in this context~\cite{BaBar:2005hhc}. Various resonant interpretations have been proposed, including assignments as a conventional charmonium, a multiquark state, or a hybrid meson~\cite{Chen:2016qju}. However, these explanations face a significant challenge: the $Y(4260)$ is not observed in the measured $R$ values or in open-charm decay channels~\cite{Chen:2016qju}. Consequently, a non-resonant interpretation was suggested, in which the $Y(4260)$ structure can be reproduced through the interference between the charmonia $\psi(4160)$ and $\psi(4415)$ \cite{Chen:2010nv}. This picture has also been applied to describe the $Y(4360)$ structure observed in $e^+e^-\to \psi(3686)\pi^+\pi^-$~\cite{Chen:2011kc}.

The year 2017 marked an important watershed with the accumulation of more precise experimental data. The BESIII collaboration performed high-precision measurements of the cross section for $e^{+}e^{-} \to J/\psi \pi^{+}\pi^{-}$~\cite{BESIII:2016bnd}. Their results revealed that the previously identified $Y(4260)$ actually consists of two substructures, now denoted as $Y(4220)$\footnote{This state is also referred to as $Y(4230)$ in some experimental and theoretical analyses, reflecting the slightly different central values obtained from various measurements and fitting schemes. In this review, we adopt the notation $Y(4220)$.} and $Y(4320)$~\cite{BESIII:2016bnd}. At the same time, the $Y(4008)$ state previously reported by Belle in the same process was not confirmed by the more precise BESIII data~\cite{Belle:2007dxy,BESIII:2016bnd}. As a result, the designation $Y(4260)$ is no longer used in the PDG listings~\cite{ParticleDataGroup:2024cfk}, with $Y(4220)$ being considered the relevant state. To date, ten $Y$ states have been reported experimentally: $Y(4220)$, $Y(4320)$, $Y(4360)$, $Y(4380)$, $Y(4390)$, $Y(4500)$, $Y(4630)$, $Y(4660)$, $Y(4710)$, and $Y(4790)$ (see Fig.~\ref{fig:ObservedY})~\cite{BESIII:2016bnd,BESIII:2016adj,BESIII:2017tqk,BESIII:2017dxi,BESIII:2018iea,BESIII:2019gjc,BESIII:2020bgb,BESIII:2020oph,Ablikim:2020jrn,BESIII:2020tgt,BESIII:2022joj,BESIII:2023cmv,BESIII:2021njb,BESIII:2022qal,BESIII:2022kcv,BESIII:2023tll,BESIII:2024yqi,BESIII:2025bce,BESIII:2026hie,BESIII:2026oyx,BESIII:2026lcb,BESIII:2026ont}.

The abundance of phenomena associated with these $Y$ states presents a challenging yet promising issue: how to understand their nature. This has become known as the "\textit{Y} problem", as highlighted in {\it White Paper on the Future Physics Programme of BESIII}~\cite{BESIII:2020nme}. Addressing this problem requires not only high-precision measurements and comprehensive experimental input to uncover the properties of these states but also theoretical breakthroughs.

In the present review, we examine how the unquenched picture provides a unified framework for addressing the $Y$ problem.

\subsection{$Y(4260)$ and its evolving interpretations}

\subsubsection{Experimental timeline and key results}

\begin{figure}[htbp]
    \centering
    \includegraphics[width=0.7\textwidth]{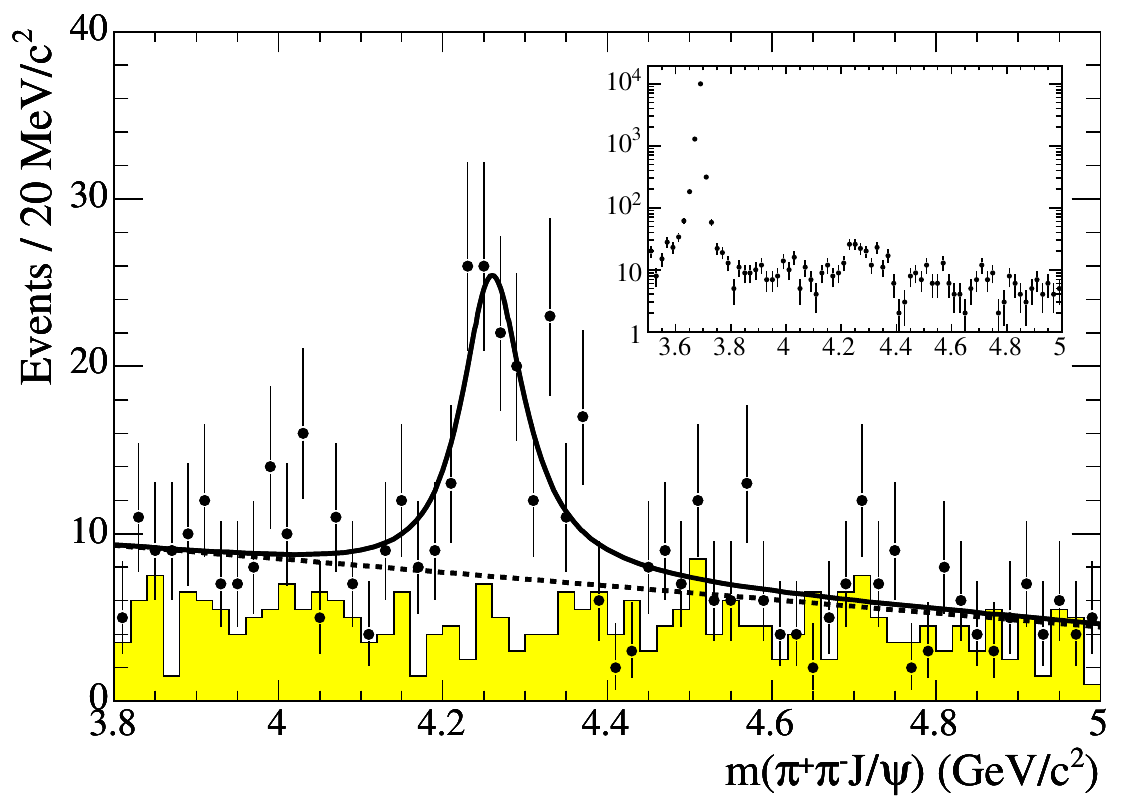}
    \caption{The observed $Y(4260)$ in the $e^{+}e^{-} \to J/\psi \pi^{+}\pi^{-}$ process from the BaBar Collaboration~\cite{BaBar:2005hhc}.}
    \label{fig:BaBar_pipiJpsigamma}
\end{figure}

The experimental journey of $Y(4260)$ resonance commenced in 2005 with the BaBar collaboration's analysis of initial-state radiation (ISR) events using 233 fb$^{-1}$ of data collected at the $\Upsilon(4S)$ resonance. Studying the process $e^+e^- \to \gamma_{\text{ISR}} \pi^+ \pi^- J/\psi$, they observed a clear accumulation of events near 4.26 GeV (see Fig. \ref{fig:BaBar_pipiJpsigamma}) in the $\pi^+\pi^-J/\psi$ invariant mass spectrum~\cite{BaBar:2005hhc}. A fit with a single Breit-Wigner function described this structure, yielding $125 \pm 23$ signal events, a mass of $M = 4259 \pm 8^{+2}_{-3}$ MeV, and a width of $\Gamma = 88 \pm 23^{+6}_{-5}$ MeV~\cite{BaBar:2005hhc}. The product $\Gamma_{e^+e^-} \times BR(Y(4260) \to \pi^+\pi^-J/\psi)$ was measured to be $5.5 \pm 1.0 ^{+0.8}_{-0.7}$ eV. Checks of the $m^2_{\text{Rec}}$ distribution and the fraction of events with a detected ISR photon confirmed the ISR production mechanism, firmly establishing the quantum numbers as $J^{PC} = 1^{--}$. BaBar designated this new structure $Y(4260)$. While the data were adequately described by a single resonance, the BaBar Collaboration also explored a two-resonance hypothesis with interference, finding suggestive but statistically inconclusive solutions involving a second state near 4.33 GeV.

The need for independent confirmation was met promptly by the CLEO collaboration in 2006~\cite{CLEO:2006tct}. Using a different data set of 13.3 fb$^{-1}$ collected at the $\Upsilon(1S-4S)$ resonances with the CLEO III detector at CESR, they performed a similar ISR search~\cite{CLEO:2006tct}. A clear signal around 4.26 GeV was observed with low background. A single-resonance fit gave a mass of $M = 4284^{+17}_{-16} \pm 4$ MeV, a width of $\Gamma = 73^{+39}_{-25} \pm 5$ MeV, and $\Gamma_{e^+e^-} \times BR(Y(4260) \to \pi^+\pi^-J/\psi)= 8.9^{+3.9}_{-3.1} \pm 1.8$ eV~\cite{CLEO:2006tct}, consistent with BaBar's results~\cite{BaBar:2005hhc} and providing a solid $5.4\sigma$ confirmation. In a complementary study using data from direct $e^+e^-$ annihilation at $\sqrt{s} = 4.26$ GeV, CLEO extended the understanding of the $Y(4260)$'s decay properties~\cite{CLEO:2006ike}. They not only confirmed the $\pi^+\pi^-J/\psi$ mode at $11\sigma$ but also reported the first observation of $Y(4260) \to \pi^0\pi^0J/\psi$ ($5.1\sigma$) and evidence for $Y(4260) \to K^+K^-J/\psi$ ($3.7\sigma$). The measured Born cross sections were $\sigma(\pi^+\pi^-J/\psi) = 58^{+12}_{-10} \pm 4$ pb, $\sigma(\pi^0\pi^0J/\psi) = 23^{+12}_{-8} \pm 1$ pb, and $\sigma(K^+K^-J/\psi) = 9^{+9}_{-5} \pm 1$ pb~\cite{CLEO:2006ike}. These findings demonstrated the resonance's significant coupling to hidden-charm final states with light hadrons, with a dipion mass spectrum consistent with an S-wave phase space model.

The narrative took a more complex turn with the high-precision measurement from the Belle Collaboration in 2007. Utilizing a much larger data sample of 548 fb$^{-1}$ collected at the $\Upsilon(4S)$, Belle measured the $e^+e^- \to \pi^+\pi^- J/\psi$ cross section via ISR from $\sqrt{s} = 3.8$ to 5.5 GeV~\cite{Belle:2007dxy}. Their results overwhelmingly confirmed the $Y(4260)$ peak near 4.25 GeV (significance $>15\sigma$) but also revealed a significant, previously unresolved cluster of events near 4.05 GeV (significance $>8\sigma$). An unbinned maximum likelihood fit to the mass spectrum showed that a model with two coherent Breit-Wigner resonances provided a significantly better description of the data than a single resonance, particularly for the lower-mass side of the 4.25 GeV enhancement. The fit yielded two solutions with interference. The parameters for the lower resonance ($R1$) were $M = 4008 \pm 40 ^{+114}_{-28}$ MeV, $\Gamma = 226 \pm 44 \pm 87$ MeV; and for the higher resonance ($R2$, associated with $Y(4260)$) $M = 4247 \pm 12 ^{+17}_{-32}$ Me, $\Gamma = 108 \pm 19 \pm 10$ MeV~\cite{Belle:2007dxy}. Belle noted that $R1$'s parameters did not match those of the known $\psi(4040)$, hinting at a more intricate underlying structure. A single-resonance fit, for comparison, yielded parameters consistent with the earlier BaBar~\cite{BaBar:2005hhc} and CLEO~\cite{CLEO:2006tct,CLEO:2006ike} results.

In summary, experimental studies from 2005 to 2007 firmly established  $Y(4260)$ resonance with $J^{PC}=1^{--}$. Its defining feature was a surprisingly strong coupling to hidden-charm final states such as $\pi\pi J/\psi$ and $K\bar{K}J/\psi$, while it remained difficult to observe in open-charm and inclusive hadronic channels. Although interpretations within the conventional (quenched) charmonium picture were explored, this distinctive decay pattern ultimately challenged such descriptions. This motivated the introduction of exotic state explanations, a topic reviewed in the following section.

\subsubsection{Resonant explanations and confronted difficulties} \label{sec:Y-1.2}

Several theoretical explanations have been proposed in response to the reported $Y(4260)$ signal; the main interpretations are summarised below (further details can be found in review articles, e.g., \cite{Brambilla:2010cs,Liu:2013waa,Hosaka:2016pey,Richard:2016eis,Chen:2016qju,Esposito:2016noz,Chen:2016spr,Lebed:2016hpi,Guo:2017jvc,Olsen:2017bmm,Ali:2017jda,Liu:2019zoy,Brambilla:2019esw,Chen:2022asf,Meng:2022ozq,Wang:2025clb}).

\begin{itemize}
    \item  Charmonium explanations: early phenomenological attempts to place $Y(4260)$ within the conventional charmonium spectrum. For example, Zhang suggested the $3^{3}D_{1}$ state assignment using Regge trajectory combined with hyperfine splitting relations~\cite{Zhang:2006td}. In Ref.~\cite{Llanes-Estrada:2005qvr},  Llanes-Estrada interpreted $Y(4260)$ as a $4S$ charmonium state, although their calculated mass of 4347 MeV for the $4S$ state is obviously larger than that of $Y(4260)$, and argued that $S$–$D$-wave interference could explain the absence of the $Y(4260)$ signal in the $R$-value scan.   Subsequently, several refined analyses of the mass spectrum and decay properties reach different conclusions~\cite{Eichten:2005ga,Segovia:2008zz}. Studies based on the Cornell coupled channel model concluded that the $2^{3}D_{1}$ assignment is incompatible with the properties of  $Y(4260)$~\cite{Eichten:2005ga}. In a constituent quark model study, Segovia \emph{et al.}~\cite{Segovia:2008zz} identified  $Y(4360)$ as the $4S$ charmonium and the $\psi(4415)$ as the $3D$ state, while finding no natural place for $Y(4260)$ within the conventional charmonium spectrum. Further analysis incorporating $\pi\pi$/$K\bar K$ final state interactions revealed a two-pole structure with strong coupling to $\omega\chi_{c0}$~\cite{Dai:2012pb}, suggesting $Y(4260)$ as a confining state with either a hybrid meson or a $3D$ charmonium component. Recent constituent quark model calculations with a Cornell-like potential likewise conclude that $Y(4260)$ does not fit the conventional meson picture~\cite{Zhao:2023hxc}. In fact, the spectroscopic overpopulation problem associated with $Y(4260)$ can be partially alleviated by the inclusion of unquenching effects. Comparisons between coupled channel approach and screened potential models confirm that both incorporate vacuum polarization from light quark pairs and yield similar global spectral features~\cite{Li:2009ad}. Within the screened potential framework~\cite{Li:2009zu}, Li and Chao predicted the mass of $\psi(4S)$ to be around 4.27~GeV, close to $Y(4260)$, a result also supported by calculations using a Martin-like potential~\cite{Shah:2012js}. Furthermore, the experimental observation that $Y(4260)$ splits into two substructures, $Y(4220)$ and $Y(4320)$ \cite{BESIII:2016bnd}, adds another layer of complexity to its assignment within the conventional charmonium spectrum.

    \item The molecular state explanations: a prominent exotic interpretation for  $Y(4260)$ is the hadronic molecular picture, motivated by the general expectation that near-threshold dynamics can generate bound states. Early proposals considered a variety of constituent combinations, such as $\rho\chi_{c1}$~\cite{Chiu:2005ey,Liu:2005ay}, $\omega\chi_{c1}$~\cite{Yuan:2005dr}, and $D_1\bar D$~\cite{Swanson:2005tq}. A distinctive scenario regarded the state as a $\Lambda_c\bar\Lambda_c$ baryononium~\cite{Qiao:2005av}, where the binding does not require color-singlet constituents and could reach several hundred MeV, representing a non-typical near-threshold molecule. With accumulating data, attention focused on the $S$-wave $D_1 \bar D$ threshold, which naturally admits $J^{PC}=1^{--}$. In this picture, $Y(4260)$ is viewed as an isospin-singlet $D_1\bar D + \text{c.c.}$ molecule~\cite{Swanson:2005tq,Ding:2008gr,Close:2009ag,Close:2010wq,Wang:2013cya,Cleven:2013mka,Wang:2013kra,Li:2013bca}, where the narrow $D_1(2420)$ belonging to the heavy-quark spin-symmetry multiplet with light angular momentum $j_l=3/2$ plays a central role. Quantitative coupled channel studies have examined the $D_1\bar D$ interaction and its coupling to related channels such as $D_0\bar D^*$~\cite{Ding:2008gr}. Analyses of the line shapes in $e^+e^- \to J/\psi\pi^+\pi^-$, $h_c\pi^+\pi^-$, and $D\bar D$, which incorporate $D_1\bar D$ scattering dynamics, extract a pole position around $4217\pm2 \text{MeV}$~\cite{Cleven:2013mka,Wang:2013kra}, consistent with  $Y(4260)$. Further combined analyses based on this assignment assumption have been presented in Refs.~\cite{Xue:2017xpu,Qin:2016spb}.

    \item Hidden-charm compact tetraquark: the compact tetraquark picture, often formulated in terms of constituent diquark-antidiquark degrees of freedom, was proposed early as a popular exotic explanation for $Y(4260)$. In such models, the mass spectrum is organized by spin-spin and spin-orbit interactions within and between the (anti)diquarks, giving rise to multiplets of states with various quantum numbers~\cite{Maiani:2005pe,Drenska:2009cd,Maiani:2014aja,Ali:2017wsf}. $Y(4260)$ has been frequently identified as a vector member of a $[cs][\bar c\bar s]$ tetraquark multiplet~\cite{Maiani:2005pe,Drenska:2009cd,Maiani:2014aja}. Based on the mass relation of Eq. (3.1) without tensor-force in Ref. \cite{Maiani:2005pe}, the spectroscopy of $P$-wave $[cs][\bar{c}\bar{s}]$ tetraquark has been updated by two different assignment scenarios~\cite{Cleven:2015era}, which assumed that $Y(4008)$/$Y(4260)$ and $Y(4260)$/$Y(4360)$ are two lowest vector tetraquark states, respectively.   
It should be emphasized that compact $[cs][\bar{c}\bar{s}]$ tetraquark interpretation of $Y(4260)$ was not supported in other theoretical frameworks. Relativistic quark model calculations based on a quasipotential formalism~\cite{Ebert:2005nc,Ebert:2008kb}, which systematically treats the diquark–antidiquark interaction via a one-gluon-exchange plus confining kernel, predict vector $[cs][\bar c\bar s]$ states substantially heavier (by about 200 MeV) than $Y(4260)$. These works instead suggest that a $[cq][\bar c\bar q]$ configuration could be a more natural candidate~\cite{Ebert:2005nc,Ebert:2008kb}, although such models tend to predict a much richer spectrum of $1^{--}$ tetraquarks than currently observed. Similarly, QCD sum rule analyses using interpolating currents for $1^{--}$ hidden-charm tetraquarks consistently yield mass estimates around 4.6-4.7 GeV for a wide range of continuum-threshold and Borel-window choices, disfavoring the identification of $Y(4260)$ with a lowest-lying compact tetraquark state~\cite{Chen:2010ze}.

\item $c\bar{c}g$ hybrid: the charmonium hybrid interpretation, in which a $c\bar{c}$ pair is coupled to an explicit gluonic excitation, provides a distinctive framework for exotic vector states. A characteristic theoretical feature is the selection rule for heavy hybrids~\cite{Zhu:1998sv,Zhu:1999wg,Close:1994hc,Close:2003mb,Kou:2005gt}: decays of a vector hybrid meson into two $S$-wave charmed mesons (e.g., $D^{(*)}\bar D^{(*)}$, $D_s^{(*)}\bar D_s^{(*)}$) are strongly suppressed, while channels involving at least one orbitally excited charmed meson (such as $D_1\bar D^{(*)}$) are favoured. Early phenomenological and lattice studies anticipated the lightest vector charmonium hybrid near 4.0-4.2 GeV~\cite{Zhu:2005hp,Close:2005iz,Barnes:1995hc,Merlin:1986tz,Lacock:1996ny,Manke:1998yg,Juge:1999aw,Drummond:1999db,Manke:1999ru}.
The Hadron Spectrum Collaboration, using a large basis of interpolators, identified a hybrid supermultiplet containing a $1^{--}$ state within $m_\pi\approx400$ MeV~\cite{HadronSpectrum:2012gic}, in which the mass difference $M_{1^{--}}-M_{\eta_c}\simeq1.3$ GeV agreed with $Y(4260)$, but subsequent studies with lighter pion masses ($m_\pi\approx240$ MeV) shift the mass upward to $\sim$4.4 GeV~\cite{Cheung:2016bym}. Independent quenched lattice studies (e.g., TWQCD~\cite{Chiu:2005ey}) also place the lightest vector hybrid about 250 MeV above $Y(4260)$. Additional insight was obtained from a lattice calculation employing nonlocal hybrid-like operators in which the $c\bar c$ pair and the chromomagnetic gluonic field are spatially separated~\cite{Chen:2016ejo}, which found a vector state at $4.33(2)$ GeV with a strongly suppressed leptonic width $\Gamma_{e^+e^-}<\mathcal{O}(40 ~\text{eV})$. Predictions from Born-Oppenheimer effective field theory place the lightest vector hybrid multiplet in the 4.0-4.3 GeV range~\cite{Berwein:2015vca}. Thus, although the hybrid picture offers an interesting explanation for $Y(4260)$, definitive support from lattice QCD remains ambiguous due to significant variations in the predicted mass across different methodologies and pion masses.
\end{itemize}

\begin{figure}[htbp]
\centering
\includegraphics[width=0.5\textwidth]{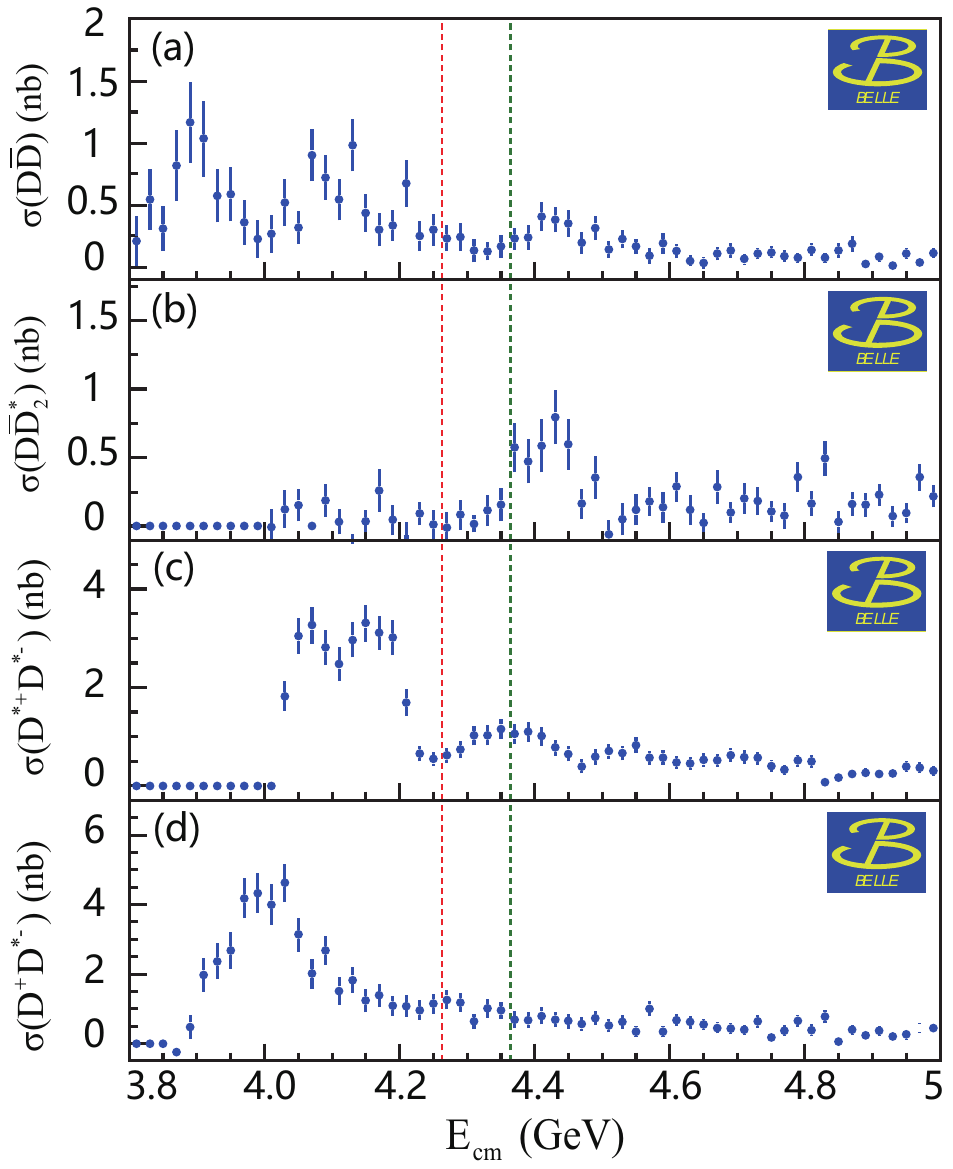}
\caption{The experimental cross sections of $e^+e^-$ annihilation from Belle Collaboration in $e^+e^-\to D\bar{D}$ (a)~\cite{Belle:2007qxm}, $e^+e^-\to D^0D^-\pi^+$ (b)~\cite{Belle:2007xvy}, $e^+e^-\to D^{*+}D^{*-}$ (c)~\cite{Belle:2006hvs}, and $e^+e^-\to D^+D^{*-}$ (d)~\cite{Belle:2006hvs} channels. The vertical red and green dashed lines represent the central masses of $Y(4260)$ and $Y(4360)$, respectively. The figure is taken from Ref.~\cite{Chen:2016qju}.}
\label{fig:belle_open_charm}
\end{figure}

\begin{figure}[htbp]
\centering
\includegraphics[width=0.9\textwidth]{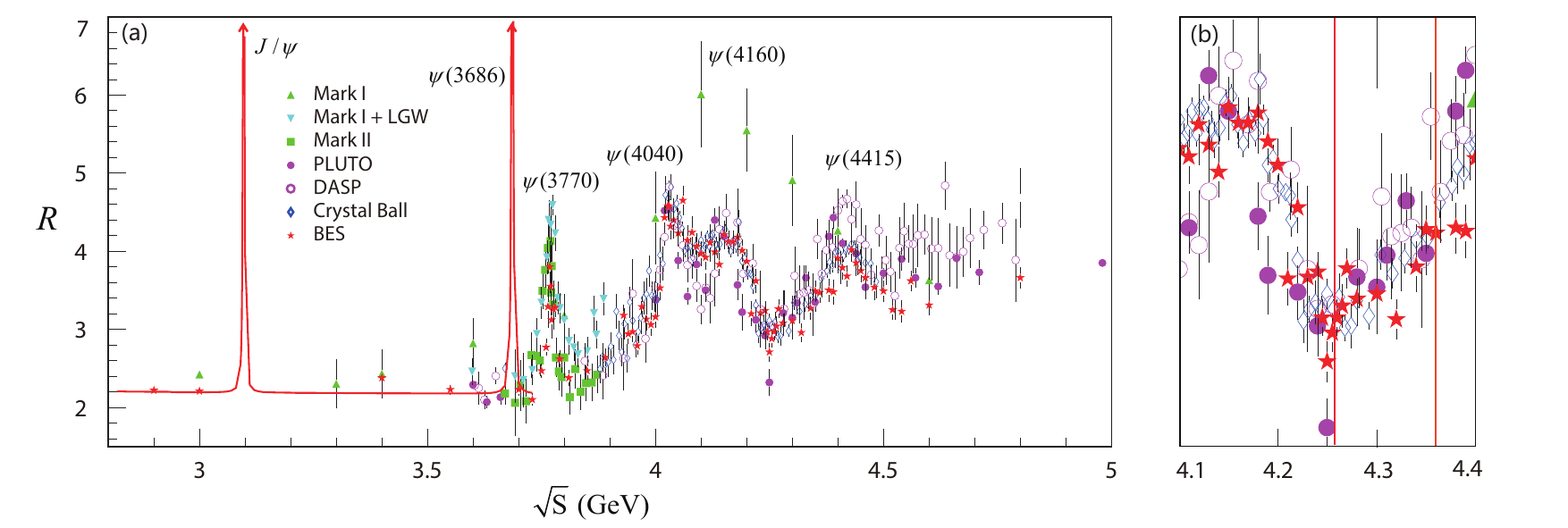}
\caption{The experimental $R$ value taken from the PDG~\cite{ParticleDataGroup:2014cgo} (a) and the detailed data in the range of $\sqrt{s}=4.1$$\sim$4.4 GeV (b). The vertical red lines in figure (b) represent the central masses of $Y(4260)$ and $Y(4360)$. The figure is taken from Ref.~\cite{Chen:2016qju}.}
\label{fig:rcc}
\end{figure}

Whether interpreting the observed $Y(4260)$ as a conventional charmonium state or as an exotic hadronic state (such as a $D_1\bar{D}$ molecule, a hidden-charm compact tetraquark, or a $c\bar{c}g$ hybrid), we face a serious problem: its hidden-charm decays cannot account for the total width of $Y(4260)$. As a resonance, its main decay channels should be open-charm modes. However, $Y(4260)$ is not observed in the measured cross sections for $e^+e^-$ annihilation into open-charm channels, such as $D\bar{D}$, $D\bar{D}^*+\text{c.c.}$, $D^*\bar{D}^*$, $D_s^+D_s^-$, $D_s^- D^{*+}_s$, $D_s^{*+}D_s^{*-}$, and $D^*\bar{D}^*\pi$~\cite{BaBar:2006qlj,Belle:2006hvs,Belle:2007xvy,Belle:2009dus,CLEO:2008ojp,BaBar:2009elc,BaBar:2010plp} (see Fig. \ref{fig:belle_open_charm}).
More importantly, $Y(4260)$ does not produce a noticeable enhancement in the measured $R$ value, as shown in Fig.~\ref{fig:rcc}. Typically, vector states such as the $\rho$, $\phi$, and $J/\psi$ exhibit clear peaks in the $R$ value. In Fig. \ref{fig:rcc}, we present the details of $R$ value around 4.2 GeV.

How to explain this puzzling phenomenon remains a major challenge when considering $Y(4260)$ as a resonance.

\subsubsection{Non-resonant picture: concept and applications}

Contrary to expectations for conventional charmonium states above open-charm threshold, neither $Y(4260)$ nor the $Y(4360)$ has been observed in exclusive open-charm decays (e.g., to $D\bar{D}$, $D^*\bar{D}$)~\cite{BaBar:2006qlj,Belle:2006hvs,Belle:2007xvy,Belle:2009dus,CLEO:2008ojp,BaBar:2009elc,BaBar:2010plp} or in the hadronic $R$-value scan~\cite{CLEO:2008ojp,PLUTO:1976jbe,DASP:1978dns,Siegrist:1981zp,BES:1999wbx,BES:2009ejh}. However, the absence in key channels poses a serious challenge to many of these resonant models. In response, a conceptually distinct explanation has been developed—the non-resonant picture—which attributes the observed peaks not to new particles but to the interference of known production amplitudes.

The core concept of the non-resonant picture is elegantly straightforward. For processes such as $e^+e^- \to J/\psi \pi^+ \pi^-$~\cite{BaBar:2005hhc} and $e^+e^- \to \psi(3686) \pi^+ \pi^-$~\cite{BaBar:2006ait}, the total production amplitude can receive contributions from distinct mechanisms. The first is a direct, non-resonant production where the virtual photon from $e^+e^-$ annihilation directly couples to the final state. The second proceeds through the formation of known intermediate resonances, such as the established charmonia $\psi(4160)$ and $\psi(4415)$, which subsequently decay into the same final state via hadronic loops. The key insight is that the coherent sum of these amplitudes, particularly when they possess a relative phase, can constructively and destructively interfere in the invariant mass ($\sqrt{s}$) spectrum. This interference can produce distinct peaks or enhancements that mimic resonant signals, even in the absence of poles in the scattering matrix. This mechanism provides a natural explanation for structures that appear prominently in specific final states (governed by the interfering pathways) but leave no imprint in other channels where the interfering terms are absent or negligible.

\begin{figure}[htbp]
\centering
\includegraphics[width=0.7\textwidth]{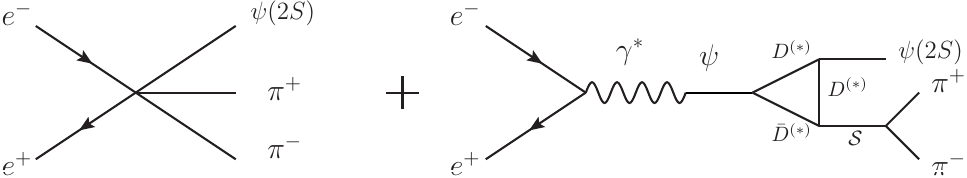}
\caption{The diagram depicting the direct production (left panel) and the production via intermediate charmonium (right panel). The figure is adapted from Ref. \cite{Chen:2011kc}.}
\label{fig:production}
\end{figure}

The work by Chen, He, and Liu provides concrete and quantitative realizations of this idea for both $Y(4260)$ \cite{Chen:2010nv} and the $Y(4360)$ \cite{Chen:2011kc}. Their model constructs the total amplitude $\mathcal{M}_{\text{tot}} = \mathcal{M}_{\text{NoR}} + e^{i\phi_1}\mathcal{A}_{\psi_1} + e^{i\phi_2}\mathcal{A}_{\psi_2}$, where $\mathcal{M}_{\text{NoR}}$ describes the direct $e^+e^- \to \text{hidden-charm} + \pi\pi$ production with an empirically parameterized form factor, and $\mathcal{A}_{\psi_i}$ represent the amplitudes for production via the intermediate states $\psi_1=\psi(4160)$ and $\psi_2=\psi(4415)$, as illusterated in Fig. \ref{fig:production}. The decay of these charmonia into the final state is calculated using a hadronic loop mechanism, which serves as a crucial non-perturbative bridge. In this framework, the charmonia couple to pairs of charmed mesons ($D^{(*)}\bar{D}^{(*)}$), which then rescatter via the exchange of charmed mesons to form the final $J/\psi$ or $\psi(3686)$ and a scalar meson ($\sigma$ or $f_0(980)$), which subsequently decays to $\pi^+ \pi^-$. This approach incorporates effective Lagrangians, form factors to account for off-shell effects and finite hadron size, and experimentally constrained couplings.

\begin{figure}[htbp]
\centering
\includegraphics[width=0.47\textwidth]{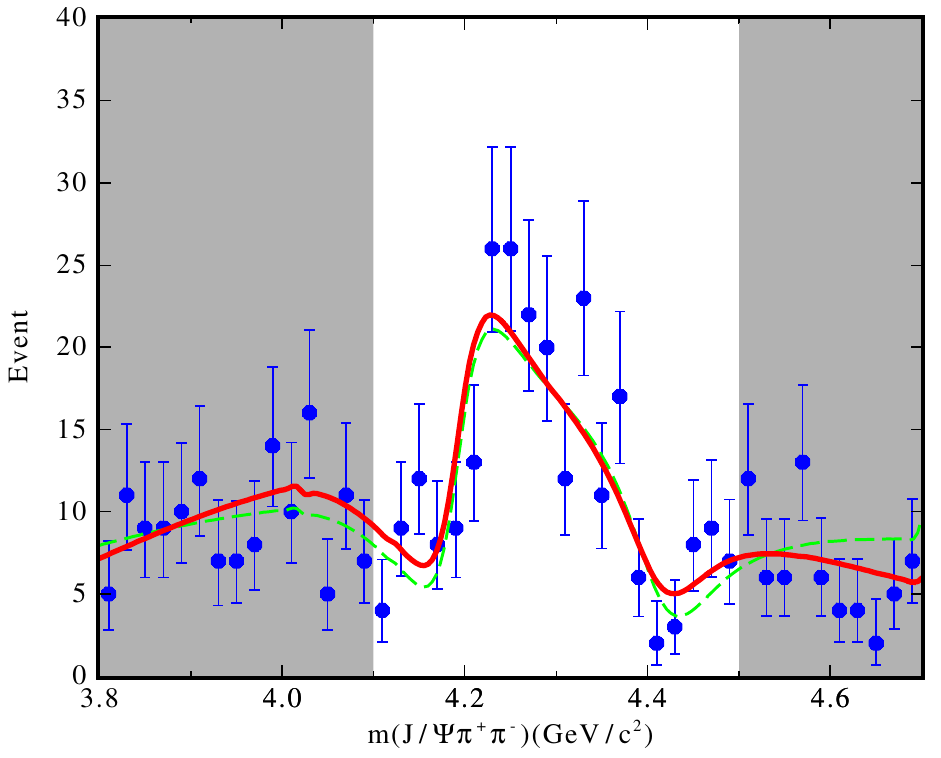}
\hspace{0.03\textwidth}
\raisebox{1ex}{\includegraphics[width=0.4\textwidth]{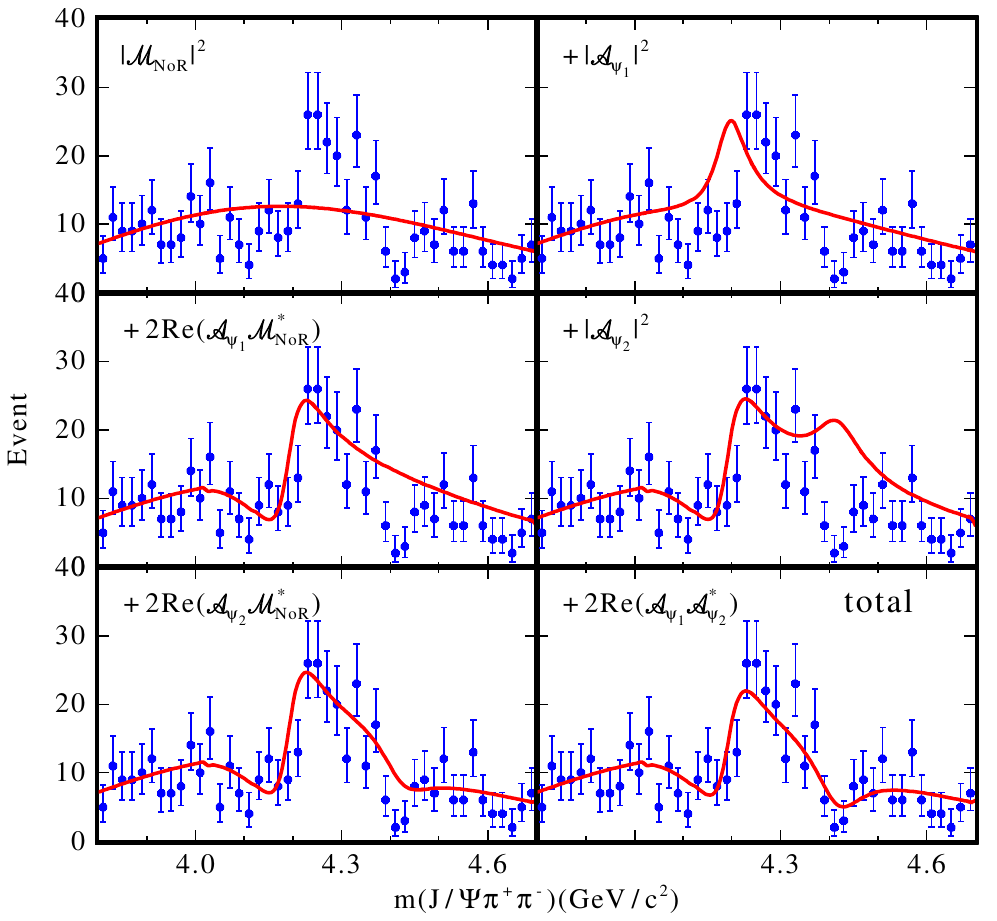}}
\caption{The fitted results of the $e^+e^-\to J/\psi\pi^+\pi^-$ events distribution (left panel), and the change of the line shape of the cross section by adding the contributions from the terms $|\mathcal{M}_{NoR}|^2$, $|\mathcal{A}_{\psi_1}|^2$,
$2\mathrm{Re}(\mathcal{A}_{\psi_1}\mathcal{M}_{NoR}^*)$,
$|\mathcal{A}_{\psi_2}|^2$,
$2\mathrm{Re}(\mathcal{A}_{\psi_2}\mathcal{M}_{NoR}^*)$, and
$2\mathrm{Re}(\mathcal{A}_{\psi_1}\mathcal{A}_{\psi_2}^*)$, step by step. The figure is adapted from Ref. \cite{Chen:2011kc}.}
\label{fig:4260}
\end{figure}

For the $Y(4260)$, the numerical analysis demonstrates that with a small set of free parameters (primarily coupling strengths and relative phases), the theoretical line shape yields an excellent fit to the $e^+e^- \to J/\psi \pi^+ \pi^-$ cross-section data from BaBar, successfully reproducing the prominent peak located between the masses of $\psi(4160)$ and $\psi(4415)$. The step-by-step decomposition of the cross-section into its components—the direct term $|\mathcal{M}_{\text{NoR}}|^2$, the individual resonant terms $|\mathcal{A}_{\psi_i}|^2$, and, most importantly, their interference terms $2\text{Re}(\mathcal{A}_{\psi_i}\mathcal{M}_{\text{NoR}}^*)$—visually illustrates how the interference between the direct and $\psi(4160)$-mediated amplitudes is primarily responsible for generating the observed enhancement, as illustrated in Fig. \ref{fig:4260}. 

\begin{figure}[htbp]
\centering
\includegraphics[width=0.85\textwidth]{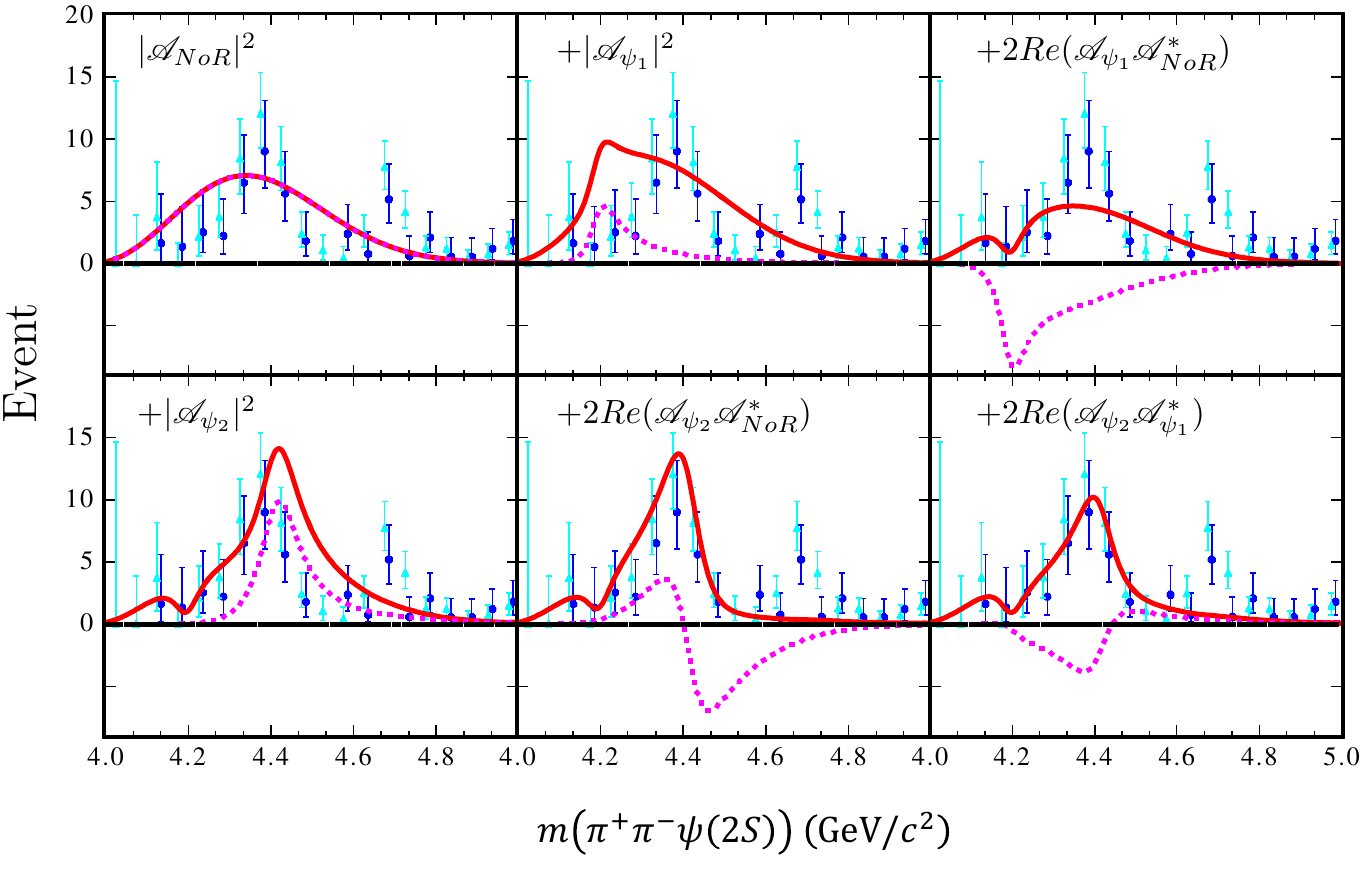}
\caption{The fitted results of the $e^+e^-\to \psi(3686)\pi^+\pi^-$ events distribution. The figure is adapted from Ref. \cite{Chen:2010nv}.}
\label{fig:4360}
\end{figure}

Encouraged by this success, the same group extended the non-resonant picture to $Y(4360)$ observed in $e^+e^- \to \psi(3686) \pi^+ \pi^-$. Employing a similar framework but with adjusted parameters to account for the different final state ($\psi(3686)$ instead of $J/\psi$), they again achieved a remarkable fit to the experimental data from both BaBar and Belle. The $Y(4360)$ structure is similarly reproduced through the interference of direct production and intermediate charmonia contributions, with the interference playing a crucial role in shaping the peak, as shown in Fig. \ref{fig:4360}. This successful application to a second, independent channel significantly strengthens the credibility of the non-resonant explanation. It demonstrates that the mechanism is not a curiosity tailored to a single dataset but a potentially general phenomenon that can account for multiple puzzling structures in the charmonium-like spectrum.

Furthermore, the model inherently explains the non-observation of $Y(4260)$ and $Y(4360)$ in other channels: since no new resonant degrees of freedom are introduced, there is no mechanism to enhance open-charm decay rates or the inclusive $R$ value near their masses. The structures are specific fingerprints of the interference patterns in their respective hidden-charm $\pi^+\pi^-$ channels alone.

The non-resonant picture offers distinct advantages and opens new avenues for investigation. It provides a parsimonious explanation that operates within the realm of known particles and established non-perturbative mechanisms (hadronic loops), avoiding the need for exotic constituents or finely tuned dynamics. Its success with both $Y(4260)$ and $Y(4360)$ suggests it should be considered a viable alternative, especially for other charmonium-like states that are observed prominently in specific hidden-charm hadronic decays from $e^+e^-$ annihilation but are elusive elsewhere.

In conclusion, the non-resonant picture represents a valuable and conceptually distinct approach in the theorist's toolkit, reminding us that not every peak in a cross-section necessitates a new particle, and that the rich interference of quantum mechanical amplitudes can itself be the source of striking structures in the spectrum of hadronic states.

\subsection{From quenched to unquenched: the evolution of charmonium potential models}

\subsubsection{1978: The Cornell potential—establishing the quenched framework}

The discovery of the $J/\psi$ in 1974 \cite{E598:1974sol,SLAC-SP-017:1974ind} opened the era of charmonium spectroscopy. It was followed by the observation of $\psi(3686)$ later that same year \cite{Abrams:1974yy}. Subsequently, $\chi_{c1}(1P)$ state was reported in 1975 \cite{Tanenbaum:1975ef}, with $\chi_{c2}(1P)$~\cite{Whitaker:1976hb} and the higher-mass $\psi(4415)$~\cite{Siegrist:1976br} identified in 1976. The year 1977 proved particularly productive, witnessing the discoveries of $\psi(3770)$ \cite{Rapidis:1977cv}, $\chi_{c0}(1P)$ \cite{Biddick:1977sv}, and $\psi(4040)$ \cite{Goldhaber:1977qn}. Moreover, the first bottomonium state, $\Upsilon(1S)$, was also observed in 1977 \cite{E288:1977xhf}, marking the parallel inception of bottomonium spectroscopy. The $\psi(4160)$ was discovered in 1978 \cite{DASP:1978dns}. Later discoveries included the $\eta_c(1S)$ in 1980 \cite{Partridge:1980vk}, the $\eta_c(2S)$ in 1982 \cite{Edwards:1982fif}, and finally the $h_c(1P)$ in 1986 \cite{R704:1986siy} (with its confirmation coming in 1992 \cite{Armstrong:1992ae}). This series of discoveries, summarized chronologically in Fig.~\ref{fig:general}, provided the foundational experimental data that stimulated the development of the quenched potential model, most notably the Cornell potential.

This systematic experimental progress established the charmonium spectrum in the late twentieth century as a central testing ground for the nonperturbative behavior of strong interaction and for phenomenological models. Consequently, a pivotal theoretical advance was the formulation of the Cornell potential \cite{Eichten:1974af, Eichten:1978tg, Eichten:1979ms}. This model, proposed shortly after the initial charmonium discoveries, provided the first quantitative framework for hadron spectroscopy. It achieved this by capturing the essential features of QCD—asymptotic freedom at short distances and confinement at long distances—in a simple phenomenological form:
\begin{align}\label{4.2.1}
\mathcal{V}(r) = -\frac{\kappa}{r} + \frac{r}{a^2},
\end{align}
where the first term represents a Coulomb-type potential originating from one-gluon exchange at short distances, and the second term is a linear confining potential that dominates at large distances, reflecting the phenomenon of quark confinement.

\begin{figure}[htbp]
    \centering 
    \begin{minipage}[c]{0.35\textwidth}
        \centering 
        \includegraphics[width=\textwidth]{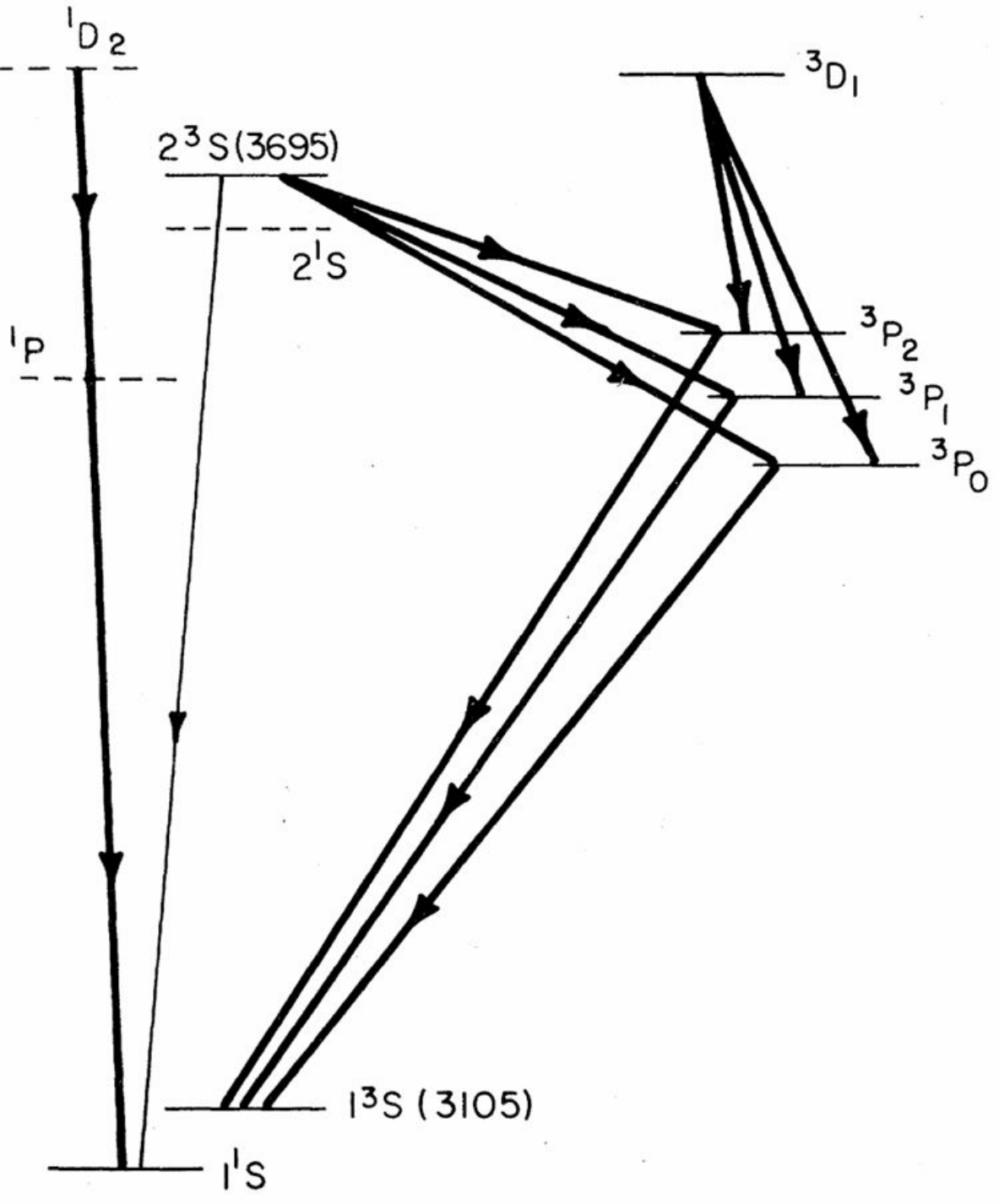}
        \caption{The charmonium spectrum obtained by Cornell group \cite{Eichten:1974af}. The vertical scale is schematic.}
        \label{Corfig}
    \end{minipage}
    \hfill
    \begin{minipage}[t]{0.60\textwidth}
        \centering 
        {
        \renewcommand\arraystretch{1.5}
\begin{tabular*}{\textwidth}{@{\extracolsep{\fill}}cccccccc}
\hline
 \multicolumn{2}{c}{\ \ \  State\ \ \  }  & \multicolumn{2}{c}{ Mass (GeV) }  & \multicolumn{2}{c}{\ \ \ \ \  Exp \ \ \ \ \  } &  \multicolumn{2}{c}{ Exp Mass (GeV) }  \\  
 \hline    
 \multicolumn{2}{c}{ $1S$ }    &   \multicolumn{2}{c}{ 3.095 }  &  \multicolumn{2}{c}{ $J/\psi$ }  & \multicolumn{2}{c}{ 3.096 }  \\
 \multicolumn{2}{c}{ $1P$ }    &   \multicolumn{2}{c}{ 3.522 }  &  \multicolumn{2}{c}{ $\chi_{c1}(1P)$ }  & \multicolumn{2}{c}{ 3.510 }  \\
 \multicolumn{2}{c}{ $2S$ }    &   \multicolumn{2}{c}{ 3.684 }  &  \multicolumn{2}{c}{ $\psi(3686)$ }  & \multicolumn{2}{c}{ 3.686 }  \\
 \multicolumn{2}{c}{ $1D$ }    &   \multicolumn{2}{c}{ 3.810 }  &  \multicolumn{2}{c}{ $\psi(3770)$ }  & \multicolumn{2}{c}{ 3.773 }  \\
 \multicolumn{2}{c}{ $3S$ }    &   \multicolumn{2}{c}{ 4.110 }  &  \multicolumn{2}{c}{ $\psi(4040)$ }  & \multicolumn{2}{c}{ 4.040 }  \\
 \multicolumn{2}{c}{ $2D$ }    &   \multicolumn{2}{c}{ 4.190 }  &  \multicolumn{2}{c}{ $\psi(4160)$ }  & \multicolumn{2}{c}{ 4.160 }  \\
 \multicolumn{2}{c}{ $4S$ }    &   \multicolumn{2}{c}{ 4.460 }  &  \multicolumn{2}{c}{ $\psi(4415)$ }  & \multicolumn{2}{c}{ 4.415 }  \\
 \multicolumn{2}{c}{ $5S$ }    &   \multicolumn{2}{c}{ 4.790 }  & \multicolumn{2}{c}{ $\times$ }   &  \multicolumn{2}{c}{ $\times$ } \\
\hline
\end{tabular*}}
       \captionof{table}{ The charmonium spectrum obtained by  Cornell model \cite{Eichten:1979ms} and is compared with the experimental data. For the state have not been observed in experiment at a period of Ref.~\cite{Eichten:1979ms} publication, a symbol "$\times$" is presented.}\label{Kangtable}
    \end{minipage}
\end{figure}

The Cornell group initially applied this model to determine the quantum numbers of $J/\psi$ and $\psi(3686)$ from their dielectron widths, and to predict properties of other low-lying $S$-, $P$-, and some $D$-wave states \cite{Eichten:1974af} (Fig.~\ref{Corfig}). The framework was later extended to incorporate a coupled-channel formalism \cite{Eichten:1978tg}. A subsequent 1980 study \cite{Eichten:1979ms} further refined the model, accurately reproducing masses and leptonic width ratios, extending it to the $\Upsilon$ and heavy-light sectors, and providing a charmonium spectrum (Table~\ref{Kangtable}).

Building upon the Cornell potential, subsequent phenomenological quark models sought to address its limitations. A key development was the 1985 Godfrey–Isgur (GI) model \cite{Godfrey:1985xj}, which introduced relativistic corrections to the kinetic term and potential, achieving a unified description of mesons from light to heavy quarkonia and successfully predicting numerous unobserved states. Despite the advancement to more sophisticated frameworks like the GI model, the original Cornell potential remains a foundational, quenched benchmark. Its successful spectrum explained the charmonia observed in the late twentieth century, and its intuitive form continues to serve as a conceptual and computational cornerstone in hadron physics.

\subsubsection{1995: From lattice QCD to the screened potential—a step toward unquenching}

\changelabel{ The Cornell potential} combines a short-range Coulomb interaction from one-gluon exchange with a long-range linearly rising confining term\footnote{The Cornell potential used here differs slightly from the standard form given in Refs.~\cite{Eichten:1974af,Eichten:1978tg,Eichten:1979ms}\changelabel{, whose Coulomb potential was not multiplied by the color factor $\frac{4}{3}$ deduced from the color operator $\frac{\mathbf{\lambda}\cdot\mathbf{\lambda^\ast}}{4}$}. For practical purposes, we retain this modified version throughout the present subsection.}, $V(r) = -\frac{4\alpha_s}{3r} + \sigma r$. This model successfully described the masses and splittings of the low-lying $c\bar{c}$ and $b\bar{b}$ states. However, as experimental data on higher excitations accumulated in the 1980s and early 1990s, significant challenges arose.

A central puzzle involved the charmonium states $\psi(4160)$ and $\psi(4415)$. In the standard linear potential model, the calculated mass of the $2D$ state fell near 4160 MeV, while the $4S$ state was predicted around 4415 MeV. This led to the natural assignments $\psi(4160) = \psi(2D)$ and $\psi(4415) = \psi(4S)$. While the mass agreement was acceptable, the predicted leptonic widths were in severe conflict with data. The leptonic width of a $^3S_1$ state is proportional to $|\psi(0)|^2$, the square of the wave function at the origin. For a $D$-wave state, $\psi(0)=0$ in the non-relativistic limit, leading to a vanishing leptonic width. Even with $S$-$D$ mixing or coupled-channel effects, the theoretical width for a $2D$-dominated state remained far too small to match the measured value of $\Gamma_{e^+e^-}(\psi(4160)) \approx 0.77$ keV, which was comparable to that of the $\psi(4040)$ ($3S$) state~\cite{Ding:1995he}. Conversely, for $\psi(4415)$ assigned as the $4S$ state, the linear potential model typically predicted $\Gamma_{e^+e^-} \sim 1.1$ keV, over a factor of two larger than the observed $0.47 \pm 0.10$ keV. These difficulties suggested that the linear confining potential might be too rigid for highly excited states, overestimating both level spacings and wave function concentrations at the origin~\cite{Ding:1995he}.

The key to resolving this impasse came from considering the dynamic nature of the QCD vacuum. In the quenched approximation, which neglects virtual quark-antiquark pair creation, the color electric flux between a static $Q\bar{Q}$ pair forms a string-like tube with constant energy density, leading to a linearly rising potential. However, when dynamical light quarks are present, as in full QCD, the string can break via the creation of a light $q\bar{q}$ pair when the stored energy is sufficient. This phenomenon, known as color screening, softens the potential at large distances, causing it to saturate. Lattice QCD simulations with dynamical fermions by Laermann \textit{et al.} in 1986 provided early evidence for this effect \cite{Laermann:1986pu,Born:1989iv}, showing a clear deviation from linearity at separations beyond roughly 1 fm.

Building on this physical insight, phenomenological studies introduced modified confining potentials that incorporated screening. A particularly instructive form, adopted by Ding, Chao, and Qin \cite{Ding:1995he}, was:
\[
V_{\text{conf}}(r) = \sigma r \left( \frac{1 - e^{-\mu r}}{\mu r} \right),
\]
which reduces to $\sigma r$ for $r \ll 1/\mu$ and approaches the constant $\sigma/\mu$ for $r \gg 1/\mu$. The parameter $\mu$ represents the inverse screening length. Combined with the Coulomb term, the total potential becomes $V(r) = -\frac{4\alpha_s}{3r} + V_{\text{conf}}(r)$. Independently, Dong \textit{et al.} employed an error-function form, $V_{\text{conf}}(r) = V_\mu \, \text{erf}(\mu_0 r)$, which exhibits similar asymptotic behavior \cite{Dong:1994zj}.

The work of Ding, Chao, and Qin performed a comprehensive analysis of both $c\bar{c}$ and $b\bar{b}$ spectra using the screened potential \cite{Ding:1993uy,Ding:1995he}. They found that a good simultaneous fit to masses and leptonic widths (including QCD radiative corrections) required a relatively large string tension, $\sigma = (0.26 - 0.32)~\text{GeV}^2$, and a screening parameter $\mu \approx 0.14 - 0.16~\text{GeV}$~\cite{Ding:1995he}. With these parameters, the calculated spectrum underwent a significant rearrangement for higher states. \changelabel{ According to the calculated mass spectrum, the assignment of $\psi(4160)$ as a predominantly $4S$ state is suggested in Ref.~\cite{Ding:1995he}, } while $\psi(4415)$ became the $5S$ state. \changelabel{ The screening effect flattens the confining potential at large distances and compresses the high-lying radial spectrum, thereby leading to a smaller effective $3S$–$4S$ mass gap than that in conventional quenched potential models.} This reassignment immediately solved the leptonic width puzzle: the $4S$ state, being an $S$-wave, naturally had a sizable $\Gamma_{e^+e^-}$, and the screened potential reduced $|\psi(0)|^2$ for higher $S$-waves, bringing $\Gamma_{e^+e^-}(\psi(4415))$ down into agreement with experiment. Their calculations also successfully reproduced the masses and leptonic widths of the $\Upsilon$ family, including the higher $6S$ state.

The study by Dong \textit{et al.} reached very similar conclusions using the error-function potential \cite{Dong:1994zj}. They obtained five $S$-wave charmonium solutions below 4.41 GeV, with the fourth and fifth matching $\psi(4160)$ and $\psi(4415)$ in both mass and leptonic width. They emphasized that the screening effect not only lowered the energies of high-lying states but also modified their wave functions, reducing their amplitude at the origin and hence the leptonic widths.

These works marked a paradigm shift. By moving from a strictly linear, quenched potential to one that incorporates the screening effect expected from dynamical quarks, they took a crucial step toward an "unquenched" phenomenological description. This approach naturally explained the anomalous mass gap between what was then considered the $3S$ and $4S$ states (now $4S$ and $5S$) and resolved long-standing discrepancies in leptonic widths. The need for a larger string tension than the canonical Regge slope value of ~0.18 GeV$^2$ was an interesting outcome, potentially reflecting the difference between the static quark-antiquark tension and the effective tension in the presence of screening.

The introduction of the screened potential also had important implications for the interpretation of other states and opened new avenues. It suggested that the $\psi(4415)$ might not be the highest $S$-wave state in that mass region and hinted at the possibility of missing states, a theme later explored in the prediction of a narrow $\psi(4S)$ around 4.26 GeV. 

\begin{figure}[htbp]
\centering
\includegraphics[width=0.6\textwidth]{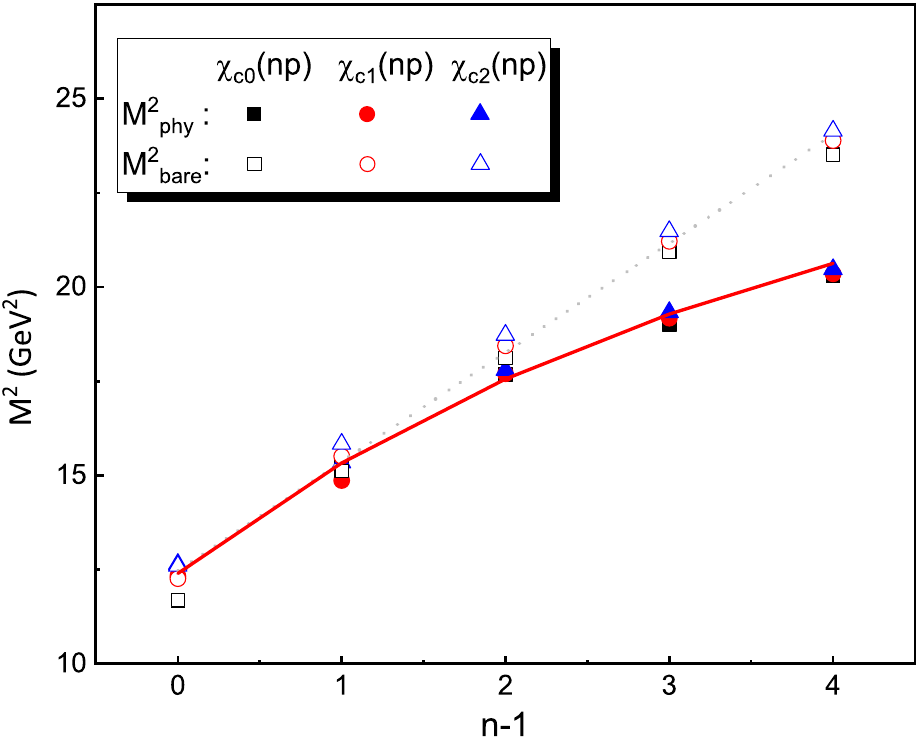}
\caption{Comparison of the physical masses of the $\chi_{cJ}(nP)\,(J=0,1,2)$ states in the coupled-channel quark model and the screening potential model (red solid curves), illustrating their equivalence. The grey dotted lines represent the predictions of the quenched GI model. Figure adapted from Ref.~\cite{Duan:2021alw}.}
\label{fig:compare_coupled_screen}
\end{figure}

Furthermore, Refs.~\cite{Li:2009ad,Duan:2021alw} demonstrated the consistency between the screening potential model and the coupled-channel quark model, emphasizing the role of coupled-channel and threshold effects, as depicting in Fig. \ref{fig:compare_coupled_screen} for $\chi_{cJ}(nP)$ cases. In this picture, the flattening of the confining potential is intrinsically linked to the opening of heavy–light meson decay channels.

In summary, research on screened potentials in the mid-1990s, driven largely by lattice QCD results, resolved key difficulties in heavy quarkonium spectroscopy. By including color screening effects from the light quark sea, it advanced beyond the quenched approximation and provided a more realistic approach for describing highly excited states. This framework enabled a systematic treatment of high-lying charmonium and can be naturally extended to bottomonium \cite{Wang:2018rjg}, heavy–light mesons \cite{Song:2015nia,Song:2015fha}, and light-flavor mesons \cite{Wang:2021gle}.

\subsubsection{2014: Mass gaps and a prediction—evidence for a narrow charmonium state near 4.2 GeV}\label{sec4.2.3}

\begin{figure}[htbp]
\centering
\includegraphics[width=0.7\textwidth]{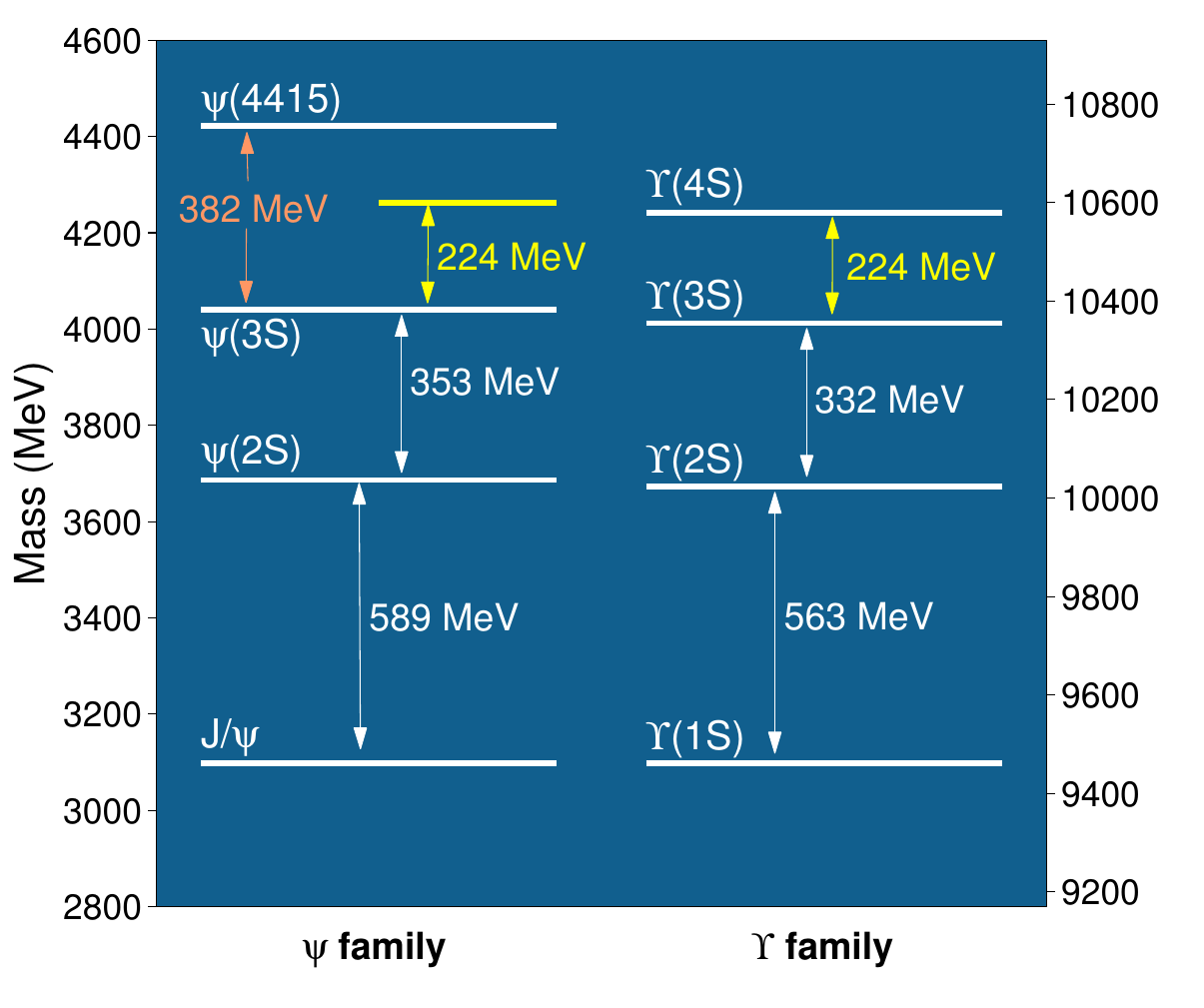}
\caption{A comparison between the $J/\psi$ and $\Upsilon$ families~\cite{He:2014xna}.}
\label{fig:ccbb}
\end{figure}

The comparison between the charmonium ($c\bar{c}$) and bottomonium ($b\bar{b}$) systems has long revealed intriguing spectroscopic regularities. A particularly clear pattern emerges when examining the mass gaps between successive radial excitations of the $S$-wave vector states ($J^{PC}=1^{--}$). For the well-established bottomonium family, the states $\Upsilon(1S)$, $\Upsilon(2S)$, $\Upsilon(3S)$, and $\Upsilon(4S)$ exhibit smoothly progressing mass differences $\Delta M_{n,n-1} = M(nS) - M((n-1)S)$. In the charmonium sector, the analogous states $J/\psi(1S)$, $\psi(2S)$ ($\psi(3686)$), and $\psi(3S)$ ($\psi(4040)$) show a strikingly similar pattern: the mass gap $\Delta M_{21}$ is nearly identical for both families ($\sim$ 589 MeV for charmonium vs. $\sim$ 563 MeV for bottomonium), and $\Delta M_{32}$ is also comparable (383 MeV vs. 333 MeV). However, this harmonious pattern appears to break for the next excitation. If the established $\psi(4415)$ is assigned as $\psi(4S)$, the resulting mass gap $\Delta M_{43}$ is about 375 MeV, which is significantly larger than the corresponding $\Upsilon(4S)$-$\Upsilon(3S)$ gap of approximately 224 MeV. This discrepancy suggests a possible misassignment or a missing state in the charmonium spectrum around 4.2 GeV, as shown in Fig. \ref{fig:ccbb}.

The 2014 work by He, Chen, Liu, and Matsuki directly confronted this puzzle by hypothesizing a missing $\psi(4S)$ state~\cite{He:2014xna}. The guiding principle was the assumed persistence of the mass-gap similarity between the two families. By adding the well-measured $\Upsilon(4S)$-$\Upsilon(3S)$ mass difference (224 MeV) to the mass of the charmonium $\psi(3S)$ (4040 MeV), they predicted the mass of the missing $\psi(4S)$ to be around 4263 MeV. This estimate is also consistent with the predictions for $\psi(4S)$ obtained in the unquenched picture \cite{Ding:1993uy,Ding:1995he,Li:2009zu,Dong:1994zj}. This \changelabel{ phenomenological spectral analogy between charmonia and bottomonia gained} theoretical credence from independent calculations. Notably, studies employing a screened potential, which incorporates the color-screening effects of dynamic light quark pairs, had already predicted the mass of the $\psi(4S)$ to be around 4273 MeV \cite{Li:2009zu}, in excellent agreement with the mass-gap estimate. Another potential model calculation yielded a similar value of 4247 MeV \cite{Dong:1994zj}. This convergence of different approaches underscored the plausibility of a missing state in this mass region.

\begin{figure}[htbp]
\centering
\includegraphics[width=0.5\textwidth]{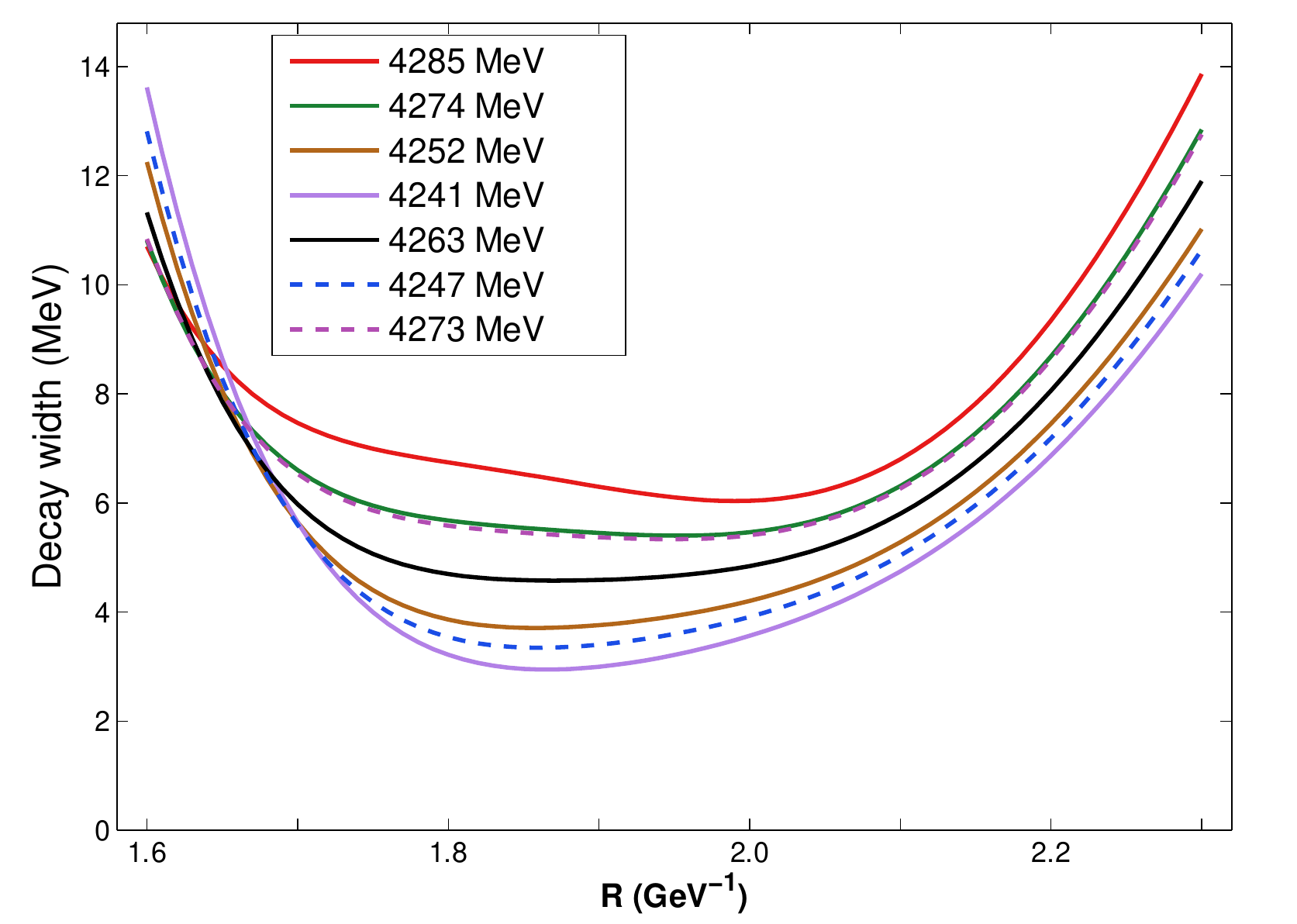}
\caption{The dependence of the total decay width of $\psi(4S)$ on the mass of $\psi(4S)$~\cite{He:2014xna}.}
\label{fig:massrelevant}
\end{figure}

The central and most surprising result of the investigation lay in the decay properties of this predicted $\psi(4S)$. Using the QPC model to calculate its OZI-allowed open-charm decays ($D\bar{D}$, $D\bar{D}^*+\text{c.c.}$, $D^*\bar{D}^*$, $D_s\bar{D}_s$, etc.), the authors found that the predicted total width was remarkably narrow, approximately 6 MeV, and stable across a reasonable range of model parameters. This narrow width persisted even when varying the input mass of the $\psi(4S)$ within a range of 4241–4285 MeV, as illusterated in Fig. \ref{fig:massrelevant}. The physical origin of this narrowness was attributed to strong node effects in the wavefunctions of higher radial excitations, which lead to significant cancellations in the decay amplitudes. This result was exceptional, as all other known charmonium states above the $D\bar{D}$ threshold—$\psi(3770)$, $\psi(4040)$, $\psi(4160)$, and $\psi(4415)$—possess widths on the order of tens to over a hundred MeV. The predicted narrow width provided a natural and compelling explanation for why such a state had escaped detection in previous experiments focused on open-charm final states and inclusive $R$-value scans, which are less sensitive to very narrow resonances.

The study naturally addressed the relationship between the predicted narrow $\psi(4S)$ and the broader charmonium-like states $Y(4260)$ and $Y(4360)$ observed in hidden-charm dipion transitions. Given the large disparity in widths (the predicted $\sim$6 MeV versus the experimental $\sim$95 MeV and $\sim$74 MeV for $Y(4260)$ and $Y(4360)$, respectively), the authors concluded that neither of these observed structures could be identified with the conventional $\psi(4S)$. This conclusion was consistent with earlier non-resonant explanations proposed for $Y(4260)$ and $Y(4360)$, which attributed their enhancements to interference effects in the production process rather than to genuine resonances.

A crucial experimental connection emerged from the analysis by Yuan in~\cite{Yuan:2013uta}. Yuan examined the BESIII data on $e^+e^- \to h_c(1P)\pi^+\pi^-$ and identified two structures: a narrow one with mass $4216 \pm 18$ MeV and width $39 \pm 22$ MeV, and a broad one around 4293 MeV. The narrow structure, denoted $Y(4216)$, bears a striking resemblance to the predicted $\psi(4S)$ in both mass and width. Although its central mass is about 47 MeV lower than the predicted 4263 MeV, this difference is within the combined theoretical and experimental uncertainties. More importantly, its narrow width aligns qualitatively with the theoretical expectation. This observation provides a tangible experimental hint that the missing $\psi(4S)$ might indeed be present in the data, awaiting further confirmation through more precise measurements and analyses.

The work also examined the properties of other states. It strongly supported the identification of $X(3940)$ with $\eta_c(3S)$, as its calculated width and dominant decay to $D\bar{D}^*$ matched the experimental data. Conversely, it decisively excluded the $\eta_c(4S)$ assignment for $X(4160)$, as the predicted $\eta_c(4S)$ was also found to be very narrow, incompatible with the broad width of $X(4160)$. Finally, the analysis explored the possibility of $\psi(4415)$ being $\psi(5S)$, finding general consistency with several measured branching ratios, though some tensions remained.

In summary, the 2014 study proposed a semi-quantitative and predictive framework from a phenomenological perspective, which resolved a spectroscopic irregularity by positing a missing, narrow $\psi(4S)$ state near 4.26 GeV. The prediction was rooted in the comparative analysis of heavy quarkonia, bolstered by potential model calculations, and received tentative support from emerging experimental data. The analysis by Yuan in~\cite{Yuan:2013uta} offered a promising experimental counterpart in the form of $Y(4216)$, reinforcing the relevance of this prediction. By highlighting the profound impact of wavefunction nodes on the decays of high radial excitations, it offered a deeper understanding of charmonium dynamics above open-charm thresholds. \changelabel{This work demonstrates} the enduring value of spectroscopic systematics and continues to motivate precise experimental investigations at facilities like BESIII, Belle, and Belle II.

\subsection{2017: Precise data reveal more charmonium-like $Y$ structures—the $Y$ problem}

In 2017, the BESIII Collaboration performed a high precise measurement of the $e^+e^- \to J/\psi \pi^+\pi^-$ process~\cite{BESIII:2016bnd}, which was a representative channel through which  $Y(4260)$ was originally discovered. The new results brought important surprises: the line shape of $Y(4260)$ enhancement in total cross section was found to be highly asymmetric, and detailed analyses revealed that it actually contains two substructures: one is a newly observed state, now called $Y(4320)$, while the other corresponds to the main enhancement previously attributed to $Y(4260)$. Compared with the resonance parameters of $Y(4260)$, it shifts downward by about $40\ \text{MeV}$ in mass and becomes narrower in width, from over $70\ \text{MeV}$ to roughly $40\ \text{MeV}$. This state has therefore been renamed as $Y(4220)$. It is worth mentioning that the updated $Y(4220)$ closely resembles a narrow peak around $4.23\ \text{GeV}$ reported by BESIII as early as in 2016 in the $\chi_{c0}\omega$ channel~\cite{BESIII:2014rja}. Additionally, $Y(4220)$ also was soon discovered in several other independent processes, including $e^+e^- \to h_c \pi^+\pi^-$~\cite{BESIII:2016adj}, $e^+e^- \to \psi(3686)\pi^+\pi^-$~\cite{BESIII:2017tqk} and $e^+e^- \to h_c \eta$~\cite{BESIII:2017dxi}.

The updated measurements of $Y(4260)$ have significantly reshaped our understanding of this state, particularly the long-standing puzzle of why no clear signal of $Y(4260)$ was observed in open-charm cross sections or in $R$-value scans. In fact, in recent years, new experimental measurements have revealed the existence of peaks or dip-like structures around $\sqrt{s}\approx 4.22$ GeV not only in several open-charm channels~\cite{BESIII:2018iea,BESIII:2023cmv} but also in the $e^+e^-\to\mu^+\mu^-$ process~\cite{Ablikim:2020jrn}, which is closely tied to the $R$-value measurement. The extracted resonance parameters in these channels are highly consistent with those of $Y(4220)$~\cite{BESIII:2016bnd,BESIII:2016adj,BESIII:2017tqk,BESIII:2017dxi}, indicating that earlier knowledge of  $Y(4260)$ indeed require substantial revision. These experimental developments suggest that many of the previous theoretical discussions regarding $Y(4260)$ may need to be reconsidered. 

\begin{figure*}[htbp]
  \centering
  \begin{tabular}{ccc}
  \subfigure{\includegraphics[width=469pt]{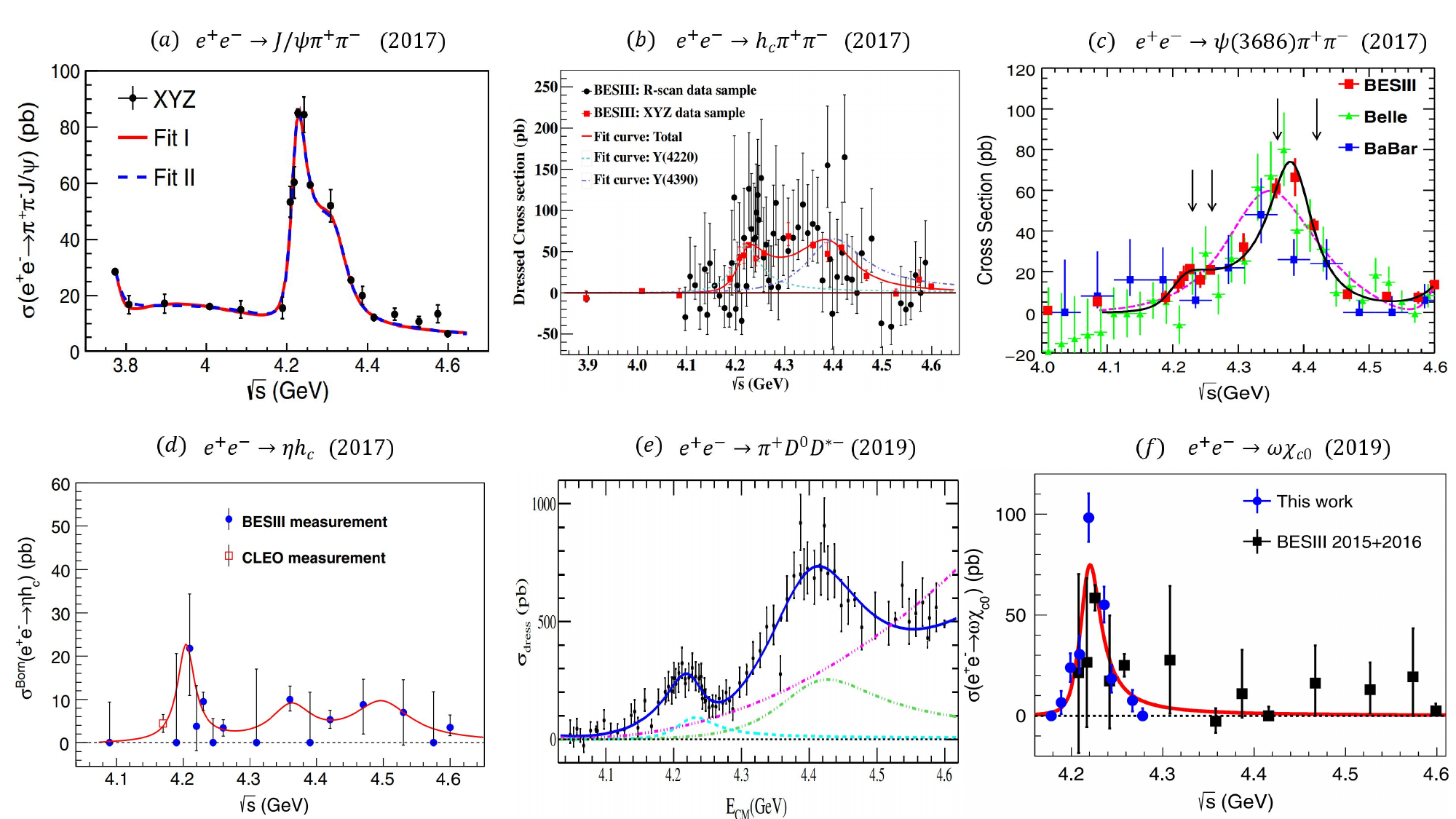}}&$\quad$\\
  \subfigure{\includegraphics[width=469pt]{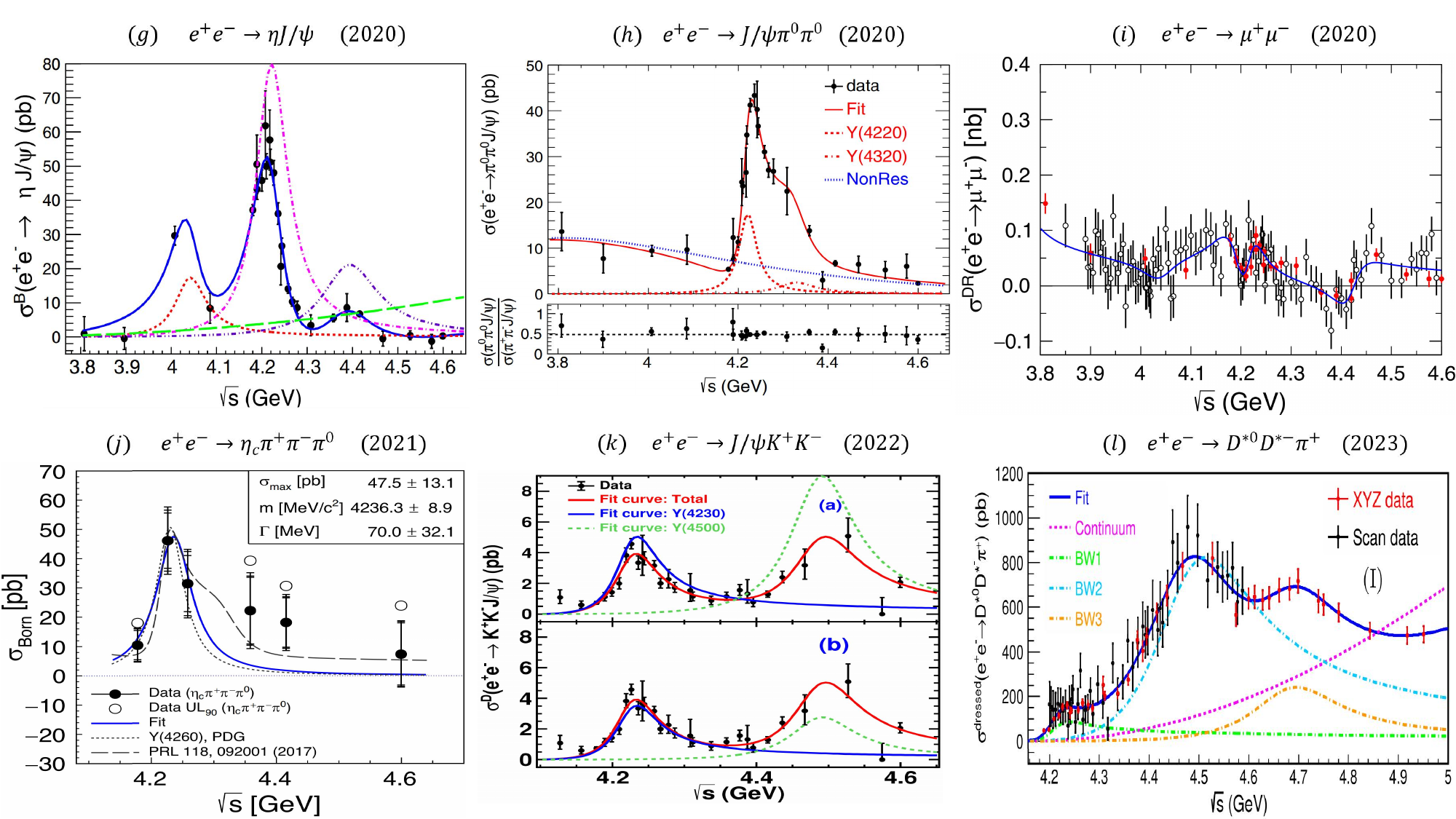}}&$\quad$\\
  \end{tabular}
  \caption{The measured cross sections of $e^+e^- \to J/\psi \pi^+\pi^-$~\cite{BESIII:2016bnd}, $e^+e^- \to h_c \pi^+\pi^-$~\cite{BESIII:2016adj}, $e^+e^- \to \psi(3686) \pi^+\pi^-$~\cite{BESIII:2017tqk},  $e^+e^- \to h_c \eta$~\cite{BESIII:2017dxi}, $e^+e^- \to  D^0D^{*-}\pi^+$~\cite{BESIII:2018iea}, $e^+e^- \to \chi_{c0}\omega$~\cite{BESIII:2019gjc},  $e^+e^- \to J/\psi \eta$~\cite{BESIII:2020bgb}, $e^+e^- \to J/\psi \pi^0\pi^0$~\cite{BESIII:2020oph}, $e^+e^- \to \mu^+\mu^-$~\cite{Ablikim:2020jrn}, $e^+e^- \to \eta_c \pi^+\pi^-\pi^0$~\cite{BESIII:2020tgt}, $e^+e^- \to J/\psi K^+K^-$~\cite{BESIII:2022joj} and $e^+e^- \to D^{*0}D^{*-}\pi^+$~\cite{BESIII:2023cmv}, in which the $Y(4220)$ state had been observed.}\label{F4-1}
\end{figure*}

The study of $Y(4220)$ has in fact become one of the central physics topics for the BESIII experiment over the past decade. Now, the existence of this state is firmly established, as it has been observed in more than ten different processes, including both hidden-charm and open-charm decay channels~\cite{BESIII:2016bnd,BESIII:2016adj,BESIII:2017tqk,BESIII:2017dxi,BESIII:2018iea,BESIII:2019gjc,BESIII:2020bgb,BESIII:2020oph,Ablikim:2020jrn,BESIII:2020tgt,BESIII:2022joj,BESIII:2023cmv,BESIII:2021njb,BESIII:2022qal,BESIII:2022kcv,BESIII:2023tll,BESIII:2024yqi,BESIII:2025bce}. A summary of the measured total cross sections of the relevant processes and corresponding $Y(4220)$ resonance signals are presented in Fig.~\ref{F4-1}. Remarkably, the fitted masses across these processes consistently point to a value around $ 4.22$ GeV, as compiled in Table~\ref{T4-1}, which also lists the resonance parameters of $Y(4220)$ and its combined product of di-lepton width and decay branching fractions in different channels. It can be seen that the decays of $Y(4220)$ are characterized by a relatively higher contribution from open-charm final states, while hidden-charm modes remain important, representing a considerable departure from some of the earlier conclusions drawn for $Y(4260)$.

\begin{figure*}[!htbp]
  \centering
  \begin{tabular}{ccc}
  \subfigure{\includegraphics[width=480pt]{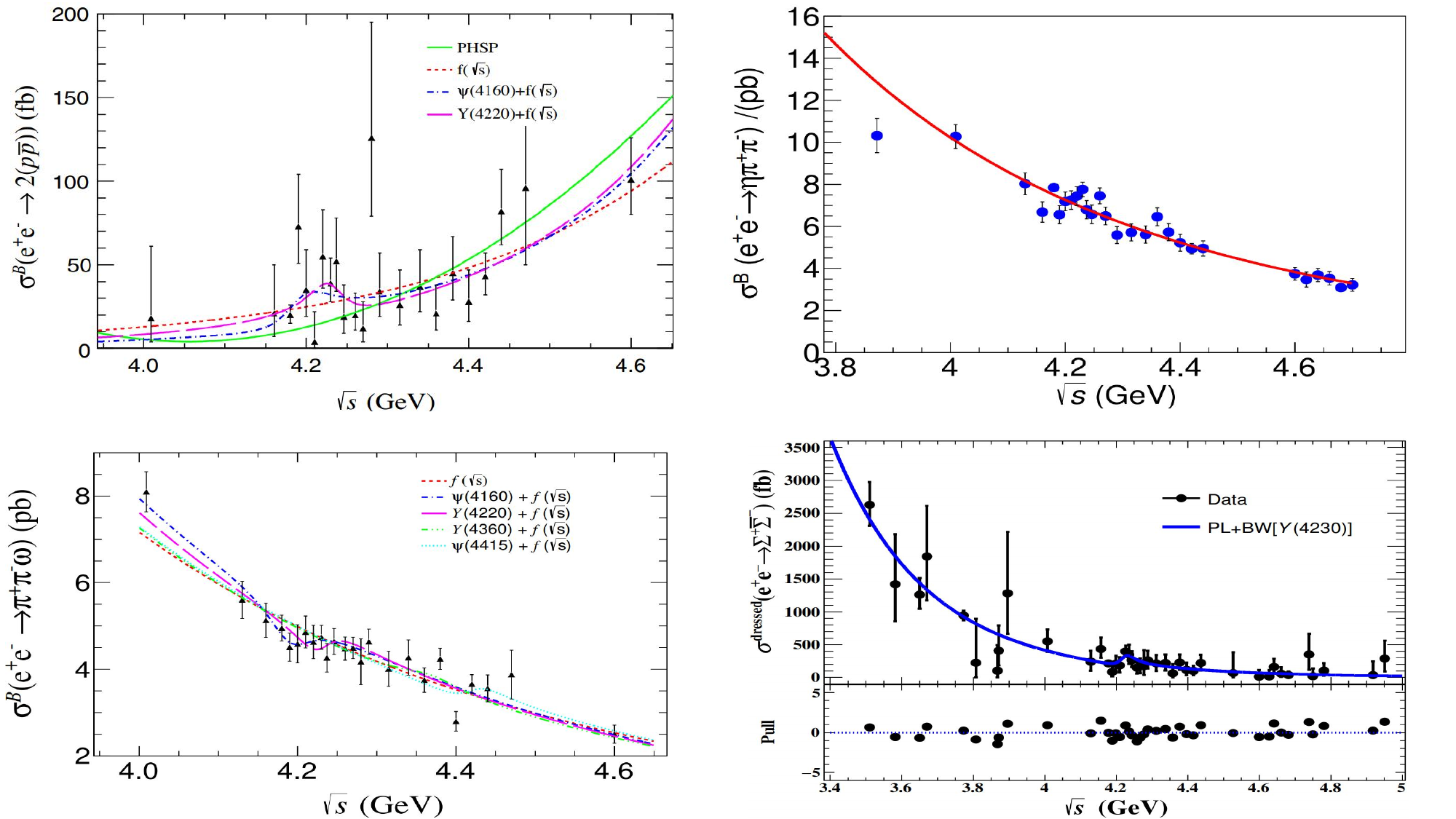}}&$\quad$\\
  \subfigure{\includegraphics[width=480pt]{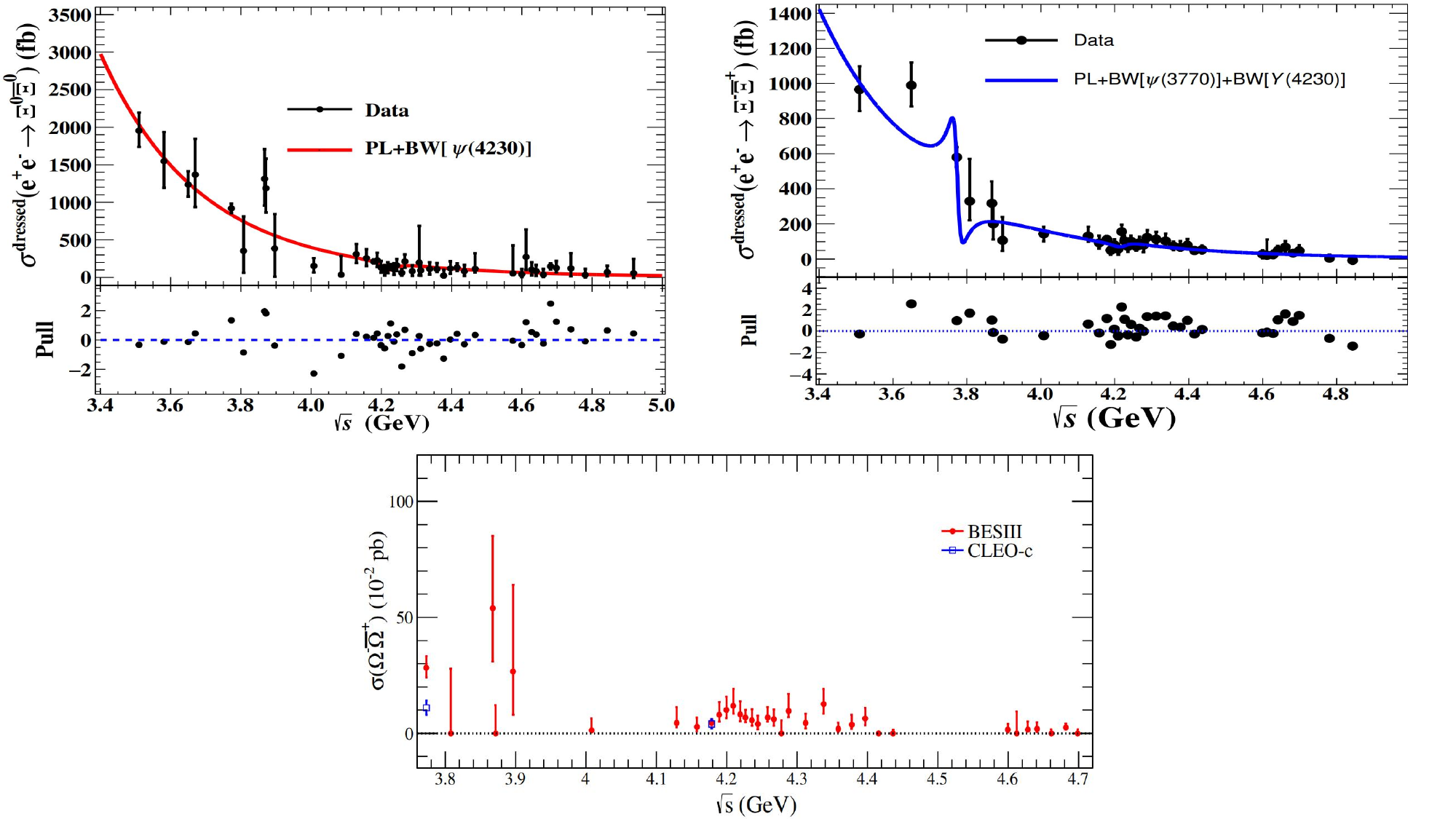}}&$\quad$\\
  \end{tabular}
  \caption{Measured cross sections of $e^{+}e^{-}$ annihilation into light hadronic final states, in which although no obvious $Y(4220)$ signal is observed, their line shapes exhibit possible structures around 4.2 GeV that warrant confirmation with higher-statistics data in the future. These processes include $e^+e^- \to 2(p\bar{p})$~\cite{BESIII:2020svk}, $e^+e^- \to \eta \pi^+\pi^-$~\cite{BESIII:2022vpa},  $e^+e^- \to \omega \pi^+\pi^- $~\cite{BESIII:2023gqy},  $e^+e^- \to \Sigma^+\bar{\Sigma}^-$~\cite{BESIII:2024umc},  $e^+e^- \to \Xi^0\bar{\Xi}^0 $~\cite{BESIII:2024ues},  $e^+e^- \to \Xi^-\bar{\Xi}^+ $~\cite{BESIII:2023rse} and  $e^+e^- \to \Omega^-\bar{\Omega}^+ $~\cite{BESIII:2025fph}. }\label{F4-2}
\end{figure*}

Additionally, the BESIII Collaboration also measured the cross sections of the processes $e^+e^- \to \pi^{+}\pi^{-}\psi(3770) $~\cite{BESIII:2019tdo}, $e^+e^- \to \eta_c \pi^+\pi^- $~\cite{BESIII:2020tgt}, $e^+e^- \to \eta_c \pi^0\gamma $~\cite{BESIII:2020tgt}, $e^+e^- \to \chi_{cJ} \pi^+ \pi^- $~\cite{BESIII:2020wzv}, $e^+e^- \to \eta\psi(3686) $~\cite{BESIII:2021fae,BESIII:2024jzt} and $e^+e^- \to \pi^{+}\pi^{-}D^{+}D^{-} $~\cite{BESIII:2022quc}, but in which they cannot find any significant signal of $Y(4220)$. Beyond hidden- and open-charm processes, channels involving purely light mesons or even baryon–antibaryon pairs are also of great interest. This is because different underlying structures of $Y(4220)$ are expected to lead to very different decay behaviors. \changelabel{ For instance, if the state contains a sizable charmonium component, one might expect that its three-gluon annihilation decay leads to abundant light-hadron final states. By contrast, such mechanisms are not naturally anticipated if $Y(4220)$ is instead a hadronic molecular state}. In recent years, BESIII has carried out extensive searches in these channels~\cite{BESIII:2020svk,BESIII:2022vpa,BESIII:2023gqy,BESIII:2024umc,BESIII:2024ues,BESIII:2023rse,BESIII:2025fph,BESIII:2018kyw,BESIII:2019cuv,BESIII:2021vkt,BESIII:2021fqx,BESIII:2021ftf,BESIII:2022zxr,BESIII:2022tjc,BESIII:2022tvj,BESIII:2023ojh,BESIII:2023tex,BESIII:2023ion,BESIII:2024ogz,BESIII:2024gql,BESIII:2025zyk,BESIII:2025lbj,BESIII:2021ccp,BESIII:2025fxf,Zhang:2025qmo,Zhang:2026qjt}. Although no definite evidence of $Y(4220)$ decays into purely light hadrons has been observed so far, largely due to limited statistics, these measurements nevertheless provide important implications for understanding its nature. The relevant cross section measurement results with possible hint of a structure around 4.2 GeV exhibiting in the line shape are summarized and presented in Fig.~\ref{F4-2}.

\begin{table*}[htbp]
  \centering
  \renewcommand\arraystretch{1.4}
  \caption{The summary of the experimental resonance parameters of $Y(4220)$ and its combined product of di-lepton width and decay branching fractions in different channels, which are obtained by a Breit-Wigner fit to total cross section. }\label{T4-1}
  {\tabcolsep0.06in
  \begin{tabular}{lcccc}
  \hline

  Resonance parameters &  Mass~(MeV) & Width~(MeV) & $\Gamma_{e^+e^-}BR_{\mathrm{Final~states}}$~(eV) \\
  \hline
  \multirow{4}{*}{$e^+e^- \to Y(4220) \to J/\psi \pi^+\pi^-$~\cite{BESIII:2016bnd}} & \multirow{4}{*}{$4222.0 \pm3.1 (4220.9\pm2.9)$} & \multirow{4}{*}{$44.1 \pm4.3 (44.1 \pm3.8)$}& $13.3 \pm 1.4 (12.0 \pm 1.0)$  \\
   &  &  & $9.2 \pm 0.7 (8.9 \pm 0.6)$   \\
    &  &  & $2.3 \pm 0.6 (2.1 \pm 0.4)$   \\
     &  &  & $1.6 \pm 0.4 (1.5 \pm 0.3)$   \\
    \hline 
  $e^+e^- \to Y(4220) \to h_c \pi^+\pi^-$~\cite{BESIII:2016adj} &  $4218.4^{+5.5}_{-4.5} \pm 0.9$ &  $66.0^{+12.3}_{-8.3} \pm 0.4$ &  $4.6^{+2.9}_{-1.4}$  \\
   \hline 
   \multirow{2}{*}{$e^+e^- \to Y(4220) \to \psi(3686) \pi^+\pi^-$~\cite{BESIII:2017tqk}} &  \multirow{2}{*}{$4209.5 \pm 7.4$} &  \multirow{2}{*}{$80.1 \pm 24.6$} &  $0.8  \pm 0.7$ \\
   &  &  & $ 0.4 \pm 0.3$   \\
    \hline 
    $e^+e^- \to Y(4220) \to h_c \eta$~\cite{BESIII:2017dxi} &  $4204 \pm 6$ &  $32 \pm 22$ &  $\cdots$ \\
     \hline 
     \multirow{4}{*}{ $e^+e^- \to Y(4220) \to D^0D^{*-}\pi^+$~\cite{BESIII:2018iea}} & \multirow{4}{*}{$4228.6 \pm 4.1$} & \multirow{4}{*}{$77.0 \pm 6.8$}& $77.4 \pm 10.1$  \\
   &  &  & $ 8.6 \pm 1.6$   \\
    &  &  & $ 99.5 \pm 14.6$   \\
     &  &  & $11.1 \pm 2.3$   \\
      \hline 
   $e^+e^- \to Y(4220) \to \chi_{c0}\omega$~\cite{BESIII:2019gjc}   &  $4218.5  \pm 1.6$ &  $28.2  \pm 3.9$ &  $2.5 \pm 0.2$ \\
    \hline 
   \multirow{3}{*}{$e^+e^- \to Y(4220) \to J/\psi \eta$~\cite{BESIII:2020bgb}} & \multirow{3}{*}{$4218.6 \pm 3.8$} & \multirow{3}{*}{$82.0 \pm 5.7$}& $8.0 \pm 1.7$  \\
   &  &  & $4.8 \pm 1.0$   \\
    &  &  & $7.0 \pm 1.5$   \\
     \hline 
     \multirow{4}{*}{$e^+e^- \to Y(4220) \to J/\psi \pi^0\pi^0$~\cite{BESIII:2020oph}} & \multirow{4}{*}{$4220.4 \pm 2.4$} & \multirow{4}{*}{$46.2 \pm 4.7$}& $0.99  \pm 0.17$  \\
   &  &  & $ 4.13 \pm 0.28$   \\
    &  &  & $ 1.38 \pm 0.30$   \\
     &  &  & $ 5.72  \pm 0.57$   \\
    \hline 
     \multirow{8}{*}{$e^+e^- \to Y(4220) \to \mu^+\mu^-$~\cite{Ablikim:2020jrn}} &$4216.7  \pm 8.9 \pm  4.1$ & $47.2  \pm 22.8   \pm10.5$& $1.53   \pm1.26   \pm0.54$\tnote{a} \\
   & $ 4213.6 \pm  7.5  \pm 4.1$ & $39.9  \pm 19.5  \pm 8.9$ & $1.28  \pm 1.09  \pm 0.46$\tnote{a}  \\
    & $4213.7  \pm 6.0  \pm 4.1$ & $38.5  \pm 12.8  \pm 8.5$ & $1.20  \pm 0.67  \pm 0.42 $\tnote{a}\\
     & $4216.2  \pm 5.7  \pm 4.1$ & $ 45.5  \pm 13.3  \pm 10.1$ & $1.46  \pm 0.89  \pm 0.52 $\tnote{a}   \\
     &$4219.4  \pm 11.2  \pm 4.1$  & $49.6  \pm 22.6  \pm 11.0$ & $ 1.50  \pm 1.03  \pm 0.53$\tnote{a} \\
    & $4212.8   \pm7.2   \pm4.0$ & $36.4  \pm 16.8   \pm8.1$ & $1.12 \pm  0.89  \pm 0.40 $\tnote{a} \\
     & $ 4216.1  \pm 7.5  \pm 4.1$ &$37.8  \pm 18.5  \pm 8.4$  & $ 1.09  \pm 0.84  \pm 0.39 $\tnote{a}  \\
      & $ 4217.3  \pm 9.1  \pm 4.1$ & $45.5  \pm 21.2  \pm 10.1$ & $ 1.40  \pm 1.08  \pm 0.50$\tnote{a}  \\
       \hline 
      $e^+e^- \to Y(4220) \to \eta_c \pi^+\pi^-\pi^0$~\cite{BESIII:2020tgt}    &  $4236.3 \pm 8.9$ &  $70.0 \pm 32.1$ &  $\cdots$ \\
       \hline 
       \multirow{2}{*}{$e^+e^- \to Y(4220) \to J/\psi K^+K^-$~\cite{BESIII:2022joj}} &  \multirow{2}{*}{$4225.3 \pm 2.3 \pm 21.5$} &  \multirow{2}{*}{$72.9 \pm 6.1 \pm 30.8$} &  $0.42 \pm 0.04 \pm 0.15$ \\
   &  &  & $0.29 \pm 0.02\pm 0.10$   \\
\hline
\end{tabular}
}
 \begin{tablenotes}
      \footnotesize
      \item[a] These values denote the partial width of $Y(4220)\to e^+e^-$ in units of keV by assuming $\Gamma_{e^+e^-}=\Gamma_{\mu^+\mu^-}$ as in the BESIII analysis~\cite{Ablikim:2020jrn}.
    \end{tablenotes}
\end{table*}

\addtocounter{table}{-1}
\begin{table*}[htbp]
  \centering
   \begin{threeparttable}
  \renewcommand\arraystretch{1.4}
  \caption{Table~\ref{T4-1} (continued).}
  {\tabcolsep0.06in
  \begin{tabular}{lcccc}
  \hline
  Resonance parameters &  Mass~(MeV) & Width~(MeV) & $\Gamma_{e^+e^-}BR_{\mathrm{Final~states}}$~(eV) \\
  \hline
  \multirow{8}{*}{$e^+e^- \to Y(4220) \to D^{*0}D^{*-}\pi^+$~\cite{BESIII:2023cmv}} & \multirow{8}{*}{$4209.6 \pm  4.7$} & \multirow{8}{*}{$81.6 \pm  17.8$}& $5.4 \pm  1.1$ \\
   & &  & $6.0 \pm  1.3$ \\
    & &  & $ 4.8 \pm  0.9$  \\
      & &  & $5.3 \pm  1.1$  \\
      & &  & $ 17.9 \pm  7.2$  \\
     & &  & $ 19.8 \pm  6.6$  \\
     & &  & $20.2 \pm  7.4$  \\
       & &  & $22.4 \pm  9.0$  \\
    \hline 
    \multirow{4}{*}{$e^+e^- \to Y(4220) \to \psi(3686) \pi^+\pi^-$~\cite{BESIII:2021njb}} &  \multirow{4}{*}{$4234.4 \pm 3.2$} &  \multirow{4}{*}{$17.6 \pm 8.1$} &  $1.59 \pm 0.75$ \\
   &  &  & $  1.63 \pm 0.78$   \\
    &  &  & $  0.02 \pm 0.01$   \\
     &  &  & $  0.02  \pm 0.01$   \\
    \hline 
     \multirow{4}{*}{$e^+e^- \to Y(4220) \to J/\psi \pi^+\pi^-$~\cite{BESIII:2022qal}} & \multirow{4}{*}{$4221.4 \pm 1.5 (4220.1 \pm 1.2)
$} & \multirow{4}{*}{$41.8 \pm2.9 (43.6 \pm 2.6)$}& $1.7 \pm 0.2(1.7 \pm 0.2)$  \\
   &  &  & $8.2  \pm0.9(8.6 \pm 0.5)$   \\
    &  &  & $3.0\pm  0.5(2.5 \pm 0.3)$   \\
     &  &  & $14.6 \pm 1.2(12.7  \pm 0.8)$   \\
      \hline 
       \multirow{4}{*}{$e^+e^- \to Y(4220) \to J/\psi K_S^0K_S^0$~\cite{BESIII:2022kcv}} &  \multirow{4}{*}{$4226.9 \pm 6.6 \pm 22.0$} &  \multirow{4}{*}{$71.7 \pm 16.2  \pm32.8$} &  $0.13 \pm 0.02\pm  0.05$ \\
        & &  & $0.14 \pm 0.03\pm  0.06$  \\
   &  &  & $ 0.18 \pm 0.05 \pm 0.07$   \\
    & &  & $0.20 \pm 0.04 \pm 0.07$  \\
    \hline 
     \multirow{4}{*}{$e^+e^- \to Y(4220) \to J/\psi \eta$~\cite{BESIII:2023tll}} & \multirow{4}{*}{$4219.7 \pm 2.5$} & \multirow{4}{*}{$80.7  \pm4.4$}& $4.0 \pm 0.5$  \\
   &  &  & $ 5.5 \pm 0.7$   \\
    &  &  & $ 8.7 \pm 1.0$   \\
    &  &  & $ 11.9 \pm 1.1$   \\
     \hline 
       $e^+e^- \to Y(4220) \to h_c \eta$~\cite{BESIII:2024yqi} &  $ 4188.8 \pm 4.7  \pm8.0$ &  $49\pm  16 \pm 19$ &  $0.80 \pm 0.19 \pm 0.45$ \\
     \hline 
      \multirow{2}{*}{$e^+e^- \to Y(4220) \to h_c \pi^+\pi^-$~\cite{BESIII:2025bce}} & \multirow{2}{*}{$4223.6^{+3.6+2.6}_{-3.7-2.9}$} & \multirow{2}{*}{$58.5^{+10.8+6.7}_{-11.4-6.5}$}& $10.2^{+1.2+1.4}_{-1.5-1.4}$  \\
   &  &  & $0.9^{+0.4+0.3}_{-0.4-0.2}$   \\
\hline
\end{tabular}
}
\end{threeparttable}
\end{table*}


Since 2017, high-statistics energy scans performed by the BESIII experiment have revealed more vector charmonium-like $Y$ states above $4.3~\mathrm{GeV}$. In the measurement of the process $e^+e^- \to \pi^+\pi^- J/\psi$~\cite{BESIII:2016bnd}, based on an integrated luminosity of about $9~\mathrm{fb}^{-1}$ collected at center-of-mass energies from $3.77~\mathrm{GeV}$ to $4.60~\mathrm{GeV}$, the observed cross section exhibits a line shape that requires more than one resonant contribution. A fit with two coherent Breit--Wigner amplitudes indicates a structure near $4.23~\mathrm{GeV}$, referred to as $Y(4220)$, together with a higher-mass enhancement around $4.32~\mathrm{GeV}$, denoted as $Y(4320)$. The resonance parameters of $Y(4320)$ are determined to be $M=(4320.0\pm10.4\pm7.0)~\mathrm{MeV}/c^2$ and $\Gamma=(101.4^{+25.3}_{-19.7}\pm10.2)~\mathrm{MeV}$, with a statistical significance exceeding $7.6\sigma$. Its measured mass is noticeably lower than that of $Y(4360)$ previously observed in the $\pi^+\pi^-\psi(3686)$ channel \cite{BaBar:2006ait,Belle:2007umv}, although the two structures have comparable widths within uncertainties. No evidence for $Y(4008)$ is found in the $\pi^+\pi^- J/\psi$ final state with highly precise data~\cite{BESIII:2016bnd}. The information of another higher $Y$ state is obtained from the study of $e^+e^- \to \pi^+\pi^- h_c$~\cite{BESIII:2016adj}, where BESIII measured the cross section at 79 center-of-mass energy points between $3.896~\mathrm{GeV}$ and $4.600~\mathrm{GeV}$. The data clearly favor a two-resonance description, with significant structures observed at $\sqrt{s}\approx4.23~\mathrm{GeV}$ and $4.39~\mathrm{GeV}$. For the higher-mass structure, referred to as $Y(4390)$, the fit yields a mass of $M=(4391.5^{+6.3}_{-6.8}\pm1.0)~\mathrm{MeV}$ and a width of $\Gamma=(139.5^{+16.2}_{-20.6}\pm0.6)~\mathrm{MeV}$, and the two-resonance hypothesis is favored over a single-resonance description with a significance exceeding $10\sigma$~\cite{BESIII:2016adj}.

Further information of vector charmonium-like states in the $4.3$–$4.4~\mathrm{GeV}$ region was provided from precise BESIII measurements of the process $e^+e^- \to \pi^+\pi^- \psi(3686)$~\cite{BESIII:2017tqk}. Over the measured energy range between $4.008~\mathrm{GeV}$ and $4.600~\mathrm{GeV}$, the measured cross section exhibits pronounced energy dependence and shows, in addition to  $Y(4220)$, a higher-mass enhancement associated with the state historically referred to as $Y(4360)$. A line-shape analysis yields resonance parameters of $M=(4383.8\pm4.2\pm0.8)~\mathrm{MeV}$ and $\Gamma=(84.2\pm12.5\pm2.1)~\mathrm{MeV}$~\cite{BESIII:2017tqk}. This state was originally observed as the $Y(4360)$ in the $\pi^+\pi^- \psi(3686)$ channel in initial-state radiation measurements \cite{BaBar:2006ait,Belle:2007umv}, however, within the current experimental uncertainties with higher precision, the resonance parameters determined by BESIII are consistent with those of $Y(4390)$ observed in $e^+e^- \to \pi^+\pi^- h_c$. Evidence for a vector state with compatible mass and width was also reported in an independent study of the process $e^+e^- \to \eta J/\psi$~\cite{BESIII:2020bgb}, where a resonance with mass $M=(4382.0\pm13.3\pm1.7)~\mathrm{MeV}$ and width $\Gamma=(135.8\pm60.8\pm22.5)~\mathrm{MeV}$ is observed. Together with the BESIII observations of $Y(4320)$ and $Y(4390)$ in other hidden-charm decay channels, these results demonstrate that multiple structures appear in the $4.3$–$4.4~\mathrm{GeV}$ energy region across different final states, while whether these structures arise from a common resonance or differ significantly among decay channels remains an open question.

The BESIII Collaboration has extended the study of vector charmonium-like states to the strange meson channel $e^+e^- \to K^+K^- J/\psi$ using data collected at center-of-mass energies between $4.127~\mathrm{GeV}$ and $4.600~\mathrm{GeV}$~\cite{BESIII:2022joj}. The measured cross sections exhibit two resonant contributions, one consistent with the established $Y(4220)$ and a second structure observed for the first time with a statistical significance exceeding $8\sigma$. This higher-mass state is denoted as $Y(4500)$, with a measured mass of $(4484.7 \pm 13.3 \pm 24.1)~\mathrm{MeV}$ and a width of $(111.1 \pm 30.1 \pm 15.2)~\mathrm{MeV}$. The observation of $Y(4500)$ in the $K^+K^- J/\psi$ final state establishes a new vector charmonium-like $Y$ state in this mass region and provides a qualitatively new experimental input beyond non-strange hidden-charm processes.

Prior to 2017, two vector charmonium-like $Y$ states, $Y(4630)$ and $Y(4660)$, had been \changelabel{reported} in the vicinity of 4.6~GeV through $e^+e^-$ annihilation.  $Y(4630)$ was first observed by Belle in the baryonic channel $e^+e^- \to \Lambda_c^+ \Lambda_c^-$, with a mass of $4634^{+8+5}_{-7-8}~\text{MeV}/c^2$ and a width of $92^{+40+10}_{-24-21}~\text{MeV}$ \cite{Belle:2008xmh}. $Y(4660)$ was discovered in the hidden-charm process $e^+e^- \to \psi(3686)\pi^+\pi^-$ by Belle, with a mass of $4664\pm11\pm5~\text{MeV}$ and a width of about $48\pm15\pm3~\text{MeV}$ \cite{Belle:2007umv}. Since 2017, the experimental landscape of $Y$ states above 4.6~GeV has also become increasingly rich. In 2019, Belle reported a vector charmonium-like state in $e^+e^- \to D_s^+ D_{s1}(2536)^-$, with a mass of $4625.9^{+6.2}_{-6.0}\pm0.4~\text{MeV}$ and a width of $49.8^{+13.9}_{-11.5}\pm4.0~\text{MeV}$ \cite{Belle:2019qoi}, whose resonant parameters are consistent with those of $Y(4630)$ within the uncertainties. \changelabel{ The structures around 4.6 GeV have been reported in different processes, while whether they correspond to a single state~\cite{Guo:2010tk} or multiple different states remains an open issue.}

Recent high-precision measurements by the BESIII Collaboration in the open-charm channel $e^+e^- \to D^{*0}D^{*-}\pi^+$ have revealed three vector charmonium-like states \cite{BESIII:2023cmv}. The third state has a mass of $4675.3 \pm 29.5 \pm 3.5~\text{MeV}/c^2$ and a width of $218.3 \pm 72.9 \pm 9.3~\text{MeV}$. While its mass is consistent with that of the earlier $Y(4660)$, its width is significantly larger, likely reflecting different interference effects or different theoretical origins. This marks the first clear observation of a $Y(4660)$-like state in an open-charm decay. Additionally, a second resonance, with mass $4469.1 \pm 26.2 \pm 3.6~\text{MeV}/c^2$ and width $246.3 \pm 36.7 \pm 9.4~\text{MeV}$, is compatible with those of the $Y(4500)$ previously observed in $e^+e^- \to K^+K^-J/\psi$ \cite{BESIII:2022joj}. If assuming them to be the same state, the substantial branching fraction of $Y(4500)$ to $D^{*}\bar{D}^{*}\pi$ compared to $K\bar{K}J/\psi$ \cite{BESIII:2023cmv}, would challenge a hidden-strangeness tetraquark interpretation for $Y(4500)$.

Very recently, the BESIII measurements of the $e^+e^- \to K_S^0 K_S^0 J/\psi$ cross section reported evidence for a new vector charmonium-like state, $Y(4710)$, with a statistical significance of 4.2$\sigma$ \cite{BESIII:2022kcv}. Its mass and width are determined to be $M = 4704.0 \pm 52.3 \pm 69.5~\text{MeV}$ and $\Gamma = 183.2 \pm 114.0 \pm 96.1~\text{MeV}$, respectively.  Notably, no significant signal is observed for  $Y(4500)$ in this decay mode \cite{BESIII:2022kcv}, in contrast to its clear appearance in $K^+K^-J/\psi$. Independently, a high-precision study of $e^+e^- \to D_s^{*+}D_s^{*-}$ reveals another distinct structure near 4.79 GeV, denoted $Y(4790)$, with a significance exceeding $6.1\sigma$ \cite{BESIII:2023wsc}. Its mass and width are found to be $4786\sim4793~\text{MeV}$ and $ 27.1\sim60~\text{MeV}$, respectively, depending on the concrete fit scenario. 
These accumulating observations above 4.6 GeV further underscore the complexity of understanding the sector of vector $Y$ states in the $XYZ$ family.

As summarized in Fig. \ref{fig:ObservedY}, a growing number of charmonium-like $Y$ states have been observed with improving experimental precision. Explaining these states---particularly within a unified framework that can systematically assign them---poses a major theoretical challenge. The so-called "$Y$ problem" was formally articulated in the {\it White Paper on the Future Physics Program of BESIII}~\cite{BESIII:2020nme}. A central issue is the overabundance of $Y$ states, which presents a significant challenge regardless of whether they are interpreted as exotic hadrons or conventional charmonia.

We need novel approaches to address the $Y$ problem. In fact, several key advances have been made over the past decade since 2011: the introduction of a Fano-like interference mechanism as a "resonance killer"~\cite{Chen:2010nv,Chen:2011kc,Chen:2015bft,Chen:2017uof}, the identification of  $Y(4220)$---a narrow structure near 4.2~GeV---as a scaling point for constructing the higher charmonium family within an unquenched framework, and the emergence of a distinctive mass spectrum for vector charmonia in the 4-4.5~GeV region, which aligns with all available experimental data and has garnered experimental support. Indeed, this resulting mass spectrum helps clarify why the mass of $\psi(4160)$ appears shifted to about 4190~MeV. In the following section, we review this progress. 

\subsection{Toward a unified theoretical framework for the $Y$ problem}

\subsubsection{Non-resonant interference revisited: killing $Y(4320)$ and $Y(4390)$ and confirming $Y(4220)$ as a genuine resonance}

\begin{figure}[htbp]
\centering
\includegraphics[width=0.43\textwidth]{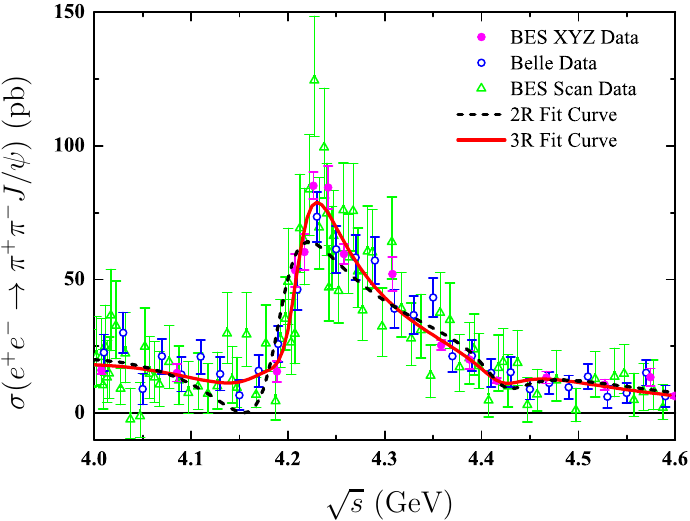}
\hspace{0.05\textwidth}
\includegraphics[width=0.43\textwidth]{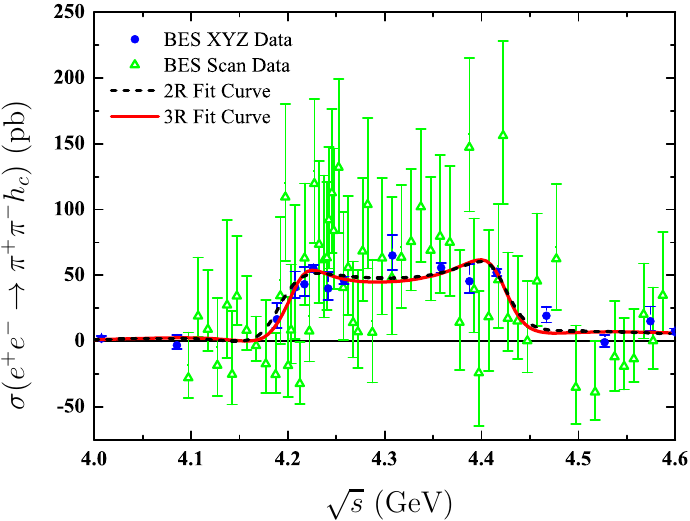}
\hspace{0.05\textwidth}
\includegraphics[width=0.43\textwidth]{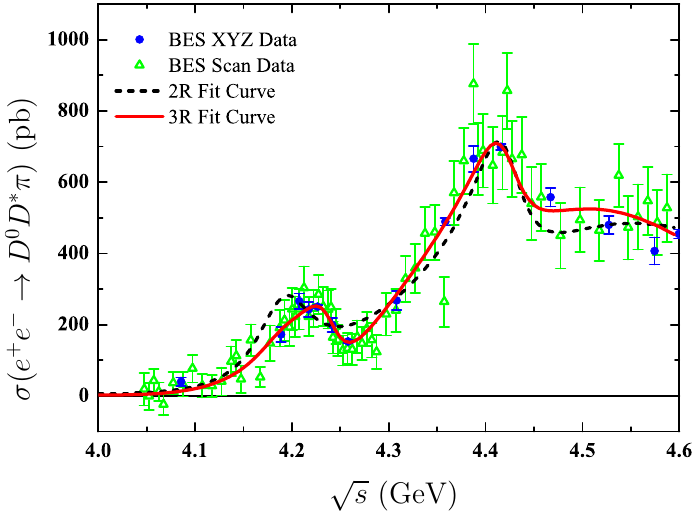}
\caption{ The Fano-like interference fit to the cross sections for the $e^+e^-\to\pi^+\pi^- J/\psi$, $e^+e^-\to \pi^+\pi^- h_c$ and $e^+e^- \to D^0D^{*-}\pi^+$ under the $2R$ and $3R$ fit schemes. Figures are from Refs. \cite{Chen:2017uof}.}
\label{fig:sec4-reskill}
\end{figure}

The non-resonant Fano-like interference mechanism has proven to be a powerful tool for understanding charmonium-like $Y$ states~\cite{Chen:2010nv,Chen:2011kc,Chen:2015bft,Chen:2017uof}. Initially, it was used to explain the broad $Y(4260)$ as an interference effect rather than a true resonance~\cite{Chen:2010nv}. Following that work, as early as in 2015, Chen \emph{et al.} pointed to the existence of a distinct narrow resonance signal near 4.22 GeV by analyzing cross section data from channels like $\pi^+\pi^-\psi(3686)$ and $\omega\chi_{c0}$ independently in the electron-positron annihilation~\cite{Chen:2014sra,Chen:2015bma}. Theoretical study of OZI-allowed two-body strong decays of higher charmonia further supported its interpretation as a missing conventional $\psi(4S)$ charmonium~\cite{He:2014xna}. Therefore, the high-precision measurements of the cross sections of $e^+e^-\to\pi^+\pi^- J/\psi$~\cite{BESIII:2016bnd} and $e^+e^-\to \pi^+\pi^- h_c$~\cite{BESIII:2016adj} as well as open-charm process $e^+e^- \to D^0D^{*-}\pi^+$~\cite{BESIII:2018iea} by BESIII allow us to re-examine the role of this non-resonant framework in understanding the observed enhancement structures. In Ref.~\cite{Chen:2017uof}, Chen, Liu and Matsuki analyzed these data with a coherent sum of amplitudes from a smooth background (parameterized phenomenologically) and the intermediate resonant contributions:
\[
\mathcal{M}_{\text{Total}} = g u^2 e^{-a u^2} + e^{i\phi_k} \sum_{k}  \frac{\sqrt{12\pi R_k \,\Gamma_k}}{s - m_k^2 + i m_k \Gamma_k} \cdot \sqrt{\frac{\Phi(s)}{\Phi(m_k^2)}},
\]
where $k$ corresponds to two established charmonium states $\psi(4160)$ and $\psi(4415)$ or other new resonance state, and $R_k \equiv \Gamma_{k}^{e^+e^-}BR(k \to \text{final states})$. \changelabel{ It is worth mentioning that introducing such a background into the amplitudes is inherently model-dependent. In particular, caution is highly warranted when fitting cross sections that feature very broad structures because it may itself generate broad peak-like shape. In practice, one should constrain the model parameters as 
much as possible to ensure that the background retains the characteristics of a smooth continuum.}  This model, with only two resonances $\psi(4160)$ and $\psi(4415)$ (2R fit), fails to describe the data near 4.2 GeV for the $e^+e^-\to\pi^+\pi^- J/\psi$ and $e^+e^- \to D^0D^{*-}\pi^+$. The fit quality dramatically improves upon introducing an additional resonance, $Y(4220)$ (3R fit), yielding $\chi^2/\text{n.d.f.}=118/153$ and $226/78$ compared to $\chi^2/\text{n.d.f.}=205/157$ and $69/74$ of the $2R$ fit scheme for $e^+e^-\to\pi^+\pi^- J/\psi$ and $e^+e^- \to D^0D^{*-}\pi^+$, respectively. The interference between $\psi(4160)$, $\psi(4415)$, and the background successfully reproduces the broad enhancements reported as $Y(4320)$ and $Y(4390)$~\cite{BESIII:2016bnd,BESIII:2016adj}, effectively removing them as necessary independent states. Simultaneously, it explains the lack of clear $\psi(4160)$ and $\psi(4415)$ peaks in these data. The key quantitative result is the consistent extraction of the $Y(4220)$ parameters: $m = 4207 \pm 12$ MeV and $\Gamma = 58 \pm 38$ MeV from $e^+e^- \to \pi^+\pi^- J/\psi$, and $m = 4211 \pm 6$ MeV and $\Gamma = 47 \pm 13$ MeV from $e^+e^- \to \pi^+\pi^- h_c$. Crucially, the same $Y(4220)$ is essential to describe the open-charm process $e^+e^- \to D^0D^{*-}\pi^+$, where it shows a large production rate consistent with dominant open-charm decays. The survival of $Y(4220)$ against this interference-based filter, which confirms its intrinsic resonant nature. The relevant fitted line shapes of cross sections for these three processes are presented in Fig. \ref{fig:sec4-reskill}. This work highlights that a comprehensive evaluation of non-resonant explanations is crucial, as it decisively distinguishes persistent spectroscopic signals from interference artifacts. Indeed, the non-resonant interpretation adheres to Occam's Razor principle, which holds that \textit{“Numquam ponenda est pluralitas sine necessitate”}—plurality should not be posited without necessity.

\subsubsection{$Y(4220)$ as a scaling point to reconstructe $J/\psi$ family within the unquenched picture}

\begin{figure}[htbp]
\centering
\includegraphics[width=0.25\textwidth]{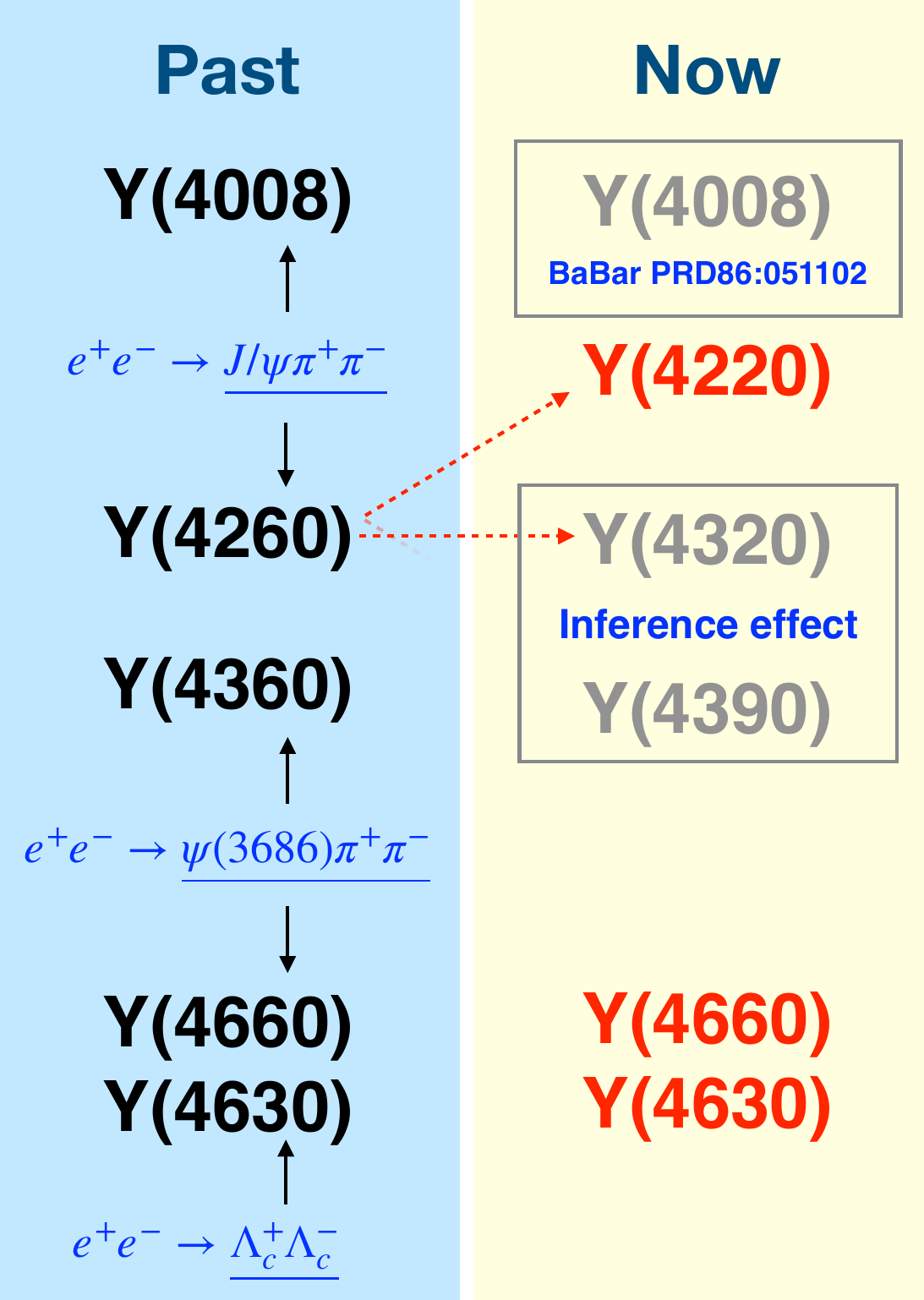}
\hspace{0.05\textwidth}
\includegraphics[width=0.5\textwidth]{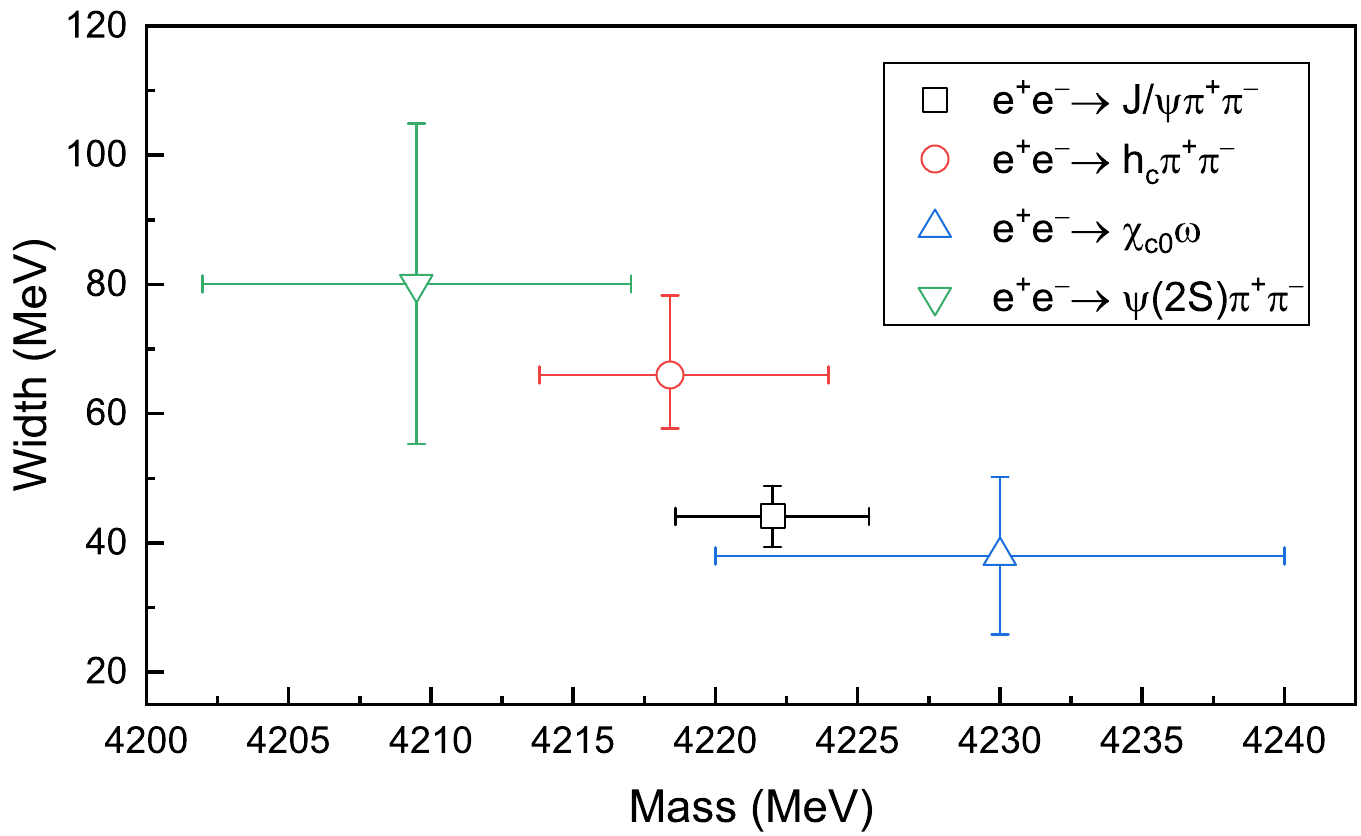}
\caption{Status of the charmonium-like $Y$ states before 2017 and their updated situation after 2017 (left panel), and the measured resonance parameters of the $Y(4220)$ state in several hidden-charm processes (right panel). Figure adapted from Refs.~\cite{Wang:2020prx,Wang:2019mhs}.}
\label{fig:4220now}
\end{figure}

As discussed in the previous subsections, once the non-resonant interference mechanism is properly incorporated, several wide structures previously assigned as members of the $Y$ family, such as $Y(4320)$, $Y(4360)$, and $Y(4390)$, can be consistently understood as the non-resonant enhancements from the interference between the background and two established charmonium states $\psi(4160)$ and $\psi(4415)$~\cite{Chen:2010nv,Chen:2011kc,Chen:2015bft,Chen:2017uof}. This reasoning follows Occam’s razor by favoring a simpler physical picture with fewer assumed resonant states, among competing descriptions that reproduce the same phenomenology.
From this perspective, as of the experimental status around 2019, the energy region between 4.0 and 4.5~GeV accommodates only the single well-established and unambiguous $Y$ particle, namely $Y(4220)$, as summarized in Fig.~\ref{fig:4220now}. It is worth emphasizing that the remaining higher $Y$ states, $Y(4630)$ and $Y(4660)$, can also be accommodated in a unified manner within the same theoretical framework, as will be discussed at the end of this subsection.
Consequently, $Y(4220)$ is the key to understanding the entire $Y$ state family.

In fact, well before  $Y(4220)$ was conclusively established by experiment in 2017~\cite{BESIII:2016bnd,BESIII:2016adj}, the existence of a narrow vector state around 4.22~GeV had been pointed out as introduced in subsection \ref{sec4.2.3}. In brief, early experimental hints for such a narrow structure already appeared in the BESIII measurement of the cross section $e^+e^-\to h_c\pi^+\pi^-$ in 2013 \cite{BESIII:2013ouc}, which suggested a possible enhancement near 4.2~GeV \cite{Yuan:2013uta}. Subsequent measurements of $e^+e^-\to \omega\chi_{cJ}$ by BESIII revealed a narrow structure with a mass of $m=4230\pm8\pm6$~MeV and a width of $\Gamma=38\pm12\pm2$~MeV \cite{BESIII:2014rja}, which was interpreted  as the long-missing $\psi(4S)$ charmonium state in Ref. \cite{Chen:2014sra}. In 2014, based on a naive comparison of the mass spectra between the charmonium $J/\psi$ and bottomonium $\Upsilon$ families, He \emph{et al.} argued that the $\psi(4S)$ mass should lie around 4.263~GeV and that this state is expected to be relatively narrow \cite{He:2014xna}, whose property is highly consistent with this narrow state around 4.2 GeV \cite{BESIII:2013ouc,BESIII:2014rja}. Further support came from the analysis of the $e^+e^-\to \pi^+\pi^- \psi(3686)$ process. In 2015, Chen and Liu \emph{et al.} pointed out that the missing $\psi(4S)$ should manifest itself in this channel \cite{Chen:2015bma}, which was finally confirmed by the precise measurement of BESIII \cite{BESIII:2017tqk}. A combined analysis of three hidden-charm channels $e^+e^-\to \psi(3686)\pi^+\pi^-$, $h_c\pi^+\pi^-$, and $\chi_{c0}\omega$ demonstrated that the narrow structures observed around 4.2~GeV can be consistently attributed to a single underlying state \cite{Chen:2015bma}.

Building upon these theoretical and experimental developments, in Ref.~\cite{Wang:2019mhs}, Wang, Chen, Liu, and Matsuki proposed that $Y(4220)$ can serve as a benchmark state for calibrating unquenched corrections to the linear confining potential of the charmonium system above 4~GeV, which provides a direct and quantitative manifestation of unquenched effects in the vector charmonium family. Based on this idea, it was found that $Y(4220)$ can be successfully described as a mixed $4S$-$3D$ unquenched charmonium state within a relativistic quark model~\cite{Wang:2019mhs}, with the $4S$ component being dominant. In the following, we present a detailed discussion of this work and its implications for the spectroscopy of higher charmonium states.

In Ref.~\cite{Wang:2019mhs}, interpreting $Y(4220)$ as a higher vector charmonium state shaped by unquenched effects leads to several nontrivial consistency requirements. First, the assignment of the well-established charmonium spectrum below 4.2~GeV must remain compatible within the unquenched theoretical framework. Second, the scenario implies the existence of an partner state of $Y(4220)$ with correlated mass and decay properties, whose presence or absence in current data provides a decisive test. Third, the well-known $\psi(4415)$ must be accommodated consistently within the same unquenched dynamics, rather than as an isolated exception. Addressing these interrelated constraints is essential to evaluate the role of  $Y(4220)$ as a scaling point for the unquenched charmonium spectrum above 4~GeV.

To quantitatively illustrate these three key points, an unquenched potential model is adopted to describe the charmonium spectrum \cite{Wang:2019mhs}. In this approach, the interaction between the quark and anti-quark can be expressed by the Hamiltonian
\begin{eqnarray}
H=E_q+ E_{\bar{q}}+V_{\mathrm{eff}}(r), \label{Eq:Htot}
\end{eqnarray}
where $E_q=(p^2+m_q^2)^{1/2}$ and $E_{\bar{q}}=(p^2+m_{\bar{q}}^2)^{1/2}$ represent the relativistic kinetic energies of the quark and antiquark, respectively, and 
$V_{\mathrm{eff}}(r)$ denotes the effective quark--antiquark interaction.
The effective potential consists of
\begin{align}
V_{\mathrm{eff}}(r)
=H^{\mathrm{conf}}_{q\bar{q}}+H^{\mathrm{SO}}_{q\bar{q}}+H^{\mathrm{hyp}}_{q\bar{q}},
\end{align}
where $H^{\mathrm{conf}}_{q\bar{q}}$ contains the short-range one-gluon-exchange interaction,
$G(r)=-\frac{4\alpha_s(r)}{3r}$, and the long-range confinement potential $S(r)$.
To incorporate unquenched effects associated with virtual quark--antiquark pair creation from the vacuum fluctuation, 
the linear confinement potential adopted in Ref. \cite{Godfrey:1985xj}
\begin{align}
S(r)=br+c,
\end{align}
is modified by a screened form,
\begin{align}
S^{\mathrm{scr}}(r)=\frac{b\left(1-e^{-\mu r}\right)}{\mu}+c,
\end{align}
which effectively accounts for the coupled channel effect at large distances and is proved to partly equivalent to the coupled-channel effect \cite{Li:2009ad}.
The remaining terms, $H^{\mathrm{SO}}_{q\bar{q}}$ and $H^{\mathrm{hyp}}_{q\bar{q}}$ represent the spin--orbit interaction and the color--hyperfine interaction, respectively \cite{Godfrey:1985xj}.
Relativistic corrections in the unquenched potential model are implemented in two aspects: the smearing transformation of the interaction and the introduction of momentum-dependent factors \cite{Godfrey:1985xj}. 
By introducing the smearing function
\begin{equation}
\rho \left(\mathbf{r}-\mathbf{r}'\right)=\frac{\sigma^3}{\pi ^{3/2}}\mathrm{exp}\left[-\sigma^2\left(\mathbf{r}-\mathbf{r}'\right)^2\right],
\end{equation}
the screened confinement potential $S^\text{scr}(r)$ and the one-gluon exchange potential $G(r)$ are smeared out according to
\begin{align}
\tilde{S}^\text{scr}(r)/\tilde{G}(r)=\int d^3\mathbf{r}'\rho(\mathbf{r}-\mathbf{r}')S^\text{scr}(r')/G(r').
\end{align}
This smearing procedure effectively incorporates the nonlocality property of the quark–antiquark interaction. Besides, in a relativistic framework the effective interaction generally depends on the momenta of the interacting quarks in the center-of-mass frame. As a result, the smeared potential $\tilde{V}_i(r)$ is modified as
\begin{equation}
{\tilde{V}_i(r)}\to\left(\frac{m_qm_{\bar{q}}}{E_qE_{\bar{q}}}\right)^{1/2+\varepsilon_{i}} {\tilde{V}_i(r)}\left(\frac{m_qm_{\bar{q}}}{E_qE_{\bar{q}}}\right)^{1/2+\varepsilon_{i}},
\end{equation}
where the parameter $\varepsilon_i$ depends on the specific type of interaction, such as contact, vector spin–orbit, and tensor terms \cite{Godfrey:1985xj}.
By fitting the masses of fourteen experimentally established charmonium states, including $J/\psi$, $\psi(3686)$, $\psi(3770)$, $\psi(4040)$, $\psi(4160)$, $\psi_2(3823)$, $\eta_c(1S)$, $\eta_c(2S)$, $h_c(1P)$, $\chi_{cJ}(1P)$ $(J=0,1,2)$, $\chi_{c0}(2P)$, and $\chi_{c2}(2P)$, together with $Y(4220)$ assigned as the $\psi(4S)$ state, all model parameters can be constrained, which induced the mass of $\psi(4S)$ as 4274 MeV. 
Additionally, the charmonium mass spectrum below 4.2 GeV can be well described, especially for
the $S$-wave ground states (see Table II of Ref.~\cite{Wang:2019mhs} for details). Thus, the first key point of constructing the unquenched charmonium spectrum by the updated $Y(4220)$ as scaling point can be achieved.

\begin{figure}[htbp]
\begin{center}
\centering
\includegraphics[width=0.43\textwidth]{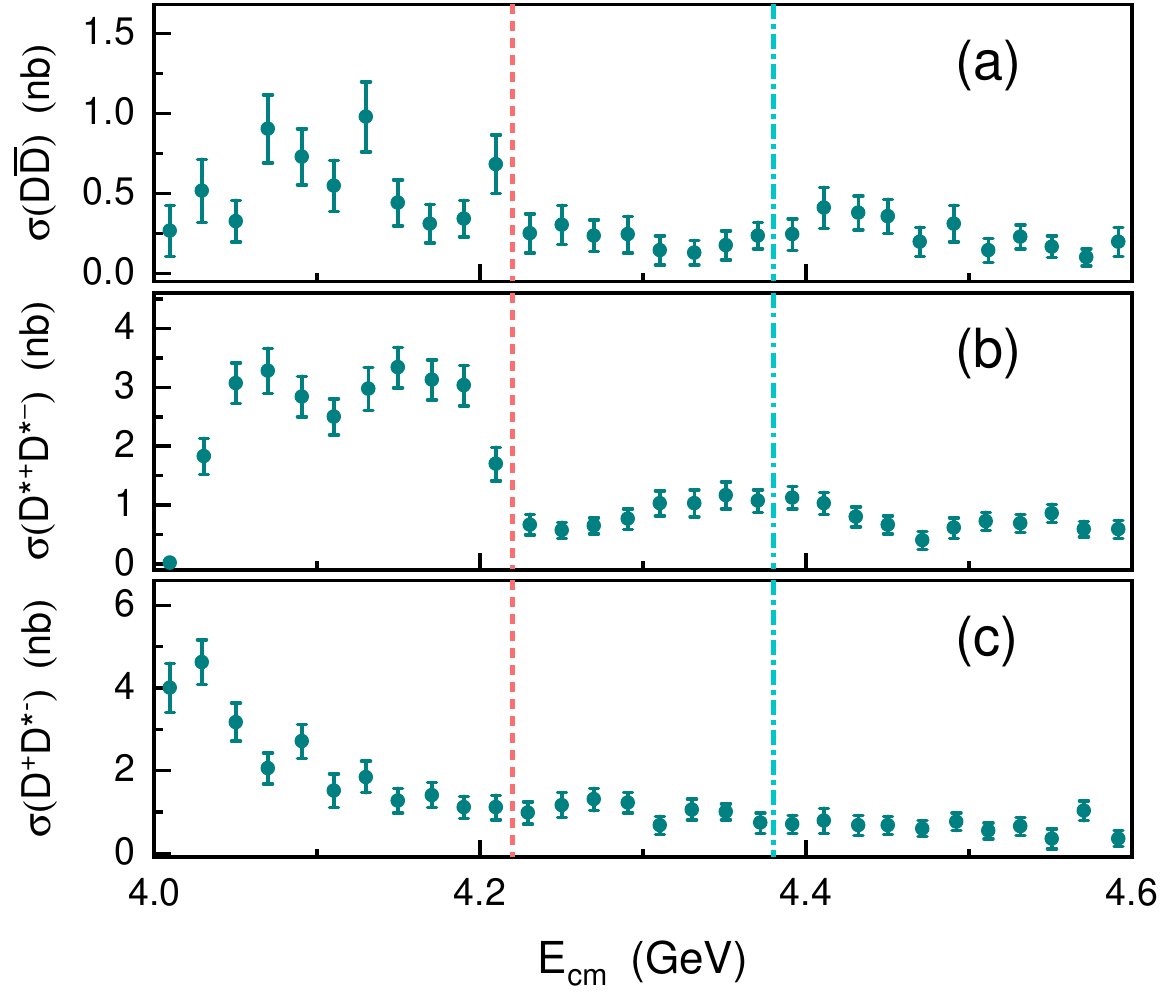}
\hspace{0.05\textwidth}
\includegraphics[width=0.43\textwidth]{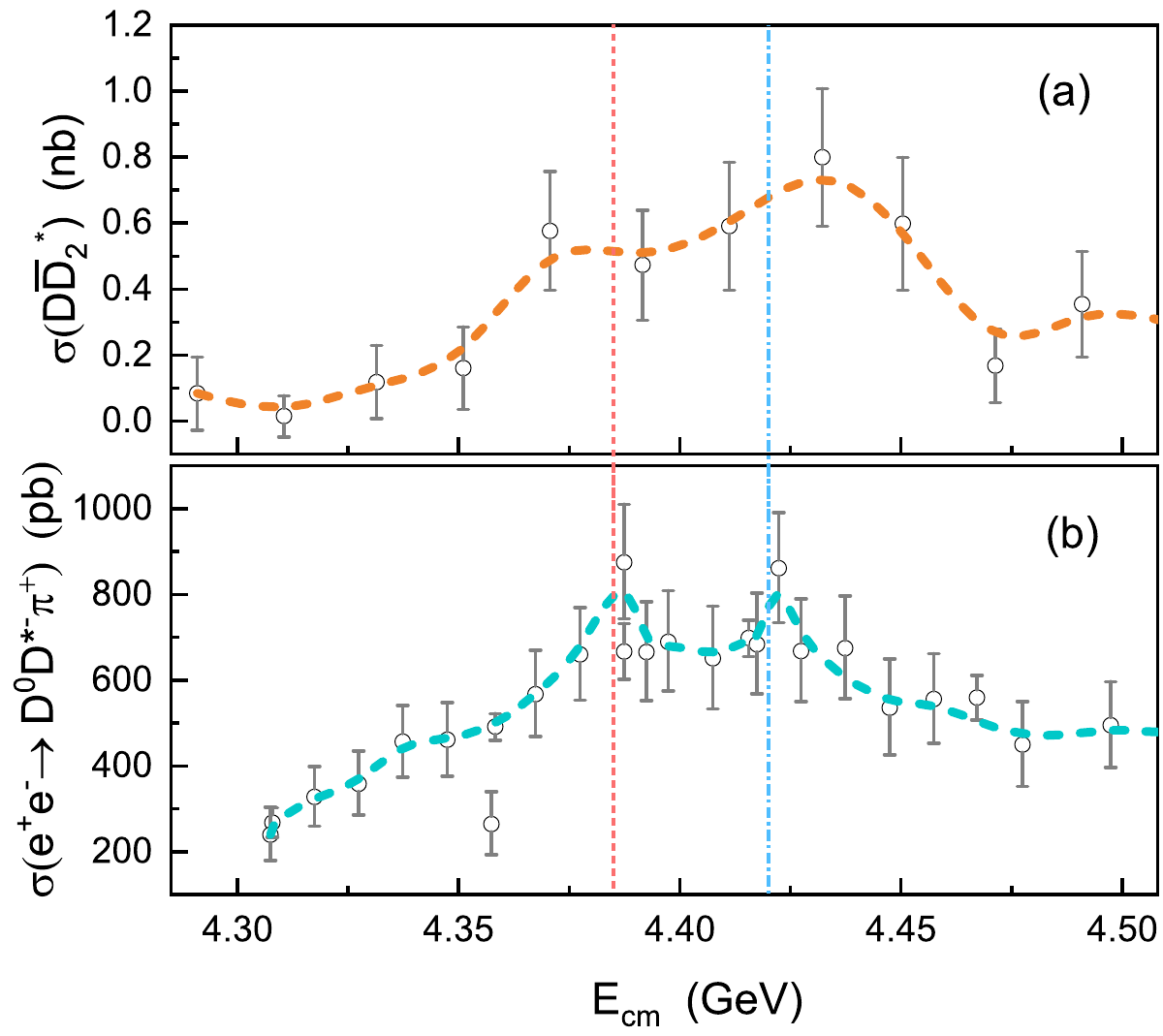}
\caption{ Experimental cross sections of selected open-charm production channels in $e^+e^-$ annihilation. 
The left panel shows two-body open-charm channels, including $e^+e^-\to D\bar D$, $D^{*+}D^{*-}$, and $D^+\bar D^{*-}$, where the vertical dashed lines indicate the center-of-mass energies at 4.22~GeV and 4.38~GeV, respectively. 
The right panel presents open-charm channels $e^+e^-\to D\bar D_2^*$ and $e^+e^-\to D^0D^{*-}\pi^+$, with dashed lines marking the energies around 4.385~GeV and 4.420~GeV. Figures are from Refs. \cite{Wang:2019mhs}. \label{4s3dopencharm}}
\end{center}
\end{figure}

After fixing $Y(4220)$ as the $\psi(4S)$ charmonium state, the authors in Ref.~\cite{Wang:2019mhs} further examine whether its decay properties are compatible with existing experimental information and investigate the expected features of its $D$-wave partner state. The analysis is performed within the quark pair creation model combined with the input of exact wave function directly solved from the unquenched potential model, with particular emphasis on open-charm decay channels.
The total width of $\psi(4S)$ is calculated to be $\Gamma_{\psi(4S)} = 27.2~\mathrm{MeV}$, which is consistent with the experimentally observed narrow width of $Y(4220)$. The predicted decay pattern is highly selective. The $D^{\ast}\bar D^{\ast}$ channel dominates, contributing about $88\%$ of the total width $\Gamma_{\psi(4S)}$, while the $D\bar D$ mode is relatively suppressed with branching ratio of $9.39\%$ and decays into strange open-charm channels and $D\bar D^{\ast}$ are found to be almost negligible, which are not contradict the collected Belle data of open-charm channels as shown in Fig. \ref{4s3dopencharm}. Although the decay $\psi(4S)\to D_1(2420)\bar D$ is not kinematically allowed, the strong $S$-wave coupling to the $D_1(2420) \bar D$ channel is emphasized as an important source of coupled-channel effects, which may influence production processes such as $e^+e^-\to D^0 D^{\ast-}\pi^+$ by the virtual $D_1(2420)^0\bar D^0$ channel as revealed by BESIII~\cite{BESIII:2018iea}.

The existence of a $\psi(4S)$ state naturally implies the presence of a partner vector charmonium $\psi(3D)$. Within the same framework, this $\psi(3D)$ state is predicted to have a mass of $M_{\psi(3D)} \simeq 4.33~\mathrm{GeV}$ and a total width of $\Gamma_{\psi(3D)} = 28.8~\mathrm{MeV}$. Its decay pattern differs qualitatively from that of $\psi(4S)$. The dominant decay mode is expected to be $D\bar D$, accounting for about $37\%$ of the total width, with sizable contributions also arising from the $D\bar D^{\ast}$ and $D^{\ast}\bar D^{\ast}$ channels. Experimentally, however, no clear enhancement corresponding to a resonance around 4.33~GeV is observed in these open-charm channels as shown in Fig. \ref{4s3dopencharm}. 


Additional experimental hints arise from the open-charm process $e^+e^-\to D\bar D_2^{\ast}(2460)$. A possible enhancement around 4.37~GeV is visible in early Belle data for the $D^0D^-\pi^+$ final state as shown in the right panel of Fig. \ref{4s3dopencharm}. However, the calculated partial width for the decay $\psi(3D)\to D \bar D_2^{\ast}(2460)$ is only $67.6~\mathrm{keV}$, corresponding to an extremely small branching fraction. Consequently, this potential structure in this channel cannot be attributed to a pure $\psi(3D)$ resonance. In addition, the electronic width of a vector $D$-wave charmonium is expected to be one to two orders of magnitude smaller than that of an $S$-wave state, implying that a pure $\psi(3D)$ resonance would be strongly suppressed in $e^+e^-$ annihilation and therefore difficult to observe through hidden-charm channels. Taken together, these observations indicate that while the decay properties of $\psi(4S)$ are broadly compatible with existing experimental information, there is no experimental evidence supporting the existence of its partner $\psi(3D)$ state. To understand this puzzling phenomenon, a new idea  of the $S$-$D$ mixing was further proposed~\cite{Wang:2019mhs}.

\subsubsection{The $4S$-$3D$ mixing scheme in charmonium}

\begin{figure}[htbp]
\begin{center}
\centering
\includegraphics[width=0.43\textwidth]{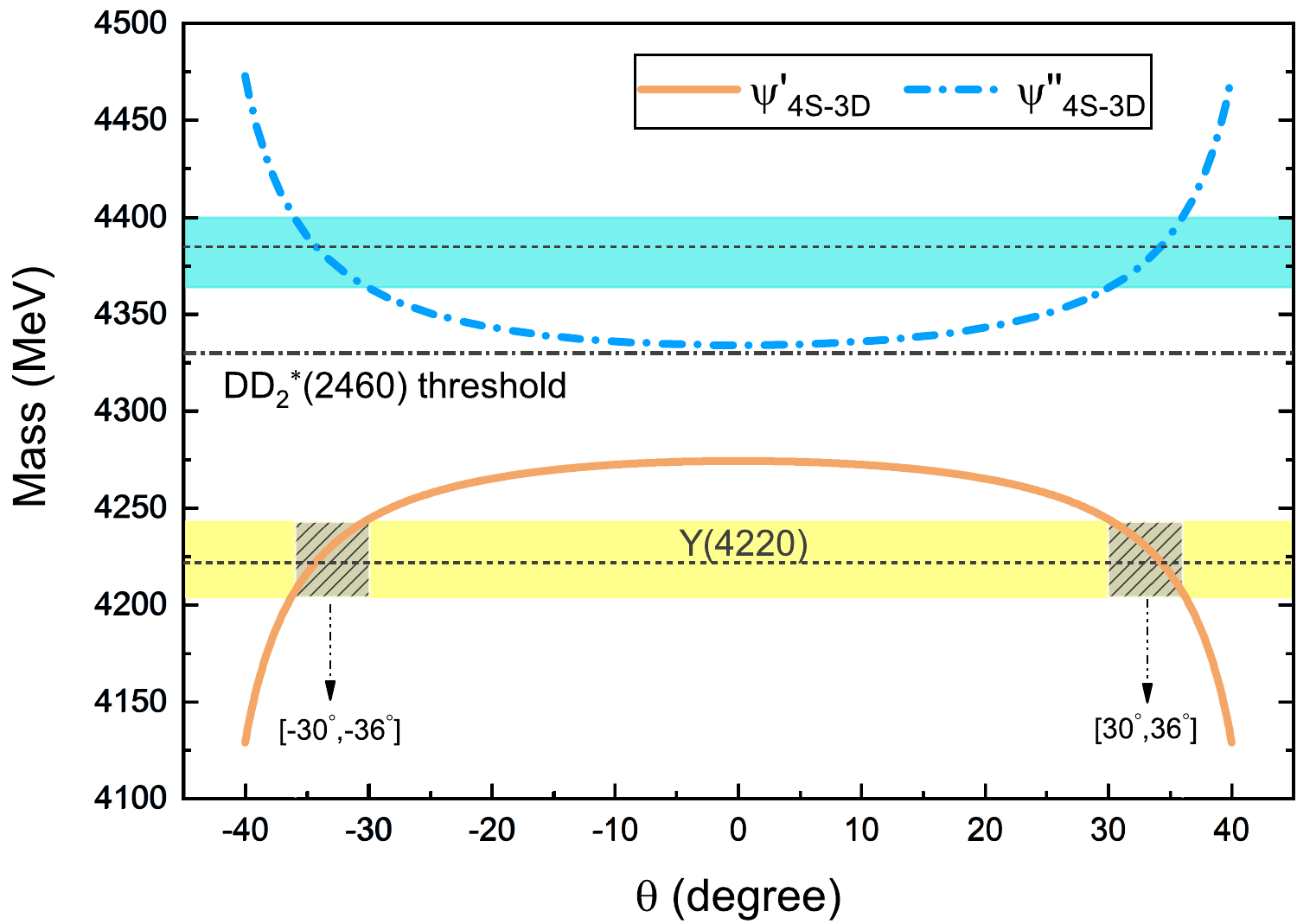}
\hspace{0.05\textwidth}
\includegraphics[width=0.43\textwidth]{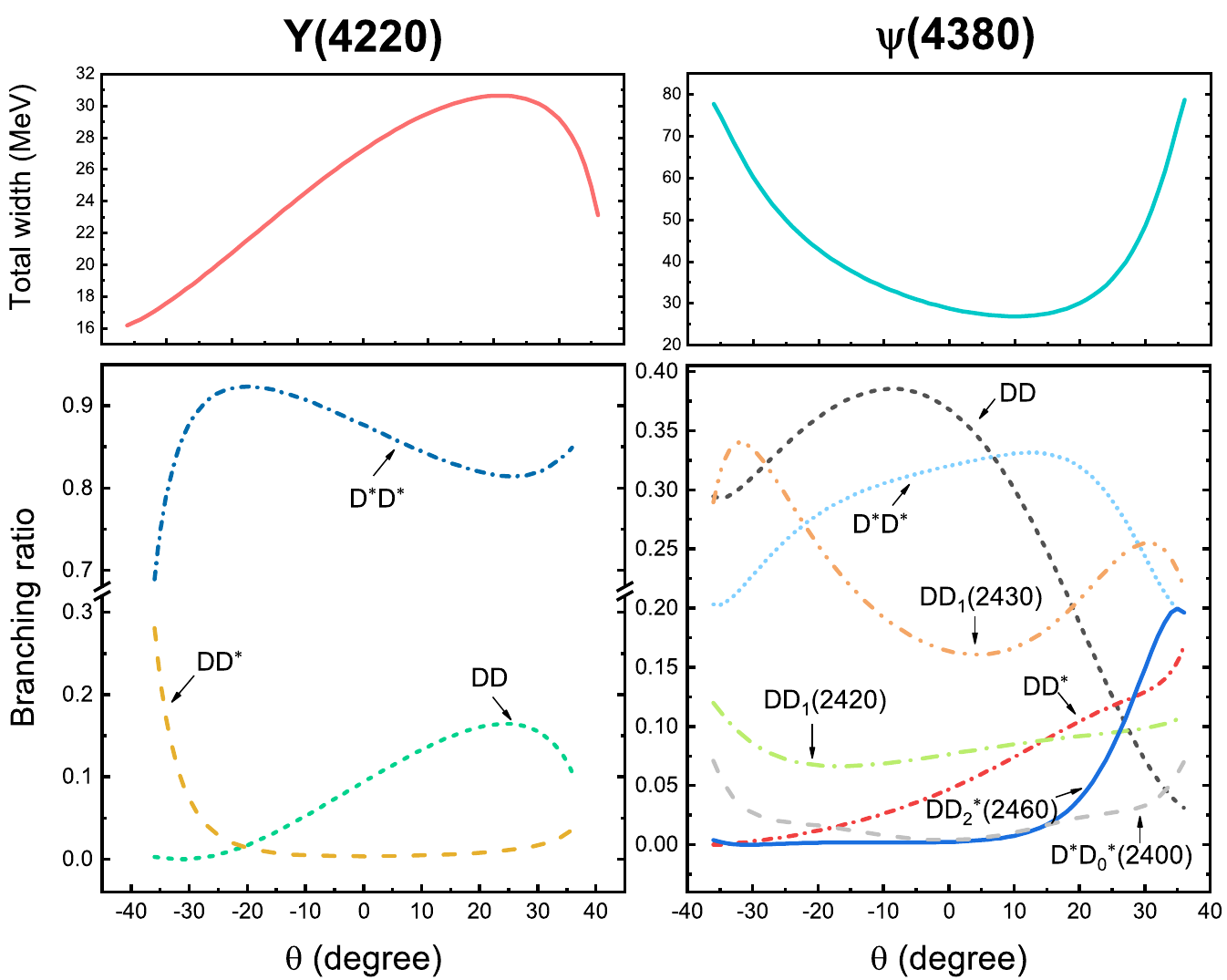}
\caption{ The masses and the open-charm decay behaviors of $\psi^{\prime}_{4S-3D}$ and $\psi^{\prime\prime}_{4S-3D}$ with dependence on the $4S$-$3D$ mixing angle $\theta$. For the left panel, the yellow and cyan bands denote the measured mass range of $Y(4220)$ and the predicted mass range of $\psi(4380)$, respectively, and the two shaded regions represent the mixing angle interval, in which the theoretical results of $\psi^{\prime}_{4S-3D}$ meet the measurements of $Y(4220)$. \label{4s3dmass}}
\end{center}
\end{figure}

To address this difficulty mentioned above, the authors of Ref.~\cite{Wang:2019mhs} further develop and explore a concrete $4S$--$3D$ mixing framework. In this scheme, the physical states are linear combinations of the pure $\psi(4S)$ and $\psi(3D)$ basis states, parameterized by a mixing angle $\theta$, i.e.,
\begin{eqnarray}
\left( \begin{array}{c}  |\psi_{4S-3D}^\prime\rangle\\  |\psi_{4S-3D}^{\prime\prime}\rangle\end{array} \right) =
\left( \begin{array}{cc} \cos{\theta} & \sin{\theta} \\
                         -\sin{\theta} & \cos{\theta} \end{array} \right)
\left( \begin{array}{c} |4^3S_1\rangle\\ |3^3D_1\rangle \end{array} \right). \label{sdmassf0}
\end{eqnarray}
Using the unquenched potential model masses $m_{4S}=4274$ MeV and $m_{3D}=4334$ MeV as input, the eigenstate masses of $\psi_{4S-3D}^\prime$ and $\psi_{4S-3D}^{\prime\prime}$ are obtained through diagonalization of the mass matrix, i.e.,
\begin{eqnarray}
m_{\psi_{4S-3D}^\prime}^2=\frac{1}{2}\left(m_{4S}^2+m_{3D}^2-\sqrt{(m_{3D}^2-m_{4S}^2)^2\sec^22\theta}\right)\label{sdmassf1},\\
m_{\psi_{4S-3D}^{\prime\prime}}^2=\frac{1}{2}\left(m_{4S}^2+m_{3D}^2+\sqrt{(m_{3D}^2-m_{4S}^2)^2\sec^22\theta}\right)\label{sdmassf2}.
\end{eqnarray}
Constraining the mixing scenario by the lower-mass eigenstate $\psi_{4S-3D}^\prime$ as experimentally observed $Y(4220)$ with mass interval of 4204$\sim$4243 MeV. This condition implies a relatively large mixing angle of $\theta \simeq \pm (30^\circ\text{--}36^\circ)$.
In Sec. \ref{coupled.S.D}, a coupled-channel dynamics mechanism explains the origin of such a large mixing angle.
Consequently, the higher-mass partner state $\psi_{4S-3D}^{\prime\prime}$ is predicted to lie in the range of 4364$\sim$4400 MeV and is tentatively labeled as $\psi(4380)$. These results are presented in the left panel of Fig. \ref{4s3dmass}.

The decay properties of the mixed $\psi_{4S-3D}^\prime$ are systematically analyzed. For the positive mixing angle, the total width of $\psi_{4S-3D}^\prime$ is calculated to be around $ 26.0$ MeV, consistent with the observed narrow width of $Y(4220)$. Its open-charm decay pattern remains dominated by the $D^{\ast}\bar D^{\ast}$ channel, closely resembling the predictions for a pure $\psi(4S)$ assignment. In contrast, a negative mixing angle leads to a significantly narrower width and an altered branching ratio structure, which seems to be disfavored by data.

Crucially, the mixing framework dramatically alters the predicted observables for the partner state $\psi(4380)$. Its total width is enhanced by approximately a factor of three compared to that of a pure $\psi(3D)$ state. The decay pattern undergoes a qualitative transformation: the dominant channels shift to $D \bar D_1(2430)$, $D^{\ast}\bar D^{\ast}$, and, most notably, a significantly enhanced $D \bar D_2^{\ast}(2460)$ mode, which can be seen in the right panel of Fig. \ref{4s3dmass}. This pronounced increased decay ratio in the $D \bar D_2^{\ast}(2460)$ channel provides a natural explanation for the potential structure around 4.37 GeV observed in the $e^+e^-\to D\bar D_2^{\ast}(2460)$ process in early Belle data as shown in Fig. \ref{4s3dopencharm}.

Furthermore, the dielectron widths of the mixed states were obtained. For the positive mixing angle solution, both $Y(4220)$ and $\psi(4380)$ are predicted to have $\Gamma_{e^+e^-}$ of $0.25\text{--}0.30$ keV. This value is substantially larger than the expected order-of-magnitude suppression for a pure $D$-wave charmonium, thereby significantly improving the prospects for observing the $\psi(4380)$ companion in $e^+e^-$ annihilation experiments. Overall, the $S$-$D$ mixing mechanism offers a unified and quantitative explanation for the key spectroscopic and decay puzzles associated with $Y(4220)$ and its partner state $\psi(4380)$.

In the proposed $4S$-$3D$ mixing framework,  $\psi(4380)$ as the partner of $Y(4220)$ is expected to have a non-negligible dilepton width, similar to $Y(4220)$ itself, suggesting that it could be observable in hidden-charm production channels. The high-precision cross section measurements of $e^{+}e^{-} \to \psi(3686)\pi^{+}\pi^{-}$ released by BESIII \cite{BESIII:2017tqk} provided an ideal opportunity to test this prediction. In Ref. \cite{Wang:2019mhs}, the Fano-like interference mechanism again was applied to analyze whether evidence for the predicted $\psi(4380)$ could be identified in this process. The analysis firstly combines a non-resonant continuum amplitude with resonant contributions from established charmonium states $\psi(4160)$ and $\psi(4415)$ together with $Y(4220)$, allowing for interference effects. A $3R$ fit including only these three resonances contributions describes much of the data, it fails to fully capture a distinct enhancement structure observed around 4.36 GeV in the cross-section distribution. Introducing a fourth resonance with free mass and width significantly improves the fit quality, yielding parameters $m = 4374 \pm 13\ \mathrm{MeV}$ and $\Gamma = 106 \pm 29\ \mathrm{MeV}$. These values are consistent with the predicted mass and enhanced width of the $\psi(4380)$ state within the $4S$-$3D$ mixing scheme, thereby providing direct evidence for its existence in the hidden-charm decay channel and further validating the mixing interpretation of $Y(4220)$.

\subsubsection{The vector charmonium mass spectrum in the unquenched picture for the 4–4.5 GeV region}

There remains one final problem in the construction of the $J/\psi$ family within the unquenched framework, i.e., the placement of the experimentally established charmonium $\psi(4415)$. In the calculated higher mass spectrum, the mass of the pure $\psi(5S)$ state is 4443~MeV, which lies closer to $\psi(4415)$ than the pure $\psi(4S)$ state. Given the compelling evidence that $Y(4220)$ structure is well described by a $4S$-$3D$ mixing scheme, it is natural to explore whether a similar mixing mechanism operates in the description of $\psi(4415)$. The $5S$-$4D$ mixing scheme is therefore proposed to interpret $\psi(4415)$, whose formula is very similar to those in Eqs. (\ref{sdmassf0})-(\ref{sdmassf2}).

Using the experimental mass range (4397$\sim$4438) MeV of $\psi(4415)$ as input, the $5S$-$4D$ mixing model yields a mixing angle of $\phi = \pm(18^\circ\text{--}36^\circ)$ and predicts a new partner state $\psi_{5S-4D}^{\prime\prime}$, denoted $\psi(4500)$, with a mass between 4489 and 4529 MeV. The calculated total width (12.6$\sim$19.9) MeV of $\psi(4415)$ is notably smaller than the PDG average of $62\pm20$ MeV. This apparent discrepancy is interpreted in the context of substantial inconsistencies in historical experimental determinations of the resonant parameters of $\psi(4415)$, as summarized in the PDG \cite{ParticleDataGroup:2024cfk}.

The analysis identifies $D^{*}\bar{D}^{*}$, $D \bar D_{1}(2420)$, and $D \bar D_{1}(2430)$ as the dominant open-charm decay channels of $\psi(4415)$. This prediction finds support in several experimental observations: 
\begin{itemize}
\item The BaBar Collaboration reported ratios $\Gamma(\psi(4415)\to D\bar{D})/\Gamma(\psi(4415)\to D^{}\bar{D}^{}) = 0.14\pm0.12\pm0.03$ and $\Gamma(\psi(4415)\to D^{*}\bar{D}+\mathrm{c.c.})/\Gamma(\psi(4415)\to D^{*}\bar{D}^{*}) = 0.17\pm0.25\pm0.03$, indicating a significantly larger $D^{*}\bar{D}^{*}$ partial width compared to other two-body open-charm modes. 
\item The BESIII measurement of $e^{+}e^{-} \to D^{0} D^{*-} \pi^{+}$ shows a structure around 4.42~GeV consistent with $\psi(4415)$ production via decay chains such as $D\bar D_{1} \to D\bar D^{*}\pi$. 
\item The Belle collaboration's observation of $\psi(4415) \to D\bar{D}_{2}^{}(2460)$ with a peak cross section of $0.74\pm0.17\pm0.08$~nb is compatible with the predicted branching fraction range of $(0.12-10.6)\%$  for this channel, given the considerable uncertainties in experimental resonance parameters.
\end{itemize}
The pattern of $\psi(4415)$ decays to final states involving charmed-strange mesons also aligns with experimental findings. While $\psi(4415) \to D_{s}^{*+} D_{s}^{*-}$ has been observed, the $D_{s}^{*+} D_{s}^{-}$ mode has not been seen, consistent with calculations showing substantially smaller partial widths for $D_{s}\bar{D}_{s}^{*}$  compared to $D_{s}^{*}\bar{D}_{s}^{*}$. Regarding the dilepton width, the predicted range of (0.147$\sim$0.344)~keV depending on the mixing angle sign encompasses the values reported by MARK~I ($0.44\pm0.14$)~keV \cite{Siegrist:1976br} and BESII ($0.35\pm0.12$)~keV \cite{BES:2007zwq} within experimental uncertainties.

For the predicted partner state $\psi(4500)$, the total width is 30$\sim$45~MeV with dominant decays into $D\bar D_{1}(2430)$, $D^{*}\bar D^{*}$, $D \bar D$, $D \bar D(2550)$, and $D \bar D_{1}(2420)$. The dilepton width of $\psi(4500)$ is predicted to be ($2.25\times10^{-3}\sim$0.189)~keV, suggesting that its discovery in $e^{+}e^{-}$ collisions will be challenging. In conclusion, the $5S$-$4D$ mixing scheme provides a \changelabel{self-consistent framework} for interpreting $\psi(4415)$ within the unquenched charmonium spectrum. While current experimental data exhibit large uncertainties that preclude definitive tests, the theoretical predictions show no contradiction with available observations and naturally lead to the prediction of a missing partner state, $\psi(4500)$. Interestingly, a prediction for this $\psi(4500)$ state exists prior to the relevant experimental discovery. Merely two years after, the BESIII experiment reported the first observation of a new vector state $Y(4500)$ around 4.5 GeV in the process $e^+e^- \to K^+K^-J/\psi$ \cite{BESIII:2022joj}. The connection between this observed $Y(4500)$ and the predicted $\psi(4500)$ has been studied in Ref. \cite{Wang:2022jxj} and will be discussed later.

\begin{figure}[htbp]
\begin{center}
\centering
\includegraphics[width=0.85\textwidth]{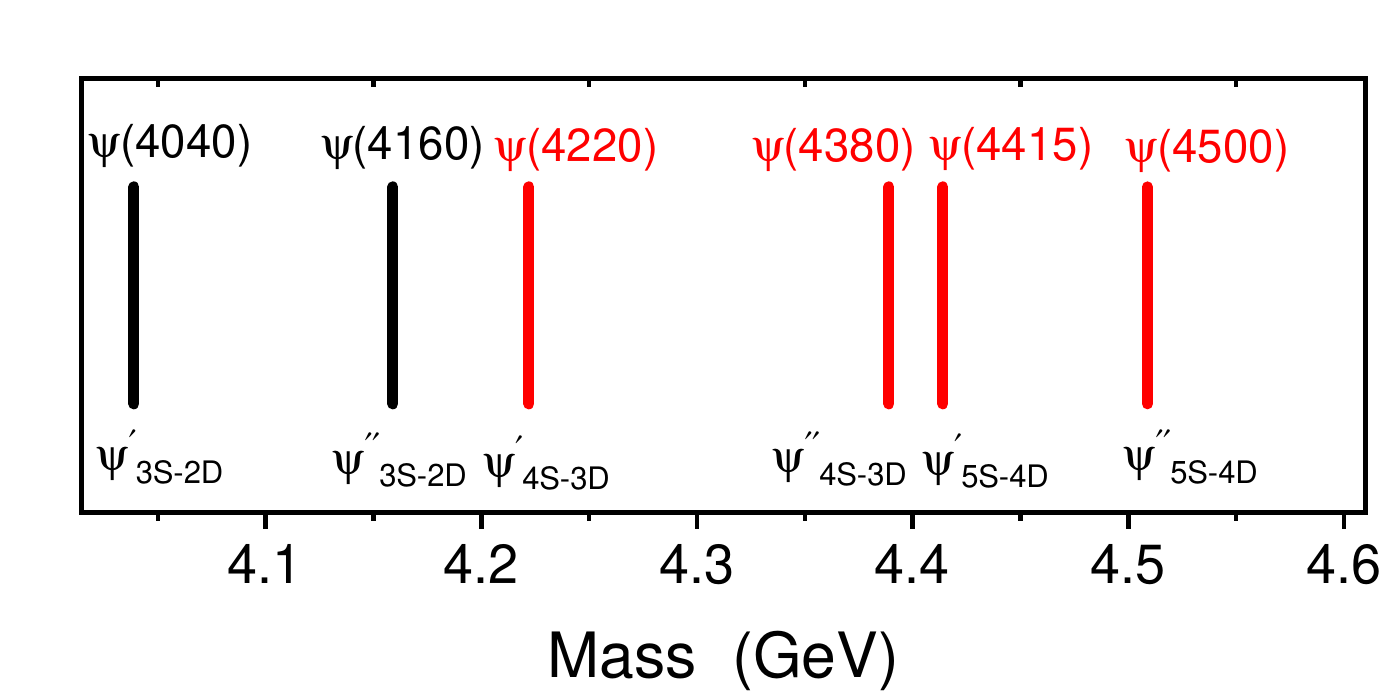}
\caption{Characteristic energy levels of the vector charmonium states in the unquenched charmonium spectrum. Figure is from Ref. \cite{Wang:2023zxj}. \label{Charaspect-2}}
\end{center}
\end{figure}

In conclusion, the unquenched framework leads to a characterized energy level of charmonium spectrum. In the energy region between 4.0 and 4.5~GeV, this spectrum accommodates the sequence $\psi(4040)$, $\psi(4160)$, $Y(4220)$, $\psi(4380)$, $\psi(4415)$, and the predicted $\psi(4500)$, as shown in Fig.~\ref{Charaspect-2}, which goes beyond the level structures anticipated in the quenched picture such as typical Cornell potential model, as summarized in Fig. \ref{Corfig} and Table \ref{Kangtable}. The emergence of such a reorganized characterized level structure indicates that the high excited charmonium spectrum encodes dynamical information beyond that of a static confining potential. From a broader perspective, these characterized energy levels provide a valuable probe of how confinement manifests itself in the presence of unquenched effects, offering insights into the nature of confinement in realistic physical systems.

\subsubsection{Are these $Y$ states around 4.6 GeV from 
$e^+e^-$ annihilation higher charmonia?}

As illustrated in Fig. \ref{fig:4220now}, two prominent states, $Y(4630)$ and $Y(4660)$, remained unclassified. The $Y(4660)$ was first observed by Belle in the hidden-charm process $e^+e^- \to \psi(3686)\pi^+\pi^-$~\cite{ Belle:2007umv}, while the $Y(4630)$ was reported in the open-charm channel $e^+e^- \to \Lambda_c^+\bar{\Lambda}_c^-$~\cite{Belle:2008xmh}. A persistent theoretical debate centered on whether these were distinct particles or different manifestations of a single structure \cite{Cotugno:2009ys,Guo:2010tk,Bugg:2008sk}. Building upon that all $Y$ states between 4.0 and 4.5 GeV and the associated puzzles could be understood within a concise and unified manner as introduced above, in Ref. \cite{Wang:2020prx}, Wang, Qian, Liu and Matsuki further extend to study the higher energy region, seeking to determine if a unified unquenched charmonium spectrum could provide a complete explanation for all observed $Y$ states.

A critical insight found in Ref. \cite{Wang:2020prx} is the inherent limitation of the model when extrapolated beyond its calibration region. It has been demonstrated that while a range of screening parameter $\mu$ values (e.g., 0.11–0.15 GeV limited by only single $Y(4220)$ state)  can equally well describe the established charmonium spectrum below 4.5 GeV, they lead to drastically different predictions for states near and above 4.6 GeV. Specifically, the predicted mass of a pure $\psi(6S)$ state varies by about 100 MeV across this $\mu$ range \cite{Wang:2020prx}. This large uncertainty indicates that the properties of the lower states do not uniquely constrain the relatively long-distance behavior of the confinment potential, and a new scaling point around the 4.6 GeV region itself is essential for a reliable description of higher mass spectrum.

With this theoretical ambiguity, the analysis first studied the possibility of interpreting $Y(4660)$ as a pure $\psi(6S)$ charmonium to play the role of this new scaling point \cite{Wang:2020prx}. However, they noted that such an assignment faces immediate difficulties. First, quark model calculations, including the results within a large range of $\mu$ value, typically predict a $\psi(6S)$ mass below 4.65 GeV, which is somewhat lower than the measured mass of $Y(4660)$. More importantly, the experimental mass difference $M_{Y(4660)} - M_{\psi(4415)} \approx 231$ MeV is found to be obviously larger than $M_{\psi(4415)} - M_{Y(4220)} \approx 191$ MeV. This trend contradicts the expected decreasing mass gap between consecutive radial excitations in a standard potential model, casting doubt on a simple $\psi(6S)$ assignment for $Y(4660)$. \changelabel{ Here, it is necessary to explain why the high-lying spectrum is compressed within a screened potential model. From the perspective of coupled-channel effects induced by intermediate charmed-meson 
loops, this behavior is natural: for low-lying states below the open-charm thresholds, the coupled-channel corrections are minimal, leaving their spectrum close to the conventional quenched quark model predictions. However, as the excitation energy increases and passes more open-charm thresholds, higher-lying states experience stronger coupled-channel shifts that significantly compress the mass spectrum. Such a phenomenon is consistent with the expected 
behavior of a screened confinement potential. Consequently, this unquenched model explains the remarkably small mass gap for $M_{\psi(4220)} - M_{\psi(4040)}$ under the $4S$--$3S$ assignment, while 
rendering the large mass gap of $M_{Y(4660)} - M_{\psi(4415)}$ somewhat unnatural if interpreted as a $6S$--$5S$ splitting within the same theoretical framework. } When moving the focus on another state $Y(4630)$, a puzzling experimental feature on the high-precision cross section of $e^+e^- \to \Lambda_c^+\bar{\Lambda}_c^-$ from BESIII was noted, which revealed a distinct "\text{platform}" enhancement near the threshold in the 4.57$\sim$4.60 GeV region \cite{BESIII:2017kqg}, starkly differing from the single peak structure $Y(4630)$ previously reported by Belle \cite{Belle:2008xmh}. This new experimental hint motivates the proposal that this novel structure near threshold may imply the existence of an additional resonance, which can be treated as candidate of the $\psi(6S)$ state and would naturally resolve the above tension.

\begin{figure}[htbp]
\centering
\begin{tabular*}{\textwidth}{@{\extracolsep{\fill}}cc}
\includegraphics[height=0.25\textheight]{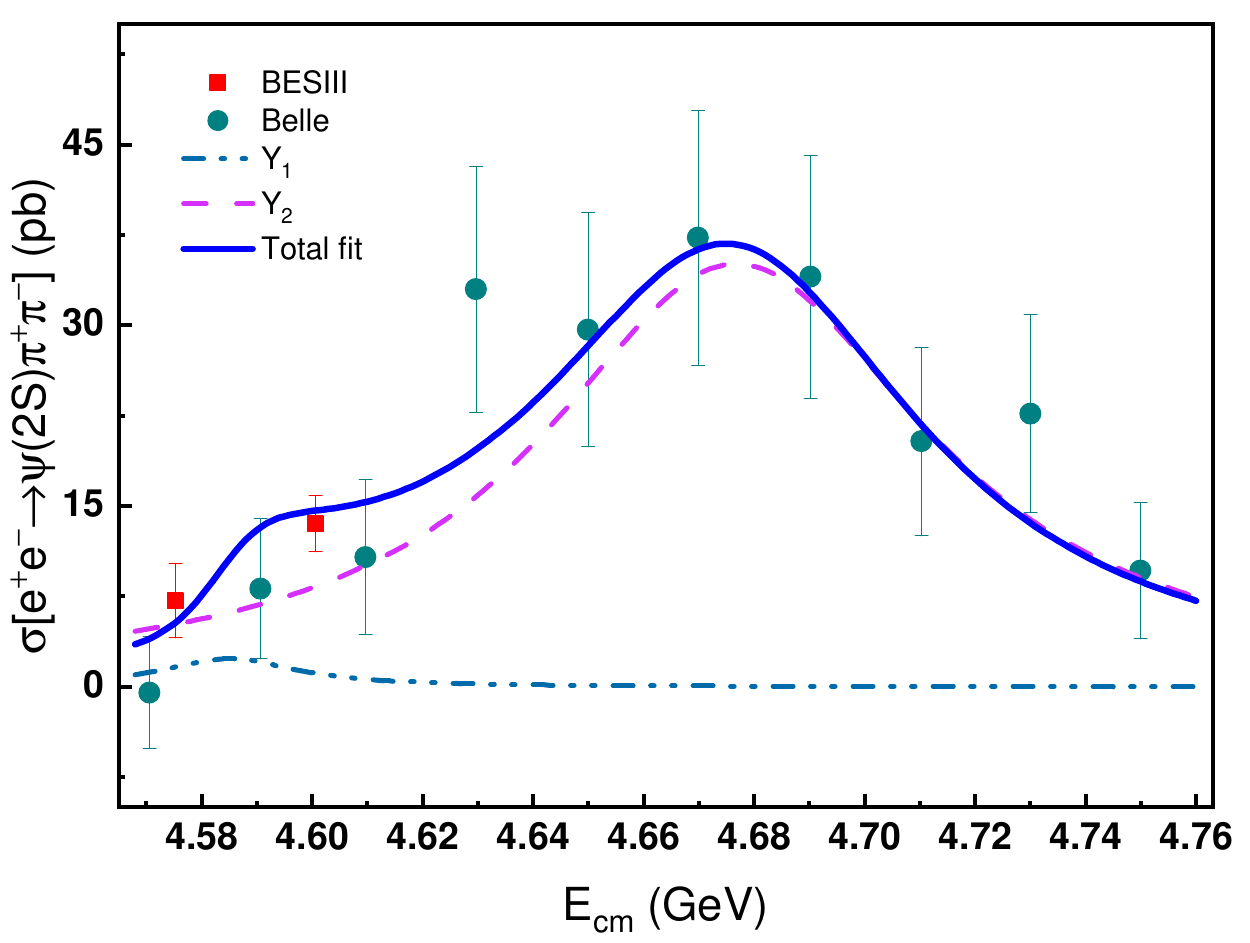}&\includegraphics[height=0.25\textheight]{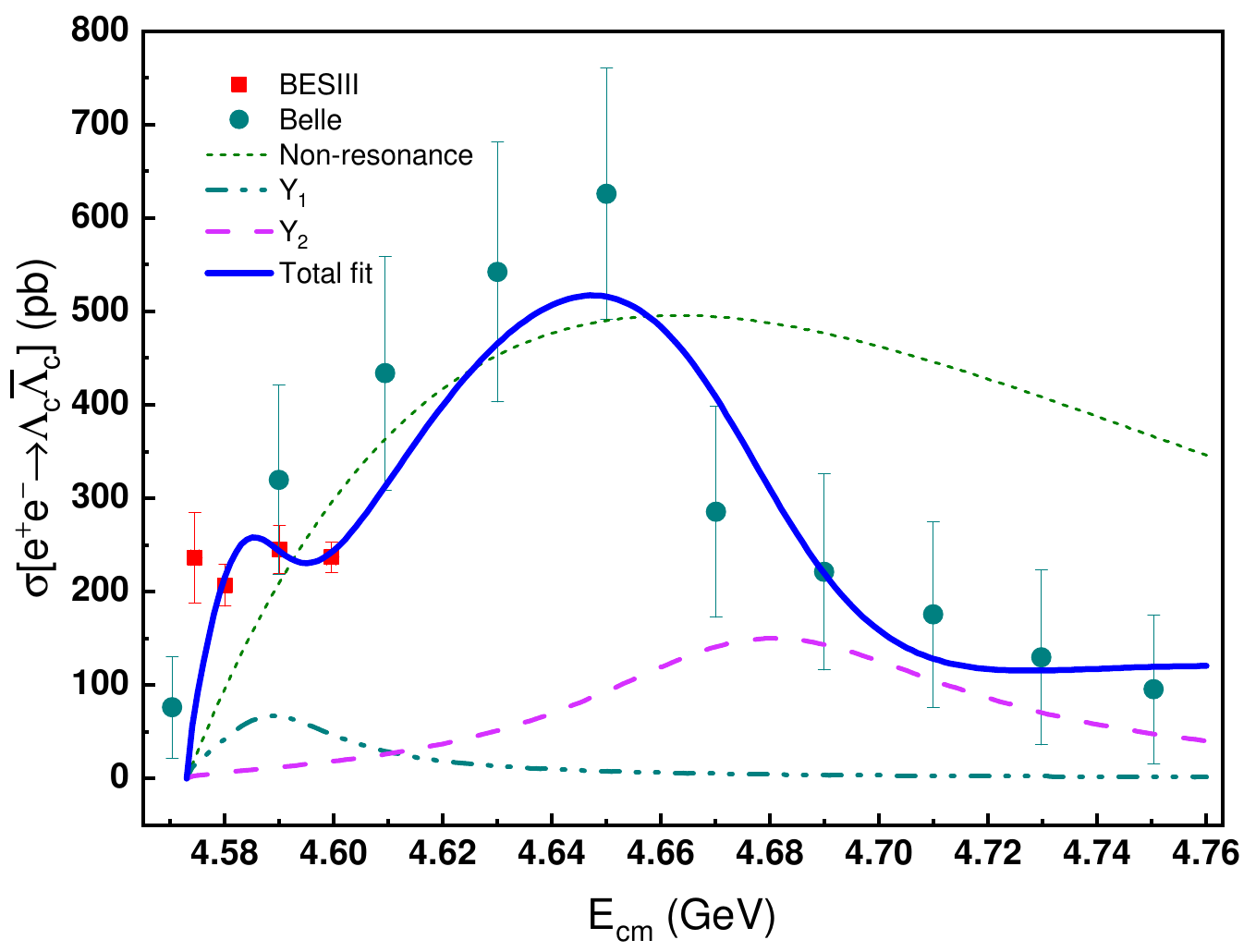}\\
(a)&\hspace{2.5em}(b)
\end{tabular*}
\caption{(a) The fitted results of the $e^+e^-\to\psi(3686)\pi^+\pi^-$ from Belle and BESIII~\cite{BESIII:2017tqk,Belle:2014wyt}.  (b) The fitted results of the $e^+e^-\to \Lambda_c\bar{\Lambda}_c $ from Belle and BESIII~\cite{Belle:2008xmh,BESIII:2017kqg}. The figures are adopted from Ref. \cite{Wang:2020prx}.}
\label{fig:Y4660}
\end{figure}

\begin{figure}[htbp]
\centering
\begin{tabular*}{\textwidth}{@{\extracolsep{\fill}}cc}
\includegraphics[height=0.34\textheight]{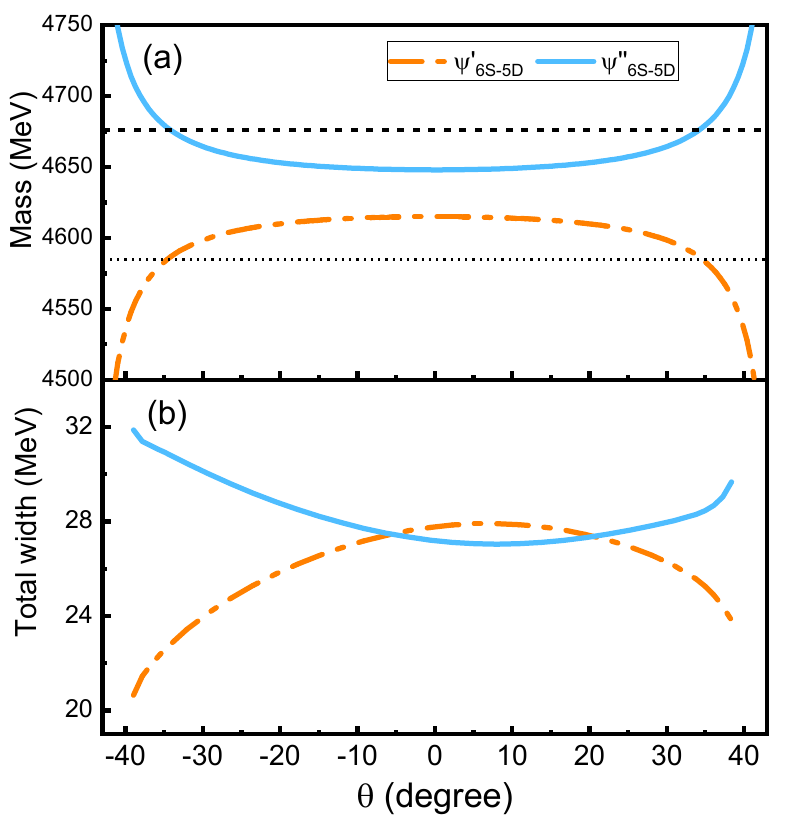}&\includegraphics[height=0.34\textheight]{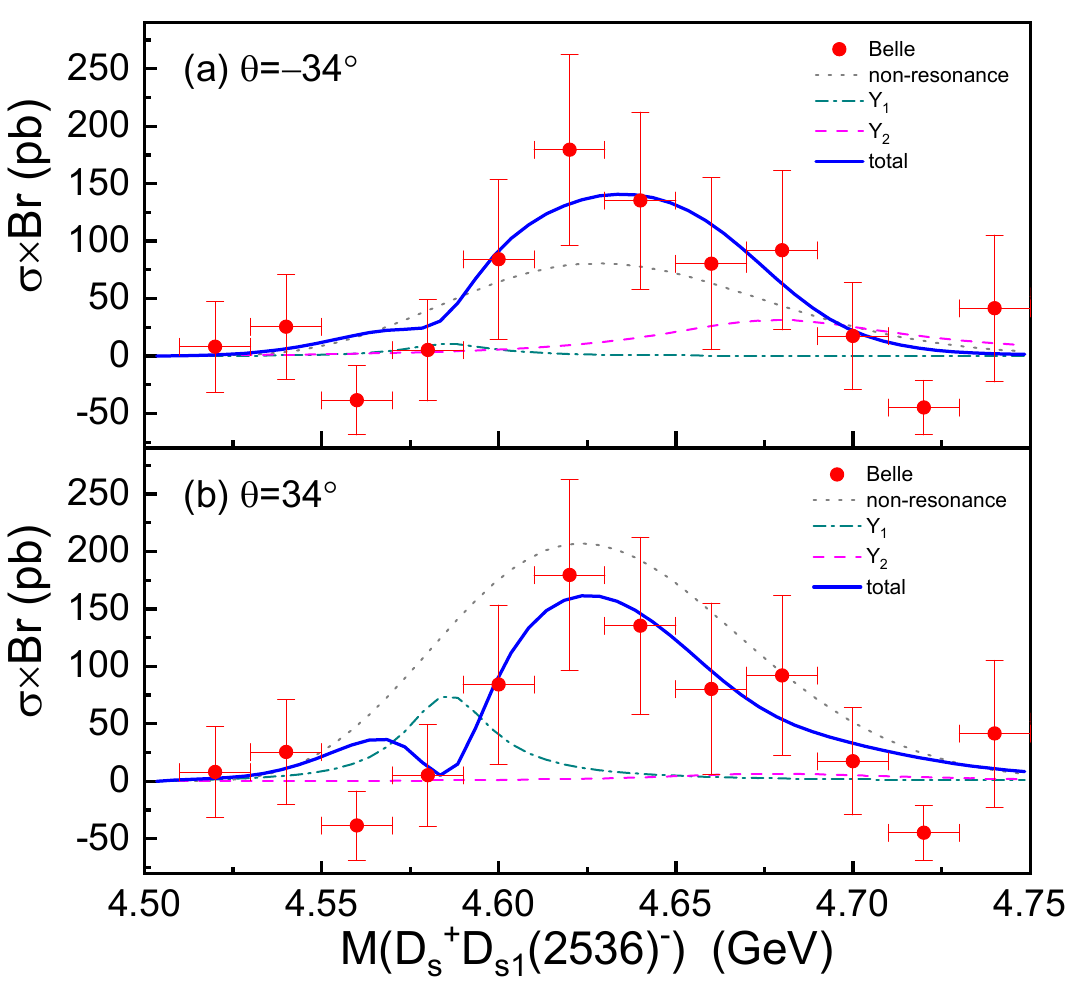}\\
(a)&\hspace{2.5em}(b)
\end{tabular*}
\caption{(a) The dependence of masses and widths of $\psi^{\prime}_{6S-5D}$ and $\psi^{\prime\prime}_{6S-5D}$ on mixing angle $\theta$ in the $6S$-$5D$.  (b) The reproduction of the experimental line shape of the $e^+e^-\to D^+_sD_{s1}(2536)^-$ from Belle \cite{Belle:2019qoi}. The figures are adopted from Ref. \cite{Wang:2020prx}.}
\label{fig:6S5D}
\end{figure}

To test this idea, a combined analysis of the processes $e^+e^- \to \psi(3686)\pi^+\pi^-$ and $e^+e^- \to \Lambda_c\bar{\Lambda}_c$ is performed \cite{Wang:2020prx}, incorporating high-precision BESIII data collected near the $\Lambda_c\bar{\Lambda}_c$ threshold \cite{BESIII:2017kqg}, as shown in Fig~\ref{fig:Y4660}. The analysis conclusively favored the existence of two vector resonances and described experimental data well with $\chi^2$/\text{d.o.f.}=1.49. The fitted resonance parameters of two vector states donated as $Y_1$ and $Y_2$ have the following masses and widths:  
\begin{align}
m_{Y_1} &= 4585 \pm 2~\text{MeV}, &
\Gamma_{Y_1} &= 29.8 \pm 8.0~\text{MeV}, \nonumber \\
m_{Y_2} &= 4676 \pm 7~\text{MeV}, &
\Gamma_{Y_2} &= 85.7 \pm 15.0~\text{MeV}.\nonumber
\end{align} From Fig.~\ref{fig:Y4660}, it can be seen that the "platform" behavior near the  $\Lambda_c\bar{\Lambda}_c$ threshold naturally emerge from the interplay of $Y_1$, $Y_2$ and non-resonant contributions. The $Y_1$ and $Y_2$ can be identified as the physical mixtures of the pure $\psi(6S)$ and $\psi(5D)$ with $\mu=0.12$. For a mixing angle of $|\theta|\approx 34^\circ$, the predicted mass of $\psi'_{6S\text{-}5D}$ (4587 MeV) and mass of $\psi''_{6S\text{-}5D}$ (4675 MeV) align strikingly with the fitted $Y_1$ and $Y_2$, respectively. Fig~\ref{fig:6S5D} (a) illustrates the dependence of the masses and total widths of $\psi'_{6S\text{-}5D}$ and $\psi''_{6S\text{-}5D}$ on the mixing angle $\theta$. In this scheme, it is argued that the reported $Y(4630)$ and $Y(4660)$ likely correspond to the same broader $Y_2$ resonance observed in different final states.

This interpretation received independent validation from the discovery of a new vector state, $Y(4626)$, in the open-charm process $e^+e^- \to D_s^+ D_{s1}(2536)^- + \text{c.c.}$ by the Belle collaboration \cite{Belle:2019qoi}. Its mass of $4625.9^{+6.2}_{-6.0}\pm 0.4$ MeV and width of $49.8^{+13.9}_{-11.5}\pm 4.0$ MeV placed it within the energy range of interest. To understand this new structure, the branching fractions $BR(\psi'_{6S5D} \to D_s \bar D_{s1}(2536))$ and $BR(\psi''_{6S5D} \to D_s \bar D_{s1}(2536))$ were calculated within the $6S$-$5D$ mixing scheme using the QPC model \cite{Wang:2020prx}. The results exhibited an obvious dependence on the sign of the mixing angle $\theta$. For $\theta = -34^\circ$, the predicted branching ratios were $0.09\%$ for the lower state $\psi'$ and $0.80\%$ for the higher state $\psi''$. Conversely, for $\theta = +34^\circ$, the pattern reversed with $\psi'$ and $\psi''$ having branching ratios of $0.64\%$ and $0.17\%$, respectively. Crucially, both scenarios yielded branching fractions of the expected order of $10^{-3}\sim10^{-2}$ for a conventional charmonium decaying into an open charm-strange pair. A subsequent fit to the measured $e^+e^- \to D_s^+ D_{s1}(2536)^-$ cross section, incorporating contributions from both $\psi'$ and $\psi''$ resonances along with a non-resonant background term, successfully reproduced the experimental line shape within both positive and negative mixing angle scenario as shown in Fig.~\ref{fig:6S5D} (b). Therefore, this interpretation not only resolves the apparent mass gap problem but provides a unified description of the relevant experimental observations.

Because the $Y_1$ state provides a new scaling point for the higher unquenched charmonium spectrum, the study proceeded to predict the properties of even higher-lying vector charmonia above 4.6 GeV \cite{Wang:2020prx}. The mass spectrum and open-charm decay widths were calculated for the next six states: $\psi(7S)$, $\psi(8S)$, $\psi(9S)$, $\psi(6D)$, $\psi(7D)$, and $\psi(8D)$. Their predicted masses lie in the range of approximately 4.7 to 4.9 GeV, indicating a dense spectrum in this region. A key and somewhat counterintuitive prediction is that these highly excited states remain relatively narrow, with total widths of only 10 to 30 MeV. This is attributed to strong node suppression effects in their decay amplitudes, i.e., the wave functions of high radial excitations possess multiple nodes, leading to significant cancellations in the overlap integrals with final-state meson wave functions during the decay process. The analysis also provided detailed predictions for their dominant open-charm decay channels. For the higher $S$-wave states, modes such as $D \bar D_1(2430)$, $D^* \bar D_0^*(2400)$, and $D^* \bar D_1(2430)$ are prominent. For the higher $D$-wave states, decays into $D^* \bar D^*$, $D \bar D$, and $D^* \bar D_1(2420)$ are among the most important. These concrete predictions serve as a valuable guide for future experimental searches at facilities like BESIII, Belle II, and a possible Super Tau-Charm \changelabel{Facility}, highlighting a rich and unexplored spectroscopy awaiting discovery in the 4.6–5.0 GeV energy range.

Collectively, this series of works presents a complete and unified solution to the long-standing "$Y$ problem" across the relevant energy region. Its success critically highlights the inclusion of unquenched effects, which are essential for correctly describing the spectrum and properties of higher excitations, and a concrete embodiment of the Occam's razor principle by the non-resonant mechanism in the research of hadron spectroscopy, achieving a comprehensive description of multiple anomalous structures without introducing exotic states. It therefore stands as a notable example of how the sustained interplay between theoretical modeling and experimental data yields a systematic resolution in spectroscopy.

\subsubsection{Accumulating experimental evidence}

\begin{figure}
\begin{tabular*}{\textwidth}{@{\extracolsep{\fill}}m{0.44\textwidth}m{0.55\textwidth}}
\includegraphics[height=0.36\textwidth]{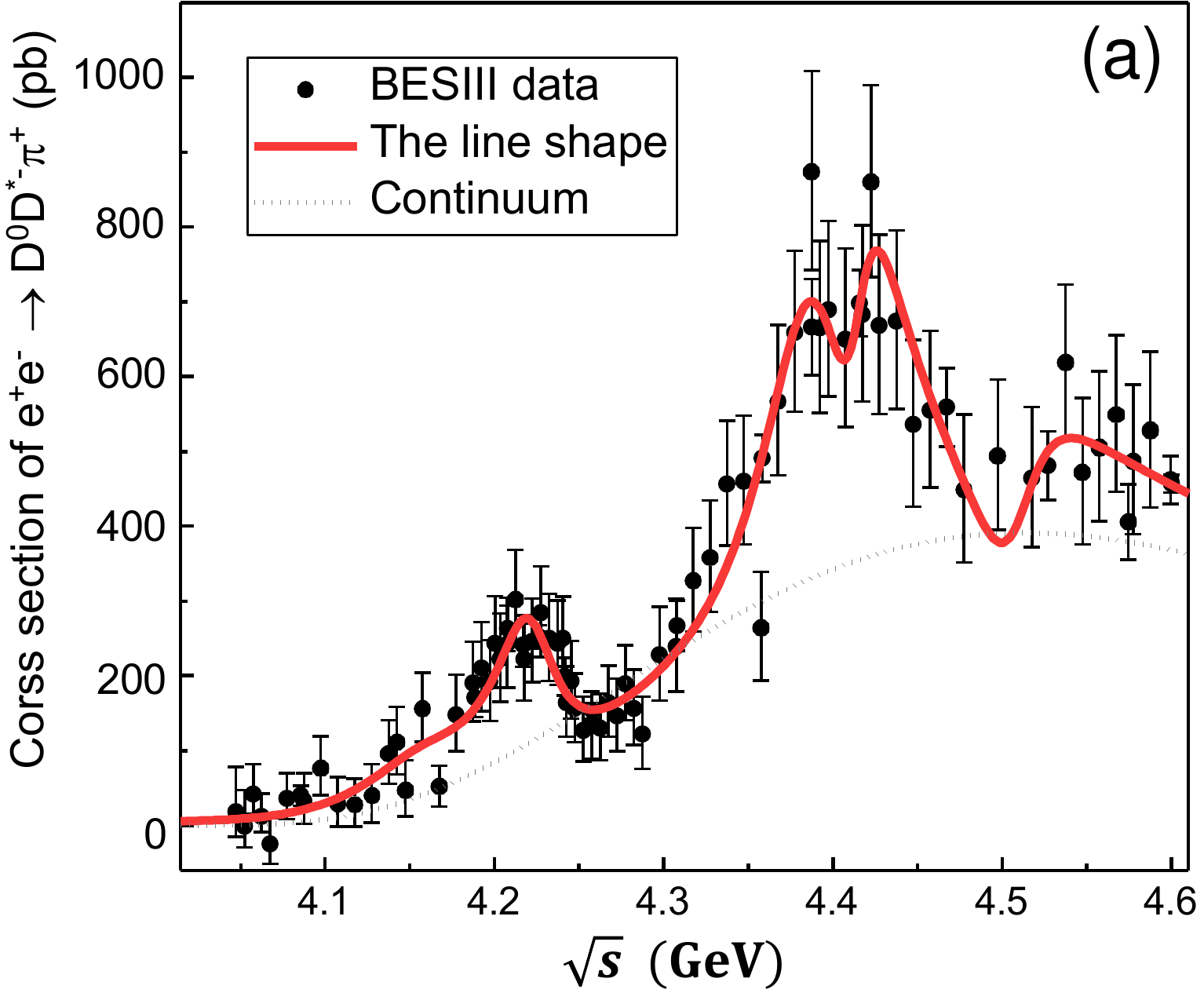}&\includegraphics[height=0.36\textwidth]{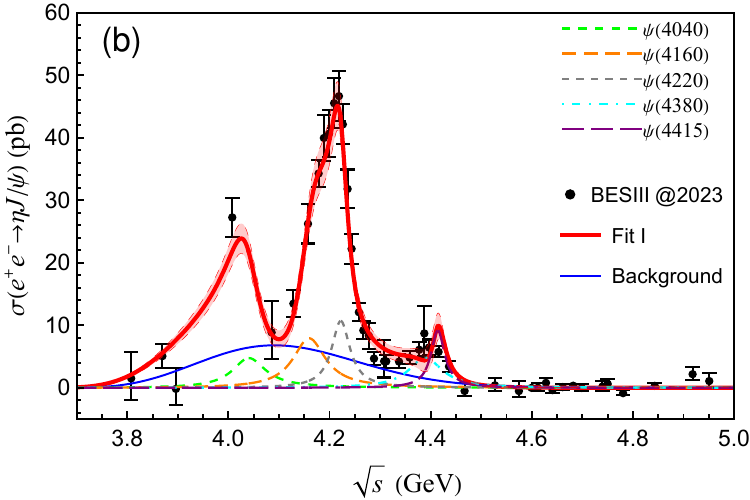}\\
\addlinespace[3em]
\includegraphics[width=0.43\textwidth]{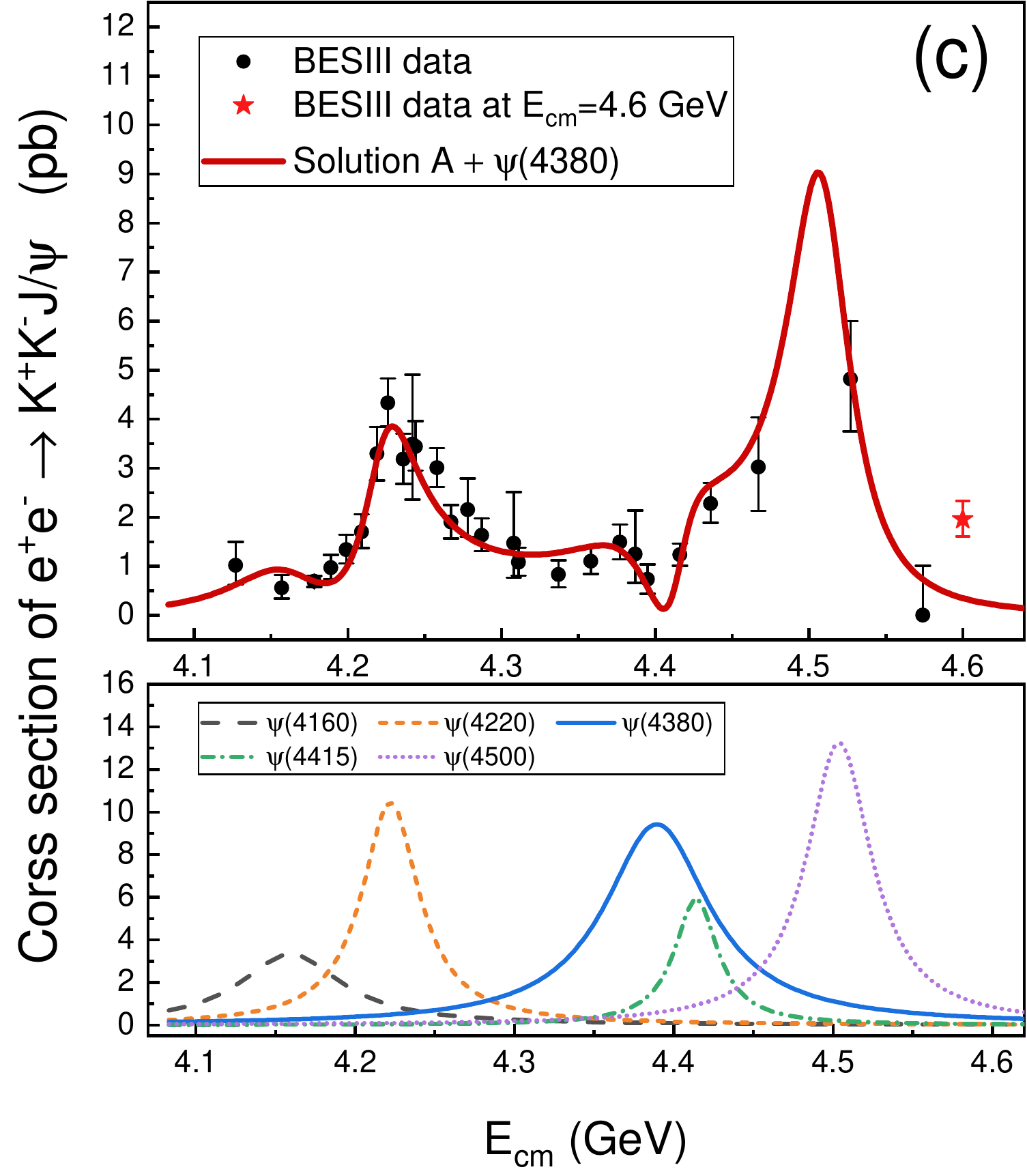}&\makecell[c]{\includegraphics[width=0.36\textwidth]{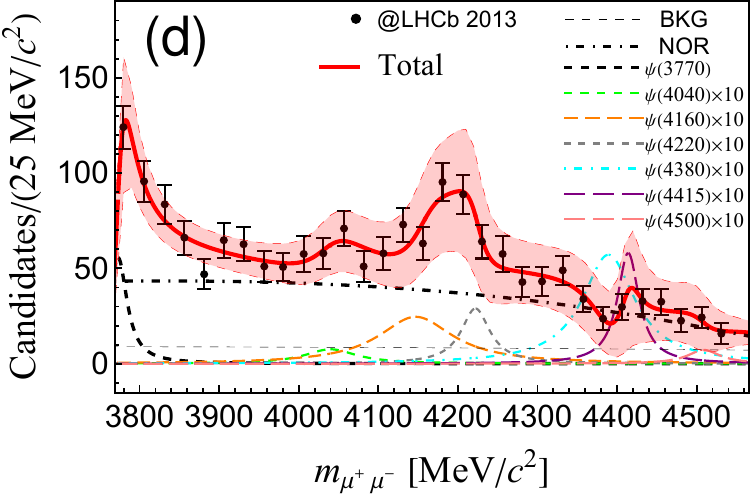}\\\includegraphics[width=0.36\textwidth]{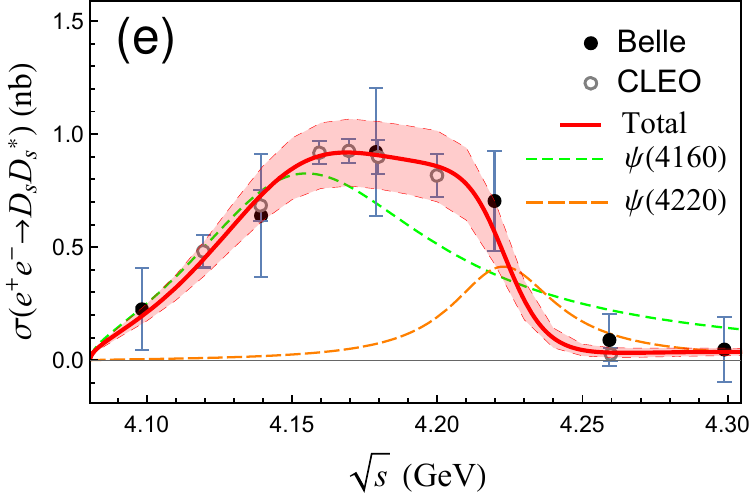}}
\end{tabular*}
    \caption{The fitted results of the $e^+e^- \to \pi^+D^0D^{*-}$ (a) \cite{Wang:2023zxj}, $e^+e^- \to \eta J/\psi$ (b) \cite{Peng:2024xui}, $e^+e^- \to K^+ K^- J/\psi$ (c) \cite{Wang:2022jxj}, dimuon invariant mass spectra in $B^+ \to K^+ \mu^+ \mu^-$ (d) \cite{Peng:2024blp} and  $e^+e^- \to D_s \bar D^*_s$ (e) \cite{Peng:2024blp} processes by the unquenched charmonium spectroscopy.}
    \label{fig:unified description}
\end{figure}

The spectroscopy of higher vector charmonium states above 4 GeV has long been a challenging problem in hadron physics. Quenched potential models predict only three states in this region: $\psi(4040)$, $\psi(4160)$, and $\psi(4415)$. However, high-precision experiments in recent years have revealed complex resonance structures that cannot be accommodated within this conventional picture. A series of studies by the Lanzhou group has proposed a characterized energy level structure consisting of six vector charmonium states in the 4–4.5 GeV range: $\psi(4040)$, $\psi(4160)$, $\psi(4220)$, $\psi(4380)$, $\psi(4415)$, and $\psi(4500)$. This framework, based on an unquenched potential model, has been systematically tested against data from multiple decay channels measured by BESIII \cite{BESIII:2018iea, BESIII:2023tll, BESIII:2022joj}, Belle \cite{Belle:2010fwv}, CLEO \cite{CLEO:2008ojp} and LHCb \cite{LHCb:2013ywr}, providing a consistent and unified description of the observed structures, which can be seen in Fig.~\ref{fig:unified description}.

\begin{enumerate}
\item The emergence of the characterized energy level structures of higher charmonium in the $e^+e^- \to D^0D^{*-}\pi^+$.
In the process $e^+e^- \to \pi^+ D^0 D^{*-}$~\cite{BESIII:2018iea}, BESIII observed two broad enhancements around 4.2 and 4.4 GeV. Earlier analyses attempted to describe these with single resonances but encountered inconsistencies in extracted widths. A combined fit incorporating the predicted states $\psi(4160)$, $\psi(4220)$, $\psi(4380)$, $\psi(4415)$, and $\psi(4500)$ yields an excellent description of the data with $\chi^2/\text{d.o.f.} = 0.74$. The enhancement near 4.4 GeV is explained not by a single wide resonance but by the interference between $\psi(4380)$ and $\psi(4415)$. Similarly, the structure around 4.2 GeV involves both $\psi(4160)$ and $\psi(4220)$. This multi-resonance interference naturally accounts for the observed line shapes and resolves the earlier discrepancy in the width of the $Y(4220)$-like structure, which is now understood as a composite signal rather than a single particle.

\item Explaining the puzzling data of the $e^+e^- \to \eta J/\psi$ by the unquenched charmonium spectroscopy.
The channel $e^+e^- \to \eta J/\psi$~\cite{BESIII:2023tll} presents another puzzle: BESIII reported a $Y(4220)$ with a width of about 80 MeV, significantly larger than the narrow width observed in other hidden-charm channels. By including the contributions of $\psi(4040)$, $\psi(4160)$, $\psi(4220)$, $\psi(4380)$, and $\psi(4415)$, and calculating their decay branching ratios to $\eta J/\psi$ via the hadronic loop mechanism, the data can be well described. The apparent large width and asymmetric line shape around 4.2 GeV arise from the overlap and interference of $\psi(4160)$ and $\psi(4220)$. Two fitting schemes were explored, one with a common cutoff parameter and another with individual parameters, both yielding good fits and demonstrating the importance of including multiple nearby resonances in the analysis.

\item Confirming the existence of a new higher charmonium $\psi(4500)$ by the newly
released data of $e^+e^- \to K^+ K^- J/\psi$. In $e^+e^- \to K^+ K^- J/\psi$ \cite{BESIII:2022joj}, BESIII observed a new structure around 4.5 GeV, denoted $Y(4500)$, with a mass consistent with the predicted $\psi(4500)$. A fit with $\psi(4220)$ and $\psi(4500)$ alone already captures the main features, but including $\psi(4160)$ and $\psi(4415)$ improves the description significantly. The extracted mass of $\psi(4500)$ is $4504 \pm 4$ MeV, and its narrower intrinsic width resolves the apparent discrepancy with the initially reported broader experimental width. Theoretical calculations of $BR(\psi \to K^+ K^- J/\psi)$ via charmed meson loops support this assignment. The analysis also indicates that $\psi(4380)$ should contribute noticeably in this channel, with a predicted branching ratio of $(4.0-6.0) \times 10^{-3}$, suggesting that future more precise data may reveal its presence.

\item Reevaluating the $\psi(4160)$ resonance parameter by the unquenched charmonium spectroscopy.
The decay $B^+ \to K^+ \mu^+ \mu^-$ \cite{LHCb:2013ywr} provides an independent probe of charmonium states through dimuon invariant mass spectra. LHCb data showed a peak around 4190 MeV, originally interpreted as $\psi(4160)$ but with a mass higher than earlier measurements. Reanalyzing the spectrum with the full set of predicted charmonium states reveals that the 4190 MeV structure is actually the result of interference between $\psi(4160)$ and $\psi(4220)$. When the contributions of both are properly accounted for, the true mass of $\psi(4160)$ is found to be around 4146 MeV, consistent with earlier experimental determinations. This clarifies a long-standing puzzle regarding the apparent mass shift of $\psi(4160)$ and underscores the importance of considering nearby resonances in line-shape analyses.

\end{enumerate}

In summary, the characterized energy level structure of vector charmonium in the 4–4.5 GeV region, comprising six states within an unquenched potential model, successfully unifies a wide range of experimental observations. It resolves previous puzzles such as the $\psi(4160)$ mass shift and the broad widths of structures in $\eta J/\psi$ and $\pi^+ D^0 D^{*-}$, and predicts new states like $\psi(4500)$ that have since been observed.

\subsection{Coupled-channel analysis for key charmonium(-like) states}

\subsubsection{$S$-$D$ mixing scheme induced by the coupled-channel mechanism}\label{coupled.S.D}

The vector charmonium spectrum above open-charm thresholds is known to be strongly influenced by unquenched effects induced by intermediate charmed meson loops. In this energy region, the conventional quark model picture becomes increasingly inadequate, as nearby open-charm thresholds can induce sizable mass shifts. In particular, the nature of  $\psi(4220)$ and $\psi(4380)$ states has remained an open issue, since their observed properties cannot be simultaneously accommodated within quenched potential models \cite{Eichten:1979ms,Godfrey:1985xj}.

Ref.~\cite{Wang:2019mhs} shows that a $4S$--$3D$ mixing scheme can naturally accommodate $Y(4220)$, reproducing its mass for a mixing angle in the range $\pm(30^\circ\sim 36^\circ)$. Nevertheless, the large mixing angle required in this phenomenological framework lacks a clear dynamical origin.  Such a sizable mixing cannot be generated by the tensor interaction in conventional potential models \cite{Godfrey:1985xj}. This motivates an investigation of the underlying dynamical mechanism responsible for the large $4S$-$3D$ mixing and a further clarification of the internal structure of $\psi(4220)$.
\begin{figure}[htbp]
\centering
\begin{tabular*}{\textwidth}{@{\extracolsep{\fill}}cc}
\includegraphics[height=0.24\textheight]{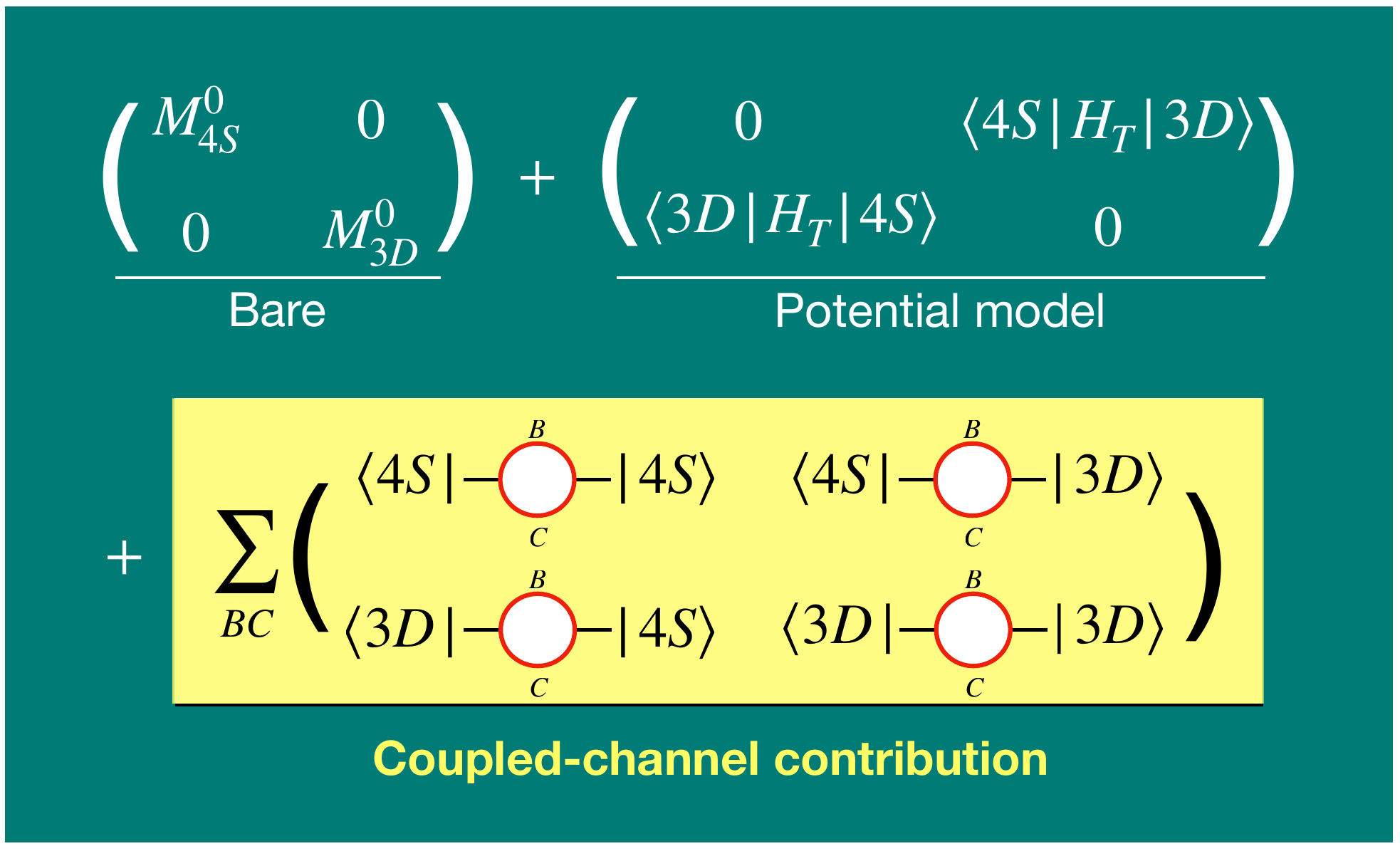}&\includegraphics[height=0.25\textheight]{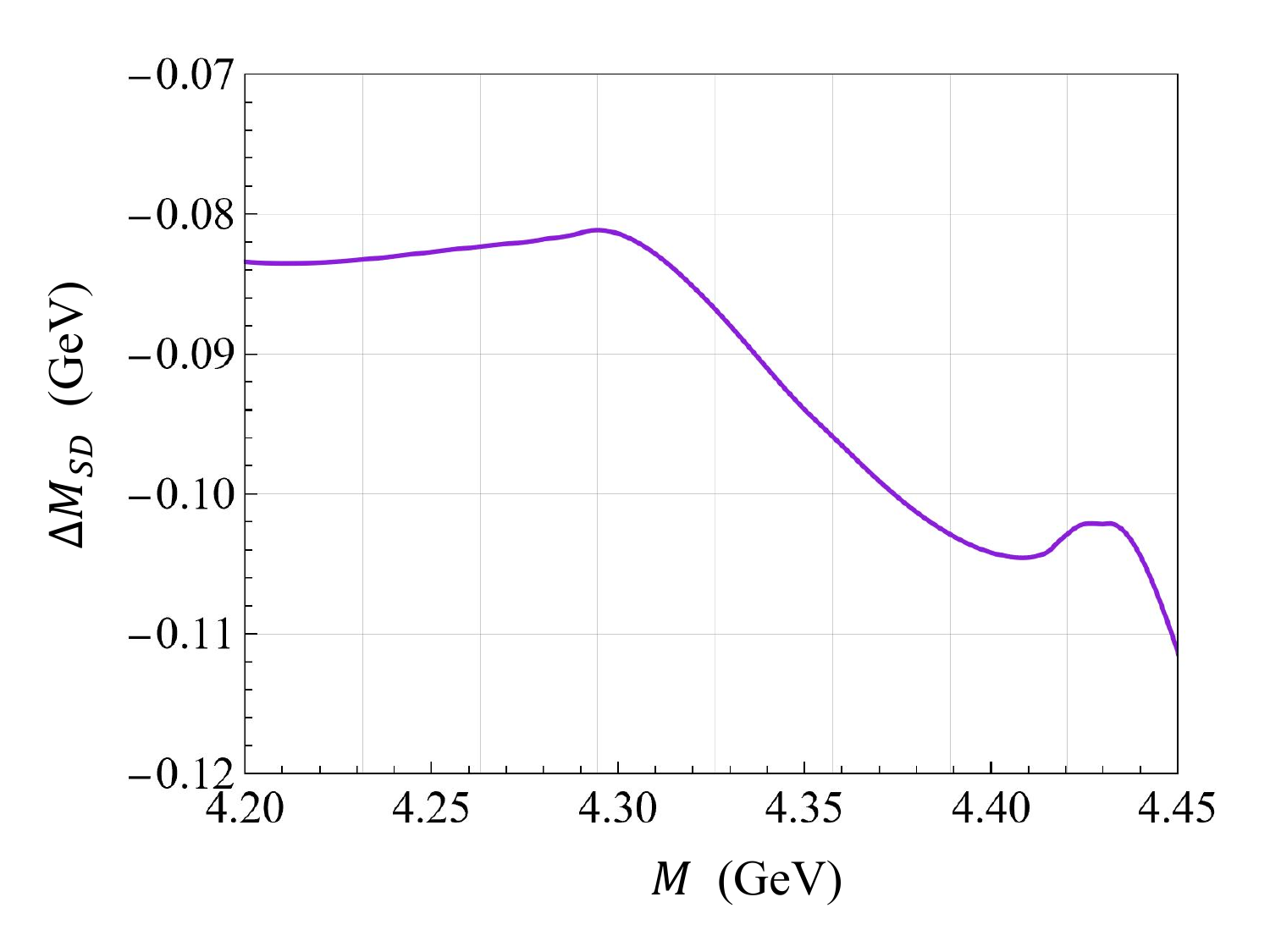}\\
(a)&\hspace{2.5em}(b)
\end{tabular*}
\caption{(a) Schematic diagram illustrating the $4S$-$3D$ mixing in the potential model and the coupled-channel mechanism. For the coupled-channel contribution, hadronic loops involving $B$ and $C$ mesons are considered in the calculation. (b) The dependence of $\Delta M_{SD}$ on $M$. The figures are adopted from Ref. \cite{Man:2025zfu}.}
\label{fig:4S3D}
\end{figure}

A systematic investigation of the $4S$-$3D$ mixing mechanism was carried out in Ref.~\cite{Man:2025zfu} within a coupled-channel framework, as illustrated schematically in Fig.~\ref{fig:4S3D} (a). Starting from the bare charmonium spectrum obtained using the GI model, the authors first examined the contribution from the tensor force. It was shown that the tensor interaction alone generates only a very small mixing between the $\psi(4S)$ and $\psi(3D)$ states, with a mixing angle well below $1^\circ$. This result confirms earlier expectations that the tensor force is insufficient to account for the large mixing angles ~\cite{Badalian:2009bu,Lu:2016mbb,Ni:2025gvx}.

The dominant source of the $S$-$D$ mixing was identified as open-charm meson loop effects.
Within this framework, the mixing angles of the relevant vector states can be determined accordingly~\cite{Heikkila:1983wd,Lu:2016mbb,Fu:2018yxq,Man:2024mvl,Ni:2025gvx}.
The mixing angle can be determined by solving the following equation:
\begin{eqnarray}\label{eq:mix_coupled_SD}
	\text{det}\begin{vmatrix}
		M_S^0 + \Delta M_{S}(M) - M &
        \langle \psi_S | H_{\text{T}} | \psi_D \rangle + \Delta M_{SD}(M) \\
		\langle \psi_D | H_{\text{T}} | \psi_S \rangle + \Delta M_{SD}(M)  &
        M_D^0 + \Delta M_{D}(M)- M
	\end{vmatrix} = 0.\nonumber\\
\end{eqnarray}
Here, $M^0_{S(D)}$ denotes the bare masses of the $\psi(3^3S_1)$ and $\psi(2^3D_1)$ states as predicted by the GI model \cite{Godfrey:1985xj}. The quantities $\Delta M_{S(D)}(M)$ correspond to the self-energy corrections arising from  meson-loop effects, while the off-diagonal term $\Delta M_{SD}(M)$ governs the mixing between the $\psi(3^3S_1)$ and $\psi(2^3D_1)$ configurations. In addition, the matrix element $\langle \psi_S | H_{\text{T}} | \psi_D \rangle$, originating from the tensor interaction in the GI model.

Using the Quark Pair Creation model to evaluate both diagonal $\Delta M_{S(D)}(M)$ and off-diagonal $\Delta M_{SD}(M)$, several relevant coupled channels were analyzed in detail. 
Fig~\ref{fig:4S3D} (b) shows the dependence of $\Delta M_{SD}$ on $M$. It is found that the magnitude of $\Delta M_{SD}$ is comparable to those of $\Delta M_{S}$ and $\Delta M_{D}$, indicating that the mixing effects provide a non-negligible contribution.
The combined effect of relevant channels results in a large dynamical mixing angle,
\begin{equation*}
\theta \simeq 35^\circ.
\end{equation*}
This value supports the large $4S$--$3D$ mixing in $\psi(4220)$ previously obtained from  fits to experimental data~\cite{Wang:2019mhs}.

After diagonalization of the coupled-channel mass matrix, two physical mixed states were obtained. The lower state, dominated by the $\psi(4S)$ component, has a mass around $4.23~\mathrm{GeV}$ and can be naturally identified with the $\psi(4220)$. The higher state, with a dominant $\psi(3D)$ component, appears near $4.39~\mathrm{GeV}$ and was associated with the $\psi(4380)$. 
Furthermore, it is found that the proximity of the $D \bar D_1$ threshold leads to a substantial mass shift of the lower mixed state, while the $D^* \bar D_1$ channel primarily affects the higher state. The predicted total widths and open-charm decay patterns of these states were shown to be compatible with available experimental data, providing further support for this assignment.

The  study  of $\psi(4220)$ further strengthens the view that unquenched dynamics are essential for a unified and self-consistent description of the charmonium spectrum above open-charm thresholds.

\begin{table}[!htbp]
\caption{The obtained $3S$-$2D$ mixing angle for $\psi(4040)$ and $\psi(4160)$.}\centering\label{mix-angle}
\renewcommand\arraystretch{1.35}
\begin{tabular*}{1.0\textwidth}{@{\extracolsep{\fill}}ccr}
\hline
$\Gamma_{e^+e^-}^{\psi(4040)}$~(keV) & $\Gamma_{e^+e^-}^{\psi(4160)}$ (keV)  &\multicolumn{1}{c}{Mixing angle}                                             \\
\hline
\multirow{6}{*}{$0.86\pm0.07$~\cite{ParticleDataGroup:2024cfk}}     &\multirow{3}{*}{$0.83\pm0.08$~\cite{Seth:2004py}}         &  $\phi=-35^{\circ},+55^{\circ}$~\cite{Chao:2007it} \footnotemark[1]        \\
                                    &                                       &  
                                    $\phi=-37^{\circ}$~\cite{Li:2009zu} \footnotemark[1] \\
                                    &                                       &      $\theta=34.8^{\circ},-55.7^{\circ}$~\cite{Badalian:2008dv} \footnotemark[1]                  \\
\cline{2-3}
                                    &\multirow{3}{*}{$0.48\pm0.22$~\cite{BES:2007zwq}}         &  $\theta=20^{\circ}$~\cite{Wang:2022jxj} \footnotemark[2]                    \\
                                    &                                       &  $\phi=45^{\circ}$~\cite{Bokade:2024tge} \footnotemark[3]                  \\
                                    &                                       &  $\phi=-21.1^{\circ},62.6^{\circ}$~\cite{Zhao:2023hxc} \footnotemark[4]    \\
\hline
\end{tabular*}
\footnotetext[1]{Extracted from the ratio (1.04) of those two di-electronic widths.}
\footnotetext[2]{Fitting the di-electronic width of the $\psi(4160)$.}
\footnotetext[3]{Extracted from the ratio (1.79) of those two di-electronic widths.}
\footnotetext[4]{Fitting the di-electronic widths of the $\psi(4040)$ and $\psi(4160)$.} 
\end{table}

Parallel to $\psi(4220)$ and $\psi(4380)$ system, $\psi(4040)$ (also denoted $\psi(3^3S_1)$ in quenched quark model classifications) represents another cornerstone in understanding unquenched effects in the vector charmonium spectrum above open-charm thresholds. First observed by the DASP Collaboration in $e^+e^-$ annihilation in 1978 \cite{DASP:1978dns}, $\psi(4040)$ is a well-established resonance listed in the Particle Data Group (PDG) reviews \cite{ParticleDataGroup:2024cfk}, yet its intrinsic structure and mixing dynamics remain incompletely resolved. A key open question revolves around the strength of the $3S$-$2D$ mixing between $\psi(4040)$ and its higher-mass counterpart $\psi(4160)$ ($\psi(2^3D_1)$), as well as the extent to which coupled-channel effects drive this mixing.

Early phenomenological studies inferred significant $3S$--$2D$ mixing for $\psi(4040)$ and $\psi(4160)$ by fitting experimental dielectronic decay widths ($\Gamma_{e^+e^-}$) or their ratios \cite{Chao:2007it,Li:2009zu, Badalian:2008dv,Anwar:2016mxo,Yang:2018mkn,Wang:2022jxj,Bokade:2024tge,Zhao:2023hxc}.
In Table \ref{mix-angle}, we summarize the current understanding of $3S$-$2D$ mixing, including the inferred mixing angles and the corresponding approaches employed to extract them.

 Using the ratio of $\Gamma_{e^+e^-}(\psi(4040)) \sim 0.86$ keV to $\Gamma_{e^+e^-}(\psi(4160)) \sim0.83$ keV, a mixing angle of about $35^\circ$  was extracted in Refs. \cite{Chao:2007it,Badalian:2008dv,Li:2009zu}. Later analyses incorporating updated BESIII Collaboration data, which reported a smaller value of $\Gamma_{e^+e^-}(\psi(4160)) \sim 0.48$ keV \cite{BES:2007zwq}, revised the mixing angle to approximately $\sim 20^\circ$ \cite{Wang:2022jxj,Zhao:2023hxc}. These relatively large mixing angles, comparable to those inferred for the $\psi(4220)$-$\psi(4380)$ system, initially suggested the presence of a universal strong $S$-$D$ mixing mechanism in higher charmonium states.

However, the dielectronic width of the $\psi(4160)$ suffers from substantial experimental uncertainties. In 2010, Mo, Yuan, and Wang reanalyzed the BES data and pointed out that the dielectronic width of the $\psi(4160)$ reported in Ref.~\cite{BES:2007zwq} represents only one of four possible solutions, with values ranging from $0.4$ to $1.1~\text{keV}$ \cite{Mo:2010bw}. It is important to emphasize that the presence of multiple possible solutions for the dielectronic width introduces significant uncertainties in extracting the $3S$-$2D$ mixing angle. Furthermore, early measurements of $\psi(4040)$ and $\psi(4160)$ were mainly based on inclusive $e^+e^- \to \text{hadrons}$ processes, with only limited information from exclusive decay channels. In particular, the dielectronic width of the $\psi(4160)$ extracted from inclusive data suffers from large uncertainties. This is primarily due to the presence of multiple open-charm thresholds in this energy region, which generate complicated structures in the measured cross sections. As a result, inclusive measurements alone are insufficient for a detailed characterization of high-mass charmonium states.
Although several exclusive processes involving $\psi(4160)$ have been reported \cite{LHCb:2013ywr,BESIII:2023wsc}, its dielectronic width has not yet been precisely determined. Consequently, approaches that rely on experimentally extracted dielectronic widths to infer the $3S$--$2D$ mixing angle may be subject to significant uncertainties and should be treated with caution.

Therefore, a more comprehensive understanding of the internal structure and mixing dynamics of $\psi(4040)$ and $\psi(4160)$ is highly desirable.

\begin{figure}
    \centering
    \includegraphics[width=0.5\textwidth]{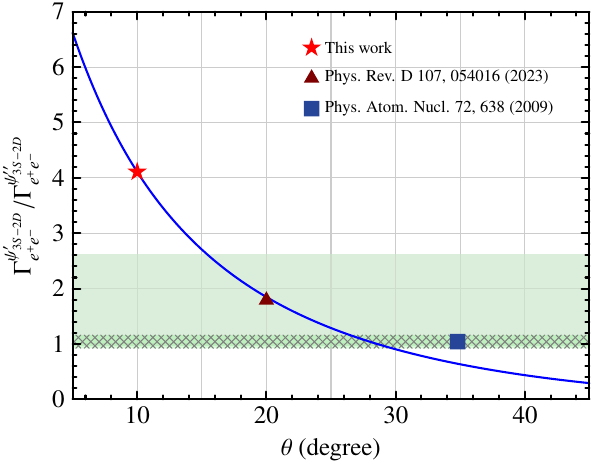}
    \caption{The ratio of dielectronic widths $\Gamma^{\psi^{\prime}_{3S\text{-}2D}}_{e^+e^-}/\Gamma^{\psi^{\prime\prime}_{3S\text{-}2D}}_{e^+e^-}$ as a function of the $3S\text{-}2D$ mixing angles. The red pentagram represents the prediction of this work ($\theta \approx 10^\circ$ and ratio $\Gamma^{\psi^{\prime}_{3S\text{-}2D}}_{e^+e^-}/\Gamma^{\psi^{\prime\prime}_{3S\text{-}2D}}_{e^+e^-}\approx 4.1$). The  light green and dark green shaded bands indicate the experimental values $1.79 \pm 0.83$ and $1.04 \pm 0.12$, respectively, from the PDG \cite {ParticleDataGroup:2024cfk}. 
 The triangle and square indicate the mixing angles $20^{\circ}$ and $34.8^{\circ}$, as obtained in Refs. \cite{Wang:2022jxj} and \cite{Badalian:2008dv}, respectively.}
    \label{fig:3S2D}
\end{figure}

Recent work by Man, Luo, and Liu \cite{Man:2025vmm} extended the coupled-channel framework to the $\psi(4040)$-$\psi(4160)$ system. Starting from the bare masses of the $\psi(3^3S_1)$ and $\psi(2^3D_1)$ states, taken as $4089$ MeV and $4172$ MeV from the GI model \cite{Godfrey:1985xj}, unquenched effects were incorporated through intermediate open-charm meson loops.
By solving Eq.~(\ref{eq:mix_coupled_SD}) within this coupled-channel formalism, the physical masses of the mixed states $\psi^{\prime}_{3S\text{-}2D}$ and $\psi^{\prime\prime}_{3S\text{-}2D}$ were obtained as
$4006.4^{+7.4}_{-8.3}$ MeV and $4089.5^{+8.8}_{-8.6}$ MeV, respectively. The corresponding eigenvectors were simultaneously determined.
The eigenvector components $(C_{S1(2)}, C_{D1(2)})$ quantify the admixtures of the $3^3S_1$ and $2^3D_1$ configurations in each physical state. Based on these coefficients, the mixing angles were extracted using the relations
$\theta_1=\arctan(-C_{D1}/C_{S1})$ and
$\theta_2=\arctan(C_{S2}/C_{D2})$ for
$\psi^{\prime}_{3S\text{-}2D}$ and $\psi^{\prime\prime}_{3S\text{-}2D}$, respectively.

As a result, the mixing angles were determined to be
\begin{equation}
\theta_1=(7.3^{+0.1}_{-0.4})^{\circ}
\quad \text{for} \quad
\psi^{\prime}_{3S\text{-}2D},
\end{equation}
and
\begin{equation}
\theta_2=(10.4^{+2.0}_{-3.5})^{\circ}
\quad \text{for} \quad
\psi^{\prime\prime}_{3S\text{-}2D}.
\end{equation}
 Unlike $\psi(4220)$ and $\psi(4380)$, which exhibit relatively large mixing angle, the mixing between the $\psi(3^3S_1)$ and $\psi(2^3D_1)$ states is found to be weaker~\cite{Man:2025vmm}.

The calculated dielectronic width of $\psi(4040)$ ($0.92$ keV) agrees with the PDG value ($0.86 \pm 0.07$ keV) \cite{ParticleDataGroup:2024cfk}, while the predicted width of $\psi(4160)$ ($0.22$ keV) lies at the lower bound of experimental measurements \cite{Mo:2010bw}. As illustrated in Fig.~\ref{fig:3S2D}, the calculation yields a ratio of approximately $4.1$, which is  larger than the experimental determinations of $1.04 \pm 0.12$ and $1.79 \pm 0.83$~\cite{ParticleDataGroup:2024cfk}. This tension between theory and experiment provides a clear and testable prediction that can be examined in future high-precision measurements.

Future experiments, particularly the upgraded BESIII facility with an expected threefold increased in luminosity~\cite{Balossino:2022ywn}, is anticipated to play a crucial role in resolving the remaining ambiguities. Precision measurements of exclusive decay channels and dielectronic widths will constrain the $3S$-$2D$ mixing angle further, while searches for the predicted $\psi(4380)$ (via $D \bar D_2^*$ decays), will offer a critical test of the universality of coupled-channel induced $S$-$D$ mixing. Collectively, these efforts will advance a unified framework for describing charmonium states above open-charm thresholds, where unquenched effects are the primary driver of mass shifts, mixing angles, and decay dynamics.

\subsubsection{Coherent coupled-channel description of near-threshold structures: $\psi(3770)$, $G(3900)$, $R(3760)$, $R(3780)$, and $R(3810)$}
As the first charmonium state above the open-charm threshold, $\psi(3770)$ displays several anomalous properties.
It exhibits a significant non-open-charm hadron (nOCH) decay fraction of approximately 15\%~\cite{BES:2007cev,BES:2008vad}, in contrast to the nearly negligible nOCH partial widths of $\psi(3686)$ and $J/\psi$.
Moreover, its line shape deviates from a standard Breit-Wigner distribution—a discrepancy that has led to proposal that two distinct structures might be present~\cite{BES:2008wyz}. Recent measurements by the BESIII collaboration have further enriched this picture. Both the hadronic cross section~\cite{BESIII:2023oql} and the nOCH cross section~\cite{BESIII:2023bed} unveil intricate structures in the $3.7-3.82$ GeV region, which are well described by the interference among three resonances:  $R(3760)$, $R(3780)$, and $R(3810)$. Clearly, only one of these can correspond to the $\psi(1D)$ state, as no additional conventional charmonia are predicted in this mass range.

In addition to the structures near the $\psi(3770)$, a resonance-like feature denoted as $G(3900)$ has been observed just above the $\psi(3770)$ peak in the $e^+e^-\to D\bar{D}$ cross section~\cite{BESIII:2024ths,BaBar:2006qlj,BaBar:2008drv,Belle:2007qxm,Wang:2025dur}. Theoretical attempts have been made from different perspectives to understand the $G(3900)$ structure, such as $P$-wave $D\bar{D}^*$ resonance states~\cite{Lin:2024qcq} and coupled-channel frameworks~\cite{Cao:2014qna,Du:2016qcr,Uglov:2016orr,Husken:2024hmi,Ye:2025ywy,Nakamura:2023obk}. 
However, these studies have yielded conflicting results; while some~\cite{Du:2016qcr,Ye:2025ywy,Nakamura:2023obk} found a pole near 3.9 GeV, others did not~\cite{Uglov:2016orr,Husken:2024hmi}.
To resolve these debates, a critical question is whether the $G(3900)$ structure can be generated from the coupled-channel dynamics of nearby $c\bar{c}$ states such as $\psi(3770)$ and $\psi(4040)$.

\begin{figure}[htbp]
  \centering
  \includegraphics[width=0.5\textwidth]{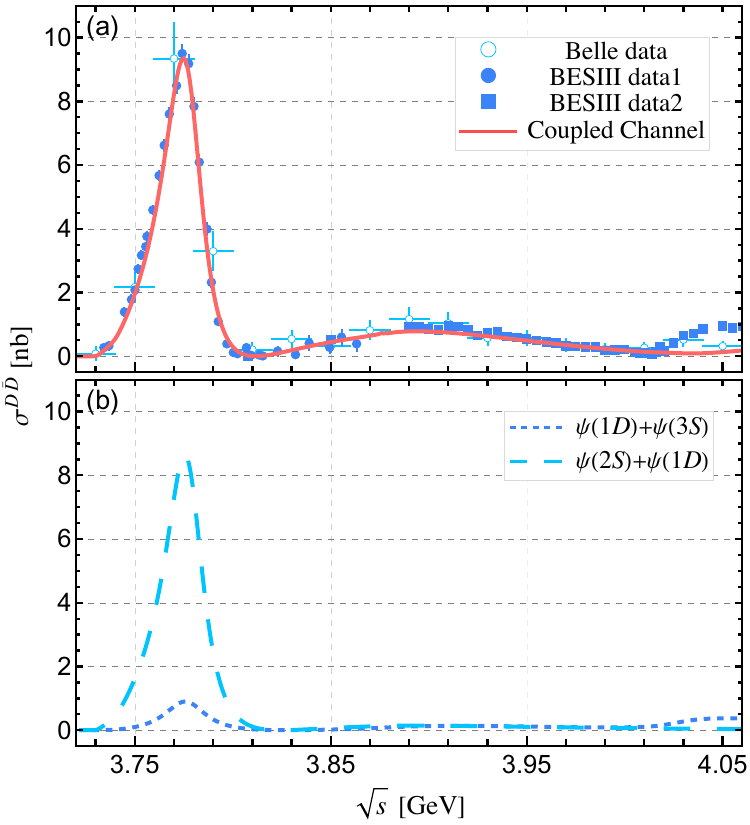}
  \caption{(a) Coupled-channel description of the $e^+e^-\to D\bar{D}$ cross section. The red solid line represents the full model including $\psi(2S)$, $\psi(1D)$, and $\psi(3S)$. Experimental data are from Belle~\cite{Belle:2007qxm} and BESIII~\cite{Husken:2024hmi,BESIII:2024ths}.
  (b) Cross section when either $\psi(2S)$ or $\psi(3S)$ is removed.\\
  Source: Taken from Ref.~\cite{Qian:2025zyp}}\label{fig:1D}
\end{figure}

\begin{figure}[htbp]
  \centering
    \includegraphics[width = 8.4 cm]{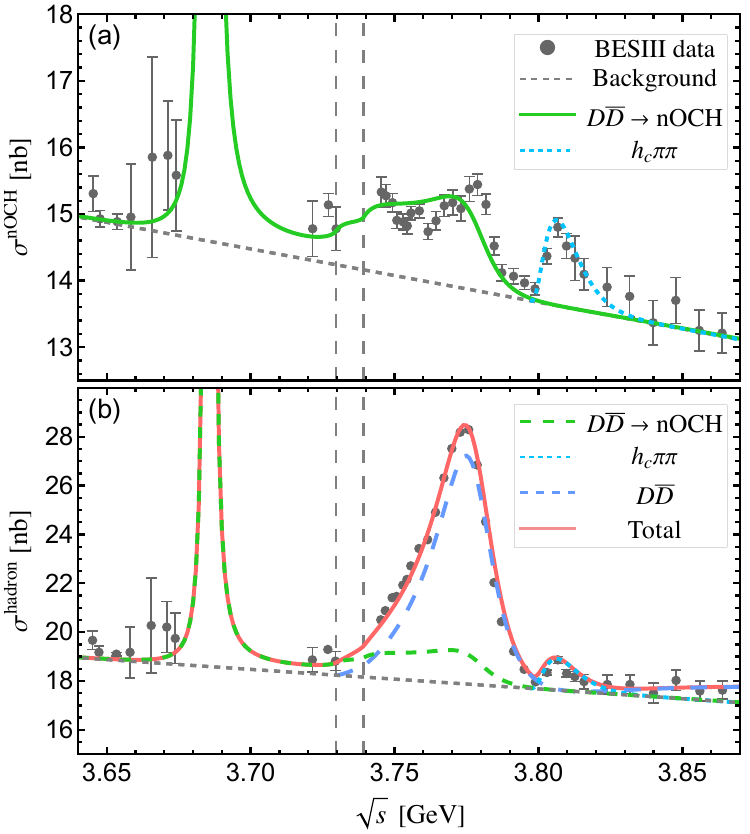}
    \caption{Coupled-channel description of (a) the $e^+e^-\to$ nonopen-charm hadron and (b) the inclusive hadronic cross sections. Vertical dashed lines indicate the $D^0\bar{D}^0$ and $D^+D^-$  thresholds. nonopen-charm hadron data are from Ref.~\cite{BESIII:2023bed}, inclusive hadronic data are from Ref.~\cite{BESIII:2023oql}.\\
    Source: Taken from Ref.~\cite{Qian:2025zyp}.}\label{fig:hadron}
\end{figure}

Interestingly, the structures near the $D\bar{D}$ threshold can be explained within a unified coupled-channel framework. In Ref.~\cite{Qian:2025zyp}, Qian and Liu propose a unified coupled-channel description to these structures. The coupled-channel framework incorporates the bare $c\bar{c}$ core, hadron-hadron kinematics, and transition interactions between bare $c\bar{c}$ states and various hadronic channels:
\begin{equation}
    H = H_{c\bar{c}} + H_{c\bar{c}\leftrightarrow hadrons} + H_{hadrons}\,.
\end{equation}
Here, the $H_{c\bar{c}}$ is modeled using a quenched model~\cite{Godfrey:1985xj}. The hadronic Hamiltonian $H_{hadrons}$ include kinematics and possibly interaction of hadronic channels. By including the $\psi(1D)$, $\psi(2S)$ and $\psi(3S)$ $c\bar{c}$ bare states, and modeling the transition from $c\bar{c}$ to $D^{(*)}\bar{D}^{(*)}$ via the QPC model, this coupled-channel model reproduce the $\psi(3770)$ and $G(3900)$ structure in $e^+e^-\to D\bar{D}$ process, as shown in Fig.~\ref{fig:1D}. Removing either $\psi(3686)$ or $\psi(3S)$ eliminates this structure, indicating  a nontrivial interference pattern. The authors found three poles near the real axis, corresponding to the $\psi(3686)$, $\psi(3770)$, and $\psi(4040)$, but no additional pole near 3.9 GeV close to the physical axis. \changelabel{However, since this model does not include direct hadron-hadron interactions. It still can not ruled out the possibilities that the introductions of additional hadron-hadron interaction, such as $D\bar{D}^{*}$ interaction, could generate the poles around 3.9 GeV.}

Beyond structures in open-charm channels, the authors of Ref.~\cite{Qian:2025zyp} find that the $R(3760)$ and $R(3780)$ resonances in non-open-charm channels can be explained by including a $D\bar{D}\to \text{nOCH}$ transition, while the $R(3810)$ can be attributed to the $\psi(1D)$-$h_c\pi\pi$ coupling, as shown in Fig.~\ref{fig:hadron}. The $D\bar{D}\to \text{nOCH}$ process corresponds to rescattering from $D\bar{D}$ to non-open-charm hadrons, modeled by
\begin{equation}
    V_{\text{nOCH},D\bar{D}}(\boldsymbol{q}) = g_{\text{nOCH}}qe^{-q/\Lambda_{\text{nOCH}}} Y_1^{1*}(\Omega_q)\,,
\end{equation} 
where $q$ is the momentum of the $D\bar{D}$ system and $Y_1^1$ is the spherical harmonic function. \changelabel{ It is worth noting that the form factor within the above potential is inherently model dependent. The authors find a characteristic momentum scale $\Lambda_{\text{nOCH}} \approx 100\; \mathrm{MeV}$ for this rescattering process, meaning the process is only significant when the $D\bar{D}$ relative momentum is small. This naturally explains the significant nOCH decay width of the $\psi(3770)$: its large nOCH branching fraction arises from its proximity to the $D\bar{D}$ threshold, enabling decay via the rescattering process $D\bar{D} \to \text{nOCH}$. Assuming the two pions are in relative $P$-wave ($L_{\pi\pi}=1$), the $\psi(1D)$-$h_c\pi\pi$ coupling is modeled as
\begin{eqnarray}
    f_{\psi(1D),h_c\pi\pi}(\boldsymbol{q},\boldsymbol{p}) &=& \frac{g_{h_c\pi\pi}}{\sqrt{4\pi}}pe^{-(q^2+p^2)/\Lambda_{h_c}^2}\frac{1}{\sqrt{2}}\left(Y_1^{0*}(\Omega_p) - Y_1^{1*}(\Omega_p) \right) \,, \nonumber
\end{eqnarray}
where $\boldsymbol{q}$ is the $h_c$ momentum in the c.m. frame, and $\boldsymbol{p}$ is the relative momentum between the pions. The authors find a characteristic momentum scale $\Lambda_{h_c} \approx 40\mathrm{MeV}$, which is easily understood because the bare mass of the $\psi(1D)$ (3.82 GeV) is only about 100 MeV above the $h_c\pi\pi$ threshold, leaving only tens of MeV for the momentum of the $h_c\pi\pi$ system.}

\changelabel{ \subsection{Complementary theoretical perspectives on the charmonium-like $Y$ structures}

The abundance of vector charmonium-like structures observed in different exclusive channels has led to various theoretical attempts to understand the $Y$ family. The framework discussed above provides a possible unified description of the $Y$ problem below 4.7 GeV by combining the non-resonant interference mechanism with the unquenched charmonium picture. This line of thought is useful for correlating several seemingly different structures within a common charmonium spectroscopic pattern. Nevertheless, many other studies have focused on specific $Y$ states or experimental channels, emphasizing different theoretical interpretations such as compact tetraquarks, hybrid charmonia, hadronic molecules, and other exotic explanations. For a balanced review, we briefly summarize below these complementary theoretical interpretations, focusing on the updated $Y$ states established after the high-precision BESIII measurements in 2017.

\subsubsection{$Y(4220)$}

After the updated BESIII measurements, the original broad $Y(4260)$ structure was resolved into the lower-mass $Y(4220)$~\cite{BESIII:2016bnd}, together with an additional structure at higher energy. Since theoretical interpretations of the $Y(4260)$ before this update have already been summarized in Sec.~\ref{sec:Y-1.2}, we here mainly focus on recent theoretical studies motivated by the updated $Y(4220)$, which are summarized as
\begin{itemize}
\item Compact tetraquark interpretations provide another possible description of the $Y(4220)$, usually formulated in terms of diquark-antidiquark degrees of freedom. In a relativized diquark model, the hidden-charm $[cq][\bar c\bar q]$ and $[cs][\bar c\bar s]$ spectra were calculated systematically, and a $[cq][\bar c\bar q]$ vector tetraquark with a mass of 4253 MeV was suggested as a possible assignment for the $Y(4220)$~\cite{Anwar:2018sol}. This calculation also predicts a rich spectrum of $1^{--}$ tetraquark states in the 4.0--5.0 GeV region, providing characteristic expectations for future experimental tests. QCD sum-rule analyses show that the conclusion is sensitive to the choice of interpolating currents: the study based on the $C\otimes\gamma_\mu C$ and $C\gamma_5\otimes\gamma_5\gamma_\mu C$ type currents disfavored the assignment of $Y(4220)$ as a pure vector tetraquark state, while a derivative current corresponding to an explicit relative $P$ wave between the diquark and antidiquark can reproduce a mass compatible with the $Y(4220)$, and a subsequent systematic analysis further supported assigning $Y(4220)$, $Y(4320/4360)$, and $Y(4390)$ to this type of $P$-wave vector tetraquark state~\cite{Wang:2018rfw,Wang:2018ntv,Wang:2018ejf}. More recently, a constituent quark model study of $1^{--}$ $P$-wave compact tetraquarks suggested that the $Y(4220)$, $Y(4360)$, and $Y(4660)$ may be assigned as $P$-wave tetraquark states, with multiple states possibly appearing around 4.23 and 4.36 GeV~\cite{Zhao:2025kno}. Since this calculation constructs the wave functions in a diquark-antidiquark basis, it would be useful to compare such assignments with future four-body studies including more general color and spatial configurations. The decay properties of compact tetraquark candidates have also been investigated through light-meson-emission and three-body strong-decay analyses~\cite{Li:2022fgd,Wang:2023dsm,Xie:2023qjg}. Overall, compact tetraquark pictures offer a direct way to account for the strong coupling of $Y(4220)$ to hidden-charm final states through quark rearrangement processes.

\item Hybrid charmonium interpretations have also been discussed for the $Y(4220)$, although dedicated studies after the updated BESIII measurements are relatively limited. In the heavy-quark sector, the separation of scales between the heavy-quark motion and the gluonic dynamics allows a systematic description of hybrids in terms of adiabatic potentials, forming the basis of the Born-Oppenheimer (BO) picture~\cite{Berwein:2024ztx}. It should be emphasized that the BO description of hybrid mesons differs from the constituent-gluon picture. In the former, the gluonic excitation is encoded in the BO potential generated by the light-field degrees of freedom at fixed heavy-quark separation, while in the latter it is treated more explicitly as an effective constituent. As a consequence, the two pictures may lead to different expectations for the spectra and decay behaviors of hybrid states. Braaten and Bruschini developed a model-independent BO analysis of hidden-heavy hadron decays into pairs of heavy hadrons~\cite{Braaten:2024stn}. In this framework, conventional quarkonia and hybrid quarkonia are characterized by BO quantum numbers associated with the light-field degrees of freedom, whose approximate conservation leads to selection rules for open-heavy-flavor decays. As summarized in Table~\ref{T4-bo}, the BO approach predicts a distinctive decay pattern for vector charmonium hybrids compared with conventional $S$-wave vector charmonia. In particular, the decay of a BO vector charmonium hybrid into the $D\bar D^*+D^*\bar D$ channel is forbidden, providing a useful test of the hybrid interpretation of the $Y(4220)$. In addition, a lattice-inspired study proposed a color-halo picture for charmonium-like hybrids, in which a relatively localized color-octet $c\bar c$ pair is surrounded by gluonic degrees of freedom~\cite{Ma:2019hsm}. In this picture, the resulting charmonium-like hybrid can naturally decay into a charmonium state accompanied by one or more light hadrons, providing a possible interpretation of the hidden-charm decay pattern of the $Y(4220)$.

\item The $D_1\bar D$ molecular interpretation has been reexamined in light of the high-statistics BESIII measurements after 2017. Recent global analyses of $e^+e^-$ annihilation into multiple hidden- and open-charm final states extracted a pole close to the $D_1\bar D$ threshold, with compatible pole positions, approximately $\sqrt{s_{\rm pole}}\simeq 4227-i25~\mathrm{MeV}$ in Ref.~\cite{vonDetten:2024eie} and $\sqrt{s_{\rm pole}}\simeq 4229.9-i23.2~\mathrm{MeV}$ in Ref.~\cite{Nakamura:2023obk}. These analyses not only include interference, but also coupled channel effects and final state interactions and indicate that the nearby $D_1\bar D$ threshold can play an important role in shaping the observed line shapes, although the physical $Y(4220)$ need not be interpreted as a purely molecular state. In the molecular picture, the proximity to the $D_1\bar D$ threshold naturally connects the $Y(4220)$ with three-body open-charm channels involving the decay of the constituent $D_1$ meson, such as $D\bar D^*\pi$. The same molecular dynamics also leads to characteristic partner-state predictions. Since the $D_1\bar D$ and $\bar D_1D$ configurations are not charge-conjugation eigenstates by themselves, their linear combinations can generate both the $J^{PC}=1^{--}$ state and an exotic $1^{-+}$ partner, as discussed in Ref.~\cite{Dong:2019ofp}. Heavy-quark spin symmetry further constrains the possible molecular partners associated with the same $S$-wave interaction between the ground-state and $P$-wave charmed-meson doublets~\cite{Anwar:2021dmg,Peng:2022nrj}. Related extensions also predicted a narrow flavor-neutral $0^{--}$ hidden-charm molecular state as a particularly clean exotic partner of this dynamics~\cite{Ji:2022blw}. The production of the $Y(4220)$ in $B$-meson decays has also been explored as a phenomenological test of its molecular component~\cite{Liu:2024hba}. Thus, the $D_1\bar D$ molecular scenario provides a near-threshold interpretation of the $Y(4220)$ line shapes and, at the same time, leads to testable predictions for exotic partner states and production mechanisms.

\item Other exotic interpretations have also been proposed for the $Y(4220)$. One representative framework is the hadro-charmonium, or hadro-quarkonium, picture proposed by Dubynskiy and Voloshin~\cite{Dubynskiy:2008mq}, in which a relatively compact charmonium state is embedded inside an extended light-hadron environment. In this scenario, the heavy $c\bar c$ pair approximately preserves its internal structure, while the surrounding light degrees of freedom help determine the quantum numbers and decay patterns of the full system. After the updated BESIII measurements, complementary analyses continued to explore this possibility for the $Y(4220)$. A calculation based on the chromo-electric polarizability in the QCD multipole expansion studied $\eta_c$- and $J/\psi$-isoscalar meson bound states and suggested that the $Y(4220)$ can be accommodated within the hadro-charmonium spectrum~\cite{Ferretti:2018kzy}. In addition, a soft-wall AdS/QCD analysis using configurational entropy as a stability criterion for non-$q\bar q$ candidates also found the hadro-quarkonium assignment to be favored for the $Y(4220)$~\cite{MartinContreras:2023oqs}. These studies support the possibility of interpreting the $Y(4220)$ as a hadro-charmonium state from different model perspectives, while further experimental information is still needed to distinguish this picture from other interpretations.
\end{itemize}

\begin{table*}[h]
\centering
 \renewcommand\arraystretch{1.3}
\caption{Relative partial decay rates for $S$-wave vector charmonia and vector charmonium hybrids into charmed mesons pairs in the Born-Oppenheimer approach~\cite{Braaten:2024stn}.} \label{T4-bo}
{\tabcolsep0.7in
\begin{tabular}{ccc}
\hline\hline
States~$(1^{--})$ & $D\bar D$~~~~~ : ~~~~~$D\bar D^*+ D^* \bar D$~~~~~ : ~~~~~$ D^*\bar D^*$ \\
\hline
$S$-wave charmonium 
& 1: 4 : 7 \\
BO charmonium hybrid 
& 1 : 0 : 3 \\
\hline\hline
\end{tabular}
}
\end{table*}

\subsubsection{$Y(4320)$, $Y(4360)$, and $Y(4390)$}

The structures around $4.3$--$4.4~\mathrm{GeV}$, usually referred to as $Y(4320)$, $Y(4360)$, and $Y(4390)$ in different experimental channels, constitute a particularly subtle part of the vector charmonium-like spectrum. Since they appear in different hidden-charm final states, it remains unclear whether they correspond to several distinct states or to different observations of the same underlying structure.

In the hadronic molecular picture, $Y(4390)$ was interpreted as an isoscalar partner of $Z(4430)$ generated from the $D^*\bar D_1(2420)$ interaction~\cite{He:2017mbh}. Its hidden-charm decays were further studied in the $D^*\bar D_1+\mathrm{h.c.}$ molecular scenario, where the sizable $h_c\pi\pi$ mode was argued to be compatible with the observed enhancement in $e^+e^-\to h_c\pi^+\pi^-$~\cite{Chen:2017abq}. Closely related molecular assignments were discussed in heavy-quark spin symmetry frameworks. In particular, the $D_1\bar D$ molecular interpretation of $Y(4220)$ can be extended to its spin partners, where the structures around $4.36$--$4.4~\mathrm{GeV}$ are associated with $D_1\bar D^*$ or $D_2^*\bar D^*$ configurations~\cite{Anwar:2021dmg,Peng:2022nrj,Ji:2022blw}. The production of $Y(4220)$ and $Y(4360)$ as a $D\bar D_1$--$D^*\bar D_1$ molecular doublet in $B$-meson decays has also been explored through triangle diagrams~\cite{Liu:2024hba}. From the data-analysis side, a global coupled-channel analysis of $e^+e^-\to c\bar c$ processes extracted several poles in this mass region, including poles around $4308$, $4346$, and $4390~\mathrm{MeV}$, which were compared with the experimentally discussed $Y(4320)$, $\psi(4360)$, and the nearby $\psi(4415)$ region, respectively~\cite{Nakamura:2023obk}. This analysis also emphasized that the $Y(4320)$ pole may be affected by the treatment of $D_1\bar D$ threshold effects, so its status remains to be further clarified. In contrast, a focused analysis of the $4.2$--$4.35~\mathrm{GeV}$ region showed that the available data can be described with a single $D_1\bar D$-dominated $Y(4220)$ state together with threshold effects and interference with $\psi(4160)$, without requiring an additional $Y(4320)$ state~\cite{vonDetten:2024eie}.

Compact tetraquark interpretations have also been applied to this mass region. A QCD sum-rule analysis using vector tetraquark currents supported assigning the $Y(4360/4320)$ to a $c\bar c q\bar q$ vector tetraquark configuration, while disfavoring the assignment of $Y(4390)$ as a pure vector tetraquark state of the same type~\cite{Wang:2018rfw}. In a relativized diquark model, the $Y(4360)$ was also accommodated within the hidden-charm tetraquark spectrum, as part of a broader set of predicted $1^{--}$ tetraquark states~\cite{Anwar:2018sol}. More recently, a constituent quark model calculation of $1^{--}$ $P$-wave compact tetraquarks suggested that $\psi(4360)$ may be assigned as a $P$-wave tetraquark state, with several nearby states possibly appearing around $4.36~\mathrm{GeV}$~\cite{Zhao:2025kno}. These complementary studies show that the nature of the $Y(4320)$, $Y(4360)$, and $Y(4390)$ structures is still not understood. More precise measurements of their masses, widths, decay patterns, and correlations among different hidden-charm and open-charm channels will be crucial for clarifying their relation.

\subsubsection{$Y(4630)$ and $Y(4660)$}

Here, we focus on theoretical studies of the $Y(4630)$ and $Y(4660)$ published after 2017, while earlier interpretations can be found in previous reviews~\cite{Klempt:2007cp,Brambilla:2010cs,Liu:2013waa,Hosaka:2016pey,Richard:2016eis,Chen:2016qju,Esposito:2016noz,Chen:2016spr,Lebed:2016hpi,Guo:2017jvc,Olsen:2017bmm,Ali:2017jda,Liu:2019zoy,Brambilla:2019esw,Chen:2022asf,Meng:2022ozq,Wang:2021aql}. Two experimental developments are relevant to the recent discussions: the BESIII measurement of the near-threshold $e^+e^-\to\Lambda_c^+\bar\Lambda_c^-$ cross section in 2017~\cite{BESIII:2017kqg}, and the Belle observation of the nearby $Y(4626)$ structure in $e^+e^-\to D_s^+D_{s1}(2536)^-+c.c.$ in 2019~\cite{Belle:2019qoi}. The central issue is whether the structures observed in $\Lambda_c\bar\Lambda_c$, $\psi(2S)\pi^+\pi^-$, and open-charm-strange channels are different manifestations of the same vector state or correspond to distinct resonances.

A reanalysis of $e^+e^-\to\Lambda_c^+\bar\Lambda_c^-$ including the $\Lambda_c\bar\Lambda_c$ final-state interaction within chiral effective field theory found resonance parameters compatible with those of the $Y(4660)$ observed in $\psi(2S)\pi^+\pi^-$, thus supporting the possibility that the $Y(4630)$ and $Y(4660)$ are the same state~\cite{Dai:2017fwx}. A different $\Lambda_c\bar\Lambda_c$-based interpretation described the $Y(4630)$ as a $^3D_1$ hadronic resonant state generated by the $\Lambda_c\bar\Lambda_c$ interaction in a one-boson-exchange potential~\cite{Mei:2022msh}. The nearby $Y(4626)$ has also motivated molecular interpretations in the open-charm-strange sector. In particular, it was proposed as a $D_s^*\bar D_{s1}(2536)$ molecular state from the coupled $D_s^*\bar D_{s1}(2536)$--$D_s\bar D_{s1}(2536)$ dynamics~\cite{He:2019csk}, and from heavy-quark spin and SU(3)-flavor symmetries~\cite{Peng:2022nrj}. A Bethe-Salpeter study of the same $D_s^*\bar D_{s1}(2536)$ system, however, found that an unrealistically large cutoff parameter would be required to generate a bound state, and therefore disfavored this molecular assignment within that framework~\cite{Ke:2020eba}. More recently, the hidden-charm decays of the $Y(4626)$ were studied under the $D_s^*\bar D_{s1}(2536)$ molecular assumption, with the channels $J/\psi\eta^{(\prime)}$ and $\chi_{cJ}\phi$ suggested as useful tests of this picture~\cite{Yue:2024bvy}.

Compact tetraquark interpretations have also been widely discussed for the $Y(4660)$ and the nearby $Y(4630)$ structure. A QCD sum-rule analysis of vector tetraquark currents supported assigning the $Y(4660/4630)$ to a $c\bar c q\bar q$ vector tetraquark configuration~\cite{Wang:2018rfw}. In another QCD sum-rule study, the $Y(4660)$ was treated as a $J^{PC}=1^{--}$ diquark-antidiquark tetraquark, and its strong decays into $J/\psi f_0(980)$, $\psi(2S)f_0(980)$, $J/\psi f_0(500)$, and $\psi(2S)f_0(500)$ were calculated, yielding a mass and width compatible with experiment~\cite{Sundu:2018toi}. The strong decays of the $Y(4660)$ were further analyzed in the vector tetraquark scenario using QCD sum rules based on solid quark-hadron duality, where the large coupling to $\psi(2S)f_0(980)$ was found to be consistent with its observation in the $\psi(2S)\pi^+\pi^-$ channel~\cite{Wang:2019iaa}. Later QCD sum-rule studies also interpreted the $Y(4660)$ as a hidden-charm hidden-strange vector tetraquark with implicit $P$ waves~\cite{Wang:2023jaw}, and a recent three-point QCD sum-rule analysis of its strong decays supported a $[sc][\bar s\bar c]$ tetraquark assignment with $J^{PC}=1^{--}$~\cite{Yang:2025lef}. In addition, a constituent quark model calculation of $1^{--}$ $P$-wave compact tetraquarks accommodated the $Y(4660)$ together with other vector charmonium-like structures, such as the $Y(4220)$ and $Y(4360)$, within the same compact-tetraquark spectrum~\cite{Zhao:2025kno}. Further measurements in these channels, together with searches for the predicted hidden-charm and hidden-strange decay modes, will be important for clarifying the relation among the $Y(4626)$, $Y(4630)$, and $Y(4660)$ structures.

\subsubsection{$Y(4500)$, $Y(4710)$, and $Y(4790)$}

The $Y(4500)$ and $Y(4710)$ were reported in $e^+e^-\to K^+K^-J/\psi$, while the $Y(4790)$ was observed in $e^+e^-\to D_s^{*+}D_s^{*-}$, as introduced in the preceding discussion. After introducing the unified interpretation based on the unquenched charmonium picture, we summarize here several complementary explanations, including hidden-charm hidden-strange molecular assignments, compact tetraquark configurations, and triangle-singularity mechanisms.

For the $Y(4500)$, one possible interpretation is a hidden-charm hidden-strange molecular state. Based on heavy-quark spin symmetry and SU(3)-flavor symmetry, the $Y(4500)$ was proposed as the strange partner of the $D\bar D_1$ molecular candidate associated with the $Y(4220)$, corresponding to a $D_s\bar D_{s1}$ configuration~\cite{Peng:2022nrj}. A QCD sum-rule analysis also tested this possibility and found that a $D_s\bar D_{s1}$ molecular current with $J^{PC}=1^{--}$ can reproduce the mass of the $Y(4500)$~\cite{Gungor:2023ksu}. A different interpretation was proposed from the triangle-singularity mechanism in $e^+e^-\to J/\psi K^+K^-$. In this analysis, the $Y(4500)$ enhancement was generated through $Z_{cs}^{(\prime)}$-mediated intermediate processes, rather than being introduced as a genuine resonance~\cite{Wang:2025zbv}. In addition, based on the unquenched charmonium assignment discussed in the previous sections, the dipionic transition $Y(4500)\to J/\psi\pi^+\pi^-$ was studied through charmed-meson loop mechanisms, including box diagrams and two types of triangle diagrams~\cite{Liu:2025bjm}.

Compact tetraquark interpretations have been developed for all three structures. The $Y(4500)$ was investigated as a $J^{PC}=1^{--}$ tetraquark state through its three-body strong decay $Y(4500)\to D^*\bar D^*\pi$ within light-cone QCD sum rules~\cite{Wang:2023hsc}. Its two-body open-charm decays were further analyzed in a vector-tetraquark picture, where several decay channels were proposed to help distinguish different possible tetraquark configurations~\cite{Wang:2024qqa}. For the higher structures, a QCD sum-rule study of hidden-charm hidden-strange vector tetraquark states with implicit $P$ waves assigned the $Y(4710)$ and $Y(4790)$ to $[sc][\bar s\bar c]$-type configurations with $J^{PC}=1^{--}$~\cite{Wang:2023jaw}. Another QCD sum-rule analysis with explicit $P$-wave tetraquark currents also discussed possible assignments of the $Y(4500)$, $Y(4710)$, and $Y(4790)$ within the vector tetraquark spectrum~\cite{Wang:2024gvf}. Most of these exotic explanations emphasize hidden-strange configurations. Confirming or excluding these structures in nonstrange hidden-charm and open-charm channels would therefore provide an especially useful test for distinguishing such interpretations from the unquenched charmonium picture.
}


\section{The initial single pion emission mechanism: explaining charged $Z_b$ and predicting charmonium-like charged $Z_c/Z_{cs}$}\label{section5}\label{sec:Z}

Of the many charmonium-like $XYZ$ states discovered over the past two decades, those that carry electric charge are collectively referred to as $Z$ states. A prominent example is provided by the bottomonium-like charged states $Z_b(10610)$ and $Z_b(10650)$, reported by the Belle Collaboration \cite{Belle:2011aa}. These $Z_b$ states have attracted extensive attention from the entire community, as their identification as hadrons necessarily implies an exotic nature beyond conventional quark models. An alternative theoretical framework for understanding their properties is the {\it Initial Single Pion Emission} (ISPE), proposed in Ref. \cite{Chen:2011pv}. As reviewed in this section, the ISPE mechanism  embodies an unquenched picture of hadron dynamics.

This section outlines the motivation behind the ISPE mechanism, which was originally developed to explain the hidden-bottom dipion decay anomaly observed in $\Upsilon(10860)$ \cite{Belle:2008nac}. The mechanism is not restricted to bottomonium systems but can also be applied to hidden-charm dipion decays of higher charmonium(-like) states. It successfully predicted the existence of charged charmonium-like $Z_c$ structures near the $D\bar{D}^*$ and $D^*\bar{D}^*$ thresholds, which should appear in the invariant mass spectra of $J/\psi\pi^\pm$, $\psi(3686)\pi^\pm$, and $h_c(1P)\pi^\pm$~\cite{Chen:2011xk}. In 2013, the BESIII and Belle collaborations observed $Z_c(3900)$ in the $J/\psi\pi^\pm$ invariant mass spectrum of the process $e^+e^-\to J/\psi\pi^+\pi^-$ at $\sqrt{s}=4.26$ GeV \cite{BESIII:2013ris,Belle:2013yex}. Since then, additional charged $Z_c$ structures have been reported in other channels \cite{BESIII:2013mhi,BESIII:2015tix,BESIII:2017tqk,BESIII:2017vtc,BESIII:2019tdo,Xiao:2013iha,BESIII:2015pqw,BESIII:2015cld,BESIII:2017bua,BESIII:2020oph,BESIII:2015ntl,BESIII:2013ouc,BESIII:2014gnk}. Furthermore, the ISPE framework has been extended to a more general initial single chiral particle emission mechanism. This extension suggests that $Z_{cs}$ structures could exist in the $J/\psi K^\pm$ invariant mass spectrum of $e^+e^-\to J/\psi K^+K^-$ at several energy points \cite{Chen:2013wca}, presenting a clear task for future experimental searches.

In the following, we review the formulation of the ISPE mechanism and its application across different quarkonium systems.

\subsection{Anomalous phenomena related to $\Upsilon(10860)$}

\subsubsection{Experimental observation of $\Upsilon(10860)$}

The bottomonia $\Upsilon(5S)$ was first reported in the $e^+ e^-$ annihilation cross section into hadrons by the CUSB Collaboration, the observed mass and width of $\Upsilon(5S)$ were measured to be $(10845\pm 20)$ and $(110 \pm 15)$ MeV, respectively~\cite{Lovelock:1985nb}. The signal of $\Upsilon(5S)$ had been successively reported by the CLEO~\cite{CLEO:1984vfn}, BaBar~\cite{BaBar:2008cmq}, and Belle~\cite{Belle:2015aea} Collaborations in the same inclusive process but with higher observed mass. The most precise mass of $\Upsilon(5S)$ from the inclusive process is $(10881.8 ^{+1.0}_{-1.1} \pm 1.2)$ MeV reported by the Belle Collaboration~\cite{Belle:2015aea}.

\begin{figure}
\centering
\scalebox{0.72}{\includegraphics{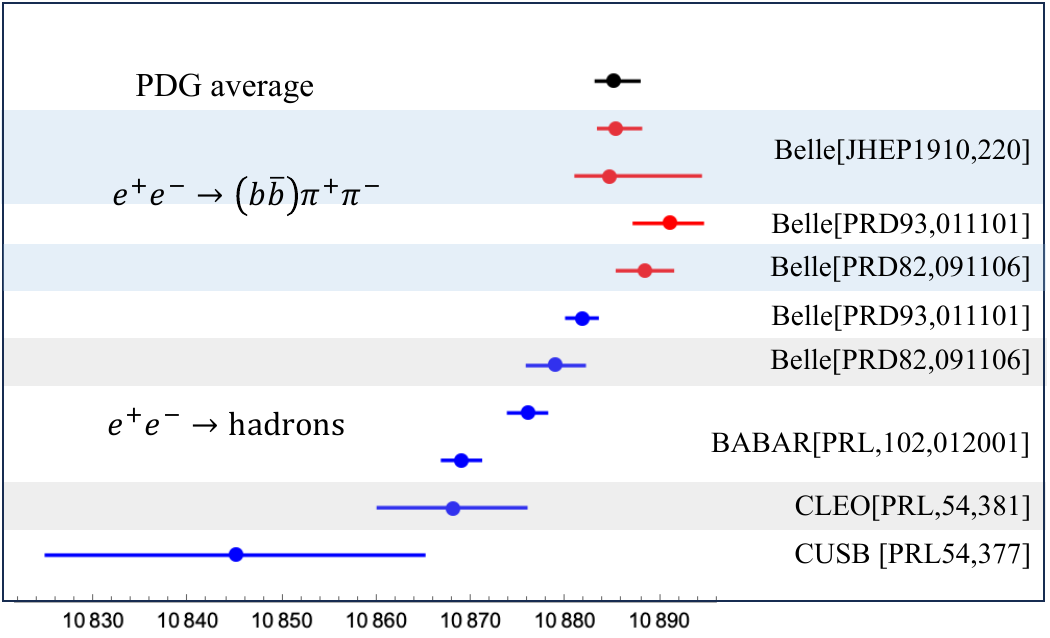}}
\caption{The mass of $\Upsilon(10860)$ in the inclusive process $e^+ e^- \to \mathrm{hadrons}$ (blue dots with error bars) and in the hidden bottom processes $e^+ e^-\to (b \bar{b}) \pi^+ pi^-$ with $\{(b\bar{b})= \Upsilon(1S,2S,3S), h_b(1P,2P)\}$ (red dots with error bars). The PDG average (black dot with error bar) is also present for comparison.\label{Fig:Ybmass}}
\end{figure}

The first exclusive production cross sections for $e^+ e^- \to \Upsilon(mS) \pi^+ \pi^-$ (with $m=1,2,3$) as a function of the center-of-mass energy $\sqrt{s}$ were reported by the Belle Collaboration~\cite{Belle:2008nac}, where the process $\Upsilon(5S) \to \Upsilon(mS) \pi^+ \pi^-$ (with $m=1,2,3$) was observed. The fitted mass and width of the resonance were determined to be $(10888.4^{+2.7}_{-2.6} \pm 1.2)$ MeV and $(30.7^{+8.3}_{-7.0} \pm 3.1)$ MeV, respectively, where the first uncertainties are statistical and the second are systematic. These values are not fully consistent with the conventional $\Upsilon(5S)$ lineshape observed in the inclusive $e^+ e^-$ annihilation cross sections~\cite{Belle:2008nac}. Subsequently, using more precise data samples, the Belle Collaboration reported further measurements of $\Upsilon(5S)$ in the processes $e^+ e^- \to \Upsilon(nS) \pi^+ \pi^-$ (with $n=1,2,3$) and $e^+ e^- \to h_b(mP) \pi^+ \pi^-$ (with $m=1,2$)~\cite{Belle:2019cbt}. The observed masses of $\Upsilon(5S)$ from the inclusive processes (blue dots with error bars) and from the exclusive hidden-bottom decay processes (red dots with error bars) are presented in Fig.~\ref{Fig:Ybmass}. From the figure, one can clearly see that the masses observed in the exclusive hidden-bottom decay processes are systematically higher than those observed in the inclusive processes. Whether the states observed in the inclusive and exclusive processes correspond to the same resonance has become an intriguing question and has stimulated considerable theoretical interest. Given that the community has not reached a definitive agreement on the nature of this state, we refer to this resonance as $\Upsilon(10860)$ in the following discussion.

\renewcommand\arraystretch{1.25}
\begin{table}
\centering
\caption{The partial widths of dipion transitions between $\Upsilon$ states~\cite{ParticleDataGroup:2024cfk, BaBar:2008xay, Belle:2007xek, BaBar:2006udk}. The partial width of $\Upsilon(3S)$ dipion decays are estimated by the PDG average of $\Upsilon(3S)$ width and the branching fractions of the  corresponding processes. \label{ISPE:UpsilonDipionWD}}
\begin{tabular}{cc|cc}
\hline
Processes  &  Width & Processes  &  Width\\
\hline 
$\Upsilon(2S) \to \Upsilon(1S) \pi^+ \pi^-$ & $(105.4 \pm 1.0 \pm 4.2)$ eV~\cite{BaBar:2008xay} &  $\Upsilon(10860) \to \Upsilon(2S) \pi^+ \pi^-$ & $(590 \pm 40 \pm 90)$ keV~\cite{Belle:2007xek}\\ 
$\Upsilon(3S) \to \Upsilon(2S) \pi^+ \pi^-$ & $(0.57 \pm 0.06)$ keV ~\cite{ParticleDataGroup:2024cfk} & $\Upsilon(10860) \to \Upsilon(1S) \pi^+ \pi^-$ & $(850 \pm 70 \pm 16)$ keV~\cite{Belle:2007xek}\\ 
$\Upsilon(3S) \to \Upsilon(1S) \pi^+ \pi^-$ & $(0.89 \pm 0.08)$ keV~ \cite{ParticleDataGroup:2024cfk}& \\ 
$\Upsilon(4S) \to \Upsilon(2S) \pi^+ \pi^-$ & $(2.7\pm 0.8 \pm 4.2)$ keV ~\cite{BaBar:2006udk}\\ 
$\Upsilon(4S) \to \Upsilon(1S) \pi^+ \pi^-$ & $(1.8 \pm 0.4)$ keV~\cite{BaBar:2006udk}\\ 
\hline
\end{tabular}
\end{table}

Besides the discrepancy in the resonance parameters observed in the inclusive $e^+ e^-$ annihilation cross sections and exclusive hidden bottom decay processes, the Belle Collaboration also reported the partial widths of $\Upsilon(1S)\pi^+ \pi^-$ and $\Upsilon(2S)\pi^+ \pi^-$, which are $(0.59\pm 0.04\stat \pm 0.09\syst)$ MeV and $(0.85 \pm 0.07 \stat \pm 0.16 \syst)$ MeV, respectively~\cite{Belle:2007xek}. The dipion transitions between the bottomonia should be suppressed by the OZI suppression rule. As shown in Table~\ref{ISPE:UpsilonDipionWD}, the partial widths of dipion decay of $\Upsilon(2S/3S/4S)$ are of order 1 keV, which are consistent with the expectations of OZI suppression rule. However, for $\Upsilon(10860)$, the partial widths of the dipion decays exceed by more than two orders of magnitude of the dipion transitions between lower $\Upsilon$ resonances as shown in Table \ref{ISPE:UpsilonDipionWD}. How to understand the anomalous large partial widths of $\Upsilon(10860)$ hidden-bottom dipion decays are also a challenge to our understanding of the properties of higher bottomonia.  

\begin{figure}
\centering
\scalebox{0.75}{\includegraphics{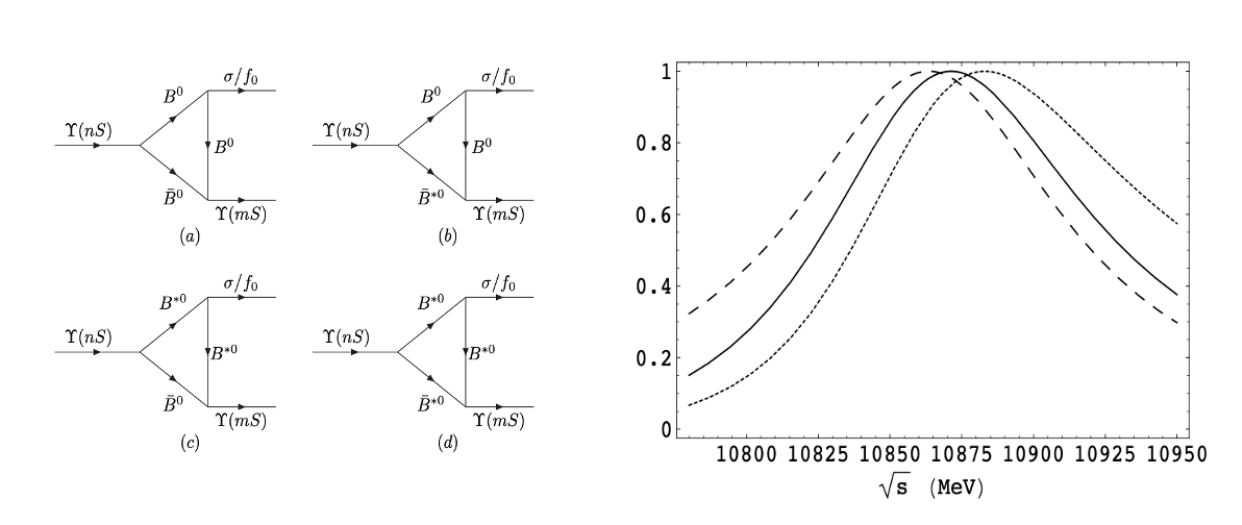}}
\caption{The diagram contributing to $\Upsilon(nS) \to \Upsilon(mS) \pi^+ \pi^-$ (left panel) and the peak shifts caused by the triangle diagrams (right panel).\\{\it Source}: Taken from~\cite{Meng:2008dd}. \label{Fig:Peakshifts}}
\end{figure}

\subsubsection{Theoretical interpretations of $\Upsilon(10860)$: bottomonium vs tetraquark}

To explain the mass difference between the inclusive $e^+ e^-$ annihilation cross section and the dipion decay process and the anomalous large hidden-bottom dipion decay width of $\Upsilon(10860)$, one should carefully check the mass spectrum of bottomonia family. For $\Upsilon(10860)$, its mass is above the thresholds of $B_{(s)}^\ast \bar{B}_{(s)}^\ast$, which indicate that $\Upsilon(5S)$ should dominantly decay into a pair of bottom mesons. Considering this fact, Meng and Chao investigated dipion transitions between $\Upsilon(5S)$ and $\Upsilon(mS)\, (m=1,2,3)$ in the final state interaction mechanism induced by unquench effect~\cite{Meng:2007tk,Meng:2008dd, Meng:2008bq}, where the hidden-bottom dipion transitions mainly proceed through the real process $\Upsilon(5S) \to B^{(\ast)} \bar{B}^{(\ast)}$ and $ B^{(\ast)} \bar{B}^{(\ast)} \to \Upsilon(mS) \sigma/f_0$ and the scalar meson $\sigma/f_0$ couples to $\pi^+\pi^-$ as shown in the left panel of Fig.~\ref{Fig:Peakshifts}. In the final state interaction mechanism, the large branching fractions of the hidden-bottom dipion transition processes could be well interpreted. Similar conclusion can be drawn from the study in Ref.~\cite{Simonov:2007cj}.
	 
In addition, Meng and Chao introduced a form factor in each $\Upsilon(5S) B^{(\ast)} \bar{B}^ {(\ast)}$ coupling to naturally balance the otherwise over-increased decay width with increased phase space in the rescattering model~\cite{Meng:2008dd}. They found that the energy distributions of $\Upsilon(mS)\pi^+ \pi^-$ markedly differ from that of  $\Upsilon(5S) \to B^{(\ast )} \bar{B}^{(\ast)}$, and as shown in the right panel of Fig.~\ref{Fig:Peakshifts}, the resonance peak is pushed up by about $7-20$ MeV for the hidden-bottom dipion transitions relative to the open-bottom decay modes~\cite{Meng:2008dd}, which naturally explain the discrepancy of the observed mass in the inclusive process and exclusive hidden-bottom dipion decay processes.

\begin{figure}
\centering
\scalebox{0.4}{\includegraphics{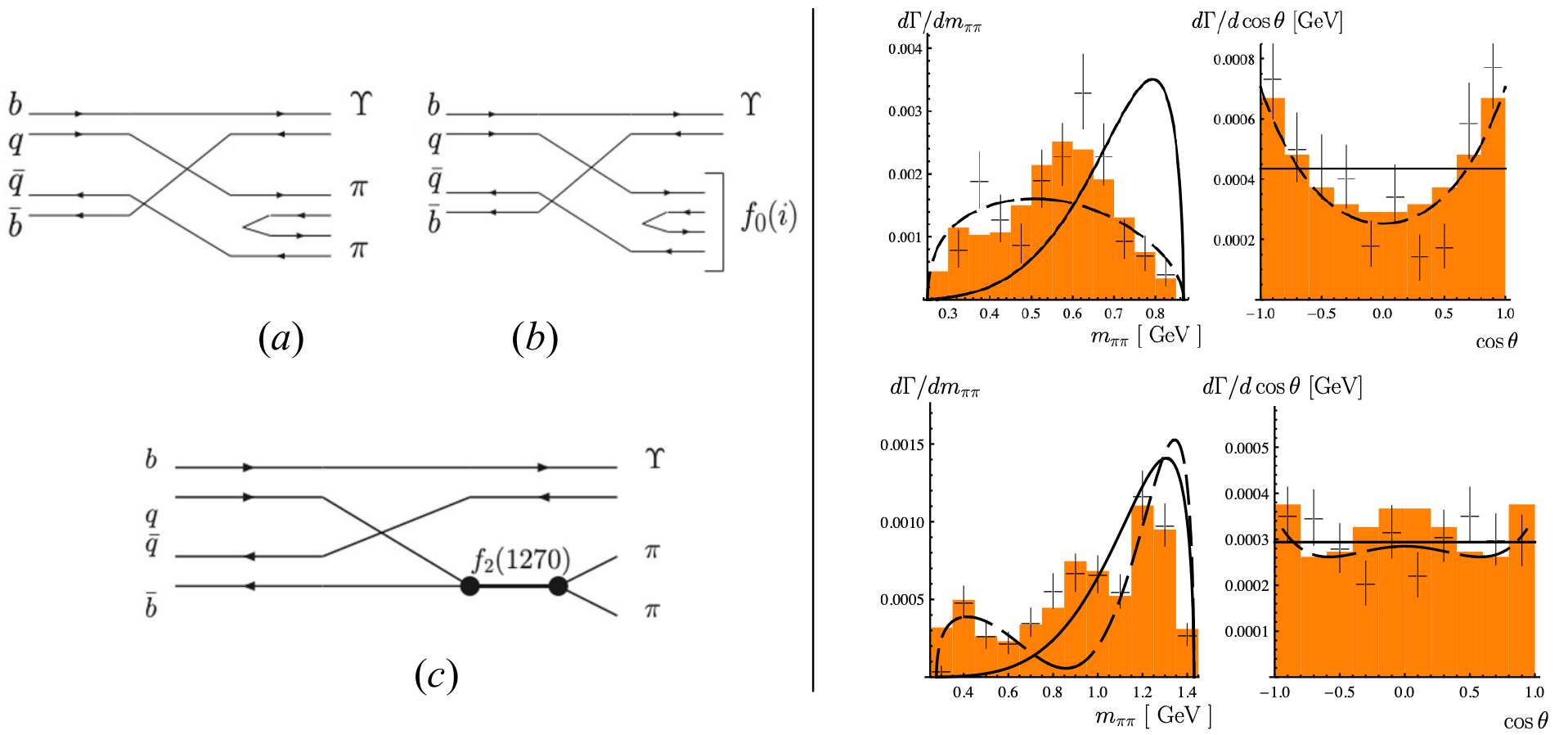}}\caption{The diagram contributing to $Y_b(10890) \to \Upsilon \pi^+ \pi^-$ in tretraquark scenario (left panel) and dipion invariant mass $(m_{\pi \pi})$ distribution and the $\cos \theta$ distributions measured by Belle~\cite{Belle:2007xek} for the final state $\Upsilon(1S ) \pi^+ \pi^-$ (upper row of right panel) and  $\Upsilon(2S ) \pi^+ \pi^-$ (lower row of right panel), and the theoretical distributions based on the tetraquark estimations (histograms)~\cite{Ali:2009es}. The solid and dashed lines show purely continuum contributions for different $\beta$. \\
{\it Source}: Taken from~\cite{Ali:2009es}.\label{Fig:Upsilon10860-tetraquark}}
\end{figure}

 Besides the conventional $\Upsilon(5S)$ interpretations, the particular properties of $\Upsilon(10860)$, observed in the cross sections for $e^+e^- \to \Upsilon(mS) \pi^+ \pi^-$, also stimulated the exotic interpretation. For example, the estimations within the constituent quark model~\cite{Ali:2009pi,Ali:2009es,Ali:2010pq,Sonnenschein:2016ibx,Patel:2016otd},  QCD sum rules~\cite{Zhang:2010mv} suggested that the mass of $P$-wave $[bq][\bar{b}\bar{q}]$ tetraquark state agrees with the experimental measurements of $\Upsilon(10860)$ in the exclusive hidden-bottom dipion decay processes. In addition to the resonance parameters, the Belle Collaboration also reported the dipion invariant mass spectra and the helicity angle distributions of the $\Upsilon(10890)\to \Upsilon(1S/2S) \pi^+ \pi^-$ processes. In the tetraquark framework, Ali {\it et al.}, considered the non-resonance contribution (Fig.~\ref{Fig:Upsilon10860-tetraquark}(a)), the resonance contributions from scalar meson $f_0(600)$ and $f_0(980)$ (Fig.~\ref{Fig:Upsilon10860-tetraquark}(b)), and tensor meson $f_2(1270)$ (Fig.~\ref{Fig:Upsilon10860-tetraquark}(c))  in $\Upsilon(10860) \to \Upsilon(1S/2S) \pi^+ \pi^-$. With these contributions, the authors in Ref.~\cite{Ali:2009es} claimed that the experimental data could be well reproduced in the tetraquark scenario as shown in the right panel of  Fig.~\ref{Fig:Upsilon10860-tetraquark}\footnote{However, in the erratum to this work, the authors indicated that the dipion invariant mass distributions and the $\cos \theta$ distributions of $\Upsilon(10860) \to \Upsilon(2S) \pi^+ \pi^-$ cannot be reproduced simultaneously, as will be discussed in the following section.}.

\begin{figure}[h]
\centering
\scalebox{0.7}{
\includegraphics{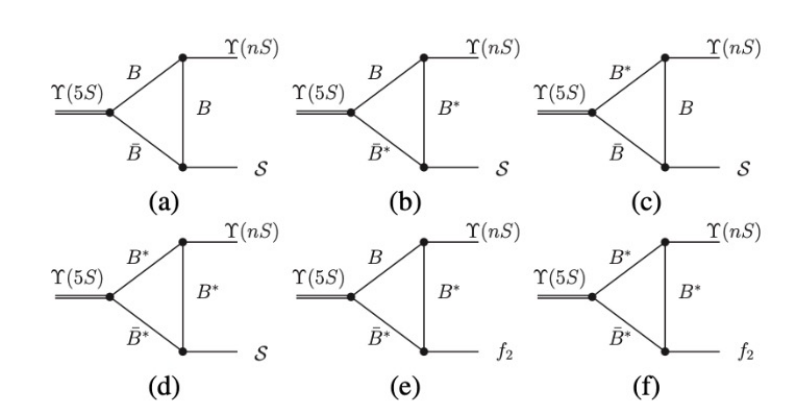}}
\caption{The schematic diagrams for $\Upsilon(5S)$ decays into $\Upsilon(nS)\mathcal{S}$ (diagrams (a)-(d)) and $\Upsilon(nS)f_2(1270)$ (diagrams (e)-(f)) $(n=1,2)$ via bottom meson loops. 
\\
{\it Source:} Taken from~\cite{Chen:2011qx}.
\label{Fig:chen-upsilon5Sdipion-ML}}
\end{figure}

\subsubsection{Puzzles in the dipion invariant mass distributions and $\cos\theta$ distributions.}

In their study of $\Upsilon(5S)$ decays, Chen {\it et al.} extended the final-state interaction mechanism developed in Refs.~\cite{Meng:2007tk, Meng:2008dd, Meng:2008bq} to investigate the dipion invariant mass spectra and helicity angle distributions of the processes $\Upsilon(5S) \to \Upsilon(1S/2S) \pi^+ \pi^-$ in Ref.~\cite{Chen:2011qx}, \changelabel{and later similar analyses were also performed in the dipion transitions of $\Upsilon(3S)$ and $\Upsilon(4S)$~\cite{Chen:2015jgl,Chen:2016mjn}}. In this approach, both non-resonance contributions and dipion resonance contributions are considered. Within the dipion resonance contributions, bottom meson loops connect the initial $\Upsilon(5S)$ to the final bottomonia and the dipion resonances. In this framework, the amplitude for $\Upsilon(5S) \to \Upsilon(nS) \pi^+ \pi^-$ takes the form,
\begin{eqnarray}
\label{model} \mathcal{M}_{\mathrm{total}}&=&\mathcal{M}
\textnormal{\Huge{$[$}}
\raisebox{-18pt}{\includegraphics[width=0.2%
\textwidth]{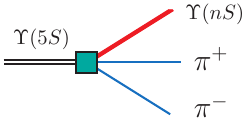}}\textnormal{\Huge{$]$}}+\sum\limits_{R}e^{i\phi_R^{(n)}}\mathcal{M}\textnormal{\Huge{$[$}}
\raisebox{-18pt}{\includegraphics[width=0.2%
\textwidth]{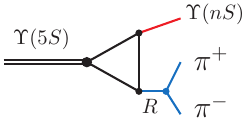}}
\textnormal{\Huge{$]$}},\label{decay}
\end{eqnarray}
where they take into account contributions from different intermediate resonances $R$ coupling to the dipion system—specifically, $R={\sigma(600),f_0(980),f_2(1270)}$ for the $\Upsilon(1S)\pi^+\pi^-$ channel and $R={\sigma(600),f_0(980)}$ for the $\Upsilon(2S)\pi^+\pi^-$ channel—as permitted by the available phase space.

In general, the decay amplitude for the non-resonance contributions to $\Upsilon(5S)\to \Upsilon(1S/2S)\pi^+\pi^-$ can be written as
\begin{eqnarray}
&&\mathcal{M}[\Upsilon(5S) \to \Upsilon(nS)(p_1)\pi^+(p_2)\pi^-(p_3)]_{\mathrm{NoR}}\nonumber \\ 
&& \qquad =\epsilon_{\Upsilon(5S)}\cdot\epsilon_{\Upsilon(nS)}^*\left\{
\left[q^2-\kappa(\Delta M)^2\left(1+\frac{2m^2_\pi}{q^2}\right)\right]_{\mathrm{S-wave}}
 +\left[\frac{3}{2}\kappa\left((\Delta M)^2-q^2\right)\left(1-\frac{4m_\pi^2}{q^2}\right)\left(\cos\theta^2-\frac{1}{3}\right)\right]_{\mathrm{D-wave}}\right\}\mathcal{A},\nonumber \\
\end{eqnarray}
This form was first proposed by Novikov and Shifman in their study of the $\psi^\prime \to J/\psi \pi^+\pi^-$ decay in Ref.~\cite{Novikov:1980fa}, where the $S$-wave and $D$-wave contributions are distinguished by the corresponding subscripts. Here, $\Delta M$ denotes the mass difference between $\Upsilon(5S)$ and $\Upsilon(nS)$. The quantity $q^2 = (p_2 + p_3)^2 \equiv m_{\pi^+\pi^-}^2$ represents the invariant mass squared of the $\pi^+\pi^-$ system, and $\mathcal{A} = F/f_\pi^2$ with $f_\pi = 130$ MeV. The angle $\theta$ is defined as the angle between the $\Upsilon(5S)$ momentum and the $\pi^-$ momentum in the $\pi^+\pi^-$ rest frame.

For the dipion resonance contributions, meson loop diagrams are introduced, as illustrated in Fig.~\ref{Fig:chen-upsilon5Sdipion-ML}. The relevant hadron-level interactions are described by an effective Lagrangian constructed in accordance with the heavy quark limit~\cite{Oh:2000qr, Casalbuoni:1996pg, Colangelo:2003sa}. After performing the loop integrals, the amplitudes for $\Upsilon(5S) \to \Upsilon(nS) \pi^+ \pi^-$ arising from such rescattering processes can be parameterized as
\begin{eqnarray}
&&\mathcal{M}[\Upsilon(5S)\to
B^{(*)}\bar{B}^{(*)}\to\Upsilon(nS)(p_1)\pi^+(p_2)\pi^-(p_3)]_S=
\left\{g^{(n)}_{0S} g_{\mu\nu}p_1\cdot
q+g_{0D}^{(n)}p_{1\mu}q_\nu\right\}\frac{\epsilon_{\Upsilon(5S)}^\mu
\epsilon_{\Upsilon(nS)}^{*\nu} g_{_{S
\pi\pi}} p_2 \cdot p_3}{q^2-m_S^2+i m_S \Gamma_S } ,\notag\\
&&\mathcal{M}[\Upsilon(5S)\to B^{(*)}\bar{B}^{(*)}\to
\Upsilon(nS)(p_1)\pi^+(p_2)\pi^-(p_3)]_{f_2(1270)}
=\left\{g_{2S}^{(n)} [g_{\mu \rho} g_{\nu
\lambda}+g_{\mu \lambda} g_{\nu \rho}] (p_1 \cdot q)^2 +
[g_{2D_1}^{(n)} g_{\mu \nu}
p_{1 \rho} q_{ \lambda} \right. \nonumber\\
&&\left.\qquad+ g_{2D_2}^{(n)} (g_{\mu \rho} q_{\nu} p_{1\lambda} +g_{\mu
\lambda} q_{\nu} p_{1\rho}) + g_{2D_3}^{(n)} (g_{\nu \lambda}
q_{\mu} p_{1\rho}+g_{\nu \rho} q_{\mu} p_{1\lambda}) ]p_1 \cdot q
+g_{2G}^{(n)}q_{\mu} q_{\nu} p_{1\rho} p_{1\lambda} \right\}\nonumber\\
&&\qquad\times
\frac{\epsilon_{\Upsilon(5S)}^\mu \epsilon_{\Upsilon(nS)}^{*\nu}
\mathcal{P}_{f_2}^{\rho \lambda \alpha \beta}(q)}{q^2 -m_{f_2}^2 + i
m_{f_2}
\Gamma_{f_2}} g_{f_2 \pi \pi}p_{2 \alpha} p_{2 \beta} ,\label{2}
\end{eqnarray}
corresponding to the contributions from the intermediate scalar states $S={\sigma(600), f_0(980)}$ and the tensor meson $f_2(1270)$, respectively. The coefficients associated with the Lorentz structures are evaluated via meson loop integrals. In the above equation, $\mathcal{P}_{f_2}^{\rho \lambda
\alpha \beta}(q)$ is defined as
\begin{eqnarray}
\mathcal{P}_{f_2}^{\rho \lambda \alpha \beta}(q) = \frac{1}{2}
(\tilde{g}^{\rho \alpha} \tilde{g}^{\lambda \beta} + \tilde{g}^{\rho
\beta} \tilde{g}^{\lambda \alpha})- \frac{1}{3} \tilde{g}^{\rho
\lambda} \tilde{g}^{\alpha \beta}\nonumber
\end{eqnarray}
with $\tilde{g}^{\alpha \beta}= g^{\alpha \beta} - q^\alpha
q^\beta/m_{f_2}^2$.

\begin{figure}[t]
\begin{center}
\scalebox{0.52}{
\includegraphics{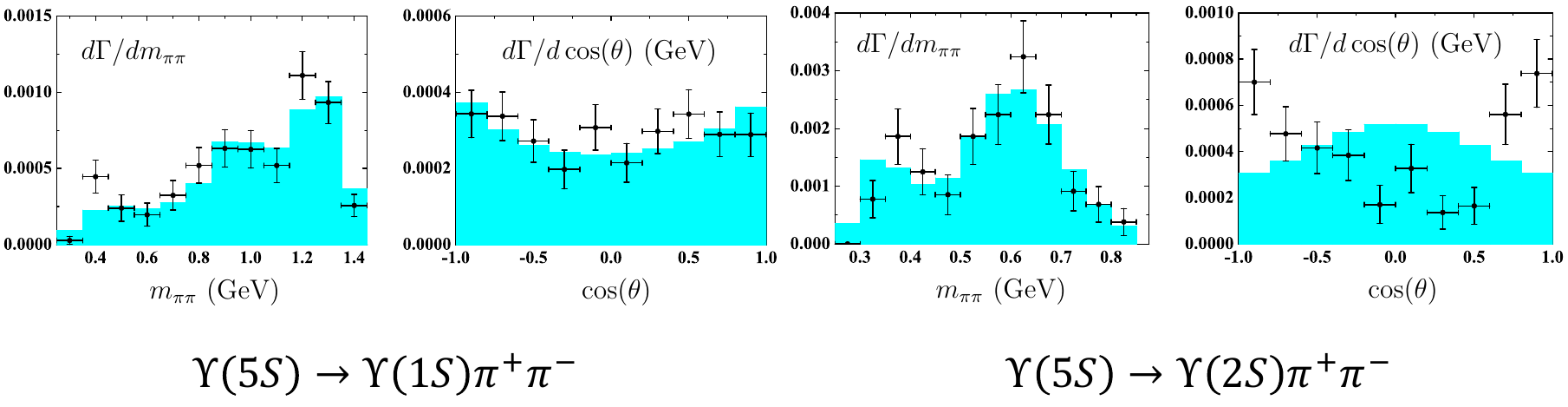}}\caption{Dipion invariant mass ($m_{\pi\pi}$) distributions (1st and 3rd frames) and the $\cos\theta$ distributions (2nd and 4th frames) measured by Belle for the final state $\Upsilon(1S) \pi^+ \pi^-$ (1st and 2nd frames, crosses) and $\Upsilon(2S) \pi^+ \pi^-$ (3rd and 4th frames, crosses) and the theoretical distributions in the conventional $\Upsilon(5S)$ frame (histograms). 
\\
{\it Source:} Taken from~\cite{Chen:2011qx}.
\label{Fig:chen-upsilon5Sdipion}}
\end{center}
\end{figure}

\begin{figure}[t]
\begin{center}
\scalebox{0.45}{
\includegraphics{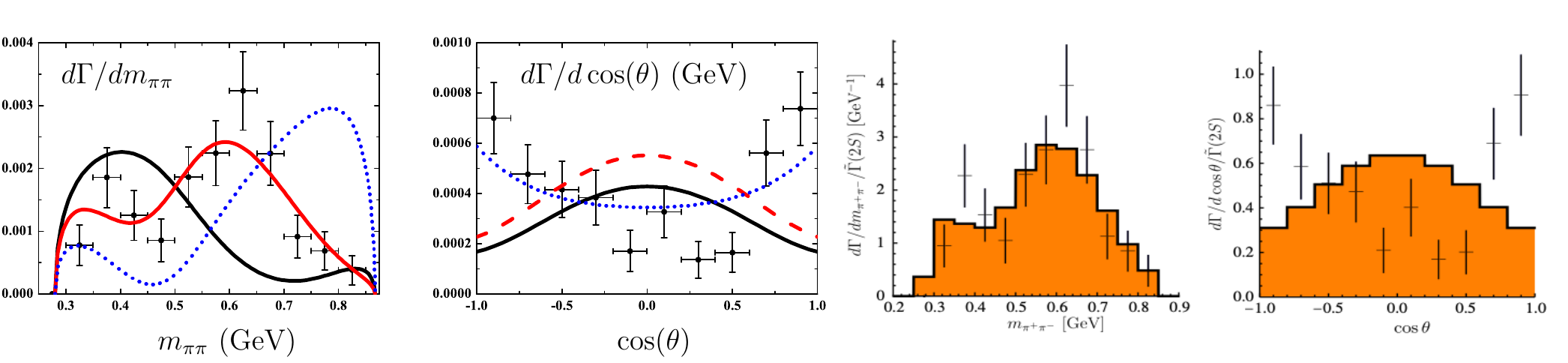}}
\caption{Dipion invariant mass ($m_{\pi\pi}$) distributions (1st and 3rd frames) and the $\cos\theta$ distributions (2nd and 4th frames) of $\Upsilon(5S) \to \Upsilon (2S) \pi^+ \pi^-$ reproduced by Chen {\it et al.} in Ref.~\cite{Chen:2011qx} (left panel) and in the erratum of Ref.~\cite{Ali:2009es}.
\\
{\it Source:} Taken from~\cite{Chen:2011qx, Ali:2009es}.
\label{Fig:Upsilon10860-tetraquark-new}
}
\end{center}
\end{figure}

In this scenario, the estimated decay widths for both $\Upsilon(10860)\to \Upsilon(2S)\pi^+\pi^-$ and $\Upsilon(10860)\to \Upsilon(1S)\pi^+\pi^-$ are consistent with the experimental measurements from the Belle Collaboration. Moreover, for the decay $\Upsilon(10860)\to \Upsilon(1S)\pi^+\pi^-$, the dipion invariant mass and $\cos\theta$ distributions can be well reproduced simultaneously, as shown in the left panel of Fig.~\ref{Fig:chen-upsilon5Sdipion}. However, for the $\Upsilon(10860)\to \Upsilon(2S)\pi^+\pi^-$ decay, the experimental data for the dipion invariant mass and $\cos\theta$ distributions cannot be described using the same set of parameters. In particular, with the parameters determined from the dipion invariant mass distributions, the resulting $d\Gamma/d\cos\theta$ for $\Upsilon(10860)\to \Upsilon(2S)\pi^+\pi^-$ exhibits a trend opposite to that of the Belle data, as illustrated in the fourth panel of Fig.~\ref{Fig:chen-upsilon5Sdipion}.

In Ref.~\cite{Chen:2011qx}, Chen and Liu also examined the dipion invariant mass spectra and $\cos\theta$ distributions of $\Upsilon(10860) \to \Upsilon(2S) \pi^+ \pi^-$ using the same formalism as in Ref.~\cite{Ali:2009es}. They found that the fitted results reported in Ref.~\cite{Ali:2009es} could not be reproduced (see left panel of Fig.~\ref{Fig:Upsilon10860-tetraquark-new}). They concluded that the dipion invariant mass spectrum and $\cos\theta$ distributions of $\Upsilon(10860) \to \Upsilon(2S) \pi^ +\pi^-$ cannot be simultaneously reproduced within either the conventional bottomonium or tetraquark frameworks, presenting a new and intriguing puzzle in the hidden-bottom dipion decays of $\Upsilon(10860)$. Subsequently, Ali {\it et al.} acknowledged this issue in the erratum associated with Ref.~\cite{Ali:2009es}, noting that one cannot simultaneously reproduce the dipion invariant mass spectrum and $\cos\theta$ distribution of $\Upsilon(5S) \to \Upsilon(2S) \pi^+ \pi^-$ using only non-resonance and dipion resonance contributions within the tetraquark framework (see right panel of Fig.~\ref{Fig:Upsilon10860-tetraquark-new}).

 \renewcommand\arraystretch{1.4}
\begin{table}[htb]
\centering
\caption{The branching fractions for $Z_b(10610)^+ $ and $Z_b(10650)^{+}$ measured by the Belle Collaboration \cite{Belle:2015upu}. The first and second uncertainties are statistical and systematic, respectively.\\
{\it Source:} Taken from~\cite{Belle:2015upu}.
\label{Tab:Zb-ratios}}
\begin{tabular}{lcc }
 \toprule[1pt]
\hspace{0.5cm}\multirow{2}{*}{Channel} &  \multicolumn{2}{c}{Branching Fraction\, $(\%)$ } \\
\hspace{2cm} & \hspace{1cm}$Z_b(10610)^+$\hspace{1cm} & \hspace{1cm}$Z_b(10650)^+$\hspace{1cm} \\
 \midrule[1pt]
 \hspace{0.5cm}$\Upsilon(1S) \pi^+$  &$0.54^{+0.16+0.11}_{-0.13-0.08}$ & $0.17^{+0.07+0.03}_{-0.06-0.02}$ \\
 \hspace{0.5cm}$\Upsilon(2S) \pi^+$  & $3.62^{+0.76+0.79}_{-0.59-0.53}$ & $1.39^{+0.48+0.34}_{-0.38-0.23}$ \\
\hspace{0.5cm} $\Upsilon(3S) \pi^+$  & $2.15^{+0.55+0.60}_{-0.42-0.43}$ & $1.63^{+0.53+0.39}_{-0.42-0.28}$ \\
\hspace{0.5cm} $h_b(1P) \pi^+$ & $3.45^{+0.87+0.86}_{-0.71-0.63}$ & $8.41^{+2.43+1.49}_{-2.12-1.06}$ \\
\hspace{0.5cm} $h_b(2P) \pi^+$ &  $4.67^{+1.24+1.18}_{-1.00-0.89}$ & $14.7^{+3.2+2.8}_{-2.8-2.3}$ \\
\hspace{0.5cm} $B^{+}\bar{B}^{\ast 0} + B^{\ast +}\bar{B}^{0}$ &$85.6^{+1.5+1.5}_{-2.0-2.1}$  &         $-$  \\
 \hspace{0.5cm}$B^{\ast +}\bar{B}^{\ast 0}$ &$-$          &$73.7^{+3.4+2.7}_{-4.4-3.5}$\\
 \bottomrule[1pt]
\end{tabular}
\end{table}

\subsubsection{Charged bottomoniumlike structures $Z_b(10610)$ and $Z_b(10650)$}

In addition to the anomalously large branching fractions of the dipion decays of $\Upsilon(10860)$ and the puzzles in the dipion invariant mass spectrum and $\cos\theta$ distribution of $\Upsilon(10860) \to \Upsilon(2S) \pi^+ \pi^-$ discussed in the previous subsection, another interesting phenomenon associated with $\Upsilon(10860)$ is the observation of charged bottomonium-like states, namely $Z_b(10610)$ and $Z_b(10650)$. In 2011, the Belle Collaboration reported these two new bottomonium-like states~\cite{Belle:2011aa}, which were identified through the invariant mass distributions of $\pi^\pm \Upsilon(nS)$ (with $n = 1, 2, 3$) and $\pi^\pm h_b(mP)$ (with $m = 1, 2$) in the decay processes $\Upsilon(5S) \to \Upsilon(nS) \pi^+ \pi^-$and $\Upsilon(5S) \to h_b(mP) \pi^+ \pi^-$. The weighted averages over all five channels yield the masses and widths of these two bottomonium-like states to be~\cite{Belle:2011aa}, 
\begin{eqnarray}
	m_{Z_b} = (10607.2 \pm 2.0) \ \mathrm{MeV}, \ \ \Gamma_{Z_b}= (18.4 \pm 2.5) \ \mathrm{MeV},\nonumber\\
	m_{Z_b^\prime} = (10652.2 \pm 1.5) \ \mathrm{MeV}, \ \ \Gamma_{Z_b^\prime}= (11.5 \pm 2.2) \ \mathrm{MeV}, 
\end{eqnarray}
respectively.

The analysis of the charged pion angular distribution indicates that the quantum numbers of the two states are consistent with $I^G(J^P) = 1^+(1^+)$~\cite{Adachi:2011mks}. Furthermore, an amplitude analysis of the process $e^+e^- \to \Upsilon(nS) \pi^+ \pi^-$ confirms $I^G(J^P) = 1^+(1^+)$ for both $Z_b(10610)$ and $Z_b(10650)$~\cite{Belle:2014vzn}. The observation of their neutral partners, $Z_b(10610)^0$ and $Z_b(10650)^0$, was achieved through a Dalitz analysis of the decay $\Upsilon(5S) \to \pi^0\pi^0\Upsilon(nS)$, as reported by the Belle Collaboration~\cite{Belle:2012glq, Belle:2013urd}, providing evidence that these two bottomonium-like states exhibit isovector characteristics. In addition to hidden-bottom decay processes, the Belle Collaboration has also observed $Z_b(10610)$ in the $B^*\bar{B}$ invariant mass spectra and $Z_b(10650)$ in the $B^*\bar{B}^*$ invariant mass distributions of the open-bottom pion decays of $\Upsilon(5S)$~\cite{Belle:2012koo, Belle:2015upu}. The branching fractions of $Z_b(10610)$ and $Z_b(10650)$ decaying into bottomonia $\Upsilon(nS) \pi$, $h_b(mP) \pi$, and $B^\ast \bar{B}^{(\ast)}$ are compiled in Table~\ref{Tab:Zb-ratios}. From these data, it can be seen that $Z_b(10610)$ and $Z_b(10650)$ decay predominantly into open-bottom channels, while the hidden-bottom channels are relatively suppressed.

\changelabel{
It is worth mentioning that in Ref.~\cite{Anisovich:1995zu}, the authors noticed that dipion invariant mass distributions of $\Upsilon(3S) \to \Upsilon(1S) \pi\pi$ has a peculiar double bump structure. To reproduce the dipion invariant mass distributions and $\Upsilon(1S)\pi$ invariant mass spectrum, a hidden bottom tetraquark states in $\Upsilon(1S) \pi$ invariant mass spectrum with $I(J^P)=1(1^+)$ and a mass in the range $10.4-10.8$ GeV should be included~\cite{Anisovich:1995zu,Guo:2004dt}. Incorporating such a tetraquark state, the dipion transitions between $\Upsilon(4S)$ and $\Upsilon(1S,2S)$ are also been investigated \cite{Guo:2006ai}. }

\subsubsection{Role of $Z_b(10610)$ and $Z_b(10650)$ in solving the puzzles of the process $\Upsilon(5S) \to \Upsilon(2S) \pi^+ \pi^-$}  
\begin{figure}[htb]
\centering
\scalebox{0.55}{\includegraphics{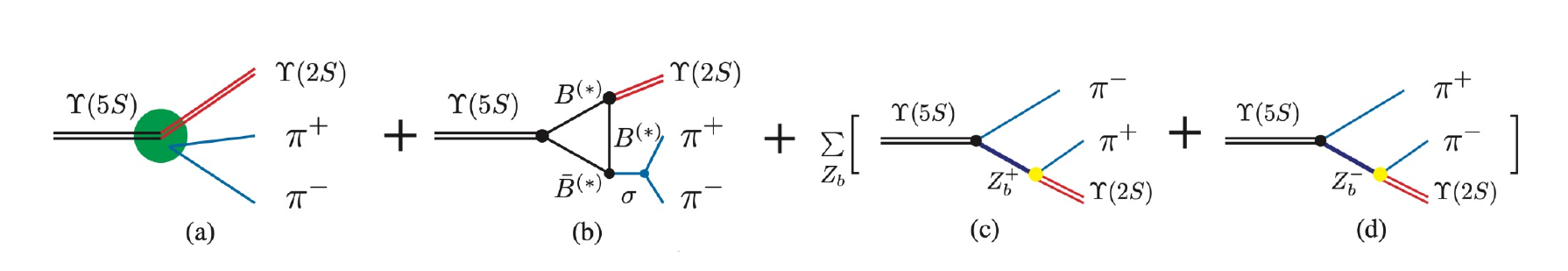}}
\caption{Diagrams contributing to $\Upsilon(5S)\to \Upsilon(2S) \pi^+ \pi^-$ .
Here, diagram (a) represents the $\Upsilon(5S)$ direct
decay into $\Upsilon(2S)\pi^+\pi^-$, while diagram (b)
denotes the intermediate hadronic loop contribution to
$\Upsilon(5S)\to \Upsilon(2S)\pi^+\pi^-$. (c) and (d) describe the
intermediate $Z_b^{\pm}$ contribution to $\Upsilon(5S)\to
\Upsilon(2S)\pi^+\pi^-$, where
$Z_b^{\pm}=\{Z_b(10610)^\pm,Z_b(10650)^\pm\}$.\\
{\it Source:} Taken from~\cite{Chen:2011zv}
 \label{Fig:chen-upsilon2s-withzb-mech}}
\end{figure}

As indicated in Ref.~\cite{Chen:2011qx}, the contributions from non-resonance and dipion resonance processes alone could not simultaneously reproduce the dipion invariant mass spectra and $\cos \theta$ distributions of the decay $\Upsilon(5S) \to \Upsilon(2S) \pi^+ \pi^-$. Following the observations of $Z_b(10610)$ and $Z_b(10650)$, Chen {\it et al.} incorporated the contributions of these two bottomonium-like states into the analysis of $\Upsilon(5S) \to \Upsilon(2S) \pi^+ \pi^-$. The contributing diagrams for this process are shown in Fig.~\ref{Fig:chen-upsilon2s-withzb-mech}. In this framework, the contributions from $Z_b(10610)$ and $Z_b(10650)$ are parameterized as\footnote{\changelabel{An alternative way to give the contribution from $Z_b$ states, especially the $\Upsilon \pi Z_{b}$ vertex, can be seen in Ref.~\cite{Wu:2018xaa}, where it arises from triangle diagrams.}}
\begin{eqnarray}
&&\mathcal{M}[\Upsilon(5S)\to Z_b^+\pi^-\to
\Upsilon(2S)\pi^+\pi^-]_{Z_b^+}= F_{Z_b^+} \epsilon_{\Upsilon(5S)}^\mu
\epsilon_{\Upsilon(2S)}^{*\nu} \frac{-g_{\mu \nu} + (p_1^\mu
+p_2^\mu) (p_1^\nu +p_2^\nu)/m_{Z_b}^2}{ (p_1+p_2)^2-m_{Z_b}^2 +
im_{Z_b}
\Gamma_{Z_b}}\notag\\
&&\mathcal{M}[\Upsilon(5S)\to Z_b^-\pi^+\to
\Upsilon(2S)\pi^-\pi^+]_{Z_b^-}= F_{Z_b^-}\epsilon_{\Upsilon(5S)}^\mu
\epsilon_{\Upsilon(2S)}^{*\nu} \frac{-g_{\mu \nu} + (p_1^\mu
+p_3^\mu) (p_1^\nu +p_3^\nu)/m_{Z_b}^2}{ (p_1+p_3)^2-m_{Z_b}^2 +
im_{Z_b} \Gamma_{Z_b}},
\end{eqnarray}
respectively, with $F_{Z_b^+}=g_{_{\Upsilon(5S)Z_b^+
\pi}}g_{_{Z_b^+ \Upsilon(2S) \pi^+}}$ and
$F_{Z_b^-}=g_{_{\Upsilon(5S)Z_b^- \pi}}g_{_{Z_b^- \Upsilon(2S)
\pi^-}}$. Since Fig. \ref{Fig:chen-upsilon2s-withzb-mech} (c) and (d) are related to each
other by charge-conjugation, thus $F_{Z_b^-}=F_{Z_b^+}=F_{Z_b}$.

The total decay amplitude of the
$\Upsilon(5S)\to \Upsilon(2S)\pi^+\pi^-$ decay is
\begin{eqnarray}
\mathcal{M}_{\mathrm{total}}&=&\mathcal{M}[\Upsilon(5S)\to
\Upsilon(2S)\pi^+\pi^-]_{\mathrm{Direct}}\nonumber\\&&+e^{i\phi_{\sigma}}
\mathcal{M}[\Upsilon(5S)\to \Upsilon(2S)\sigma(600)\to
\Upsilon(2S)\pi^+\pi^-]+e^{i\phi_{f_0}}
\mathcal{M}[\Upsilon(5S)\to \Upsilon(2S)f_0(980)\to
\Upsilon(2S)\pi^+\pi^-]\nonumber\\&&+\sum_{_{Z_b}}e^{i\varphi_{_{Z_b}}}\bigg\{
\mathcal{M}[\Upsilon(5S)\to Z_b^+\pi^-\to
\Upsilon(2S)\pi^+\pi^-]_{Z_b^+}+\mathcal{M}[\Upsilon(5S)\to
Z_b^-\pi^+\to \Upsilon(2S)\pi^+\pi^-]_{Z_b^-}\bigg\},
\end{eqnarray}
with these introduced phase angles $\phi_\sigma$, $\phi_{f_0}$,
$\varphi_{_{Z_b(10610)}}$ and $\varphi_{_{Z_b(10650)}}$.

\begin{figure}[htb]
\begin{center}
\scalebox{0.7}{
\includegraphics{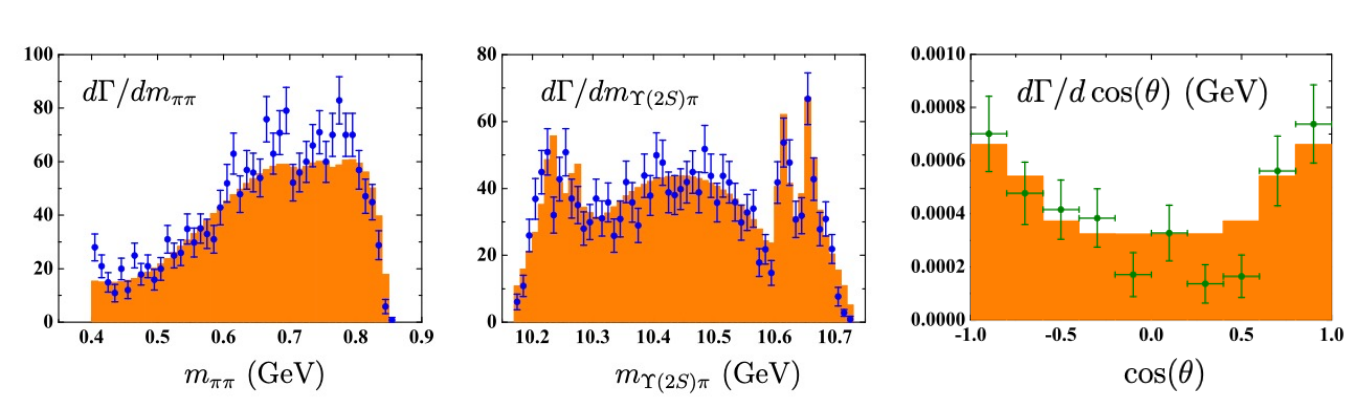}}
\caption{The $m_{\pi \pi}$ (left frame), $m_{\Upsilon(2S) \pi}$ (middle frame)
 invariant mass spectra, and $\cos\theta$ distribution (right frame)  
 for $\Upsilon(5S) \to \Upsilon(2S) \pi^+
\pi^-$ process. The histograms are
the fitting results with the contributions from $Z_b(10610)$ and $Z_b(10650)$, and the dots with errors correspond to the Belle
data \cite{Belle:2007xek, Belle:2011aa}.
\\
{\it Source:} Taken from~\cite{Chen:2011zv}.
\label{Fig:chen-reproduce-withzb}}
\end{center}
\end{figure}

In Ref.~\cite{Chen:2011zv}, following the inclusion of contributions from $Z_b(10610)$ and $Z_b(10650)$, it was found that the dipion invariant mass spectrum, the $\cos\theta$ distributions, and the $\Upsilon(2S)\pi$ spectrum could all be well reproduced simultaneously, as shown in Fig.~\ref{Fig:chen-reproduce-withzb}. The $Z_b(10610)$ and $Z_b(10650)$ states were found to be related to the anomalous features observed in $\Upsilon(2S)\pi^+\pi^-$ production near the $\Upsilon(5S)$ resonance, previously reported by Belle~\cite{Belle:2007xek}. By comparing the fitting results obtained with and without the contributions from these newly observed bottomonium-like states, the analysis in Ref.~\cite{Chen:2011zv} demonstrated that the intermediate $Z_b(10610)$ and $Z_b(10650)$ play a crucial role in shaping the behavior of $d\Gamma(\Upsilon(5S) \to \Upsilon(2S) \pi^+ \pi^-)/d\cos\theta$. The inclusion of $Z_b(10610)$ and $Z_b(10650)$ contributions to $\Upsilon(5S) \to \Upsilon(2S)\pi^+\pi^-$ thus provides a unique mechanism for understanding the puzzling $\cos\theta$ distribution observed in $\Upsilon(2S)\pi^+\pi^-$ production near the $\Upsilon(5S)$ resonance~\cite{Belle:2007xek}.

\begin{figure}[htb]
\centering
\scalebox{0.6}{\includegraphics{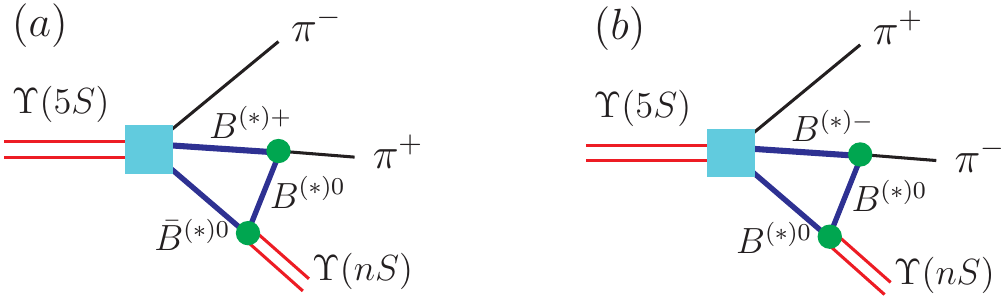}}
\caption{The schematic diagrams for $\Upsilon(5S)\to
\Upsilon(nS)\pi^+\pi^-$ by the ISPE mechanism. Here, diagrams (a)
and (b) are related to each other by particle antiparticle conjugation, i.e.,
$B^{(*)}\rightleftharpoons\bar{B}^{(*)}$ and $\pi^+\rightleftharpoons \pi^-$. After performing the transformations $B^{(*)+}\rightleftharpoons B^{(*)0}$, $B^{(*)-}\rightleftharpoons \bar{B}^{(*)0}$ and $\pi^+\rightleftharpoons \pi^-$, we obtain the remaining
diagrams. By replacing $\Upsilon(nS)$ with $h_b(mP)$, one obtains
the diagrams for $\Upsilon(5S)\to h_b(mP)\pi^+\pi^-$. \\
{\it Source:} Taken from~\cite{Chen:2011pv}.
\label{Fig:Chen-ISPE-Zb}}
\end{figure}

\begin{figure}[htb]
\centering
\scalebox{0.8}{\includegraphics{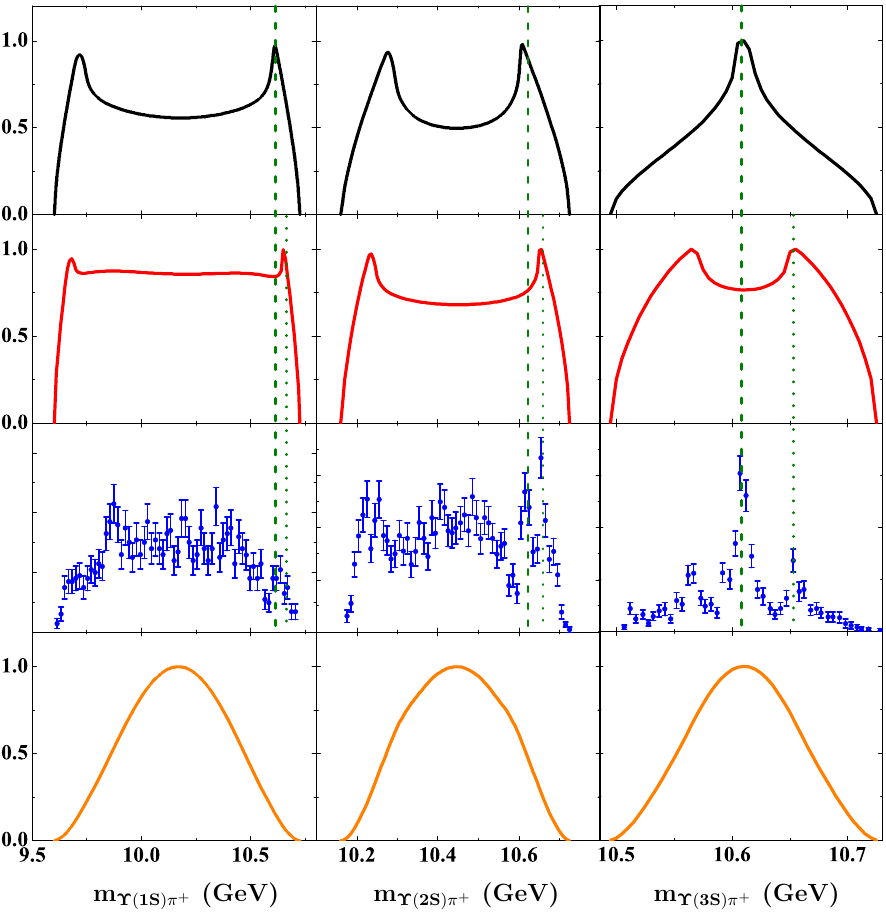}}
\caption{The obtained theoretical line shapes of $d\Gamma(\Upsilon(5S)\to\Upsilon(nS)\pi^+\pi^-)/d m_{\Upsilon(nS)\pi^+}$, and the comparison of our result with the Belle data (the third column) ~\cite{Belle:2011aa}. The first, the second and the fourth columns correspond to the numerical result considering $B\bar{B}^*+h.c.$, $B^*\bar{B}^*$ and $B\bar{B}$ intermediate state contributions respectively, while the first, the second and the third rows are the results corresponding to the distributions of the $\Upsilon(1S)\pi^+$, $\Upsilon(2S)\pi^+$ and $\Upsilon(3S)\pi^+$ invariant mass spectra. We use the vertical dashed and dotted lines to mark the masses of $Z_b(10610)$ and $Z_b(10650)$, respectively. Here, the maximum of the theoretical line shape is normalized to 1. \\
{\it Source:} Taken from~\cite{Chen:2011pv}
\label{Fig:chen-ISPE-Zb-Upsilon}}
\end{figure}

\begin{figure}[htb]
\centering
\scalebox{0.85}{\includegraphics{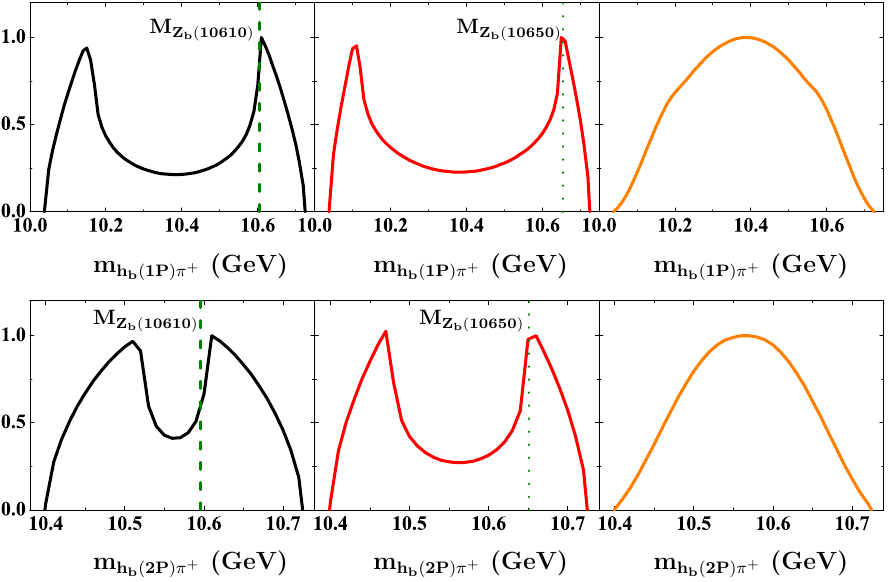}} 
\caption{The theoretical curves of $d\Gamma(\Upsilon(5S)\to h_b(1P)\pi^+\pi^-)/d m_{h_b(1P)\pi^+}$ (the first column) and $d\Gamma(\Upsilon(5S)\to h_b(2P)\pi^+\pi^-)/d m_{h_b(2P)\pi^+}$ (the second column). For easily comparing our result with the experimental data, one adopts the vertical dashed and dotted lines to denote the masses of $Z_b(10610)$ and $Z_b(10650)$ respectively. The first, the second and the third rows correspond to the numerical result respectively considering $B\bar{B}^*+h.c.$, $B^*\bar{B}^*$ and $B\bar{B}$ intermediate state contributions in Fig. \ref{Fig:Chen-ISPE-Zb}. Here, the maximum of the theoretical line shape is normalized to 1.\\
{\it Source:} Taken from~\cite{Chen:2011pv}
 \label{Fig:chen-ISPE-Zb-hb}}
\end{figure}

\subsection{Initial single pion emission mechanism}

\subsubsection{Charged $Z_b$ states and the proposal of the initial single pion emission mechanism}

The proximity of their masses to the thresholds and their nonzero electric charges have stimulated considerable theoretical interest in understanding the nature of the $Z_b(10610)$ and $Z_b(10650)$ states. Motivated by their distinctive properties, various interpretations have been proposed, including descriptions as $B^{\ast}\bar{B}^{(\ast)}$ molecular states~\cite{Prelovsek:2019ywc, Zhang:2011jja, Yang:2011rp, Nieves:2011zz, Sun:2011uh, Cleven:2011gp, Mehen:2011yh, Ohkoda:2011vj, Li:2012wf, Liu:2017mrh, Zhao:2014gqa, Chen:2015ata, Wang:2018jlv, Bondar:2011ev, Dong:2012hc, Li:2012uc, Wu:2020edh, Xiao:2017uve, Wu:2022hck, Wu:2018xaa}, $[bq][\bar{b}\bar{q}]$ tetraquark states~\cite{Wang:2013zra, Wang:2019mxn, Ke:2012gm, Wang:2013daa, Gupta:2012gba, Ali:2011ug}, and kinematic effects~\cite{Bugg:2011jr, Swanson:2014tra}.

In addition to these QCD-exotic and kinematic interpretations, the authors of Ref.~\cite{Chen:2011pv} extended the description of the $Z_b(10610)/Z_b(10650)$ contributions in $\Upsilon(5S)$ dipion decays (as shown in Figs.~\ref{Fig:chen-upsilon2s-withzb-mech}(c) and (d)) to the diagrams in Fig.~\ref{Fig:Chen-ISPE-Zb}. In this extended picture, the initial $\Upsilon(5S)$ transitions into a pair of $B^{(*)}$ and $\bar{B}^{(*)}$ mesons accompanied by the emission of a single pion. Owing to the continuous energy distribution of the emitted pion, the $B^{(*)}$ and $\bar{B}^{(*)}$ mesons, when produced with low relative momentum, can interact with each other and subsequently transition into $\Upsilon(nS)\pi$ via the exchange of an appropriate $B^{(*)}$ meson. This specific mechanism in $\Upsilon(5S)$ dipion decays was termed the ISPE mechanism, \changelabel{a kinematical effect that can generate near threshold enhancements on the invariant mass spectra}, by the authors of Ref.~\cite{Chen:2011pv}. To some extent, the role of the pion in the ISPE mechanism within $\Upsilon$ decays is analogous to that of the photon in the well-known Initial State Radiation (ISR) mechanism in $e^+e^-$ collisions, \changelabel{but with a distinction that ISR is a physical electromagnetic process with a photon emitted before the hard scattering, while ISPE is part of a strong decay amplitude and cannot be separately from other parts in a clean way. Anyway, such a mechanism} has enabled a series of observations of charmonium-like $XYZ$ states in recent years.

Using the effective Lagrangian approach~\cite{Oh:2000qr, Casalbuoni:1996pg, Colangelo:2003sa}, the general expressions for the hadron-level amplitudes corresponding to Figs.~\ref{Fig:Chen-ISPE-Zb}(a) and (b) are given by
\begin{eqnarray}
&&\mathcal{M}\big\{\Upsilon(5S)\to \pi^-+[B^{(*)+}\bar{B}^{(*)0}\rightarrowtail\Upsilon(nS)\pi^+]_{B^{(*)0}}\big\} \notag\\
&&\qquad \qquad =\prod_i g_i\int\frac{d^4 q}{(2\pi)^4}\frac{\big[p_1,p_2,p_3,q\big]_{\mu\nu}\epsilon_{\Upsilon(5S)}^{\mu}
\epsilon_{\Upsilon(nS)}^{\nu}}{\big[(p_2+q)^2-m_{B^{(*)}}^2\big]\big[(p_1-q)^2-m_{B^{(*)}}^2\big]}\nonumber
\frac{1}{q^2-m_{B^{(*)}}^2}\mathcal{F}^2(q^2,m_{B^{(*)}}^2),\label{h1}\\
&&\mathcal{M}\big\{\Upsilon(5S)\to \pi^++[B^{(*)-}{B}^{(*)0}\rightarrowtail\Upsilon(nS)\pi^-]_{B^{(*)0}}\big\}\nonumber\\
&&\qquad \qquad =\prod_i g_i\int\frac{d^4
q}{(2\pi)^4}\frac{\big[p_1,p_2,p_3,q\big]_{\mu\nu}\epsilon_{\Upsilon(5S)}^{\mu}
\epsilon_{\Upsilon(nS)}^{\nu}}{\big[(p_3+q)^2-m_{B^{(*)}}^2\big]\big[(p_1-q)^2-m_{B^{(*)}}^2\big]}
\frac{1}{q^2-m_{B^{(*)}}^2}\mathcal{F}^2(q^2,m_{B^{(*)}}^2),\label{h2}
\end{eqnarray}
where $\big[p_1,p_2,p_3,q\big]_{\mu\nu}$ denotes the Lorentz structures constructed by four-momenta $p_1$, $p_2$, $p_3$ and $q$. In the above amplitude, a form factor is introduced to depict the internal structure of the involved bottom mesons and make the loop integral finite in the ultraviolet region, the concrete form of the form factor is, 
\begin{eqnarray}
	\mathcal{F}(q^2,m_{B^{(\ast)}}^2) = \frac{\Lambda^2-m_{B^{(\ast)}}^2}{\Lambda^2-q^2},
\end{eqnarray}
where $\Lambda=m_{B^{(\ast)}} +\alpha \Lambda_{\mathrm{QCD}}$ with $\Lambda_{\mathrm{QCD}}=220$ MeV. $\alpha$ is a model parameter, which should be of order unity~\cite{Cheng:2004ru}.

The estimations in Ref.~\cite{Chen:2011pv} found that lineshapes of $d\Gamma(\Upsilon(5S)\to\Upsilon(nS)\pi^+\pi^-)/d m_{\Upsilon(nS)\pi^+}$ generated by the ISPE mechanism are almost independent on the model parameter $\alpha$ introduced by the form factor in the amplitudes. From the estimated lineshapes with different intermediate states, one can obtain,
\begin{itemize}
	\item For the $\Upsilon(5S) \to \Upsilon(nS) \pi^+\pi^-\,  (n=1,2,3)$ processes,  just shown in the first and the second columns of Fig. \ref{Fig:chen-ISPE-Zb-Upsilon}, combined with the corresponding reflections, the sharp peaks around $B\bar{B}^*$ and $B^*\bar{B}^*$ thresholds appear in the $m_{\Upsilon(1S)\pi^+}$ and $m_{\Upsilon(2S)\pi^+}$ distributions of $d\Gamma(\Upsilon(5S)\to\Upsilon(1S)\pi^+\pi^-)/d m_{\Upsilon(1S)\pi^+}$ and $d\Gamma(\Upsilon(5S)\to\Upsilon(2S)\pi^+\pi^-)/d m_{\Upsilon(2S)\pi^+}$. The comparison of these results with the Belle data \cite{Belle:2011aa} indicates that we indeed can mimic the peak structures similar to the $Z_{b}(10610)$ and $Z_{b}(10650)$ reported by Belle if introducing the ISPE mechanism. The theoretical result of the $\Upsilon(5S)\to\Upsilon(3S)\pi^+\pi^-$ decay further indicates that there also exists a peak around 10610 MeV, which combines with its reflection in the $m_{\Upsilon(3S)\pi^+}$ distribution to form a broad structure. In addition, a structure at $\sim 10650$ MeV and its reflection are reproduced. These results qualitatively and naturally explain why there are three structures appearing in the $\Upsilon(3S)\pi^+$ invariant mass spectrum just announced by Belle~\cite{Belle:2011aa}. 
	\item For the $\Upsilon(5S) \to h_b(mP) \pi^+\pi^-\,  (m=1,2)$ processes, the ISPE mechanism has also been extend to  study the $\Upsilon(5S)\to h_b(mP)\pi^+\pi^-$ ($m=1,2$) decays. Similar to the situation of $\Upsilon(5S)\to \Upsilon(1S,2S)\pi^+\pi^-$, one can find two structures around 10610 MeV and 10650 MeV and their reflections in the theoretical line shape of $d\Upsilon(5S)\to h_b(mP)\pi^+\pi^-/dm_{h_b(mP)\pi^+}$ as shown in Fig.~\ref{Fig:chen-ISPE-Zb-hb}. These structures are well consistent with the centers of peaks reported by the Belle Collaboration~\cite{Belle:2011aa}. 
	\item In addition, the Belle data also give a very intriguing phenomenon, i.e., there does not exist the structure near the $B\bar{B}$ threshold. The ISPE mechanism can provide a direct explanation to it. If only considering the $B\bar{B}$ contribution in Fig. \ref{Fig:Chen-ISPE-Zb}, the calculation shows that there are no peak structure close to the $B\bar{B}$ threshold in the $\Upsilon(nS)\pi^+$ and $h_b(mP)\pi^+$ invariant mass spectra. Alternately, the smooth line shapes similar to phase space of corresponding decay processes appear in the invariant mass spectra of $\Upsilon(nS)\pi^+$ and $h_b(mP)\pi^+$.
\end{itemize}

Besides the $\Upsilon(5S)$ hidden-bottom dipion decays, the ISPE mechanism has also been applied to investigate the hidden-bottom dipion decays of $\Upsilon(11020)$~\cite{Chen:2011pu}, where clear peak structures near $B^\ast \bar{B}$ and $B^\ast \bar{B}^\ast$ thresholds are predicted in the $\Upsilon(nS) \pi^+\, (n=1,2,3)$ and $h_b(mP) \pi^+\,  (m=1,2)$ invariant mass distributions. In addition to the hidden-bottom decay process, in the open bottom pion decays of $\Upsilon(5S)$, the lineshapes of $B^\ast \bar{B}$ and $B^\ast \bar{B}^\ast$ invariant mass distributions estimated by ISPE mechanism also emerge peak structures in the vicinity of  $B^\ast \bar{B}$ and $B^\ast \bar{B}^\ast$ thresholds, which well correspond to $Z_b(10610)$ and $Z_b(10610)$ observed by the Belle Collaboration~\cite{Chen:2012yr}.

\begin{figure}[htb]
\centering 
\scalebox{0.75}{\includegraphics{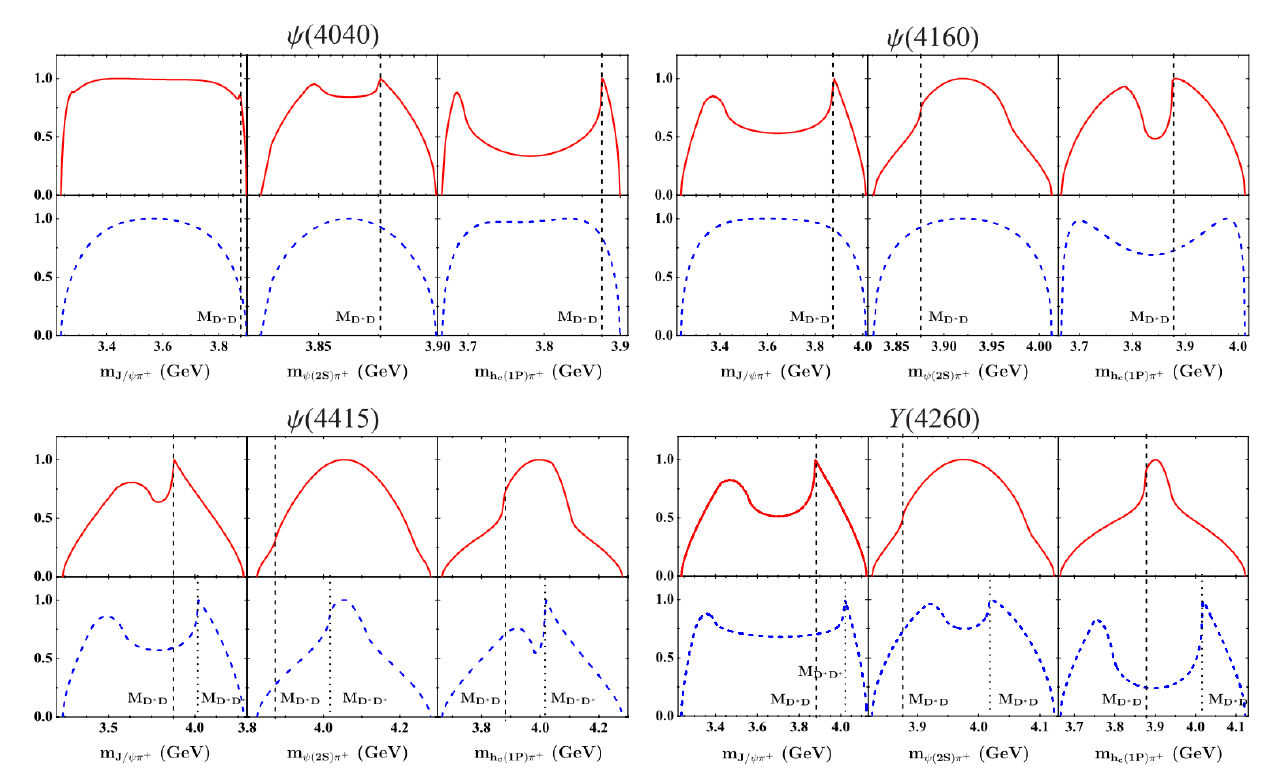}}
\caption{(Color online.) The invariant mass spectra of $J/\psi\pi^+$, $\psi(3686)\pi^+$ and $h_c(1P)\pi^+$ for the $\psi(4040)$, $\psi(4160)$, $\psi(4415)$ and $Y(4260)$ decays into $J/\psi\pi^+\pi^-$, $\psi(3686)\pi^+\pi^-$ and $h_c(1P)\pi^+\pi^-$. Here, the solid, dashed correspond to the results considering intermediate  $D\bar{D}^*+h.c.$ and $D^*\bar{D}^*$ respectively. The vertical dashed lines and the dotted lines denote the threshhold of $D^\ast \bar{D}$ and $D^\ast \bar{D}^\ast$ respectively. Here, the maximum of the line shape is normalized to 1. \\ 
{\it Source:} Taken from~\cite{Chen:2011xk} 
\label{Fig:chen-ISPE-Zc}}
\end{figure}

\begin{figure}[htb]
\centering 
\scalebox{1}{\includegraphics{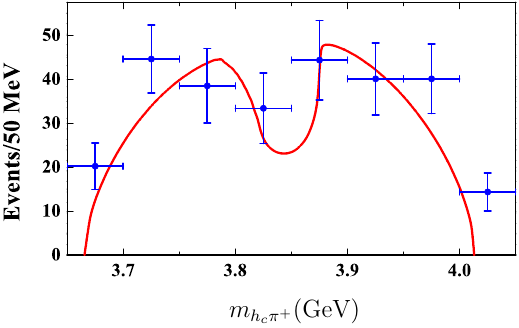}}
\caption{(Color online.) A comparison of the $h_c\pi^\pm$ mass distribution of $\psi(4160)\to h_c(1P) \pi^+\pi^-$ (solid line) predicted in this work and measurement by CLEO-c (points with errors) \cite{CLEO:2011aa}. Here, CLEO-c measured the $h_c(1P)\pi^\pm$ mass distribution from $e^+e^-\to h_c(1P) \pi^+\pi^-$ at $E_{CM}=4170$ MeV \cite{CLEO:2011aa}. We normalize our numbers for a real comparison with the available CLEO-c data. \\ 
{\it Source:} Taken from~\cite{Chen:2011xk} 
\label{Fig:ISPE-hcpi-CLEOc}}
\end{figure}

\subsubsection{Prediction of charmonium-like charged $Z_c$ structure}

Considering the ISPE mechanism is a universal mechanism existing in heavy quarkonium decay, one can naturally extend such physical picture to study hidden-charm decays of higher vector charmonia due to the similarity between charmonium and bottomonium families, and predict some novel phenomena similar to the $Z_b$ structures. In Ref.~\cite{Chen:2011xk}, the dipion decays of the charmonia $\psi(4040)$, $\psi(4160)$ and $\psi(4415)$ and the charmonium-like state $Y(4260)$ have been investigated by ISPE mechanism. In Fig. \ref{Fig:chen-ISPE-Zc}, the results of $d\Gamma/d m_{J/\psi\pi^+}$, $d\Gamma/d m_{\psi(3686)\pi^+}$ and $d\Gamma/d m_{h_c(1P)\pi^+}$ of $\psi(4040)$, $\psi(4160)$, $\psi(4415)$, $Y(4260)$ decays into $J/\psi\pi^+\pi^-$, $\psi(3686)\pi^+\pi^-$, $h_c(1P)\pi^+\pi^-$ are presented, from the lineshapes, one can find the following intriguing properties. 
\begin{itemize}
\item{
For the dipion decays of $\psi(4040)$, there exit sharp peak structures close to the $D^\ast \bar{D}$ threshold and the corresponding reflections in the distributions of $d\Gamma/d m_{J/\psi\pi^+}$, $d\Gamma/d m_{\psi(3686)\pi^+}$ and $d\Gamma/d m_{h_c(1P)\pi^+}$ of $\psi(4040)\to J/\psi\pi^+\pi^-$, $\psi(4040)\to \psi(3686)\pi^+\pi^-$ and $\psi(4040)\to h_c(1P)\pi^+\pi^-$ decays. While this structure appearing in $d\Gamma(\psi(4040)\to J/\psi\pi^+\pi^-)/d m_{J/\psi\pi^+}$ is not obvious comparing with the structure in the $\psi(3686)\pi^+$ and $h_c(1P)\pi^+$ invariant mass spectra in the processes  $\psi(4040)\to \psi(3686)\pi^+\pi^-$ and $\psi(4040)\to h_c(1P)\pi^+\pi^-$, respectively. }

\item{For the dipion decay of $\psi(4160)$, two sharp peaks appear in the distributions of  $d\Gamma(\psi(4160)\to J/\psi\pi^+\pi^-)/d m_{J/\psi\pi^+}$ and $d\Gamma(\psi(4160)\to h_c(1P)\pi^+\pi^-)/d m_{h_c(1P)\pi^+}$, which are close the $D^\ast \bar{D}$ threshold. The structure in the $J/\psi\pi^+$ invariant mass spectrum is more narrow than that in the $h_c(1P)\pi^+$ invariant mass spectrum. 
}

\item{
In the hidden-charm dipion decays of $\psi(4415)$, we find two sharp peak structures around the $D^\ast \bar{D}$ and $D^\ast \bar{D}^\ast$ thresholds appearing in the $J/\psi\pi^+$ invariant mass spectra. In addition, a sharp peak close the $D^\ast \bar{D}^\ast$ threshold is observed in the $h_c(1P)\pi^+$ invariant mass spectrum distribution. In the $d\Gamma(\psi(4415)\to \psi(3686)\pi^+\pi^-)/d m_{\psi(3686)\pi^+}$ distribution, a peak near $D^\ast\bar{D}^\ast$ with its reflection form a broad structure. Under the ISPE mechanism, the intermediate $D^*\bar{D}$ can result in a very broad structure in the $h_c(1P)\pi^+$ invariant mass spectrum distribution.}

\item{
In the hidden-charm dipion decays of $Y(4260)$, there exist the sharp peaks close to $D^\ast \bar{D}$ threshold in the $d\Gamma(\psi(4260)\to J/\psi\pi^+\pi^-)/d m_{J/\psi\pi^+}$ and $d\Gamma(\psi(4260)\to h_c(1P)\pi^+\pi^-)/d m_{h_c(1P)\pi^+}$ distributions. In the distributions of $d\Gamma(\psi(4260)\to \psi(3686)\pi^+\pi^-)/d m_{\psi(3686)\pi^+}$ and $d\Gamma(\psi(4260)\to h_c(1P)\pi^+\pi^-)/d m_{h_c(1P)\pi^+}$, the ISPE mechanism also lead to the structures around $D^\ast \bar{D}^\ast$ threshold. As for $Y(4626) \to h_c(1P) \pi^+ \pi^- $ process, the peak close \changelabel{to} the $D^\ast \bar{D}$ threshold and its reflection overlap with itself to form a broad structure in the $h_c(1P)\pi^+$ invariant mass spectrum. }
\end{itemize}

In 2011, the CLEO-c Collaboration announced the preliminary measurements of the $h_c(1P)\pi^\pm$ mass distribution from $e^+e^-\to h_c(1P) \pi^+\pi^-$ at $E_{CM}=4170$ MeV (the points with errors in Fig. 4 (b) of Ref.~\cite{CLEO:2011aa}). In Ref. ~\cite{Chen:2011xk}, Chen and Liu compared the predicted $h_c\pi^\pm$ mass distribution of $\psi(4160)\to h_c(1P) \pi^+\pi^-$ (see Fig. \ref{Fig:chen-ISPE-Zc}) with the CLEO-c data. They noticed that the predicted theoretical line shape of the $h_c(1P)\pi^\pm$ mass distribution of $\psi(4160)\to h_c(1P) \pi^+\pi^-$ is consistent with the data measured by CLEO-c, which is listed in Fig. \ref{Fig:ISPE-hcpi-CLEOc}, where there indeed exist a broad structure around $D\bar{D}^*$ threshold and its reflection. To some extent, this fact provides a direct test to ISPE prediction in Ref.~\cite{Chen:2011xk}.

In Ref.~\cite{Belle:2007dxy}, the Belle Collaboration measured the cross sections for $e^+ e^- \to  \pi^+ \pi^- J/\psi$ via initial state radiation, the $\pi^+ \pi^-$ invariant mass distributions were reported. At the conference "Hadron Structure and Interactions in 2011", Yuan reported peak structures at $12~\mathrm{GeV}^2$ and $15~\mathrm{GeV}^2$ in the $\pi^\pm J/\psi$ invariant mass squared ($M^2_{\pi^\pm J/\psi}$) distribution from Belle data~\cite{Yuan:2021talk}; these could be evidence of the enhancement predicted by the ISPE mechanism~\cite{Chen:2011xk}.    

In addition to the dipion decays of $Y(4260)$, Chen {\it et al.} also investigated the hidden-charm dipion decays of the charmonium-like state $Y(4360)$ by the ISPE mechanism in Ref.~\cite{Chen:2013bha}. The estimations shows that there also exist charge charmonium-like structure near $D\bar{D}^\ast$ and $D^\ast \bar{D}^\ast$ threshold in the $J/\psi \pi^+$, $\psi(3686) \pi^+$ and $h_c(1P) \pi^+ $ invariant mass spectra of the corresponding hidden-charm dipion decays processes. Using the ISPE mechanism, the authors in Refs.~\cite{Chen:2011xk, Chen:2013bha} not only predicted existences of the charmonium-like structures near the $D^{(\ast)} \bar{D}^\ast$ thresholds, \changelabel{which are kinematically generated enhancements instead of real physical states}, but also provided specific schemes for observing these structures experimentally, such as the hidden-charm dipion decays processes, $Y(4260) \to J/\psi \pi^+ \pi^-$,  $Y(4260) \to \psi(3686) \pi^+ \pi^-$, and  $Y(4260) \to h_c \pi^+ \pi^-$. 

\changelabel{ In addition to ISPE mechanism, the possible existence of near $D^\ast \bar{D}^{(\ast)}$ structures had also been discussed from the perspective of mass spectra or decay. For examples, the authors in Ref.~\cite{Maiani:2004vq} taking the $X(3872)$ as input, predicted the existence of charmonium-like state with $J^{PC}=1^{+-}$ around 3882 MeV in the constituent quark model with diquark-antidiquark interactions, and in the same tetraquark hypothesis, the decays properties of these tetraquark states were investigated~\cite{Ali:2011ug}. In the one-boson-exchange model, the charm analoges of $Z_b(10610)/Z_b(10650)$ were predicted with the masses around the thresholds of $D^\ast \bar{D}/D^\ast \bar{D}^{\ast}$~\cite{Sun:2011uh}.}

\begin{figure}[hbtp]
\centering 
\scalebox{0.55}{\includegraphics{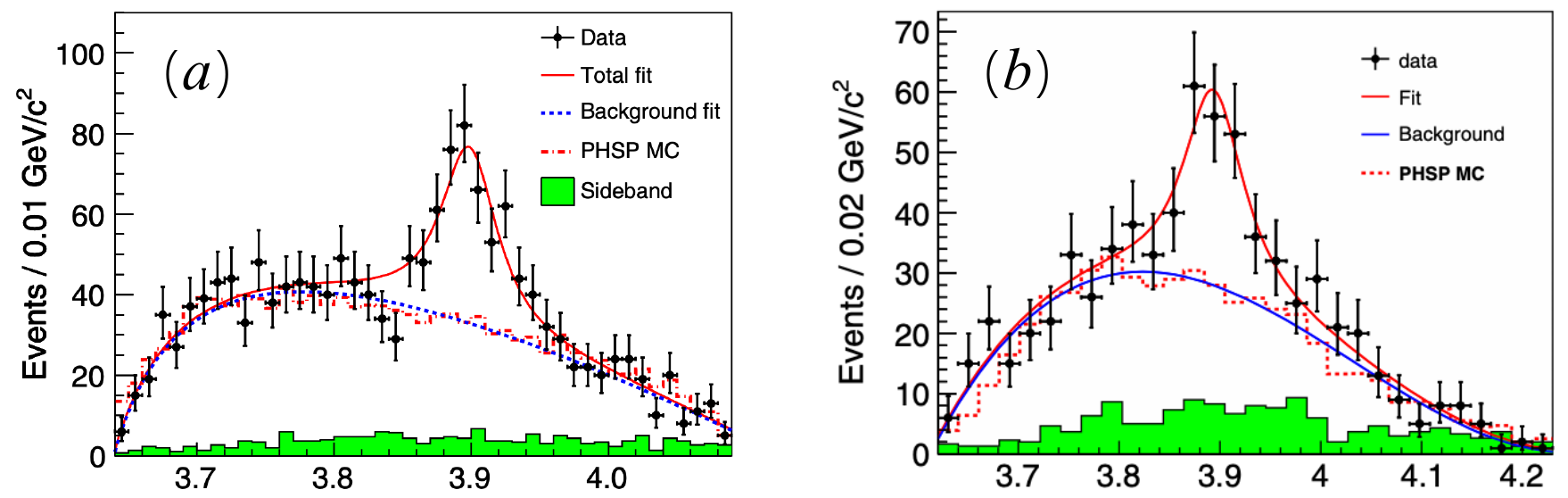}}\caption{Fit to the $M_{\mathrm{max}}(\pi^\pm J/\psi)$ distributions of $e^+ e^- \to \pi^+ \pi^- J/\psi$ at $\sqrt{s}=4.26$ GeV form the BESIII (diagram (a))~\cite{BESIII:2013ris} and Belle (diagrma (b))~\cite{Belle:2013yex} Collaborations. Dots  with error bars are experimental data from BESIII and Belle Collaborations~\cite{BESIII:2013ris,Belle:2013yex}, the red solid curves show the total fit, and the blue dotted curves show the results of a phase space Mento Carlo simulation and the green shaded histogram shows the normalized $J/\psi$ sideband events.  \\ 
{\it Source:} Taken from~\cite{BESIII:2013ris,Belle:2013yex}. 
\label{Fig:exp-Zc}
}
\end{figure}
\subsubsection{Experimental observations of the $Z_c$ structures}

In 2013, the BESIII Collaboration studied the process $e^+ e^- \to \pi^+ \pi^- J/\psi$ at a center-of-mass energy of 4.260 GeV using a data sample of 525 pb$^{-1}$ collected with the BESIII detector. The cross section was measured to be $62.9 \pm 1.9 \pm 3.7$ pb, which is in good agreement with previous measurements from the BaBar~\cite{BaBar:2012vyb}, Belle~\cite{Belle:2007dxy}, and CLEO~\cite{CLEO:2006ike} collaborations. In addition, a new structure at around 3.9 GeV in the $\pi^\pm J/\psi$ mass spectrum, subsequently named $Z_c(3900)$, was observed, as shown in Fig.~\ref{Fig:exp-Zc} (a). The signal shape was parameterized using an $S$-wave Breit-Wigner function. Neglecting interference between the Breit-Wigner signal and the background, the mass and width were reported to be~\cite{BESIII:2013ris}
\begin{eqnarray*}
M = (3899.0 \pm 3.6 \pm 4.9) \ \mathrm{MeV},\quad \Gamma = 46 \pm 10 \pm 20 \ \mathrm{MeV}.
\end{eqnarray*}

At nearly the same time, the Belle Collaboration reported cross section measurements for $e^+ e^- \to \pi^+\pi^- J/\psi$ between 3.8 GeV and 5.5 GeV, using a $967~\mathrm{fb}^{-1}$ data sample collected by the Belle detector at or near the $\Upsilon(nS)$ (with $n=1$ to $5$) resonances via the ISR technique~\cite{Belle:2013yex}. A structure in the $\pi^\pm J/\psi$ mass spectrum was observed with a significance of $5.2\sigma$, as shown in Fig.~\ref{Fig:exp-Zc}(b). The resonance parameters were determined to be~\cite{Belle:2013yex}
\begin{eqnarray*}
M = (3894.5 \pm 6.6 \pm 4.5) \ \mathrm{MeV},\quad \Gamma = (63 \pm 24 \pm 26) \ \mathrm{MeV}.
\end{eqnarray*}

Subsequently, former members of the CLEO-c collaboration, Xiao {\it et al.}, reanalyzed data collected by the CLEO-c detector at the $\psi(4160)$ resonance and observed the charged $Z_c(3900)$ structure~\cite{Xiao:2013iha}, consistent with the findings of the BESIII~\cite{BESIII:2013ris} and Belle~\cite{Belle:2013yex} collaborations. In addition, evidence for the neutral $Z_c(3900)$ was first reported at a significance of $3\sigma$~\cite{Xiao:2013iha}, and was later firmly established by the BESIII Collaboration in the process $e^+ e^- \to \pi^0 \pi^0 J/\psi$ with a significance of $10.4\sigma$~\cite{BESIII:2015cld}. A partial wave analysis of the processes $e^+ e^- \to \pi^{+} \pi^{-} J/\psi$ and $\pi^{0} \pi^{0} J/\psi$ determined the spin and parity of the $Z_c(3900)$ to be $J^P = 1^+$~\cite{BESIII:2017bua, BESIII:2020oph}. In addition to hidden-charm decays, the BESIII Collaboration also reported a near-threshold structure in the $D\bar{D}^\ast$ system, denoted $Z_c(3885)$, observed in the processes $e^+ e^- \to (D\bar{D}^\ast)^\pm \pi^\mp$~\cite{BESIII:2015pqw} and $e^+ e^- \to (D\bar{D}^\ast)^0 \pi^0$~\cite{BESIII:2015ntl}.

Furthermore, in the $\pi^{\pm,0} h_c$ invariant mass distributions of the processes $e^+ e^- \to \pi^{+,0} \pi^{-,0} h_c$, the BESIII Collaboration reported another charged charmonium-like structure, $Z_c(4020)$, with masses of $4022.9 \pm 0.8 \pm 2.7$ MeV and $4023.0 \pm 2.2 \pm 3.8$ MeV for the charged and neutral states, respectively~\cite{BESIII:2013ouc, BESIII:2014gnk}. In the open-charm sector, a near-threshold structure in the $D^\ast \bar{D}^\ast$ system, referred to as $Z_c(4025)$, was observed in the $(D^\ast\bar{D}^\ast)^{\pm}$ and $(D^\ast\bar{D}^\ast)^0$ invariant mass spectra of the processes $e^+ e^- \to (D^\ast\bar{D}^\ast)^{\pm} \pi^\mp$~\cite{BESIII:2013mhi} and $e^+ e^- \to (D^\ast\bar{D}^\ast)^{0} \pi^0$~\cite{BESIII:2015tix}. Beyond these channels, the BESIII Collaboration has also searched for charmonium-like structures in the processes $e^+e^-\to \pi^+ \pi^- \psi(3686)$~\cite{BESIII:2017tqk}, $e^+e^-\to \pi^0 \pi^0 \psi(3686)$~\cite{BESIII:2017vtc}, and $e^+e^- \to \pi^+ \pi^- \psi(3770)$~\cite{BESIII:2019tdo}.

In addition to studies in $e^+e^-$ annihilation, the charmonium-like structures $Z_c(3900)$ and $Z_c(4020)$ have also been searched for in $B$ meson decays. For example, the Belle Collaboration searched for $Z_c(3900)$ in the decay $\bar{B}^0 \to J/\psi K^-\pi^+$ and found no significant signal~\cite{Belle:2014nuw}. Using $p\bar{p}$ collision data collected by the D\O\, experiment at the Fermilab Tevatron collider, the D\O\, Collaboration reported evidence for $Z_c(3900)^\pm$ decaying to $J/\psi \pi^\pm$ in semi-inclusive weak decays of $b$-flavored hadrons~\cite{D0:2018wyb}.
\begin{figure}[htbp]
\centering%
\scalebox{0.76}{\includegraphics{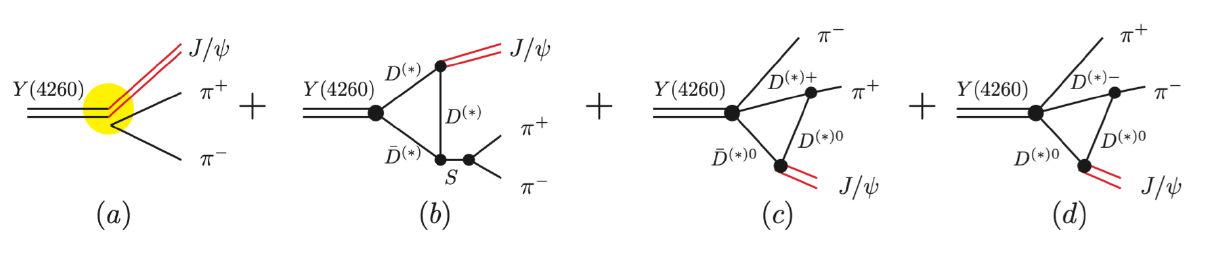}}%
\caption{The typical diagrams depicting $Y(4260) \to
J/\psi\pi^+\pi^-$ decay. Here, diagram (a) denotes
$Y(4260)$ direct decay into $J/\psi\pi^+\pi^-$, diagram (b) describes the intermediate hadronic loop
contribution to $Y(4260)\to J/\psi\pi^+\pi^-$, and diagrams (c) and (d) are from the ISPE mechanism.\\
{\it Source:} Taken from~\cite{Chen:2013coa}
\label{Fig:ISPE-reproduceZc-mech}}
\end{figure}

\subsubsection{Reproduce $Z_c(3900)$ in the $\pi^\pm J/\psi $ invariant mass spectra of $e^+e^- \to \pi^+ \pi^- J/\psi$ at $\sqrt{s}=4.260$ GeV}

The observations of $Z_c$ structures in the hidden-charm decay processes suggested by the ISPE mechanism provide important evidence in support of the ISPE mechanism. With abundant experimental information on the distributions of the $J/\psi\pi^\pm$ and $\pi^+\pi^-$ invariant mass spectra in $Y(4260)\to \pi^+\pi^- J/\psi$~\cite{BESIII:2013ris,Belle:2013yex, Xiao:2013iha}, the authors of Refs.~\cite{Chen:2011xk, Chen:2013bha} conducted further in-depth studies of these processes using the ISPE mechanism to gain a deeper understanding of these charmonium-like structures.

Following the observations of  $Z_c(3900)$ in the $\pi^+J/\psi$ invariant mass distributions of the process $e^+ e^- \to \pi^+ \pi^- J/\psi$ at $\sqrt{s}=4.260$ GeV by the BESIII and Belle collaborations~\cite{BESIII:2013ris,Belle:2013yex}, Chen, Liu, and Matsuki considered contributions from the direct decay of $Y(4260)$, the dipion scalar resonance contributions, and the ISPE contributions, as illustrated in Fig.~\ref{Fig:ISPE-reproduceZc-mech} of Ref.~\cite{Chen:2013coa}. The corresponding amplitudes are expressed as,
\begin{eqnarray}
\mathcal{M}_{\mathrm{Direct}} &=&
{{F}\over{f_\pi^2}}\epsilon_{Y(4260)}\cdot
\epsilon_{J/\psi}\Big\{\Big[q^2-\kappa (\Delta
M)^2\Big(1+\frac{2m^2_\pi}{q^2}\Big)\Big]_{\mathrm{S-wave}}+\Big[\frac{3}{2}\kappa\big((\Delta M)^2-q^2\big)
\Big(1-\frac{4m_\pi^2}{q^2}\Big)
\Big(\cos\theta^2-\frac{1}{3}\Big)\Big]_{\mathrm{D-wave}}\Big\}\nonumber\\
\mathcal{M}_{\mathcal{S}} 
&=& \epsilon_{Y(4260)}^\mu \epsilon_{J/\psi}^\nu\; g_{\mu \nu}\;
\frac{F_{\mathcal{S}}}{(p_1+p_2)^2-m_{\mathcal{S}}^2 +
im_{\mathcal{S}} \Gamma_{\mathcal{S}}},\notag\\
\mathcal{M}_{\mathrm{ISPE}}^{D\bar{D}}&=&g_{Y(4260)D D\pi}\epsilon_{Y(4260)}^{\mu}
\epsilon_{J/\psi}^\nu \big[ A_0  g_{\mu \nu} + \big(A_1 p_{1 \mu}
p_{1 \nu}  +A_2 p_{1 \mu} p_{2 \nu} +A_3  p_{2
\mu} p_{1 \nu}+ A_4 p_{2 \mu} p_{2 \nu}\big)\big],\notag\\
\mathcal{M}_{\mathrm{ISPE}}^{D\bar{D}^*}&=& g_{Y(4260)D^*D\pi}\epsilon_{Y(4260)}^{\mu}
\epsilon_{J/\psi}^\nu \big[ B_0  g_{\mu \nu} + \big(B_1 p_{1 \mu}
p_{1 \nu} +B_2 p_{1 \mu} p_{2 \nu} +B_3  p_{2
\mu} p_{1 \nu}+ B_4 p_{2 \mu} p_{2 \nu}\big)\big],\notag
\\
\mathcal{M}_{\mathrm{ISPE}}^{D^*\bar{D}^*} &=&g_{Y(4260)D^*D^*\pi}\epsilon_{Y(4260)}^{\mu}
\epsilon_{J/\psi}^\nu \big[ C_0  g_{\mu \nu} + \big(C_1 p_{1 \mu}
p_{1 \nu}+ C_2  p_{1 \mu} p_{2 \nu} +C_3 p_{2
\mu} p_{1 \nu}+ C_4  p_{2 \mu} p_{2 \nu}\big)\big],
\end{eqnarray}
where the coefficients $A_i$, $B_i$, and $C_i$ (with $i = 0$ to $4$) associated with the various Lorentz structures are evaluated from the loop integrals in the ISPE mechanism (see Ref.~\cite{Chen:2011xk} for further details). In this framework, the total decay amplitude for $Y(4260) \to \pi^+ \pi^- J/\psi$ is given by the sum of these sub-amplitudes, i.e.,
\begin{eqnarray}
&&\mathcal{M}^{\mathrm{Total}}[Y(4260) \to \pi^+ \pi^-
J/\psi] = \mathcal{M}_{\mathrm{Direct}} +
e^{i\phi_\sigma} \mathcal{M}_{\sigma} + e^{i\phi_{f_0(980)}}
\mathcal{M}_{f_0(980)}+ e^{i\phi_{\mathrm{ISPE}}}
\left(\mathcal{M}_{\mathrm{ISPE}}^{D\bar{D}}
+\mathcal{M}_{\mathrm{ISPE}}^{D\bar{D}^*}
+\mathcal{M}_{\mathrm{ISPE}}^{D^\ast \bar{D}^\ast} \right), \qquad 
\end{eqnarray}
where three phase angles $\phi_\sigma$, $\phi_{f_0(980)}$, and $\phi_{\mathrm{ISPE}}$ are introduced. With the amplitude given above, the differential decay width for $Y(4260) \to \pi^+ \pi^- J/\psi$ is expressed as
\begin{eqnarray}
&&d\Gamma\left(Y(4260)\to \pi^+\pi^-
J/\psi\right)=\frac{1}{3}\frac{1}{(2\pi)^3}\frac{1}{32m^3_{Y(4260)}}\big|\mathcal{M}^{\mathrm{Total}}\big|^2
d{m^2_{J/\psi\pi^\pm}}d{m^2_{\pi^+\pi^-}},
\end{eqnarray}
where $m_{J/\psi\pi^\pm}^2=(p_2+p_3)^2$ and $m_{\pi^+\pi^-}^2=(p_1+p_2)^2$. Using the above differential decay width, the theoretical distributions for the $J/\psi \pi^\pm$ and $\pi^+\pi^-$ invariant mass spectra can be obtained and subsequently applied in fitting to the BESIII and Belle data~\cite{BESIII:2013ris,Belle:2013yex}.

\begin{figure}[t]
\centering%
\scalebox{0.9}{\includegraphics{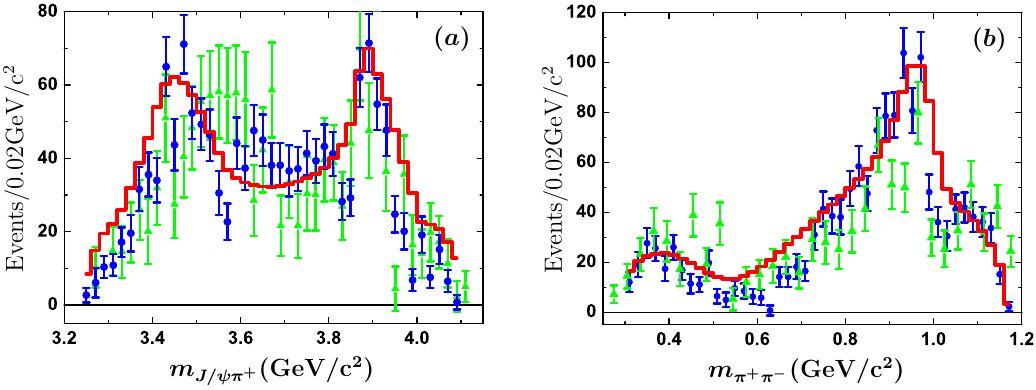}}\caption{(color online). The distributions of the $J/\psi\pi^+$ and $\pi^+\pi^-$ invariant mass spectra of  $Y(4260) \to \pi^+ \pi^- J/\psi$. The blue and green dots with error bars are the experimental data given by BESIII \cite{BESIII:2013ris} and Belle \cite{Belle:2013yex}, respectively. The red histograms are our results considering contributions of the ISPE mechanism to the $Y(4260) \to \pi^+ \pi^- J/\psi$ decay. \\
{\it Source:} Taken from~\cite{Chen:2013coa}
\label{Fig:ISPE-reproduceZc-results}}
\end{figure}

In Fig.~\ref{Fig:ISPE-reproduceZc-results}, the $J/\psi \pi^+$ (left panel) and $\pi^+ \pi^-$ (right panel) invariant mass spectra for $Y(4260) \to \pi^+ \pi^- J/\psi$ are shown. The two red histograms represent the fit results to the BESIII data using the ISPE mechanism. For comparison, the data from the Belle Collaboration~\cite{Belle:2013yex} are also included. From the figure, one can see that the present fit agrees well with the experimental data reported by both the BESIII and Belle Collaborations~\cite{BESIII:2013ris, Belle:2013yex}. In particular, two peaks—the $Z_c(3900)$ structure and its reflection—are nicely reproduced, with the fitting results clearly reflecting that $Z_c(3900)$ is much sharper than its reflection. In addition, the double-peak structures in the dipion invariant mass distributions are also well described.

As shown in Fig.~\ref{Fig:ISPE-reproduceZc-results}, the $Z_c(3900)$ structure is remarkably reproduced within the ISPE framework, which incorporates interference effects among different intermediate processes. This result affirmatively confirms that $Z_c(3900)$ corresponds to the charged charmonium-like structure near the $D\bar{D}^*$ threshold predicted via the ISPE mechanism in Ref.~\cite{Chen:2011xk} \changelabel{as a kinematical generated enhancement}.

\subsubsection{Reproduce the enhancement structures in the $\pi^\pm \psi(3686)$ invariant mass spectra of $e^+ e^- \to \pi^+ \pi^- \psi(3686)$}

In Ref.~\cite{BESIII:2017tqk}, the BESIII Collaboration investigated the process $e^+e^- \to \pi^+ \pi^- \psi(3686)$ using a $5.1~\mathrm{fb}^{-1}$ data sample collected with the BESIII detector at 16 center-of-mass energy points ranging from 4.008 to 4.600 GeV. The cross sections were measured with improved precision and found to be consistent with previous measurements from the Belle and BaBar collaborations. In addition, a charged charmonium-like structure was observed in the $\pi^\pm \psi(3686)$ invariant mass spectrum for the data at $\sqrt{s}=4.416$ GeV, with a significance of $9.2\sigma$. The resonance parameters of this structure were fitted to be~\cite{BESIII:2017tqk},
\begin{eqnarray}
	M=(4032.1 \pm 2.4)\ \mathrm{MeV}, \ \ \Gamma=(26.1 \pm 5.3)~\mathrm{MeV},
\end{eqnarray}	
respectively. It should be mentioned that this process is also one of the processes suggested by the ISPE mechanism for searching charged charmonium-like structures, as proposed in Ref.~\cite{Chen:2011xk}.

Moreover, unlike the $Z_c(3900)$ observed in $e^+e^-\to \pi^+ \pi^- J/\psi$ at $\sqrt{s}=4.260$ GeV, the BESIII Collaboration has reported on the dipion and $\pi^\pm \psi(3686)$ invariant mass spectra in the process $e^+ e^-\to \pi^+ \pi^- \psi(3686)$ at six different center-of-mass energies: $\sqrt{s}=4.226,\ 4.258,\ 4.358,\ 4.387,\ 4.416,\ 4.600$ GeV. The lineshapes of the $\pi^\pm \psi(3686)$ distributions vary significantly across these energy points. Therefore, reproducing these invariant mass distributions within a unified ISPE mechanism would provide even stronger evidence in support of the ISPE mechanism. 

\begin{figure}[t]
\centering%
\scalebox{0.7}{\includegraphics{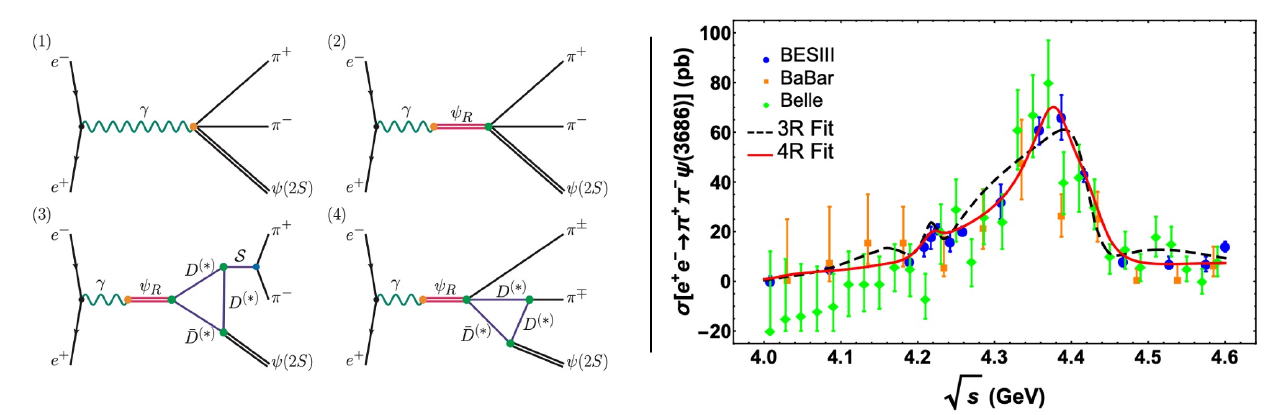}}
\caption{(color online). Feynman diagrams contributing to the process $e^+ e^-\to \pi^+ \pi^- \psi(3686)$ (left panel) and the fitted cross sections for $e^+ e^-\to \pi^+ \pi^- \psi(3686)$ given by the BESIII Collaboration in Ref.~\cite{BESIII:2017tqk}. The data from the BaBar~\cite{BaBar:2006ait, BaBar:2012hpr} and Belle~\cite{Belle:2007umv, Belle:2014wyt} collaborations are also presented for comparison. 
  \\
{\it Source:} Taken from~\cite{Huang:2019agb}
\label{Fig:ispe-psi2s}}
\end{figure}

In the left panel of Fig.~\ref{Fig:ispe-psi2s}, the Feynman diagrams contributing to the process $e^+e^- \to \pi^+ \pi^- \psi(3686)$ are presented. Fig.~\ref{Fig:ispe-psi2s}-(1) shows the non-resonance production process, where the virtual photon directly couples to $\pi^+ \pi^-\psi(3686)$. Fig.~\ref{Fig:ispe-psi2s}-(2) depicts the direct production process, in which the dipion is produced directly via gluon emission. Fig.~\ref{Fig:ispe-psi2s}-(3) illustrates the dipion resonance production process, where the dipion is produced through a scalar meson, and Fig.~\ref{Fig:ispe-psi2s}-(4) represents the production process involving the ISPE mechanism. In the three-charmonium scenario, $\psi(4160)$, $\psi(4220)$, and $\psi(4415)$ were included in the fit. As shown in the right panel of Fig.~\ref{Fig:ispe-psi2s}, this three-charmonium fit accurately reproduces the structure in the vicinity of 4.2 GeV and the broad enhancement near 4.4 GeV. However, the fitted curve lies above the experimental data around 4.3 GeV, which indicates the need for an additional resonance contribution. In the 4R fit scenario, an additional resonance, $\psi(4380)$, was taken into consideration, and one finds that the cross sections can be better described than in the three-charmonium scenario.

\begin{figure}[t]
\centering%
\scalebox{0.73}{\includegraphics{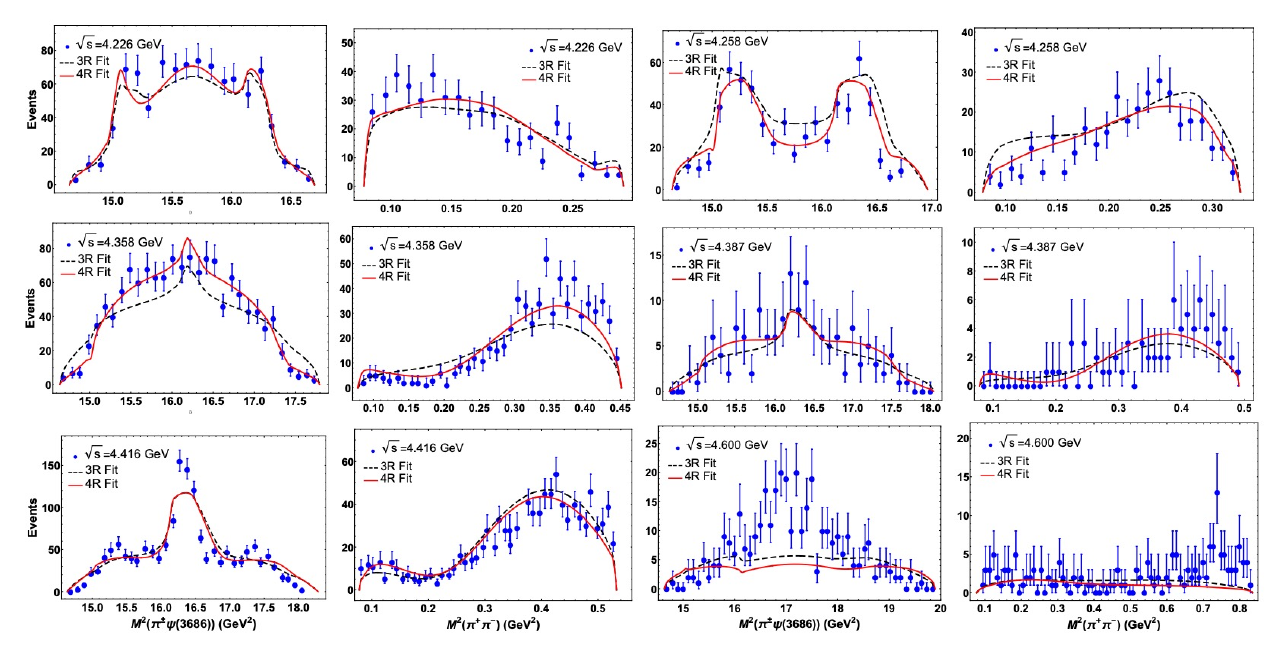}}
\caption{(color online). Combined fit for the $\psi(3686) \pi^\pm$ and $\pi^+ \pi^-$ invariant mass spectra at different center-of-mass energies reported by the BESIII Collaboration in Ref.~\cite{BESIII:2017tqk}. The black dashed and red solid curves are the fitted  results under three-charmonium and four-charmonium scenarios, respectively.  \\
{\it Source:} Taken from~\cite{Huang:2019agb}
\label{Fig:ISPE-reproduce-psi2spi-results}}
\end{figure}

In addition to the cross sections, the dipion and $\pi^\pm \psi(3686)$ invariant mass distributions are also fitted simultaneously. As shown in Fig.~\ref{Fig:ISPE-reproduce-psi2spi-results}, the clear peak structures in the $\psi(3686) \pi^+$ invariant mass spectra at $\sqrt{s}=4.226, \ 4.258$, and $4.416$ GeV, as reported by the BESIII Collaboration~\cite{BESIII:2017tqk}, are well described by the ISPE mechanism. For the data at 4.600 GeV, however, the fitted curves obtained from both fit scenarios lie below the experimental data for the $\pi^\pm \psi(3686)$ invariant mass distributions, suggesting possible contributions from a charmonium state around 4.6 GeV. 

The predictions of charmonium-like structures near the $D^\ast \bar{D}$ and $D^\ast \bar{D}^\ast$ thresholds, along with subsequent successful descriptions of various processes related to $Z_c(3900)$ and $Z_c(4020)$ using the ISPE mechanism after their experimental observation, demonstrate that the ISPE mechanism is universal in heavy quarkonium decay processes. The ISPE mechanism is crucial for understanding the observed near-threshold structures.

It is worth mentioning again that in the $e^+ e^- \to \pi^+ \pi^- \psi(3686)$ process, the BESIII Collaboration reported the $\psi(3686) \pi^\pm$ and $\pi^+ \pi^-$ invariant mass spectra at different center-of-mass energies. From Fig.~\ref{Fig:ISPE-reproduce-psi2spi-results}, one can notice that the lineshapes at different center-of-mass energies are rather different. This distinct property of the lineshape poses a great challenge for the aforementioned tetraquark and molecular interpretations, which are generally expected to behave similarly across different channels and center-of-mass energies. Unlike a "genuine state," and as indicated in Fig.~\ref{Fig:chen-ISPE-Zc}, the predicted lineshapes of the $J/\psi \pi$ invariant mass distributions vary with the initial state energy, which is consistent with the experimental measurements from the BESIII Collaboration. The reproduction of experimental data at different center-of-mass energy points within a unified framework in Ref.~\cite{Huang:2019agb} further proves the validity of the ISPE mechanism.

\changelabel{
Apart from the ISPE mechanism, following the experimental observation of these charged charmonium-like exotic structures, diverse theoretical frameworks have been proposed to interpret these states, largely motivated by their proximity to the $D^{\ast}\bar{D}^{(\ast)}$ mass thresholds and their isovector nature. Within these formalisms, multiple physical observables, such as mass spectra, decay properties, and production in various processes have been extensively explored in existing literature, as summarized below:
\begin{itemize}
    \item Compact tetraquark states of $[cq][\bar{c}\bar{q}]$ configuration. The mass spectra of charged charmonium-like resonances have been systematically studied within the compact tetraquark picture via QCD sum rules in Refs.~\cite{Wang:2022fdu,Wang:2013exa,Wang:2013vex,Wang:2019tlw,Wang:2019hnw, Qiao:2013dda, Dias:2013xfa}. Combining complementary results from lattice NRQCD and Born–Oppenheimer potential calculations, the authors of Ref.~\cite{Braaten:2013boa} proposed that $Z_c(3900)$ can be assigned as a tetraquark state with quantum numbers $J^P=1^+$. Furthermore, the decay patterns of the $Z_c$ family were analyzed within the compact tetraquark scheme in Ref.~\cite{Maiani:2013nmn}, which predicts hidden-charm decay channels to dominate. This prediction, however, conflicts with available experimental measurements.

    \item Bound states, resonances, or virtual states originating from $D^\ast \bar{D}/D^\ast \bar{D}^\ast$ hadronic interactions. QCD sum rule calculations reported in Refs.~\cite{Wang:2020dgr,Wang:2013daa, Cui:2013xla,Zhang:2013aoa, Cui:2013yva,Chen:2013omd,Khemchandani:2013iwa} demonstrate that the measured masses of $Z_c^{(\prime)}$ can be naturally reproduced under the hadronic molecular ansatz for $D^{\ast}\bar{D}^{(\ast)}$ systems. Consistent conclusions have also been reached using heavy-quark spin symmetry arguments~\cite{Guo:2013sya,Ji:2022uie} and phenomenological potential models~\cite{He:2013nwa}. By contrast, coupled-channel quasipotential Bethe–Salpeter equation calculations in Ref.~\cite{He:2017lhy} suggest that $Z_c(3900)$ and $Z_c(3985)$ correspond to virtual states. Additional coupled-channel analyses that simultaneously fit the $D^\ast\bar{D}$ and $J/\psi\pi$ invariant mass distributions further support the interpretation that $Z_c(3900)/Z_c(3985)$ are virtual states or resonances rather than stable bound states~\cite{Albaladejo:2015lob,Du:2022jjv}. Beyond mass spectroscopy, researchers have investigated the decay properties~\cite{Wilbring:2013cha,Ke:2013gia,Dong:2013iqa,Li:2013xia,Xiao:2018kfx,Wang:2022aiu,Chen:2015igx} and production mechanisms~\cite{Wang:2013cya, Wu:2023rrp,Qi:2023kwc, Wu:2019vbk,Chen:2017abq,Liu:2024ziu} of $Z_c^{(\prime)}$ within the molecular assumption.

\item Kinematic enhancements driven by triangle singularities. Apart from resonant exotic state interpretations, kinematic effects—most notably triangle singularities—have also been advanced to explain the observed near-threshold enhancements. Calculations in Ref.~\cite{Wang:2013hga} show that triangle singularities can generate threshold bumps within a narrow kinematic window. As elaborated in Ref.~\cite{Liu:2013vfa}, triangle singularities emerge in rescattering amplitudes under specific kinematic configurations and drastically modify threshold spectral behavior. A dedicated coupled-channel analysis covering the $D\bar{D}^\ast$, $\eta_c \rho$, and $J/\psi \pi$ channels concluded that the prominent $Z_c(3900)/Z_c(3985)$ peaks observed in hidden-charm spectra mainly arise from triangle-loop kinematic effects~\cite{Yu:2024sqv}. In Ref.~\cite{Chen:2023def}, the authors performed a comprehensive fit to experimental $\pi^+\pi^-$ and $J/\psi \pi^\pm$ invariant mass spectra, incorporating charmed-meson loops with triangle singularities, coupled-channel $J/\psi \pi-D\bar{D}^\ast$ rescattering, and dispersive treatments of final-state $\pi\pi-KK$ interactions. Their analysis implies that molecular and non-molecular contributions to the internal structure of $Z_c(3900)$ are comparably significant.
\end{itemize}

Despite extensive theoretical efforts dedicated to interpreting $Z_c(3900)$ from all the above perspectives, its fundamental underlying nature remains unresolved. In Ref.~\cite{Pilloni:2016obd}, the authors carried out an amplitude analysis on experimental data to probe the microscopic composition of $Z_c(3900)$, and they argued that contemporary experimental precision is insufficient to discriminate between the competing pictures of compact QCD tetraquark states, virtual hadronic states, and pure kinematic enhancements. Separately, finite-temperature properties of $Z_c(3900)$ have been explored in Ref.~\cite{Zhang:2025fcv} within both the hadronic-molecule and triangle-singularity formalisms, but the behaviors of $Z_c(3900)$ turn out to be similar in these two different interpretations. In Refs.~\cite{Ermolina:2024uln,Danilkin:2020kce}, the authors provide a simultaneous and accurate description of the $\pi^+\pi^-$ and $K^+K^-$ invariant mass distributions  at specific energies, utilizing the Muskhelishvili-Omnès formalism with $\pi\pi$ rescattering effects and $K\bar{K}$ coupled-channel unitarity, which offers a crucial theoretical framework for experimental measurements of $Z_c/Z_{cs}$ states at BESIII and Belle II.
}

In addition, the nature of these charged charmonium-like structures has been comprehensively investigated using lattice QCD. In Refs.~\cite{Prelovsek:2013xba,Prelovsek:2014swa}, the authors searched for charged charmonium-like structure with masses below 4.2 GeV in the channel $I^G(J^{PC})=1^+(1^{+-})$ using lattice QCD, where the two-meson channels $J/\psi \pi$, $\psi(3686)\pi$, $\psi(1D)\pi$, $D\bar{D}^\ast$, $D^\ast\bar{D}^\ast$, and $\eta_c \rho$ were included. The spectrum of eigenstates was extracted using a number of meson-meson and diquark-antidiquark interpolating fields, but no additional candidate for $Z_c(3900)$ below 4.2 GeV was found with a pion mass of 266 MeV. The HAL QCD Collaboration studied $Z_c(3900)$ using the method of coupled-channel scattering in lattice QCD, where the interactions among $\pi J/\psi$, $\rho \eta_c$, and $D\bar{D}^\ast$ channels were derived from (2+1)-flavor QCD simulations at $m_\pi=410-700$ MeV~\cite{HALQCD:2016ofq,Ikeda:2017mee}. Their analysis found that the interactions were dominated by the off-diagonal $\pi J/\psi-\bar{D}D^\ast$ and $\rho \eta_c-\bar{D}D^\ast$ couplings, suggesting that $Z_c(3900)$ is not a genuine resonance but rather a threshold cusp. Similarly, using the coupled-channel Lüscher finite-size formalism, the CLQCD Collaboration investigated the charmonium-like structure $Z_c(3900)$ within a two-channel scattering model~\cite{CLQCD:2019npr}. Combined with the two-channel Ross-Shaw theory, the scattering parameters were extracted from the energy levels by solving a generalized eigenvalue problem. However, the scattering length parameters suggested that the best-fit parameters for the particular lattice setup do not correspond to a peak in the elastic scattering cross-section near the threshold. \changelabel{ Comparison the $T$-matrix analysis in a finite box with the Lattice QCD simulation in Ref.~\cite{Prelovsek:2014swa} indicates that both resonance and virtual state interpretations are acceptable~\cite{Albaladejo:2016jsg}. However, both the joint analysis of the experimental data for $J/\psi \pi$ and $D \bar{D}^\ast$ invariant mass distributions~
\cite{Yan:2023bwt} and the lattice finite-volume energy levels resulting from the coupled-channel simulations~\cite{Cheung:2017tnt,CLQCD:2019npr} do not preclude the existence of a physical $Z_c(3900)$ state. 
}

\begin{figure}[htb] 
\centering %
\scalebox{0.75}{\includegraphics{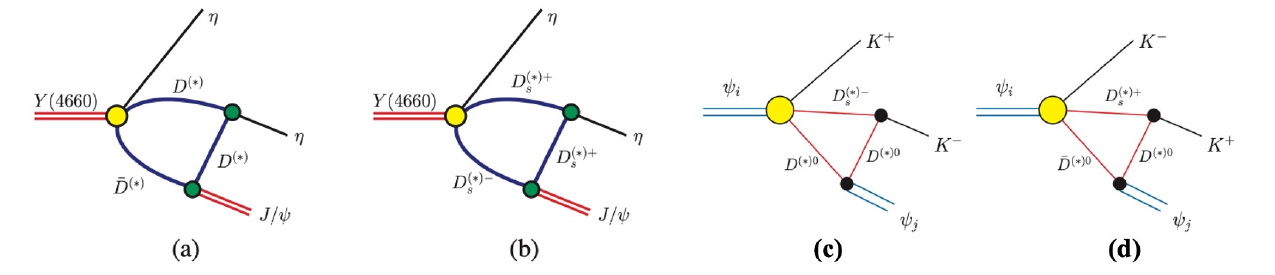}}
\caption{The di-$\eta$ (diagrams (a) and (b))and di-kaon (diagrams (c) and (d)) decays in the ISChE mechanism. It's worth mentioning that in the di-$\eta$ decay process, the initial and final states could be connected by both the charmed meson loops (diagram (a)) and charmed-strange meson loops (diagram (b)). \\
{\it Source:} Taken from~\cite{Chen:2013axa,Chen:2013wca}.
\label{Fig:di-eta-di-K}} 
\end{figure}

\begin{figure}[htb] 
\centering %
\scalebox{0.5}{\includegraphics{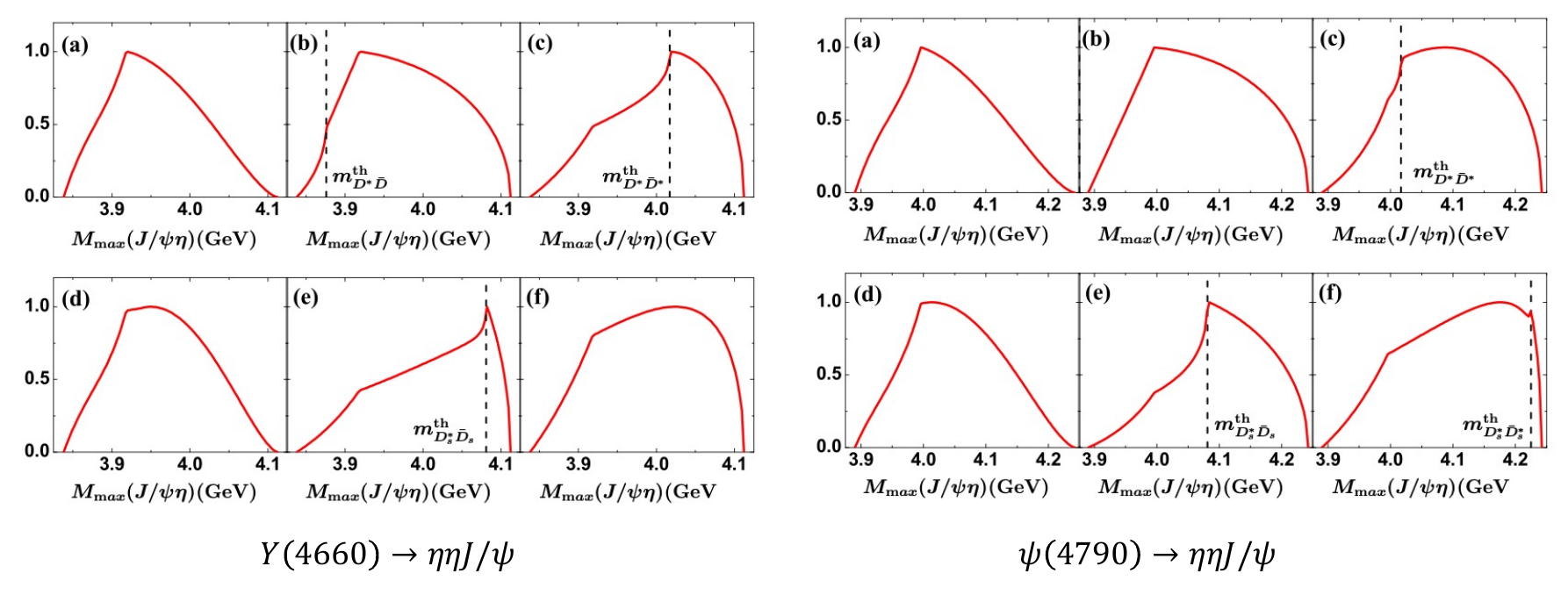}}
\caption{ The $M_{\mathrm{max}}({J/\psi \eta})$ distributions of $Y(4660) \to \eta \eta J/\psi$ (left panel) and $\psi(4790) \to \eta \eta J/\psi$ (right panel). Here, the diagrams (a), (b) and (c) are the lineshapes resulted from the intermediate $D\bar{D}$, $D^\ast \bar{D} +h.c.$ and $D^\ast \bar{D}^\ast$, respectively, while the digrams (d), (e) and (f) are the results considering the intermediate $D_s\bar{D}_s$, $D_s^\ast \bar{D}_s +h.c.$ and $D_s^\ast \bar{D}_s^\ast$ contributions, respectively. The thresholds of $D\bar{D}^*$, $D^*\bar{D}^*$ and $D_s^*\bar{D}_s$ are marked by the vertical dashed lines. The maxima of these lineshapes are normalized to 1.\\
{\it Source:} Taken from~\cite{Chen:2013axa}.
\label{Fig:ISChE-dieta}} 
\end{figure}

\subsection{Initial single chiral particle emission mechanism and the predicted $Z_{cs}$ structures}

Recognizing that the pion is a chiral particle, the authors in Ref.~\cite{Chen:2013wca} generalized the ISPE mechanism to the Initial Single Chiral particle Emission (ISChE) mechanism, wherein the pion in the ISPE framework can be replaced by any other chiral particle, such as a kaon or an $\eta$ meson. Within the ISChE mechanism, the di-$\eta$ decay~\cite{Chen:2013axa} and dikaon decay processes of higher charmonia have been investigated~\cite{Chen:2013wca}. These studies predicted the existence of isoscalar partners of the $Z_c$ states near the $D^\ast \bar{D}^{(\ast)}/D_s^{\ast +}D_{s}^{(\ast)-}$ thresholds, as well as charmonium-like states with strangeness, i.e., $Z_{cs}$ structures. Moreover, analogous to the predictions made for the $Z_c$ structures, the authors in Refs.~\cite{Chen:2013axa,Chen:2013wca} not only anticipated the presence of these charmonium-like structures but also clearly identified the specific channels in which they could be experimentally observed.

\subsubsection{Prediction of isoscalar partner of $Z_c$}
In Ref.~\cite{Chen:2013axa}, the hidden-charm di-$\eta$ decay processes of $Y(4660)$ and $\psi(4790)$ were investigated using the ISChE mechanism, where the decays $Y(4660)/\psi(4790) \to J/\psi \eta \eta$ proceed through intermediate states involving $\eta$ coupled to $D^{(\ast)} \bar{D}^{(\ast)}$ and/or $\eta$ coupled to $D_s^{(\ast)+} D_s^{(\ast)-}$, as illustrated in Fig. \ref{Fig:di-eta-di-K}. Through this process, one can address the important question of whether isoscalar charmonium-like structures exist near the $D\bar{D}^{\ast}$, $D^\ast \bar{D}^\ast$, $D_s^+ D_s^{\ast-}$, and $D_s^{\ast +}D_{s}^{\ast-}$ thresholds. From the ISChE-based analysis of the $Y(4660)/\psi(4790) \to J/\psi \eta \eta$ processes, the following key insights were obtained:
\begin{itemize}
\item For the $Y(4660) \to \eta \eta J/\psi$ process, a clear structure around 3.92 GeV appears in the $M_{\mathrm{max}}(\eta J/\psi)$ distributions across all diagrams shown in the left panel of Fig.~\ref{Fig:ISChE-dieta}. This structure arises from the phase space integral of $M_{\mathrm{max}}(J/\psi \eta)$. In addition, peak structures are visible near the $D^\ast \bar{D}$, $D^\ast \bar{D}^\ast$, and $D_s^+ D_s^{\ast-}$ thresholds. However, these structures are relatively broad, and the enhancement near the $D^\ast \bar{D}$ threshold is less pronounced compared to those observed in the $\pi^\pm J/\psi$ invariant mass distributions.

\item For the $\psi(4790) \to \eta \eta J/\psi$ process, clear peak structures near the $D^\ast \bar{D}^\ast$ and $D_s^+ D_s^{\ast-}$ thresholds were predicted in the $M_{\mathrm{max}}(\eta J/\psi)$ distributions. In contrast, the structures near the $D\bar{D}^\ast$ and $D_s^{\ast +} D_s^{\ast-}$ thresholds are less prominent, as illustrated in the right panel of Fig.~\ref{Fig:ISChE-dieta}. The peak resulting from the phase space integral for $\psi(4790) \to \eta \eta J/\psi$ is centered around 4.0 GeV.

\end{itemize}

In 2026, the BESIII Collaboration searched for the isospin partner of $Z_c(3900)$ in the $J/\psi \eta$ invariant mass distributions of $e^+ e^- \to \eta \eta J/\psi$ at center-of-mass energies ranging from $\sqrt{s}=4.226$ to $4.690$ GeV~\cite{BESIII:2026kja}. Due to the low statistical significance observed at nearly all energy points, no evidence of an isospin partner of $Z_c(3900)$ decaying to $\eta J/\psi$ was found.

\begin{figure}[htb] 
\centering %
\scalebox{1.15}{\includegraphics{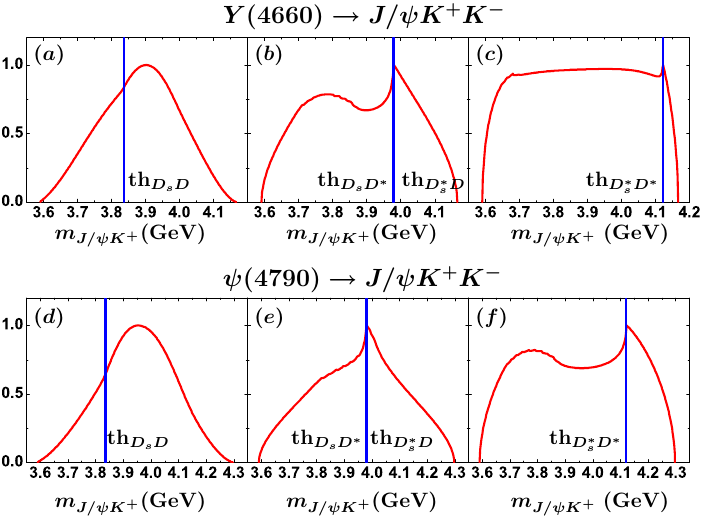}}
\caption{Dependence of the distribution of $d\Gamma/dm_{J/\psi K^+}$ on the $J/\psi K^+$ invariant mass spectrum (red solid curves). The diagrams (a) and (d), diagrams (b) and (e), and diagrams (c) and (f), are the results considering the intermediate $D \bar{D}_s+\text{h.c.}$, $ {D}^{*}\bar{D}_s+D  \bar{D}_s^\ast +\text{h.c.}$, and $D^*\bar{D}_s^* +\text{h.c.}$ contributions, respectively. Here, the line shape of distribution of $d\Gamma/dm_{J/\psi K^+}$ is normalized to 1. \\
{\it Source:} Taken from~\cite{Chen:2013wca}.
\label{Fig:ISChE-Zcs-prediction}} 
\end{figure}

\subsubsection{Predictions of charged $Z_{cs}$ structures}

Similar to the hidden-charm dipion decay processes, the hidden-charm di-kaon decay processes of $\psi(4415)$, $Y(4660)$, and $\psi(4790)$ were investigated in Ref.~\cite{Chen:2013wca}, where structures with hidden charm and open strangeness were predicted. Within the ISChE mechanism, the following results were obtained:
\begin{itemize}
\item The vector charmonium $\psi(4415)$ lies below the thresholds of $D_s^{\ast} \bar{D} \bar{K} / D_s \bar{D}^{\ast} \bar{K}$ and $D_s^\ast \bar{D}^\ast \bar{K}$. Consequently, under the ISChE mechanism, $\psi(4415)\to J/\psi K^+K^-$ proceeds only via the intermediate $D\bar{D}_s + \text{h.c.}$ The lineshape of $d\Gamma/dm{J/\psi K^+}$ is smooth and does not exhibit a sharp peak near the $D\bar{D}_s+\text{h.c.}$ threshold. This behavior also holds for the decays $Y(4660)\to J/\psi K^+K^-$ and $\psi(4790)\to J/\psi K^+K^-$, as shown in Fig.~\ref{Fig:ISChE-Zcs-prediction} (a) and (d).

\item When considering the intermediate $ {D}^{*}\bar{D}_s + D \bar{D}_s^\ast + \text{h.c.}$ contribution to $Y(4660)\to J/\psi K^+K^-$, two structures are observed in the lineshape presented in Fig.~\ref{Fig:ISChE-Zcs-prediction} (b). The higher one is a sharp peak near the $D_s\bar{D}^{*}/\bar{D}_s^{*}{D}$ thresholds, while the lower one is a broad structure that appears as its reflection. If only the intermediate $D^*\bar{D}_s^* + \text{h.c.}$ is included, the resulting $d\Gamma/dm_{J/\psi K^+}$ distribution for $Y(4660)\to J/\psi K^+K^-$ exhibits a small, sharp peak around the $D^*\bar{D}_s^*/\bar{D}^* D_s^*$ threshold. In addition, its reflection, appearing around $3.7$ GeV, is less pronounced (see Fig.~\ref{Fig:ISChE-Zcs-prediction} (c) for further details).

\item For the $\psi(4790)\to J/\psi K^+K^-$ decay, the corresponding $d\Gamma/dm_{J/\psi K^+}$ lineshapes are presented, considering the intermediate $ {D}^{*}\bar{D}_s + D \bar{D}_s^\ast + \text{h.c.}$ and $D^*\bar{D}_s^* + \text{h.c.}$ contributions in diagrams (e) and (f) of Fig.~\ref{Fig:ISChE-Zcs-prediction}. Figure~\ref{Fig:ISChE-Zcs-prediction} (e) shows an enhancement structure in the $d\Gamma/dm_{J/\psi K^+}$ distribution, composed of a sharp peak near the $D_s\bar{D}^{*}/\bar{D}_s^{*}{D}$ threshold and its reflection. This situation differs from that of $Y(4660)\to J/\psi K^+K^-$ discussed above. Figure~\ref{Fig:ISChE-Zcs-prediction} (f) reveals a sharp peak close to the $D^*\bar{D}_s^*/\bar{D}^* D_s^*$ threshold and a corresponding broad structure from its reflection, both of which are more distinct compared to the features shown in Fig.~\ref{Fig:ISChE-Zcs-prediction} (c).
\end{itemize}

It should be mentioned that, besides the ISChE mechanism, the existence of hidden-charm and open-strange states has also been predicted within the relativistic diquark-antidiquark framework~\cite{Ebert:2008kb}. Estimates from QCD sum rules have also indicated the possible existence of a $D_s^- D^\ast + D_s^{\ast-} D$ molecular state~\cite{Lee:2008uy}\footnote{It is worth noting that these studies primarily focus on the calculation of mass spectra.}. In Ref.~\cite{Dias:2013qga}, the decay properties of charged states near the $D_s\bar{D}^\ast/D_s^\ast \bar{D}$ threshold were investigated using QCD sum rules.

\begin{figure}[htb]
\centering%
\scalebox{0.55}{\includegraphics{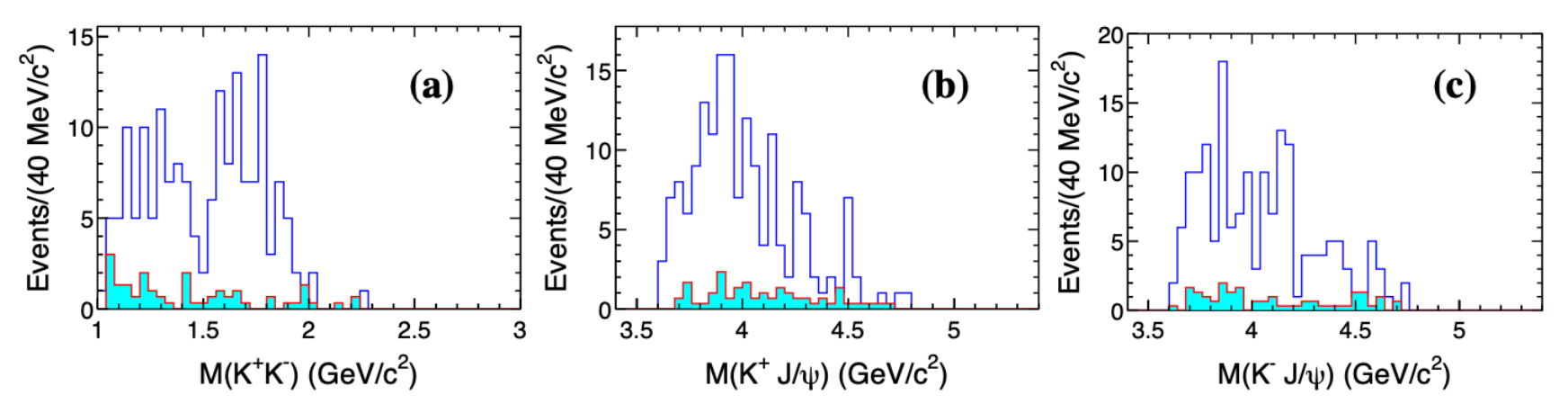}}
\caption{Invariant mass distributions of (a) $K^+ K^-$, (b) $K^+ J/\psi$, and (c) $K^- J/\psi$ for $K^+ K^- J/\psi$ events with $4.4 <M(K^+ K^-J/\psi)<5.5\ \mathrm{GeV}/c^2$. Solid histograms are for events in the $J/\psi$ signal region, and the shaded histograms are normalized background from the $J/\psi$ mass sidebands. \\
{\it Source:} Taken from~\cite{Belle:2014fgf}
\label{Fig:exp-Zcs-Belle}}
\end{figure}

\begin{figure}[htb]
\centering%
\scalebox{0.7}{\includegraphics{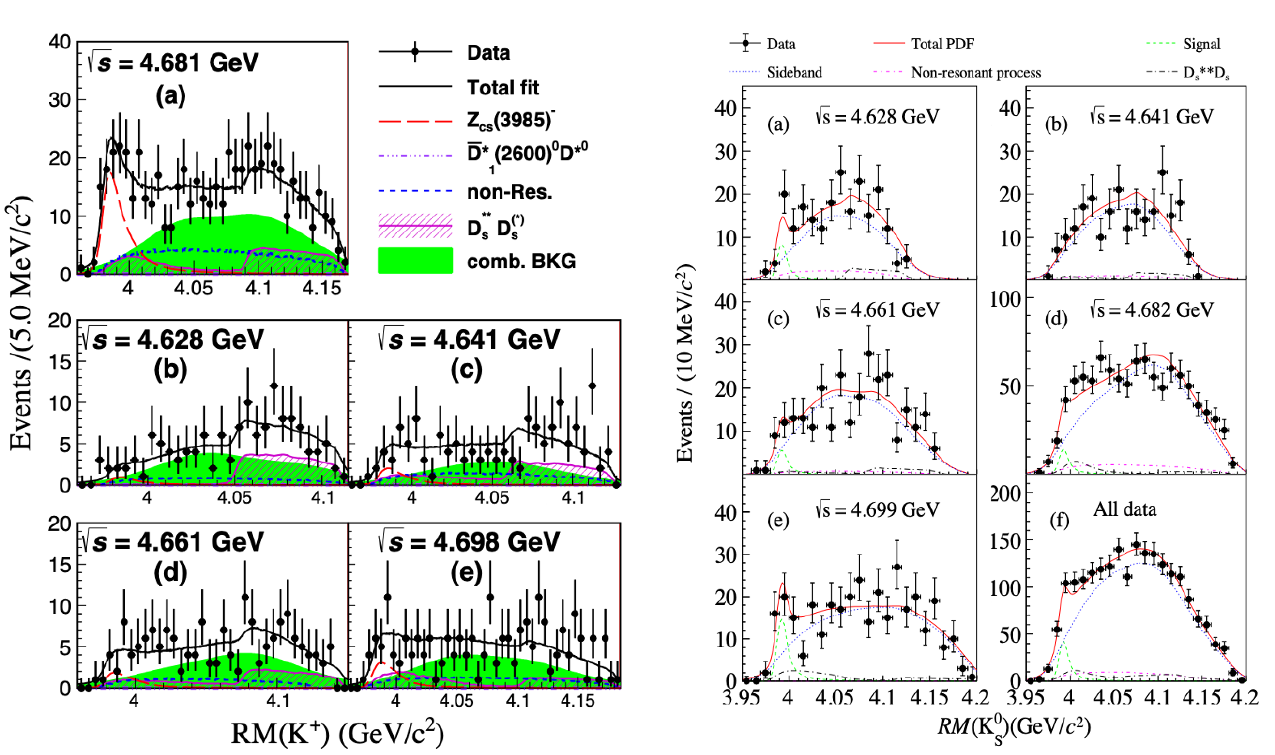}}
\caption{The experimental observation of $Z_{cs}^-$ (left panel) in the $K^+$ recoil mass spectra of $e^+e^- \to K^+ (D_s^- D^{\ast 0} +D_s^{\ast -} D^0) $~\cite{BESIII:2020qkh} and $Z_{cs}^0$ (right panel) in the $K_s^0$ recoil mass spectra of $e^+e^- \to K^0_s (D_s^+ D^{\ast-} +D_s^{\ast+} D^-)$~\cite{BESIII:2022qzr} by the BESIII Collaboration.\\
{\it Source:} Taken from~\cite{BESIII:2020qkh,BESIII:2022qzr}
\label{Fig:Zcs-BESIII}}
\end{figure}

\begin{figure}[htb]
\centering%
\scalebox{1}{\includegraphics{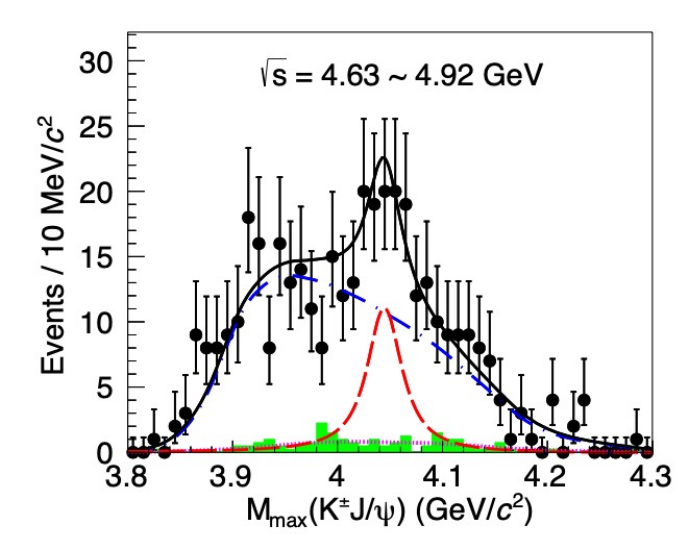}}
\caption{Distributions of $M_{\mathrm{max}}(K^\pm J/\psi)$ of $e^+ e^- \to K^+ K^- J/\psi$ reported by BESIII Collaboration~\cite{BESIII:2023wqy}. The red dashed, blue dash-dotted, and pink dotted lines are the signal component of $Z_{cs}$, the phase space and the combinatorial background, respectively.\\
{\it Source:} Taken from~\cite{BESIII:2023wqy}
\label{Fig:Zcs-BESIII-1s}}
\end{figure}

\begin{figure}[htb]
\centering%
\scalebox{0.45}{\includegraphics{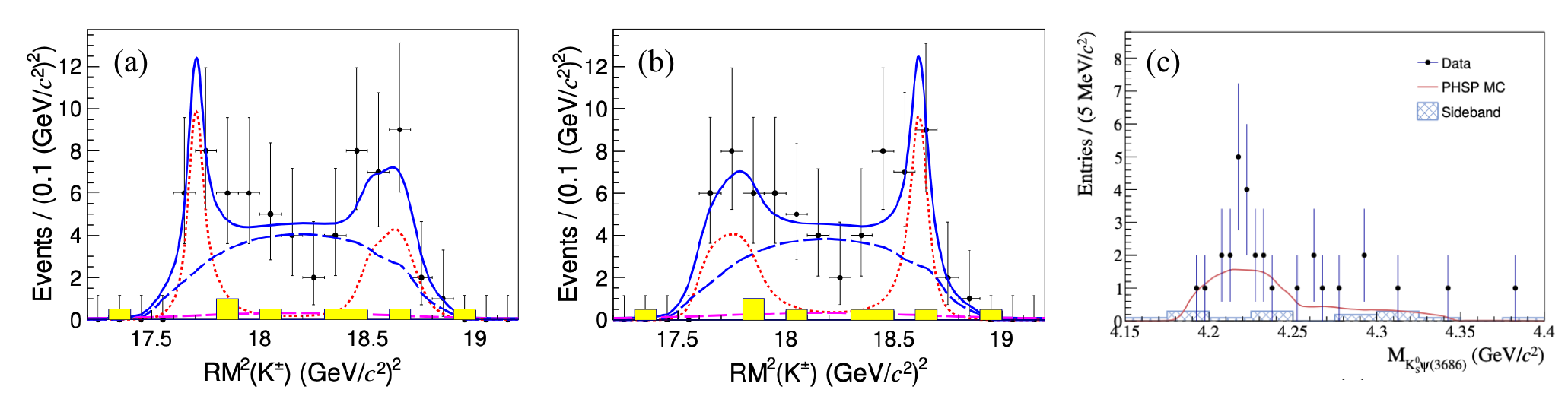}}
\caption{Distributions of $RM^2(K^\pm)$ of $e^+ e^- \to K^+ K^- \psi(3686)$ (diagram (a) and (b)) and $m_{K_S^0 \psi(3686)}$ of $e^+ e^- \to K_S^0 K_S^0 \psi(3686)$ reported by BESIII Collaboration~\cite{BESIII:2024vjf,BESIII:2024aya} .\\
{\it Source:} Taken from~\cite{BESIII:2024vjf,BESIII:2024aya}
\label{Fig:Zcs-BESIII-2s}}
\end{figure} 

\subsubsection{Experimental searches for $Z_{cs}$ structures}
As discussed above, charged hidden-charm four quark matters with strangeness near the $D_s^- D^{\ast 0}/D_s^{\ast -} D^0$ thresholds have been theoretically predicted by various models, including the compact tetraquark scenario~\cite{Ebert:2008kb}, the hadronic molecular model~\cite{Lee:2008uy}, and the ISChE mechanism~\cite{Chen:2013wca}. Experimental searches for these charged charmonium-like structures provide a crucial test of the above predictions. In the following, we present a brief review of the experimental progress on charmonium-like structures with strangeness.

\begin{itemize}
\item  Search for $Z_{cs}$ structures in $e^+e^- \to K^+ K^- J/\psi$ by the Belle Collaboration. In 2014, the Belle Collaboration accumulated a sample of $e^+ e^- \to K^+ K^- J/\psi$ events via initial-state radiation at center-of-mass energies between the threshold and 6.0 GeV, using a data sample of $673\ \mathrm{fb}^{-1}$ collected at or near $\sqrt{s}=10.58$ GeV~\cite{Belle:2014fgf}. The cross sections for $e^+ e^- \to K^+ K^- J/\psi$ and $K_S^0 K_L^0 J/\psi$ were measured, and possible intermediate states in the selected $K^+ K^- J/\psi$ events were investigated by examining the Dalitz plot. As shown in Fig.~\ref{Fig:exp-Zcs-Belle}, no clear structure was observed in the $K^\pm J/\psi$ system at that level of statistical precision. A larger data sample is required to obtain more detailed information about possible structures in the $K^\pm J/\psi$ system.

\item Observation of $Z_{cs}$ by the BESIII Collaboration. Based on $e^+ e^-$ annihilation data samples collected by the BESIII detector at five center-of-mass energies ranging from 4.628 to 4.698 GeV, with a total integrated luminosity of $3.7\ \mathrm{fb}^{-1}$, the BESIII Collaboration analyzed the processes $e^+ e^- \to K^+ D_s^- D^{\ast 0}$ and $K^+ D_s^{\ast -} D^0$~\cite{BESIII:2020qkh}. As shown in the left panel of Fig.~\ref{Fig:Zcs-BESIII}, the $K^+$ recoil-mass distributions for events collected at $\sqrt{s}=4.681$ GeV exhibit an excess of events over the known contributions from conventional charmed mesons, near the $D_s^- D^{\ast }$ and $D_s^{\ast-} D^0 $ thresholds. Fitting this near-threshold structure with a mass-dependent-width Breit-Wigner function yields a pole mass and width of~\cite{BESIII:2020qkh}
\begin{eqnarray*}
m =(3982.5^{+1.8}_{-2.6} \pm 2.1) \ \mathrm{MeV} ,\quad 	\Gamma = (12.8^{+5.3}_{-4.4} \pm 3.0) \ \mathrm{MeV}, 
\end{eqnarray*}
where the first uncertainties are statistical and the second are systematic. The significance of the resonance hypothesis over contributions from conventional charmed mesons alone is about $5.3\sigma$. This marks the first observation of a candidate charged hidden-charm structure with strangeness, consistent with predictions from the ISChE mechanism~\cite{Chen:2013wca} and other QCD exotic estimates~\cite{Ebert:2008kb, Lee:2008uy, Dias:2013qga}.

Subsequently, the BESIII Collaboration studied the neutral component of charmonium-like structures with strangeness, i.e., $Z_{cs}^0$, in the process $e^+e^- \to K_S^0 (D_s^+ D^{\ast-} + D_s^{\ast+} D^{-})$ and its charge conjugate, at five center-of-mass energies between 4.628 and 4.699 GeV. In the $K_S^0$ recoil-mass spectrum, evidence for a structure near the $D_s^+ D^{\ast-}$ and $D_s^{\ast+ }D^-$ thresholds was reported (see right panel of Fig.~\ref{Fig:Zcs-BESIII}). Fitting with a Breit-Wigner lineshape, the BESIII Collaboration obtained the mass and width of this structure as
\begin{eqnarray*}
m = (3992.2 \pm 1.7 \pm 1.6) \ \mathrm{MeV}, \quad \Gamma = (7.7^{+4.1}_{-3.8} \pm 4.3) \ \mathrm{MeV},
\end{eqnarray*} 
where the first uncertainties are statistical and the second are systematic. The significance of the signal, including both statistical and systematic uncertainties, is $4.6\sigma$. The measured mass of $Z_{cs}(3985)^0$ is close to that of $Z_{cs}(3985)^+$, suggesting that $Z_{cs}(3985)^0$ is the neutral isospin partner of $Z_{cs}(3985)^+$.

In the hidden-charm decay channel, the BESIII Collaboration searched for charged charmonium-like structures $Z_{cs}$ in the $M_{\mathrm{max}}(K^\pm J/\psi)$ distributions of the process $e^+ e^- \to K^+ K^- J/\psi$, using data samples with an integrated luminosity of $5.85\ \mathrm{fb}^{-1}$ collected at center-of-mass energies from 4.61 to 4.95 GeV. As shown in Fig.~\ref{Fig:Zcs-BESIII-1s}, a small excess corresponding to a $Z_{cs}$ signal is observed above other components. The mass and width are fitted to be $4044 \pm 6$ MeV and $36 \pm 16$ MeV, respectively. The uncertainties are statistical only, and the significance is $2.3\sigma$.

In addition, the BESIII Collaboration searched for $Z_{cs}$ structures in the processes $e^+e^- \to K^+K^- \psi(3686)$~\cite{BESIII:2024vjf} and $e^+e^- \to K_S^0K_S^0 \psi(3686)$~\cite{BESIII:2024aya}. In Ref.~\cite{BESIII:2024vjf}, the BESIII Collaboration fitted the $M^2_{K^\pm}$ distributions using two different schemes. The mass of $Z_{cs}$ was assumed to be around 4.205 GeV in Fit I (as shown in Fig.~\ref{Fig:Zcs-BESIII-2s} (a)) and around 4.315 GeV in Fit II (as shown in Fig.~\ref{Fig:Zcs-BESIII-2s} (b)). In Fit I, the mass and width for $Z_{cs}^\pm$ were fitted to be $(4208 \pm 3.1)$ MeV and $6.1 \pm 5.7$ MeV, respectively, with a global significance of $1.2\sigma$. In Fit II, the resonance parameters of $Z_{cs}^\pm$ were determined to be $M = (4316 \pm 2.7)$ MeV and $\Gamma = (9.0 \pm 8.6)$ MeV, with a global significance of $1.1\sigma$. The BESIII Collaboration also searched for the $Z_{cs}^0$ in the $K_{S}^0\psi(3686)$ invariant mass distributions of the process $e^+e^- \to K_S^0K_S^0 \psi(3686)$~\cite{BESIII:2024aya}. As shown in Fig.~\ref{Fig:Zcs-BESIII-2s} (c), the $K_{S}^0\psi(3686)$ invariant mass distribution is consistent with three-body phase space, and no obvious structure is observed.
\end{itemize}

As indicated in Ref.~\cite{Chen:2013wca}, the lineshapes of the $Z_{cs}$ structures depend on the initial state. For example, the structure near the $D_s D^\ast/D_s^\ast D$ threshold is prominent in the di-kaon decay of $Y(4660)$, whereas the structure near the $D_s^\ast D^\ast$ threshold is expected to be more evident in the $\psi(4790)$ decay. \changelabel{Such a lineshape dependence on the intial state is also find by Ref.~\cite{Du:2022jjv,Baru:2021ddn,Yang:2020nrt}, where it can be related to the triangle singularity effect}. Thus, to further investigate the $Z_{cs}$ structures in the hidden-charm di-kaon decays of higher charmonia, larger data samples are required, which should be accessible to the BESIII and Belle II Collaborations.

It should also be noted that the LHCb Collaboration has reported two charmonium-like structures with strangeness in $B$ meson decays~\cite{LHCb:2021uow, LHCb:2023hxg}. Specifically, in the decay $B^+ \to J/\psi \phi K^+$, two $Z_{cs}$ structures were observed in the $J/\psi K^+$ invariant mass spectrum~\cite{LHCb:2021uow}. The resonance parameters of the lower-mass state, denoted $Z_{cs}(4000)$, were measured to be $m = (4003 \pm 6^{+4}_{-14}) \ \mathrm{MeV}$ and $\Gamma = (131 \pm 15 \pm 26) \ \mathrm{MeV}$~\cite{LHCb:2021uow}. While its mass is consistent with the structure observed by the BESIII Collaboration, there is a significant discrepancy in the width. For the higher-mass structure, referred to as $Z_{cs}(4200)$, the mass and width were reported as $(4216 \pm 24^{+43}_{-30}) \ \mathrm{MeV}$ and $(233 \pm 52^{+97}_{-73}) \ \mathrm{MeV}$, respectively~\cite{LHCb:2021uow}. In addition, a neutral partner of the $Z_{cs}(4000)^+$ was observed in the $J/\psi K_S^0$ invariant mass distribution of the decay $B^0 \to J/\psi \phi K_S^0$, with a mass of $(3991^{+12+9}_{-10-17}) \ \mathrm{MeV}$ and a width of $(105^{+29+17}_{-25-23}) \ \mathrm{MeV}$~\cite{LHCb:2023hxg}. Given the substantial difference in width between the $Z_{cs}(3985)$ and $Z_{cs}(4000)$, it remains an open question whether these two structures share the same origin \changelabel{as Ref.~\cite{Ortega:2021enc} indicated, where these two structures are possibly related to the same virtual-state pole}.

\changelabel{ Following the experimental observations of the $Z_{cs}$ structures, their unique properties have aroused intense theoretical investigations. The mass spectra, decay behaviors and production mechanisms of these hidden-charm strange states have been comprehensively studied in various theoretical frameworks, which can be mainly classified into the compact tetraquark scenario, the hadronic molecular scenario, and kinematic effects as summarized below.

\begin{itemize}
    \item Compact tetraquark scenario. In the framework of QCD sum rules, the $Z_{cs}(3985)$ state was studied as the diquark–antidiquark type axialvector tetraquark state with $c\bar{c}u\bar{s}$ configuration, and the predicted mass is highly consistent with the experimental data from BESIII~\cite{Wang:2020iqt}, while the  analysis in Ref.~\cite{Wan:2020oxt} indicated that $Z_{cs}(3985)$ can be regarded as the mixing state of different color multiplet structures. By considering the meson–diquark interaction and introducing both diquark–antidiquark and molecular components, the $Z_{cs}(4000)$ state can be explained as the mixed state of $\bar{D}^{\ast 0} D_s^+$ and $\bar{A}_{cs}S_{cu}$ components via solving the Bethe–Salpeter equation~\cite{Cao:2022rjp}. Based on the chiral quark model with real- and complex-scaling methods, the $Z_{cs}$ states observed by BESIII and LHCb can be well described as either compact tetraquark states or hadronic molecular states~\cite{Yang:2021zhe}. In the framework of the dynamical diquark model with Hamiltonian formalism, the large mass splitting between $Z_{cs}(4000)$ and $Z_{cs}(4220)$ can be well explained by the $\mathrm{SU(3)_{flavor}}$ mixing of all $J^P=1^+$ states in the compact tetraquark frame~\cite{Giron:2021sla}. In Ref.~\cite{Karliner:2021qok}, the authors argued that the $Z_{cs}$ states are regarded as the analogues of $K_1(1270)$ and $K_1(1400)$ with an extra $c\bar{c}$ pair. Using the effective Hamiltonian with color, spin and flavor dependent interactions, $Z_{cs}(3985)$ is assigned as the $[sc][\bar{q}\bar{c}]$ state with $J^P=0^+$ or $1^+$, while $Z_{cs}(4000)$ and $Z_{cs}(4200)$ are identified as the $[qc][\bar{s}\bar{c}]$ states with $J^P=1^+$~\cite{Shi:2021jyr}. Furthermore, the observed $Z_{cs}$ states can be well organized into two $\mathrm{SU(3)}_f$ nonets with $J^P=1^+$ in the tetraquark picture~\cite{Maiani:2021tri}.

    \item Hadronic molecular and virtual state resulted from $\bar{D}^{(\ast)}_s D^{(\ast)}$ interactions. The properties of $Z_{cs}$ state has been extensively studied in the $\bar{D}^{(\ast)}_s D^{(\ast)}$ molecular frame in various theoretical approaches. In QCD sum rules, the mass and width of $Z_{cs}(3985)$ calculated with the $1^+$ $\bar{D}^{(\ast)}_s D^{(\ast)}$ molecular interpolating currents are in good agreement with experimental measurements~\cite{Xu:2020evn}. Combined with effective range expansion and compositeness relation, the decay properties of $Z_{cs}(3985)$ were analyzed, revealing that the width of open-charm channel is about one order of magnitude larger than that of hidden-charm channel~\cite{Guo:2020vmu}. In addition, based on heavy quark spin symmetry and $\mathrm{SU(3)}$ flavor symmetry, the pole structure extracted from experimental data suggests that $Z_{cs}(3985)$ is a genuine state, which can be either a virtual state or a bound state~\cite{Du:2020vwb,Baru:2021ddn}. Under the $\mathrm{SU(3)}_F$ symmetry and heavy quark spin symmetry, $Z_{cs}(3985)$ is regarded as the $U$-spin partner of $Z_c(3900)$ in the coupled-channel calculation~\cite{Meng:2020ihj}. Similarly, $Z_{cs}(3985)$ is considered as the strange partner of $Z_c(3900)$, which can be a virtual state or a resonance near the $D^\ast D_s - DD_s^{\ast}$ threshold, and the inclusion of triangle singularity contribution improves the consistency with BESIII data~\cite{Yang:2020nrt}. In chiral effective field theory, the resonance parameters and event distributions of $Z_{cs}(3985)$ are well reproduced by using the parameters fitted from $Z_c(3900)$ data, which strongly supports the molecular nature of $Z_{cs}(3985)$~\cite{Wang:2020htx,Zhai:2022ied}. The quark-exchange model calculations of hidden-charm decay modes also favor $Z_{cs}(3985)$ and $Z_{cs}(4000)$ as molecular states~\cite{Han:2022fup}, while the potential model with the axial meson exchange calculation supports $Z_{cs}(3985)$ as a bound state~\cite{Yan:2021tcp}. A combined analysis of $Z_c(3900)$ and $Z_{cs}(3985)$ shows that both states can be uniformly explained as $D^{(\ast)}\bar{D}_{(s)}^{(\ast)}$ resonances~\cite{Du:2022jjv}. However, the one-boson-exchange model with coupled-channel effects indicates that $Z_{cs}(3985)$ cannot be identified as a $D^{\ast 0} D_s^-/D^0 D_s^{\ast -}/D^{\ast 0}D_s^{\ast -}$ resonance~\cite{Chen:2020yvq}. Additionally, the production mechanisms of $Z_{cs}$ states have been widely investigated with the molecular assumption, including the productions in $B^+$, $B_s$ decays~\cite{Wu:2021cyc,Wu:2023fyh,Yu:2025pyu}, $K^-p$ scattering~\cite{Liu:2021ojf}, and the hidden-charm decay properties~\cite{Wu:2021ezz}.

  \item Kinematic threshold enhancements. Same as $Z_c(3900)$, kinematic effect can also reproduce the position of $Z_{cs}(3985)$. The estimations in Ref.~\cite{Ge:2021sdq} indicate that the structures $Z_{cs}(4000)$, $Z_{cs}(4200)$, and $X(4700)$ reported by the LHCb Collaboration can be simulated by the $J/\psi K^\ast$, $\psi^\prime K$ and $\psi^\prime \phi$ threshold cusp without introducing genuine exotic resonances. In Ref.~\cite{Ikeno:2020mra}, the authors found that the interactions is not strong enough to produce a bound state or resonance, but the large accumulation of strength at the $D^-_s \bar{D}^\ast $ threshold in the $e^+ e^- \to K^+ (D_s^{\ast -} D^0 +D_s^- D^{-\ast})$ reaction can be well reproduced with coupled channels effect. While the estimations in Ref.~\cite{Wang:2020kej} proposed a reflection picture to understand the nature of $Z_{cs}(3985)$, where the enhancement structure can be a reflection of $D_{s2}^\ast (2573)$.

\end{itemize}
}

\section{Interaction between hadrons involved in charmonia}

\subsection{Experimental observation of di-$J/\psi$ invariant mass spectrum from $pp$ collisions}

\begin{figure}[htb!]
\begin{center}
   \subfigure[]{\includegraphics[width=0.32991\linewidth]{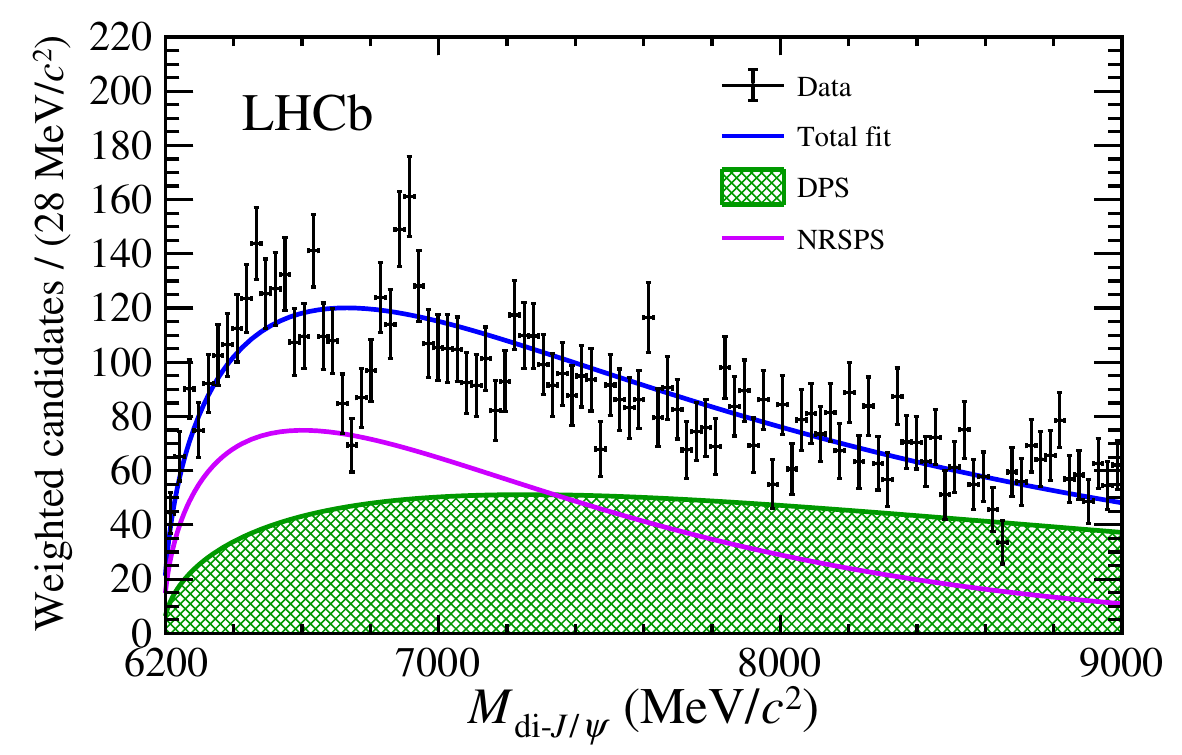}}
   \subfigure[]{\includegraphics[width=0.32991\linewidth]{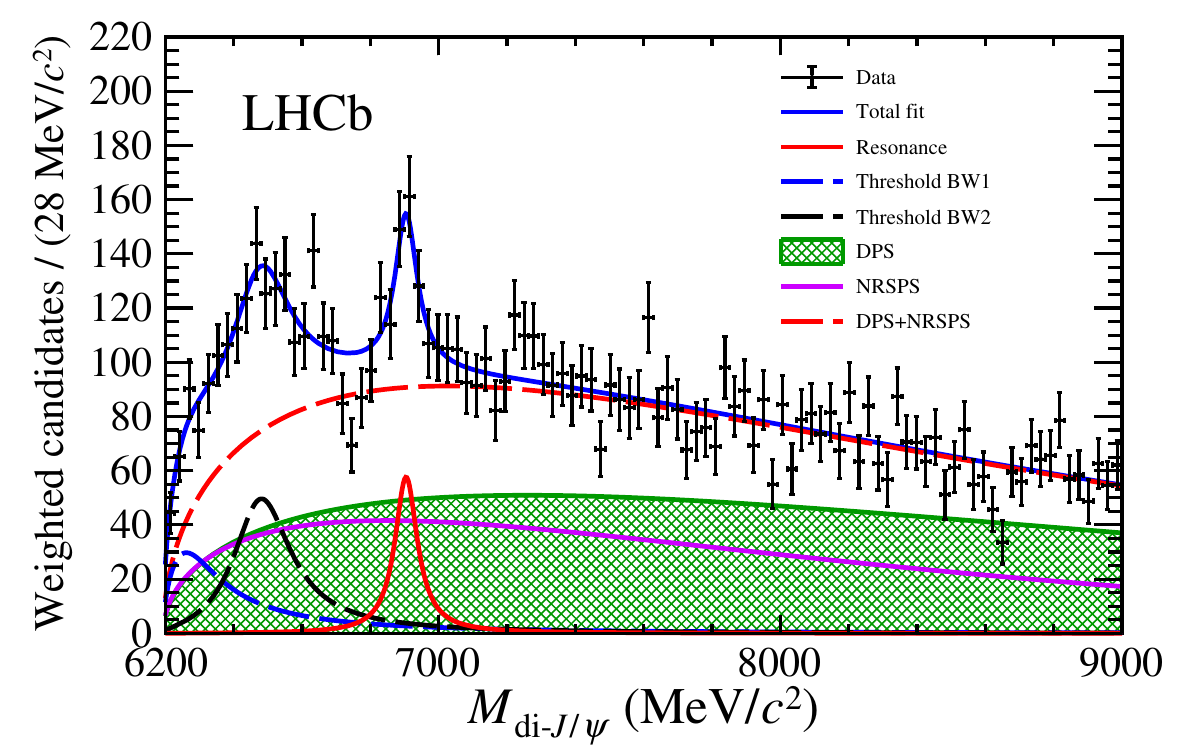}}
   \subfigure[]{\includegraphics[width=0.32991\linewidth]{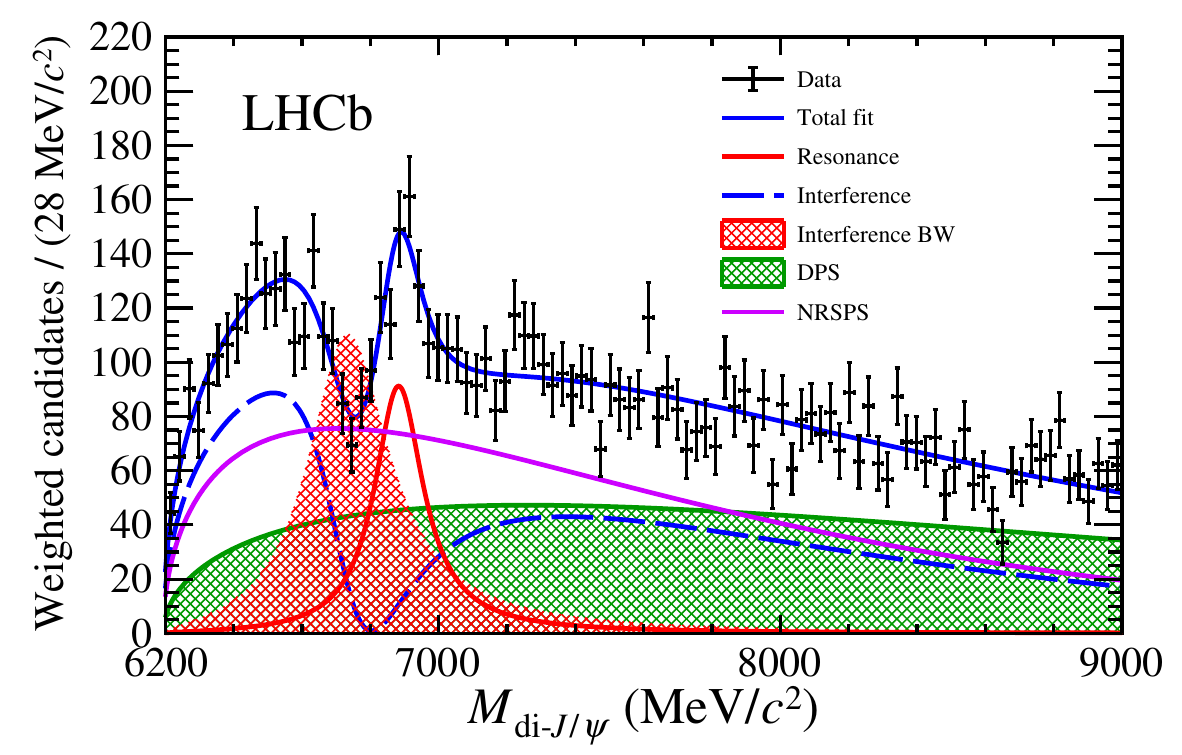}}
\vspace*{-0.1cm}
\end{center}
\caption{Di-$J/\psi$ invariant mass spectra with $p_T^{\mathrm{di-}J/\psi}>5.2$ GeV from LHCb, showing fits with (a) NRSPS+DPS, (b) model I, and (c) model II~\cite{LHCb:2020bwg}.}\label{fig:LHCbTccccfit}
\end{figure}

In 2020, the LHCb Collaboration reported the observation of a narrow structure near 6.9 GeV and a broad enhancement just above the di-$J/\psi$ threshold, using $pp$ collision data at $\sqrt{s}=7,8,13$ TeV (9 fb$^{-1}$)~\cite{LHCb:2020bwg}. A fit with only non-resonant single-parton scattering (NRSPS) and double-parton scattering (DPS) failed to describe the spectrum (Fig.~\ref{fig:LHCbTccccfit}(a)). Including two threshold Breit-Wigner functions and one resonance near 6.9 GeV (model I) gave a good description, with the resonance parameters $M=6905\pm11\pm7$ MeV, $\Gamma=80\pm19$ MeV (Fig.~\ref{fig:LHCbTccccfit}(b)). An alternative model II introducing interference between the NRSPS and a threshold resonance improved the description of a dip near 6.75 GeV, yielding $M=6886\pm11$ MeV, $\Gamma=168\pm33$ MeV (Fig.~\ref{fig:LHCbTccccfit}(c)). Both models confirmed a new structure, denoted $X(6900)$.

\begin{figure*}[!htbp]  
\centering   
\includegraphics[width=0.49\textwidth]{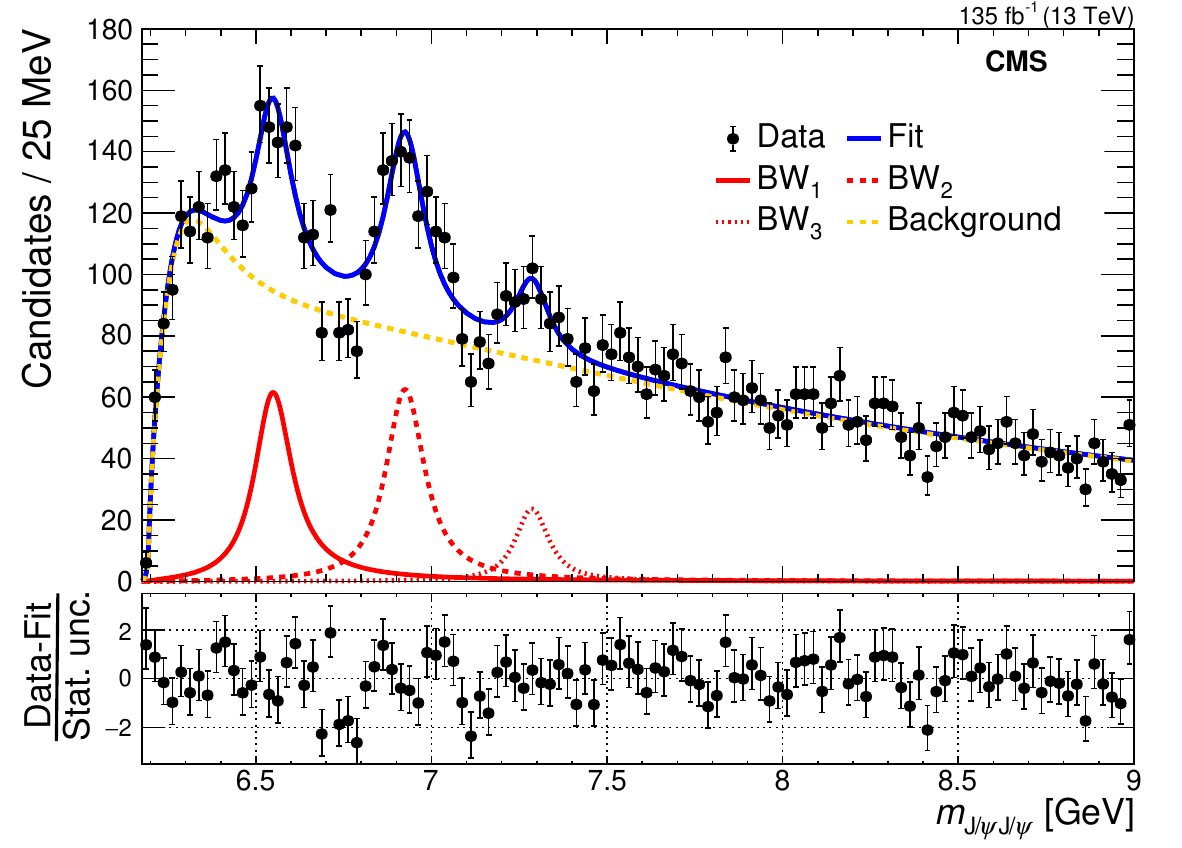} 
\includegraphics[width=0.49\textwidth]{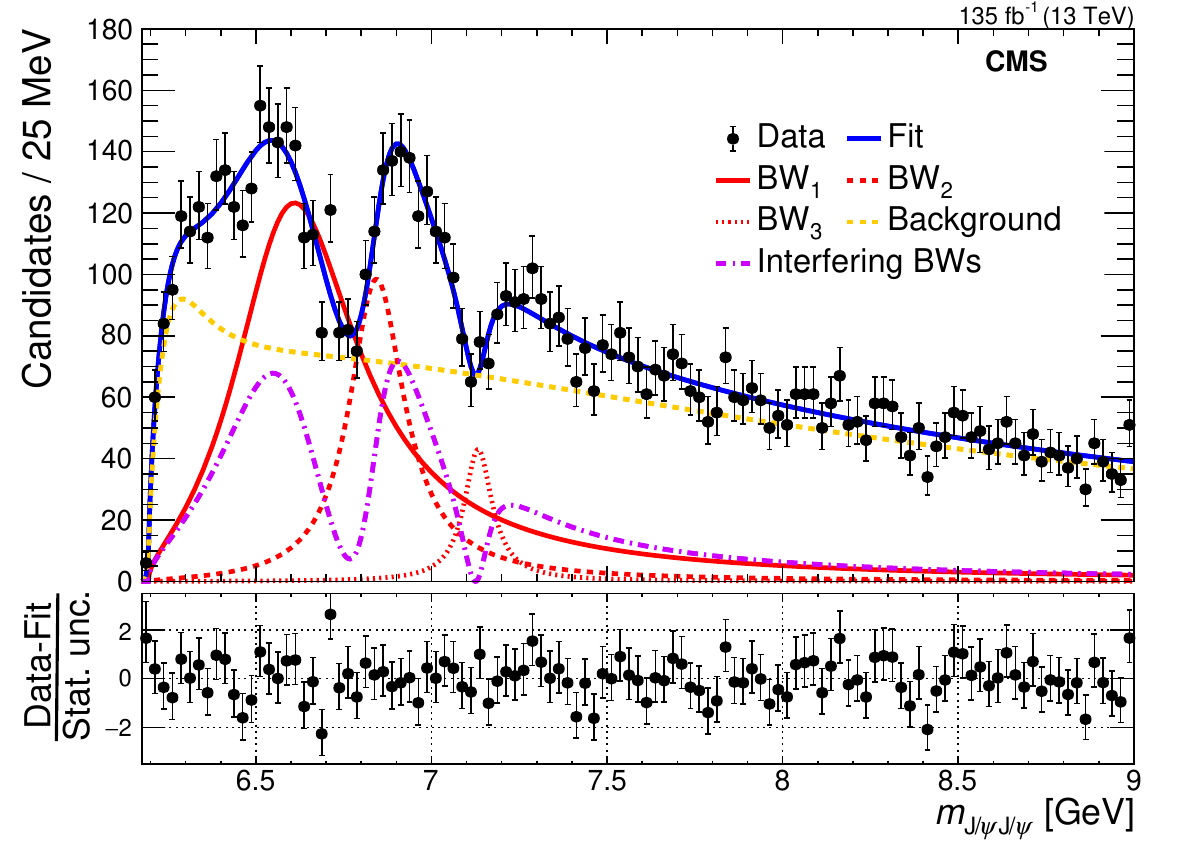} 
\caption{CMS di-$J/\psi$ mass spectrum fitted with three Breit-Wigners without (left) and with (right) interference~\cite{CMS:2023owd}.}
\label{fig:CMSTccccfit}
\end{figure*}

Subsequently, CMS and ATLAS confirmed the existence of $X(6900)$. In 2022, CMS analyzed 135 fb$^{-1}$ of 13 TeV data and reported three structures: $X(6600)$, $X(6900)$ and $X(7100)$~\cite{CMS:2023owd}, which are presented in Fig.~\ref{fig:CMSTccccfit}. Two fit scenarios were used: one with three incoherent Breit-Wigners plus background, and another including interference among them. The resonant parameters are listed in Table~\ref{tab:CMSTccccfit}. Interference significantly improved the description of dips near 6.75 and 7.15 GeV, though it altered the extracted masses and widths.

\begin{table}[htb]
\centering
\caption{Fitted masses and widths (MeV) of the three structures from CMS~\cite{CMS:2023owd}.}
\renewcommand{\arraystretch}{1.3}
\begin{tabular*}{1.0\textwidth}{l@{\extracolsep{\fill}}ccccc}
\hline
                &                       & $\mathrm{BW_1}$               & $\mathrm{BW_2}$           & $\mathrm{BW_3}$           \\ 
\hline
No interference & $m$ (MeV)       & $6552\pm10\pm12$              & $6927\pm9\pm4$            & $7287^{+20}_{-18}\pm5$    \\   
                & $\Gamma$ (MeV)   & $124^{+32}_{-26}\pm33$        & $122^{+24}_{-21}\pm18$    & $95^{+59}_{-40}\pm19$     \\ 
                & Significance & $6.5\sigma$ & $9.4\sigma$ & $4.1\sigma$ \\[1ex]
Interference    & $m$ (MeV)        & $6638^{+43+16}_{-38-31}$      & $6847^{+44+48}_{-28-20}$  & $7134^{+48+41}_{-25-15}$  \\    
                & $\Gamma$ (MeV)   & $440^{+230+110}_{-200-240}$   & $191^{+66+25}_{-49-17}$   & $97^{+40+29}_{-29-26}$    \\
                & Significance & $7.9\sigma$ & $9.8\sigma$ & $4.7\sigma$ \\
                \hline
\end{tabular*}
\label{tab:CMSTccccfit}
\end{table}

\begin{figure}[htbp]
\centering
\includegraphics[width=0.40\textwidth]{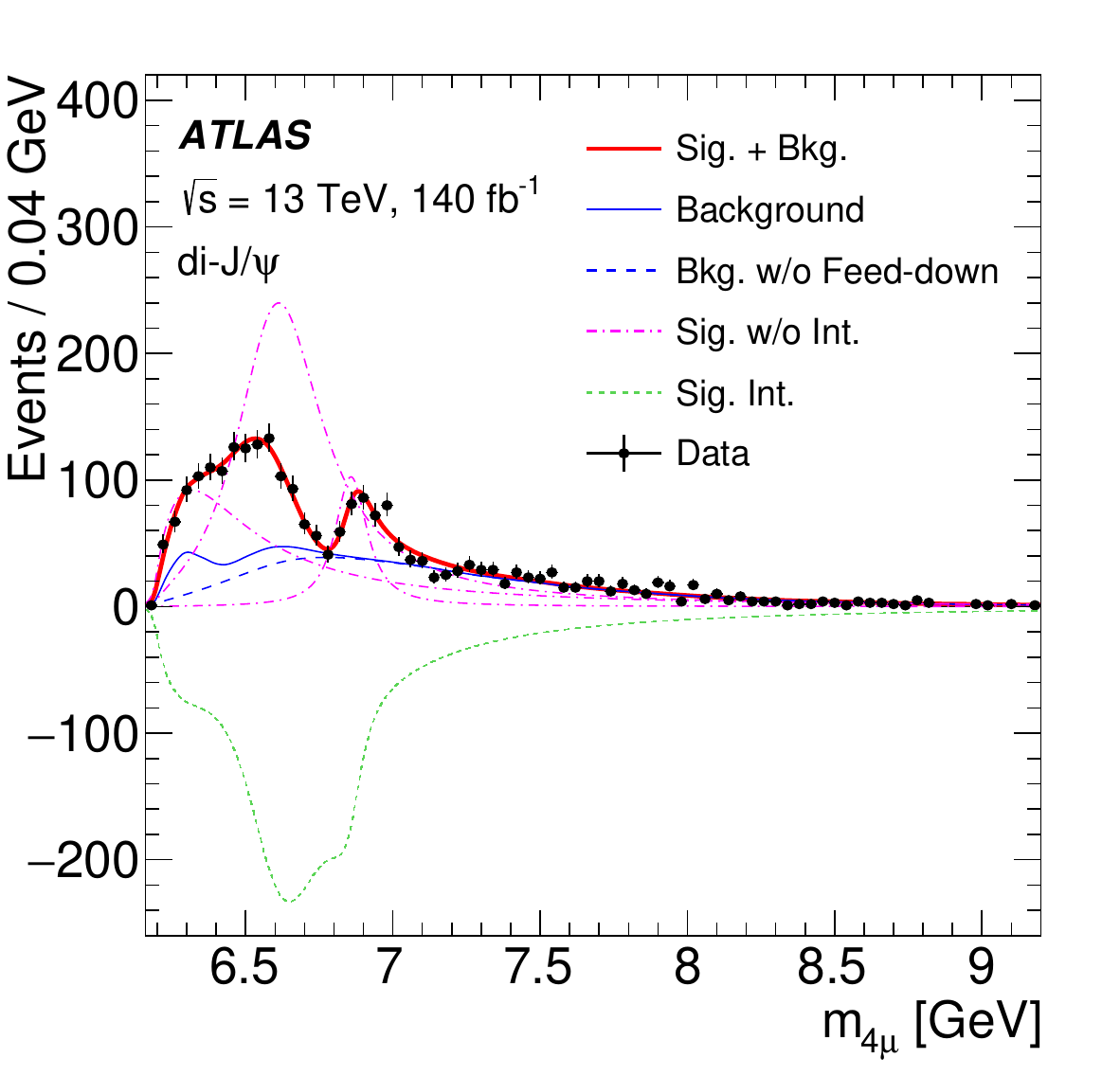}
\put(-50, 40){\textbf{(a)}}
\includegraphics[width=0.40\textwidth]{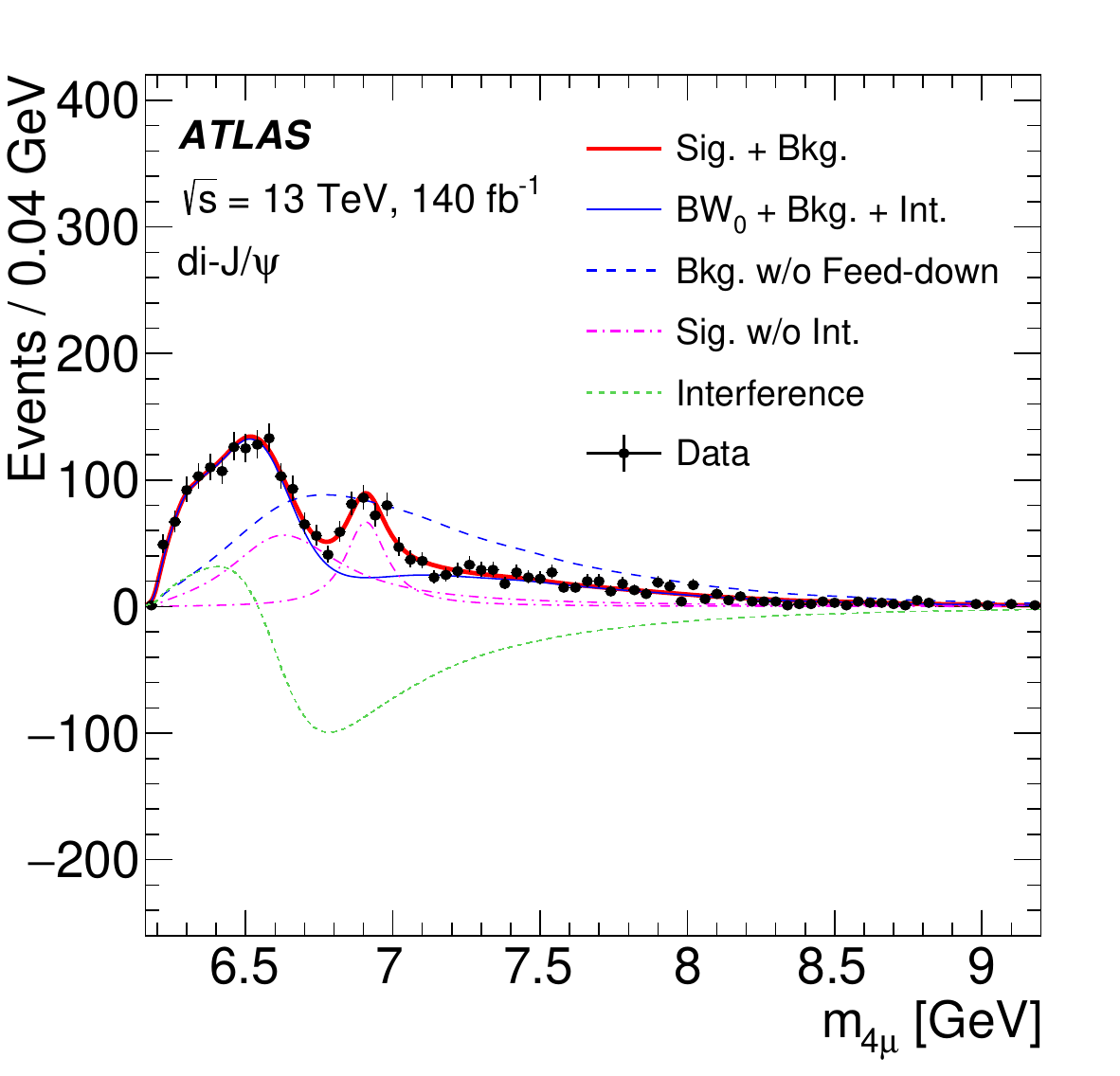}
\put(-50, 40){\textbf{(b)}}\\
\includegraphics[width=0.40\textwidth]{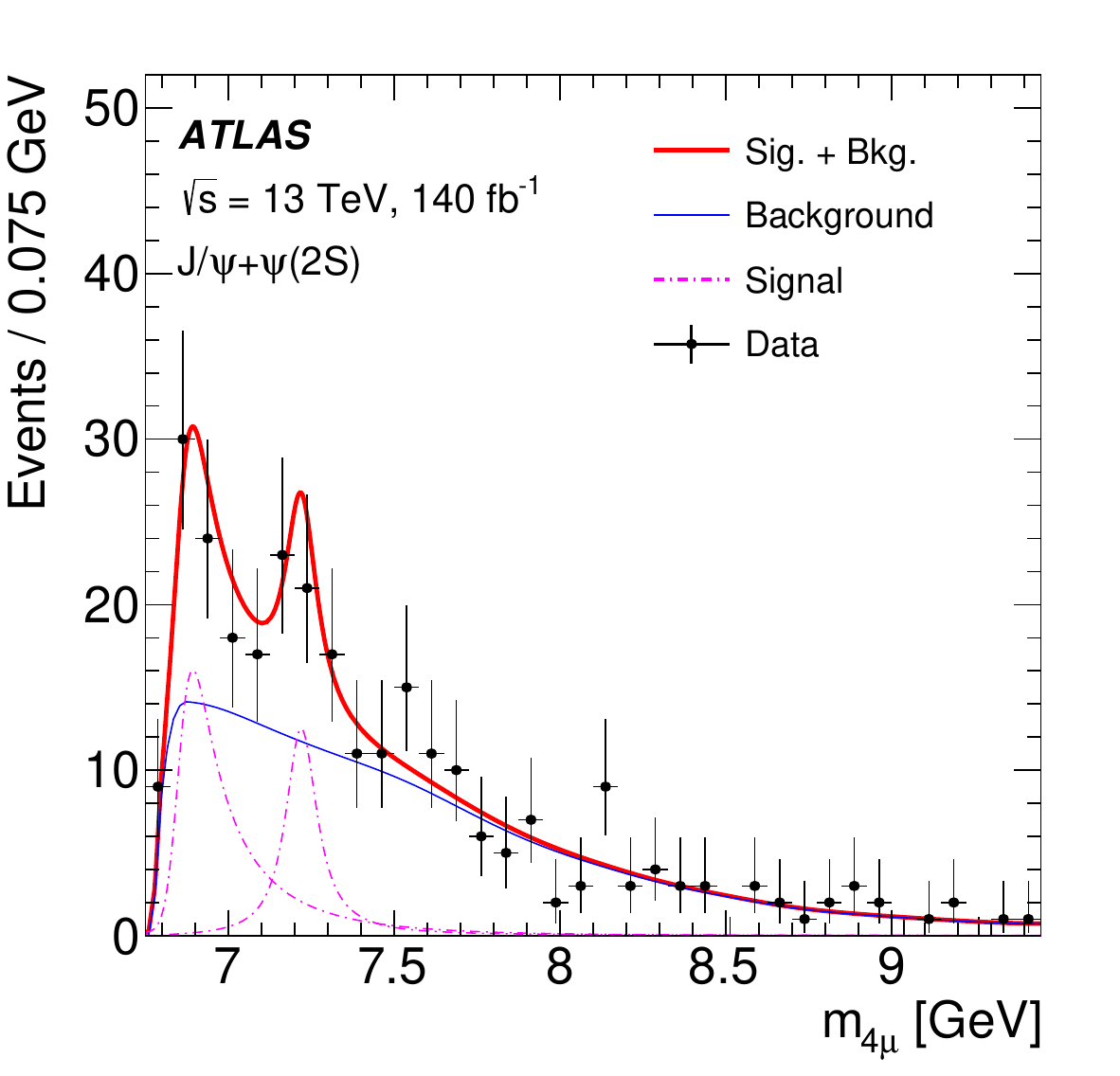}
\put(-50, 60){\textbf{(c)}}
\includegraphics[width=0.40\textwidth]{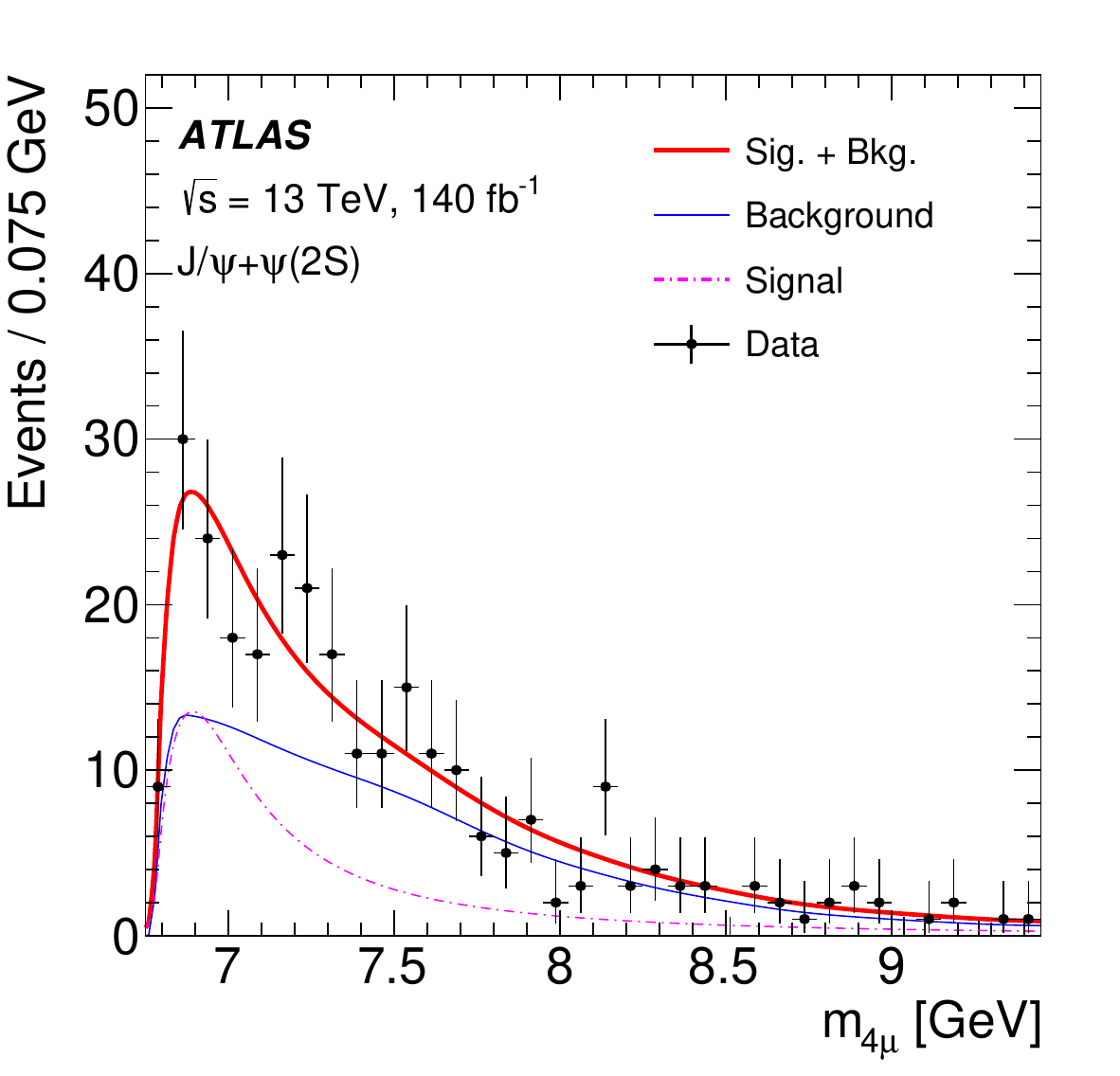}
\put(-50, 60){\textbf{(d)}}
\caption{Fitted spectra in di-$J/\psi$ channel with models A (a), B (b), and in $J/\psi+\psi(2S)$ channel with models $\alpha$ (c) and $\beta$ (d), by ATLAS~\cite{ATLAS:2023bft}. \iffalse Fit results for models A (a), B (b), $\alpha$ (c) and $\beta$ (d) are shown. The purple dash-dotted lines represent the components of individual resonances, and the green short dashed ones represent the interferences among them. \fi}
\label{fig:ATLASTccccfit}
\end{figure}

ATLAS also analyzed di-$J/\psi$ and $J/\psi+\psi(2S)$ channels using 140 fb$^{-1}$ at 13 TeV~\cite{ATLAS:2023bft}, whose invariant mass spectra are shown in Fig.~\ref{fig:ATLASTccccfit}. For di-$J/\psi$, two models were employed: model A included three interfering resonances; model B considered interference between one resonance and the SPS background, requiring only two resonances. Both models confirmed $X(6600)$ and $X(6900)$, but $X(7100)$ was not clearly seen. In the $J/\psi+\psi(2S)$ channel, two similar fit models $\alpha$ and $\beta$ yielded modest significances ($\sim 4\sigma$) for structures near 6.9–7.2 GeV (Table~\ref{tab:ATLASTccccfit}).

\begin{table}[htbp]
\caption{ATLAS fit results on di-$J/\psi$ and $J/\psi\psi(2S)$ spectra in GeV~\cite{ATLAS:2023bft}.}
\label{tab:ATLASTccccfit}
\centering
\renewcommand{\arraystretch}{1.2}
\begin{tabular*}{1.0\textwidth}{l@{\extracolsep{\fill}}ccc}
\hline
di-$J/\psi$ & model A & model B \\ 
\hline
$m_0$ & $6.41\pm 0.08_{-0.03}^{+0.08}$ & $6.65\pm 0.02_{-0.02}^{+0.03}$ \\
$\Gamma_0$ & $0.59\pm 0.35_{-0.20}^{+0.12}$ & $0.44\pm 0.05_{-0.05}^{+0.06}$ \\
$m_1$ & $6.63\pm 0.05_{-0.01}^{+0.08}$ & \multirow{2}{*}{---} \\
$\Gamma_1$ & $0.35\pm 0.11_{-0.04}^{+0.11}$&  \\
$m_2$ & $6.86\pm 0.03_{-0.02}^{+0.01}$ & $6.91\pm 0.01\pm 0.01$ \\
$\Gamma_2$ & $0.11\pm 0.05_{-0.01}^{+0.02}$ & $0.15\pm 0.03\pm 0.01$ \\
\hline
$J/\psi$+$\psi$(2S) & model $\alpha$ & model $\beta$ \\ 
\hline
$m_3$ & $7.22\pm 0.03_{-0.04}^{+0.01}$ & $6.96\pm 0.05\pm 0.03$ \\
$\Gamma_3$ & $0.09\pm 0.06_{-0.05}^{+0.06}$ & $0.51\pm 0.17_{-0.10}^{+0.11}$ \\
\hline
\end{tabular*}
\end{table}
\begin{table}[h!]
\centering
\caption{Updated fit results of the $J/\psi J/\psi$ mass spectrum with interference (Run~2+Run~3) in MeV
(uncertainties are statistical followed by systematic), from CMS~\cite{CMS:2025xwt}.}
\renewcommand{\arraystretch}{1.3}
\begin{tabular*}{1.0\textwidth}{l@{\extracolsep{\fill}}ccccc}
\hline
                &                       & $\mathrm{BW_1}$               & $\mathrm{BW_2}$           & $\mathrm{BW_3}$           \\ 
\hline
Interference    & $m$ (MeV)        & $6593^{+15}_{-14}\pm 25$      & $6847^{+10}_{-10}\pm 15$  & $7173^{+9}_{-10}\pm 13$  \\    
(Run 2 + Run 3) & $\Gamma$ (MeV)   & $446^{+66}_{-54}\pm 87$   & $135^{+16}_{-14}\pm 14$   & $73^{+18}_{-15}\pm 10$    \\
                & Significance & $15.2\sigma$ & $16.7\sigma$ & $7.7\sigma$ \\
                \hline
\end{tabular*}
\label{tab:CMSTccccfit3}
\end{table}

In 2024, CMS updated the di-$J/\psi$ analysis with 315 fb$^{-1}$ (Run~2+Run~3)~\cite{CMS:2025xwt}. Using an interference model similar to the earlier one, they obtained consistent but more precise parameters (Table~\ref{tab:CMSTccccfit3}), with all three structures now exceeding $5\sigma$ significance. The dips near 6.75 and 7.15 GeV were confirmed with high significance ($9.7\sigma$ and $6.5\sigma$), and the interference pattern suggests the three states likely share the same $J^{PC}$.

In 2025, CMS also reported the $J/\psi+\psi(2S)$ channel using 314 fb$^{-1}$~\cite{CMS:2025vnq}. An interference fit with two resonances gave $m_{X(6900)}=6876^{+46+110}_{-29-110}$ MeV, $\Gamma_{X(6900)}=253^{+290+120}_{-100-120}$ MeV and $m_{X(7100)}=7169^{+26+74}_{-52-70}$ MeV, $\Gamma_{X(7100)}=154^{+110+140}_{-82-160}$ MeV, consistent with the di-$J/\psi$ results within uncertainties. The significances were $7.9\sigma$ for $X(6900)$ and $4.0\sigma$ for $X(7100)$.
Furthermore, a CMS analysis of angular distributions using a matrix-element likelihood approach favored the quantum numbers $2^{++}$ for $X(6600)$, $X(6900)$ and $X(7100)$ over other $J^{PC}$ hypotheses~\cite{CMS:2025fpt}.

Through analyses conducted by different experiments, the extracted resonance parameters for structures observed in the di-$J/\psi$ invariant mass spectrum differ, as presented in Tables \ref{tab:CMSTccccfit}, \ref{tab:ATLASTccccfit}, and \ref{tab:CMSTccccfit3}\footnote{We need to clarify that they represent typical data analysis, which is distinct from physics analysis. As noted in Ref. \cite{Wang:2020axi}, as theorists, we only trust the experimental data provided by experimentalists. To a certain extent, exploring the underlying mechanism behind the experimental data should be left to theorists.}. This leads to ambiguity to make conclusion when only comparing theoretical results with the experimental resonance parameters.

\subsection{Theoretical status and challenges in identifying genuine multiquark states}

The theoretical exploration of heavy flavor tetraquarks, particularly the fully heavy systems with the \(QQ\bar{Q}\bar{Q}\) configuration, has a history spanning nearly five decades. Initial speculations emerged soon after the "November Revolution" of particle physics \cite{E598:1974sol,SLAC-SP-017:1974ind}, with early works attempting to address the possible existence of such states, albeit with primitive numerical results due to the lack of experimental guidance \cite{Iwasaki:1975pv, Iwasaki:1976cn, Chao:1980dv, Badalian:1985es, Ader:1981db, Li:1983ru}. The landscape transformed with the rapid advancement of charm physics, where precise measurements of charmonium spectra across various \(J^{PC}\) quantum numbers enabled the refinement of theoretical models. This progress ignited a renewed wave of interest in fully heavy exotic states~\cite{Lloyd:2003yc, Liu:2019zuc, Wang:2019rdo, Chen:2020lgj,Debastiani:2017msn, Bedolla:2019zwg,Richard:2017vry, Berezhnoy:2011xn, Wu:2016vtq, Karliner:2016zzc,Chen:2016jxd, Wang:2018poa,Anwar:2017toa,Bai:2016int,Debastiani:2017xcr}, as models and their parameters could now be more reliably constrained.

A significant early contribution came from Chao {\it et al.}, who proposed that peculiar structures in \(R(e^{+}e^{-}\rightarrow \text{hadrons})\) around \(\sqrt{s}=6-7\ \text{GeV}\) might be attributed to predicted \(P\)-wave \((cc)\)-\((\bar{c}\bar{c})\) states produced between 6.4 and 6.8 GeV, predominantly decaying into charmed mesons \cite{Chao:1980dv}. Subsequent foundational studies employed the potential model \cite{Badalian:1985es,Ader:1981db} and the MIT bag model within the Born-Oppenheimer approximation \cite{Heller:1985cb}. While a non-relativistic potential analysis initially found no \(QQ\bar{Q}\bar{Q}\) bound state \cite{Zouzou:1986qh}, Lloyd {\it et al.} later reported several closely-lying states using a parameterized Hamiltonian with a large oscillator basis \cite{Lloyd:2003yc}.

Starting around 2003, systematic calculations of the fully heavy tetraquark spectrum proliferated. These employed diverse frameworks including constituent quark models \changelabel{that based on the one-gluon-exchange in addition with a phenomenological confinement}~\cite{Lloyd:2003yc, Liu:2019zuc, Wang:2019rdo, Chen:2020lgj}, diquark\changelabel{-antidiquark models}~\cite{Debastiani:2017msn, Bedolla:2019zwg}, analyses of hyperfine splitting and color-magnetic interactions \cite{Richard:2017vry, Berezhnoy:2011xn, Wu:2016vtq, Karliner:2016zzc}, and QCD sum rules \cite{Wang:2017jtz,Chen:2016jxd, Wang:2018poa}. Mass predictions for the fully charmed tetraquark \changelabel{in these approaches} varied widely from about 5.8 to 7.4 GeV, \changelabel{which reflects the very strong model dependence of the conclusion. However, there still exists} a common feature from quark model approaches: the lowest-lying state was typically an \(S\)-wave configuration, and due to the Pauli principle, it was almost invariably assigned the quantum numbers \(J^{PC}=0^{++}\). Representative mass predictions from this era include those by Karliner {\it et al.} (\(\mathrm{M}(X_{cc\bar{c}\bar{c}})=6192\pm25\ \text{MeV}\), \(\mathrm{M}(X_{bb\bar{b}\bar{b}})=18826\pm25\ \text{MeV}\) for \(0^{++}\)) \cite{Karliner:2016zzc}, Anwar {\it et al.} (\((18.72\pm0.02)\ \text{GeV}\) for \(bb\bar{b}\bar{b}\)) \cite{Anwar:2017toa}, and Bai {\it et al.} (using diffusion Monte Carlo for \(bb\bar{b}\bar{b}\)) \cite{Bai:2016int}. Further insights came from an updated Cornell model applied to a diquark-antidiquark configuration \cite{Debastiani:2017xcr} and moment QCD sum rules supporting compact states, suggesting searches in the \(J/\psi J/\psi\) and \(\eta_{c}(1S)\eta_{c}(1S)\) channels \cite{Chen:2016jxd}. Silva {\it et al.} studied the fully-heavy tetraquarks in the vacuum and in a hot environment~\cite{Silva:2025bdg}. In the addendum of Ref.~\cite{Silva:2025bdg}, the result of the $J^{PC}=2^{++}$ $S$-wave tetraquark is consistent with the measurement from the CMS Collaboration~\cite{CMS:2025fpt}.

The experimental quest for these states intensified with accumulating data. Early hints came from CMS, which reported a structure near 18.4 GeV in the four-lepton channel (global significance \(3.6\sigma\)), proposed as a fully-bottom tetraquark candidate \cite{CMS:2016liw}, though it was not confirmed later \cite{CMS:2020qwa}. LHCb also found no significant \(bb\bar{b}\bar{b}\) signal in the \(\Upsilon(1S)\mu^{+}\mu^{-}\) spectrum between 17.5 and 20.0 GeV \cite{LHCb:2018uwm}. A major breakthrough arrived in 2020 with the LHCb discovery of the narrow \(X(6900)\) resonance in the di-\(J/\psi\) spectrum, alongside a broad structure (6.2–6.8 GeV) and a potential peak near 7.3 GeV \cite{LHCb:2020bwg}. Subsequent measurements by ATLAS and CMS confirmed \(X(6900)\) and revealed additional structures \cite{CMS:2023owd, Xu:2022rnl, Zhang:2022toq}, sparking extensive theoretical discussion on their nature 
.

Post-discovery, a significant volume of theoretical work aimed to interpret the observed di-\(J/\psi\) structures. Since these states lie above the di-\(J/\psi\) threshold, many studies started to focus on their internal dynamics and resonant production mechanisms. \changelabel{Apart from the possible $\mathbf{8_{c\bar{c}}} \otimes \bar{\mathbf{8}}_{c\bar{c}}$ assignments~\cite{Wang:2021mma,Yang:2020wkh,Tang:2024zvf}, a prominent interpretation is to posit them as excited compact tetraquark states~\iffalse\cite{Deng:2020iqw, Wang:2020ols, Lu:2020cns, Sonnenschein:2020nwn, Giron:2020wpx, Karliner:2020dta, Wang:2020dlo, Zhu:2020xni, Zhao:2020jvl, Ke:2021iyh, Mutuk:2021hmi, Li:2021ygk, Wang:2021kfv, Wang:2021mma, Liu:2021rtn, Santowsky:2021bhy, Wang:2022xja, Zhang:2022qtp, Faustov:2022mvs, Dong:2022sef, Yu:2022lak, Agaev:2023wua, Chen:2024bpz, Lin:2024olg, Galkin:2023wox}\fi. For example, for the most experimentally determined state $X(6900)$, it not only can be interpreted as ground $L=1$ or $L=2$ orbital excited tetraquark states with possible quantum numbers $0^{\pm}$, $1^{\pm}$, $2^{\pm}$~\cite{Deng:2020iqw,Sonnenschein:2020nwn,Wang:2021kfv,Liu:2021rtn,Zhang:2022qtp,Faustov:2022mvs,Yu:2022lak,Chen:2024bpz,Lin:2024olg,Galkin:2023wox}, but also can be explained as the first or second radial excited state of the scalar, axialvector, or tensor all charmed tetraquark state~\cite{Wang:2020ols,Lu:2020cns,Sonnenschein:2020nwn,Giron:2020wpx,Karliner:2020dta,Wang:2020dlo,Zhao:2020jvl,Ke:2021iyh,Mutuk:2021hmi,Li:2021ygk,Santowsky:2021bhy,Wang:2022xja}, or even as a radial and orbital simultaneously excited $2P$ state~\cite{Zhu:2020xni,Dong:2022sef} \iffalse\cite{Deng:2020iqw, Wang:2020ols, Lu:2020cns, Sonnenschein:2020nwn, Giron:2020wpx, Karliner:2020dta, Wang:2020dlo, Zhu:2020xni, Zhao:2020jvl, Ke:2021iyh, Mutuk:2021hmi, Li:2021ygk, Wang:2021kfv, Wang:2021mma, Liu:2021rtn, Santowsky:2021bhy, Wang:2022xja, An:2022qpt, Zhang:2022qtp, Faustov:2022mvs, Dong:2022sef, Yu:2022lak, Agaev:2023wua, Chen:2024bpz, Lin:2024olg, NgaOngodo:2025mkt,Galkin:2023wox}\fi. Also, a}lternative ideas include the interpretation of \(X(6900)\) as a tetracharm hybrid \cite{Wan:2020fsk,Tang:2024kmh,Tang:2025ept} or a $\chi_{c0}\chi_{c0}$ molecule~\cite{Agaev:2023wua}. Crucially, to reproduce resonant masses in the experimental range, studies highlight the essential role of coupled-channel effects. \changelabel{In tetraquark picture, it is not only found that the single $K-$type space configuration can generate states with masses vary from 6.4 to 6.9 GeV and thus may give contributions to the formations of the experimental observed $T_{cc\bar{c}\bar{c}}$~\cite{Yang:2021hrb}, but also found that hidden-color channels in addition with meson-meson channels can provide additional resonant mechanisms because they can generate states near those in diquark-antidiquark picture~\cite{Jin:2020jfc,Wang:2023jqs,Wu:2024tif}. Such a conclusion can be especially reflected in an analysis based on the Kohn-type variational method, where only meson-meson channels are considered~\cite{Ortega:2023pmr}, and the result do show that coupled channels that contain excited charmonia can generate states with experimental masses. Therefore,} it underscores that the formation of these states, mostly interpreted as compact tetraquark states, likely involves complex interactions among components with different structures.

However, the fundamental question of stability for the ground-state fully heavy tetraquark remains contentious. Some studies, such as those using an updated Cornell model \cite{Debastiani:2017msn}, an extended chromomagnetic model \cite{Weng:2020jao}, or the quark delocalization color screening model \cite{Jin:2020jfc}, suggest bound states can exist. Conversely, other comprehensive analyses conclude the opposite. For instance, \changelabel{after using also the extended chromomagnetic model, There is no stable $S-$wave all charmed tetraquark state~\cite{An:2022qpt}}. Similar conclusion are also obtained trough nonrelativistic quark models, where the masses of ground states are above the lowest meson-meson thresholds in nonrelativistic quark models \cite{Wang:2019rdo,Meng:2024yhu}. Lattice nonrelativistic QCD calculations found no \(bb\bar{b}\bar{b}\) state below the bottomonium-pair threshold \cite{Hughes:2017xie}, and Richard {\it et al.} argued that a properly treated four-body problem in a standard quark model does not yield a stable \(QQ\bar{Q}\bar{Q}\) configuration \cite{Richard:2018yrm}. Wu {\it et al.} performed a benchmark calculations of fully heavy compact tetraquark states using three different quark
potential models, which does not find the existence of any bound states for both $cc\bar{c}\bar{c}$ and $bb\bar{b}\bar{b}$ systems \cite{Wu:2024euj}. In light of the current lack of a theoretical consensus on stability, it would be valuable to engage in more discussion \changelabel{with more precise and reasonable model parameters in the future}.

This situation highlights the profound challenges in identifying genuine multiquark states. The plethora of theoretical investigations within quark models faces a critical problem: most interpretations heavily rely on matching predicted masses to experimentally extracted resonance parameters, which themselves vary significantly across different experimental analyses. Furthermore, theoretical predictions are sensitive to model details and parameter fitting schemes, leading to substantial discrepancies. Also, the eigenvalue problem approach in quark-level calculations often yields a dense spectrum of excited states with similar masses but different configurations, complicating unambiguous state assignment. Therefore, confirming the nature of these states demands more than just mass comparisons. Future progress will require richer experimental input—such as detailed decay properties and production rates—coupled with a deeper theoretical understanding of the interactions between heavy quarkonia and the non-perturbative dynamics governing multiquark systems.

\subsection{Employing scattering between charmonia to understand these observed enhancements in the di-$J/\psi$ invariant mass spectrum}

Experimental analyses on the enhancement structures observed in the di-$J/\psi$ invariant mass spectrum typically extract masses and widths of possible resonances by fitting these peaks with phenomenological functions. However, the resulting resonance parameters vary noticeably across different experiments and depend sensitively on the specific fitting assumptions, as illustrated in the above subsection. Quark-model interpretations then rely on these extracted parameters to test whether compact $cc\bar{c}\bar{c}$ tetraquark configurations can accommodate the observed states, rendering their conclusions contingent upon the reliability of the resonance parameter extraction procedure. An alternative and more direct approach is to interpret the experimental line shape through dynamical scattering mechanisms. Rather than invoking pre-extracted resonance parameters, this perspective considers the rescattering of intermediate charmonium pairs directly produced in $pp$ collisions, which subsequently transform into the final $J/\psi J/\psi$ state. By connecting the observed invariant mass distribution directly to the underlying scattering amplitudes, this approach circumvents the intermediate step of resonance parameter extraction from experiments and its associated model dependence, thereby offering a more immediate way to explore the nature and structure of these fully-charm enhancements.


\begin{figure}[htbp]
\begin{center}
   \includegraphics[width=0.66\linewidth]{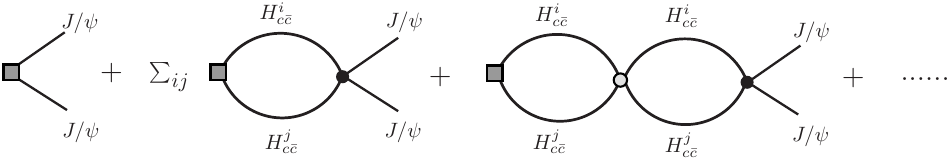}
\end{center}
\caption{The schematic diagrams for the production mechanism of a double charmonium $J/\psi J/\psi$, where $H_{c\bar{c}}$ stands for allowed intermediate charmonium states, such as $\eta_c$, $J/\psi$, $\chi_{cJ}$ with $J = 0, 1, 2$, etc. Here, the gray rectangle corresponds to direct production of a double charmonium in hadron collisions, fit I or II means the cutoff parameters are set as the same or independent~\cite{Wang:2022jmb}.}\label{fig:wangfitdiagram}
\end{figure}

A preliminary study was carried out by Refs.~\cite{Wang:2020wrp,Wang:2022jmb} through the rescattering mechanism, where the final $J/\psi J/\psi$ invariant mass spectrum were produced by the rescattering of the intermediate charmonia $H_{c\bar{c}}$ as given in Fig.~\ref{fig:wangfitdiagram}.  The key dynamical quantity is the scalar two-point loop integral describing the rescattering of an intermediate charmonium pair $H_{c\bar{c}}^i H_{c\bar{c}}^j$ into the same pair. In the nonrelativistic limit, this loop function takes the form
\[
L_{ij}(m_{J/\psi J/\psi}) = \frac{-1}{4m_i m_j} \left( -\frac{2\mu\alpha}{(2\pi)^{3/2}} + \frac{2\mu\sqrt{2\mu m_0}\left( \mathrm{erf}\left[\frac{\sqrt{8\mu m_0}}{\alpha}\right] - i \right)}{2\pi e^{\frac{8\mu m_0}{\alpha^2}}} \right),
\]
where $\mu = m_i m_j/(m_i+m_j)$, $m_0 = m_{J/\psi J/\psi} - m_i - m_j$, and $\alpha$ is a cutoff parameter of regulating the ultraviolet behavior.
By summing the infinite series of loop diagrams (see Fig.~\ref{fig:wangfitdiagram}), the full production amplitude for a given intermediate channel $ij$ is obtained as
\[
\mathcal{A}_{ij}^2(m_{J/\psi J/\psi}) = \frac{g_{ij}^2 L_{ij}(m_{J/\psi J/\psi})^2}{\bigl(1 + C_{ij}L_{ij}(m_{J/\psi J/\psi})\bigr)^2} \frac{e^{c_0 m_{J/\psi J/\psi}} p_{J/\psi}^{2L+1}}{m_{J/\psi J/\psi}},
\]
where $p_{J/\psi}$ is the momentum of the final $J/\psi$, $L=0$ for parity-even systems and $L=1$ for parity-odd ones, and $C_{ij}$ represents the coupling strength of the rescattering interaction. The total scattering amplitude is then constructed by summing coherently over the continuum term and all relevant channels with appropriate phase factors.

\begin{figure}[htb!]
\centering
\begin{minipage}{0.42\linewidth}
\centering
(I)\quad\includegraphics[width=\linewidth]{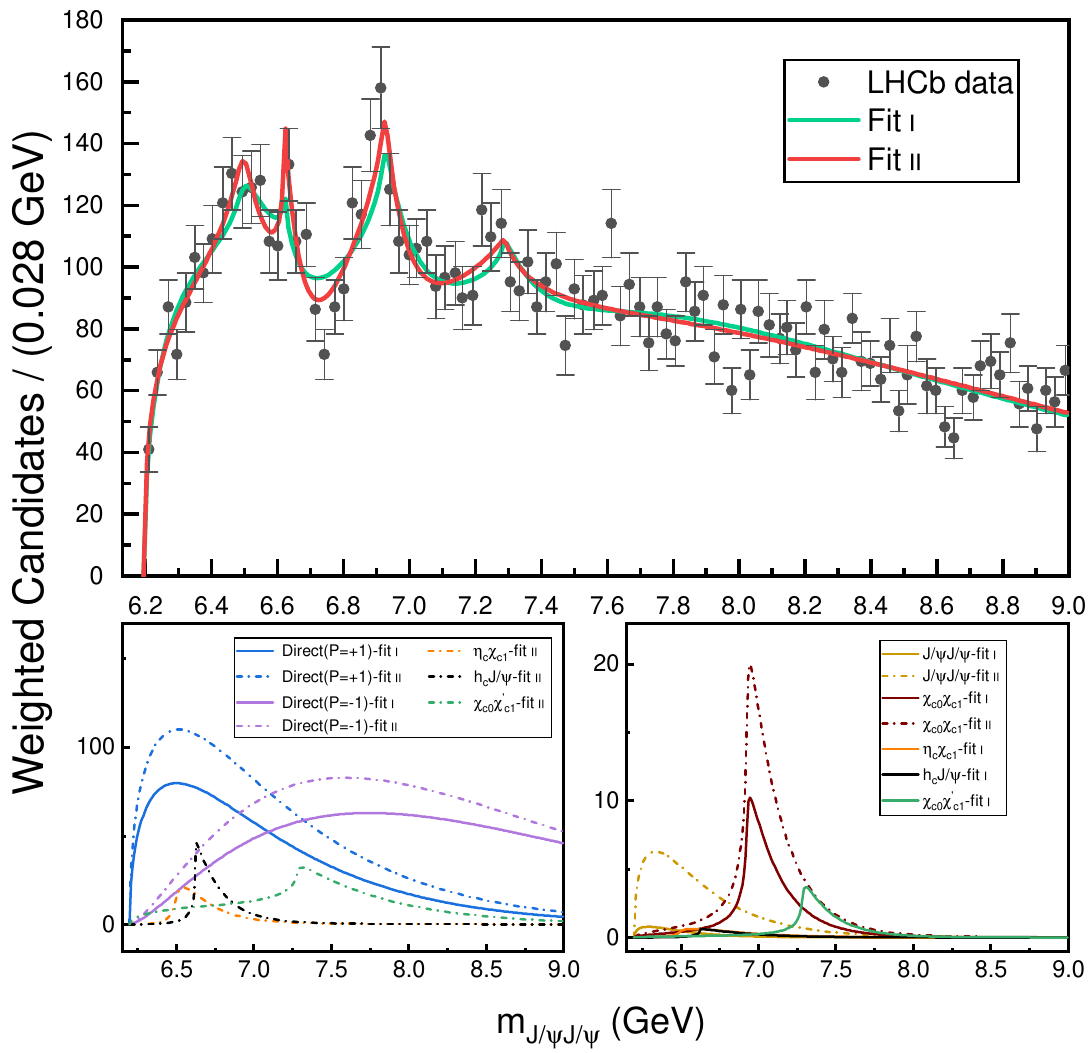}
\end{minipage}
\hfill
\begin{minipage}{0.48\linewidth}
\centering
(II)\quad\includegraphics[width=\linewidth]{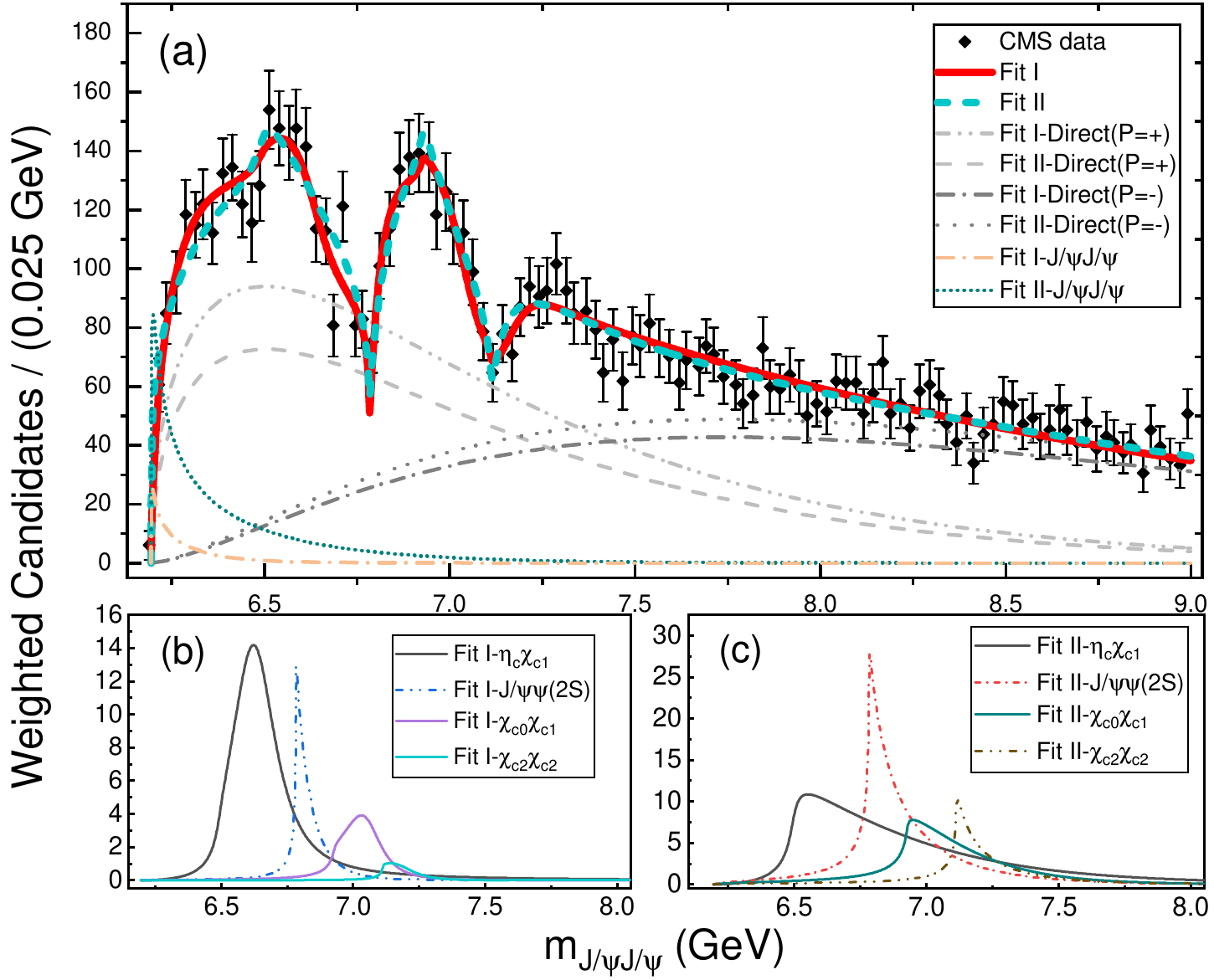}
\end{minipage}
\caption{Fit results to the experimental results, where (I) for LHCb data fit~\cite{Wang:2020wrp} and (II) for CMS data fit~\cite{Wang:2022jmb}.}
\label{fig:wangfitresult}
\end{figure}

After considering the intermediate states as $\eta_c\chi_{c1}$, $h_c J/\psi$, $\chi_{c0}\chi_{c1}$, and $J/\psi J/\psi$ itself, Ref.~\cite{Wang:2020wrp} successfully reproduced the LHCb data as in Fig.~\ref{fig:wangfitresult}~(I), and then, a improved work in Ref.~\cite{Wang:2022jmb} also reproduced the CMS data by considering the $\eta_c\chi_{c1}$, $J/\psi \psi(3686)$, $\chi_{c0}\chi_{c1}$, $\chi_{c2}\chi_{c2}$, and $J/\psi J/\psi$ channels as in Fig.~\ref{fig:wangfitresult}~(II). Especially, during the fit on the CMS data, Ref.~\cite{Wang:2022jmb} found three resonant poles that mainly related to the $\chi_{c1}\eta_c$, $\chi_{c0}\chi_{c1}$, and $\chi_{c2}\chi_{c2}$ channels, respectively, which can be assigned to the experimental enhancements around 6.6 and 7.1 GeV. In addition, several virtual poles below the $J/\psi J/\psi$, $J/\psi \psi(3686)$, $\eta_c\chi_{c1}$, and $\chi_{c2} \chi_{c2}$ thresholds were also found, which indicates that the interaction between these charmonia should be attractive, but not strong enough to form bound states. Meanwhile, the extracted positive scattering lengths in Ref.~\cite{Wang:2022jmb} with the following scattering amplitude $A_0$, i.e., 
$
A_0^{-1}=\frac{1}{a_0}-i\sqrt{2\mu E}$,
where $E$ is the binding energy, and $\mu$ is the reduced mass of the intermediate two body channel, also proves that the attractive property of the interaction between these charmonia.

Furthermore, Ref.~\cite{Wang:2022jmb} provides predictions for the $J/\psi\psi(3686)$ invariant mass spectrum, which are compared with preliminary ATLAS data, showing good agreement and lending additional support to the proposed dynamical mechanism. The authors in Ref.~\cite{Wang:2022jmb} emphasize that future lattice QCD calculations will be essential to quantitatively determine the coupling strengths of double charmonium scattering and to clarify the nature of these structures.

\begin{figure}[htb!]
\begin{center}
   \subfigure[]{\includegraphics[width=0.48\linewidth]{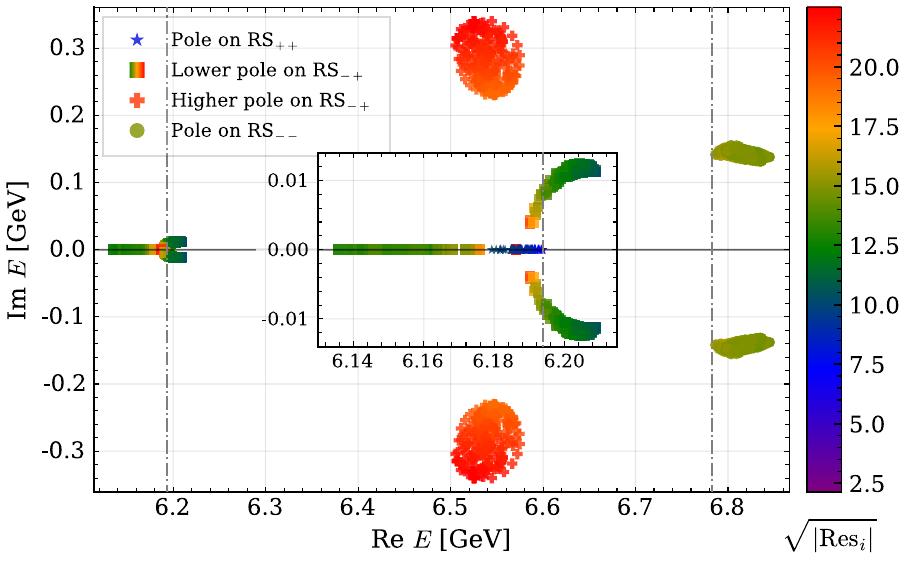}}
   \subfigure[]{\includegraphics[width=0.48\linewidth]{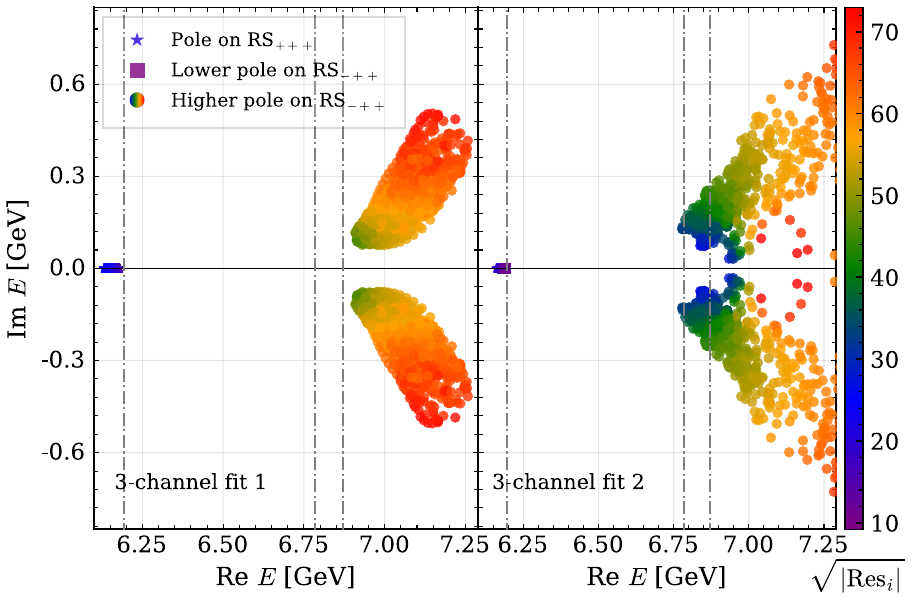}}\\
\vspace*{-0.1cm}
\end{center}
\caption{Extracted pole positions from Refs.~\cite{Dong:2020nwy}, where left is for the two channel model, right is for the three channel model~\cite{Dong:2020nwy}.}\label{fig:dongfitresult}
\end{figure}

\begin{figure}[htb!]
\begin{center}
   \subfigure[]{\includegraphics[width=0.48\linewidth]{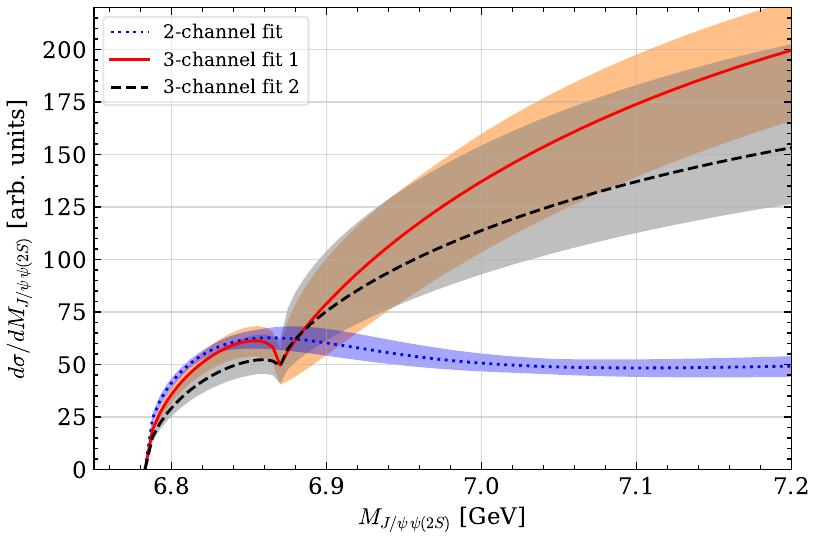}}
   \subfigure[]{\includegraphics[width=0.505\linewidth]{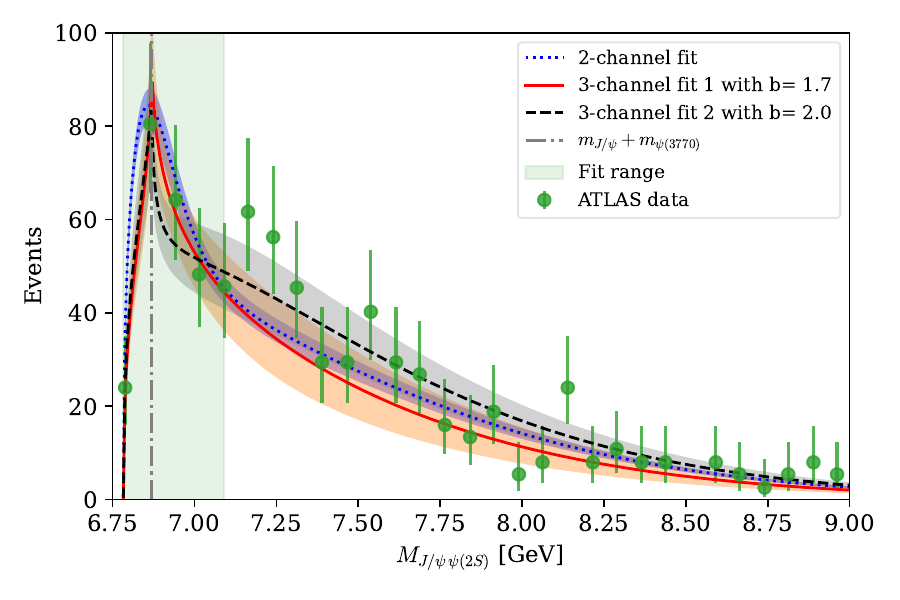}}\\
\vspace*{-0.1cm}
\end{center}
\caption{The $J/\psi \psi(2S)$ invariant mass spectrum, where (a) is the predictions given by Ref.~\cite{Dong:2020nwy}, (b) is a comparison with ATLAS data presented in Ref.~\cite{Song:2024ykq}.}\label{fig:songfitresult}
\end{figure}

Concurrently, Ref.~\cite{Dong:2020nwy} also performed an analysis of the di-$J/\psi$ spectrum measured by LHCb. Different from Refs.~\cite{Wang:2020wrp,Wang:2022jmb}, where the scattering amplitude is directly parameterized, Ref.~\cite{Dong:2020nwy} obtained the transition matrix through the Bethe-Salpeter equation, where the interactions between charmonia were described by contact terms, and only $\{J/\psi J/\psi, J/\psi \psi(3686)\}$ channels were considered. Two fit schemes were considered, one is a two channel model employing $\{J/\psi J/\psi, J/\psi \psi(3686)\}$ as parameterized potential as
\begin{eqnarray}
        V_{2 \mathrm{ch}}(E)=\left(\begin{array}{cc}
        a_1+b_1 k_1^2 & c \\
        c & a_2+b_2 k_2^2
        \end{array}\right),
\end{eqnarray}
where $a_i$, $b_i$, and $c$ are free parameters, and the introduction of the momentum dependence in the potential is to produce nontrivial structures above the higher threshold~\cite{Dong:2020nwy}. The other is a $\{J/\psi J/\psi ,J/\psi \psi(3686),J/\psi \psi(3770)\}$ three channel model, and the potential contains only energy independent contact terms as
\begin{eqnarray}
        V_{3 \mathrm{ch}}(E)=\left(\begin{array}{lll}
a_{11} & a_{12} & a_{13} \\
a_{12} & a_{22} & a_{23} \\
a_{13} & a_{23} & a_{33}
\end{array}\right).
\end{eqnarray}
It turns out that although much less channels are considered, the fit results on the LHCb data are also quiet well. In addition, the extracted pole positions are also given as in Fig.~\ref{fig:dongfitresult} with the two upper panels.
Obviously, in these two fit schemes, the extracted pole positions are quite different, which is reflected by the existence of a pole with wide width around 6.6 GeV, along with the description on the structures around 6.8 GeV, since either the positions and the stabilities of the poles there are different. This means, if there actually exists $X(6600)$, and the description on the dip and $X(6900)$ in the experimental data, is model dependent here. However, in both models, a pole below $J/\psi J/\psi$ threshold is also found, and the extracted scattering parameters also indicate that the interaction between $J/\psi J/\psi$ is attractive, which is consistent with the results given by Ref.~\cite{Wang:2022jmb}, although the compositeness analysis through $\bar{X}_A$ that defined as
$\bar{X}_A=\left(1+2\left|r_0 / a_0\right|\right)^{-1 / 2}$
are a little different, where the two channel model indicates that the pole around 6.2 GeV is a compact state, while the three channel model supports it as a molecule. Such a phenomena indicates that the current experimental data is still insufficient to determine the properties of the experimental enhancements, i.e., fitting the experimental data only is far from enough to understand these states. After the release of CMS and ATLAS data, Ref.~\cite{Song:2024ykq} fit the experimental data again with the same procedure, and the results of poles are quite similar to those given by Ref.~\cite{Dong:2020nwy}. However, as presented in Fig.~\ref{fig:songfitresult}, the descriptions on the $J/\psi \psi(3686)$ invariant mass spectrum are too different, even in the same work with different models.

\begin{figure}[htb!]
\begin{center}
   \subfigure[]{\includegraphics[width=0.48\linewidth]{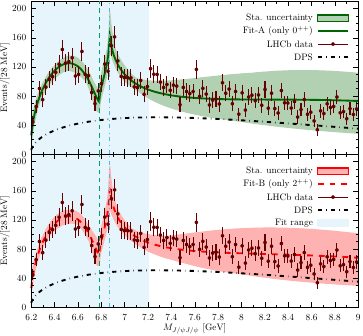}}
   \subfigure[]{\includegraphics[width=0.48\linewidth]{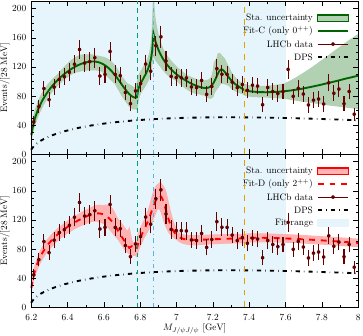}}\\
   \subfigure[]{\includegraphics[width=0.48\linewidth]{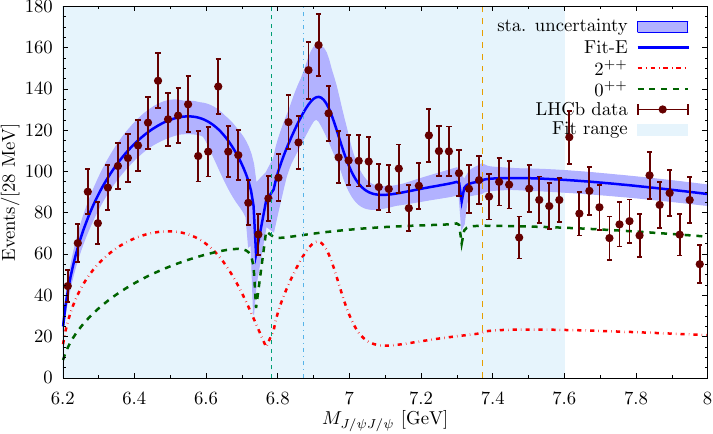}}
   \subfigure[]{\includegraphics[width=0.48\linewidth]{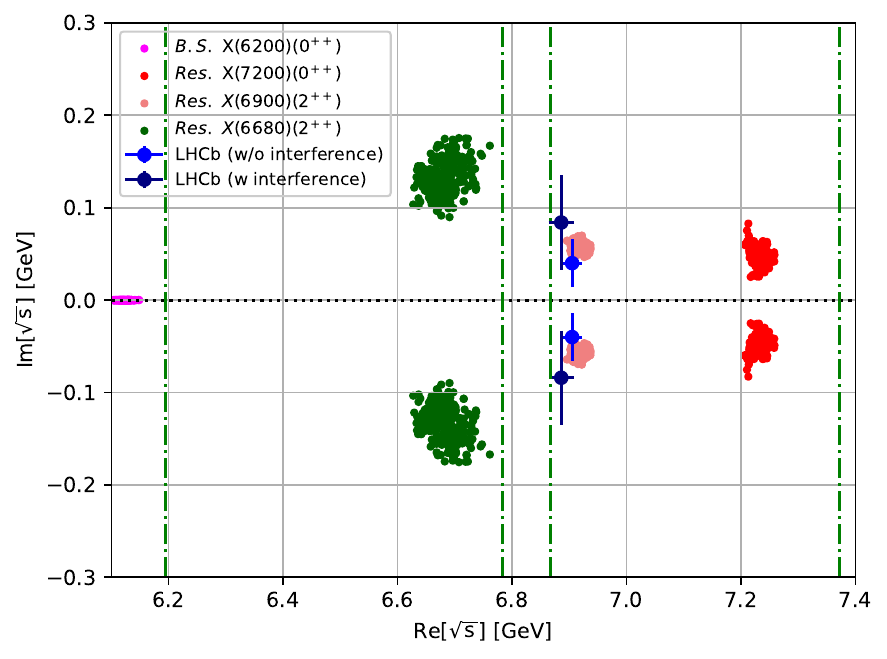}}\\
\vspace*{-0.1cm}
\end{center}
\caption{Fit results to the LHCb data. Here, the top left figure considers $\{J/\psi J/\psi ,J/\psi \psi(3686),J/\psi \psi(3770)\}$, the top right one is for $\{J/\psi J/\psi ,J/\psi \psi(3686),J/\psi \psi(3770),\psi(3686) \psi(3686)\}$ channels, the bottom left result is a combined fit with both $0^{++}$ and $2^{++}$ potentials in $\{J/\psi J/\psi ,J/\psi \psi(3686),\psi(3686) \psi(3686)\}$ channels, and the one located at bottom right is the poles for the four channel model~\cite{Liang:2021fzr}.}\label{fig:liangfitresult}
\end{figure}

\begin{figure}[htb!]
\begin{center}
   \includegraphics[width=0.48\linewidth]{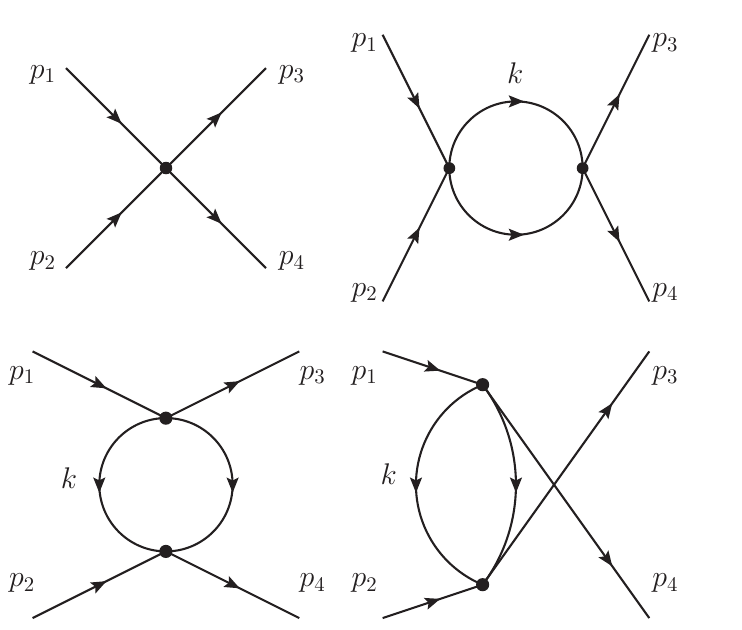}
\end{center}
\caption{Feynman diagrams of the scattering amplitudes of charmonia given in Ref.~\cite{Zhou:2022xpd}.}\label{zhoudiagram}
\end{figure}

Then, to further study the quantum numbers of the experimental observed structures in di-$J/\psi$ spectrum, Ref.~\cite{Liang:2021fzr} performed a partial wave analysis on the LHCb data, where the analysis scheme is similar to Ref.~\cite{Dong:2020nwy} as presented in Fig.~\ref{fig:liangfitresult}, but replacing the potential from just free parameters to the following form, where for a $V_1(p_1,\epsilon_1)V_2(p_2,\epsilon_2)\to V_3(p_3,\epsilon_3)V_4(p_4,\epsilon_4)$ process, as
$        V_{i j}=\mathcal{C}_1 \epsilon_1 \cdot \epsilon_2 \epsilon_3^{\dagger} \cdot \epsilon_4^{\dagger}+\mathcal{C}_2 \epsilon_1 \cdot \epsilon_3^{\dagger} \epsilon_2 \cdot \epsilon_4^{\dagger}+\mathcal{C}_3 \epsilon_1 \cdot \epsilon_4^{\dagger} \epsilon_2 \cdot \epsilon_3^{\dagger}$,
with $\mathcal{C}_i$ being constants that related to the undetermined coupling constants of the 4-vector contact interactions. After doing a partial wave projection on the potential to the $0^{++}$ and $2^{++}$ quantum numbers, Ref.~\cite{Liang:2021fzr} performed a series of fits on the LHCb data, and the corresponding pole positions were extracted, as presented in Fig.~\ref{fig:liangfitresult}.
Similarly, the LHCb data can also be well described in both $0^{++}$ and $2^{++}$ quantum numbers. In addition, a state below the $J/\psi J/\psi$ threshold is also found, whose quantum number is $0^{++}$, i.e., the $J=0$ partial wave potential between $J/\psi J/\psi$ should be attractive. However, it immediately arises the same question as in Ref.~\cite{Dong:2020nwy,Song:2024ykq} that what is exactly the interpretation of $X(6600)$, since it appears as a pole in $2^{++}$ fit but disappears in $0^{++}$ configuration, but whether it is a pole or not, the experimental data can be nicely fit. Another one is from the comprehensive comparisons between the figures in Fig.~\ref{fig:liangfitresult}, which indicates that the $J/\psi \psi(3770)$ channel, about 330 MeV away from the location of $X(7200)$, will have a significant influence on the formation of $X(7200)$, since it is so clear in the four channel fit under $0^{++}$, but disappears in the combined fit that without $J/\psi \psi(3770)$ channel. Whatsmore, the $X(6900)$ here is found dominantly contributed by a $2^{++}$ resonance. However, Refs.~\cite{Zhou:2022xpd,Kuang:2023vac} found that, after considering the scattering amplitudes up to next-to-leading order as shown in Fig.~\ref{zhoudiagram}, under pole counting rule, $X(6900)$ changes to a $0^{++}$ compact state.

To decode the nature of $X(7200)$, Ref.~\cite{Niu:2022jqp} carried out an analysis on the experimental data, and pointed out that the controversy around 7.2 GeV among the three collaborations might be the key to further understand the fully charmed tetraquarks. Thus, considering$X(6600)$ is near the $\eta_c \eta_c^\prime$ threshold and $X(7200)$ is close to the $\eta_c^\prime \eta_c^\prime$ threshold, Ref.~\cite{Niu:2022jqp} carried out a more complex coupled channel analysis on the experimental data, where six channels as $\{\eta_c \eta_c, J/\psi J/\psi ,\eta_c \eta_c^\prime,J/\psi \psi(3686),\eta_c^\prime \eta_c^\prime,\psi(3686) \psi(3686)\}$ were considered. By adopting different fit schemes with $0^{++}$ and $2^{++}$ quantum numbers, although Ref.~\cite{Niu:2022jqp} could obtain good fit quality on the experimental data in all the cases, the extracted pole positions changed a lot, where for $X(6600)$, it appears as a pole only in one kind of $2^{++}$ fit scheme so that its existence is still questionable. For $X(6900)$, its pole appears in all situations, i.e., it can be either a $0^{++}$ or $2^{++}$ state, but the pole positions are a little different in these two quantum numbers. As for $X(7200)$, it appears only in the $0^{++}$ fit scheme, that is because to form a $2^{++}$ state in $\eta_c^\prime \eta_c^\prime$ channel, the partial wave must be $D$-wave, which is not easy to form a resonance. Furthermore, it turns out that after considering so many channels, the binding energy of the $0^{++}$ state mentioned above that below $J/\psi J/\psi$ threshold will become much deeper. additionally, its $2^{++}$ partner also shows up. 

Obviously, for $X(6600)$ and $X(7200)$, not only their quantum numbers are not determined, but also their existences are still questionable. Actually, even for the firstly observed $X(6900)$, by just using the fit approach, similar disagreements also existed. For example, apart from the controversy on its quantum number, Refs.~\cite{Wang:2020wrp,Zhuang:2021pci} proposed that $X(6900)$ might be only a cusp effect. In addition, for the pole property, except the compact state interpretation~\cite{Zhou:2022xpd,Cao:2020gul,Liu:2024pio,Guo:2020pvt}, molecular state interpretation might also be permitted~\cite{Lu:2023ccs}. Such a mess situation, which can spark to an entire inspection on the whole experimental spectra, indicates that, the current experimental data is still far from enough to allow us to just integrate all the possible degrees of freedom out to pinch the interactions between charmonia as contact terms, i.e., to truly understand the nature of these fully charmed states, we must try our best to make deeper exploration on the interaction mechanisms between charmonia under a more microscopical and detailed level. At least, we should find some reliable constraints on the fitted parameters from the first principle, if we use contact interactions.

\subsection{Exploration on microscopic mechanism of the interaction between charmonia}

\begin{figure}[h]
\centering
\subfigure[ ~$t$ channel process]{
\includegraphics[width=0.45\linewidth]{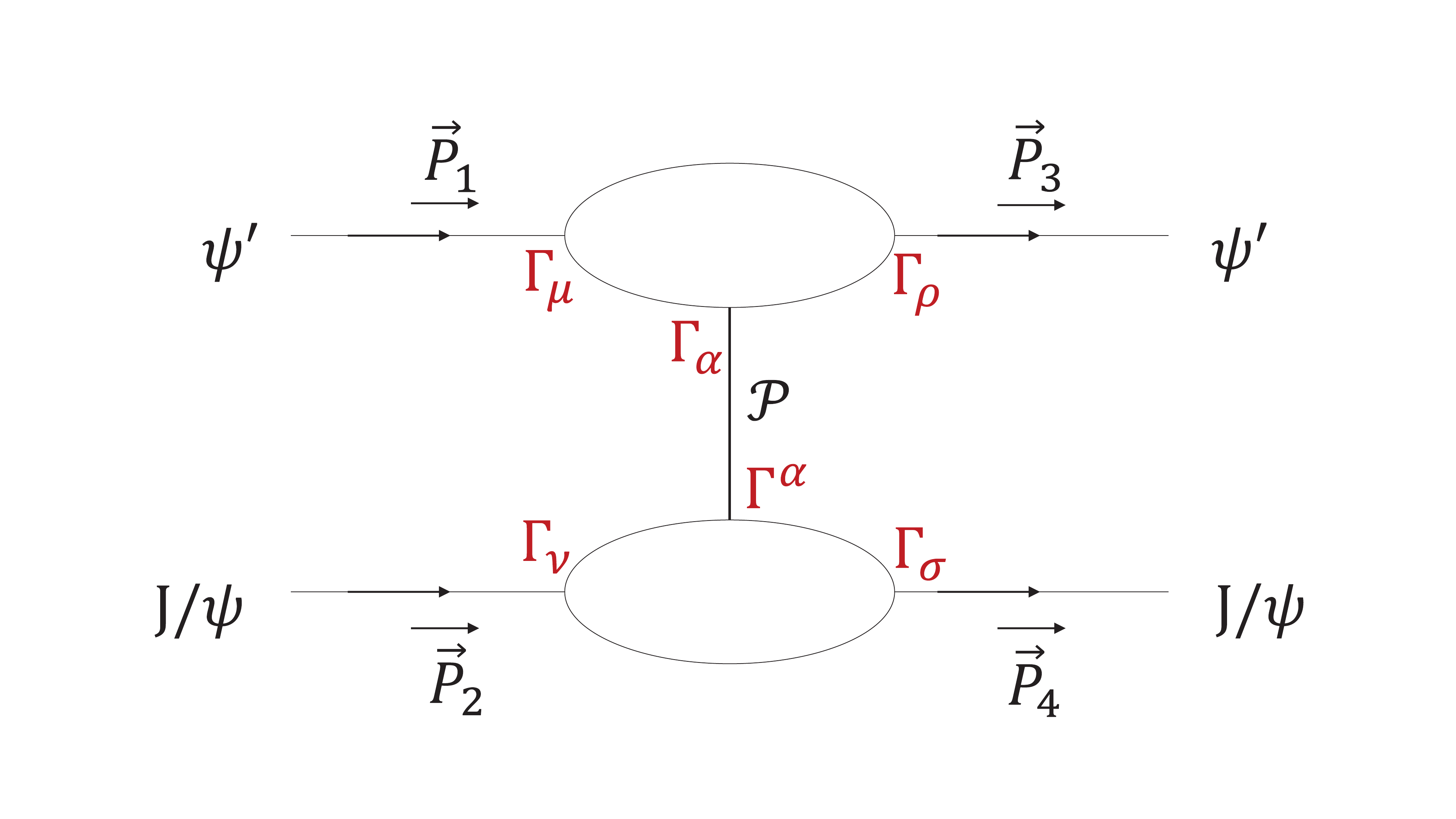}
}
\quad
\subfigure[ ~$u$ channel process]{
\includegraphics[width=0.45\linewidth]{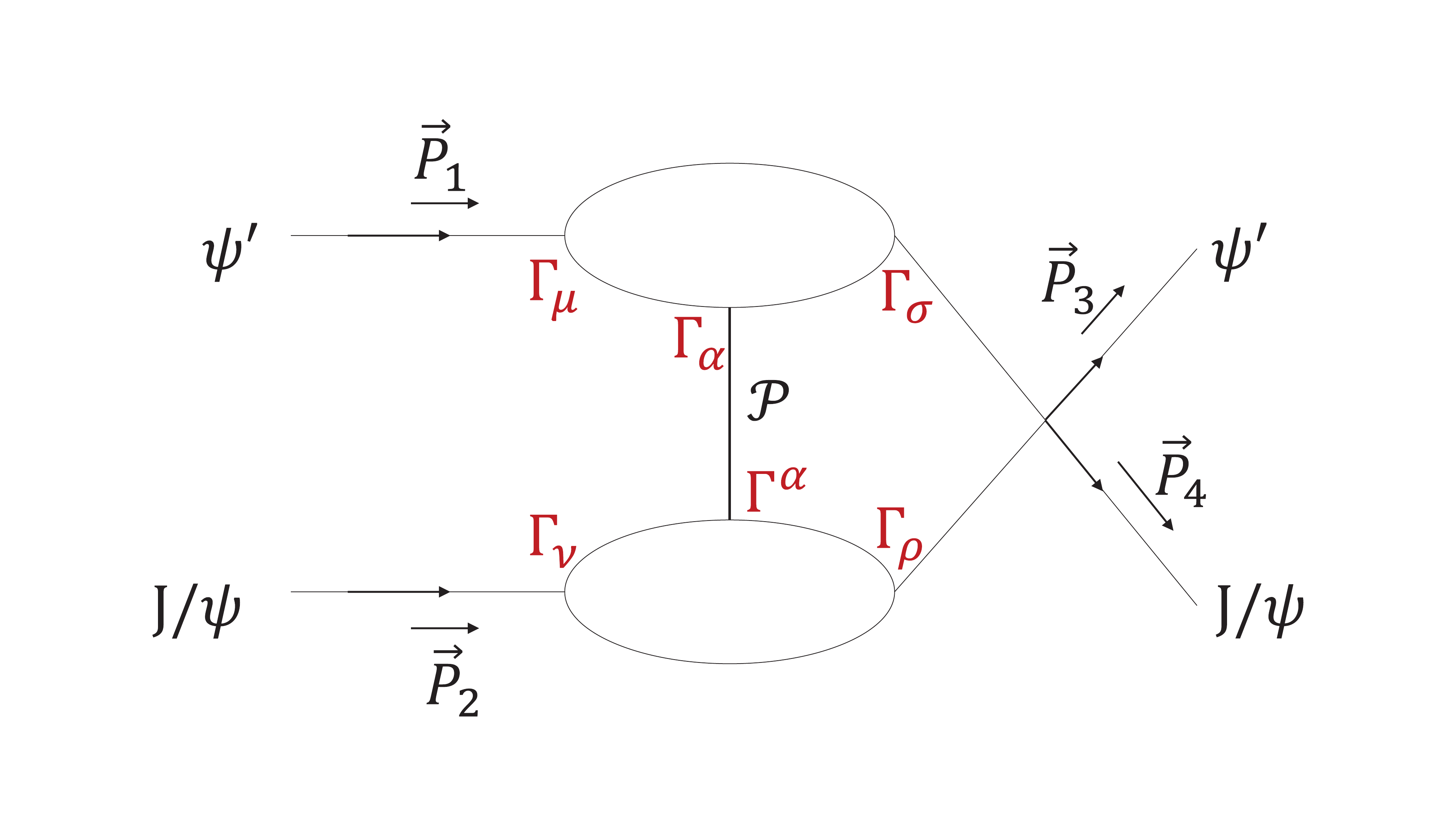}
}
\caption{Illustrative diagrams for (a) $t$ channel and (b) $u$ channel Pomeron exchanges in $J/ \psi \  \psi' \to J/\psi \  \psi' $~\cite{Gong:2020bmg}. Here, $\psi^\prime$ denotes $\psi(3686)$}
\label{fig:pomerondiagram}
\end{figure}

Unlike hadronic systems containing light quarks, where long-range interactions can be mediated by exchanges of light mesons such as pions, the fully heavy charmonium pairs lack such light degrees of freedom. This feature suggests that the interaction between charmonia may naturally arise from short-distance dynamics, although the full interaction picture could be more complex. A common effective field theory treatment in this context is to integrate out all possible degrees of freedom and parametrize the interaction via contact terms as introduced in the above subsection. However, integrating out all possible degrees of freedom to capture the interactions between charmonia as contact terms may not be sufficient to reveal the interaction mechanisms responsible for producing the fully charmed tetraquark structures, since this treatment reflects nearly no clear dynamics at this stage. 
Thus, further explorations are still needed, and this subsection is devoted to exploring various theoretical attempts in this direction. For this purpose, Ref.~\cite{Gong:2020bmg} proposed a microscopic model, where charmonia scatter in short distance through Pomeron exchange as in Fig.~\ref{fig:pomerondiagram}. After using the same scattering equation as Ref.~\cite{Dong:2020nwy}, but replacing the contact terms with the partial wave potentials derived from Fig.~\ref{fig:pomerondiagram}, it is found that this model can provide very strong contributions to the interactions between vector charmonia, since there all exists poles near thresholds, including $J/\psi J/\psi$, $J/\psi \psi(2S)$, and $\psi(2S) \psi(2S)$ channels. Furthermore, a comparison of the scattering cross section to the experimental data of LHCb is performed, in which the peak around 6.9 GeV is nicely reproduced, demonstrating the effectiveness of this model. In addition, it indicates that there may exist a sate around 6.3 GeV, although it maybe covered by the very wide structure. Furthermore, the $J/\psi \psi(3686)$ spectrum is also predicted, which shows a significant enhancement caused by the pole around $J/\psi \psi(3686)$ threshold.

Notably, in this model, the parameters, such as cutoffs, that relate to the interaction between charmonia and Pomeron are previously fixed by the typical size of charmonia, i.e., $\sim \hbar c / \Lambda \approx 0.3$ fm, what are changed just the relative ratios that channels contribute to the scattering. Absolutely, such an approach obtain a new insight into the dynamical interaction between charmonia. However, the Pomeron itself is a phenomenological object, in addition, such a model is still a short-distance interaction, which may compete with, even be absorbed by the model parameters, gluon exchange, makes it hard to examine.

\begin{figure}[t]
 \centering
 \includegraphics[width=0.66\linewidth]{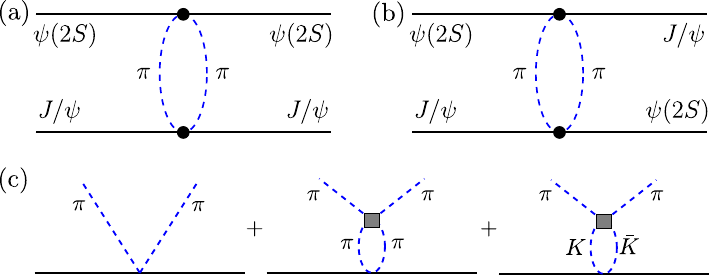}
 \caption{Two-pion exchanges between the $\psi(2S)$ and $J/\psi$, where (a) is for $t$-channel and (b) is for $u$-channel. (c) The contributions to the amplitude for the transition $\psi_\alpha\to \psi_\beta\pi\pi$ including the coupled-channel ($\pi\pi$ and $K\bar K$) FSI. In (c), the solid and dashed lines label charmonia and light mesons, respectively, the filled squares denote the $\pi\pi$ and $K\bar K$ interactions and the black dots mean that the interactions between the light mesons, as depicted in (c),  have been taken into account~\cite{Dong:2021lkh}.}
 \label{fig:twopiondiagram}
\end{figure}

To reveal the possible midrange or long-distance interaction between charmonia and explore the plausibility of the predicted $X(6200)$ in Ref.~\cite{Dong:2020nwy}, Ref.~\cite{Dong:2021lkh} proposed that two pion or two kaon exchange can also give considerable contributions in the scattering of charmonia. \changelabel{It starts from the point that quarkonia can interact with gluonic fields through the instantaneous dipole moments, which are created when the quarkonium emits a gluon transitioning into a virtual color-octet state followed by an emission of a second gluon and a return to the original quarkonium state~\cite{Brambilla:2015rqa}. Then, this two gluons transition can be matched onto the chiral theory as a two-pion exchange in the isoscalar-scalar channel, which will result into long range forces in QCD~\cite{Brambilla:2015rqa,Fujii:1999xn}. After generalizing this point to the SU(3) symmetry, two-kaon exchange should also be possible. Thus, the scattering between two charmonia can be constructed }as in Fig.~\ref{fig:twopiondiagram}, \changelabel{and the reason why $\eta\eta$ channel is neglected is that it is not essential for the $f_0$ states, while two-kaon exchange can be related to $f_0(980)$ and two-pion exchange can be related to both $f_0(980)$ and $\sigma~(f_0(500))$~\cite{Dong:2021lkh}.}

\begin{figure}
\centering
\includegraphics[width=0.66\linewidth]{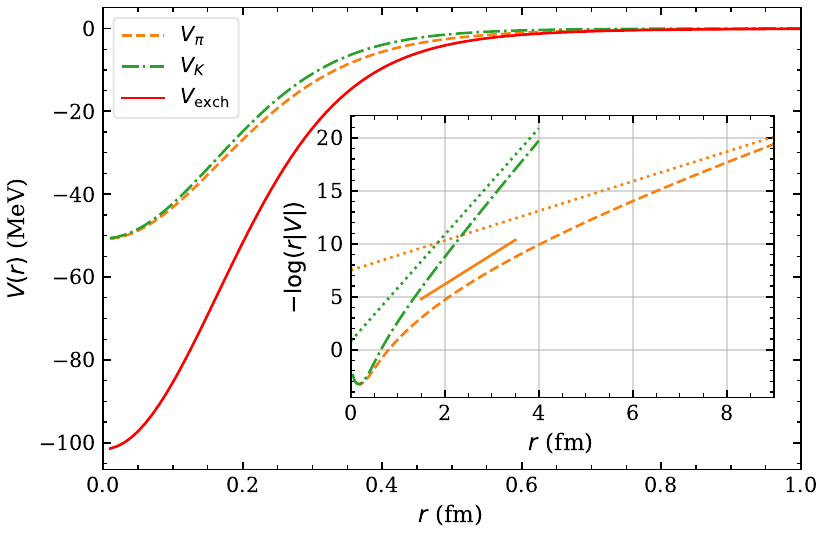}
\caption{The behaviour of the regularised potentials $V_\pi(r,\Lambda)$ and $V_K(r,\Lambda)$ as functions of $r$ for the cut-off $\Lambda=2$~GeV. The inlay demonstrates the mid- and long-range patterns for both potentials. To guide the eye, the dotted lines show the asymptotic slopes in the long-range:
$2m_\pi$ and $2m_K$ for the two-pion and two-kaon potential, respectively. The solid orange line in the inlay with a slope of $0.55$ GeV implies that the behaviour of the contribution from the $f_0(500)$ dominates in the mid-range~\cite{Dong:2021lkh}.}
\label{fig:twopionpotential}
\end{figure}

After the determinations on the coupling constants, such as $\alpha_{\psi(3686)J/\psi}$ and $\alpha_{J/\psi J/\psi}$, Ref.~\cite{Dong:2021lkh} gives the potential for $J/\psi J/\psi$ scattering \changelabel{as in Fig.~\ref{fig:twopionpotential}. Apparently,} such potentials indicate that both two pion and two kaon exchanges can provide considerably attractive contributions, \changelabel{and their orders of magnitudes are almostly the same}. Especially, the result finds that the contribution from $f_0(500)$, i.e., $\sigma$ particle, dominates in the mid-range, which is consistent with the usual experiences in one boson exchange model. Then, combining the adopted potential with contact term, Ref.~\cite{Dong:2021lkh} further investigate the ratio of the light meson exchange contribution to the total potential, and it shows that this ratio can indeed reach above 50\%, makes such a mechanism plausible in explaining the existence of the state below $J/\psi J/\psi$ threshold, although there also exists some uncertainties, i.e., (1) there still exists uncertainties in the determination on the $\alpha_{J/\psi J/\psi}$ due to the lack of the experimental information on the $J/\psi J/\psi$ system, thus it can only start from an educated guess that the meson-exchange picture may be important; (2) although the ratio implies that the meson-exchange contribution may be important, the value of it varies a lot with different model parameters and assumptions, \changelabel{which will vary the nature of $X(6200)$ from bound to virtual}. Thus, Ref.~\cite{Dong:2021lkh} calls for a lattice QCD calculation, which will help in more robust predictions. \changelabel{Nevertheless, it does provide an origin of the long range scattering between heavy quarknonia, and it does indicate that long range forces can play very important roles in the charmonia scattering.}

\begin{figure}
    \centering
    \includegraphics[width=0.7\linewidth]{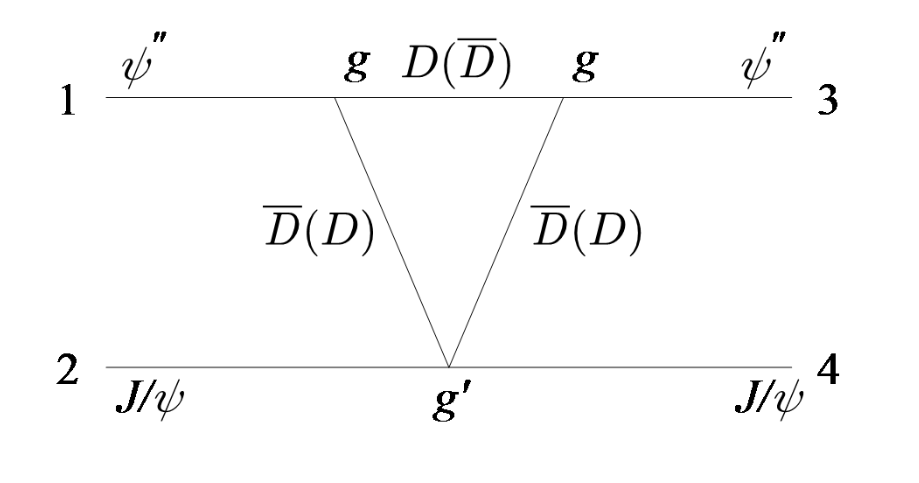}
   \caption{The $D\bar D$ triangle diagram in $J/\psi \psi(3770)$ scattering process, where $\psi^{\prime\prime}$ denotes the $\psi(3770)$ meson~\cite{Lu:2023aal}.}
    \label{fig:ludiagram}
\end{figure}

\begin{figure}
    \centering
    \subfigure[ ]{
    \includegraphics[width=0.48\linewidth]{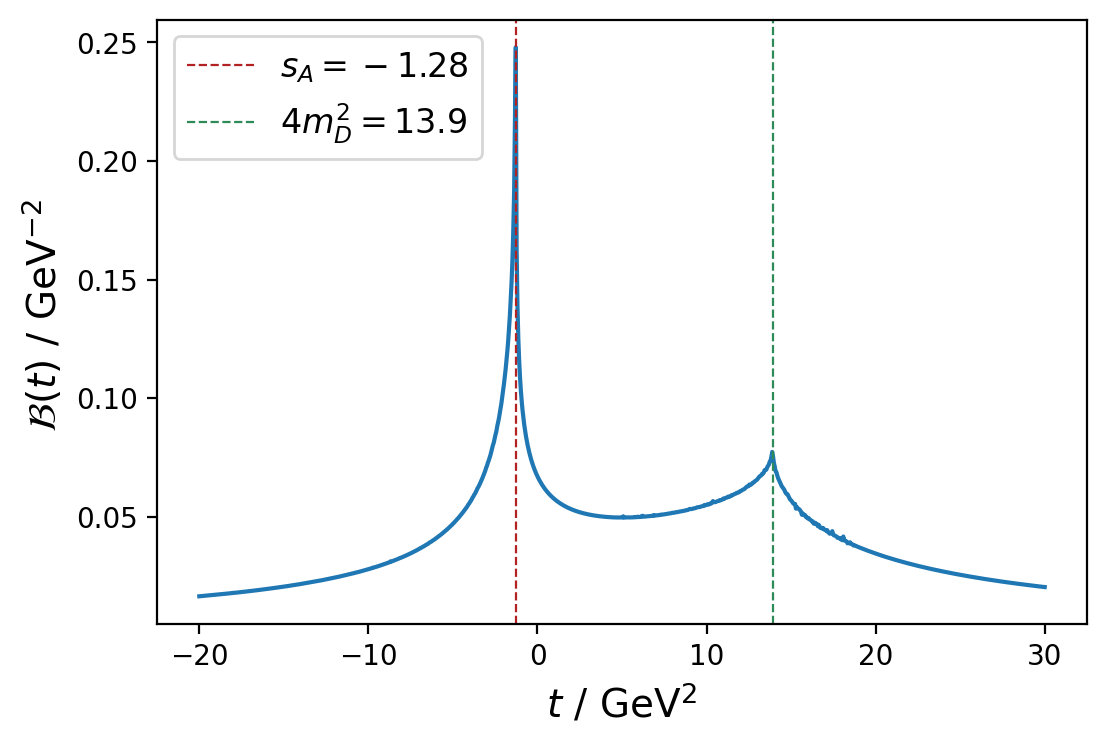}
    }
    \subfigure[ ]{
    \includegraphics[width=0.48\linewidth]{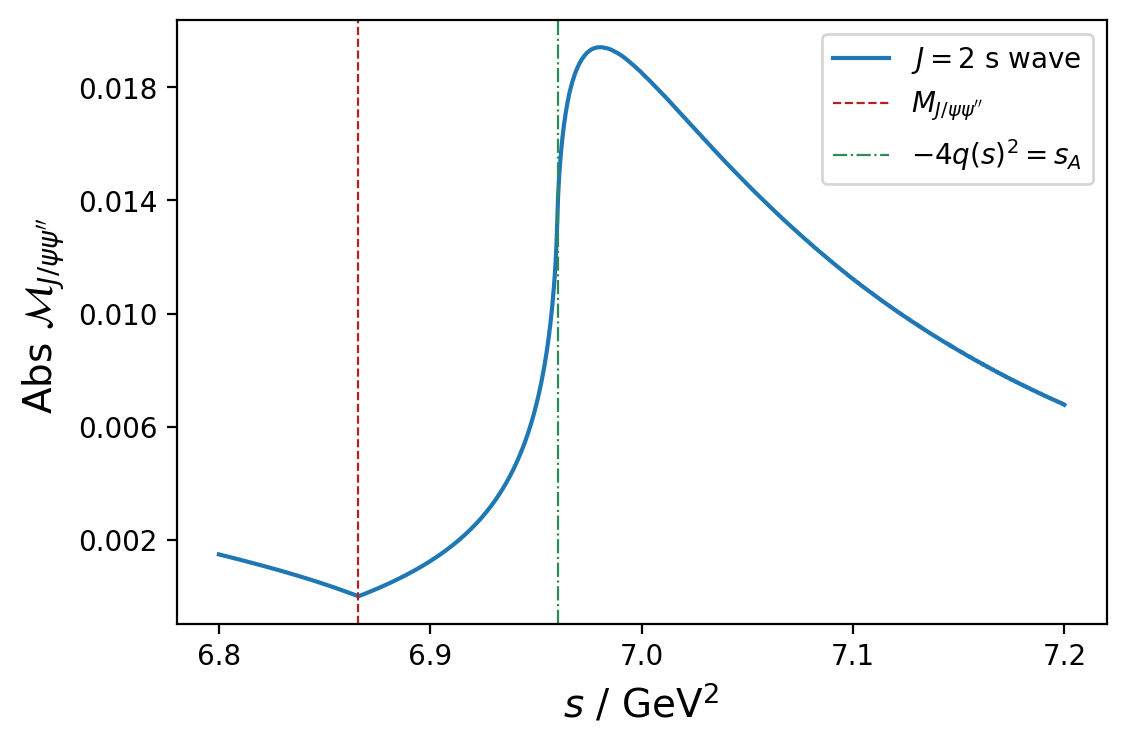}
    }
    \caption{(a): Triangle diagram contribution, with $t$ as the Mandelstam variable. Here, the left sharp peak is caused by the anomalous threshold, while the small one on the right is due to normal threshold effect. (b): The enhancement of the $J/\psi\psi(3770)$ scattering amplitude from the triangle diagram.~\cite{Lu:2023aal}.}
    \label{fig:luresult}
\end{figure}

Another mechanism to explore the nature of $X(6900)$ is proposed by Ref.~\cite{Lu:2023aal}. Since this peak is closed to the $J/\psi \psi(3770)$ threshold, considering that $\psi(3770)$ can strongly couple to a pair of charmed mesons $D\bar{D}$, Ref.~\cite{Lu:2023aal} proposed a possible mechanism for the $J/\psi \psi(3770) \to J/\psi \psi(3770)$ scattering process as Fig.~\ref{fig:ludiagram}.
After transiting such a diagram into the scattering amplitude, Ref.~\cite{Lu:2023aal} pointed out that the loop integral will contain the effect of anomalous triangle singularity, whose position is located at
\begin{equation}
    s_A = 4m^2-\frac{(M^2-2m^2 )^2 }{m^2},
\end{equation}
with $M$ and $m$ as the masses of $\psi(3770)$ and $D$ meson, respectively. As a result, since the mass of $\psi(3770)$ is above the $D\bar{D}$ threshold, these two charmed mesons in the loop can be real particles as long as the scattering energy is above the $J/\psi \psi(3770)$ threshold. At this time, apart from the normal threshold effect caused by $D\bar{D}$, anomalous threshold may also give a considerable contribution as in Fig.~\ref{fig:luresult} (a), which will lead to an enhancement around 6.98 GeV as in Fig.~\ref{fig:luresult} (b).

However, when substituting this mechanism into the fit, Ref.~\cite{Lu:2023aal} found that the triangle diagram contributions are small, and the reason might be that the peak position through anomalous threshold contribution is roughly 60 -- 80MeV above the $X(6900)$ peak, and hence the fit does not like it. In addition, \changelabel{using the same estimation method on the effective interaction range proposed by Yukawa}~\cite{Yukawa:1935xg}, the direct exchanges of charmed mesons between charmonia is still a short-distance interaction with effective range about $\hbar c / m_D \approx 0.1$ fm, which, in our view, is a little unnatural to be used under the hadronic level. A similar discussion can be also used for Ref.~\cite{Liu:2024pio}, where exchanged bosons there are charmonia. While if not, the contact four particle interaction, i.e., $J/\psi J/\psi D D$ in Fig.~\ref{fig:ludiagram} will still have detailed information loss on the interaction mechanism.

\begin{figure}[htbp]
    \centering
\begin{tabular}{cc}
\includegraphics[width=0.44\linewidth]{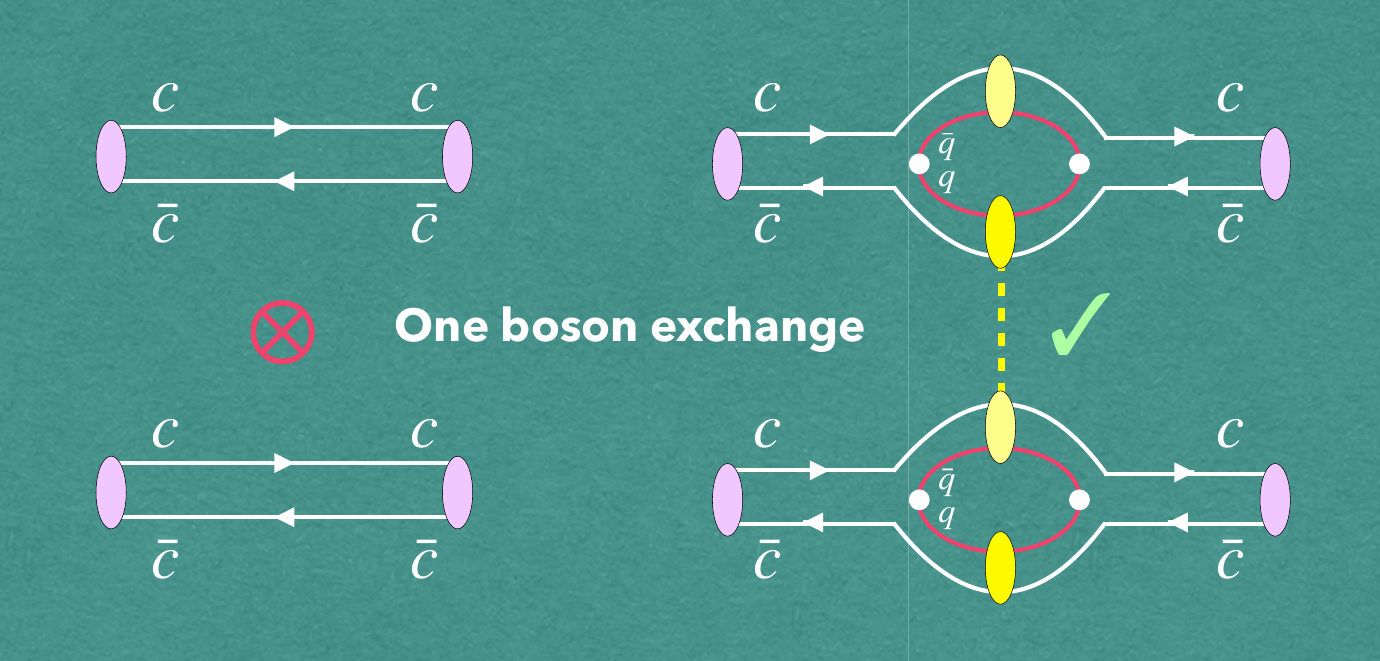}     &  \includegraphics[width=0.44\linewidth]{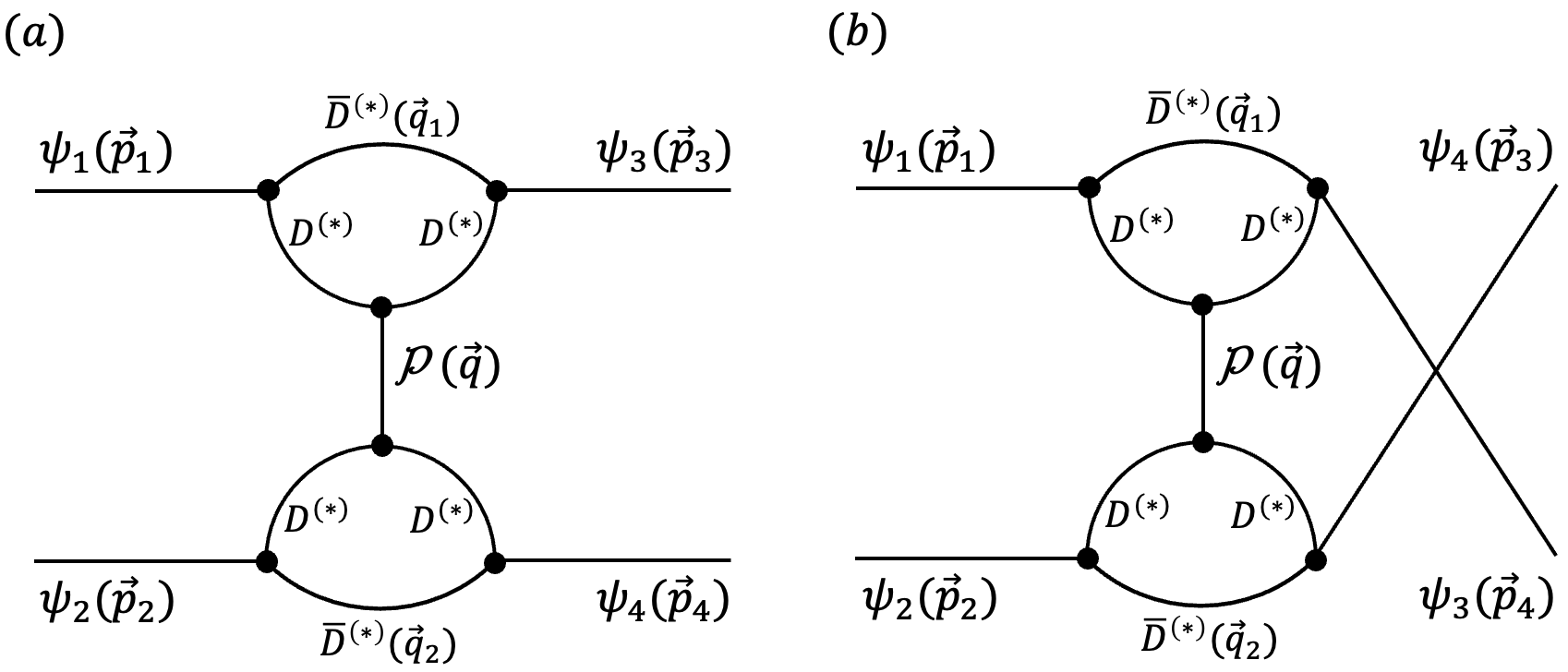}\\
  (I)   & (II) 
\end{tabular}
    \caption{(I) \changelabel{An approach enabling one-boson exchange for effective charmonium scattering based on the unquenched picture.} The cross symbol on the left indicates that direct light flavor one-boson exchange ($\pi,\rho,\sigma,~\cdots$) cannot occur. (II) The scattering mechanisms of the $\psi_1 \psi_2 \to \psi_3 \psi_4$ process involve $D$ meson loop·s, with diagram $(a)$ representing the $t$-channel contribution and diagram $(b)$ representing the $u$-channel contribution. Here, $\psi_i$ can be $J/\psi$, $\psi(3686)$, or $\psi(3770)$, and $\mathcal{P}$ can be $\eta$ or $\sigma$. Furthermore, when $\psi_1 = \psi_2$ and $\psi_3 = \psi_4$, there is no $u$-channel contribution~\cite{Huang:2024jin}.}\label{fig:huangdiagram}
\end{figure}

Nevertheless, in our view, such $D$ loop mechanism \changelabel{can be viewed as a direct application} of the unquenched effect. Since the fully charmed tetraquark states are observed in the di-$J/\psi$ spectrum, it is natural to image that in the experimental detector, the amount of charmonia is extremely abundant. These charmonia, especially the ones above the open-charm threshold, may have large probabilities to oscillate between itself and a pair of charmed mesons in the unquenched picture. Thus, there indeed exists environment for the happening of such a mechanism, and the contribution of it may be not negligible. \changelabel{In addition, the appearance of the light sea quarks makes the exchanges of light hadrons, which have much larger interaction range, possible between these charmonia. Furthermore, such a kind of oscillation has already becomes a hadronic-scale process, which may further enlarge the interaction space. Thus, guided} by this motivation, Ref.~\cite{Huang:2024jin} proposed a new mechanism for the interactions between charmonia, where both the charmonia can oscillate into a pair of charmed mesons, and light mesons can exchange between these intermediate charmed mesons, as in Fig.~\ref{fig:huangdiagram}. At this time, due to conservation of $C$-parity, the exchanged light mesons can only be $\eta$ and $\sigma$. In addition, due to the heavy quark symmetry, all the coupling constants can be previously determined. 

\begin{figure}[htbp]
    \centering
    \includegraphics[width=0.68\linewidth]{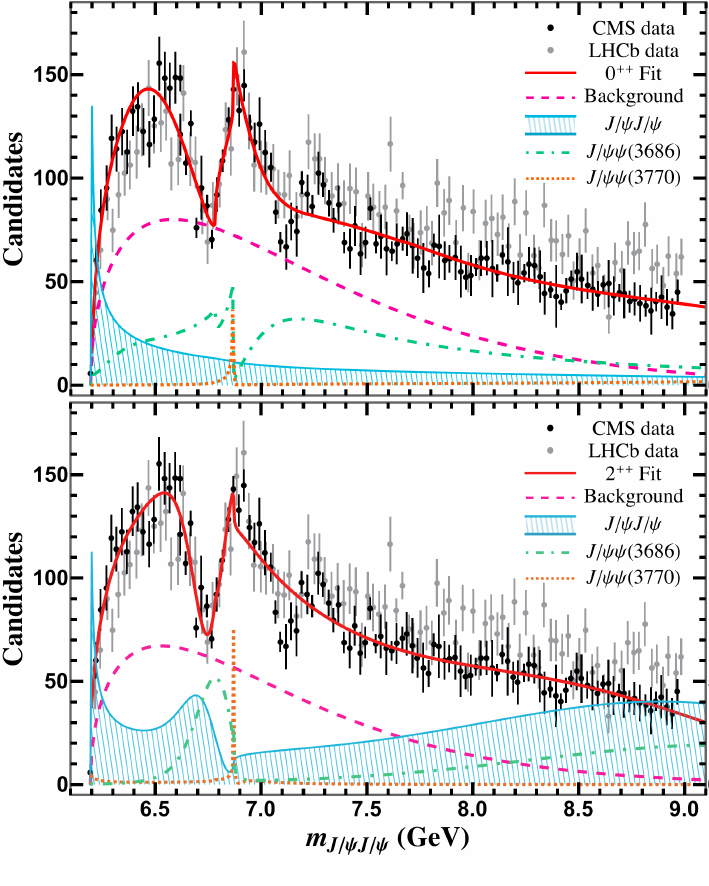}
    \caption{The fit results on experimental data are presented, with the top panel for $0^{++}$ configuration and the bottom panel for $2^{++}$ configuration. In each panel, the dark and light grey error bars represent the CMS data and LHCb data respectively, and the red solid line indicates the fit result. The magenta dashed line, light blue line filled with shadows, light green dot-dashed line, and orange dotted line show the contributions of background, $J/\psi J/\psi$ channel, $J/\psi \psi(3686)$ channel, and $J/\psi \psi(3770)$ channel, respectively~\cite{Huang:2024jin}.}
  \label{fig:huangfit}
\end{figure}

To illustrate the effectiveness of this mechanism, Ref.~\cite{Huang:2024jin} applies this mechanism on the experimental data from CMS with both $0^{++}$ and $2^{++}$ configurations, and the result is presented in Fig.~\ref{fig:huangfit}.
Obviously, for either $0^{++}$ or $2^{++}$ configuration, the fit results are good, while the $\chi^2/\text{d.o.f.}$ suggests a slight preference for the $0^{++}$ configuration. In addition, for the peak around 6.9 GeV, an explicit twist can be found, and the reason is easy to see from the sharp but narrow peak around 6.865 GeV in the $J/\psi \psi(3770)$ channel. Such an effect generates because when the scattering energy reach up to the $J/\psi \psi(3770)$ threshold, rescattering between coupled channels exceed the branch point located at the $J/\psi \psi(3770)$ threshold, which can cause a cusp at this threshold~\cite{Wang:2020wrp}. In addition, since $\psi(3770)$ can be a real particle at this time, when the two charmed mesons linked with it are $D\bar{D}$, the branch point located at $D\bar{D}$ threshold is simultaneously triggered, which may result into further enhancements, and consequently modifies the shape of the peak near 6.9~GeV and becomes a crucial feature of this mechanism. Furthermore, from the extracted pole positions, it is found that for either $0^{++}$ or $2^{++}$ configurations, there always exists a states below the $J/\psi J/\psi$ threshold. In addition, $X(6600)$ is not a real physical state but an effect of interference, while the dip around 6.8 GeV correspond to a pole mainly generated by the $J/\psi\psi(3686)$ channel. Also, the peak around 6.9~GeV is dominantly generated by a physical state emerging from the $J/\psi\psi(3770)$ channel. Last but not least, after analyzing the contributions from different light meson exchanges, it is found that the $\sigma$ exchange dominants, which is consistent with the usual experience on one boson exchange model.

Absolutely, the mechanism proposed by Ref.~\cite{Huang:2024jin} answers a critical issue that how traditional one boson exchange model is introduced in understanding the charmonia scattering. Also, it gives a more refined description on the interaction vertex between charmonia and light mesons, which can make all the coupling constants previously determined. Furthermore, due to the heavy quark symmetry, such a mechanism should also exist in $\Upsilon\Upsilon$, $J/\psi\Upsilon$, $\chi_{bJ}\chi_{bJ}$, $\chi_{cJ}\chi_{bJ}$ channels, which can result into the generations of near threshold states. However, such a mechanism still have some problems to be solved, one is for the suppression on the loop, especially in Fig.~\ref{fig:huangdiagram} the diagrams contain two loops, which may be further suppressed and make the practical contribution not considerable, another is that, there still exists many undetermined parameters in this model, especially the cutoff parameters from the form factors in both loops and exchanged light mesons, which may also lead to large uncertainties on the contributions. Nevertheless, such a mechanism also make a step to a deeper understanding on the scattering mechanisms in fully heavy hadron systems, although further investigations are eager to be carried on to clarify the importance of this mechanism.

\subsection{Lattice QCD result of interactions involved in fully heavy hadrons}

After observing these enhancement structures in the di-$J/\psi$ invariant mass spectrum \cite{LHCb:2020bwg,CMS:2023owd,ATLAS:2023bft,CMS:2025xwt}, some lattice QCD research groups joined the effort. For the scattering between $S$-wave charmonia, Ref.~\cite{Meng:2024czd} performed a lattice QCD calculation of the single-channel $S$-wave scattering length in the $0^{++}$ sector of $\eta_c \eta_c$ and the $2^{++}$ sector of $J/\psi J/\psi$. However, the result indicates that the interactions between the two charmonium pairs are repulsive in both cases, making $2^{++}$ states below the $J/\psi J/\psi$ threshold disfavored in lattice QCD. This finding was subsequently confirmed by another lattice QCD study in Refs.~\cite{Li:2025vbd,Li:2025ftn}, where an additional investigation on the $0^{++}$ sector of $J/\psi J/\psi$ scattering reveals an attractive interaction, suggesting a $0^{++}$ virtual state located 20-40 MeV below the threshold. In addition, although the $2^{++}$ sector of $J/\psi J/\psi$ scattering is repulsive, a $2^{++}$ resonance compatible with the $X(6600)$ is found~\cite{Li:2025vbd,Li:2025ftn}. Although there is still a long way to go before drawing a conclusion on charmonium scattering from lattice QCD results, this information can provide useful hints for understanding charmonium scattering.

\begin{figure}[htbp]
    \centering
    \subfigure[]{\includegraphics[width=0.48\linewidth]{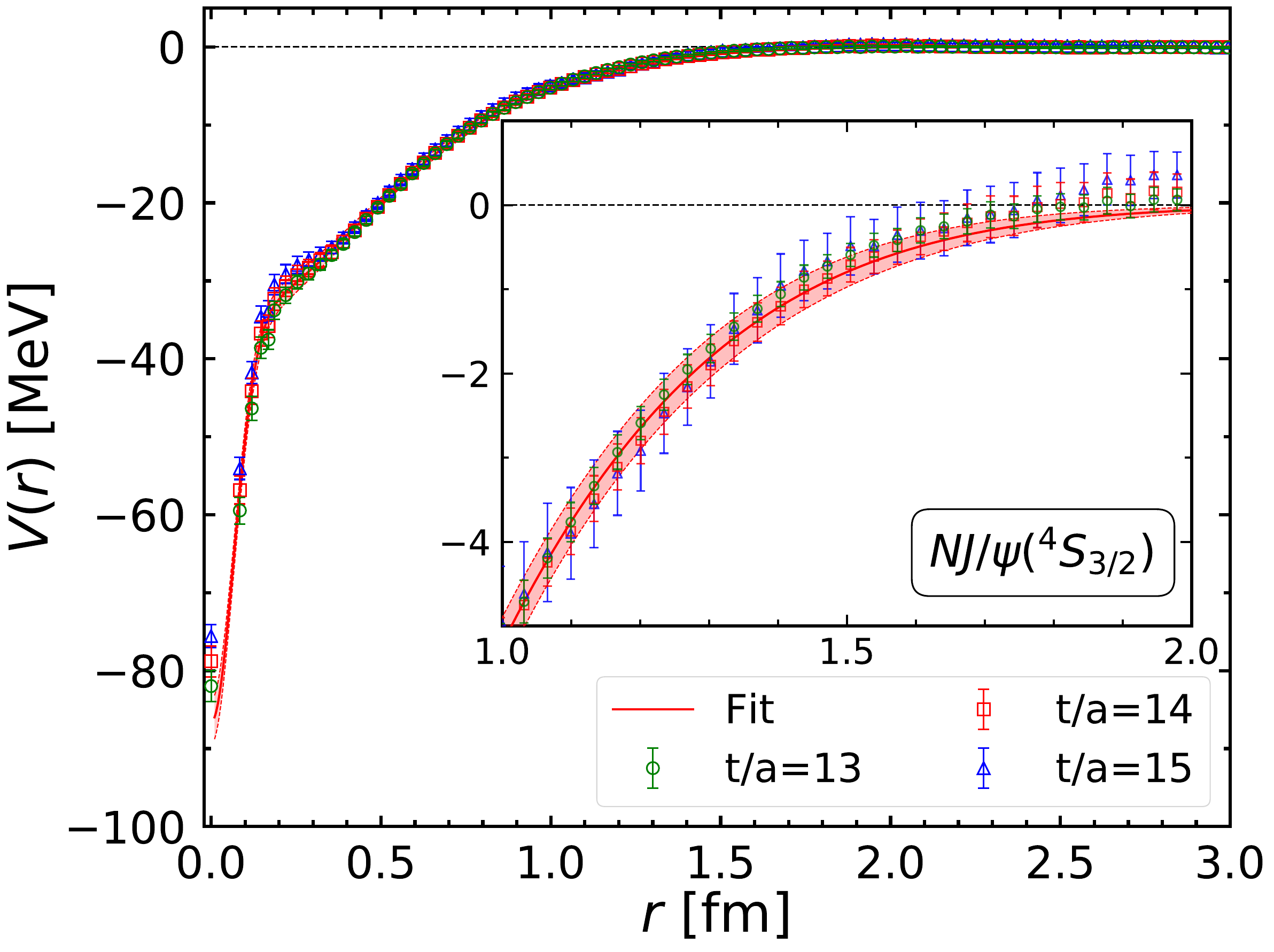}}
    \subfigure[]{\includegraphics[width=0.48\linewidth]{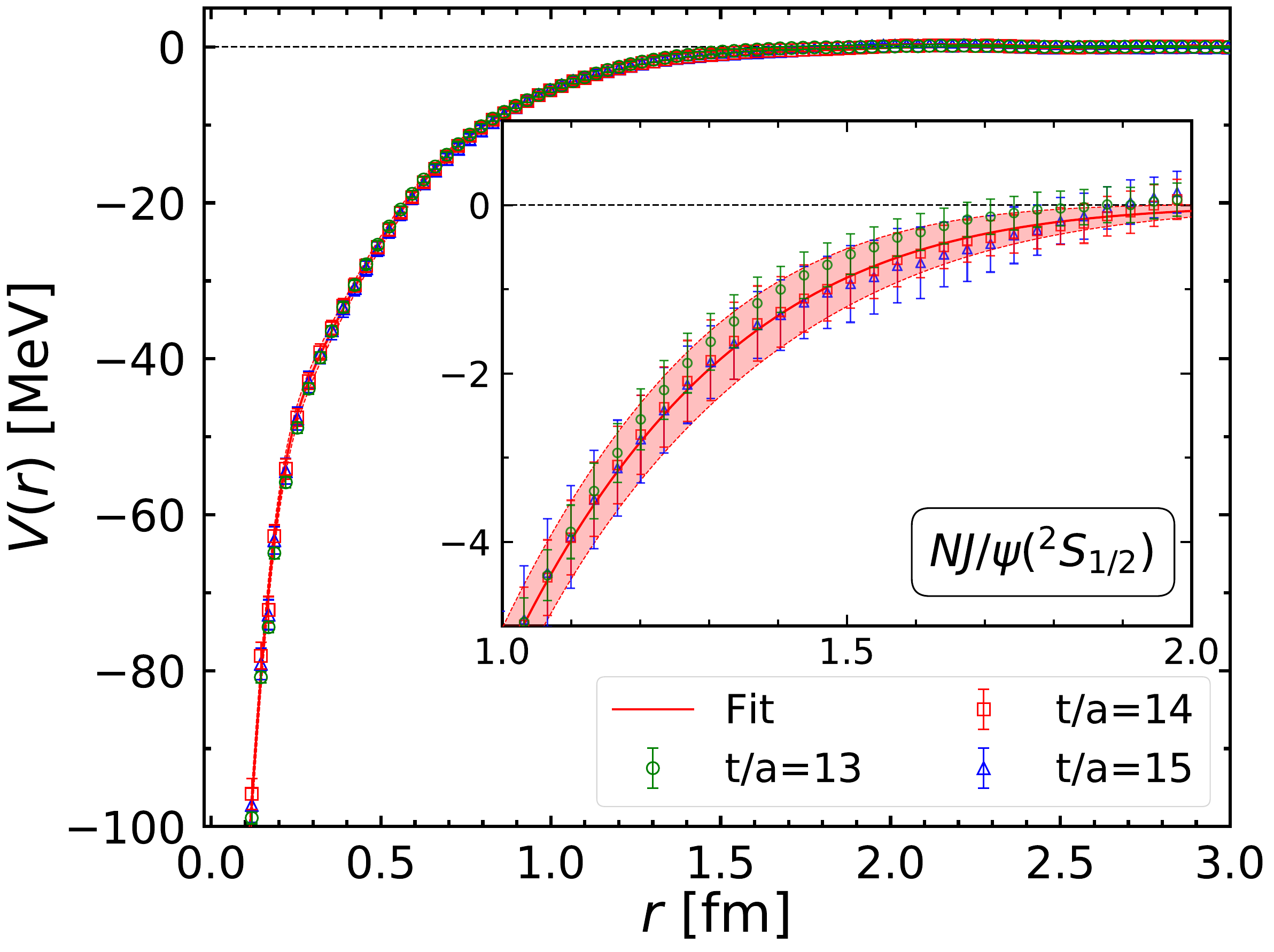}}
    \subfigure[]{\includegraphics[width=0.48\linewidth]{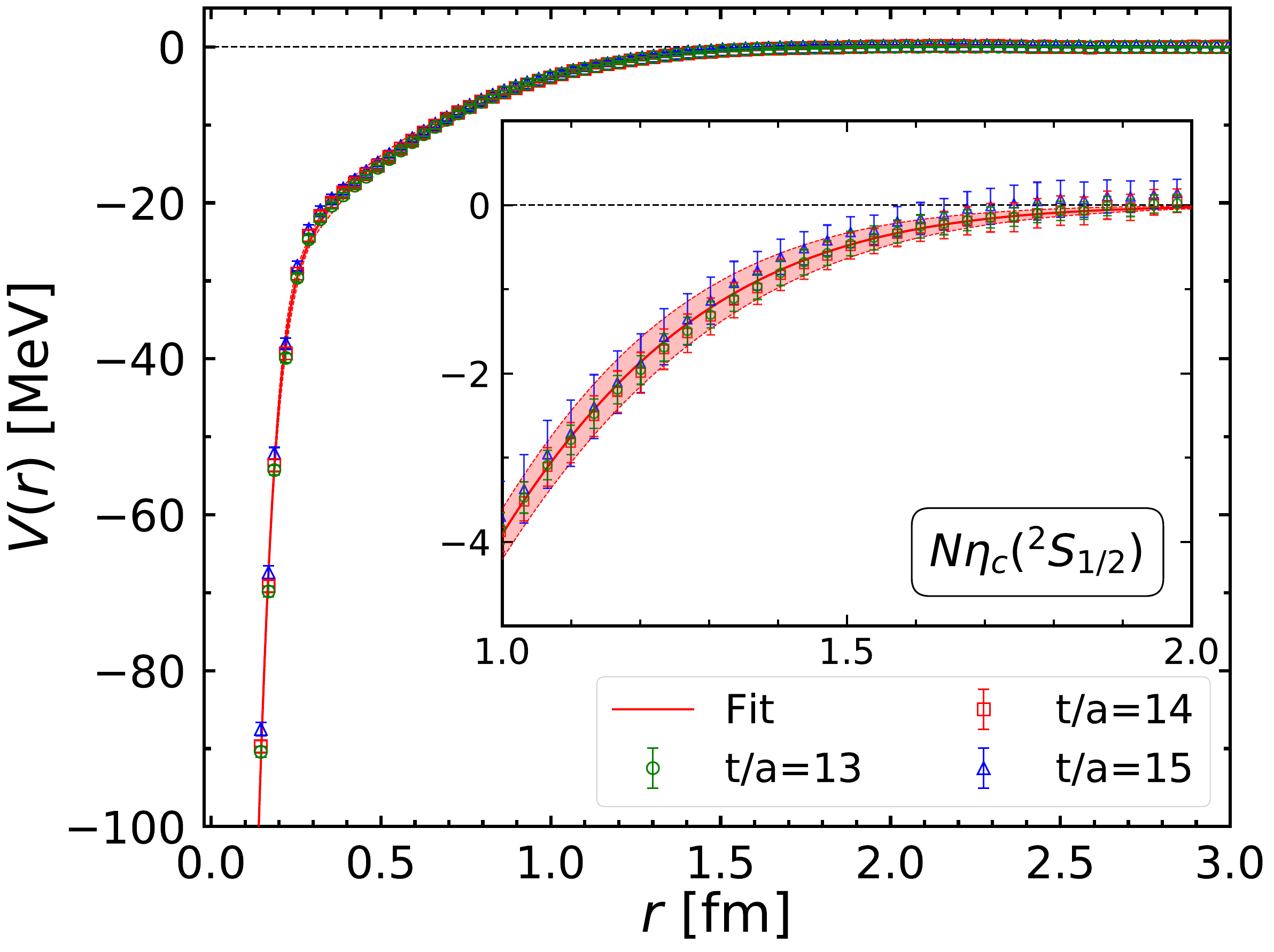}}
    \subfigure[]{\includegraphics[width=0.48\linewidth]{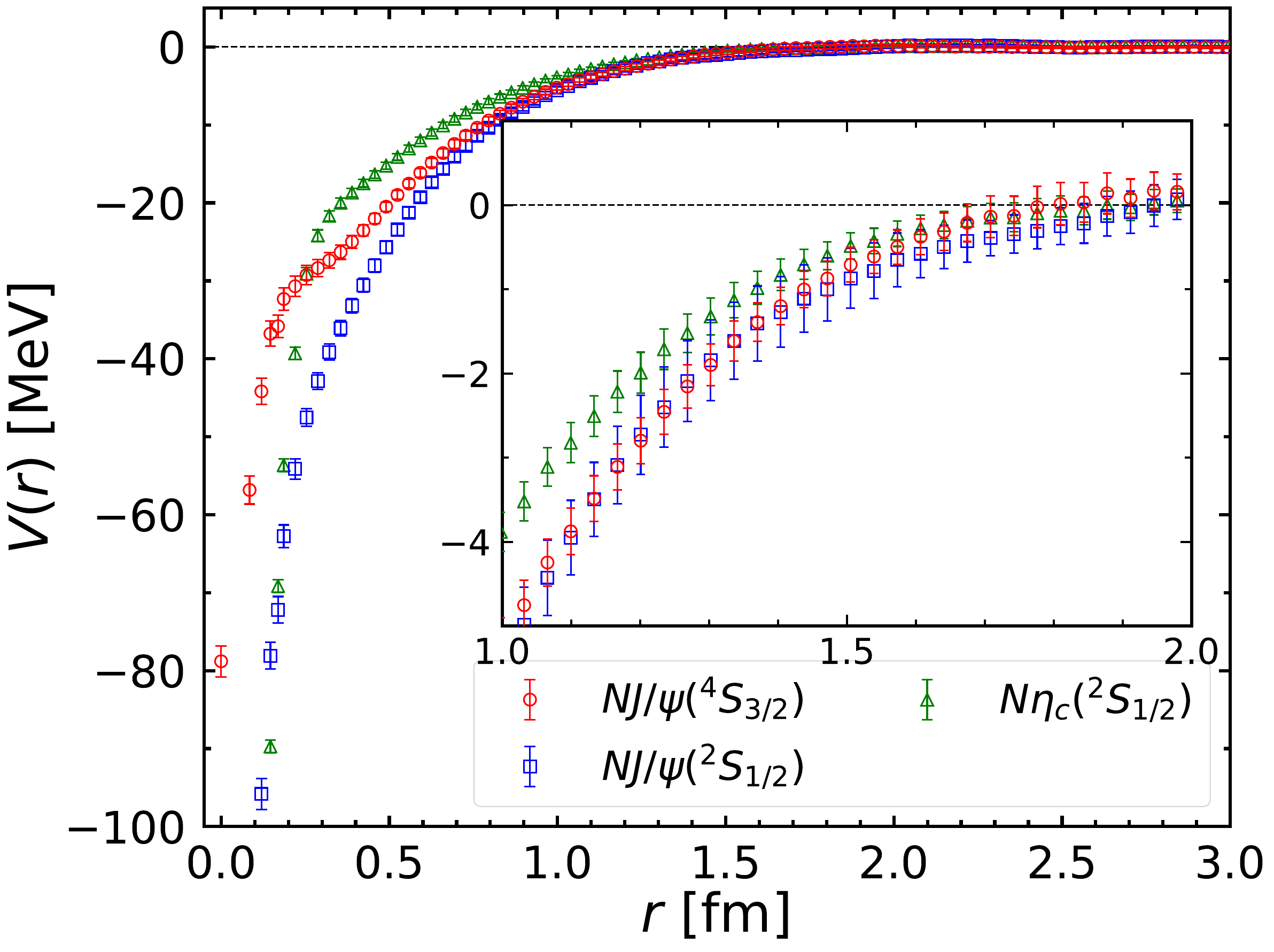}}
    \caption{The $S$-wave $N$-$c\bar c$ potential extracted at $t/a=13$, $14$, and $15$ for $N$-$J/\psi$ with $^4S_{3/2}$ (a), with $^2S_{1/2}$ (b), and $N$-$\eta_c$ with $^2S_{1/2}$ (c). The red bands show the fit results with phenomenological three-range Gaussians at $t/a=14$.
    The three potentials at $t/a=14$ are also shown in (d) for a direct comparison.
    A magnification is shown in the inset for each panel~\cite{Lyu:2024ttm}.
    }
    \label{fig:ccN-lattice}
\end{figure}

\begin{figure}[htbp]
  \centering
  \includegraphics[width=0.66\linewidth]{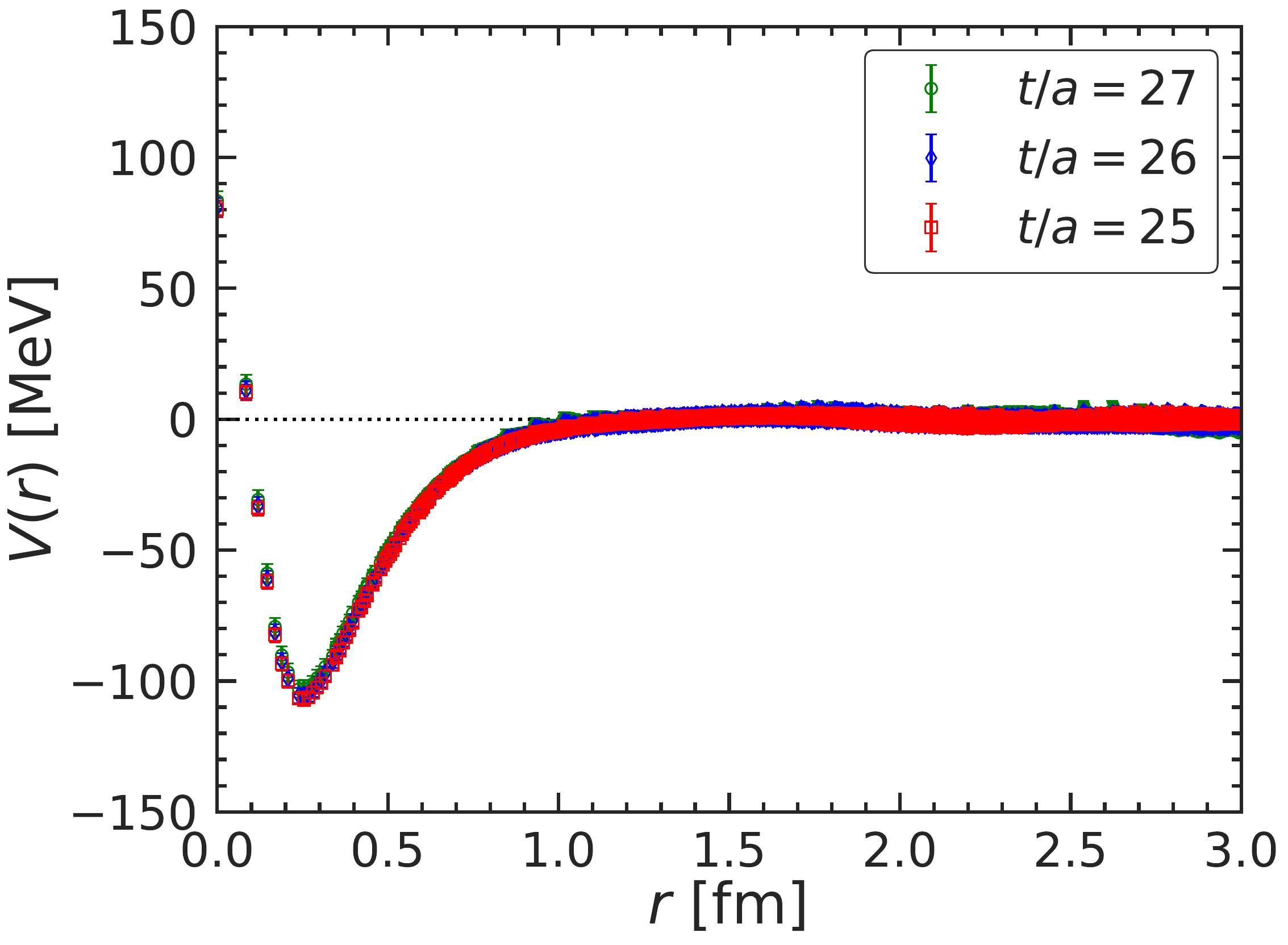}
  \caption{(Color online).
  The $\Omega_{ccc}$-$\Omega_{ccc}$ potential $V(r)$ in the ${^1S_0}$ channel as a function of separation $r$ at Euclidean time $t/a= 25$ (red square), $26$ (blue diamond) and $27$ (green circle).
  }
  \label{fig:cccccc-lattice}
\end{figure}

The interactions involving fully heavy hadrons have recently become a topic of interest for lattice QCD research groups \cite{Lyu:2024ttm,Lyu:2021qsh,Meng:2024czd,Li:2025vbd,Li:2025ftn,Zhang:2025zaa,Mathur:2022ovu,Dhindsa:2025zjk}. As two primary approaches in lattice QCD for studying hadron-hadron interactions, the HAL QCD method is applied to obtain an energy-independent, non-local potential by analyzing the spatiotemporal correlations between two hadrons \cite{Lyu:2024ttm,Lyu:2021qsh,Zhang:2025zaa}. For example, the $S$-wave $c\bar{c}$-$N$ potentials were extracted at $t/a=13$, 14, and 15 for $J/\psi$-$N$ with $^4S_{3/2}$ and $^4S_{1/2}$ and $\eta_c$-$N$ with $^3S_{1/2}$ (see Fig.~\ref{fig:ccN-lattice} for the details of these potentials) \cite{Lyu:2024ttm}. The HAL QCD method has also been employed to extract the potential of $\Omega_{ccc}$-$\Omega_{ccc}$ in the $^1S_0$ channel \cite{Lyu:2021qsh} as illustrated in Fig.~\ref{fig:cccccc-lattice}. These findings provide a window into understanding how fully heavy hadrons interact with each other or with other hadrons.

Returning to the theme of this review—unquenched charmonium—may offer a unique perspective for understanding the underlying mechanisms. Based on the experience from nuclear force studies, the one-boson exchange model \cite{Yukawa:1935xg,Lattes:1947mw,10.1143/PTP.27.1199} has been regarded as an effective approach for investigating hadronic interactions, with numerous applications over the past years in the study of new hadronic states~ (see review papers \cite{Liu:2013waa, Chen:2016qju, Chen:2016spr, Hosaka:2016pey,Guo:2017jvc, Liu:2019zoy,Dong:2021juy, Chen:2022asf, Meng:2022ozq,Liu:2024uxn} for more details). However, when applying this model to interactions involving fully heavy hadrons, the one-boson exchange model encounters a difficulty: no light bosons can be exchanged to mediate the interaction between hadrons. In such cases, one often attempts to introduce extremely short-range exchange contributions \cite{Wang:2020wrp,Wang:2022jmb,Dong:2020nwy,Song:2024ykq,Liang:2021fzr,Zhou:2022xpd,Kuang:2023vac,Niu:2022jqp}. However, this picture is inconsistent with our understanding of low-energy strong interactions. In studying the interactions of charmonia, as introduced in the previous section, an interaction mechanism based on the unquenched picture \cite{Huang:2024jin} offers a promising alternative. More importantly, such an unquenched mechanism may serve as \changelabel{\iffalse\sout{universal framework}\fi another universal approach} for studying interactions involving all types of fully heavy hadrons \changelabel{in the future}.

\begin{figure}[htbp]
  \centering
\includegraphics[width=0.9\linewidth]{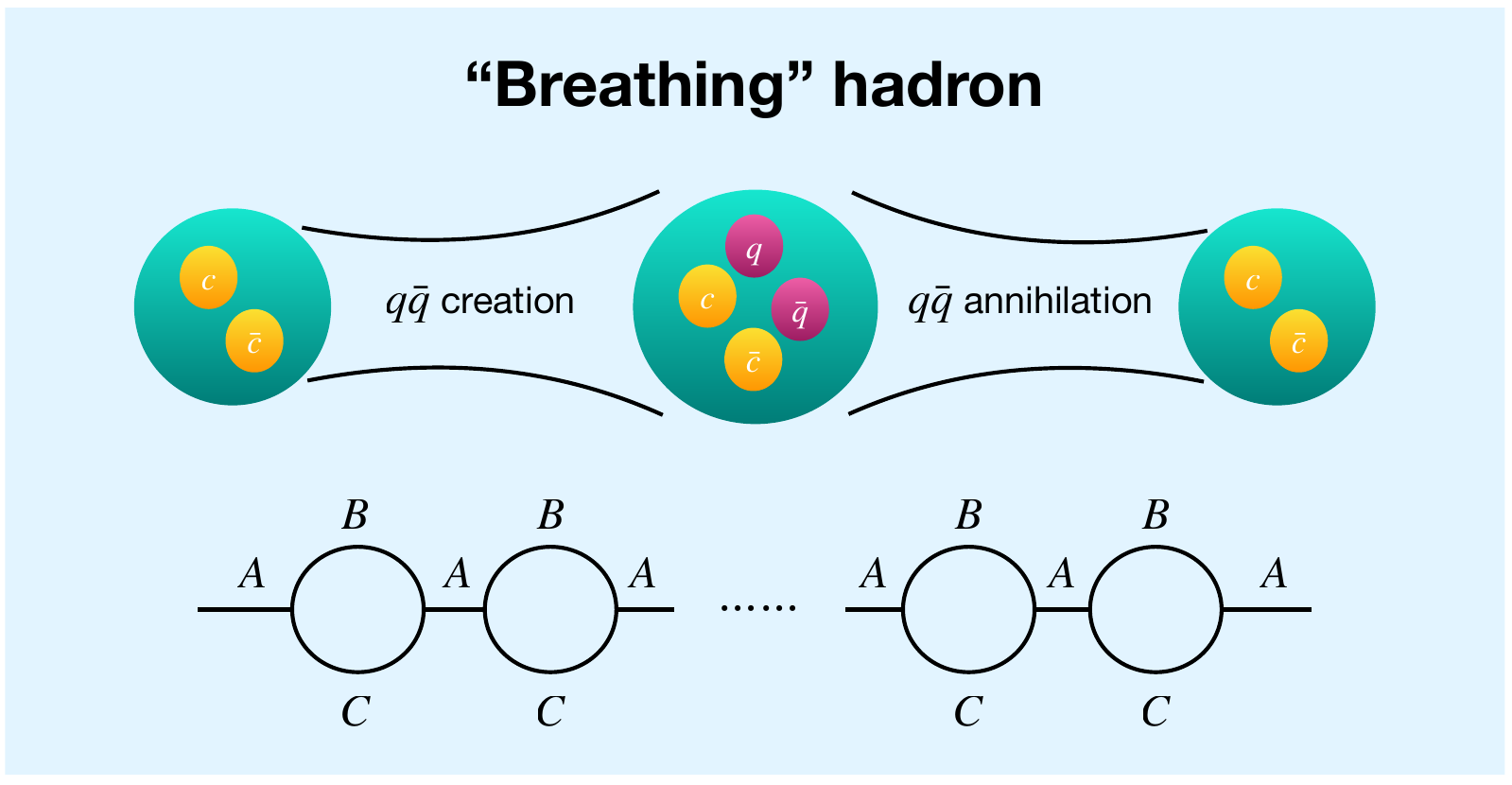}
  \caption{Breathing hadron inspired by unquenched picture, which corresponds to the standard hadronic self-energy in quantum field theory.
  }
  \label{fig:breathing}
\end{figure}

Thus, for fully heavy hadrons, we cannot treat them as static objects. According to the unquenched picture, fully heavy hadrons can couple to corresponding hadron channels via the heavy quarks capturing the sea quarks created from the vacuum. Of course, the light quarks in these hadron channels can also annihilate, transitioning back to a fully heavy hadron state. Such a process can occur repeatedly, much like a "breathing" object, as illustrated in Fig.~\ref{fig:breathing}. This picture, \changelabel{which can be related to the concept of quantum fluctuation and self-energy in quantum field theory}, provides \changelabel{another} clear and universally applicable microscopic mechanism for the hadronic interactions involving fully heavy hadrons. Since the intermediate hadron channels contain light quarks, the one-boson exchange mechanism can still play a role in understanding these interactions involving fully heavy hadrons (see Fig.~\ref{fig:huangdiagram} for the application of this mechanism to charmonium interactions). With more experimental data, lattice QCD results, and corresponding phenomenological studies, we believe this research issue is valuable and merits sustained attention.

\section{Extension to $\Upsilon$ family and light flavor vector mesonic states}\label{section7}

\subsection{Universality of vector mesonic structures in $e^+e^-$ annihilation}

In particle physics, the $R$ value is a fundamental observable that has played a pivotal historical role. It is defined as the ratio of the total cross section for electron-positron annihilation into hadrons to the cross section for annihilation into a muon pair at the same center-of-mass energy, i.e.,
\begin{equation}
R \equiv \frac{\sigma(e^+e^- \rightarrow \text{hadrons})}{\sigma(e^+e^- \rightarrow \mu^+\mu^-)}.
\label{eq:R_def}
\end{equation}

\begin{figure}[htbp]
    \centering
    \includegraphics[width=0.9\textwidth]{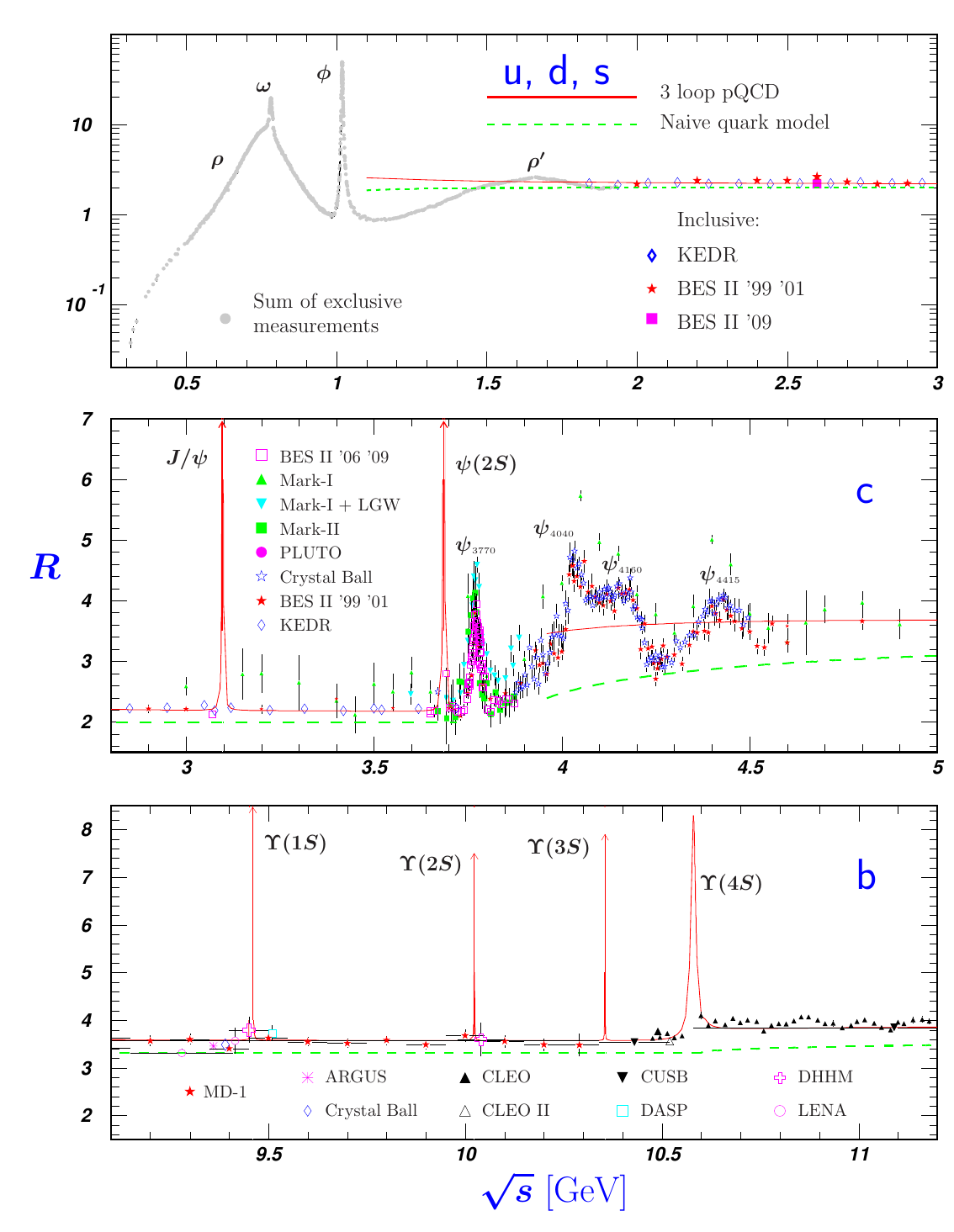}
    \caption{The $R$ value as a function of center-of-mass energy $\sqrt{s}$ from various experiments. The step-like structure corresponds to the opening of new quark flavor thresholds, while sharp peaks indicate vector meson resonances. Notable structures include the $\phi(1020)$, $J/\psi$, $\psi(2S)$, $\Upsilon$ family, and the $Z$ boson. The regions containing the anomalous $Y$ structures discussed in this work are highlighted. Figure adapted from \cite{ParticleDataGroup:2020ssz}.}
    \label{fig:Rvalue}
\end{figure}

Fig.~\ref{fig:Rvalue} displays the measured $R$ values as a function of the center-of-mass energy $\sqrt{s}$, ranging from low energies (around 1 GeV) to high energies (up to several hundred GeV). The data exhibit the characteristic step-like structure predicted by the quark-parton model, with distinct steps at the charm and bottom quark thresholds. Narrow resonances such as the $J/\psi$ and $\Upsilon$ families appear as sharp peaks. Above the $Z$ boson resonance, the ratio is dominated by $Z$ boson exchange and by the interference between photon and $Z$ exchange.

Section~\ref{section4} reviews how unquenched effects play a crucial role in the charmonium sector for resolving the so-called $Y$ problem \cite{Barnes:2005pb}. A unified framework can be established to understand the puzzling phenomena associated with charmonium-like $Y$ structures produced directly in $e^+e^-$ collisions. Indeed, a remarkable universality emerges across different flavor sectors when examining vector mesonic structures in $e^+e^-$ annihilation. 

Prominent examples include the anomalous $Y(4260)$ observed in $e^+e^-\to J/\psi\pi^+\pi^-$ \cite{BaBar:2005hhc} (later resolved into two substructures by BESIII \cite{BESIII:2016bnd}), the $Y(2175)$ seen in $e^+e^-\to\phi\pi^+\pi^-$ \cite{Belle:2008kuo} as a typical light-flavor strangeonium-like state, and the $\Upsilon(10860)$ produced in $e^+e^-$ annihilation into $\Upsilon(nS)\pi^+\pi^-$ ($n=1,2,3$) \cite{Belle:2007xek,Belle:2015aea,Belle:2019cbt}. These states have been the subject of extensive theoretical investigations \cite{Chen:2008qw,Ding:2009vj,Guo:2016fgl,Meng:2007tk,Meng:2008dd,Wang:2019mhs}.

Their similar phenomenological behaviors and analogous decay patterns suggest a possible universal dynamical mechanism related to unquenched effects. Building on insights gained from addressing the $Y$ problem in the charmonium sector, it is plausible that a unified approach based on unquenched effects can be extended to explain these vector mesonic structures across light-flavor, charmonium, and bottomonium sectors. The following sections extend this unquenched analysis systematically to the bottomonium family and light-flavor vector mesonic states.

\subsection{Higher bottomonia in the unquenched picture}

\subsubsection{Higher bottomonium}\label{higher bottomonium}

As a prominent member of the heavy quarkonium family, bottomonium holds a unique position in hadron spectroscopy, offering crucial insights into the strong-interaction dynamics governed by QCD. Over the past decades, concerted experimental and theoretical efforts have been dedicated to studying bottomonium, resulting in a series of discoveries that have greatly enriched our knowledge of its spectral properties. The discovery of the $\Upsilon(1S)$ state in 1977~\cite{E288:1977xhf}, which first established the existence of the bottom quark, was followed by the gradual mapping of a rich spectrum of bottomonium resonances. Subsequent experiments identified a sequence of radially excited $\Upsilon(nS)$ states ($n = 2$--$6$)~\cite{E288:1977efs,CLEO:1984vfn,Lovelock:1985nb}, as well as the $P$-wave spin-triplet states $\chi_{bJ}(1P)$ and $\chi_{bJ}(2P)$ ($J=0,1,2$) almost four decades ago~\cite{Klopfenstein:1983nx,Pauss:1983pa,Han:1982zk,Eigen:1982zm}.

The experimental landscape of bottomonium physics has advanced considerably in recent years, thanks to improved capabilities at the $B$-factories BaBar and Belle, the upgraded Belle~II experiment, and the high-energy collisions at the LHC. A number of new states have been firmly established, including the pseudoscalar ground state $\eta_b(1S)$~\cite{BaBar:2008dae,CLEO:2009nxu,Dobbs:2012zn,Belle:2012fkf}, its first radial excitation $\eta_b(2S)$~\cite{BaBar:2011xka,Dobbs:2012zn,Belle:2012fkf}, the $P$-wave excitations $\chi_{b1}(3P)$~\cite{D0:2012pig,BaBar:2011ljf}, $h_b(1P)$~\cite{BaBar:2011ljf,Belle:2011wqq}, $h_b(2P)$~\cite{Belle:2011wqq}, the $D$-wave state $\Upsilon(1^3D_2)$~\cite{CLEO:2004npj,BaBar:2010tqb}, and the puzzling $\Upsilon(10753)$~\cite{Belle:2019cbt}. This wealth of experimental data offers an excellent testing ground for phenomenological models of bottomonium spectroscopy~\cite{Richardson:1978bt,Buchmuller:1980su,Martin:1980jx,Godfrey:1985xj,Gupta:1986xt,Beyer:1992nd,Ding:1993uy,Ding:1995he,Motyka:1997di,Ebert:2002pp,Gonzalez:2003gx,Vijande:2004he,Radford:2007vd,Li:2009nr,Ferretti:2013vua,Wei-Zhao:2013sta,Gonzalez:2014nka,Li:2015zda,Godfrey:2015dia,Segovia:2016xqb,Deng:2016ktl,Wang:2018rjg,Ni:2025gvx,Kaushal:2025kbz,Bokade:2025voh}.

As in the charmonium sector, quenched potential models---notably the GI  model~\cite{Godfrey:1985xj,Godfrey:2015dia}---provide a satisfactory description of the low-lying bottomonium spectrum. However, they systematically overpredict the masses of higher-lying states when compared with experimental measurements. A clear example is the $\Upsilon(6S)$ state, whose predicted mass in the GI model exceeds the measured mass of the $\Upsilon(11020)$ by about 100~MeV, as illustrated in Table~\ref{tab:11020mass}. Such discrepancies highlight the increasing relevance of unquenched effects, which become more pronounced for excited states with larger spatial extensions.

\begin{table}[htbp]
\centering
\caption{A comparison of the mass of the $\Upsilon(11020)$ state among the modified GI (MGI) model, the original GI model, other screened potential models and experimental measurements.}
\renewcommand\arraystretch{1.3}
\setlength{\tabcolsep}{4pt}
\begin{tabular*}{1.0\textwidth}{l@{\extracolsep{\fill}}ccccccccc}
            \hline
Model &MGI \cite{Wang:2018rjg} &GI \cite{Godfrey:1985xj,Godfrey:2015dia} & Ref.~\cite{Li:2009nr} & Ref.~\cite{Deng:2016ktl} & Ref.~\cite{Kaushal:2025kbz}& Experiment \cite{ParticleDataGroup:2024cfk}\\
\hline
Mass (MeV)&11001 &11102 & 11023& 10997 & 11022& $11000\pm4$\\
\hline
\end{tabular*}
\label{tab:11020mass}
\end{table}

In Sec.~\ref{section4}, it has been demonstrated that a series of $Y$ structures in the 4.0--4.7~GeV region can be successfully described within the charmonium spectrum once unquenched effects are incorporated into the GI model. Specifically, the linear confining term is modified as $b r\to {b(1 - e^{-\mu r})}/{\mu}$, which softens the potential at large distances due to vacuum polarization induced by light-quark pair creation~\cite{Laermann:1986pu,Born:1989iv}. Although meson--meson channels are not explicitly included, this phenomenological treatment effectively mimics coupled-channel dynamics and thus represents an unquenched correction to the conventional quenched potential. 

Inspired by the success of this screened potential in resolving the long-standing ``$Y$ problem'' in charmonium(-like) sector, it is natural to expect analogous unquenched effects to be important for high-lying bottomonium spectroscopy. Following this motivation, in Ref.~\cite{Wang:2018rjg}, Wang, Sun, Liu, and Matsuki introduced the same screening mechanism into a modified GI model to reanalyze the bottomonium spectrum. The screening parameter $\mu = 0.075$~GeV and other parameters are determined by a $\chi^2$ fit to well-established experimental bottomonium states. \changelabel{ It is also worth noting that the screened-potential picture has been applied to the bottomonium spectrum in other quark-model studies. In a pioneering study, Li and Chao~\cite{Li:2009nr} systematically investigated the mass spectrum, electromagnetic decays, and E1 radiative transitions of bottomonium states by formulating a color confinement potential as ${b(1 - e^{-\mu r})}/{\mu}$ within the non-relativistic quark model. Deng {\it et al.} performed a systematic calculation in a nonrelativistic screened-potential quark model and discussed how the screening effect modifies the highly excited bottomonium spectrum~\cite{Deng:2016ktl}. More recently, Kaushal and Bhaghyesh analyzed the beauty-hadron spectrum in a phenomenological screened-potential model, where the screened confinement interaction was also used to describe bottomonium states~\cite{Kaushal:2025kbz}. The predictions for the mass of the $\Upsilon(11020)$ from different screened-potential models are summarized in Table~\ref{tab:11020mass}. It is worth emphasizing that Ref.~\cite{Wang:2018rjg} evaluates the screened potential within the framework of the semi-relativistic GI quark model, whereas other investigations were based on a non-relativistic quark model. }


\begin{figure}[htbp]
\centering
\includegraphics[width=0.9\textwidth]{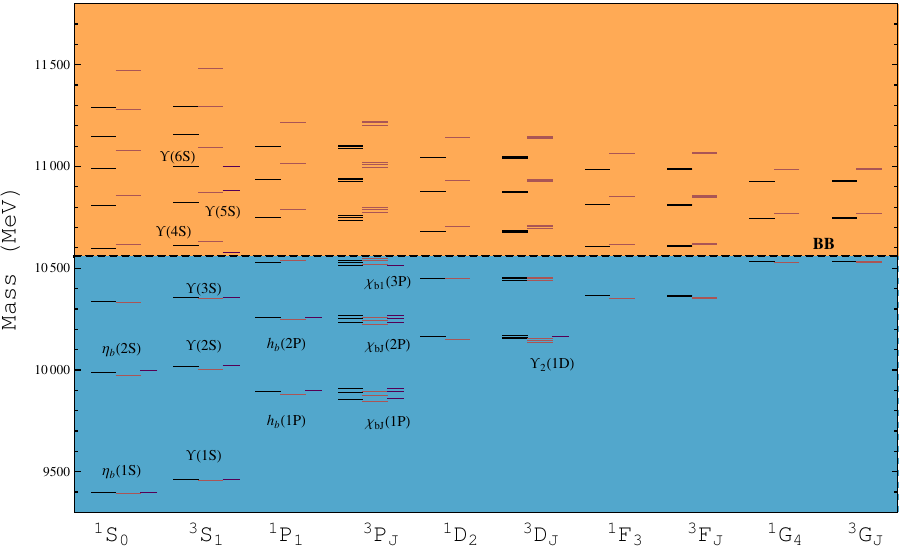}
\caption{Bottomonium mass spectrum. From left to right: predictions of the modified GI model~\cite{Wang:2018rjg} (black lines), original GI model~\cite{Godfrey:1985xj,Godfrey:2015dia} (pink lines), and experimental values from the PDG~\cite{ParticleDataGroup:2016lqr} (purple lines). The open-bottom threshold is indicated by the dashed line. Figure adapted from Ref.~\cite{Wang:2018rjg}.}
\label{fig:bb_spectrum}
\end{figure}

\begin{figure}[htbp]
\centering
\includegraphics[width=0.7\textwidth]{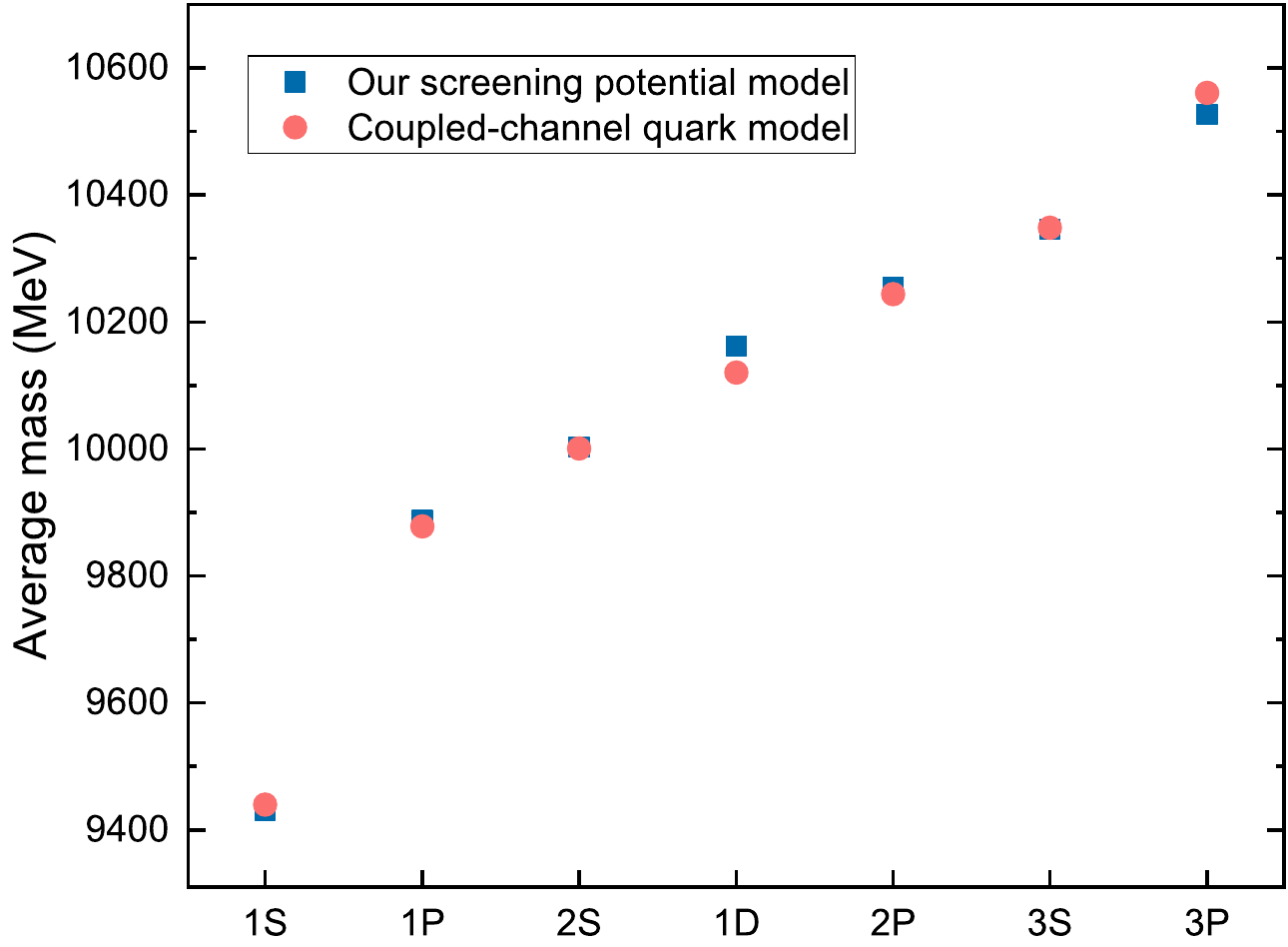}
\caption{A comparison of the results of screening potential model
and those obtained by a coupled-channel quark model~\cite{Ferretti:2013vua} for bottomonium spectra. The figure is adapted from Ref.~\cite{Wang:2018rjg}.}
\label{fig:comparison-MGI_cc}
\end{figure}

Using the fitted parameters of the modified GI model, the bottomonium spectrum can be calculated systematically. The resulting masses, together with the original GI predictions and available experimental data, are displayed in Fig.~\ref{fig:bb_spectrum}. For low-lying states, the GI model typically yields masses of $10\sim20$~MeV below the experimental values, whereas the modified GI model shows noticeably better agreement. In contrast, for higher-lying states, the quenched GI model systematically overestimates the masses, in which a representative case is $\Upsilon(11020)$ and its measured mass is nearly 100~MeV lower than the GI prediction. This discrepancy is largely resolved once screening effect is included. Notably, the recently observed $\Upsilon(10753)$~\cite{Belle:2019cbt} can be naturally interpreted within this framework as a $4S$-$3D$ mixed state with dominant $3D$ character, consistent with expectations from the unquenched potential model. A detailed discussion of this assignment will be given in subsequent sections. Overall, these results indicate that the unquenched potential model effectively remedies the shortcomings of the quenched approach and provides a more accurate and unified description of the bottomonium spectrum. Moreover, as shown in Sec.~\ref{section4}, the results from the screening potential model and the coupled-channel model are compatible in the charmonium sector. This compatibility also holds in the bottomonium sector, as illustrated in Fig.~\ref{fig:comparison-MGI_cc}, further underscoring the equivalence of these two unquenched formulations.

An exception is $\Upsilon(10860)$, for which the predicted mass of $\Upsilon(5S)$ in the present work lies about 60~MeV below the experimental value. It is noteworthy that measurements reported by BaBar and Belle after 2010 tend to be roughly 30~MeV higher than earlier determinations~\cite{ParticleDataGroup:2024cfk}, as illustrated in Fig. \ref{Fig:Ybmass}. In addition, the total width of $\Upsilon(10860)$ measured by Belle differs from earlier results that suggested a rather broad state. Furthermore, the cross section data for $e^+e^- \to \Upsilon(nS)\pi^+\pi^-$ $(n=1,2,3)$ around 10.88~GeV from Belle do not appear to be fully described by a single $\Upsilon(10860)$ resonance~\cite{Belle:2019cbt}. Similar difficulties in reproducing the $\Upsilon(5S)$ mass are encountered in other non-relativistic potential models that incorporate screening effects~\cite{Ding:1993uy,Li:2009nr,Deng:2016ktl}, indicating that a complete understanding of this state remains an open issue.

By taking the obtained mass spectra and the corresponding spatial wave function as input, Ref.~\cite{Wang:2018rjg} further systematically calculated partial widths for allowed radiative transitions, annihilation decays, hidden-bottom hadronic transitions, and open-bottom two-body strong decays for each bottomonium state, providing useful benchmarks for future experimental searches.

\subsubsection{Anomalous hidden-bottom decay phenomena of several high-lying bottomonia}

While the rich spectrum of higher bottomonium states reflects the importance of unquenched effects in spectroscopy, their decay properties provide an equally complementary window into the underlying dynamics. In particular, several anomalous exclusive hidden-bottom transitions of high-lying bottomonium states have been observed, highlighting the role of unquenched effects in their decay dynamics.

\begin{figure}[htbp]
\centering
\includegraphics[width=0.4\textwidth]{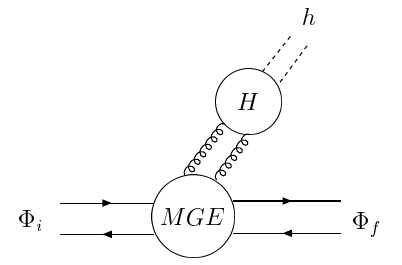}
\caption{Schematic diagram for a typical hadronic transition in the QCDME approach. Here, MGE denotes multipole gluon emission from the heavy quark--antiquark pair, and $H$ represents the hadronization of the emitted gluons into light hadron(s). Figure adapted from Ref.~\cite{Kuang:2006me}.}
\label{fig:qcdme}
\end{figure}

Prior to the high-precision measurements from the $B$-factories, the QCD multipole expansion (QCDME) approach had been established as a \changelabel{good} theoretical framework for studying hadronic transitions among low-lying heavy quarkonia \changelabel{with the limited phase space, which usually can suppress the contribution of the $f_0(500)/\sigma$ resonance from the $S$-wave $\pi\pi$ final state interaction}~\cite{Gottfried:1977gp,Bhanot:1979af,Peskin:1979va,Bhanot:1979vb,Voloshin:1978hc,Yan:1980uh,Kuang:1981se,Kuang:1988bz,Kuang:1989ub,Kuang:2006me}. This method relies on the clear separation of energy scales in heavy-quark systems: the spatial size of the quarkonium is small compared with the wavelength of the emitted soft gluons, allowing the heavy quarkonium system to be regarded as a compact color source emitting soft gluons, and thereby permitting a multipole expansion of the gluon field analogous to that in classical electrodynamics. As depicted in Fig.~\ref{fig:qcdme}, the transition amplitude factorizes into a multipole gluon emission (MGE) part---which depends on the quarkonium wave function---and a subsequent hadronization ($H$) part, where the emitted gluons form light hadron(s) at the scale of light-flavor dynamics.

However, with the accumulation of high-precision data from Belle and BaBar, several unexpected features have emerged in the hadronic transitions of heavy quarkonia, especially above the open-flavor thresholds. These observations pose serious challenges to the conventional QCDME picture.

\begin{figure}[htbp]
\centering
\includegraphics[width=0.3\textwidth]{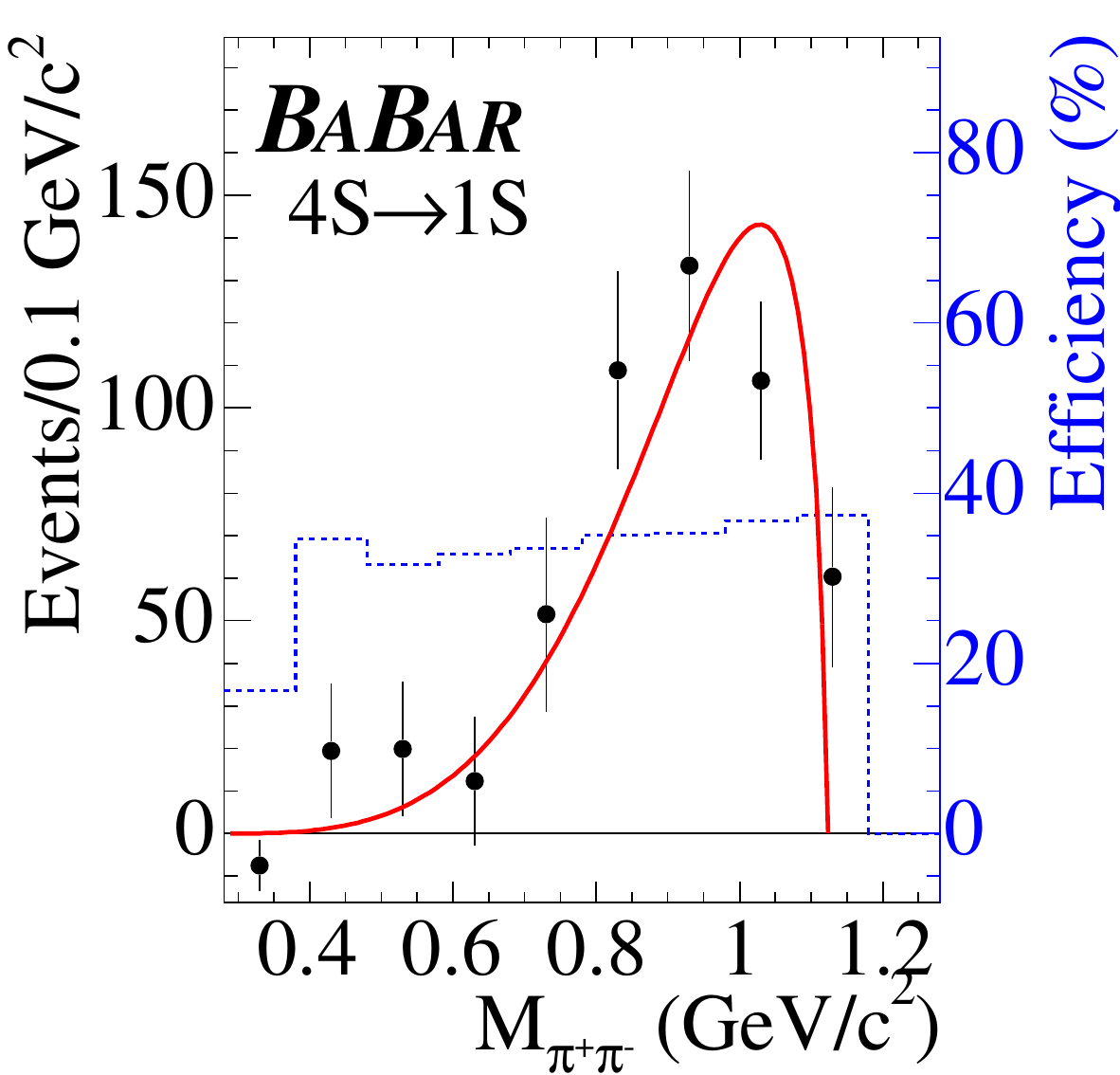}
\hspace{0.05\textwidth}
\includegraphics[width=0.3\textwidth]{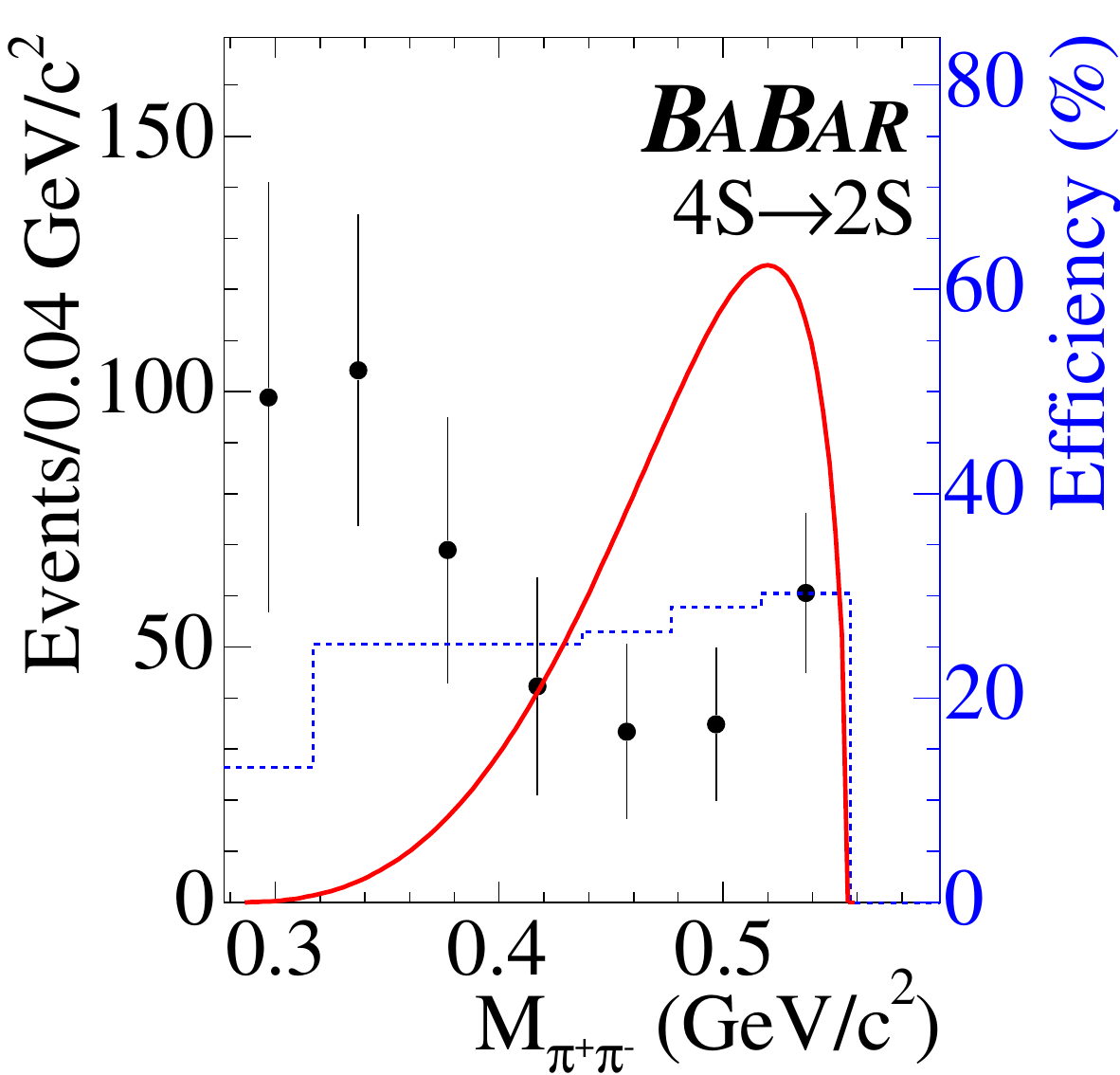}
\caption{Measured dipion invariant mass distributions for the processes $\Upsilon(10580)\to\Upsilon(nS)\pi^+\pi^-$ $(n=1,2)$, compared with the predictions of the QCDME approach. The figure is adapted from Ref.~\cite{BaBar:2006udk}.}
\label{fig:Mpipi-4S-1S-Pap2-Model}
\end{figure}

Anomalous decay features have been observed in several transitions of the $\Upsilon(10580)$ state. One representative manifestation is found in the dipion invariant mass spectrum measured by the BaBar Collaboration \cite{BaBar:2006udk}. As shown in Fig.~\ref{fig:Mpipi-4S-1S-Pap2-Model}, the measured $m_{\pi^+\pi^-}$ distribution for $\Upsilon(10580)\to\Upsilon(1S)\pi^+\pi^-$ is broadly consistent with the line shape predicted by the QCDME approach \cite{Kuang:1981se}. However, the corresponding distribution for $\Upsilon(10580)\to\Upsilon(2S)\pi^+\pi^-$ exhibits a pronounced deviation from the QCDME predictions.

Another class of anomalies arises in the transitions $\Upsilon(nS)\to\Upsilon(1S)\eta$ $(n\geq 2)$. Within the QCDME framework, such processes are expected to be highly suppressed due to the spin-flip effects of the heavy quarks \cite{Kuang:1981se}. The ratio
\begin{align}
R(n)=\frac{BR[\Upsilon(nS)\to\Upsilon(1S)\eta]}{BR[\Upsilon(nS)\to\Upsilon(1S)\pi^+\pi^-]},
\end{align}
is typically of the order ${\cal O}(10^{-3})$, in good agreement with the QCDME expectations for $n=2,3$. However, the Belle Collaboration reported a significantly enhanced decay rate for $\Upsilon(4S)\to\Upsilon(1S)\eta$ \cite{BaBar:2008xay}, with the corresponding ratio determined to be $R(4)=2.41\pm0.40\pm0.12$, which is about two orders of magnitude larger than the QCDME prediction.

In addition, the OZI-suppressed spin-flip transition $\Upsilon(4S)\to h_b(1P)\eta$ has been found to have the largest non-$B\bar B$ branching fraction among known $\Upsilon(4S)$ hadronic transitions \cite{Belle:2015hnh}, even though spin-flipping processes are generally expected to be power suppressed \cite{Kuang:2006me,Voloshin:2007dx}.

Anomalous decay behaviors have also been reported in the transitions of the $\Upsilon(10860)$ state, as discussed in Sec.~\ref{section5}. In particular, the hidden-bottom dipion transitions $\Upsilon(10860)\to \Upsilon(nS)\pi^+\pi^-$ $(n=1,2,3)$ exhibit anomalously large partial widths. As summarized in Table~\ref{ISPE:UpsilonDipionWD}, the decay rates for $\Upsilon(10860)\to \Upsilon(1S,\,2S)\pi^+\pi^-$ are approximately two orders of magnitude larger than those of conventional dipion transitions $\Upsilon(mS)\to \Upsilon(nS)\pi^+\pi^-$ with $n<m\leq 4$. For lower-lying bottomonium states, both the magnitudes of the dipion transition rates and the associated invariant mass distributions are well described within the QCDME framework. By contrast, the exceptionally large widths of $\Upsilon(10860)\to \Upsilon(nS)\pi^+\pi^-$ $(n=1,2,3)$ clearly exceed the expectations of QCDME. Moreover, similar to the situation encountered in the $\Upsilon(10580)$ transitions, the experimentally measured dipion invariant mass spectra in the $\Upsilon(10860)$ decays exhibit pronounced deviations from the typical QCDME patterns. Anomalous behaviors are also observed in spin-flipping transitions. The Belle Collaboration reported the processes $\Upsilon(10860)\to h_b(nP)\pi^+\pi^-$ $(n=1,2)$ with branching fractions at the level of ${\cal O}(10^{-3})$~\cite{Belle:2011wqq}, which are significantly larger than the corresponding upper limits for $\Upsilon(3S)\to h_b(nP)\pi^+\pi^-$ $(n=1,2)$ measured by the BaBar Collaboration~\cite{BaBar:2011krt}. As discussed in Sec.~\ref{section5}, these enhanced transition rates are closely correlated with the experimentally observed $Z_b$ structures, which can be naturally understood within the ISPE mechanism. In addition, the Belle Collaboration observed the OZI-suppressed transitions $\Upsilon(10860)\to \chi_{bJ}\omega$ $(J=0,1,2)$~\cite{Belle:2014sys}, whose measured branching fractions are also at the level of ${\cal O}(10^{-3})$, far exceeding naive expectations.

\subsubsection{A global description of the anomalous bottomonium decays via the hadronic loop mechanism}

\changelabel{ In light of these anomalous transitions observed in experiments, various mechanisms have been proposed. Within the QCDME framework, improved descriptions of the two-gluon hadronization matrix elements and the dipion transition amplitudes were developed to account for hadronic transitions between heavy quarkonia more systematically~\cite{Kuang:2006me}. However, the unusually large $\Upsilon(10860)\to\Upsilon(nS)\pi^+\pi^-$ rates suggest that additional dynamics may be important for states above the open-bottom thresholds. An alternative interpretation identifies the enhancement near $10.89~\mathrm{GeV}$ as a hidden-bottom tetraquark state, $Y_b(10890)$~\cite{Ali:2009pi}. In this picture, the Belle data for $\Upsilon(1S,2S)\pi^+\pi^-$ were analyzed by including intermediate light-meson contributions such as $f_0(600)$, $f_0(980)$, and $f_2(1270)$~\cite{Ali:2009es}, and related channels including $\Upsilon(1S)K^+K^-$ and $\Upsilon(1S)\eta\pi^0$ were proposed as further tests of the tetraquark scenario~\cite{Ali:2010pq}.} 

Beyond such extensions, the hadronic loop mechanism, as outlined in Sec.~\ref{section2}, offers a global framework to account for all the anomalous transitions discussed above. An important reason why the QCDME works well for low-lying states but fails for high-lying ones may be attributed to the increasing importance of long-distance final state interactions associated with unquenched effects. For low-lying bottomonium, such contributions are not obvious because their masses lie far below the open-flavor $B\bar B$ threshold~\cite{Zhou:1990ik}. In contrast, for high-lying bottomonium states close to or above the open-flavor threshold, long-distance effects will become dominant, and some of the power counting rules underlying the QCDME approach may break down. In the following, a concise review  within this mechanism will be presented.

\paragraph{Dipion transitions of $\Upsilon(10580)$ and $\Upsilon(10860)$}
Analyses of the transitions $\Upsilon(10580/10860)\to\Upsilon(1S/2S)\pi^+\pi^-$ within the hadronic loop mechanism have been carried out in Ref. \cite{Meng:2007tk,Meng:2008dd,Chen:2011jp,Chen:2011qx}. As illustrated in Fig. \ref{Fig:Peakshifts}, the initial states $\Upsilon(10580)$ and $\Upsilon(10860)$ first decay into a $B^{(*)}\bar B^{(*)}$ pair, which subsequently undergoes a rescattering process. Through the exchange of a $B^{(*)}$ meson, the intermediate $B^{(*)}\bar B^{(*)}$ converts into a lower $\Upsilon$ state accompanied by a scalar meson, $\sigma/f_0$, which then decays into a $\pi^+\pi^-$ pair. In this picture, the initial and final states are dynamically connected by a hadronic loop composed of $B^{(*)}$ mesons.

As shown in Table~\ref{Tab:Predictions-Width}, the obtained branching ratios for $\Upsilon(10580,10860)\to\Upsilon(1S,2S)\pi^+\pi^-$ are in good agreement with the available experimental measurements. A consistent description is also obtained by incorporating the coupled-channel effects \cite{Simonov:2008qy,Simonov:2008ci}. Moreover, the much larger transition rates of $\Upsilon(10860)$ relative to those of $\Upsilon(10580)$ can be mainly attributed to the difference in the available phase space for the open-bottom processes $\Upsilon(10580,10860)\to B^{(*)}\bar B^{(*)}$, which is characterized by the three-momentum $\vec p_1$ of the $B^{(*)}$ mesons in the initial-state rest frame. With reasonable choices of the model parameters, the ratios of the transition rates are estimated to lie in the range of $200$–$600$ (see Refs.~\cite{Meng:2007tk,Meng:2008dd} for details).

\begin{center}\begin{table}
\caption{Transition widths, in units of keV, for the processes
$\Upsilon(4S,5S)\to\Upsilon(nS)\,\pi^+\pi^-$ and $K^+K^-$. The table is adapted from Ref.~\cite{Meng:2007tk}.}
\centering
\begin{tabular*}{1.0\textwidth}{l@{\extracolsep{\fill}}ccccc}
\hline
   & from $\sigma$ & from $f_0(980)$ &
   total           & Experimental data  \\
\hline
$\Upsilon(4S)\to\Upsilon(1S)\pi^+\pi^-$   & $0.54^{+0.00}_{-0.00}$ & $0.93^{+0.03}_{-0.03}$ &  $1.47\pm0.03$  & $1.8\pm0.4$         \\
$\Upsilon(4S)\to\Upsilon(2S)\pi^+\pi^-$   & $1.09^{+0.23}_{-0.21}$ & $0.05^{+0.00}_{-0.01}$ &  $1.14^{+0.23}_{-0.21}$  & $2.7\pm0.8$  \\
\hline $\Upsilon(5S)\to\Upsilon(1S)\pi^+\pi^-$  &
$102^{+1+42+21}_{-0-35-9}$ & $225^{+1+93+47}_{-1-77-43}$ &
$327^{+114}_{-97}$ & $590\pm 40 \pm 90$\\
$\Upsilon(5S)\to\Upsilon(2S)\pi^+\pi^-$  &
$385^{+10+164+87}_{-11-135-78}$ & $37^{+4+16+9}_{-3-13-7}$ &
$422^{+187}_{-157}$ & $850\pm 70 \pm 160$\\
$\Upsilon(5S)\to\Upsilon(3S)\pi^+\pi^-$  &
$306^{+78+133+73}_{-64-108-64}$ & $13^{+1+6+4}_{-1-4-2}$ &
$319^{+171}_{-141}$ & $520^{+200}_{-170}\pm 100$\\
\hline $\Upsilon(5S)\to\Upsilon(1S)K^+K^-$  &
 & $32^{+5+13+5}_{-5-11-6}$ &
$32^{+15}_{-13}$ & $67^{+17}_{-13}\pm 13$\\
\hline \label{Tab:Predictions-Width}\end{tabular*}
\end{table}\end{center}

In particular, Ref.~\cite{Meng:2008dd} studied the energy distributions of the dipion transitions $\Upsilon(10860)\to\Upsilon(1S,\,2S,\,3S)\pi^+\pi^-$ within the hadronic loop framework. It was found that this mechanism can shift the resonance peak of the $\Upsilon(10860)$ upward by about $7$–$20$~MeV in the processes $e^+e^-\to\Upsilon(1S,\,2S,\,3S)\pi^+\pi^-$ relative to that observed in the dominant open-bottom channels $e^+e^-\to B^{(*)}\bar B^{(*)}$, as shown in Fig. \ref{Fig:Peakshifts}. Such an effect \changelabel{provides} a possible clue to why the masses of $\Upsilon(10860)$ extracted from exclusive hidden-bottom decay modes by the BaBar and Belle Collaborations are systematically higher than those obtained from earlier inclusive measurements, and may also be relevant for understanding the difficulties in assigning $\Upsilon(10860)$ to the previously discussed unquenched spectra.

\begin{figure}[htbp]
\centering
\includegraphics[width=0.7\textwidth]{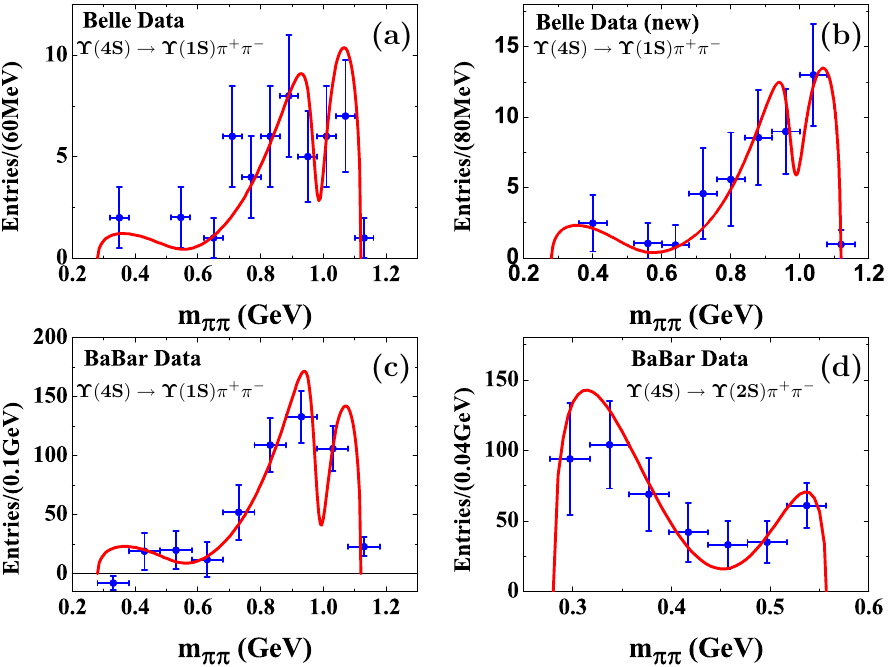}
\includegraphics[width=0.7\textwidth]{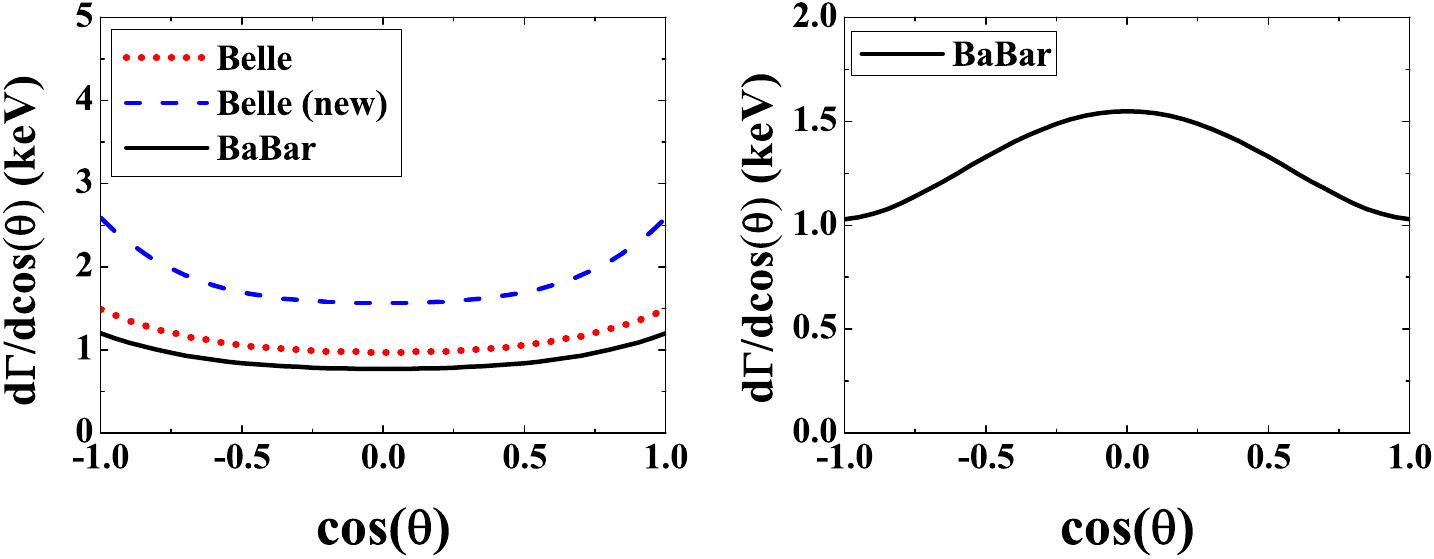}
\caption{The fitted dipion invariant mass distributions and the corresponding predictions for the angular distributions $d\Gamma/d\cos\theta$ for $\Upsilon(10580)\to \Upsilon(1S)\pi^+\pi^-$ (left panel) and $\Upsilon(4S)\to \Upsilon(2S)\pi^+\pi^-$ (right panel). Figure adapted from Ref.~\cite{Chen:2011jp}.}
\label{fig:Upsilon4S}
\end{figure}

Besides the magnitudes of the branching ratios, the dipion invariant mass spectra provide crucial information on the underlying decay mechanisms. The $\pi^+\pi^-$ mass distributions for the processes $\Upsilon(10580)\to\Upsilon(1S,\,2S)\pi^+\pi^-$ and $\Upsilon(10860)\to\Upsilon(1S,\,2S)\pi^+\pi^-$ have been analyzed in Refs.~\cite{Chen:2011jp,Chen:2011qx}. In these studies, both the dominant hadronic loop contributions and the direct decay mechanisms were taken into account, as exemplified by Eq.~\eqref{decay} for the case of $\Upsilon(10860)$. By incorporating the interference between the corresponding decay amplitudes, the experimental dipion invariant mass spectra were well reproduced, as shown in Figs.~\ref{fig:Upsilon4S} and \ref{Fig:chen-upsilon5Sdipion-ML}. In addition, the resulting angular distributions for the $\Upsilon(10860)\to \Upsilon(1S)\pi^+\pi^-$ transitions are consistent with the available experimental data (the disagreement for $\Upsilon(10860)\to \Upsilon(2S)\pi^+\pi^-$ process is related to the $Z_b$ problem, see Sec.~\ref{section5} for details), while those for the $\Upsilon(10580)$ transitions constitute clear theoretical predictions awaiting future experimental tests.

\changelabel{ Additionally, it is worth emphasizing that Guo, Shen, Chiang and Ping in Ref.~\cite{Guo:2006ai} also analyzed the $\pi^+\pi^-$ invariant-mass spectra in $\Upsilon(10580)\to\Upsilon(1S,2S)\pi^+\pi^-$ and predicted the corresponding $\cos\theta_\pi^\ast$ distributions. In that work, the observed dipion spectra were described by introducing an additional sequential decay mechanism, $\Upsilon(10580)\to\pi X\to\Upsilon(1S,2S)\pi^+\pi^-$, together with the $\pi\pi$ final-state interaction, where $X$ denotes an intermediate $b\bar b q\bar q$ configuration. This provides an early alternative description of the nontrivial dipion spectra, complementary to the hadronic loop mechanism reviewed above.}

\paragraph{$\Upsilon(10860)$ and $\Upsilon(10580)$ transitions with $\eta$ and $\omega$ emission} 

\begin{figure}[htbp]
\centering
\includegraphics[width=0.8\textwidth]{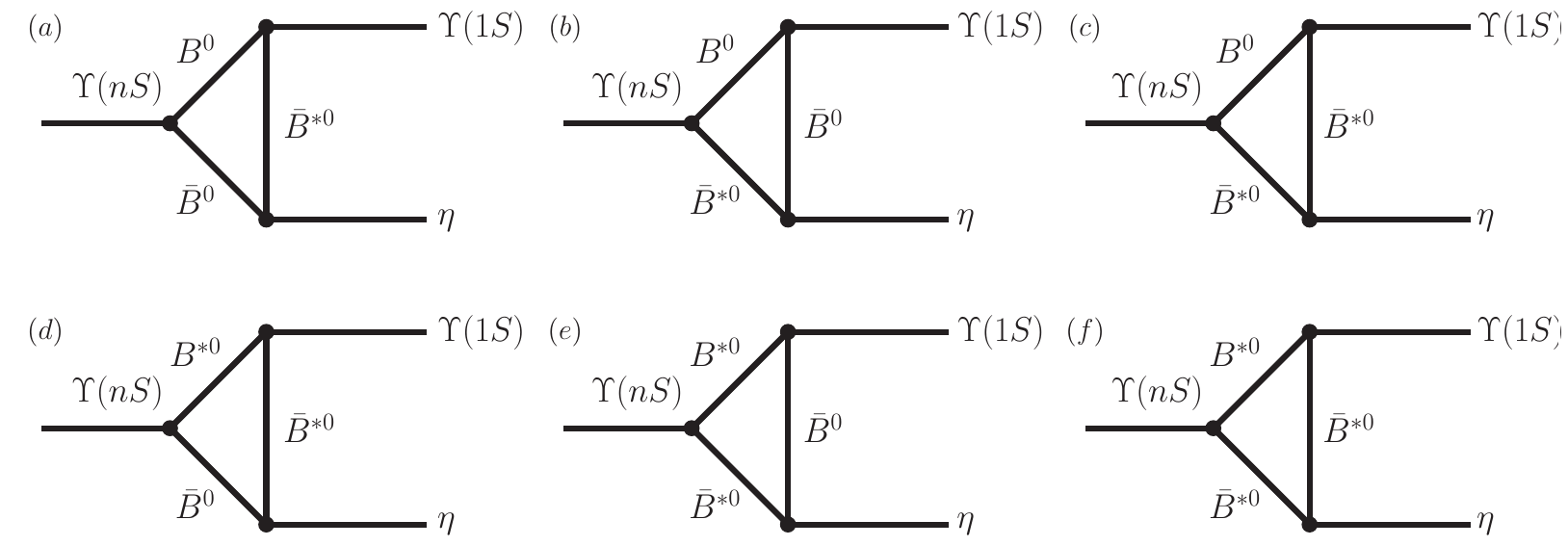}
\caption{Feynman diagram illustrating the hadronic loop contributions to the transitions $\Upsilon(10580),\,\Upsilon(10860)\to \Upsilon(1S)\eta$. Figure adapted from Ref.~\cite{Meng:2008bq}.}
\label{fig:Y-Yeta}
\end{figure}

The transitions $\Upsilon(10580),\,\Upsilon(10860)\to \Upsilon(1S,\,2S)\eta$ were studied within the hadronic loop mechanism in Ref.~\cite{Meng:2008bq}. The corresponding diagram is illustrated in Fig.~\ref{fig:Y-Yeta}. The resulting partial width for $\Upsilon(10580)\to \Upsilon(1S)\eta$ is found to be $2.94~\text{keV}$. Combined with the partial width of $\Upsilon(10580)\to \Upsilon(1S)\pi^+\pi^-$ obtained in the same framework and listed in Table~\ref{Tab:Predictions-Width}, the ratio
\begin{align}
R(4)=\frac{BR[\Upsilon(10580)\to\Upsilon(1S)\eta]}{BR[\Upsilon(10580)\to\Upsilon(1S)\pi^+\pi^-]}
\simeq 2.0,
\end{align}
is consistent with the experimental measurement. 

For the $\eta$ transitions of the $\Upsilon(10860)$, the widths are highly sensitive to the coupling constants $g_{\Upsilon(5S)B^{(*)}B^{(*)}}$, due to a substantial cancellation among the contributions from the $B\bar{B}$, $B\bar{B}^*+\text{c.c.}$, and $B^*\bar{B}^*$ channels. Consequently, only a rough estimate can be provided, yielding $\Gamma(\Upsilon(10860)\to\Upsilon(1S,2S)\eta)=10-200$ keV, which correspond to branching fractions in the range $(0.3$–$5.4)\times10^{-3}$ by take $\Gamma(\Upsilon(10860))=37$ MeV as a reference \cite{ParticleDataGroup:2024cfk}. While recent Belle measurements reported $BR(\Upsilon(10860)\to\Upsilon(1S)\eta)=(0.85\pm0.15\pm0.08)\times10^{-3}$, and $BR(\Upsilon(10860)\to\Upsilon(2S)\eta)=(4.13\pm0.41\pm0.37)\times10^{-3}$ \cite{Belle:2021gws}, which are in good agreement with the above estimates.
One may further consider the ratio
\begin{align}
R(5)=\frac{BR[\Upsilon(10860)\to\Upsilon(1S)\eta]}{BR[\Upsilon(10860)\to\Upsilon(1S)\pi^+\pi^-]}
\simeq 0.03-0.30,
\end{align}
which also encompasses the value measured by the Belle Collaboration,
$R(5)=0.19\pm0.04\pm0.01$ \cite{Belle:2021gws}.

The estimated contributions from virtual hadronic loops to the transitions $\Upsilon(2S,3S)\to\Upsilon(1S)\eta$ are small, with the contribution for $n=3$ roughly consistent with the experimental upper limit and that for $n=2$ an order of magnitude below the measured value \cite{CLEO:2008dfe}. This indicates that the QCDME mechanism remains dominant for the $\Upsilon(2S)\to\Upsilon(1S)\eta$ transition, as expected for a state lying far below the $B\bar B$ threshold.

\begin{figure}[htbp]
\centering
\includegraphics[width=1.0\textwidth]{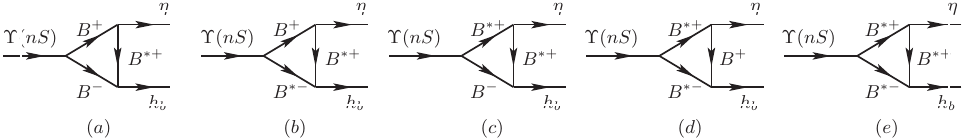}
\caption{Feynman diagram illustrating the hadronic loop contributions to the transitions $\Upsilon(10580),\,\Upsilon(10860),\,\Upsilon(11020)\to h_b(1P)\eta$. Figure adapted from Ref.~\cite{Zhang:2018eeo}.}
\label{fig:fig_diagram}
\end{figure}

Ref.~\cite{Zhang:2018eeo} investigated the processes
$\Upsilon(10580),\,\Upsilon(10860),\,\Upsilon(11020)\to h_b(1P)\eta$
within the hadronic loop mechanism. The corresponding loop diagrams are shown in Fig.~\ref{fig:fig_diagram}. The calculated branching ratios for $\Upsilon(10580)$ were found to be consistent with the experimental measurements, indicating that hadronic loop effects can substantially enhance spin-flipping hidden-bottom transitions. It was further shown that the predicted branching ratios for $\Upsilon(10860)$ and $\Upsilon(11020)$ are approximately one and two orders of magnitude smaller, respectively, than that for $\Upsilon(10580)$. This suppression may partially account for the absence of experimental signals for these two processes so far.

\begin{figure}[htbp]
\centering
\begin{tabular}{c @{\hspace{6pt}\vrule\hspace{6pt}} c 
                @{\hspace{6pt}\vrule\hspace{6pt}} c}
  \includegraphics[width=0.30\textwidth]{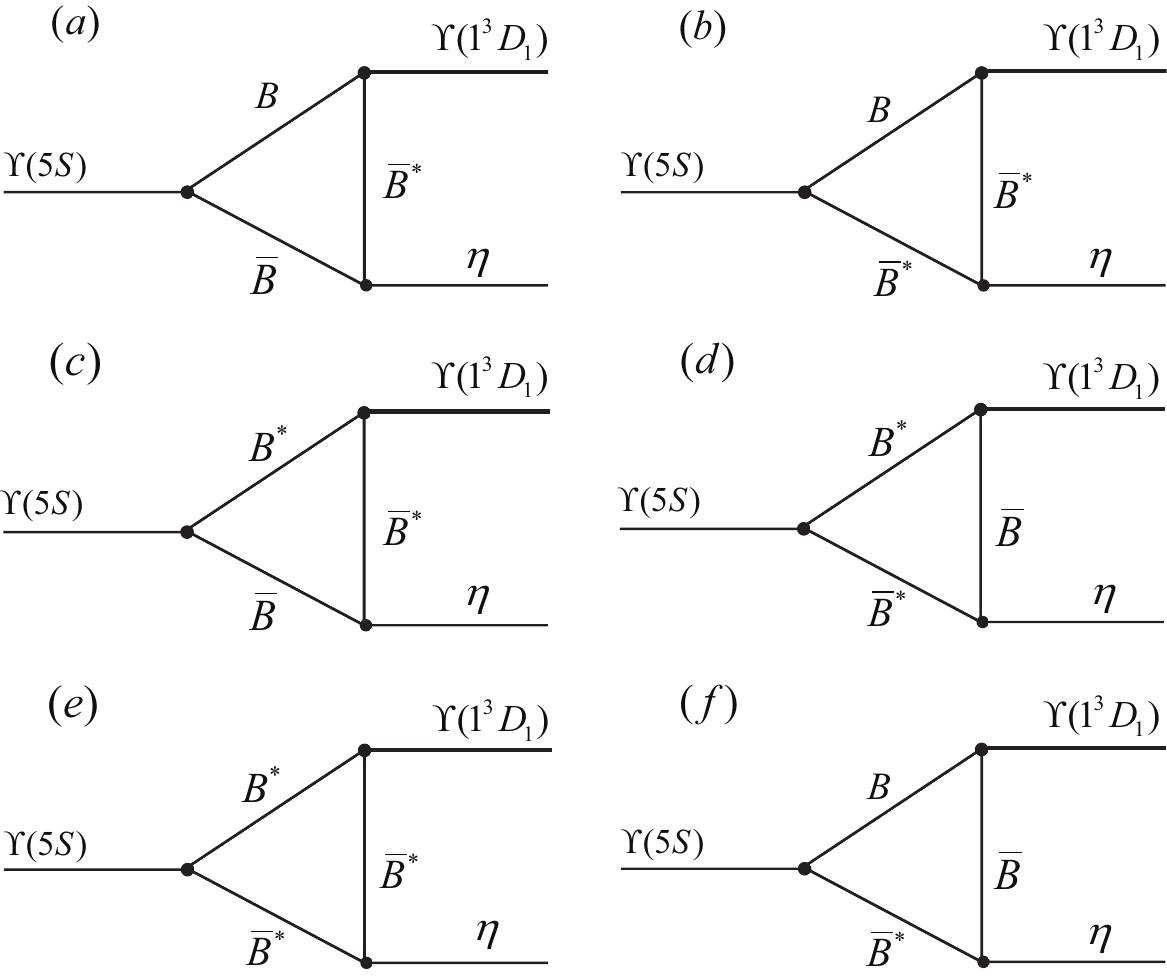} &
  \includegraphics[width=0.30\textwidth]{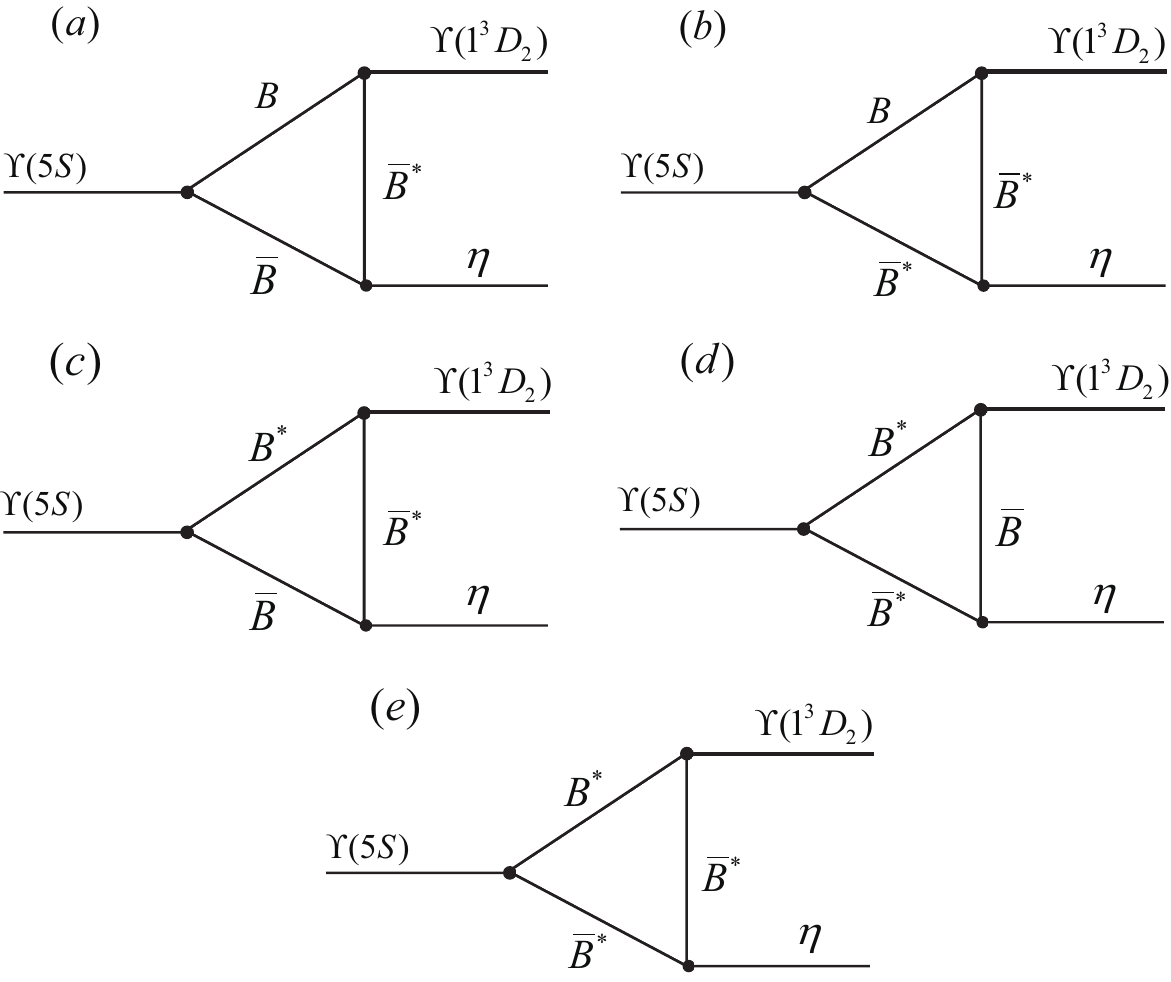} &
  \raisebox{42pt}{\includegraphics[width=0.30\textwidth]{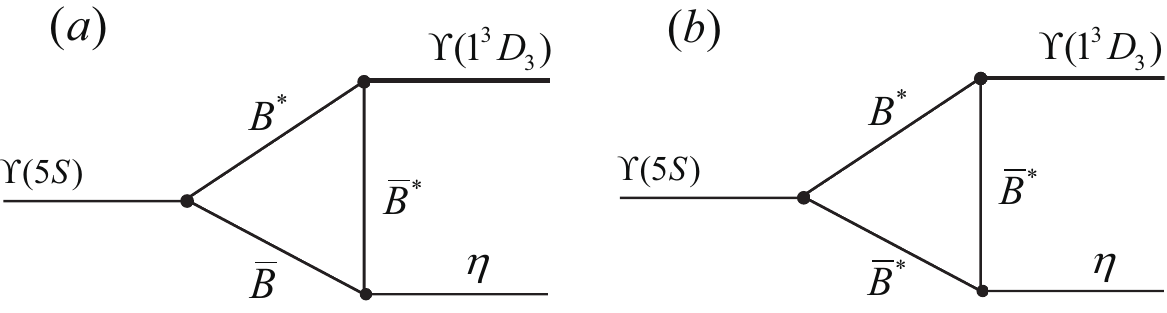}}
\end{tabular}
\caption{Feynman diagrams illustrating the $\eta$ transitions of
$\Upsilon(10860)\to\Upsilon(1^3D_J)\eta$ with $J=1,2,3$.
Adapted from Ref.~\cite{Wang:2016qmz}.}
\label{fig:feyn123}
\end{figure}

\begin{figure}[htbp]
\centering
\includegraphics[width=0.4\textwidth]{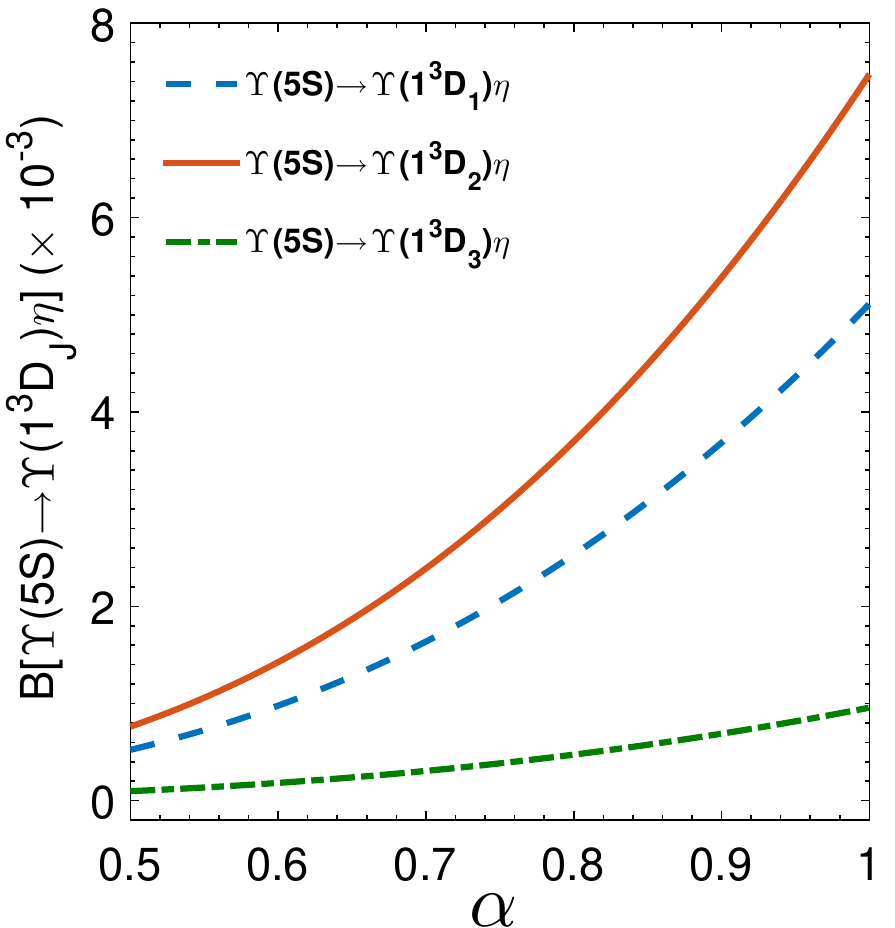}
\hspace{0.05\textwidth}
\includegraphics[width=0.415\textwidth]{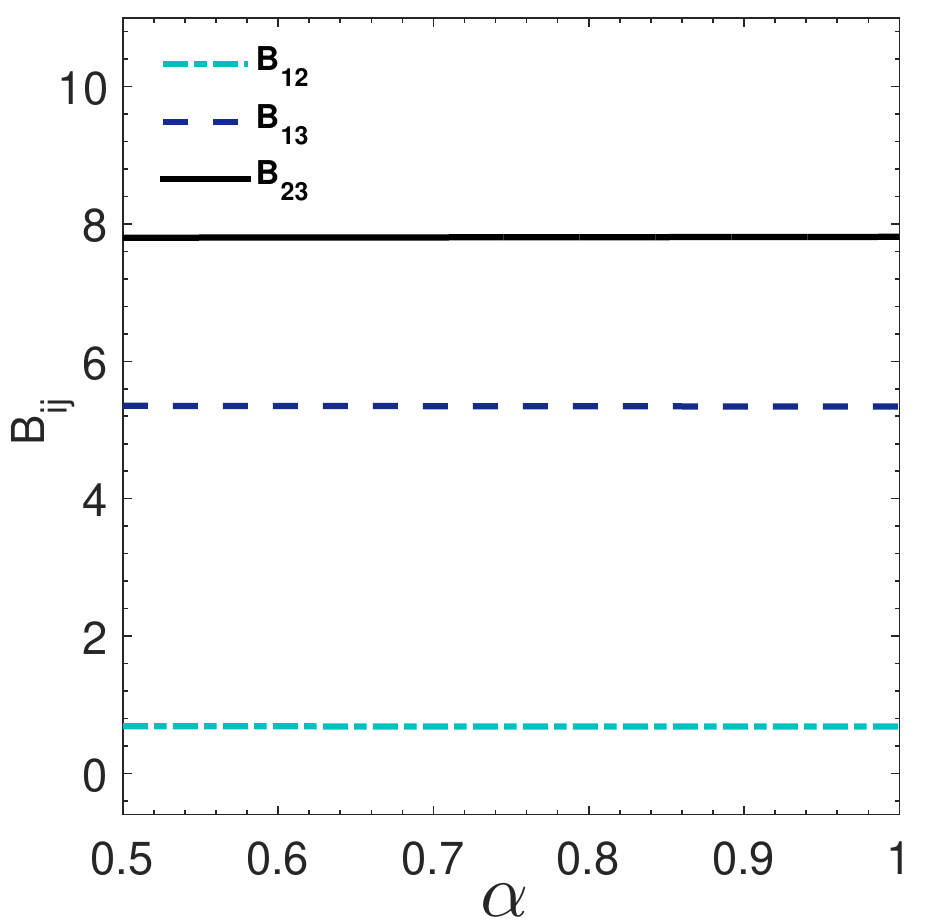}
\caption{The predicted branching ratios of $\Upsilon(10860)\to\Upsilon(1^3D_J)\eta$ (left panel), as well as the ratios between them (right panel). Figure adapted from Ref.~\cite{Wang:2016qmz}.}
\label{fig:BR123}
\end{figure}

Furthermore, the $\eta$ transitions of $\Upsilon(10860)$ into the $D$-wave bottomonium states
$\Upsilon(1^3D_J)$ with $J=1,2,3$ were systematically investigated in Ref.~\cite{Wang:2016qmz}. Experimentally, only the $\Upsilon(1^3D_2)$ state has been firmly established so far~\cite{CLEO:2004npj,BaBar:2010tqb}, while the $\Upsilon(1^3D_1)$ and $\Upsilon(1^3D_3)$ states remain unobserved. The corresponding hadronic loop diagrams are illustrated in Fig.~\ref{fig:feyn123}. The predicted branching ratios, together with their
characteristic ratios, are presented in Fig.~\ref{fig:BR123}.

It is found that the branching ratios for $\Upsilon(10860)\to \Upsilon(1^3D_J)\eta$ are of the order of $\mathcal{O}(10^{-3})$, which are comparable to those of the experimentally
observed dipion transitions $\Upsilon(10860)\to \Upsilon(nS)\pi^+\pi^-\,(n=1,2,3)$.
This indicates that these $\eta$ transitions possess considerable discovery potential in future experiments. Moreover, the characteristic ratios reveal that the $\eta$ transitions into
$\Upsilon(1^3D_1)$ and $\Upsilon(1^3D_2)$ occur at a similar magnitude, whereas the decay $\Upsilon(10860)\to \Upsilon(1^3D_3)\eta$ is suppressed by approximately one order of magnitude relative to the former two channels.

Following the publication of the theoretical work, the Belle Collaboration reported the first observation of the process
$\Upsilon(10860) \to \Upsilon_J(1D)$~\cite{Belle:2018hjt}.
The measured branching fraction,
$(4.82 \pm 0.92 \pm 0.67) \times 10^{-3}$, agrees well with the expectation from the hadronic loop mechanism~\cite{Wang:2016qmz}.
However, no direct experimental evidence has been established for the simultaneous presence of all three members of the $\Upsilon_J(1D)$ triplet.
Consequently, only $90\%$ confidence-level upper limits are set on the production fractions of the $J=1$ and $J=3$ states relative to the $J=2$ state.

These overall agreements provide further support for the hadronic loop mechanism in describing the anomalous $\eta$ transitions of  $\Upsilon(10580)$ and $\Upsilon(10860)$.

\begin{figure}[htbp]
\centering
\includegraphics[width=0.8\textwidth]{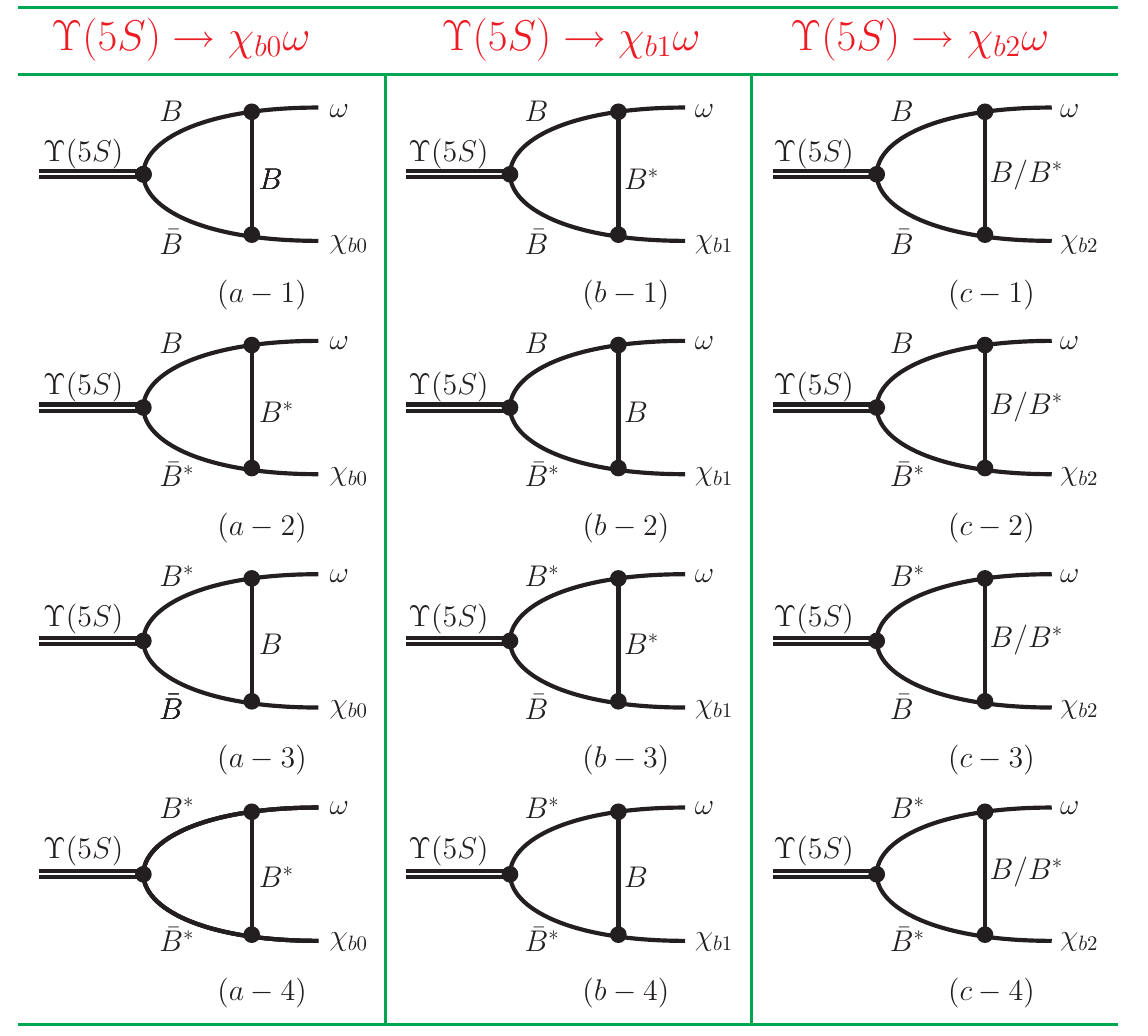}
\caption{Feynman diagram depicting $\Upsilon(10860)\to \chi_{bJ}\omega$ decay within the hadronic loop mechanism. Figure adapted from Ref.~\cite{Chen:2014ccr}.}
\label{fig:chibJV}
\end{figure}

\begin{figure}[htbp]
\centering
\includegraphics[width=0.7\textwidth]{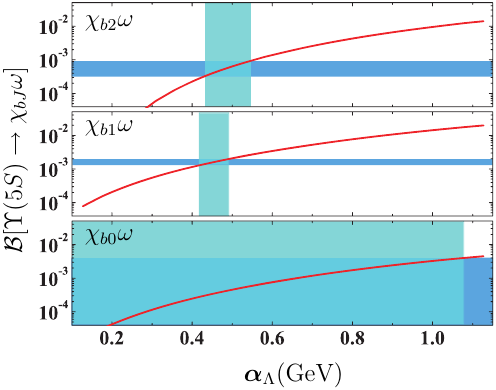}
\caption{The branching ratios of $\Upsilon(10860) \to \chi_{bJ}\omega$ as functions of the parameter $\alpha_\Lambda$. The horizontal bands represent the experimental measurements reported by the Belle Collaboration \cite{Belle:2014sys}, while the vertical bands indicate the reasonable ranges of $\alpha_\Lambda$ where the theoretical results overlap with the Belle data. The figure is adapted from Ref. \cite{Chen:2014ccr}.}
\label{fig:omega}
\end{figure}

The hadronic loop mechanism also provides a natural explanation for the large branching ratios of $\Upsilon(10860)\to \chi_{bJ}\omega$ $(J=0,1,2)$ ~\cite{Chen:2014ccr}. The corresponding Feynman diagrams are presented in Fig.~\ref{fig:chibJV}. The resulting branching ratios are shown in Fig.~\ref{fig:omega}, where the experimentally measured values are well reproduced with reasonable choices of parameters, indicating that hadronic loop effects give dominant contributions to these decay processes.

\changelabel{ The $\chi_{bJ}\omega$ hadronic transitions have also been discussed from the viewpoint of HQSS~\cite{Guo:2014qra,Luo:2025kid}. If $\Upsilon(10860)$ is treated as a pure $5S$ bottomonium state, HQSS leads to the characteristic relation $\Gamma[\chi_{b0}\omega]:\Gamma[\chi_{b1}\omega]:\Gamma[\chi_{b2}\omega]=1:2.8:4.4$ for $\Upsilon(10860)\to\chi_{bJ}\omega$~\cite{Guo:2014qra}, while the Belle data give $\mathcal B[\chi_{b1}\omega]/\mathcal B[\chi_{b2}\omega]=2.62\pm1.30$~\cite{Belle:2014sys}, indicating a sizable deviation from this simple expectation. The authors in Ref.~\cite{Guo:2014qra} pointed out that a small $S$-$D$ mixing in the $\Upsilon(10860)$ wave function can induce an enhanced breaking of these HQSS ratios, especially when the decay of the $D$-wave component is dominated by coupled-channel effects. More recently, Ref.~\cite{Luo:2025kid} studied $\Upsilon(10753)$ and $\Upsilon(10860)$ as $4S$-$3D$ and $5S$-$4D$ mixed states, respectively, and treated the open bottom loop contributions to the $D$-wave components within a nonrelativistic effective field theory framework. These studies provide a useful perspective of HQSS and bottom meson loop on the $\chi_{bJ}\omega$ hadronic transitions of higher bottomonium states.}

In short, within the unquenched picture, the bottomonium spectrum can be described more accurately, and the observed anomalous hidden-bottom decays of high-lying bottomonium states can be naturally understood by incorporating hadronic loop effects. Moreover, several corresponding predictions have been confirmed by subsequent experiments. These results suggest that the observed anomalous phenomena can be accommodated within the unquenched quarkonium framework, and that exotic interpretations may not be necessary.

\paragraph{Investigation of hidden-bottom decays of $\Upsilon(11020)$}

In the above discussions, we have shown that a series of anomalous hidden-bottom decay behaviors observed in $\Upsilon(10580)$ and $\Upsilon(10860)$ can be well understood within a global hadronic loop mechanism. These studies indicate that unquenched hadronic loop effects play a dominant role in the hidden-bottom decays of high-lying bottomonium states.

As the highest established vector bottomonium state, $\Upsilon(11020)$ is also located above the open-flavor $B^{*}\bar B^{*}$ thresholds, where similar hadronic loop effects are expected to be significant. Given that our current knowledge of $\Upsilon(11020)$ remains rather limited due to the scarcity of relevant experimental information \cite{ParticleDataGroup:2024cfk}, it is therefore worthwhile to extend the hadronic loop framework to $\Upsilon(11020)$ and to explore its possible hidden-bottom decay behaviors. Since the corresponding loop diagrams are analogous to those involved in the $\Upsilon(10860)$ transitions, we do not repeat the detailed diagrams here, but instead summarize the main results from the related studies in Table \ref{tab:upsilon11020hidden}. These results may provide useful guidance for future experimental investigations of this state.

\begin{table}[htbp]
\centering
\caption{Predicted branching ratios for the hidden-bottom decays of $\Upsilon(11020)$, together with several characteristic ratios among them. The results are summarized from Refs.~\cite{Huang:2018pmk,Huang:2018cco,Huang:2017kkg,Zhang:2018eeo}.}
\renewcommand\arraystretch{1.3}
\begin{tabular*}{1.0\textwidth}{l@{\extracolsep{\fill}}cccc}
\hline
\multicolumn{2}{c}{Final states} & $BR\,(\times10^{-3})$ &\multicolumn{2}{c}{characteristic ratios} \\
\hline
\multirow{3}{*}{$\Upsilon(nS)\eta$ \cite{Huang:2018pmk}} &$\Upsilon(1S)\eta$& 3.238 &$BR[(\Upsilon(2S)\eta)]/BR[(\Upsilon(1S)\eta)]$ &0.579--0.626\\
{}&$\Upsilon(2S)\eta$ &1.947 &$BR[(\Upsilon(3S)\eta)]/BR[(\Upsilon(1S)\eta)]$ &0.039--0.045\\
{}&$\Upsilon(3S)\eta$& 0.135 &$BR[(\Upsilon(3S)\eta)]/BR[(\Upsilon(2S)\eta)]$ &0.067--0.072\\
\hline
\multirow{2}{*}{$\Upsilon(nS)\eta^\prime$ \cite{Huang:2018pmk}} &$\Upsilon(1S)\eta^\prime$& 1.448 &$BR[(\Upsilon(1S)\eta)]/BR[(\Upsilon(1S)\eta^\prime)]$ &2.231--2.241\\
{}&$\Upsilon(2S)\eta^\prime$ &$1.305\times10^{-3}$ &$BR[(\Upsilon(2S)\eta)]/BR[(\Upsilon(2S)\eta^\prime)]$ &$(1.478\text{--}1.508)\times10^3$\\
\hline
\multirow{3}{*}{$\Upsilon(1^3D_J)\eta$ \cite{Huang:2018cco}} &$\Upsilon(1^3D_1)\eta$& 0.174--1.720 &$BR[(\Upsilon(1^3D_2)\eta)]/BR[(\Upsilon(1^3D_1)\eta)]$ &0.841--0.853\\
{}&$\Upsilon(1^3D_2)\eta$ &0.149--1.440 &$BR[(\Upsilon(1^3D_3)\eta)]/BR[(\Upsilon(1^3D_1)\eta)]$ &1.289--1.306\\
{}&$\Upsilon(1^3D_3)\eta$&0.228--2.210 &$BR[(\Upsilon(1^3D_3)\eta)]/BR[(\Upsilon(1^3D_2)\eta)]$ &1.531--1.533\\
\hline
\multirow{3}{*}{$\chi_{bJ}\phi$ \cite{Huang:2017kkg}} &$\chi_{b0}\phi$&$(0.68\text{--}4.62)\times 10^{-3}$ &$BR[(\chi_{b1}\phi)]/BR[(\chi_{b0}\phi)]$ &0.74\\
{}&$\chi_{b1}\phi$ &$(0.50\text{--}3.43)\times10^{-3}$ &$BR[(\chi_{b2}\phi)]/BR[(\chi_{b0}\phi)]$ &3.28\\
{}&$\chi_{b2}\phi$ &$(2.22\text{--}15.18)\times10^{-3}$ &$BR[(\chi_{b2}\phi)]/BR[(\chi_{b1}\phi)]$ &4.43\\
\hline
\multirow{3}{*}{$\chi_{bJ}\omega$ \cite{Huang:2017kkg}} &$\chi_{b0}\omega$&0.15--2.81 &$BR[(\chi_{b1}\omega)]/BR[(\chi_{b0}\omega)]$ &4.11\\
{}&$\chi_{b1}\omega$ &0.63--11.68 &$BR[(\chi_{b2}\omega)]/BR[(\chi_{b0}\omega)]$ &7.06\\
{}&$\chi_{b2}\omega$ &1.08--20.02 &$BR[(\chi_{b2}\omega)]/BR[(\chi_{b1}\omega)]$ &1.72\\
\hline
\multicolumn{2}{c}{$h_b(1P)\eta$ \cite{Zhang:2018eeo}} &0.1--10.0 &\multicolumn{2}{c}{$\cdots$}\\
\hline
\end{tabular*}
\label{tab:upsilon11020hidden}
\end{table}

As summarized in Table~\ref{tab:upsilon11020hidden}, several decay channels, including $\Upsilon(nS)\eta$, $\Upsilon(1S)\eta^\prime$, $\Upsilon(1^3D_J)\eta$, and $\chi_{bJ}\omega$, are predicted to possess relatively large branching ratios and thus represent promising modes for experimental investigation. Subsequently, experimental analyses have reported evidence for the decay $\Upsilon(11020)\to \chi_{bJ}\omega$, although the spin assignment ($J=1$ or $2$) could not be unambiguously determined. This signal was observed via the process $\Upsilon(11020)\to \chi_{bJ}\pi^+\pi^-\pi^0$, with a measured branching ratio of $(8.7\pm4.3\pm6.1^{+4.5}_{-2.5})\times10^{-3}$, which is compatible with the theoretical expectations obtained in Ref.~\cite{Huang:2017kkg}.

\subsubsection{A consistent assignment of $\Upsilon(10753)$ into the bottomonium family}
\paragraph{Observation of $\Upsilon(10753)$ and subsequent theoretical interpretations}

\begin{figure}[htbp]
\centering
\includegraphics[width=0.6\textwidth]{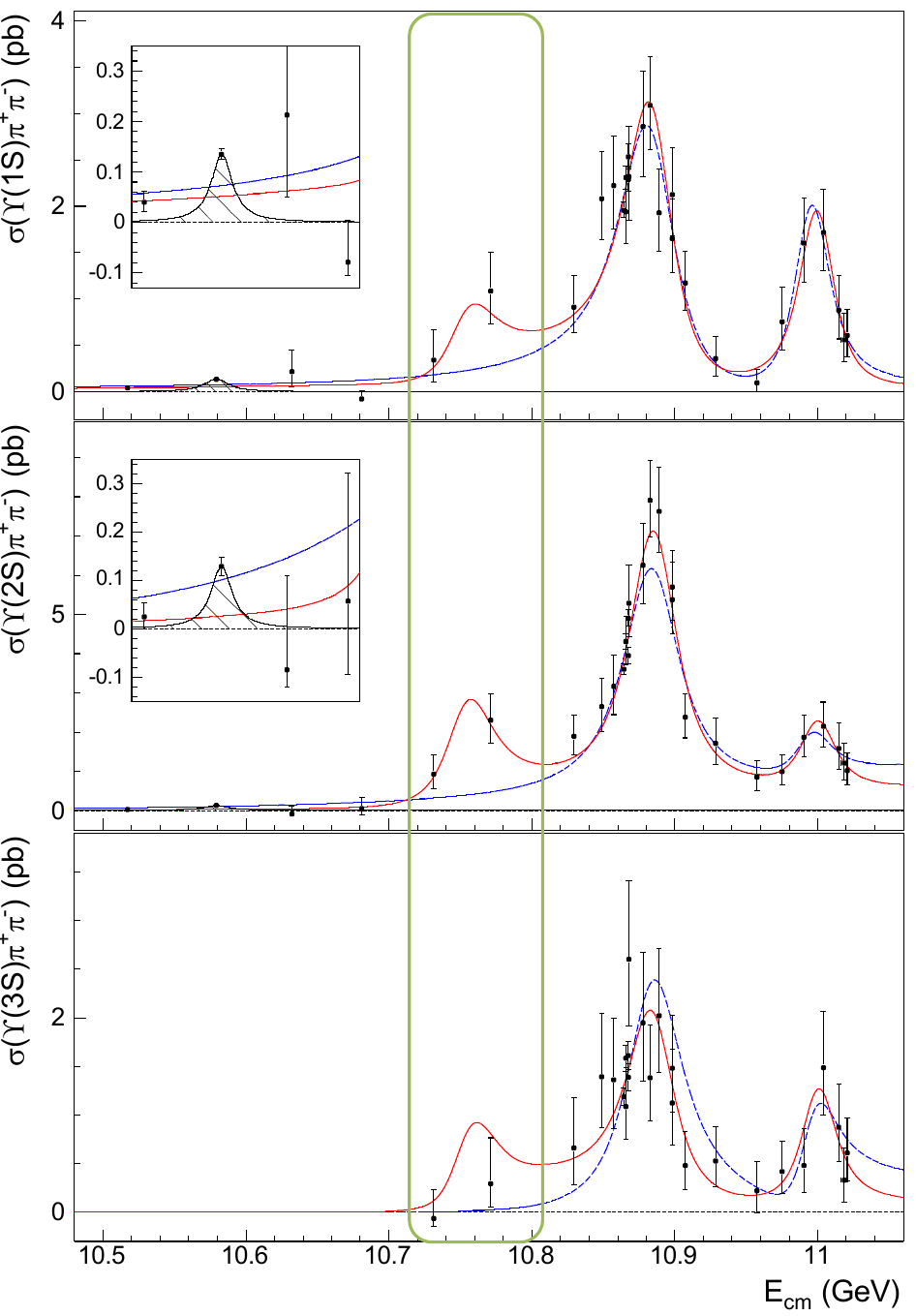}
\caption{Cross sections for $e^{+}e^{-}\to\Upsilon(nS)\pi^{+}\pi^{-}$
measured by the Belle Collaboration~\cite{Belle:2019cbt}, showing the
structure of $\Upsilon(10753)$. The figure is adapted from Ref.~\cite{Belle:2019cbt}.}
\label{fig:observe10753}
\end{figure}

In 2019, the Belle Collaboration updated its analysis of the processes $e^{+}e^{-}\to\Upsilon(nS)\pi^{+}\pi^{-}$ $(n=1,2,3)$ \cite{Belle:2019cbt}, superseding earlier measurements~\cite{Belle:2015aea}. Besides the known resonances $\Upsilon(10860)$ and $\Upsilon(11020)$, a new structure with a statistical significance of $5.2\sigma$ was observed near 10.75~GeV, as shown in Fig. \ref{fig:observe10753}. This state is now listed by the PDG as $\Upsilon(10753)$ \cite{ParticleDataGroup:2024cfk}. The measured resonance parameters are

\begin{align}
M = 10752.7 \pm 5.9^{+0.7}_{-1.1}~\mathrm{MeV}\qquad \Gamma = 35.5^{+17.6+3.9}_{-11.3-3.3}~\mathrm{MeV}
\end{align}

Following its experimental observation, this bottomonium-like structure has stimulated extensive theoretical investigations into its nature. A variety of interpretations have been proposed, including its assignment as a conventional bottomonium state~\cite{Chen:2019uzm,Li:2019qsg,Kher:2022gbz,Li:2021jjt,Kaushal:2025kbz,Bokade:2025voh,Luo:2025kid}, as well as several exotic scenarios, such as a tetraquark configuration~\cite{Wang:2019veq,Ali:2019okl,Zhao:2025kno} and a hybrid state~\cite{TarrusCastella:2019lyq,TarrusCastella:2021pld}. In Ref.~\cite{Ortega:2024rrv}, $\Upsilon(10753)$ is interpreted as a dressed hadronic resonance, whose structure consists of an approximately equal admixture of a conventional $b\bar b$ bottomonium component and a $B^*\bar B^*$ molecular component. In addition, Refs.~\cite{Bicudo:2019ymo,Bicudo:2020qhp} described $\Upsilon(10753)$ as a dynamical meson–meson resonance with about 76\% meson–meson content, while Ref.~\cite{Ni:2025gvx} proposed that the observed structure may instead arise from a threshold effect associated with the coupling of the $B^*\bar B^*$ channel to  $\Upsilon(4S)$, rather than corresponding to a genuine resonance.

The production of $\Upsilon(10753)$ in $e^{+}e^{-}$ annihilation identifies it as a vector state, with quantum numbers $J^{PC}=1^{--}$. Its mass lies between the experimentally observed $\Upsilon(10580)$ and $\Upsilon(10860)$, commonly identified as the $\Upsilon(4S)$ and $\Upsilon(5S)$ states, respectively. 
A natural preliminary interpretation is therefore to assign $\Upsilon(10753)$ as a $\Upsilon(3D)$ bottomonium state \cite{Chen:2019uzm}. However, such an assignment encounters serious difficulties.

First, the measured mass of $\Upsilon(10753)$ is significantly higher than theoretical expectations for the $\Upsilon(3D)$ state \cite{Chen:2019uzm}. Specifically, it exceeds the prediction of the unquenched model discussed above by about $80~\mathrm{MeV}$~\cite{Wang:2018rjg}, is roughly $100~\mathrm{MeV}$ higher than the nonrelativistic constituent quark model result~\cite{Segovia:2016xqb}, and is about $55~\mathrm{MeV}$ and $35$-$55~\mathrm{MeV}$ above the predictions of the quenched GI model~\cite{Godfrey:1985xj,Godfrey:2015dia} and the relativistic string Hamiltonian approach~\cite{Badalian:2008ik,Badalian:2009bu}, respectively. Second, the dielectron width expected for a conventional $\Upsilon(3D)$ state is extremely small. Various theoretical studies estimate $\Gamma_{ee}[\Upsilon(3D)]$ to be only $1$--$3~\mathrm{eV}$
~\cite{Godfrey:1985xj,Godfrey:2015dia,Wang:2018rjg,Badalian:2008ik,Badalian:2009bu}. Compared with the dielectron widths of the $\Upsilon(4S)$ and $\Upsilon(5S)$, this represents a suppression of nearly two orders of magnitude, making the $\Upsilon(3D)$ state difficult to observe directly in the $e^{+}e^{-}$ annihilation process. This expectation is clearly at odds with the prominent signal of $\Upsilon(10753)$ observed in the $e^{+}e^{-}\to\Upsilon(nS)\pi^{+}\pi^{-}$ processes~\cite{Belle:2019cbt}.

\paragraph{$4S$-$3D$ mixing scheme}

To address these difficulties, it is instructive to recall some established experience in constructing the charmonium family. In Ref.~\cite{Rosner:2001nm}, Rosner introduced a mixing scheme between the $2S$ and $1D$ charmonium states to account for puzzling phenomena such as the ``$\rho\pi$ puzzle’’ associated with the $\psi(3686)$ and $\psi(3770)$ states. When extending to higher charmonia, the $4S$–$3D$ and $5S$–$4D$ mixing schemes were further introduced to construct a coherent vector charmonium mass spectrum in the 4.0–4.5~GeV region within the unquenched picture, as illustrated in Sec.~\ref{section4}. This framework successfully accommodates a wide range of experimental observations. From these previous experiences in constructing the charmonium family, it can be concluded that the $S$–$D$ mixing effect is rather universal and should not be neglected in a realistic description of heavy quarkonium spectroscopy.

Motivated by the studies on charmonium spectroscopy~\cite{Rosner:2001nm,Wang:2019mhs}, it has been suggested that an analogous $S$–$D$ wave mixing mechanism may also play an important role in resolving the mass puzzle associated with the $\Upsilon(10753)$~\cite{Li:2021jjt}. According to the PDG compilation~\cite{ParticleDataGroup:2024cfk}, the experimentally measured mass of $\Upsilon(10580)$, $10579.4$~MeV, is noticeably lower than the mass range predicted for the $\Upsilon(4S)$ state, $10607$–$10640$~MeV, in various potential model calculations~\cite{Segovia:2016xqb,Wang:2018rjg,Godfrey:2015dia,Badalian:2008ik,Badalian:2009bu}. Meanwhile, the observed mass of $\Upsilon(10753)$ is located above the predicted mass region of the $\Upsilon(3D)$ state, $10653$–$10717$~MeV.
Such a mass ordering provides a typical environment for level repulsion induced by $S$–$D$ mixing and thus naturally motivates the introduction of a $4S$–$3D$ mixing scheme, within which the mass discrepancy associated with $\Upsilon(10753)$ can be effectively accommodated. In particular, the physical vector bottomonium states $\Upsilon(10580)$ and $\Upsilon(10753)$, resulting from the $4S$–$3D$ mixing can be written as
\begin{align}  
\left(\begin{array}{c} |\Upsilon_{4S-3D}^\prime\rangle \\ |\Upsilon_{4S-3D}^{\prime\prime}\rangle \end{array}\right)= \left(\begin{array}{cc}\cos\,\theta&\sin\,\theta\\ 
-\sin\,\theta&\cos\,\theta\end{array}\right) \left(\begin{array}{c}|\Upsilon(4S)\rangle\\ |\Upsilon(3D)\rangle \end{array}\right), 
\end{align} 
where $\theta$ denotes the mixing angle.

\begin{figure}[htbp]
\centering
\raisebox{0.015\textwidth}{%
    \includegraphics[width=0.45\textwidth]{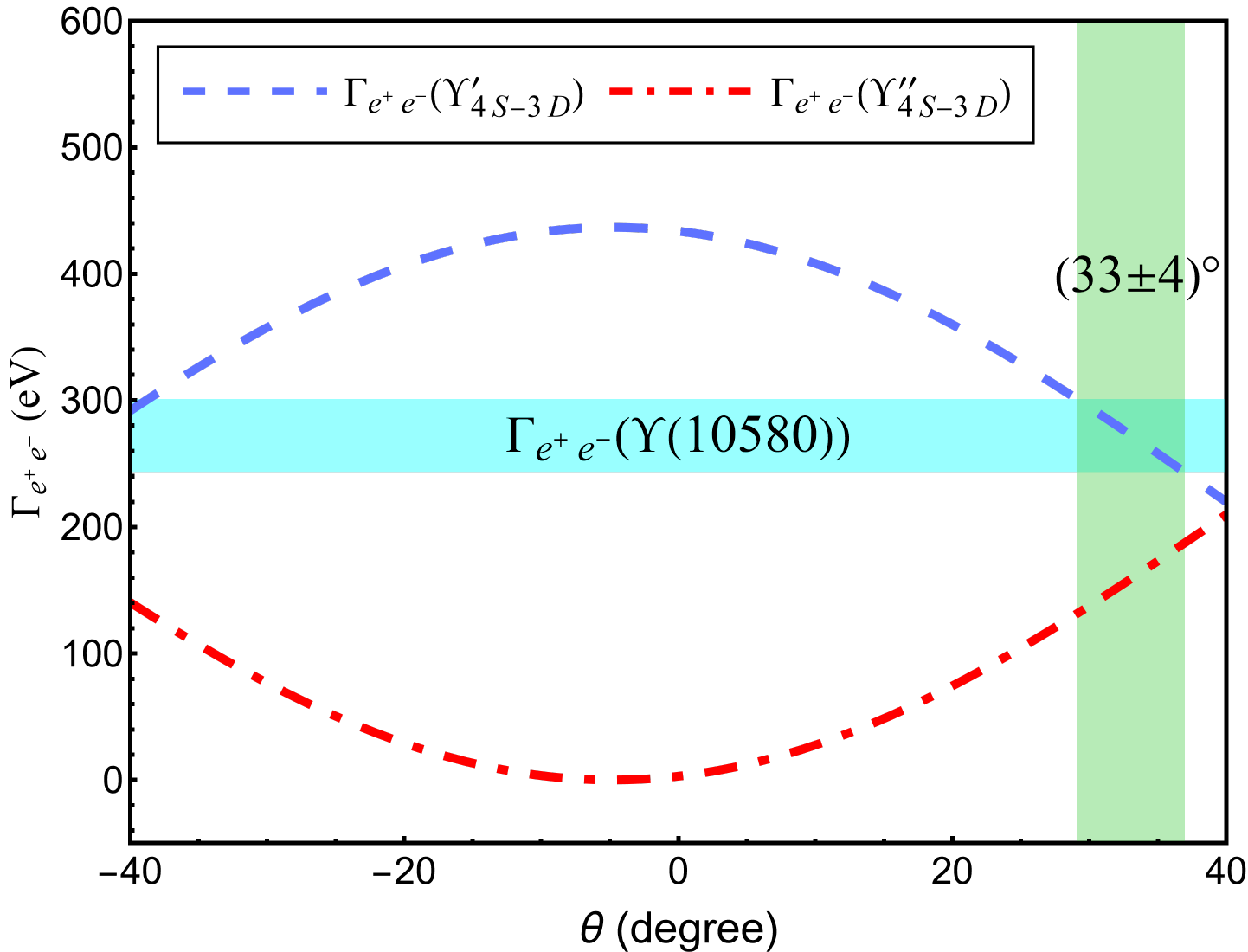}}
  \hspace{0.04\textwidth}
\includegraphics[width=0.45\textwidth]{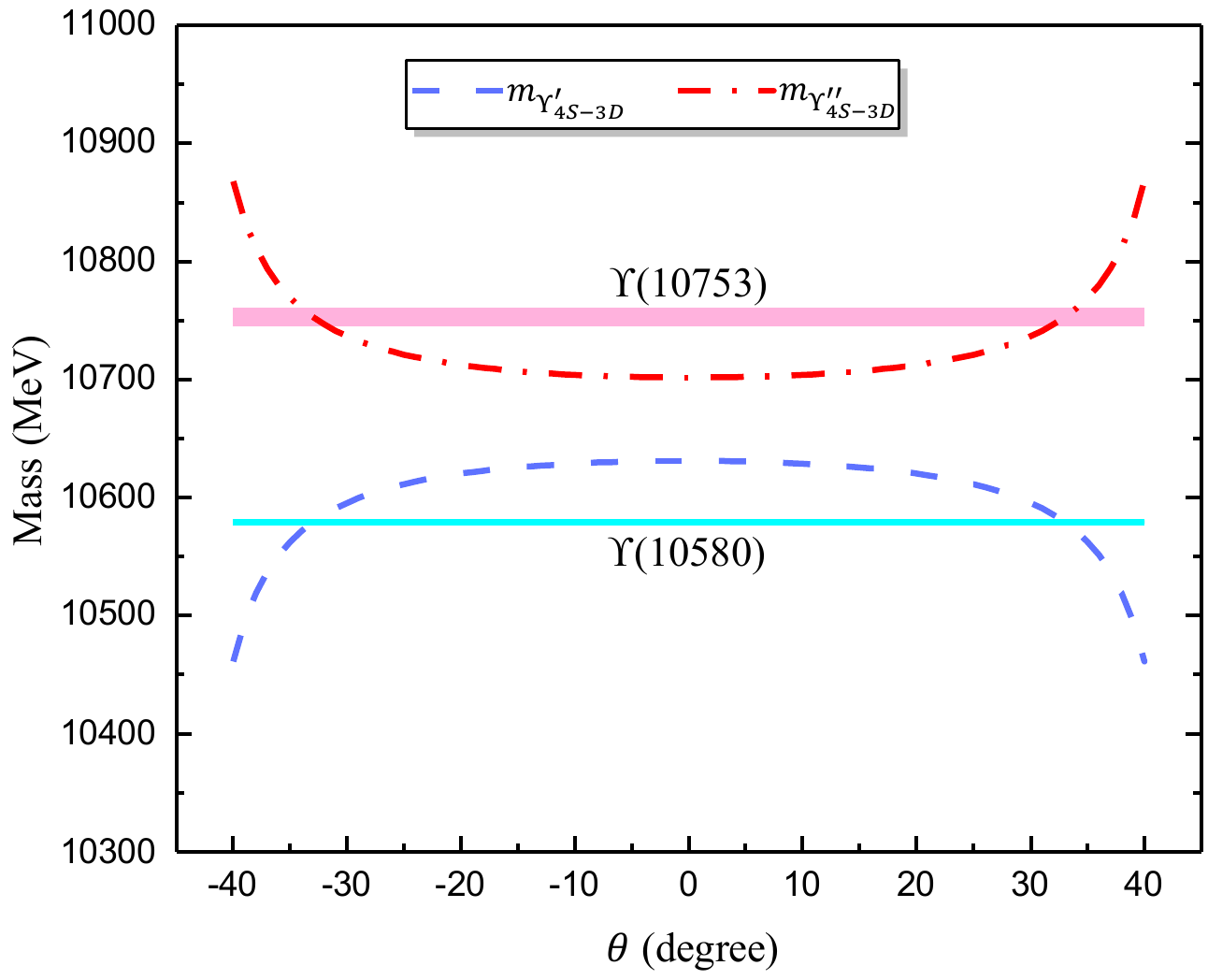}
\caption{The predicted masses (left panel) and dielectron widths (right panel) of the mixed bottomonium states $\Upsilon_{4S-3D}^\prime$ and $\Upsilon_{4S-3D}^{\prime\prime}$ as functions of the mixing angle $\theta$ \cite{Li:2021jjt}. The blue horizontal bands denote the experimental values, while the green vertical band indicates the favored mixing angle.}
\label{fig:upsilon10753mix}
\end{figure}

After introducing the $4S$--$3D$ mixing scheme, the predicted masses and dielectron widths of the mixed states as functions of the mixing angle $\theta$ are shown in Fig.~\ref{fig:upsilon10753mix}. 
The mixing angle can be determined by matching the calculated dielectron width $\Gamma_{e^+e^-}(\Upsilon^{\prime}_{4S\text{--}3D})$ to the experimentally measured value of $\Upsilon(10580)$, yielding \cite{Li:2021jjt} 
\begin{align}
\theta = (33 \pm 4)^\circ.
\end{align}
This result is consistent with the estimation in Refs.~\cite{Badalian:2008ik,Badalian:2009bu,Luo:2025kid}.
With the above mixing angle, the masses of $\Upsilon(10580)$ and the $\Upsilon(10753)$ are simultaneously reproduced. Meanwhile, the dielectron width of $\Upsilon(10753)$ is enhanced by nearly two orders of magnitude relative to the pure $3D$ assignment, becoming comparable to that of an $S$-wave state, which makes its observation in $e^{+}e^{-}$ annihilation feasible, as shown in Fig.~\ref{fig:upsilon10753mix}.

\paragraph{Hidden-bottom decay propertis}

\begin{figure}[htbp]
\centering
\includegraphics[width=0.9\textwidth]{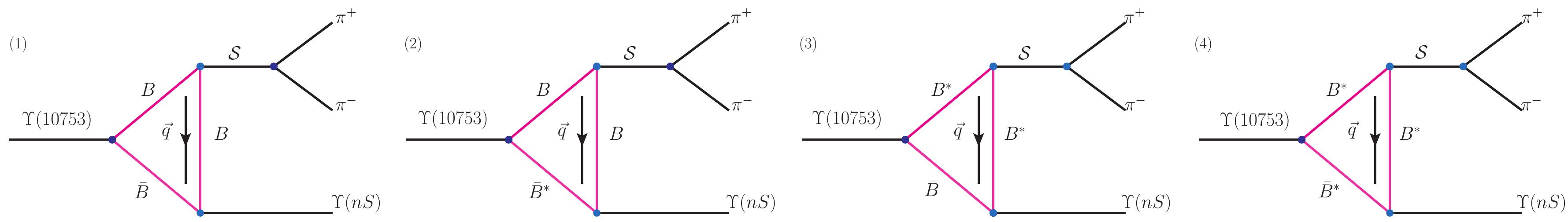}
\caption{Schematic diagrams for the $\Upsilon(10753)\to\Upsilon(nS)\pi^{+}\pi^{-}$ ($n=1,2,3$) processes within the hadronic loop mechanism. Here, $\mathcal{S}$ represents the scalar mesons $\sigma$ and $f_0(980)$. The figure is adapted from Ref.~\cite{Bai:2022cfz}.}
\label{fig:fig_FSI}
\end{figure}

To further assess the plausibility of assigning the newly observed $\Upsilon(10753)$ to the mixed $\Upsilon_{4S-3D}^{\prime\prime}$ state, its hidden-bottom decay channel $\Upsilon(10753)\to\Upsilon(nS)\pi^{+}\pi^{-}$ $(n=1,2,3)$—which serves as the discovery mode of this state—was examined in Ref.~\cite{Bai:2022cfz}.
Since $\Upsilon(10753)$ lies above the open-flavor thresholds of $B^{(*)}\bar B^{(*)}$, hadronic loop effects are expected to play a dominate role in its hidden decay dynamics, as already illustrated in the cases of $\Upsilon(10580)$ and $\Upsilon(10860)$ discussed above.
Within this framework, Ref.~\cite{Bai:2022cfz} employed the hadronic loop mechanism to investigate the transitions $\Upsilon(10753)\to\Upsilon(nS)\pi^{+}\pi^{-}$, in which the dipion system is generated through intermediate scalar meson contributions, as shown in Fig.~\ref{fig:fig_FSI}.
It is worth noting that, when  $\Upsilon(10753)$ is treated as the mixed $4S$–$3D$ state $\Upsilon_{4S-3D}^{\prime\prime}$, the decay amplitudes receive contributions from both the $4S$-wave and $3D$-wave components; further technical details can be found in Ref.~\cite{Bai:2022cfz}.

\begin{figure}[htbp]
\centering
\includegraphics[width=0.72\textwidth]{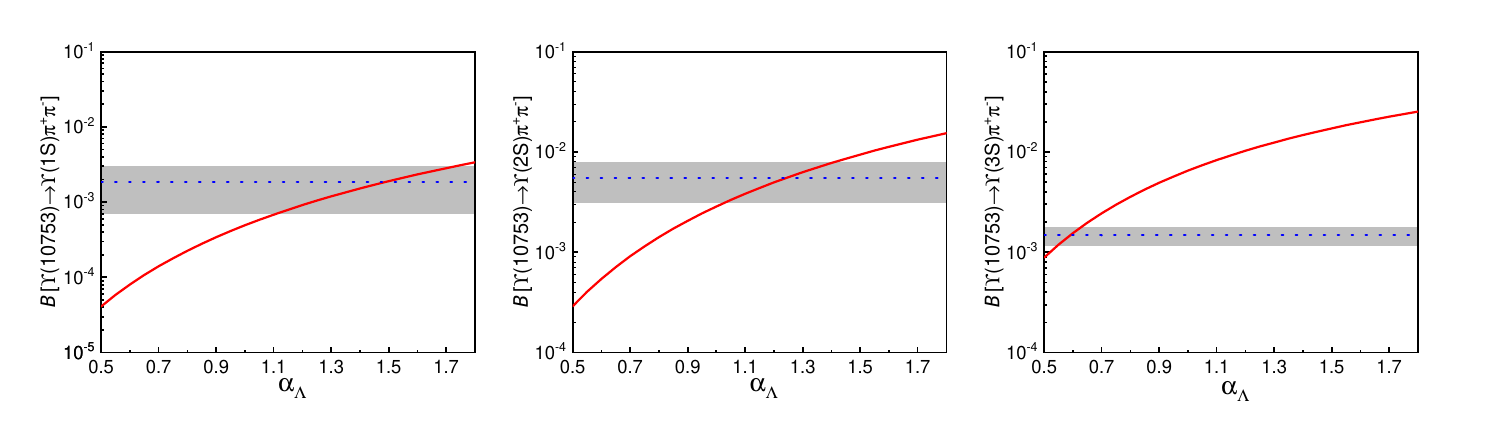}
\raisebox{0.011\textwidth}{\includegraphics[width=0.19\textwidth]{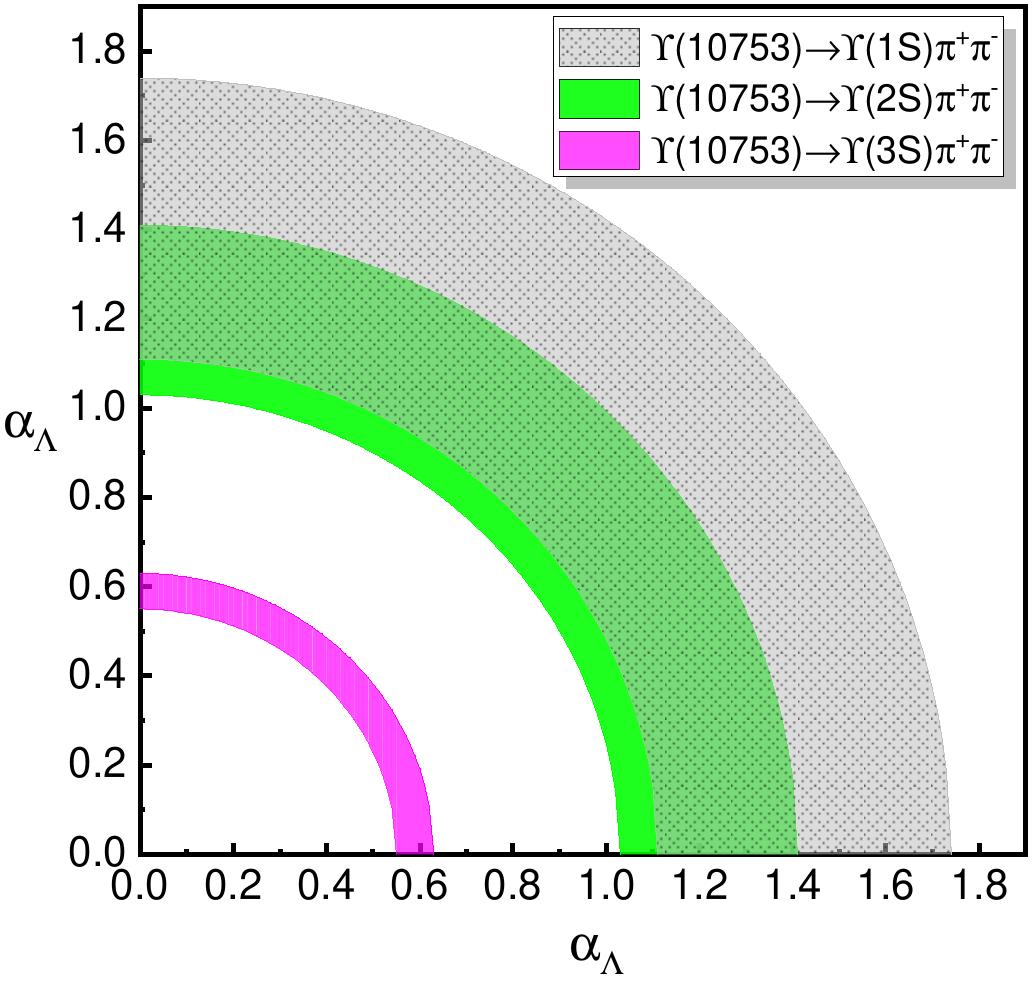}}
\caption{Dependence of the branching ratios $BR[\Upsilon(10753)\to\Upsilon(nS)\pi^+\pi^-]$ ($n=1,2,3$) on the parameter $\alpha_{\Lambda}$. The red solid lines represent the predictions from the hadronic loop mechanism \cite{Bai:2022cfz}, while the gray bands with blue dotted lines indicate the experimental values extracted by the Belle Collaboration \cite{Belle:2019cbt}. The rightmost panel shows the ranges of $\alpha_{\Lambda}$ for which the theoretical predictions are consistent with the experiment, showing that these values are of order unity and thus lie within a phenomenologically reasonable range, as suggested in Ref.~\cite{Cheng:2004ru}. Figure Adapted from Ref.~\cite{Bai:2022cfz}.}
\label{fig:br_alpha}
\end{figure}

The calculated branching ratios of the processes
$\Upsilon(10753)\to\Upsilon(nS)\pi^{+}\pi^{-}$ $(n=1,2,3)$
are presented in Fig.~\ref{fig:br_alpha}.
On the experimental side, the Belle Collaboration has extracted the physical quantities
\begin{equation}
\mathcal{R}_{n}=\Gamma_{e^+e^-}(\Upsilon(10753))\times BR[\Upsilon(10753)\to\Upsilon(nS)\pi^+\pi^-]\;(n=1,2,3),
\end{equation}
from the measured cross sections~\cite{Belle:2019cbt}.
These quantities are directly proportional to the transition amplitudes of the processes
$e^{+}e^{-}\to\Upsilon(nS)\pi^{+}\pi^{-}$ mediated by the $\Upsilon(10753)$ resonance.

Treating $\Upsilon(10753)$ as the mixed
$\Upsilon_{4S-3D}^{\prime\prime}$ state with the mixing angle $\theta=(33\pm4)^\circ$,
the corresponding dielectron width
$\Gamma_{e^{+}e^{-}}(\Upsilon(10753))$
can be estimated, as illustrated in the left panel of Fig. \ref{fig:upsilon10753mix}.
Combining this result with the experimentally extracted
$\mathcal{R}_{n}$ values,
the branching ratios
$BR\left[\Upsilon(10753)\to\Upsilon(nS)\pi^{+}\pi^{-}\right]$
can be deduced.
The resulting experimental ranges are shown as the grey bands in
Fig.~\ref{fig:br_alpha}.

\begin{figure}[htbp]
\centering
\includegraphics[width=1.02\textwidth]{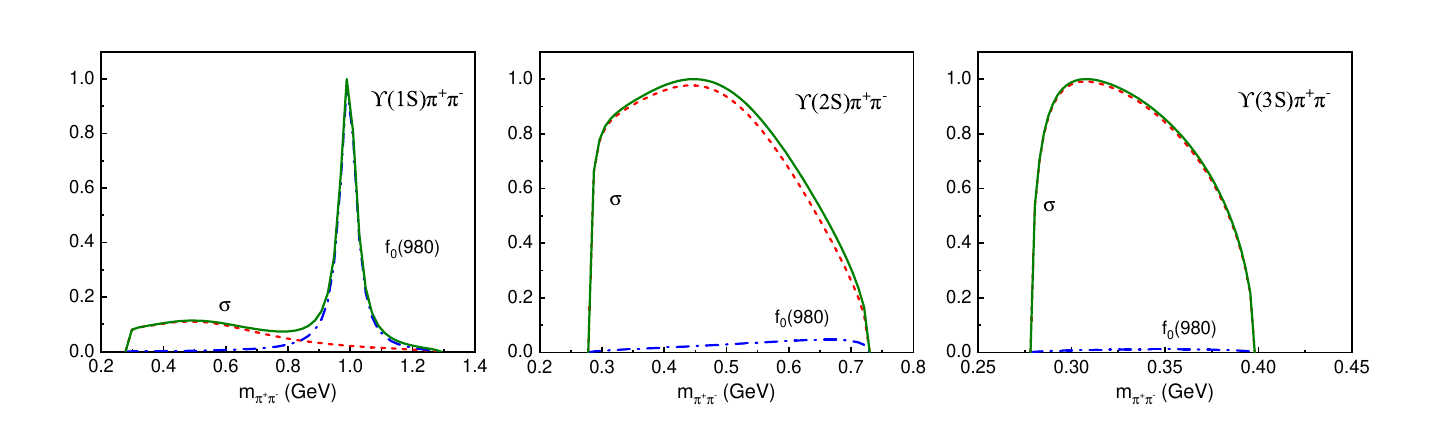}
\caption{Predicted dipion invariant-mass spectra for the decays $\Upsilon(10753)\to\Upsilon(nS)\pi^+\pi^-$ ($n=1,2,3$), normalized to unity at their respective maxima. Adapted from Ref.~\cite{Bai:2022cfz}.}
\label{fig:Upsilon10753dipion}
\end{figure}

As illustrated in Fig.~\ref{fig:br_alpha}, for reasonable choices of the model parameters, the calculated branching ratios are in good agreement with the available experimental constraints, providing further support for assigning the $\Upsilon(10753)$ as a $4S$--$3D$ mixed state. In addition, the dipion invariant mass distributions for the decays $\Upsilon(10753)\to\Upsilon(nS)\pi^{+}\pi^{-}$ ($n=1,2,3$) are predicted, as shown in Fig.~\ref{fig:Upsilon10753dipion}.

\begin{figure}[htbp]
\centering
\includegraphics[width=0.4\textwidth]{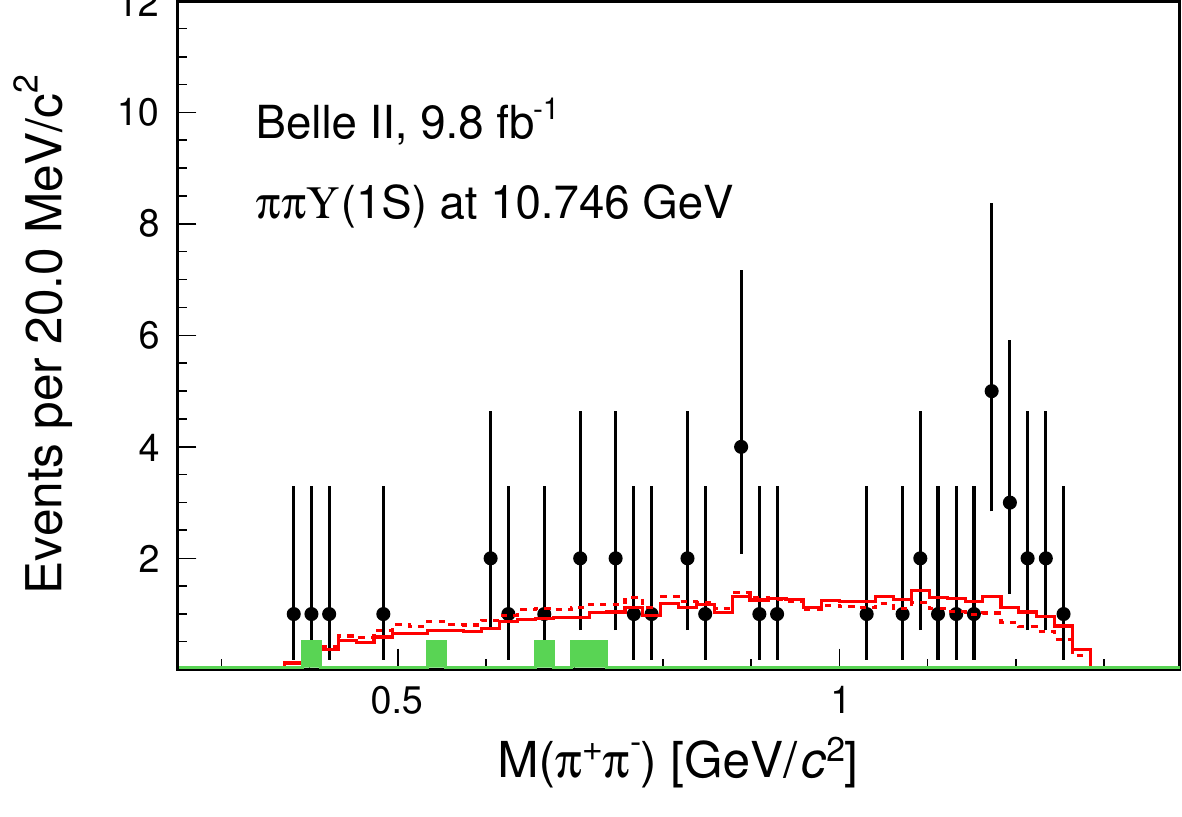}
\hspace{0.05\textwidth}
\includegraphics[width=0.4\textwidth]{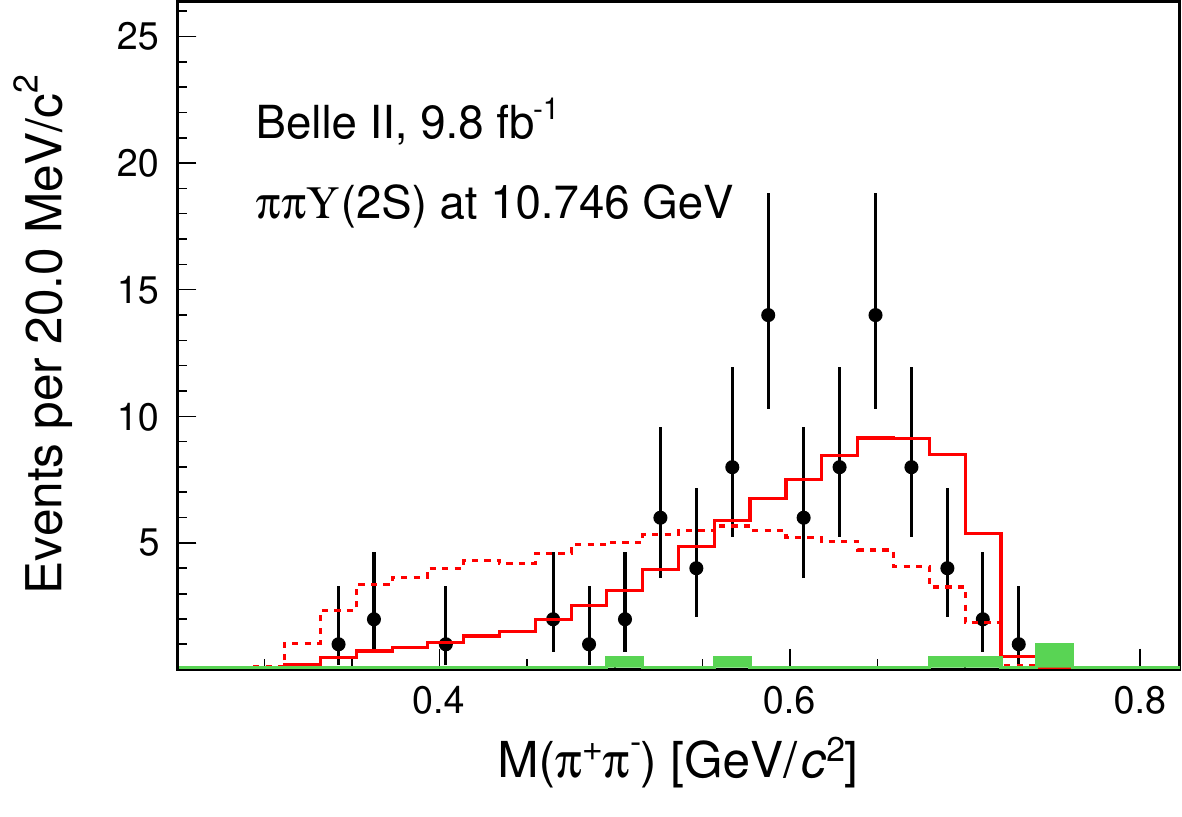}
\caption{Distributions of dipion mass at $\sqrt s=10.746$ GeV for $\Upsilon(1S)\pi^+\pi^-$ (left panel) and $\Upsilon(2S)\pi^+\pi^-$ (right panel). Adapted from Ref.~\cite{Belle-II:2024mjm}.}
\label{fig:Mpipi_reWeighted}
\end{figure}

The forthcoming Belle II experiments have confirmed the $\Upsilon(10753)$ signals in the
$e^+e^-\to \Upsilon(1S,2S)\pi^+\pi^-$ processes, while no evidence is observed in the
$\Upsilon(3S)\pi^+\pi^-$ channel \cite{Belle-II:2024mjm}. The measured dipion invariant mass distributions are shown in Fig.~\ref{fig:Mpipi_reWeighted}.

At the current level of experimental precision, the measured distributions do not appear to be fully consistent with the theoretical expectations based on the hadronic loop mechanism shown in Fig.~\ref{fig:Upsilon10753dipion}. On the theoretical side, although the hadronic-loop contributions are expected to be dominant, additional effects—such as direct decay contributions and interference among different amplitudes—may slightly modify the line shapes, as discussed above in the $\Upsilon(10580)$ and $\Upsilon(10860)$ analyses. On the experimental side, the present measurements still suffer from relatively large uncertainties. As a result, no definite conclusion can be drawn at this stage. More precise measurements with improved statistics from the Belle II Collaboration are expected.

\begin{figure}[htbp]
\centering
\includegraphics[width=0.3\textwidth]{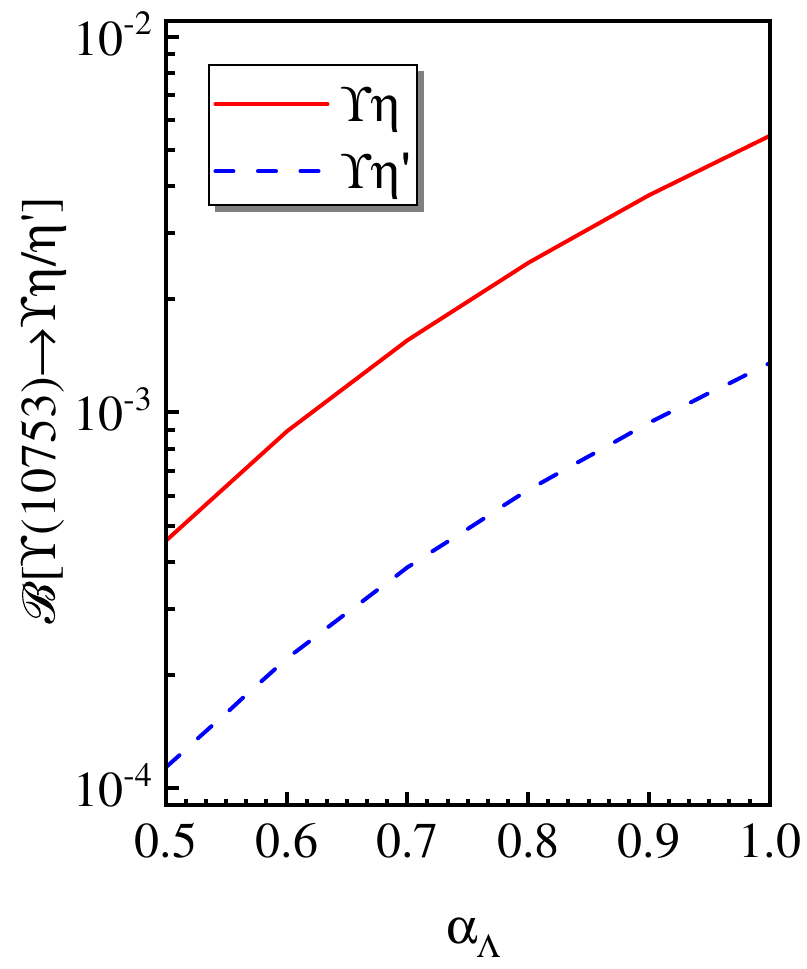}
\includegraphics[width=0.3\textwidth]{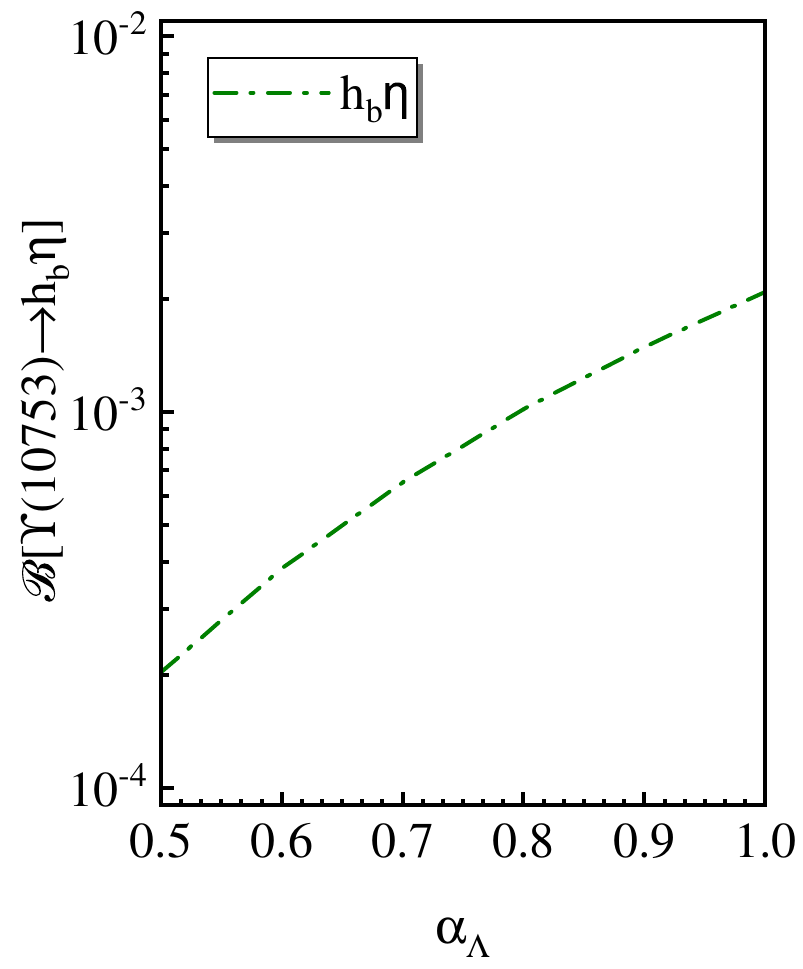}
\includegraphics[width=0.3\textwidth]{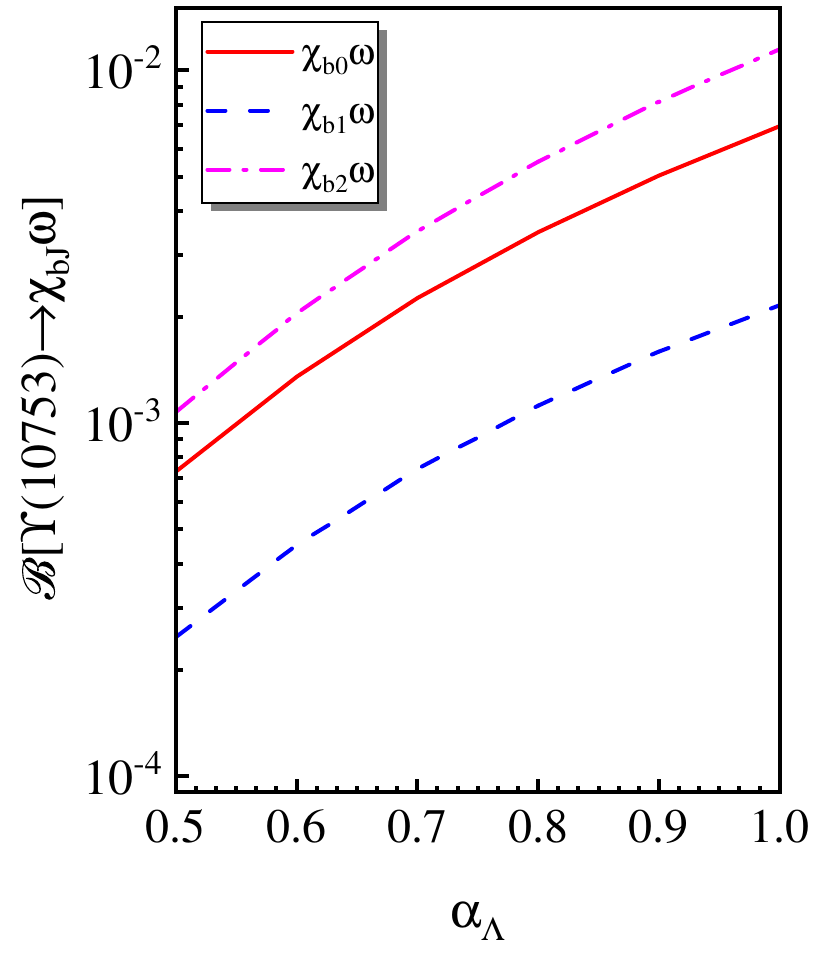}\\
\includegraphics[width=0.3\textwidth]{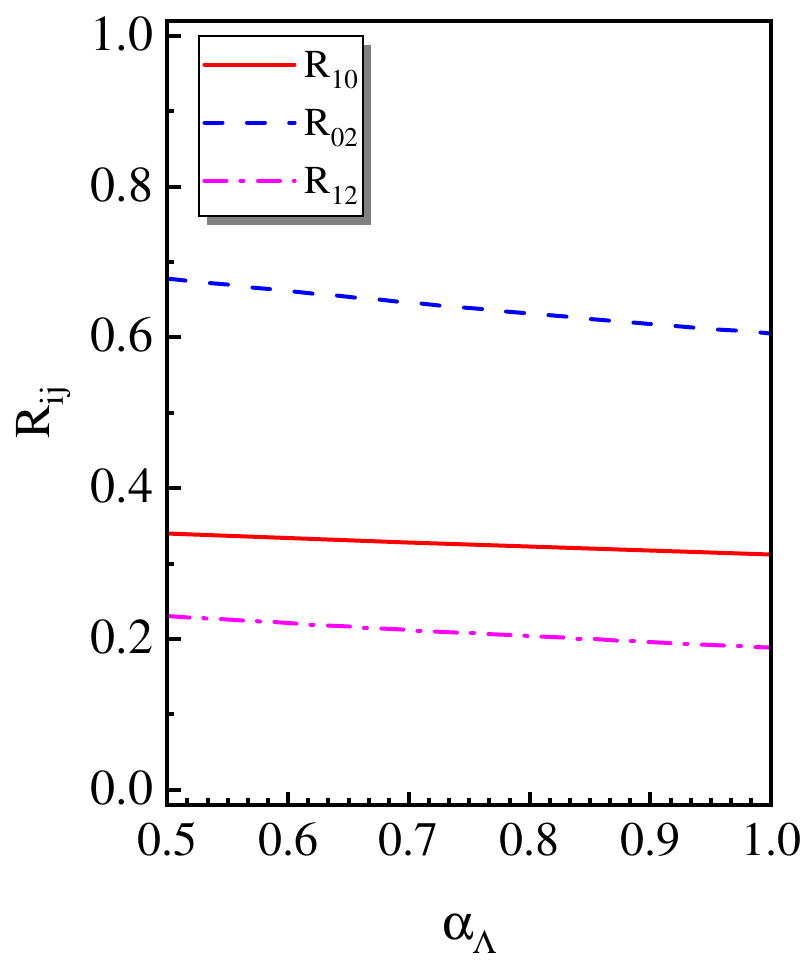}
\includegraphics[width=0.3\textwidth]{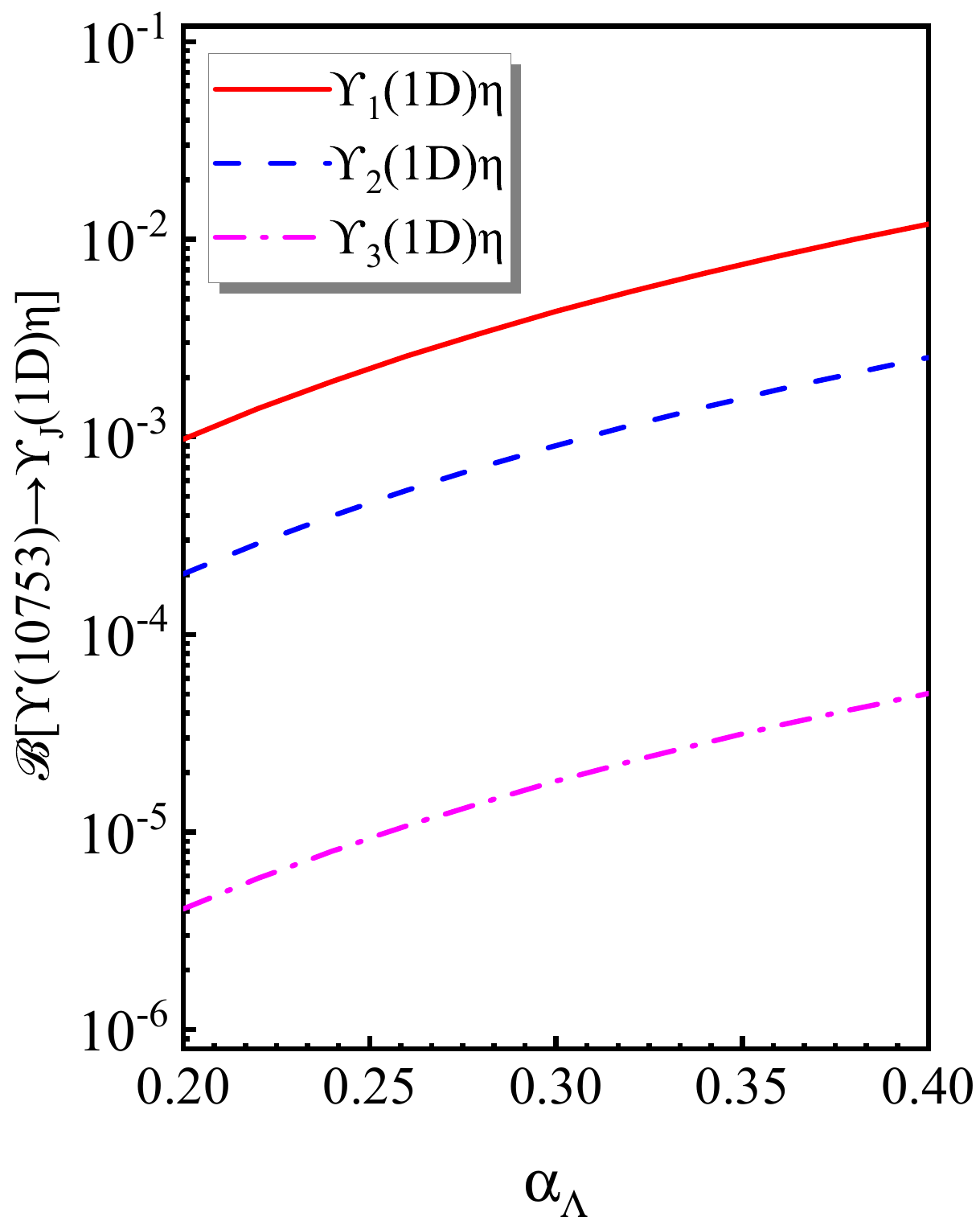}
\includegraphics[width=0.3\textwidth]{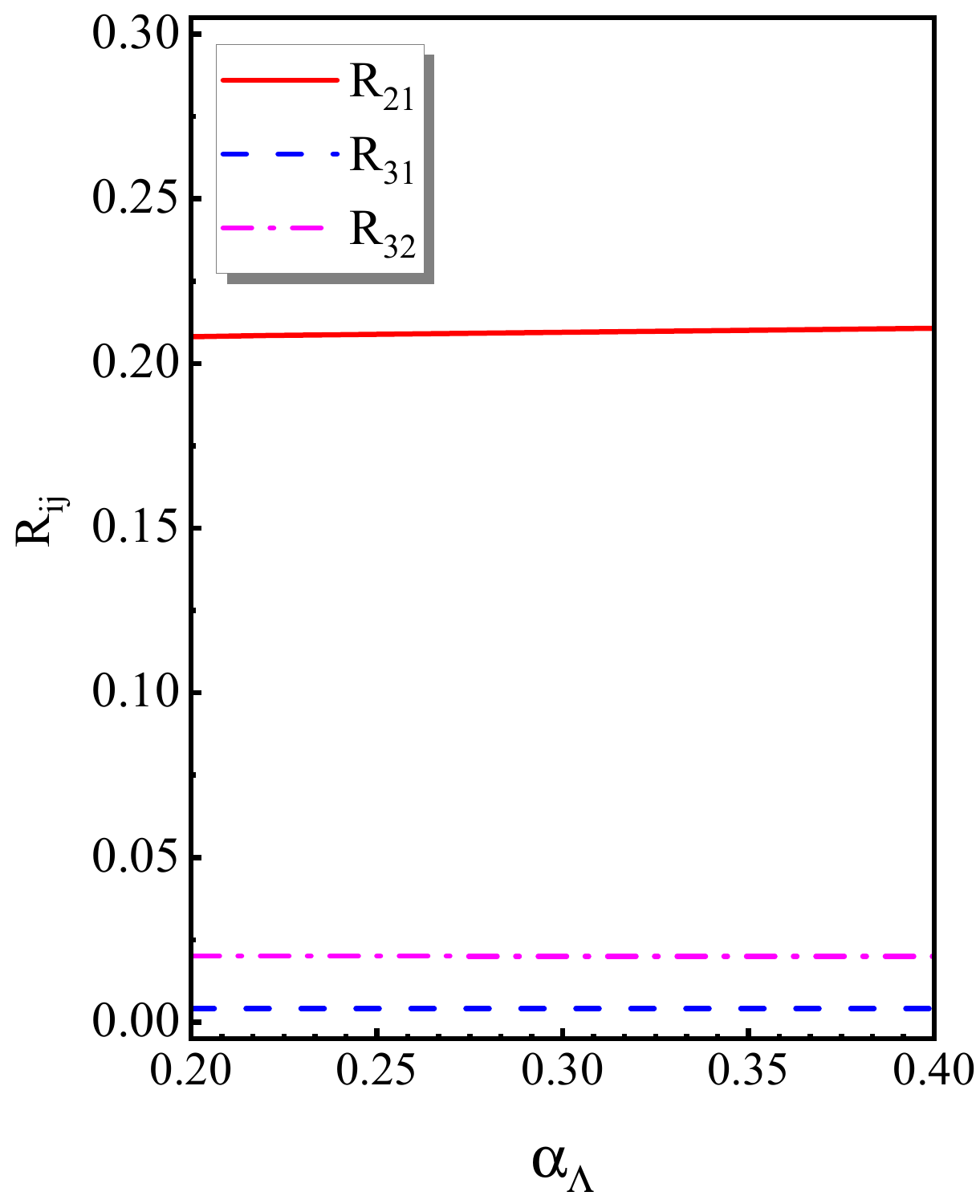}
\caption{The predicted branching ratios and characteristic ratios for the two-body hidden-bottom decay channels of $\Upsilon(10753)$, interpreted as the $4S$-$3D$ mixed state  $\Upsilon_{4S-3D}^{\prime\prime}$, within the hadronic loop mechanism. The quantities shown include:
(1) $BR[\Upsilon(10753)\to \Upsilon(1S)\eta^{(\prime)}]$; 
(2) $BR[\Upsilon(10753)\to h_b\eta]$; 
(3) $BR[\Upsilon(10753)\to \chi_{bJ}\omega]$ with $J=0,1,2$; 
(4) the ratios $R_{ij}^{\chi_b}\equiv BR[\Upsilon(10753)\to \chi_{b i}\omega]/BR[\Upsilon(10753)\to \chi_{b j}\omega]$; 
(5) $BR[\Upsilon(10753)\to \Upsilon(1^3D_J)\eta]$ with $J=1,2,3$; 
and 
(6) the ratios $R_{ij}^{D}\equiv BR[\Upsilon(10753)\to \Upsilon(1^3D_i)\eta]/BR[\Upsilon(10753)\to \Upsilon(1^3D_j)\eta]$. These figures are adapted from Ref. \cite{Li:2021jjt,Li:2022leg}.}
\label{fig:brupsilon10753}
\end{figure}

\begin{figure}[htbp]
\centering
\includegraphics[width=0.3\textwidth]{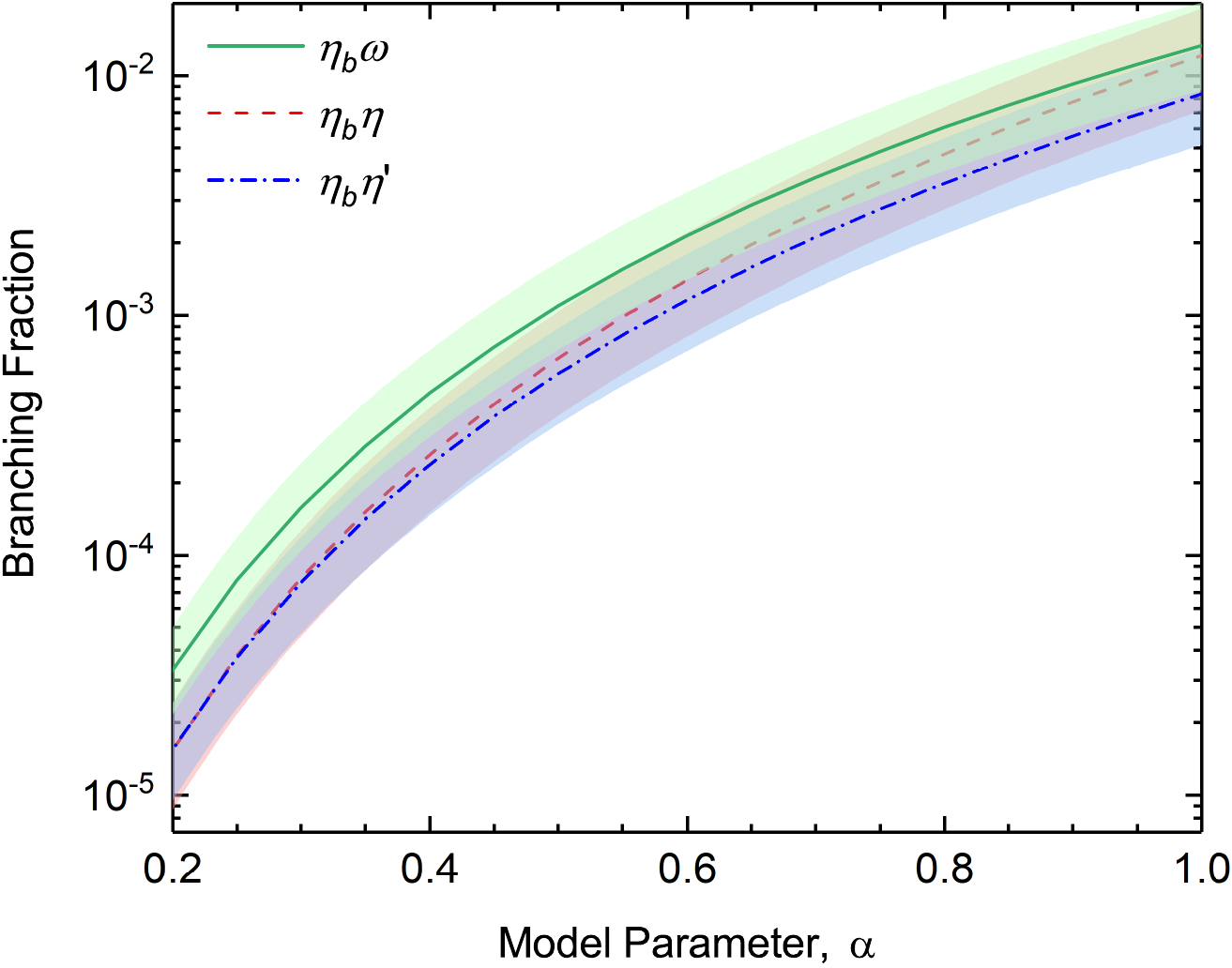}
\includegraphics[width=0.3\textwidth]{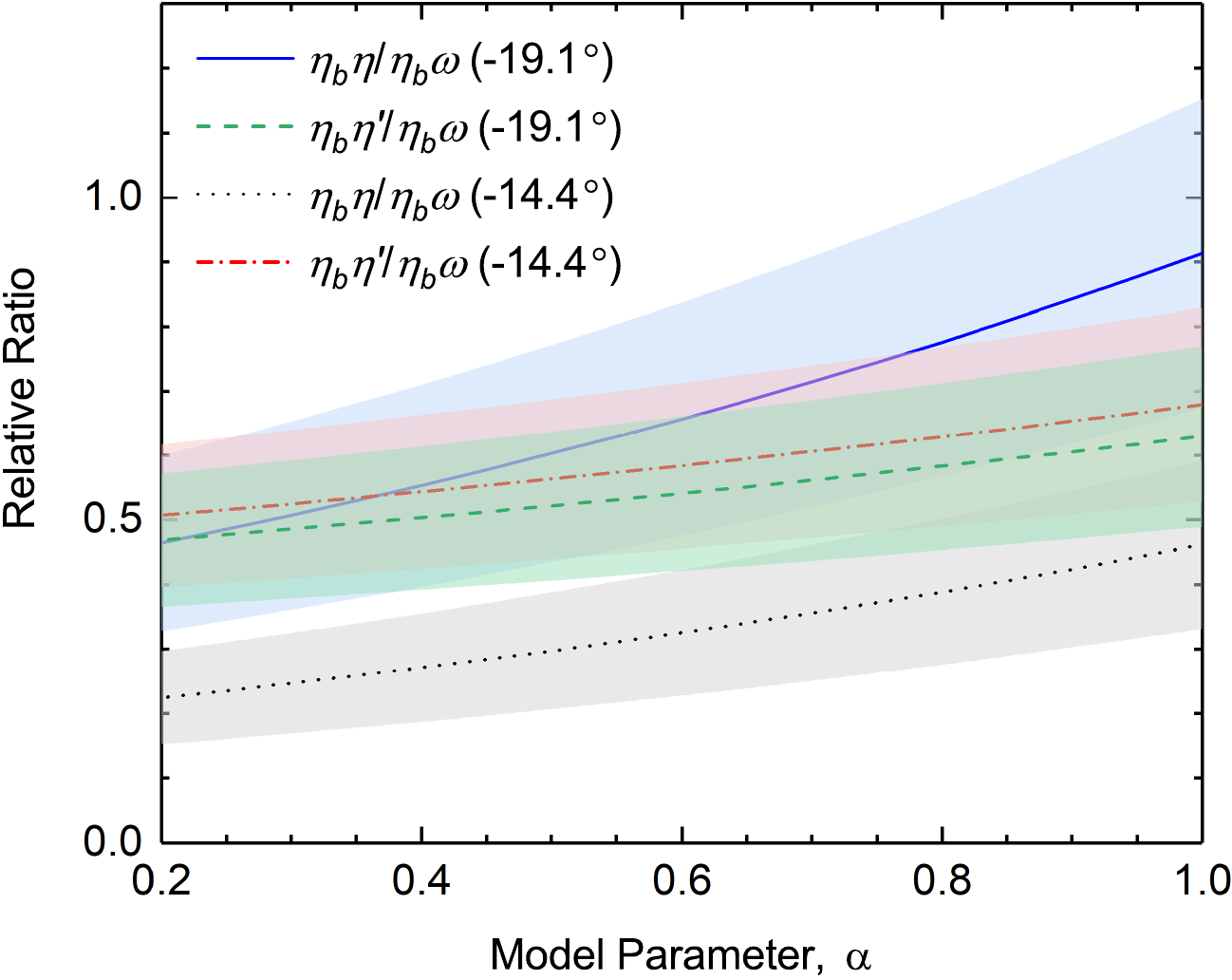}
\includegraphics[width=0.3\textwidth]{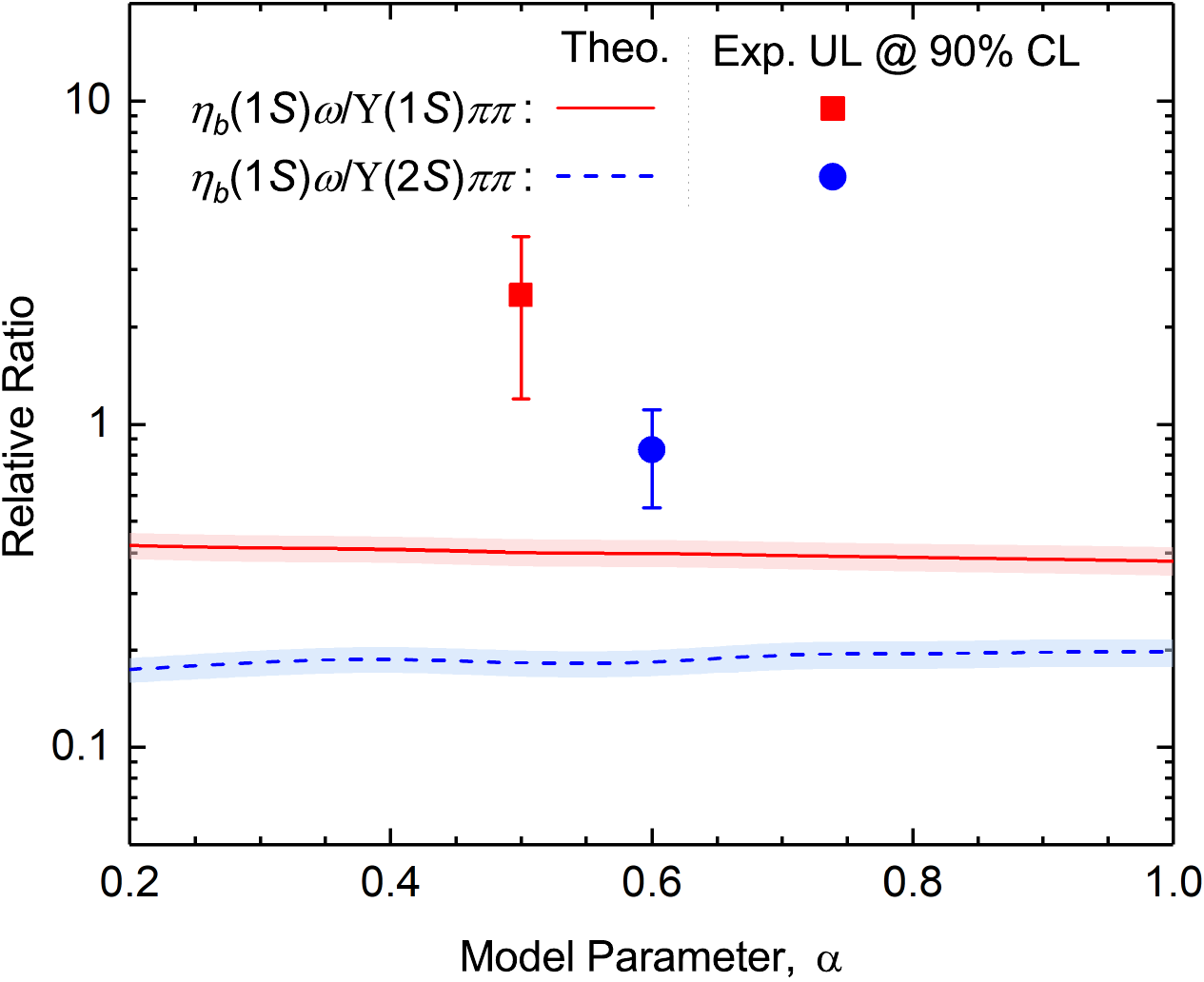}
\caption{(1) The predicted branching ratios for $\Upsilon(10753)\to \eta_b\eta^{(\prime)}$ and $\Upsilon(10753)\to \eta_b(1S)\omega$ within the hadronic loop mechanism;
(2) the characteristic ratios between $BR[\Upsilon(10753)\to \eta_b\eta^{(\prime)}]$ and $BR[\Upsilon(10753)\to \eta_b\omega]$, evaluated with the $\eta$--$\eta^\prime$ mixing angle taken as $-19.1^\circ$ or $-14.4^\circ$;
(3) the characteristic ratio $BR[\Upsilon(10753)\to \eta_b\eta]/BR[\Upsilon(10753)\to \Upsilon(1S,\,2S)\pi^+\pi^-]$, where the points denote the upper limits extracted from experimental measurements~\cite{Belle:2019cbt,Belle-II:2023twj}.
In the calculations, $\Upsilon(10753)$ is interpreted as the $4S$--$3D$ mixed state $\Upsilon_{4S-3D}^{\prime\prime}$, with the favored $\alpha$ parameter in the range $0.3$--$0.5$.
The figures are adapted from Ref.~\cite{Liu:2023gtx}.}
\label{fig:upsilon10753etab}
\end{figure}

The other two-body hidden-bottom decay channels of $\Upsilon(10753)$ have also been systematically studied within the hadronic loop mechanism, where $\Upsilon(10753)$ is interpreted as the $4S$--$3D$ mixed state $\Upsilon_{4S-3D}^{\prime\prime}$ \cite{Li:2021jjt,Li:2022leg,Liu:2023gtx}. The corresponding results are summarized in Figs.~\ref{fig:brupsilon10753} and \ref{fig:upsilon10753etab}, and the main features can be outlined as follows:

\begin{itemize}
\item The branching ratios of the transitions $\Upsilon(10753)\to \Upsilon(1S)\eta^{(\prime)}$, $\Upsilon(1^3D_J)\eta$ ($J=1,2$), $h_b(1P)\eta$, and $\chi_{bJ}(1P)\omega$ ($J=0,1,2$) are found to be of order $\mathcal{O}(10^{-4}$--$10^{-3})$, indicating that these modes should be accessible at the Belle~II experiment.

\item The $\Upsilon(1^3D_1)$ and $\Upsilon(1^3D_2)$ states may be accessible via the $\eta$ transitions of $\Upsilon(10753)$. In contrast, the observation of the $\Upsilon(1^3D_3)$ state is not expected to be favorable in this process due to its small decay rate~\cite{Li:2022leg}. At present, neither the $\Upsilon(1^3D_1)$ nor the $\Upsilon(1^3D_3)$ state has been experimentally observed, rendering this channel a potentially viable avenue for searching for the as-yet unestablished $\Upsilon(1^3D_1)$ state.

\item The theoretical ratios $BR[\Upsilon(10753)\to \eta_b(1S)\eta]/BR[\Upsilon(10753)\to \Upsilon(1S,\,2S)\pi^+\pi^-]$ are compatible with the experimental upper limits, as shown in the
rightmost panel of Fig.~\ref{fig:upsilon10753etab}. This consistency, to some extent, supports the interpretation of the $\Upsilon(10753)$ as the $4S$--$3D$ mixed state $\Upsilon_{4S-3D}^{\prime\prime}$~\cite{Liu:2023gtx}.

\item Within the $4S$--$3D$ mixing scenario, the partial decay width of $\Upsilon(10753)\to \eta_b(1S)\omega$ is predicted to be in the range $(5.6$--$38.8)\,\mathrm{keV}$, which is significantly smaller than the value expected from the tetraquark assignment, $2.46^{+4.70}_{-1.60}\,\mathrm{MeV}$~\cite{Wang:2019veq}. This marked difference provides an important discriminator for probing the internal structure of the $\Upsilon(10753)$~\cite{Liu:2023gtx}.
\end{itemize}

Following the theoretical study in Ref.~\cite{Li:2021jjt}, the Belle~II Collaboration reported the first observation of signals for $e^+e^-\to\chi_{b1}\omega$ and evidence for $e^+e^-\to\chi_{b2}\omega$ at $\sqrt{s}=10.745~\mathrm{GeV}$ \cite{Belle-II:2022xdi}. The measured energy dependences of the Born cross sections are consistent with the shape of the $\Upsilon(10753)$ state. In a more recent update, Belle~II reported measurements of the products of the dielectron width and the corresponding branching ratios, yielding $(1.46\pm0.25\pm0.20)\,\mathrm{eV}$ and $(1.29\pm0.38\pm0.31)\,\mathrm{eV}$ for $e^+e^-\to\Upsilon(10753)\to\chi_{b1}\omega$ and $e^+e^-\to\Upsilon(10753)\to\chi_{b2}\omega$, respectively \cite{Belle-II:2025jus}. Using the estimated dielectron width $\Gamma_{e^+e^-}=159\pm30~\mathrm{eV}$ \cite{Li:2021jjt,Bai:2022cfz}, the corresponding branching ratios are inferred to be $(9.2\pm2.7)\times10^{-3}$ and $(8.1\pm3.4)\times10^{-3}$ for the $\Upsilon(10753)\to\chi_{b1}\omega$ and $\Upsilon(10753)\to\chi_{b2}\omega$ decays, respectively. The former slightly exceeds the theoretical expectation of $(0.25$–$2.16)\times10^{-3}$, whereas the latter is compatible with the predicted range $(1.08$–$11.5)\times10^{-3}$ \cite{Li:2021jjt}. Furthermore, the experimentally determined ratio
\[
\frac{BR[\Upsilon(10753)\to\chi_{b1}\omega]}{BR[\Upsilon(10753)\to\chi_{b2}\omega]}
=1.13\pm0.38\pm0.34
\]
is consistent with the theoretical prediction of $0.2$ \cite{Li:2021jjt} at the $1.8\sigma$ level.

Ref.~\cite{Liu:2024ets} also investigated the radiative transitions $\Upsilon(10753)\to \gamma X_b$, where $X_b$ denotes the heavy quark flavor symmetry counterpart of $X(3872)$ in the bottomonium sector. It was found that the corresponding partial decay width is of the order of $10~\mathrm{keV}$. Based on this result, the authors suggested that the Belle~II experiment search for the $X_b$ state via the process $e^+e^-\to \gamma X_b$, mediated by $\Upsilon(10753)$, with the subsequent decay $X_b\to\chi_{b1}\pi\pi$.

In summary, a consistent assignment of $\Upsilon(10753)$ into the bottomonium family can be achieved by interpreting it as the $4S$--$3D$ mixed state $\Upsilon_{4S-3D}^{\prime\prime}$ in the unquenched picture. This scenario provides a coherent description of its mass spectrum, production properties, and several observed decay behaviors. The predicted decay patterns and the mixing-induced features offer clear experimental signatures, which can be stringently tested by future high-luminosity $e^+e^-$ facilities, such as Belle II.

\subsection{Construction of light-flavor vector mesons around the 2 GeV mass region}
\subsubsection{Status of light-flavor meson spectroscopy}

Light hadron spectroscopy occupies an important position in hadron physics owing to both its historical significance and the richness of the observed spectrum. The systematic study of light-flavor hadrons played a central role in the establishment of SU(3) flavor symmetry and ultimately led to the formulation of the quark model \cite{Gell-Mann:1964ewy,Zweig:1964ruk}. Since then, the light-flavor sector has continued to dominate the experimental landscape, exhibiting a dense spectrum and a wide variety of production and decay channels.

Over the past decade, the construction of the light-flavor meson spectrum has primarily relied on systematic comparisons between theoretically estimated masses and total decay widths, and the corresponding experimental measurements, following an initial classification based on fundamental quantum numbers. Among them, Regge trajectory analyses \cite{Anisovich:2000kxa} have provided a simple and widely used framework for organizing experimentally observed states, particularly in the low-lying region where the spectrum appears relatively sparse. Following this methodology, the Lanzhou research group has systematically constructed the spectroscopy of light-flavor meson families \cite{Yu:2011ta,Wang:2012wa,Ye:2012gu,He:2013ttg,Pang:2014laa,Wang:2014sea,Chen:2015iqa,Pang:2015eha,Pang:2019ovr,Zhou:2022ark}.


This conventional construction strategy has proven to be relatively effective for the identification of low-lying light meson states. This effectiveness can be traced to a generic feature of hadron spectra governed by color confinement: with increasing excitation energy, the level spacing between successive states gradually decreases. As a consequence, states in the low-energy region remain relatively well separated, allowing experimentally observed resonances to be matched to theoretical expectations with limited ambiguity. 
In the highly excited region, the energy level of spectrum is usually compressed beyond the expectations of simple Regge systematics, enhancing correlations among neighboring excited states. As a result, different production and decay processes often exhibit markedly different line-shape structures in the same mass region. If these structures are naively interpreted as originating from a single excited state, the extracted resonance parameters such as masses and widths can differ substantially from one reaction to another. This problem is most clearly illustrated in the spectroscopy of light-flavor vector mesons around 2.0~GeV, where the identification and classification of excited states remain highly controversial. As illustrated in Fig.~\ref{fig:MassSpectrum}, the low-lying vector states show good correspondence with the experimental data, whereas the \changelabel{highly excited} states above 2.0 GeV remain far from being established.

A typical example is the strangeonium-like state $Y(2175)$ (referred to as $\phi(2170)$ in the PDG \cite{ParticleDataGroup:2024cfk}), which was first reported by the $BABAR$ Collaboration in the process $e^+e^-\to Y(2175)\to \phi f_0(980)$ via the initial-state radiation technique \cite{BaBar:2006gsq}. This observation was subsequently confirmed by the Belle \cite{Belle:2008kuo}, $BABAR$ \cite{BaBar:2007ptr,BaBar:2011btv}, and BESIII \cite{BESIII:2021aet} Collaborations in the same channel, as well as in the processes $e^+e^-\to \eta Y(2175)\to \eta \phi f_0(980)$ \cite{BES:2007sqy,BESIII:2014ybv,BESIII:2017qkh}, $e^+e^-\to Y(2175)\to\phi\eta^{(\prime)}$ \cite{BESIII:2020gnc,BESIII:2021bjn} reported by the BES and BESIII Collaborations. The $Y(2175)$ has been suggested as the strange-sector partner of the $Y(4260)$, since both states are observed prominently in hidden-flavor final states. 

\changelabel{ Apart from the unquenched light-meson interpretation combined with interference mechanisms that will be discussed later in this section, various other theoretical interpretations have been proposed for the $Y(2175)$:
\begin{itemize}
\item In the conventional strangeonium picture, the $Y(2175)$ has been assigned either to the $2^3D_1$ or $3^3S_1$ $s\bar s$ state. The $2D$ assignment was investigated using decay models and Regge-trajectory arguments, with open-strangeness decay modes such as $K\bar K$, $K^*\bar K^*$, $K(1460)\bar K$, and $h_1(1380)\eta$ proposed as important tests to distinguish it from a hybrid or a $3S$ strangeonium assignment~\cite{Ding:2007pc,Wang:2012wa,Pang:2019ttv,Li:2020xzs}. The $3S$ assignment has also been discussed in quark-model and phenomenological analyses of the vector strangeonium spectrum~\cite{Barnes:2002mu,Afonin:2014nya,Li:2020xzs}. Many of these predictions were made before the recent measurements in open-strangeness channels. These new data therefore provide important tests of the conventional strangeonium assignments and motivate the unquenched spectral analysis discussed later in this section.

\item The strangeonium-hybrid interpretation treats the $Y(2175)$ as a $1^{--}$ $s\bar s g$ state. In the flux-tube picture, the hybrid assignment was shown to have decay properties different from those of a higher $s\bar s$ state, especially in open-strangeness channels~\cite{Ding:2006ya}. QCD Gaussian-sum-rule analyses found that the relative resonance strength of the $Y(2175)$ is much smaller than that of a heavier state around $2.9~\mathrm{GeV}$, disfavoring its interpretation as a predominantly hybrid state~\cite{Govaerts:1985fx,Ho:2019org}. However, a recent comprehensive next-to-leading-order (NLO) sum-rule analysis has revised the hybrid-mass prediction downward, making it compatible with the observed $Y(2175)$~\cite{Li:2025hsp}. 

\item Compact tetraquark interpretations have been investigated in several frameworks. QCD sum-rule studies based on color-octet or diquark-antidiquark currents found masses compatible with the observed $Y(2175)$ within uncertainties and discussed its possible coupling to $\phi f_0(980)$ and related hidden-strangeness channels~\cite{Wang:2006ri,Chen:2008ej}. Diquark-model analyses also considered the $Y(2175)$ as a member of a higher tetraquark multiplet~\cite{Drenska:2008gr}. More recent studies further examined the fully strange tetraquark interpretation. A fall-apart decay analysis of an $ss\bar s\bar s$ tetraquark suggested sizable couplings to $\phi f_0(980)$, $h_1\eta^{(\prime)}$, and $K_1K$ channels~\cite{Ke:2018evd}, while a light-cone QCD sum-rule study of the decays into $\phi f_0(980)$, $\phi\eta$, and $\phi\eta^\prime$ obtained a mass and total width compatible with experiment~\cite{Agaev:2019coa}. On the other hand, a nonrelativistic potential-model study of the fully strange $ss\bar s\bar s$ spectrum predicted the lowest $1^{--}$ tetraquark states to be substantially heavier, which challenges a simple fully strange compact-tetraquark assignment for the $Y(2175)$~\cite{Liu:2020lpw}.

\item Baryon-antibaryon molecular interpretations identify the $Y(2175)$ with a near-threshold $\Lambda\bar\Lambda$ bound state or resonance. In one-boson-exchange and color-flux-tube models, the $Y(2175)$ was suggested as a $\Lambda\bar\Lambda({}^3S_1)$ bound state, emphasizing the attraction in the baryon-antibaryon system and the proximity to the $\Lambda\bar\Lambda$ threshold~\cite{Zhao:2013ffn,Deng:2013aca}. The decay properties of the $\phi(2170)$ were also studied under the $\Lambda\bar\Lambda$ molecular assumption, where selected strong decay modes were used to test this assignment~\cite{Dong:2017rmg}.

\item Nonresonant explanations have also been discussed. It was pointed out that an $S$-wave threshold cusp, or a related kinematic threshold effect, could not be excluded as a possible origin of the observed enhancement~\cite{Zhu:2007wz}. In the related $e^+e^-\to\phi K^+K^-$ process, nonresonant contributions, kaon form factors, and meson-meson rescattering effects were analyzed together with a possible $X(2175)$ contribution, showing that interference can play an important role in describing the data~\cite{Gomez-Avila:2007pgn}.

\item A three-body hadronic-molecule interpretation has also been proposed. In this approach, the $Y(2175)$ is generated dynamically from the $\phi K\bar K$ interaction and appears as an isoscalar $\phi f_0(980)$-like resonance in the three-body system~\cite{MartinezTorres:2008gy}. Its decays into final states involving kaonic resonances, such as $K(1460)\bar K$, $K_1(1270)\bar K$, and $K_1(1400)\bar K$, provide important tests of this three-body molecular picture~\cite{Li:2025hsp}.
\end{itemize}
 }

\begin{figure}[htbp]
\centering
\includegraphics[width=0.45\textwidth]{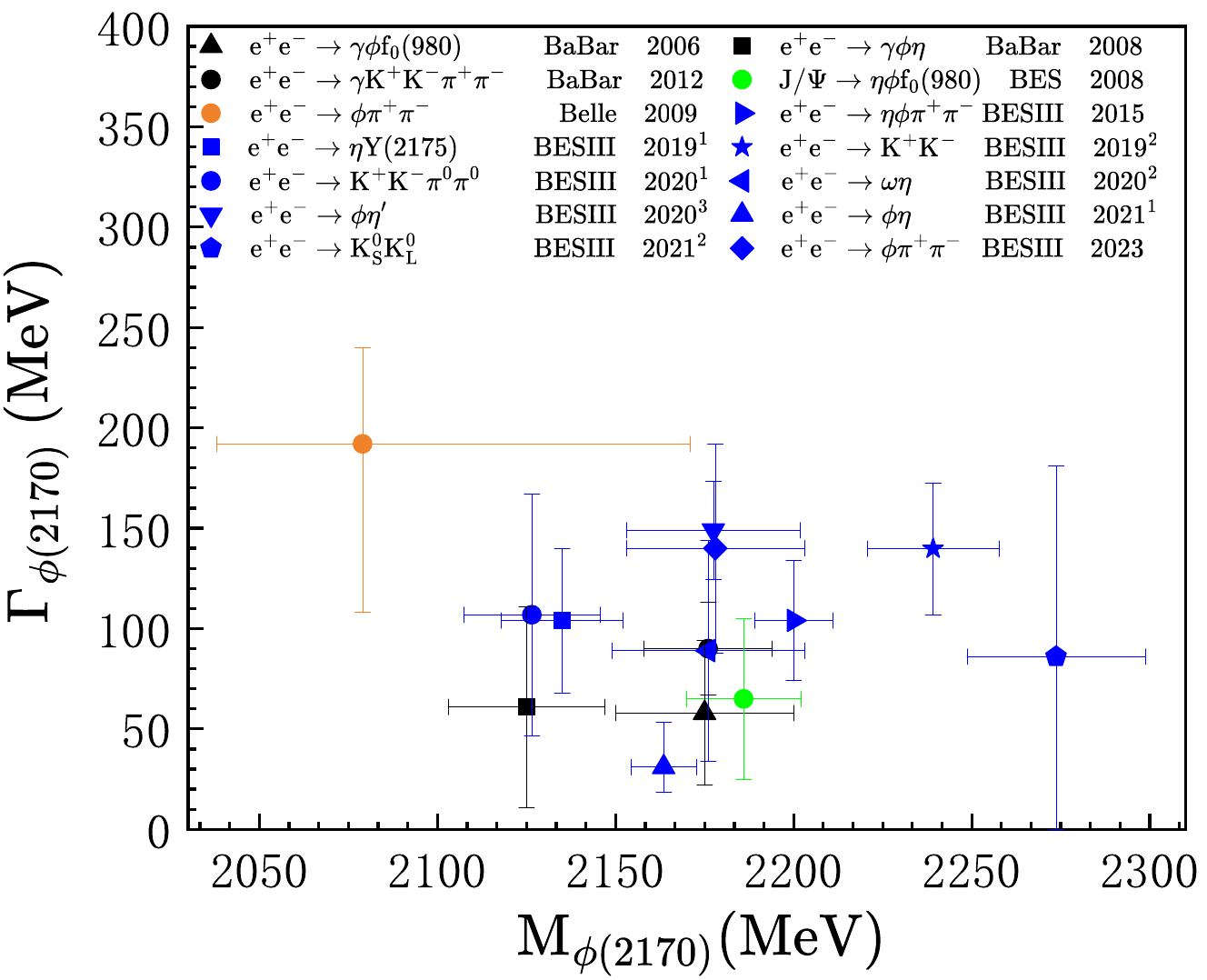}
\hspace{0.04\textwidth}
\includegraphics[width=0.45\textwidth]{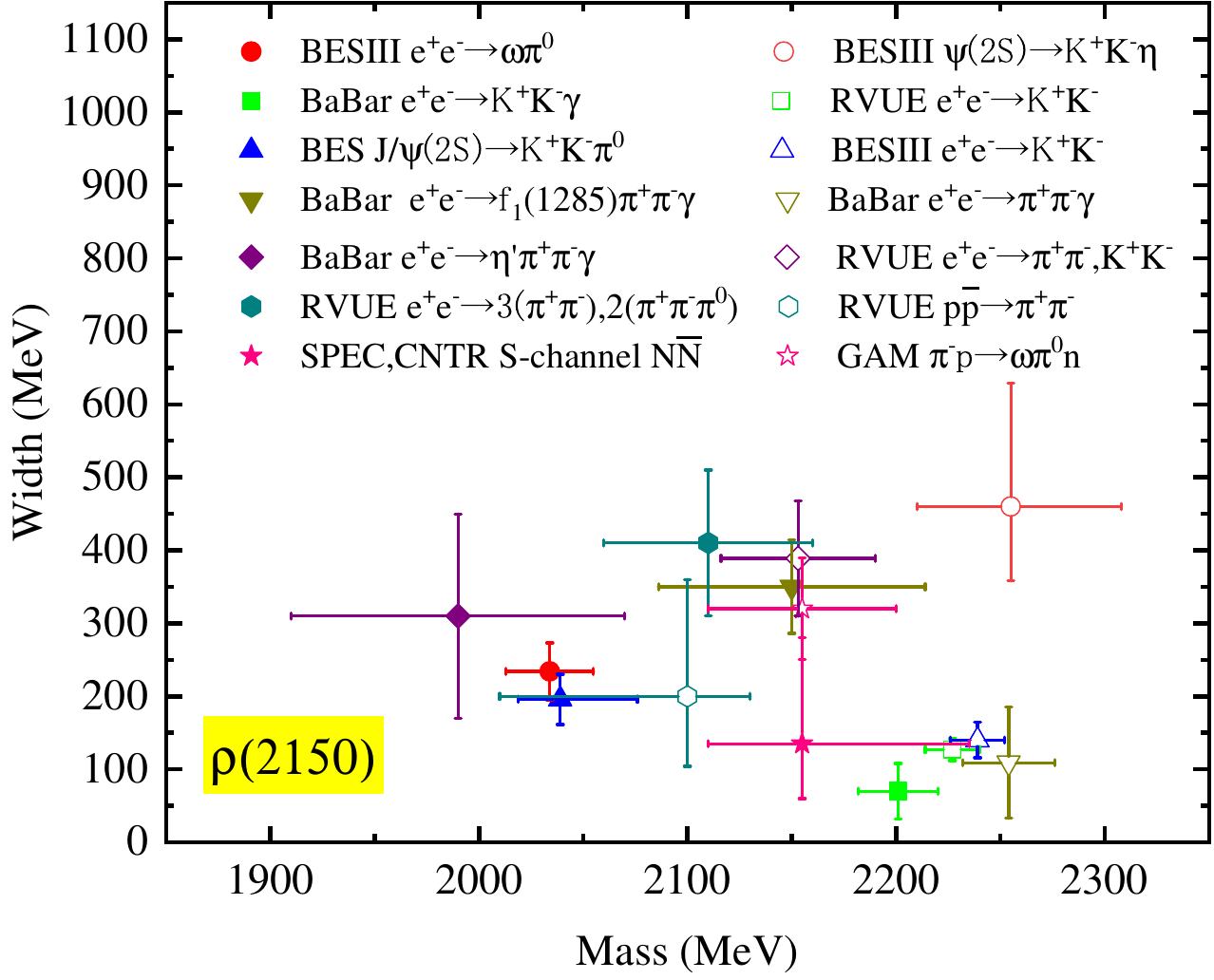}
\caption{Measured resonance parameters of the $Y(2175)$ (left panel) and the $\rho(2150)$ (right panel) from different experimental processes. The figures are adapted from Refs. \cite{Wei:2025ejv,Zhou:2022ark}.}
\label{fig:Y2175res}
\end{figure}

As shown in Fig.~\ref{fig:Y2175res}, the resonance parameters reported by different experimental groups exhibit substantial discrepancies, a situation similar to that observed for the $\rho(2150)$.

The current difficulties encountered in the construction of \changelabel{highly excited} light-flavor mesons are closely related to the analysis way of experimental cross-section data, in which different production or decay processes are studied independently, and the observed structures are interpreted through fits to individual data sets without the theoretical support of spectroscopy. As a result, the extracted resonance parameters tend to be highly process dependent, obscuring the underlying structure of the \changelabel{highly excited} spectrum. Thus, it is not a simple task to solve this messy situation associated with the vector meson spectrum around 2.0 GeV.

Here, a new perspective of constructing \changelabel{highly excited} light-flavor meson spectroscopy was provided in Refs. \cite{Wang:2021gle,Wang:2020kte}.  A guiding principle comes from the well-established success of unquenched approaches in the spectroscopy of charmonium and bottomonium, as introduced in Sec.~\ref{section4} and Sec.~\ref{section7}, respectively. \changelabel{ It should be clarified that the "unquenched" discussed in this subsection are effectively realized through a screened potential to describe the mass spectrum. This screening approach serves as an effective description of 
continuum-induced unquenching, which is distinct from explicit coupled-channel calculations. Of course, a full coupled-channel treatment in the light-flavor meson sector is considerably more challenging than that in heavy 
quarkonium, primarily because the contributions from a large number of open intermediate channels must be simultaneously considered, which warrants systematic experimental and theoretical investigations in the future. } Motivated by these achievements, it is natural to extend the same unquenched spectroscopy description to the light-flavor meson sector, where such effects are expected to be even more pronounced.  Compared with traditional spectroscopic analysis based on Regge trajectory, this approach can offer a more precise description of the \changelabel{highly excited} spectrum and provide exact spatial wave functions for the physical states. These wave functions enable more reliable calculations of decay amplitudes and production strengths, which are indispensable for connecting spectroscopy with experimental observables. With such spectroscopic support, the cross sections measured in different processes can be analyzed in a unified and consistent manner, rather than through isolated fits. The strategy outlined above actually provides a systematic and generally applicable framework for the construction of \changelabel{highly excited} spectra. In the following, we illustrate the application of this strategy by decoding the nature of the well-known $Y(2175)$ through a combined analysis of scattering cross sections in open-strange channels from electron-positron collision.

\subsubsection{The unquenched spectrum of light vector meson and their contribution in the cross sections of the $e^+e^-$ annihilations into open-strange channels}

Recent measurements of open-strange processes in $e^+e^-$ annihilation have provided a series of intriguing cross-section data in the $\sqrt{s} \sim 2.2\;\text{GeV}$ region~\cite{BESIII:2018ldc,BESIII:2020vtu}. The BESIII collaboration reported a clear enhancement in the cross section for $e^+e^- \to K^+K^-$ with a mass of $2239.2 \pm 7.1 \pm 11.3\;\text{MeV}$ and width of $139.8 \pm 12.3 \pm 20.6\;\text{MeV}$~\cite{BESIII:2018ldc}. Subsequently, a combined analysis of the process $e^+e^- \to K^+K^-\pi^0\pi^0$ through its sub-processes $e^+e^- \to K^+(1460)K^-$, $K_1^+(1400)K^-$, $K_1^+(1270)K^-$, and $K^{*+}K^{*-}$ extracted a structure with a significantly lower mass of $2126.5 \pm 16.8 \pm 12.4\;\text{MeV}$ and width of $106.9 \pm 32.1 \pm 28.1\;\text{MeV}$~\cite{BESIII:2020vtu}. It is worth mentioning that open-strange decay channels are usually dominant for exciting light vector meson states, so these reaction processes actually provide very key information to construct \changelabel{highly excited} $\phi$, $\rho$ and $\omega$-meson spectra. Although both observed structures lie near the mass region of $Y(2175)$, the mass difference of about $110\;\text{MeV}$ poses a challenge to a simple single-resonance interpretation. The observed discrepancies of resonance parameters in different open-strange processes suggest that a more comprehensive analysis, incorporating the possible contributions from multiple light vector mesons, is necessary. In addition, the BESIII measurement shows the absence of a clear signal in the typical open-strange mode $K^{*+}K^{*-}$ channel~\cite{BESIII:2020vtu}, which also should be appropriately understood. 

\begin{figure}[htbp]
\centering
\includegraphics[width=0.8\textwidth]{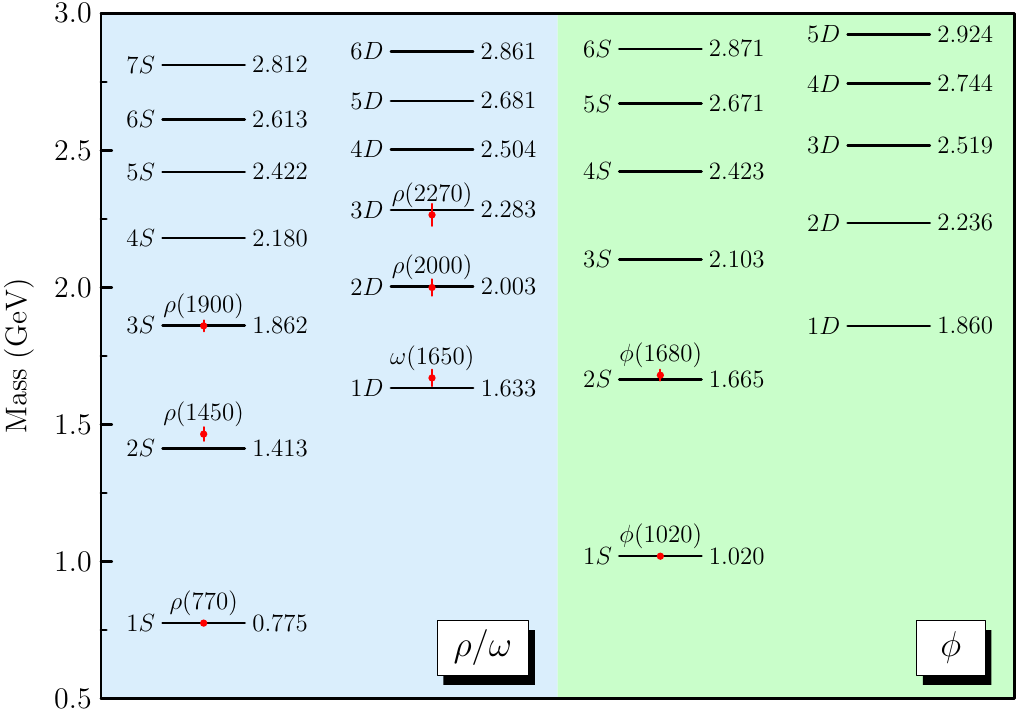}
\caption{Comparison between the theoretical predictions and experimental measurements for the masses of light-flavor vector mesons. Adapted from Ref.~\cite{Wang:2021abg}.}
\label{fig:MassSpectrum}
\end{figure}

To clarify this situation, in Ref. \cite{Wang:2021gle}, Wang, Wang, Liu and Matsuki first systematically study the spectrum and decay properties of light vector mesons above $2\;\text{GeV}$ using an unquenched relativized potential model, whose formula have been introduced in Sec.~\ref{section4}. The model parameters, particularly the color screening constant $\mu$, are constrained by fitting to the experimental masses of established higher excitations in other light meson families (such as $\pi(2360)$ \cite{Anisovich:2001pn}, $\pi_2(2285)$ \cite{Anisovich:2010nh}, $\rho_2(2285)$ \cite{Anisovich:2002su}, and $\rho_3(2250)$ \cite{VES:2000xjq}) up to about $2.3\;\text{GeV}$. This calibration yields $\mu = 0.081\;\text{GeV}$, indicating significant unquenched effects. The mass spectra of light-flavor mesons are systematically predicted and it is found that the global features of the $S$-wave and $D$-wave light-flavor meson spectra are in good overall agreement with the available experimental data \cite{Wang:2021gle}. The predicted spectrum for the vector mesons is presented in Fig. \ref{fig:MassSpectrum}. Using the resulting wave functions, the open-strange strong decay branching ratios are calculated via the quark pair creation model, and the dilepton widths $\Gamma_{e^+e^-}$ are derived from wave-function overlaps. As shown in Table~\ref{tab:val}, di-lepton width predictions for well-known low-lying vector mesons agree reasonably with experimental averages, supporting the theoretical extrapolation to that of higher states.

\begin{table}[ht]
\centering
\caption{Comparison of theoretical predictions with experimental data for selected decays of established light vector mesons. The di-lepton width $\Gamma_{e^+e^-}$ is in keV. Table data is adapted from Ref. \cite{Wang:2021gle}.}
\begin{tabular}{lcc}
\hline
\hline
Observable quantity & Theory & Experiment (PDG) \cite{ParticleDataGroup:2024cfk} \\
\hline
\hline
$\Gamma_{e^+e^-}(\rho(770))$ & 6.98 & $7.04 \pm 0.06$ \\
$BR_{e^+e^-}(\rho(1450)) \cdot BR_{\omega\pi}(\rho(1450)) \;(10^{-6})$ & 4.4 & $3.7 \pm 0.4$ \\
$\Gamma_{e^+e^-}(\omega(782))$ & 0.78 & $0.60 \pm 0.02$ \\
$BR_{e^+e^-}(\omega(1420)) \cdot BR_{\omega\pi}(\omega(1420)) \;(10^{-7})$ & 2.74 & $6.58 \pm 1.49$ \\
$\Gamma_{e^+e^-}(\phi(1020))$ & 3.19 & $1.27 \pm 0.04$ \\
$BR_{e^+e^-}(\phi(1680)) \cdot BR_{K\bar{K}}(\phi(1680)) \;(10^{-7})$ & 20.4 & $22.2 \pm 8.7$ \\
\hline
\hline
\end{tabular}
\label{tab:val}
\end{table}

Crucially, this unquenched spectroscopic study reveals that the electron-positron annihilation into the open-strange channels are dominated by strangeonium states $\phi(3S)$, $\phi(2D)$, and $\phi(4S)$. The calculated dilepton widths for excited $\omega$ states are about an order of magnitude smaller than those for $\rho$ states, and both have branching ratios into open-strange final states typically below a few percent. In contrast, the $\phi$-meson resonances have comparable or larger $\Gamma_{e^+e^-}$ and significantly larger branching ratios into open-strange channels, often exceeding $10\%$. This systematic difference justifies neglecting the numerous possible $\rho$ and $\omega$ contributions, which would otherwise make a combined analysis intractable due to the large number of free parameters.

\begin{figure}[htbp]
\centering
\includegraphics[width=0.85\textwidth]{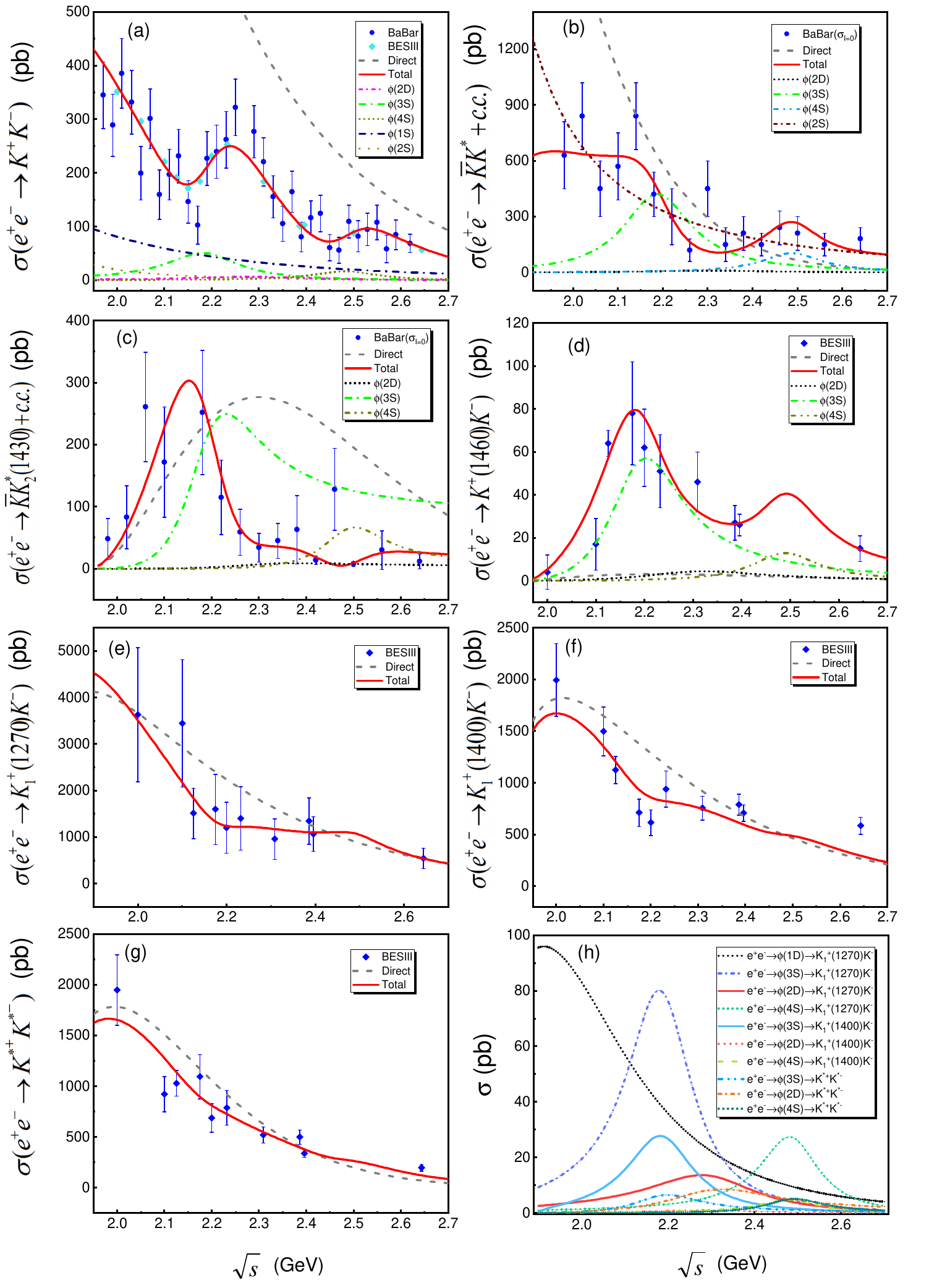}
\caption{Simultaneous description of the cross sections for the seven open-strange processes $e^+e^- \to K^+K^-$, $K\bar{K}^*+\text{c.c.}$, $K^{*+}K^{*-}$, $K_1(1270)^+K^-$, $K_1(1400)^+K^-$, $K_2^*(1430)\bar{K}+\text{c.c.}$, and $K(1460)^+K^-$. Figure adapted from Ref. \cite{Wang:2021gle}. }
\label{fig:kkfit}
\end{figure}

With this strong support from the theoretical spectroscopy, a combined analysis of the seven measured open-strange cross sections ($e^+e^- \to K^+K^-$, $K\bar{K}^*+\text{c.c.}$, $K^{*+}K^{*-}$, $K_1(1270)^+K^-$, $K_1(1400)^+K^-$, $K_2^*(1430)\bar{K}+\text{c.c.}$, and $K(1460)^+K^-$) is performed in Ref. \cite{Wang:2021gle}. The theoretical description for each process $e^+e^- \to K_{(J)}^{(*)}\bar{K}_{(J)}^{(*)}$ includes a direct production amplitude and resonant contributions from selected light vector mesons:
\[
\mathcal{M}_{\text{Total}} = \mathcal{M}_{\text{Direct}} + e^{i\theta_n} \sum_{V_n} \mathcal{M}_{V_n},
\]
where the sum runs over well-known $\phi(1020)$, $\phi(1680)$, and predicted $\phi(1D)$, $\phi(3S)$, $\phi(2D)$, and $\phi(4S)$. The corresponding cross section is given by
\[
\sigma = \int \frac{1}{64\pi s} \frac{1}{|\mathbf{p}_{3\text{cm}}|^2} \bigl| \mathcal{M}_{\text{Total}} \bigr|^2 \, dt .
\]
It is worth emphasizing that all relevant coupling constants in resonance contributions of $\phi(1D)$, $\phi(3S)$, $\phi(2D)$, and $\phi(4S)$ have been fixed by the theoretical calculations. The simultaneous fit to all seven channels achieves a good description with $\chi^2/\text{d.o.f.}=2.29$. The combined fitted cross section for open-strange channels are shown in Fig. \ref{fig:kkfit}. The fitted resonance parameters for the higher strangeonia are:
\[
\begin{aligned}
m_{\phi(3S)} &= 2183 \pm 1\;\text{MeV}, &\Gamma_{\phi(3S)} &= 185 \pm 4\;\text{MeV},\\
m_{\phi(2D)} &= 2290 \pm 3\;\text{MeV}, &\Gamma_{\phi(2D)} &= 312 \pm 6\;\text{MeV},\\
m_{\phi(4S)} &= 2485 \pm 5\;\text{MeV}, &\Gamma_{\phi(4S)} &= 165 \pm 3\;\text{MeV}.
\end{aligned}
\]
These masses are consistent with expectations for higher radial and orbital excitations in an unquenched potential model. The widths, particularly of $\phi(3S)$ and $\phi(2D)$, are broad, in line with theoretical estimates but notably larger than the $~100\;\text{MeV}$ width extracted from simple Breit-Wigner fits to individual channels~\cite{BESIII:2018ldc,BESIII:2020vtu}. This discrepancy is precisely resolved by the interference mechanism: the $\phi(3S)$ and $\phi(2D)$ states, with their substantial intrinsic widths, interfere constructively and destructively with each other and with the non-resonant background across different final states. This interference pattern naturally produces the apparent peak near $2.2\;\text{GeV}$, but with a lineshape that is narrower than either individual resonance, mimicking a single narrower state. It also provides a direct explanation for the mass difference of about $ 110\;\text{MeV}$ observed between the $K^+K^-$ and the $K^+K^-\pi^0\pi^0$ sub-processes, which means that the interference result, and hence the peak position, is sensitive to the relative phases and amplitudes specific to each final state.

A particularly instructive outcome concerns the $e^+e^- \to K^{*+}K^{*-}$ channel, where no significant resonant enhancement was observed. The analysis traces this absence to a specific dynamical suppression: the branching ratio $BR(\phi(3S) \to K^{*}\bar{K}^{*})$ is predicted to be only $ \sim 0.7\%$, an order of magnitude smaller than its decays to other open-strange modes like $K_1(1270)K$ or $K(1460)K$. This suppression originates from a nodal structure in the overlap integral of the $\phi(3S)$ wave function with the $K^{*}\bar{K}^{*}$ final state, a direct consequence of the radial excitation. Thus, what might have seemed like a puzzling null result becomes a successful postdiction and a non-trivial test of the unquenched model.

Finally, this analysis predicted the existence of a new resonance, $\phi(4S)$, with a mass around $2.485\;\text{GeV}$. Its contribution, though less prominent than the lower $\phi$ states in the current data, is discernible in several channels, most notably $e^+e^- \to K\bar{K}(1460)$, offering a clear target for future high-statistics experiments at BESIII and Belle~II. In conclusion, this analysis demonstrates that a global analysis of production cross sections, guided by spectroscopic calculations, can coherently resolve experimental puzzles and constrain the \changelabel{highly excited} light vector meson spectrum. It thereby provides a new interpretation of $Y(2175)$: rather than an isolated particle, it manifests as a pronounced interference structure between the broad, overlapping $\phi(3S)$ and $\phi(2D)$ resonances.

\subsubsection{Unified description of the cross sections for more relevant reaction processes within the support of the unquenched spectroscopy}

\paragraph{High-lying $\phi$-related process}

\begin{figure}[htbp]
\centering
\includegraphics[width=0.6\textwidth]{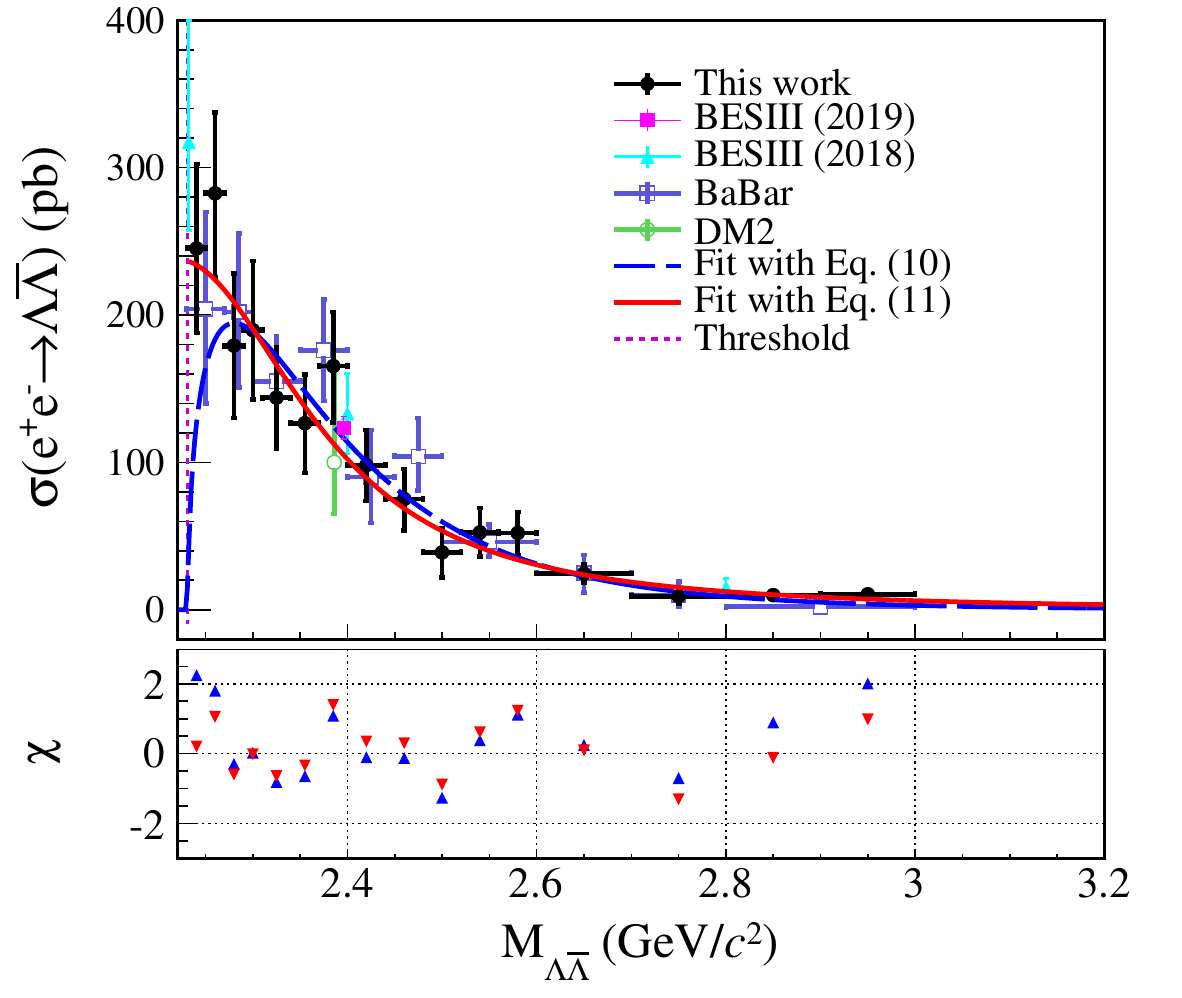}
\caption{Measured cross section for the process $e^+e^-\to \Lambda\bar{\Lambda}$ reported by the BESIII Collaboration in 2023~\cite{BESIII:2023ioy}, together with previous measurements \cite{BaBar:2007fsu,BESIII:2017hyw,BESIII:2019nep,DM2:1990tut}. Adapted from Ref.~\cite{BESIII:2023ioy}.}
\label{fig:mLambda}
\end{figure}

In 2023, the BESIII Collaboration measured the cross section of the process $e^+e^-\to \Lambda\bar{\Lambda}$ at center-of-mass energies from the production threshold up to 3.0~GeV \cite{BESIII:2023ioy}, as shown in Fig.~\ref{fig:mLambda}. It is observed that the cross section data in the vicinity of 2.4~GeV are not well accommodated by the experimental fit. Notably, this energy region overlaps with the masses of several predicted high-lying strangeonium states. This feature naturally motivates the investigation of the decays of high-lying strangeonia into $\Lambda\bar{\Lambda}$.

\begin{figure}[htbp]
\centering
\raisebox{0.045\textwidth}{\includegraphics[width=0.45\textwidth]{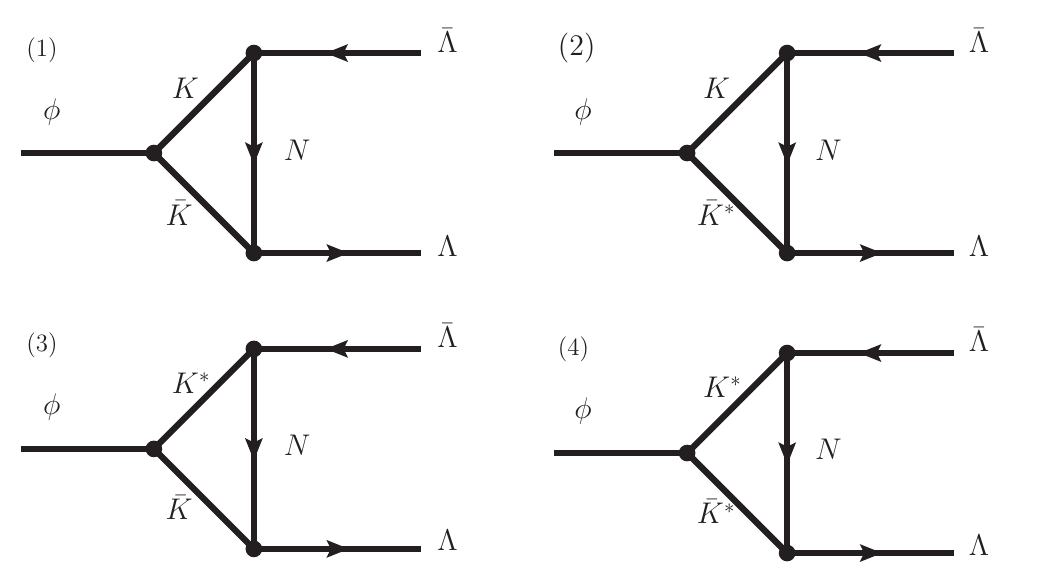}}
\hspace{0.02\textwidth}
\includegraphics[width=0.45\textwidth]{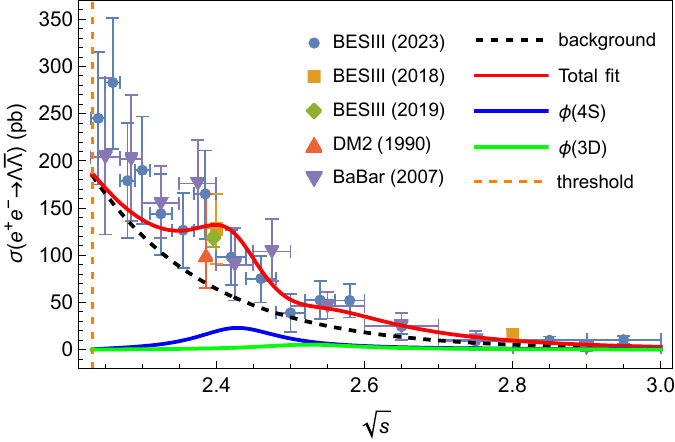}
\caption{Left panel: Feynman diagrams for the processes $\phi(4S)/\phi(3D)\to \Lambda\bar{\Lambda}$ within the hadronic loop mechanism. Right panel: Fit to the cross section of $e^+e^-\to \Lambda\bar{\Lambda}$ incorporating the contributions from $\phi(4S)$ and $\phi(3D)$. Figures adapted from Ref.~\cite{Bai:2023dhc}.}
\label{fig:tLambda}
\end{figure}

In Ref.~\cite{Bai:2023dhc}, Bai, Zhou, and Liu investigated the decays of $\phi(4S)$ and $\phi(3D)$ into $\Lambda\bar{\Lambda}$ via the hadronic loop mechanism, as illustrated in the left panel of Fig.~\ref{fig:tLambda}. In the unquenched spectroscopy, the masses of these states are predicted to be around 2.4~GeV. A refit to the measured cross section of $e^+e^-\to \Lambda\bar{\Lambda}$ was performed by including the contributions from $\phi(4S)$, $\phi(3D)$, and a continuum background, together with their mutual interference. The fit result, shown in the right panel of Fig.~\ref{fig:tLambda}, demonstrates that the cross section data are well reproduced. In particular, the deviation around 2.4~GeV can be mainly attributed to the contribution from the $\phi(4S)$ state. The extracted branching ratios of $\phi(4S)$/$\phi(3D)\to\Lambda\bar{\Lambda}$ are in good agreement with the theoretical predictions based on the hadronic loop mechanism.

Motivated by these results, the decays of higher strangeonia ($\phi(nS)$ with $n=4,5,6$ and $\phi(mD)$ with $m=3,4,5$) into light-flavor baryon–antibaryon pairs, such as $\Lambda\bar{\Lambda}$, $\Sigma\bar{\Sigma}$, and $\Xi\bar{\Xi}$, were further investigated within the same framework. The corresponding branching ratios are found to lie in the range of $10^{-5}$–$10^{-2}$, indicating that these processes may be accessible in future experimental measurements.

\paragraph{High-lying $\omega$-related processes}

\begin{figure}[htbp]
\centering
\includegraphics[width=0.6\textwidth]{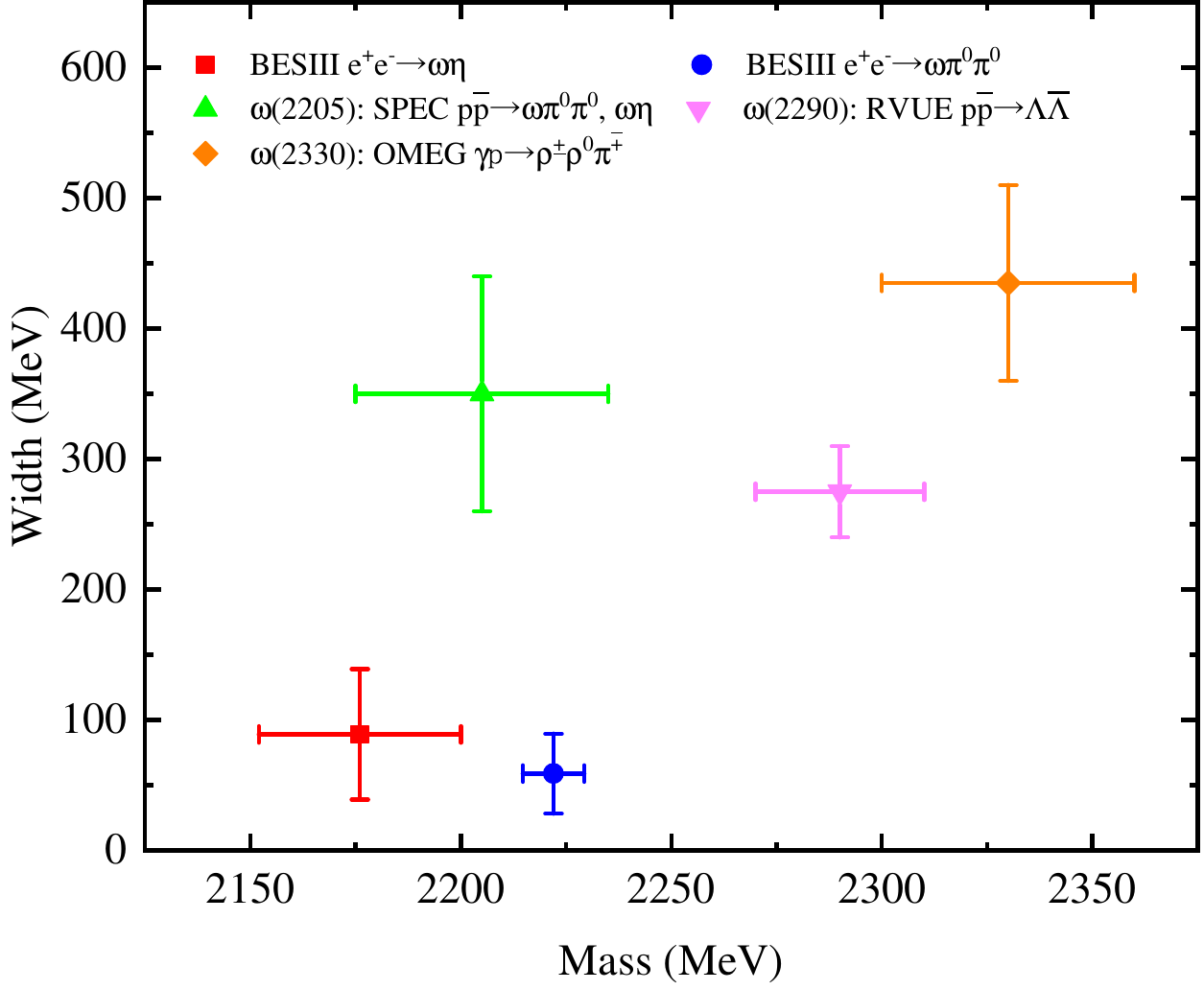}
\caption{A comparison of resonance parameters of the
reported $\omega$ states with masses around 2.2 GeV \cite{ParticleDataGroup:2024cfk,BESIII:2020xmw,BESIII:2021uni}. Adapted from Ref.~\cite{Zhou:2022wwk}.}
\label{fig:RP}
\end{figure}

Recently, the BESIII Collaboration reported measurements of the cross sections for the processes
$e^+e^-\to\omega\eta$ \cite{BESIII:2020xmw} and
$e^+e^-\to\omega\pi^0\pi^0$ \cite{BESIII:2021uni}
at center-of-mass energies ranging from 2.00 to 3.08~GeV.
These channels are related to high-lying $\omega$ meson states due to isospin conservation.
Two structures were observed and interpreted in terms of resonances with the following parameters:
$m_1 = 2222 \pm 7 \pm 2$~MeV, $\Gamma_1 = 59 \pm 30 \pm 6$~MeV, and
$m_2 = 2179 \pm 21 \pm 3$~MeV, $\Gamma_2 = 89 \pm 28 \pm 5$~MeV.
Notably, the resonance parameters $(m_1, \Gamma_1)$ and $(m_2, \Gamma_2)$
extracted from the two processes show noticeable differences. When comparing the two reported $\omega$-like states with $\omega(2205)$, $\omega(2290)$, and
$\omega(2330)$ listed as "further states" in the PDG \cite{ParticleDataGroup:2024cfk},
as illustrated in Fig.~\ref{fig:RP},
their resonance parameters are found to differ significantly.
It is evident that the accumulation of the five $\omega$ states within the same energy region is puzzling from a spectroscopic point of view.

\begin{figure*}[htbp]
  \centering
  \begin{tabular}{@{}cc@{}}
    \includegraphics[width=0.3\textwidth]{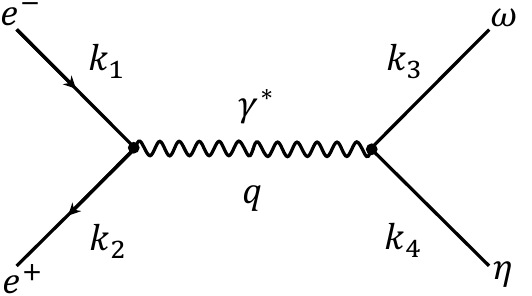} \hspace{0.04\textwidth} &
    \includegraphics[width=0.3\textwidth]{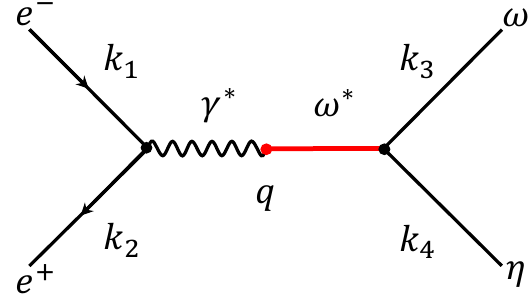} \\
    (a) & (b)
  \end{tabular}

  \vspace{0.3cm}

  \begin{tabular}{@{}ccc@{}}
    \includegraphics[width=0.3\textwidth]{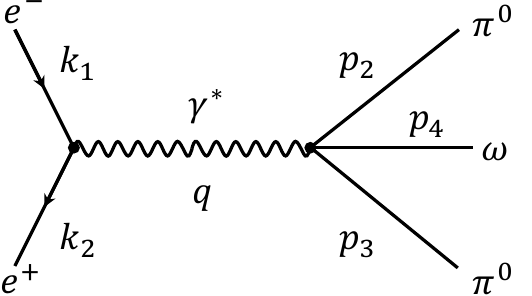} &
    \includegraphics[width=0.3\textwidth]{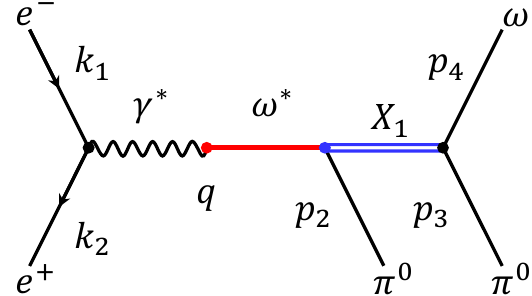} &
    \includegraphics[width=0.3\textwidth]{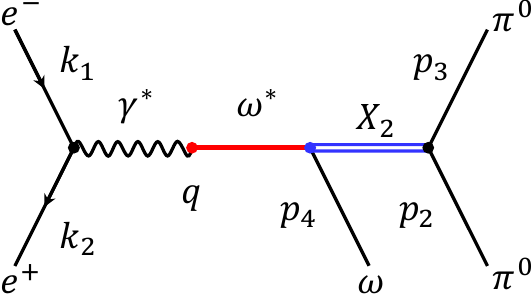} \\
    (c) & (d) & (e)
  \end{tabular}

  \caption{Feynman diagrams for the reactions 
  $e^+e^-\to\omega\eta$ and $e^+e^-\to\omega\pi^0\pi^0$.
  Panels (a) and (b) correspond to $e^+e^-\to\omega\eta$.
  Panels (c)--(e) depict the mechanisms of $e^+e^-\to\omega\pi^0\pi^0$.
  Here, $\omega^*$ denotes the $\omega(4S)$ and $\omega(3D)$ intermediate states,
  while $X_1$ denotes $\rho$, $\rho(1450)$, and $b_1(1235)$, and
  $X_2$ represents $f_2(1270)$.
  Adapted from Ref.~\cite{Zhou:2022wwk}.}
  \label{Feyomega_combined}
\end{figure*}

To clarify this puzzling situation, in Ref.~\cite{Zhou:2022wwk},
Zhou, Wang, and Liu introduced the contributions from the intermediate
$\omega(4S)$ and $\omega(3D)$ states, whose masses are around 2.2~GeV
according to the unquenched spectrum of light vector mesons,
to reproduce the cross section data of
$e^+e^-\to\omega\eta$ and $e^+e^-\to\omega\pi^0\pi^0$. Specifically, two reaction mechanisms are considered here. One corresponds to the direct annihilation of the initial $e^+e^-$ pair into the final state via a virtual photon, which provides a nonresonant
background contribution.
The other proceeds via the intermediate $\omega(4S)$ and $\omega(3D)$ states, as illustrated in Fig.~\ref{Feyomega_combined}.

\begin{figure*}[htbp]
\centering
\begin{tabular}{@{}cc@{}}
\includegraphics[width=0.35\textwidth]{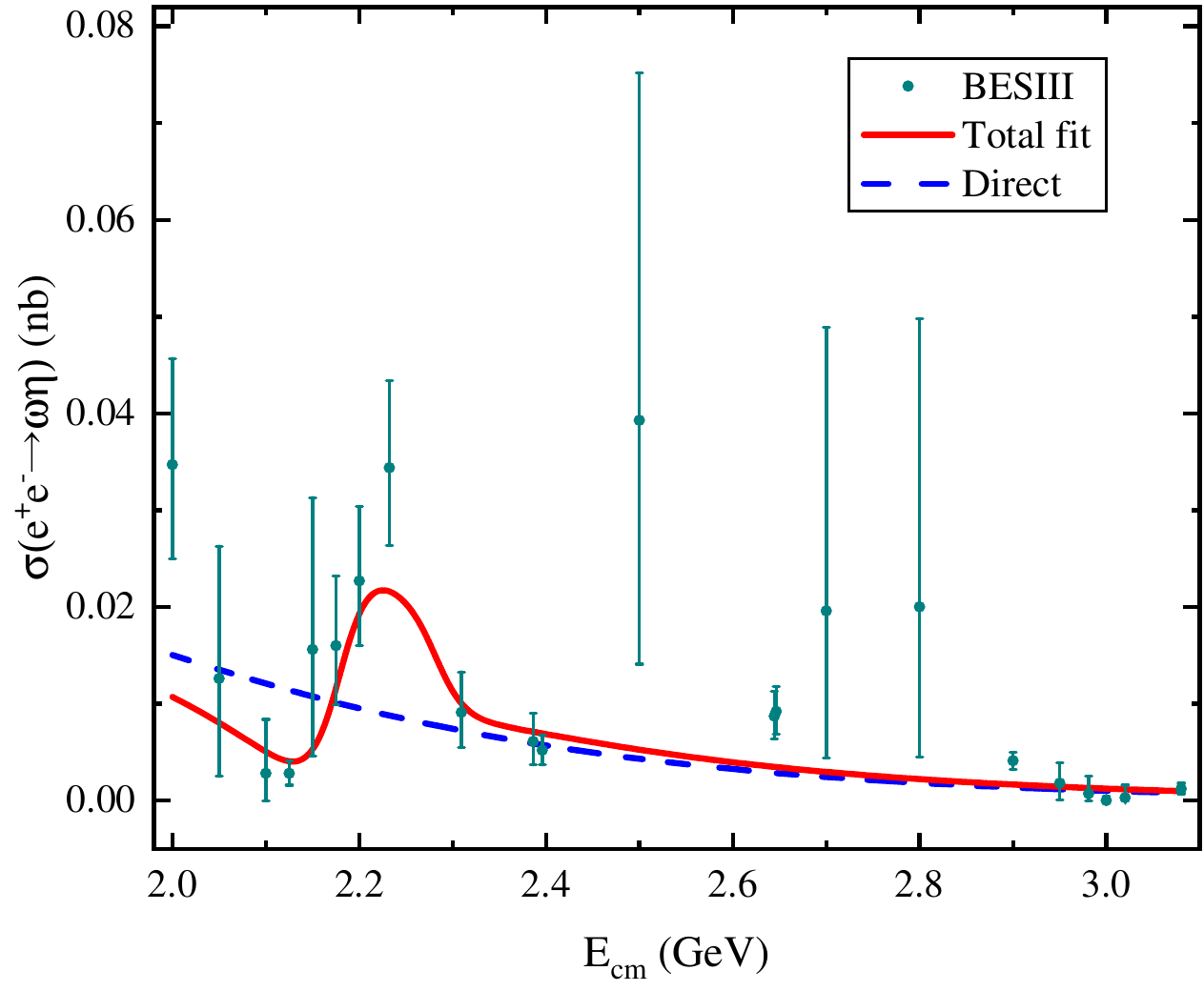} \hspace{0.04\textwidth} &
\includegraphics[width=0.35\textwidth]{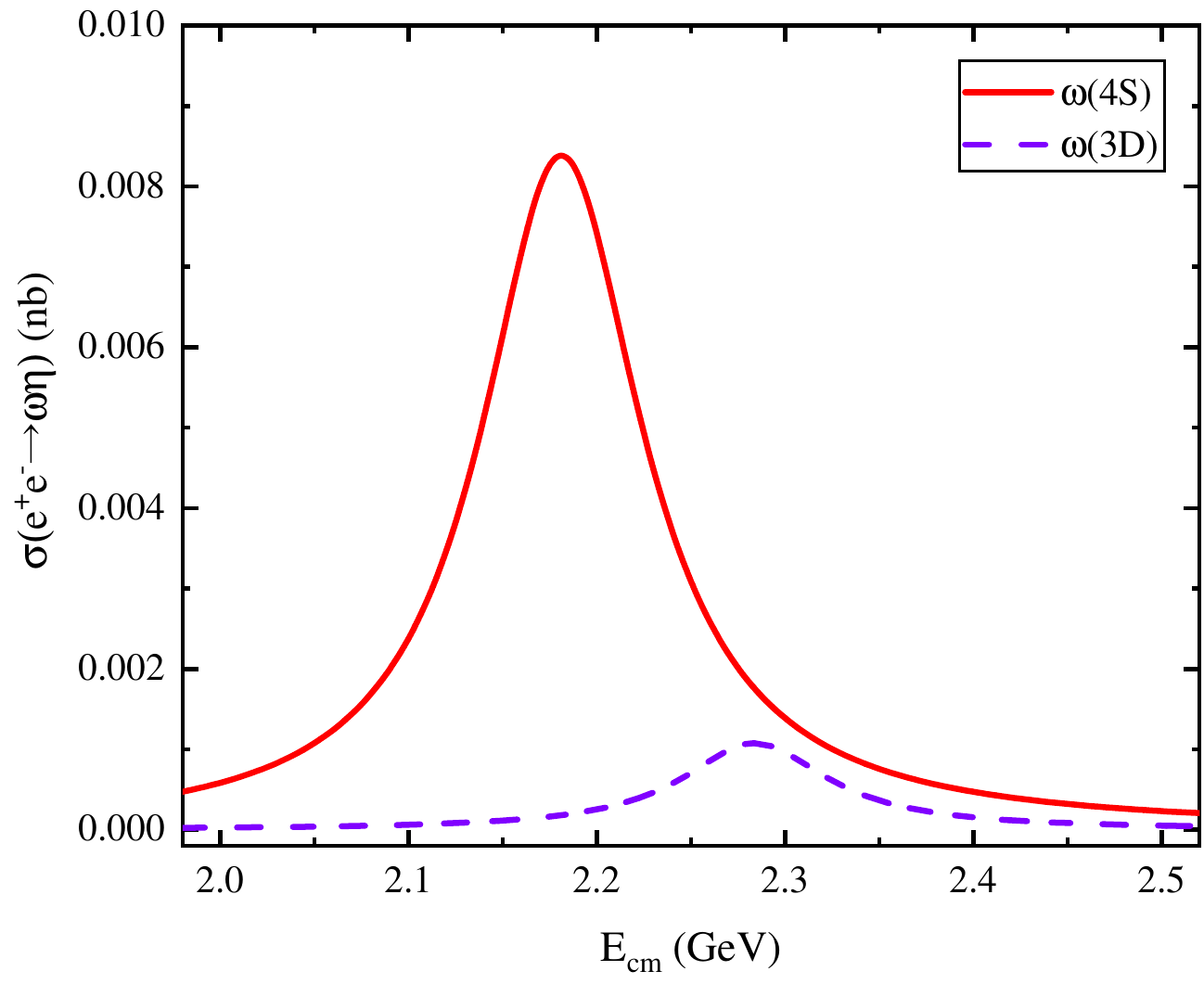} \\
(a) & (b)
\end{tabular}

\vspace{0.3cm}

\begin{tabular}{@{}ccc@{}}
\includegraphics[width=0.3\textwidth]{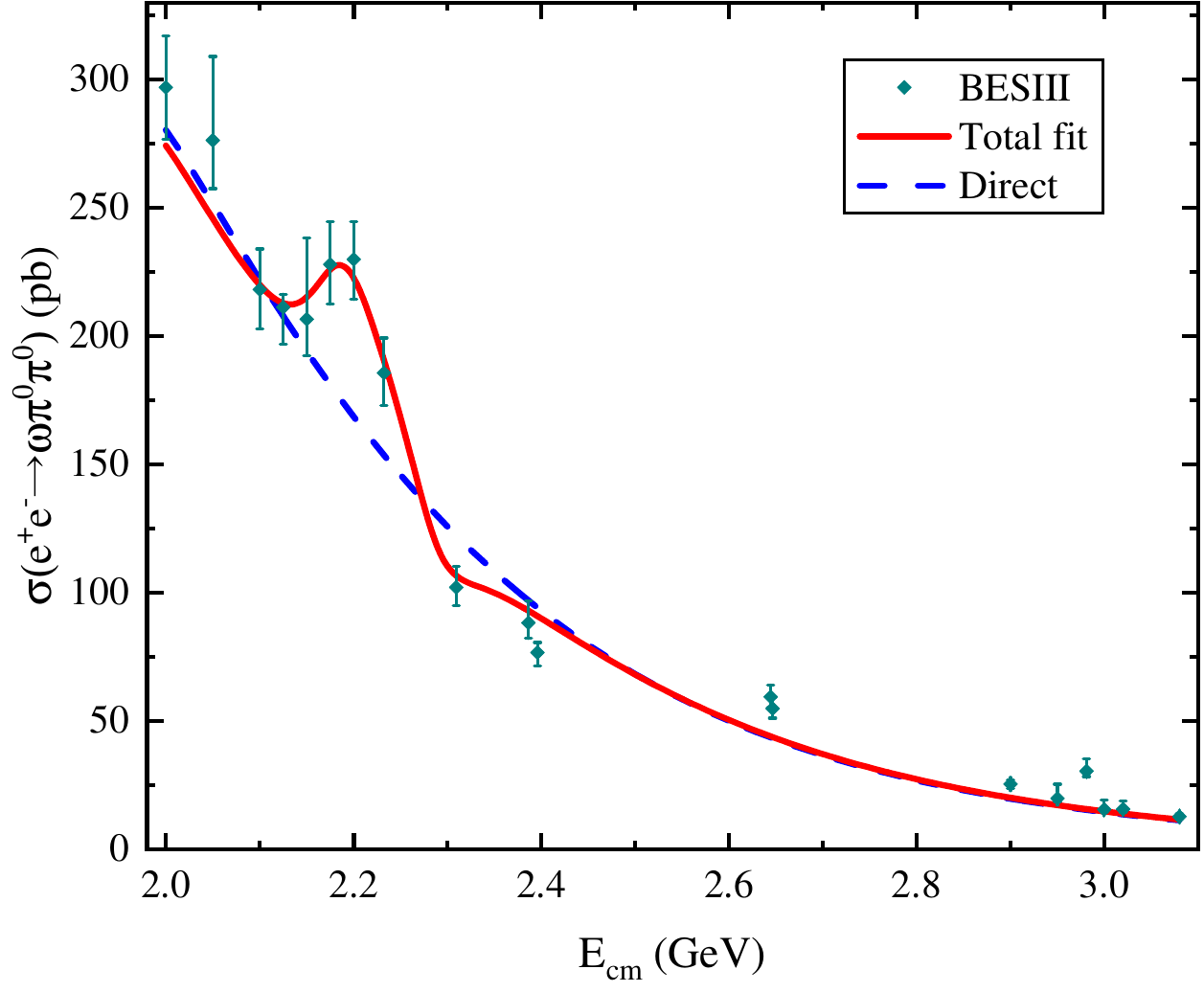} &
\includegraphics[width=0.3\textwidth]{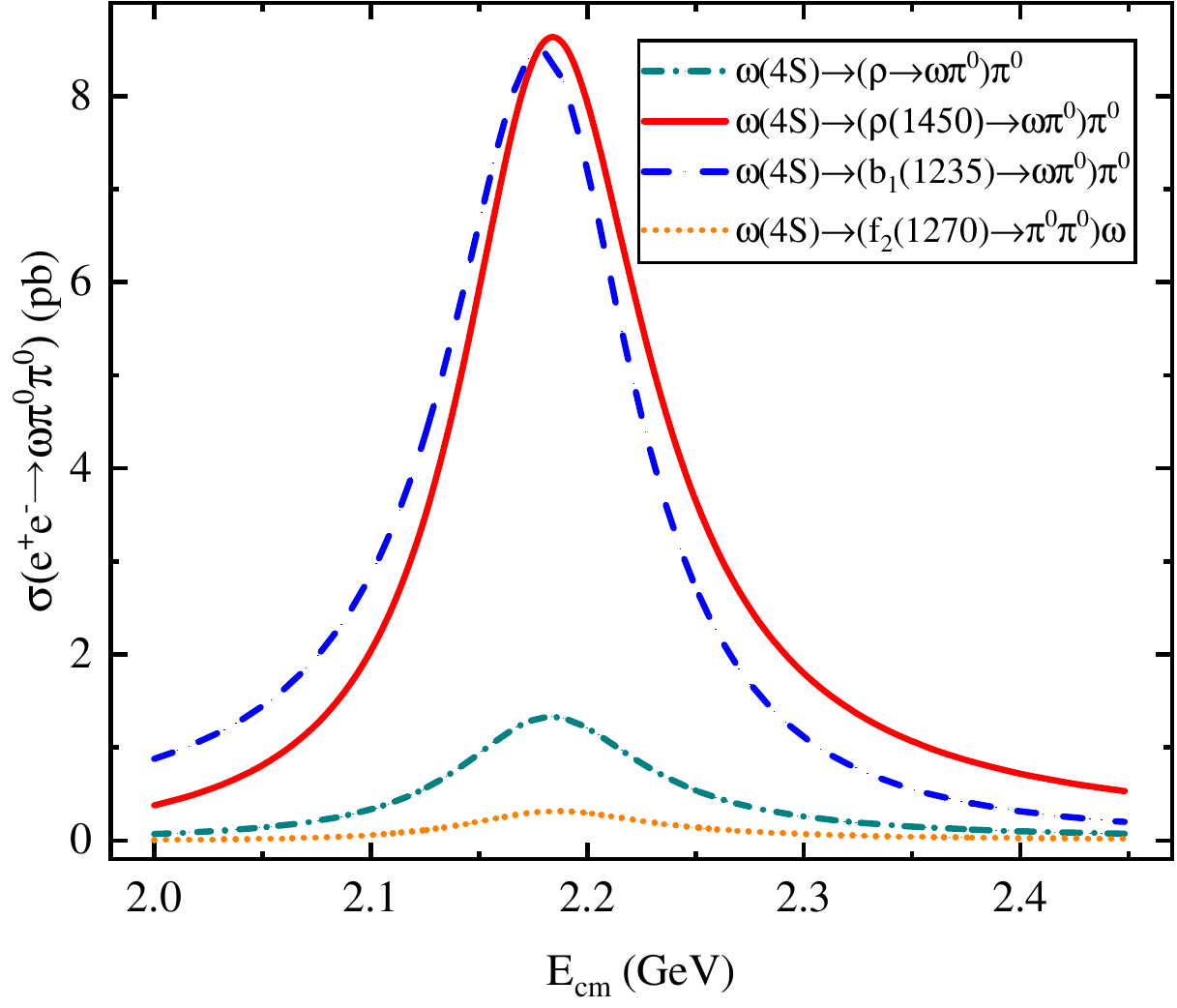} &
\includegraphics[width=0.3\textwidth]{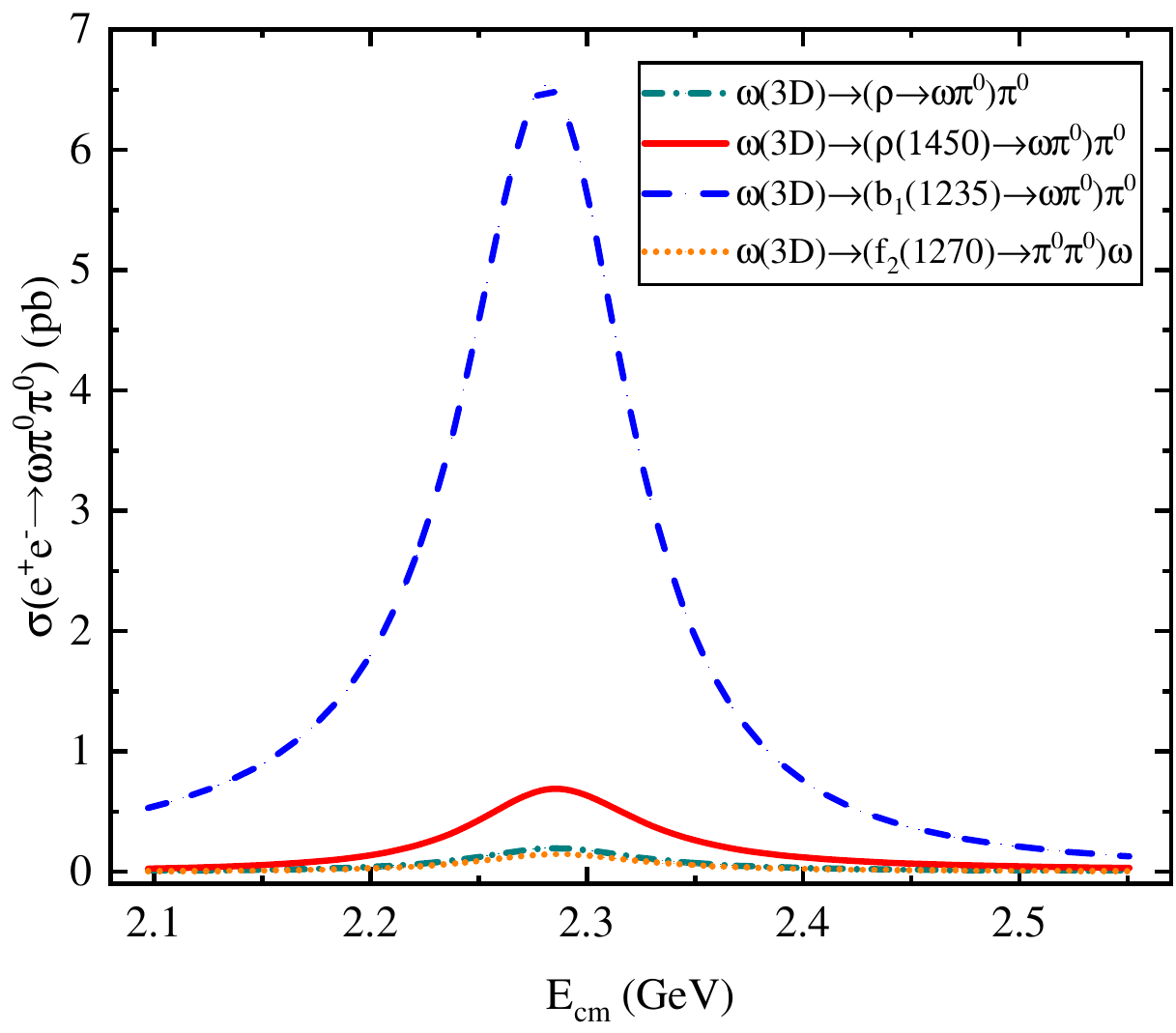} \\
(c) & (d) & (e)
\end{tabular}

\caption{Fitted results to the experimental cross section data and the corresponding resonance contributions for 
$e^+e^-\to\omega\eta$ (Panels (a) and (b)) and $e^+e^-\to\omega\pi^0\pi^0$ (Panels (c)--(e)).
Figures adapted from Ref.~\cite{Zhou:2022wwk}.}
\label{fig:oe_opp_combined}
\end{figure*}

The coupling constants entering the resonance amplitudes are fixed
by the decay properties of the $\omega(4S)$ and $\omega(3D)$ states
as provided by the unquenched spectroscopy.
By taking into account the interference between the direct amplitude
and the corresponding resonance amplitudes,
the measured cross section data of $e^+e^-\to\omega\eta$ and
$e^+e^-\to\omega\pi^0\pi^0$ can be well reproduced,
as shown in Fig.~\ref{fig:oe_opp_combined}. Notably, $\omega(4S)$ dominates in $e^+e^-\to\omega\eta$, 
whereas in $e^+e^-\to\omega\pi^0\pi^0$, both $\omega(4S)$ and $\omega(3D)$ 
contribute with comparable significance.

\begin{figure}[htbp]
\centering
\includegraphics[width=0.8\textwidth]{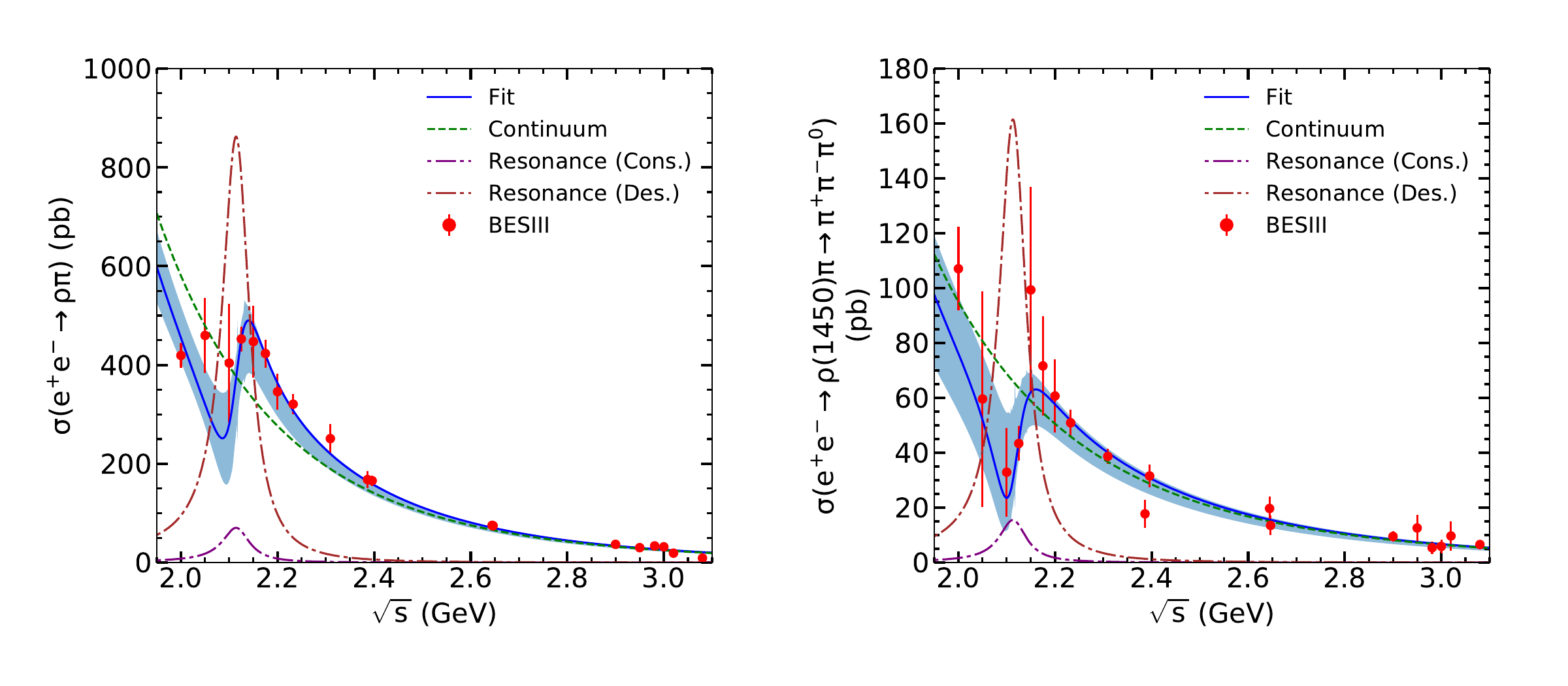}
\caption{The measured cross sections for (a) $e^+e^-\to \rho\pi$ and (b) $e^+e^-\to \rho(1450)\pi\to\pi^+\pi^-\pi^0$. The red points with error bars denote the experimental data. The blue solid curve with the shaded band represents the total fit, while the green dashed curve shows the continuum contribution. The purple dot-dashed and brown dot-dashed curves correspond to the resonance amplitudes for the constructive and destructive interference solutions, respectively. Adapted from Ref. \cite{BESIII:2024okl}.}
\label{fig:observeY2119}
\end{figure}

Subsequently, in 2024, the BESIII Collaboration measured the cross sections for the processes $e^+e^-\to \rho\pi$ and $e^+e^-\to \rho(1450)\pi$ through a partial wave analysis of $e^+e^-\to \pi^+\pi^-\pi^0$ process, and reported a resonance structure with a mass of $M = 2119 \pm 11 \pm 15$ MeV and a width of $\Gamma = 69 \pm 30 \pm 5$ MeV \cite{BESIII:2024okl}, as shown in Fig.~\ref{fig:observeY2119}. This resonance, with a significance of $5.9\sigma$, is hereafter referred to as the $Y(2119)$. The observed decay channels of the $Y(2119)$ indicate that it is a typical $\omega$-like mesonic state with isospin $I=0$.

However, a comparison of the observed $Y(2119)$ with the predicted mass spectrum of the $\omega$-meson family reveals a significant discrepancy, as illustrated in Table~\ref{MGI}. Furthermore, the structure cannot be attributed to interference effects, since its measured mass lies far from the dominant $S$-wave states, while the contributions from nearby $D$-wave states are expected to be negligible due to the suppression of their dielectron widths. These considerations suggest that the $Y(2119)$ cannot be readily accommodated within the conventional $\omega$-meson family.

\begin{table}[htbp]
\centering
\caption{The predicted mass spectrum of the selected high-lying $\omega$ meson states from Ref.~\cite{Wang:2021gle}, together with a comparison to the observed $Y(2119)$ \cite{BESIII:2024okl}. Table adapted from Ref. \cite{Bai:2025knk}}	\label{MGI}
		\renewcommand\arraystretch{1.3}
  \begin{tabular*}{1.0\textwidth}{l@{\extracolsep{\fill}}cccc}
			 \hline
          \hline
			$\omega$ states &$\omega(2D)$ &$\omega(4S)$ &$\omega(3D)$ &$Y(2119)$\\
			\hline
			Mass (MeV)      &2003         &2180         &2283        &$2119\pm11\pm15$\\
                Width (MeV)     &181          &104          &94          &$69\pm30\pm5$\\
                $\Gamma_{e^+e^-}$ (eV)  &2.2     &7.0          &1.8         &$\cdots$\\
			\hline
            \hline
		\end{tabular*}
	\end{table}	

\begin{figure}[htbp]
\centering
\includegraphics[width=0.6\textwidth]{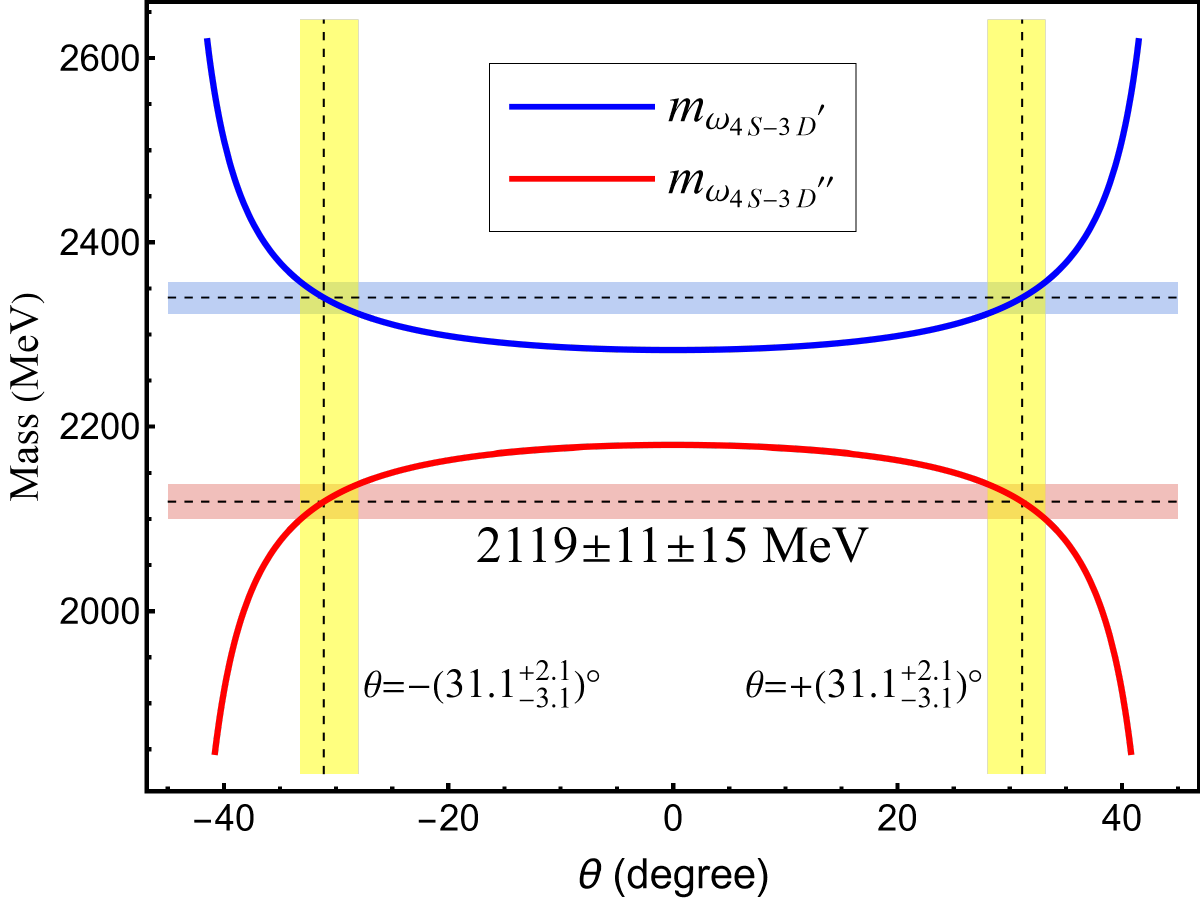}
\caption{The masses of the mixed state $\omega_{4S-3D}^\prime$ and $\omega_{4S-3D}^{\prime\prime}$ as function of the $4S$-$3D$ mixing angle $\theta$. The light red horizontal band represents the measured mass of the $Y(2119)$, while the yellow vertical bands indicate the mixing angles at which the theoretical mass of $\omega_{4S-3D}^\prime$ coincides with the $Y(2119)$. Additionally, the light blue horizontal band represents the predicted mass of $\omega_{4S-3D}^{\prime\prime}$ at these mixing angles. Adapted from Ref. \cite{Bai:2025knk}.}
\label{fig:mixY2119}
\end{figure}

Upon examining the details of the above comparison, we observe that the mass of the $\omega(4S)$ is higher than that of the $Y(2119)$. To address this discrepancy, it is essential to identify a mechanism capable of lowering the mass of the $\omega(4S)$ to align with that of the $Y(2119)$. A natural solution is to introduce a $4S$-$3D$ state mixing scheme.
The $4S$--$3D$ mixing scheme has been employed in the analyses of the charmoniumlike state $Y(4220)$ and the bottomoniumlike state $Y(10753)$, as reviewed in the previous sections. In view of these applications, it is natural to consider extending this framework to the $\omega$-meson family. Specifically, in Ref.~\cite{Bai:2025knk}, Bai, Zhou, and Liu proposed the following mixing scheme:
\begin{align}
	\left(\begin{array}{c}
	|\omega_{4S-3D}^\prime\rangle \\
	|\omega_{4S-3D}^{\prime\prime}\rangle
	\end{array}\right)=
	\left(\begin{array}{cc}
	\cos\theta & \sin\theta\\
	-\sin\theta & \cos\theta
	\end{array}\right)
	\left(\begin{array}{c}
	|\omega(4S)\rangle\\
	|\omega(3D)\rangle
	\end{array}\right),
\end{align}
and extracted a mixing angle of $\theta=\pm(31.1^{+2.1}_{-3.1})^\circ$ by associating the observed $Y(2119)$ with the $\omega_{4S-3D}^\prime$ state, as shown in Fig.~\ref{fig:mixY2119}.

Then, the decay properties of $Y(2119)$ were studied by assuming $Y(2119)$ as the mixed state $\omega_{4S\text{-}3D}^\prime$, and a comparison between the theoretical calculations and the corresponding experimental measurements for the relevant decay channels is presented in Table~\ref{tab:comparison}.

\begin{table}[htbp]
\centering
\caption{Theoretical predictions for the mixed $\omega_{4S\text{-}3D}^\prime$ state's mass, width ($\Gamma$), dielectron width ($\Gamma^{\mathcal{R}}_{e^+e^-}$), partial widths of $\rho\pi$ and $\rho(1450)\pi$ channels, and products $\Gamma_{e^+e^-}^\mathcal{R} \times BR_{\mathcal{R}\to\rho\pi}$ and $\Gamma_{e^+e^-}^\mathcal{R}\times BR_{\mathcal{R}\to\rho(1450)\pi\to\pi^+\pi^-\pi^0}$, calculated for both negative and positive angles. Comparison with BESIII Collaboration measurements of the observed $Y(2119)$ for constructive and destructive interference scenarios \cite{BESIII:2024okl} is included. Adapted from Ref. \cite{Bai:2025knk}.} 
\label{tab:comparison}
\renewcommand\arraystretch{1.5}
\setlength{\tabcolsep}{4.0pt}
\begin{tabular*}{1.0\textwidth}{l@{\extracolsep{\fill}}cccc}
\hline
\hline
Intermediate state $\mathcal{R}$ &\multicolumn{2}{c}{$\omega_{4S-3D}^\prime$}  &\multicolumn{2}{c}{$Y(2119)$  \cite{BESIII:2024okl}} \\
\hline
\multirow{2}{*}{Status}  &\multirow{2}{*}{$\theta$=$-(31.1^{+2.1}_{-3.1})^\circ$} &\multirow{2}{*}{$\theta$=$+(31.1^{+2.1}_{-3.1})^\circ$} &Constructive interference  &Destructive interference \\
&{} &{} &scenario &scenario\\
\hline
Mass (MeV) &$2119\pm11\pm15$  &$2119\pm11\pm15$   &$2119\pm11\pm15$  &$2119\pm11\pm15$ \\
$\Gamma$ (MeV)     &$66.9^{+11.4}_{-9.9}$ &$52.9^{+11.5}_{-9.5}$  &$69\pm30\pm5$ &$69\pm30\pm5$\\
$\Gamma_{e^+e^-}^{\mathcal{R}}$ (eV) &$10.0\pm0.2$ &$2.6^{+0.4}_{-0.2}$  &$\cdots$  &$\cdots$\\
$\Gamma_{\mathcal{R}\to\rho\pi}$ (MeV) &$12.8^{+2.2}_{-1.9}$ &$4.6^{+1.4}_{-1.0}$ &$\cdots$  &$\cdots$\\
$\Gamma_{\mathcal{R}\to\rho(1450)\pi}$ (MeV) &$24.8^{+4.4}_{-4.1}$ &$8.5^{+3.0}_{-2.0}$ &$\cdots$  &$\cdots$\\
$\Gamma_{e^+e^-}^{\mathcal{R}}\times BR_{\mathcal{R}\to\rho\pi}\,(\text{eV})$ &$1.9\pm0.5$ &$0.2\pm0.1$ &$1.5\pm0.7$ &$17\pm12$\\
$\Gamma_{e^+e^-}^{\mathcal{R}}\times BR_{\mathcal{R}\to\rho(1450)\pi\to\pi^+\pi^-\pi^0}\,(\text{eV})$ &$0.4\pm0.1$ &$0.04\pm0.02$ &$0.3\pm0.2$ &$3\pm2$ \\
\hline
\hline
\end{tabular*}
\end{table}

As shown in Table~\ref{tab:comparison}, the calculated masses and total widths are consistent with the measured parameters of the $Y(2119)$ for both positive and negative mixing angles. A further comparison between the experimental measurements and the theoretical calculations of the products of the dielectron width and branching ratios for the $\rho\pi$ and $\rho(1450)\pi$ channels is performed. These quantities are directly proportional to the amplitudes for $e^+e^-$ annihilation into the corresponding final states via the $Y(2119)$ resonance. It is found that the $\omega_{4S\text{-}3D}^\prime$ state with a negative mixing angle yields $\Gamma_{e^+e^-}^{\mathcal{R}}\times BR_{\mathcal{R}\to\rho\pi}$ and $\Gamma_{e^+e^-}^{\mathcal{R}}\times BR_{\mathcal{R}\to\rho(1450)\pi\to\pi^+\pi^-\pi^0}$, that are in excellent agreement with those of the measured $Y(2119)$ in the constructive interference scenario. This supports the assignment of the $Y(2119)$ to the mixed state $\omega_{4S-3D}^\prime$ with a negative mixing angle $\theta = -(31.1^{+2.1}_{-3.1})^\circ$. Moreover, the experimental analysis provides two interference solutions, corresponding to constructive and destructive interference between the resonance amplitude and the continuum background for the $Y(2119)$, as shown in Fig. \ref{fig:observeY2119}. The present analysis indicates that only the constructive interference solution is physically viable, whereas the destructive interference solution does not correspond to a viable physical scenario.

Furthermore, the dominant decay channels of the $Y(2119)$ are investigated within the mixed state scenario with the negative mixing angle. In addition to the $\rho\pi$ and $\rho(1450)\pi$ modes, the $\rho a_0(980)$ and $b_1(1235)\pi$ channels are identified as promising for future experimental searches for the $Y(2119)$. In this framework, a $D$-wave dominated partner of the $Y(2119)$ is also expected, with a mass around 2340~MeV. However, there are significant experimental challenges in observing this state in $e^+e^-$ collisions, since its dielectron width is predicted to be extremely small. Such a state may instead be accessible at other experimental facilities, such as meson beam experiments.
\paragraph{High-lying $\rho$-related processes}

Theoretically, the dielectron widths of the $\rho$-mesonic states are generally about one order of magnitude larger than those of the $\omega$-mesonic states \cite{Wang:2021gle}, implying that $\rho$ mesons are produced more copiously at $e^+e^-$ collision facilities. In recent years, several structures associated with high-lying $\rho$ states have been reported in experimental measurements of exclusive cross sections around 2 GeV, including $e^+e^-\to\pi^+\pi^-$ \cite{BaBar:2012bdw,BaBar:2019kds}, $e^+e^-\to f_1(1285)\pi^+\pi^-$ \cite{BaBar:2007qju,BaBar:2022ahi}, $e^+e^-\to\omega\pi^0$ \cite{BESIII:2020xmw,Achasov:2016zvn}, $e^+e^-\to\eta^\prime\pi^+\pi^-$ \cite{BESIII:2020kpr}, $e^+ e^-\to \eta \pi^+ \pi^- $ \cite{BESIII:2023sbq}, and $e^+e^-\to p\bar{p}\pi^0$ \cite{Guo:2025igf}.

In the mass region around 2~GeV, three $\rho$-mesonic states, namely $\rho(1900)$, $\rho(2000)$, and $\rho(2150)$, have been collected by the PDG \cite{ParticleDataGroup:2024cfk}. In Refs.~\cite{Wang:2020kte,Liu:2022yrt,Zhou:2022ark}, the Lanzhou group pointed out that interference effects play an essential role in understanding these reported structures and the corresponding measured cross-section data. In particular, the measured cross sections for $e^+e^-\to\pi^+\pi^-$, $e^+e^-\to f_1(1285)\pi^+\pi^-$, and $e^+e^-\to\omega\pi^0$ can be well reproduced by the intermediate $\rho(1900)$ and $\rho(2150)$ contributions, together with the interference effect between them. Meanwhile, the cross section for the $e^+e^-\to\eta^\prime\pi^+\pi^-$ process can be well described by the coherent contributions of $\rho(2000)$ and $\rho(2150)$. Within these interpretations, $\rho(1900)$, $\rho(2000)$, and $\rho(2150)$ are assigned as the $\rho(3S)$, $\rho(2D)$, and $\rho(4S)$ states, respectively.

More recently, the BESIII Collaboration reported precise measurements of the cross sections for the process $e^+e^- \to \eta\pi^+\pi^-$ at center-of-mass energies ranging from 2.00 to 3.08 GeV~\cite{BESIII:2023sbq}. Through a partial wave analysis, a $\rho$-like structure, denoted as $Y(2044)$, was observed in the cross section distribution of the subprocess $e^+e^- \to a_2(1320)\pi$, with a fitted mass $M = 2044 \pm 21 \pm 4$ MeV and a fitted width $\Gamma = 163 \pm 69 \pm 24$ MeV. Furthermore, the product of the dielectron width and the branching ratio, $\Gamma_{e^+e^-} , BR_{a_2(1320)\pi}$, was determined to be $(34.6 \pm 17.1 \pm 6.0)$ eV or $(137.1 \pm 73.3 \pm 2.1)$ eV, depending on whether constructive or destructive interference was assumed in the cross section analysis.

The extracted resonance parameters of $Y(2044)$ are broadly consistent with the theoretical expectations for the $\rho(2D)$ state. Nevertheless, assigning $Y(2044)$ as the $\rho(2D)$ state is challenged by a notable discrepancy. In the unquenched spectroscopy framework, the theoretical predictions for the dielectron width $\Gamma_{e^+e^-}$ and the branching ratio $BR_{a_2(1320)\pi}$ of the pure $\rho(2D)$ state are about 20 eV and 3.6\%, respectively. Consequently, the resulting product $\Gamma_{e^+e^-}  BR_{a_2(1320)\pi}$ is nearly two orders of magnitude smaller than the corresponding experimental values reported for $Y(2044)$. This inconsistency disfavors an interpretation of $Y(2044)$ as a pure $\rho(2D)$ state.

\begin{figure}[htbp]
\centering
\includegraphics[width=0.6\textwidth]{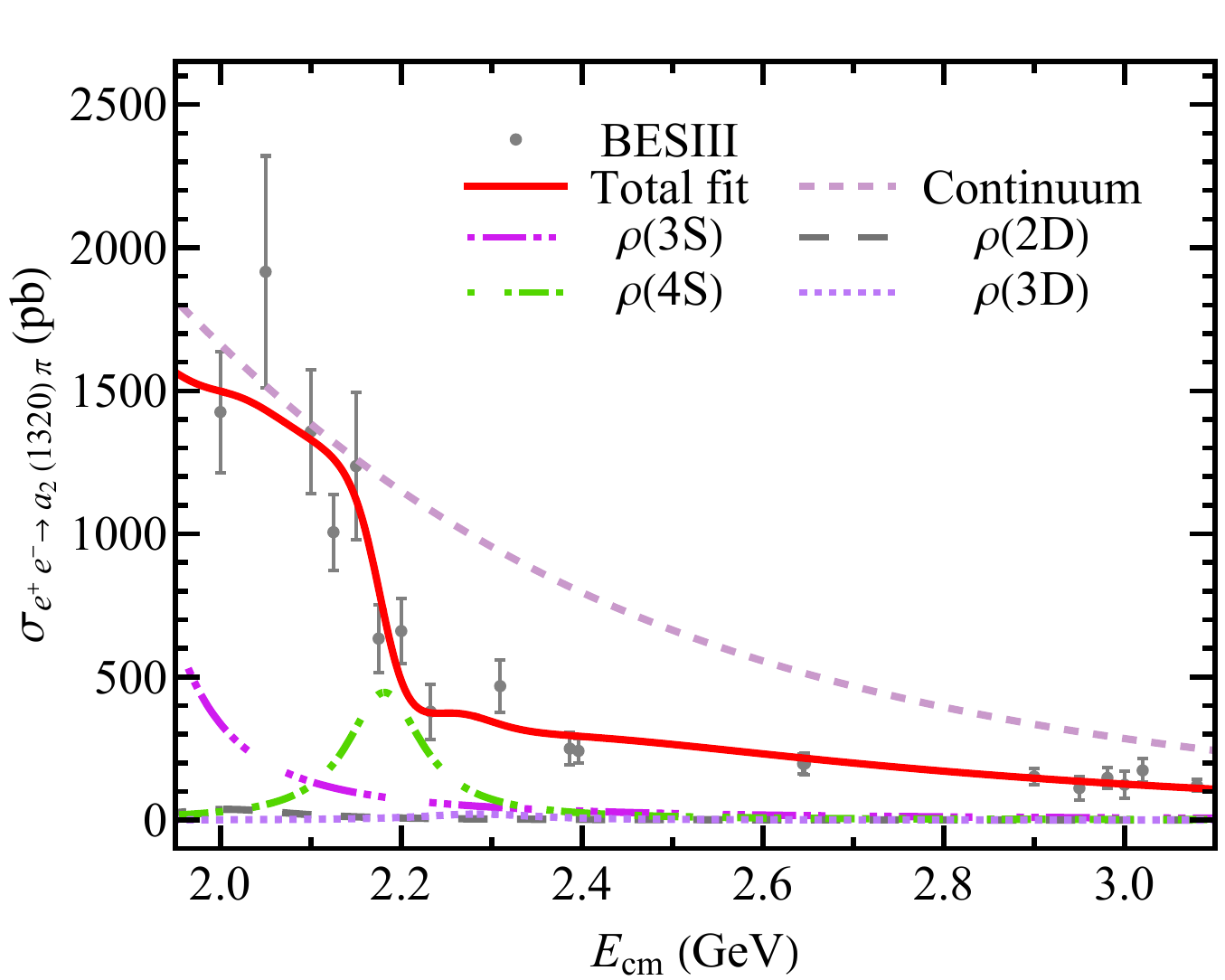}
\caption{Fit to the measured cross section of the process $e^+e^- \to a_2(1320)\pi$ including only the interference among the $\rho(3S)$, $\rho(2D)$, $\rho(4S)$, and $\rho(3D)$ states~\cite{BESIII:2023sbq}, where the resonance contributions are fixed by the unquenched spectroscopy framework. The figure is adapted from Ref.~\cite{Zhou:2025rxb}.}
\label{fig:00a}
\end{figure}

To understand the $Y(2044)$ structure observed in the measured cross section of $e^+e^- \to a_2(1320)\pi$, a first attempt was made in Ref.~\cite{Zhou:2025rxb} by considering the interference effects among the four high-lying $\rho$ meson states around 2 GeV. However, the fitting results for the $e^+e^- \to a_2(1320)\pi$ cross section indicate that the peak structure around 2044 MeV cannot be reproduced by taking into account only the interference among these states, as shown in Fig.~\ref{fig:00a}. This result suggests that the unsuccessful fit may be attributed to the limited contribution of the $\rho(2D)$ state to the $e^+e^- \to a_2(1320)\pi$ cross section.

It has been shown that the $S$-$D$ mixing effect plays a crucial role in understanding the spectroscopy of meson families in both the heavy- and light-flavor sectors, as discussed above. In Ref.~\cite{Zhou:2025rxb}, Zhou, Bai, Wang, Xu, and Liu proposed the $3S$-$2D$ and $4S$-$3D$ mixing schemes to reanalyze the cross sections of $e^+e^- \to a_2(1320)\pi$, as well as several additional high-lying $\rho$-related processes, including $e^+e^-\to\omega\pi^0$, $e^+e^-\to f_1(1285)\pi^+\pi^-$, $e^+e^-\to\pi^+\pi^-$, $e^+e^-\to\rho\eta$, and $e^+e^-\to\eta^\prime\pi^+\pi^-$.  

\begin{figure}[htbp]
  \centering
  \includegraphics[width=0.45\textwidth]{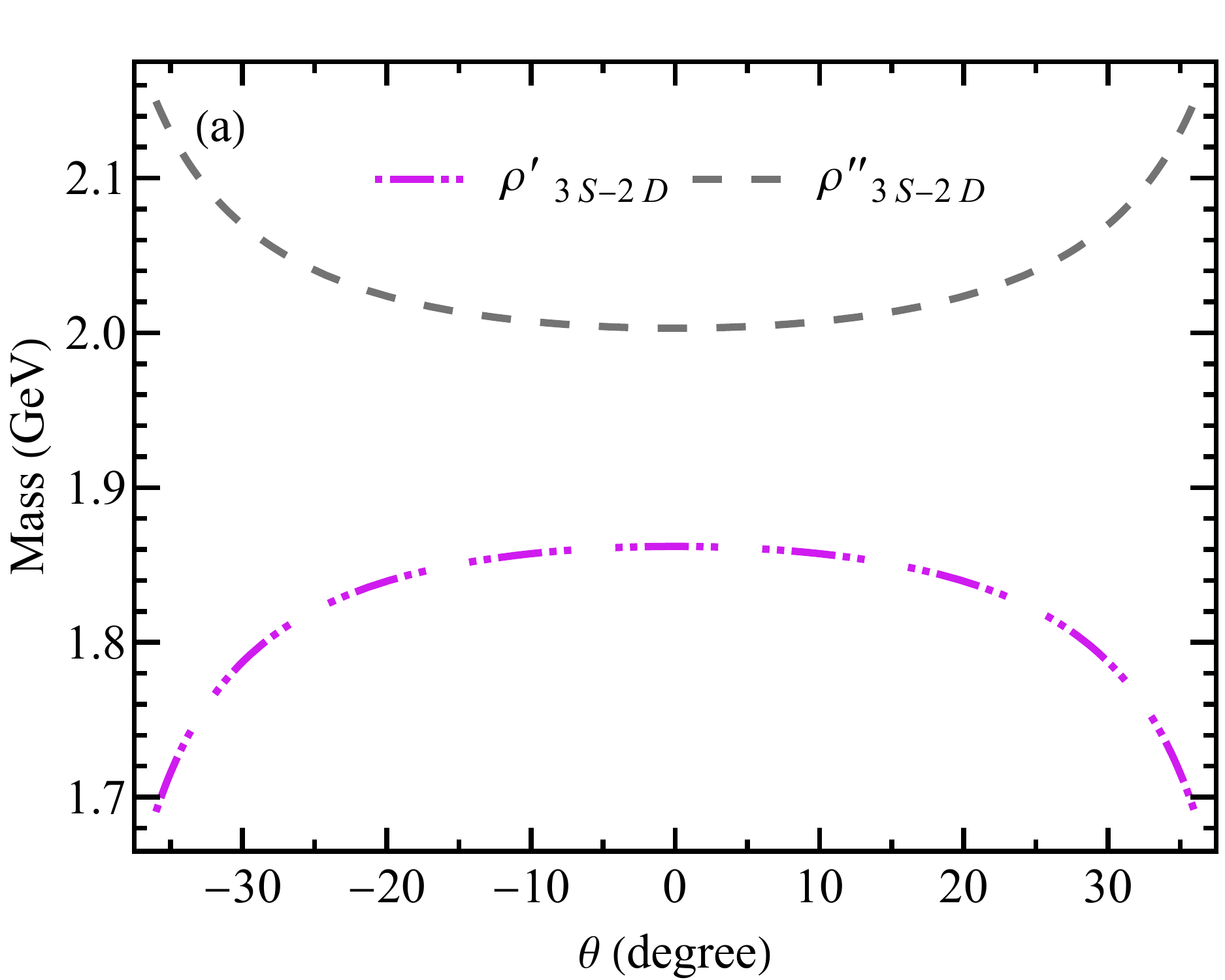}
  \hspace{0.04\textwidth}
  \includegraphics[width=0.45\textwidth]{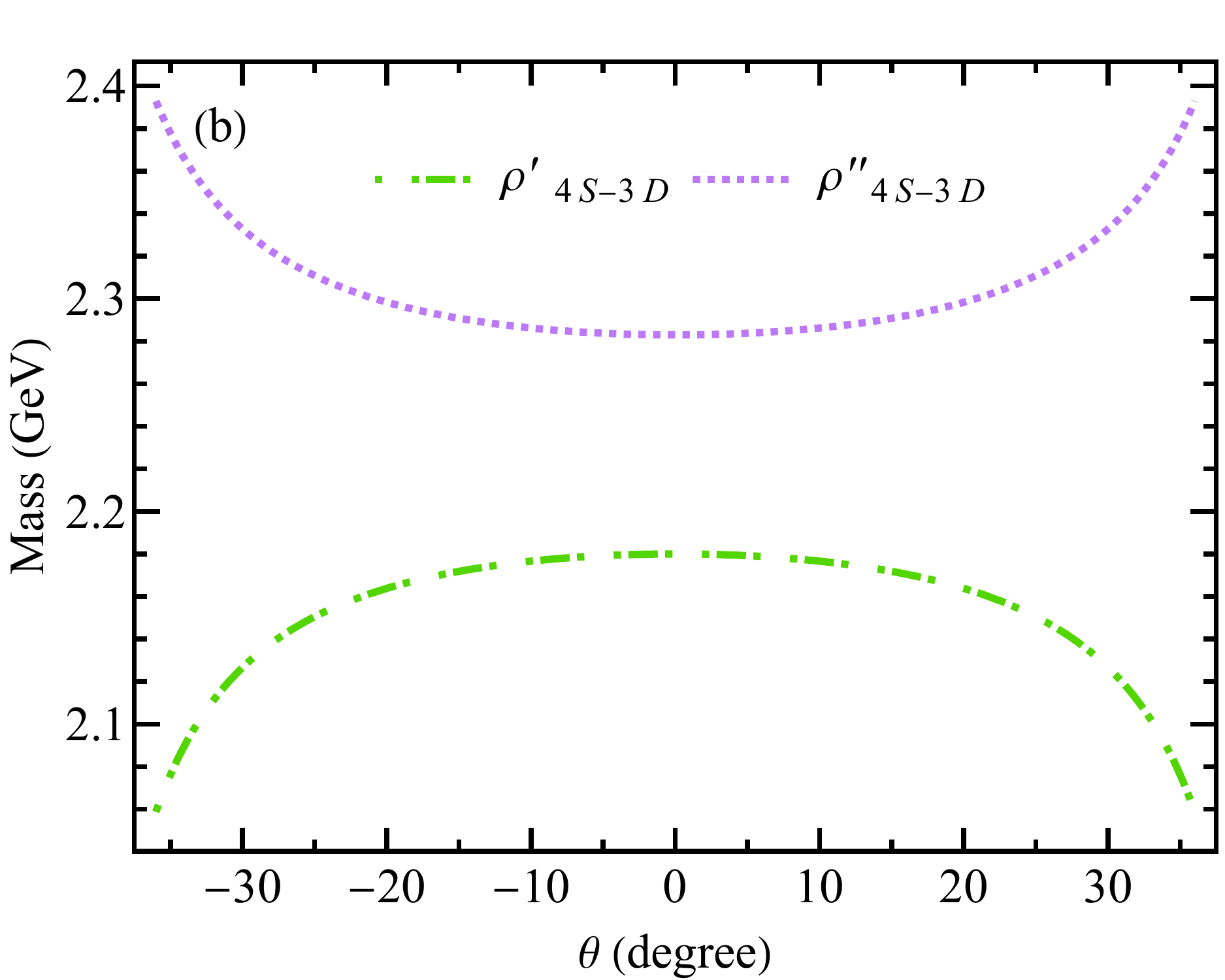}
  \caption{Masses of the mixed states as functions of the mixing angle $\theta$. Panel (a) shows the masses of the mixed states $\rho_{3S-2D}^{\prime}$ and $\rho_{3S-2D}^{\prime\prime}$, while panel (b) shows those of $\rho_{4S-3D}^{\prime}$ and $\rho_{4S-3D}^{\prime\prime}$. Adapted from Ref. \cite{Zhou:2025rxb}.}
  \label{fig:rhomass}
\end{figure}

\begin{figure}[htbp]
  \centering
  \includegraphics[width=0.245\textwidth]{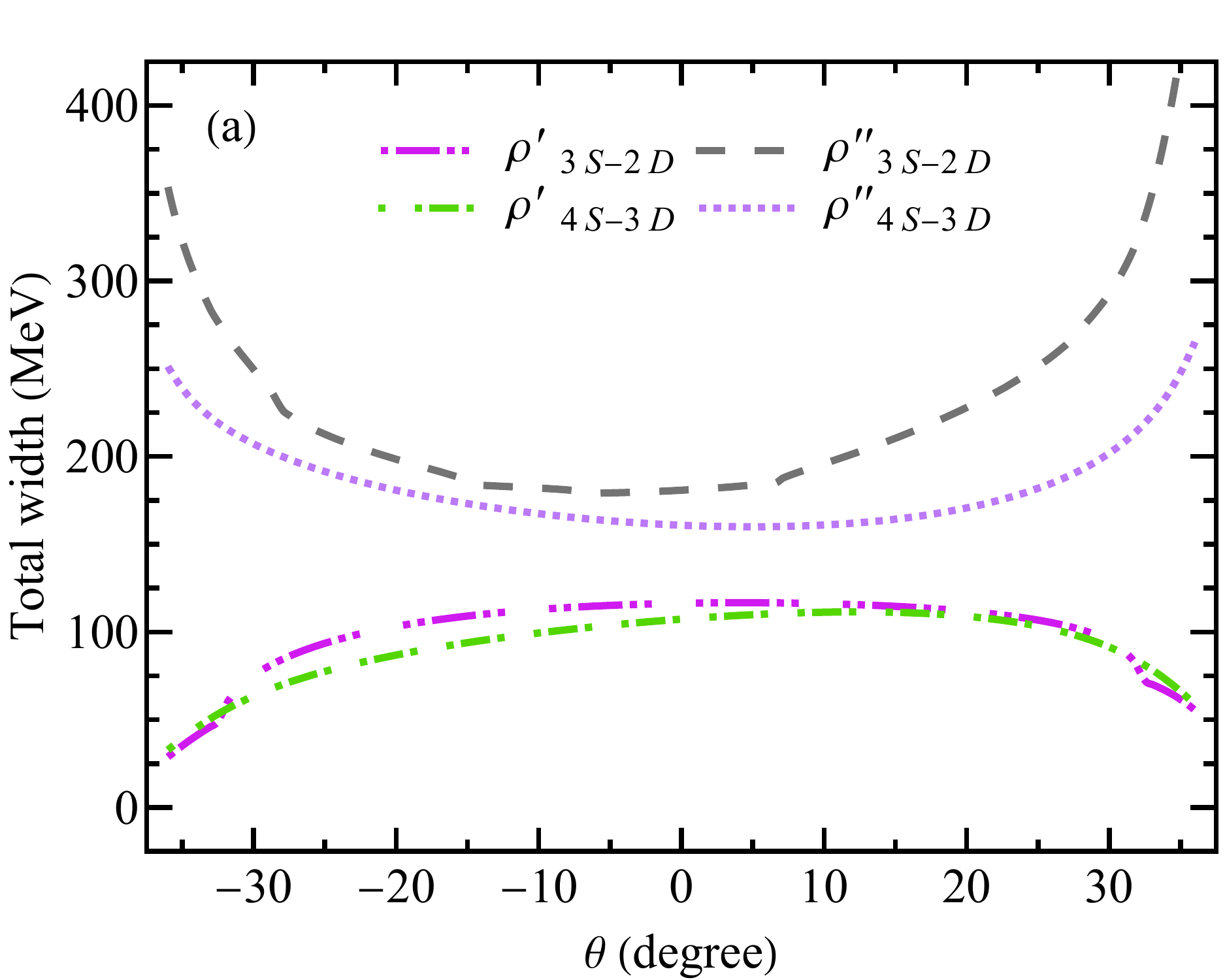}
  \includegraphics[width=0.245\textwidth]{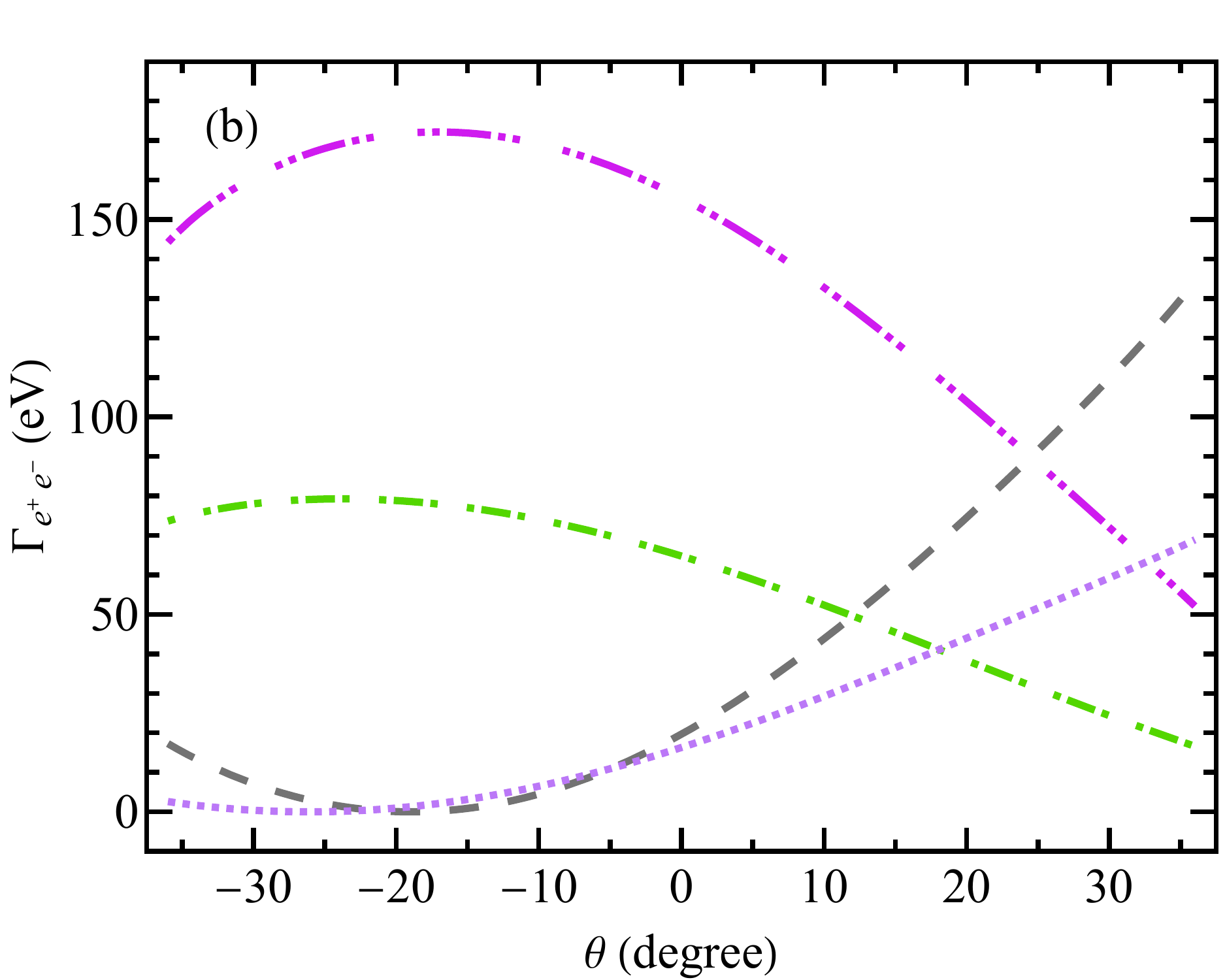}
  \includegraphics[width=0.245\textwidth]{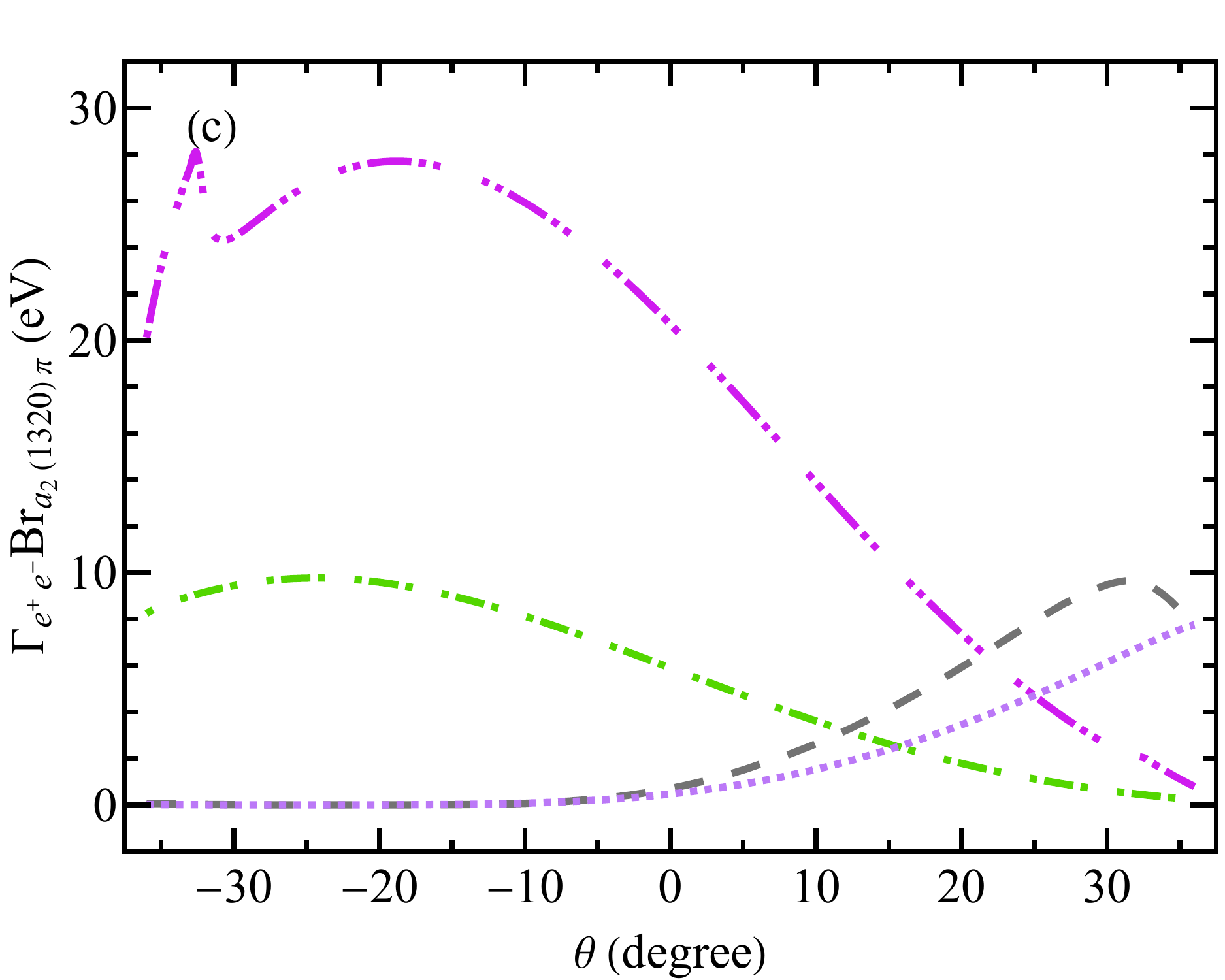}
  \includegraphics[width=0.245\textwidth]{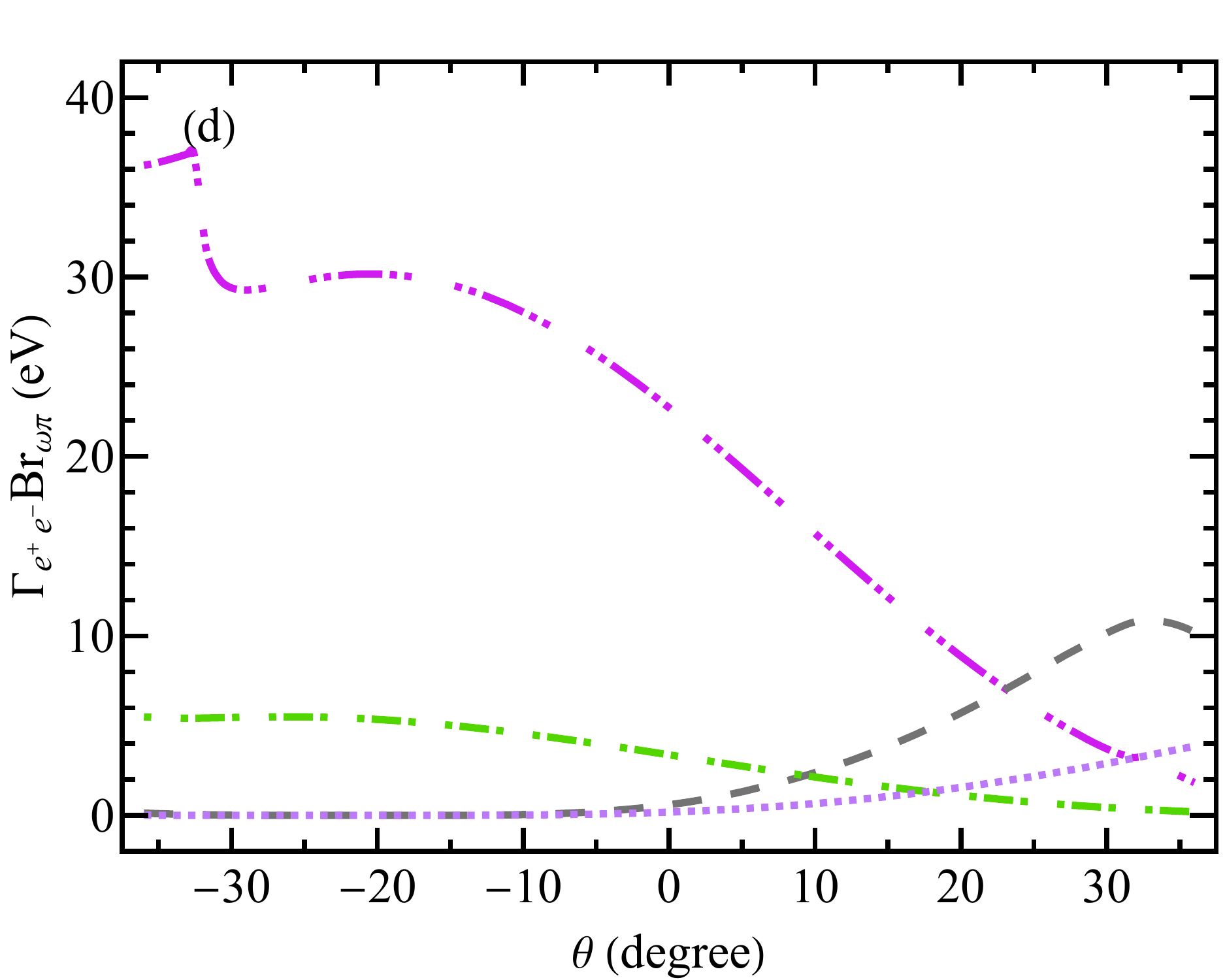}\\
  \includegraphics[width=0.245\textwidth]{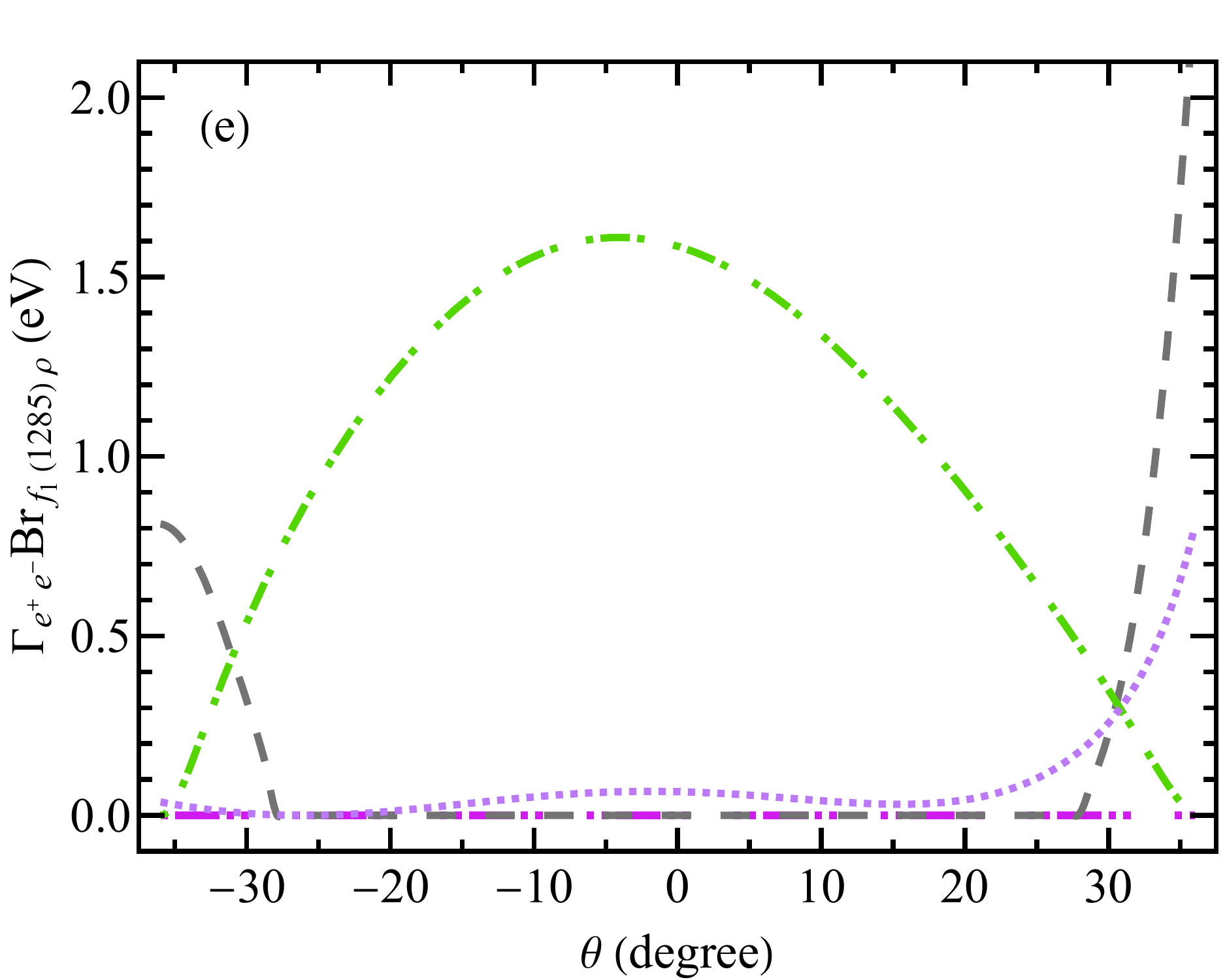}
  \includegraphics[width=0.245\textwidth]{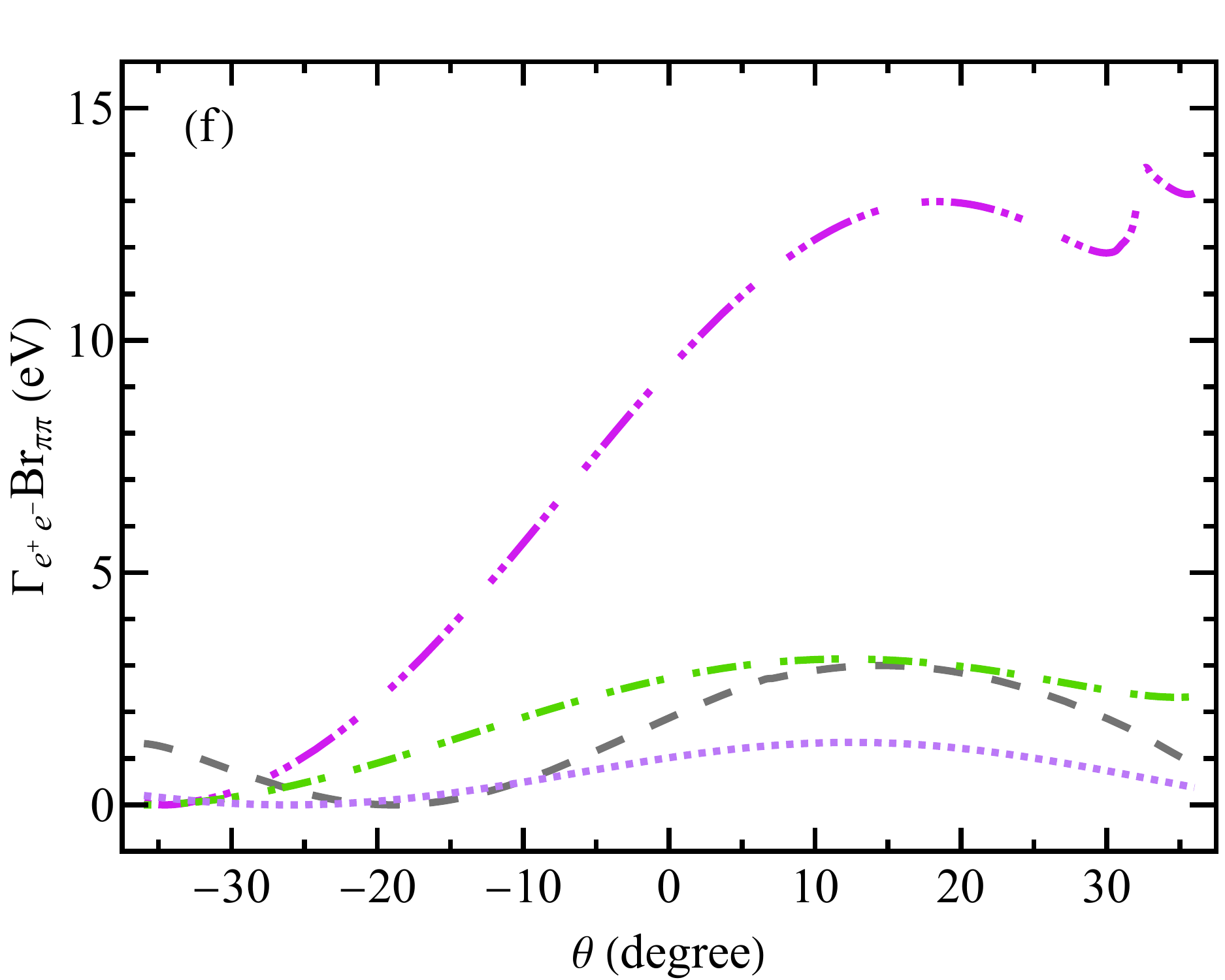}
  \includegraphics[width=0.245\textwidth]{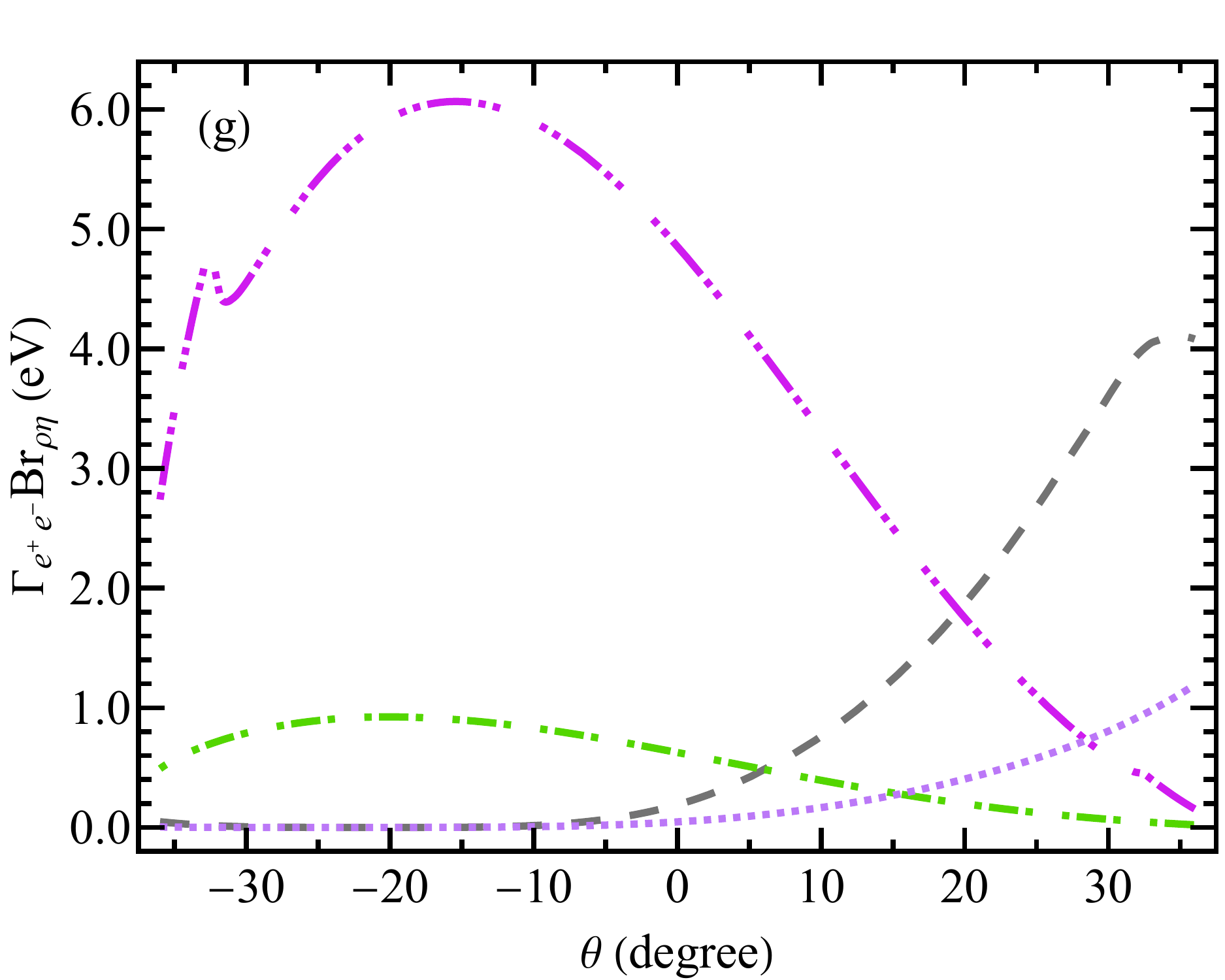}
  \includegraphics[width=0.245\textwidth]{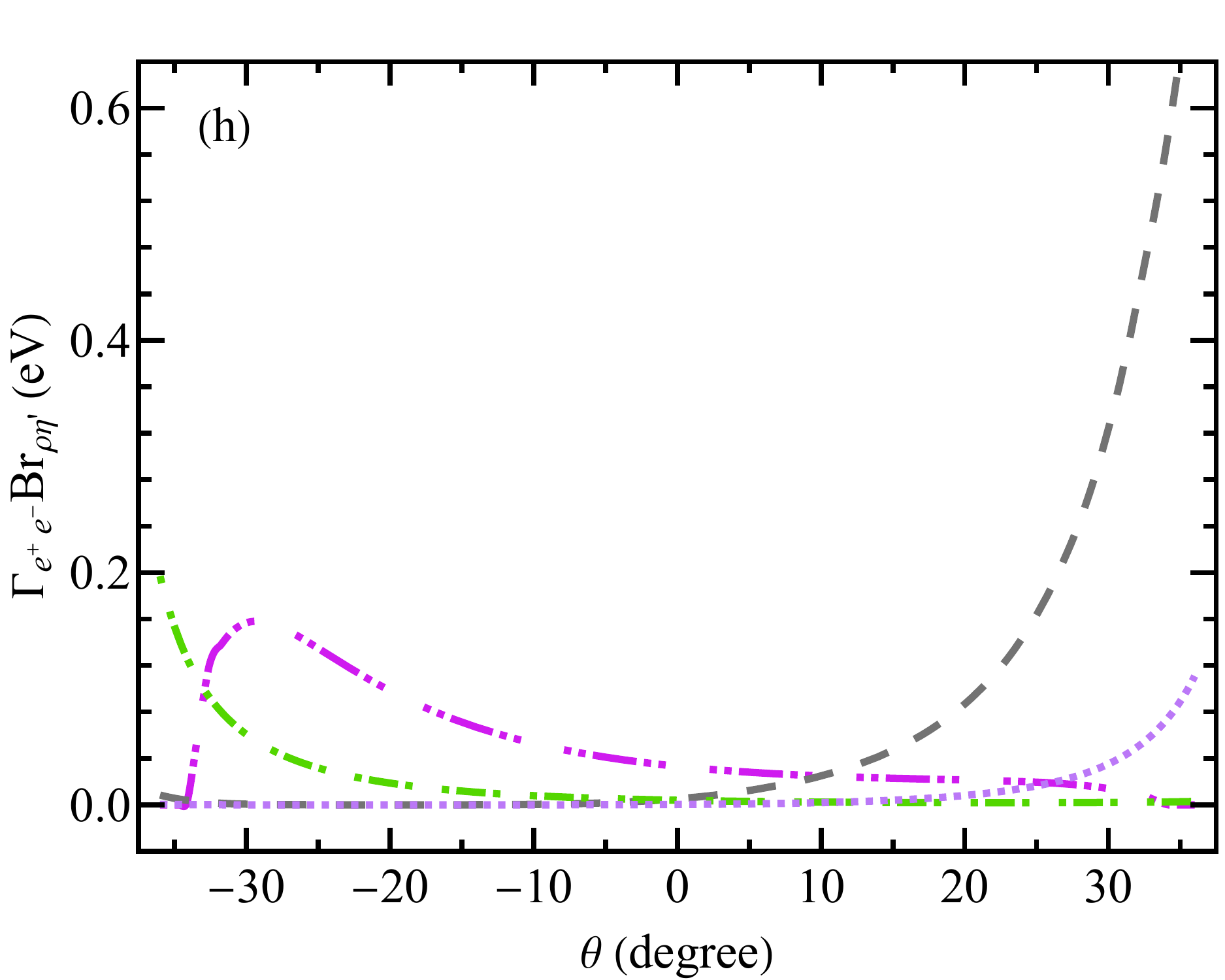}
  \caption{Decay behaviors of the mixed states $\rho_{3S-2D}^{\prime}$, $\rho_{3S-2D}^{\prime\prime}$, $\rho_{4S-3D}^{\prime}$, and $\rho_{4S-3D}^{\prime\prime}$ as functions of the mixing angle $\theta$. Panel (a) shows the total widths, panel (b) shows the dielectron widths, and panels (c)--(h) show the products of the dielectron width and branching ratios for $a_2(1320)\pi$ (panel (c)), $\omega\pi$ (panel (d)), $f_1(1285)\rho$ (panel (e)), $\pi\pi$ (panel (f)), $\rho\eta$ (panel (g)), and $\rho\eta^\prime$ (panel (h)). Figures adapted from Ref. \cite{Zhou:2025rxb}.}
  \label{fig:rhodecay}
\end{figure}

Within the $S$-$D$ mixing framework, the $\rho(nS)$-$\rho((n-1)D)$ mixing can be expressed as~\cite{Zhou:2025rxb}
\begin{eqnarray}
\begin{pmatrix}
|\rho_{nS-(n-1)D}^\prime\rangle \\
|\rho_{nS-(n-1)D}^{\prime\prime}\rangle
\end{pmatrix}=
\begin{pmatrix}
\cos\theta & \sin\theta\\
-\sin\theta & \cos\theta
\end{pmatrix}
\begin{pmatrix}
|\rho(nS)\rangle\\
|\rho((n-1)D)\rangle
\end{pmatrix},
\end{eqnarray}
where $\theta$ denotes the mixing angle between $\rho(nS)$ and $\rho((n-1)D)$. The masses and decay properties of the mixed states $\rho_{3S-2D}^{\prime}$, $\rho_{3S-2D}^{\prime\prime}$, $\rho_{4S-3D}^{\prime}$, and $\rho_{4S-3D}^{\prime\prime}$ as functions of the mixing angle $\theta$ are shown in Figs.~\ref{fig:rhomass} and \ref{fig:rhodecay}, respectively.

From Fig.~\ref{fig:rhodecay}, it is evident that the $S$-$D$ mixing effects have a significant impact on the decay behaviors of these high-lying $\rho$ states. The central idea for determining the mixing angles is to identify the ideal processes in which a specific intermediate $\rho$ state dominates over other $\rho$ states in the 2~GeV energy region, such that the corresponding interference effects can be neglected.

Following this strategy, the $3S$-$2D$ mixing angle $\theta_{3S-2D}$ is first constrained to be positive in order to enhance the contribution of $\rho_{3S-2D}^{\prime\prime}$ to the cross section of $e^+e^- \to a_2(1320)\pi$, as shown in Fig.~\ref{fig:rhodecay}(c). In this positive $\theta_{3S-2D}$ scenario, the contribution of $\rho_{3S-2D}^{\prime\prime}$ dominates the $Y(2034)$ structure observed in the $e^+e^-\to \omega\pi^0$ process~\cite{BESIII:2020xmw}, while the contribution from the adjacent $\rho_{4S-3D}^{\prime}$ state is relatively small, as shown in Fig.~\ref{fig:rhodecay}(d). Moreover, the masses of the other two mixed states are considerably distant from 2034~MeV, so their interference effects are expected to be negligible. Therefore, the $Y(2034)$ structure can be regarded as a good candidate for $\rho_{3S-2D}^{\prime\prime}$, leading to a determination of the mixing angle $\theta_{3S-2D}=23.4^\circ$. The theoretical width of $\rho_{3S\text{-}2D}^{\prime\prime}$ at this mixing angle is also consistent with the measured width of $Y(2034)$.

The process $e^+e^-\to f_1(1285)\rho$ then provides an ideal channel to determine the $4S$-$3D$ mixing angle. The study by the BaBar Collaboration on $e^+e^-\to f_1(1285)\pi^+\pi^-$~\cite{BaBar:2022ahi} shows that the $\pi^+\pi^-$ system is dominantly produced via $\rho(770)$. Consequently, when $\rho_{3S-2D}^{\prime}$ and $\rho_{3S-2D}^{\prime\prime}$ serve as intermediate states, their contributions are suppressed because the $\rho(770)$ is off shell. In addition, the $BABAR$ data reveal an enhancement only around 2150~MeV, with no evidence of additional structures at higher energies, and the mixing angle $\theta_{4S-3D}$ is unlikely to be very large. This further indicates that the contribution of $\rho_{4S-3D}^{\prime\prime}$ should be not significant in this channel. Hence, the $\rho(2150)$ structure can be safely assigned to $\rho_{4S-3D}^{\prime}$, yielding $\theta_{4S-3D}=\pm 25.1^\circ$. 

\begin{figure}[htbp]
  \centering
  \includegraphics[width=0.32\textwidth]{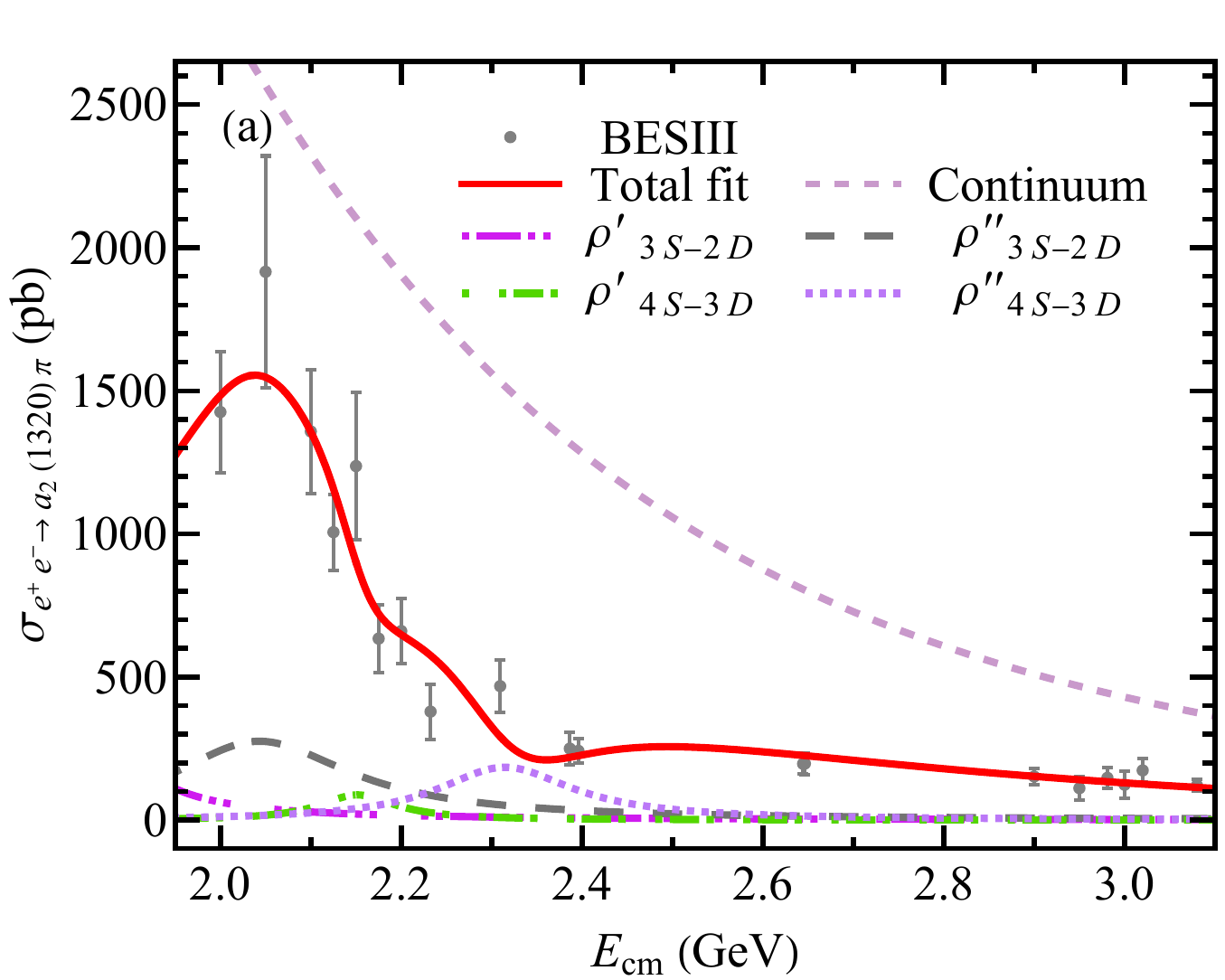}
  \includegraphics[width=0.32\textwidth]{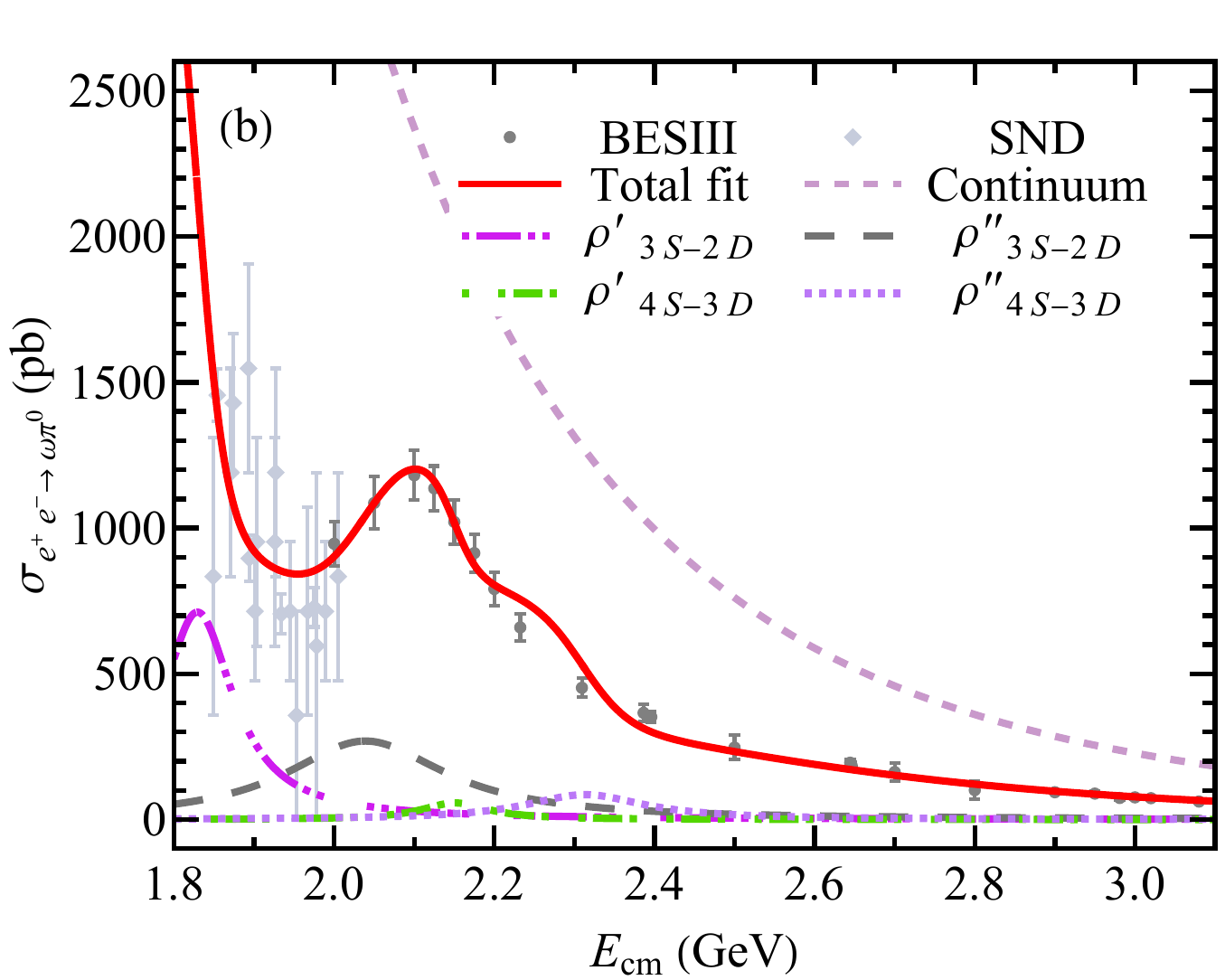}
  \includegraphics[width=0.32\textwidth]{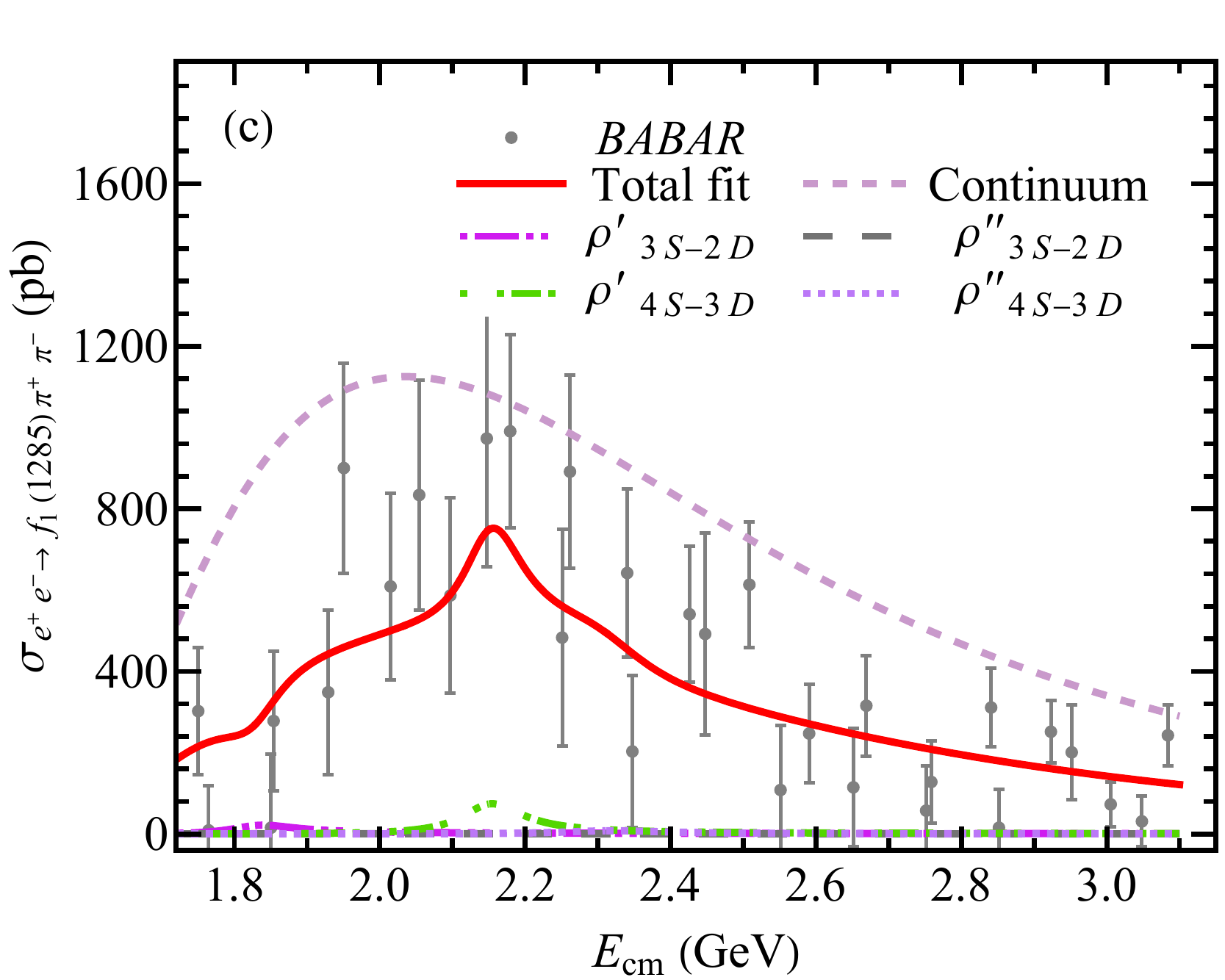}\\
  \includegraphics[width=0.32\textwidth]{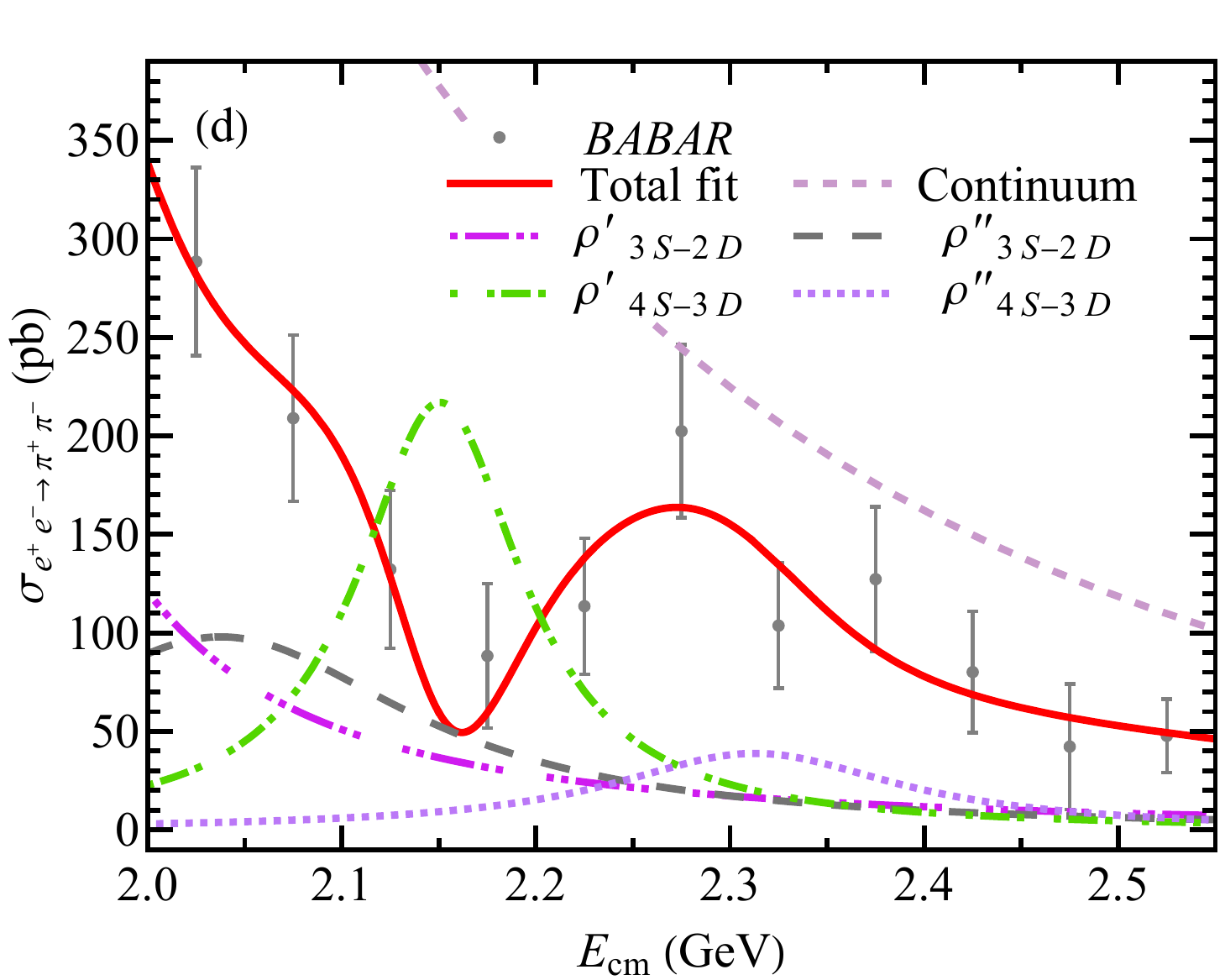}
  \includegraphics[width=0.32\textwidth]{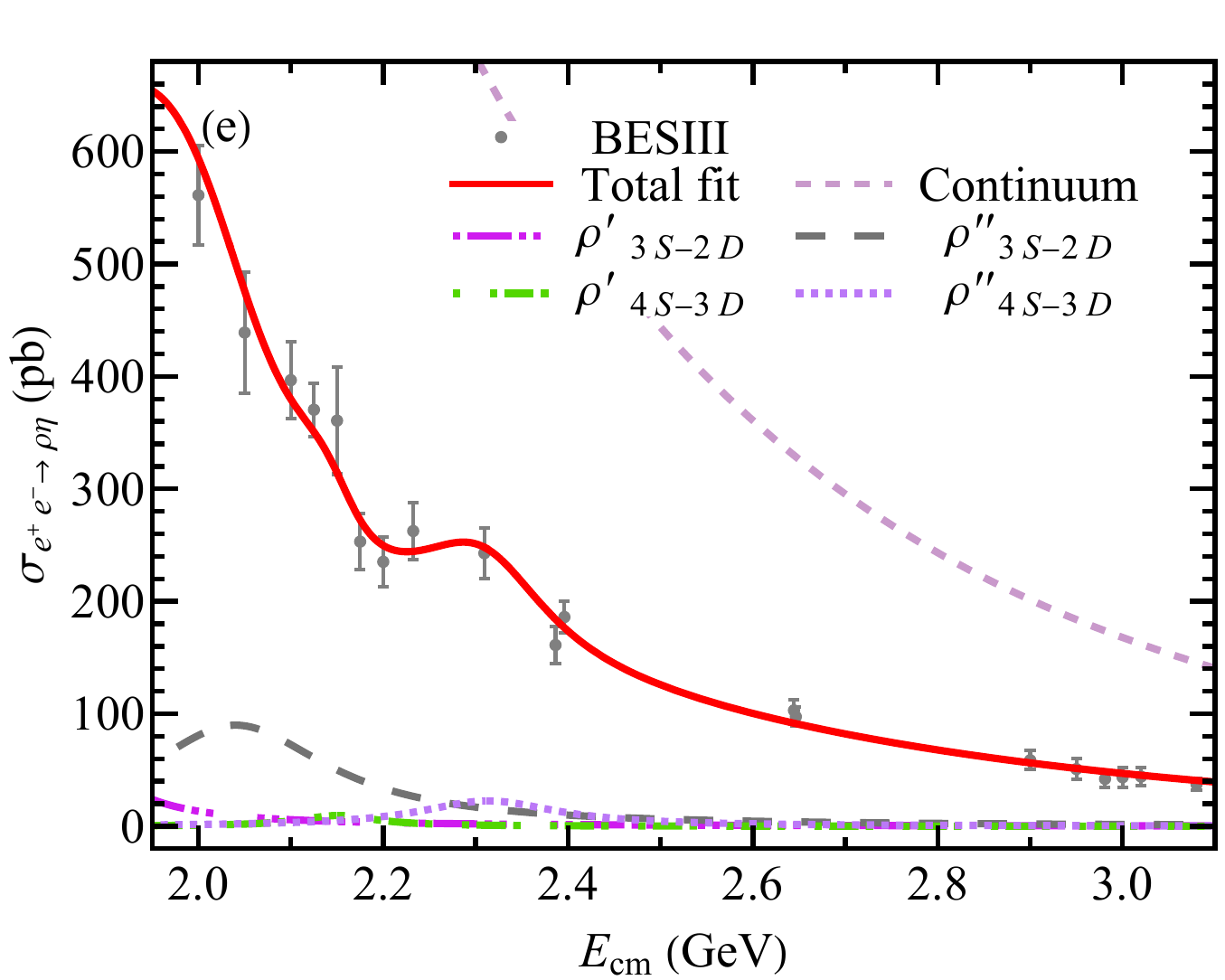}
  \includegraphics[width=0.32\textwidth]{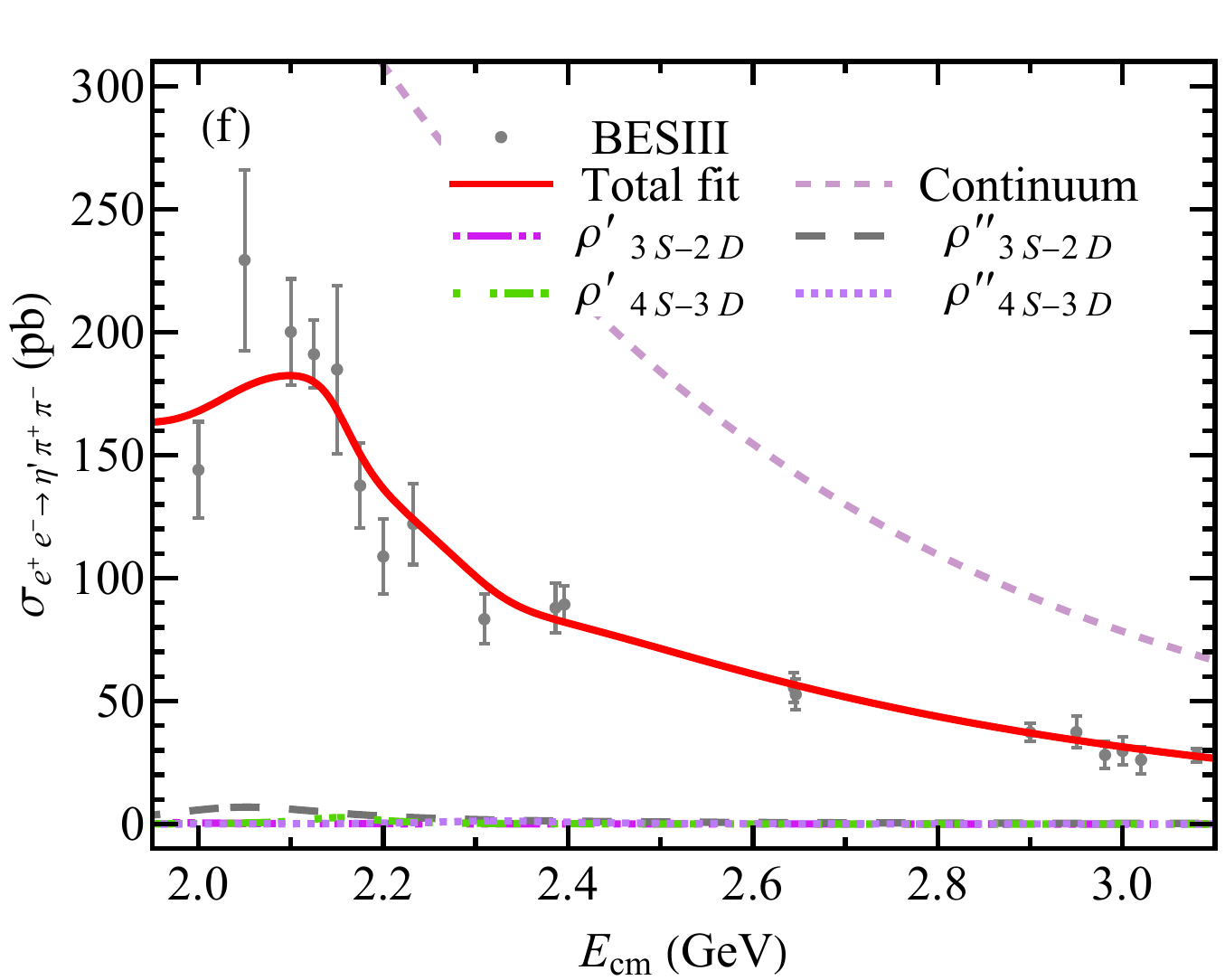}
  \caption{Combined fit to the measured cross sections for the processes $e^+e^-\to a_2(1320)\pi$ \cite{BESIII:2023sbq} (panel (a)), $e^+e^-\to\omega\pi^0$ \cite{BESIII:2020xmw,Achasov:2016zvn} (panel (b)), $e^+e^-\to f_1(1285)\pi^+\pi^-$ \cite{BaBar:2007qju,BaBar:2022ahi} (panel (c)), $e^+e^-\to\pi^+\pi^-$ \cite{BaBar:2007qju,BaBar:2022ahi} (panel (d)), $e^+e^-\to\rho\eta$ \cite{BESIII:2023sbq} (panel (e)), and $e^+e^-\to\eta^\prime\pi^+\pi^-$ \cite{BESIII:2020kpr} (panel (f)). The figures are adapted from Ref.~\cite{Zhou:2025rxb}.}
  \label{fig:combinefit}
\end{figure}

Based on the proposed mixing scheme, a combined fit to the cross section data of the processes
$e^+e^-\to a_2(1320)\pi$, $e^+e^-\to\omega\pi^0$, $e^+e^-\to f_1(1285)\pi^+\pi^-$, $e^+e^-\to\pi^+\pi^-$, $e^+e^-\to\rho\eta$, and $e^+e^-\to\eta^\prime\pi^+\pi^-$ is performed, where four mixed states around 2~GeV, $\rho_{3S\text{-}2D}^{\prime}$, $\rho_{3S\text{-}2D}^{\prime\prime}$, $\rho_{4S\text{-}3D}^{\prime}$, and $\rho_{4S\text{-}3D}^{\prime\prime}$, are taken as intermediate resonances. The contributions of these resonances to the corresponding cross sections are fixed by the unquenched spectroscopy framework. Two initial schemes, $\theta_{4S\text{-}3D}=-25.1^\circ$ and $\theta_{4S\text{-}3D}=25.1^\circ$, are adopted in the fit. It is found that the negative-$\theta_{4S\text{-}3D}$ scheme fails to reproduce the measured cross section data, whereas the positive-$\theta_{4S\text{-}3D}$ scheme provides a good description of all six processes, as shown in Fig.~\ref{fig:combinefit}. These results indicate that the $\rho(4S)$–$\rho(3D)$ mixing angle favors a positive value, $\theta_{4S\text{-}3D}=25.1^\circ$. In this framework, the $Y(2044)$ structure observed in $e^+e^-\to a_2(1320)\pi$ \cite{BESIII:2023sbq} is reproduced by the four $S$–$D$ mixed $\rho$ states, with the dominant contribution arising from $\rho_{3S\text{-}2D}^{\prime\prime}$ and a non-negligible interference effect among the resonances.

\begin{table}[htbp]
\centering
\caption{Masses, total widths, and dielectron widths of the mixed states $\rho_{3S\text{-}2D}^{\prime}$, $\rho_{3S\text{-}2D}^{\prime\prime}$, $\rho_{4S\text{-}3D}^{\prime}$, and $\rho_{4S\text{-}3D}^{\prime\prime}$ in Ref.~\cite{Zhou:2025rxb}, evaluated with the mixing angles $\theta_{3S\text{-}2D}=23.4^\circ$ and $\theta_{4S\text{-}3D}=25.1^\circ$.}	\label{MGIrho}
		\renewcommand\arraystretch{1.3}
  \begin{tabular*}{1.0\textwidth}{l@{\extracolsep{\fill}}cccc}
			 \hline
          \hline
			$\rho$ states &$\rho^{\prime}_{3S-2D}$ &$\rho^{\prime\prime}_{3S-2D}$ &$\rho^{\prime}_{4S-3D}$   & $\rho^{\prime\prime}_{4S-3D}$\\
			\hline
			Mass (MeV)  & 1828      & 2034      & 2150  &2311\\
                Width (MeV)    & 109       & 243       & 103       & 182\\
                $\Gamma_{e^+e^-}$ (eV) & 93.4     & 85.9    & 31.0     & 51.8\\
			\hline
            \hline
		\end{tabular*}
	\end{table}	

Overall, by introducing four theoretically constructed mixed states, $\rho_{3S\text{-}2D}^{\prime}$, $\rho_{3S\text{-}2D}^{\prime\prime}$, $\rho_{4S\text{-}3D}^{\prime}$, and $\rho_{4S\text{-}3D}^{\prime\prime}$ (listed in Table~\ref{MGIrho}), the measured cross section data for the high-lying $\rho$-related processes can be well described within a unified theoretical framework.

Collectively, the puzzles of the apparent inconsistencies among the reported resonance parameters in various final states—spanning the $\rho$, $\omega$, and $\phi$ sectors around 2 GeV—are coherently solved within a unified framework that combines the same unquenched spectroscopy with the interference effects. This series of concrete applications validates the proposed strategy for constructing the high-lying spectra via spectroscopy-guided cross section analysis, demonstrating its viability for studying the light-flavor vector meson family and its promising potential for extension to other light hadron sectors with different quantum numbers.

\section{Summary}

Driven by sustained advances in experimental techniques and the continual accumulation of data, hadron spectroscopy has undergone rapid development over the past two decades. This progress is evident from the multitude of high-quality reviews and related literature. It has not only fueled the extensive search for exotic hadrons—a cornerstone in building the "Particle Zoo 2.0"—but also established itself as a major direction in contemporary hadron spectroscopy research.

Confronted with a wealth of recent experimental discoveries, it is instructive to recall pivotal moments in the history of particle physics. In the 1950s and 1960s, the discovery of light hadrons spurred the study of hadron classification, leading to the establishment of SU(3) flavor symmetry. In the 1970s and 1980s, the subsequent discoveries of charmonia gave rise to the Cornell potential, a model describing quark interactions. This model not only successfully explained the charmonium spectrum at the time but also laid the quantitative foundation for hadron spectroscopy, becoming the progenitor of many subsequent spectroscopic models. Following this historical logic, the continuous emergence of high-statistics data over the last twenty years—primarily represented by charmonium-like $XYZ$ states—gives us strong reason to anticipate new theoretical breakthroughs, just as history has repeatedly demonstrated. We find ourselves both witnesses and participants in this ongoing process.

While the search for exotic hadrons commands significant attention, an equally important parallel aspect must be emphasized: these high-precision data provide stringent tests for the reliability and applicability of various phenomenological models. This presents a rare opportunity. Due to the non-perturbative nature of the strong interaction, phenomenological models remain indispensable tools, on par with experiment and lattice QCD, for understanding non-perturbative strong dynamics. These three pillars can be likened to the three interlocked rings of a Borromean ring: intimately connected and mutually supportive, together they form a robust framework for hadron physics research.

The solution to the low-mass puzzles associated with new hadrons such as \(X(3872)\), \(D_{s0}^*(2317)\), \(D_{s1}^\prime(2460)\), and \(\Lambda_c(2940)\) has highlighted the critical importance of "unquenched effects" in hadron spectroscopy. Before the year 2000, the construction of the hadron spectrum focused predominantly on low-lying states. The quenched model achieved remarkable success in describing the spectroscopy of these states, primarily because most low-lying states lie below the thresholds of their allowed hadronic channels, thus experiencing negligible coupled-channel effects. Consequently, the physical states largely coincide with the bare states.

However, since 2003, a large number of newly discovered hadrons are closely related to high-lying states. Within the framework of coupled-channel dynamics, these states often exhibit strong couplings to open hadronic channels. This leads to significant modifications of the physical states, resulting in various novel phenomena. These phenomena appear "anomalous" precisely because they deviate from the predictions and calculations of the quenched model. This context makes the two statements quoted in Fig. 1 self-evident:

\begin{quote}
``The discovered charmonia prompted the development of the quenched potential model.''\\
``The quenched potential model has become outdated in light of the discovery of new hadronic states.''
\end{quote}

Therefore, we are experiencing a "{\it paradigm shift from the quenched to the unquenched picture}". \changelabel{ This viewpoint is also in line with the perspective advocated in Ref.~\cite{Du:2017zvv}, which argues that the continuum dynamics playing an essential role in shaping the observed hadron spectrum.}

This review adopts this very perspective to examine hadron spectroscopy. Through a series of concrete examples, it systematically demonstrates how unquenched effects manifest in spectroscopic studies. These effects are not only prominently displayed in (charmonium-like) charmonium spectroscopy but are also universally relevant to bottomonium and light hadron spectroscopy. This universality underscores that unquenched effects are not isolated curiosities but a general rule. Such pervasiveness typically indicates that we are touching upon the essence of the phenomena—an insight that cannot be overlooked.

The research trajectory outlined here spans the past two decades. We take satisfaction in using this review to illustrate the enduring vitality of the unquenched picture in hadron spectroscopy. Our sentiment at this juncture is aptly captured by the ancient Chinese poetic line: \textbf{``The falling leaf reveals the coming of autumn (\begin{CJK}{UTF8}{gbsn}一叶落而知天下秋\end{CJK})''}—perceiving the overarching shift from a subtle sign.

Standing at the vantage point of 2026, this review has looked back on the past, but we must also look forward. A natural question arises: \textbf{What is the future direction of hadron spectroscopy?}

Our answer is clear: hadron spectroscopy is entering a "high-precision era". This is reflected not only in the continually improving experimental precision at facilities like BESIII, LHCb, and Belle II but also inevitably in the enhanced precision of theoretical studies. Now that the crucial role of unquenched effects is recognized, we must take this as our starting point to usher in a new phase of high-precision hadron spectroscopy. This entails developing more sophisticated unquenched models to deepen our understanding of non-perturbative strong interaction dynamics. This is not only our responsibility but also the shared mission of the new generation of particle physicists.

\section*{Acknowledgements}

We would like to thank Ri-Qing Qian, Zi-Long Man, Qin-Song Zhou, Li-Ming Wang, Tian-Cai Peng, and Xue-Qian Li for their collaboration.
This review is also dedicated to the memory of two esteemed mentors.
One of the authors, X.L., wishes to honor Professor Eef van Beveren, under whose guidance X.L. conducted postdoctoral research at the University of Coimbra with funding from FCT. Professor van Beveren was among the first internationally to recognize the importance of unquenched effects in hadron spectroscopy, as exemplified by his pioneering explanation for the low-mass puzzle of the $D_{s0}^*(2317)$. Beyond his scientific insight, he instilled in me a deep respect for experimental data and an enduring optimism in facing challenges.
X.L. also dedicates this work to Professor Takayuki Matsuki. Many of the results presented here are the fruit of our collaboration that spanned over a decade.
This work is also supported by the National Natural Science Foundation of China under Grant Nos. 12335001, 12305087, 12247101, 12547101, 12405088, and 12405098, the ‘111 Center’ under Grant No. B20063, National Key Research and Development Program under Grant No.2024YFA1610503, the Natural Science Foundation of Gansu Province (No. 26RCKA012, No. 25JRRA799), the fundamental Research Funds for the Central Universities (No. lzujbky-2023-stlt01), the project for top-notch innovative talents of Gansu province, Lanzhou City High-Level Talent Funding, the Postdoctoral Fellowship Program of CPSF under Grant Number GZB20260801, and the Talent Scientific Fund of Lanzhou University.


\appendix






\bibliographystyle{elsarticle-num-collab}
\bibliography{ref}

\begin{thebibliography}{1000}
\expandafter\ifx\csname url\endcsname\relax
  \def\url#1{\texttt{#1}}\fi
\expandafter\ifx\csname urlprefix\endcsname\relax\def\urlprefix{URL }\fi
\expandafter\ifx\csname href\endcsname\relax
  \def\href#1#2{#2} \def\path#1{#1}\fi

\bibitem{SLAC:2024-11-20}
{SLAC National Accelerator Laboratory},
  \href{https://www6.slac.stanford.edu/news/2024-11-20-slac-celebrates-50-years-nobel-winning-discovery-particle-physics}{{SLAC}
  celebrates 50 years of a nobel-winning discovery in particle physics}
  (Nov.~20 2024).
\newline\urlprefix\url{https://www6.slac.stanford.edu/news/2024-11-20-slac-celebrates-50-years-nobel-winning-discovery-particle-physics}

\bibitem{CDS:2935663}
IHEP--CERN Joint Symposium, \href{https://cds.cern.ch/record/2935663}{50 years
  of the $J/\psi$ discovery: from the november revolution to the precision
  frontier}.
\newline\urlprefix\url{https://cds.cern.ch/record/2935663}

\bibitem{E598:1974sol}
J.~J. Aubert, et~al., E598 Collaboration, {Experimental Observation of a Heavy
  Particle $J$}, Phys. Rev. Lett. 33 (1974) 1404--1406.
\newblock \href {https://doi.org/10.1103/PhysRevLett.33.1404}
  {\path{doi:10.1103/PhysRevLett.33.1404}}.

\bibitem{SLAC-SP-017:1974ind}
J.~E. Augustin, et~al., SLAC-SP-017 Collaboration, {Discovery of a Narrow
  Resonance in $e^+ e^-$ Annihilation}, Phys. Rev. Lett. 33 (1974) 1406--1408.
\newblock \href {https://doi.org/10.1103/PhysRevLett.33.1406}
  {\path{doi:10.1103/PhysRevLett.33.1406}}.

\bibitem{Glashow:1970gm}
S.~L. Glashow, J.~Iliopoulos, L.~Maiani, {Weak interactions with lepton-hadron
  symmetry}, Phys. Rev. D 2 (1970) 1285--1292.
\newblock \href {https://doi.org/10.1103/PhysRevD.2.1285}
  {\path{doi:10.1103/PhysRevD.2.1285}}.

\bibitem{Gell-Mann:1961omu}
M.~Gell-Mann, {The eightfold way: A theory of strong interaction symmetry} (3
  1961).
\newblock \href {https://doi.org/10.2172/4008239} {\path{doi:10.2172/4008239}}.

\bibitem{Neeman:1961jhl}
Y.~Ne'eman, {Derivation of strong interactions from a gauge invariance}, Nucl.
  Phys. 26 (1961) 222--229.
\newblock \href {https://doi.org/10.1016/0029-5582(61)90134-1}
  {\path{doi:10.1016/0029-5582(61)90134-1}}.

\bibitem{Gell-Mann:1964ewy}
M.~Gell-Mann, {A schematic model of baryons and mesons}, Phys. Lett. 8 (1964)
  214--215.
\newblock \href {https://doi.org/10.1016/S0031-9163(64)92001-3}
  {\path{doi:10.1016/S0031-9163(64)92001-3}}.

\bibitem{Zweig:1964ruk}
G.~Zweig, {An SU(3) model for strong interaction symmetry and its breaking.
  Version 1} (1 1964).
\newblock \href {https://doi.org/10.17181/CERN-TH-401}
  {\path{doi:10.17181/CERN-TH-401}}.

\bibitem{Eichten:1974af}
E.~Eichten, K.~Gottfried, T.~Kinoshita, J.~B. Kogut, K.~D. Lane, T.-M. Yan,
  {The spectrum of charmonium}, Phys. Rev. Lett. 34 (1975) 369--372, [Erratum:
  Phys.Rev.Lett. 36, 1276 (1976)].
\newblock \href {https://doi.org/10.1103/PhysRevLett.34.369}
  {\path{doi:10.1103/PhysRevLett.34.369}}.

\bibitem{Eichten:1978tg}
E.~Eichten, K.~Gottfried, T.~Kinoshita, K.~D. Lane, T.-M. Yan, {Charmonium: The
  model}, Phys. Rev. D 17 (1978) 3090, [Erratum: Phys.Rev.D 21, 313 (1980)].
\newblock \href {https://doi.org/10.1103/PhysRevD.17.3090}
  {\path{doi:10.1103/PhysRevD.17.3090}}.

\bibitem{Eichten:1979ms}
E.~Eichten, K.~Gottfried, T.~Kinoshita, K.~D. Lane, T.-M. Yan, {Charmonium:
  Comparison with experiment}, Phys. Rev. D 21 (1980) 203.
\newblock \href {https://doi.org/10.1103/PhysRevD.21.203}
  {\path{doi:10.1103/PhysRevD.21.203}}.

\bibitem{Godfrey:1985xj}
S.~Godfrey, N.~Isgur, {Mesons in a relativized quark model with
  chromodynamics}, Phys. Rev. D 32 (1985) 189--231.
\newblock \href {https://doi.org/10.1103/PhysRevD.32.189}
  {\path{doi:10.1103/PhysRevD.32.189}}.

\bibitem{Klempt:2007cp}
E.~Klempt, A.~Zaitsev, {Glueballs, Hybrids, Multiquarks. Experimental facts
  versus QCD inspired concepts}, Phys. Rept. 454 (2007) 1--202.
\newblock \href {http://arxiv.org/abs/0708.4016} {\path{arXiv:0708.4016}},
  \href {https://doi.org/10.1016/j.physrep.2007.07.006}
  {\path{doi:10.1016/j.physrep.2007.07.006}}.

\bibitem{Brambilla:2010cs}
N.~Brambilla, et~al., {Heavy quarkonium: Progress, puzzles, and opportunities},
  Eur. Phys. J. C 71 (2011) 1534.
\newblock \href {http://arxiv.org/abs/1010.5827} {\path{arXiv:1010.5827}},
  \href {https://doi.org/10.1140/epjc/s10052-010-1534-9}
  {\path{doi:10.1140/epjc/s10052-010-1534-9}}.

\bibitem{Liu:2013waa}
X.~Liu, {An overview of $XYZ$ new particles}, Chin. Sci. Bull. 59 (2014)
  3815--3830.
\newblock \href {http://arxiv.org/abs/1312.7408} {\path{arXiv:1312.7408}},
  \href {https://doi.org/10.1007/s11434-014-0407-2}
  {\path{doi:10.1007/s11434-014-0407-2}}.

\bibitem{Hosaka:2016pey}
A.~Hosaka, T.~Iijima, K.~Miyabayashi, Y.~Sakai, S.~Yasui, {Exotic hadrons with
  heavy flavors: $X$, $Y$, $Z$, and related states}, PTEP 2016 (2016) 062C01.
\newblock \href {http://arxiv.org/abs/1603.09229} {\path{arXiv:1603.09229}},
  \href {https://doi.org/10.1093/ptep/ptw045} {\path{doi:10.1093/ptep/ptw045}}.

\bibitem{Richard:2016eis}
J.-M. Richard, {Exotic hadrons: review and perspectives}, Few Body Syst. 57
  (2016) 1185--1212.
\newblock \href {http://arxiv.org/abs/1606.08593} {\path{arXiv:1606.08593}},
  \href {https://doi.org/10.1007/s00601-016-1159-0}
  {\path{doi:10.1007/s00601-016-1159-0}}.

\bibitem{Chen:2016qju}
H.-X. Chen, W.~Chen, X.~Liu, S.-L. Zhu, {The hidden-charm pentaquark and
  tetraquark states}, Phys. Rept. 639 (2016) 1--121.
\newblock \href {http://arxiv.org/abs/1601.02092} {\path{arXiv:1601.02092}},
  \href {https://doi.org/10.1016/j.physrep.2016.05.004}
  {\path{doi:10.1016/j.physrep.2016.05.004}}.

\bibitem{Esposito:2016noz}
A.~Esposito, A.~Pilloni, A.~D. Polosa, {Multiquark resonances}, Phys. Rept. 668
  (2017) 1--97.
\newblock \href {http://arxiv.org/abs/1611.07920} {\path{arXiv:1611.07920}},
  \href {https://doi.org/10.1016/j.physrep.2016.11.002}
  {\path{doi:10.1016/j.physrep.2016.11.002}}.

\bibitem{Chen:2016spr}
H.-X. Chen, W.~Chen, X.~Liu, Y.-R. Liu, S.-L. Zhu, {A review of the open charm
  and open bottom systems}, Rept. Prog. Phys. 80 (2017) 076201.
\newblock \href {http://arxiv.org/abs/1609.08928} {\path{arXiv:1609.08928}},
  \href {https://doi.org/10.1088/1361-6633/aa6420}
  {\path{doi:10.1088/1361-6633/aa6420}}.

\bibitem{Lebed:2016hpi}
R.~F. Lebed, R.~E. Mitchell, E.~S. Swanson, {Heavy-quark QCD exotica}, Prog.
  Part. Nucl. Phys. 93 (2017) 143--194.
\newblock \href {http://arxiv.org/abs/1610.04528} {\path{arXiv:1610.04528}},
  \href {https://doi.org/10.1016/j.ppnp.2016.11.003}
  {\path{doi:10.1016/j.ppnp.2016.11.003}}.

\bibitem{Guo:2017jvc}
F.-K. Guo, C.~Hanhart, U.-G. Mei{\ss}ner, Q.~Wang, Q.~Zhao, B.-S. Zou,
  {Hadronic molecules}, Rev. Mod. Phys. 90 (2018) 015004, [Erratum:
  Rev.Mod.Phys. 94, 029901 (2022)].
\newblock \href {http://arxiv.org/abs/1705.00141} {\path{arXiv:1705.00141}},
  \href {https://doi.org/10.1103/RevModPhys.90.015004}
  {\path{doi:10.1103/RevModPhys.90.015004}}.

\bibitem{Olsen:2017bmm}
S.~L. Olsen, T.~Skwarnicki, D.~Zieminska, {Nonstandard heavy mesons and
  baryons: Experimental evidence}, Rev. Mod. Phys. 90 (2018) 015003.
\newblock \href {http://arxiv.org/abs/1708.04012} {\path{arXiv:1708.04012}},
  \href {https://doi.org/10.1103/RevModPhys.90.015003}
  {\path{doi:10.1103/RevModPhys.90.015003}}.

\bibitem{Ali:2017jda}
A.~Ali, J.~S. Lange, S.~Stone, {Exotics: Heavy pentaquarks and tetraquarks},
  Prog. Part. Nucl. Phys. 97 (2017) 123--198.
\newblock \href {http://arxiv.org/abs/1706.00610} {\path{arXiv:1706.00610}},
  \href {https://doi.org/10.1016/j.ppnp.2017.08.003}
  {\path{doi:10.1016/j.ppnp.2017.08.003}}.

\bibitem{Liu:2019zoy}
Y.-R. Liu, H.-X. Chen, W.~Chen, X.~Liu, S.-L. Zhu, {Pentaquark and tetraquark
  states}, Prog. Part. Nucl. Phys. 107 (2019) 237--320.
\newblock \href {http://arxiv.org/abs/1903.11976} {\path{arXiv:1903.11976}},
  \href {https://doi.org/10.1016/j.ppnp.2019.04.003}
  {\path{doi:10.1016/j.ppnp.2019.04.003}}.

\bibitem{Brambilla:2019esw}
N.~Brambilla, S.~Eidelman, C.~Hanhart, A.~Nefediev, C.-P. Shen, C.~E. Thomas,
  A.~Vairo, C.-Z. Yuan, {The $XYZ$ states: Experimental and theoretical status
  and perspectives}, Phys. Rept. 873 (2020) 1--154.
\newblock \href {http://arxiv.org/abs/1907.07583} {\path{arXiv:1907.07583}},
  \href {https://doi.org/10.1016/j.physrep.2020.05.001}
  {\path{doi:10.1016/j.physrep.2020.05.001}}.

\bibitem{Chen:2022asf}
H.-X. Chen, W.~Chen, X.~Liu, Y.-R. Liu, S.-L. Zhu, {An updated review of the
  new hadron states}, Rept. Prog. Phys. 86 (2023) 026201.
\newblock \href {http://arxiv.org/abs/2204.02649} {\path{arXiv:2204.02649}},
  \href {https://doi.org/10.1088/1361-6633/aca3b6}
  {\path{doi:10.1088/1361-6633/aca3b6}}.

\bibitem{Meng:2022ozq}
L.~Meng, B.~Wang, G.-J. Wang, S.-L. Zhu, {Chiral perturbation theory for heavy
  hadrons and chiral effective field theory for heavy hadronic molecules},
  Phys. Rept. 1019 (2023) 1--149.
\newblock \href {http://arxiv.org/abs/2204.08716} {\path{arXiv:2204.08716}},
  \href {https://doi.org/10.1016/j.physrep.2023.04.003}
  {\path{doi:10.1016/j.physrep.2023.04.003}}.

\bibitem{Wang:2021aql}
F.-L. Wang, X.-D. Yang, R.~Chen, X.~Liu, {Correlation of the hidden-charm
  molecular tetraquarks and the charmoniumlike structures existing in the $B\to
  XYZ+K$ process}, Phys. Rev. D 104 (2021) 094010.
\newblock \href {http://arxiv.org/abs/2103.04698} {\path{arXiv:2103.04698}},
  \href {https://doi.org/10.1103/PhysRevD.104.094010}
  {\path{doi:10.1103/PhysRevD.104.094010}}.

\bibitem{Wang:2025dur}
X.~Wang, X.~Liu, Y.~Gao, {Colloquium: Hadron Production in Open-charm Meson
  Pair at $e^+e^-$ Collider} (2 2025).
\newblock \href {http://arxiv.org/abs/2502.15117} {\path{arXiv:2502.15117}}.

\bibitem{Belle:2003nnu}
S.~K. Choi, et~al., Belle Collaboration, {Observation of a narrow
  charmonium-like state in exclusive $B^\pm \to K^\pm \pi^+ \pi^- J/\psi$
  decays}, Phys. Rev. Lett. 91 (2003) 262001.
\newblock \href {http://arxiv.org/abs/hep-ex/0309032}
  {\path{arXiv:hep-ex/0309032}}, \href
  {https://doi.org/10.1103/PhysRevLett.91.262001}
  {\path{doi:10.1103/PhysRevLett.91.262001}}.

\bibitem{BaBar:2003oey}
B.~Aubert, et~al., BaBar Collaboration, {Observation of a narrow meson decaying
  to $D_s^+ \pi^0$ at a mass of 2.32 GeV/$c^2$}, Phys. Rev. Lett. 90 (2003)
  242001.
\newblock \href {http://arxiv.org/abs/hep-ex/0304021}
  {\path{arXiv:hep-ex/0304021}}, \href
  {https://doi.org/10.1103/PhysRevLett.90.242001}
  {\path{doi:10.1103/PhysRevLett.90.242001}}.

\bibitem{CLEO:2003ggt}
D.~Besson, et~al., CLEO Collaboration, {Observation of a narrow resonance of
  mass 2.46 GeV/$c^2$ decaying to $D^{*+}_s\pi^0$ and confirmation of the
  $D^*_{sJ}(2317)$ state}, Phys. Rev. D 68 (2003) 032002, [Erratum: Phys.Rev.D
  75, 119908 (2007)].
\newblock \href {http://arxiv.org/abs/hep-ex/0305100}
  {\path{arXiv:hep-ex/0305100}}, \href
  {https://doi.org/10.1103/PhysRevD.68.032002}
  {\path{doi:10.1103/PhysRevD.68.032002}}.

\bibitem{BaBar:2006itc}
B.~Aubert, et~al., BaBar Collaboration, {Observation of a charmed baryon
  decaying to $D^0p$ at a mass near 2.94 GeV/$c^2$}, Phys. Rev. Lett. 98 (2007)
  012001.
\newblock \href {http://arxiv.org/abs/hep-ex/0603052}
  {\path{arXiv:hep-ex/0603052}}, \href
  {https://doi.org/10.1103/PhysRevLett.98.012001}
  {\path{doi:10.1103/PhysRevLett.98.012001}}.

\bibitem{Silvestre-Brac:1991qqx}
B.~Silvestre-Brac, C.~Gignoux, {Unitary effects in spin orbit splitting of
  $P$-wave baryons}, Phys. Rev. D 43 (1991) 3699--3708.
\newblock \href {https://doi.org/10.1103/PhysRevD.43.3699}
  {\path{doi:10.1103/PhysRevD.43.3699}}.

\bibitem{vanBeveren:2003kd}
E.~van Beveren, G.~Rupp, {Observed $D_s(2317)$ and tentative
  $D(2100\text{--}2300)$ as the charmed cousins of the light scalar nonet},
  Phys. Rev. Lett. 91 (2003) 012003.
\newblock \href {http://arxiv.org/abs/hep-ph/0305035}
  {\path{arXiv:hep-ph/0305035}}, \href
  {https://doi.org/10.1103/PhysRevLett.91.012003}
  {\path{doi:10.1103/PhysRevLett.91.012003}}.

\bibitem{Hwang:2004cd}
D.~S. Hwang, D.-W. Kim, {Mass of $D^*_{sJ}(2317)$ and coupled channel effect},
  Phys. Lett. B 601 (2004) 137--143.
\newblock \href {http://arxiv.org/abs/hep-ph/0408154}
  {\path{arXiv:hep-ph/0408154}}, \href
  {https://doi.org/10.1016/j.physletb.2004.09.040}
  {\path{doi:10.1016/j.physletb.2004.09.040}}.

\bibitem{Kalashnikova:2005ui}
Y.~S. Kalashnikova, {Coupled-channel model for charmonium levels and an option
  for $X(3872)$}, Phys. Rev. D 72 (2005) 034010.
\newblock \href {http://arxiv.org/abs/hep-ph/0506270}
  {\path{arXiv:hep-ph/0506270}}, \href
  {https://doi.org/10.1103/PhysRevD.72.034010}
  {\path{doi:10.1103/PhysRevD.72.034010}}.

\bibitem{Li:2009ad}
B.-Q. Li, C.~Meng, K.-T. Chao, {Coupled-channel and screening effects in
  charmonium spectrum}, Phys. Rev. D 80 (2009) 014012.
\newblock \href {http://arxiv.org/abs/0904.4068} {\path{arXiv:0904.4068}},
  \href {https://doi.org/10.1103/PhysRevD.80.014012}
  {\path{doi:10.1103/PhysRevD.80.014012}}.

\bibitem{Zhang:2009bv}
O.~Zhang, C.~Meng, H.~Q. Zheng, {Ambiversion of $X(3872)$}, Phys. Lett. B 680
  (2009) 453--458.
\newblock \href {http://arxiv.org/abs/0901.1553} {\path{arXiv:0901.1553}},
  \href {https://doi.org/10.1016/j.physletb.2009.09.033}
  {\path{doi:10.1016/j.physletb.2009.09.033}}.

\bibitem{Ortega:2009hj}
P.~G. Ortega, J.~Segovia, D.~R. Entem, F.~Fernandez, {Coupled channel approach
  to the structure of the $X(3872)$}, Phys. Rev. D 81 (2010) 054023.
\newblock \href {http://arxiv.org/abs/0907.3997} {\path{arXiv:0907.3997}},
  \href {https://doi.org/10.1103/PhysRevD.81.054023}
  {\path{doi:10.1103/PhysRevD.81.054023}}.

\bibitem{Danilkin:2010cc}
I.~V. Danilkin, Y.~A. Simonov, {Dynamical origin and the pole structure of
  $X(3872)$}, Phys. Rev. Lett. 105 (2010) 102002.
\newblock \href {http://arxiv.org/abs/1006.0211} {\path{arXiv:1006.0211}},
  \href {https://doi.org/10.1103/PhysRevLett.105.102002}
  {\path{doi:10.1103/PhysRevLett.105.102002}}.

\bibitem{Song:2015nia}
Q.-T. Song, D.-Y. Chen, X.~Liu, T.~Matsuki, {Charmed-strange mesons revisited:
  mass spectra and strong decays}, Phys. Rev. D 91 (2015) 054031.
\newblock \href {http://arxiv.org/abs/1501.03575} {\path{arXiv:1501.03575}},
  \href {https://doi.org/10.1103/PhysRevD.91.054031}
  {\path{doi:10.1103/PhysRevD.91.054031}}.

\bibitem{Luo:2019qkm}
S.-Q. Luo, B.~Chen, Z.-W. Liu, X.~Liu, {Resolving the low mass puzzle of
  $\Lambda_c(2940)^+$}, Eur. Phys. J. C 80 (2020) 301.
\newblock \href {http://arxiv.org/abs/1910.14545} {\path{arXiv:1910.14545}},
  \href {https://doi.org/10.1140/epjc/s10052-020-7874-1}
  {\path{doi:10.1140/epjc/s10052-020-7874-1}}.

\bibitem{Luo:2021dvj}
S.-Q. Luo, B.~Chen, X.~Liu, T.~Matsuki, {Predicting a new resonance as
  charmed-strange baryonic analog of $D^*_{s0}$(2317)}, Phys. Rev. D 103 (2021)
  074027.
\newblock \href {http://arxiv.org/abs/2102.00679} {\path{arXiv:2102.00679}},
  \href {https://doi.org/10.1103/PhysRevD.103.074027}
  {\path{doi:10.1103/PhysRevD.103.074027}}.

\bibitem{Zhang:2022pxc}
Z.-L. Zhang, Z.-W. Liu, S.-Q. Luo, F.-L. Wang, B.~Wang, H.~Xu,
  {$\Lambda_c(2910)$ and $\Lambda_c(2940)$ as conventional baryons dressed with
  the $D^*N$ channel}, Phys. Rev. D 107 (2023) 034036.
\newblock \href {http://arxiv.org/abs/2210.17188} {\path{arXiv:2210.17188}},
  \href {https://doi.org/10.1103/PhysRevD.107.034036}
  {\path{doi:10.1103/PhysRevD.107.034036}}.

\bibitem{Zhang:2024usz}
Z.-L. Zhang, Z.-W. Liu, S.-Q. Luo, P.~Chen, Z.-H. Guo, {Masses and radiative
  decay widths of $D_{s0}^*(2317)$ and $D_{s1}^\prime(2460)$ and their bottom
  analogs}, Phys. Rev. D 110 (2024) 094037.
\newblock \href {http://arxiv.org/abs/2409.05337} {\path{arXiv:2409.05337}},
  \href {https://doi.org/10.1103/PhysRevD.110.094037}
  {\path{doi:10.1103/PhysRevD.110.094037}}.

\bibitem{Belle:2009and}
S.~Uehara, et~al., Belle Collaboration, {Observation of a charmonium-like
  enhancement in the $\gamma\gamma\to\omega J/\psi$ process}, Phys. Rev. Lett.
  104 (2010) 092001.
\newblock \href {http://arxiv.org/abs/0912.4451} {\path{arXiv:0912.4451}},
  \href {https://doi.org/10.1103/PhysRevLett.104.092001}
  {\path{doi:10.1103/PhysRevLett.104.092001}}.

\bibitem{Belle:2005rte}
S.~Uehara, et~al., Belle Collaboration, {Observation of a $\chi_{c2}^\prime$
  candidate in $\gamma\gamma\to D\bar D$ production at Belle}, Phys. Rev. Lett.
  96 (2006) 082003.
\newblock \href {http://arxiv.org/abs/hep-ex/0512035}
  {\path{arXiv:hep-ex/0512035}}, \href
  {https://doi.org/10.1103/PhysRevLett.96.082003}
  {\path{doi:10.1103/PhysRevLett.96.082003}}.

\bibitem{BaBar:2005hhc}
B.~Aubert, et~al., BaBar Collaboration, {Observation of a broad structure in
  the $\pi^+ \pi^- J/\psi$ mass spectrum around 4.26 GeV/$c^2$}, Phys. Rev.
  Lett. 95 (2005) 142001.
\newblock \href {http://arxiv.org/abs/hep-ex/0506081}
  {\path{arXiv:hep-ex/0506081}}, \href
  {https://doi.org/10.1103/PhysRevLett.95.142001}
  {\path{doi:10.1103/PhysRevLett.95.142001}}.

\bibitem{BaBar:2006ait}
B.~Aubert, et~al., BaBar Collaboration, {Evidence of a broad structure at an
  invariant mass of 4.32 GeV$/c^{2}$ in the reaction $e^{+} e^{-} \to \pi^{+}
  \pi^{-} \psi(2S)$ measured at BaBar}, Phys. Rev. Lett. 98 (2007) 212001.
\newblock \href {http://arxiv.org/abs/hep-ex/0610057}
  {\path{arXiv:hep-ex/0610057}}, \href
  {https://doi.org/10.1103/PhysRevLett.98.212001}
  {\path{doi:10.1103/PhysRevLett.98.212001}}.

\bibitem{Belle:2007umv}
X.~L. Wang, et~al., Belle Collaboration, {Observation of two resonant
  structures in $e^+e^-\to\pi^+\pi^-\psi(2S)$ via initial state radiation at
  Belle}, Phys. Rev. Lett. 99 (2007) 142002.
\newblock \href {http://arxiv.org/abs/0707.3699} {\path{arXiv:0707.3699}},
  \href {https://doi.org/10.1103/PhysRevLett.99.142002}
  {\path{doi:10.1103/PhysRevLett.99.142002}}.

\bibitem{Belle:2008xmh}
G.~Pakhlova, et~al., Belle Collaboration, {Observation of a near-threshold
  enhancement in the $e^+e^-\to\Lambda_c^+\Lambda_c^-$ cross section using
  initial-state radiation}, Phys. Rev. Lett. 101 (2008) 172001.
\newblock \href {http://arxiv.org/abs/0807.4458} {\path{arXiv:0807.4458}},
  \href {https://doi.org/10.1103/PhysRevLett.101.172001}
  {\path{doi:10.1103/PhysRevLett.101.172001}}.

\bibitem{BESIII:2016bnd}
M.~Ablikim, et~al., BESIII Collaboration, {Precise measurement of the
  $e^+e^-\to \pi^+\pi^-J/\psi$ cross section at center-of-mass energies from
  3.77 to 4.60 GeV}, Phys. Rev. Lett. 118 (2017) 092001.
\newblock \href {http://arxiv.org/abs/1611.01317} {\path{arXiv:1611.01317}},
  \href {https://doi.org/10.1103/PhysRevLett.118.092001}
  {\path{doi:10.1103/PhysRevLett.118.092001}}.

\bibitem{BESIII:2016adj}
M.~Ablikim, et~al., BESIII Collaboration, {Evidence of two resonant structures
  in $e^+ e^- \to \pi^+ \pi^- h_c$}, Phys. Rev. Lett. 118 (2017) 092002.
\newblock \href {http://arxiv.org/abs/1610.07044} {\path{arXiv:1610.07044}},
  \href {https://doi.org/10.1103/PhysRevLett.118.092002}
  {\path{doi:10.1103/PhysRevLett.118.092002}}.

\bibitem{BESIII:2020nme}
M.~Ablikim, et~al., BESIII Collaboration, {Future physics programme of BESIII},
  Chin. Phys. C 44 (2020) 040001.
\newblock \href {http://arxiv.org/abs/1912.05983} {\path{arXiv:1912.05983}},
  \href {https://doi.org/10.1088/1674-1137/44/4/040001}
  {\path{doi:10.1088/1674-1137/44/4/040001}}.

\bibitem{Harris:1999wn}
F.~A. Harris, {Recent charmonium results from BES}, in: {American Physical
  Society (APS) Meeting of the Division of Particles and Fields (DPF 99)},
  1999.
\newblock \href {http://arxiv.org/abs/hep-ex/9903036}
  {\path{arXiv:hep-ex/9903036}}.

\bibitem{Liu:2006dq}
X.~Liu, X.-Q. Zeng, X.-Q. Li, {Study on contributions of hadronic loops to
  decays of $J/\psi \to$ vector + pseudoscalar mesons}, Phys. Rev. D 74 (2006)
  074003.
\newblock \href {http://arxiv.org/abs/hep-ph/0606191}
  {\path{arXiv:hep-ph/0606191}}, \href
  {https://doi.org/10.1103/PhysRevD.74.074003}
  {\path{doi:10.1103/PhysRevD.74.074003}}.

\bibitem{Liu:2009dr}
X.~Liu, B.~Zhang, X.-Q. Li, {The puzzle of excessive non-$D\bar{D}$ component
  of the inclusive $\psi(3770)$ decay and the long-distant contribution}, Phys.
  Lett. B 675 (2009) 441--445.
\newblock \href {http://arxiv.org/abs/0902.0480} {\path{arXiv:0902.0480}},
  \href {https://doi.org/10.1016/j.physletb.2009.04.047}
  {\path{doi:10.1016/j.physletb.2009.04.047}}.

\bibitem{Li:2013zcr}
G.~Li, X.-h. Liu, Q.~Wang, Q.~Zhao, {Further understanding of the
  non-$D\bar{D}$ decays of $\psi(3770)$}, Phys. Rev. D 88 (2013) 014010.
\newblock \href {http://arxiv.org/abs/1302.1745} {\path{arXiv:1302.1745}},
  \href {https://doi.org/10.1103/PhysRevD.88.014010}
  {\path{doi:10.1103/PhysRevD.88.014010}}.

\bibitem{Zhang:2009kr}
Y.-J. Zhang, G.~Li, Q.~Zhao, {Towards a dynamical understanding of the
  non-$D\bar{D}$ decay of $\psi(3770)$}, Phys. Rev. Lett. 102 (2009) 172001.
\newblock \href {http://arxiv.org/abs/0902.1300} {\path{arXiv:0902.1300}},
  \href {https://doi.org/10.1103/PhysRevLett.102.172001}
  {\path{doi:10.1103/PhysRevLett.102.172001}}.

\bibitem{Chen:2009ah}
D.-Y. Chen, J.~He, X.-Q. Li, X.~Liu, {Understanding the branching ratios of
  $\chi_{c1}\to\phi\phi$, $\omega\omega$, $\omega\phi$ observed at BES-III},
  Phys. Rev. D 81 (2010) 074006.
\newblock \href {http://arxiv.org/abs/0912.4860} {\path{arXiv:0912.4860}},
  \href {https://doi.org/10.1103/PhysRevD.81.074006}
  {\path{doi:10.1103/PhysRevD.81.074006}}.

\bibitem{Liu:2009vv}
X.-H. Liu, Q.~Zhao, {The Evasion of helicity selection rule in $\chi_{c1} \to
  VV$ and $\chi_{c2} \to VP$ via intermediate charmed meson loops}, Phys. Rev.
  D 81 (2010) 014017.
\newblock \href {http://arxiv.org/abs/0912.1508} {\path{arXiv:0912.1508}},
  \href {https://doi.org/10.1103/PhysRevD.81.014017}
  {\path{doi:10.1103/PhysRevD.81.014017}}.

\bibitem{Chen:2010re}
D.-Y. Chen, Y.-B. Dong, X.~Liu, {Long-distant contribution and $\chi_{c1}$
  radiative decays to light vector meson}, Eur. Phys. J. C 70 (2010) 177--182.
\newblock \href {http://arxiv.org/abs/1005.0066} {\path{arXiv:1005.0066}},
  \href {https://doi.org/10.1140/epjc/s10052-010-1449-5}
  {\path{doi:10.1140/epjc/s10052-010-1449-5}}.

\bibitem{Huang:2021kfm}
Q.~Huang, J.-Z. Wang, R.-G. Ping, X.~Liu, {Detecting the polarization in
  $\chi_{cJ} \to \phi \phi $ decays to probe hadronic loop effect}, Phys. Rev.
  D 103 (2021) 096006.
\newblock \href {http://arxiv.org/abs/2102.07104} {\path{arXiv:2102.07104}},
  \href {https://doi.org/10.1103/PhysRevD.103.096006}
  {\path{doi:10.1103/PhysRevD.103.096006}}.

\bibitem{CLEO:2005mpm}
D.~Besson, et~al., CLEO Collaboration, {Measurement of $\sigma(e^+e^- \to
  \psi(3770) \to {\rm hadrons})$ at $E_{\rm c.m.}$ = 3773 MeV}, Phys. Rev.
  Lett. 96 (2006) 092002, [Erratum: Phys.Rev.Lett. 104, 159901 (2010)].
\newblock \href {http://arxiv.org/abs/1004.1358} {\path{arXiv:1004.1358}},
  \href {https://doi.org/10.1103/PhysRevLett.96.092002}
  {\path{doi:10.1103/PhysRevLett.96.092002}}.

\bibitem{BES:2006fpf}
M.~Ablikim, et~al., BES Collaboration, {Measurements of the braching fractions
  for $\psi(3770) \to D^0 \bar{D}^0, D^+ D^-, D \bar{D}$ and the resonance
  parameters of $\psi(3770)$ and $\psi(2S)$}, Phys. Rev. Lett. 97 (2006)
  121801.
\newblock \href {http://arxiv.org/abs/hep-ex/0605107}
  {\path{arXiv:hep-ex/0605107}}, \href
  {https://doi.org/10.1103/PhysRevLett.97.121801}
  {\path{doi:10.1103/PhysRevLett.97.121801}}.

\bibitem{BES:2006dso}
M.~Ablikim, et~al., BES Collaboration, {Measurements of the cross-sections for
  $e^+ e^- \to$ hadrons at 3.650 GeV, 3.6648 GeV, 3.773 GeV and the branching
  fraction for $\psi(3770) \to {\rm non} - D \bar{D}$}, Phys. Lett. B 641
  (2006) 145--155.
\newblock \href {http://arxiv.org/abs/hep-ex/0605105}
  {\path{arXiv:hep-ex/0605105}}, \href
  {https://doi.org/10.1016/j.physletb.2006.08.049}
  {\path{doi:10.1016/j.physletb.2006.08.049}}.

\bibitem{BES:2007cev}
M.~Ablikim, et~al., BES Collaboration, {Direct measurements of the
  non-$D\bar{D}$ cross section $\sigma_{\psi(3770)\to \mathrm{non}-D\bar{D}}$
  at $E_{cm} = 3.773$ GeV and the branching fraction for $\psi(3770)\to
  \mathrm{non}-D\bar{D}$}, Phys. Rev. D 76 (2007) 122002.
\newblock \href {https://doi.org/10.1103/PhysRevD.76.122002}
  {\path{doi:10.1103/PhysRevD.76.122002}}.

\bibitem{BES:2008vad}
M.~Ablikim, et~al., BES Collaboration, {Direct measurements of the cross
  sections for $e^+ e^- \to$ hadrons (non-$D \bar{D}$) in the range from 3.65
  GeV to 3.87 GeV and the branching fraction for $\psi(3770) \to
  \mathrm{non}-D\bar{D}$}, Phys. Lett. B 659 (2008) 74--79.
\newblock \href {https://doi.org/10.1016/j.physletb.2007.11.078}
  {\path{doi:10.1016/j.physletb.2007.11.078}}.

\bibitem{CLEO:2008sah}
J.~V. Bennett, et~al., CLEO Collaboration, {Observation of $\chi_{cJ}$
  radiative decays to light vector mesons}, Phys. Rev. Lett. 101 (2008) 151801.
\newblock \href {http://arxiv.org/abs/0807.3718} {\path{arXiv:0807.3718}},
  \href {https://doi.org/10.1103/PhysRevLett.101.151801}
  {\path{doi:10.1103/PhysRevLett.101.151801}}.

\bibitem{BESIII:2011ysp}
M.~Ablikim, et~al., BESIII Collaboration, {Study of $\chi_{cJ}$ radiative
  decays into a vector meson}, Phys. Rev. D 83 (2011) 112005.
\newblock \href {http://arxiv.org/abs/1103.5564} {\path{arXiv:1103.5564}},
  \href {https://doi.org/10.1103/PhysRevD.83.112005}
  {\path{doi:10.1103/PhysRevD.83.112005}}.

\bibitem{BES:2004imp}
M.~Ablikim, et~al., BES Collaboration, {Observation of $K^* (892)^0 \bar{K}^*
  (892)^0$ in $\chi_{cJ}$ decays}, Phys. Rev. D 70 (2004) 092003.
\newblock \href {http://arxiv.org/abs/hep-ex/0408012}
  {\path{arXiv:hep-ex/0408012}}, \href
  {https://doi.org/10.1103/PhysRevD.70.092003}
  {\path{doi:10.1103/PhysRevD.70.092003}}.

\bibitem{BESII:2011hcd}
M.~Ablikim, et~al., BESII Collaboration, {Observation of $\chi_{c1}$ decays
  into vector meson pairs $\phi\phi$, $\omega\omega$, and $\omega\phi$}, Phys.
  Rev. Lett. 107 (2011) 092001.
\newblock \href {http://arxiv.org/abs/1104.5068} {\path{arXiv:1104.5068}},
  \href {https://doi.org/10.1103/PhysRevLett.107.092001}
  {\path{doi:10.1103/PhysRevLett.107.092001}}.

\bibitem{CLEO:2005zky}
N.~E. Adam, et~al., CLEO Collaboration, {Observation of $\psi(3770)\to\pi\pi
  J/\psi$ and measurement of $\Gamma_{ee}[\psi(2S)]$}, Phys. Rev. Lett. 96
  (2006) 082004.
\newblock \href {http://arxiv.org/abs/hep-ex/0508023}
  {\path{arXiv:hep-ex/0508023}}, \href
  {https://doi.org/10.1103/PhysRevLett.96.082004}
  {\path{doi:10.1103/PhysRevLett.96.082004}}.

\bibitem{Kuang:1989ub}
Y.-P. Kuang, T.-M. Yan, {Hadronic Transitions of $D$ Wave Quarkonium and
  $\psi(3770) \to J/\psi \pi \pi$}, Phys. Rev. D 41 (1990) 155.
\newblock \href {https://doi.org/10.1103/PhysRevD.41.155}
  {\path{doi:10.1103/PhysRevD.41.155}}.

\bibitem{Chen:2011jp}
D.-Y. Chen, X.~Liu, X.-Q. Li, {Anomalous dipion invariant mass distribution of
  the $\Upsilon(4S)$ decays into $\Upsilon(1S) \pi^{+} \pi^{-}$ and
  $\Upsilon(2S) \pi^{+} \pi^{-}$}, Eur. Phys. J. C 71 (2011) 1808.
\newblock \href {http://arxiv.org/abs/1109.1406} {\path{arXiv:1109.1406}},
  \href {https://doi.org/10.1140/epjc/s10052-011-1808-x}
  {\path{doi:10.1140/epjc/s10052-011-1808-x}}.

\bibitem{Belle:2007xek}
K.~F. Chen, et~al., Belle Collaboration, {Observation of anomalous
  $\Upsilon(1S) \pi^+ \pi^-$ and $\Upsilon(2S) \pi^+ \pi^-$ production near the
  $\Upsilon(5S)$ resonance}, Phys. Rev. Lett. 100 (2008) 112001.
\newblock \href {http://arxiv.org/abs/0710.2577} {\path{arXiv:0710.2577}},
  \href {https://doi.org/10.1103/PhysRevLett.100.112001}
  {\path{doi:10.1103/PhysRevLett.100.112001}}.

\bibitem{Belle:2011aa}
A.~Bondar, et~al., Belle Collaboration, {Observation of two charged
  bottomonium-like resonances in $\Upsilon(5S)$ decays}, Phys. Rev. Lett. 108
  (2012) 122001.
\newblock \href {http://arxiv.org/abs/1110.2251} {\path{arXiv:1110.2251}},
  \href {https://doi.org/10.1103/PhysRevLett.108.122001}
  {\path{doi:10.1103/PhysRevLett.108.122001}}.

\bibitem{Chen:2011pv}
D.-Y. Chen, X.~Liu, {$Z_b(10610)$ and $Z_b(10650)$ structures produced by the
  initial single pion emission in the $\Upsilon(5S)$ decays}, Phys. Rev. D 84
  (2011) 094003.
\newblock \href {http://arxiv.org/abs/1106.3798} {\path{arXiv:1106.3798}},
  \href {https://doi.org/10.1103/PhysRevD.84.094003}
  {\path{doi:10.1103/PhysRevD.84.094003}}.

\bibitem{Chen:2011xk}
D.-Y. Chen, X.~Liu, {Predicted charged charmonium-like structures in the
  hidden-charm dipion decay of higher charmonia}, Phys. Rev. D 84 (2011)
  034032.
\newblock \href {http://arxiv.org/abs/1106.5290} {\path{arXiv:1106.5290}},
  \href {https://doi.org/10.1103/PhysRevD.84.034032}
  {\path{doi:10.1103/PhysRevD.84.034032}}.

\bibitem{BESIII:2013qmu}
M.~Ablikim, et~al., BESIII Collaboration, {Observation of a charged
  $(D\bar{D}^{*})^\pm$ mass peak in $e^{+}e^{-} \to \pi D\bar{D}^{*}$ at
  $\sqrt{s} =4.26$ GeV}, Phys. Rev. Lett. 112 (2014) 022001.
\newblock \href {http://arxiv.org/abs/1310.1163} {\path{arXiv:1310.1163}},
  \href {https://doi.org/10.1103/PhysRevLett.112.022001}
  {\path{doi:10.1103/PhysRevLett.112.022001}}.

\bibitem{LHCb:2020bwg}
R.~Aaij, et~al., LHCb Collaboration, {Observation of structure in the $J
  /\psi$-pair mass spectrum}, Sci. Bull. 65 (2020) 1983--1993.
\newblock \href {http://arxiv.org/abs/2006.16957} {\path{arXiv:2006.16957}},
  \href {https://doi.org/10.1016/j.scib.2020.08.032}
  {\path{doi:10.1016/j.scib.2020.08.032}}.

\bibitem{ATLAS:2023bft}
G.~Aad, et~al., ATLAS Collaboration, {Observation of an excess of dicharmonium
  events in the four-muon final state with the ATLAS detector}, Phys. Rev.
  Lett. 131 (2023) 151902.
\newblock \href {http://arxiv.org/abs/2304.08962} {\path{arXiv:2304.08962}},
  \href {https://doi.org/10.1103/PhysRevLett.131.151902}
  {\path{doi:10.1103/PhysRevLett.131.151902}}.

\bibitem{CMS:2023owd}
A.~Hayrapetyan, et~al., CMS Collaboration, {New structures in the $J/\psi
  J/\psi$ mass spectrum in proton-proton collisions at $\sqrt{s}=13$ TeV},
  Phys. Rev. Lett. 132 (2024) 111901.
\newblock \href {http://arxiv.org/abs/2306.07164} {\path{arXiv:2306.07164}},
  \href {https://doi.org/10.1103/PhysRevLett.132.111901}
  {\path{doi:10.1103/PhysRevLett.132.111901}}.

\bibitem{ATLAS:2025nsd}
G.~Aad, et~al., ATLAS Collaboration, {Observation of structures in the
  $J/\psi+\psi(2S)$ mass spectrum with the ATLAS detector} (9 2025).
\newblock \href {http://arxiv.org/abs/2509.13101} {\path{arXiv:2509.13101}}.

\bibitem{CMS:2025xwt}
CMS Collaboration, {The CMS Collaboration, Observation of a family of all-charm
  tetraquark candidates at the LHC, CMS PAS BPH-24-003;
  \href{https://cds.cern.ch/record/2929472/files/BPH-24-003-pas.pdf}{https://cds.cern.ch/record/2929472/files/BPH-24-003-pas.pdf}}.

\bibitem{CMS:2025vnq}
CMS Collaboration, {The CMS Collaboration, Observation of $X(6900)$ and
  evidence of $X(7100)$ in the $J/\psi\psi(2S)\to\mu^+\mu^-\mu^+\mu^-$ mass
  spectrum in $pp$ collisions at CMS, CMS PAS BPH-22-004;
  \href{https://cds.cern.ch/record/2929529/files/BPH-22-004-pas.pdf}{https://cds.cern.ch/record/2929529/files/BPH-22-004-pas.pdf}}.

\bibitem{CMS:2025fpt}
A.~Hayrapetyan, et~al., CMS Collaboration, {Determination of the spin and
  parity of all-charm tetraquarks}, Nature 648 (2025) 58--63.
\newblock \href {http://arxiv.org/abs/2506.07944} {\path{arXiv:2506.07944}},
  \href {https://doi.org/10.1038/s41586-025-09711-7}
  {\path{doi:10.1038/s41586-025-09711-7}}.

\bibitem{Lyu:2024ttm}
Y.~Lyu, T.~Doi, T.~Hatsuda, T.~Sugiura, {Nucleon-charmonium interactions from
  lattice QCD}, Phys. Lett. B 860 (2025) 139178.
\newblock \href {http://arxiv.org/abs/2410.22755} {\path{arXiv:2410.22755}},
  \href {https://doi.org/10.1016/j.physletb.2024.139178}
  {\path{doi:10.1016/j.physletb.2024.139178}}.

\bibitem{Lyu:2021qsh}
Y.~Lyu, H.~Tong, T.~Sugiura, S.~Aoki, T.~Doi, T.~Hatsuda, J.~Meng, T.~Miyamoto,
  {Dibaryon with highest charm number near unitarity from lattice QCD}, Phys.
  Rev. Lett. 127 (2021) 072003.
\newblock \href {http://arxiv.org/abs/2102.00181} {\path{arXiv:2102.00181}},
  \href {https://doi.org/10.1103/PhysRevLett.127.072003}
  {\path{doi:10.1103/PhysRevLett.127.072003}}.

\bibitem{Appelquist:1974zd}
T.~Appelquist, H.~D. Politzer, {Orthocharmonium and $e^+ e^-$ annihilation},
  Phys. Rev. Lett. 34 (1975) 43.
\newblock \href {https://doi.org/10.1103/PhysRevLett.34.43}
  {\path{doi:10.1103/PhysRevLett.34.43}}.

\bibitem{ParticleDataGroup:2024cfk}
S.~Navas, et~al., Particle Data Group Collaboration, {Review of particle
  physics}, Phys. Rev. D 110 (2024) 030001.
\newblock \href {https://doi.org/10.1103/PhysRevD.110.030001}
  {\path{doi:10.1103/PhysRevD.110.030001}}.

\bibitem{Hou:1982kh}
W.-S. Hou, A.~Soni, {Vector gluonium as a possible explanation for anomalous
  $\psi$ decays}, Phys. Rev. Lett. 50 (1983) 569.
\newblock \href {https://doi.org/10.1103/PhysRevLett.50.569}
  {\path{doi:10.1103/PhysRevLett.50.569}}.

\bibitem{Brodsky:1987bb}
S.~J. Brodsky, G.~P. Lepage, S.~F. Tuan, {Exclusive charmonium decays: The
  $J/\psi (\psi^\prime$) $\to \rho \pi$, $K^* \bar{K}$ puzzle}, Phys. Rev.
  Lett. 59 (1987) 621.
\newblock \href {https://doi.org/10.1103/PhysRevLett.59.621}
  {\path{doi:10.1103/PhysRevLett.59.621}}.

\bibitem{Brodsky:1997fj}
S.~J. Brodsky, M.~Karliner, Intrinsic charm of vector mesons: A possible
  solution of the "$\rho \pi$ puzzle", Phys. Rev. Lett. 78 (1997) 4682--4685.
\newblock \href {http://arxiv.org/abs/hep-ph/9704379}
  {\path{arXiv:hep-ph/9704379}}, \href
  {https://doi.org/10.1103/PhysRevLett.78.4682}
  {\path{doi:10.1103/PhysRevLett.78.4682}}.

\bibitem{Feldmann:2000hs}
T.~Feldmann, P.~Kroll, {Implications of light quark admixtures on charmonium
  decays into meson pairs}, Phys. Rev. D 62 (2000) 074006.
\newblock \href {http://arxiv.org/abs/hep-ph/0003096}
  {\path{arXiv:hep-ph/0003096}}, \href
  {https://doi.org/10.1103/PhysRevD.62.074006}
  {\path{doi:10.1103/PhysRevD.62.074006}}.

\bibitem{Guo:2023igo}
X.-D. Guo, D.-Y. Chen, X.-Q. Li, Z.-Y. Yuan, S.~Sang, {Molecular components in
  $J/\psi$ and $\rho$-$\pi$ puzzle}, Chin. Phys. C 48 (2024) 053110.
\newblock \href {http://arxiv.org/abs/2308.06121} {\path{arXiv:2308.06121}},
  \href {https://doi.org/10.1088/1674-1137/ad2a65}
  {\path{doi:10.1088/1674-1137/ad2a65}}.

\bibitem{Clavelli:1983rk}
L.~J. Clavelli, G.~W. Intemann, {Vector meson mixing and hadronic decays of
  $\psi$}, Phys. Rev. D 28 (1983) 2767.
\newblock \href {https://doi.org/10.1103/PhysRevD.28.2767}
  {\path{doi:10.1103/PhysRevD.28.2767}}.

\bibitem{Li:1996yn}
X.-Q. Li, D.~V. Bugg, B.-S. Zou, {A Possible explanation of the "$\rho \pi$
  puzzle" in $J/\psi$, $\psi^\prime$ decays}, Phys. Rev. D 55 (1997)
  1421--1424.
\newblock \href {https://doi.org/10.1103/PhysRevD.55.1421}
  {\path{doi:10.1103/PhysRevD.55.1421}}.

\bibitem{Suzuki:1998ea}
M.~Suzuki, {Long distance final state interactions and $J/\psi$ decay}, Phys.
  Rev. D 57 (1998) 5717--5722.
\newblock \href {http://arxiv.org/abs/hep-ph/9801284}
  {\path{arXiv:hep-ph/9801284}}, \href
  {https://doi.org/10.1103/PhysRevD.57.5717}
  {\path{doi:10.1103/PhysRevD.57.5717}}.

\bibitem{Suzuki:2000yq}
M.~Suzuki, {Anomalies in $\psi(2S)$ decay and the $\rho-\pi$ puzzle}, Phys.
  Rev. D 63 (2001) 054021.
\newblock \href {http://arxiv.org/abs/hep-ph/0006296}
  {\path{arXiv:hep-ph/0006296}}, \href
  {https://doi.org/10.1103/PhysRevD.63.054021}
  {\path{doi:10.1103/PhysRevD.63.054021}}.

\bibitem{Rosner:2001nm}
J.~L. Rosner, {Charmless final states and $S$-$D$ wave mixing in the
  $\psi^{\prime \prime}$}, Phys. Rev. D 64 (2001) 094002.
\newblock \href {http://arxiv.org/abs/hep-ph/0105327}
  {\path{arXiv:hep-ph/0105327}}, \href
  {https://doi.org/10.1103/PhysRevD.64.094002}
  {\path{doi:10.1103/PhysRevD.64.094002}}.

\bibitem{Gerard:1999uf}
J.~M. Gerard, J.~Weyers, {Phases and amplitudes in inclusive $\psi$ and
  $\psi^\prime$ decays}, Phys. Lett. B 462 (1999) 324--328.
\newblock \href {http://arxiv.org/abs/hep-ph/9906357}
  {\path{arXiv:hep-ph/9906357}}, \href
  {https://doi.org/10.1016/S0370-2693(99)00849-7}
  {\path{doi:10.1016/S0370-2693(99)00849-7}}.

\bibitem{Chen:1998ma}
Y.-Q. Chen, E.~Braaten, {An explanation for the $\rho \pi$ puzzle of $J / \psi$
  and $\psi^\prime$ decays}, Phys. Rev. Lett. 80 (1998) 5060--5063.
\newblock \href {http://arxiv.org/abs/hep-ph/9801226}
  {\path{arXiv:hep-ph/9801226}}, \href
  {https://doi.org/10.1103/PhysRevLett.80.5060}
  {\path{doi:10.1103/PhysRevLett.80.5060}}.

\bibitem{Karl:1984en}
G.~Karl, W.~Roberts, {Sequential fragmentation and rate oscillations in
  exclusive decays}, Phys. Lett. B 144 (1984) 263--265.
\newblock \href {https://doi.org/10.1016/0370-2693(84)91817-3}
  {\path{doi:10.1016/0370-2693(84)91817-3}}.

\bibitem{Pinsky:1989ue}
S.~S. Pinsky, {The $\psi^\prime$ to $J/\psi$ hadronic decay puzzle}, Phys.
  Lett. B 236 (1990) 479--482.
\newblock \href {https://doi.org/10.1016/0370-2693(90)90387-L}
  {\path{doi:10.1016/0370-2693(90)90387-L}}.

\bibitem{Chaichian:1988kn}
M.~Chaichian, N.~A. Tornqvist, {Preasymptotic versus asymptotic {QCD} relations
  for the hadronic form-factor and the anomalous $J/\psi$ and $\psi^\prime$
  decays}, Nucl. Phys. B 323 (1989) 75--89.
\newblock \href {https://doi.org/10.1016/0550-3213(89)90588-9}
  {\path{doi:10.1016/0550-3213(89)90588-9}}.

\bibitem{Okubo:1963fa}
S.~Okubo, {Phi meson and unitary symmetry model}, Phys. Lett. 5 (1963)
  165--168.
\newblock \href {https://doi.org/10.1016/S0375-9601(63)92548-9}
  {\path{doi:10.1016/S0375-9601(63)92548-9}}.

\bibitem{Zweig:1964jf}
G.~Zweig, {An SU(3) model for strong interaction symmetry and its breaking.
  Version 2}, 1964, pp. 22--101.
\newblock \href {https://doi.org/10.17181/CERN-TH-412}
  {\path{doi:10.17181/CERN-TH-412}}.

\bibitem{Iizuka:1966fk}
J.~Iizuka, {Systematics and phenomenology of meson family}, Prog. Theor. Phys.
  Suppl. 37 (1966) 21--34.
\newblock \href {https://doi.org/10.1143/PTPS.37.21}
  {\path{doi:10.1143/PTPS.37.21}}.

\bibitem{Lipkin:1991bf}
H.~J. Lipkin, {The OZI rule and nucleons}, Int. J. Mod. Phys. E 1 (1992)
  603--628.
\newblock \href {https://doi.org/10.1142/S0218301392000291}
  {\path{doi:10.1142/S0218301392000291}}.

\bibitem{Nomokonov:2002jb}
V.~P. Nomokonov, M.~G. Sapozhnikov, {Experimental tests of the
  Okubo-Zweig-Iizuka rule in hadron interactions}, Phys. Part. Nucl. 34 (2003)
  94--123.
\newblock \href {http://arxiv.org/abs/hep-ph/0204259}
  {\path{arXiv:hep-ph/0204259}}.

\bibitem{LeYaouanc:1977gm}
A.~Le~Yaouanc, L.~Oliver, O.~Pene, J.~C. Raynal, {Why is $\psi(4.414)$ so
  narrow?}, Phys. Lett. B 72 (1977) 57--61.
\newblock \href {https://doi.org/10.1016/0370-2693(77)90062-4}
  {\path{doi:10.1016/0370-2693(77)90062-4}}.

\bibitem{Barnes:2005pb}
T.~Barnes, S.~Godfrey, E.~S. Swanson, {Higher charmonia}, Phys. Rev. D 72
  (2005) 054026.
\newblock \href {http://arxiv.org/abs/hep-ph/0505002}
  {\path{arXiv:hep-ph/0505002}}, \href
  {https://doi.org/10.1103/PhysRevD.72.054026}
  {\path{doi:10.1103/PhysRevD.72.054026}}.

\bibitem{Rong:2005it}
G.~Rong, D.-h. Zhang, J.-c. Chen, {Determination of the branching fractions for
  $\psi(3770) \to D^0 \bar{D}^0, D^+ D^-, D \bar{D}$ and $\psi(3770) \to {\rm
  non}-D \bar{D}$} (6 2005).
\newblock \href {http://arxiv.org/abs/hep-ex/0506051}
  {\path{arXiv:hep-ex/0506051}}.

\bibitem{Lipkin:1986av}
H.~J. Lipkin, {The {OZI} rule in charmonium decays above $D \bar{D}$
  threshold}, Phys. Lett. B 179 (1986) 278.
\newblock \href {https://doi.org/10.1016/0370-2693(86)90580-0}
  {\path{doi:10.1016/0370-2693(86)90580-0}}.

\bibitem{Ding:1991vu}
Y.-B. Ding, D.-H. Qin, K.-T. Chao, {Electric dipole transitions of $\psi
  (3770)$ and $S$-$D$ mixing between $\psi (3686)$ and $\psi (3770)$}, Phys.
  Rev. D 44 (1991) 3562--3566.
\newblock \href {https://doi.org/10.1103/PhysRevD.44.3562}
  {\path{doi:10.1103/PhysRevD.44.3562}}.

\bibitem{Achasov:1990gt}
N.~N. Achasov, A.~A. Kozhevnikov, {Direct decays of heavy quarkonia}, Phys.
  Lett. B 260 (1991) 425--428.
\newblock \href {https://doi.org/10.1016/0370-2693(91)91637-B}
  {\path{doi:10.1016/0370-2693(91)91637-B}}.

\bibitem{Achasov:1991qp}
N.~N. Achasov, A.~A. Kozhevnikov, {Decays of heavy quarkonia which violate the
  OZI rule}, JETP Lett. 54 (1991) 193--196.

\bibitem{Achasov:1994vh}
N.~N. Achasov, A.~A. Kozhevnikov, {Dynamical violation of the OZI rule and $G$
  parity in the decays of heavy quarkonia}, Phys. Rev. D 49 (1994) 275--282.
\newblock \href {https://doi.org/10.1103/PhysRevD.49.275}
  {\path{doi:10.1103/PhysRevD.49.275}}.

\bibitem{Achasov:2005qb}
N.~N. Achasov, A.~A. Kozhevnikov, {Branching ratios for the decays of
  $\psi(3770)$ and $\Upsilon(10580)$ mesons to a pair of light hadrons}, Phys.
  Atom. Nucl. 69 (2006) 988--998.
\newblock \href {http://arxiv.org/abs/hep-ph/0505146}
  {\path{arXiv:hep-ph/0505146}}, \href
  {https://doi.org/10.1134/S1063778806060093}
  {\path{doi:10.1134/S1063778806060093}}.

\bibitem{Rosner:2004wy}
J.~L. Rosner, {$\psi^{\prime\prime}$ decays to charmless final states}, Annals
  Phys. 319 (2005) 1--12.
\newblock \href {http://arxiv.org/abs/hep-ph/0411003}
  {\path{arXiv:hep-ph/0411003}}, \href
  {https://doi.org/10.1016/j.aop.2005.02.004}
  {\path{doi:10.1016/j.aop.2005.02.004}}.

\bibitem{Voloshin:2005sd}
M.~B. Voloshin, {The $\bar{c} c$ purity of $\psi(3770)$ and $\psi^\prime$
  challenged}, Phys. Rev. D 71 (2005) 114003.
\newblock \href {http://arxiv.org/abs/hep-ph/0504197}
  {\path{arXiv:hep-ph/0504197}}, \href
  {https://doi.org/10.1103/PhysRevD.71.114003}
  {\path{doi:10.1103/PhysRevD.71.114003}}.

\bibitem{Eichten:2007qx}
E.~Eichten, S.~Godfrey, H.~Mahlke, J.~L. Rosner, {Quarkonia and their
  transitions}, Rev. Mod. Phys. 80 (2008) 1161--1193.
\newblock \href {http://arxiv.org/abs/hep-ph/0701208}
  {\path{arXiv:hep-ph/0701208}}, \href
  {https://doi.org/10.1103/RevModPhys.80.1161}
  {\path{doi:10.1103/RevModPhys.80.1161}}.

\bibitem{He:2008xb}
Z.-G. He, Y.~Fan, K.-T. Chao, {QCD prediction for the non-$D\bar{D}$
  annihilation decay of $\psi(3770)$}, Phys. Rev. Lett. 101 (2008) 112001.
\newblock \href {http://arxiv.org/abs/0802.1849} {\path{arXiv:0802.1849}},
  \href {https://doi.org/10.1103/PhysRevLett.101.112001}
  {\path{doi:10.1103/PhysRevLett.101.112001}}.

\bibitem{Qi:2025xkz}
X.-Y. Qi, X.-D. Guo, D.-Y. Chen, {Light meson decays of $D$ wave charmonia},
  Phys. Rev. D 112 (2025) 054003.
\newblock \href {http://arxiv.org/abs/2505.05960} {\path{arXiv:2505.05960}},
  \href {https://doi.org/10.1103/hdmz-b2xq} {\path{doi:10.1103/hdmz-b2xq}}.

\bibitem{Chernyak:1981zz}
V.~L. Chernyak, A.~R. Zhitnitsky, {Exclusive decays of heavy mesons}, Nucl.
  Phys. B 201 (1982) 492, [Erratum: Nucl.Phys.B 214, 547 (1983)].
\newblock \href {https://doi.org/10.1016/0550-3213(83)90251-1}
  {\path{doi:10.1016/0550-3213(83)90251-1}}.

\bibitem{Chernyak:1983ej}
V.~L. Chernyak, A.~R. Zhitnitsky, {Asymptotic behavior of exclusive processes
  in QCD}, Phys. Rept. 112 (1984) 173.
\newblock \href {https://doi.org/10.1016/0370-1573(84)90126-1}
  {\path{doi:10.1016/0370-1573(84)90126-1}}.

\bibitem{Anisovich:1995zu}
V.~V. Anisovich, D.~V. Bugg, A.~V. Sarantsev, B.~S. Zou, {$\Upsilon (3S) \to
  \Upsilon (1S) \pi \pi$ decay: Is the $\pi \pi$ spectrum puzzle an indication
  of a $b \bar{b} q \bar{q}$ resonance?}, Phys. Rev. D 51 (1995) R4619--R4622.
\newblock \href {https://doi.org/10.1103/PhysRevD.51.R4619}
  {\path{doi:10.1103/PhysRevD.51.R4619}}.

\bibitem{Guo:2004dt}
F.~K. Guo, P.~N. Shen, H.~C. Chiang, R.~G. Ping, {Heavy quarkonium pi+ pi-
  transitions and a possible b anti-b q anti-q state}, Nucl. Phys. A 761 (2005)
  269--282.
\newblock \href {http://arxiv.org/abs/hep-ph/0410204}
  {\path{arXiv:hep-ph/0410204}}, \href
  {https://doi.org/10.1016/j.nuclphysa.2005.07.019}
  {\path{doi:10.1016/j.nuclphysa.2005.07.019}}.

\bibitem{Belle:2006wip}
A.~Sokolov, et~al., Belle Collaboration, {Observation of the decay
  $\Upsilon(4S)\to\Upsilon(1S)\pi^+\pi^-$}, Phys. Rev. D 75 (2007) 071103.
\newblock \href {http://arxiv.org/abs/hep-ex/0611026}
  {\path{arXiv:hep-ex/0611026}}, \href
  {https://doi.org/10.1103/PhysRevD.75.071103}
  {\path{doi:10.1103/PhysRevD.75.071103}}.

\bibitem{BaBar:2006udk}
B.~Aubert, et~al., BaBar Collaboration, {Observation of $\Upsilon(4S)$ decays
  to $\pi^+ \pi^- \Upsilon(1S)$ and $\pi^+ \pi^- \Upsilon(2S)$}, Phys. Rev.
  Lett. 96 (2006) 232001.
\newblock \href {http://arxiv.org/abs/hep-ex/0604031}
  {\path{arXiv:hep-ex/0604031}}, \href
  {https://doi.org/10.1103/PhysRevLett.96.232001}
  {\path{doi:10.1103/PhysRevLett.96.232001}}.

\bibitem{Kuang:1981se}
Y.-P. Kuang, T.-M. Yan, {Predictions for hadronic transitions in the $b\bar b$
  system}, Phys. Rev. D 24 (1981) 2874.
\newblock \href {https://doi.org/10.1103/PhysRevD.24.2874}
  {\path{doi:10.1103/PhysRevD.24.2874}}.

\bibitem{Guo:2006ai}
F.-K. Guo, P.-N. Shen, H.-C. Chiang, R.-G. Ping, {On the structure of the
  $\pi\pi$ invariant mass spectra of the $\Upsilon(4S)\to
  \Upsilon(1S,\,2S)\pi^+\pi^-$.}, Phys. Lett. B 658 (2007) 27--32.
\newblock \href {http://arxiv.org/abs/hep-ph/0601120}
  {\path{arXiv:hep-ph/0601120}}, \href
  {https://doi.org/10.1016/j.physletb.2007.10.021}
  {\path{doi:10.1016/j.physletb.2007.10.021}}.

\bibitem{Gao:2006bc}
Y.-J. Gao, Y.-J. Zhang, K.-T. Chao, {Radiative decays of charmonium into light
  mesons}, Chin. Phys. Lett. 23 (2006) 2376--2378.
\newblock \href {http://arxiv.org/abs/hep-ph/0607278}
  {\path{arXiv:hep-ph/0607278}}, \href
  {https://doi.org/10.1088/0256-307X/23/9/008}
  {\path{doi:10.1088/0256-307X/23/9/008}}.

\bibitem{Belle:2012fkf}
R.~Mizuk, et~al., Belle Collaboration, {Evidence for the $\eta_b(2S)$ and
  observation of $h_b(1P) \to \eta_b(1S) \gamma$ and $h_b(2P) \to \eta_b(1S)
  \gamma$}, Phys. Rev. Lett. 109 (2012) 232002.
\newblock \href {http://arxiv.org/abs/1205.6351} {\path{arXiv:1205.6351}},
  \href {https://doi.org/10.1103/PhysRevLett.109.232002}
  {\path{doi:10.1103/PhysRevLett.109.232002}}.

\bibitem{Godfrey:2002rp}
S.~Godfrey, J.~L. Rosner, {Production of singlet $P$-wave $c\bar c$ and $b\bar
  b$ states}, Phys. Rev. D 66 (2002) 014012.
\newblock \href {http://arxiv.org/abs/hep-ph/0205255}
  {\path{arXiv:hep-ph/0205255}}, \href
  {https://doi.org/10.1103/PhysRevD.66.014012}
  {\path{doi:10.1103/PhysRevD.66.014012}}.

\bibitem{Kinnunen:1978qm}
R.~Kinnunen, N.~A. Tornqvist, {Prediction of the $A_1$-$B$, the $Q_1$-$Q_2$
  mass difference and the $Q_1$-$Q_2$ mixing}, Lett. Nuovo Cim. 23 (1978) 517.
\newblock \href {https://doi.org/10.1007/BF02770285}
  {\path{doi:10.1007/BF02770285}}.

\bibitem{Tornqvist:1979hx}
N.~A. Tornqvist, {The meson mass spectrum and unitarity}, Annals Phys. 123
  (1979) 1.
\newblock \href {https://doi.org/10.1016/0003-4916(79)90262-8}
  {\path{doi:10.1016/0003-4916(79)90262-8}}.

\bibitem{Ono:1983rd}
S.~Ono, N.~A. Tornqvist, {Continuum mixing and coupled channel effects in $c
  \bar{c}$ and $b \bar{b}$ quarkonium}, Z. Phys. C 23 (1984) 59.
\newblock \href {https://doi.org/10.1007/BF01558041}
  {\path{doi:10.1007/BF01558041}}.

\bibitem{Ding:1993uy}
Y.-B. Ding, K.-T. Chao, D.-H. Qin, {Screened $Q \bar{Q}$ potential and spectrum
  of heavy quarkonium}, Chin. Phys. Lett. 10 (1993) 460--463.
\newblock \href {https://doi.org/10.1088/0256-307X/10/8/004}
  {\path{doi:10.1088/0256-307X/10/8/004}}.

\bibitem{Ding:1995he}
Y.-B. Ding, K.-T. Chao, D.-H. Qin, {Possible effects of color screening and
  large string tension in heavy quarkonium spectra}, Phys. Rev. D 51 (1995)
  5064--5068.
\newblock \href {http://arxiv.org/abs/hep-ph/9502409}
  {\path{arXiv:hep-ph/9502409}}, \href
  {https://doi.org/10.1103/PhysRevD.51.5064}
  {\path{doi:10.1103/PhysRevD.51.5064}}.

\bibitem{Badalian:1981xj}
A.~M. Badalian, L.~P. Kok, M.~I. Polikarpov, Y.~A. Simonov, {Resonances in
  coupled channels in nuclear and particle physics}, Phys. Rept. 82 (1982)
  31--177.
\newblock \href {https://doi.org/10.1016/0370-1573(82)90014-X}
  {\path{doi:10.1016/0370-1573(82)90014-X}}.

\bibitem{Heikkila:1983wd}
K.~Heikkila, S.~Ono, N.~A. Tornqvist, {Heavy $c \bar{c}$ and $b \bar{b}$
  quarkonium states and unitarity effects}, Phys. Rev. D 29 (1984) 110,
  [Erratum: Phys.Rev.D 29, 2136 (1984)].
\newblock \href {https://doi.org/10.1103/PhysRevD.29.2136}
  {\path{doi:10.1103/PhysRevD.29.2136}}.

\bibitem{Zhou:1990ik}
H.-Y. Zhou, Y.-P. Kuang, {Coupled channel effects in hadronic transitions in
  heavy quarkonium systems}, Phys. Rev. D 44 (1991) 756--769.
\newblock \href {https://doi.org/10.1103/PhysRevD.44.756}
  {\path{doi:10.1103/PhysRevD.44.756}}.

\bibitem{Geiger:1989yc}
P.~Geiger, N.~Isgur, {The quenched approximation in the quark model}, Phys.
  Rev. D 41 (1990) 1595.
\newblock \href {https://doi.org/10.1103/PhysRevD.41.1595}
  {\path{doi:10.1103/PhysRevD.41.1595}}.

\bibitem{Geiger:1991qe}
P.~Geiger, N.~Isgur, {How the Okubo-Zweig-Iizuka rule evades large loop
  corrections}, Phys. Rev. Lett. 67 (1991) 1066--1069.
\newblock \href {https://doi.org/10.1103/PhysRevLett.67.1066}
  {\path{doi:10.1103/PhysRevLett.67.1066}}.

\bibitem{Geiger:1991ab}
P.~Geiger, N.~Isgur, {Reconciling the OZI rule with strong pair creation},
  Phys. Rev. D 44 (1991) 799--808.
\newblock \href {https://doi.org/10.1103/PhysRevD.44.799}
  {\path{doi:10.1103/PhysRevD.44.799}}.

\bibitem{Geiger:1992va}
P.~Geiger, N.~Isgur, {When can hadronic loops scuttle the OZI rule?}, Phys.
  Rev. D 47 (1993) 5050--5059.
\newblock \href {https://doi.org/10.1103/PhysRevD.47.5050}
  {\path{doi:10.1103/PhysRevD.47.5050}}.

\bibitem{Isgur:1998kr}
N.~Isgur, {Spin orbit inversion of excited heavy quark mesons}, Phys. Rev. D 57
  (1998) 4041--4053.
\newblock \href {https://doi.org/10.1103/PhysRevD.57.4041}
  {\path{doi:10.1103/PhysRevD.57.4041}}.

\bibitem{Maiani:2004qj}
L.~Maiani, F.~Piccinini, A.~D. Polosa, V.~Riquer, {$J/\psi$ absorption in heavy
  ion collisions II}, Nucl. Phys. A 748 (2005) 209--225.
\newblock \href {http://arxiv.org/abs/hep-ph/0408150}
  {\path{arXiv:hep-ph/0408150}}, \href
  {https://doi.org/10.1016/j.nuclphysa.2004.10.023}
  {\path{doi:10.1016/j.nuclphysa.2004.10.023}}.

\bibitem{Franklin:1983ve}
M.~E.~B. Franklin, et~al., {Measurement of $\psi(3097)$ and $\psi^\prime$(3686)
  decays into selected hadronic modes}, Phys. Rev. Lett. 51 (1983) 963--966.
\newblock \href {https://doi.org/10.1103/PhysRevLett.51.963}
  {\path{doi:10.1103/PhysRevLett.51.963}}.

\bibitem{Zhu:2006uc}
Y.-S. Zhu, BES Collaboration, {New results from $\psi^\prime$ decays and
  $\psi^{\prime\prime}\to\rho\pi$ decay}, AIP Conf. Proc. 814 (2006) 580--586.
\newblock \href {https://doi.org/10.1063/1.2176546}
  {\path{doi:10.1063/1.2176546}}.

\bibitem{Guo:2007up}
F.-K. Guo, S.~Krewald, U.-G. Meissner, {Hadronic-loop induced mass shifts in
  scalar heavy-light mesons}, Phys. Lett. B 665 (2008) 157--163.
\newblock \href {http://arxiv.org/abs/0712.2953} {\path{arXiv:0712.2953}},
  \href {https://doi.org/10.1016/j.physletb.2008.06.008}
  {\path{doi:10.1016/j.physletb.2008.06.008}}.

\bibitem{Wang:2012mf}
Q.~Wang, G.~Li, Q.~Zhao, {Open charm effects in the explanation of the
  long-standing `$\rho\pi$ puzzle'}, Phys. Rev. D 85 (2012) 074015.
\newblock \href {http://arxiv.org/abs/1201.1681} {\path{arXiv:1201.1681}},
  \href {https://doi.org/10.1103/PhysRevD.85.074015}
  {\path{doi:10.1103/PhysRevD.85.074015}}.

\bibitem{Meng:2007tk}
C.~Meng, K.-T. Chao, {Scalar resonance contributions to the dipion transition
  rates of $\Upsilon(4S,5S)$ in the re-scattering model}, Phys. Rev. D 77
  (2008) 074003.
\newblock \href {http://arxiv.org/abs/0712.3595} {\path{arXiv:0712.3595}},
  \href {https://doi.org/10.1103/PhysRevD.77.074003}
  {\path{doi:10.1103/PhysRevD.77.074003}}.

\bibitem{Meng:2008bq}
C.~Meng, K.-T. Chao, {$\Upsilon(4S,5S) \to \Upsilon(1S) \eta$ transitions in
  the rescattering model and the new BaBar measurement}, Phys. Rev. D 78 (2008)
  074001.
\newblock \href {http://arxiv.org/abs/0806.3259} {\path{arXiv:0806.3259}},
  \href {https://doi.org/10.1103/PhysRevD.78.074001}
  {\path{doi:10.1103/PhysRevD.78.074001}}.

\bibitem{Cheng:2004ru}
H.-Y. Cheng, C.-K. Chua, A.~Soni, {Final state interactions in hadronic $B$
  decays}, Phys. Rev. D 71 (2005) 014030.
\newblock \href {http://arxiv.org/abs/hep-ph/0409317}
  {\path{arXiv:hep-ph/0409317}}, \href
  {https://doi.org/10.1103/PhysRevD.71.014030}
  {\path{doi:10.1103/PhysRevD.71.014030}}.

\bibitem{ParticleDataGroup:2006fqo}
W.~M. Yao, et~al., Particle Data Group Collaboration, {Review of particle
  physics}, J. Phys. G 33 (2006) 1--1232.
\newblock \href {https://doi.org/10.1088/0954-3899/33/1/001}
  {\path{doi:10.1088/0954-3899/33/1/001}}.

\bibitem{Meng:2007cx}
C.~Meng, K.-T. Chao, {Decays of the $X(3872) $ and $\chi_{c1} (2P)$
  charmonium}, Phys. Rev. D 75 (2007) 114002.
\newblock \href {http://arxiv.org/abs/hep-ph/0703205}
  {\path{arXiv:hep-ph/0703205}}, \href
  {https://doi.org/10.1103/PhysRevD.75.114002}
  {\path{doi:10.1103/PhysRevD.75.114002}}.

\bibitem{Liu:2007qi}
X.~Liu, B.~Zhang, L.-L. Shen, S.-L. Zhu, {$X(1576)$ and the final state
  interaction effect}, Phys. Rev. D 75 (2007) 074017.
\newblock \href {http://arxiv.org/abs/hep-ph/0701022}
  {\path{arXiv:hep-ph/0701022}}, \href
  {https://doi.org/10.1103/PhysRevD.75.074017}
  {\path{doi:10.1103/PhysRevD.75.074017}}.

\bibitem{Zhang:2007su}
B.~Zhang, X.~Liu, S.-L. Zhu, {The Dispersive contribution of $\rho(1450,1700)$
  decays and $X(1576)$}, Chin. Phys. Lett. 24 (2007) 2537--2539.
\newblock \href {http://arxiv.org/abs/0705.3082} {\path{arXiv:0705.3082}},
  \href {https://doi.org/10.1088/0256-307X/24/9/020}
  {\path{doi:10.1088/0256-307X/24/9/020}}.

\bibitem{Pennington:2007xr}
M.~R. Pennington, D.~J. Wilson, {Decay channels and charmonium mass-shifts},
  Phys. Rev. D 76 (2007) 077502.
\newblock \href {http://arxiv.org/abs/0704.3384} {\path{arXiv:0704.3384}},
  \href {https://doi.org/10.1103/PhysRevD.76.077502}
  {\path{doi:10.1103/PhysRevD.76.077502}}.

\bibitem{vanBeveren:2007cb}
E.~van Beveren, G.~Rupp, {Relating multichannel scattering and production
  amplitudes in a microscopic OZI-based model}, Annals Phys. 323 (2008)
  1215--1229.
\newblock \href {http://arxiv.org/abs/0706.4119} {\path{arXiv:0706.4119}},
  \href {https://doi.org/10.1016/j.aop.2007.11.012}
  {\path{doi:10.1016/j.aop.2007.11.012}}.

\bibitem{BES:2005sch}
M.~Ablikim, et~al., BES Collaboration, {Search for $\psi^{\prime \prime} \to
  \rho \pi$ at BESII}, Phys. Rev. D 72 (2005) 072007.
\newblock \href {http://arxiv.org/abs/hep-ex/0507092}
  {\path{arXiv:hep-ex/0507092}}, \href
  {https://doi.org/10.1103/PhysRevD.72.072007}
  {\path{doi:10.1103/PhysRevD.72.072007}}.

\bibitem{CLEO:2005zrs}
G.~S. Adams, et~al., CLEO Collaboration, {Decay of the $\psi(3770)$ to light
  hadrons}, Phys. Rev. D 73 (2006) 012002.
\newblock \href {http://arxiv.org/abs/hep-ex/0509011}
  {\path{arXiv:hep-ex/0509011}}, \href
  {https://doi.org/10.1103/PhysRevD.73.012002}
  {\path{doi:10.1103/PhysRevD.73.012002}}.

\bibitem{Close:2000yk}
F.~E. Close, A.~Kirk, {The Mixing of the $f_0(1370)$, $f_0(1500)$ and
  $f_0(1710)$ and the search for the scalar glueball}, Phys. Lett. B 483 (2000)
  345--352.
\newblock \href {http://arxiv.org/abs/hep-ph/0004241}
  {\path{arXiv:hep-ph/0004241}}, \href
  {https://doi.org/10.1016/S0370-2693(00)00623-7}
  {\path{doi:10.1016/S0370-2693(00)00623-7}}.

\bibitem{Li:2007ky}
G.~Li, Q.~Zhao, C.-H. Chang, {Decays of $J/ \psi$ and $\psi^\prime$ into vector
  and pseudoscalar meson and the pseudoscalar glueball-$q \bar{q}$ mixing}, J.
  Phys. G 35 (2008) 055002.
\newblock \href {http://arxiv.org/abs/hep-ph/0701020}
  {\path{arXiv:hep-ph/0701020}}, \href
  {https://doi.org/10.1088/0954-3899/35/5/055002}
  {\path{doi:10.1088/0954-3899/35/5/055002}}.

\bibitem{ParticleDataGroup:2008zun}
C.~Amsler, et~al., Particle Data Group Collaboration, {Review of particle
  physics}, Phys. Lett. B 667 (2008) 1--1340.
\newblock \href {https://doi.org/10.1016/j.physletb.2008.07.018}
  {\path{doi:10.1016/j.physletb.2008.07.018}}.

\bibitem{Dolinsky:1991vq}
S.~I. Dolinsky, et~al., {Summary of experiments with the neutral detector at
  the $e^+ e^-$ storage ring VEPP-2M}, Phys. Rept. 202 (1991) 99--170.
\newblock \href {https://doi.org/10.1016/0370-1573(91)90127-8}
  {\path{doi:10.1016/0370-1573(91)90127-8}}.

\bibitem{Benayoun:1999fv}
M.~Benayoun, L.~DelBuono, S.~Eidelman, V.~N. Ivanchenko, H.~B. O'Connell,
  {Radiative decays, nonet symmetry and SU(3) breaking}, Phys. Rev. D 59 (1999)
  114027.
\newblock \href {http://arxiv.org/abs/hep-ph/9902326}
  {\path{arXiv:hep-ph/9902326}}, \href
  {https://doi.org/10.1103/PhysRevD.59.114027}
  {\path{doi:10.1103/PhysRevD.59.114027}}.

\bibitem{Kucukarslan:2006wk}
A.~Kucukarslan, U.-G. Meissner, {Omega-phi mixing in chiral perturbation
  theory}, Mod. Phys. Lett. A 21 (2006) 1423--1430.
\newblock \href {http://arxiv.org/abs/hep-ph/0603061}
  {\path{arXiv:hep-ph/0603061}}, \href
  {https://doi.org/10.1142/S0217732306020743}
  {\path{doi:10.1142/S0217732306020743}}.

\bibitem{Li:2007xr}
G.~Li, Q.~Zhao, {Hadronic loop contributions to $J/\psi$ and $\psi^\prime$
  radiative decays into $\gamma \eta_c$ or $\gamma \eta_c^\prime$}, Phys. Lett.
  B 670 (2008) 55--60.
\newblock \href {http://arxiv.org/abs/0709.4639} {\path{arXiv:0709.4639}},
  \href {https://doi.org/10.1016/j.physletb.2008.10.033}
  {\path{doi:10.1016/j.physletb.2008.10.033}}.

\bibitem{Li:2011ssa}
G.~Li, Q.~Zhao, {Revisit the radiative decays of $J/\psi$ and $\psi^\prime\to
  \gamma\eta_c (\gamma\eta_c^\prime)$}, Phys. Rev. D 84 (2011) 074005.
\newblock \href {http://arxiv.org/abs/1107.2037} {\path{arXiv:1107.2037}},
  \href {https://doi.org/10.1103/PhysRevD.84.074005}
  {\path{doi:10.1103/PhysRevD.84.074005}}.

\bibitem{ParticleDataGroup:2010dbb}
K.~Nakamura, et~al., Particle Data Group Collaboration, {Review of particle
  physics}, J. Phys. G 37 (2010) 075021.
\newblock \href {https://doi.org/10.1088/0954-3899/37/7A/075021}
  {\path{doi:10.1088/0954-3899/37/7A/075021}}.

\bibitem{BES:2012uhz}
M.~Ablikim, et~al., BES Collaboration, {First observation of the M1 transition
  $\psi(3686)\to \gamma\eta_c(2S)$}, Phys. Rev. Lett. 109 (2012) 042003.
\newblock \href {http://arxiv.org/abs/1205.5103} {\path{arXiv:1205.5103}},
  \href {https://doi.org/10.1103/PhysRevLett.109.042003}
  {\path{doi:10.1103/PhysRevLett.109.042003}}.

\bibitem{CLEO:2008pln}
R.~E. Mitchell, et~al., CLEO Collaboration, {$J/\psi$ and $\psi(2S)$ radiative
  decays to $\eta_c$}, Phys. Rev. Lett. 102 (2009) 011801, [Erratum:
  Phys.Rev.Lett. 106, 159903 (2011)].
\newblock \href {http://arxiv.org/abs/0805.0252} {\path{arXiv:0805.0252}},
  \href {https://doi.org/10.1103/PhysRevLett.102.011801}
  {\path{doi:10.1103/PhysRevLett.102.011801}}.

\bibitem{Dudek:2009kk}
J.~J. Dudek, R.~Edwards, C.~E. Thomas, {Exotic and excited-state radiative
  transitions in charmonium from lattice QCD}, Phys. Rev. D 79 (2009) 094504.
\newblock \href {http://arxiv.org/abs/0902.2241} {\path{arXiv:0902.2241}},
  \href {https://doi.org/10.1103/PhysRevD.79.094504}
  {\path{doi:10.1103/PhysRevD.79.094504}}.

\bibitem{Chen:2013cpa}
D.-Y. Chen, X.~Liu, T.~Matsuki, {Anomalous radiative transitions between
  $h_b(nP)$ and $\eta_b(mS)$ and hadronic loop effect}, Phys. Rev. D 87 (2013)
  094010.
\newblock \href {http://arxiv.org/abs/1304.0372} {\path{arXiv:1304.0372}},
  \href {https://doi.org/10.1103/PhysRevD.87.094010}
  {\path{doi:10.1103/PhysRevD.87.094010}}.

\bibitem{Guo:2009wr}
F.-K. Guo, C.~Hanhart, U.-G. Meissner, {On the extraction of the light quark
  mass ratio from the decays $\psi^\prime\to J/\psi\pi^0(\eta)$}, Phys. Rev.
  Lett. 103 (2009) 082003, [Erratum: Phys.Rev.Lett. 104, 109901 (2010)].
\newblock \href {http://arxiv.org/abs/0907.0521} {\path{arXiv:0907.0521}},
  \href {https://doi.org/10.1103/PhysRevLett.103.082003}
  {\path{doi:10.1103/PhysRevLett.103.082003}}.

\bibitem{Guo:2010ak}
F.-K. Guo, C.~Hanhart, G.~Li, U.-G. Meissner, Q.~Zhao, {Effect of charmed meson
  loops on charmonium transitions}, Phys. Rev. D 83 (2011) 034013.
\newblock \href {http://arxiv.org/abs/1008.3632} {\path{arXiv:1008.3632}},
  \href {https://doi.org/10.1103/PhysRevD.83.034013}
  {\path{doi:10.1103/PhysRevD.83.034013}}.

\bibitem{Ioffe:1979rv}
B.~L. Ioffe, {Masses of light quarks and interaction of low-energy $eta$
  mesons. (In Russian)}, Yad. Fiz. 29 (1979) 1611--1619.

\bibitem{Ioffe:1980mx}
B.~L. Ioffe, M.~A. Shifman, {The decays $\psi^\prime\to J/\psi+\pi^0(\eta)$ and
  quark masses}, Phys. Lett. B 95 (1980) 99--102.
\newblock \href {https://doi.org/10.1016/0370-2693(80)90409-8}
  {\path{doi:10.1016/0370-2693(80)90409-8}}.

\bibitem{Donoghue:1989sj}
J.~F. Donoghue, {Light quark masses and chiral symmetry}, Ann. Rev. Nucl. Part.
  Sci. 39 (1989) 1--17.
\newblock \href {https://doi.org/10.1146/annurev.ns.39.120189.000245}
  {\path{doi:10.1146/annurev.ns.39.120189.000245}}.

\bibitem{Meissner:1993ah}
U.~G. Meissner, {Recent developments in chiral perturbation theory}, Rept.
  Prog. Phys. 56 (1993) 903--996.
\newblock \href {http://arxiv.org/abs/hep-ph/9302247}
  {\path{arXiv:hep-ph/9302247}}, \href
  {https://doi.org/10.1088/0034-4885/56/8/001}
  {\path{doi:10.1088/0034-4885/56/8/001}}.

\bibitem{Leutwyler:1996eq}
H.~Leutwyler, {Light quark masses}, NATO Sci. Ser. B 363 (1997) 149--164.
\newblock \href {http://arxiv.org/abs/hep-ph/9609467}
  {\path{arXiv:hep-ph/9609467}}.

\bibitem{Weinberg:1977hb}
S.~Weinberg, {The problem of mass}, Trans. New York Acad. Sci. 38 (1977)
  185--201.
\newblock \href {https://doi.org/10.1111/j.2164-0947.1977.tb02958.x}
  {\path{doi:10.1111/j.2164-0947.1977.tb02958.x}}.

\bibitem{Leutwyler:1996sa}
H.~Leutwyler, {Bounds on the light quark masses}, Phys. Lett. B 374 (1996)
  163--168.
\newblock \href {http://arxiv.org/abs/hep-ph/9601234}
  {\path{arXiv:hep-ph/9601234}}, \href
  {https://doi.org/10.1016/0370-2693(96)85876-X}
  {\path{doi:10.1016/0370-2693(96)85876-X}}.

\bibitem{Leutwyler:1996qg}
H.~Leutwyler, {The ratios of the light quark masses}, Phys. Lett. B 378 (1996)
  313--318.
\newblock \href {http://arxiv.org/abs/hep-ph/9602366}
  {\path{arXiv:hep-ph/9602366}}, \href
  {https://doi.org/10.1016/0370-2693(96)00386-3}
  {\path{doi:10.1016/0370-2693(96)00386-3}}.

\bibitem{Guo:2010zk}
F.-K. Guo, C.~Hanhart, G.~Li, U.-G. Meissner, Q.~Zhao, {Novel analysis of the
  decays $\psi^\prime \to h_{c} \pi^0$ and $\eta^\prime_c \to \chi_{c0}
  \pi^0$}, Phys. Rev. D 82 (2010) 034025.
\newblock \href {http://arxiv.org/abs/1002.2712} {\path{arXiv:1002.2712}},
  \href {https://doi.org/10.1103/PhysRevD.82.034025}
  {\path{doi:10.1103/PhysRevD.82.034025}}.

\bibitem{Guo:2010ca}
F.-K. Guo, C.~Hanhart, U.-G. Meissner, {Extracting the light quark mass ratio
  $m_u/m_d$ from bottomonia transitions}, Phys. Rev. Lett. 105 (2010) 162001.
\newblock \href {http://arxiv.org/abs/1007.4682} {\path{arXiv:1007.4682}},
  \href {https://doi.org/10.1103/PhysRevLett.105.162001}
  {\path{doi:10.1103/PhysRevLett.105.162001}}.

\bibitem{Guo:2011dv}
F.-K. Guo, U.-G. Meissner, {Examining coupled-channel effects in radiative
  charmonium transitions}, Phys. Rev. Lett. 108 (2012) 112002.
\newblock \href {http://arxiv.org/abs/1111.1151} {\path{arXiv:1111.1151}},
  \href {https://doi.org/10.1103/PhysRevLett.108.112002}
  {\path{doi:10.1103/PhysRevLett.108.112002}}.

\bibitem{Guo:2012tg}
F.-K. Guo, U.-G. Meissner, {Light quark mass dependence in heavy quarkonium
  physics}, Phys. Rev. Lett. 109 (2012) 062001.
\newblock \href {http://arxiv.org/abs/1203.1116} {\path{arXiv:1203.1116}},
  \href {https://doi.org/10.1103/PhysRevLett.109.062001}
  {\path{doi:10.1103/PhysRevLett.109.062001}}.

\bibitem{Mehen:2011tp}
T.~Mehen, D.-L. Yang, {On the role of charmed meson loops in charmonium
  decays}, Phys. Rev. D 85 (2012) 014002.
\newblock \href {http://arxiv.org/abs/1111.3884} {\path{arXiv:1111.3884}},
  \href {https://doi.org/10.1103/PhysRevD.85.014002}
  {\path{doi:10.1103/PhysRevD.85.014002}}.

\bibitem{Guo:2014qra}
F.-K. Guo, U.-G. Mei{\ss}ner, C.-P. Shen, {Enhanced breaking of heavy quark
  spin symmetry}, Phys. Lett. B 738 (2014) 172--177.
\newblock \href {http://arxiv.org/abs/1406.6543} {\path{arXiv:1406.6543}},
  \href {https://doi.org/10.1016/j.physletb.2014.09.043}
  {\path{doi:10.1016/j.physletb.2014.09.043}}.

\bibitem{Cleven:2011gp}
M.~Cleven, F.-K. Guo, C.~Hanhart, U.-G. Meissner, {Bound state nature of the
  exotic $Z_b$ states}, Eur. Phys. J. A 47 (2011) 120.
\newblock \href {http://arxiv.org/abs/1107.0254} {\path{arXiv:1107.0254}},
  \href {https://doi.org/10.1140/epja/i2011-11120-6}
  {\path{doi:10.1140/epja/i2011-11120-6}}.

\bibitem{Cleven:2013sq}
M.~Cleven, Q.~Wang, F.-K. Guo, C.~Hanhart, U.-G. Meissner, Q.~Zhao, {Confirming
  the molecular nature of the $Z_b(10610)$ and the $Z_b(10650)$}, Phys. Rev. D
  87~(7) (2013) 074006.
\newblock \href {http://arxiv.org/abs/1301.6461} {\path{arXiv:1301.6461}},
  \href {https://doi.org/10.1103/PhysRevD.87.074006}
  {\path{doi:10.1103/PhysRevD.87.074006}}.

\bibitem{Guo:2013zbw}
F.-K. Guo, C.~Hanhart, U.-G. Mei{\ss}ner, Q.~Wang, Q.~Zhao, {Production of the
  X(3872) in charmonia radiative decays}, Phys. Lett. B 725 (2013) 127--133.
\newblock \href {http://arxiv.org/abs/1306.3096} {\path{arXiv:1306.3096}},
  \href {https://doi.org/10.1016/j.physletb.2013.06.053}
  {\path{doi:10.1016/j.physletb.2013.06.053}}.

\bibitem{Esposito:2014hsa}
A.~Esposito, A.~L. Guerrieri, A.~Pilloni, {Probing the nature of
  $Z_c^{(\prime)}$ states via the $\eta_c\rho$ decay}, Phys. Lett. B 746 (2015)
  194--201.
\newblock \href {http://arxiv.org/abs/1409.3551} {\path{arXiv:1409.3551}},
  \href {https://doi.org/10.1016/j.physletb.2015.04.057}
  {\path{doi:10.1016/j.physletb.2015.04.057}}.

\bibitem{Mehen:2015efa}
T.~Mehen, {Hadronic loops versus factorization in effective field theory
  calculations of $X(3872)\to\chi_{cJ}\pi^0$}, Phys. Rev. D 92~(3) (2015)
  034019.
\newblock \href {http://arxiv.org/abs/1503.02719} {\path{arXiv:1503.02719}},
  \href {https://doi.org/10.1103/PhysRevD.92.034019}
  {\path{doi:10.1103/PhysRevD.92.034019}}.

\bibitem{Abreu:2016xlr}
L.~M. Abreu, A.~Lafayette~Vasconcellos, {Production of $Z_b^{(')}$ states in
  heavy-meson effective theory}, Phys. Rev. D 94~(9) (2016) 096009.
\newblock \href {https://doi.org/10.1103/PhysRevD.94.096009}
  {\path{doi:10.1103/PhysRevD.94.096009}}.

\bibitem{Huo:2015uka}
W.-S. Huo, G.-Y. Chen, {The nature of $Z_b$ states from a combined analysis of
  $\Upsilon (5S)\rightarrow h_b(mP) \pi ^+ \pi ^-$ and $\Upsilon
  (5S)\rightarrow B^{(*)}\bar{B}^{(*)}\pi $}, Eur. Phys. J. C 76~(3) (2016)
  172.
\newblock \href {http://arxiv.org/abs/1501.02189} {\path{arXiv:1501.02189}},
  \href {https://doi.org/10.1140/epjc/s10052-016-4013-0}
  {\path{doi:10.1140/epjc/s10052-016-4013-0}}.

\bibitem{Wu:2016dws}
Q.~Wu, G.~Li, F.~Shao, Q.~Wang, R.~Wang, Y.~Zhang, Y.~Zheng, {Production of
  $X_b$ in $\Upsilon(5S, 6S)\to \gamma X_b$ Radiative Decays}, Adv. High Energy
  Phys. 2016 (2016) 3729050.
\newblock \href {http://arxiv.org/abs/1606.05118} {\path{arXiv:1606.05118}},
  \href {https://doi.org/10.1155/2016/3729050}
  {\path{doi:10.1155/2016/3729050}}.

\bibitem{Chen:2016mjn}
Y.-H. Chen, M.~Cleven, J.~T. Daub, F.-K. Guo, C.~Hanhart, B.~Kubis, U.-G.
  Mei{\ss}ner, B.-S. Zou, {Effects of $Z_b$ states and bottom meson loops on
  $\Upsilon(4S) \to \Upsilon(1S,2S) \pi^+\pi^-$ transitions}, Phys. Rev. D 95
  (2017) 034022.
\newblock \href {http://arxiv.org/abs/1611.00913} {\path{arXiv:1611.00913}},
  \href {https://doi.org/10.1103/PhysRevD.95.034022}
  {\path{doi:10.1103/PhysRevD.95.034022}}.

\bibitem{Guo:2016yxl}
F.-K. Guo, U.-G. Mei{\ss}ner, Z.~Yang, {Hindered magnetic dipole transitions
  between P-wave bottomonia and coupled-channel effects}, Phys. Lett. B 760
  (2016) 417--421.
\newblock \href {http://arxiv.org/abs/1604.00770} {\path{arXiv:1604.00770}},
  \href {https://doi.org/10.1016/j.physletb.2016.07.023}
  {\path{doi:10.1016/j.physletb.2016.07.023}}.

\bibitem{Colangelo:1989gi}
P.~Colangelo, G.~Nardulli, N.~Paver, Riazuddin, {Long distance effects in $b
  \to s$ exclusive decays}, Z. Phys. C 45 (1990) 575.
\newblock \href {https://doi.org/10.1007/BF01556270}
  {\path{doi:10.1007/BF01556270}}.

\bibitem{Beneke:1999br}
M.~Beneke, G.~Buchalla, M.~Neubert, C.~T. Sachrajda, {QCD factorization for $B
  \to \pi \pi$ decays: Strong phases and $CP$ violation in the heavy quark
  limit}, Phys. Rev. Lett. 83 (1999) 1914--1917.
\newblock \href {http://arxiv.org/abs/hep-ph/9905312}
  {\path{arXiv:hep-ph/9905312}}, \href
  {https://doi.org/10.1103/PhysRevLett.83.1914}
  {\path{doi:10.1103/PhysRevLett.83.1914}}.

\bibitem{Beneke:2000ry}
M.~Beneke, G.~Buchalla, M.~Neubert, C.~T. Sachrajda, {QCD factorization for
  exclusive, nonleptonic $B$ meson decays: General arguments and the case of
  heavy light final states}, Nucl. Phys. B 591 (2000) 313--418.
\newblock \href {http://arxiv.org/abs/hep-ph/0006124}
  {\path{arXiv:hep-ph/0006124}}, \href
  {https://doi.org/10.1016/S0550-3213(00)00559-9}
  {\path{doi:10.1016/S0550-3213(00)00559-9}}.

\bibitem{Cheng:2000kt}
H.-Y. Cheng, K.-C. Yang, {$B \to J / \psi K$ decays in QCD factorization},
  Phys. Rev. D 63 (2001) 074011.
\newblock \href {http://arxiv.org/abs/hep-ph/0011179}
  {\path{arXiv:hep-ph/0011179}}, \href
  {https://doi.org/10.1103/PhysRevD.63.074011}
  {\path{doi:10.1103/PhysRevD.63.074011}}.

\bibitem{Keum:2000wi}
Y.~Y. Keum, H.-N. Li, A.~I. Sanda, {Penguin enhancement and $B \to K \pi$
  decays in perturbative QCD}, Phys. Rev. D 63 (2001) 054008.
\newblock \href {http://arxiv.org/abs/hep-ph/0004173}
  {\path{arXiv:hep-ph/0004173}}, \href
  {https://doi.org/10.1103/PhysRevD.63.054008}
  {\path{doi:10.1103/PhysRevD.63.054008}}.

\bibitem{Keum:2000ph}
Y.-Y. Keum, H.-n. Li, A.~I. Sanda, {Fat penguins and imaginary penguins in
  perturbative QCD}, Phys. Lett. B 504 (2001) 6--14.
\newblock \href {http://arxiv.org/abs/hep-ph/0004004}
  {\path{arXiv:hep-ph/0004004}}, \href
  {https://doi.org/10.1016/S0370-2693(01)00247-7}
  {\path{doi:10.1016/S0370-2693(01)00247-7}}.

\bibitem{Lu:2000em}
C.-D. Lu, K.~Ukai, M.-Z. Yang, {Branching ratio and $CP$ violation of $B \to
  \pi \pi$ decays in perturbative QCD approach}, Phys. Rev. D 63 (2001) 074009.
\newblock \href {http://arxiv.org/abs/hep-ph/0004213}
  {\path{arXiv:hep-ph/0004213}}, \href
  {https://doi.org/10.1103/PhysRevD.63.074009}
  {\path{doi:10.1103/PhysRevD.63.074009}}.

\bibitem{Beneke:2001ev}
M.~Beneke, G.~Buchalla, M.~Neubert, C.~T. Sachrajda, {QCD factorization in $B
  \to \pi K, \pi \pi$ decays and extraction of Wolfenstein parameters}, Nucl.
  Phys. B 606 (2001) 245--321.
\newblock \href {http://arxiv.org/abs/hep-ph/0104110}
  {\path{arXiv:hep-ph/0104110}}, \href
  {https://doi.org/10.1016/S0550-3213(01)00251-6}
  {\path{doi:10.1016/S0550-3213(01)00251-6}}.

\bibitem{Song:2002mh}
Z.-z. Song, K.-T. Chao, {Problems of QCD factorization in exclusive decays of
  $B$ meson to charmonium}, Phys. Lett. B 568 (2003) 127--134.
\newblock \href {http://arxiv.org/abs/hep-ph/0206253}
  {\path{arXiv:hep-ph/0206253}}, \href
  {https://doi.org/10.1016/j.physletb.2003.06.062}
  {\path{doi:10.1016/j.physletb.2003.06.062}}.

\bibitem{Beneke:2003zv}
M.~Beneke, M.~Neubert, {QCD factorization for $B \to PP$ and $B \to PV$
  decays}, Nucl. Phys. B 675 (2003) 333--415.
\newblock \href {http://arxiv.org/abs/hep-ph/0308039}
  {\path{arXiv:hep-ph/0308039}}, \href
  {https://doi.org/10.1016/j.nuclphysb.2003.09.026}
  {\path{doi:10.1016/j.nuclphysb.2003.09.026}}.

\bibitem{Colangelo:2002mj}
P.~Colangelo, F.~De~Fazio, T.~N. Pham, {$B^- \to K^- \chi_{c0}$ decay from
  charmed meson rescattering}, Phys. Lett. B 542 (2002) 71--79.
\newblock \href {http://arxiv.org/abs/hep-ph/0207061}
  {\path{arXiv:hep-ph/0207061}}, \href
  {https://doi.org/10.1016/S0370-2693(02)02306-7}
  {\path{doi:10.1016/S0370-2693(02)02306-7}}.

\bibitem{Colangelo:2003sa}
P.~Colangelo, F.~De~Fazio, T.~N. Pham, {Nonfactorizable contributions in $B$
  decays to charmonium: The Case of $B^- \to K^- h_c$}, Phys. Rev. D 69 (2004)
  054023.
\newblock \href {http://arxiv.org/abs/hep-ph/0310084}
  {\path{arXiv:hep-ph/0310084}}, \href
  {https://doi.org/10.1103/PhysRevD.69.054023}
  {\path{doi:10.1103/PhysRevD.69.054023}}.

\bibitem{Liu:2007qs}
X.~Liu, X.-Q. Li, {Effects of hadronic loops on the direct $CP$ violation of
  $B_c$}, Phys. Rev. D 77 (2008) 096010.
\newblock \href {http://arxiv.org/abs/0707.0919} {\path{arXiv:0707.0919}},
  \href {https://doi.org/10.1103/PhysRevD.77.096010}
  {\path{doi:10.1103/PhysRevD.77.096010}}.

\bibitem{Fajfer:2003ag}
S.~Fajfer, A.~Prapotnik, P.~Singer, J.~Zupan, {Final state interactions in the
  $D^+_s \to \omega \pi^+$ and $D^+_s \to \rho^0 \pi^+$ decays}, Phys. Rev. D
  68 (2003) 094012.
\newblock \href {http://arxiv.org/abs/hep-ph/0308100}
  {\path{arXiv:hep-ph/0308100}}, \href
  {https://doi.org/10.1103/PhysRevD.68.094012}
  {\path{doi:10.1103/PhysRevD.68.094012}}.

\bibitem{Yu:2021euw}
Y.~Yu, Y.-K. Hsiao, B.-C. Ke, {Study of the $D_s^+ \to a_0(980) \rho $ and
  $a_0(980) \omega $ decays}, Eur. Phys. J. C 81 (2021) 1093.
\newblock \href {http://arxiv.org/abs/2108.02936} {\path{arXiv:2108.02936}},
  \href {https://doi.org/10.1140/epjc/s10052-021-09895-y}
  {\path{doi:10.1140/epjc/s10052-021-09895-y}}.

\bibitem{Hsiao:2019ait}
Y.-K. Hsiao, Y.~Yu, B.-C. Ke, {Resonant $a_0(980)$ state in triangle
  rescattering $D_s^+\to \pi ^+\pi ^0\eta $ decays}, Eur. Phys. J. C 80 (2020)
  895.
\newblock \href {http://arxiv.org/abs/1909.07327} {\path{arXiv:1909.07327}},
  \href {https://doi.org/10.1140/epjc/s10052-020-08468-9}
  {\path{doi:10.1140/epjc/s10052-020-08468-9}}.

\bibitem{Ling:2021qzl}
X.-Z. Ling, M.-Z. Liu, J.-X. Lu, L.-S. Geng, J.-J. Xie, {Can the nature of
  $a_0(980)$ be tested in the $D_s^{+}\to \pi^{+}\pi^0 \eta$ decay?}, Phys.
  Rev. D 103 (2021) 116016.
\newblock \href {http://arxiv.org/abs/2102.05349} {\path{arXiv:2102.05349}},
  \href {https://doi.org/10.1103/PhysRevD.103.116016}
  {\path{doi:10.1103/PhysRevD.103.116016}}.

\bibitem{Bediaga:2022sxw}
I.~Bediaga, T.~Frederico, P.~C. Magalh{\~a}es, {Enhanced charm $CP$ asymmetries
  from final state interactions}, Phys. Rev. Lett. 131 (2023) 051802.
\newblock \href {http://arxiv.org/abs/2203.04056} {\path{arXiv:2203.04056}},
  \href {https://doi.org/10.1103/PhysRevLett.131.051802}
  {\path{doi:10.1103/PhysRevLett.131.051802}}.

\bibitem{Pich:2023kim}
A.~Pich, E.~Solomonidi, L.~Vale~Silva, {Final-state interactions in the $CP$
  asymmetries of charm-meson two-body decays}, Phys. Rev. D 108 (2023) 036026.
\newblock \href {http://arxiv.org/abs/2305.11951} {\path{arXiv:2305.11951}},
  \href {https://doi.org/10.1103/PhysRevD.108.036026}
  {\path{doi:10.1103/PhysRevD.108.036026}}.

\bibitem{Geng:2024uxp}
C.-Q. Geng, X.-N. Jin, C.-W. Liu, X.~Yu, {Hidden strangeness in meson weak
  decays to baryon pair}, Phys. Rev. D 110 (2024) 113008.
\newblock \href {http://arxiv.org/abs/2409.11374} {\path{arXiv:2409.11374}},
  \href {https://doi.org/10.1103/PhysRevD.110.113008}
  {\path{doi:10.1103/PhysRevD.110.113008}}.

\bibitem{Wang:2025mdn}
Y.-L. Wang, S.-T. Cai, Y.-K. Hsiao, {Rescattering-induced $D\to SS$ weak
  decays} (9 2025).
\newblock \href {http://arxiv.org/abs/2509.16879} {\path{arXiv:2509.16879}}.

\bibitem{Yu:2020vlt}
Y.~Yu, Y.-K. Hsiao, {Cabibbo-favored $\Lambda_c^+ \to \Lambda a_0(980)^+$ decay
  in the final state interaction}, Phys. Lett. B 820 (2021) 136586.
\newblock \href {http://arxiv.org/abs/2012.14575} {\path{arXiv:2012.14575}},
  \href {https://doi.org/10.1016/j.physletb.2021.136586}
  {\path{doi:10.1016/j.physletb.2021.136586}}.

\bibitem{Jia:2024pyb}
C.-P. Jia, H.-Y. Jiang, J.-P. Wang, F.-S. Yu, {Final-state rescattering
  mechanism of charmed baryon decays}, JHEP 11 (2024) 072.
\newblock \href {http://arxiv.org/abs/2408.14959} {\path{arXiv:2408.14959}},
  \href {https://doi.org/10.1007/JHEP11(2024)072}
  {\path{doi:10.1007/JHEP11(2024)072}}.

\bibitem{He:2024unv}
X.-G. He, C.-W. Liu, {Large $CP$ violation in charmed baryon decays}, Sci.
  Bull. 70 (2025) 2598--2603.
\newblock \href {http://arxiv.org/abs/2404.19166} {\path{arXiv:2404.19166}},
  \href {https://doi.org/10.1016/j.scib.2025.05.045}
  {\path{doi:10.1016/j.scib.2025.05.045}}.

\bibitem{Wang:2024oyi}
J.-P. Wang, F.-S. Yu, {$CP$ violation of baryon decays with $N \pi$
  rescatterings}, Chin. Phys. C 48 (2024) 101002.
\newblock \href {http://arxiv.org/abs/2407.04110} {\path{arXiv:2407.04110}},
  \href {https://doi.org/10.1088/1674-1137/ad75f4}
  {\path{doi:10.1088/1674-1137/ad75f4}}.

\bibitem{Duan:2024zjv}
Z.-D. Duan, J.-P. Wang, R.-H. Li, C.-D. L{\"u}, F.-S. Yu, {Final-state
  rescattering mechanism in bottom-baryon decays}, JHEP 09 (2025) 160.
\newblock \href {http://arxiv.org/abs/2412.20458} {\path{arXiv:2412.20458}},
  \href {https://doi.org/10.1007/JHEP09(2025)160}
  {\path{doi:10.1007/JHEP09(2025)160}}.

\bibitem{Hsiao:2024szt}
Y.-K. Hsiao, S.-T. Cai, Y.-L. Wang, {Hidden charm $P_{cs}(4338)^0$ production
  in baryonic $B^-\to J/\psi \Lambda \bar{p}$ decay}, Phys. Rev. D 111 (2025)
  076020.
\newblock \href {http://arxiv.org/abs/2409.04951} {\path{arXiv:2409.04951}},
  \href {https://doi.org/10.1103/PhysRevD.111.076020}
  {\path{doi:10.1103/PhysRevD.111.076020}}.

\bibitem{Yu:2017zst}
F.-S. Yu, H.-Y. Jiang, R.-H. Li, C.-D. L{\"u}, W.~Wang, Z.-X. Zhao, {Discovery
  potentials of doubly charmed baryons}, Chin. Phys. C 42 (2018) 051001.
\newblock \href {http://arxiv.org/abs/1703.09086} {\path{arXiv:1703.09086}},
  \href {https://doi.org/10.1088/1674-1137/42/5/051001}
  {\path{doi:10.1088/1674-1137/42/5/051001}}.

\bibitem{Li:2020qrh}
R.-H. Li, J.-J. Hou, B.~He, Y.-R. Wang, {Weak decays of doubly heavy baryons:
  ${\cal B}_{cc}\to {\cal B} D^{(*)}$}, Chin. Phys. C 45 (2021) 043108.
\newblock \href {http://arxiv.org/abs/2010.09362} {\path{arXiv:2010.09362}},
  \href {https://doi.org/10.1088/1674-1137/abe0bc}
  {\path{doi:10.1088/1674-1137/abe0bc}}.

\bibitem{Han:2021azw}
J.-J. Han, H.-Y. Jiang, W.~Liu, Z.-J. Xiao, F.-S. Yu, {Rescattering mechanism
  of weak decays of double-charm baryons}, Chin. Phys. C 45 (2021) 053105.
\newblock \href {http://arxiv.org/abs/2101.12019} {\path{arXiv:2101.12019}},
  \href {https://doi.org/10.1088/1674-1137/abec68}
  {\path{doi:10.1088/1674-1137/abec68}}.

\bibitem{Hu:2024uia}
X.-H. Hu, C.-P. Jia, Y.~Xing, F.-S. Yu, {Final-state rescattering mechanism of
  double-charm baryon decays: $B_{cc} \to B_cP$}, Phys. Rev. D 111 (2025)
  076002.
\newblock \href {http://arxiv.org/abs/2403.09511} {\path{arXiv:2403.09511}},
  \href {https://doi.org/10.1103/PhysRevD.111.076002}
  {\path{doi:10.1103/PhysRevD.111.076002}}.

\bibitem{Geng:2025gtd}
C.-Q. Geng, X.-N. Jin, C.-W. Liu, X.-Y. Liu, X.~Yu, {Potential of discovering
  $\Xi_{cc}^+$ in $\Lambda_b $ decays} (10 2025).
\newblock \href {http://arxiv.org/abs/2510.22181} {\path{arXiv:2510.22181}}.

\bibitem{Xu:2016kbn}
H.~Xu, X.~Liu, T.~Matsuki, {Understanding $B^- \to X(3823)K^-$ via rescattering
  mechanism and predicting $B^-\to \eta_{c2} (^1D_2)/\psi_3(^3D_3)K^-$}, Phys.
  Rev. D 94 (2016) 034005.
\newblock \href {http://arxiv.org/abs/1605.04776} {\path{arXiv:1605.04776}},
  \href {https://doi.org/10.1103/PhysRevD.94.034005}
  {\path{doi:10.1103/PhysRevD.94.034005}}.

\bibitem{Duan:2021bna}
M.-X. Duan, J.-Z. Wang, Y.-S. Li, X.~Liu, {Role of the newly measured $B \to
  KD\bar{D}$ process to establish $\chi_{c0}(2P)$ state}, Phys. Rev. D 104
  (2021) 034035.
\newblock \href {http://arxiv.org/abs/2104.09132} {\path{arXiv:2104.09132}},
  \href {https://doi.org/10.1103/PhysRevD.104.034035}
  {\path{doi:10.1103/PhysRevD.104.034035}}.

\bibitem{Yuan:2025pnt}
Q.-W. Yuan, Q.~Wu, M.-Z. Liu, {$B$-meson decays to vector charmoniumlike states
  and a $K$ meson: The role of final-state interactions}, Phys. Rev. D 111
  (2025) 114035.
\newblock \href {http://arxiv.org/abs/2504.11121} {\path{arXiv:2504.11121}},
  \href {https://doi.org/10.1103/xrjr-zxpz} {\path{doi:10.1103/xrjr-zxpz}}.

\bibitem{Liu:2010um}
X.-H. Liu, Q.~Zhao, {Further study of the helicity selection rule evading
  mechanism in $\eta_c$, $\chi_{c0}$ and $h_c$ decaying to baryon anti-baryon
  pairs}, J. Phys. G 38 (2011) 035007.
\newblock \href {http://arxiv.org/abs/1004.0496} {\path{arXiv:1004.0496}},
  \href {https://doi.org/10.1088/0954-3899/38/3/035007}
  {\path{doi:10.1088/0954-3899/38/3/035007}}.

\bibitem{Chen:2010nv}
D.-Y. Chen, J.~He, X.~Liu, {Nonresonant explanation for the $Y(4260)$ structure
  observed in the $e^+e^-\to J/\psi\pi^+\pi^-$ process}, Phys. Rev. D 83 (2011)
  054021.
\newblock \href {http://arxiv.org/abs/1012.5362} {\path{arXiv:1012.5362}},
  \href {https://doi.org/10.1103/PhysRevD.83.054021}
  {\path{doi:10.1103/PhysRevD.83.054021}}.

\bibitem{Chen:2011kc}
D.-Y. Chen, J.~He, X.~Liu, {A Novel explanation of charmonium-like structure in
  $e^+e^-\to \psi(2S)\pi^+\pi^-$}, Phys. Rev. D 83 (2011) 074012.
\newblock \href {http://arxiv.org/abs/1101.2474} {\path{arXiv:1101.2474}},
  \href {https://doi.org/10.1103/PhysRevD.83.074012}
  {\path{doi:10.1103/PhysRevD.83.074012}}.

\bibitem{Wang:2011yh}
Q.~Wang, X.-H. Liu, Q.~Zhao, {Open charm effects in $e^+e^-\to J/\psi \eta$,
  $J/\psi \pi^0$ and $\phi\eta_c$}, Phys. Rev. D 84 (2011) 014007.
\newblock \href {http://arxiv.org/abs/1103.1095} {\path{arXiv:1103.1095}},
  \href {https://doi.org/10.1103/PhysRevD.84.014007}
  {\path{doi:10.1103/PhysRevD.84.014007}}.

\bibitem{Wang:2012wj}
Q.~Wang, X.-H. Liu, Q.~Zhao, {Updated study of the $\eta_c$ and $\eta_c^\prime$
  decays into light vector mesons}, Phys. Lett. B 711 (2012) 364--370.
\newblock \href {http://arxiv.org/abs/1202.3026} {\path{arXiv:1202.3026}},
  \href {https://doi.org/10.1016/j.physletb.2012.04.022}
  {\path{doi:10.1016/j.physletb.2012.04.022}}.

\bibitem{Chen:2012nva}
D.-Y. Chen, X.~Liu, T.~Matsuki, {$\eta$ transitions between charmonia with
  meson loop contributions}, Phys. Rev. D 87 (2013) 054006.
\newblock \href {http://arxiv.org/abs/1209.0064} {\path{arXiv:1209.0064}},
  \href {https://doi.org/10.1103/PhysRevD.87.054006}
  {\path{doi:10.1103/PhysRevD.87.054006}}.

\bibitem{Guo:2012tj}
Z.-k. Guo, S.~Narison, J.-M. Richard, Q.~Zhao, {Isospin violating decay of
  $\psi(3770) \to J/\psi + \pi^0$}, Phys. Rev. D 85 (2012) 114007.
\newblock \href {http://arxiv.org/abs/1204.1448} {\path{arXiv:1204.1448}},
  \href {https://doi.org/10.1103/PhysRevD.85.114007}
  {\path{doi:10.1103/PhysRevD.85.114007}}.

\bibitem{Li:2013jma}
G.~Li, X.-H. Liu, Q.~Zhao, {Evasion of HSR in $S$-wave charmonium decaying to
  $P$-wave light hadrons}, Eur. Phys. J. C 73 (2013) 2576.
\newblock \href {https://doi.org/10.1140/epjc/s10052-013-2576-6}
  {\path{doi:10.1140/epjc/s10052-013-2576-6}}.

\bibitem{Li:2013xia}
G.~Li, {Hidden-charmonium decays of $Z_c(3900)$ and $Z_c(4025)$ in intermediate
  meson loops model}, Eur. Phys. J. C 73 (2013) 2621.
\newblock \href {http://arxiv.org/abs/1304.4458} {\path{arXiv:1304.4458}},
  \href {https://doi.org/10.1140/epjc/s10052-013-2621-5}
  {\path{doi:10.1140/epjc/s10052-013-2621-5}}.

\bibitem{Li:2013yla}
G.~Li, X.-H. Liu, {Investigating possible decay modes of $Y(4260)$ under
  $D_1(2420)\bar{D}$ + $c.c.$ molecular state ansatz}, Phys. Rev. D 88 (2013)
  094008.
\newblock \href {http://arxiv.org/abs/1307.2622} {\path{arXiv:1307.2622}},
  \href {https://doi.org/10.1103/PhysRevD.88.094008}
  {\path{doi:10.1103/PhysRevD.88.094008}}.

\bibitem{Chen:2013yxa}
D.-Y. Chen, X.~Liu, T.~Matsuki, {Hidden-charm decays of $X(3915)$ and $Z(3930)$
  as the $P$-wave charmonia}, PTEP 2015 (2015) 043B05.
\newblock \href {http://arxiv.org/abs/1311.6274} {\path{arXiv:1311.6274}},
  \href {https://doi.org/10.1093/ptep/ptv038} {\path{doi:10.1093/ptep/ptv038}}.

\bibitem{Chen:2014sra}
D.-Y. Chen, X.~Liu, T.~Matsuki, {Observation of $e^+e^-\to \chi_{c0}\omega$ and
  missing higher charmonium $\psi(4S)$}, Phys. Rev. D 91 (2015) 094023.
\newblock \href {http://arxiv.org/abs/1411.5136} {\path{arXiv:1411.5136}},
  \href {https://doi.org/10.1103/PhysRevD.91.094023}
  {\path{doi:10.1103/PhysRevD.91.094023}}.

\bibitem{Dong:2013kta}
Y.~Dong, A.~Faessler, T.~Gutsche, V.~E. Lyubovitskij, {Selected strong decay
  modes of $Y(4260)$}, Phys. Rev. D 89 (2014) 034018.
\newblock \href {http://arxiv.org/abs/1310.4373} {\path{arXiv:1310.4373}},
  \href {https://doi.org/10.1103/PhysRevD.89.034018}
  {\path{doi:10.1103/PhysRevD.89.034018}}.

\bibitem{Li:2014gxa}
G.~Li, C.-S. An, P.-Y. Li, D.~Liu, X.~Zhang, Z.~Zhou, {Investigations on the
  charmless decays of $Y(4260)$}, Chin. Phys. C 39 (2015) 063102.
\newblock \href {http://arxiv.org/abs/1412.3221} {\path{arXiv:1412.3221}},
  \href {https://doi.org/10.1088/1674-1137/39/6/063102}
  {\path{doi:10.1088/1674-1137/39/6/063102}}.

\bibitem{Chen:2015bma}
D.-Y. Chen, X.~Liu, T.~Matsuki, {Search for missing $\psi(4S)$ in the
  $e^+e^-\to \pi^+\pi^-\psi(2S)$ process}, Phys. Rev. D 93 (2016) 034028.
\newblock \href {http://arxiv.org/abs/1509.00736} {\path{arXiv:1509.00736}},
  \href {https://doi.org/10.1103/PhysRevD.93.034028}
  {\path{doi:10.1103/PhysRevD.93.034028}}.

\bibitem{Wang:2015xsa}
B.~Wang, H.~Xu, X.~Liu, D.-Y. Chen, S.~Coito, E.~Eichten, {Using $X(3823)\to
  J/\psi\pi^+\pi^-$ to identify coupled-channel effects}, Front. Phys.
  (Beijing) 11 (2016) 111402.
\newblock \href {http://arxiv.org/abs/1507.07985} {\path{arXiv:1507.07985}},
  \href {https://doi.org/10.1007/s11467-016-0564-7}
  {\path{doi:10.1007/s11467-016-0564-7}}.

\bibitem{Guo:2016iej}
X.-D. Guo, D.-Y. Chen, H.-W. Ke, X.~Liu, X.-Q. Li, {Study on the rare decays of
  $Y(4630)$ induced by final state interactions}, Phys. Rev. D 93 (2016)
  054009.
\newblock \href {http://arxiv.org/abs/1602.02222} {\path{arXiv:1602.02222}},
  \href {https://doi.org/10.1103/PhysRevD.93.054009}
  {\path{doi:10.1103/PhysRevD.93.054009}}.

\bibitem{Qian:2021neg}
R.-Q. Qian, Q.~Huang, X.~Liu, {Predicted $\Lambda \bar{\Lambda}$ and $\Xi^-
  \Xi^+$ decay modes of the charmoniumlike $Y(4230)$}, Phys. Lett. B 833 (2022)
  137292.
\newblock \href {http://arxiv.org/abs/2111.13821} {\path{arXiv:2111.13821}},
  \href {https://doi.org/10.1016/j.physletb.2022.137292}
  {\path{doi:10.1016/j.physletb.2022.137292}}.

\bibitem{Qian:2021gby}
R.-Q. Qian, J.-Z. Wang, X.~Liu, T.~Matsuki, {Charmonium decays into $\Lambda_c
  \bar{\Lambda}_c$ pair governed by the hadronic loop mechanism}, Phys. Rev. D
  104 (2021) 094001.
\newblock \href {http://arxiv.org/abs/2104.13270} {\path{arXiv:2104.13270}},
  \href {https://doi.org/10.1103/PhysRevD.104.094001}
  {\path{doi:10.1103/PhysRevD.104.094001}}.

\bibitem{Wang:2022jxj}
J.-Z. Wang, X.~Liu, {Confirming the existence of a new higher charmonium
  $\psi(4500)$ by the newly released data of $e^+e^- \to K^+K^-J/\psi$}, Phys.
  Rev. D 107 (2023) 054016.
\newblock \href {http://arxiv.org/abs/2212.13512} {\path{arXiv:2212.13512}},
  \href {https://doi.org/10.1103/PhysRevD.107.054016}
  {\path{doi:10.1103/PhysRevD.107.054016}}.

\bibitem{Liu:2023sdw}
S.~Liu, Z.~Cai, Y.~Zheng, G.~Li, {Hidden charmonium decays of $\psi (nS)$
  through charmed meson loops}, Eur. Phys. J. C 83 (2023) 820.
\newblock \href {https://doi.org/10.1140/epjc/s10052-023-12012-w}
  {\path{doi:10.1140/epjc/s10052-023-12012-w}}.

\bibitem{Qian:2023taw}
R.-Q. Qian, X.~Liu, {Production of charmonium $\chi_{cJ}(2P)$ plus one $\omega$
  meson by $e^+e^-$ annihilation}, Phys. Rev. D 108 (2023) 094046.
\newblock \href {http://arxiv.org/abs/2308.14072} {\path{arXiv:2308.14072}},
  \href {https://doi.org/10.1103/PhysRevD.108.094046}
  {\path{doi:10.1103/PhysRevD.108.094046}}.

\bibitem{Peng:2024xui}
T.-C. Peng, Z.-Y. Bai, J.-Z. Wang, X.~Liu, {How higher charmonia shape the
  puzzling data of the $e^+e^- \to \eta J/\psi$ cross section}, Phys. Rev. D
  109 (2024) 094048.
\newblock \href {http://arxiv.org/abs/2403.03705} {\path{arXiv:2403.03705}},
  \href {https://doi.org/10.1103/PhysRevD.109.094048}
  {\path{doi:10.1103/PhysRevD.109.094048}}.

\bibitem{Gao:2024qth}
T.-L. Gao, R.-Q. Qian, X.~Liu, {Discovery potential of charmonium $2P$ states
  through the $e^+e^- \to \gamma D\bar{D}$ processes}, Phys. Rev. D 111 (2025)
  054021.
\newblock \href {http://arxiv.org/abs/2412.06400} {\path{arXiv:2412.06400}},
  \href {https://doi.org/10.1103/PhysRevD.111.054021}
  {\path{doi:10.1103/PhysRevD.111.054021}}.

\bibitem{Zheng:2024eia}
Y.~Zheng, Z.~Cai, G.~Li, S.~Liu, J.~Wu, Q.~Wu, {Hidden charmonium decays of
  spin-2 partner of $X(3872)$}, Phys. Rev. D 109 (2024) 014027.
\newblock \href {http://arxiv.org/abs/2401.03219} {\path{arXiv:2401.03219}},
  \href {https://doi.org/10.1103/PhysRevD.109.014027}
  {\path{doi:10.1103/PhysRevD.109.014027}}.

\bibitem{Bai:2024lps}
Z.-Y. Bai, B.-J. Lai, Q.-S. Zhou, X.~Liu, {Non-$D{\bar{D}}$ decays into light
  meson pairs of the $D$-wave charmonium $\psi _3(3842)$}, Eur. Phys. J. C 85
  (2025) 649.
\newblock \href {http://arxiv.org/abs/2412.09408} {\path{arXiv:2412.09408}},
  \href {https://doi.org/10.1140/epjc/s10052-025-14376-7}
  {\path{doi:10.1140/epjc/s10052-025-14376-7}}.

\bibitem{Qi:2025hwd}
X.-Y. Qi, Q.~Wu, X.-D. Guo, D.-Y. Chen, {Two-body hidden charm decays of $D$
  wave charmonia}, Eur. Phys. J. C 85 (2025) 96.
\newblock \href {http://arxiv.org/abs/2501.16124} {\path{arXiv:2501.16124}},
  \href {https://doi.org/10.1140/epjc/s10052-025-13808-8}
  {\path{doi:10.1140/epjc/s10052-025-13808-8}}.

\bibitem{Laermann:1986pu}
E.~Laermann, F.~Langhammer, I.~Schmitt, P.~M. Zerwas, {The interquark
  potential: SU(2) color gauge theory with fermions}, Phys. Lett. B 173 (1986)
  437--442.
\newblock \href {https://doi.org/10.1016/0370-2693(86)90411-9}
  {\path{doi:10.1016/0370-2693(86)90411-9}}.

\bibitem{Born:1989iv}
K.~D. Born, E.~Laermann, N.~Pirch, T.~F. Walsh, P.~M. Zerwas, {Hadron
  properties in lattice QCD with dynamical fermions}, Phys. Rev. D 40 (1989)
  1653--1663.
\newblock \href {https://doi.org/10.1103/PhysRevD.40.1653}
  {\path{doi:10.1103/PhysRevD.40.1653}}.

\bibitem{Otto:1984qr}
S.~W. Otto, J.~D. Stack, {The SU(3) Heavy Quark Potential with High
  Statistics}, Phys. Rev. Lett. 52 (1984) 2328.
\newblock \href {https://doi.org/10.1103/PhysRevLett.52.2328}
  {\path{doi:10.1103/PhysRevLett.52.2328}}.

\bibitem{Barkai:1984ca}
D.~Barkai, K.~J.~M. Moriarty, C.~Rebbi, {The Force Between Static Quarks},
  Phys. Rev. D 30 (1984) 1293.
\newblock \href {https://doi.org/10.1103/PhysRevD.30.1293}
  {\path{doi:10.1103/PhysRevD.30.1293}}.

\bibitem{Barkai:1984pz}
D.~Barkai, K.~J.~M. Moriarty, C.~Rebbi, {Force Between Static Charges and
  Universality in Lattice {QCD}}, Phys. Rev. D 30 (1984) 2201.
\newblock \href {https://doi.org/10.1103/PhysRevD.30.2201}
  {\path{doi:10.1103/PhysRevD.30.2201}}.

\bibitem{Isgur:1977ef}
N.~Isgur, G.~Karl, {Hyperfine Interactions in Negative Parity Baryons}, Phys.
  Lett. B 72 (1977) 109.
\newblock \href {https://doi.org/10.1016/0370-2693(77)90074-0}
  {\path{doi:10.1016/0370-2693(77)90074-0}}.

\bibitem{Isgur:1978wd}
N.~Isgur, G.~Karl, {Positive Parity Excited Baryons in a Quark Model with
  Hyperfine Interactions}, Phys. Rev. D 19 (1979) 2653, [Erratum: Phys.Rev.D
  23, 817 (1981)].
\newblock \href {https://doi.org/10.1103/PhysRevD.19.2653}
  {\path{doi:10.1103/PhysRevD.19.2653}}.

\bibitem{Isgur:1978xj}
N.~Isgur, G.~Karl, {$P$-wave baryons in the quark model}, Phys. Rev. D 18
  (1978) 4187.
\newblock \href {https://doi.org/10.1103/PhysRevD.18.4187}
  {\path{doi:10.1103/PhysRevD.18.4187}}.

\bibitem{Capstick:1986ter}
S.~Capstick, N.~Isgur, {Baryons in a relativized quark model with
  chromodynamics}, Phys. Rev. D 34 (1986) 2809--2835.
\newblock \href {https://doi.org/10.1103/physrevd.34.2809}
  {\path{doi:10.1103/physrevd.34.2809}}.

\bibitem{Tan:2024pqs}
Y.~Tan, X.~Liu, X.~Chen, Y.~Wu, H.~Huang, J.~Ping, {Equivalence among
  color-singlet, color-octet, and diquark structures in a chiral quark model},
  Phys. Rev. D 109 (2024) 076026.
\newblock \href {http://arxiv.org/abs/2402.16697} {\path{arXiv:2402.16697}},
  \href {https://doi.org/10.1103/PhysRevD.109.076026}
  {\path{doi:10.1103/PhysRevD.109.076026}}.

\bibitem{Hu:2022zdh}
J.~Hu, B.-R. He, J.-L. Ping, {Investigating full-heavy tetraquarks composed of
  $cc{\bar{c}}{\bar{b}}$ and $bb{\bar{b}}{\bar{c}}$}, Eur. Phys. J. C 83 (2023)
  559.
\newblock \href {http://arxiv.org/abs/2202.10380} {\path{arXiv:2202.10380}},
  \href {https://doi.org/10.1140/epjc/s10052-023-11715-4}
  {\path{doi:10.1140/epjc/s10052-023-11715-4}}.

\bibitem{Duan:2021alw}
M.-X. Duan, X.~Liu, {Where are $3P$ and higher $P$-wave states in the
  charmonium family?}, Phys. Rev. D 104 (2021) 074010.
\newblock \href {http://arxiv.org/abs/2107.14438} {\path{arXiv:2107.14438}},
  \href {https://doi.org/10.1103/PhysRevD.104.074010}
  {\path{doi:10.1103/PhysRevD.104.074010}}.

\bibitem{vanBeveren:1979bd}
E.~van Beveren, C.~Dullemond, G.~Rupp, {Spectra and strong decays of $c\bar{c}$
  and $b\bar{b}$ states}, Phys. Rev. D 21 (1980) 772, [Erratum: Phys.Rev.D 22,
  787 (1980)].
\newblock \href {https://doi.org/10.1103/PhysRevD.21.772}
  {\path{doi:10.1103/PhysRevD.21.772}}.

\bibitem{vanBeveren:1986ea}
E.~van Beveren, T.~A. Rijken, K.~Metzger, C.~Dullemond, G.~Rupp, J.~E. Ribeiro,
  {A Low Lying Scalar Meson Nonet in a Unitarized Meson Model}, Z. Phys. C 30
  (1986) 615--620.
\newblock \href {http://arxiv.org/abs/0710.4067} {\path{arXiv:0710.4067}},
  \href {https://doi.org/10.1007/BF01571811} {\path{doi:10.1007/BF01571811}}.

\bibitem{Ono:1985eu}
S.~Ono, A.~I. Sanda, N.~A. Tornqvist, {$B$ Meson Production Between the
  $\Upsilon (4s)$ and $\Upsilon (6s)$ and the Possibility of Detecting $B
  \bar{B}$ Mixing}, Phys. Rev. D 34 (1986) 186.
\newblock \href {https://doi.org/10.1103/PhysRevD.34.186}
  {\path{doi:10.1103/PhysRevD.34.186}}.

\bibitem{Tornqvist:1984fy}
N.~A. Tornqvist, P.~Zenczykowski, {Ground State Baryon Mass Splittings From
  Unitarity}, Phys. Rev. D 29 (1984) 2139.
\newblock \href {https://doi.org/10.1103/PhysRevD.29.2139}
  {\path{doi:10.1103/PhysRevD.29.2139}}.

\bibitem{Tornqvist:1985fi}
N.~A. Tornqvist, P.~Zenczykowski, {The Spectrum of the $P$ Wave Baryons and
  Hadronic Loops}, Z. Phys. C 30 (1986) 83.
\newblock \href {https://doi.org/10.1007/BF01560681}
  {\path{doi:10.1007/BF01560681}}.

\bibitem{Zenczykowski:1985uh}
P.~Zenczykowski, {Baryon Spectroscopy: Symmetries, Symmetry Breaking and
  Hadronic Loops}, Annals Phys. 169 (1986) 453.
\newblock \href {https://doi.org/10.1016/0003-4916(86)90176-4}
  {\path{doi:10.1016/0003-4916(86)90176-4}}.

\bibitem{Tornqvist:1995kr}
N.~A. Tornqvist, {Understanding the scalar meson $q \bar{q}$ nonet}, Z. Phys. C
  68 (1995) 647--660.
\newblock \href {http://arxiv.org/abs/hep-ph/9504372}
  {\path{arXiv:hep-ph/9504372}}, \href {https://doi.org/10.1007/BF01565264}
  {\path{doi:10.1007/BF01565264}}.

\bibitem{Tornqvist:1982yv}
N.~A. Tornqvist, {Scalar mesons in the unitarized quark model}, Phys. Rev.
  Lett. 49 (1982) 624--627.
\newblock \href {https://doi.org/10.1103/PhysRevLett.49.624}
  {\path{doi:10.1103/PhysRevLett.49.624}}.

\bibitem{Tornqvist:1984xz}
N.~A. Tornqvist, {Quarkonium and Quark Loops}, Acta Phys. Polon. B 16 (1985)
  503, [Erratum: Acta Phys.Polon.B 16, 683 (1985)].

\bibitem{Klempt:2009pi}
E.~Klempt, J.-M. Richard, {Baryon spectroscopy}, Rev. Mod. Phys. 82 (2010)
  1095--1153.
\newblock \href {http://arxiv.org/abs/0901.2055} {\path{arXiv:0901.2055}},
  \href {https://doi.org/10.1103/RevModPhys.82.1095}
  {\path{doi:10.1103/RevModPhys.82.1095}}.

\bibitem{Cheng:2015iom}
H.-Y. Cheng, {Charmed baryons circa 2015}, Front. Phys. (Beijing) 10 (2015)
  101406.
\newblock \href {https://doi.org/10.1007/s11467-015-0483-z}
  {\path{doi:10.1007/s11467-015-0483-z}}.

\bibitem{Yuan:2018inv}
C.-Z. Yuan, {The $XYZ$ states revisited}, Int. J. Mod. Phys. A 33 (2018)
  1830018.
\newblock \href {http://arxiv.org/abs/1808.01570} {\path{arXiv:1808.01570}},
  \href {https://doi.org/10.1142/S0217751X18300181}
  {\path{doi:10.1142/S0217751X18300181}}.

\bibitem{Dong:2021juy}
X.-K. Dong, F.-K. Guo, B.-S. Zou, {A survey of heavy-antiheavy hadronic
  molecules}, Progr. Phys. 41 (2021) 65--93.
\newblock \href {http://arxiv.org/abs/2101.01021} {\path{arXiv:2101.01021}},
  \href {https://doi.org/10.13725/j.cnki.pip.2021.02.001}
  {\path{doi:10.13725/j.cnki.pip.2021.02.001}}.

\bibitem{Cheng:2021qpd}
H.-Y. Cheng, {Charmed baryon physics circa 2021}, Chin. J. Phys. 78 (2022)
  324--362.
\newblock \href {http://arxiv.org/abs/2109.01216} {\path{arXiv:2109.01216}},
  \href {https://doi.org/10.1016/j.cjph.2022.06.021}
  {\path{doi:10.1016/j.cjph.2022.06.021}}.

\bibitem{Gross:2022hyw}
F.~Gross, et~al., {50 Years of quantum chromodynamics}, Eur. Phys. J. C 83
  (2023) 1125.
\newblock \href {http://arxiv.org/abs/2212.11107} {\path{arXiv:2212.11107}},
  \href {https://doi.org/10.1140/epjc/s10052-023-11949-2}
  {\path{doi:10.1140/epjc/s10052-023-11949-2}}.

\bibitem{Liu:2024uxn}
M.-Z. Liu, Y.-W. Pan, Z.-W. Liu, T.-W. Wu, J.-X. Lu, L.-S. Geng, {Three ways to
  decipher the nature of exotic hadrons: Multiplets, three-body hadronic
  molecules, and correlation functions}, Phys. Rept. 1108 (2025) 1--108.
\newblock \href {http://arxiv.org/abs/2404.06399} {\path{arXiv:2404.06399}},
  \href {https://doi.org/10.1016/j.physrep.2024.12.001}
  {\path{doi:10.1016/j.physrep.2024.12.001}}.

\bibitem{Wang:2025sic}
Z.-G. Wang, {Review of the QCD sum rules for exotic states}, Front. Phys.
  (Beijing) 21 (2026) 016300.
\newblock \href {http://arxiv.org/abs/2502.11351} {\path{arXiv:2502.11351}},
  \href {https://doi.org/10.15302/frontphys.2026.016300}
  {\path{doi:10.15302/frontphys.2026.016300}}.

\bibitem{Belle:2003guh}
P.~Krokovny, et~al., Belle Collaboration, {Observation of the $D_{sJ}(2317)$
  and $D_{sJ}(2457)$ in $B$ decays}, Phys. Rev. Lett. 91 (2003) 262002.
\newblock \href {http://arxiv.org/abs/hep-ex/0308019}
  {\path{arXiv:hep-ex/0308019}}, \href
  {https://doi.org/10.1103/PhysRevLett.91.262002}
  {\path{doi:10.1103/PhysRevLett.91.262002}}.

\bibitem{BaBar:2006eep}
B.~Aubert, et~al., BaBar Collaboration, {A study of the $D^*_{sJ}(2317)$ and
  $D_{sJ}(2460)$ mesons in inclusive $c \bar{c}$ production near $\sqrt{s}$ =
  10.6 GeV}, Phys. Rev. D 74 (2006) 032007.
\newblock \href {http://arxiv.org/abs/hep-ex/0604030}
  {\path{arXiv:hep-ex/0604030}}, \href
  {https://doi.org/10.1103/PhysRevD.74.032007}
  {\path{doi:10.1103/PhysRevD.74.032007}}.

\bibitem{Belle:2003kup}
Y.~Mikami, et~al., Belle Collaboration, {Measurements of the $D_{sJ}$ resonance
  properties}, Phys. Rev. Lett. 92 (2004) 012002.
\newblock \href {http://arxiv.org/abs/hep-ex/0307052}
  {\path{arXiv:hep-ex/0307052}}, \href
  {https://doi.org/10.1103/PhysRevLett.92.012002}
  {\path{doi:10.1103/PhysRevLett.92.012002}}.

\bibitem{BESIII:2017vdm}
M.~Ablikim, et~al., BESIII Collaboration, {Measurement of the absolute
  branching fraction of $D_{s0}^{*\pm}(2317)\to \pi^0 D_{s}^{\pm}$}, Phys. Rev.
  D 97 (2018) 051103.
\newblock \href {http://arxiv.org/abs/1711.08293} {\path{arXiv:1711.08293}},
  \href {https://doi.org/10.1103/PhysRevD.97.051103}
  {\path{doi:10.1103/PhysRevD.97.051103}}.

\bibitem{BaBar:2004yux}
B.~Aubert, et~al., BaBar Collaboration, {Study of $B \to D_{sJ}^{(*)+}
  \bar{D}^{(*)}$ decays}, Phys. Rev. Lett. 93 (2004) 181801.
\newblock \href {http://arxiv.org/abs/hep-ex/0408041}
  {\path{arXiv:hep-ex/0408041}}, \href
  {https://doi.org/10.1103/PhysRevLett.93.181801}
  {\path{doi:10.1103/PhysRevLett.93.181801}}.

\bibitem{Barnes:2003dj}
T.~Barnes, F.~E. Close, H.~J. Lipkin, {Implications of a $DK$ molecule at 2.32
  GeV}, Phys. Rev. D 68 (2003) 054006.
\newblock \href {http://arxiv.org/abs/hep-ph/0305025}
  {\path{arXiv:hep-ph/0305025}}, \href
  {https://doi.org/10.1103/PhysRevD.68.054006}
  {\path{doi:10.1103/PhysRevD.68.054006}}.

\bibitem{Cheng:2003kg}
H.-Y. Cheng, W.-S. Hou, {$B$ decays as spectroscope for charmed four quark
  states}, Phys. Lett. B 566 (2003) 193--200.
\newblock \href {http://arxiv.org/abs/hep-ph/0305038}
  {\path{arXiv:hep-ph/0305038}}, \href
  {https://doi.org/10.1016/S0370-2693(03)00834-7}
  {\path{doi:10.1016/S0370-2693(03)00834-7}}.

\bibitem{Kolomeitsev:2003ac}
E.~E. Kolomeitsev, M.~F.~M. Lutz, {On Heavy light meson resonances and chiral
  symmetry}, Phys. Lett. B 582 (2004) 39--48.
\newblock \href {http://arxiv.org/abs/hep-ph/0307133}
  {\path{arXiv:hep-ph/0307133}}, \href
  {https://doi.org/10.1016/j.physletb.2003.10.118}
  {\path{doi:10.1016/j.physletb.2003.10.118}}.

\bibitem{Hayashigaki:2004st}
A.~Hayashigaki, K.~Terasaki, {Isospin quantum number of $D^+_{s0}(2317)$},
  Prog. Theor. Phys. 114 (2006) 1191--1200.
\newblock \href {http://arxiv.org/abs/hep-ph/0410393}
  {\path{arXiv:hep-ph/0410393}}, \href {https://doi.org/10.1143/PTP.114.1191}
  {\path{doi:10.1143/PTP.114.1191}}.

\bibitem{Chen:2004dy}
Y.-Q. Chen, X.-Q. Li, {A Comprehensive four-quark interpretation of
  $D_s(2317)$, $D_s(2457)$ and $D_s(2632)$}, Phys. Rev. Lett. 93 (2004) 232001.
\newblock \href {http://arxiv.org/abs/hep-ph/0407062}
  {\path{arXiv:hep-ph/0407062}}, \href
  {https://doi.org/10.1103/PhysRevLett.93.232001}
  {\path{doi:10.1103/PhysRevLett.93.232001}}.

\bibitem{Dmitrasinovic:2005gc}
V.~Dmitrasinovic, {$D_{s0}^+(2317)^ - D_0(2308)$ mass difference as evidence
  for tetraquarks}, Phys. Rev. Lett. 94 (2005) 162002.
\newblock \href {https://doi.org/10.1103/PhysRevLett.94.162002}
  {\path{doi:10.1103/PhysRevLett.94.162002}}.

\bibitem{Kim:2005gt}
H.~Kim, Y.~Oh, {$D_s(2317)$ as a four-quark state in QCD sum rules}, Phys. Rev.
  D 72 (2005) 074012.
\newblock \href {http://arxiv.org/abs/hep-ph/0508251}
  {\path{arXiv:hep-ph/0508251}}, \href
  {https://doi.org/10.1103/PhysRevD.72.074012}
  {\path{doi:10.1103/PhysRevD.72.074012}}.

\bibitem{Terasaki:2006wm}
K.~Terasaki, {$D^+_{s0}(2317)$ as an iso-triplet four-quark meson}, Eur. Phys.
  J. A 31 (2007) 676.
\newblock \href {http://arxiv.org/abs/hep-ph/0609223}
  {\path{arXiv:hep-ph/0609223}}, \href
  {https://doi.org/10.1140/epja/i2006-10176-7}
  {\path{doi:10.1140/epja/i2006-10176-7}}.

\bibitem{Guo:2006fu}
F.-K. Guo, P.-N. Shen, H.-C. Chiang, R.-G. Ping, B.-S. Zou, {Dynamically
  generated $0^+$ heavy mesons in a heavy chiral unitary approach}, Phys. Lett.
  B 641 (2006) 278--285.
\newblock \href {http://arxiv.org/abs/hep-ph/0603072}
  {\path{arXiv:hep-ph/0603072}}, \href
  {https://doi.org/10.1016/j.physletb.2006.08.064}
  {\path{doi:10.1016/j.physletb.2006.08.064}}.

\bibitem{Guo:2006rp}
F.-K. Guo, P.-N. Shen, H.-C. Chiang, {Dynamically generated 1+ heavy mesons},
  Phys. Lett. B 647 (2007) 133--139.
\newblock \href {http://arxiv.org/abs/hep-ph/0610008}
  {\path{arXiv:hep-ph/0610008}}, \href
  {https://doi.org/10.1016/j.physletb.2007.01.050}
  {\path{doi:10.1016/j.physletb.2007.01.050}}.

\bibitem{Faessler:2007gv}
A.~Faessler, T.~Gutsche, V.~E. Lyubovitskij, Y.-L. Ma, {Strong and radiative
  decays of the $D_{s0}^*(2317)$ meson in the $DK$-molecule picture}, Phys.
  Rev. D 76 (2007) 014005.
\newblock \href {http://arxiv.org/abs/0705.0254} {\path{arXiv:0705.0254}},
  \href {https://doi.org/10.1103/PhysRevD.76.014005}
  {\path{doi:10.1103/PhysRevD.76.014005}}.

\bibitem{Xie:2010zza}
Z.-X. Xie, G.-Q. Feng, X.-H. Guo, {Analyzing $D_{s0}^*(2317)^+$ in the $DK$
  molecule picture in the Beth-Salpeter approach}, Phys. Rev. D 81 (2010)
  036014.
\newblock \href {https://doi.org/10.1103/PhysRevD.81.036014}
  {\path{doi:10.1103/PhysRevD.81.036014}}.

\bibitem{Guo:2023wkv}
F.-K. Guo, {Exotic hadrons from an effective field theory perspective}, PoS
  LATTICE2022 (2023) 232.
\newblock \href {https://doi.org/10.22323/1.430.0232}
  {\path{doi:10.22323/1.430.0232}}.

\bibitem{Close:2003sg}
F.~E. Close, P.~R. Page, {The $D^{*0} \bar{D}^0$ threshold resonance}, Phys.
  Lett. B 578 (2004) 119--123.
\newblock \href {http://arxiv.org/abs/hep-ph/0309253}
  {\path{arXiv:hep-ph/0309253}}, \href
  {https://doi.org/10.1016/j.physletb.2003.10.032}
  {\path{doi:10.1016/j.physletb.2003.10.032}}.

\bibitem{Voloshin:2003nt}
M.~B. Voloshin, {Interference and binding effects in decays of possible
  molecular component of $X(3872)$}, Phys. Lett. B 579 (2004) 316--320.
\newblock \href {http://arxiv.org/abs/hep-ph/0309307}
  {\path{arXiv:hep-ph/0309307}}, \href
  {https://doi.org/10.1016/j.physletb.2003.11.014}
  {\path{doi:10.1016/j.physletb.2003.11.014}}.

\bibitem{Wong:2003xk}
C.-Y. Wong, {Molecular states of heavy quark mesons}, Phys. Rev. C 69 (2004)
  055202.
\newblock \href {http://arxiv.org/abs/hep-ph/0311088}
  {\path{arXiv:hep-ph/0311088}}, \href
  {https://doi.org/10.1103/PhysRevC.69.055202}
  {\path{doi:10.1103/PhysRevC.69.055202}}.

\bibitem{Swanson:2003tb}
E.~S. Swanson, {Short range structure in the $X(3872)$}, Phys. Lett. B 588
  (2004) 189--195.
\newblock \href {http://arxiv.org/abs/hep-ph/0311229}
  {\path{arXiv:hep-ph/0311229}}, \href
  {https://doi.org/10.1016/j.physletb.2004.03.033}
  {\path{doi:10.1016/j.physletb.2004.03.033}}.

\bibitem{Tornqvist:2004qy}
N.~A. Tornqvist, {Isospin breaking of the narrow charmonium state of Belle at
  3872 MeV as a deuson}, Phys. Lett. B 590 (2004) 209--215.
\newblock \href {http://arxiv.org/abs/hep-ph/0402237}
  {\path{arXiv:hep-ph/0402237}}, \href
  {https://doi.org/10.1016/j.physletb.2004.03.077}
  {\path{doi:10.1016/j.physletb.2004.03.077}}.

\bibitem{AlFiky:2005jd}
M.~T. AlFiky, F.~Gabbiani, A.~A. Petrov, {$X(3872)$: Hadronic molecules in
  effective field theory}, Phys. Lett. B 640 (2006) 238--245.
\newblock \href {http://arxiv.org/abs/hep-ph/0506141}
  {\path{arXiv:hep-ph/0506141}}, \href
  {https://doi.org/10.1016/j.physletb.2006.07.069}
  {\path{doi:10.1016/j.physletb.2006.07.069}}.

\bibitem{Liu:2008fh}
Y.-R. Liu, X.~Liu, W.-Z. Deng, S.-L. Zhu, {Is $X(3872)$ really a molecular
  state?}, Eur. Phys. J. C 56 (2008) 63--73.
\newblock \href {http://arxiv.org/abs/0801.3540} {\path{arXiv:0801.3540}},
  \href {https://doi.org/10.1140/epjc/s10052-008-0640-4}
  {\path{doi:10.1140/epjc/s10052-008-0640-4}}.

\bibitem{Thomas:2008ja}
C.~E. Thomas, F.~E. Close, {Is $X(3872)$ a molecule?}, Phys. Rev. D 78 (2008)
  034007.
\newblock \href {http://arxiv.org/abs/0805.3653} {\path{arXiv:0805.3653}},
  \href {https://doi.org/10.1103/PhysRevD.78.034007}
  {\path{doi:10.1103/PhysRevD.78.034007}}.

\bibitem{Liu:2009qhy}
X.~Liu, Z.-G. Luo, Y.-R. Liu, S.-L. Zhu, {$X(3872)$ and other possible heavy
  molecular states}, Eur. Phys. J. C 61 (2009) 411--428.
\newblock \href {http://arxiv.org/abs/0808.0073} {\path{arXiv:0808.0073}},
  \href {https://doi.org/10.1140/epjc/s10052-009-1020-4}
  {\path{doi:10.1140/epjc/s10052-009-1020-4}}.

\bibitem{Lee:2009hy}
I.~W. Lee, A.~Faessler, T.~Gutsche, V.~E. Lyubovitskij, {$X(3872)$ as a
  molecular $DD^*$ state in a potential model}, Phys. Rev. D 80 (2009) 094005.
\newblock \href {http://arxiv.org/abs/0910.1009} {\path{arXiv:0910.1009}},
  \href {https://doi.org/10.1103/PhysRevD.80.094005}
  {\path{doi:10.1103/PhysRevD.80.094005}}.

\bibitem{Gamermann:2009uq}
D.~Gamermann, J.~Nieves, E.~Oset, E.~Ruiz~Arriola, {Couplings in coupled
  channels versus wave functions: application to the $X(3872)$ resonance},
  Phys. Rev. D 81 (2010) 014029.
\newblock \href {http://arxiv.org/abs/0911.4407} {\path{arXiv:0911.4407}},
  \href {https://doi.org/10.1103/PhysRevD.81.014029}
  {\path{doi:10.1103/PhysRevD.81.014029}}.

\bibitem{Braaten:2010mg}
E.~Braaten, H.~W. Hammer, T.~Mehen, {Scattering of an ultrasoft pion and the
  $X(3872)$}, Phys. Rev. D 82 (2010) 034018.
\newblock \href {http://arxiv.org/abs/1005.1688} {\path{arXiv:1005.1688}},
  \href {https://doi.org/10.1103/PhysRevD.82.034018}
  {\path{doi:10.1103/PhysRevD.82.034018}}.

\bibitem{Wang:2013kva}
P.~Wang, X.~G. Wang, {Study on $X(3872)$ from effective field theory with pion
  exchange interaction}, Phys. Rev. Lett. 111 (2013) 042002.
\newblock \href {http://arxiv.org/abs/1304.0846} {\path{arXiv:1304.0846}},
  \href {https://doi.org/10.1103/PhysRevLett.111.042002}
  {\path{doi:10.1103/PhysRevLett.111.042002}}.

\bibitem{Baru:2013rta}
V.~Baru, E.~Epelbaum, A.~A. Filin, C.~Hanhart, U.~G. Meissner, A.~V. Nefediev,
  {Quark mass dependence of the $X(3872)$ binding energy}, Phys. Lett. B 726
  (2013) 537--543.
\newblock \href {http://arxiv.org/abs/1306.4108} {\path{arXiv:1306.4108}},
  \href {https://doi.org/10.1016/j.physletb.2013.08.073}
  {\path{doi:10.1016/j.physletb.2013.08.073}}.

\bibitem{Belle:2006xni}
K.~Abe, et~al., Belle Collaboration, {Experimental constraints on the possible
  $J^P$ quantum numbers of the $\Lambda_c(2880)^+$}, Phys. Rev. Lett. 98 (2007)
  262001.
\newblock \href {http://arxiv.org/abs/hep-ex/0608043}
  {\path{arXiv:hep-ex/0608043}}, \href
  {https://doi.org/10.1103/PhysRevLett.98.262001}
  {\path{doi:10.1103/PhysRevLett.98.262001}}.

\bibitem{LHCb:2017jym}
R.~Aaij, et~al., LHCb Collaboration, {Study of the $D^0 p$ amplitude in
  $\Lambda_b^0\to D^0 p \pi^-$ decays}, JHEP 05 (2017) 030.
\newblock \href {http://arxiv.org/abs/1701.07873} {\path{arXiv:1701.07873}},
  \href {https://doi.org/10.1007/JHEP05(2017)030}
  {\path{doi:10.1007/JHEP05(2017)030}}.

\bibitem{Yu:2022ymb}
G.-L. Yu, Z.-Y. Li, Z.-G. Wang, J.~Lu, M.~Yan, {Systematic analysis of single
  heavy baryons $\Lambda_Q$, $\Sigma_Q$ and $\Omega_Q$}, Nucl. Phys. B 990
  (2023) 116183.
\newblock \href {http://arxiv.org/abs/2206.08128} {\path{arXiv:2206.08128}},
  \href {https://doi.org/10.1016/j.nuclphysb.2023.116183}
  {\path{doi:10.1016/j.nuclphysb.2023.116183}}.

\bibitem{Ebert:2007nw}
D.~Ebert, R.~N. Faustov, V.~O. Galkin, {Masses of excited heavy baryons in the
  relativistic quark model}, Phys. Lett. B 659 (2008) 612--620.
\newblock \href {http://arxiv.org/abs/0705.2957} {\path{arXiv:0705.2957}},
  \href {https://doi.org/10.1016/j.physletb.2007.11.037}
  {\path{doi:10.1016/j.physletb.2007.11.037}}.

\bibitem{Ebert:2011kk}
D.~Ebert, R.~N. Faustov, V.~O. Galkin, {Spectroscopy and Regge trajectories of
  heavy baryons in the relativistic quark-diquark picture}, Phys. Rev. D 84
  (2011) 014025.
\newblock \href {http://arxiv.org/abs/1105.0583} {\path{arXiv:1105.0583}},
  \href {https://doi.org/10.1103/PhysRevD.84.014025}
  {\path{doi:10.1103/PhysRevD.84.014025}}.

\bibitem{Chen:2014nyo}
B.~Chen, K.-W. Wei, A.~Zhang, {Assignments of $\Lambda_Q$ and $\Xi_Q$ baryons
  in the heavy quark-light diquark picture}, Eur. Phys. J. A 51 (2015) 82.
\newblock \href {http://arxiv.org/abs/1406.6561} {\path{arXiv:1406.6561}},
  \href {https://doi.org/10.1140/epja/i2015-15082-3}
  {\path{doi:10.1140/epja/i2015-15082-3}}.

\bibitem{Chen:2016iyi}
B.~Chen, K.-W. Wei, X.~Liu, T.~Matsuki, {Low-lying charmed and charmed-strange
  baryon states}, Eur. Phys. J. C 77 (2017) 154.
\newblock \href {http://arxiv.org/abs/1609.07967} {\path{arXiv:1609.07967}},
  \href {https://doi.org/10.1140/epjc/s10052-017-4708-x}
  {\path{doi:10.1140/epjc/s10052-017-4708-x}}.

\bibitem{Cheng:2017ove}
H.-Y. Cheng, C.-W. Chiang, {Quantum numbers of $\Omega_c$ states and other
  charmed baryons}, Phys. Rev. D 95 (2017) 094018.
\newblock \href {http://arxiv.org/abs/1704.00396} {\path{arXiv:1704.00396}},
  \href {https://doi.org/10.1103/PhysRevD.95.094018}
  {\path{doi:10.1103/PhysRevD.95.094018}}.

\bibitem{BaBar:2003cdx}
B.~Aubert, et~al., BaBar Collaboration, {Observation of a narrow meson decaying
  to $D_s^+ \pi^0 \gamma$ at a mass of 2.458-GeV/c$^2$}, Phys. Rev. D 69 (2004)
  031101.
\newblock \href {http://arxiv.org/abs/hep-ex/0310050}
  {\path{arXiv:hep-ex/0310050}}, \href
  {https://doi.org/10.1103/PhysRevD.69.031101}
  {\path{doi:10.1103/PhysRevD.69.031101}}.

\bibitem{BESIII:2022bse}
M.~Ablikim, et~al., BESIII Collaboration, {Observation of a new $X(3872)$
  production process $e^+e^-\to\omega X(3872)$}, Phys. Rev. Lett. 130 (2023)
  151904.
\newblock \href {http://arxiv.org/abs/2212.07291} {\path{arXiv:2212.07291}},
  \href {https://doi.org/10.1103/PhysRevLett.130.151904}
  {\path{doi:10.1103/PhysRevLett.130.151904}}.

\bibitem{LHCb:2020fvo}
R.~Aaij, et~al., LHCb Collaboration, {Study of the $\psi_2(3823)$ and
  $\chi_{c1}(3872)$ states in $B^+ \rightarrow \left(
  J\psi\pi^+\pi^-\right)K^+$ decays}, JHEP 08 (2020) 123.
\newblock \href {http://arxiv.org/abs/2005.13422} {\path{arXiv:2005.13422}},
  \href {https://doi.org/10.1007/JHEP08(2020)123}
  {\path{doi:10.1007/JHEP08(2020)123}}.

\bibitem{BESIII:2013fnz}
M.~Ablikim, et~al., BESIII Collaboration, {Observation of $e^+e^-\to\gamma
  X(3872)$ at BESIII}, Phys. Rev. Lett. 112 (2014) 092001.
\newblock \href {http://arxiv.org/abs/1310.4101} {\path{arXiv:1310.4101}},
  \href {https://doi.org/10.1103/PhysRevLett.112.092001}
  {\path{doi:10.1103/PhysRevLett.112.092001}}.

\bibitem{LHCb:2011zzp}
R.~Aaij, et~al., LHCb Collaboration, {Observation of $X(3872)$ production in
  $pp$ collisions at $\sqrt{s}=7$ TeV}, Eur. Phys. J. C 72 (2012) 1972.
\newblock \href {http://arxiv.org/abs/1112.5310} {\path{arXiv:1112.5310}},
  \href {https://doi.org/10.1140/epjc/s10052-012-1972-7}
  {\path{doi:10.1140/epjc/s10052-012-1972-7}}.

\bibitem{Godfrey:2015dva}
S.~Godfrey, K.~Moats, {Properties of excited charm and charm-strange mesons},
  Phys. Rev. D 93 (2016) 034035.
\newblock \href {http://arxiv.org/abs/1510.08305} {\path{arXiv:1510.08305}},
  \href {https://doi.org/10.1103/PhysRevD.93.034035}
  {\path{doi:10.1103/PhysRevD.93.034035}}.

\bibitem{Ni:2021pce}
R.-H. Ni, Q.~Li, X.-H. Zhong, {Mass spectra and strong decays of charmed and
  charmed-strange mesons}, Phys. Rev. D 105 (2022) 056006.
\newblock \href {http://arxiv.org/abs/2110.05024} {\path{arXiv:2110.05024}},
  \href {https://doi.org/10.1103/PhysRevD.105.056006}
  {\path{doi:10.1103/PhysRevD.105.056006}}.

\bibitem{Ebert:2009ua}
D.~Ebert, R.~N. Faustov, V.~O. Galkin, {Heavy-light meson spectroscopy and
  Regge trajectories in the relativistic quark model}, Eur. Phys. J. C 66
  (2010) 197--206.
\newblock \href {http://arxiv.org/abs/0910.5612} {\path{arXiv:0910.5612}},
  \href {https://doi.org/10.1140/epjc/s10052-010-1233-6}
  {\path{doi:10.1140/epjc/s10052-010-1233-6}}.

\bibitem{Li:2010vx}
D.-M. Li, P.-F. Ji, B.~Ma, {The newly observed open-charm states in quark
  model}, Eur. Phys. J. C 71 (2011) 1582.
\newblock \href {http://arxiv.org/abs/1011.1548} {\path{arXiv:1011.1548}},
  \href {https://doi.org/10.1140/epjc/s10052-011-1582-9}
  {\path{doi:10.1140/epjc/s10052-011-1582-9}}.

\bibitem{Zeng:1994vj}
J.~Zeng, J.~W. Van~Orden, W.~Roberts, {Heavy mesons in a relativistic model},
  Phys. Rev. D 52 (1995) 5229--5241.
\newblock \href {http://arxiv.org/abs/hep-ph/9412269}
  {\path{arXiv:hep-ph/9412269}}, \href
  {https://doi.org/10.1103/PhysRevD.52.5229}
  {\path{doi:10.1103/PhysRevD.52.5229}}.

\bibitem{Lahde:1999ih}
T.~A. Lahde, C.~J. Nyfalt, D.~O. Riska, {Spectra and M1 decay widths of heavy
  light mesons}, Nucl. Phys. A 674 (2000) 141--167.
\newblock \href {http://arxiv.org/abs/hep-ph/9908485}
  {\path{arXiv:hep-ph/9908485}}, \href
  {https://doi.org/10.1016/S0375-9474(00)00154-8}
  {\path{doi:10.1016/S0375-9474(00)00154-8}}.

\bibitem{Deng:2016stx}
W.-J. Deng, H.~Liu, L.-C. Gui, X.-H. Zhong, {Charmonium spectrum and their
  electromagnetic transitions with higher multipole contributions}, Phys. Rev.
  D 95 (2017) 034026.
\newblock \href {http://arxiv.org/abs/1608.00287} {\path{arXiv:1608.00287}},
  \href {https://doi.org/10.1103/PhysRevD.95.034026}
  {\path{doi:10.1103/PhysRevD.95.034026}}.

\bibitem{Ebert:2011jc}
D.~Ebert, R.~N. Faustov, V.~O. Galkin, {Spectroscopy and Regge trajectories of
  heavy quarkonia and $B_c$ mesons}, Eur. Phys. J. C 71 (2011) 1825.
\newblock \href {http://arxiv.org/abs/1111.0454} {\path{arXiv:1111.0454}},
  \href {https://doi.org/10.1140/epjc/s10052-011-1825-9}
  {\path{doi:10.1140/epjc/s10052-011-1825-9}}.

\bibitem{Li:2009zu}
B.-Q. Li, K.-T. Chao, {Higher Charmonia and $X$, $Y$, $Z$ states with Screened
  Potential}, Phys. Rev. D 79 (2009) 094004.
\newblock \href {http://arxiv.org/abs/0903.5506} {\path{arXiv:0903.5506}},
  \href {https://doi.org/10.1103/PhysRevD.79.094004}
  {\path{doi:10.1103/PhysRevD.79.094004}}.

\bibitem{Wang:2019mhs}
J.-Z. Wang, D.-Y. Chen, X.~Liu, T.~Matsuki, {Constructing $J/\psi$ family with
  updated data of charmoniumlike $Y$ states}, Phys. Rev. D 99 (2019) 114003.
\newblock \href {http://arxiv.org/abs/1903.07115} {\path{arXiv:1903.07115}},
  \href {https://doi.org/10.1103/PhysRevD.99.114003}
  {\path{doi:10.1103/PhysRevD.99.114003}}.

\bibitem{Weng:2024roa}
X.-Z. Weng, W.-Z. Deng, S.-L. Zhu, {Heavy baryons in the relativized quark
  model with chromodynamics}, Phys. Rev. D 110 (2024) 056052.
\newblock \href {http://arxiv.org/abs/2405.19039} {\path{arXiv:2405.19039}},
  \href {https://doi.org/10.1103/PhysRevD.110.056052}
  {\path{doi:10.1103/PhysRevD.110.056052}}.

\bibitem{vanBeveren:2005ha}
E.~van Beveren, J.~E. G.~N. Costa, F.~Kleefeld, G.~Rupp, {From the $\kappa$ via
  the $D^*_{s0}(2317)$ to the $\chi_{c0}$: Connecting light and heavy scalar
  mesons}, Phys. Rev. D 74 (2006) 037501.
\newblock \href {http://arxiv.org/abs/hep-ph/0509351}
  {\path{arXiv:hep-ph/0509351}}, \href
  {https://doi.org/10.1103/PhysRevD.74.037501}
  {\path{doi:10.1103/PhysRevD.74.037501}}.

\bibitem{Simonov:2004ar}
Y.~A. Simonov, J.~A. Tjon, {The Coupled-channel analysis of the $D$ and $D_s$
  mesons}, Phys. Rev. D 70 (2004) 114013.
\newblock \href {http://arxiv.org/abs/hep-ph/0409361}
  {\path{arXiv:hep-ph/0409361}}, \href
  {https://doi.org/10.1103/PhysRevD.70.114013}
  {\path{doi:10.1103/PhysRevD.70.114013}}.

\bibitem{Ortega:2016mms}
P.~G. Ortega, J.~Segovia, D.~R. Entem, F.~Fernandez, {Molecular components in
  $P$-wave charmed-strange mesons}, Phys. Rev. D 94 (2016) 074037.
\newblock \href {http://arxiv.org/abs/1603.07000} {\path{arXiv:1603.07000}},
  \href {https://doi.org/10.1103/PhysRevD.94.074037}
  {\path{doi:10.1103/PhysRevD.94.074037}}.

\bibitem{Cheng:2014bca}
H.-Y. Cheng, F.-S. Yu, {Near mass degeneracy in the scalar meson sector:
  Implications for $B^*_{(s)0}$ and $B^\prime_{(s)1}$ mesons}, Phys. Rev. D 89
  (2014) 114017.
\newblock \href {http://arxiv.org/abs/1404.3771} {\path{arXiv:1404.3771}},
  \href {https://doi.org/10.1103/PhysRevD.89.114017}
  {\path{doi:10.1103/PhysRevD.89.114017}}.

\bibitem{Cheng:2017oqh}
H.-Y. Cheng, F.-S. Yu, {Masses of scalar and axial-vector $B$ Mesons
  revisited}, Eur. Phys. J. C 77 (2017) 668.
\newblock \href {http://arxiv.org/abs/1704.01208} {\path{arXiv:1704.01208}},
  \href {https://doi.org/10.1140/epjc/s10052-017-5252-4}
  {\path{doi:10.1140/epjc/s10052-017-5252-4}}.

\bibitem{Dai:2003yg}
Y.-B. Dai, C.-S. Huang, C.~Liu, S.-L. Zhu, {Understanding the $D^+_{sJ}(2317)$
  and $D^+_{sJ}(2460)$ with sum rules in HQET}, Phys. Rev. D 68 (2003) 114011.
\newblock \href {http://arxiv.org/abs/hep-ph/0306274}
  {\path{arXiv:hep-ph/0306274}}, \href
  {https://doi.org/10.1103/PhysRevD.68.114011}
  {\path{doi:10.1103/PhysRevD.68.114011}}.

\bibitem{Colangelo:2003vg}
P.~Colangelo, F.~De~Fazio, {Understanding $D_{sJ}(2317)$}, Phys. Lett. B 570
  (2003) 180--184.
\newblock \href {http://arxiv.org/abs/hep-ph/0305140}
  {\path{arXiv:hep-ph/0305140}}, \href
  {https://doi.org/10.1016/j.physletb.2003.08.003}
  {\path{doi:10.1016/j.physletb.2003.08.003}}.

\bibitem{Coito:2011qn}
S.~Coito, G.~Rupp, E.~van Beveren, {Quasi-bound states in the continuum: a
  dynamical coupled-channel calculation of axial-vector charmed mesons}, Phys.
  Rev. D 84 (2011) 094020.
\newblock \href {http://arxiv.org/abs/1106.2760} {\path{arXiv:1106.2760}},
  \href {https://doi.org/10.1103/PhysRevD.84.094020}
  {\path{doi:10.1103/PhysRevD.84.094020}}.

\bibitem{Barnes:2007xu}
T.~Barnes, E.~S. Swanson, {Hadron loops: General theorems and application to
  charmonium}, Phys. Rev. C 77 (2008) 055206.
\newblock \href {http://arxiv.org/abs/0711.2080} {\path{arXiv:0711.2080}},
  \href {https://doi.org/10.1103/PhysRevC.77.055206}
  {\path{doi:10.1103/PhysRevC.77.055206}}.

\bibitem{Liu:2011yp}
J.-F. Liu, G.-J. Ding, {Bottomonium spectrum with coupled-channel effects},
  Eur. Phys. J. C 72 (2012) 1981.
\newblock \href {http://arxiv.org/abs/1105.0855} {\path{arXiv:1105.0855}},
  \href {https://doi.org/10.1140/epjc/s10052-012-1981-6}
  {\path{doi:10.1140/epjc/s10052-012-1981-6}}.

\bibitem{Ni:2025gvx}
R.-H. Ni, Q.~Deng, J.-J. Wu, X.-H. Zhong, {Bottomonia in an unquenched quark
  model}, Phys. Rev. D 111 (2025) 114027.
\newblock \href {http://arxiv.org/abs/2501.15110} {\path{arXiv:2501.15110}},
  \href {https://doi.org/10.1103/6x3t-x15s} {\path{doi:10.1103/6x3t-x15s}}.

\bibitem{Sultan:2025dfe}
M.~A. Sultan, W.~Hao, E.~S. Swanson, L.~Chang, {Bottomonium meson spectrum with
  quenched and unquenched quark models}, Eur. Phys. J. A 61 (2025) 137.
\newblock \href {http://arxiv.org/abs/2503.10178} {\path{arXiv:2503.10178}},
  \href {https://doi.org/10.1140/epja/s10050-025-01607-4}
  {\path{doi:10.1140/epja/s10050-025-01607-4}}.

\bibitem{Lu:2016mbb}
Y.~Lu, M.~N. Anwar, B.-S. Zou, {Coupled-Channel Effects for the Bottomonium
  with Realistic Wave Functions}, Phys. Rev. D 94 (2016) 034021.
\newblock \href {http://arxiv.org/abs/1606.06927} {\path{arXiv:1606.06927}},
  \href {https://doi.org/10.1103/PhysRevD.94.034021}
  {\path{doi:10.1103/PhysRevD.94.034021}}.

\bibitem{Lu:2017hma}
Y.~Lu, M.~N. Anwar, B.-S. Zou, {How large is the contribution of excited mesons
  in coupled-channel effects?}, Phys. Rev. D 95 (2017) 034018.
\newblock \href {http://arxiv.org/abs/1701.00692} {\path{arXiv:1701.00692}},
  \href {https://doi.org/10.1103/PhysRevD.95.034018}
  {\path{doi:10.1103/PhysRevD.95.034018}}.

\bibitem{Ferretti:2012zz}
J.~Ferretti, G.~Galata, E.~Santopinto, A.~Vassallo, {Bottomonium self-energies
  due to the coupling to the meson-meson continuum}, Phys. Rev. C 86 (2012)
  015204.
\newblock \href {https://doi.org/10.1103/PhysRevC.86.015204}
  {\path{doi:10.1103/PhysRevC.86.015204}}.

\bibitem{Ferretti:2013vua}
J.~Ferretti, E.~Santopinto, {Higher mass bottomonia}, Phys. Rev. D 90 (2014)
  094022.
\newblock \href {http://arxiv.org/abs/1306.2874} {\path{arXiv:1306.2874}},
  \href {https://doi.org/10.1103/PhysRevD.90.094022}
  {\path{doi:10.1103/PhysRevD.90.094022}}.

\bibitem{Coito:2010if}
S.~Coito, G.~Rupp, E.~van Beveren, {Delicate interplay between the $D^0-D*0$,
  $\rho^0-J/\psi$, and $\omega-J/\psi$ channels in the $X(3872)$ resonance},
  Eur. Phys. J. C 71 (2011) 1762.
\newblock \href {http://arxiv.org/abs/1008.5100} {\path{arXiv:1008.5100}},
  \href {https://doi.org/10.1140/epjc/s10052-011-1762-7}
  {\path{doi:10.1140/epjc/s10052-011-1762-7}}.

\bibitem{Coito:2012vf}
S.~Coito, G.~Rupp, E.~van Beveren, {$X(3872)$ is not a true molecule}, Eur.
  Phys. J. C 73 (2013) 2351.
\newblock \href {http://arxiv.org/abs/1212.0648} {\path{arXiv:1212.0648}},
  \href {https://doi.org/10.1140/epjc/s10052-013-2351-8}
  {\path{doi:10.1140/epjc/s10052-013-2351-8}}.

\bibitem{Cardoso:2014xda}
M.~Cardoso, G.~Rupp, E.~van Beveren, {Unquenched quark-model calculation of
  $X(3872)$ electromagnetic decays}, Eur. Phys. J. C 75 (2015) 26.
\newblock \href {http://arxiv.org/abs/1411.1654} {\path{arXiv:1411.1654}},
  \href {https://doi.org/10.1140/epjc/s10052-014-3254-z}
  {\path{doi:10.1140/epjc/s10052-014-3254-z}}.

\bibitem{Ortega:2012rs}
P.~G. Ortega, D.~R. Entem, F.~Fernandez, {Molecular structures in charmonium
  spectrum: The $XYZ$ puzzle}, J. Phys. G 40 (2013) 065107.
\newblock \href {http://arxiv.org/abs/1205.1699} {\path{arXiv:1205.1699}},
  \href {https://doi.org/10.1088/0954-3899/40/6/065107}
  {\path{doi:10.1088/0954-3899/40/6/065107}}.

\bibitem{Ferretti:2013faa}
J.~Ferretti, G.~Galat{\`a}, E.~Santopinto, {Interpretation of the $X(3872)$ as
  a charmonium state plus an extra component due to the coupling to the
  meson-meson continuum}, Phys. Rev. C 88 (2013) 015207.
\newblock \href {http://arxiv.org/abs/1302.6857} {\path{arXiv:1302.6857}},
  \href {https://doi.org/10.1103/PhysRevC.88.015207}
  {\path{doi:10.1103/PhysRevC.88.015207}}.

\bibitem{Zhou:2011sp}
Z.-Y. Zhou, Z.~Xiao, {Hadron loops effect on mass shifts of the charmed and
  charmed-strange spectra}, Phys. Rev. D 84 (2011) 034023.
\newblock \href {http://arxiv.org/abs/1105.6025} {\path{arXiv:1105.6025}},
  \href {https://doi.org/10.1103/PhysRevD.84.034023}
  {\path{doi:10.1103/PhysRevD.84.034023}}.

\bibitem{Zhou:2013ada}
Z.-Y. Zhou, Z.~Xiao, {Comprehending heavy charmonia and their decays by hadron
  loop effects}, Eur. Phys. J. A 50 (2014) 165.
\newblock \href {http://arxiv.org/abs/1309.1949} {\path{arXiv:1309.1949}},
  \href {https://doi.org/10.1140/epja/i2014-14165-y}
  {\path{doi:10.1140/epja/i2014-14165-y}}.

\bibitem{Duan:2020tsx}
M.-X. Duan, S.-Q. Luo, X.~Liu, T.~Matsuki, {Possibility of charmoniumlike state
  $X(3915)$ as $\chi_{c0}(2P)$ state}, Phys. Rev. D 101 (2020) 054029.
\newblock \href {http://arxiv.org/abs/2002.03311} {\path{arXiv:2002.03311}},
  \href {https://doi.org/10.1103/PhysRevD.101.054029}
  {\path{doi:10.1103/PhysRevD.101.054029}}.

\bibitem{Man:2025zfu}
Z.-L. Man, S.-Q. Luo, Z.-Y. Bai, X.~Liu, {Coupled-channel study of $4S$-$3D$
  mixing dynamics in $\psi(4220)$ and $\psi(4380)$}, Phys. Lett. B 868 (2025)
  139644.
\newblock \href {http://arxiv.org/abs/2502.08072} {\path{arXiv:2502.08072}},
  \href {https://doi.org/10.1016/j.physletb.2025.139644}
  {\path{doi:10.1016/j.physletb.2025.139644}}.

\bibitem{Man:2025vmm}
Z.-L. Man, S.-Q. Luo, X.~Liu, {Is the $3S$-$2D$ mixing strong for the charmonia
  $\psi(4040)$ and $\psi(4160)$?}, Phys. Rev. D 112 (2025) 074025.
\newblock \href {http://arxiv.org/abs/2507.18536} {\path{arXiv:2507.18536}},
  \href {https://doi.org/10.1103/76fd-njsy} {\path{doi:10.1103/76fd-njsy}}.

\bibitem{Danilkin:2009hr}
I.~V. Danilkin, Y.~A. Simonov, {Channel coupling in heavy quarkonia: Energy
  levels, mixing, widths and new states}, Phys. Rev. D 81 (2010) 074027.
\newblock \href {http://arxiv.org/abs/0907.1088} {\path{arXiv:0907.1088}},
  \href {https://doi.org/10.1103/PhysRevD.81.074027}
  {\path{doi:10.1103/PhysRevD.81.074027}}.

\bibitem{Wang:2024ytk}
J.-Z. Wang, Z.-Y. Lin, Y.-K. Chen, L.~Meng, S.-L. Zhu, {Uncovering the mystery
  of $X(3872)$ with the coupled-channel dynamics}, Phys. Rev. D 111 (2025)
  L111502.
\newblock \href {http://arxiv.org/abs/2404.16575} {\path{arXiv:2404.16575}},
  \href {https://doi.org/10.1103/929g-cv15} {\path{doi:10.1103/929g-cv15}}.

\bibitem{Colangelo:2025uhs}
P.~Colangelo, F.~De~Fazio, G.~Roselli, {Charming case of $X(3872)$ and
  $\chi_{c1}(2P)$}, Phys. Rev. D 111 (2025) 074014.
\newblock \href {http://arxiv.org/abs/2501.15888} {\path{arXiv:2501.15888}},
  \href {https://doi.org/10.1103/PhysRevD.111.074014}
  {\path{doi:10.1103/PhysRevD.111.074014}}.

\bibitem{LHCb:2024tpv}
R.~Aaij, et~al., LHCb Collaboration, {Probing the nature of the
  $\chi_{c1}(3872)$ state using radiative decays}, JHEP 11 (2024) 121.
\newblock \href {http://arxiv.org/abs/2406.17006} {\path{arXiv:2406.17006}},
  \href {https://doi.org/10.1007/JHEP11(2024)121}
  {\path{doi:10.1007/JHEP11(2024)121}}.

\bibitem{Roberts:1992esl}
W.~Roberts, B.~Silvestre-Brac, {General method of calculation of any hadronic
  decay in the $^3P_0$ model}, Few Body Syst. 11~(4) (1992) 171--193.
\newblock \href {https://doi.org/10.1007/bf01641821}
  {\path{doi:10.1007/bf01641821}}.

\bibitem{Blundell:1996as}
H.~G. Blundell, {Meson properties in the quark model: A look at some
  outstanding problems}, Other thesis (7 1996).
\newblock \href {http://arxiv.org/abs/hep-ph/9608473}
  {\path{arXiv:hep-ph/9608473}}.

\bibitem{Chen:2017gnu}
B.~Chen, X.~Liu, {New $\Omega_c^0$ baryons discovered by LHCb as the members of
  $1P$ and $2S$ states}, Phys. Rev. D 96~(9) (2017) 094015.
\newblock \href {http://arxiv.org/abs/1704.02583} {\path{arXiv:1704.02583}},
  \href {https://doi.org/10.1103/PhysRevD.96.094015}
  {\path{doi:10.1103/PhysRevD.96.094015}}.

\bibitem{vanBeveren:2020eis}
E.~van Beveren, G.~Rupp, {Modern meson spectroscopy: the fundamental role of
  unitarity}, Prog. Part. Nucl. Phys. 117 (2021) 103845.
\newblock \href {http://arxiv.org/abs/2012.03693} {\path{arXiv:2012.03693}},
  \href {https://doi.org/10.1016/j.ppnp.2020.103845}
  {\path{doi:10.1016/j.ppnp.2020.103845}}.

\bibitem{Guo:2019twa}
F.-K. Guo, X.-H. Liu, S.~Sakai, {Threshold cusps and triangle singularities in
  hadronic reactions}, Prog. Part. Nucl. Phys. 112 (2020) 103757.
\newblock \href {http://arxiv.org/abs/1912.07030} {\path{arXiv:1912.07030}},
  \href {https://doi.org/10.1016/j.ppnp.2020.103757}
  {\path{doi:10.1016/j.ppnp.2020.103757}}.

\bibitem{Belle:2009rkh}
C.~P. Shen, et~al., Belle Collaboration, {Evidence for a new resonance and
  search for the $Y(4140)$ in the $\gamma \gamma \to \phi J/\psi$ process},
  Phys. Rev. Lett. 104 (2010) 112004.
\newblock \href {http://arxiv.org/abs/0912.2383} {\path{arXiv:0912.2383}},
  \href {https://doi.org/10.1103/PhysRevLett.104.112004}
  {\path{doi:10.1103/PhysRevLett.104.112004}}.

\bibitem{CDF:2009jgo}
T.~Aaltonen, et~al., CDF Collaboration, {Evidence for a narrow near-threshold
  structure in the $J/\psi\phi$ mass spectrum in $B^+\to J/\psi\phi K^+$
  decays}, Phys. Rev. Lett. 102 (2009) 242002.
\newblock \href {http://arxiv.org/abs/0903.2229} {\path{arXiv:0903.2229}},
  \href {https://doi.org/10.1103/PhysRevLett.102.242002}
  {\path{doi:10.1103/PhysRevLett.102.242002}}.

\bibitem{Liu:2009fe}
X.~Liu, Z.-G. Luo, Z.-F. Sun, {$X(3915)$ and $X(4350)$ as new members in
  $P$-wave charmonium family}, Phys. Rev. Lett. 104 (2010) 122001.
\newblock \href {http://arxiv.org/abs/0911.3694} {\path{arXiv:0911.3694}},
  \href {https://doi.org/10.1103/PhysRevLett.104.122001}
  {\path{doi:10.1103/PhysRevLett.104.122001}}.

\bibitem{Landau:1948kw}
L.~D. Landau, {On the angular momentum of a system of two photons}, Dokl. Akad.
  Nauk SSSR 60 (1948) 207--209.
\newblock \href {https://doi.org/10.1016/B978-0-08-010586-4.50070-5}
  {\path{doi:10.1016/B978-0-08-010586-4.50070-5}}.

\bibitem{Yang:1950rg}
C.-N. Yang, {Selection rules for the dematerialization of a particle into two
  photons}, Phys. Rev. 77 (1950) 242--245.
\newblock \href {https://doi.org/10.1103/PhysRev.77.242}
  {\path{doi:10.1103/PhysRev.77.242}}.

\bibitem{Belle:2007qae}
S.~Uehara, et~al., Belle Collaboration, {Study of charmonia in four-meson final
  states produced in two-photon collisions}, Eur. Phys. J. C 53 (2008) 1--14.
\newblock \href {http://arxiv.org/abs/0706.3955} {\path{arXiv:0706.3955}},
  \href {https://doi.org/10.1140/epjc/s10052-007-0451-z}
  {\path{doi:10.1140/epjc/s10052-007-0451-z}}.

\bibitem{Belle:2013eck}
S.~Uehara, et~al., Belle Collaboration, {High-statistics study of $K^0_S$ pair
  production in two-photon collisions}, PTEP 2013 (2013) 123C01.
\newblock \href {http://arxiv.org/abs/1307.7457} {\path{arXiv:1307.7457}},
  \href {https://doi.org/10.1093/ptep/ptt097} {\path{doi:10.1093/ptep/ptt097}}.

\bibitem{BaBar:2012nxg}
J.~P. Lees, et~al., BaBar Collaboration, {Study of $X(3915) \to J/\psi \omega$
  in two-photon collisions}, Phys. Rev. D 86 (2012) 072002.
\newblock \href {http://arxiv.org/abs/1207.2651} {\path{arXiv:1207.2651}},
  \href {https://doi.org/10.1103/PhysRevD.86.072002}
  {\path{doi:10.1103/PhysRevD.86.072002}}.

\bibitem{ParticleDataGroup:2012pjm}
J.~Beringer, et~al., Particle Data Group Collaboration, {Review of particle
  physics (RPP)}, Phys. Rev. D 86 (2012) 010001.
\newblock \href {https://doi.org/10.1103/PhysRevD.86.010001}
  {\path{doi:10.1103/PhysRevD.86.010001}}.

\bibitem{Guo:2012tv}
F.-K. Guo, U.-G. Meissner, {Where is the $\chi_{c0}(2P)$?}, Phys. Rev. D 86
  (2012) 091501.
\newblock \href {http://arxiv.org/abs/1208.1134} {\path{arXiv:1208.1134}},
  \href {https://doi.org/10.1103/PhysRevD.86.091501}
  {\path{doi:10.1103/PhysRevD.86.091501}}.

\bibitem{Olsen:2014maa}
S.~L. Olsen, {Is the $X$(3915) the $\chi_{c0}(2P)$?}, Phys. Rev. D 91 (2015)
  057501.
\newblock \href {http://arxiv.org/abs/1410.6534} {\path{arXiv:1410.6534}},
  \href {https://doi.org/10.1103/PhysRevD.91.057501}
  {\path{doi:10.1103/PhysRevD.91.057501}}.

\bibitem{Shi:2024llv}
P.-P. Shi, M.~Albaladejo, M.-L. Du, F.-K. Guo, J.~Nieves, {$P$-wave charmonium
  contribution to hidden-charm states from a reanalysis of lattice QCD data},
  Phys. Rev. D 111 (2025) 074043.
\newblock \href {http://arxiv.org/abs/2410.19563} {\path{arXiv:2410.19563}},
  \href {https://doi.org/10.1103/PhysRevD.111.074043}
  {\path{doi:10.1103/PhysRevD.111.074043}}.

\bibitem{Prelovsek:2020eiw}
S.~Prelovsek, S.~Collins, D.~Mohler, M.~Padmanath, S.~Piemonte,
  {Charmonium-like resonances with J$^{PC}$ = 0$^{++}$, 2$^{++}$ in coupled $
  \mathrm{D}\overline{\mathrm{D}} $, $
  {\mathrm{D}}_{\mathrm{s}}{\overline{\mathrm{D}}}_{\mathrm{s}} $ scattering on
  the lattice}, JHEP 06 (2021) 035.
\newblock \href {http://arxiv.org/abs/2011.02542} {\path{arXiv:2011.02542}},
  \href {https://doi.org/10.1007/JHEP06(2021)035}
  {\path{doi:10.1007/JHEP06(2021)035}}.

\bibitem{Prelovsek:2013cra}
S.~Prelovsek, L.~Leskovec, {Evidence for X(3872) from DD* scattering on the
  lattice}, Phys. Rev. Lett. 111 (2013) 192001.
\newblock \href {http://arxiv.org/abs/1307.5172} {\path{arXiv:1307.5172}},
  \href {https://doi.org/10.1103/PhysRevLett.111.192001}
  {\path{doi:10.1103/PhysRevLett.111.192001}}.

\bibitem{Belle:2017egg}
K.~Chilikin, et~al., Belle Collaboration, {Observation of an alternative
  $\chi_{c0}(2P)$ candidate in $e^+ e^- \rightarrow J/\psi D \bar{D}$}, Phys.
  Rev. D 95 (2017) 112003.
\newblock \href {http://arxiv.org/abs/1704.01872} {\path{arXiv:1704.01872}},
  \href {https://doi.org/10.1103/PhysRevD.95.112003}
  {\path{doi:10.1103/PhysRevD.95.112003}}.

\bibitem{Zhou:2017dwj}
Z.-Y. Zhou, Z.~Xiao, {Understanding $X(3862)$, $X(3872)$, and $X(3930)$ in a
  Friedrichs-model-like scheme}, Phys. Rev. D 96 (2017) 054031, [Erratum:
  Phys.Rev.D 96, 099905 (2017)].
\newblock \href {http://arxiv.org/abs/1704.04438} {\path{arXiv:1704.04438}},
  \href {https://doi.org/10.1103/PhysRevD.96.054031}
  {\path{doi:10.1103/PhysRevD.96.054031}}.

\bibitem{Yu:2017bsj}
G.~L. Yu, Z.~G. Wang, Z.~Y. Li, {The analysis of the charmonium-like states
  $X^{*}(3860)$,$X(3872)$, $X(3915)$, $X(3930)$ and $X(3940)$ according to its
  strong decay behaviors}, Chin. Phys. C 42 (2018) 4.
\newblock \href {http://arxiv.org/abs/1704.06763} {\path{arXiv:1704.06763}},
  \href {https://doi.org/10.1088/1674-1137/42/4/043107}
  {\path{doi:10.1088/1674-1137/42/4/043107}}.

\bibitem{Ortega:2017qmg}
P.~G. Ortega, J.~Segovia, D.~R. Entem, F.~Fern{\'a}ndez, {Charmonium resonances
  in the 3.9 GeV/$c^2$ energy region and the $X(3915)/X(3930)$ puzzle}, Phys.
  Lett. B 778 (2018) 1--5.
\newblock \href {http://arxiv.org/abs/1706.02639} {\path{arXiv:1706.02639}},
  \href {https://doi.org/10.1016/j.physletb.2018.01.005}
  {\path{doi:10.1016/j.physletb.2018.01.005}}.

\bibitem{BaBar:2010jfn}
B.~Aubert, et~al., BaBar Collaboration, {Observation of the $\chi_{c2}(2p)$
  meson in the reaction $\gamma \gamma \to D \bar{D}$ at {BaBar}}, Phys. Rev. D
  81 (2010) 092003.
\newblock \href {http://arxiv.org/abs/1002.0281} {\path{arXiv:1002.0281}},
  \href {https://doi.org/10.1103/PhysRevD.81.092003}
  {\path{doi:10.1103/PhysRevD.81.092003}}.

\bibitem{Munz:1996hb}
C.~R. Munz, {Two photon decays of mesons in a relativistic quark model}, Nucl.
  Phys. A 609 (1996) 364--376.
\newblock \href {http://arxiv.org/abs/hep-ph/9601206}
  {\path{arXiv:hep-ph/9601206}}, \href
  {https://doi.org/10.1016/S0375-9474(96)00265-5}
  {\path{doi:10.1016/S0375-9474(96)00265-5}}.

\bibitem{Ebert:2003mu}
D.~Ebert, R.~N. Faustov, V.~O. Galkin, {Two photon decay rates of heavy
  quarkonia in the relativistic quark model}, Mod. Phys. Lett. A 18 (2003)
  601--608.
\newblock \href {http://arxiv.org/abs/hep-ph/0302044}
  {\path{arXiv:hep-ph/0302044}}, \href
  {https://doi.org/10.1142/S021773230300971X}
  {\path{doi:10.1142/S021773230300971X}}.

\bibitem{Hwang:2010iq}
C.-W. Hwang, R.-S. Guo, {Two-photon and two-gluon decays of p-wave heavy
  quarkonium using a covariant light-front approach}, Phys. Rev. D 82 (2010)
  034021.
\newblock \href {http://arxiv.org/abs/1005.2811} {\path{arXiv:1005.2811}},
  \href {https://doi.org/10.1103/PhysRevD.82.034021}
  {\path{doi:10.1103/PhysRevD.82.034021}}.

\bibitem{Wang:2007nb}
G.-L. Wang, {Annihilation rate of heavy $0^{++}$ P-wave quarkonium in
  relativistic Salpeter method}, Phys. Lett. B 653 (2007) 206--209.
\newblock \href {http://arxiv.org/abs/0708.3516} {\path{arXiv:0708.3516}},
  \href {https://doi.org/10.1016/j.physletb.2007.08.017}
  {\path{doi:10.1016/j.physletb.2007.08.017}}.

\bibitem{Chen:2012wy}
D.-Y. Chen, J.~He, X.~Liu, T.~Matsuki, T.~Matsuki, {Does the enhancement
  observed in $\gamma\gamma\to D\bar{D}$ contain two $P$-wave higher
  charmonia?}, Eur. Phys. J. C 72 (2012) 2226.
\newblock \href {http://arxiv.org/abs/1207.3561} {\path{arXiv:1207.3561}},
  \href {https://doi.org/10.1140/epjc/s10052-012-2226-4}
  {\path{doi:10.1140/epjc/s10052-012-2226-4}}.

\bibitem{Zhou:2015uva}
Z.-Y. Zhou, Z.~Xiao, H.-Q. Zhou, {Could the $X(3915)$ and the $X(3930)$ Be the
  Same Tensor State?}, Phys. Rev. Lett. 115 (2015) 022001.
\newblock \href {http://arxiv.org/abs/1501.00879} {\path{arXiv:1501.00879}},
  \href {https://doi.org/10.1103/PhysRevLett.115.022001}
  {\path{doi:10.1103/PhysRevLett.115.022001}}.

\bibitem{Baru:2017fgv}
V.~Baru, C.~Hanhart, A.~V. Nefediev, {Can $X(3915)$ be the tensor partner of
  the $X(3872)$?}, JHEP 06 (2017) 010.
\newblock \href {http://arxiv.org/abs/1703.01230} {\path{arXiv:1703.01230}},
  \href {https://doi.org/10.1007/JHEP06(2017)010}
  {\path{doi:10.1007/JHEP06(2017)010}}.

\bibitem{LHCb:2020pxc}
R.~Aaij, et~al., LHCb Collaboration, {Amplitude analysis of the $B^+\to
  D^+D^-K^+$ decay}, Phys. Rev. D 102 (2020) 112003.
\newblock \href {http://arxiv.org/abs/2009.00026} {\path{arXiv:2009.00026}},
  \href {https://doi.org/10.1103/PhysRevD.102.112003}
  {\path{doi:10.1103/PhysRevD.102.112003}}.

\bibitem{LHCb:2020bls}
R.~Aaij, et~al., LHCb Collaboration, {A model-independent study of resonant
  structure in $B^+\to D^+D^-K^+$ decays}, Phys. Rev. Lett. 125 (2020) 242001.
\newblock \href {http://arxiv.org/abs/2009.00025} {\path{arXiv:2009.00025}},
  \href {https://doi.org/10.1103/PhysRevLett.125.242001}
  {\path{doi:10.1103/PhysRevLett.125.242001}}.

\bibitem{Yu:2017pmn}
F.-S. Yu, {Weak-decay searches for $Qs{\bar{u}}{\bar{d}}$ tetraquarks}, Eur.
  Phys. J. C 82 (2022) 641.
\newblock \href {http://arxiv.org/abs/1709.02571} {\path{arXiv:1709.02571}},
  \href {https://doi.org/10.1140/epjc/s10052-022-10567-8}
  {\path{doi:10.1140/epjc/s10052-022-10567-8}}.

\bibitem{Liu:2020orv}
X.-H. Liu, M.-J. Yan, H.-W. Ke, G.~Li, J.-J. Xie, {Triangle singularity as the
  origin of $X_0(2900)$ and $X_1(2900)$ observed in $B^+\to D^+ D^- K^+$}, Eur.
  Phys. J. C 80 (2020) 1178.
\newblock \href {http://arxiv.org/abs/2008.07190} {\path{arXiv:2008.07190}},
  \href {https://doi.org/10.1140/epjc/s10052-020-08762-6}
  {\path{doi:10.1140/epjc/s10052-020-08762-6}}.

\bibitem{Zhang:2020oze}
J.-R. Zhang, {Open-charm tetraquark candidate: Note on $X_0$(2900)}, Phys. Rev.
  D 103 (2021) 054019.
\newblock \href {http://arxiv.org/abs/2008.07295} {\path{arXiv:2008.07295}},
  \href {https://doi.org/10.1103/PhysRevD.103.054019}
  {\path{doi:10.1103/PhysRevD.103.054019}}.

\bibitem{He:2020jna}
X.-G. He, W.~Wang, R.~Zhu, {Open-charm tetraquark $X_c$ and open-bottom
  tetraquark $X_b$}, Eur. Phys. J. C 80 (2020) 1026.
\newblock \href {http://arxiv.org/abs/2008.07145} {\path{arXiv:2008.07145}},
  \href {https://doi.org/10.1140/epjc/s10052-020-08597-1}
  {\path{doi:10.1140/epjc/s10052-020-08597-1}}.

\bibitem{Chen:2020aos}
H.-X. Chen, W.~Chen, R.-R. Dong, N.~Su, {$X_0$(2900) and $X_1$(2900): hadronic
  molecules or compact tetraquarks}, Chin. Phys. Lett. 37 (2020) 101201.
\newblock \href {http://arxiv.org/abs/2008.07516} {\path{arXiv:2008.07516}},
  \href {https://doi.org/10.1088/0256-307X/37/10/101201}
  {\path{doi:10.1088/0256-307X/37/10/101201}}.

\bibitem{Lu:2020qmp}
Q.-F. L{\"u}, D.-Y. Chen, Y.-B. Dong, {Open charm and bottom tetraquarks in an
  extended relativized quark model}, Phys. Rev. D 102 (2020) 074021.
\newblock \href {http://arxiv.org/abs/2008.07340} {\path{arXiv:2008.07340}},
  \href {https://doi.org/10.1103/PhysRevD.102.074021}
  {\path{doi:10.1103/PhysRevD.102.074021}}.

\bibitem{Huang:2020ptc}
Y.~Huang, J.-X. Lu, J.-J. Xie, L.-S. Geng, {Strong decays of
  ${\bar{D}}^{*}K^{*}$ molecules and the newly observed $X_{0,1}$ states}, Eur.
  Phys. J. C 80 (2020) 973.
\newblock \href {http://arxiv.org/abs/2008.07959} {\path{arXiv:2008.07959}},
  \href {https://doi.org/10.1140/epjc/s10052-020-08516-4}
  {\path{doi:10.1140/epjc/s10052-020-08516-4}}.

\bibitem{Wang:2020xyc}
Z.-G. Wang, {Analysis of the $X_0(2900)$ as the scalar tetraquark state via the
  QCD sum rules}, Int. J. Mod. Phys. A 35 (2020) 2050187.
\newblock \href {http://arxiv.org/abs/2008.07833} {\path{arXiv:2008.07833}},
  \href {https://doi.org/10.1142/S0217751X20501870}
  {\path{doi:10.1142/S0217751X20501870}}.

\bibitem{Hu:2020mxp}
M.-W. Hu, X.-Y. Lao, P.~Ling, Q.~Wang, {$X_0$(2900) and its heavy quark spin
  partners in molecular picture}, Chin. Phys. C 45 (2021) 021003.
\newblock \href {http://arxiv.org/abs/2008.06894} {\path{arXiv:2008.06894}},
  \href {https://doi.org/10.1088/1674-1137/abcfaa}
  {\path{doi:10.1088/1674-1137/abcfaa}}.

\bibitem{Karliner:2020vsi}
M.~Karliner, J.~L. Rosner, {First exotic hadron with open heavy flavor: $cs\bar
  u\bar d$ tetraquark}, Phys. Rev. D 102 (2020) 094016.
\newblock \href {http://arxiv.org/abs/2008.05993} {\path{arXiv:2008.05993}},
  \href {https://doi.org/10.1103/PhysRevD.102.094016}
  {\path{doi:10.1103/PhysRevD.102.094016}}.

\bibitem{Agaev:2020nrc}
S.~S. Agaev, K.~Azizi, H.~Sundu, {New scalar resonance $X_ 0(2900)$ as a
  molecule: mass and width}, J. Phys. G 48 (2021) 085012.
\newblock \href {http://arxiv.org/abs/2008.13027} {\path{arXiv:2008.13027}},
  \href {https://doi.org/10.1088/1361-6471/ac0b31}
  {\path{doi:10.1088/1361-6471/ac0b31}}.

\bibitem{Yang:2020atz}
G.~Yang, J.~Ping, J.~Segovia, {Tetra- and penta-quark structures in the
  constituent quark model}, Symmetry 12 (2020) 1869.
\newblock \href {http://arxiv.org/abs/2009.00238} {\path{arXiv:2009.00238}},
  \href {https://doi.org/10.3390/sym12111869} {\path{doi:10.3390/sym12111869}}.

\bibitem{Chen:2020eyu}
Y.-K. Chen, J.-J. Han, Q.-F. L{\"u}, J.-P. Wang, F.-S. Yu, {Branching fractions
  of $B^-\rightarrow D^-X_{0,1}(2900)$ and their implications}, Eur. Phys. J. C
  81 (2021) 71.
\newblock \href {http://arxiv.org/abs/2009.01182} {\path{arXiv:2009.01182}},
  \href {https://doi.org/10.1140/epjc/s10052-021-08857-8}
  {\path{doi:10.1140/epjc/s10052-021-08857-8}}.

\bibitem{Mutuk:2020igv}
H.~Mutuk, {Monte-Carlo based QCD sum rules analysis of $X_0$(2900) and
  $X_1$(2900)}, J. Phys. G 48 (2021) 055007.
\newblock \href {http://arxiv.org/abs/2009.02492} {\path{arXiv:2009.02492}},
  \href {https://doi.org/10.1088/1361-6471/abeb7f}
  {\path{doi:10.1088/1361-6471/abeb7f}}.

\bibitem{Burns:2020xne}
T.~J. Burns, E.~S. Swanson, {Discriminating among interpretations for the
  $X(2900)$ states}, Phys. Rev. D 103 (2021) 014004.
\newblock \href {http://arxiv.org/abs/2009.05352} {\path{arXiv:2009.05352}},
  \href {https://doi.org/10.1103/PhysRevD.103.014004}
  {\path{doi:10.1103/PhysRevD.103.014004}}.

\bibitem{Dong:2020rgs}
X.-K. Dong, B.-S. Zou, {Prediction of possible $DK_1$ bound states}, Eur. Phys.
  J. A 57 (2021) 139.
\newblock \href {http://arxiv.org/abs/2009.11619} {\path{arXiv:2009.11619}},
  \href {https://doi.org/10.1140/epja/s10050-021-00442-7}
  {\path{doi:10.1140/epja/s10050-021-00442-7}}.

\bibitem{Xiao:2020ltm}
C.-J. Xiao, D.-Y. Chen, Y.-B. Dong, G.-W. Meng, {Study of the decays of
  $S-$wave $\bar D^\ast K^\ast$ hadronic molecules: The scalar $X_0(2900)$ and
  its spin partners $X_{J(J=1,2)}$}, Phys. Rev. D 103 (2021) 034004.
\newblock \href {http://arxiv.org/abs/2009.14538} {\path{arXiv:2009.14538}},
  \href {https://doi.org/10.1103/PhysRevD.103.034004}
  {\path{doi:10.1103/PhysRevD.103.034004}}.

\bibitem{Tan:2020cpu}
Y.~Tan, J.~Ping, {$X(2900)$ in a chiral quark model}, Chin. Phys. C 45 (2021)
  093104.
\newblock \href {http://arxiv.org/abs/2010.04045} {\path{arXiv:2010.04045}},
  \href {https://doi.org/10.1088/1674-1137/ac0ba4}
  {\path{doi:10.1088/1674-1137/ac0ba4}}.

\bibitem{An:2020vku}
H.-T. An, K.~Chen, X.~Liu, {Manifestly exotic pentaquarks with a single heavy
  quark}, Phys. Rev. D 105 (2022) 034018.
\newblock \href {http://arxiv.org/abs/2010.05014} {\path{arXiv:2010.05014}},
  \href {https://doi.org/10.1103/PhysRevD.105.034018}
  {\path{doi:10.1103/PhysRevD.105.034018}}.

\bibitem{Wang:2020prk}
G.-J. Wang, L.~Meng, L.-Y. Xiao, M.~Oka, S.-L. Zhu, {Mass spectrum and strong
  decays of tetraquark ${\bar{c}}{\bar{s}} qq$ states}, Eur. Phys. J. C 81
  (2021) 188.
\newblock \href {http://arxiv.org/abs/2010.09395} {\path{arXiv:2010.09395}},
  \href {https://doi.org/10.1140/epjc/s10052-021-08978-0}
  {\path{doi:10.1140/epjc/s10052-021-08978-0}}.

\bibitem{Abreu:2020ony}
L.~M. Abreu, {$X_J(2900)$ states in a hot hadronic medium}, Phys. Rev. D 103
  (2021) 036013.
\newblock \href {http://arxiv.org/abs/2010.14955} {\path{arXiv:2010.14955}},
  \href {https://doi.org/10.1103/PhysRevD.103.036013}
  {\path{doi:10.1103/PhysRevD.103.036013}}.

\bibitem{Yang:2021izl}
G.~Yang, J.~Ping, J.~Segovia, {${sQ\bar{q}\bar{q}}$ ${(q=u,\,d;\, Q=c,\,b)}$
  tetraquarks in the chiral quark model}, Phys. Rev. D 103 (2021) 074011.
\newblock \href {http://arxiv.org/abs/2101.04933} {\path{arXiv:2101.04933}},
  \href {https://doi.org/10.1103/PhysRevD.103.074011}
  {\path{doi:10.1103/PhysRevD.103.074011}}.

\bibitem{Qi:2021iyv}
J.-J. Qi, Z.-Y. Wang, Z.-F. Zhang, X.-H. Guo, {Studying the ${\bar{D}}_1K$
  molecule in the Bethe{\textendash}Salpeter equation approach}, Eur. Phys. J.
  C 81 (2021) 639.
\newblock \href {http://arxiv.org/abs/2101.06688} {\path{arXiv:2101.06688}},
  \href {https://doi.org/10.1140/epjc/s10052-021-09422-z}
  {\path{doi:10.1140/epjc/s10052-021-09422-z}}.

\bibitem{Albuquerque:2021svg}
R.~M. Albuquerque, S.~Narison, D.~Rabetiarivony, G.~Randriamanatrika, {The new
  charm-strange resonances in the $D^- K^+$ channel}, Nucl. Part. Phys. Proc.
  312-317 (2021) 125--129.
\newblock \href {http://arxiv.org/abs/2102.04622} {\path{arXiv:2102.04622}},
  \href {https://doi.org/10.1016/j.nuclphysbps.2021.05.032}
  {\path{doi:10.1016/j.nuclphysbps.2021.05.032}}.

\bibitem{Wang:2021lwy}
B.~Wang, S.-L. Zhu, {How to understand the $X(2900)$?}, Eur. Phys. J. C 82
  (2022) 419.
\newblock \href {http://arxiv.org/abs/2107.09275} {\path{arXiv:2107.09275}},
  \href {https://doi.org/10.1140/epjc/s10052-022-10396-9}
  {\path{doi:10.1140/epjc/s10052-022-10396-9}}.

\bibitem{Chen:2021tad}
H.~Chen, H.-R. Qi, H.-Q. Zheng, {$X_1(2900)$ as a $\bar{D}_1 K$ molecule}, Eur.
  Phys. J. C 81 (2021) 812.
\newblock \href {http://arxiv.org/abs/2108.02387} {\path{arXiv:2108.02387}},
  \href {https://doi.org/10.1140/epjc/s10052-021-09603-w}
  {\path{doi:10.1140/epjc/s10052-021-09603-w}}.

\bibitem{ParticleDataGroup:2022pth}
R.~L. Workman, et~al., Particle Data Group Collaboration, {Review of particle
  physics}, PTEP 2022 (2022) 083C01.
\newblock \href {https://doi.org/10.1093/ptep/ptac097}
  {\path{doi:10.1093/ptep/ptac097}}.

\bibitem{Belle:2007dxy}
C.~Z. Yuan, et~al., Belle Collaboration, {Measurement of
  $e^+e^-\to\pi^+\pi^-J/\psi$ cross-section via initial state radiation at
  Belle}, Phys. Rev. Lett. 99 (2007) 182004.
\newblock \href {http://arxiv.org/abs/0707.2541} {\path{arXiv:0707.2541}},
  \href {https://doi.org/10.1103/PhysRevLett.99.182004}
  {\path{doi:10.1103/PhysRevLett.99.182004}}.

\bibitem{BESIII:2017tqk}
M.~Ablikim, et~al., BESIII Collaboration, {Measurement of
  $e^{+}e^{-}\rightarrow \pi^{+}\pi^{-}\psi(3686)$ from 4.008 to 4.600 GeV and
  observation of a charged structure in the $\pi^{\pm}\psi(3686)$ mass
  spectrum}, Phys. Rev. D 96 (2017) 032004, [Erratum: Phys.Rev.D 99, 019903
  (2019)].
\newblock \href {http://arxiv.org/abs/1703.08787} {\path{arXiv:1703.08787}},
  \href {https://doi.org/10.1103/PhysRevD.96.032004}
  {\path{doi:10.1103/PhysRevD.96.032004}}.

\bibitem{BESIII:2017dxi}
M.~Ablikim, et~al., BESIII Collaboration, {Observation of $e^{+}e^{-} \to \eta
  h_{c}$ at center-of-mass energies from 4.085 to 4.600 GeV}, Phys. Rev. D 96
  (2017) 012001.
\newblock \href {http://arxiv.org/abs/1704.08033} {\path{arXiv:1704.08033}},
  \href {https://doi.org/10.1103/PhysRevD.96.012001}
  {\path{doi:10.1103/PhysRevD.96.012001}}.

\bibitem{BESIII:2018iea}
M.~Ablikim, et~al., BESIII Collaboration, {Evidence of a resonant structure in
  the $e^+e^-\to \pi^+D^0D^{*-}$ cross section between 4.05 and 4.60 GeV},
  Phys. Rev. Lett. 122 (2019) 102002.
\newblock \href {http://arxiv.org/abs/1808.02847} {\path{arXiv:1808.02847}},
  \href {https://doi.org/10.1103/PhysRevLett.122.102002}
  {\path{doi:10.1103/PhysRevLett.122.102002}}.

\bibitem{BESIII:2019gjc}
M.~Ablikim, et~al., BESIII Collaboration, {Cross section measurements of $e^+
  e^-\to\omega\chi_{c0}$ form $\sqrt{s}=$ 4.178 to 4.278 GeV}, Phys. Rev. D 99
  (2019) 091103.
\newblock \href {http://arxiv.org/abs/1903.02359} {\path{arXiv:1903.02359}},
  \href {https://doi.org/10.1103/PhysRevD.99.091103}
  {\path{doi:10.1103/PhysRevD.99.091103}}.

\bibitem{BESIII:2020bgb}
M.~Ablikim, et~al., BESIII Collaboration, {Observation of the $Y(4220)$ and
  $Y(4360)$ in the process $e^{+}e^{-} \to \eta J/\psi$}, Phys. Rev. D 102
  (2020) 031101.
\newblock \href {http://arxiv.org/abs/2003.03705} {\path{arXiv:2003.03705}},
  \href {https://doi.org/10.1103/PhysRevD.102.031101}
  {\path{doi:10.1103/PhysRevD.102.031101}}.

\bibitem{BESIII:2020oph}
M.~Ablikim, et~al., BESIII Collaboration, {Study of the process
  $e^+e^-\to\pi^0\pi^0 J/\psi$ and neutral charmonium-like state
  $Z_c(3900)^0$}, Phys. Rev. D 102 (2020) 012009.
\newblock \href {http://arxiv.org/abs/2004.13788} {\path{arXiv:2004.13788}},
  \href {https://doi.org/10.1103/PhysRevD.102.012009}
  {\path{doi:10.1103/PhysRevD.102.012009}}.

\bibitem{Ablikim:2020jrn}
M.~Ablikim, BESIII Collaboration, {Measurement of cross sections for
  $e^{+}e^{-} \rightarrow \mu^+\mu^-$ at center-of-mass energies from 3.80 to
  4.60 GeV}, Phys. Rev. D 102 (2020) 112009.
\newblock \href {http://arxiv.org/abs/2007.12872} {\path{arXiv:2007.12872}},
  \href {https://doi.org/10.1103/PhysRevD.102.112009}
  {\path{doi:10.1103/PhysRevD.102.112009}}.

\bibitem{BESIII:2020tgt}
M.~Ablikim, et~al., BESIII Collaboration, {Measurements of $e^+e^-\rightarrow
  \eta_{\rm c}\pi^+ \pi^-\pi^0$, $\eta_{\rm c}\pi^+ \pi^-$, and $\eta_{\rm
  c}\pi^0\gamma$ at $\sqrt{s}$ from 4.18 to 4.60 GeV, and search for a $Z_{\rm
  c}$ state close to the $D\bar{D}$ threshold decaying to $\eta_{\rm c}\pi$ at
  $\sqrt{s}$ = 4.23 GeV}, Phys. Rev. D 103 (2021) 032006.
\newblock \href {http://arxiv.org/abs/2010.14415} {\path{arXiv:2010.14415}},
  \href {https://doi.org/10.1103/PhysRevD.103.032006}
  {\path{doi:10.1103/PhysRevD.103.032006}}.

\bibitem{BESIII:2022joj}
M.~Ablikim, et~al., BESIII Collaboration, {Observation of the $Y(4230)$ and a
  new structure in $e^+e^-\to K^+K^-J/\psi$}, Chin. Phys. C 46 (2022) 111002.
\newblock \href {http://arxiv.org/abs/2204.07800} {\path{arXiv:2204.07800}},
  \href {https://doi.org/10.1088/1674-1137/ac945c}
  {\path{doi:10.1088/1674-1137/ac945c}}.

\bibitem{BESIII:2023cmv}
M.~Ablikim, et~al., BESIII Collaboration, {Observation of three charmoniumlike
  States with $J^{PC}=1^{--}$ in $e^+e^-\to D^{*0}D^{*-}\pi^+$}, Phys. Rev.
  Lett. 130 (2023) 121901.
\newblock \href {http://arxiv.org/abs/2301.07321} {\path{arXiv:2301.07321}},
  \href {https://doi.org/10.1103/PhysRevLett.130.121901}
  {\path{doi:10.1103/PhysRevLett.130.121901}}.

\bibitem{BESIII:2021njb}
M.~Ablikim, et~al., BESIII Collaboration, {Cross section measurement of
  $e^+e^-\rightarrow\pi^+\pi^-(3686)$ from $\sqrt{s}=4.0076$ to 4.6984 GeV},
  Phys. Rev. D 104 (2021) 052012.
\newblock \href {http://arxiv.org/abs/2107.09210} {\path{arXiv:2107.09210}},
  \href {https://doi.org/10.1103/PhysRevD.104.052012}
  {\path{doi:10.1103/PhysRevD.104.052012}}.

\bibitem{BESIII:2022qal}
M.~Ablikim, et~al., BESIII Collaboration, {Study of the resonance structures in
  the process $e^+e^+ \to \pi^+\pi^- J/ \psi$}, Phys. Rev. D 106 (2022) 072001.
\newblock \href {http://arxiv.org/abs/2206.08554} {\path{arXiv:2206.08554}},
  \href {https://doi.org/10.1103/PhysRevD.106.072001}
  {\path{doi:10.1103/PhysRevD.106.072001}}.

\bibitem{BESIII:2022kcv}
M.~Ablikim, et~al., BESIII Collaboration, {Observation of the $Y(4230)$ and
  evidence for a new vector charmoniumlike state $Y(4710)$ in $e^+e^-\to
  K_S^0K_S^0 J/\psi$}, Phys. Rev. D 107 (2023) 092005.
\newblock \href {http://arxiv.org/abs/2211.08561} {\path{arXiv:2211.08561}},
  \href {https://doi.org/10.1103/PhysRevD.107.092005}
  {\path{doi:10.1103/PhysRevD.107.092005}}.

\bibitem{BESIII:2023tll}
M.~Ablikim, et~al., BESIII Collaboration, {Measurement of $e^+e^-\to\eta J\psi$
  cross section from $\sqrt s=3.808$ GeV to 4.951 GeV}, Phys. Rev. D 109 (2024)
  092012.
\newblock \href {http://arxiv.org/abs/2310.03361} {\path{arXiv:2310.03361}},
  \href {https://doi.org/10.1103/PhysRevD.109.092012}
  {\path{doi:10.1103/PhysRevD.109.092012}}.

\bibitem{BESIII:2024yqi}
M.~Ablikim, et~al., BESIII Collaboration, {Measurement of the born cross
  section for $e^+e^-\to\eta h_c$ at center-of-mass energies between 4.1 and
  4.6~GeV}, Phys. Rev. D 111 (2025) L011101.
\newblock \href {http://arxiv.org/abs/2404.06718} {\path{arXiv:2404.06718}},
  \href {https://doi.org/10.1103/PhysRevD.111.L011101}
  {\path{doi:10.1103/PhysRevD.111.L011101}}.

\bibitem{BESIII:2025bce}
M.~Ablikim, et~al., BESIII Collaboration, {Observation of three resonance
  structure in the cross section of $e^+e^-\to\pi^+\pi^- h_c$} (4 2025).
\newblock \href {http://arxiv.org/abs/2504.04096} {\path{arXiv:2504.04096}}.

\bibitem{BESIII:2026hie}
M.~Ablikim, et~al., BESIII Collaboration, {Measurement of born cross sections
  for $e^+e^-\to\Sigma^-\bar{\Sigma}^+$ at $\sqrt{s}=3.51-4.95$ GeV and
  observation of $\psi(3770)\to\Sigma^-\bar{\Sigma}^+$} (2 2026).
\newblock \href {http://arxiv.org/abs/2602.23835} {\path{arXiv:2602.23835}}.

\bibitem{BESIII:2026oyx}
M.~Ablikim, et~al., BESIII Collaboration, {Measurement of born cross sections
  for $e^+e^- \to K^+\Xi^0\bar{\Sigma}^-$ at $\sqrt{s} = 3.51-4.95$ GeV and
  observation of $\psi(3770) \to K^+\Xi^0\bar{\Sigma}^-$} (5 2026).
\newblock \href {http://arxiv.org/abs/2605.19780} {\path{arXiv:2605.19780}}.

\bibitem{BESIII:2026lcb}
M.~Ablikim, et~al., BESIII Collaboration, {Observation of $\psi(3770)\to p\bar
  p$ and measurement of electromagnetic form factors of proton at $\sqrt{s} =
  3.510-4.946$ GeV} (6 2026).
\newblock \href {http://arxiv.org/abs/2606.22759} {\path{arXiv:2606.22759}}.

\bibitem{BESIII:2026ont}
M.~Ablikim, et~al., BESIII Collaboration, {Measurement of born cross sections
  for $e^+e^-\to p\bar p$ at $\sqrt{s} =3.510-4.946$ GeV} (6 2026).
\newblock \href {http://arxiv.org/abs/2606.22762} {\path{arXiv:2606.22762}}.

\bibitem{BESIII:2023wsc}
M.~Ablikim, et~al., BESIII Collaboration, {Precise Measurement of the $e^+e^-
  \to D_s^{*+}D_s^{*-}$ Cross Sections at Center-of-Mass Energies from
  Threshold to 4.95~GeV}, Phys. Rev. Lett. 131 (2023) 151903.
\newblock \href {http://arxiv.org/abs/2305.10789} {\path{arXiv:2305.10789}},
  \href {https://doi.org/10.1103/PhysRevLett.131.151903}
  {\path{doi:10.1103/PhysRevLett.131.151903}}.

\bibitem{BESIII:2024jzg}
M.~Ablikim, et~al., BESIII Collaboration, {Observation of Structures in the
  Processes $e^+e^- \to \omega \chi_{c1}$ and $\omega \chi_{c2}$}, Phys. Rev.
  Lett. 132 (2024) 161901.
\newblock \href {http://arxiv.org/abs/2401.14720} {\path{arXiv:2401.14720}},
  \href {https://doi.org/10.1103/PhysRevLett.132.161901}
  {\path{doi:10.1103/PhysRevLett.132.161901}}.

\bibitem{CLEO:2006tct}
Q.~He, et~al., CLEO Collaboration, {Confirmation of the $Y(4260)$ resonance
  production in initial state radiation}, Phys. Rev. D 74 (2006) 091104.
\newblock \href {http://arxiv.org/abs/hep-ex/0611021}
  {\path{arXiv:hep-ex/0611021}}, \href
  {https://doi.org/10.1103/PhysRevD.74.091104}
  {\path{doi:10.1103/PhysRevD.74.091104}}.

\bibitem{CLEO:2006ike}
T.~E. Coan, et~al., CLEO Collaboration, {Charmonium decays of $Y(4260)$,
  $\psi(4160)$ and $\psi(4040)$}, Phys. Rev. Lett. 96 (2006) 162003.
\newblock \href {http://arxiv.org/abs/hep-ex/0602034}
  {\path{arXiv:hep-ex/0602034}}, \href
  {https://doi.org/10.1103/PhysRevLett.96.162003}
  {\path{doi:10.1103/PhysRevLett.96.162003}}.

\bibitem{Wang:2025clb}
Q.~Wang, Q.~Zhao, {A Short Review of the Vector Charmonium-Like State
  {\ensuremath{\psi}}(4230)}, Chin. Phys. Lett. 42 (2025) 110201.
\newblock \href {http://arxiv.org/abs/2508.05304} {\path{arXiv:2508.05304}},
  \href {https://doi.org/10.1088/0256-307X/42/11/110201}
  {\path{doi:10.1088/0256-307X/42/11/110201}}.

\bibitem{Zhang:2006td}
A.~Zhang, {Charmonium spectrum and new observed states}, Phys. Lett. B 647
  (2007) 140--144.
\newblock \href {http://arxiv.org/abs/hep-ph/0603093}
  {\path{arXiv:hep-ph/0603093}}, \href
  {https://doi.org/10.1016/j.physletb.2007.01.062}
  {\path{doi:10.1016/j.physletb.2007.01.062}}.

\bibitem{Llanes-Estrada:2005qvr}
F.~J. Llanes-Estrada, {$Y(4260)$ and possible charmonium assignment}, Phys.
  Rev. D 72 (2005) 031503.
\newblock \href {http://arxiv.org/abs/hep-ph/0507035}
  {\path{arXiv:hep-ph/0507035}}, \href
  {https://doi.org/10.1103/PhysRevD.72.031503}
  {\path{doi:10.1103/PhysRevD.72.031503}}.

\bibitem{Eichten:2005ga}
E.~J. Eichten, K.~Lane, C.~Quigg, {New states above charm threshold}, Phys.
  Rev. D 73 (2006) 014014, [Erratum: Phys.Rev.D 73, 079903 (2006)].
\newblock \href {http://arxiv.org/abs/hep-ph/0511179}
  {\path{arXiv:hep-ph/0511179}}, \href
  {https://doi.org/10.1103/PhysRevD.73.014014}
  {\path{doi:10.1103/PhysRevD.73.014014}}.

\bibitem{Segovia:2008zz}
J.~Segovia, A.~M. Yasser, D.~R. Entem, F.~Fernandez, {$J^{PC}=1^{--}$ hidden
  charm resonances}, Phys. Rev. D 78 (2008) 114033.
\newblock \href {https://doi.org/10.1103/PhysRevD.78.114033}
  {\path{doi:10.1103/PhysRevD.78.114033}}.

\bibitem{Dai:2012pb}
L.~Y. Dai, M.~Shi, G.-Y. Tang, H.~Q. Zheng, {Nature of $X(4260)$}, Phys. Rev. D
  92 (2015) 014020.
\newblock \href {http://arxiv.org/abs/1206.6911} {\path{arXiv:1206.6911}},
  \href {https://doi.org/10.1103/PhysRevD.92.014020}
  {\path{doi:10.1103/PhysRevD.92.014020}}.

\bibitem{Zhao:2023hxc}
Z.~Zhao, K.~Xu, A.~Limphirat, W.~Sreethawong, N.~Tagsinsit, A.~Kaewsnod,
  X.~Liu, K.~Khosonthongkee, S.~Cheedket, Y.~Yan, {Mass spectrum of $1^{--}$
  heavy quarkonium}, Phys. Rev. D 109 (2024) 016012.
\newblock \href {http://arxiv.org/abs/2304.06243} {\path{arXiv:2304.06243}},
  \href {https://doi.org/10.1103/PhysRevD.109.016012}
  {\path{doi:10.1103/PhysRevD.109.016012}}.

\bibitem{Shah:2012js}
M.~Shah, A.~Parmar, P.~C. Vinodkumar, {Leptonic and Digamma decay Properties of
  $S$-wave quarkonia states}, Phys. Rev. D 86 (2012) 034015.
\newblock \href {http://arxiv.org/abs/1203.6184} {\path{arXiv:1203.6184}},
  \href {https://doi.org/10.1103/PhysRevD.86.034015}
  {\path{doi:10.1103/PhysRevD.86.034015}}.

\bibitem{Chiu:2005ey}
T.-W. Chiu, T.-H. Hsieh, TWQCD Collaboration, {$Y(4260)$ on the lattice}, Phys.
  Rev. D 73 (2006) 094510.
\newblock \href {http://arxiv.org/abs/hep-lat/0512029}
  {\path{arXiv:hep-lat/0512029}}, \href
  {https://doi.org/10.1103/PhysRevD.73.094510}
  {\path{doi:10.1103/PhysRevD.73.094510}}.

\bibitem{Liu:2005ay}
X.~Liu, X.-Q. Zeng, X.-Q. Li, {Possible molecular structure of the newly
  observed $Y(4260)$}, Phys. Rev. D 72 (2005) 054023.
\newblock \href {http://arxiv.org/abs/hep-ph/0507177}
  {\path{arXiv:hep-ph/0507177}}, \href
  {https://doi.org/10.1103/PhysRevD.72.054023}
  {\path{doi:10.1103/PhysRevD.72.054023}}.

\bibitem{Yuan:2005dr}
C.~Z. Yuan, P.~Wang, X.~H. Mo, {The $Y(4260)$ as an $\omega \chi_{c1}$
  molecular state}, Phys. Lett. B 634 (2006) 399--402.
\newblock \href {http://arxiv.org/abs/hep-ph/0511107}
  {\path{arXiv:hep-ph/0511107}}, \href
  {https://doi.org/10.1016/j.physletb.2006.01.031}
  {\path{doi:10.1016/j.physletb.2006.01.031}}.

\bibitem{Swanson:2005tq}
E.~Swanson, {Review of heavy hadron spectroscopy}, Int. J. Mod. Phys. A 21
  (2006) 733--738.
\newblock \href {http://arxiv.org/abs/hep-ph/0509327}
  {\path{arXiv:hep-ph/0509327}}, \href
  {https://doi.org/10.1142/S0217751X0603196X}
  {\path{doi:10.1142/S0217751X0603196X}}.

\bibitem{Qiao:2005av}
C.-F. Qiao, {One explanation for the exotic state $Y(4260)$}, Phys. Lett. B 639
  (2006) 263--265.
\newblock \href {http://arxiv.org/abs/hep-ph/0510228}
  {\path{arXiv:hep-ph/0510228}}, \href
  {https://doi.org/10.1016/j.physletb.2006.06.038}
  {\path{doi:10.1016/j.physletb.2006.06.038}}.

\bibitem{Ding:2008gr}
G.-J. Ding, {Are $Y(4260)$ and $Z^+_2$ are $D_1 D$ or $D_0 D^*$ Hadronic
  Molecules?}, Phys. Rev. D 79 (2009) 014001.
\newblock \href {http://arxiv.org/abs/0809.4818} {\path{arXiv:0809.4818}},
  \href {https://doi.org/10.1103/PhysRevD.79.014001}
  {\path{doi:10.1103/PhysRevD.79.014001}}.

\bibitem{Close:2009ag}
F.~Close, C.~Downum, {On the possibility of Deeply Bound Hadronic Molecules
  from single Pion Exchange}, Phys. Rev. Lett. 102 (2009) 242003.
\newblock \href {http://arxiv.org/abs/0905.2687} {\path{arXiv:0905.2687}},
  \href {https://doi.org/10.1103/PhysRevLett.102.242003}
  {\path{doi:10.1103/PhysRevLett.102.242003}}.

\bibitem{Close:2010wq}
F.~Close, C.~Downum, C.~E. Thomas, {Novel Charmonium and Bottomonium
  Spectroscopies due to Deeply Bound Hadronic Molecules from Single Pion
  Exchange}, Phys. Rev. D 81 (2010) 074033.
\newblock \href {http://arxiv.org/abs/1001.2553} {\path{arXiv:1001.2553}},
  \href {https://doi.org/10.1103/PhysRevD.81.074033}
  {\path{doi:10.1103/PhysRevD.81.074033}}.

\bibitem{Wang:2013cya}
Q.~Wang, C.~Hanhart, Q.~Zhao, {Decoding the riddle of $Y(4260)$ and
  $Z_c(3900)$}, Phys. Rev. Lett. 111 (2013) 132003.
\newblock \href {http://arxiv.org/abs/1303.6355} {\path{arXiv:1303.6355}},
  \href {https://doi.org/10.1103/PhysRevLett.111.132003}
  {\path{doi:10.1103/PhysRevLett.111.132003}}.

\bibitem{Cleven:2013mka}
M.~Cleven, Q.~Wang, F.-K. Guo, C.~Hanhart, U.-G. Mei{\ss}ner, Q.~Zhao,
  {$Y(4260)$ as the first $S$-wave open charm vector molecular state?}, Phys.
  Rev. D 90 (2014) 074039.
\newblock \href {http://arxiv.org/abs/1310.2190} {\path{arXiv:1310.2190}},
  \href {https://doi.org/10.1103/PhysRevD.90.074039}
  {\path{doi:10.1103/PhysRevD.90.074039}}.

\bibitem{Wang:2013kra}
Q.~Wang, M.~Cleven, F.-K. Guo, C.~Hanhart, U.-G. Mei{\ss}ner, X.-G. Wu,
  Q.~Zhao, {$Y(4260)$: hadronic molecule versus hadro-charmonium
  interpretation}, Phys. Rev. D 89 (2014) 034001.
\newblock \href {http://arxiv.org/abs/1309.4303} {\path{arXiv:1309.4303}},
  \href {https://doi.org/10.1103/PhysRevD.89.034001}
  {\path{doi:10.1103/PhysRevD.89.034001}}.

\bibitem{Li:2013bca}
M.-T. Li, W.-L. Wang, Y.-B. Dong, Z.-Y. Zhang, {A Study of $P$-wave Heavy Meson
  Interactions in A Chiral Quark Model} (3 2013).
\newblock \href {http://arxiv.org/abs/1303.4140} {\path{arXiv:1303.4140}}.

\bibitem{Xue:2017xpu}
S.-R. Xue, H.-J. Jing, F.-K. Guo, Q.~Zhao, {Disentangling the role of the
  $Y(4260)$ in $e^+e^-\to D^*\bar{D}^*$ and $D_s^*\bar{D}_s^*$ via line shape
  studies}, Phys. Lett. B 779 (2018) 402--408.
\newblock \href {http://arxiv.org/abs/1708.06961} {\path{arXiv:1708.06961}},
  \href {https://doi.org/10.1016/j.physletb.2018.02.027}
  {\path{doi:10.1016/j.physletb.2018.02.027}}.

\bibitem{Qin:2016spb}
W.~Qin, S.-R. Xue, Q.~Zhao, {Production of $Y(4260)$ as a hadronic molecule
  state of $\bar{D}D_1 +c.c.$ in $e^+e^-$ annihilations}, Phys. Rev. D 94
  (2016) 054035.
\newblock \href {http://arxiv.org/abs/1605.02407} {\path{arXiv:1605.02407}},
  \href {https://doi.org/10.1103/PhysRevD.94.054035}
  {\path{doi:10.1103/PhysRevD.94.054035}}.

\bibitem{Maiani:2005pe}
L.~Maiani, V.~Riquer, F.~Piccinini, A.~D. Polosa, {Four quark interpretation of
  $Y(4260)$}, Phys. Rev. D 72 (2005) 031502.
\newblock \href {http://arxiv.org/abs/hep-ph/0507062}
  {\path{arXiv:hep-ph/0507062}}, \href
  {https://doi.org/10.1103/PhysRevD.72.031502}
  {\path{doi:10.1103/PhysRevD.72.031502}}.

\bibitem{Drenska:2009cd}
N.~V. Drenska, R.~Faccini, A.~D. Polosa, {Exotic Hadrons with Hidden Charm and
  Strangeness}, Phys. Rev. D 79 (2009) 077502.
\newblock \href {http://arxiv.org/abs/0902.2803} {\path{arXiv:0902.2803}},
  \href {https://doi.org/10.1103/PhysRevD.79.077502}
  {\path{doi:10.1103/PhysRevD.79.077502}}.

\bibitem{Maiani:2014aja}
L.~Maiani, F.~Piccinini, A.~D. Polosa, V.~Riquer, {The $Z(4430)$ and a New
  Paradigm for Spin Interactions in Tetraquarks}, Phys. Rev. D 89 (2014)
  114010.
\newblock \href {http://arxiv.org/abs/1405.1551} {\path{arXiv:1405.1551}},
  \href {https://doi.org/10.1103/PhysRevD.89.114010}
  {\path{doi:10.1103/PhysRevD.89.114010}}.

\bibitem{Ali:2017wsf}
A.~Ali, L.~Maiani, A.~V. Borisov, I.~Ahmed, M.~Jamil~Aslam, A.~Y. Parkhomenko,
  A.~D. Polosa, A.~Rehman, {A new look at the $Y$ tetraquarks and $\Omega _c$
  baryons in the diquark model}, Eur. Phys. J. C 78 (2018) 29.
\newblock \href {http://arxiv.org/abs/1708.04650} {\path{arXiv:1708.04650}},
  \href {https://doi.org/10.1140/epjc/s10052-017-5501-6}
  {\path{doi:10.1140/epjc/s10052-017-5501-6}}.

\bibitem{Cleven:2015era}
M.~Cleven, F.-K. Guo, C.~Hanhart, Q.~Wang, Q.~Zhao, {Employing spin symmetry to
  disentangle different models for the $XYZ$ states}, Phys. Rev. D 92 (2015)
  014005.
\newblock \href {http://arxiv.org/abs/1505.01771} {\path{arXiv:1505.01771}},
  \href {https://doi.org/10.1103/PhysRevD.92.014005}
  {\path{doi:10.1103/PhysRevD.92.014005}}.

\bibitem{Ebert:2005nc}
D.~Ebert, R.~N. Faustov, V.~O. Galkin, {Masses of heavy tetraquarks in the
  relativistic quark model}, Phys. Lett. B 634 (2006) 214--219.
\newblock \href {http://arxiv.org/abs/hep-ph/0512230}
  {\path{arXiv:hep-ph/0512230}}, \href
  {https://doi.org/10.1016/j.physletb.2006.01.026}
  {\path{doi:10.1016/j.physletb.2006.01.026}}.

\bibitem{Ebert:2008kb}
D.~Ebert, R.~N. Faustov, V.~O. Galkin, {Excited heavy tetraquarks with hidden
  charm}, Eur. Phys. J. C 58 (2008) 399--405.
\newblock \href {http://arxiv.org/abs/0808.3912} {\path{arXiv:0808.3912}},
  \href {https://doi.org/10.1140/epjc/s10052-008-0754-8}
  {\path{doi:10.1140/epjc/s10052-008-0754-8}}.

\bibitem{Chen:2010ze}
W.~Chen, S.-L. Zhu, {The Vector and Axial-Vector Charmonium-like States}, Phys.
  Rev. D 83 (2011) 034010.
\newblock \href {http://arxiv.org/abs/1010.3397} {\path{arXiv:1010.3397}},
  \href {https://doi.org/10.1103/PhysRevD.83.034010}
  {\path{doi:10.1103/PhysRevD.83.034010}}.

\bibitem{Zhu:1998sv}
S.-L. Zhu, {Masses and decay widths of heavy hybrid mesons}, Phys. Rev. D 60
  (1999) 014008.
\newblock \href {http://arxiv.org/abs/hep-ph/9812405}
  {\path{arXiv:hep-ph/9812405}}, \href
  {https://doi.org/10.1103/PhysRevD.60.014008}
  {\path{doi:10.1103/PhysRevD.60.014008}}.

\bibitem{Zhu:1999wg}
S.-L. Zhu, {Some decay modes of the $1^{-+}$ hybrid meson in QCD sum rules
  revisited}, Phys. Rev. D 60 (1999) 097502.
\newblock \href {http://arxiv.org/abs/hep-ph/9903537}
  {\path{arXiv:hep-ph/9903537}}, \href
  {https://doi.org/10.1103/PhysRevD.60.097502}
  {\path{doi:10.1103/PhysRevD.60.097502}}.

\bibitem{Close:1994hc}
F.~E. Close, P.~R. Page, {The Production and decay of hybrid mesons by flux
  tube breaking}, Nucl. Phys. B 443 (1995) 233--254.
\newblock \href {http://arxiv.org/abs/hep-ph/9411301}
  {\path{arXiv:hep-ph/9411301}}, \href
  {https://doi.org/10.1016/0550-3213(95)00085-7}
  {\path{doi:10.1016/0550-3213(95)00085-7}}.

\bibitem{Close:2003mb}
F.~E. Close, S.~Godfrey, {Charmonium hybrid production in exclusive $B$ meson
  decays}, Phys. Lett. B 574 (2003) 210--216.
\newblock \href {http://arxiv.org/abs/hep-ph/0305285}
  {\path{arXiv:hep-ph/0305285}}, \href
  {https://doi.org/10.1016/j.physletb.2003.09.011}
  {\path{doi:10.1016/j.physletb.2003.09.011}}.

\bibitem{Kou:2005gt}
E.~Kou, O.~Pene, {Suppressed decay into open charm for the $Y(4260)$ being an
  hybrid}, Phys. Lett. B 631 (2005) 164--169.
\newblock \href {http://arxiv.org/abs/hep-ph/0507119}
  {\path{arXiv:hep-ph/0507119}}, \href
  {https://doi.org/10.1016/j.physletb.2005.09.013}
  {\path{doi:10.1016/j.physletb.2005.09.013}}.

\bibitem{Zhu:2005hp}
S.-L. Zhu, {The Possible interpretations of $Y(4260)$}, Phys. Lett. B 625
  (2005) 212.
\newblock \href {http://arxiv.org/abs/hep-ph/0507025}
  {\path{arXiv:hep-ph/0507025}}, \href
  {https://doi.org/10.1016/j.physletb.2005.08.068}
  {\path{doi:10.1016/j.physletb.2005.08.068}}.

\bibitem{Close:2005iz}
F.~E. Close, P.~R. Page, {Gluonic charmonium resonances at BaBar and BELLE?},
  Phys. Lett. B 628 (2005) 215--222.
\newblock \href {http://arxiv.org/abs/hep-ph/0507199}
  {\path{arXiv:hep-ph/0507199}}, \href
  {https://doi.org/10.1016/j.physletb.2005.09.016}
  {\path{doi:10.1016/j.physletb.2005.09.016}}.

\bibitem{Barnes:1995hc}
T.~Barnes, F.~E. Close, E.~S. Swanson, {Hybrid and conventional mesons in the
  flux tube model: Numerical studies and their phenomenological implications},
  Phys. Rev. D 52 (1995) 5242--5256.
\newblock \href {http://arxiv.org/abs/hep-ph/9501405}
  {\path{arXiv:hep-ph/9501405}}, \href
  {https://doi.org/10.1103/PhysRevD.52.5242}
  {\path{doi:10.1103/PhysRevD.52.5242}}.

\bibitem{Merlin:1986tz}
J.~Merlin, J.~E. Paton, {Spin Interactions in the Flux Tube Model and Hybrid
  Meson Masses}, Phys. Rev. D 35 (1987) 1668.
\newblock \href {https://doi.org/10.1103/PhysRevD.35.1668}
  {\path{doi:10.1103/PhysRevD.35.1668}}.

\bibitem{Lacock:1996ny}
P.~Lacock, C.~Michael, P.~Boyle, P.~Rowland, UKQCD Collaboration, {Hybrid
  mesons from quenched QCD}, Phys. Lett. B 401 (1997) 308--312.
\newblock \href {http://arxiv.org/abs/hep-lat/9611011}
  {\path{arXiv:hep-lat/9611011}}, \href
  {https://doi.org/10.1016/S0370-2693(97)00384-5}
  {\path{doi:10.1016/S0370-2693(97)00384-5}}.

\bibitem{Manke:1998yg}
T.~Manke, I.~T. Drummond, R.~R. Horgan, H.~P. Shanahan, UKQCD Collaboration,
  {Heavy hybrids from NRQCD}, Phys. Rev. D 57 (1998) 3829--3832.
\newblock \href {http://arxiv.org/abs/hep-lat/9710083}
  {\path{arXiv:hep-lat/9710083}}, \href
  {https://doi.org/10.1103/PhysRevD.57.R3829}
  {\path{doi:10.1103/PhysRevD.57.R3829}}.

\bibitem{Juge:1999aw}
K.~J. Juge, J.~Kuti, C.~J. Morningstar, {The Heavy hybrid spectrum from NRQCD
  and the Born-Oppenheimer approximation}, Nucl. Phys. B Proc. Suppl. 83 (2000)
  304--306.
\newblock \href {http://arxiv.org/abs/hep-lat/9909165}
  {\path{arXiv:hep-lat/9909165}}, \href
  {https://doi.org/10.1016/S0920-5632(00)91655-4}
  {\path{doi:10.1016/S0920-5632(00)91655-4}}.

\bibitem{Drummond:1999db}
I.~T. Drummond, N.~A. Goodman, R.~R. Horgan, H.~P. Shanahan, L.~C. Storoni,
  {Spin effects in heavy hybrid mesons on an anisotropic lattice}, Phys. Lett.
  B 478 (2000) 151--160.
\newblock \href {http://arxiv.org/abs/hep-lat/9912041}
  {\path{arXiv:hep-lat/9912041}}, \href
  {https://doi.org/10.1016/S0370-2693(00)00225-2}
  {\path{doi:10.1016/S0370-2693(00)00225-2}}.

\bibitem{Manke:1999ru}
T.~Manke, CP-PACS Collaboration, {Exotic quarkonia from anisotropic lattices},
  Nucl. Phys. B Proc. Suppl. 86 (2000) 397--400.
\newblock \href {http://arxiv.org/abs/hep-lat/9909038}
  {\path{arXiv:hep-lat/9909038}}, \href
  {https://doi.org/10.1016/S0920-5632(00)00593-4}
  {\path{doi:10.1016/S0920-5632(00)00593-4}}.

\bibitem{HadronSpectrum:2012gic}
L.~Liu, G.~Moir, M.~Peardon, S.~M. Ryan, C.~E. Thomas, P.~Vilaseca, J.~J.
  Dudek, R.~G. Edwards, B.~Joo, D.~G. Richards, Hadron Spectrum Collaboration,
  {Excited and exotic charmonium spectroscopy from lattice QCD}, JHEP 07 (2012)
  126.
\newblock \href {http://arxiv.org/abs/1204.5425} {\path{arXiv:1204.5425}},
  \href {https://doi.org/10.1007/JHEP07(2012)126}
  {\path{doi:10.1007/JHEP07(2012)126}}.

\bibitem{Cheung:2016bym}
G.~K.~C. Cheung, C.~O'Hara, G.~Moir, M.~Peardon, S.~M. Ryan, C.~E. Thomas,
  D.~Tims, Hadron Spectrum Collaboration, {Excited and exotic charmonium, $D_s$
  and $D$ meson spectra for two light quark masses from lattice QCD}, JHEP 12
  (2016) 089.
\newblock \href {http://arxiv.org/abs/1610.01073} {\path{arXiv:1610.01073}},
  \href {https://doi.org/10.1007/JHEP12(2016)089}
  {\path{doi:10.1007/JHEP12(2016)089}}.

\bibitem{Chen:2016ejo}
Y.~Chen, W.-F. Chiu, M.~Gong, L.-C. Gui, Z.~Liu, {Exotic vector charmonium and
  its leptonic decay width}, Chin. Phys. C 40 (2016) 081002.
\newblock \href {http://arxiv.org/abs/1604.03401} {\path{arXiv:1604.03401}},
  \href {https://doi.org/10.1088/1674-1137/40/8/081002}
  {\path{doi:10.1088/1674-1137/40/8/081002}}.

\bibitem{Berwein:2015vca}
M.~Berwein, N.~Brambilla, J.~Tarr{\'u}s~Castell{\`a}, A.~Vairo, {Quarkonium
  Hybrids with Nonrelativistic Effective Field Theories}, Phys. Rev. D 92
  (2015) 114019.
\newblock \href {http://arxiv.org/abs/1510.04299} {\path{arXiv:1510.04299}},
  \href {https://doi.org/10.1103/PhysRevD.92.114019}
  {\path{doi:10.1103/PhysRevD.92.114019}}.

\bibitem{BaBar:2006qlj}
B.~Aubert, et~al., BaBar Collaboration, {Study of the exclusive initial-state
  radiation production of the $D\bar D$ System}, Phys. Rev. D 76 (2007) 111105.
\newblock \href {http://arxiv.org/abs/hep-ex/0607083}
  {\path{arXiv:hep-ex/0607083}}, \href
  {https://doi.org/10.1103/PhysRevD.76.111105}
  {\path{doi:10.1103/PhysRevD.76.111105}}.

\bibitem{Belle:2006hvs}
K.~Abe, et~al., Belle Collaboration, {Measurement of the near-threshold
  $e^+e^-\to D^{(*)\pm}D^{(*)\mp}$ cross section using initial-state
  radiation}, Phys. Rev. Lett. 98 (2007) 092001.
\newblock \href {http://arxiv.org/abs/hep-ex/0608018}
  {\path{arXiv:hep-ex/0608018}}, \href
  {https://doi.org/10.1103/PhysRevLett.98.092001}
  {\path{doi:10.1103/PhysRevLett.98.092001}}.

\bibitem{Belle:2007xvy}
G.~Pakhlova, et~al., Belle Collaboration, {Observation of $\psi(4415)\to D\bar
  D_2^*(2460)$ decay using initial-state radiation}, Phys. Rev. Lett. 100
  (2008) 062001.
\newblock \href {http://arxiv.org/abs/0708.3313} {\path{arXiv:0708.3313}},
  \href {https://doi.org/10.1103/PhysRevLett.100.062001}
  {\path{doi:10.1103/PhysRevLett.100.062001}}.

\bibitem{Belle:2009dus}
G.~Pakhlova, et~al., Belle Collaboration, {Measurement of the $e^+ e^-\to D^0
  D^{*-} \pi^+$ cross section using initial-state radiation}, Phys. Rev. D 80
  (2009) 091101.
\newblock \href {http://arxiv.org/abs/0908.0231} {\path{arXiv:0908.0231}},
  \href {https://doi.org/10.1103/PhysRevD.80.091101}
  {\path{doi:10.1103/PhysRevD.80.091101}}.

\bibitem{CLEO:2008ojp}
D.~Cronin-Hennessy, et~al., CLEO Collaboration, {Measurement of charm
  production cross sections in $e^+e^-$ annihilation at energies between 3.97
  and 4.26 GeV}, Phys. Rev. D 80 (2009) 072001.
\newblock \href {http://arxiv.org/abs/0801.3418} {\path{arXiv:0801.3418}},
  \href {https://doi.org/10.1103/PhysRevD.80.072001}
  {\path{doi:10.1103/PhysRevD.80.072001}}.

\bibitem{BaBar:2009elc}
B.~Aubert, et~al., BaBar Collaboration, {Exclusive initial-state-radiation
  production of the $D\bar D$, $D^*\bar D$, and $D^*\bar D^*$ systems}, Phys.
  Rev. D 79 (2009) 092001.
\newblock \href {http://arxiv.org/abs/0903.1597} {\path{arXiv:0903.1597}},
  \href {https://doi.org/10.1103/PhysRevD.79.092001}
  {\path{doi:10.1103/PhysRevD.79.092001}}.

\bibitem{BaBar:2010plp}
P.~del Amo~Sanchez, et~al., BaBar Collaboration, {Exclusive production of
  $D^+_s D^-_s$, $D^{*+}_s D^-_s$, and $D^{*+}_s D^{*-}_s$ via $e^+ e^-$
  annihilation with initial-state-radiation}, Phys. Rev. D 82 (2010) 052004.
\newblock \href {http://arxiv.org/abs/1008.0338} {\path{arXiv:1008.0338}},
  \href {https://doi.org/10.1103/PhysRevD.82.052004}
  {\path{doi:10.1103/PhysRevD.82.052004}}.

\bibitem{Belle:2007qxm}
G.~Pakhlova, et~al., Belle Collaboration, {Measurement of the near-threshold
  $e^+ e^- \to D \bar{D}$ cross section using initial-state radiation}, Phys.
  Rev. D 77 (2008) 011103.
\newblock \href {http://arxiv.org/abs/0708.0082} {\path{arXiv:0708.0082}},
  \href {https://doi.org/10.1103/PhysRevD.77.011103}
  {\path{doi:10.1103/PhysRevD.77.011103}}.

\bibitem{ParticleDataGroup:2014cgo}
K.~A. Olive, et~al., Particle Data Group Collaboration, {Review of particle
  physics}, Chin. Phys. C 38 (2014) 090001.
\newblock \href {https://doi.org/10.1088/1674-1137/38/9/090001}
  {\path{doi:10.1088/1674-1137/38/9/090001}}.

\bibitem{PLUTO:1976jbe}
J.~Burmester, et~al., PLUTO Collaboration, {The total hadronic cross-section
  for $e^+e^-$ annihilation between 3.1 GeV and 4.8 GeV center-of-mass energy},
  Phys. Lett. B 66 (1977) 395--400.
\newblock \href {https://doi.org/10.1016/0370-2693(77)90023-5}
  {\path{doi:10.1016/0370-2693(77)90023-5}}.

\bibitem{DASP:1978dns}
R.~Brandelik, et~al., DASP Collaboration, {Total cross-section for hadron
  production by $e^+ e^-$ annihilation at center-of-mass energies between 3.6
  and 5.2 GeV}, Phys. Lett. B 76 (1978) 361.
\newblock \href {https://doi.org/10.1016/0370-2693(78)90807-9}
  {\path{doi:10.1016/0370-2693(78)90807-9}}.

\bibitem{Siegrist:1981zp}
J.~Siegrist, et~al., {Hadron production by $e^+ e^-$ annihilation at
  center-of-mass energies between 2.6 GeV and 7.8 GeV: Part 1. total cross
  section, multiplicities and inclusive momentum distributions}, Phys. Rev. D
  26 (1982) 969.
\newblock \href {https://doi.org/10.1103/PhysRevD.26.969}
  {\path{doi:10.1103/PhysRevD.26.969}}.

\bibitem{BES:1999wbx}
J.~Z. Bai, et~al., BES Collaboration, {Measurement of the total cross-section
  for hadronic production by $e^+e^-$ annihilation at energies between 2.6--5
  GeV}, Phys. Rev. Lett. 84 (2000) 594--597.
\newblock \href {http://arxiv.org/abs/hep-ex/9908046}
  {\path{arXiv:hep-ex/9908046}}, \href
  {https://doi.org/10.1103/PhysRevLett.84.594}
  {\path{doi:10.1103/PhysRevLett.84.594}}.

\bibitem{BES:2009ejh}
M.~Ablikim, et~al., BES Collaboration, {$R$ value measurements for $e^+e^-$
  annihilation at 2.60 GeV, 3.07 GeV and 3.65 GeV}, Phys. Lett. B 677 (2009)
  239--245.
\newblock \href {http://arxiv.org/abs/0903.0900} {\path{arXiv:0903.0900}},
  \href {https://doi.org/10.1016/j.physletb.2009.05.055}
  {\path{doi:10.1016/j.physletb.2009.05.055}}.

\bibitem{Abrams:1974yy}
G.~S. Abrams, et~al., {The discovery of a second narrow resonance in $e^+ e^-$
  annihilation}, Phys. Rev. Lett. 33 (1974) 1453--1455.
\newblock \href {https://doi.org/10.1103/PhysRevLett.33.1453}
  {\path{doi:10.1103/PhysRevLett.33.1453}}.

\bibitem{Tanenbaum:1975ef}
W.~M. Tanenbaum, et~al., {Observation of an intermediate state in $\psi^\prime
  (3684)$ radiative cascade decay}, Phys. Rev. Lett. 35 (1975) 1323.
\newblock \href {https://doi.org/10.1103/PhysRevLett.35.1323}
  {\path{doi:10.1103/PhysRevLett.35.1323}}.

\bibitem{Whitaker:1976hb}
J.~S. Whitaker, et~al., {Radiative Decays of $\psi (3095)$ and $\psi^\prime
  (3684)$}, Phys. Rev. Lett. 37 (1976) 1596.
\newblock \href {https://doi.org/10.1103/PhysRevLett.37.1596}
  {\path{doi:10.1103/PhysRevLett.37.1596}}.

\bibitem{Siegrist:1976br}
J.~Siegrist, et~al., {Observation of a Resonance at 4.4 GeV and Additional
  Structure Near 4.1 GeV in $e^+ e^-$ Annihilation}, Phys. Rev. Lett. 36 (1976)
  700.
\newblock \href {https://doi.org/10.1103/PhysRevLett.36.700}
  {\path{doi:10.1103/PhysRevLett.36.700}}.

\bibitem{Rapidis:1977cv}
P.~A. Rapidis, et~al., {Observation of a resonance in $e^+e^-$ annihilation
  just above charm threshold}, Phys. Rev. Lett. 39 (1977) 526, [Erratum:
  Phys.Rev.Lett. 39, 974 (1977)].
\newblock \href {https://doi.org/10.1103/PhysRevLett.39.526}
  {\path{doi:10.1103/PhysRevLett.39.526}}.

\bibitem{Biddick:1977sv}
C.~J. Biddick, et~al., {Inclusive gamma-ray spectra from $\psi (3095)$ and
  $\psi^\prime (3684)$}, Phys. Rev. Lett. 38 (1977) 1324.
\newblock \href {https://doi.org/10.1103/PhysRevLett.38.1324}
  {\path{doi:10.1103/PhysRevLett.38.1324}}.

\bibitem{Goldhaber:1977qn}
G.~Goldhaber, et~al., {$D$ and $D^*$ Meson production near 4 GeV in $e^+ e^-$
  annihilation}, Phys. Lett. B 69 (1977) 503--507.
\newblock \href {https://doi.org/10.1016/0370-2693(77)90855-3}
  {\path{doi:10.1016/0370-2693(77)90855-3}}.

\bibitem{E288:1977xhf}
S.~W. Herb, et~al., E288 Collaboration, {Observation of a dimuon resonance at
  9.5 GeV in 400 GeV proton-nucleus collisions}, Phys. Rev. Lett. 39 (1977)
  252--255.
\newblock \href {https://doi.org/10.1103/PhysRevLett.39.252}
  {\path{doi:10.1103/PhysRevLett.39.252}}.

\bibitem{Partridge:1980vk}
R.~Partridge, et~al., {Observation of an $\eta_c$ candidate state with mass
  2978 MeV $\pm$9 MeV}, Phys. Rev. Lett. 45 (1980) 1150--1153.
\newblock \href {https://doi.org/10.1103/PhysRevLett.45.1150}
  {\path{doi:10.1103/PhysRevLett.45.1150}}.

\bibitem{Edwards:1982fif}
C.~Edwards, et~al., {Observation of an $\eta_c^\prime$ candidate state with
  mass 3592 MeV $\pm$5 MeV}, Phys. Rev. Lett. 48 (1982) 70.
\newblock \href {https://doi.org/10.1103/PhysRevLett.48.70}
  {\path{doi:10.1103/PhysRevLett.48.70}}.

\bibitem{R704:1986siy}
C.~Baglin, et~al., R704,
  Annecy(LAPP)-CERN-Genoa-Lyon-Oslo-Rome-Strasbourg-Turin Collaboration,
  {Search for the $p$ wave singlet charmonium state in $\bar{p} p$
  annihilations at the {CERN} intersecting storage rings}, Phys. Lett. B 171
  (1986) 135--141.
\newblock \href {https://doi.org/10.1016/0370-2693(86)91013-0}
  {\path{doi:10.1016/0370-2693(86)91013-0}}.

\bibitem{Armstrong:1992ae}
T.~A. Armstrong, et~al., {Observation of the $p$ wave singlet state of
  charmonium}, Phys. Rev. Lett. 69 (1992) 2337--2340.
\newblock \href {https://doi.org/10.1103/PhysRevLett.69.2337}
  {\path{doi:10.1103/PhysRevLett.69.2337}}.

\bibitem{Dong:1994zj}
Y.-B. Dong, Y.-W. Yu, Z.-Y. Zhang, P.-N. Shen, {Leptonic decay of charmonium},
  Phys. Rev. D 49 (1994) 1642--1644.
\newblock \href {https://doi.org/10.1103/PhysRevD.49.1642}
  {\path{doi:10.1103/PhysRevD.49.1642}}.

\bibitem{Wang:2018rjg}
J.-Z. Wang, Z.-F. Sun, X.~Liu, T.~Matsuki, {Higher bottomonium zoo}, Eur. Phys.
  J. C 78 (2018) 915.
\newblock \href {http://arxiv.org/abs/1802.04938} {\path{arXiv:1802.04938}},
  \href {https://doi.org/10.1140/epjc/s10052-018-6372-1}
  {\path{doi:10.1140/epjc/s10052-018-6372-1}}.

\bibitem{Song:2015fha}
Q.-T. Song, D.-Y. Chen, X.~Liu, T.~Matsuki, {Higher radial and orbital
  excitations in the charmed meson family}, Phys. Rev. D 92 (2015) 074011.
\newblock \href {http://arxiv.org/abs/1503.05728} {\path{arXiv:1503.05728}},
  \href {https://doi.org/10.1103/PhysRevD.92.074011}
  {\path{doi:10.1103/PhysRevD.92.074011}}.

\bibitem{Wang:2021gle}
J.-Z. Wang, L.-M. Wang, X.~Liu, T.~Matsuki, {Deciphering the light vector meson
  contribution to the cross sections of $e^+e^-$ annihilations into the
  open-strange channels through a combined analysis}, Phys. Rev. D 104 (2021)
  054045.
\newblock \href {http://arxiv.org/abs/2106.14582} {\path{arXiv:2106.14582}},
  \href {https://doi.org/10.1103/PhysRevD.104.054045}
  {\path{doi:10.1103/PhysRevD.104.054045}}.

\bibitem{He:2014xna}
L.-P. He, D.-Y. Chen, X.~Liu, T.~Matsuki, {Prediction of a missing higher
  charmonium around 4.26 GeV in $J/\psi$ family}, Eur. Phys. J. C 74 (2014)
  3208.
\newblock \href {http://arxiv.org/abs/1405.3831} {\path{arXiv:1405.3831}},
  \href {https://doi.org/10.1140/epjc/s10052-014-3208-5}
  {\path{doi:10.1140/epjc/s10052-014-3208-5}}.

\bibitem{Yuan:2013uta}
C.-Z. Yuan, {Evidence for resonant structures in $e^{+}e^{-} \to
  \pi^{+}\pi^{-}h_c$}, Chin. Phys. C 38 (2014) 043001.
\newblock \href {http://arxiv.org/abs/1312.6399} {\path{arXiv:1312.6399}},
  \href {https://doi.org/10.1088/1674-1137/38/4/043001}
  {\path{doi:10.1088/1674-1137/38/4/043001}}.

\bibitem{BESIII:2014rja}
M.~Ablikim, et~al., BESIII Collaboration, {Study of $e^+e^-\to\omega\chi_{cJ}$
  at center-of-mass energies from 4.21 to 4.42 GeV}, Phys. Rev. Lett. 114
  (2015) 092003.
\newblock \href {http://arxiv.org/abs/1410.6538} {\path{arXiv:1410.6538}},
  \href {https://doi.org/10.1103/PhysRevLett.114.092003}
  {\path{doi:10.1103/PhysRevLett.114.092003}}.

\bibitem{BESIII:2019tdo}
M.~Ablikim, et~al., BESIII Collaboration, {Observation of
  $e^{+}e^{-}\rightarrow \pi^{+}\pi^{-}\psi(3770)$ and
  $D_{1}(2420)^{0}\bar{D}^{0} +\text{c.c.} $}, Phys. Rev. D 100 (2019) 032005.
\newblock \href {http://arxiv.org/abs/1903.08126} {\path{arXiv:1903.08126}},
  \href {https://doi.org/10.1103/PhysRevD.100.032005}
  {\path{doi:10.1103/PhysRevD.100.032005}}.

\bibitem{BESIII:2020wzv}
M.~Ablikim, et~al., BESIII Collaboration, {Search for the reaction $e^{+}e^{-}
  \rightarrow \chi_{cJ} \pi^+ \pi^-$ and a charmonium-like structure decaying
  to $\chi_{cJ}\pi ^{\pm}$ between 4.18 and 4.60 GeV}, Phys. Rev. D 103 (2021)
  052010.
\newblock \href {http://arxiv.org/abs/2012.02682} {\path{arXiv:2012.02682}},
  \href {https://doi.org/10.1103/PhysRevD.103.052010}
  {\path{doi:10.1103/PhysRevD.103.052010}}.

\bibitem{BESIII:2021fae}
M.~Ablikim, et~al., BESIII Collaboration, {Observation of
  $e^{+}e^{-}\rightarrow\eta\psi(2S)$~at center-of-mass energies from 4.236 to
  4.600 GeV}, JHEP 10 (2021) 177.
\newblock \href {http://arxiv.org/abs/2103.01480} {\path{arXiv:2103.01480}},
  \href {https://doi.org/10.1007/JHEP10(2021)177}
  {\path{doi:10.1007/JHEP10(2021)177}}.

\bibitem{BESIII:2024jzt}
M.~Ablikim, et~al., BESIII Collaboration, {Cross section measurement of $e^+e^-
  \to \eta \psi(2S)$ and search for $e^+e^- \to \eta \tilde{X}(3872)$}, Phys.
  Rev. D 109 (2024) 112004.
\newblock \href {http://arxiv.org/abs/2403.16811} {\path{arXiv:2403.16811}},
  \href {https://doi.org/10.1103/PhysRevD.109.112004}
  {\path{doi:10.1103/PhysRevD.109.112004}}.

\bibitem{BESIII:2022quc}
M.~Ablikim, et~al., BESIII Collaboration, {Measurement of
  $e^{+}e^{-}\rightarrow\pi^{+}\pi^{-}D^{+}D^{-}$ cross sections at
  center-of-mass energies from 4.190 to 4.946 GeV}, Phys. Rev. D 106 (2022)
  052012.
\newblock \href {http://arxiv.org/abs/2208.00099} {\path{arXiv:2208.00099}},
  \href {https://doi.org/10.1103/PhysRevD.106.052012}
  {\path{doi:10.1103/PhysRevD.106.052012}}.

\bibitem{BESIII:2020svk}
M.~Ablikim, et~al., BESIII Collaboration, {Study of $e^+e^- \to 2(p\bar{p})$ at
  center-of-mass energies between 4.0 and 4.6 GeV}, Phys. Rev. D 103 (2021)
  052003.
\newblock \href {http://arxiv.org/abs/2012.11079} {\path{arXiv:2012.11079}},
  \href {https://doi.org/10.1103/PhysRevD.103.052003}
  {\path{doi:10.1103/PhysRevD.103.052003}}.

\bibitem{BESIII:2022vpa}
M.~Ablikim, et~al., BESIII Collaboration, {Measurement of the cross section of
  $e^{+}e^{-}\to\eta\pi^+\pi^-$ at center-of-mass energies from 3.872 GeV to
  4.700 GeV}, JHEP 12 (2022) 153.
\newblock \href {http://arxiv.org/abs/2202.12748} {\path{arXiv:2202.12748}},
  \href {https://doi.org/10.1007/JHEP12(2022)153}
  {\path{doi:10.1007/JHEP12(2022)153}}.

\bibitem{BESIII:2023gqy}
M.~Ablikim, et~al., BESIII Collaboration, {Study of the
  $e^+e^-\to\pi^+\pi^-\omega$ process at center-of-mass energies between 4.0
  and 4.6 GeV}, JHEP 08 (2023) 159.
\newblock \href {http://arxiv.org/abs/2303.09718} {\path{arXiv:2303.09718}},
  \href {https://doi.org/10.1007/JHEP08(2023)159}
  {\path{doi:10.1007/JHEP08(2023)159}}.

\bibitem{BESIII:2024umc}
M.~Ablikim, et~al., BESIII Collaboration, {Measurement of born cross section of
  $ e^{+}e^{-}\to \Sigma^{+}\bar\Sigma^- $ at center-of-mass energies between
  3.510 and 4.951 GeV}, JHEP 05 (2024) 022.
\newblock \href {http://arxiv.org/abs/2401.09468} {\path{arXiv:2401.09468}},
  \href {https://doi.org/10.1007/JHEP05(2024)022}
  {\path{doi:10.1007/JHEP05(2024)022}}.

\bibitem{BESIII:2024ues}
M.~Ablikim, et~al., BESIII Collaboration, {Measurement of Born cross sections
  of $e^+e^-\to \Xi^0\bar\Xi^0$ and search for charmonium(-like) states at $
  \sqrt{s} $ = 3.51{\textendash}4.95 GeV}, JHEP 11 (2024) 062.
\newblock \href {http://arxiv.org/abs/2409.00427} {\path{arXiv:2409.00427}},
  \href {https://doi.org/10.1007/JHEP11(2024)062}
  {\path{doi:10.1007/JHEP11(2024)062}}.

\bibitem{BESIII:2023rse}
M.~Ablikim, et~al., BESIII Collaboration, {Measurement of the cross section of
  $e^+e^-\to\Xi^-\bar\Xi^+$ at center-of-mass energies between 3.510 and 4.843
  GeV}, JHEP 11 (2023) 228.
\newblock \href {http://arxiv.org/abs/2309.04215} {\path{arXiv:2309.04215}},
  \href {https://doi.org/10.1007/JHEP11(2023)228}
  {\path{doi:10.1007/JHEP11(2023)228}}.

\bibitem{BESIII:2025fph}
M.~Ablikim, et~al., BESIII Collaboration, {Measurement of Born cross sections
  and effective form factors of $e^+e^-\to \Omega^{-}\bar\Omega^{+}$ from
  $\sqrt{s}= 3.7$ to 4.7 GeV} (8 2025).
\newblock \href {http://arxiv.org/abs/2508.01359} {\path{arXiv:2508.01359}}.

\bibitem{BESIII:2018kyw}
M.~Ablikim, et~al., BESIII Collaboration, {Measurements of $e^+e^- \to
  K_{S}^{0}K^{\pm}\pi^{\mp}\pi^0$ and $K_{S}^{0}K^{\pm}\pi^{\mp}\eta$ at
  center-of-mass energies from $3.90$ to $4.60~\mathrm{GeV}$}, Phys. Rev. D 99
  (2019) 012003.
\newblock \href {http://arxiv.org/abs/1810.09395} {\path{arXiv:1810.09395}},
  \href {https://doi.org/10.1103/PhysRevD.99.012003}
  {\path{doi:10.1103/PhysRevD.99.012003}}.

\bibitem{BESIII:2019cuv}
M.~Ablikim, et~al., BESIII Collaboration, {Measurement of the cross section for
  $e^{+}e^{-}\rightarrow\Xi^{-}\bar\Xi^{+}$ and observation of an excited $\Xi$
  baryon}, Phys. Rev. Lett. 124 (2020) 032002.
\newblock \href {http://arxiv.org/abs/1910.04921} {\path{arXiv:1910.04921}},
  \href {https://doi.org/10.1103/PhysRevLett.124.032002}
  {\path{doi:10.1103/PhysRevLett.124.032002}}.

\bibitem{BESIII:2021vkt}
M.~Ablikim, et~al., BESIII Collaboration, {Cross section measurement of $e^+e^-
  \to p\bar{p}\eta$ and $e^+e^- \to p\bar{p}\omega$ at center-of-mass energies
  between 3.773~GeV and 4.6~GeV}, Phys. Rev. D 104 (2021) 092008.
\newblock \href {http://arxiv.org/abs/2102.04268} {\path{arXiv:2102.04268}},
  \href {https://doi.org/10.1103/PhysRevD.104.092008}
  {\path{doi:10.1103/PhysRevD.104.092008}}.

\bibitem{BESIII:2021fqx}
M.~Ablikim, et~al., BESIII Collaboration, {Observation of a near-threshold
  enhancement in the $\Lambda\bar{\Lambda}$ mass spectrum from
  $e^+e^-\to\phi\Lambda\bar{\Lambda}$ at $\sqrt{s}$ from 3.51 to 4.60 GeV},
  Phys. Rev. D 104 (2021) 052006.
\newblock \href {http://arxiv.org/abs/2104.08754} {\path{arXiv:2104.08754}},
  \href {https://doi.org/10.1103/PhysRevD.104.052006}
  {\path{doi:10.1103/PhysRevD.104.052006}}.

\bibitem{BESIII:2021ftf}
M.~Ablikim, et~al., BESIII Collaboration, {Cross sections for the reactions
  $e^+e^-\rightarrow K^+K^-\pi^+\pi^-(\pi^0)$, $K^+K^-K^+K^-(\pi^0)$,
  $\pi^+\pi^-\pi^+\pi^-(\pi^0)$, $p\bar{p}\pi^+\pi^-(\pi^0)$ in the energy
  region between 3.773 and 4.600 GeV}, Phys. Rev. D 104 (2021) 112009.
\newblock \href {http://arxiv.org/abs/2109.12751} {\path{arXiv:2109.12751}},
  \href {https://doi.org/10.1103/PhysRevD.104.112009}
  {\path{doi:10.1103/PhysRevD.104.112009}}.

\bibitem{BESIII:2022zxr}
M.~Ablikim, et~al., BESIII Collaboration, {Cross section measurements of the
  processes~$e^+e^- \rightarrow \omega\pi^{0}$ and $\omega\eta$~~at
  center-of-mass energies between 3.773 and 4.701 GeV}, JHEP 07 (2022) 064.
\newblock \href {http://arxiv.org/abs/2203.02682} {\path{arXiv:2203.02682}},
  \href {https://doi.org/10.1007/JHEP07(2022)064}
  {\path{doi:10.1007/JHEP07(2022)064}}.

\bibitem{BESIII:2022tjc}
M.~Ablikim, et~al., BESIII Collaboration, {Measurement of $e^+e^- \to \phi
  \eta^\prime$ cross sections at center-of-mass energies between 3.508 and
  4.600~GeV}, Phys. Rev. D 107 (2023) 072003.
\newblock \href {http://arxiv.org/abs/2210.06988} {\path{arXiv:2210.06988}},
  \href {https://doi.org/10.1103/PhysRevD.107.072003}
  {\path{doi:10.1103/PhysRevD.107.072003}}.

\bibitem{BESIII:2022tvj}
M.~Ablikim, et~al., BESIII Collaboration, {Measurement of
  $e^+e^-\rightarrow\Lambda\bar{\Lambda}\eta$ from 3.5106 to 4.6988 GeV and
  study of $\Lambda\bar{\Lambda}$ mass threshold enhancement}, Phys. Rev. D 107
  (2023) 112001.
\newblock \href {http://arxiv.org/abs/2211.10755} {\path{arXiv:2211.10755}},
  \href {https://doi.org/10.1103/PhysRevD.107.112001}
  {\path{doi:10.1103/PhysRevD.107.112001}}.

\bibitem{BESIII:2023ojh}
M.~Ablikim, et~al., BESIII Collaboration, {Measurement of $ {e}^{+}{e}^{-}\to
  p{K}^{-}\overline{\Lambda}+c.c. $ cross sections between 4.009 GeV and 4.951
  GeV}, JHEP 12 (2023) 027.
\newblock \href {http://arxiv.org/abs/2307.02328} {\path{arXiv:2307.02328}},
  \href {https://doi.org/10.1007/JHEP12(2023)027}
  {\path{doi:10.1007/JHEP12(2023)027}}.

\bibitem{BESIII:2023tex}
M.~Ablikim, et~al., BESIII Collaboration, {Measurement of $e^+e^- \to \phi
  \eta^\prime$ cross sections at center-of-mass energies from 3.508 to
  4.951~GeV and search for the decay $\psi(3770) \to \phi \eta^\prime$}, Phys.
  Rev. D 108 (2023) 052015.
\newblock \href {http://arxiv.org/abs/2307.12736} {\path{arXiv:2307.12736}},
  \href {https://doi.org/10.1103/PhysRevD.108.052015}
  {\path{doi:10.1103/PhysRevD.108.052015}}.

\bibitem{BESIII:2023ion}
M.~Ablikim, et~al., BESIII Collaboration, {Study of $e^+e^- \to \eta \phi$ at
  center-of-mass energies from 3.773 to 4.600~GeV}, Phys. Rev. D 108 (2023)
  112011.
\newblock \href {http://arxiv.org/abs/2308.08161} {\path{arXiv:2308.08161}},
  \href {https://doi.org/10.1103/PhysRevD.108.112011}
  {\path{doi:10.1103/PhysRevD.108.112011}}.

\bibitem{BESIII:2024ogz}
M.~Ablikim, et~al., BESIII Collaboration, {Measurement of the cross sections of
  $ {e}^{+}{e}^{-}\to {K}^{-}{\overline{\Xi}}^{+}\Lambda /{\Sigma}^0 $ at
  center-of-mass energies between 3.510 and 4.914 GeV}, JHEP 07 (2024) 258.
\newblock \href {http://arxiv.org/abs/2406.18183} {\path{arXiv:2406.18183}},
  \href {https://doi.org/10.1007/JHEP07(2024)258}
  {\path{doi:10.1007/JHEP07(2024)258}}.

\bibitem{BESIII:2024gql}
M.~Ablikim, et~al., BESIII Collaboration, {Measurement of Born cross section of
  $e^+e^- \to \Sigma^0 \bar{\Sigma}^0$ at $\sqrt{s}$=3.50-4.95 GeV}, Phys. Rev.
  D 111 (2025) L051502.
\newblock \href {http://arxiv.org/abs/2412.20305} {\path{arXiv:2412.20305}},
  \href {https://doi.org/10.1103/PhysRevD.111.L051502}
  {\path{doi:10.1103/PhysRevD.111.L051502}}.

\bibitem{BESIII:2025zyk}
M.~Ablikim, et~al., BESIII Collaboration, {Cross section measurement of
  $e^{+}e^{-} \to f_{1}(1285)\pi^{+}\pi^{-}$ at center-of-mass energies between
  $3.808$ and $4.951\rm GeV$} (1 2025).
\newblock \href {http://arxiv.org/abs/2501.14206} {\path{arXiv:2501.14206}}.

\bibitem{BESIII:2025lbj}
M.~Ablikim, et~al., BESIII Collaboration, {Measurement of the Born cross
  section for $e^+e^- \to p K^- K^- \bar \Sigma^+$ at $\sqrt{s} =$ 3.5-4.9 GeV}
  (8 2025).
\newblock \href {http://arxiv.org/abs/2508.11276} {\path{arXiv:2508.11276}}.

\bibitem{BESIII:2021ccp}
M.~Ablikim, et~al., BESIII Collaboration, {Measurement of the cross section for
  $e^{+}e^{-}\rightarrow\Lambda\bar\Lambda$~and evidence of the decay
  $\psi(3770)\rightarrow\Lambda\bar\Lambda$}, Phys. Rev. D 104 (2021) L091104.
\newblock \href {http://arxiv.org/abs/2108.02410} {\path{arXiv:2108.02410}},
  \href {https://doi.org/10.1103/PhysRevD.104.L091104}
  {\path{doi:10.1103/PhysRevD.104.L091104}}.

\bibitem{BESIII:2025fxf}
M.~Ablikim, et~al., BESIII Collaboration, {Cross sections measurement of
  $e^+e^-\to \Xi(1530)^0\bar\Xi^0 + c.c.$ and search for $\psi(3770)\to
  \Xi(1530)^0\bar\Xi^0 + c.c.$} (12 2025).
\newblock \href {http://arxiv.org/abs/2512.17275} {\path{arXiv:2512.17275}}.

\bibitem{Zhang:2025qmo}
R.~Zhang, X.~Wang, {Search for charmonium(-like) states decaying into the
  $\Omega^-\bar\Omega^+$ final states} (8 2025).
\newblock \href {http://arxiv.org/abs/2508.03454} {\path{arXiv:2508.03454}}.

\bibitem{Zhang:2026qjt}
R.~Zhang, X.~Wang, {A review of hyperon physics at BESIII experiment}, Symmetry
  18 (2026) 200.
\newblock \href {http://arxiv.org/abs/2601.15116} {\path{arXiv:2601.15116}},
  \href {https://doi.org/10.3390/sym18010200} {\path{doi:10.3390/sym18010200}}.

\bibitem{Belle:2019qoi}
S.~Jia, et~al., Belle Collaboration, {Observation of a vector charmoniumlike
  state in $e^+e^- \to D^+_sD_{s1}(2536)^-+c.c.$}, Phys. Rev. D 100 (2019)
  111103.
\newblock \href {http://arxiv.org/abs/1911.00671} {\path{arXiv:1911.00671}},
  \href {https://doi.org/10.1103/PhysRevD.100.111103}
  {\path{doi:10.1103/PhysRevD.100.111103}}.

\bibitem{Guo:2010tk}
F.-K. Guo, J.~Haidenbauer, C.~Hanhart, U.-G. Meissner, {Reconciling the
  $X(4630)$ with the $Y(4660)$}, Phys. Rev. D 82 (2010) 094008.
\newblock \href {http://arxiv.org/abs/1005.2055} {\path{arXiv:1005.2055}},
  \href {https://doi.org/10.1103/PhysRevD.82.094008}
  {\path{doi:10.1103/PhysRevD.82.094008}}.

\bibitem{Chen:2015bft}
D.-Y. Chen, X.~Liu, X.-Q. Li, H.-W. Ke, {Unified Fano-like interference picture
  for charmoniumlike states $Y(4008)$, $Y(4260)$ and $Y(4360)$}, Phys. Rev. D
  93 (2016) 014011.
\newblock \href {http://arxiv.org/abs/1512.04157} {\path{arXiv:1512.04157}},
  \href {https://doi.org/10.1103/PhysRevD.93.014011}
  {\path{doi:10.1103/PhysRevD.93.014011}}.

\bibitem{Chen:2017uof}
D.-Y. Chen, X.~Liu, T.~Matsuki, {Interference effect as resonance killer of
  newly observed charmoniumlike states $Y(4320)$ and $Y(4390)$}, Eur. Phys. J.
  C 78 (2018) 136.
\newblock \href {http://arxiv.org/abs/1708.01954} {\path{arXiv:1708.01954}},
  \href {https://doi.org/10.1140/epjc/s10052-018-5635-1}
  {\path{doi:10.1140/epjc/s10052-018-5635-1}}.

\bibitem{BESIII:2013ouc}
M.~Ablikim, et~al., BESIII Collaboration, {Observation of a charged
  charmoniumlike structure $Z_c$(4020) and search for the $Z_c$(3900) in
  $e^+e^- \to \pi^+\pi^-h_c$}, Phys. Rev. Lett. 111 (2013) 242001.
\newblock \href {http://arxiv.org/abs/1309.1896} {\path{arXiv:1309.1896}},
  \href {https://doi.org/10.1103/PhysRevLett.111.242001}
  {\path{doi:10.1103/PhysRevLett.111.242001}}.

\bibitem{Wang:2020prx}
J.-Z. Wang, R.-Q. Qian, X.~Liu, T.~Matsuki, {Are the $Y$ states around 4.6 GeV
  from $e^+e^-$ annihilation higher charmonia?}, Phys. Rev. D 101 (2020)
  034001.
\newblock \href {http://arxiv.org/abs/2001.00175} {\path{arXiv:2001.00175}},
  \href {https://doi.org/10.1103/PhysRevD.101.034001}
  {\path{doi:10.1103/PhysRevD.101.034001}}.

\bibitem{BES:2007zwq}
M.~Ablikim, et~al., BES Collaboration, {Determination of the $\psi(3770)$,
  $\psi(4040)$, $\psi(4160)$ and $\psi(4415)$ resonance parameters}, eConf
  C070805 (2007) 02.
\newblock \href {http://arxiv.org/abs/0705.4500} {\path{arXiv:0705.4500}},
  \href {https://doi.org/10.1016/j.physletb.2007.11.100}
  {\path{doi:10.1016/j.physletb.2007.11.100}}.

\bibitem{Wang:2023zxj}
J.-Z. Wang, X.~Liu, {Identifying a characterized energy level structure of
  higher charmonium well matched to the peak structures in $e^+e^- \to
  \pi^+D^0D^{*-}$}, Phys. Lett. B 849 (2024) 138456.
\newblock \href {http://arxiv.org/abs/2306.14695} {\path{arXiv:2306.14695}},
  \href {https://doi.org/10.1016/j.physletb.2024.138456}
  {\path{doi:10.1016/j.physletb.2024.138456}}.

\bibitem{Cotugno:2009ys}
G.~Cotugno, R.~Faccini, A.~D. Polosa, C.~Sabelli, {Charmed baryonium}, Phys.
  Rev. Lett. 104 (2010) 132005.
\newblock \href {http://arxiv.org/abs/0911.2178} {\path{arXiv:0911.2178}},
  \href {https://doi.org/10.1103/PhysRevLett.104.132005}
  {\path{doi:10.1103/PhysRevLett.104.132005}}.

\bibitem{Bugg:2008sk}
D.~V. Bugg, {An alternative fit to Belle mass spectra for $D\bar{D}$, $D^*
  \bar{D}^*$ and $\Lambda_c \bar{\Lambda}_c$}, J. Phys. G 36 (2009) 075002.
\newblock \href {http://arxiv.org/abs/0811.2559} {\path{arXiv:0811.2559}},
  \href {https://doi.org/10.1088/0954-3899/36/7/075002}
  {\path{doi:10.1088/0954-3899/36/7/075002}}.

\bibitem{BESIII:2017kqg}
M.~Ablikim, et~al., BESIII Collaboration, {Precision measurement of the
  $e^{+}e^{-}~\rightarrow~\Lambda_{c}^{+} \bar{\Lambda}_{c}^{-}$ cross section
  near threshold}, Phys. Rev. Lett. 120 (2018) 132001.
\newblock \href {http://arxiv.org/abs/1710.00150} {\path{arXiv:1710.00150}},
  \href {https://doi.org/10.1103/PhysRevLett.120.132001}
  {\path{doi:10.1103/PhysRevLett.120.132001}}.

\bibitem{Belle:2014wyt}
X.~L. Wang, et~al., Belle Collaboration, {Measurement of $e^+e^- \to
  \pi^+\pi^-\psi(2S)$ via Initial State Radiation at Belle}, Phys. Rev. D 91
  (2015) 112007.
\newblock \href {http://arxiv.org/abs/1410.7641} {\path{arXiv:1410.7641}},
  \href {https://doi.org/10.1103/PhysRevD.91.112007}
  {\path{doi:10.1103/PhysRevD.91.112007}}.

\bibitem{Belle:2010fwv}
G.~Pakhlova, et~al., Belle Collaboration, {Measurement of $e^+e^-\to D_s^{(*)+}
  D_s^{(*)-}$ cross sections near threshold using initial-state radiation},
  Phys. Rev. D 83 (2011) 011101.
\newblock \href {http://arxiv.org/abs/1011.4397} {\path{arXiv:1011.4397}},
  \href {https://doi.org/10.1103/PhysRevD.83.011101}
  {\path{doi:10.1103/PhysRevD.83.011101}}.

\bibitem{LHCb:2013ywr}
R.~Aaij, et~al., LHCb Collaboration, {Observation of a resonance in $B^+ \to
  K^+ \mu^+\mu^-$ decays at low recoil}, Phys. Rev. Lett. 111 (2013) 112003.
\newblock \href {http://arxiv.org/abs/1307.7595} {\path{arXiv:1307.7595}},
  \href {https://doi.org/10.1103/PhysRevLett.111.112003}
  {\path{doi:10.1103/PhysRevLett.111.112003}}.

\bibitem{Peng:2024blp}
T.-C. Peng, Z.-Y. Bai, J.-Z. Wang, X.~Liu, {Reevaluating the $\psi(4160)$
  resonance parameter using $B^+ \to K^+ \mu^+ \mu^-$ data in the context of
  unquenched charmonium spectroscopy}, Phys. Rev. D 111 (2025) 054023.
\newblock \href {http://arxiv.org/abs/2412.11096} {\path{arXiv:2412.11096}},
  \href {https://doi.org/10.1103/PhysRevD.111.054023}
  {\path{doi:10.1103/PhysRevD.111.054023}}.

\bibitem{Badalian:2009bu}
A.~M. Badalian, B.~L.~G. Bakker, I.~V. Danilkin, {Dielectron widths of the
  $S$-, $D$-vector bottomonium states}, Phys. Atom. Nucl. 73 (2010) 138--149.
\newblock \href {http://arxiv.org/abs/0903.3643} {\path{arXiv:0903.3643}},
  \href {https://doi.org/10.1134/S1063778810010163}
  {\path{doi:10.1134/S1063778810010163}}.

\bibitem{Fu:2018yxq}
H.-F. Fu, L.~Jiang, {Coupled-channel-induced $S{-}D$ mixing of Charmonia and
  testing possible assignments for $Y$(4260) and $Y$(4360)}, Eur. Phys. J. C 79
  (2019) 460.
\newblock \href {http://arxiv.org/abs/1812.00179} {\path{arXiv:1812.00179}},
  \href {https://doi.org/10.1140/epjc/s10052-019-6976-0}
  {\path{doi:10.1140/epjc/s10052-019-6976-0}}.

\bibitem{Man:2024mvl}
Z.-L. Man, C.-R. Shu, Y.-R. Liu, H.~Chen, {Charmonium states in a
  coupled-channel model}, Eur. Phys. J. C 84 (2024) 810.
\newblock \href {http://arxiv.org/abs/2402.02765} {\path{arXiv:2402.02765}},
  \href {https://doi.org/10.1140/epjc/s10052-024-13132-7}
  {\path{doi:10.1140/epjc/s10052-024-13132-7}}.

\bibitem{Seth:2004py}
K.~K. Seth, {Alternative analysis of the $R$ measurements: Resonance parameters
  of the higher vector states of charmonium}, Phys. Rev. D 72 (2005) 017501.
\newblock \href {http://arxiv.org/abs/hep-ex/0405007}
  {\path{arXiv:hep-ex/0405007}}, \href
  {https://doi.org/10.1103/PhysRevD.72.017501}
  {\path{doi:10.1103/PhysRevD.72.017501}}.

\bibitem{Chao:2007it}
K.-T. Chao, {Interpretations for the $X(4160)$ observed in the double charm
  production at $B$ factories}, Phys. Lett. B 661 (2008) 348--353.
\newblock \href {http://arxiv.org/abs/0707.3982} {\path{arXiv:0707.3982}},
  \href {https://doi.org/10.1016/j.physletb.2008.02.039}
  {\path{doi:10.1016/j.physletb.2008.02.039}}.

\bibitem{Badalian:2008dv}
A.~M. Badalian, B.~L.~G. Bakker, I.~V. Danilkin, {The $S$-$D$ mixing and
  di-electron widths of higher charmonium $1^{--}$ states}, Phys. Atom. Nucl.
  72 (2009) 638--646.
\newblock \href {http://arxiv.org/abs/0805.2291} {\path{arXiv:0805.2291}},
  \href {https://doi.org/10.1134/S1063778809040085}
  {\path{doi:10.1134/S1063778809040085}}.

\bibitem{Bokade:2024tge}
C.~A. Bokade, Bhaghyesh, {Conventional and $XYZ$ charmonium states in a
  relativistic screened potential model}, Phys. Rev. D 111 (2025) 014030.
\newblock \href {http://arxiv.org/abs/2408.06759} {\path{arXiv:2408.06759}},
  \href {https://doi.org/10.1103/PhysRevD.111.014030}
  {\path{doi:10.1103/PhysRevD.111.014030}}.

\bibitem{Anwar:2016mxo}
M.~N. Anwar, Y.~Lu, B.-S. Zou, {Modeling charmonium-$\eta$ decays of
  $J^{PC}=1^{--}$ Higher charmonia}, Phys. Rev. D 95 (2017) 114031.
\newblock \href {http://arxiv.org/abs/1612.05396} {\path{arXiv:1612.05396}},
  \href {https://doi.org/10.1103/PhysRevD.95.114031}
  {\path{doi:10.1103/PhysRevD.95.114031}}.

\bibitem{Yang:2018mkn}
Y.~Yang, Y.~Guo, J.~Sun, N.~Wang, Q.~Chang, G.~Lu, {$B_{u}$ ${\to}$ ${\psi}M$
  decays and $S$-$D$ wave mixing effects}, Chin. Phys. C 42 (2018) 113102.
\newblock \href {http://arxiv.org/abs/1808.07605} {\path{arXiv:1808.07605}},
  \href {https://doi.org/10.1088/1674-1137/42/11/113102}
  {\path{doi:10.1088/1674-1137/42/11/113102}}.

\bibitem{Mo:2010bw}
X.~H. Mo, C.~Z. Yuan, P.~Wang, {On the leptonic partial widths of the excited
  $\psi$ states}, Phys. Rev. D 82 (2010) 077501.
\newblock \href {http://arxiv.org/abs/1007.0084} {\path{arXiv:1007.0084}},
  \href {https://doi.org/10.1103/PhysRevD.82.077501}
  {\path{doi:10.1103/PhysRevD.82.077501}}.

\bibitem{Balossino:2022ywn}
I.~Balossino, F.~Cossio, R.~Farinelli, L.~Lavezzi, {The CGEM-IT: An upgrade for
  the BESIII experiment}, Symmetry 14 (2022) 905.
\newblock \href {https://doi.org/10.3390/sym14050905}
  {\path{doi:10.3390/sym14050905}}.

\bibitem{BES:2008wyz}
M.~Ablikim, et~al., BES Collaboration, {Anomalous line shape of the cross
  section for $e^+e^- \to$ hadrons in the center-of-mass energy region between
  3.650 and 3.872 GeV}, Phys. Rev. Lett. 101 (2008) 102004.
\newblock \href {https://doi.org/10.1103/PhysRevLett.101.102004}
  {\path{doi:10.1103/PhysRevLett.101.102004}}.

\bibitem{BESIII:2023oql}
M.~Ablikim, et~al., BESIII Collaboration, {$\mathcal R(3780)$ resonance
  interpreted as the $1^3D_1$-wave dominant state of charmonium from precise
  measurements of the cross section of $e^+e^-\rightarrow$~hadrons}, Phys. Rev.
  Lett. 133 (2024) 241902.
\newblock \href {http://arxiv.org/abs/2401.00878} {\path{arXiv:2401.00878}},
  \href {https://doi.org/10.1103/PhysRevLett.133.241902}
  {\path{doi:10.1103/PhysRevLett.133.241902}}.

\bibitem{BESIII:2023bed}
M.~Ablikim, et~al., BESIII Collaboration, {First observation of a
  three-resonance structure in $e^+e^-\rightarrow$ nonopen charm hadrons},
  Phys. Rev. Lett. 132 (2024) 191902.
\newblock \href {http://arxiv.org/abs/2307.10948} {\path{arXiv:2307.10948}},
  \href {https://doi.org/10.1103/PhysRevLett.132.191902}
  {\path{doi:10.1103/PhysRevLett.132.191902}}.

\bibitem{BESIII:2024ths}
M.~Ablikim, et~al., BESIII Collaboration, {Precise measurement of Born cross
  sections for $e^+e^-\to D\bar{D}$ at $\sqrt{s} =$ 3.80--4.95 GeV}, Phys. Rev.
  Lett. 133 (2024) 081901.
\newblock \href {http://arxiv.org/abs/2402.03829} {\path{arXiv:2402.03829}},
  \href {https://doi.org/10.1103/PhysRevLett.133.081901}
  {\path{doi:10.1103/PhysRevLett.133.081901}}.

\bibitem{BaBar:2008drv}
B.~Aubert, et~al., BaBar Collaboration, {Study of the Exclusive
  Initial-State-Radiation Production of the $D \bar{D}$ System} (2 2008).
\newblock \href {http://arxiv.org/abs/0710.1371} {\path{arXiv:0710.1371}}.

\bibitem{Lin:2024qcq}
Z.-Y. Lin, J.-Z. Wang, J.-B. Cheng, L.~Meng, S.-L. Zhu, {Identification of the
  $G(3900)$ as the $P$-wave $D\bar{D}^*/\bar{D}D^*$ resonance}, Phys. Rev.
  Lett. 133 (2024) 241903.
\newblock \href {http://arxiv.org/abs/2403.01727} {\path{arXiv:2403.01727}},
  \href {https://doi.org/10.1103/PhysRevLett.133.241903}
  {\path{doi:10.1103/PhysRevLett.133.241903}}.

\bibitem{Cao:2014qna}
X.~Cao, H.~Lenske, {The nature and line shapes of charmonium in the $e^+e^- \to
  D\bar{D}$ reactions} (10 2014).
\newblock \href {http://arxiv.org/abs/1410.1375} {\path{arXiv:1410.1375}}.

\bibitem{Du:2016qcr}
M.-L. Du, U.-G. Mei{\ss}ner, Q.~Wang, {$P$-wave coupled channel effects in
  electron-positron annihilation}, Phys. Rev. D 94 (2016) 096006.
\newblock \href {http://arxiv.org/abs/1608.02537} {\path{arXiv:1608.02537}},
  \href {https://doi.org/10.1103/PhysRevD.94.096006}
  {\path{doi:10.1103/PhysRevD.94.096006}}.

\bibitem{Uglov:2016orr}
T.~V. Uglov, Y.~S. Kalashnikova, A.~V. Nefediev, G.~V. Pakhlova, P.~N. Pakhlov,
  {Exclusive open-charm near-threshold cross sections in a coupled-channel
  approach}, JETP Lett. 105 (2017) 1--7.
\newblock \href {http://arxiv.org/abs/1611.07582} {\path{arXiv:1611.07582}},
  \href {https://doi.org/10.1134/S0021364017010064}
  {\path{doi:10.1134/S0021364017010064}}.

\bibitem{Husken:2024hmi}
N.~H{\"u}sken, R.~F. Lebed, R.~E. Mitchell, E.~S. Swanson, Y.-Q. Wang, C.-Z.
  Yuan, {Poles and poltergeists in $e^+e^-\to D\bar{D}$ data}, Phys. Rev. D 109
  (2024) 114010.
\newblock \href {http://arxiv.org/abs/2404.03896} {\path{arXiv:2404.03896}},
  \href {https://doi.org/10.1103/PhysRevD.109.114010}
  {\path{doi:10.1103/PhysRevD.109.114010}}.

\bibitem{Ye:2025ywy}
Q.~Ye, Z.~Zhang, M.-L. Du, U.-G. Mei{\ss}ner, P.-Y. Niu, Q.~Wang, {Resonance
  parameters of the vector charmoniumlike state $G(3900)$}, Phys. Rev. D 112
  (2025) 016015.
\newblock \href {http://arxiv.org/abs/2504.17431} {\path{arXiv:2504.17431}},
  \href {https://doi.org/10.1103/qq61-ncln} {\path{doi:10.1103/qq61-ncln}}.

\bibitem{Nakamura:2023obk}
S.~X. Nakamura, X.~H. Li, H.~P. Peng, Z.~T. Sun, X.~R. Zhou, {Global
  coupled-channel analysis of $e^+e^-\to c\bar{c}$ processes in
  $\sqrt{s}=3.75-4.7$ GeV} (12 2023).
\newblock \href {http://arxiv.org/abs/2312.17658} {\path{arXiv:2312.17658}}.

\bibitem{Qian:2025zyp}
R.-Q. Qian, X.~Liu, {Unified coupled-channel description for the five
  near-threshold structures $\psi(3770)$, $G(3900)$, $R(3760)$, $R(3780)$ and
  $R(3810)$ from $e^+e^-$ annihilation}, Phys. Rev. D 112 (2025) L091502.
\newblock \href {http://arxiv.org/abs/2509.17679} {\path{arXiv:2509.17679}},
  \href {https://doi.org/10.1103/c5vs-2l1k} {\path{doi:10.1103/c5vs-2l1k}}.

\bibitem{Anwar:2018sol}
M.~N. Anwar, J.~Ferretti, E.~Santopinto, {Spectroscopy of the hidden-charm
  $[qc][\bar q \bar c]$ and $[sc][\bar s \bar c]$ tetraquarks in the
  relativized diquark model}, Phys. Rev. D 98 (2018) 094015.
\newblock \href {http://arxiv.org/abs/1805.06276} {\path{arXiv:1805.06276}},
  \href {https://doi.org/10.1103/PhysRevD.98.094015}
  {\path{doi:10.1103/PhysRevD.98.094015}}.

\bibitem{Wang:2018rfw}
Z.-G. Wang, {Vector tetraquark state candidates: $Y(4260/4220)$,
  $Y(4360/4320)$, $Y(4390)$ and $Y(4660/4630)$}, Eur. Phys. J. C 78 (2018) 518.
\newblock \href {http://arxiv.org/abs/1803.05749} {\path{arXiv:1803.05749}},
  \href {https://doi.org/10.1140/epjc/s10052-018-5996-5}
  {\path{doi:10.1140/epjc/s10052-018-5996-5}}.

\bibitem{Wang:2018ntv}
Z.-G. Wang, {Lowest vector tetraquark states: $Y(4260/4220)$ or $Z_c(4100)$},
  Eur. Phys. J. C 78 (2018) 933.
\newblock \href {http://arxiv.org/abs/1809.10299} {\path{arXiv:1809.10299}},
  \href {https://doi.org/10.1140/epjc/s10052-018-6417-5}
  {\path{doi:10.1140/epjc/s10052-018-6417-5}}.

\bibitem{Wang:2018ejf}
Z.-G. Wang, {Analysis of the vector tetraquark states with P-waves between the
  diquarks and antidiquarks via the QCD sum rules}, Eur. Phys. J. C 79 (2019)
  29.
\newblock \href {http://arxiv.org/abs/1811.02726} {\path{arXiv:1811.02726}},
  \href {https://doi.org/10.1140/epjc/s10052-019-6568-z}
  {\path{doi:10.1140/epjc/s10052-019-6568-z}}.

\bibitem{Zhao:2025kno}
Z.~Zhao, A.~Kaewsnod, K.~Xu, N.~Tagsinsit, X.~Liu, A.~Limphirat, Y.~Yan, {Study
  of $1^{--}$ $P$ wave charmoniumlike and bottomoniumlike tetraquark
  spectroscopy} (3 2025).
\newblock \href {http://arxiv.org/abs/2503.00552} {\path{arXiv:2503.00552}}.

\bibitem{Li:2022fgd}
N.~Li, H.-Z. He, W.~Liang, Q.-F. L{\"u}, D.-Y. Chen, Y.-B. Dong, {Light meson
  emissions of selected charmonium-like states within compact tetraquark
  configurations*}, Chin. Phys. C 47 (2023) 063102.
\newblock \href {http://arxiv.org/abs/2210.17148} {\path{arXiv:2210.17148}},
  \href {https://doi.org/10.1088/1674-1137/acc648}
  {\path{doi:10.1088/1674-1137/acc648}}.

\bibitem{Wang:2023dsm}
Z.-G. Wang, {Three-body strong decays of the Y(4230) via the light-cone QCD sum
  rules}, Int. J. Mod. Phys. A 38 (2023) 2350175.
\newblock \href {http://arxiv.org/abs/2309.01337} {\path{arXiv:2309.01337}},
  \href {https://doi.org/10.1142/S0217751X23501750}
  {\path{doi:10.1142/S0217751X23501750}}.

\bibitem{Xie:2023qjg}
Y.~Xie, H.~Sun, {Strong decay of in light cone sum rules*}, Chin. Phys. C 48
  (2024) 023105.
\newblock \href {http://arxiv.org/abs/2306.06741} {\path{arXiv:2306.06741}},
  \href {https://doi.org/10.1088/1674-1137/ad13f9}
  {\path{doi:10.1088/1674-1137/ad13f9}}.

\bibitem{Berwein:2024ztx}
M.~Berwein, N.~Brambilla, A.~Mohapatra, A.~Vairo, {Hybrids, tetraquarks,
  pentaquarks, doubly heavy baryons, and quarkonia in Born-Oppenheimer
  effective theory}, Phys. Rev. D 110 (2024) 094040.
\newblock \href {http://arxiv.org/abs/2408.04719} {\path{arXiv:2408.04719}},
  \href {https://doi.org/10.1103/PhysRevD.110.094040}
  {\path{doi:10.1103/PhysRevD.110.094040}}.

\bibitem{Braaten:2024stn}
E.~Braaten, R.~Bruschini, {Model-independent predictions for decays of
  hidden-heavy hadrons into pairs of heavy hadrons}, Phys. Rev. D 109 (2024)
  094051.
\newblock \href {http://arxiv.org/abs/2403.12868} {\path{arXiv:2403.12868}},
  \href {https://doi.org/10.1103/PhysRevD.109.094051}
  {\path{doi:10.1103/PhysRevD.109.094051}}.

\bibitem{Ma:2019hsm}
Y.~Ma, W.~Sun, Y.~Chen, M.~Gong, Z.~Liu, {Color halo scenario of
  charmonium-like hybrids}, Chin. Phys. C 45 (2021) 093111.
\newblock \href {http://arxiv.org/abs/1910.09819} {\path{arXiv:1910.09819}},
  \href {https://doi.org/10.1088/1674-1137/ac0ee2}
  {\path{doi:10.1088/1674-1137/ac0ee2}}.

\bibitem{vonDetten:2024eie}
L.~von Detten, V.~Baru, C.~Hanhart, Q.~Wang, D.~Winney, Q.~Zhao, {How many
  vector charmoniumlike states lie in the mass range
  4.2{\textendash}4.35~GeV?}, Phys. Rev. D 109 (2024) 116002.
\newblock \href {http://arxiv.org/abs/2402.03057} {\path{arXiv:2402.03057}},
  \href {https://doi.org/10.1103/PhysRevD.109.116002}
  {\path{doi:10.1103/PhysRevD.109.116002}}.

\bibitem{Dong:2019ofp}
X.-K. Dong, Y.-H. Lin, B.-S. Zou, {Prediction of an exotic state around 4240
  MeV with $J^{PC}=1^{-+}$ as C-parity partner of Y(4260) in molecular
  picture}, Phys. Rev. D 101 (2020) 076003.
\newblock \href {http://arxiv.org/abs/1910.14455} {\path{arXiv:1910.14455}},
  \href {https://doi.org/10.1103/PhysRevD.101.076003}
  {\path{doi:10.1103/PhysRevD.101.076003}}.

\bibitem{Anwar:2021dmg}
M.~N. Anwar, Y.~Lu, {Heavy quark spin partners of the Y(4260) in
  coupled-channel formalism}, Phys. Rev. D 104 (2021) 094006.
\newblock \href {http://arxiv.org/abs/2109.02539} {\path{arXiv:2109.02539}},
  \href {https://doi.org/10.1103/PhysRevD.104.094006}
  {\path{doi:10.1103/PhysRevD.104.094006}}.

\bibitem{Peng:2022nrj}
F.-Z. Peng, M.-J. Yan, M.~S{\'a}nchez~S{\'a}nchez, M.~Pavon~Valderrama, {Light-
  and heavy-quark symmetries and the $Y(4230)$, $Y(4360)$, $Y(4500)$,
  $Y(4620)$, and $X(4630)$ resonances}, Phys. Rev. D 107 (2023) 016001.
\newblock \href {http://arxiv.org/abs/2205.13590} {\path{arXiv:2205.13590}},
  \href {https://doi.org/10.1103/PhysRevD.107.016001}
  {\path{doi:10.1103/PhysRevD.107.016001}}.

\bibitem{Ji:2022blw}
T.~Ji, X.-K. Dong, F.-K. Guo, B.-S. Zou, {Prediction of a Narrow Exotic
  Hadronic State with Quantum Numbers $J^{PC}=0^{--}$}, Phys. Rev. Lett. 129
  (2022) 102002.
\newblock \href {http://arxiv.org/abs/2205.10994} {\path{arXiv:2205.10994}},
  \href {https://doi.org/10.1103/PhysRevLett.129.102002}
  {\path{doi:10.1103/PhysRevLett.129.102002}}.

\bibitem{Liu:2024hba}
M.-Z. Liu, Q.~Wu, {Exploring the nature of $Y(4230)$ and $Y(4360)$ in $B$
  decays}, Eur. Phys. J. C 85 (2025) 188.
\newblock \href {http://arxiv.org/abs/2409.06539} {\path{arXiv:2409.06539}},
  \href {https://doi.org/10.1140/epjc/s10052-025-13838-2}
  {\path{doi:10.1140/epjc/s10052-025-13838-2}}.

\bibitem{Dubynskiy:2008mq}
S.~Dubynskiy, M.~B. Voloshin, {Hadro-Charmonium}, Phys. Lett. B 666 (2008)
  344--346.
\newblock \href {http://arxiv.org/abs/0803.2224} {\path{arXiv:0803.2224}},
  \href {https://doi.org/10.1016/j.physletb.2008.07.086}
  {\path{doi:10.1016/j.physletb.2008.07.086}}.

\bibitem{Ferretti:2018kzy}
J.~Ferretti, {$\eta_{\rm c}$- and $J/\psi$-isoscalar meson bound states in the
  hadro-charmonium picture}, Phys. Lett. B 782 (2018) 702--706.
\newblock \href {http://arxiv.org/abs/1805.04717} {\path{arXiv:1805.04717}},
  \href {https://doi.org/10.1016/j.physletb.2018.06.032}
  {\path{doi:10.1016/j.physletb.2018.06.032}}.

\bibitem{MartinContreras:2023oqs}
M.~A. Martin~Contreras, A.~Vega, {Holographic stability for non-$q\bar{q}$
  candidates}, Phys. Rev. D 108 (2023) 126024.
\newblock \href {http://arxiv.org/abs/2309.02905} {\path{arXiv:2309.02905}},
  \href {https://doi.org/10.1103/PhysRevD.108.126024}
  {\path{doi:10.1103/PhysRevD.108.126024}}.

\bibitem{He:2017mbh}
J.~He, D.-Y. Chen, {Interpretation of $Y(4390)$ as an isoscalar partner of
  $Z(4430)$ from $D^*(2010)\bar{D}_1(2420)$ interaction}, Eur. Phys. J. C 77
  (2017) 398.
\newblock \href {http://arxiv.org/abs/1704.08776} {\path{arXiv:1704.08776}},
  \href {https://doi.org/10.1140/epjc/s10052-017-4973-8}
  {\path{doi:10.1140/epjc/s10052-017-4973-8}}.

\bibitem{Chen:2017abq}
D.-Y. Chen, C.-J. Xiao, J.~He, {Hidden-charm decays of $Y(4390)$ in a hadronic
  molecular scenario}, Phys. Rev. D 96 (2017) 054017.
\newblock \href {https://doi.org/10.1103/PhysRevD.96.054017}
  {\path{doi:10.1103/PhysRevD.96.054017}}.

\bibitem{Dai:2017fwx}
L.-Y. Dai, J.~Haidenbauer, U.~G. Mei{\ss}ner, {Re-examining the $X(4630)$
  resonance in the reaction $e^+e^-\rightarrow \Lambda^+_c\bar\Lambda^-_c$},
  Phys. Rev. D 96 (2017) 116001.
\newblock \href {http://arxiv.org/abs/1710.03142} {\path{arXiv:1710.03142}},
  \href {https://doi.org/10.1103/PhysRevD.96.116001}
  {\path{doi:10.1103/PhysRevD.96.116001}}.

\bibitem{Mei:2022msh}
X.-H. Mei, Z.~Yu, M.~Song, J.-Y. Guo, G.~Li, X.~Luo, {Explanation of $Y(4630)$
  as a hadronic resonant state*}, Chin. Phys. C 47 (2023) 033104.
\newblock \href {http://arxiv.org/abs/2212.02218} {\path{arXiv:2212.02218}},
  \href {https://doi.org/10.1088/1674-1137/aca959}
  {\path{doi:10.1088/1674-1137/aca959}}.

\bibitem{He:2019csk}
J.~He, Y.~Liu, J.-T. Zhu, D.-Y. Chen, {Y(4626) as a molecular state from
  interaction ${D}^*_s{\bar{D}}_{s1}(2536)-{D}_s{\bar{D}}_{s1}(2536)$}, Eur.
  Phys. J. C 80 (2020) 246.
\newblock \href {http://arxiv.org/abs/1912.08420} {\path{arXiv:1912.08420}},
  \href {https://doi.org/10.1140/epjc/s10052-020-7820-2}
  {\path{doi:10.1140/epjc/s10052-020-7820-2}}.

\bibitem{Ke:2020eba}
H.-W. Ke, X.-H. Liu, X.-Q. Li, {Study on the possible molecular state composed
  of $D^*_s\bar D_{s1} $ within the Bethe-Salpeter framework}, Chin. Phys. C 44
  (2020) 093104.
\newblock \href {http://arxiv.org/abs/2004.03167} {\path{arXiv:2004.03167}},
  \href {https://doi.org/10.1088/1674-1137/44/9/093104}
  {\path{doi:10.1088/1674-1137/44/9/093104}}.

\bibitem{Yue:2024bvy}
Z.-L. Yue, Y.~Pan, D.-Y. Chen, {Hidden charm decays of $Y(4626)$ in a
  $D_s^{*+}D_{s1}(2536)^-$ molecular frame}, Phys. Rev. D 110 (2024) 074013.
\newblock \href {http://arxiv.org/abs/2408.08546} {\path{arXiv:2408.08546}},
  \href {https://doi.org/10.1103/PhysRevD.110.074013}
  {\path{doi:10.1103/PhysRevD.110.074013}}.

\bibitem{Sundu:2018toi}
H.~Sundu, S.~S. Agaev, K.~Azizi, {Resonance $Y(4660)$ as a vector tetraquark
  and its strong decay channels}, Phys. Rev. D 98 (2018) 054021.
\newblock \href {http://arxiv.org/abs/1805.04705} {\path{arXiv:1805.04705}},
  \href {https://doi.org/10.1103/PhysRevD.98.054021}
  {\path{doi:10.1103/PhysRevD.98.054021}}.

\bibitem{Wang:2019iaa}
Z.-G. Wang, {Strong decays of the $Y(4660)$ as a vector tetraquark state in
  solid quark-hadron duality}, Eur. Phys. J. C 79 (2019) 184.
\newblock \href {http://arxiv.org/abs/1901.02177} {\path{arXiv:1901.02177}},
  \href {https://doi.org/10.1140/epjc/s10052-019-6708-5}
  {\path{doi:10.1140/epjc/s10052-019-6708-5}}.

\bibitem{Wang:2023jaw}
Z.-G. Wang, {Analysis of the vector hidden-charm-hidden-strange tetraquark
  states with implicit P-waves via the QCD sum rules}, Nucl. Phys. B 1002
  (2024) 116514.
\newblock \href {http://arxiv.org/abs/2312.10292} {\path{arXiv:2312.10292}},
  \href {https://doi.org/10.1016/j.nuclphysb.2024.116514}
  {\path{doi:10.1016/j.nuclphysb.2024.116514}}.

\bibitem{Yang:2025lef}
X.-S. Yang, Z.-G. Wang, {Analysis of the strong decays of the $Y(4660)$ in
  tetraquark scenario via the QCD sum rules} (11 2025).
\newblock \href {http://arxiv.org/abs/2511.09098} {\path{arXiv:2511.09098}}.

\bibitem{Gungor:2023ksu}
E.~G{\"u}ng{\"o}r, H.~Sundu, J.~Y. S{\"u}ng{\"u}, E.~V. Veliev, {Possible
  Molecular Explanation for the Resonance $Y$(4500)}, Few Body Syst. 64 (2023)
  53.
\newblock \href {http://arxiv.org/abs/2303.05756} {\path{arXiv:2303.05756}},
  \href {https://doi.org/10.1007/s00601-023-01807-y}
  {\path{doi:10.1007/s00601-023-01807-y}}.

\bibitem{Wang:2025zbv}
X.-Y. Wang, L.-L. Wang, X.-H. Liu, Q.~Zhao, {Threshold effects as the origin of
  $Y(4500)$ observed in $e^+e^- \to J/\psi K^+K^-$}, Phys. Rev. D 113 (2026)
  014030.
\newblock \href {http://arxiv.org/abs/2510.23099} {\path{arXiv:2510.23099}},
  \href {https://doi.org/10.1103/xrdw-64rv} {\path{doi:10.1103/xrdw-64rv}}.

\bibitem{Liu:2025bjm}
S.-D. Liu, Q.~Wu, G.~Li, {Dipionic transitions of Y(4500) to
  J/{\ensuremath{\psi}}}, Phys. Rev. D 112 (2025) 074036.
\newblock \href {http://arxiv.org/abs/2504.14792} {\path{arXiv:2504.14792}},
  \href {https://doi.org/10.1103/fz1m-3mdg} {\path{doi:10.1103/fz1m-3mdg}}.

\bibitem{Wang:2023hsc}
Z.-G. Wang, {Analysis of the decay $Y(4500) \to D^* \bar{D}^* \pi$ with the
  light-cone QCD sum rules}, Nucl. Phys. B 993 (2023) 116265.
\newblock \href {http://arxiv.org/abs/2304.14153} {\path{arXiv:2304.14153}},
  \href {https://doi.org/10.1016/j.nuclphysb.2023.116265}
  {\path{doi:10.1016/j.nuclphysb.2023.116265}}.

\bibitem{Wang:2024qqa}
Z.-G. Wang, {Strong decays of the vector tetraquark states with the masses
  about 4.5 GeV via the QCD sum rules}, Nucl. Phys. B 1005 (2024) 116580.
\newblock \href {http://arxiv.org/abs/2404.05328} {\path{arXiv:2404.05328}},
  \href {https://doi.org/10.1016/j.nuclphysb.2024.116580}
  {\path{doi:10.1016/j.nuclphysb.2024.116580}}.

\bibitem{Wang:2024gvf}
Z.-G. Wang, {Ground states and first radial excitations of vector tetraquark
  states with explicit P-waves via QCD sum rules*}, Chin. Phys. C 48 (2024)
  103103.
\newblock \href {http://arxiv.org/abs/2405.04145} {\path{arXiv:2405.04145}},
  \href {https://doi.org/10.1088/1674-1137/ad5ae5}
  {\path{doi:10.1088/1674-1137/ad5ae5}}.

\bibitem{Belle:2008nac}
K.~F. Chen, et~al., Belle Collaboration, {Observation of an enhancement in
  $e^+e^- \to \Upsilon(1S)\pi^+ \pi^-$, $\Upsilon(2S)\pi^+ \pi^-$, and
  $\Upsilon(3S)\pi^+ \pi^-$ production around $\sqrt{s}=10.89$ GeV at Belle},
  Phys. Rev. D 82 (2010) 091106.
\newblock \href {http://arxiv.org/abs/0810.3829} {\path{arXiv:0810.3829}},
  \href {https://doi.org/10.1103/PhysRevD.82.091106}
  {\path{doi:10.1103/PhysRevD.82.091106}}.

\bibitem{BESIII:2013ris}
M.~Ablikim, et~al., BESIII Collaboration, {Observation of a charged
  charmoniumlike structure in $e^+e^- \to \pi^+\pi^- J/\psi$ at $\sqrt{s}$
  =4.26 GeV}, Phys. Rev. Lett. 110 (2013) 252001.
\newblock \href {http://arxiv.org/abs/1303.5949} {\path{arXiv:1303.5949}},
  \href {https://doi.org/10.1103/PhysRevLett.110.252001}
  {\path{doi:10.1103/PhysRevLett.110.252001}}.

\bibitem{Belle:2013yex}
Z.~Q. Liu, et~al., Belle Collaboration, {Study of $e^+e^- \to \pi^+ \pi^-
  J/\psi$ and Observation of a charged charmoniumlike state at Belle}, Phys.
  Rev. Lett. 110 (2013) 252002, [Erratum: Phys.Rev.Lett. 111, 019901 (2013)].
\newblock \href {http://arxiv.org/abs/1304.0121} {\path{arXiv:1304.0121}},
  \href {https://doi.org/10.1103/PhysRevLett.110.252002}
  {\path{doi:10.1103/PhysRevLett.110.252002}}.

\bibitem{BESIII:2013mhi}
M.~Ablikim, et~al., BESIII Collaboration, {Observation of a charged
  charmoniumlike structure in $e^+e^- \to (D^{*} \bar{D}^{*})^{\pm} \pi^\mp$ at
  $\sqrt{s}=4.26$GeV}, Phys. Rev. Lett. 112 (2014) 132001.
\newblock \href {http://arxiv.org/abs/1308.2760} {\path{arXiv:1308.2760}},
  \href {https://doi.org/10.1103/PhysRevLett.112.132001}
  {\path{doi:10.1103/PhysRevLett.112.132001}}.

\bibitem{BESIII:2015tix}
M.~Ablikim, et~al., BESIII Collaboration, {Observation of a neutral
  charmoniumlike state $Z_c(4025)^0$ in $e^{+} e^{-} \to (D^{*}
  \bar{D}^{*})^{0} \pi^0$}, Phys. Rev. Lett. 115 (2015) 182002.
\newblock \href {http://arxiv.org/abs/1507.02404} {\path{arXiv:1507.02404}},
  \href {https://doi.org/10.1103/PhysRevLett.115.182002}
  {\path{doi:10.1103/PhysRevLett.115.182002}}.

\bibitem{BESIII:2017vtc}
M.~Ablikim, et~al., BESIII Collaboration, {Measurement of $e^{+}e^{-} \to
  \pi^{0}\pi^{0}\psi(3686)$ at $\sqrt{s}$ from 4.009 to 4.600 GeV and
  observation of a neutral charmoniumlike structure}, Phys. Rev. D 97 (2018)
  052001.
\newblock \href {http://arxiv.org/abs/1710.10740} {\path{arXiv:1710.10740}},
  \href {https://doi.org/10.1103/PhysRevD.97.052001}
  {\path{doi:10.1103/PhysRevD.97.052001}}.

\bibitem{Xiao:2013iha}
T.~Xiao, S.~Dobbs, A.~Tomaradze, K.~K. Seth, {Observation of the charged hadron
  $Z_c^{\pm}(3900)$ and evidence for the neutral $Z_c^0(3900)$ in $e^+e^-\to
  \pi\pi J/\psi$ at $\sqrt{s}=4170$ MeV}, Phys. Lett. B 727 (2013) 366--370.
\newblock \href {http://arxiv.org/abs/1304.3036} {\path{arXiv:1304.3036}},
  \href {https://doi.org/10.1016/j.physletb.2013.10.041}
  {\path{doi:10.1016/j.physletb.2013.10.041}}.

\bibitem{BESIII:2015pqw}
M.~Ablikim, et~al., BESIII Collaboration, {Confirmation of a charged
  charmoniumlike state $Z_c(3885)^{\mp}$ in
  $e^+e^-\to\pi^{\pm}(D\bar{D}^*)^\mp$ with double $D$ tag}, Phys. Rev. D 92
  (2015) 092006.
\newblock \href {http://arxiv.org/abs/1509.01398} {\path{arXiv:1509.01398}},
  \href {https://doi.org/10.1103/PhysRevD.92.092006}
  {\path{doi:10.1103/PhysRevD.92.092006}}.

\bibitem{BESIII:2015cld}
M.~Ablikim, et~al., BESIII Collaboration, {Observation of $Z_c(3900)^{0}$ in
  $e^+e^-\to\pi^0\pi^0 J/\psi$}, Phys. Rev. Lett. 115 (2015) 112003.
\newblock \href {http://arxiv.org/abs/1506.06018} {\path{arXiv:1506.06018}},
  \href {https://doi.org/10.1103/PhysRevLett.115.112003}
  {\path{doi:10.1103/PhysRevLett.115.112003}}.

\bibitem{BESIII:2017bua}
M.~Ablikim, et~al., BESIII Collaboration, {Determination of the Spin and Parity
  of the $Z_c(3900)$}, Phys. Rev. Lett. 119 (2017) 072001.
\newblock \href {http://arxiv.org/abs/1706.04100} {\path{arXiv:1706.04100}},
  \href {https://doi.org/10.1103/PhysRevLett.119.072001}
  {\path{doi:10.1103/PhysRevLett.119.072001}}.

\bibitem{BESIII:2015ntl}
M.~Ablikim, et~al., BESIII Collaboration, {Observation of a Neutral Structure
  near the $D\bar{D}^{*}$ Mass Threshold in $e^{+}e^{-}\to (D
  \bar{D}^*)^0\pi^0$ at $\sqrt{s}$ = 4.226 and 4.257 GeV}, Phys. Rev. Lett. 115
  (2015) 222002.
\newblock \href {http://arxiv.org/abs/1509.05620} {\path{arXiv:1509.05620}},
  \href {https://doi.org/10.1103/PhysRevLett.115.222002}
  {\path{doi:10.1103/PhysRevLett.115.222002}}.

\bibitem{BESIII:2014gnk}
M.~Ablikim, et~al., BESIII Collaboration, {Observation of $e^+e^- \to \pi^0
  \pi^0 h_c$ and a Neutral Charmoniumlike Structure $Z_c(4020)^0$}, Phys. Rev.
  Lett. 113 (2014) 212002.
\newblock \href {http://arxiv.org/abs/1409.6577} {\path{arXiv:1409.6577}},
  \href {https://doi.org/10.1103/PhysRevLett.113.212002}
  {\path{doi:10.1103/PhysRevLett.113.212002}}.

\bibitem{Chen:2013wca}
D.-Y. Chen, X.~Liu, T.~Matsuki, {Predictions of charged charmoniumlike
  structures with hidden-charm and open-strange channels}, Phys. Rev. Lett. 110
  (2013) 232001.
\newblock \href {http://arxiv.org/abs/1303.6842} {\path{arXiv:1303.6842}},
  \href {https://doi.org/10.1103/PhysRevLett.110.232001}
  {\path{doi:10.1103/PhysRevLett.110.232001}}.

\bibitem{Lovelock:1985nb}
D.~M.~J. Lovelock, et~al., {Masses, widths, and leptonic widths of the higher
  upsilon resonances}, Phys. Rev. Lett. 54 (1985) 377--380.
\newblock \href {https://doi.org/10.1103/PhysRevLett.54.377}
  {\path{doi:10.1103/PhysRevLett.54.377}}.

\bibitem{CLEO:1984vfn}
D.~Besson, et~al., CLEO Collaboration, {Observation of new structure in the
  $e^+e^-$ annihilation cross-section above $B\bar B$ threshold}, Phys. Rev.
  Lett. 54 (1985) 381.
\newblock \href {https://doi.org/10.1103/PhysRevLett.54.381}
  {\path{doi:10.1103/PhysRevLett.54.381}}.

\bibitem{BaBar:2008cmq}
B.~Aubert, et~al., BaBar Collaboration, {Measurement of the $e^{+} e^{-} \to b
  \bar{b}$ cross section between $\sqrt{s}$ = 10.54-GeV and 11.20-GeV}, Phys.
  Rev. Lett. 102 (2009) 012001.
\newblock \href {http://arxiv.org/abs/0809.4120} {\path{arXiv:0809.4120}},
  \href {https://doi.org/10.1103/PhysRevLett.102.012001}
  {\path{doi:10.1103/PhysRevLett.102.012001}}.

\bibitem{Belle:2015aea}
D.~Santel, et~al., Belle Collaboration, {Measurements of the $\Upsilon$(10860)
  and $\Upsilon$(11020) resonances via $\sigma(e^+e^-\to \Upsilon(nS)\pi^+
  \pi^-)$}, Phys. Rev. D 93 (2016) 011101.
\newblock \href {http://arxiv.org/abs/1501.01137} {\path{arXiv:1501.01137}},
  \href {https://doi.org/10.1103/PhysRevD.93.011101}
  {\path{doi:10.1103/PhysRevD.93.011101}}.

\bibitem{Belle:2019cbt}
R.~Mizuk, et~al., Belle Collaboration, {Observation of a new structure near
  10.75 GeV in the energy dependence of the $e^{+}e^{-}\to
  \Upsilon(nS)\pi^+\pi-$ (n = 1, 2, 3) cross sections}, JHEP 10 (2019) 220.
\newblock \href {http://arxiv.org/abs/1905.05521} {\path{arXiv:1905.05521}},
  \href {https://doi.org/10.1007/JHEP10(2019)220}
  {\path{doi:10.1007/JHEP10(2019)220}}.

\bibitem{BaBar:2008xay}
B.~Aubert, et~al., BaBar Collaboration, {Study of hadronic transitions between
  $\Upsilon$ states and observation of $\Upsilon(4S)\to\eta \Upsilon(1S)$
  decay}, Phys. Rev. D 78 (2008) 112002.
\newblock \href {http://arxiv.org/abs/0807.2014} {\path{arXiv:0807.2014}},
  \href {https://doi.org/10.1103/PhysRevD.78.112002}
  {\path{doi:10.1103/PhysRevD.78.112002}}.

\bibitem{Meng:2008dd}
C.~Meng, K.-T. Chao, {Peak shifts due to $B^{(*)}-\bar B^{(*)}$ rescattering in
  $\Upsilon(5S)$ dipion transitions}, Phys. Rev. D 78 (2008) 034022.
\newblock \href {http://arxiv.org/abs/0805.0143} {\path{arXiv:0805.0143}},
  \href {https://doi.org/10.1103/PhysRevD.78.034022}
  {\path{doi:10.1103/PhysRevD.78.034022}}.

\bibitem{Simonov:2007cj}
Y.~A. Simonov, {Di-pion emission in heavy quarkonia decays}, JETP Lett. 87
  (2008) 121--123.
\newblock \href {http://arxiv.org/abs/0712.2197} {\path{arXiv:0712.2197}},
  \href {https://doi.org/10.1007/s11448-008-3001-5}
  {\path{doi:10.1007/s11448-008-3001-5}}.

\bibitem{Ali:2009pi}
A.~Ali, C.~Hambrock, I.~Ahmed, M.~J. Aslam, {A case for hidden $b\bar{b}$
  tetraquarks based on $e^+e^- \to b\bar{b}$ cross section between
  $\sqrt{s}=10.54$ and 11.20 GeV}, Phys. Lett. B 684 (2010) 28--39.
\newblock \href {http://arxiv.org/abs/0911.2787} {\path{arXiv:0911.2787}},
  \href {https://doi.org/10.1016/j.physletb.2009.12.053}
  {\path{doi:10.1016/j.physletb.2009.12.053}}.

\bibitem{Ali:2009es}
A.~Ali, C.~Hambrock, M.~J. Aslam, {Tetraquark interpretation of the BELLE data
  on the anomalous $\Upsilon(1S)\pi^+\pi^-$ and $\Upsilon(2S)\pi^+\pi^-$
  production near the $\Upsilon(5S)$ resonance}, Phys. Rev. Lett. 104 (2010)
  162001, [Erratum: Phys.Rev.Lett. 107, 049903 (2011)].
\newblock \href {http://arxiv.org/abs/0912.5016} {\path{arXiv:0912.5016}},
  \href {https://doi.org/10.1103/PhysRevLett.104.162001}
  {\path{doi:10.1103/PhysRevLett.104.162001}}.

\bibitem{Ali:2010pq}
A.~Ali, C.~Hambrock, S.~Mishima, {Tetraquark-based analysis and predictions of
  the cross sections and distributions for the processes $e^+ e^- \to
  \Upsilon(1S) (\pi^+ \pi^-, K^+ K^-, \eta \pi^0)$ near $\Upsilon(5S)$}, Phys.
  Rev. Lett. 106 (2011) 092002.
\newblock \href {http://arxiv.org/abs/1011.4856} {\path{arXiv:1011.4856}},
  \href {https://doi.org/10.1103/PhysRevLett.106.092002}
  {\path{doi:10.1103/PhysRevLett.106.092002}}.

\bibitem{Sonnenschein:2016ibx}
J.~Sonnenschein, D.~Weissman, {A tetraquark or not a tetraquark? A holography
  inspired stringy hadron (HISH) perspective}, Nucl. Phys. B 920 (2017)
  319--344.
\newblock \href {http://arxiv.org/abs/1606.02732} {\path{arXiv:1606.02732}},
  \href {https://doi.org/10.1016/j.nuclphysb.2017.04.016}
  {\path{doi:10.1016/j.nuclphysb.2017.04.016}}.

\bibitem{Patel:2016otd}
S.~Patel, P.~C. Vinodkumar, {Tetraquark states in the bottom sector and the
  status of the $Y_b$ (10890) state}, Eur. Phys. J. C 76 (2016) 356.
\newblock \href {http://arxiv.org/abs/1606.01047} {\path{arXiv:1606.01047}},
  \href {https://doi.org/10.1140/epjc/s10052-016-4186-6}
  {\path{doi:10.1140/epjc/s10052-016-4186-6}}.

\bibitem{Zhang:2010mv}
J.-R. Zhang, M.-Q. Huang, {Could $Y_{b}(10890)$ be the $P$-wave
  $[bq][\bar{b}\bar{q}]$ tetraquark state?}, JHEP 11 (2010) 057.
\newblock \href {http://arxiv.org/abs/1011.2815} {\path{arXiv:1011.2815}},
  \href {https://doi.org/10.1007/JHEP11(2010)057}
  {\path{doi:10.1007/JHEP11(2010)057}}.

\bibitem{Chen:2011qx}
D.-Y. Chen, J.~He, X.-Q. Li, X.~Liu, {Dipion invariant mass distribution of the
  anomalous $\Upsilon(1S) \pi^{+} \pi^{-}$ and $\Upsilon(2S) \pi^{+} \pi^{-}$
  production near the peak of $\Upsilon(10860)$}, Phys. Rev. D 84 (2011)
  074006.
\newblock \href {http://arxiv.org/abs/1105.1672} {\path{arXiv:1105.1672}},
  \href {https://doi.org/10.1103/PhysRevD.84.074006}
  {\path{doi:10.1103/PhysRevD.84.074006}}.

\bibitem{Chen:2015jgl}
Y.-H. Chen, J.~T. Daub, F.-K. Guo, B.~Kubis, U.-G. Mei{\ss}ner, B.-S. Zou,
  {Effect of $Z_b$ states on $\Upsilon(3S)\to\Upsilon(1S)\pi\pi$ decays}, Phys.
  Rev. D 93 (2016) 034030.
\newblock \href {http://arxiv.org/abs/1512.03583} {\path{arXiv:1512.03583}},
  \href {https://doi.org/10.1103/PhysRevD.93.034030}
  {\path{doi:10.1103/PhysRevD.93.034030}}.

\bibitem{Novikov:1980fa}
V.~A. Novikov, M.~A. Shifman, {Comment on the $\psi^\prime\to J/\psi\pi\pi$
  decay}, Z. Phys. C 8 (1981) 43.
\newblock \href {https://doi.org/10.1007/BF01429829}
  {\path{doi:10.1007/BF01429829}}.

\bibitem{Oh:2000qr}
Y.-s. Oh, T.~Song, S.~H. Lee, {$J / \psi$ absorption by $\pi$ and $\rho$ mesons
  in meson exchange model with anomalous parity interactions}, Phys. Rev. C 63
  (2001) 034901.
\newblock \href {http://arxiv.org/abs/nucl-th/0010064}
  {\path{arXiv:nucl-th/0010064}}, \href
  {https://doi.org/10.1103/PhysRevC.63.034901}
  {\path{doi:10.1103/PhysRevC.63.034901}}.

\bibitem{Casalbuoni:1996pg}
R.~Casalbuoni, A.~Deandrea, N.~Di~Bartolomeo, R.~Gatto, F.~Feruglio,
  G.~Nardulli, {Phenomenology of heavy meson chiral Lagrangians}, Phys. Rept.
  281 (1997) 145--238.
\newblock \href {http://arxiv.org/abs/hep-ph/9605342}
  {\path{arXiv:hep-ph/9605342}}, \href
  {https://doi.org/10.1016/S0370-1573(96)00027-0}
  {\path{doi:10.1016/S0370-1573(96)00027-0}}.

\bibitem{Belle:2015upu}
A.~Garmash, et~al., Belle Collaboration, {Observation of $Z_b(10610)$ and
  $Z_b(10650)$ decaying to $B$ mesons}, Phys. Rev. Lett. 116 (2016) 212001.
\newblock \href {http://arxiv.org/abs/1512.07419} {\path{arXiv:1512.07419}},
  \href {https://doi.org/10.1103/PhysRevLett.116.212001}
  {\path{doi:10.1103/PhysRevLett.116.212001}}.

\bibitem{Adachi:2011mks}
I.~Adachi, Belle Collaboration, {Observation of two charged bottomonium-like
  resonances}, in: {9th Conference on Flavor Physics and CP Violation}, 2011.
\newblock \href {http://arxiv.org/abs/1105.4583} {\path{arXiv:1105.4583}}.

\bibitem{Belle:2014vzn}
A.~Garmash, et~al., Belle Collaboration, {Amplitude analysis of $e^+e^- \to
  \Upsilon(nS) \pi^+\pi^-$ at $\sqrt{s}=10.865$ GeV}, Phys. Rev. D 91 (2015)
  072003.
\newblock \href {http://arxiv.org/abs/1403.0992} {\path{arXiv:1403.0992}},
  \href {https://doi.org/10.1103/PhysRevD.91.072003}
  {\path{doi:10.1103/PhysRevD.91.072003}}.

\bibitem{Belle:2012glq}
I.~Adachi, et~al., Belle Collaboration, {Evidence for a $Z_b^0(10610)$ in
  Dalitz analysis of $\Upsilon(5S)\to\Upsilon(nS)\pi^0\pi^0$} (7 2012).
\newblock \href {http://arxiv.org/abs/1207.4345} {\path{arXiv:1207.4345}}.

\bibitem{Belle:2013urd}
P.~Krokovny, et~al., Belle Collaboration, {First observation of the
  $Z^{0}_{b}$(10610) in a Dalitz analysis of $\Upsilon$(10860) $\to
  \Upsilon$(nS)$\pi^0 \pi^0$}, Phys. Rev. D 88 (2013) 052016.
\newblock \href {http://arxiv.org/abs/1308.2646} {\path{arXiv:1308.2646}},
  \href {https://doi.org/10.1103/PhysRevD.88.052016}
  {\path{doi:10.1103/PhysRevD.88.052016}}.

\bibitem{Belle:2012koo}
I.~Adachi, et~al., Belle Collaboration, {Study of three-body $\Upsilon(10860)$
  decays}, 2012.
\newblock \href {http://arxiv.org/abs/1209.6450} {\path{arXiv:1209.6450}}.

\bibitem{Wu:2018xaa}
Q.~Wu, D.-Y. Chen, F.-K. Guo, {Production of the $Z_b^{(\prime)}$ states from
  the $\Upsilon(5S,6S)$ decays}, Phys. Rev. D 99 (2019) 034022.
\newblock \href {http://arxiv.org/abs/1810.09696} {\path{arXiv:1810.09696}},
  \href {https://doi.org/10.1103/PhysRevD.99.034022}
  {\path{doi:10.1103/PhysRevD.99.034022}}.

\bibitem{Chen:2011zv}
D.-Y. Chen, X.~Liu, S.-L. Zhu, {Charged bottomonium-like states $Z_b(10610)$
  and $Z_b(10650)$ and the $\Upsilon(5S)\to \Upsilon(2S)\pi^+\pi^-$ decay},
  Phys. Rev. D 84 (2011) 074016.
\newblock \href {http://arxiv.org/abs/1105.5193} {\path{arXiv:1105.5193}},
  \href {https://doi.org/10.1103/PhysRevD.84.074016}
  {\path{doi:10.1103/PhysRevD.84.074016}}.

\bibitem{Prelovsek:2019ywc}
S.~Prelovsek, H.~Bahtiyar, J.~Petkovic, {$Z_b$ tetraquark channel from lattice
  QCD and Born-Oppenheimer approximation}, Phys. Lett. B 805 (2020) 135467.
\newblock \href {http://arxiv.org/abs/1912.02656} {\path{arXiv:1912.02656}},
  \href {https://doi.org/10.1016/j.physletb.2020.135467}
  {\path{doi:10.1016/j.physletb.2020.135467}}.

\bibitem{Zhang:2011jja}
J.-R. Zhang, M.~Zhong, M.-Q. Huang, {Could $Z_{b}(10610)$ be a $B^{*}\bar{B}$
  molecular state?}, Phys. Lett. B 704 (2011) 312--315.
\newblock \href {http://arxiv.org/abs/1105.5472} {\path{arXiv:1105.5472}},
  \href {https://doi.org/10.1016/j.physletb.2011.09.039}
  {\path{doi:10.1016/j.physletb.2011.09.039}}.

\bibitem{Yang:2011rp}
Y.~Yang, J.~Ping, C.~Deng, H.-S. Zong, {Possible interpretation of the
  $Z_b$(10610) and $Z_b$(10650) in a chiral quark model}, J. Phys. G 39 (2012)
  105001.
\newblock \href {http://arxiv.org/abs/1105.5935} {\path{arXiv:1105.5935}},
  \href {https://doi.org/10.1088/0954-3899/39/10/105001}
  {\path{doi:10.1088/0954-3899/39/10/105001}}.

\bibitem{Nieves:2011zz}
J.~Nieves, M.~P. Valderrama, {Deriving the existence of $B\bar{B}^*$ bound
  states from the $X(3872)$ and heavy quark symmetry}, Phys. Rev. D 84 (2011)
  056015.
\newblock \href {http://arxiv.org/abs/1106.0600} {\path{arXiv:1106.0600}},
  \href {https://doi.org/10.1103/PhysRevD.84.056015}
  {\path{doi:10.1103/PhysRevD.84.056015}}.

\bibitem{Sun:2011uh}
Z.-F. Sun, J.~He, X.~Liu, Z.-G. Luo, S.-L. Zhu, {$Z_b(10610)^\pm$ and
  $Z_b(10650)^\pm$ as the $B^*\bar{B}$ and $B^*\bar{B}^{*}$ molecular states},
  Phys. Rev. D 84 (2011) 054002.
\newblock \href {http://arxiv.org/abs/1106.2968} {\path{arXiv:1106.2968}},
  \href {https://doi.org/10.1103/PhysRevD.84.054002}
  {\path{doi:10.1103/PhysRevD.84.054002}}.

\bibitem{Mehen:2011yh}
T.~Mehen, J.~W. Powell, {Heavy quark symmetry predictions for weakly bound
  $B$-meson molecules}, Phys. Rev. D 84 (2011) 114013.
\newblock \href {http://arxiv.org/abs/1109.3479} {\path{arXiv:1109.3479}},
  \href {https://doi.org/10.1103/PhysRevD.84.114013}
  {\path{doi:10.1103/PhysRevD.84.114013}}.

\bibitem{Ohkoda:2011vj}
S.~Ohkoda, Y.~Yamaguchi, S.~Yasui, K.~Sudoh, A.~Hosaka, {Exotic mesons with
  hidden bottom near thresholds}, Phys. Rev. D 86 (2012) 014004.
\newblock \href {http://arxiv.org/abs/1111.2921} {\path{arXiv:1111.2921}},
  \href {https://doi.org/10.1103/PhysRevD.86.014004}
  {\path{doi:10.1103/PhysRevD.86.014004}}.

\bibitem{Li:2012wf}
M.~T. Li, W.~L. Wang, Y.~B. Dong, Z.~Y. Zhang, {$Z_b(10650)$ and $Z_b(10610)$
  states in a chiral quark model}, J. Phys. G 40 (2013) 015003.
\newblock \href {http://arxiv.org/abs/1204.3959} {\path{arXiv:1204.3959}},
  \href {https://doi.org/10.1088/0954-3899/40/1/015003}
  {\path{doi:10.1088/0954-3899/40/1/015003}}.

\bibitem{Liu:2017mrh}
M.-Z. Liu, D.-J. Jia, D.-Y. Chen, {Possible hadronic molecular states composed
  of $S$-wave heavy-light mesons}, Chin. Phys. C 41 (2017) 053105.
\newblock \href {http://arxiv.org/abs/1702.04440} {\path{arXiv:1702.04440}},
  \href {https://doi.org/10.1088/1674-1137/41/5/053105}
  {\path{doi:10.1088/1674-1137/41/5/053105}}.

\bibitem{Zhao:2014gqa}
L.~Zhao, L.~Ma, S.-L. Zhu, {Spin-orbit force, recoil corrections, and possible
  $B \bar{B}^{*}$ and $D \bar{D}^{*}$ molecular states}, Phys. Rev. D 89 (2014)
  094026.
\newblock \href {http://arxiv.org/abs/1403.4043} {\path{arXiv:1403.4043}},
  \href {https://doi.org/10.1103/PhysRevD.89.094026}
  {\path{doi:10.1103/PhysRevD.89.094026}}.

\bibitem{Chen:2015ata}
W.~Chen, T.~G. Steele, H.-X. Chen, S.-L. Zhu, {Mass spectra of $Z_c$ and $Z_b$
  exotic states as hadron molecules}, Phys. Rev. D 92 (2015) 054002.
\newblock \href {http://arxiv.org/abs/1505.05619} {\path{arXiv:1505.05619}},
  \href {https://doi.org/10.1103/PhysRevD.92.054002}
  {\path{doi:10.1103/PhysRevD.92.054002}}.

\bibitem{Wang:2018jlv}
Q.~Wang, V.~Baru, A.~A. Filin, C.~Hanhart, A.~V. Nefediev, J.~L. Wynen, {Line
  shapes of the $Z_b(10610)$ and $Z_b(10650)$ in the elastic and inelastic
  channels revisited}, Phys. Rev. D 98 (2018) 074023.
\newblock \href {http://arxiv.org/abs/1805.07453} {\path{arXiv:1805.07453}},
  \href {https://doi.org/10.1103/PhysRevD.98.074023}
  {\path{doi:10.1103/PhysRevD.98.074023}}.

\bibitem{Bondar:2011ev}
A.~E. Bondar, A.~Garmash, A.~I. Milstein, R.~Mizuk, M.~B. Voloshin, {Heavy
  quark spin structure in $Z_b$ resonances}, Phys. Rev. D 84 (2011) 054010.
\newblock \href {http://arxiv.org/abs/1105.4473} {\path{arXiv:1105.4473}},
  \href {https://doi.org/10.1103/PhysRevD.84.054010}
  {\path{doi:10.1103/PhysRevD.84.054010}}.

\bibitem{Dong:2012hc}
Y.~Dong, A.~Faessler, T.~Gutsche, V.~E. Lyubovitskij, {Decays of $Z_b^+$ and
  $Z_b^{\prime+}$ as hadronic molecules}, J. Phys. G 40 (2013) 015002.
\newblock \href {http://arxiv.org/abs/1203.1894} {\path{arXiv:1203.1894}},
  \href {https://doi.org/10.1088/0954-3899/40/1/015002}
  {\path{doi:10.1088/0954-3899/40/1/015002}}.

\bibitem{Li:2012uc}
X.~Li, M.~B. Voloshin, {$Z_b(10610)$ and $Z_b(10650)$ decays to bottomonium
  plus pion}, Phys. Rev. D 86 (2012) 077502.
\newblock \href {http://arxiv.org/abs/1207.2425} {\path{arXiv:1207.2425}},
  \href {https://doi.org/10.1103/PhysRevD.86.077502}
  {\path{doi:10.1103/PhysRevD.86.077502}}.

\bibitem{Wu:2020edh}
Q.~Wu, D.-Y. Chen, T.~Matsuki, {$D$ wave bottomonia production from
  $Z_b^{(\prime)}$ decay}, Phys. Rev. D 102 (2020) 114037.
\newblock \href {http://arxiv.org/abs/2012.02940} {\path{arXiv:2012.02940}},
  \href {https://doi.org/10.1103/PhysRevD.102.114037}
  {\path{doi:10.1103/PhysRevD.102.114037}}.

\bibitem{Xiao:2017uve}
C.-J. Xiao, D.-Y. Chen, {Analysis of the hidden bottom decays of $Z_b(10610)$
  and $Z_b(10650)$ via final state interaction}, Phys. Rev. D 96 (2017) 014035.
\newblock \href {https://doi.org/10.1103/PhysRevD.96.014035}
  {\path{doi:10.1103/PhysRevD.96.014035}}.

\bibitem{Wu:2022hck}
Q.~Wu, Y.~Zheng, S.~Liu, G.~Li, {Investigations on the light hadron decays of
  $Z_b(10610)$ and $Z_b(10650)$}, Phys. Rev. D 107 (2023) 034028.
\newblock \href {http://arxiv.org/abs/2211.02896} {\path{arXiv:2211.02896}},
  \href {https://doi.org/10.1103/PhysRevD.107.034028}
  {\path{doi:10.1103/PhysRevD.107.034028}}.

\bibitem{Wang:2013zra}
Z.-G. Wang, T.~Huang, {The $Z_b(10610)$ and $Z_b(10650)$ as axial-vector
  tetraquark states in the QCD sum rules}, Nucl. Phys. A 930 (2014) 63--85.
\newblock \href {http://arxiv.org/abs/1312.2652} {\path{arXiv:1312.2652}},
  \href {https://doi.org/10.1016/j.nuclphysa.2014.08.084}
  {\path{doi:10.1016/j.nuclphysa.2014.08.084}}.

\bibitem{Wang:2019mxn}
Z.-G. Wang, {Analysis of the hidden-bottom tetraquark mass spectrum with the
  QCD sum rules}, Eur. Phys. J. C 79 (2019) 489.
\newblock \href {http://arxiv.org/abs/1903.10895} {\path{arXiv:1903.10895}},
  \href {https://doi.org/10.1140/epjc/s10052-019-7019-6}
  {\path{doi:10.1140/epjc/s10052-019-7019-6}}.

\bibitem{Ke:2012gm}
H.-W. Ke, X.-Q. Li, Y.-L. Shi, G.-L. Wang, X.-H. Yuan, {Is $Z_b(10610)$ a
  molecular state?}, JHEP 04 (2012) 056.
\newblock \href {http://arxiv.org/abs/1202.2178} {\path{arXiv:1202.2178}},
  \href {https://doi.org/10.1007/JHEP04(2012)056}
  {\path{doi:10.1007/JHEP04(2012)056}}.

\bibitem{Wang:2013daa}
Z.-G. Wang, T.~Huang, {Possible assignments of the $X(3872)$, $Z_c(3900)$ and
  $Z_b(10610)$ as axial-vector molecular states}, Eur. Phys. J. C 74 (2014)
  2891.
\newblock \href {http://arxiv.org/abs/1312.7489} {\path{arXiv:1312.7489}},
  \href {https://doi.org/10.1140/epjc/s10052-014-2891-6}
  {\path{doi:10.1140/epjc/s10052-014-2891-6}}.

\bibitem{Gupta:2012gba}
V.~Gupta, G.~Sanchez-Colon, S.~Rajpoot, {Sum rules for tetraquark decay
  coupling constants with broken SU(3) symmetry}, Mod. Phys. Lett. A 27 (2012)
  1250165.
\newblock \href {http://arxiv.org/abs/1304.1177} {\path{arXiv:1304.1177}},
  \href {https://doi.org/10.1142/S0217732312501659}
  {\path{doi:10.1142/S0217732312501659}}.

\bibitem{Ali:2011ug}
A.~Ali, C.~Hambrock, W.~Wang, {Tetraquark interpretation of the charged
  bottomonium-like states $Z_b^{+-}(10610)$ and $Z_b^{+-}(10650)$ and
  implications}, Phys. Rev. D 85 (2012) 054011.
\newblock \href {http://arxiv.org/abs/1110.1333} {\path{arXiv:1110.1333}},
  \href {https://doi.org/10.1103/PhysRevD.85.054011}
  {\path{doi:10.1103/PhysRevD.85.054011}}.

\bibitem{Bugg:2011jr}
D.~V. Bugg, {An Explanation of Belle states $Z_b(10610)$ and $Z_b(10650)$}, EPL
  96 (2011) 11002.
\newblock \href {http://arxiv.org/abs/1105.5492} {\path{arXiv:1105.5492}},
  \href {https://doi.org/10.1209/0295-5075/96/11002}
  {\path{doi:10.1209/0295-5075/96/11002}}.

\bibitem{Swanson:2014tra}
E.~S. Swanson, {$Z_b$ and $Z_c$ exotic states as coupled channel cusps}, Phys.
  Rev. D 91 (2015) 034009.
\newblock \href {http://arxiv.org/abs/1409.3291} {\path{arXiv:1409.3291}},
  \href {https://doi.org/10.1103/PhysRevD.91.034009}
  {\path{doi:10.1103/PhysRevD.91.034009}}.

\bibitem{Chen:2011pu}
D.-Y. Chen, X.~Liu, T.~Matsuki, {Charged bottomonium-like structures in the
  hidden-bottom dipion decays of $\Upsilon(11020)$}, Phys. Rev. D 84 (2011)
  074032.
\newblock \href {http://arxiv.org/abs/1108.4458} {\path{arXiv:1108.4458}},
  \href {https://doi.org/10.1103/PhysRevD.84.074032}
  {\path{doi:10.1103/PhysRevD.84.074032}}.

\bibitem{Chen:2012yr}
D.-Y. Chen, X.~Liu, T.~Matsuki, {Interpretation of $Z_b$(10610) and
  $Z_b$(10650) in the ISPE mechanism and the charmonium counterpart}, Chin.
  Phys. C 38 (2014) 053102.
\newblock \href {http://arxiv.org/abs/1208.2411} {\path{arXiv:1208.2411}},
  \href {https://doi.org/10.1088/1674-1137/38/5/053102}
  {\path{doi:10.1088/1674-1137/38/5/053102}}.

\bibitem{CLEO:2011aa}
T.~K. Pedlar, et~al., CLEO Collaboration, {Observation of the $h_c(1P)$ using
  $e^+e^-$ collisions above $D\bar{D}$ threshold}, Phys. Rev. Lett. 107 (2011)
  041803.
\newblock \href {http://arxiv.org/abs/1104.2025} {\path{arXiv:1104.2025}},
  \href {https://doi.org/10.1103/PhysRevLett.107.041803}
  {\path{doi:10.1103/PhysRevLett.107.041803}}.

\bibitem{Yuan:2021talk}
C.-Z. Yuan, {Talk given in Hadron Structure and Interactions in 2011, held by
  Research Center for Nuclear Physics of Osaka University} (2011).

\bibitem{Chen:2013bha}
D.-Y. Chen, X.~Liu, T.~Matsuki, {Novel charged charmoniumlike structures in the
  hidden-charm dipion decays of $Y(4360)$}, Phys. Rev. D 88 (2013) 014034.
\newblock \href {http://arxiv.org/abs/1306.2080} {\path{arXiv:1306.2080}},
  \href {https://doi.org/10.1103/PhysRevD.88.014034}
  {\path{doi:10.1103/PhysRevD.88.014034}}.

\bibitem{Maiani:2004vq}
L.~Maiani, F.~Piccinini, A.~D. Polosa, V.~Riquer, {Diquark-antidiquarks with
  hidden or open charm and the nature of X(3872)}, Phys. Rev. D 71 (2005)
  014028.
\newblock \href {http://arxiv.org/abs/hep-ph/0412098}
  {\path{arXiv:hep-ph/0412098}}, \href
  {https://doi.org/10.1103/PhysRevD.71.014028}
  {\path{doi:10.1103/PhysRevD.71.014028}}.

\bibitem{BaBar:2012vyb}
J.~P. Lees, et~al., BaBar Collaboration, {Study of the reaction $e^{+}e^{-} \to
  J/\psi\pi^{+}\pi^{-}$ via initial-state radiation at BaBar}, Phys. Rev. D 86
  (2012) 051102.
\newblock \href {http://arxiv.org/abs/1204.2158} {\path{arXiv:1204.2158}},
  \href {https://doi.org/10.1103/PhysRevD.86.051102}
  {\path{doi:10.1103/PhysRevD.86.051102}}.

\bibitem{Belle:2014nuw}
K.~Chilikin, et~al., Belle Collaboration, {Observation of a new charged
  charmoniumlike state in $\bar{B}^0\to J/\psi K^-\pi^+$ decays}, Phys. Rev. D
  90 (2014) 112009.
\newblock \href {http://arxiv.org/abs/1408.6457} {\path{arXiv:1408.6457}},
  \href {https://doi.org/10.1103/PhysRevD.90.112009}
  {\path{doi:10.1103/PhysRevD.90.112009}}.

\bibitem{D0:2018wyb}
V.~M. Abazov, et~al., D0 Collaboration, {Evidence for $Z_c^{\pm}(3900)$ in
  semi-inclusive decays of $b$-flavored hadrons}, Phys. Rev. D 98 (2018)
  052010.
\newblock \href {http://arxiv.org/abs/1807.00183} {\path{arXiv:1807.00183}},
  \href {https://doi.org/10.1103/PhysRevD.98.052010}
  {\path{doi:10.1103/PhysRevD.98.052010}}.

\bibitem{Chen:2013coa}
D.-Y. Chen, X.~Liu, T.~Matsuki, {Reproducing the $Z_c(3900)$ structure through
  the initial-single-pion-emission mechanism}, Phys. Rev. D 88 (2013) 036008.
\newblock \href {http://arxiv.org/abs/1304.5845} {\path{arXiv:1304.5845}},
  \href {https://doi.org/10.1103/PhysRevD.88.036008}
  {\path{doi:10.1103/PhysRevD.88.036008}}.

\bibitem{BaBar:2012hpr}
J.~P. Lees, et~al., BaBar Collaboration, {Study of the reaction $e^{+}e^{-}\to
  \psi(2S)\pi^{+}\pi^{-}$ via initial-state radiation at BaBar}, Phys. Rev. D
  89 (2014) 111103.
\newblock \href {http://arxiv.org/abs/1211.6271} {\path{arXiv:1211.6271}},
  \href {https://doi.org/10.1103/PhysRevD.89.111103}
  {\path{doi:10.1103/PhysRevD.89.111103}}.

\bibitem{Huang:2019agb}
Q.~Huang, D.-Y. Chen, X.~Liu, T.~Matsuki, {Charged charmoniumlike structures in
  the $e^+ e^- \to \psi (3686) \pi ^+ \pi ^-$ process based on the ISPE
  mechanism}, Eur. Phys. J. C 79 (2019) 613.
\newblock \href {http://arxiv.org/abs/1905.05650} {\path{arXiv:1905.05650}},
  \href {https://doi.org/10.1140/epjc/s10052-019-7121-9}
  {\path{doi:10.1140/epjc/s10052-019-7121-9}}.

\bibitem{Wang:2022fdu}
Z.-G. Wang, {Strange cousin of Z $_{c}$(4020/4025) as a tetraquark state*},
  Chin. Phys. C 46 (2022) 123106.
\newblock \href {http://arxiv.org/abs/2207.00947} {\path{arXiv:2207.00947}},
  \href {https://doi.org/10.1088/1674-1137/ac8c21}
  {\path{doi:10.1088/1674-1137/ac8c21}}.

\bibitem{Wang:2013exa}
Z.-G. Wang, {Analysis of the $Z_c(4020)$, $Z_c(4025)$, $Y(4360)$ and $Y(4660)$
  as vector tetraquark states with QCD sum rules}, Eur. Phys. J. C 74 (2014)
  2874.
\newblock \href {http://arxiv.org/abs/1311.1046} {\path{arXiv:1311.1046}},
  \href {https://doi.org/10.1140/epjc/s10052-014-2874-7}
  {\path{doi:10.1140/epjc/s10052-014-2874-7}}.

\bibitem{Wang:2013vex}
Z.-G. Wang, T.~Huang, {Analysis of the $X(3872)$, $Z_c(3900)$ and $Z_c(3885)$
  as axial-vector tetraquark states with QCD sum rules}, Phys. Rev. D 89 (2014)
  054019.
\newblock \href {http://arxiv.org/abs/1310.2422} {\path{arXiv:1310.2422}},
  \href {https://doi.org/10.1103/PhysRevD.89.054019}
  {\path{doi:10.1103/PhysRevD.89.054019}}.

\bibitem{Wang:2019tlw}
Z.-G. Wang, {Analysis of the hidden-charm tetraquark mass spectrum with the QCD
  sum rules}, Phys. Rev. D 102 (2020) 014018.
\newblock \href {http://arxiv.org/abs/1908.07914} {\path{arXiv:1908.07914}},
  \href {https://doi.org/10.1103/PhysRevD.102.014018}
  {\path{doi:10.1103/PhysRevD.102.014018}}.

\bibitem{Wang:2019hnw}
Z.-G. Wang, {Axialvector tetraquark candidates for $Z_c(3900)$, $Z_c(4020)$,
  $Z_c(4430)$, $Z_c(4600)$}, Chin. Phys. C 44 (2020) 063105.
\newblock \href {http://arxiv.org/abs/1901.10741} {\path{arXiv:1901.10741}},
  \href {https://doi.org/10.1088/1674-1137/44/6/063105}
  {\path{doi:10.1088/1674-1137/44/6/063105}}.

\bibitem{Qiao:2013dda}
C.-F. Qiao, L.~Tang, {Interpretation of $Z_c(4025)$ as the hidden charm
  tetraquark states via QCD Sum Rules}, Eur. Phys. J. C 74 (2014) 2810.
\newblock \href {http://arxiv.org/abs/1308.3439} {\path{arXiv:1308.3439}},
  \href {https://doi.org/10.1140/epjc/s10052-014-2810-x}
  {\path{doi:10.1140/epjc/s10052-014-2810-x}}.

\bibitem{Dias:2013xfa}
J.~M. Dias, F.~S. Navarra, M.~Nielsen, C.~M. Zanetti, {$Z^+_c$(3900) decay
  width in QCD sum rules}, Phys. Rev. D 88 (2013) 016004.
\newblock \href {http://arxiv.org/abs/1304.6433} {\path{arXiv:1304.6433}},
  \href {https://doi.org/10.1103/PhysRevD.88.016004}
  {\path{doi:10.1103/PhysRevD.88.016004}}.

\bibitem{Braaten:2013boa}
E.~Braaten, {How the $Z_c$(3900) Reveals the Spectra of Quarkonium Hybrid and
  Tetraquark Mesons}, Phys. Rev. Lett. 111 (2013) 162003.
\newblock \href {http://arxiv.org/abs/1305.6905} {\path{arXiv:1305.6905}},
  \href {https://doi.org/10.1103/PhysRevLett.111.162003}
  {\path{doi:10.1103/PhysRevLett.111.162003}}.

\bibitem{Maiani:2013nmn}
L.~Maiani, V.~Riquer, R.~Faccini, F.~Piccinini, A.~Pilloni, A.~D. Polosa, {A
  $J^{PG}=1^{++}$ Charged Resonance in the $Y(4260) \to \pi^+ \pi^- J/\psi$
  Decay?}, Phys. Rev. D 87 (2013) 111102.
\newblock \href {http://arxiv.org/abs/1303.6857} {\path{arXiv:1303.6857}},
  \href {https://doi.org/10.1103/PhysRevD.87.111102}
  {\path{doi:10.1103/PhysRevD.87.111102}}.

\bibitem{Wang:2020dgr}
Z.-G. Wang, {Analysis of the Hidden-charm Tetraquark molecule mass spectrum
  with the QCD sum rules}, Int. J. Mod. Phys. A 36 (2021) 2150107.
\newblock \href {http://arxiv.org/abs/2012.11869} {\path{arXiv:2012.11869}},
  \href {https://doi.org/10.1142/S0217751X21501074}
  {\path{doi:10.1142/S0217751X21501074}}.

\bibitem{Cui:2013xla}
C.-Y. Cui, Y.-L. Liu, M.-Q. Huang, {Could $Z_c$(4025) be a $J^P$ = $1^+ D^*
  \bar{D^*}$ molecular state?}, Eur. Phys. J. C 73 (2013) 2661.
\newblock \href {http://arxiv.org/abs/1308.3625} {\path{arXiv:1308.3625}},
  \href {https://doi.org/10.1140/epjc/s10052-013-2661-x}
  {\path{doi:10.1140/epjc/s10052-013-2661-x}}.

\bibitem{Zhang:2013aoa}
J.-R. Zhang, {Improved QCD sum rule study of $Z_{c}(3900)$ as a $\bar{D}D^{*}$
  molecular state}, Phys. Rev. D 87 (2013) 116004.
\newblock \href {http://arxiv.org/abs/1304.5748} {\path{arXiv:1304.5748}},
  \href {https://doi.org/10.1103/PhysRevD.87.116004}
  {\path{doi:10.1103/PhysRevD.87.116004}}.

\bibitem{Cui:2013yva}
C.-Y. Cui, Y.-L. Liu, W.-B. Chen, M.-Q. Huang, {Could $Z_{c}(3900)$ be a
  $I^{G}J^{P}=1^{+}1^{+}$ $D^{*}\bar{D}$ molecular state?}, J. Phys. G 41
  (2014) 075003.
\newblock \href {http://arxiv.org/abs/1304.1850} {\path{arXiv:1304.1850}},
  \href {https://doi.org/10.1088/0954-3899/41/7/075003}
  {\path{doi:10.1088/0954-3899/41/7/075003}}.

\bibitem{Chen:2013omd}
W.~Chen, T.~G. Steele, M.-L. Du, S.-L. Zhu, {$D^*\bar D^*$ molecule
  interpretation of $Z_c(4025)$}, Eur. Phys. J. C 74 (2014) 2773.
\newblock \href {http://arxiv.org/abs/1308.5060} {\path{arXiv:1308.5060}},
  \href {https://doi.org/10.1140/epjc/s10052-014-2773-y}
  {\path{doi:10.1140/epjc/s10052-014-2773-y}}.

\bibitem{Khemchandani:2013iwa}
K.~P. Khemchandani, A.~Martinez~Torres, M.~Nielsen, F.~S. Navarra, {Relating
  $D^* \bar{D}^*$ currents with $J^\pi= 0^+,1^+$ and $2^+$ to $Z_c$ states},
  Phys. Rev. D 89 (2014) 014029.
\newblock \href {http://arxiv.org/abs/1310.0862} {\path{arXiv:1310.0862}},
  \href {https://doi.org/10.1103/PhysRevD.89.014029}
  {\path{doi:10.1103/PhysRevD.89.014029}}.

\bibitem{Guo:2013sya}
F.-K. Guo, C.~Hidalgo-Duque, J.~Nieves, M.~P. Valderrama, {Consequences of
  heavy quark symmetries for hadronic molecules}, Phys. Rev. D 88 (2013)
  054007.
\newblock \href {http://arxiv.org/abs/1303.6608} {\path{arXiv:1303.6608}},
  \href {https://doi.org/10.1103/PhysRevD.88.054007}
  {\path{doi:10.1103/PhysRevD.88.054007}}.

\bibitem{Ji:2022uie}
T.~Ji, X.-K. Dong, M.~Albaladejo, M.-L. Du, F.-K. Guo, J.~Nieves, {Establishing
  the heavy quark spin and light flavor molecular multiplets of the $X(3872)$,
  $Z_c(3900)$, and $X(3960)$}, Phys. Rev. D 106 (2022) 094002.
\newblock \href {http://arxiv.org/abs/2207.08563} {\path{arXiv:2207.08563}},
  \href {https://doi.org/10.1103/PhysRevD.106.094002}
  {\path{doi:10.1103/PhysRevD.106.094002}}.

\bibitem{He:2013nwa}
J.~He, X.~Liu, Z.-F. Sun, S.-L. Zhu, {$Z_c(4025)$ as the hadronic molecule with
  hidden charm}, Eur. Phys. J. C 73 (2013) 2635.
\newblock \href {http://arxiv.org/abs/1308.2999} {\path{arXiv:1308.2999}},
  \href {https://doi.org/10.1140/epjc/s10052-013-2635-z}
  {\path{doi:10.1140/epjc/s10052-013-2635-z}}.

\bibitem{He:2017lhy}
J.~He, D.-Y. Chen, {$Z_c(3900)/Z_c(3885)$ as a virtual state from $\pi
  J/\psi-\bar{D}^*D$ interaction}, Eur. Phys. J. C 78 (2018) 94.
\newblock \href {http://arxiv.org/abs/1712.05653} {\path{arXiv:1712.05653}},
  \href {https://doi.org/10.1140/epjc/s10052-018-5580-z}
  {\path{doi:10.1140/epjc/s10052-018-5580-z}}.

\bibitem{Albaladejo:2015lob}
M.~Albaladejo, F.-K. Guo, C.~Hidalgo-Duque, J.~Nieves, {$Z_c(3900)$: What has
  been really seen?}, Phys. Lett. B 755 (2016) 337--342.
\newblock \href {http://arxiv.org/abs/1512.03638} {\path{arXiv:1512.03638}},
  \href {https://doi.org/10.1016/j.physletb.2016.02.025}
  {\path{doi:10.1016/j.physletb.2016.02.025}}.

\bibitem{Du:2022jjv}
M.-L. Du, M.~Albaladejo, F.-K. Guo, J.~Nieves, {Combined analysis of the
  $Z_c(3900)$ and the $Z_{cs}(3985)$ exotic states}, Phys. Rev. D 105 (2022)
  074018.
\newblock \href {http://arxiv.org/abs/2201.08253} {\path{arXiv:2201.08253}},
  \href {https://doi.org/10.1103/PhysRevD.105.074018}
  {\path{doi:10.1103/PhysRevD.105.074018}}.

\bibitem{Wilbring:2013cha}
E.~Wilbring, H.~W. Hammer, U.~G. Mei{\ss}ner, {Electromagnetic structure of the
  $Z_c(3900)$}, Phys. Lett. B 726 (2013) 326--329.
\newblock \href {http://arxiv.org/abs/1304.2882} {\path{arXiv:1304.2882}},
  \href {https://doi.org/10.1016/j.physletb.2013.08.059}
  {\path{doi:10.1016/j.physletb.2013.08.059}}.

\bibitem{Ke:2013gia}
H.-W. Ke, Z.-T. Wei, X.-Q. Li, {Is $Z_c(3900)$ a molecular state}, Eur. Phys.
  J. C 73 (2013) 2561.
\newblock \href {http://arxiv.org/abs/1307.2414} {\path{arXiv:1307.2414}},
  \href {https://doi.org/10.1140/epjc/s10052-013-2561-0}
  {\path{doi:10.1140/epjc/s10052-013-2561-0}}.

\bibitem{Dong:2013iqa}
Y.~Dong, A.~Faessler, T.~Gutsche, V.~E. Lyubovitskij, {Strong decays of
  molecular states Z$_{c}^{+}$ and Z$_{c}^{'+}$}, Phys. Rev. D 88 (2013)
  014030.
\newblock \href {http://arxiv.org/abs/1306.0824} {\path{arXiv:1306.0824}},
  \href {https://doi.org/10.1103/PhysRevD.88.014030}
  {\path{doi:10.1103/PhysRevD.88.014030}}.

\bibitem{Xiao:2018kfx}
C.-J. Xiao, D.-Y. Chen, Y.-B. Dong, W.~Zuo, T.~Matsuki, {Understanding the
  $\eta_c\rho$ decay mode of $Z_c^{(\prime)}$ via the triangle loop mechanism},
  Phys. Rev. D 99 (2019) 074003.
\newblock \href {http://arxiv.org/abs/1811.04688} {\path{arXiv:1811.04688}},
  \href {https://doi.org/10.1103/PhysRevD.99.074003}
  {\path{doi:10.1103/PhysRevD.99.074003}}.

\bibitem{Wang:2022aiu}
X.-Y. Wang, G.~Li, C.-S. An, J.-J. Xie, {Radiative decays of the neutral
  $Z_c(3900)$ and $Z_c(4020)$}, Phys. Rev. D 106 (2022) 074026.
\newblock \href {http://arxiv.org/abs/2210.06783} {\path{arXiv:2210.06783}},
  \href {https://doi.org/10.1103/PhysRevD.106.074026}
  {\path{doi:10.1103/PhysRevD.106.074026}}.

\bibitem{Chen:2015igx}
D.-Y. Chen, Y.-B. Dong, {Radiative decays of the neutral $Z_c(3900)$}, Phys.
  Rev. D 93 (2016) 014003.
\newblock \href {http://arxiv.org/abs/1510.00829} {\path{arXiv:1510.00829}},
  \href {https://doi.org/10.1103/PhysRevD.93.014003}
  {\path{doi:10.1103/PhysRevD.93.014003}}.

\bibitem{Wu:2023rrp}
Q.~Wu, M.-Z. Liu, L.-S. Geng, {Productions of $X(3872)$, $Z_c(3900)$,
  $X_2(4013)$, and $Z_c(4020)$ in $B_{(s)}$ decays offer strong clues on their
  molecular nature}, Eur. Phys. J. C 84 (2024) 147.
\newblock \href {http://arxiv.org/abs/2304.05269} {\path{arXiv:2304.05269}},
  \href {https://doi.org/10.1140/epjc/s10052-024-12501-6}
  {\path{doi:10.1140/epjc/s10052-024-12501-6}}.

\bibitem{Qi:2023kwc}
X.-Y. Qi, Q.~Wu, D.-Y. Chen, {Pionic transitions from $Z_c(4020)$ to $D$ wave
  charmonia}, Eur. Phys. J. C 83 (2023) 1006.
\newblock \href {http://arxiv.org/abs/2302.10050} {\path{arXiv:2302.10050}},
  \href {https://doi.org/10.1140/epjc/s10052-023-12151-0}
  {\path{doi:10.1140/epjc/s10052-023-12151-0}}.

\bibitem{Wu:2019vbk}
Q.~Wu, D.-Y. Chen, X.-J. Fan, G.~Li, {Production of $Z_c(3900)$ and $Z_c(4020)$
  in $B_c$ decay}, Eur. Phys. J. C 79 (2019) 265.
\newblock \href {http://arxiv.org/abs/1902.05737} {\path{arXiv:1902.05737}},
  \href {https://doi.org/10.1140/epjc/s10052-019-6784-6}
  {\path{doi:10.1140/epjc/s10052-019-6784-6}}.

\bibitem{Liu:2024ziu}
M.-Z. Liu, X.-Z. Ling, L.-S. Geng, {Productions of $X(3872)/Z_c(3900)$ and
  $X_2(4013)/Z_c(4020)$ in $Y(4220)$ and $Y(4360)$ decays}, Phys. Rev. D 110
  (2024) 054035.
\newblock \href {http://arxiv.org/abs/2404.07681} {\path{arXiv:2404.07681}},
  \href {https://doi.org/10.1103/PhysRevD.110.054035}
  {\path{doi:10.1103/PhysRevD.110.054035}}.

\bibitem{Wang:2013hga}
Q.~Wang, C.~Hanhart, Q.~Zhao, {Systematic study of the singularity mechanism in
  heavy quarkonium decays}, Phys. Lett. B 725 (2013) 106--110.
\newblock \href {http://arxiv.org/abs/1305.1997} {\path{arXiv:1305.1997}},
  \href {https://doi.org/10.1016/j.physletb.2013.06.049}
  {\path{doi:10.1016/j.physletb.2013.06.049}}.

\bibitem{Liu:2013vfa}
X.-H. Liu, G.~Li, {Exploring the threshold behavior and implications on the
  nature of $Y(4260)$ and $Z_c(3900)$}, Phys. Rev. D 88 (2013) 014013.
\newblock \href {http://arxiv.org/abs/1306.1384} {\path{arXiv:1306.1384}},
  \href {https://doi.org/10.1103/PhysRevD.88.014013}
  {\path{doi:10.1103/PhysRevD.88.014013}}.

\bibitem{Yu:2024sqv}
K.~Yu, G.-J. Wang, J.-J. Wu, Z.~Yang, {Three-coupled-channel analysis of
  $Z_c(3900)$ involving $D\bar{D}^\ast$, $\pi J/\psi$, and $\rho \eta_c$},
  Phys. Rev. D 110 (2024) 114029.
\newblock \href {http://arxiv.org/abs/2409.10865} {\path{arXiv:2409.10865}},
  \href {https://doi.org/10.1103/PhysRevD.110.114029}
  {\path{doi:10.1103/PhysRevD.110.114029}}.

\bibitem{Chen:2023def}
Y.-H. Chen, M.-L. Du, F.-K. Guo, {Precise determination of the pole position of
  the exotic Z$_{c}$(3900)}, Sci. China Phys. Mech. Astron. 67 (2024) 291011.
\newblock \href {http://arxiv.org/abs/2310.15965} {\path{arXiv:2310.15965}},
  \href {https://doi.org/10.1007/s11433-023-2408-1}
  {\path{doi:10.1007/s11433-023-2408-1}}.

\bibitem{Pilloni:2016obd}
A.~Pilloni, C.~Fernandez-Ramirez, A.~Jackura, V.~Mathieu, M.~Mikhasenko,
  J.~Nys, A.~P. Szczepaniak, JPAC Collaboration, {Amplitude analysis and the
  nature of the Z$_c$(3900)}, Phys. Lett. B 772 (2017) 200--209.
\newblock \href {http://arxiv.org/abs/1612.06490} {\path{arXiv:1612.06490}},
  \href {https://doi.org/10.1016/j.physletb.2017.06.030}
  {\path{doi:10.1016/j.physletb.2017.06.030}}.

\bibitem{Zhang:2025fcv}
Y.~Zhang, A.~Hosaka, Q.~Wang, S.~Yasui, {$Z_c(3900)$ in a hadronic molecule and
  a triangle singularity approach at finite temperature}, Phys. Rev. D 112
  (2025) 016014.
\newblock \href {http://arxiv.org/abs/2503.19374} {\path{arXiv:2503.19374}},
  \href {https://doi.org/10.1103/5nlp-vwrc} {\path{doi:10.1103/5nlp-vwrc}}.

\bibitem{Ermolina:2024uln}
V.~Ermolina, I.~Danilkin, M.~Vanderhaeghen, {Dalitz-plot decomposition for the
  $e^+ e^- \to J/\psi \pi \pi (KK)$ and $e^+ e^- \to \pi \pi h_c$ processes},
  Phys. Lett. B 864 (2025) 139450.
\newblock \href {http://arxiv.org/abs/2410.19946} {\path{arXiv:2410.19946}},
  \href {https://doi.org/10.1016/j.physletb.2025.139450}
  {\path{doi:10.1016/j.physletb.2025.139450}}.

\bibitem{Danilkin:2020kce}
I.~Danilkin, D.~A.~S. Molnar, M.~Vanderhaeghen, {Simultaneous description of
  the $e^+e^- \to J/ \psi \, \pi \pi\, (K \bar{K})$ processes}, Phys. Rev. D
  102~(1) (2020) 016019.
\newblock \href {http://arxiv.org/abs/2004.13499} {\path{arXiv:2004.13499}},
  \href {https://doi.org/10.1103/PhysRevD.102.016019}
  {\path{doi:10.1103/PhysRevD.102.016019}}.

\bibitem{Prelovsek:2013xba}
S.~Prelovsek, L.~Leskovec, {Search for $Z^{+}_{c}(3900)$ in the $1^{+-}$
  channel on the lattice}, Phys. Lett. B 727 (2013) 172--176.
\newblock \href {http://arxiv.org/abs/1308.2097} {\path{arXiv:1308.2097}},
  \href {https://doi.org/10.1016/j.physletb.2013.10.009}
  {\path{doi:10.1016/j.physletb.2013.10.009}}.

\bibitem{Prelovsek:2014swa}
S.~Prelovsek, C.~B. Lang, L.~Leskovec, D.~Mohler, {Study of the $Z_c^+$ channel
  using lattice QCD}, Phys. Rev. D 91 (2015) 014504.
\newblock \href {http://arxiv.org/abs/1405.7623} {\path{arXiv:1405.7623}},
  \href {https://doi.org/10.1103/PhysRevD.91.014504}
  {\path{doi:10.1103/PhysRevD.91.014504}}.

\bibitem{HALQCD:2016ofq}
Y.~Ikeda, S.~Aoki, T.~Doi, S.~Gongyo, T.~Hatsuda, T.~Inoue, T.~Iritani,
  N.~Ishii, K.~Murano, K.~Sasaki, HAL QCD Collaboration, {Fate of the
  tetraquark candidate $Z_c$(3900) from lattice QCD}, Phys. Rev. Lett. 117
  (2016) 242001.
\newblock \href {http://arxiv.org/abs/1602.03465} {\path{arXiv:1602.03465}},
  \href {https://doi.org/10.1103/PhysRevLett.117.242001}
  {\path{doi:10.1103/PhysRevLett.117.242001}}.

\bibitem{Ikeda:2017mee}
Y.~Ikeda, HAL QCD Collaboration, {The tetraquark candidate $Z_c$(3900) from
  dynamical lattice QCD simulations}, J. Phys. G 45 (2018) 024002.
\newblock \href {http://arxiv.org/abs/1706.07300} {\path{arXiv:1706.07300}},
  \href {https://doi.org/10.1088/1361-6471/aa9afd}
  {\path{doi:10.1088/1361-6471/aa9afd}}.

\bibitem{CLQCD:2019npr}
T.~Chen, Y.~Chen, M.~Gong, C.~Liu, L.~Liu, Y.-B. Liu, Z.~Liu, J.-P. Ma,
  M.~Werner, J.-B. Zhang, CLQCD Collaboration, {A coupled-channel lattice study
  on the resonance-like structure $Z_c(3900)$}, Chin. Phys. C 43 (2019) 103103.
\newblock \href {http://arxiv.org/abs/1907.03371} {\path{arXiv:1907.03371}},
  \href {https://doi.org/10.1088/1674-1137/43/10/103103}
  {\path{doi:10.1088/1674-1137/43/10/103103}}.

\bibitem{Albaladejo:2016jsg}
M.~Albaladejo, P.~Fernandez-Soler, J.~Nieves, {$Z_c(3900)$: Confronting theory
  and lattice simulations}, Eur. Phys. J. C 76~(10) (2016) 573.
\newblock \href {http://arxiv.org/abs/1606.03008} {\path{arXiv:1606.03008}},
  \href {https://doi.org/10.1140/epjc/s10052-016-4427-8}
  {\path{doi:10.1140/epjc/s10052-016-4427-8}}.

\bibitem{Yan:2023bwt}
L.-W. Yan, Z.-H. Guo, F.-K. Guo, D.-L. Yao, Z.-Y. Zhou, {Reconciling
  experimental and lattice data of Zc(3900) in a
  J/{\ensuremath{\psi}}{\ensuremath{\pi}}-DD{\textasciimacron}* coupled-channel
  analysis}, Phys. Rev. D 109 (2024) 014026.
\newblock \href {http://arxiv.org/abs/2307.12283} {\path{arXiv:2307.12283}},
  \href {https://doi.org/10.1103/PhysRevD.109.014026}
  {\path{doi:10.1103/PhysRevD.109.014026}}.

\bibitem{Cheung:2017tnt}
G.~K.~C. Cheung, C.~E. Thomas, J.~J. Dudek, R.~G. Edwards, Hadron Spectrum
  Collaboration, {Tetraquark operators in lattice QCD and exotic flavour states
  in the charm sector}, JHEP 11 (2017) 033.
\newblock \href {http://arxiv.org/abs/1709.01417} {\path{arXiv:1709.01417}},
  \href {https://doi.org/10.1007/JHEP11(2017)033}
  {\path{doi:10.1007/JHEP11(2017)033}}.

\bibitem{Chen:2013axa}
D.-Y. Chen, X.~Liu, T.~Matsuki, {Prediction of isoscalar charmoniumlike
  structures in the hidden-charm di-eta decays of higher charmonia}, J. Phys. G
  42 (2015) 015002.
\newblock \href {http://arxiv.org/abs/1309.4528} {\path{arXiv:1309.4528}},
  \href {https://doi.org/10.1088/0954-3899/42/1/015002}
  {\path{doi:10.1088/0954-3899/42/1/015002}}.

\bibitem{BESIII:2026kja}
M.~Ablikim, et~al., BESIII Collaboration, {Search for the reaction channel $e^+
  e^- \to \eta \eta \,J/\psi$ and the isospin partner of the $Z_c(3900)$ at
  center-of-mass energies $\sqrt{s} = 4.226-4.950$ GeV} (1 2026).
\newblock \href {http://arxiv.org/abs/2601.15882} {\path{arXiv:2601.15882}}.

\bibitem{Lee:2008uy}
S.~H. Lee, M.~Nielsen, U.~Wiedner, {$D_sD^*$ molecule as an axial meson}, J.
  Korean Phys. Soc. 55 (2009) 424.
\newblock \href {http://arxiv.org/abs/0803.1168} {\path{arXiv:0803.1168}},
  \href {https://doi.org/10.3938/jkps.55.424} {\path{doi:10.3938/jkps.55.424}}.

\bibitem{Dias:2013qga}
J.~M. Dias, X.~Liu, M.~Nielsen, {Predicition for the decay width of a charged
  state near the $D_s\bar{D}^*/D^*_s\bar{D}$ threshold}, Phys. Rev. D 88 (2013)
  096014.
\newblock \href {http://arxiv.org/abs/1307.7100} {\path{arXiv:1307.7100}},
  \href {https://doi.org/10.1103/PhysRevD.88.096014}
  {\path{doi:10.1103/PhysRevD.88.096014}}.

\bibitem{Belle:2014fgf}
C.~P. Shen, et~al., Belle Collaboration, {Updated cross section measurement of
  $e^+ e^- \to K^+ K^- J/\psi$ and $K_S^0K_S^0J/\psi$ via initial state
  radiation at Belle}, Phys. Rev. D 89 (2014) 072015.
\newblock \href {http://arxiv.org/abs/1402.6578} {\path{arXiv:1402.6578}},
  \href {https://doi.org/10.1103/PhysRevD.89.072015}
  {\path{doi:10.1103/PhysRevD.89.072015}}.

\bibitem{BESIII:2020qkh}
M.~Ablikim, et~al., BESIII Collaboration, {Observation of a near-threshold
  structure in the $K^+$ recoil-mass spectra in $e^+e^- \rightarrow
  K^+(D_s^-D^{*0}+D_s^{*-}D^0$)}, Phys. Rev. Lett. 126 (2021) 102001.
\newblock \href {http://arxiv.org/abs/2011.07855} {\path{arXiv:2011.07855}},
  \href {https://doi.org/10.1103/PhysRevLett.126.102001}
  {\path{doi:10.1103/PhysRevLett.126.102001}}.

\bibitem{BESIII:2022qzr}
M.~Ablikim, et~al., BESIII Collaboration, {Evidence for a neutral
  near-threshold structure in the $K^{0}_{S}$ recoil-mass spectra in $e^+e^-\to
  K^{0}_{S}D_s^+D^{*-}$ and $e^+e^-\to K^{0}_{S}D_s^{*+}D^{-}$}, Phys. Rev.
  Lett. 129 (2022) 112003.
\newblock \href {http://arxiv.org/abs/2204.13703} {\path{arXiv:2204.13703}},
  \href {https://doi.org/10.1103/PhysRevLett.129.112003}
  {\path{doi:10.1103/PhysRevLett.129.112003}}.

\bibitem{BESIII:2023wqy}
M.~Ablikim, et~al., BESIII Collaboration, {Observation of a vector
  charmoniumlike state at 4.7 $\mathrm{GeV/c}^2$ and search for $Z_{cs}$ in
  $e^+e^-\to K^+K^-J/\psi$}, Phys. Rev. Lett. 131 (2023) 211902.
\newblock \href {http://arxiv.org/abs/2308.15362} {\path{arXiv:2308.15362}},
  \href {https://doi.org/10.1103/PhysRevLett.131.211902}
  {\path{doi:10.1103/PhysRevLett.131.211902}}.

\bibitem{BESIII:2024vjf}
M.~Ablikim, et~al., BESIII Collaboration, {Measurement of the $e^+e^-\to
  K^+K^-\psi(2S)$ cross section at center-of-mass energies from 4.699 to 4.951
  GeV and search for $Z_{cs}^{\pm}$ in the $Z_{cs}^\pm\to K^\pm\psi(2S)$ decay}
  (7 2024).
\newblock \href {http://arxiv.org/abs/2407.20009} {\path{arXiv:2407.20009}}.

\bibitem{BESIII:2024aya}
M.~Ablikim, et~al., BESIII Collaboration, {Measurement of cross sections of
  $e^+e^-\to K^0_S K^0_S \psi(3686)$ from $\sqrt{s}=$ 4.682 to 4.951 GeV}, JHEP
  02 (2025) 120.
\newblock \href {http://arxiv.org/abs/2411.15752} {\path{arXiv:2411.15752}},
  \href {https://doi.org/10.1007/JHEP02(2025)120}
  {\path{doi:10.1007/JHEP02(2025)120}}.

\bibitem{Baru:2021ddn}
V.~Baru, E.~Epelbaum, A.~A. Filin, C.~Hanhart, A.~V. Nefediev, {Is
  $Z_{cs}(3982)$ a molecular partner of $Z_c(3900)$ and $Z_c(4020)$ states?},
  Phys. Rev. D 105 (2022) 034014.
\newblock \href {http://arxiv.org/abs/2110.00398} {\path{arXiv:2110.00398}},
  \href {https://doi.org/10.1103/PhysRevD.105.034014}
  {\path{doi:10.1103/PhysRevD.105.034014}}.

\bibitem{Yang:2020nrt}
Z.~Yang, X.~Cao, F.-K. Guo, J.~Nieves, M.~P. Valderrama, {Strange molecular
  partners of the $Z_c$(3900) and $Z_c$(4020)}, Phys. Rev. D 103 (2021) 074029.
\newblock \href {http://arxiv.org/abs/2011.08725} {\path{arXiv:2011.08725}},
  \href {https://doi.org/10.1103/PhysRevD.103.074029}
  {\path{doi:10.1103/PhysRevD.103.074029}}.

\bibitem{LHCb:2021uow}
R.~Aaij, et~al., LHCb Collaboration, {Observation of new resonances decaying to
  $J/\psi K^+$ and $J/\psi \phi$}, Phys. Rev. Lett. 127 (2021) 082001.
\newblock \href {http://arxiv.org/abs/2103.01803} {\path{arXiv:2103.01803}},
  \href {https://doi.org/10.1103/PhysRevLett.127.082001}
  {\path{doi:10.1103/PhysRevLett.127.082001}}.

\bibitem{LHCb:2023hxg}
R.~Aaij, et~al., LHCb Collaboration, {Evidence of a $J/\psi K_S^0$ structure in
  $B^0\to J/\psi \phi K_S^0$ decays}, Phys. Rev. Lett. 131 (2023) 131901.
\newblock \href {http://arxiv.org/abs/2301.04899} {\path{arXiv:2301.04899}},
  \href {https://doi.org/10.1103/PhysRevLett.131.131901}
  {\path{doi:10.1103/PhysRevLett.131.131901}}.

\bibitem{Ortega:2021enc}
P.~G. Ortega, D.~R. Entem, F.~Fernandez, {The strange partner of the $Z_c$
  structures in a coupled-channels model}, Phys. Lett. B 818 (2021) 136382.
\newblock \href {http://arxiv.org/abs/2103.07871} {\path{arXiv:2103.07871}},
  \href {https://doi.org/10.1016/j.physletb.2021.136382}
  {\path{doi:10.1016/j.physletb.2021.136382}}.

\bibitem{Wang:2020iqt}
Z.-G. Wang, {Analysis of $Z_{cs}(3985)$ as the axialvector tetraquark state},
  Chin. Phys. C 45 (2021) 073107.
\newblock \href {http://arxiv.org/abs/2011.10959} {\path{arXiv:2011.10959}},
  \href {https://doi.org/10.1088/1674-1137/abfa83}
  {\path{doi:10.1088/1674-1137/abfa83}}.

\bibitem{Wan:2020oxt}
B.-D. Wan, C.-F. Qiao, {About the exotic structure of $Z_{cs}$}, Nucl. Phys. B
  968 (2021) 115450.
\newblock \href {http://arxiv.org/abs/2011.08747} {\path{arXiv:2011.08747}},
  \href {https://doi.org/10.1016/j.nuclphysb.2021.115450}
  {\path{doi:10.1016/j.nuclphysb.2021.115450}}.

\bibitem{Cao:2022rjp}
Z.-H. Cao, W.~He, Z.-F. Sun, {$Z_{cs}$ states and the mixture of hadronic
  molecule and diquark-anti-diquark components within effective field theory},
  Phys. Rev. D 107 (2023) 014017.
\newblock \href {http://arxiv.org/abs/2205.11150} {\path{arXiv:2205.11150}},
  \href {https://doi.org/10.1103/PhysRevD.107.014017}
  {\path{doi:10.1103/PhysRevD.107.014017}}.

\bibitem{Yang:2021zhe}
G.~Yang, J.~Ping, J.~Segovia, {Hidden-charm tetraquarks with strangeness in the
  chiral quark model}, Phys. Rev. D 104 (2021) 094035.
\newblock \href {http://arxiv.org/abs/2109.04311} {\path{arXiv:2109.04311}},
  \href {https://doi.org/10.1103/PhysRevD.104.094035}
  {\path{doi:10.1103/PhysRevD.104.094035}}.

\bibitem{Giron:2021sla}
J.~F. Giron, R.~F. Lebed, S.~R. Martinez, {Spectrum of hidden-charm,
  open-strange exotics in the dynamical diquark model}, Phys. Rev. D 104 (2021)
  054001.
\newblock \href {http://arxiv.org/abs/2106.05883} {\path{arXiv:2106.05883}},
  \href {https://doi.org/10.1103/PhysRevD.104.054001}
  {\path{doi:10.1103/PhysRevD.104.054001}}.

\bibitem{Karliner:2021qok}
M.~Karliner, J.~L. Rosner, {Configuration mixing in strange tetraquarks
  $Z_{cs}$}, Phys. Rev. D 104 (2021) 034033.
\newblock \href {http://arxiv.org/abs/2107.04915} {\path{arXiv:2107.04915}},
  \href {https://doi.org/10.1103/PhysRevD.104.034033}
  {\path{doi:10.1103/PhysRevD.104.034033}}.

\bibitem{Shi:2021jyr}
P.-P. Shi, F.~Huang, W.-L. Wang, {Hidden charm tetraquark states in a diquark
  model}, Phys. Rev. D 103 (2021) 094038.
\newblock \href {http://arxiv.org/abs/2105.02397} {\path{arXiv:2105.02397}},
  \href {https://doi.org/10.1103/PhysRevD.103.094038}
  {\path{doi:10.1103/PhysRevD.103.094038}}.

\bibitem{Maiani:2021tri}
L.~Maiani, A.~D. Polosa, V.~Riquer, {The new resonances $Z_{cs}(3985)$ and
  $Z_{cs}(4003)$ (almost) fill two tetraquark nonets of broken
  $\text{SU}(3)_f$}, Sci. Bull. 66 (2021) 1616--1619.
\newblock \href {http://arxiv.org/abs/2103.08331} {\path{arXiv:2103.08331}},
  \href {https://doi.org/10.1016/j.scib.2021.04.040}
  {\path{doi:10.1016/j.scib.2021.04.040}}.

\bibitem{Xu:2020evn}
Y.-J. Xu, Y.-L. Liu, C.-Y. Cui, M.-Q. Huang, {$\bar D^{(*)}_s D^{(*)}$
  molecular state with $J^P$= $1^+$}, Phys. Rev. D 104 (2021) 094028.
\newblock \href {http://arxiv.org/abs/2011.14313} {\path{arXiv:2011.14313}},
  \href {https://doi.org/10.1103/PhysRevD.104.094028}
  {\path{doi:10.1103/PhysRevD.104.094028}}.

\bibitem{Guo:2020vmu}
Z.-H. Guo, J.~A. Oller, {Unified description of the hidden-charm tetraquark
  states $Z_{cs}(3985), Z_c(3900)$, and $X(4020)$}, Phys. Rev. D 103 (2021)
  054021.
\newblock \href {http://arxiv.org/abs/2012.11904} {\path{arXiv:2012.11904}},
  \href {https://doi.org/10.1103/PhysRevD.103.054021}
  {\path{doi:10.1103/PhysRevD.103.054021}}.

\bibitem{Du:2020vwb}
M.-C. Du, Q.~Wang, Q.~Zhao, {The nature of charged charmonium-like states
  $Z_c(3900)$ and its strange partner $Z_{cs}(3982)$} (11 2020).
\newblock \href {http://arxiv.org/abs/2011.09225} {\path{arXiv:2011.09225}}.

\bibitem{Meng:2020ihj}
L.~Meng, B.~Wang, S.-L. Zhu, {$Z_{cs}(3985)^-$ as the $U$-spin partner of
  $Z_c(3900)^-$ and implication of other states in the $\text{SU(3)}_F$
  symmetry and heavy quark symmetry}, Phys. Rev. D 102 (2020) 111502.
\newblock \href {http://arxiv.org/abs/2011.08656} {\path{arXiv:2011.08656}},
  \href {https://doi.org/10.1103/PhysRevD.102.111502}
  {\path{doi:10.1103/PhysRevD.102.111502}}.

\bibitem{Wang:2020htx}
B.~Wang, L.~Meng, S.-L. Zhu, {Decoding the nature of $Z_{cs}(3985)$ and
  establishing the spectrum of charged heavy quarkoniumlike states in chiral
  effective field theory}, Phys. Rev. D 103 (2021) L021501.
\newblock \href {http://arxiv.org/abs/2011.10922} {\path{arXiv:2011.10922}},
  \href {https://doi.org/10.1103/PhysRevD.103.L021501}
  {\path{doi:10.1103/PhysRevD.103.L021501}}.

\bibitem{Zhai:2022ied}
Q.-Y. Zhai, M.-Z. Liu, J.-X. Lu, L.-S. Geng, {$Z_{cs}(3985)$ in
  next-to-leading-order chiral effective field theory: The first truncation
  uncertainty analysis}, Phys. Rev. D 106 (2022) 034026.
\newblock \href {http://arxiv.org/abs/2205.03878} {\path{arXiv:2205.03878}},
  \href {https://doi.org/10.1103/PhysRevD.106.034026}
  {\path{doi:10.1103/PhysRevD.106.034026}}.

\bibitem{Han:2022fup}
S.~Han, L.-Y. Xiao, {Aspects of $Z_{cs}(3985)$ and $Z_{cs}(4000)$}, Phys. Rev.
  D 105 (2022) 054008.
\newblock \href {http://arxiv.org/abs/2203.00168} {\path{arXiv:2203.00168}},
  \href {https://doi.org/10.1103/PhysRevD.105.054008}
  {\path{doi:10.1103/PhysRevD.105.054008}}.

\bibitem{Yan:2021tcp}
M.-J. Yan, F.-Z. Peng, M.~S{\'a}nchez~S{\'a}nchez, M.~Pavon~Valderrama, {Axial
  meson exchange and the $Z_c(3900)$ and $Z_{cs}(3985)$ resonances as heavy
  hadron molecules}, Phys. Rev. D 104 (2021) 114025.
\newblock \href {http://arxiv.org/abs/2102.13058} {\path{arXiv:2102.13058}},
  \href {https://doi.org/10.1103/PhysRevD.104.114025}
  {\path{doi:10.1103/PhysRevD.104.114025}}.

\bibitem{Chen:2020yvq}
R.~Chen, Q.~Huang, {$Z_{cs}(3985)^-$: A strange hidden-charm tetraquark
  resonance or not?}, Phys. Rev. D 103 (2021) 034008.
\newblock \href {http://arxiv.org/abs/2011.09156} {\path{arXiv:2011.09156}},
  \href {https://doi.org/10.1103/PhysRevD.103.034008}
  {\path{doi:10.1103/PhysRevD.103.034008}}.

\bibitem{Wu:2021cyc}
Q.~Wu, D.-Y. Chen, W.-H. Qin, G.~Li, {Production of $Z_{cs}$ in $B$ and $B_s$
  decays}, Eur. Phys. J. C 82 (2022) 520.
\newblock \href {http://arxiv.org/abs/2111.13347} {\path{arXiv:2111.13347}},
  \href {https://doi.org/10.1140/epjc/s10052-022-10465-z}
  {\path{doi:10.1140/epjc/s10052-022-10465-z}}.

\bibitem{Wu:2023fyh}
Q.~Wu, Y.-K. Chen, G.~Li, S.-D. Liu, D.-Y. Chen, {Hunting for the hidden-charm
  molecular states with strange quarks in $B$ and $B_s$ decays}, Phys. Rev. D
  107 (2023) 054044.
\newblock \href {http://arxiv.org/abs/2302.01696} {\path{arXiv:2302.01696}},
  \href {https://doi.org/10.1103/PhysRevD.107.054044}
  {\path{doi:10.1103/PhysRevD.107.054044}}.

\bibitem{Yu:2025pyu}
Z.~Yu, Q.~Wu, D.-Y. Chen, {$Z_{cs}^+$ production in $B^+$ decays}, Phys. Rev. D
  112 (2025) 094054.
\newblock \href {http://arxiv.org/abs/2506.06702} {\path{arXiv:2506.06702}},
  \href {https://doi.org/10.1103/fjsv-fxpm} {\path{doi:10.1103/fjsv-fxpm}}.

\bibitem{Liu:2021ojf}
J.~Liu, D.-Y. Chen, J.~He, {Double exotic state productions in pion and kaon
  induced reactions}, Eur. Phys. J. C 81 (2021) 965.
\newblock \href {http://arxiv.org/abs/2108.00148} {\path{arXiv:2108.00148}},
  \href {https://doi.org/10.1140/epjc/s10052-021-09766-6}
  {\path{doi:10.1140/epjc/s10052-021-09766-6}}.

\bibitem{Wu:2021ezz}
Q.~Wu, D.-Y. Chen, {Exploration of the hidden charm decays of $Z_{cs}(3985)$},
  Phys. Rev. D 104 (2021) 074011.
\newblock \href {http://arxiv.org/abs/2108.06700} {\path{arXiv:2108.06700}},
  \href {https://doi.org/10.1103/PhysRevD.104.074011}
  {\path{doi:10.1103/PhysRevD.104.074011}}.

\bibitem{Ge:2021sdq}
Y.-H. Ge, X.-H. Liu, H.-W. Ke, {Threshold effects as the origin of
  $Z_{cs}(4000)$, $Z_{cs}(4220)$ and $X(4700)$ observed in $B^+\to J/\psi \phi
  K^+$}, Eur. Phys. J. C 81 (2021) 854.
\newblock \href {http://arxiv.org/abs/2103.05282} {\path{arXiv:2103.05282}},
  \href {https://doi.org/10.1140/epjc/s10052-021-09590-y}
  {\path{doi:10.1140/epjc/s10052-021-09590-y}}.

\bibitem{Ikeno:2020mra}
N.~Ikeno, R.~Molina, E.~Oset, {The $Z_{cs}(3985)$ as a threshold effect from
  the $\bar D_s^* D + \bar D_s D^*$ interaction}, Phys. Lett. B 814 (2021)
  136120.
\newblock \href {http://arxiv.org/abs/2011.13425} {\path{arXiv:2011.13425}},
  \href {https://doi.org/10.1016/j.physletb.2021.136120}
  {\path{doi:10.1016/j.physletb.2021.136120}}.

\bibitem{Wang:2020kej}
J.-Z. Wang, Q.-S. Zhou, X.~Liu, T.~Matsuki, {Toward charged $Z_{cs}(3985)$
  structure under a reflection mechanism}, Eur. Phys. J. C 81 (2021) 51.
\newblock \href {http://arxiv.org/abs/2011.08628} {\path{arXiv:2011.08628}},
  \href {https://doi.org/10.1140/epjc/s10052-021-08877-4}
  {\path{doi:10.1140/epjc/s10052-021-08877-4}}.

\bibitem{Wang:2020axi}
J.-Z. Wang, D.-Y. Chen, X.~Liu, T.~Matsuki, {Universal non-resonant explanation
  to charmoniumlike structures $Z_c(3885)$ and $Z_c(4025)$}, Eur. Phys. J. C 80
  (2020) 1040.
\newblock \href {http://arxiv.org/abs/2007.02263} {\path{arXiv:2007.02263}},
  \href {https://doi.org/10.1140/epjc/s10052-020-08621-4}
  {\path{doi:10.1140/epjc/s10052-020-08621-4}}.

\bibitem{Iwasaki:1975pv}
Y.~Iwasaki, {A possible model for new resonances-exotics and hidden charm},
  Prog. Theor. Phys. 54 (1975) 492.
\newblock \href {https://doi.org/10.1143/PTP.54.492}
  {\path{doi:10.1143/PTP.54.492}}.

\bibitem{Iwasaki:1976cn}
Y.~Iwasaki, {Is a state $c\bar{c}c\bar{c}$ found at 6.0 GeV?}, Phys. Rev. Lett.
  36 (1976) 1266.
\newblock \href {https://doi.org/10.1103/PhysRevLett.36.1266}
  {\path{doi:10.1103/PhysRevLett.36.1266}}.

\bibitem{Chao:1980dv}
K.-T. Chao, {The ($cc$) - ($\bar{c}\bar{c}$) (diquark-anti-diquark) states in
  $e^+ e^-$ annihilation}, Z. Phys. C 7 (1981) 317.
\newblock \href {https://doi.org/10.1007/BF01431564}
  {\path{doi:10.1007/BF01431564}}.

\bibitem{Badalian:1985es}
A.~M. Badalian, B.~L. Ioffe, A.~V. Smilga, {Four quark states in the heavy
  quark system}, Nucl. Phys. B 281 (1987) 85.
\newblock \href {https://doi.org/10.1016/0550-3213(87)90248-3}
  {\path{doi:10.1016/0550-3213(87)90248-3}}.

\bibitem{Ader:1981db}
J.~P. Ader, J.~M. Richard, P.~Taxil, {Do narrow heavy heavy multi-quark state
  exist?}, Phys. Rev. D 25 (1982) 2370.
\newblock \href {https://doi.org/10.1103/PhysRevD.25.2370}
  {\path{doi:10.1103/PhysRevD.25.2370}}.

\bibitem{Li:1983ru}
B.-A. Li, K.-F. Liu, {$J/\psi$ pair production in hadronic collisions}, Phys.
  Rev. D 29 (1984) 426.
\newblock \href {https://doi.org/10.1103/PhysRevD.29.426}
  {\path{doi:10.1103/PhysRevD.29.426}}.

\bibitem{Lloyd:2003yc}
R.~J. Lloyd, J.~P. Vary, {All charm tetraquarks}, Phys. Rev. D 70 (2004)
  014009.
\newblock \href {http://arxiv.org/abs/hep-ph/0311179}
  {\path{arXiv:hep-ph/0311179}}, \href
  {https://doi.org/10.1103/PhysRevD.70.014009}
  {\path{doi:10.1103/PhysRevD.70.014009}}.

\bibitem{Liu:2019zuc}
M.-S. Liu, Q.-F. L{\"u}, X.-H. Zhong, Q.~Zhao, {All-heavy tetraquarks}, Phys.
  Rev. D 100 (2019) 016006.
\newblock \href {http://arxiv.org/abs/1901.02564} {\path{arXiv:1901.02564}},
  \href {https://doi.org/10.1103/PhysRevD.100.016006}
  {\path{doi:10.1103/PhysRevD.100.016006}}.

\bibitem{Wang:2019rdo}
G.-J. Wang, L.~Meng, S.-L. Zhu, {Spectrum of the fully-heavy tetraquark state
  $QQ\bar Q' \bar Q'$}, Phys. Rev. D 100 (2019) 096013.
\newblock \href {http://arxiv.org/abs/1907.05177} {\path{arXiv:1907.05177}},
  \href {https://doi.org/10.1103/PhysRevD.100.096013}
  {\path{doi:10.1103/PhysRevD.100.096013}}.

\bibitem{Chen:2020lgj}
X.~Chen, {Fully-charm tetraquarks: $cc\bar{c}\bar{c}$} (1 2020).
\newblock \href {http://arxiv.org/abs/2001.06755} {\path{arXiv:2001.06755}}.

\bibitem{Debastiani:2017msn}
V.~R. Debastiani, F.~S. Navarra, {A non-relativistic model for the
  $[cc][\bar{c}\bar{c}]$ tetraquark}, Chin. Phys. C 43 (2019) 013105.
\newblock \href {http://arxiv.org/abs/1706.07553} {\path{arXiv:1706.07553}},
  \href {https://doi.org/10.1088/1674-1137/43/1/013105}
  {\path{doi:10.1088/1674-1137/43/1/013105}}.

\bibitem{Bedolla:2019zwg}
M.~A. Bedolla, J.~Ferretti, C.~D. Roberts, E.~Santopinto, {Spectrum of
  fully-heavy tetraquarks from a diquark+antidiquark perspective}, Eur. Phys.
  J. C 80 (2020) 1004.
\newblock \href {http://arxiv.org/abs/1911.00960} {\path{arXiv:1911.00960}},
  \href {https://doi.org/10.1140/epjc/s10052-020-08579-3}
  {\path{doi:10.1140/epjc/s10052-020-08579-3}}.

\bibitem{Richard:2017vry}
J.-M. Richard, A.~Valcarce, J.~Vijande, {String dynamics and metastability of
  all-heavy tetraquarks}, Phys. Rev. D 95 (2017) 054019.
\newblock \href {http://arxiv.org/abs/1703.00783} {\path{arXiv:1703.00783}},
  \href {https://doi.org/10.1103/PhysRevD.95.054019}
  {\path{doi:10.1103/PhysRevD.95.054019}}.

\bibitem{Berezhnoy:2011xn}
A.~V. Berezhnoy, A.~V. Luchinsky, A.~A. Novoselov, {Tetraquarks composed of 4
  heavy quarks}, Phys. Rev. D 86 (2012) 034004.
\newblock \href {http://arxiv.org/abs/1111.1867} {\path{arXiv:1111.1867}},
  \href {https://doi.org/10.1103/PhysRevD.86.034004}
  {\path{doi:10.1103/PhysRevD.86.034004}}.

\bibitem{Wu:2016vtq}
J.~Wu, Y.-R. Liu, K.~Chen, X.~Liu, S.-L. Zhu, {Heavy-flavored tetraquark states
  with the $QQ\bar{Q}\bar{Q}$ configuration}, Phys. Rev. D 97 (2018) 094015.
\newblock \href {http://arxiv.org/abs/1605.01134} {\path{arXiv:1605.01134}},
  \href {https://doi.org/10.1103/PhysRevD.97.094015}
  {\path{doi:10.1103/PhysRevD.97.094015}}.

\bibitem{Karliner:2016zzc}
M.~Karliner, S.~Nussinov, J.~L. Rosner, {$Q Q \bar Q \bar Q$ states: masses,
  production, and decays}, Phys. Rev. D 95 (2017) 034011.
\newblock \href {http://arxiv.org/abs/1611.00348} {\path{arXiv:1611.00348}},
  \href {https://doi.org/10.1103/PhysRevD.95.034011}
  {\path{doi:10.1103/PhysRevD.95.034011}}.

\bibitem{Chen:2016jxd}
W.~Chen, H.-X. Chen, X.~Liu, T.~G. Steele, S.-L. Zhu, {Hunting for exotic
  doubly hidden-charm/bottom tetraquark states}, Phys. Lett. B 773 (2017)
  247--251.
\newblock \href {http://arxiv.org/abs/1605.01647} {\path{arXiv:1605.01647}},
  \href {https://doi.org/10.1016/j.physletb.2017.08.034}
  {\path{doi:10.1016/j.physletb.2017.08.034}}.

\bibitem{Wang:2018poa}
Z.-G. Wang, Z.-Y. Di, {Analysis of the vector and axialvector
  $QQ\bar{Q}\bar{Q}$ tetraquark states with QCD sum rules}, Acta Phys. Polon. B
  50 (2019) 1335.
\newblock \href {http://arxiv.org/abs/1807.08520} {\path{arXiv:1807.08520}},
  \href {https://doi.org/10.5506/APhysPolB.50.1335}
  {\path{doi:10.5506/APhysPolB.50.1335}}.

\bibitem{Anwar:2017toa}
M.~N. Anwar, J.~Ferretti, F.-K. Guo, E.~Santopinto, B.-S. Zou, {Spectroscopy
  and decays of the fully-heavy tetraquarks}, Eur. Phys. J. C 78 (2018) 647.
\newblock \href {http://arxiv.org/abs/1710.02540} {\path{arXiv:1710.02540}},
  \href {https://doi.org/10.1140/epjc/s10052-018-6073-9}
  {\path{doi:10.1140/epjc/s10052-018-6073-9}}.

\bibitem{Bai:2016int}
Y.~Bai, S.~Lu, J.~Osborne, {Beauty-full Tetraquarks}, Phys. Lett. B 798 (2019)
  134930.
\newblock \href {http://arxiv.org/abs/1612.00012} {\path{arXiv:1612.00012}},
  \href {https://doi.org/10.1016/j.physletb.2019.134930}
  {\path{doi:10.1016/j.physletb.2019.134930}}.

\bibitem{Debastiani:2017xcr}
V.~R. Debastiani, F.~S. Navarra, {Spectroscopy of the All-Charm Tetraquark},
  PoS Hadron2017 (2018) 238.
\newblock \href {http://arxiv.org/abs/1711.11495} {\path{arXiv:1711.11495}},
  \href {https://doi.org/10.22323/1.310.0238} {\path{doi:10.22323/1.310.0238}}.

\bibitem{Heller:1985cb}
L.~Heller, J.~A. Tjon, {On Bound States of Heavy $Q^2 \bar{Q}^2$ Systems},
  Phys. Rev. D 32 (1985) 755.
\newblock \href {https://doi.org/10.1103/PhysRevD.32.755}
  {\path{doi:10.1103/PhysRevD.32.755}}.

\bibitem{Zouzou:1986qh}
S.~Zouzou, B.~Silvestre-Brac, C.~Gignoux, J.~M. Richard, {Four quark bound
  states}, Z. Phys. C 30 (1986) 457.
\newblock \href {https://doi.org/10.1007/BF01557611}
  {\path{doi:10.1007/BF01557611}}.

\bibitem{Wang:2017jtz}
Z.-G. Wang, {Analysis of the $QQ\bar{Q}\bar{Q}$ tetraquark states with QCD sum
  rules}, Eur. Phys. J. C 77 (2017) 432.
\newblock \href {http://arxiv.org/abs/1701.04285} {\path{arXiv:1701.04285}},
  \href {https://doi.org/10.1140/epjc/s10052-017-4997-0}
  {\path{doi:10.1140/epjc/s10052-017-4997-0}}.

\bibitem{Silva:2025bdg}
V.~S. Silva, C.~Pigozzo, L.~M. Abreu, {Addendum to: Fully-heavy tetraquarks in
  the vacuum and in a hot environment}, Eur. Phys. J. C 85 (2025) 1154.
\newblock \href {http://arxiv.org/abs/2503.12160} {\path{arXiv:2503.12160}},
  \href {https://doi.org/10.1140/epjc/s10052-025-14871-x}
  {\path{doi:10.1140/epjc/s10052-025-14871-x}}.

\bibitem{CMS:2016liw}
V.~Khachatryan, et~al., CMS Collaboration, {Observation of $\Upsilon$(1S) pair
  production in proton-proton collisions at $ \sqrt{s}=8 $ TeV}, JHEP 05 (2017)
  013.
\newblock \href {http://arxiv.org/abs/1610.07095} {\path{arXiv:1610.07095}},
  \href {https://doi.org/10.1007/JHEP05(2017)013}
  {\path{doi:10.1007/JHEP05(2017)013}}.

\bibitem{CMS:2020qwa}
A.~M. Sirunyan, et~al., CMS Collaboration, {Measurement of the $\Upsilon$(1S)
  pair production cross section and search for resonances decaying to
  $\Upsilon$(1S)$\mu^+\mu^-$ in proton-proton collisions at $\sqrt{s} =$ 13
  TeV}, Phys. Lett. B 808 (2020) 135578.
\newblock \href {http://arxiv.org/abs/2002.06393} {\path{arXiv:2002.06393}},
  \href {https://doi.org/10.1016/j.physletb.2020.135578}
  {\path{doi:10.1016/j.physletb.2020.135578}}.

\bibitem{LHCb:2018uwm}
R.~Aaij, et~al., LHCb Collaboration, {Search for beautiful tetraquarks in the
  $\Upsilon(1S)\mu^+\mu^-$ invariant-mass spectrum}, JHEP 10 (2018) 086.
\newblock \href {http://arxiv.org/abs/1806.09707} {\path{arXiv:1806.09707}},
  \href {https://doi.org/10.1007/JHEP10(2018)086}
  {\path{doi:10.1007/JHEP10(2018)086}}.

\bibitem{Xu:2022rnl}
Y.~Xu, ATLAS Collaboration, {ATLAS results on exotic hadronic resonances}, Acta
  Phys. Polon. Supp. 16 (2023) 21.
\newblock \href {http://arxiv.org/abs/2209.12173} {\path{arXiv:2209.12173}},
  \href {https://doi.org/10.21468/SciPostPhysProc.15.013}
  {\path{doi:10.21468/SciPostPhysProc.15.013}}.

\bibitem{Zhang:2022toq}
J.~Zhang, K.~Yi, CMS Collaboration, {Recent CMS results on exotic resonances},
  PoS ICHEP2022 (2022) 775.
\newblock \href {http://arxiv.org/abs/2212.00504} {\path{arXiv:2212.00504}},
  \href {https://doi.org/10.22323/1.414.0775} {\path{doi:10.22323/1.414.0775}}.

\bibitem{Wang:2021mma}
Q.-N. Wang, Z.-Y. Yang, W.~Chen, {Exotic fully-heavy $Q\bar QQ\bar Q$
  tetraquark states in $\mathbf{8}_{[Q\bar{Q}]}\otimes \mathbf{8}_{[Q\bar{Q}]}$
  color configuration}, Phys. Rev. D 104 (2021) 114037.
\newblock \href {http://arxiv.org/abs/2109.08091} {\path{arXiv:2109.08091}},
  \href {https://doi.org/10.1103/PhysRevD.104.114037}
  {\path{doi:10.1103/PhysRevD.104.114037}}.

\bibitem{Yang:2020wkh}
B.-C. Yang, L.~Tang, C.-F. Qiao, {Scalar fully-heavy tetraquark states
  $QQ^\prime {\bar{Q}} \bar{Q^\prime }$ in QCD sum rules}, Eur. Phys. J. C 81
  (2021) 324.
\newblock \href {http://arxiv.org/abs/2012.04463} {\path{arXiv:2012.04463}},
  \href {https://doi.org/10.1140/epjc/s10052-021-09096-7}
  {\path{doi:10.1140/epjc/s10052-021-09096-7}}.

\bibitem{Tang:2024zvf}
C.-M. Tang, C.-G. Duan, L.~Tang, {Fully charmed tetraquark states in
  $8_{[c\bar{c}]}\otimes 8_{[c\bar{c}]}$ color structure via QCD sum rules},
  Eur. Phys. J. C 84 (2024) 743.
\newblock \href {http://arxiv.org/abs/2405.05042} {\path{arXiv:2405.05042}},
  \href {https://doi.org/10.1140/epjc/s10052-024-13102-z}
  {\path{doi:10.1140/epjc/s10052-024-13102-z}}.

\bibitem{Deng:2020iqw}
C.~Deng, H.~Chen, J.~Ping, {Towards the understanding of fully-heavy tetraquark
  states from various models}, Phys. Rev. D 103 (2021) 014001.
\newblock \href {http://arxiv.org/abs/2003.05154} {\path{arXiv:2003.05154}},
  \href {https://doi.org/10.1103/PhysRevD.103.014001}
  {\path{doi:10.1103/PhysRevD.103.014001}}.

\bibitem{Sonnenschein:2020nwn}
J.~Sonnenschein, D.~Weissman, {Deciphering the recently discovered tetraquark
  candidates around 6.9 GeV}, Eur. Phys. J. C 81 (2021) 25.
\newblock \href {http://arxiv.org/abs/2008.01095} {\path{arXiv:2008.01095}},
  \href {https://doi.org/10.1140/epjc/s10052-020-08818-7}
  {\path{doi:10.1140/epjc/s10052-020-08818-7}}.

\bibitem{Wang:2021kfv}
G.-J. Wang, L.~Meng, M.~Oka, S.-L. Zhu, {Higher fully charmed tetraquarks:
  Radial excitations and $P$-wave states}, Phys. Rev. D 104 (2021) 036016.
\newblock \href {http://arxiv.org/abs/2105.13109} {\path{arXiv:2105.13109}},
  \href {https://doi.org/10.1103/PhysRevD.104.036016}
  {\path{doi:10.1103/PhysRevD.104.036016}}.

\bibitem{Liu:2021rtn}
F.-X. Liu, M.-S. Liu, X.-H. Zhong, Q.~Zhao, {Higher mass spectra of the
  fully-charmed and fully-bottom tetraquarks}, Phys. Rev. D 104 (2021) 116029.
\newblock \href {http://arxiv.org/abs/2110.09052} {\path{arXiv:2110.09052}},
  \href {https://doi.org/10.1103/PhysRevD.104.116029}
  {\path{doi:10.1103/PhysRevD.104.116029}}.

\bibitem{Zhang:2022qtp}
J.~Zhang, J.-B. Wang, G.~Li, C.-S. An, C.-R. Deng, J.-J. Xie, {Spectrum of the
  $S$-wave fully-heavy tetraquark states}, Eur. Phys. J. C 82 (2022) 1126.
\newblock \href {http://arxiv.org/abs/2209.13856} {\path{arXiv:2209.13856}},
  \href {https://doi.org/10.1140/epjc/s10052-022-11111-4}
  {\path{doi:10.1140/epjc/s10052-022-11111-4}}.

\bibitem{Faustov:2022mvs}
R.~N. Faustov, V.~O. Galkin, E.~M. Savchenko, {Fully Heavy Tetraquark
  Spectroscopy in the Relativistic Quark Model}, Symmetry 14 (2022) 2504.
\newblock \href {http://arxiv.org/abs/2210.16015} {\path{arXiv:2210.16015}},
  \href {https://doi.org/10.3390/sym14122504} {\path{doi:10.3390/sym14122504}}.

\bibitem{Yu:2022lak}
G.-L. Yu, Z.-Y. Li, Z.-G. Wang, J.~Lu, M.~Yan, {The $S$- and $P$-wave fully
  charmed tetraquark states and their radial excitations}, Eur. Phys. J. C 83
  (2023) 416.
\newblock \href {http://arxiv.org/abs/2212.14339} {\path{arXiv:2212.14339}},
  \href {https://doi.org/10.1140/epjc/s10052-023-11445-7}
  {\path{doi:10.1140/epjc/s10052-023-11445-7}}.

\bibitem{Chen:2024bpz}
Z.-Z. Chen, X.-L. Chen, P.-F. Yang, W.~Chen, {$P$-wave fully charm and fully
  bottom tetraquark states}, Phys. Rev. D 109 (2024) 094011.
\newblock \href {http://arxiv.org/abs/2402.03117} {\path{arXiv:2402.03117}},
  \href {https://doi.org/10.1103/PhysRevD.109.094011}
  {\path{doi:10.1103/PhysRevD.109.094011}}.

\bibitem{Lin:2024olg}
Y.-Y. Lin, J.-Y. Wang, A.~Zhang, {Mass spectrum of fully charmed
  $[cc][\bar{c}\bar{c}]$ tetraquarks}, Eur. Phys. J. Plus 139 (2024) 707.
\newblock \href {http://arxiv.org/abs/2404.08971} {\path{arXiv:2404.08971}},
  \href {https://doi.org/10.1140/epjp/s13360-024-05480-w}
  {\path{doi:10.1140/epjp/s13360-024-05480-w}}.

\bibitem{Galkin:2023wox}
V.~O. Galkin, E.~M. Savchenko, {Relativistic description of asymmetric fully
  heavy tetraquarks in the diquark{\textendash}antidiquark model}, Eur. Phys.
  J. A 60 (2024) 96.
\newblock \href {http://arxiv.org/abs/2310.20247} {\path{arXiv:2310.20247}},
  \href {https://doi.org/10.1140/epja/s10050-024-01311-9}
  {\path{doi:10.1140/epja/s10050-024-01311-9}}.

\bibitem{Wang:2020ols}
Z.-G. Wang, {Tetraquark candidates in the LHCb's di-$J/\psi$ mass spectrum},
  Chin. Phys. C 44 (2020) 113106.
\newblock \href {http://arxiv.org/abs/2006.13028} {\path{arXiv:2006.13028}},
  \href {https://doi.org/10.1088/1674-1137/abb080}
  {\path{doi:10.1088/1674-1137/abb080}}.

\bibitem{Lu:2020cns}
Q.-F. L{\"u}, D.-Y. Chen, Y.-B. Dong, {Masses of fully heavy tetraquarks $QQ
  {\bar{Q}} {\bar{Q}}$ in an extended relativized quark model}, Eur. Phys. J. C
  80 (2020) 871.
\newblock \href {http://arxiv.org/abs/2006.14445} {\path{arXiv:2006.14445}},
  \href {https://doi.org/10.1140/epjc/s10052-020-08454-1}
  {\path{doi:10.1140/epjc/s10052-020-08454-1}}.

\bibitem{Giron:2020wpx}
J.~F. Giron, R.~F. Lebed, {Simple spectrum of $c\bar c c\bar c$ states in the
  dynamical diquark model}, Phys. Rev. D 102 (2020) 074003.
\newblock \href {http://arxiv.org/abs/2008.01631} {\path{arXiv:2008.01631}},
  \href {https://doi.org/10.1103/PhysRevD.102.074003}
  {\path{doi:10.1103/PhysRevD.102.074003}}.

\bibitem{Karliner:2020dta}
M.~Karliner, J.~L. Rosner, {Interpretation of structure in the di- $J/\psi$
  spectrum}, Phys. Rev. D 102 (2020) 114039.
\newblock \href {http://arxiv.org/abs/2009.04429} {\path{arXiv:2009.04429}},
  \href {https://doi.org/10.1103/PhysRevD.102.114039}
  {\path{doi:10.1103/PhysRevD.102.114039}}.

\bibitem{Wang:2020dlo}
Z.-G. Wang, {Revisit the tetraquark candidates in the $J/\psi J/\psi$ mass
  spectrum}, Int. J. Mod. Phys. A 36 (2021) 2150014.
\newblock \href {http://arxiv.org/abs/2009.05371} {\path{arXiv:2009.05371}},
  \href {https://doi.org/10.1142/S0217751X21500147}
  {\path{doi:10.1142/S0217751X21500147}}.

\bibitem{Zhao:2020jvl}
Z.~Zhao, K.~Xu, A.~Kaewsnod, X.~Liu, A.~Limphirat, Y.~Yan, {Study of
  charmoniumlike and fully-charm tetraquark spectroscopy}, Phys. Rev. D 103
  (2021) 116027.
\newblock \href {http://arxiv.org/abs/2012.15554} {\path{arXiv:2012.15554}},
  \href {https://doi.org/10.1103/PhysRevD.103.116027}
  {\path{doi:10.1103/PhysRevD.103.116027}}.

\bibitem{Ke:2021iyh}
H.-W. Ke, X.~Han, X.-H. Liu, Y.-L. Shi, {Tetraquark state $X(6900)$ and the
  interaction between diquark and antidiquark}, Eur. Phys. J. C 81 (2021) 427.
\newblock \href {http://arxiv.org/abs/2103.13140} {\path{arXiv:2103.13140}},
  \href {https://doi.org/10.1140/epjc/s10052-021-09229-y}
  {\path{doi:10.1140/epjc/s10052-021-09229-y}}.

\bibitem{Mutuk:2021hmi}
H.~Mutuk, {Nonrelativistic treatment of fully-heavy tetraquarks as
  diquark-antidiquark states}, Eur. Phys. J. C 81 (2021) 367.
\newblock \href {http://arxiv.org/abs/2104.11823} {\path{arXiv:2104.11823}},
  \href {https://doi.org/10.1140/epjc/s10052-021-09176-8}
  {\path{doi:10.1140/epjc/s10052-021-09176-8}}.

\bibitem{Li:2021ygk}
Q.~Li, C.-H. Chang, G.-L. Wang, T.~Wang, {Mass spectra and wave functions of
  $T_{QQ\bar{Q}\bar{Q}}$ tetraquarks}, Phys. Rev. D 104 (2021) 014018.
\newblock \href {http://arxiv.org/abs/2104.12372} {\path{arXiv:2104.12372}},
  \href {https://doi.org/10.1103/PhysRevD.104.014018}
  {\path{doi:10.1103/PhysRevD.104.014018}}.

\bibitem{Santowsky:2021bhy}
N.~Santowsky, C.~S. Fischer, {Four-quark states with charm quarks in a two-body
  Bethe{\textendash}Salpeter approach}, Eur. Phys. J. C 82 (2022) 313.
\newblock \href {http://arxiv.org/abs/2111.15310} {\path{arXiv:2111.15310}},
  \href {https://doi.org/10.1140/epjc/s10052-022-10272-6}
  {\path{doi:10.1140/epjc/s10052-022-10272-6}}.

\bibitem{Wang:2022xja}
Z.-G. Wang, {Analysis of the $X(6600)$, $X(6900)$, $X(7300)$ and related
  tetraquark states with the QCD sum rules}, Nucl. Phys. B 985 (2022) 115983.
\newblock \href {http://arxiv.org/abs/2207.08059} {\path{arXiv:2207.08059}},
  \href {https://doi.org/10.1016/j.nuclphysb.2022.115983}
  {\path{doi:10.1016/j.nuclphysb.2022.115983}}.

\bibitem{Zhu:2020xni}
R.~Zhu, {Fully-heavy tetraquark spectra and production at hadron colliders},
  Nucl. Phys. B 966 (2021) 115393.
\newblock \href {http://arxiv.org/abs/2010.09082} {\path{arXiv:2010.09082}},
  \href {https://doi.org/10.1016/j.nuclphysb.2021.115393}
  {\path{doi:10.1016/j.nuclphysb.2021.115393}}.

\bibitem{Dong:2022sef}
W.-C. Dong, Z.-G. Wang, {Going in quest of potential tetraquark interpretations
  for the newly observed $T_{\psi\psi}$ states in light of the
  diquark-antidiquark scenarios}, Phys. Rev. D 107 (2023) 074010.
\newblock \href {http://arxiv.org/abs/2211.11989} {\path{arXiv:2211.11989}},
  \href {https://doi.org/10.1103/PhysRevD.107.074010}
  {\path{doi:10.1103/PhysRevD.107.074010}}.

\bibitem{Wan:2020fsk}
B.-D. Wan, C.-F. Qiao, {Gluonic tetracharm configuration of $X (6900)$}, Phys.
  Lett. B 817 (2021) 136339.
\newblock \href {http://arxiv.org/abs/2012.00454} {\path{arXiv:2012.00454}},
  \href {https://doi.org/10.1016/j.physletb.2021.136339}
  {\path{doi:10.1016/j.physletb.2021.136339}}.

\bibitem{Tang:2024kmh}
C.-M. Tang, C.-G. Duan, L.~Tang, C.-F. Qiao, {A novel configuration of gluonic
  tetraquark state}, Eur. Phys. J. C 85 (2025) 396.
\newblock \href {http://arxiv.org/abs/2411.11433} {\path{arXiv:2411.11433}},
  \href {https://doi.org/10.1140/epjc/s10052-025-14106-z}
  {\path{doi:10.1140/epjc/s10052-025-14106-z}}.

\bibitem{Tang:2025ept}
C.-M. Tang, C.-G. Duan, L.~Tang, C.-F. Qiao, {QCD sum rule predictions on
  gluonic tetraquark states with $J^{PC}=0^{+-},0^{--}$ and $1^{\pm \pm}$} (11
  2025).
\newblock \href {http://arxiv.org/abs/2511.18807} {\path{arXiv:2511.18807}}.

\bibitem{Agaev:2023wua}
S.~S. Agaev, K.~Azizi, B.~Barsbay, H.~Sundu, {Exploring fully heavy scalar
  tetraquarks $QQ\bar{Q}\bar{Q}$}, Phys. Lett. B 844 (2023) 138089.
\newblock \href {http://arxiv.org/abs/2304.03244} {\path{arXiv:2304.03244}},
  \href {https://doi.org/10.1016/j.physletb.2023.138089}
  {\path{doi:10.1016/j.physletb.2023.138089}}.

\bibitem{Yang:2021hrb}
G.~Yang, J.~Ping, J.~Segovia, {Exotic resonances of fully-heavy tetraquarks in
  a lattice-QCD insipired quark model}, Phys. Rev. D 104 (2021) 014006.
\newblock \href {http://arxiv.org/abs/2104.08814} {\path{arXiv:2104.08814}},
  \href {https://doi.org/10.1103/PhysRevD.104.014006}
  {\path{doi:10.1103/PhysRevD.104.014006}}.

\bibitem{Jin:2020jfc}
X.~Jin, Y.~Xue, H.~Huang, J.~Ping, {Full-heavy tetraquarks in constituent quark
  models}, Eur. Phys. J. C 80 (2020) 1083.
\newblock \href {http://arxiv.org/abs/2006.13745} {\path{arXiv:2006.13745}},
  \href {https://doi.org/10.1140/epjc/s10052-020-08650-z}
  {\path{doi:10.1140/epjc/s10052-020-08650-z}}.

\bibitem{Wang:2023jqs}
G.-J. Wang, M.~Oka, D.~Jido, {Quark confinement for multiquark systems:
  Application to fully charmed tetraquarks}, Phys. Rev. D 108 (2023) L071501.
\newblock \href {http://arxiv.org/abs/2307.04310} {\path{arXiv:2307.04310}},
  \href {https://doi.org/10.1103/PhysRevD.108.L071501}
  {\path{doi:10.1103/PhysRevD.108.L071501}}.

\bibitem{Wu:2024tif}
Y.~Wu, X.~Liu, J.~Ping, H.~Huang, Y.~Tan, {Further study of $c\bar{c}c\bar{c}$
  system within a chiral quark model}, Eur. Phys. J. C 85 (2025) 147.
\newblock \href {http://arxiv.org/abs/2403.10375} {\path{arXiv:2403.10375}},
  \href {https://doi.org/10.1140/epjc/s10052-025-13822-w}
  {\path{doi:10.1140/epjc/s10052-025-13822-w}}.

\bibitem{Ortega:2023pmr}
P.~G. Ortega, D.~R. Entem, F.~Fern{\'a}ndez, {Exploring $T_{\psi\psi}$
  tetraquark candidates in a coupled-channels formalism}, Phys. Rev. D 108
  (2023) 094023.
\newblock \href {http://arxiv.org/abs/2307.00532} {\path{arXiv:2307.00532}},
  \href {https://doi.org/10.1103/PhysRevD.108.094023}
  {\path{doi:10.1103/PhysRevD.108.094023}}.

\bibitem{Weng:2020jao}
X.-Z. Weng, X.-L. Chen, W.-Z. Deng, S.-L. Zhu, {Systematics of fully heavy
  tetraquarks}, Phys. Rev. D 103 (2021) 034001.
\newblock \href {http://arxiv.org/abs/2010.05163} {\path{arXiv:2010.05163}},
  \href {https://doi.org/10.1103/PhysRevD.103.034001}
  {\path{doi:10.1103/PhysRevD.103.034001}}.

\bibitem{An:2022qpt}
H.-T. An, S.-Q. Luo, Z.-W. Liu, X.~Liu, {Spectroscopic behavior of fully heavy
  tetraquarks}, Eur. Phys. J. C 83 (2023) 740.
\newblock \href {http://arxiv.org/abs/2208.03899} {\path{arXiv:2208.03899}},
  \href {https://doi.org/10.1140/epjc/s10052-023-11847-7}
  {\path{doi:10.1140/epjc/s10052-023-11847-7}}.

\bibitem{Meng:2024yhu}
Q.~Meng, G.-J. Wang, M.~Oka, {Mass spectra of full-heavy and double-heavy
  tetraquark states in the conventional quark model}, Phys. Rev. D 111 (2025)
  014018.
\newblock \href {http://arxiv.org/abs/2404.01238} {\path{arXiv:2404.01238}},
  \href {https://doi.org/10.1103/PhysRevD.111.014018}
  {\path{doi:10.1103/PhysRevD.111.014018}}.

\bibitem{Hughes:2017xie}
C.~Hughes, E.~Eichten, C.~T.~H. Davies, {Searching for beauty-fully bound
  tetraquarks using lattice nonrelativistic QCD}, Phys. Rev. D 97 (2018)
  054505.
\newblock \href {http://arxiv.org/abs/1710.03236} {\path{arXiv:1710.03236}},
  \href {https://doi.org/10.1103/PhysRevD.97.054505}
  {\path{doi:10.1103/PhysRevD.97.054505}}.

\bibitem{Richard:2018yrm}
J.-M. Richard, A.~Valcarce, J.~Vijande, {Few-body quark dynamics for doubly
  heavy baryons and tetraquarks}, Phys. Rev. C 97 (2018) 035211.
\newblock \href {http://arxiv.org/abs/1803.06155} {\path{arXiv:1803.06155}},
  \href {https://doi.org/10.1103/PhysRevC.97.035211}
  {\path{doi:10.1103/PhysRevC.97.035211}}.

\bibitem{Wu:2024euj}
W.-L. Wu, Y.-K. Chen, L.~Meng, S.-L. Zhu, {Benchmark calculations of fully
  heavy compact and molecular tetraquark states}, Phys. Rev. D 109 (2024)
  054034.
\newblock \href {http://arxiv.org/abs/2401.14899} {\path{arXiv:2401.14899}},
  \href {https://doi.org/10.1103/PhysRevD.109.054034}
  {\path{doi:10.1103/PhysRevD.109.054034}}.

\bibitem{Wang:2022jmb}
J.-Z. Wang, X.~Liu, {Improved understanding of the peaking phenomenon existing
  in the new di-$J/\psi$ invariant mass spectrum from the CMS Collaboration},
  Phys. Rev. D 106 (2022) 054015.
\newblock \href {http://arxiv.org/abs/2207.04893} {\path{arXiv:2207.04893}},
  \href {https://doi.org/10.1103/PhysRevD.106.054015}
  {\path{doi:10.1103/PhysRevD.106.054015}}.

\bibitem{Wang:2020wrp}
J.-Z. Wang, D.-Y. Chen, X.~Liu, T.~Matsuki, {Producing fully charm structures
  in the $J/\psi$ -pair invariant mass spectrum}, Phys. Rev. D 103 (2021)
  071503.
\newblock \href {http://arxiv.org/abs/2008.07430} {\path{arXiv:2008.07430}},
  \href {https://doi.org/10.1103/PhysRevD.103.L071503}
  {\path{doi:10.1103/PhysRevD.103.L071503}}.

\bibitem{Dong:2020nwy}
X.-K. Dong, V.~Baru, F.-K. Guo, C.~Hanhart, A.~Nefediev, {Coupled-Channel
  Interpretation of the LHCb Double-~$J/\psi$~Spectrum and Hints of a New State
  Near the~ $J/\psi J/\psi$~~Threshold}, Phys. Rev. Lett. 126 (2021) 132001,
  [Erratum: Phys.Rev.Lett. 127, 119901 (2021)].
\newblock \href {http://arxiv.org/abs/2009.07795} {\path{arXiv:2009.07795}},
  \href {https://doi.org/10.1103/PhysRevLett.127.119901}
  {\path{doi:10.1103/PhysRevLett.127.119901}}.

\bibitem{Song:2024ykq}
Y.-L. Song, Y.~Zhang, V.~Baru, F.-K. Guo, C.~Hanhart, A.~Nefediev, {Toward a
  precision determination of the $X(6200)$ parameters from data}, Phys. Rev. D
  111 (2025) 034038.
\newblock \href {http://arxiv.org/abs/2411.12062} {\path{arXiv:2411.12062}},
  \href {https://doi.org/10.1103/PhysRevD.111.034038}
  {\path{doi:10.1103/PhysRevD.111.034038}}.

\bibitem{Liang:2021fzr}
Z.-R. Liang, X.-Y. Wu, D.-L. Yao, {Hunting for states in the recent LHCb
  di-$J/\psi$ invariant mass spectrum}, Phys. Rev. D 104 (2021) 034034.
\newblock \href {http://arxiv.org/abs/2104.08589} {\path{arXiv:2104.08589}},
  \href {https://doi.org/10.1103/PhysRevD.104.034034}
  {\path{doi:10.1103/PhysRevD.104.034034}}.

\bibitem{Zhou:2022xpd}
Q.~Zhou, D.~Guo, S.-Q. Kuang, Q.-H. Yang, L.-Y. Dai, {Nature of the $X(6900)$
  in partial wave decomposition of $J/\psi J/\psi$ scattering}, Phys. Rev. D
  106 (2022) L111502.
\newblock \href {http://arxiv.org/abs/2207.07537} {\path{arXiv:2207.07537}},
  \href {https://doi.org/10.1103/PhysRevD.106.L111502}
  {\path{doi:10.1103/PhysRevD.106.L111502}}.

\bibitem{Kuang:2023vac}
S.-Q. Kuang, Q.~Zhou, D.~Guo, Q.-H. Yang, L.-Y. Dai, {Study of $X(6900)$ with
  unitarized coupled channel scattering amplitudes}, Eur. Phys. J. C 83 (2023)
  383.
\newblock \href {http://arxiv.org/abs/2302.03968} {\path{arXiv:2302.03968}},
  \href {https://doi.org/10.1140/epjc/s10052-023-11473-3}
  {\path{doi:10.1140/epjc/s10052-023-11473-3}}.

\bibitem{Niu:2022jqp}
P.~Niu, Z.~Zhang, Q.~Wang, M.-L. Du, {The third peak structure in the double
  $J/\psi$ spectrum}, Sci. Bull. 68 (2023) 800--803.
\newblock \href {http://arxiv.org/abs/2212.06535} {\path{arXiv:2212.06535}},
  \href {https://doi.org/10.1016/j.scib.2023.03.025}
  {\path{doi:10.1016/j.scib.2023.03.025}}.

\bibitem{Zhuang:2021pci}
Z.~Zhuang, Y.~Zhang, Y.~Ma, Q.~Wang, {Lineshape of the compact fully heavy
  tetraquark}, Phys. Rev. D 105 (2022) 054026.
\newblock \href {http://arxiv.org/abs/2111.14028} {\path{arXiv:2111.14028}},
  \href {https://doi.org/10.1103/PhysRevD.105.054026}
  {\path{doi:10.1103/PhysRevD.105.054026}}.

\bibitem{Cao:2020gul}
Q.-F. Cao, H.~Chen, H.-R. Qi, H.-Q. Zheng, {Some remarks on $X(6900)$}, Chin.
  Phys. C 45 (2021) 103102.
\newblock \href {http://arxiv.org/abs/2011.04347} {\path{arXiv:2011.04347}},
  \href {https://doi.org/10.1088/1674-1137/ac0ee5}
  {\path{doi:10.1088/1674-1137/ac0ee5}}.

\bibitem{Liu:2024pio}
W.-Y. Liu, H.-X. Chen, {Hadronic Molecules with Four Charm or Beauty Quarks},
  Universe 11 (2025) 36.
\newblock \href {http://arxiv.org/abs/2405.14404} {\path{arXiv:2405.14404}},
  \href {https://doi.org/10.3390/universe11020036}
  {\path{doi:10.3390/universe11020036}}.

\bibitem{Guo:2020pvt}
Z.-H. Guo, J.~A. Oller, {Insights into the inner structures of the fully
  charmed tetraquark state $X(6900)$}, Phys. Rev. D 103 (2021) 034024.
\newblock \href {http://arxiv.org/abs/2011.00978} {\path{arXiv:2011.00978}},
  \href {https://doi.org/10.1103/PhysRevD.103.034024}
  {\path{doi:10.1103/PhysRevD.103.034024}}.

\bibitem{Lu:2023ccs}
Y.~Lu, C.~Chen, K.-G. Kang, G.-y. Qin, H.-Q. Zheng, {X(6900) peak could be a
  molecular state}, Phys. Rev. D 107 (2023) 094006.
\newblock \href {http://arxiv.org/abs/2302.04150} {\path{arXiv:2302.04150}},
  \href {https://doi.org/10.1103/PhysRevD.107.094006}
  {\path{doi:10.1103/PhysRevD.107.094006}}.

\bibitem{Gong:2020bmg}
C.~Gong, M.-C. Du, Q.~Zhao, X.-H. Zhong, B.~Zhou, {Nature of $X(6900)$ and its
  production mechanism at LHCb}, Phys. Lett. B 824 (2022) 136794.
\newblock \href {http://arxiv.org/abs/2011.11374} {\path{arXiv:2011.11374}},
  \href {https://doi.org/10.1016/j.physletb.2021.136794}
  {\path{doi:10.1016/j.physletb.2021.136794}}.

\bibitem{Dong:2021lkh}
X.-K. Dong, V.~Baru, F.-K. Guo, C.~Hanhart, A.~Nefediev, B.-S. Zou, {Is the
  existence of a $J/\psi J/\psi$ bound state plausible?}, Sci. Bull. 66 (2021)
  2462--2470.
\newblock \href {http://arxiv.org/abs/2107.03946} {\path{arXiv:2107.03946}},
  \href {https://doi.org/10.1016/j.scib.2021.09.009}
  {\path{doi:10.1016/j.scib.2021.09.009}}.

\bibitem{Brambilla:2015rqa}
N.~Brambilla, G.~Krein, J.~Tarr{\'u}s~Castell{\`a}, A.~Vairo, {Long-range
  properties of $1S$ bottomonium states}, Phys. Rev. D 93 (2016) 054002.
\newblock \href {http://arxiv.org/abs/1510.05895} {\path{arXiv:1510.05895}},
  \href {https://doi.org/10.1103/PhysRevD.93.054002}
  {\path{doi:10.1103/PhysRevD.93.054002}}.

\bibitem{Fujii:1999xn}
H.~Fujii, D.~Kharzeev, {Long range forces of QCD}, Phys. Rev. D 60 (1999)
  114039.
\newblock \href {http://arxiv.org/abs/hep-ph/9903495}
  {\path{arXiv:hep-ph/9903495}}, \href
  {https://doi.org/10.1103/PhysRevD.60.114039}
  {\path{doi:10.1103/PhysRevD.60.114039}}.

\bibitem{Lu:2023aal}
Y.~Lu, C.~Chen, G.-y. Qin, H.-Q. Zheng, {A discussion on the anomalous
  threshold enhancement of $J/\psi$--$\psi(3770)$ couplings and $X(6900)$
  peak}, Chin. Phys. C 48 (2024) 041001.
\newblock \href {http://arxiv.org/abs/2312.10711} {\path{arXiv:2312.10711}},
  \href {https://doi.org/10.1088/1674-1137/ad2361}
  {\path{doi:10.1088/1674-1137/ad2361}}.

\bibitem{Yukawa:1935xg}
H.~Yukawa, {On the Interaction of Elementary Particles I}, Proc. Phys. Math.
  Soc. Jap. 17 (1935) 48--57.
\newblock \href {https://doi.org/10.1143/PTPS.1.1}
  {\path{doi:10.1143/PTPS.1.1}}.

\bibitem{Huang:2024jin}
Q.~Huang, R.~Chen, J.~He, X.~Liu, {Discovering a Novel Dynamics Mechanism for
  Charmonium Scattering} (7 2024).
\newblock \href {http://arxiv.org/abs/2407.16316} {\path{arXiv:2407.16316}}.

\bibitem{Meng:2024czd}
Y.~Meng, C.~Liu, X.-Y. Tuo, H.~Yan, Z.~Zhang, {Lattice calculation of the $\eta
  _c\eta _c$ and $J/\psi J/\psi $ s-wave scattering length}, Eur. Phys. J. C 85
  (2025) 458.
\newblock \href {http://arxiv.org/abs/2411.11533} {\path{arXiv:2411.11533}},
  \href {https://doi.org/10.1140/epjc/s10052-025-14192-z}
  {\path{doi:10.1140/epjc/s10052-025-14192-z}}.

\bibitem{Li:2025vbd}
G.~Li, C.~Shi, Y.~Chen, W.~Sun, {$\eta_c\eta_c$ and $J/\psi J/\psi$ scatterings
  from lattice QCD} (5 2025).
\newblock \href {http://arxiv.org/abs/2505.23220} {\path{arXiv:2505.23220}}.

\bibitem{Li:2025ftn}
G.~Li, C.~Shi, Y.~Chen, W.~Sun, {Tensor Resonance in $J/\psi J/\psi$ Scattering
  from Lattice QCD} (5 2025).
\newblock \href {http://arxiv.org/abs/2505.24213} {\path{arXiv:2505.24213}}.

\bibitem{Zhang:2025zaa}
L.~Zhang, T.~Doi, Y.~Lyu, T.~Hatsuda, Y.-G. Ma, {Probing
  nucleon-$\bar{\Omega}_{ccc}$ interaction via lattice QCD at physical quark
  masses}, Phys. Lett. B 871 (2025) 139998.
\newblock \href {http://arxiv.org/abs/2508.10388} {\path{arXiv:2508.10388}},
  \href {https://doi.org/10.1016/j.physletb.2025.139998}
  {\path{doi:10.1016/j.physletb.2025.139998}}.

\bibitem{Mathur:2022ovu}
N.~Mathur, M.~Padmanath, D.~Chakraborty, {Strongly Bound Dibaryon with Maximal
  Beauty Flavor from Lattice QCD}, Phys. Rev. Lett. 130 (2023) 111901.
\newblock \href {http://arxiv.org/abs/2205.02862} {\path{arXiv:2205.02862}},
  \href {https://doi.org/10.1103/PhysRevLett.130.111901}
  {\path{doi:10.1103/PhysRevLett.130.111901}}.

\bibitem{Dhindsa:2025zjk}
N.~S. Dhindsa, D.~Chakraborty, A.~Radhakrishnan, N.~Mathur, M.~Padmanath,
  {Precisely determining the ground state mass of Spin-3/2 $\Omega_{ccc}$
  baryon from Lattice QCD}, 2025.
\newblock \href {http://arxiv.org/abs/2512.23417} {\path{arXiv:2512.23417}}.

\bibitem{Lattes:1947mw}
C.~M.~G. Lattes, H.~Muirhead, G.~P.~S. Occhialini, C.~F. Powell, {Processes
  involving charged mesons}, Nature 159 (1947) 694--697.
\newblock \href {https://doi.org/10.1038/159694a0}
  {\path{doi:10.1038/159694a0}}.

\bibitem{10.1143/PTP.27.1199}
N.~Hoshizaki, S.~Otsuki, W.~Watari, M.~Yonezawa,
  \href{https://doi.org/10.1143/PTP.27.1199}{Nuclear forces and bosons: The
  sakata model and one-boson-exchange-potentials}, Progress of Theoretical
  Physics 27~(6) (1962) 1199--1220.
\newblock \href
  {http://arxiv.org/abs/https://academic.oup.com/ptp/article-pdf/27/6/1199/5382274/27-6-1199.pdf}
  {\path{arXiv:https://academic.oup.com/ptp/article-pdf/27/6/1199/5382274/27-6-1199.pdf}},
  \href {https://doi.org/10.1143/PTP.27.1199} {\path{doi:10.1143/PTP.27.1199}}.
\newline\urlprefix\url{https://doi.org/10.1143/PTP.27.1199}

\bibitem{ParticleDataGroup:2020ssz}
P.~A. Zyla, et~al., Particle Data Group Collaboration, {Review of particle
  physics}, PTEP 2020 (2020) 083C01.
\newblock \href {https://doi.org/10.1093/ptep/ptaa104}
  {\path{doi:10.1093/ptep/ptaa104}}.

\bibitem{Belle:2008kuo}
C.~P. Shen, et~al., Belle Collaboration, {Observation of the $\phi(1680)$ and
  the $Y(2175)$ in $e^+e^-\to\phi\pi^+\pi^-$}, Phys. Rev. D 80 (2009) 031101.
\newblock \href {http://arxiv.org/abs/0808.0006} {\path{arXiv:0808.0006}},
  \href {https://doi.org/10.1103/PhysRevD.80.031101}
  {\path{doi:10.1103/PhysRevD.80.031101}}.

\bibitem{Chen:2008qw}
H.-X. Chen, A.~Hosaka, S.-L. Zhu, {The $I^GJ^{PC}=1^-1^{-+}$ tetraquark
  states}, Phys. Rev. D 78 (2008) 054017.
\newblock \href {http://arxiv.org/abs/0806.1998} {\path{arXiv:0806.1998}},
  \href {https://doi.org/10.1103/PhysRevD.78.054017}
  {\path{doi:10.1103/PhysRevD.78.054017}}.

\bibitem{Ding:2009vj}
G.-J. Ding, J.-F. Liu, M.-L. Yan, {Dynamics of hadronic molecule in one-boson
  exchange approach and possible heavy flavor molecules}, Phys. Rev. D 79
  (2009) 054005.
\newblock \href {http://arxiv.org/abs/0901.0426} {\path{arXiv:0901.0426}},
  \href {https://doi.org/10.1103/PhysRevD.79.054005}
  {\path{doi:10.1103/PhysRevD.79.054005}}.

\bibitem{Guo:2016fgl}
P.~Guo, {One spatial dimensional finite volume three-body interaction for a
  short-range potential}, Phys. Rev. D 95 (2017) 054508.
\newblock \href {http://arxiv.org/abs/1607.03184} {\path{arXiv:1607.03184}},
  \href {https://doi.org/10.1103/PhysRevD.95.054508}
  {\path{doi:10.1103/PhysRevD.95.054508}}.

\bibitem{E288:1977efs}
W.~R. Innes, et~al., E288 Collaboration, {Observation of structure in the
  $\Upsilon$ region}, Phys. Rev. Lett. 39 (1977) 1240--1242, [Erratum:
  Phys.Rev.Lett. 39, 1640 (1977)].
\newblock \href {https://doi.org/10.1103/PhysRevLett.39.1240}
  {\path{doi:10.1103/PhysRevLett.39.1240}}.

\bibitem{Klopfenstein:1983nx}
C.~Klopfenstein, et~al., {Observation of the lowest $P$ wave $b \bar{b}$ bound
  states}, Phys. Rev. Lett. 51 (1983) 160.
\newblock \href {https://doi.org/10.1103/PhysRevLett.51.160}
  {\path{doi:10.1103/PhysRevLett.51.160}}.

\bibitem{Pauss:1983pa}
F.~Pauss, et~al., {Observation of $\chi_b$ production in the exclusive reaction
  $\Upsilon^\prime \to \gamma \chi_b \to \gamma \gamma \Upsilon \to \gamma
  \gamma$ ($e^+ e^-$ or $\mu^+ \mu^-$)}, Phys. Lett. B 130 (1983) 439.
\newblock \href {https://doi.org/10.1016/0370-2693(83)91539-3}
  {\path{doi:10.1016/0370-2693(83)91539-3}}.

\bibitem{Han:1982zk}
K.~Han, et~al., {Observation of $P$-wave $b\bar b$ bound states}, Phys. Rev.
  Lett. 49 (1982) 1612--1616.
\newblock \href {https://doi.org/10.1103/PhysRevLett.49.1612}
  {\path{doi:10.1103/PhysRevLett.49.1612}}.

\bibitem{Eigen:1982zm}
G.~Eigen, et~al., {Evidence for $\chi_b^\prime$ production in the exclusive
  reaction
  $\Upsilon^{\prime\prime}\to\gamma\chi_{b}^\prime\to$($\gamma\gamma\Upsilon^\prime$
  or $\gamma\gamma\Upsilon$)}, Phys. Rev. Lett. 49 (1982) 1616--1619.
\newblock \href {https://doi.org/10.1103/PhysRevLett.49.1616}
  {\path{doi:10.1103/PhysRevLett.49.1616}}.

\bibitem{BaBar:2008dae}
B.~Aubert, et~al., BaBar Collaboration, {Observation of the bottomonium ground
  state in the decay $\Upsilon(3S) \to \gamma \eta_b$}, Phys. Rev. Lett. 101
  (2008) 071801, [Erratum: Phys.Rev.Lett. 102, 029901 (2009)].
\newblock \href {http://arxiv.org/abs/0807.1086} {\path{arXiv:0807.1086}},
  \href {https://doi.org/10.1103/PhysRevLett.101.071801}
  {\path{doi:10.1103/PhysRevLett.101.071801}}.

\bibitem{CLEO:2009nxu}
G.~Bonvicini, et~al., CLEO Collaboration, {Measurement of the $\eta_b(1S)$ mass
  and the branching fraction for $\Upsilon(3S)\to\gamma\eta_b(1S)$}, Phys. Rev.
  D 81 (2010) 031104.
\newblock \href {http://arxiv.org/abs/0909.5474} {\path{arXiv:0909.5474}},
  \href {https://doi.org/10.1103/PhysRevD.81.031104}
  {\path{doi:10.1103/PhysRevD.81.031104}}.

\bibitem{Dobbs:2012zn}
S.~Dobbs, Z.~Metreveli, K.~K. Seth, A.~Tomaradze, T.~Xiao, {Observation of the
  $\eta_b(2S)$ meson in $\Upsilon(2S) \to \gamma \eta_b(2S)$, $\eta_b(2S) \to $
  hadrons, and confirmation of $\eta_b(1S)$ meson}, Phys. Rev. Lett. 109 (2012)
  082001.
\newblock \href {http://arxiv.org/abs/1204.4205} {\path{arXiv:1204.4205}},
  \href {https://doi.org/10.1103/PhysRevLett.109.082001}
  {\path{doi:10.1103/PhysRevLett.109.082001}}.

\bibitem{BaBar:2011xka}
J.~P. Lees, et~al., BaBar Collaboration, {Study of radiative bottomonium
  transitions using converted photons}, Phys. Rev. D 84 (2011) 072002.
\newblock \href {http://arxiv.org/abs/1104.5254} {\path{arXiv:1104.5254}},
  \href {https://doi.org/10.1103/PhysRevD.84.072002}
  {\path{doi:10.1103/PhysRevD.84.072002}}.

\bibitem{D0:2012pig}
V.~M. Abazov, et~al., D0 Collaboration, {Observation of a narrow mass state
  decaying into $\Upsilon(1S) + \gamma$ in $p\bar{p}$ collisions at $\sqrt{s} =
  1.96$ TeV}, Phys. Rev. D 86 (2012) 031103.
\newblock \href {http://arxiv.org/abs/1203.6034} {\path{arXiv:1203.6034}},
  \href {https://doi.org/10.1103/PhysRevD.86.031103}
  {\path{doi:10.1103/PhysRevD.86.031103}}.

\bibitem{BaBar:2011ljf}
J.~P. Lees, et~al., BaBar Collaboration, {Evidence for the $h_b(1P)$ meson in
  the decay $\Upsilon(3S) \to \pi^0 h_b(1P)$}, Phys. Rev. D 84 (2011) 091101.
\newblock \href {http://arxiv.org/abs/1102.4565} {\path{arXiv:1102.4565}},
  \href {https://doi.org/10.1103/PhysRevD.84.091101}
  {\path{doi:10.1103/PhysRevD.84.091101}}.

\bibitem{Belle:2011wqq}
I.~Adachi, et~al., Belle Collaboration, {First observation of the $P$-wave
  spin-singlet bottomonium states $h_b(1P)$ and $h_b(2P)$}, Phys. Rev. Lett.
  108 (2012) 032001.
\newblock \href {http://arxiv.org/abs/1103.3419} {\path{arXiv:1103.3419}},
  \href {https://doi.org/10.1103/PhysRevLett.108.032001}
  {\path{doi:10.1103/PhysRevLett.108.032001}}.

\bibitem{CLEO:2004npj}
G.~Bonvicini, et~al., CLEO Collaboration, {First observation of a
  $\Upsilon(1D)$ state}, Phys. Rev. D 70 (2004) 032001.
\newblock \href {http://arxiv.org/abs/hep-ex/0404021}
  {\path{arXiv:hep-ex/0404021}}, \href
  {https://doi.org/10.1103/PhysRevD.70.032001}
  {\path{doi:10.1103/PhysRevD.70.032001}}.

\bibitem{BaBar:2010tqb}
P.~del Amo~Sanchez, et~al., BaBar Collaboration, {Observation of the
  $\Upsilon(1^3D_J)$ bottomonium state through decays to
  $\pi^+\pi^-\Upsilon(1S)$}, Phys. Rev. D 82 (2010) 111102.
\newblock \href {http://arxiv.org/abs/1004.0175} {\path{arXiv:1004.0175}},
  \href {https://doi.org/10.1103/PhysRevD.82.111102}
  {\path{doi:10.1103/PhysRevD.82.111102}}.

\bibitem{Richardson:1978bt}
J.~L. Richardson, {The heavy quark potential and the $\Upsilon$, $J/\psi$
  systems}, Phys. Lett. B 82 (1979) 272--274.
\newblock \href {https://doi.org/10.1016/0370-2693(79)90753-6}
  {\path{doi:10.1016/0370-2693(79)90753-6}}.

\bibitem{Buchmuller:1980su}
W.~Buchmuller, S.~H.~H. Tye, {Quarkonia and quantum chromodynamics}, Phys. Rev.
  D 24 (1981) 132.
\newblock \href {https://doi.org/10.1103/PhysRevD.24.132}
  {\path{doi:10.1103/PhysRevD.24.132}}.

\bibitem{Martin:1980jx}
A.~Martin, {A fit of upsilon and charmonium spectra}, Phys. Lett. B 93 (1980)
  338--342.
\newblock \href {https://doi.org/10.1016/0370-2693(80)90527-4}
  {\path{doi:10.1016/0370-2693(80)90527-4}}.

\bibitem{Gupta:1986xt}
S.~N. Gupta, S.~F. Radford, W.~W. Repko, {Semirelativistic potential model for
  heavy quarkonia}, Phys. Rev. D 34 (1986) 201--206.
\newblock \href {https://doi.org/10.1103/PhysRevD.34.201}
  {\path{doi:10.1103/PhysRevD.34.201}}.

\bibitem{Beyer:1992nd}
M.~Beyer, U.~Bohn, M.~G. Huber, B.~C. Metsch, J.~Resag, {Relativistic effects
  and the constituent quark model of heavy quarkonia}, Z. Phys. C 55 (1992)
  307--315.
\newblock \href {https://doi.org/10.1007/BF01482594}
  {\path{doi:10.1007/BF01482594}}.

\bibitem{Motyka:1997di}
L.~Motyka, K.~Zalewski, {Mass spectra and leptonic decay widths of heavy
  quarkonia}, Eur. Phys. J. C 4 (1998) 107--114.
\newblock \href {http://arxiv.org/abs/hep-ph/9709254}
  {\path{arXiv:hep-ph/9709254}}, \href {https://doi.org/10.1007/s100520050190}
  {\path{doi:10.1007/s100520050190}}.

\bibitem{Ebert:2002pp}
D.~Ebert, R.~N. Faustov, V.~O. Galkin, {Properties of heavy quarkonia and $B_c$
  mesons in the relativistic quark model}, Phys. Rev. D 67 (2003) 014027.
\newblock \href {http://arxiv.org/abs/hep-ph/0210381}
  {\path{arXiv:hep-ph/0210381}}, \href
  {https://doi.org/10.1103/PhysRevD.67.014027}
  {\path{doi:10.1103/PhysRevD.67.014027}}.

\bibitem{Gonzalez:2003gx}
P.~Gonzalez, A.~Valcarce, H.~Garcilazo, J.~Vijande, {Heavy meson description
  with a screened potential}, Phys. Rev. D 68 (2003) 034007.
\newblock \href {http://arxiv.org/abs/hep-ph/0307310}
  {\path{arXiv:hep-ph/0307310}}, \href
  {https://doi.org/10.1103/PhysRevD.68.034007}
  {\path{doi:10.1103/PhysRevD.68.034007}}.

\bibitem{Vijande:2004he}
J.~Vijande, F.~Fernandez, A.~Valcarce, {Constituent quark model study of the
  meson spectra}, J. Phys. G 31 (2005) 481.
\newblock \href {http://arxiv.org/abs/hep-ph/0411299}
  {\path{arXiv:hep-ph/0411299}}, \href
  {https://doi.org/10.1088/0954-3899/31/5/017}
  {\path{doi:10.1088/0954-3899/31/5/017}}.

\bibitem{Radford:2007vd}
S.~F. Radford, W.~W. Repko, {Potential model calculations and predictions for
  heavy quarkonium}, Phys. Rev. D 75 (2007) 074031.
\newblock \href {http://arxiv.org/abs/hep-ph/0701117}
  {\path{arXiv:hep-ph/0701117}}, \href
  {https://doi.org/10.1103/PhysRevD.75.074031}
  {\path{doi:10.1103/PhysRevD.75.074031}}.

\bibitem{Li:2009nr}
B.-Q. Li, K.-T. Chao, {Bottomonium spectrum with screened potential}, Commun.
  Theor. Phys. 52 (2009) 653--661.
\newblock \href {http://arxiv.org/abs/0909.1369} {\path{arXiv:0909.1369}},
  \href {https://doi.org/10.1088/0253-6102/52/4/20}
  {\path{doi:10.1088/0253-6102/52/4/20}}.

\bibitem{Wei-Zhao:2013sta}
T.~Wei-Zhao, C.~Lu, Y.~You-Chang, C.~Hong, {Bottomonium states versus recent
  experimental observations in the QCD-inspired potential model}, Chin. Phys. C
  37 (2013) 083101.
\newblock \href {http://arxiv.org/abs/1308.0960} {\path{arXiv:1308.0960}},
  \href {https://doi.org/10.1088/1674-1137/37/8/083101}
  {\path{doi:10.1088/1674-1137/37/8/083101}}.

\bibitem{Gonzalez:2014nka}
P.~Gonzalez, {Generalized screened potential model}, J. Phys. G 41 (2014)
  095001.
\newblock \href {http://arxiv.org/abs/1406.5025} {\path{arXiv:1406.5025}},
  \href {https://doi.org/10.1088/0954-3899/41/9/095001}
  {\path{doi:10.1088/0954-3899/41/9/095001}}.

\bibitem{Li:2015zda}
Y.~Li, P.~Maris, X.~Zhao, J.~P. Vary, {Heavy quarkonium in a holographic
  basis}, Phys. Lett. B 758 (2016) 118--124.
\newblock \href {http://arxiv.org/abs/1509.07212} {\path{arXiv:1509.07212}},
  \href {https://doi.org/10.1016/j.physletb.2016.04.065}
  {\path{doi:10.1016/j.physletb.2016.04.065}}.

\bibitem{Godfrey:2015dia}
S.~Godfrey, K.~Moats, {Bottomonium mesons and strategies for their
  observation}, Phys. Rev. D 92 (2015) 054034.
\newblock \href {http://arxiv.org/abs/1507.00024} {\path{arXiv:1507.00024}},
  \href {https://doi.org/10.1103/PhysRevD.92.054034}
  {\path{doi:10.1103/PhysRevD.92.054034}}.

\bibitem{Segovia:2016xqb}
J.~Segovia, P.~G. Ortega, D.~R. Entem, F.~Fern{\'a}ndez, {Bottomonium spectrum
  revisited}, Phys. Rev. D 93 (2016) 074027.
\newblock \href {http://arxiv.org/abs/1601.05093} {\path{arXiv:1601.05093}},
  \href {https://doi.org/10.1103/PhysRevD.93.074027}
  {\path{doi:10.1103/PhysRevD.93.074027}}.

\bibitem{Deng:2016ktl}
W.-J. Deng, H.~Liu, L.-C. Gui, X.-H. Zhong, {Spectrum and electromagnetic
  transitions of bottomonium}, Phys. Rev. D 95 (2017) 074002.
\newblock \href {http://arxiv.org/abs/1607.04696} {\path{arXiv:1607.04696}},
  \href {https://doi.org/10.1103/PhysRevD.95.074002}
  {\path{doi:10.1103/PhysRevD.95.074002}}.

\bibitem{Kaushal:2025kbz}
R.~Kaushal, Bhaghyesh, {Beauty hadron spectrum in a screened potential model},
  Eur. Phys. J. C 85 (2025) 814.
\newblock \href {http://arxiv.org/abs/2505.13987} {\path{arXiv:2505.13987}},
  \href {https://doi.org/10.1140/epjc/s10052-025-14538-7}
  {\path{doi:10.1140/epjc/s10052-025-14538-7}}.

\bibitem{Bokade:2025voh}
C.~A. Bokade, A.~Bhaghyesh, {Predictions for bottomonium from a relativistic
  screened potential model}, Chin. Phys. 49 (2025) 073102.
\newblock \href {http://arxiv.org/abs/2501.03147} {\path{arXiv:2501.03147}},
  \href {https://doi.org/10.1088/1674-1137/adc084}
  {\path{doi:10.1088/1674-1137/adc084}}.

\bibitem{ParticleDataGroup:2016lqr}
C.~Patrignani, et~al., Particle Data Group Collaboration, {Review of particle
  physics}, Chin. Phys. C 40 (2016) 100001.
\newblock \href {https://doi.org/10.1088/1674-1137/40/10/100001}
  {\path{doi:10.1088/1674-1137/40/10/100001}}.

\bibitem{Gottfried:1977gp}
K.~Gottfried, {Hadronic transitions between quark-antiquark bound states},
  Phys. Rev. Lett. 40 (1978) 598.
\newblock \href {https://doi.org/10.1103/PhysRevLett.40.598}
  {\path{doi:10.1103/PhysRevLett.40.598}}.

\bibitem{Bhanot:1979af}
G.~Bhanot, W.~Fischler, S.~Rudaz, {A multipole expansion and the Casimir-Polder
  effect in quantum chromodynamics}, Nucl. Phys. B 155 (1979) 208--236.
\newblock \href {https://doi.org/10.1016/0550-3213(79)90363-8}
  {\path{doi:10.1016/0550-3213(79)90363-8}}.

\bibitem{Peskin:1979va}
M.~E. Peskin, {Short distance analysis for heavy quark systems. 1.
  Diagrammatics}, Nucl. Phys. B 156 (1979) 365--390.
\newblock \href {https://doi.org/10.1016/0550-3213(79)90199-8}
  {\path{doi:10.1016/0550-3213(79)90199-8}}.

\bibitem{Bhanot:1979vb}
G.~Bhanot, M.~E. Peskin, {Short distance analysis for heavy quark systems. 2.
  Applications}, Nucl. Phys. B 156 (1979) 391--416.
\newblock \href {https://doi.org/10.1016/0550-3213(79)90200-1}
  {\path{doi:10.1016/0550-3213(79)90200-1}}.

\bibitem{Voloshin:1978hc}
M.~B. Voloshin, {On dynamics of heavy quarks in nonperturbative QCD vacuum},
  Nucl. Phys. B 154 (1979) 365--380.
\newblock \href {https://doi.org/10.1016/0550-3213(79)90037-3}
  {\path{doi:10.1016/0550-3213(79)90037-3}}.

\bibitem{Yan:1980uh}
T.-M. Yan, {Hadronic transitions between heavy quark states in quantum
  chromodynamics}, Phys. Rev. D 22 (1980) 1652.
\newblock \href {https://doi.org/10.1103/PhysRevD.22.1652}
  {\path{doi:10.1103/PhysRevD.22.1652}}.

\bibitem{Kuang:1988bz}
Y.-P. Kuang, S.~F. Tuan, T.-M. Yan, {Hadronic transitions and $P$ wave singlet
  states of heavy quarkonia}, Phys. Rev. D 37 (1988) 1210--1219.
\newblock \href {https://doi.org/10.1103/PhysRevD.37.1210}
  {\path{doi:10.1103/PhysRevD.37.1210}}.

\bibitem{Kuang:2006me}
Y.-P. Kuang, {QCD multipole expansion and hadronic transitions in heavy
  quarkonium systems}, Front. Phys. China 1 (2006) 19--37.
\newblock \href {http://arxiv.org/abs/hep-ph/0601044}
  {\path{arXiv:hep-ph/0601044}}, \href
  {https://doi.org/10.1007/s11467-005-0012-6}
  {\path{doi:10.1007/s11467-005-0012-6}}.

\bibitem{Belle:2015hnh}
U.~Tamponi, et~al., Belle Collaboration, {First observation of the hadronic
  transition $ \Upsilon(4S) \to \eta h_{b}(1P)$ and new measurement of the
  $h_b(1P)$ and $\eta_b(1S)$ parameters}, Phys. Rev. Lett. 115 (2015) 142001.
\newblock \href {http://arxiv.org/abs/1506.08914} {\path{arXiv:1506.08914}},
  \href {https://doi.org/10.1103/PhysRevLett.115.142001}
  {\path{doi:10.1103/PhysRevLett.115.142001}}.

\bibitem{Voloshin:2007dx}
M.~B. Voloshin, {Charmonium}, Prog. Part. Nucl. Phys. 61 (2008) 455--511.
\newblock \href {http://arxiv.org/abs/0711.4556} {\path{arXiv:0711.4556}},
  \href {https://doi.org/10.1016/j.ppnp.2008.02.001}
  {\path{doi:10.1016/j.ppnp.2008.02.001}}.

\bibitem{BaBar:2011krt}
J.~P. Lees, et~al., BaBar Collaboration, {Study of di-pion bottomonium
  transitions and search for the $h_b(1P)$ state}, Phys. Rev. D 84 (2011)
  011104.
\newblock \href {http://arxiv.org/abs/1105.4234} {\path{arXiv:1105.4234}},
  \href {https://doi.org/10.1103/PhysRevD.84.011104}
  {\path{doi:10.1103/PhysRevD.84.011104}}.

\bibitem{Belle:2014sys}
X.~H. He, et~al., Belle Collaboration, {Observation of $e^+e^- \to \pi^+ \pi^-
  \pi^0 \chi_{bJ}$ and Search for $X_b \to \omega \Upsilon(1S)$ at
  $\sqrt{s}=10.867$ GeV}, Phys. Rev. Lett. 113 (2014) 142001.
\newblock \href {http://arxiv.org/abs/1408.0504} {\path{arXiv:1408.0504}},
  \href {https://doi.org/10.1103/PhysRevLett.113.142001}
  {\path{doi:10.1103/PhysRevLett.113.142001}}.

\bibitem{Simonov:2008qy}
Y.~A. Simonov, A.~I. Veselov, {Bottomonium dipion transitions}, Phys. Rev. D 79
  (2009) 034024.
\newblock \href {http://arxiv.org/abs/0804.4635} {\path{arXiv:0804.4635}},
  \href {https://doi.org/10.1103/PhysRevD.79.034024}
  {\path{doi:10.1103/PhysRevD.79.034024}}.

\bibitem{Simonov:2008ci}
Y.~A. Simonov, A.~I. Veselov, {Strong decays and dipion transitions of
  $\Upsilon(5S)$}, Phys. Lett. B 671 (2009) 55--59.
\newblock \href {http://arxiv.org/abs/0805.4499} {\path{arXiv:0805.4499}},
  \href {https://doi.org/10.1016/j.physletb.2008.12.001}
  {\path{doi:10.1016/j.physletb.2008.12.001}}.

\bibitem{Belle:2021gws}
E.~Kovalenko, et~al., Belle Collaboration, {Study of $e^+e^- \rightarrow
  \Upsilon(1S,2S)\eta$ and $e^+e^-\rightarrow \Upsilon(1S)\eta^{\prime}$ at
  $\sqrt{s}=10.866$ GeV with the Belle detector}, Phys. Rev. D 104 (2021)
  112006.
\newblock \href {http://arxiv.org/abs/2108.04426} {\path{arXiv:2108.04426}},
  \href {https://doi.org/10.1103/PhysRevD.104.112006}
  {\path{doi:10.1103/PhysRevD.104.112006}}.

\bibitem{CLEO:2008dfe}
Q.~He, et~al., CLEO Collaboration, {Observation of $\Upsilon(2S)\to\eta
  \Upsilon(1S)$ and search for related transitions}, Phys. Rev. Lett. 101
  (2008) 192001.
\newblock \href {http://arxiv.org/abs/0806.3027} {\path{arXiv:0806.3027}},
  \href {https://doi.org/10.1103/PhysRevLett.101.192001}
  {\path{doi:10.1103/PhysRevLett.101.192001}}.

\bibitem{Zhang:2018eeo}
Y.~Zhang, G.~Li, {Exploring the $\Upsilon(4S,5S,6S) \to h_b(1P)\eta$
  hidden-bottom hadronic transitions}, Phys. Rev. D 97 (2018) 014018.
\newblock \href {http://arxiv.org/abs/1801.03582} {\path{arXiv:1801.03582}},
  \href {https://doi.org/10.1103/PhysRevD.97.014018}
  {\path{doi:10.1103/PhysRevD.97.014018}}.

\bibitem{Wang:2016qmz}
B.~Wang, X.~Liu, D.-Y. Chen, {Prediction of anomalous
  $\Upsilon(5S)\to\Upsilon(1^3D_J)\eta$ transitions}, Phys. Rev. D 94 (2016)
  094039.
\newblock \href {http://arxiv.org/abs/1611.02369} {\path{arXiv:1611.02369}},
  \href {https://doi.org/10.1103/PhysRevD.94.094039}
  {\path{doi:10.1103/PhysRevD.94.094039}}.

\bibitem{Belle:2018hjt}
U.~Tamponi, et~al., Belle Collaboration, {Inclusive study of bottomonium
  production in association with an $\eta $ meson in $e^+e^-$ annihilations
  near $\Upsilon(5S)$}, Eur. Phys. J. C 78 (2018) 633.
\newblock \href {http://arxiv.org/abs/1803.03225} {\path{arXiv:1803.03225}},
  \href {https://doi.org/10.1140/epjc/s10052-018-6086-4}
  {\path{doi:10.1140/epjc/s10052-018-6086-4}}.

\bibitem{Chen:2014ccr}
D.-Y. Chen, X.~Liu, T.~Matsuki, {Explaining the anomalous $\Upsilon(5S)\to
  \chi_{bJ}\omega$ decays through the hadronic loop effect}, Phys. Rev. D 90
  (2014) 034019.
\newblock \href {http://arxiv.org/abs/1406.6763} {\path{arXiv:1406.6763}},
  \href {https://doi.org/10.1103/PhysRevD.90.034019}
  {\path{doi:10.1103/PhysRevD.90.034019}}.

\bibitem{Luo:2025kid}
Z.-L. Luo, Y.-L. Song, F.-K. Guo, {Decays of $\Upsilon(10860)$ and
  $\Upsilon(10753)$ into $\omega\chi_{bJ}$}, Phys. Lett. B 870 (2025) 139960.
\newblock \href {http://arxiv.org/abs/2508.18720} {\path{arXiv:2508.18720}},
  \href {https://doi.org/10.1016/j.physletb.2025.139960}
  {\path{doi:10.1016/j.physletb.2025.139960}}.

\bibitem{Huang:2018pmk}
Q.~Huang, X.~Liu, T.~Matsuki, {Proposal of searching for the $\Upsilon(6S)$
  hadronic decays into $\Upsilon(nS)$ plus $\eta^{(\prime)}$}, Phys. Rev. D 98
  (2018) 054008.
\newblock \href {http://arxiv.org/abs/1807.09137} {\path{arXiv:1807.09137}},
  \href {https://doi.org/10.1103/PhysRevD.98.054008}
  {\path{doi:10.1103/PhysRevD.98.054008}}.

\bibitem{Huang:2018cco}
Q.~Huang, H.~Xu, X.~Liu, T.~Matsuki, {Potential observation of the
  $\Upsilon(6S) \to \Upsilon(1^3D_J) \eta$ transitions at Belle II}, Phys. Rev.
  D 97 (2018) 094018.
\newblock \href {http://arxiv.org/abs/1804.01017} {\path{arXiv:1804.01017}},
  \href {https://doi.org/10.1103/PhysRevD.97.094018}
  {\path{doi:10.1103/PhysRevD.97.094018}}.

\bibitem{Huang:2017kkg}
Q.~Huang, B.~Wang, X.~Liu, D.-Y. Chen, T.~Matsuki, {Exploring the $\Upsilon
  (6S)\rightarrow \chi _{bJ}\phi $ and $\Upsilon (6S)\rightarrow \chi
  _{bJ}\omega $ hidden-bottom hadronic transitions}, Eur. Phys. J. C 77 (2017)
  165.
\newblock \href {http://arxiv.org/abs/1701.00894} {\path{arXiv:1701.00894}},
  \href {https://doi.org/10.1140/epjc/s10052-017-4726-8}
  {\path{doi:10.1140/epjc/s10052-017-4726-8}}.

\bibitem{Chen:2019uzm}
B.~Chen, A.~Zhang, J.~He, {Bottomonium spectrum in the relativistic flux tube
  model}, Phys. Rev. D 101 (2020) 014020.
\newblock \href {http://arxiv.org/abs/1910.06065} {\path{arXiv:1910.06065}},
  \href {https://doi.org/10.1103/PhysRevD.101.014020}
  {\path{doi:10.1103/PhysRevD.101.014020}}.

\bibitem{Li:2019qsg}
Q.~Li, M.-S. Liu, Q.-F. L{\"u}, L.-C. Gui, X.-H. Zhong, {Canonical
  interpretation of $Y(10750)$ and $\Upsilon(10860)$ in the $\Upsilon$ family},
  Eur. Phys. J. C 80 (2020) 59.
\newblock \href {http://arxiv.org/abs/1905.10344} {\path{arXiv:1905.10344}},
  \href {https://doi.org/10.1140/epjc/s10052-020-7626-2}
  {\path{doi:10.1140/epjc/s10052-020-7626-2}}.

\bibitem{Kher:2022gbz}
V.~Kher, R.~Chaturvedi, N.~Devlani, A.~K. Rai, {Bottomonium spectroscopy using
  Coulomb plus linear (Cornell) potential}, Eur. Phys. J. Plus 137 (2022) 357.
\newblock \href {http://arxiv.org/abs/2201.08317} {\path{arXiv:2201.08317}},
  \href {https://doi.org/10.1140/epjp/s13360-022-02538-5}
  {\path{doi:10.1140/epjp/s13360-022-02538-5}}.

\bibitem{Li:2021jjt}
Y.-S. Li, Z.-Y. Bai, Q.~Huang, X.~Liu, {Hidden-bottom hadronic decays of
  $\Upsilon(10753)$ with a $\eta^{(\prime)}$ or $\omega$ emission}, Phys. Rev.
  D 104 (2021) 034036.
\newblock \href {http://arxiv.org/abs/2106.14123} {\path{arXiv:2106.14123}},
  \href {https://doi.org/10.1103/PhysRevD.104.034036}
  {\path{doi:10.1103/PhysRevD.104.034036}}.

\bibitem{Wang:2019veq}
Z.-G. Wang, {Vector hidden-bottom tetraquark candidate: $Y(10750)$}, Chin.
  Phys. C 43 (2019) 123102.
\newblock \href {http://arxiv.org/abs/1905.06610} {\path{arXiv:1905.06610}},
  \href {https://doi.org/10.1088/1674-1137/43/12/123102}
  {\path{doi:10.1088/1674-1137/43/12/123102}}.

\bibitem{Ali:2019okl}
A.~Ali, L.~Maiani, A.~Y. Parkhomenko, W.~Wang, {Interpretation of $Y_b (10753)$
  as a tetraquark and its production mechanism}, Phys. Lett. B 802 (2020)
  135217.
\newblock \href {http://arxiv.org/abs/1910.07671} {\path{arXiv:1910.07671}},
  \href {https://doi.org/10.1016/j.physletb.2020.135217}
  {\path{doi:10.1016/j.physletb.2020.135217}}.

\bibitem{TarrusCastella:2019lyq}
J.~Tarr{\'u}s~Castell{\`a}, {Spin structure of heavy-quark hybrids}, AIP Conf.
  Proc. 2249 (2020) 020008.
\newblock \href {http://arxiv.org/abs/1908.05179} {\path{arXiv:1908.05179}},
  \href {https://doi.org/10.1063/5.0008570} {\path{doi:10.1063/5.0008570}}.

\bibitem{TarrusCastella:2021pld}
J.~Tarr{\'u}s~Castell{\`a}, E.~Passemar, {Exotic to standard bottomonium
  transitions}, Phys. Rev. D 104 (2021) 034019.
\newblock \href {http://arxiv.org/abs/2104.03975} {\path{arXiv:2104.03975}},
  \href {https://doi.org/10.1103/PhysRevD.104.034019}
  {\path{doi:10.1103/PhysRevD.104.034019}}.

\bibitem{Ortega:2024rrv}
P.~G. Ortega, D.~R. Entem, F.~Fern{\'a}ndez, J.~Segovia, {Assessment of
  $\Upsilon$-states above $B\bar B$-threshold using a constituent-quark-model
  based meson-meson coupled-channels framework}, Phys. Rev. D 109 (2024)
  114007.
\newblock \href {http://arxiv.org/abs/2403.03770} {\path{arXiv:2403.03770}},
  \href {https://doi.org/10.1103/PhysRevD.109.114007}
  {\path{doi:10.1103/PhysRevD.109.114007}}.

\bibitem{Bicudo:2019ymo}
P.~Bicudo, M.~Cardoso, N.~Cardoso, M.~Wagner, {Bottomonium resonances with $I =
  0$ from lattice QCD correlation functions with static and light quarks},
  Phys. Rev. D 101 (2020) 034503.
\newblock \href {http://arxiv.org/abs/1910.04827} {\path{arXiv:1910.04827}},
  \href {https://doi.org/10.1103/PhysRevD.101.034503}
  {\path{doi:10.1103/PhysRevD.101.034503}}.

\bibitem{Bicudo:2020qhp}
P.~Bicudo, N.~Cardoso, L.~Mueller, M.~Wagner, {Computation of the quarkonium
  and meson-meson composition of the $\Upsilon(nS)$ states and of the new
  $\Upsilon(10753)$ Belle resonance from lattice QCD static potentials}, Phys.
  Rev. D 103 (2021) 074507.
\newblock \href {http://arxiv.org/abs/2008.05605} {\path{arXiv:2008.05605}},
  \href {https://doi.org/10.1103/PhysRevD.103.074507}
  {\path{doi:10.1103/PhysRevD.103.074507}}.

\bibitem{Badalian:2008ik}
A.~M. Badalian, B.~L.~G. Bakker, I.~V. Danilkin, {On the possibility to observe
  higher $n^3D_1$ bottomonium states in the $e^+e^-$ processes}, Phys. Rev. D
  79 (2009) 037505.
\newblock \href {http://arxiv.org/abs/0812.2136} {\path{arXiv:0812.2136}},
  \href {https://doi.org/10.1103/PhysRevD.79.037505}
  {\path{doi:10.1103/PhysRevD.79.037505}}.

\bibitem{Bai:2022cfz}
Z.-Y. Bai, Y.-S. Li, Q.~Huang, X.~Liu, T.~Matsuki,
  {$\Upsilon(10753)\to\Upsilon(nS)\pi^+\pi^-$ decays induced by hadronic loop
  mechanism}, Phys. Rev. D 105 (2022) 074007.
\newblock \href {http://arxiv.org/abs/2201.12715} {\path{arXiv:2201.12715}},
  \href {https://doi.org/10.1103/PhysRevD.105.074007}
  {\path{doi:10.1103/PhysRevD.105.074007}}.

\bibitem{Belle-II:2024mjm}
I.~Adachi, et~al., Belle-II Collaboration, {Study of $\Upsilon(10753)$ decays
  to $\pi^{+}\pi^{-}\Upsilon(nS)$ final states at Belle II}, JHEP 07 (2024)
  116.
\newblock \href {http://arxiv.org/abs/2401.12021} {\path{arXiv:2401.12021}},
  \href {https://doi.org/10.1007/JHEP07(2024)116}
  {\path{doi:10.1007/JHEP07(2024)116}}.

\bibitem{Li:2022leg}
Y.-S. Li, Z.-Y. Bai, X.~Liu, {Investigating the
  $\Upsilon(10753)\to\Upsilon(1^3D_J)\eta$ transitions}, Phys. Rev. D 105
  (2022) 114041.
\newblock \href {http://arxiv.org/abs/2205.04049} {\path{arXiv:2205.04049}},
  \href {https://doi.org/10.1103/PhysRevD.105.114041}
  {\path{doi:10.1103/PhysRevD.105.114041}}.

\bibitem{Liu:2023gtx}
S.-D. Liu, Z.-X. Cai, Z.-S. Jia, G.~Li, J.-J. Xie, {Hidden-bottom hadronic
  transitions of $\Upsilon(10753)$}, Phys. Rev. D 109 (2024) 014039.
\newblock \href {http://arxiv.org/abs/2312.02761} {\path{arXiv:2312.02761}},
  \href {https://doi.org/10.1103/PhysRevD.109.014039}
  {\path{doi:10.1103/PhysRevD.109.014039}}.

\bibitem{Belle-II:2023twj}
I.~Adachi, et~al., Belle-II Collaboration, {Search for the
  $e^+e^-\to\eta_b(1S)\omega$ and $e^+e^-\to\chi_{b0}(1P)\omega$ processes at
  $\sqrt s=10.745$ GeV}, Phys. Rev. D 109 (2024) 072013.
\newblock \href {http://arxiv.org/abs/2312.13043} {\path{arXiv:2312.13043}},
  \href {https://doi.org/10.1103/PhysRevD.109.072013}
  {\path{doi:10.1103/PhysRevD.109.072013}}.

\bibitem{Belle-II:2022xdi}
I.~Adachi, et~al., Belle-II Collaboration, {Observation of
  $e^+e^-\to\omega\chi_{bJ}(1P)$ and search for $X_b\to\omega\Upsilon(1S)$ at
  $\sqrt s$ near 10.75~GeV}, Phys. Rev. Lett. 130 (2023) 091902.
\newblock \href {http://arxiv.org/abs/2208.13189} {\path{arXiv:2208.13189}},
  \href {https://doi.org/10.1103/PhysRevLett.130.091902}
  {\path{doi:10.1103/PhysRevLett.130.091902}}.

\bibitem{Belle-II:2025jus}
I.~Adachi, et~al., Belle-II, Belle Collaboration, {Improved measurement of Born
  cross sections for $\chi_{bJ}\omega$ and
  $\chi_{bJ}(\pi^+\pi^-\pi^0)_{\text{non}-\omega}$at Belle and Belle II} (10
  2025).
\newblock \href {http://arxiv.org/abs/2510.25461} {\path{arXiv:2510.25461}}.

\bibitem{Liu:2024ets}
S.-D. Liu, H.-D. Cai, Z.-X. Cai, H.-S. Gao, G.~Li, F.~Wang, J.-J. Xie,
  {Production of $X_b$ via radiative transition of $\Upsilon(10753)$}, Phys.
  Rev. D 109 (2024) 094045.
\newblock \href {http://arxiv.org/abs/2403.01676} {\path{arXiv:2403.01676}},
  \href {https://doi.org/10.1103/PhysRevD.109.094045}
  {\path{doi:10.1103/PhysRevD.109.094045}}.

\bibitem{Anisovich:2000kxa}
A.~V. Anisovich, V.~V. Anisovich, A.~V. Sarantsev, {Systematics of $q\bar q$
  states in the $(n,M^2)$ and $(J,M^2)$ planes}, Phys. Rev. D 62 (2000) 051502.
\newblock \href {http://arxiv.org/abs/hep-ph/0003113}
  {\path{arXiv:hep-ph/0003113}}, \href
  {https://doi.org/10.1103/PhysRevD.62.051502}
  {\path{doi:10.1103/PhysRevD.62.051502}}.

\bibitem{Yu:2011ta}
J.-S. Yu, Z.-F. Sun, X.~Liu, Q.~Zhao, {Categorizing resonances $X(1835)$,
  $X(2120)$ and $X(2370)$ in the pseudoscalar meson family}, Phys. Rev. D 83
  (2011) 114007.
\newblock \href {http://arxiv.org/abs/1104.3064} {\path{arXiv:1104.3064}},
  \href {https://doi.org/10.1103/PhysRevD.83.114007}
  {\path{doi:10.1103/PhysRevD.83.114007}}.

\bibitem{Wang:2012wa}
X.~Wang, Z.-F. Sun, D.-Y. Chen, X.~Liu, T.~Matsuki, {Non-strange partner of
  strangeonium-like state $Y(2175)$}, Phys. Rev. D 85 (2012) 074024.
\newblock \href {http://arxiv.org/abs/1202.4139} {\path{arXiv:1202.4139}},
  \href {https://doi.org/10.1103/PhysRevD.85.074024}
  {\path{doi:10.1103/PhysRevD.85.074024}}.

\bibitem{Ye:2012gu}
Z.-C. Ye, X.~Wang, X.~Liu, Q.~Zhao, {The mass spectrum and strong decays of
  isoscalar tensor mesons}, Phys. Rev. D 86 (2012) 054025.
\newblock \href {http://arxiv.org/abs/1206.0097} {\path{arXiv:1206.0097}},
  \href {https://doi.org/10.1103/PhysRevD.86.054025}
  {\path{doi:10.1103/PhysRevD.86.054025}}.

\bibitem{He:2013ttg}
L.-P. He, X.~Wang, X.~Liu, {Towards two-body strong decay behavior of higher
  $\rho$ and $\rho_3$ mesons}, Phys. Rev. D 88 (2013) 034008.
\newblock \href {http://arxiv.org/abs/1306.5562} {\path{arXiv:1306.5562}},
  \href {https://doi.org/10.1103/PhysRevD.88.034008}
  {\path{doi:10.1103/PhysRevD.88.034008}}.

\bibitem{Pang:2014laa}
C.-Q. Pang, L.-P. He, X.~Liu, T.~Matsuki, {Phenomenological study of the
  isovector tensor meson family}, Phys. Rev. D 90 (2014) 014001.
\newblock \href {http://arxiv.org/abs/1405.3189} {\path{arXiv:1405.3189}},
  \href {https://doi.org/10.1103/PhysRevD.90.014001}
  {\path{doi:10.1103/PhysRevD.90.014001}}.

\bibitem{Wang:2014sea}
B.~Wang, C.-Q. Pang, X.~Liu, T.~Matsuki, {Pseudotensor meson family}, Phys.
  Rev. D 91 (2015) 014025.
\newblock \href {http://arxiv.org/abs/1410.3930} {\path{arXiv:1410.3930}},
  \href {https://doi.org/10.1103/PhysRevD.91.014025}
  {\path{doi:10.1103/PhysRevD.91.014025}}.

\bibitem{Chen:2015iqa}
K.~Chen, C.-Q. Pang, X.~Liu, T.~Matsuki, {Light axial vector mesons}, Phys.
  Rev. D 91 (2015) 074025.
\newblock \href {http://arxiv.org/abs/1501.07766} {\path{arXiv:1501.07766}},
  \href {https://doi.org/10.1103/PhysRevD.91.074025}
  {\path{doi:10.1103/PhysRevD.91.074025}}.

\bibitem{Pang:2015eha}
C.-Q. Pang, B.~Wang, X.~Liu, T.~Matsuki, {High-spin mesons below 3 GeV}, Phys.
  Rev. D 92 (2015) 014012.
\newblock \href {http://arxiv.org/abs/1505.04105} {\path{arXiv:1505.04105}},
  \href {https://doi.org/10.1103/PhysRevD.92.014012}
  {\path{doi:10.1103/PhysRevD.92.014012}}.

\bibitem{Pang:2019ovr}
C.-Q. Pang, Y.-R. Wang, J.-F. Hu, T.-J. Zhang, X.~Liu, {Study of the $\omega$
  meson family and newly observed $\omega$-like state $X(2240)$}, Phys. Rev. D
  101 (2020) 074022.
\newblock \href {http://arxiv.org/abs/1910.12408} {\path{arXiv:1910.12408}},
  \href {https://doi.org/10.1103/PhysRevD.101.074022}
  {\path{doi:10.1103/PhysRevD.101.074022}}.

\bibitem{Zhou:2022ark}
Q.-S. Zhou, J.-Z. Wang, X.~Liu, T.~Matsuki, {Identifying the contribution of
  higher $\rho$ mesons around 2~GeV in the $e^+e^-\to\omega\pi^0$ and
  $e^+e^-\to\rho\eta^\prime$ processes}, Phys. Rev. D 105 (2022) 074035.
\newblock \href {http://arxiv.org/abs/2201.06393} {\path{arXiv:2201.06393}},
  \href {https://doi.org/10.1103/PhysRevD.105.074035}
  {\path{doi:10.1103/PhysRevD.105.074035}}.

\bibitem{BaBar:2006gsq}
B.~Aubert, et~al., BaBar Collaboration, {A structure at 2175 MeV in $e^{+}
  e^{-} \to \phi f_0(980)$ observed via initial-state radiation}, Phys. Rev. D
  74 (2006) 091103.
\newblock \href {http://arxiv.org/abs/hep-ex/0610018}
  {\path{arXiv:hep-ex/0610018}}, \href
  {https://doi.org/10.1103/PhysRevD.74.091103}
  {\path{doi:10.1103/PhysRevD.74.091103}}.

\bibitem{BaBar:2007ptr}
B.~Aubert, et~al., BaBar Collaboration, {The $e^+e^-\to K^+K^-\pi^+\pi^-$,
  $K^+K^-\pi^0\pi^0$ and $K^+K^-K^+K^-$ cross sections measured with
  initial-state radiation}, Phys. Rev. D 76 (2007) 012008.
\newblock \href {http://arxiv.org/abs/0704.0630} {\path{arXiv:0704.0630}},
  \href {https://doi.org/10.1103/PhysRevD.76.012008}
  {\path{doi:10.1103/PhysRevD.76.012008}}.

\bibitem{BaBar:2011btv}
J.~P. Lees, et~al., BaBar Collaboration, {Cross sections for the reactions
  $e^+e^-\to K^+K^-\pi^+\pi^-$, $K^+K^-\pi^0\pi^0$, and $K^+K^-K^+K^-$ measured
  using initial-state radiation events}, Phys. Rev. D 86 (2012) 012008.
\newblock \href {http://arxiv.org/abs/1103.3001} {\path{arXiv:1103.3001}},
  \href {https://doi.org/10.1103/PhysRevD.86.012008}
  {\path{doi:10.1103/PhysRevD.86.012008}}.

\bibitem{BESIII:2021aet}
M.~Ablikim, et~al., BESIII Collaboration, {Measurement of
  $e^+e^-\to\phi\pi^+\pi^-$ cross sections at center-of-mass energies from 2.00
  to 3.08~GeV}, Phys. Rev. D 108 (2023) 032011.
\newblock \href {http://arxiv.org/abs/2112.13219} {\path{arXiv:2112.13219}},
  \href {https://doi.org/10.1103/PhysRevD.108.032011}
  {\path{doi:10.1103/PhysRevD.108.032011}}.

\bibitem{BES:2007sqy}
M.~Ablikim, et~al., BES Collaboration, {Observation of $Y(2175)$ in $J / \psi
  \to\eta \phi f_0(980)$}, Phys. Rev. Lett. 100 (2008) 102003.
\newblock \href {http://arxiv.org/abs/0712.1143} {\path{arXiv:0712.1143}},
  \href {https://doi.org/10.1103/PhysRevLett.100.102003}
  {\path{doi:10.1103/PhysRevLett.100.102003}}.

\bibitem{BESIII:2014ybv}
M.~Ablikim, et~al., BESIII Collaboration, {Study of $J/\psi \to \eta \phi \pi^+
  \pi^-$ at BESIII}, Phys. Rev. D 91 (2015) 052017.
\newblock \href {http://arxiv.org/abs/1412.5258} {\path{arXiv:1412.5258}},
  \href {https://doi.org/10.1103/PhysRevD.91.052017}
  {\path{doi:10.1103/PhysRevD.91.052017}}.

\bibitem{BESIII:2017qkh}
M.~Ablikim, et~al., BESIII Collaboration, {Observation of $e^+ e^- \to \eta
  Y(2175)$ at center-of-mass energies above 3.7 GeV}, Phys. Rev. D 99 (2019)
  012014.
\newblock \href {http://arxiv.org/abs/1709.04323} {\path{arXiv:1709.04323}},
  \href {https://doi.org/10.1103/PhysRevD.99.012014}
  {\path{doi:10.1103/PhysRevD.99.012014}}.

\bibitem{BESIII:2020gnc}
M.~Ablikim, et~al., BESIII Collaboration, {Observation of a structure in
  $e^{+}e^{-} \to \phi \eta^{\prime}$ at $\sqrt{s}$ from 2.05 to 3.08 GeV},
  Phys. Rev. D 102 (2020) 012008.
\newblock \href {http://arxiv.org/abs/2003.13064} {\path{arXiv:2003.13064}},
  \href {https://doi.org/10.1103/PhysRevD.102.012008}
  {\path{doi:10.1103/PhysRevD.102.012008}}.

\bibitem{BESIII:2021bjn}
M.~Ablikim, et~al., BESIII Collaboration, {Study of the process
  $e^{+}e^{-}\rightarrow\phi\eta$ at center-of-mass energies between 2.00 and
  3.08 GeV}, Phys. Rev. D 104 (2021) 032007.
\newblock \href {http://arxiv.org/abs/2104.05549} {\path{arXiv:2104.05549}},
  \href {https://doi.org/10.1103/PhysRevD.104.032007}
  {\path{doi:10.1103/PhysRevD.104.032007}}.

\bibitem{Ding:2007pc}
G.-J. Ding, M.-L. Yan, {$Y(2175)$: Distinguish hybrid state from higher
  quarkonium}, Phys. Lett. B 657 (2007) 49--54.
\newblock \href {http://arxiv.org/abs/hep-ph/0701047}
  {\path{arXiv:hep-ph/0701047}}, \href
  {https://doi.org/10.1016/j.physletb.2007.10.020}
  {\path{doi:10.1016/j.physletb.2007.10.020}}.

\bibitem{Pang:2019ttv}
C.-Q. Pang, {Excited states of $\phi$ meson}, Phys. Rev. D 99 (2019) 074015.
\newblock \href {http://arxiv.org/abs/1902.02206} {\path{arXiv:1902.02206}},
  \href {https://doi.org/10.1103/PhysRevD.99.074015}
  {\path{doi:10.1103/PhysRevD.99.074015}}.

\bibitem{Li:2020xzs}
Q.~Li, L.-C. Gui, M.-S. Liu, Q.-F. L{\"u}, X.-H. Zhong, {Mass spectrum and
  strong decays of strangeonium in a constituent quark model}, Chin. Phys. C 45
  (2021) 023116.
\newblock \href {http://arxiv.org/abs/2004.05786} {\path{arXiv:2004.05786}},
  \href {https://doi.org/10.1088/1674-1137/abcf22}
  {\path{doi:10.1088/1674-1137/abcf22}}.

\bibitem{Barnes:2002mu}
T.~Barnes, N.~Black, P.~R. Page, {Strong decays of strange quarkonia}, Phys.
  Rev. D 68 (2003) 054014.
\newblock \href {http://arxiv.org/abs/nucl-th/0208072}
  {\path{arXiv:nucl-th/0208072}}, \href
  {https://doi.org/10.1103/PhysRevD.68.054014}
  {\path{doi:10.1103/PhysRevD.68.054014}}.

\bibitem{Afonin:2014nya}
S.~S. Afonin, I.~V. Pusenkov, {Universal description of radially excited heavy
  and light vector mesons}, Phys. Rev. D 90 (2014) 094020.
\newblock \href {http://arxiv.org/abs/1411.2390} {\path{arXiv:1411.2390}},
  \href {https://doi.org/10.1103/PhysRevD.90.094020}
  {\path{doi:10.1103/PhysRevD.90.094020}}.

\bibitem{Ding:2006ya}
G.-J. Ding, M.-L. Yan, {A Candidate for $1^{--}$ strangeonium hybrid}, Phys.
  Lett. B 650 (2007) 390--400.
\newblock \href {http://arxiv.org/abs/hep-ph/0611319}
  {\path{arXiv:hep-ph/0611319}}, \href
  {https://doi.org/10.1016/j.physletb.2007.05.026}
  {\path{doi:10.1016/j.physletb.2007.05.026}}.

\bibitem{Govaerts:1985fx}
J.~Govaerts, L.~J. Reinders, J.~Weyers, {Radial excitations and exotic mesons
  via QCD sum rules}, Nucl. Phys. B 262 (1985) 575--592.
\newblock \href {https://doi.org/10.1016/0550-3213(85)90505-X}
  {\path{doi:10.1016/0550-3213(85)90505-X}}.

\bibitem{Ho:2019org}
J.~Ho, R.~Berg, T.~G. Steele, W.~Chen, D.~Harnett, {Is the $Y(2175)$ a
  strangeonium hybrid meson?}, Phys. Rev. D 100 (2019) 034012.
\newblock \href {http://arxiv.org/abs/1905.12779} {\path{arXiv:1905.12779}},
  \href {https://doi.org/10.1103/PhysRevD.100.034012}
  {\path{doi:10.1103/PhysRevD.100.034012}}.

\bibitem{Li:2025hsp}
S.-H. Li, Z.-R. Huang, W.~Chen, H.-Y. Jin, {Revising the mass of light hybrid
  mesons: NLO QCD sum rules point to $\phi(2170)$ as a prime candidate}, JHEP
  03 (2026) 087.
\newblock \href {http://arxiv.org/abs/2506.22412} {\path{arXiv:2506.22412}},
  \href {https://doi.org/10.1007/JHEP03(2026)087}
  {\path{doi:10.1007/JHEP03(2026)087}}.

\bibitem{Wang:2006ri}
Z.-G. Wang, {Analysis of the $Y(2175)$ as a tetraquark state with QCD sum
  rules}, Nucl. Phys. A 791 (2007) 106--116.
\newblock \href {http://arxiv.org/abs/hep-ph/0610171}
  {\path{arXiv:hep-ph/0610171}}, \href
  {https://doi.org/10.1016/j.nuclphysa.2007.04.012}
  {\path{doi:10.1016/j.nuclphysa.2007.04.012}}.

\bibitem{Chen:2008ej}
H.-X. Chen, X.~Liu, A.~Hosaka, S.-L. Zhu, {The $Y(2175)$ state in the QCD sum
  rule}, Phys. Rev. D 78 (2008) 034012.
\newblock \href {http://arxiv.org/abs/0801.4603} {\path{arXiv:0801.4603}},
  \href {https://doi.org/10.1103/PhysRevD.78.034012}
  {\path{doi:10.1103/PhysRevD.78.034012}}.

\bibitem{Drenska:2008gr}
N.~V. Drenska, R.~Faccini, A.~D. Polosa, {Higher tetraquark particles}, Phys.
  Lett. B 669 (2008) 160--166.
\newblock \href {http://arxiv.org/abs/0807.0593} {\path{arXiv:0807.0593}},
  \href {https://doi.org/10.1016/j.physletb.2008.09.038}
  {\path{doi:10.1016/j.physletb.2008.09.038}}.

\bibitem{Ke:2018evd}
H.-W. Ke, X.-Q. Li, {Study of the strong decays of $\phi(2170)$ and the future
  charm-tau factory}, Phys. Rev. D 99 (2019) 036014.
\newblock \href {http://arxiv.org/abs/1810.07912} {\path{arXiv:1810.07912}},
  \href {https://doi.org/10.1103/PhysRevD.99.036014}
  {\path{doi:10.1103/PhysRevD.99.036014}}.

\bibitem{Agaev:2019coa}
S.~S. Agaev, K.~Azizi, H.~Sundu, {Nature of the vector resonance $Y(2175)$},
  Phys. Rev. D 101 (2020) 074012.
\newblock \href {http://arxiv.org/abs/1911.09743} {\path{arXiv:1911.09743}},
  \href {https://doi.org/10.1103/PhysRevD.101.074012}
  {\path{doi:10.1103/PhysRevD.101.074012}}.

\bibitem{Liu:2020lpw}
F.-X. Liu, M.-S. Liu, X.-H. Zhong, Q.~Zhao, {Fully-strange tetraquark
  $ss\bar{s}\bar{s}$ spectrum and possible experimental evidence}, Phys. Rev. D
  103 (2021) 016016.
\newblock \href {http://arxiv.org/abs/2008.01372} {\path{arXiv:2008.01372}},
  \href {https://doi.org/10.1103/PhysRevD.103.016016}
  {\path{doi:10.1103/PhysRevD.103.016016}}.

\bibitem{Zhao:2013ffn}
L.~Zhao, N.~Li, S.-L. Zhu, B.-S. Zou, {Meson-exchange model for the
  $\Lambda\bar{\Lambda}$ interaction}, Phys. Rev. D 87 (2013) 054034.
\newblock \href {http://arxiv.org/abs/1302.1770} {\path{arXiv:1302.1770}},
  \href {https://doi.org/10.1103/PhysRevD.87.054034}
  {\path{doi:10.1103/PhysRevD.87.054034}}.

\bibitem{Deng:2013aca}
C.~Deng, J.~Ping, Y.~Yang, F.~Wang, {Baryonia and near-threshold enhancements},
  Phys. Rev. D 88 (2013) 074007.
\newblock \href {http://arxiv.org/abs/1306.6725} {\path{arXiv:1306.6725}},
  \href {https://doi.org/10.1103/PhysRevD.88.074007}
  {\path{doi:10.1103/PhysRevD.88.074007}}.

\bibitem{Dong:2017rmg}
Y.~Dong, A.~Faessler, T.~Gutsche, Q.~L{\"u}, V.~E. Lyubovitskij, {Selected
  strong decays of $\eta(2225)$ and $\phi(2170)$ as $\Lambda \bar\Lambda$ bound
  states}, Phys. Rev. D 96 (2017) 074027.
\newblock \href {http://arxiv.org/abs/1705.09631} {\path{arXiv:1705.09631}},
  \href {https://doi.org/10.1103/PhysRevD.96.074027}
  {\path{doi:10.1103/PhysRevD.96.074027}}.

\bibitem{Zhu:2007wz}
S.-L. Zhu, {New hadron states}, Int. J. Mod. Phys. E 17 (2008) 283--322.
\newblock \href {http://arxiv.org/abs/hep-ph/0703225}
  {\path{arXiv:hep-ph/0703225}}, \href
  {https://doi.org/10.1142/S0218301308009446}
  {\path{doi:10.1142/S0218301308009446}}.

\bibitem{Gomez-Avila:2007pgn}
S.~Gomez-Avila, M.~Napsuciale, E.~Oset, {$\phi K^{+}K^{-}$ production in
  electron-positron annihilation}, Phys. Rev. D 79 (2009) 034018.
\newblock \href {http://arxiv.org/abs/0711.4147} {\path{arXiv:0711.4147}},
  \href {https://doi.org/10.1103/PhysRevD.79.034018}
  {\path{doi:10.1103/PhysRevD.79.034018}}.

\bibitem{MartinezTorres:2008gy}
A.~Martinez~Torres, K.~P. Khemchandani, L.~S. Geng, M.~Napsuciale, E.~Oset,
  {The $X(2175)$ as a resonant state of the $\phi K\bar K$ system}, Phys. Rev.
  D 78 (2008) 074031.
\newblock \href {http://arxiv.org/abs/0801.3635} {\path{arXiv:0801.3635}},
  \href {https://doi.org/10.1103/PhysRevD.78.074031}
  {\path{doi:10.1103/PhysRevD.78.074031}}.

\bibitem{Wei:2025ejv}
X.~Wei, Q.-H. Shen, X.-H. Liu, J.-J. Xie, {Shedding light on the nature of the
  $\phi(2170)$ state in the $e^+e^-\to\phi\pi^+\pi^-$ reaction}, Phys. Rev. D
  113 (2026) 014022.
\newblock \href {http://arxiv.org/abs/2509.12552} {\path{arXiv:2509.12552}},
  \href {https://doi.org/10.1103/frs3-zpt2} {\path{doi:10.1103/frs3-zpt2}}.

\bibitem{Wang:2020kte}
L.-M. Wang, J.-Z. Wang, X.~Liu, {Toward $e^+e^-\to \pi^+\pi^-$ annihilation
  inspired by higher $\rho$ mesonic states around 2.2 GeV}, Phys. Rev. D 102
  (2020) 034037.
\newblock \href {http://arxiv.org/abs/2007.03118} {\path{arXiv:2007.03118}},
  \href {https://doi.org/10.1103/PhysRevD.102.034037}
  {\path{doi:10.1103/PhysRevD.102.034037}}.

\bibitem{BESIII:2018ldc}
M.~Ablikim, et~al., BESIII Collaboration, {Measurement of $e^{+} e^{-}
  \rightarrow K^{+} K^{-}$ cross section at $\sqrt{s} = 2.00 - 3.08$ GeV},
  Phys. Rev. D 99 (2019) 032001.
\newblock \href {http://arxiv.org/abs/1811.08742} {\path{arXiv:1811.08742}},
  \href {https://doi.org/10.1103/PhysRevD.99.032001}
  {\path{doi:10.1103/PhysRevD.99.032001}}.

\bibitem{BESIII:2020vtu}
M.~Ablikim, et~al., BESIII Collaboration, {Observation of a Resonant Structure
  in $e^{+}e^{-} \to K^{+}K^{-}\pi^{0}\pi^{0}$}, Phys. Rev. Lett. 124 (2020)
  112001.
\newblock \href {http://arxiv.org/abs/2001.04131} {\path{arXiv:2001.04131}},
  \href {https://doi.org/10.1103/PhysRevLett.124.112001}
  {\path{doi:10.1103/PhysRevLett.124.112001}}.

\bibitem{Anisovich:2001pn}
A.~V. Anisovich, C.~A. Baker, C.~J. Batty, D.~V. Bugg, V.~A. Nikonov, A.~V.
  Sarantsev, V.~V. Sarantsev, B.~S. Zou, {Partial wave analysis of $\bar pp$
  annihilation channels in flight with $I = 1$, $C = +1$}, Phys. Lett. B 517
  (2001) 261--272.
\newblock \href {http://arxiv.org/abs/1110.0278} {\path{arXiv:1110.0278}},
  \href {https://doi.org/10.1016/S0370-2693(01)01017-6}
  {\path{doi:10.1016/S0370-2693(01)01017-6}}.

\bibitem{Anisovich:2010nh}
A.~V. Anisovich, C.~J. Batty, D.~V. Bugg, V.~A. Nikonov, A.~V. Sarantsev, {A
  fresh look at $\eta_2(1645), \eta_2(1870), \eta_2(2030)$ and $f_2(1910)$ in
  $\bar{p} p \to \eta 3\pi^0$}, Eur. Phys. J. C 71 (2011) 1511.
\newblock \href {http://arxiv.org/abs/1009.1781} {\path{arXiv:1009.1781}},
  \href {https://doi.org/10.1140/epjc/s10052-010-1511-3}
  {\path{doi:10.1140/epjc/s10052-010-1511-3}}.

\bibitem{Anisovich:2002su}
A.~V. Anisovich, C.~A. Baker, C.~J. Batty, D.~V. Bugg, L.~Montanet, V.~A.
  Nikonov, A.~V. Sarantsev, V.~V. Sarantsev, B.~S. Zou, {Combined analysis of
  meson channels with $I = 1$, $C = -1$ from 1940 to 2410 MeV}, Phys. Lett. B
  542 (2002) 8--18.
\newblock \href {http://arxiv.org/abs/1109.5247} {\path{arXiv:1109.5247}},
  \href {https://doi.org/10.1016/S0370-2693(02)02302-X}
  {\path{doi:10.1016/S0370-2693(02)02302-X}}.

\bibitem{VES:2000xjq}
D.~V. Amelin, et~al., VES Collaboration, {Natural parity resonances in $\eta
  \pi^+ \pi^-$}, Nucl. Phys. A 668 (2000) 83--96.
\newblock \href {https://doi.org/10.1016/S0375-9474(99)00370-X}
  {\path{doi:10.1016/S0375-9474(99)00370-X}}.

\bibitem{Wang:2021abg}
L.-M. Wang, S.-Q. Luo, X.~Liu, {Light unflavored vector meson spectroscopy
  around the mass range of $2.4\sim3$ GeV and possible experimental evidence},
  Phys. Rev. D 105 (2022) 034011.
\newblock \href {http://arxiv.org/abs/2109.06617} {\path{arXiv:2109.06617}},
  \href {https://doi.org/10.1103/PhysRevD.105.034011}
  {\path{doi:10.1103/PhysRevD.105.034011}}.

\bibitem{BESIII:2023ioy}
M.~Ablikim, et~al., BESIII Collaboration, {Measurement of the $e^+e^-\to
  \Lambda\bar{\Lambda}$ cross section from threshold to 3.00 GeV using events
  with initial-state radiation}, Phys. Rev. D 107 (2023) 072005.
\newblock \href {http://arxiv.org/abs/2303.07629} {\path{arXiv:2303.07629}},
  \href {https://doi.org/10.1103/PhysRevD.107.072005}
  {\path{doi:10.1103/PhysRevD.107.072005}}.

\bibitem{BaBar:2007fsu}
B.~Aubert, et~al., BaBar Collaboration, {Study of $e^{+} e^{-} \to \Lambda
  \bar{\Lambda}$, $\Lambda \bar{\Sigma}^0$, $\Sigma^0 \bar{\Sigma}^0$ using
  initial state radiation with $BABAR$}, Phys. Rev. D 76 (2007) 092006.
\newblock \href {http://arxiv.org/abs/0709.1988} {\path{arXiv:0709.1988}},
  \href {https://doi.org/10.1103/PhysRevD.76.092006}
  {\path{doi:10.1103/PhysRevD.76.092006}}.

\bibitem{BESIII:2017hyw}
M.~Ablikim, et~al., BESIII Collaboration, {Observation of a cross-section
  enhancement near mass threshold in
  $e^{+}e^{-}\rightarrow\Lambda\bar{\Lambda}$}, Phys. Rev. D 97 (2018) 032013.
\newblock \href {http://arxiv.org/abs/1709.10236} {\path{arXiv:1709.10236}},
  \href {https://doi.org/10.1103/PhysRevD.97.032013}
  {\path{doi:10.1103/PhysRevD.97.032013}}.

\bibitem{BESIII:2019nep}
M.~Ablikim, et~al., BESIII Collaboration, {Complete measurement of the
  $\Lambda$ electromagnetic form factors}, Phys. Rev. Lett. 123 (2019) 122003.
\newblock \href {http://arxiv.org/abs/1903.09421} {\path{arXiv:1903.09421}},
  \href {https://doi.org/10.1103/PhysRevLett.123.122003}
  {\path{doi:10.1103/PhysRevLett.123.122003}}.

\bibitem{DM2:1990tut}
D.~Bisello, et~al., DM2 Collaboration, {Baryon pair production in $e^+e^-$
  annihilation at $\sqrt s=2.4$ GeV}, Z. Phys. C 48 (1990) 23--28.
\newblock \href {https://doi.org/10.1007/BF01565602}
  {\path{doi:10.1007/BF01565602}}.

\bibitem{Bai:2023dhc}
Z.-Y. Bai, Q.-S. Zhou, X.~Liu, {Higher strangeonium decays into light flavor
  baryon pairs like $\Lambda\bar{\Lambda}$, $\Sigma\bar{\Sigma}$, and
  $\Xi\bar{\Xi}$}, Phys. Rev. D 108 (2023) 094036.
\newblock \href {http://arxiv.org/abs/2307.16255} {\path{arXiv:2307.16255}},
  \href {https://doi.org/10.1103/PhysRevD.108.094036}
  {\path{doi:10.1103/PhysRevD.108.094036}}.

\bibitem{BESIII:2020xmw}
M.~Ablikim, et~al., BESIII Collaboration, {Observation of a resonant structure
  in $e^{+}e^{-} \to \omega\eta$ and another in $e^{+}e^{-} \to \omega\pi^{0}$
  at center-of-mass energies between 2.00 and 3.08 GeV}, Phys. Lett. B 813
  (2021) 136059.
\newblock \href {http://arxiv.org/abs/2009.08099} {\path{arXiv:2009.08099}},
  \href {https://doi.org/10.1016/j.physletb.2020.136059}
  {\path{doi:10.1016/j.physletb.2020.136059}}.

\bibitem{BESIII:2021uni}
M.~Ablikim, et~al., BESIII Collaboration, {Measurement of the
  $e^{+}e^{-}\rightarrow\omega\pi^{0}\pi^{0}$ cross section at center-of-mass
  energies from 2.0 to 3.08~GeV}, Phys. Rev. D 105 (2022) 032005.
\newblock \href {http://arxiv.org/abs/2112.15076} {\path{arXiv:2112.15076}},
  \href {https://doi.org/10.1103/PhysRevD.105.032005}
  {\path{doi:10.1103/PhysRevD.105.032005}}.

\bibitem{Zhou:2022wwk}
Q.-S. Zhou, J.-Z. Wang, X.~Liu, {Role of the $\omega(4S)$ and $\omega(3D)$
  states in mediating the $e^+e^-\to\omega\eta$ and $e^+e^-\to\omega\pi^0\pi^0$
  processes}, Phys. Rev. D 106 (2022) 034010.
\newblock \href {http://arxiv.org/abs/2207.00276} {\path{arXiv:2207.00276}},
  \href {https://doi.org/10.1103/PhysRevD.106.034010}
  {\path{doi:10.1103/PhysRevD.106.034010}}.

\bibitem{BESIII:2024okl}
M.~Ablikim, et~al., BESIII Collaboration, {Study of $e^+e^-\to\pi^+\pi^-\pi^0$
  at $\sqrt s$ from 2.00 to 3.08~GeV at BESIII}, Phys. Rev. D 110 (2024)
  032005.
\newblock \href {http://arxiv.org/abs/2401.14711} {\path{arXiv:2401.14711}},
  \href {https://doi.org/10.1103/PhysRevD.110.032005}
  {\path{doi:10.1103/PhysRevD.110.032005}}.

\bibitem{Bai:2025knk}
Z.-Y. Bai, Q.-S. Zhou, X.~Liu, {Role of $4S$-$3D$ mixing in explaining the
  $\omega$-like $Y(2119)$ observed in $e^+e^-\to \rho\pi$ and $e^+e^-\to
  \rho(1450)\pi$}, Phys. Rev. D 111 (2025) 054013.
\newblock \href {http://arxiv.org/abs/2502.05754} {\path{arXiv:2502.05754}},
  \href {https://doi.org/10.1103/PhysRevD.111.054013}
  {\path{doi:10.1103/PhysRevD.111.054013}}.

\bibitem{BaBar:2012bdw}
J.~P. Lees, et~al., BaBar Collaboration, {Precise measurement of the $e^+ e^-
  \to \pi^+\pi^- (\gamma)$ cross section with the initial-state radiation
  method at $BABAR$}, Phys. Rev. D 86 (2012) 032013.
\newblock \href {http://arxiv.org/abs/1205.2228} {\path{arXiv:1205.2228}},
  \href {https://doi.org/10.1103/PhysRevD.86.032013}
  {\path{doi:10.1103/PhysRevD.86.032013}}.

\bibitem{BaBar:2019kds}
J.~P. Lees, et~al., BaBar Collaboration, {Resonances in $e^+e^-$ annihilation
  near 2.2 GeV}, Phys. Rev. D 101 (2020) 012011.
\newblock \href {http://arxiv.org/abs/1912.04512} {\path{arXiv:1912.04512}},
  \href {https://doi.org/10.1103/PhysRevD.101.012011}
  {\path{doi:10.1103/PhysRevD.101.012011}}.

\bibitem{BaBar:2007qju}
B.~Aubert, et~al., BaBar Collaboration, {The $e^+e^-\to 2(\pi^+\pi^-)\pi^0$,
  $2(\pi^+ \pi^-) \eta$, $K^+ K^- \pi^+ \pi^- \pi^0$ and $K^+ K^- \pi^+ \pi^-
  \eta$ cross sections measured with initial-state radiation}, Phys. Rev. D 76
  (2007) 092005, [Erratum: Phys.Rev.D 77, 119902 (2008)].
\newblock \href {http://arxiv.org/abs/0708.2461} {\path{arXiv:0708.2461}},
  \href {https://doi.org/10.1103/PhysRevD.76.092005}
  {\path{doi:10.1103/PhysRevD.76.092005}}.

\bibitem{BaBar:2022ahi}
J.~P. Lees, et~al., BaBar Collaboration, {Study of the reactions $e^+e^-\to
  K^+K^-\pi^0\pi^0\pi^0$, $e^+e^-\to K_S^0K^\pm\pi^\pm\pi^0\pi^0$, and
  $e^+e^-\to K_S^0K^\pm\pi^\pm\pi^+\pi^-$ at center-of-mass energies from
  threshold to 4.5~GeV using initial-state radiation}, Phys. Rev. D 107 (2023)
  072001.
\newblock \href {http://arxiv.org/abs/2207.10340} {\path{arXiv:2207.10340}},
  \href {https://doi.org/10.1103/PhysRevD.107.072001}
  {\path{doi:10.1103/PhysRevD.107.072001}}.

\bibitem{Achasov:2016zvn}
M.~N. Achasov, et~al., {Updated measurement of the $e^+e^- \to \omega \pi^0 \to
  \pi^0\pi^0\gamma$ cross section with the SND detector}, Phys. Rev. D 94
  (2016) 112001.
\newblock \href {http://arxiv.org/abs/1610.00235} {\path{arXiv:1610.00235}},
  \href {https://doi.org/10.1103/PhysRevD.94.112001}
  {\path{doi:10.1103/PhysRevD.94.112001}}.

\bibitem{BESIII:2020kpr}
M.~Ablikim, et~al., BESIII Collaboration, {Measurement of the Born cross
  sections for $e^+e^- \to \eta^\prime \pi^{+}\pi^{-}$ at center-of-mass
  energies between $2.00$ and $3.08${\textasciitilde}GeV}, Phys. Rev. D 103
  (2021) 072007.
\newblock \href {http://arxiv.org/abs/2012.07360} {\path{arXiv:2012.07360}},
  \href {https://doi.org/10.1103/PhysRevD.103.072007}
  {\path{doi:10.1103/PhysRevD.103.072007}}.

\bibitem{BESIII:2023sbq}
M.~Ablikim, et~al., BESIII Collaboration, {Measurement of the cross sections
  for $e^+ e^-\to \eta \pi^+ \pi^-$ at center-of-mass energies between 2.00 and
  3.08~GeV}, Phys. Rev. D 108 (2023) L111101.
\newblock \href {http://arxiv.org/abs/2310.10452} {\path{arXiv:2310.10452}},
  \href {https://doi.org/10.1103/PhysRevD.108.L111101}
  {\path{doi:10.1103/PhysRevD.108.L111101}}.

\bibitem{Guo:2025igf}
D.~Guo, J.-Z. Wang, Q.-S. Zhou, {Potential of the reaction $e^+e^-\to
  p\bar{p}\pi^0$ for constructing higher $\rho$-meson spectroscopy above
  2.4~GeV}, Phys. Rev. D 112~(11) (2025) 114029.
\newblock \href {http://arxiv.org/abs/2509.07821} {\path{arXiv:2509.07821}},
  \href {https://doi.org/10.1103/w5gh-lq88} {\path{doi:10.1103/w5gh-lq88}}.

\bibitem{Liu:2022yrt}
X.~Liu, Q.-S. Zhou, L.-M. Wang, {Broad resonance structure in $e^+e^-\to
  f_1(1285)\pi^+\pi^-$ and higher $\rho$-meson excitations}, Phys. Rev. D 106
  (2022) 094012.
\newblock \href {http://arxiv.org/abs/2209.11525} {\path{arXiv:2209.11525}},
  \href {https://doi.org/10.1103/PhysRevD.106.094012}
  {\path{doi:10.1103/PhysRevD.106.094012}}.

\bibitem{Zhou:2025rxb}
Q.-S. Zhou, Z.-Y. Bai, J.-Z. Wang, H.~Xu, X.~Liu, {Decoding the role
  of~$\rho$~mesonic states for elucidating the~$e^+e^-\to a_2(1320)\pi$~data
  and other reactions}, Phys. Rev. D 112 (2025) 074021.
\newblock \href {http://arxiv.org/abs/2507.21554} {\path{arXiv:2507.21554}},
  \href {https://doi.org/10.1103/pdfh-mk79} {\path{doi:10.1103/pdfh-mk79}}.

\bibitem{Du:2017zvv}
M.-L. Du, M.~Albaladejo, P.~Fern{\'a}ndez-Soler, F.-K. Guo, C.~Hanhart, U.-G.
  Mei{\ss}ner, J.~Nieves, D.-L. Yao, {Towards a new paradigm for heavy-light
  meson spectroscopy}, Phys. Rev. D 98 (2018) 094018.
\newblock \href {http://arxiv.org/abs/1712.07957} {\path{arXiv:1712.07957}},
  \href {https://doi.org/10.1103/PhysRevD.98.094018}
  {\path{doi:10.1103/PhysRevD.98.094018}}.

\end{thebibliography}

\end{document}